\documentclass[12pt,twoside,a4paper,hidelinks]{book}

\usepackage[utf8]{inputenc}
\usepackage[T1]{fontenc}
\usepackage{amsmath,amsfonts,amssymb}
\usepackage{graphicx}
\usepackage[english]{babel}
\usepackage{csquotes}
\usepackage[a4paper,inner=3cm,outer=2.5cm,top=2.4cm,bottom=1.5cm,includehead,includefoot]{geometry}
\usepackage{caption}
\usepackage{subcaption}
\usepackage{fancyhdr}
\usepackage{nameref}
\usepackage{hyperref}
\usepackage{multirow}
\usepackage{appendix}
\usepackage{setspace}
\usepackage{titlesec}
\usepackage{subfiles} 
\usepackage{indentfirst}
\usepackage{minitoc}
\usepackage{verbatim}
\usepackage{sfmath}
\usepackage{siunitx}
\usepackage{url}  
\usepackage{doi}
\usepackage{tabularx}
\usepackage{array} 
\usepackage{makecell} 
\usepackage{microtype}
\usepackage{float}
\usepackage[table,xcdraw]{xcolor}
\usepackage{rotating} 
\usepackage{acronym} 
\usepackage{etoolbox}
\usepackage{sectsty}
\usepackage{multicol}
\usepackage{ragged2e} 
\usepackage[none]{hyphenat}

\hypersetup{
    colorlinks=true,
    linkcolor=mipurple,   
    citecolor=mipurple,
    urlcolor=mipurple,
    filecolor=mipurple,
}

\definecolor{mipurple}{RGB}{186,85,211}

\mtcsetfeature{minitoc}{before}{\vspace{20pt}}
\mtcsetfeature{minitoc}{after}{\vspace{20pt}}
\mtcsetfont{minitoc}{section}{\color{mipurple}\normalfont\small\bfseries}
\mtcsetfont{minitoc}{subsection}{\color{mipurple}\normalfont\small}

\fancypagestyle{onlypagenumber}{
  \fancyhf{} 
  \fancyhead[RO,LE]{\thepage} 
   
}

\fancypagestyle{rest}{
  \fancyhf{} 
  \fancyhead[LE,RO]{\MakeLowercase{\rightmark}} 
  \fancyhead[LO,RE]{\thepage} 
}

\titleformat{\chapter}[display]
  {\normalfont\huge\bfseries}
  {\centering\hrule\vspace{0.25ex}\hrule\vspace{1.5ex}\MakeUppercase{\chaptertitlename}\ \thechapter\vspace{1.5ex}\hrule}
  {0.1em}
  {\centering}

\titlespacing*{\chapter}{0pt}{0pt}{0pt}
\titlespacing*{\section}{0pt}{0pt}{0pt}
\titlespacing*{\subsection}{0pt}{0pt}{0pt}
\titlespacing*{\subsubsection}{0pt}{0pt}{0pt}

\makeatletter
\AtBeginDocument{%
  \renewcommand*{\AC@hyperlink}[2]{%
    \begingroup
      \hypersetup{hidelinks}%
      \hyperlink{#1}{#2}%
    \endgroup
  }%
}
\makeatother

\author{
\large
Memoria presentada por \\
\textbf{Tamara Pardo Yanguas} \\
para optar al grado de \\
Doctora en Física \\
\vspace{1cm}
y dirigida por \\
\textbf{María Luisa Sarsa Sarsa} \\
\textbf{María Martínez Pérez} \\
\vspace{3.5cm}
Grupo de Física Nuclear y Astropartículas \\
Área de Física Atómica, Molecular y Nuclear \\
Departamento de Física Teórica \\
Centro de Astropartículas y Física de Altas Energías (CAPA) \\
\textbf{Universidad de Zaragoza} \\
\vspace{0.5cm}
Septiembre 2025
}

\title{\textbf{Advances in detector response and background modelling in the ANAIS-112 experiment for annual modulation and other rare event searches}}

\makeatletter
\renewcommand{\maketitle}{
  \vspace*{1cm}
  \begin{center}
    \hrule height 0.5pt
    \vspace{1ex}
    \hrule height 0.5pt
    \vspace{2ex}
    {\normalfont\LARGE \bfseries \@title \par}
    \vspace{2ex}
    \hrule height 0.5pt
    \vspace{1ex}
    \vspace{2cm}
    \parbox{0.8\textwidth}{\centering \normalfont\normalsize \@author}
  \end{center}
}
\makeatother

\begin{document}

\dominitoc

\maketitle
\thispagestyle{empty} 

\clearpage
\thispagestyle{empty}
\null
\clearpage

\pagenumbering{roman}  
\setcounter{page}{3}  

\tableofcontents
\clearpage

\clearpage\vspace{1.5ex}
\pagestyle{plain}  


\pagestyle{plain}       
{\begin{center}
  \normalfont\huge\bfseries
  \vspace{1.5ex}
  \textbf{List of Abbreviations}
\end{center}}
\phantomsection
\addcontentsline{toc}{section}{List of Abbreviations} 
\setcounter{page}{7}  

\vspace{0.2cm}
{\normalsize

\begin{multicols}{2}
\RaggedRight 

\noindent
\makebox[5em][l]{\textbf{ALP}} Axion-Like Particle\\[6pt]
\makebox[5em][l]{\textbf{BAO}} Baryon Acoustic Oscillations\\[6pt]
\makebox[5em][l]{\textbf{BBN}} Big Bang Nucleosynthesis\\[6pt]
\makebox[5em][l]{\textbf{BDT}} Boosted Decision Tree\\[6pt]
\makebox[5em][l]{\textbf{BR}} Branching Ratio\\[6pt]
\makebox[5em][l]{\textbf{BSM}} Beyond the Standard Model\\[6pt]
\makebox[5em][l]{\textbf{CCD}} Charge Coupled Device\\[6pt]
\makebox[5em][l]{\textbf{CDM}} Cold Dark Matter\\[6pt]
\makebox[5em][l]{\textbf{CFD}} \parbox[t]{\dimexpr\linewidth-5em}{Constant Fraction Discriminator}\\[6pt]
\makebox[5em][l]{\textbf{CJPL}} \parbox[t]{\dimexpr\linewidth-5em}{China Jinping Underground Laboratory}\\[6pt]
\makebox[5em][l]{\textbf{C.L.}} Confidence Level\\[6pt]
\makebox[5em][l]{\textbf{CMB}} \parbox[t]{\dimexpr\linewidth-5em}{Cosmic Microwave Background}\\[6pt]
\makebox[5em][l]{\textbf{CP}} Charge-Parity\\[6pt]
\makebox[5em][l]{\textbf{CR}} Cosmic Ray\\[6pt]
\makebox[5em][l]{\textbf{DAQ}} Data Acquisition\\[6pt]
\makebox[5em][l]{\textbf{DD}} Direct Detection\\[6pt]
\makebox[5em][l]{\textbf{DM}} Dark Matter\\[6pt]
\makebox[5em][l]{\textbf{EC}} Electron Capture\\[6pt]
\makebox[5em][l]{\textbf{EFT}} Effective Field Theory\\[6pt]
\makebox[5em][l]{\textbf{ER}} Electron Recoil\\[6pt]
\makebox[5em][l]{\textbf{FLRW}} \parbox[t]{\dimexpr\linewidth-5em}{Friedmann - Lemaître - Robertson - Walker metric}\\[6pt]
\makebox[5em][l]{\textbf{FM}} First Moment\\[6pt]
\makebox[5em][l]{\textbf{GPS}} \parbox[t]{\dimexpr\linewidth-5em}{G4GeneralParticleSource}\\[6pt]
\makebox[5em][l]{\textbf{G4NDL}} \parbox[t]{\dimexpr\linewidth-5em}{Geant4 Neutron Data Library}\\[6pt]
\makebox[5em][l]{\textbf{GR}} General Relativity\\[6pt]
\makebox[5em][l]{\textbf{HPGe}} High Precision Germanium\\[6pt]
\makebox[5em][l]{\textbf{HV}} High Voltage\\[6pt]
\makebox[5em][l]{\textbf{ICP-MS}} \parbox[t]{\dimexpr\linewidth-5em}{Inductively Coupled Plasma - Mass Spectrometry}\\[6pt]
\makebox[5em][l]{\textbf{INFN}} \parbox[t]{\dimexpr\linewidth-5em}{Istituto Nazionale di Fisica Nucleare}\\[6pt]
\makebox[5em][l]{\textbf{LM}} Likelihood Method \\[6pt]
\makebox[5em][l]{\textbf{LNGS}} \parbox[t]{\dimexpr\linewidth-5em}{Laboratori Nazionali del Gran Sasso}\\[6pt]
\makebox[5em][l]{\textbf{LSC}} \parbox[t]{\dimexpr\linewidth-5em}{Laboratorio Subterráneo de Canfranc}\\[6pt]
\makebox[5em][l]{\textbf{LSM}} \parbox[t]{\dimexpr\linewidth-5em}{Laboratoire Souterrain de Modane}\\[6pt]
\makebox[5em][l]{\textbf{LY}} Light Yield\\[6pt]
\makebox[5em][l]{\textbf{MACHO}} \parbox[t]{\dimexpr\linewidth-5em}{Massive Compact Halo Object}\\[6pt]
\makebox[5em][l]{\textbf{MC}} Monte Carlo\\[6pt]
\makebox[5em][l]{\textbf{ML}} Machine Learning\\[6pt]
\makebox[5em][l]{\textbf{MOND}} \parbox[t]{\dimexpr\linewidth-5em}{Modified Newtonian Dynamics}\\[6pt]
\makebox[5em][l]{\textbf{NAA}} Neutron Activation Analysis\\[6pt]
\makebox[5em][l]{\textbf{nphe}} Number of Photoelectrons\\[6pt]
\makebox[5em][l]{\textbf{NR}} Nuclear Recoil\\[6pt]
\makebox[5em][l]{\textbf{OFE}} Oxygen-Free Electronic\\[6pt]
\makebox[5em][l]{\textbf{OFHC}} \parbox[t]{\dimexpr\linewidth-5em}{Oxygen-Free High Conductivity}\\[6pt]
\makebox[5em][l]{\textbf{PBH}} Primordial Black Hole\\[6pt]
\makebox[5em][l]{\textbf{PDF}} Probability Density Function\\[6pt]
\makebox[5em][l]{\textbf{phe}} Photoelectron\\[6pt]
\makebox[5em][l]{\textbf{PK}} Photocathode\\[6pt]
\makebox[5em][l]{\textbf{PMT}} Photomultiplier Tube\\[6pt]
\makebox[5em][l]{\textbf{PQ}} Peccei-Quinn\\[6pt]
\makebox[5em][l]{\textbf{pNG}} \parbox[t]{\dimexpr\linewidth-5em}{pseudo-Nambu–Goldstone}\\[6pt]
\makebox[5em][l]{\textbf{PSA}} Pulse Shape Analysis\\[6pt]
\makebox[5em][l]{\textbf{PSD}} Pulse Shape Discrimination\\[6pt]
\makebox[5em][l]{\textbf{PSV}} Pulse Shape Variable\\[6pt]
\makebox[5em][l]{\textbf{QE}} Quantum Efficiency\\[6pt]
\makebox[5em][l]{\textbf{QED}} Quantum Electrodynamics\\[6pt]
\makebox[5em][l]{\textbf{QCD}} Quantum Chromodynamics\\[6pt]
\makebox[5em][l]{\textbf{QDC}} Charge-to-Digital Converters\\[6pt]
\makebox[5em][l]{\textbf{QF}} Quenching Factor\\[6pt]
\makebox[5em][l]{\textbf{RFA}} Radon Free Air\\[6pt]
\makebox[5em][l]{\textbf{ROI}} Region Of Interest\\[6pt]
\makebox[5em][l]{\textbf{SD}} Spin-Dependent\\[6pt]
\makebox[5em][l]{\textbf{SER}} Single Electron Response\\[6pt]
\makebox[5em][l]{\textbf{SF}} Spontaneous Fission\\[6pt]
\makebox[5em][l]{\textbf{SHM}} Standard Halo Model\\[6pt]
\makebox[5em][l]{\textbf{SI}} Spin-Independent\\[6pt]
\makebox[5em][l]{\textbf{SiPMs}} Silicon Photomultiplier\\[6pt]
\makebox[5em][l]{\textbf{SM}} Standard Model\\[6pt]
\makebox[5em][l]{\textbf{SURF}} \parbox[t]{\dimexpr\linewidth-5em}{Stanford Underground Research Facility}\\[6pt]
\makebox[5em][l]{\textbf{SUSY}} Supersymmetry\\[6pt]
\makebox[5em][l]{\textbf{TES}} Transition Edge Sensor\\[6pt]
\makebox[5em][l]{\textbf{TPC}} Time Projection Chamber\\[6pt]
\makebox[5em][l]{\textbf{TUNL}} \parbox[t]{\dimexpr\linewidth-5em}{Triangle Universities Nuclear Laboratory}\\[6pt]
\makebox[5em][l]{\textbf{WIMP}} \parbox[t]{\dimexpr\linewidth-5em}{Weakly Interacting Massive Particle}\\[6pt]

\end{multicols}
}

\mainmatter
\pagestyle{headings} 

\chapter*{} 
\vspace{-2cm}
\label{Chapter:Abstract}
\phantomsection
\addcontentsline{toc}{section}{Abstract} 
{\begin{center}

    {\normalfont\LARGE \bfseries Abstract}

\end{center}}

\pagenumbering{roman}  
\setcounter{page}{9}  

Numerous astronomical and cosmological observations point to the existence of dark matter, which constitutes about 27\% of the Universe. Despite extensive experimental efforts, the nature of DM remains unknown. Among the preferred dark matter candidates are axions and Weakly Interacting Massive Particles, which can be searched for using ground-based detectors. Only the DAMA/LIBRA experiment, using NaI(Tl) detectors at Gran Sasso National Laboratory, has reported a positive dark matter signal: an annual modulation in its detection rates compatible with the signal that would be produced by dark matter particles distributed in the galactic halo following the most standard halo models. To independently verify this result without relying on specific DM or halo models, using the same NaI target is essential. This is the goal of ANAIS-112, operating with 112.5 kg of NaI(Tl) scintillators since August 3, 2017, at the Canfranc Underground Laboratory (LSC). This thesis presents work conducted within the ANAIS-112 experiment, focusing on data analysis and the development of Geant4-based simulations, with the goal of reducing systematic effects and increasing the experimental sensitivity. Firstly, efforts have been directed towards improving the understanding of the ANAIS-112 crystals' response to nuclear recoils. Six years of ANAIS data challenge the dark matter interpretation of the DAMA/LIBRA signal. However, systematic uncertainties, particularly related to the scintillation quenching factors (QFs) of sodium and iodine nuclei recoiling in NaI(Tl), must be addressed to ensure reliable comparisons between ANAIS, DAMA/LIBRA, and other experiments. For that purpose, onsite neutron calibrations in ANAIS-112 have been performed since 2021 at LSC. The calibration procedure is based on the exposure of the full detector array to \textsuperscript{252}Cf neutron sources. This work has compared the data from these calibrations with dedicated Geant4-based neutron simulations. Moreover, the neutron simulations have revealed deficiencies in the Geant4 implementation of certain decay processes or cross sections in some versions. This study has enabled the evaluation of different QF models. The simulations have proven highly sensitive to the QF used, favoring models in which the QF increases with energy, and disfavouring the values assumed by the DAMA collaboration. In parallel, this thesis has contributed to the improvement of the ANAIS-112 background model through a multiparametric fit of its various background components. An exhaustive review of the contributions considered in the previous background model has been performed, which has enabled a significantly better agreement in all the analyzed observables. Subsequently, new physics searches have been conducted using ANAIS-112 data. Specifically, a reanalysis of the annual modulation signal and a search for solar axions have been carried out employing the experiment’s six-year accumulated exposure. Finally, this thesis also includes the work conducted within the COSINUS experiment, focused on background modelling and the analysis of the impact of internal background components on the experiment’s sensitivity.

\clearpage
\thispagestyle{plain} 
\pagenumbering{roman} 
\setcounter{page}{10} 
\null 
\clearpage

\chapter*{} 
\vspace{-2cm}
\label{Chapter:Resumen}
\phantomsection
\addcontentsline{toc}{section}{Resumen} 
{\begin{center}

    {\normalfont\LARGE \bfseries Resumen}

\end{center}}

\thispagestyle{plain}

\pagenumbering{roman}  
\setcounter{page}{11}  

Numerosas observaciones astronómicas y cosmológicas correspondientes a distintas escalas y tiempos de la historia del Universo apuntan a la existencia de la materia oscura, la cual constituye aproximadamente el 27\% del Universo. A pesar de los grandes esfuerzos experimentales, la naturaleza de la materia oscura sigue siendo desconocida. Entre los candidatos favoritos a materia oscura se encuentran los axiones y las partículas masivas débilmente interactuantes (WIMPs, por sus siglas en inglés), cuya detección puede ser abordada haciendo uso de técnicas diferentes mediante detectores situados en la Tierra. Sólo el experimento DAMA/LIBRA, que utiliza detectores de NaI(Tl) en el Laboratorio Nacional del Gran Sasso, ha reportado una señal positiva de materia oscura: una modulación anual en sus ritmos de detección compatible con la esperada para partículas de materia oscura del halo galáctico. Para verificar de forma independiente este resultado sin depender de modelos específicos de materia oscura o del halo, es esencial utilizar el mismo material blanco de yoduro de sodio (NaI). Este es el objetivo de ANAIS-112, que opera con 112,5 kg de centelleadores de NaI(Tl) desde el 3 de agosto de 2017 en el Laboratorio Subterráneo de Canfranc (LSC). Esta tesis presenta el trabajo realizado dentro del experimento ANAIS-112, centrado en el análisis de datos y el desarrollo de simulaciones basadas en Geant4, con el objetivo de mejorar el entendimiento de los efectos sistemáticos que afectan a la comparación entre DAMA/LIBRA y ANAIS-112, y de aumentar la sensibilidad del experimento. En primer lugar, se ha trabajado en mejorar la comprensión de la respuesta de los cristales de ANAIS-112 a los retrocesos nucleares. Seis años de datos de ANAIS cuestionan de manera significativa la interpretación en términos de materia oscura de la señal observada por DAMA/LIBRA. Sin embargo, es necesario abordar las incertidumbres sistemáticas, especialmente las relacionadas con los factores de quenching (QFs) para el centelleo de los núcleos de sodio e yodo en NaI(Tl), para garantizar comparaciones rigurosas entre ANAIS, DAMA/LIBRA y otros experimentos. Para ello, desde 2021 se han realizado calibraciones con neutrones in situ en ANAIS-112 en el LSC. El procedimiento de calibración se basa en la exposición de todo el conjunto de detectores de ANAIS a fuentes de neutrones de \textsuperscript{252}Cf. Este trabajo ha comparado los datos obtenidos en dichas calibraciones con simulaciones específicas de neutrones basadas en Geant4. Además, las simulaciones de neutrones han puesto de manifiesto deficiencias en la implementación de ciertos procesos de decaimiento o secciones eficaces de neutrones en algunas versiones de Geant4. Este estudio ha permitido evaluar distintos modelos de QF. Las simulaciones han demostrado una alta sensibilidad al modelo de QF empleado, favoreciendo aquellos en los que el QF aumenta con la energía y desfavoreciendo los valores de QF asumidos por DAMA/LIBRA. Paralelamente, esta tesis ha contribuido a la mejora del modelo de fondo de ANAIS-112 mediante un ajuste multiparamétrico de sus diferentes componentes del fondo. Se ha realizado una revisión exhaustiva de las contribuciones consideradas en el modelo de fondo previo, lo que ha permitido lograr un acuerdo significativamente mejor en todos los observables analizados. Posteriormente, se han realizado nuevas búsquedas de física con los datos de ANAIS-112. En concreto, se ha llevado a cabo un reanálisis de la señal de modulación anual y una búsqueda de axiones solares utilizando la exposición acumulada de seis años del experimento. Finalmente, esta tesis también incorpora el trabajo realizado durante una estancia de investigación en el experimento COSINUS, centrado en el modelado de fondo y en el análisis del impacto de las componentes del fondo interno en la sensibilidad del experimento.

\thispagestyle{plain}

\clearpage
  \thispagestyle{empty}
  \null
  \clearpage

\pagenumbering{arabic} 
\setcounter{chapter}{0} 
\chapter{Understanding the Universe}\label{Chapter:Intro}
\minitoc
\vspace{-0.2cm}
\vspace{0.5cm}

Human understanding of what the Universe is and how it works has changed significantly throughout the millennia. Even though knowing the Universe has always been a subject of interest, it was until recently a matter reserved to philosophers. In fact, the nature of cosmos is a topic where scientific and philosophical narratives have typically got blurred. From the undeniable ancient greek contributions to the great milestones of the 20$\mathrm{^{th}}$ century astrophysics, there have been astonishing discoveries that have exponentially improved how mankind understands the Universe. 

While this rate of progress still advances at great pace, when looking up into the night sky, most of the energy and mass of the Universe remain of unknown nature. Nowadays, the model of Universe favored by most recent measurements \cite{Planck:2018jri} is a spatially-flat Universe dominated by a 68\% of the so-called dark energy, responsible for the accelerating expansion of the Universe (see Section \ref{StandardCosmologicalModelSection}). Baryonic matter accounts for only 5\% of the total energy content, and less than 1\% corresponds to luminous baryonic matter. The remaining 27\% corresponds to a stable (or very long-lived), non-baryonic, non-luminous, cold matter component known by the name of dark matter (DM). Compelling astrophysical and cosmological observations obtained using tracers of its gravitational interactions support the existence of this elusive ingredient of the Universe (see Section~\ref{DarkMatterNeedSection}). 

As a long-standing open question in particle physics, cosmology, and astrophysics, many DM candidates have been proposed to solve this puzzle (see Section \ref{DMCandidatesSection}). Considering that the preferred scenario supports that DM has a particle nature, none of the particles of the Standard Model of particle physics (SM) has the properties required to be a valid candidate. Consequently, a large number of hypothetical new DM particles beyond the Standard Model (BSM) have been proposed \cite{Baudis2016}. Among the most popular candidates are the Weakly Interacting Massive Particles (WIMPs) and axions/ALPs (Axion-Like Particles).

In order to unveil the ultimate nature of DM, increasingly sensitive experiments have been designed to try to detect or produce DM for decades. There are three classes of complementary DM detection techniques: direct detection, by identifying the interaction of DM particles with a convenient detector; indirect detection, by studying SM particle fluxes produced in the annihilation or decay of DM particles; and production at colliders, by searching for the DM particle as the final product of the collision of very energetic SM particles (see Section \ref{DMDetectionSection}). Any indication of new physics that points towards a viable BSM theory could also shed light on the DM problem. Despite the overwhelming global experimental effort, no unequivocal positive signal coming from DM particles has been found to date. 

Nevertheless, DAMA/LIBRA, a DM direct detection experiment which uses NaI(Tl) crystals at the Gran Sasso National Laboratory (LNGS), in Italy, has been reporting since end of the 1990s a long-standing positive result: the observation of a highly statistically significant annual modulation in its detection rate compatible with that expected for DM particles following the most
standard halo models \cite{DAMA2008FirstResults,DAMA:2010gpn,DAMA2013,DAMA2018}. This claim of discovery is in strong tension with the many limits set by other experiments with different target materials and detection techniques, like those from CDMS \cite{SuperCDMS:2015eexAbdelhameed}, CRESST \cite{CRESST:2019jnq}, Darkside-50 \cite{agnes2023search}, EDELWEISS \cite{EDELWEISS:2019vjv}, KIMS \cite{Lee:2014zsa}, LUX \cite{LUX:2016ggv}, PICO \cite{Krauss:2020ofg} or XENON collaborations \cite{XENON:2020kmp}, who have been ruling out for years the most plausible compatibility scenarios. Nevertheless, these experiments do not employ NaI(Tl) as target material. As a result, they cannot conclusively confirm or refute the DAMA/LIBRA claim due to many model-related uncertainties when comparing results obtained with different detector target.

Thus, the DAMA/LIBRA result remains a puzzling mystery. The ANAIS (Annual modulation with NaI Scintillators) project \cite{Amare:2018ndh} is a direct DM detection experiment whose goal is to confirm or refute in a model independent way the positive annual modulation signal reported by DAMA/LIBRA collaboration using the same target and technique, but different experimental conditons. Consisting of 112.5 kg of ultrapure NaI(Tl) detectors, ANAIS-112 is taking data at the Canfranc Underground Laboratory (LSC) in Spain since the 3$^{\textnormal{rd}}$ of August 2017. 

\section{The Standard Cosmological Model} \label{StandardCosmologicalModelSection}

Cosmology is the science that aims at studying the origin, structure, evolution and ultimate fate of the Universe. Throughout the centuries, notable contributions have been made towards a mathematical formulation of the Universe, built upon observations of heavenly objects. Early geocentric models placed the Earth at the center of the Universe. This model was later replaced in the 16th century by the heliocentric model proposed by Copernicus, which suggested that the Earth and the other planets of the Solar System orbited around the Sun, later supported by Galileo observations.


Newton provided a new understanding of the forces ruling the Universe by formulating the universal theory of gravitation and the laws of motion. It was not until the beginning of the 20$\mathrm{^{th}}$ century that the Milky Way was discovered to be one among many others galaxies of the Universe. The development of Einstein's Theory of General Relativity (GR) in 1915 \cite{GALE1996279}, together with high-precision observational data acquired during the last decade of the past century, led to the birth of modern cosmology.

The most complete, data-supported accepted theory of the Universe is the so-called Lambda-Cold DM ($\Lambda$CDM) \cite{Carroll:2001,Peebles:2003}. It is considered the Standard Model of Cosmology due to its simplicity and great predictability. This consistent framework is based not only on GR, but also on the Cosmological Principle, which asserts that at sufficiently large scales ($\gtrsim$~100 Mpc), the Universe is statistically homogeneous and isotropic, implying that all spatial locations are fundamentally equivalent in their physical properties.

According to the $\Lambda$CDM model, the Universe is composed of four fundamental components: dark energy (denoted by $\Lambda$) \cite{Copeland:2006wr}, non-baryonic cold dark matter (CDM) \cite{Arbey:2021}, ordinary matter (baryons and leptons), and radiation (photons and neutrinos). A wide range of cosmological observations have tested the model to a very high degree of precision. These include the accelerating expansion of the Universe, the anisotropies of the Cosmic Microwave Background (CMB) \cite{Page_2003}, the spectrum and statistical properties of large-scale structures of the Universe \cite{BERNARDEAU20021}, and the observed abundances of different light elements \cite{Steigman:2007xt}.  

In 1929, Edwin Hubble changed the understanding of the cosmos providing the first observational evidence of the expansion of the Universe \cite{Hubble:1929, Hubble:1931}. Contrary to the previously held view of a static Universe, Hubble observations ultimately proved that galaxies are moving away in all directions at velocities directly proportional to their distance from the Earth following the Hubble-Lema\^itre law:

\begin{equation}
 v = H_0 r,
\end{equation}

with $H_0$ the Hubble parameter, $v$ the galaxy velocity, and $r$ its distance to the observer.

Following this breakthrough discovery, in 1931 George Lema\^itre realized that
projecting the observed expansion of the Universe back in time implied that, at some finite moment in the past, the entire Universe must have been smaller and hotter, concentrated at a single point of infinite density at the initial instant: the Big Bang \cite{LEMAÎTRE1931,LEMAÎTRE1950}. Based on the Lema\^itre hypothesis, George Gamow formulated the Big Bang theory during the 1940s \cite{Gamow:1946eb, Alpher:1948gsu}.

The homogeneity and isotropy of the Universe is introduced in the theory by the Friedmann-Lema\^itre-Robertson- Walker (FLRW) metric, describing the space-time \cite{Arbey:2021}. Later, the CMB radiation would provide compelling observational evidence supporting such homogeneous and isotropic Universe. In the Standard Cosmological model, the overall geometry and evolution of the Universe are described in terms of two cosmological parameters: the spatial curvature, $k$, and a time-dependent scale factor of the Universe that determines the expansion (or contraction) of the Universe, $a(t)$. The FLRW line element expressed in terms of the comoving coordinates is:

\begin{equation}
ds^2 = g_{\mu\nu}dx^{\mu}dx^{\nu} = c^2 dt^2 - a^2(t) \left(\frac{dr^2}{1-kr^2} +r^2 d\theta + r^2 \sin^2\theta d\phi^2 \right),
\end{equation}

where $g_{\mu\nu}$ is the metric tensor and $x^{\mu} = (ct, r, \theta, \phi)$. The possible values of the curvature parameter $k$ are +1, 0 or -1 for a closed, spatially flat or open Universe. 

In order to determine how the evolution of the Universe occurs, GR equations must be employed, which relates the spacetime geometry and its mass/energy distribution:

\begin{equation}
R_{\mu\nu} - \frac{1}{2} R g_{\mu\nu} + \Lambda g_{\mu\nu} = \frac{8\pi G}{c^4} T_{\mu\nu},
\end{equation}

where \( R_{\mu\nu} \) is the Ricci curvature tensor, \( R = g_{\mu\nu} R^{\mu\nu} \) the Ricci scalar, \( \Lambda \) the cosmological constant, \( G \) the Newtonian gravitational constant, and \( T_{\mu\nu} \) the energy- momentum tensor. Einstein originally introduced \( \Lambda \) to allow for a static Universe, but after the discovery of cosmic expansion by Hubble, he regarded it as a serious error and labelled the term his “biggest blunder”. However, \( \Lambda \) regained significance as it became the most successful model to explain the observation of the accelerated expansion of the Universe \cite{schmidt2012nobel}.

Considering the Universe as a perfect fluid (without viscosity or heat conduction), the energy-momentum tensor can be written in the form: 

\begin{equation}
T_{\mu\nu} = \left(\rho+\frac{P}{c^2}\right)u_\mu u_\nu + Pg_{\mu\nu},
\end{equation}

with $u_\mu$ the four-velocity of the fluid element, $P$ the pressure and $\rho$ the density. 

By applying the Einstein equations with the FLRW metric under the ideal fluid assumption, the Friedmann equations are obtained. While the time component of this basic set of cosmic equations becomes the first Friedmann equation and governs the evolution of the Universe, the spatial component gives rise to the second Friedmann equation, which describes whether the expansion of the Universe is accelerating or decelerating:

\begin{equation}
H^2(t) \equiv \left(\frac{\dot{a}}{a}\right)^2 = \frac{8\pi G}{3} \rho_{tot} - \frac{kc^2}{a^2},
\label{1stFriedmann}
\end{equation}
and
\begin{equation}
\frac{\ddot{a}}{a} = -\frac{4\pi G}{3c^2} \left(\rho_{tot} c^2 +3P\right),
\end{equation}

with $H(t)$ the Hubble parameter, being $H(t_0)$ = $H_0$ its present value, and $\rho_{tot}$ the total matter and energy density of the Universe, in which matter, $\rho_{m}$, radiation, $\rho_{rad}$, and dark energy $\rho_{\Lambda} = \frac{\Lambda}{8\pi G}$ contributes. 

Since the different components have different equations of state, the temporal evolution of each energy density is: 

\begin{equation}
\frac{d{\rho_i}}{dt} = -3\left(\rho_i + \frac{P_i}{c^2}\right) \frac{\dot{a}}{a}.
\label{1p7}
\end{equation}

Under the assumption of a perfect fluid, the equation of state that relates linearly the pressure to the density can be expressed as: 

\begin{equation}
    P_i=\omega_i\rho_ic^2,
    \label{1p8}
\end{equation}

where $i$ accounts for the matter, radiation, dark energy or any other component contributing
to the energy density of the Universe. 

Combining Equations \ref{1p7} and \ref{1p8}, the temporal evolution of the energy density for each component is obtained:

\begin{equation}
\rho_i(t) \propto a(t)^{-3(\omega_i+1)}.
\end{equation}

\begin{itemize}

\item Non-relativistic matter is represented by a set of non-interacting particles with zero pressure ($\omega_m$ = 0), leading to $\rho_m \propto a^{-3}$. 

\item For radiation, which is defined as a very hot gas of relativistic particles (mainly photons and neutrinos), the energy-momentum tensor of Quantum Electrodynamics (QED) indicates $\omega_r = 1/3$, resulting in its evolution as $\rho_r \propto a^{-4}$. This behavior corresponds to the redshift of radiation energy as the Universe expands, and it strictly applies to massless particles. For particles with very small mass, such as neutrinos, $\omega_r$ tends to zero as the Universe expands.

\item Regarding dark energy, there are two different models: its energy density remains constant, $\omega_\Lambda = -1 $, which corresponds to a cosmological constant, $\rho_\Lambda$ = const.; or, more generally,
any $\omega_\Lambda < -1/3$ value, being possible a time dependence in $\omega_\Lambda (t)$, and then, a complex evolution in time of the $\rho_\Lambda$.

\end{itemize}

From the first Friedmann equation (Equation \ref{1stFriedmann}), it is convenient to define a critical density $\rho_c$ , which corresponds to the energy density condition with $k$ = 0 (flat Universe), being:

\begin{equation}
    \rho_c = \frac{3H_o^2}{8\pi G} \\
\end{equation}

The condition $\rho_{tot} > \rho_c$ indicates a closed Universe, $\rho_{tot} = \rho_c$ implies a flat Universe, and $\rho_{tot} < \rho_c$ corresponds to an open Universe. 

Consequently, a dimensionless density parameter can be built as the energy density relative to the critical density, indicating the contribution of each component to the Universe energy:

\begin{equation}
    \Omega_i = \frac{\rho_i}{\rho_c}
\end{equation}

According to this definition, $\sum_{i}\Omega_i$ = 1, <1 or >1 for a flat, open and closed Universe, respectively. In some cases, although it does not account for a real contribution to the Universe energy, it can be interesting to consider curvature as a contribution to the density. In this case, $\sum_{i}\Omega_i$ = 1 by construction. 


The Standard Cosmological Model is tremendously successful at providing a self- contained
picture of the composition and evolution of the Universe. Yet, there are questions that are not properly addressed by the $\Lambda$CDM framework as introduced so far, mainly why is the Universe so isotropic, homogeneous and flat, as suggested by observations. These are the so-called horizon (how can it be that apparently disjoint patches of space have nearly the same densities and temperatures) and the flatness problems. 
A process of accelerated expansion in the very early stages of the Universe can elegantly address both issues. This process is known as inflation and it was proposed by Alan Guth in 1981 \cite{Guth:1980zm}. Indeed, during the inflationary epoch, tiny primordial energy density fluctuations would have been amplified, acting as seeds for the formation of large-scale structures that otherwise would not have been able to form.

Inflation provides a structure formation scenario that completes the Standard Cosmological Model: a six-parameter
model based on a flat Universe dominated by dark energy and CDM, with initial gaussian, adiabatic fluctuations seeded by inflation. In spite of its success explaining the observations of the Universe at all scales (see Section \ref{evidences}), this model includes unknown ingredients according to the present knowledge of particle physics: the particle responsible of the CDM, the source of the dark energy, or the inflaton field responsible of the inflation.

\subsection{CMB and cosmological parameters}\label{CMB_cosmoPara}

In the last three decades,  precision measurements of the CMB temperature and polarisation anisotropies have incredibly boosted the understanding of the Universe and have allowed to extensively test the $\Lambda$CDM cosmological model.

The Big Bang theory states that the early Universe consisted of a hot soup of matter and radiation in thermal equilibrium. As the Universe expanded, the particles forming this plasma cooled down, which eventually led to the formation of light nuclei allowing the Big Bang Nucleosynthesis (BBN) \cite{Steigman:2007xt}. It took place between 3 and 20 minutes
after the Big Bang, when the temperature of the Universe was $\simeq$ 10$^{10}$ K. Later, recombination of electrons and light nuclei forming the first neutral atoms occurred when the temperature of the Universe fell down below 3000 K about 380000 years after the Big Bang. The Universe, which previously consisted of a highly-ionized plasma opaque to electromagnetic radiation, changed then into a gas of neutral atoms, transparent to photons. From that moment, photons decoupled from the rest of the Universe contents and started free-streaming.

Those emitted photons formed the cosmic microwave background (CMB), which provides the oldest picture of the Universe. These relic photons have remained almost unalterable until today, filling the Universe and travelling freely through space, while losing energy with the expansion of the Universe. This radiation was first observed in 1964 by Arno Penzias and Robert Wilson \cite{Penzias:1965apj}. At the cosmological scale, the CMB is considered the first tangible evidence of the of the Bing Bang cosmology and gives very valuable information of the early times of the Universe.

\begin{figure}[t!]
\begin{center}
\includegraphics[width=0.65\textwidth]{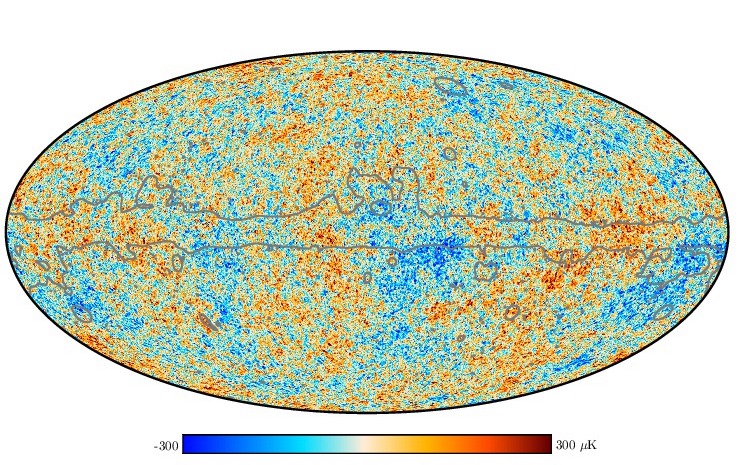}
\caption{\label{PlanckCMB} Full-sky Planck 2018 map of the CMB temperature anisotropies over an average temperature of 2.725 K. The figure shows temperature fluctuations corresponding to regions of varying densities which would later led to the formation of all the structures in the Universe. The gray lines/contours show the position of the microwave
emissions in the Milky Way and local structures, which have been removed from the
data \cite{Planck2018}.}
\end{center}
\end{figure}
The CMB presents an almost perfect black body isotropic and homogeneous distribution corresponding to a temperature of 2.73 K. However, dedicated measurements from the last decades have shown very small temperature anisotropies ($\Delta$T/T $\sim$ 10$^{-5}$). CMB radiation has been consistently mapped at increasing levels of sensitivity and angular resolution. The latest and most accurate map of the CMB fluctuations has been obtained by the European Planck satellite (Figure \ref{PlanckCMB}) \cite{Planck2018}. These measurements offer improved angular resolution, broader frequency coverage, and a more complete characterization of polarization modes compared to its predecessor, WMAP \cite{WMAP:2012fli,WMAP:2012nax}. This map illustrates the temperature distribution of the early Universe at the time of recombination, with blue spots corresponding to colder, denser regions, and red spots to hotter, less dense regions.

The angular distribution of the CMB temperature fluctuations can
be expressed in terms of the spherical harmonics:

\begin{figure}[t!]
\begin{center}
\includegraphics[width=0.65\textwidth]{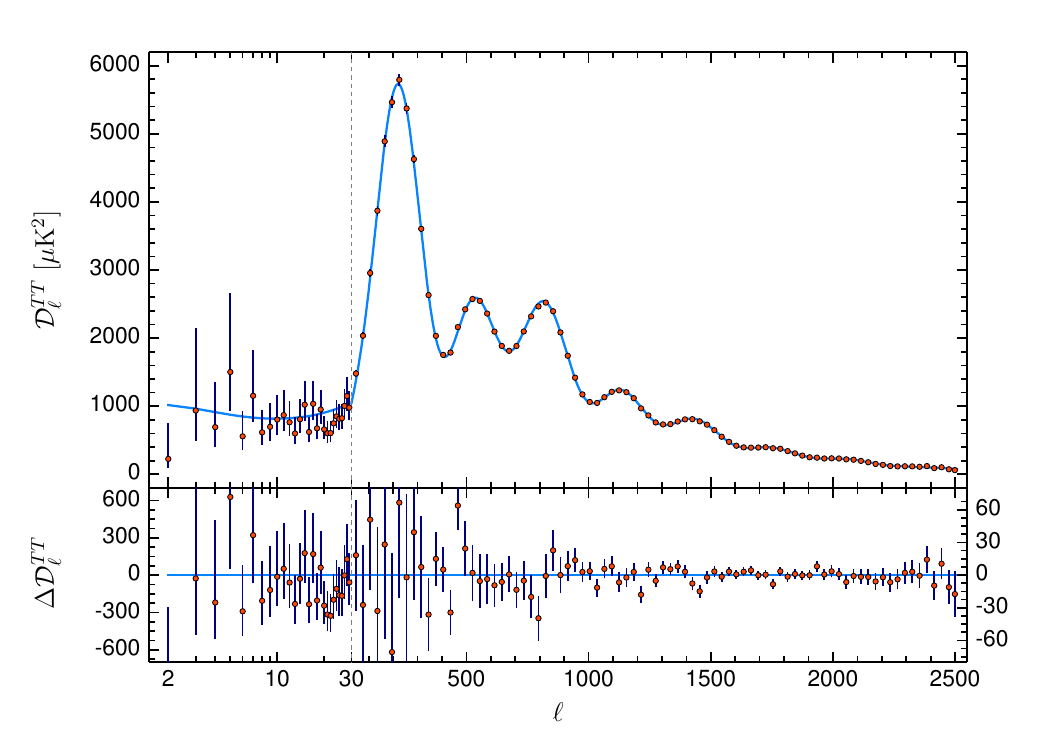}
\caption{\label{PlanckFit} Planck 2018 measured power spectrum (orange dots) compared with the
predictions of the $\Lambda$CDM cosmological model (blue line). The temperature (TT) modes are represented, although there are also measurements of the polarization modes and the combination of temperature and polarization modes. It can be observed that there is a very good agreement between
the experimental data and the predictions of the $\Lambda$CDM model at high multipoles. Residuals are also shown \cite{akrami2020planck}.}
\end{center}
\vspace{-0.5cm}
\end{figure}

\begin{equation}
\frac{\Delta T}{T}(\theta,\phi) = \sum_{lm}{a_{lm}Y_{lm}(\theta,\phi)},
\end{equation}
where $l$ and $m$ correspond to a specific angular scale.

The anisotropy of the CMB is typically quantified by its angular power spectra $D_l$ derived from the analysis of the temperature map, which is determined by the square of the amplitude of the
temperature fluctuations for each $l$:

\begin{equation}
    D_l = \frac{l(l+1)}{2\pi} C_l ,
\end{equation}

where $C_l$ = $\frac{1}{2l+1}$$\sum_{m} |a_{lm}|^2$ measures the power of the multipole $l$. The features of the density profile of the early Universe are crucial for cosmological parameters estimation.

The angular power spectrum of the CMB temperature collected by Planck is shown in Figure~\ref{PlanckFit}. The temperature (TT) modes are represented, although there are also measurements of the polarization modes and the combination of temperature and polarization modes. The height of the first peak in the power spectrum is linked to the density of all matter (dark and baryonic), whereas the second peak is associated with the density of baryonic matter.

The best fit predictions of the $\Lambda$CDM cosmological model parameterized with six independent (free) parameters are also displayed in Figure~\ref{PlanckFit}. Similar good agreement between observations and predictions is obtained for other modes. In particular, the six fundamental cosmological parameters are: the present baryon ($\Omega_b$$h^2$) and CDM ($\Omega_c$$h^2$) densities multiplied by $h^2$, where $h$ = $H_0$/(100 km s$^{-1}$Mpc$^{-1}$) and $H_0$ is the Hubble constant, the sound horizon at last scattering ($\theta_{MC}$); the reionization optical depth ($\tau$); the amplitude of the primordial comoving curvature power spectrum ($A_S$); and the power-law index of the scalar spectrum  of primordial curvature perturbations~($n_S$).

It is worth highlighting the very good agreement between
the experimental data and the predictions of the $\Lambda$CDM model at high angular scales. However, $\Lambda$CDM also faces
several challenges, including inconsistencies in measurements of the Hubble constant derived from Planck with respect to the value coming from other probes. 

According to the latest measurements from DESI in combination with BBN priors~\cite{karim2025desi}, 
\ensuremath{H_0 = (68.51 \pm 0.58)\,\text{km\,s}^{-1}\,\text{Mpc}^{-1}}, 
which is mostly compatible with the Planck value~\cite{Planck:2018vyg}, 
\ensuremath{(67.66 \pm 0.42)\,\text{km\,s}^{-1}\,\text{Mpc}^{-1}}, 
while the estimates based on local measurements are noticeably higher, 
\ensuremath{(72.3 \pm 1.4)\,\text{km\,s}^{-1}\,\text{Mpc}^{-1}}~\cite{galbany2023updated}, 
reinforcing the Hubble tension between local and cosmological determinations.

By fitting the $\Lambda$CDM model to the peaks of the CMB power spectrum, and combining information from CMB lensing reconstruction
and Baryon Acoustic Oscillations (BAO), an estimation of the six parameters of the cosmological model with unprecedented accuracy is obtained. In particular, the most recent estimation of $\Omega_x$, where $x~=~b, \chi, m$ are the cosmic densities of baryons, DM and total matter, respectively, is \cite{Planck:2018vyg}:

\begin{equation}
\begin{split}
\Omega_bh^2 & = 0.02242 \pm 0.00014 \\
\Omega_{\chi}h^2 & = 0.11933 \pm 0.00091 \\
\Omega_m & = 0.3111 \pm 0.0056
\end{split}
\end{equation}

According to these results, the total mass-energy budget of the Universe is made up of about 68\% of unknown dark energy, 27\% of non-baryonic DM, and 5\% of ordinary matter. Applying BBN priors, DESI yields values of \(\Omega_m =0.2977 \pm 0.0086\) \cite{karim2025desi}. Both measurements are consistent, with values around 30\% of matter.

Furthermore, DESI results supports a flat Universe undergoing accelerated expansion, $\Omega_k$~=~0.0023~$\pm$~0.0011~\cite{karim2025desi}. Concerning the baryonic matter, the value reported by Planck is in agreement with the estimations derived from the comparison of the measured abundances of the light nuclei and the predictions of the BBN. 

In addition, the current contribution of radiation to the total mass-energy density of the Universe is negligible. Constraints on the possible contributions of neutrinos and other light particles behaving as radiation (i.e., relativistic) at the time of recombination can be bound. From the combination of DESI and CMB, upper limits on the sum of neutrino masses have been derived, finding \( \Sigma m_\nu < 0.064 \, \text{eV} \) with a 95\% confidence level (C.L.), assuming the \(\Lambda\)CDM model \cite{karim2025desi}. This result rules out the inverted mass hierarchy and, notably, identifies zero as the most likely value for the total neutrino mass, thereby almost excluding the normal ordering as well. The derived upper bound is in clear tension with neutrino oscillation experiment results, which report $\Sigma m_\nu > 0.059 \, \text{eV}$, favouring normal ordering. This discrepancy can be considered as a hint that the current model for interpreting the observational data has some drawbacks or limitations.


\section{Dark matter need} \label{DarkMatterNeedSection}

The existence of an as-yet-unknown kind of matter, which does not interact with the electromagnetic radiation and constitutes a significant contribution to the Universe density, is now a crucial element in understanding the thermal history and evolution of the cosmos. This enigmatic substance is referred to as dark matter.

\subsection{Dark matter evidences}\label{evidences}

Derived from cosmological and astrophysical observations, the evidence supporting its existence has been accumulating for almost a century, and relies on tracers that unveil its gravitational interactions. What might have initially seemed as a collection of unrelated cosmological or gravitational hints, has converged into a list of essential properties that any potential DM candidate must fulfill. One of the first solid hints of the existence of this dark component goes back to the 1930s, when Fritz Zwicky used the virial theorem to infer the existence of unseen matter in the Coma galaxy cluster \cite{Zwicky:1933gu}. Thereafter, numerous observations have emerged, indicating the presence of DM at a wide range of scales, spanning from galactic to cosmological scales. 

\subsubsection{Evidences at the galactic and galaxy cluster scales}

On the galactic scale, one of the most compelling evidence for the existence of DM comes from the study of the galactic rotation curves, which are the rotational velocities
of objects formed by stars, hot gas and cosmic dust
 in a galaxy, as a function of their distance from the center of the galaxy. The prediction from Newtonian mechanics is clear: given the distribution of
mass observed in a galaxy, the rotational velocity $v(r)$ of objects in the outer regions
of the galaxy must decrease with the distance $r$ from the center of the galaxy as: 

\begin{equation}
    v(r) = \sqrt{\frac{G M(r)}{r}},
\end{equation}

where a spherical mass distribution $M(r)$ is assumed for the galaxy, being $G$ the Newtonian gravitational constant. 

\begin{figure}[t!]
\begin{center}
\includegraphics[width=0.55\textwidth]{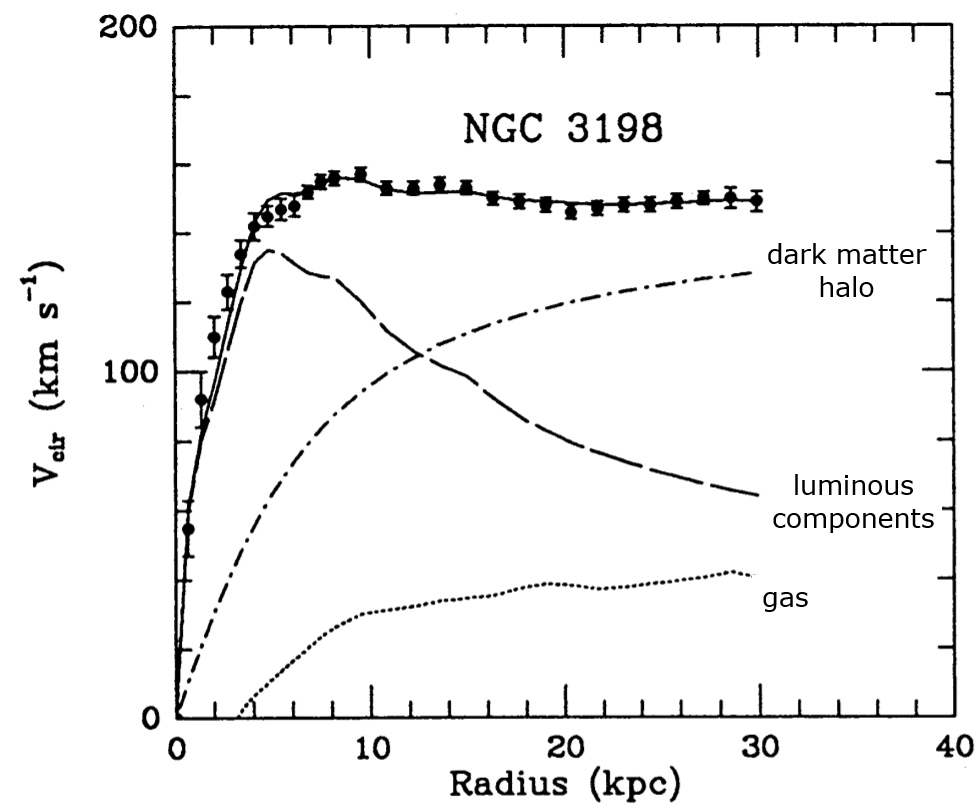}
\caption{\label{curvaDeRot} Example of galactic rotation curve. The dotted, the dashed and the dashed-dotted lines correspond to gas, luminous and DM halo estimated contributions to rotation velocity, respectively \cite{Rubin:1978kmz}. }
\vspace{-0.8cm}
\end{center}
\end{figure}

Despite the fact that spiral galaxies are not spherical distributions, a decrease of the rotational velocity as $v(r) \sim 1/\sqrt{r}$ beyond the visible galactic radius is expected. On the contrary, observations made by Vera
Rubin, Kent Ford and Albert Bosma in the 1970s indicated that rotation curves as determined in \cite{Rubin:1978kmz} showed a flattening, i.e. rotational velocities remained constant at large distances from
the galactic center. As an example, Figure~\ref{curvaDeRot} shows the rotational curve for galaxy NGC 3198. As it can be observed, up to the reach of the measurements, the rotational velocity does not decrease. Assuming that galaxies are surrounded by a DM halo with mass distribution $M(r) \propto r$, extending well beyond the regions occupied by visible baryonic matter, constitutes the most widely accepted explanation for this observed behavior. In addition, such DM haloes around the galaxies are consistent with the results from many-body simulations carried out on galaxy evolution, playing a crucial role by stabilizing the galactic disk in spiral galaxies.

Galaxy clusters stand as the most largest gravitationally bound structures within the observable Universe, with masses between 10$^{14}$ to 10$^{15}$\( \mathrm{M}_\odot \). These objects constitute excellent targets not only to evidence the existence of DM, but also to test GR and the Standard Cosmological Model. 

As done by Zwicky with the Coma cluster, measurements of the velocity dispersion of galaxies by applying the virial theorem can be employed to infer the mass of the object.
Nowadays, more precise methods are available to determine the mass of a galaxy cluster, depending on its location. One widely used approach for estimating the mass of a cluster involves measuring its X-ray surface brightness profile. In particular, X-ray measurements indicate that the temperature of the observed hot gas in galaxy clusters cannot be explained by considering only the gravitational attraction arising from the visible matter within the clusters.

\begin{figure}[t!]
\begin{center}
\includegraphics[width=1.\textwidth]{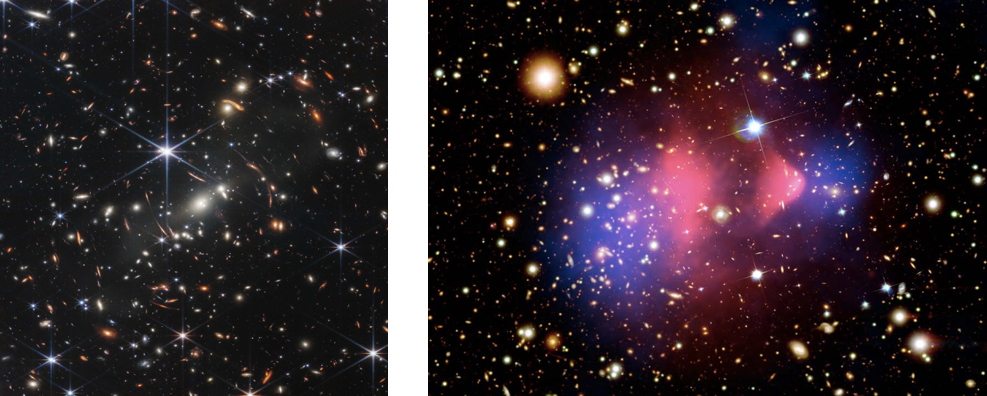}

\caption{\label{lensing} \textbf{Left panel:} Near-infrared image captured by NASA James Webb Space Telescope of the galaxy cluster SMACS-0723. The cluster is so massive that it bends the light coming from more distant galaxies, resulting in the creation of multiple distorted images (the orangish arc structures) of the same object (the brightest galaxy near the center of the image) \cite{JamesWebbTelescope}. \textbf{Right panel:} Composite image of the Bullet Cluster. Hot gas detected by Chandra X-ray telescope (in pink) is superimposed on the optical image of the galaxies from Magellan and Hubble Space Telescope (in orange and white). The blue region of the image shows the mass distribution in the cluster, as deduced from observations using gravitational lensing. The image depicts the separation between DM and hot gas within the cluster \cite{Bullet}.}
\vspace{-0.5cm}
\end{center}
\end{figure}

In addition to X-rays, the mass of a cluster can also be determined by gravitational lensing. GR predicts that massive objects bend light as it travels from the source to the observer, leading to distortions in the image of the source \cite{Tyson2009}. The more massive the lensing object, the stronger the effect. Hence, gravitational lensing can be used to estimate the mass distribution of
the object acting as a lens, as for example a cluster of galaxies. This effect can be observed in the left panel of Figure \ref{lensing}, where multiple images of the source are obtained.

\setlength{\belowcaptionskip}{-10pt}

While clusters are generally assumed to be stable and virialized objects, they can be found in various dynamical states. Clusters can have undergone merger events, as identified in Coma and Virgo clusters, and collisions, with the Bullet Cluster (1E0657-558) being the most well-known case. The Bullet Cluster is considered the most direct observational evidence to date for DM since, unlike galactic rotation curves, this evidence can be interpreted without requiring a Dynamics/Gravitation theoretical framework.

The Bullet Cluster is a very illustrative example of
how gravitational lensing can point towards the existence of DM \cite{Clowe:2003tk}. This cluster consists of two large colliding clusters of
galaxies, whose collision seems to have resulted in the separation of DM and baryonic matter (see right panel of Figure \ref{lensing}). 

On the one hand, X-ray observations reveal that a significant portion of the baryonic (gas) matter in the cluster is concentrated at the center of the system, as depicted in pink in the figure. During the collision, electromagnetic interactions among gas particles caused them to decelerate and to lose energy and to be left behind each cluster. 

On the other hand, weak gravitational lensing observations of the same system indicate that the majority of the mass is situated outside the central baryonic gas region, in blue. If most of the invisible matter of the cluster was not slowed down, it means it must be collisionless (or almost collisionless) matter. Consequently, DM should consist of a particle with gravitational interactions and possibly, very weakly interacting, both with itself and with baryonic matter. In fact, the study of the Bullet Cluster has set an upper limit on the DM self-interaction cross-section, specifically $\sigma_{\text{DM}}/m < 1.25 \, \text{cm}^2/\text{g}$, although more stringent limits have been derived from analyses with more clusters. Then, the DM components of the two clusters would pass through each other without substantial deceleration, leaving behind the baryonic gas components. The existence of a DM component in both clusters accounts for the observed separation.

\subsubsection{Evidences at cosmological scale}

On the cosmological scale, additional evidence is derived from large-scales structures. According to the $\Lambda$CDM model, as recombination took place, baryonic matter decoupled from photons, and thereafter its evolution was exclusively driven by gravity. Matter then began to aggregate, ultimately giving rise to the structures observed in the Universe today.

DM, making up the majority of the mass of the Universe, clearly played a pivotal role in the formation of these structures. The structure formation models predict that if the DM component is hot, i.e. relativistic, the formation of structures would have occured later and larger structures would have formed before smaller ones. This contrasts with the current hierarchical model of structure formation (from smaller to larger scales) supported by observations \cite{Bergstrom:2000pn}. Comparison between observations and simulations reveals that much more matter than visible baryonic matter is present in the Universe, but also that this DM component has to be mainly cold, i.e. non-relativistic when structures formed, as it is also derived from the CMB anisotropies analysis.

While large-scale structure observations reveal the preference of the Universe for cold DM, BBN provides strong evidence of its
non-baryonic nature. BBN corresponds to the brief period in the early Universe, beginning when the Universe was about 10 seconds old and lasting for about 20 minutes, during which the light
nuclides up to $^7$Li were formed \cite{Steigman:2007xt}. The comparison between the predictions of the BBN and the measurements of the light nuclei abundances in the Universe provides an estimation of the density of baryons ($\Omega_b$), which is fully compatible with the value derived from CMB anisotropies analysis. The consistency of such two independent results, from two different moments of the history of the Universe and dependent on very different assumptions and data, can be considered a strong support to the current model of the Universe.

BBN has proven to be an indispensable ingredient in the understanding of the current composition of the Universe. In the first moments of the Universe, when the temperature was of the order of $k_BT~\sim$ 100~MeV, the primordial plasma consisted of a soup of photons, neutrinos, electrons, and positrons in thermal equilibrium with baryons. The Universe continued to expand and cool, and when the temperature dropped below $k_BT~\sim0.8$ MeV, the weak interactions keeping neutrons and protons in chemical equilibrium stopped being efficient and the neutron-to-proton ratio froze at a constant value, except for a small slow due to neutron decay (877.75 $\pm$ 0.34 s \cite{gonzalez2021improved}). When $k_BT~\sim$ 80~keV, all neutrons were bounded off in
light nuclei. Once primordial nucleosynthesis of light elements had taken place, the Universe was left primarily with $^1$H and $^4$He, plus trace amounts of deuterium, $^3$He, and $^7$Li. Heavier elements were produced in very small quantities and quickly decayed. 

During this BBN epoch,  the production of nuclides depended solely on the baryon-to- photon ratio. Thus, given that stellar evolution has only marginally altered the relative proportions of these elements, measurements of their primordial abundances allow for an accurate inference of the baryon-to-photon ratio in the early Universe. Excellent agreement with predicted values is found for deuterium, \( ^3\mathrm{He} \), and \( ^4\mathrm{He} \). However, a notable discrepancy remains for $^7\mathrm{Li}$, whose predicted abundance, $^7\mathrm{Li}/\mathrm{H} = (5.38 \pm 0.35) \times 10^{-10}$, exceeds the observed value, $(1.6 \pm 0.3) \times 10^{-10}$, by approximately a factor of three. This tension, known as the lithium problem, may point to unresolved astrophysical processes or the need for new physics BSM. 

The comparison between the calculated abundance of light elements derived from BBN and observational data indicates that ordinary matter comprises approximately 5\% of the critical density. Consequently, combining this result with the other many indications that there is a larger amount of mass in clusters of galaxies and galaxies, it can be inferred that DM should be of non-baryonic origin, which is also a remarkable concordance with the results from CMB anisotropies analysis.

\section{Dark matter candidates} \label{DMCandidatesSection}

In the quest for the nature of DM, little is known beyond what is observed from its gravitational interactions. Nevertheless, in order to account for all the gathered observational evidence supporting its existence, potential candidates for DM must share a set of fundamental properties:

\begin{itemize}
    \item The absence of its detection through standard astrophysical observations implies that DM must be electrically neutral, or at most extremely weakly charged, to be dissipationless.
    \item In order to reproduce the structures at the
scales observed in the Universe and the matter distribution within clusters like Bullet Cluster, DM is required to interact either only gravitationally or via weak-scale interactions with ordinary matter.
    \item From the observations of the large-scale structures, DM has to be cold (or at least warm), that is, non-relativistic at the onset of galaxy formation.
    \item Primordial nucleosynthesis and CMB data indicate that DM must be composed of non-baryonic material.

    \item DM should be stable or very long-lived on cosmological timescales.

    \item All the evidence of the DM come from gravitational origin, thus DM must be massive. 
\end{itemize}

Numerous theories have been proposed to identify particle candidates for DM, aiming for a hypothetical candidate that meets the requirements outlined above, aligns perfectly with experimental constraints, minimizes arbitrary parameter choices, offers testable predictions or even ideally solves other existing problems within the Standard Model of particle physics. 

\subsection{Dark matter candidates zoo}

The baryonic DM hypothesis was explored in the late 20th century. In this context, a very popular candidate consisted of normal baryonic matter in form of Massive Compact Halo Objects (MACHOs). MACHOs~\cite{paczynski1986gravitational}, which encompass non-luminous objects like brown dwarfs or black holes, contribute to the overall mass content of the galaxy but not to the non-baryonic DM. Searches for MACHOs within our galaxy at the end of the 20th century enabled the quantification of their contribution to the galactic mass, revealing it to be significantly lower than the necessary to explain the observed dynamics. 
\begin{figure}[b!]
\begin{center}
\includegraphics[width=0.9\textwidth]{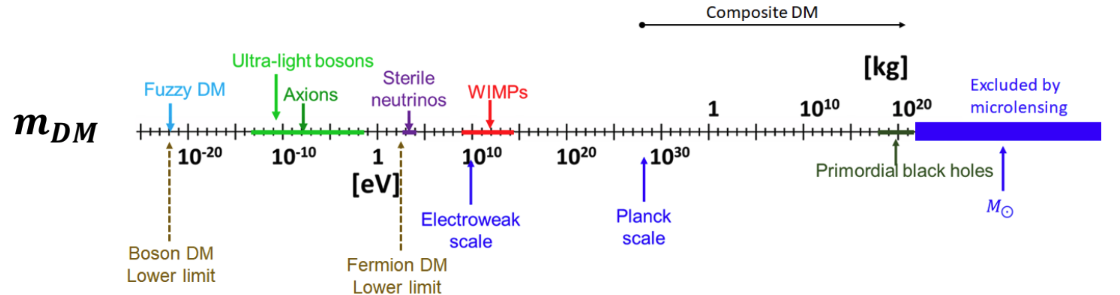}
\caption{\label{massscale} Graphical representation of various DM candidates according to their typical mass, shown on a logarithmic scale. Particle DM is found in the middle mass range, while at very low masses candidates can be considered wave-like, and near or beyond the Planck scale, composite or macroscopic DM candidates emerge. }
\end{center}
\end{figure}

Standard model neutrinos were one of the first particles considered as DM candidates even before that neutrino oscillation experiments confirmed that neutrinos possess mass. However, the mass of the neutrinos has not yet been measured. The most stringent upper limit on the electron neutrino mass is 0.8 eV at 90\% C.L., as reported by the KATRIN collaboration \cite{onillon2023neutrino}. In addition, as alredy mentioned, current cosmological observations constrain the sum of the masses of the three neutrino eigenstates to be \(\lesssim\)~0.1~eV~\cite{particle2020review,karim2025desi}. Consequently, their low mass prevents SM neutrinos from accounting for the observed DM density in the Universe. Most importantly, neutrinos emerge from the early Universe with relativistic velocities. Hence, as hot DM, they are not able to reproduce the cosmic structures, showing the greatest difficulties in particular at small scales. 

Having ruled out the neutrinos as the exclusive candidate for DM, the search for candidates beyond the SM becomes imperative as no other particle within the SM fulfills the required properties. Figure \ref{massscale} illustrates the most relevant categories of viable DM candidates, covering a range of masses which spans more than 90 orders of magnitude. Their possible interaction cross sections also extend across many orders of magnitude. Axions and WIMPs stand out as the preferred  particle-like candidates, since they were not proposed ad-hoc but to solve problems in particle physics not related to DM \cite{particle2022review,freese2017status}. This work will focus on particle-like candidates, which also include sterile neutrinos as other potential candidates for DM. Nonetheless, there also exist other macroscopic or composite candidates that, despite certain limitations, could still represent a significant component of the solution to the DM problem, such as primordial black holes (PBHs).



\begin{itemize}
     
 \item{\textbf{Axions and ALPs.}} 
 
 The SM of particle physics, while remarkably successful, remains incomplete, as it fails to account for key phenomena such as DM, the matter-antimatter asymmetry, and the hierarchy problem. 

Another significant shortcoming lies in the absence of a natural mechanism accounting for the lack of charge-parity (CP) violation in strong interactions, a symmetry that is violated in weak interactions. Although quantum chromodynamics (QCD) permits CP violation via the \(\theta\) parameter, such a violation has never been experimentally observed. In particular, the non-observation of a neutron electric dipole moment imposes a stringent upper bound on this parameter, constraining it to \(\theta < 10^{-10}\) \cite{baker2006improved,pendlebury2015revised}. Since the SM predicts that this dimensionless parameter should take values $\sim$ 1, its extremely small observed value requires a fine-tuning, commonly referred to as the strong CP problem.

The solution proposed in 1977 by Peccei and Quinn (PQ) involves the introduction of a global U(1)\(_\text{PQ}\) symmetry, spontaneously broken at a high energy scale $f_A$ \cite{peccei1977cp,peccei1977constraints}. The pseudo-Nambu–Goldstone (pNG) boson resulting from this symmetry breaking, subsequently named the axion by Weinberg \cite{weinberg1978new} and Wilczek \cite{wilczek1978problem}, dynamically relaxes the \(\theta\) parameter toward zero, thereby offering a natural solution to the strong CP problem. The particular pNG boson that solves the strong CP problem via the PQ mechanism is called the QCD axion, while the broader class of such pNG bosons are referred to as axion-like particles (ALPs). From the detection perspective, the main difference between the two lies in the fact that, for QCD axions, both the mass and the interactions are determined by the symmetry-breaking scale $f_A$, leading to a fixed relation between coupling and mass. In contrast, no such relation exists for ALPs.


The first proposed axion model is the Peccei-Quinn-Weinberg-Wilczek axion~\cite{peccei1977cp,weinberg1978new,wilczek1978problem}. In this scenario, the symmetry breaking occurs close to the electroweak scale $v_{\text{EW}} \approx 250\,\mathrm{GeV} \approx f_A$. This made the model quickly excluded by a
number of experimental constraints \cite{bardeen1987constraints}. Observations therefore require that $f_A \gg v_{\text{EW}}$, which implies extremely small axion masses and couplings, as these properties scale inversely with the symmetry-breaking scale. This class of models is commonly referred to as invisible axions \cite{dine1981simple}. The two most well-known invisible axion models are the KSVZ model \cite{shifman1980can}, which introduces a new heavy, electrically neutral quark carrying PQ charge, and the DFSZ model \cite{dine1981simple,zhitnitsky1980possible}, where two Higgs doublets are present and the PQ charge is carried by SM quarks and leptons. In addition to these, there are more elaborate models that combine features of both.

Axions and ALPs are nearly collisionless, with feeble interactions with SM particles, neutral,
non-baryonic, and may be present in sufficient quantities
to account for the entirety or an arbitrary fraction of the observed DM density. These axions could have been produced in the early Universe  through both thermal and non-thermal mechanism, would play a role in the formation of cosmic structures and are expected to form halos around galaxies in general, and the Milky Way in particular. These axions are referred to as galactic axions. Consequently, their detection can be pursued using haloscopes, which are resonant cavities designed to detect axions from the galactic DM halo as they traverse the Earth by converting them into detectable signals within a strong magnetic field.


Should axions exist, they could have also significant implications in astrophysics, for instance in the stellar evolution. Axions could be produced within stars and, due to their weak interaction with baryonic matter, they might serve as an efficient mechanism for stellar cooling. Some of the most stringent experimental constraints on axion existence have been derived from the study of white dwarfs \cite{gill2011constraining}.

    At present, several experiments are actively searching for both axions and ALPs, mainly exploiting the Primakoff effect, which is the axion-to-photon conversion in the presence of a very strong electromagnetic field. Among the different experimental
strategies, it is worth highlighting haloscopes, which aim to detect axions in the galactic halo, such as the Axion DM Experiment (ADMX) \cite{ADMX:2018gho,ADMX:2019uok}. Another category of experiments, referred to as helioscopes, focuses on detecting axions originating from the Sun. Notable examples are the CERN Axion Solar Telescope (CAST) \cite{CAST:2004gzq,CAST:2017uph} or the next-generation International Axion Observatory (IAXO) \cite{IAXO:2012eqj,IAXO:2019mpb}.  Additionally, the search for ALPs includes experiments employing the light shining through a wall technique, with the ALPS (Any Light Particle Search) collaboration \cite{isleif2022any} at DESY (German Electron Synchrotron), as an example.
Nevertheless, beyond experiments specifically designed for axion searches, WIMP detection experiments could also be sensitive to this hypothetical particle. This thesis will present a search for solar axions using data from the ANAIS-112 experiment (see Chapter \ref{Chapter:annual}).

In addition to axions and ALPs, ultralight bosons also include dark, or hidden, photons \cite{essig2013working,fabbrichesi2021physics}. Given the increasingly constrained DM parameter space, the scientific community has shifted its interest toward alternative candidates, such as those arising from the hidden sector, hidden because they are not charged under the SM gauge groups. The hidden sector would be parallel to the ordinary one, with interactions between the two mediated by a portal.

Such dark sector can contain few or many states, like fermions
or scalars or both, depending on the model. The most motivated candidate within this sector is the dark photon \cite{battaglieri2017us}. Typically featuring a sub-GeV mass range, the dark photon can act either as a mediator within the dark sector or as a DM candidate itself depending on the model. It becomes experimentally accessible through kinetic mixing with the ordinary photon, which serves as the portal between the dark and visible sectors.

This makes the dark photon an attractive target for direct detection experiments, such as DAMIC-M, which are sensitive to hidden-sector DM-electron interactions and optimized for low-mass searches \cite{aggarwal2025probing}. It could also be explored in ANAIS+, the next phase of the ANAIS-112 experiment, which aims to lower the energy threshold by using SiPMs at cryogenic temperatures, thus reducing noise and enhancing light yield~(LY) (see Section~\ref{experimentstestingDAMA}).\\

 \item \textbf{Sterile neutrinos.} 
 
SM neutrinos, due to their extremely small masses, cannot account for cold DM. However, a hypothetical fourth type of neutrino, called sterile because it lacks electroweak interactions with SM particles, could serve as a candidate for warm or cold DM. 

Sterile neutrinos, whose mass lies in the range of a few keV to several MeV, are neutral leptons
with no ordinary weak interactions except those induced by mixing with
active neutrinos, but could have interactions involving new physics. The discovery of neutrino oscillations showed that at least two of the three neutrinos must be massive, contradicting the SM prediction of massless neutrinos \cite{glashow1961partial}. This motivated the introduction of sterile neutrinos, which arise naturally in the most widely studied mechanism for generating neutrino masses: the seesaw mechanism. Depending on the model, the properties of these sterile neutrinos make them compelling and viable DM candidates, provided their lifetime is sufficiently long (comparable to the age of the Universe) and their relic density is large enough \cite{dodelson1994sterile}.\\

 \item \textbf{PBHs.} 
 
 PBHs would have formed by the primordial density perturbations in the radiation- dominated era of the early Universe \cite{Carr:2016drx}. Unlike astrophysical black holes, which are considered baryonic DM, PBHs should therefore be classified as non-baryonic and behave like any other form of CDM, since they were assembled before the BBN and the CMB.
PBHs continuously lose mass because of their evaporation due to Hawking radiation. For masses larger than 5~$\times10^{11}$ kg, PBHs are stable at the scale of the age of the Universe. There are several constraints on their existence \cite{Carr:2016drx}, yet they remain as potential candidates for DM.\\
\end{itemize}

\subsection{Weakly Interacting Massive Particles (WIMPs)}

\begin{figure}[b!]
\begin{center}
\includegraphics[width=0.6\textwidth]{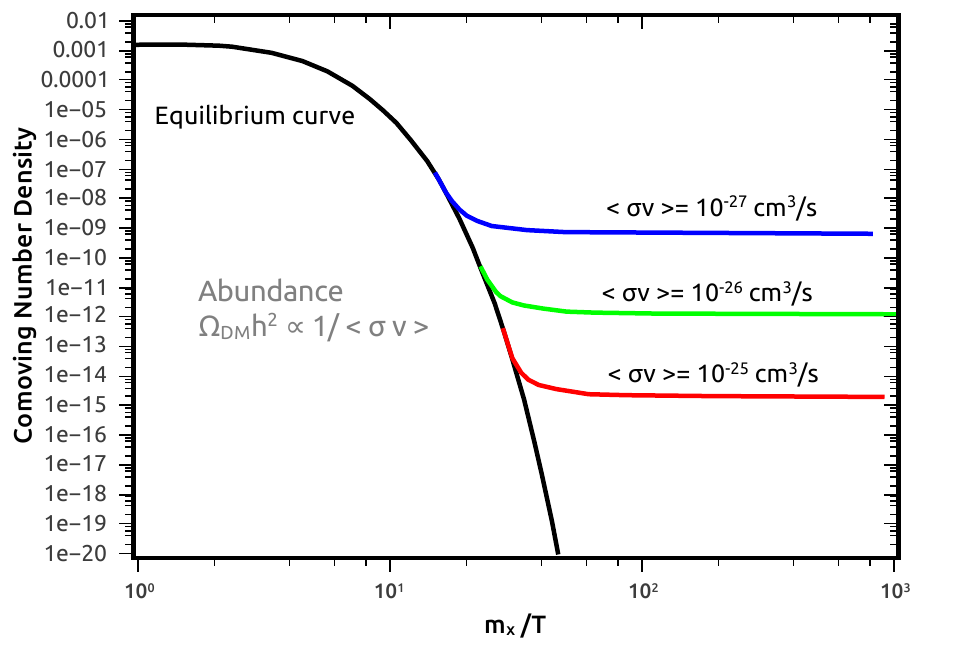}
\caption{\label{freezeout} Evolution of the number density of thermal relics with time. At early times, the relic particle is in chemical equilibrium with the primordial plasma (black).
Its abundance is settled by rates of particle creation and annihilation. At later times, the
expanding Universe cools and annihilation dominates the evolution of the density until it becomes freeze-out because the expansion rate prevents the particles from effective annihilation. Colored curves indicate
the relic abundance for different values of the velocity averaged annihilation cross-section, <$\sigma v $ >. To reproduce the current DM abundance, <$\sigma v $> is of the order of the
weak interaction scale, $\sim$~3~x~10$^{-26}$~cm$^3$/s~\cite{arcadi2018waning}.}
\end{center}
\end{figure}

Weakly Interacting Massive Particles (WIMPs) have been for decades the most attractive candidate to explain DM \cite{Steigman:1984ac}. They are defined to be particles that would interact with SM particles via weak-scale interactions, with cross sections comparable to those of weak interactions, and have masses in the range between 10 GeV and 1~TeV. WIMPs with these characteristics would have been produced thermally and their relic cosmological abundance set by the equilibrium between thermal production and annihilation, as it is shown in Figure~\ref{freezeout}. 

Consider a WIMP $\chi$ with mass m$_\chi$. In the early Universe, such a particle is in chemical equilibrium at
temperatures T $\geqslant$ m$_\chi$ with the primordial hot plasma. Equilibrium is maintained by annihilation processes to SM particles and antiparticles. However, as the Universe expands and cools below the DM mass m$_\chi$, the particle-production is inhibited, but the annihilation process is still efficient. Eventually, the averaged annihilation rate becomes slower than the Hubble expansion rate. Then, these particles fall out from the chemical equilibrium. If they are stable, they leave as an imprint of their existence a fixed density, becoming a thermal relic. This chemical decoupling mechanism is usually called freeze-out. 

The final WIMP relic abundance is inversely proportional to the annihilation cross section velocity averaged. Most interestingly, a weak-scale annihilation cross section, <$\sigma v $>~$\sim$~3~x~10$^{-26}$~cm$^3$/s, naturally gives a thermally produced WIMP relic abundance that matches shockingly the observed DM density. This striking cosmological coincidence is known as the “WIMP miracle”, and it is the reason why WIMPs are by far the most searched for non-baryonic DM candidates. In addition, if DM is mainly composed by WIMPs, interactions with SM particles were required for their production in the early Universe. Consequently, WIMPs offer the advantage that strategies for their direct and indirect detection, as well as for their production, can be developed, since they are expected to interact, though weakly, with SM particles.

There is a plethora of theoretical models, but the most motivated candidates for WIMPs come from supersymmetric (SUSY) theories \cite{nilles1984supersymmetry}. These extensions of the SM solve, among other things, the hierarchy problem. This problem is related to the enormous difference between the electroweak scale and the Planck scale, which appears in radiative corrections to the mass of the Higgs boson. SUSY introduces a symmetry such as for every particle of the SM, there exists a supersymmetric  partner, with the same mass and quantum numbers, but differing in the spin by one-half. However, these new particles have not been found, so this symmetry must be broken at some scale. When breaking that symmetry, the superpartners acquire masses greater than the SM ones and related with the breaking scale. SUSY provides neutral,
massive and stable candidates which can play the role of DM, such as the lightest neutralino. However, up to date no hint on supersymmetric particles has been found in accelerators.

\subsection{Alternatives to dark matter}

Given that all observational evidence supporting the existence of DM arises from its gravitational interactions, several alternative theories have been proposed over the years to account for these observations \cite{bertone2018history}. Modified Newtonian Dynamics
(MOND) \cite{milgrom1983modification} and modified gravity theories, which correspond to modifications of GR, are amongst the most well-known. However, so far none of them has proven capable of explaining all the different mismatches in observations that support the DM hypothesis. The reason is that evidence comes from so many different observables, spatial and temporal scales of the Universe evolution that, even if these theories can explain individual phenomena, it is very difficult to address all of the evidence
simultaneously without invoking the existence of DM.

\section{Dark matter detection} \label{DMDetectionSection}
The strategies followed to detect this elusive cosmological component vary significantly depending on the properties of the candidates, being very different for PBHs, for instance, than for
particle candidates, but also for axions with respect to WIMPs \cite{irastorza2018new}. In this section, the methods to detect WIMPs will be reviewed. 

\begin{figure}[b!]
\begin{center}
\hspace{1cm}\includegraphics[width=0.45\textwidth]{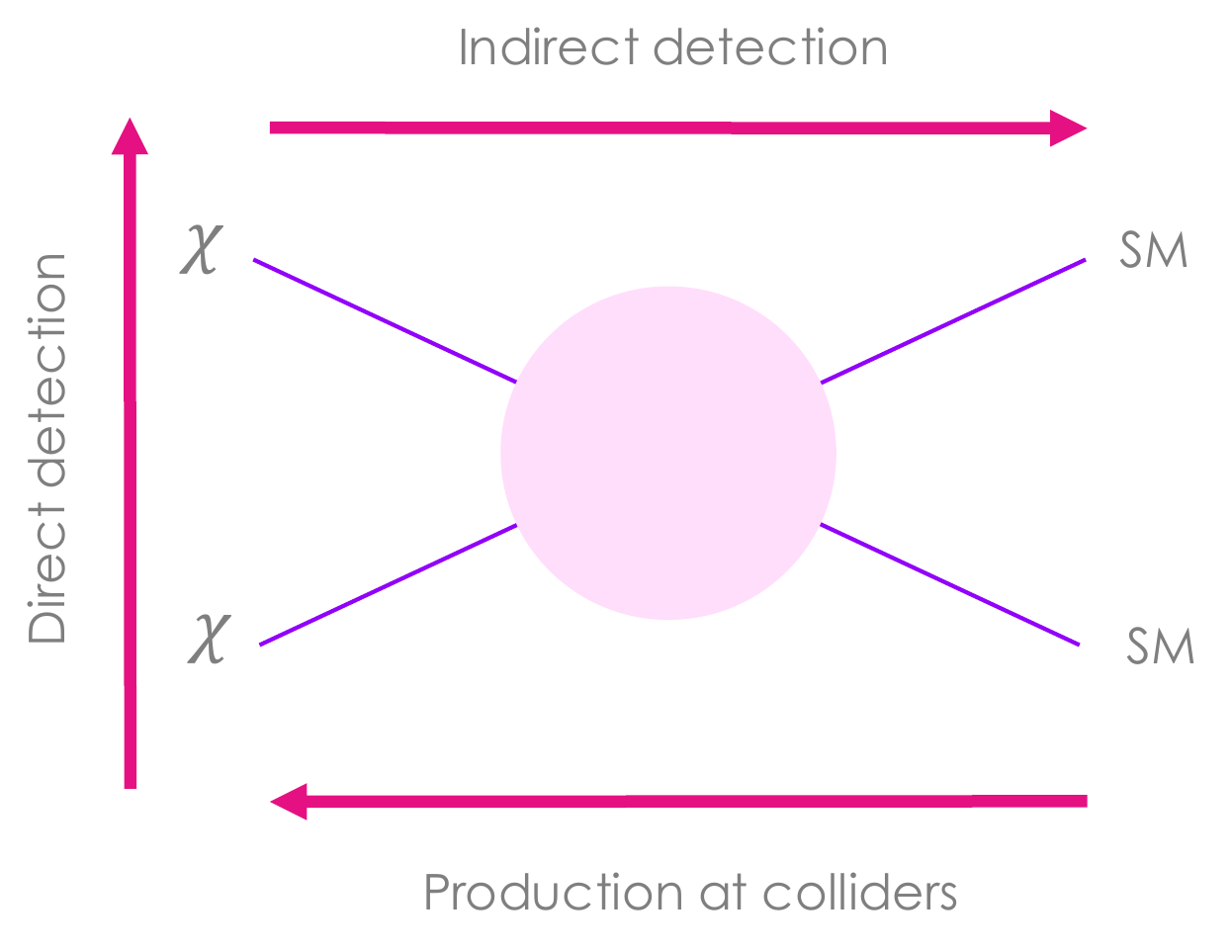}
\caption{\label{stratetegies} Diagram of the three different DM detection channels. Production at colliders are
represented from right to left; direct searches from bottom to top, and indirect searches
from left to right.} 
\end{center}
\end{figure}

A scheme of the different WIMP search strategies, assuming some coupling between WIMP and SM particles, is displayed in Figure~\ref{stratetegies}. Experimental
searches for WIMP DM can be broken down into three categories based on the different
ways WIMPs can interact with SM particles: direct detection, indirect detection and production at colliders. It is worth highlighting the complementarity among them, as results from all three should be combined to achieve a comprehensive understanding, although such integration is subject to strong model dependencies.

\subsection{Dark matter searches at colliders}

Many compelling theories beyond the SM predict that DM is light
enough to be produced in colliders such as the Large Hadron Collider (LHC) \cite{Evans:2008zzb}. Given its feeble interactions with ordinary matter, the produced DM would not be detected. However, it would still carry away energy and momentum from collisions. Collider searches
probe DM models by searching for missing energy and momentum from collisions exploiting the so-called mono-X or mediator searches \cite{Boveia:2018yeb,Buchmueller:2014yoa}. Similarly, collider experiments play a central role in probing new particles and new physics, providing valuable guidance in identifying the theoretical models that should be explored in the search for viable DM candidates.

ATLAS \cite{ATLAS:2008xda} and CMS \cite{CMS:2008euu} collaborations at LHC have been searching for DM in proton-proton collisions. Since no significant deviations from
the predictions of the SM have been observed, constraints
on the coupling of DM mediators or limits on the masses of the
mediator and DM particles have been settled. On the other hand, fixed target experiments such as the NA64 experiment
at CERN \cite{andreev2025searching} is very sensitive to sub-GeV
dark-sector particles that interact with the SM ones via light mediators. 

Collider searches are particularly competitive for low DM particles masses, although the validity of this comparison relies strongly on the selection of couplings between DM and SM particles.

\subsection{Indirect dark matter detection}
Indirect detection methods rely on the self-annihilation or decay of WIMP particles in DM dominated astrophysical objects, such as the galactic center, galactic halos, and subhalos. These processes lead to intermediate states (leptons, quarks, gluons, gauge bosons..), which then decay or hadronize into stable particles. The observation of an excess in the fluxes of secondary particles with respect to that expected by
the known sources of background can point to the the presence of DM \cite{Gaskins:2016cha}. 

In the field of indirect detection, increasingly stringent limits are being set, and for certain annihilation channels and WIMP mass ranges, current sensitivities have already reached the thermal relic cross section. However, in this DM detection channel, setting meaningful constraints depends not only on the particle properties of the DM candidate, but also on the assumed spatial distribution of DM, which determines the expected fluxes, and is subject to significant uncertainties. Moreover, astrophysical backgrounds are not fully
understood and difficult to model.

\begin{itemize}
    \item  \textbf{$\gamma$-ray searches}

Cosmic $\gamma$-rays are the most energetic photons of the Universe. They travel from the source to the detector without suffering deflection from the magnetic fields, and pointing back directly to their source, but can be affected by absorption in the interstellar medium. WIMPs with typical masses in the GeV to TeV range would eventually produce photons with energies in the $\gamma$-ray range. These $\gamma$-rays from DM annihilation may exhibit specific energy distributions, with peaks at distinctive energies that would allow to reject astrophysical backgrounds \cite{Bergstrom:1997fj}. In the last years, many observations in the $\gamma$-ray window have been conducted in different energy ranges, with both instruments in space and on ground. 

$\gamma$-rays can be observed directly from space with satellites.  Fermi-LAT \cite{Fermi-LAT:2009ihh} is a $\gamma$-ray observatory that has been
operating since 2008. It is sensitive to energies between 20~MeV and 300 GeV. Fermi-LAT detected an
excess from the galactic center around a few GeV \cite{Goodenough:2009gk}. In addition to a plausible
DM annihilation signal, this excess could be produced by a
large number of sources (e.g.~pulsars \cite{Gordon:2013vta}) with such a small angular distance
to the Galactic Center that they could not be resolved by Fermi-LAT. Today, the ultimate nature of this excess remains still unclear. 

There has been another excess observed in the 3.5 keV X-ray line, discovered in 2014 in galaxy clusters and the Andromeda galaxy, which has sparked speculation about its origin \cite{bhargava2020xmm}. One hypothesis suggests it could be a signature of decaying DM, specifically from sterile neutrinos, potentially explaining small-scale structures in the Local Group. While some studies have detected this line in additional objects, others have found no evidence, and conventional astrophysical explanations have been proposed. Future experiments may test the sterile neutrino hypothesis.

Higher-energy $\gamma$ rays (> 100 GeV) must be detected indirectly with ground-based telescopes via the Cherenkov light emitted by the showers of secondary particles produced by the interaction of very high energetic $\gamma$ rays hitting the upper part of the atmosphere.  Cherenkov telescopes often target satellite galaxies of the Milky Way because of their high DM content and low astrophysical backgrounds. HESS \cite{HESS:2011zpk}, MAGIC \cite{MAGIC:2009tyk} and VERITAS \cite{VERITAS:2017tif} are Cherenkov telescopes sensitive
to high-energy $\gamma$ rays. The future Cherenkov Telescope Array
(CTA) \cite{CTAConsortium:2012fwj} is an upcoming next-generation Cherenkov observatory that could significantly improve the sensitivity in this search. \\

\item \textbf{Neutrino searches}

Neutrino production is an attractive signature
for DM searches. They are also neutrally charged, and they can travel to Earth pointing back directly to their source, just as $\gamma$-rays. Nevertheless, unlike the latter, they do not lose energy or are absorbed through their propagation over cosmic distances. One of the most typical DM searches from neutrino telescopes is to search for neutrinos coming from a heavy celestial object like the Sun, where DM could be gravitationally bound and then, annihilate. 

An intrinsic limitation of experiments searching for neutrinos is their extremely low interaction cross section with ordinary
matter. Thus, their detection is a challenge, requiring large exposure times and enormous amounts of target material, typically ice or water. Examples of neutrino telescopes include ANTARES \cite{ANTARES:2016xuh}, located
in the Mediterranean Sea,  or IceCube \cite{IceCube-Gen2:2020qha}, under the ice
of the South Pole. In the future, larger neutrino telescopes such as KM3Net \cite{KM3NeT:2024xca}, which is already producing results with its first installed modules, or Hyper-Kamiokande \cite{Hyper-Kamiokande:2016srs} will improve the sensitivity of the current telescopes and therefore, either provide a robust detection or stronger constraints on the WIMP properties.\\

\item \textbf{Cosmic-ray searches}

Hadrons and leptons produced by WIMP annihilation or decay can arrive to Earth, but their fluxes are expected to be isotropic due to the strong deflections caused by interstellar magnetic fields during propagation. This deflection, together with interactions with the interstellar medium that cause energy losses, makes tracing these charged particles back to their sources very challenging. Moreover, there is still many uncertainties on sources of cosmic rays (CRs) and propagation, and numerous astrophysical processes also generate CR fluxes that can mimic potential DM signals. A CR flux of antiparticles could be more easily discriminated, because antimatter fluxes from astrophysical sources are relatively
rare \cite{Ibarra:2008qg}. For this reason, DM is primarily searched for in antimatter products, although these searches are strongly dependent on modelling of backgrounds
to extract information from DM annihilation. 


An excess of a positron flux at GeV energies was first detected by PAMELA \cite{PAMELA:2017bna}, and after confirmed by Fermi-LAT \cite{Fermi-LAT:2009ihh} and AMS \cite{AMS:2019rhg}. The origin of this excess is still a matter of debate within the community \cite{cuoco2017novel,luque2021combined}, as long as it could favour a DM origin or other astrophysical sources, such as emission from pulsars or supernova remnants. A better modelling of those backgrounds is required to improve the sensitivity of these searches.

\end{itemize}

\subsection{Direct dark matter detection}

Direct DM detection (DDD) experiments aim to detect DM particles of the Milky Way dark halo through their interactions with
ordinary matter in convenient detection systems. Particular characteristics of the generated signal,
or the lack thereof, can then be used to infer properties of DM when compared with the expected
radioactive background of the detector. Therefore, in order to unequivocally
identify such interactions, a good understanding of the signal signatures and backgrounds, as well as a thorough understanding of the detection technique to be employed, is required.

Although this thesis focuses on WIMP scattering off nuclei, it is worth highlighting that DM interactions with electrons are also theoretically well motivated. As already mentioned, sub-GeV DM candidates, such as those from the hidden sector, are receiving increasing attention from the scientific community, as their parameter space remains relatively unexplored. Current direct detection experiments are not sensitive to such particles via elastic nuclear scattering, since the resulting nuclear recoil energies would fall below detection thresholds due to the low DM mass.

To overcome this limitation, alternative interaction channels must be considered. One such possibility is inelastic scattering off bound electrons, which can result in detectable energy transfers down to DM masses on the order of MeV. In this context, semiconductors are particularly promising targets thanks to their 1 eV band gaps, allowing for the detection of small ionization signals. Additionally, their lower detection thresholds provide enhanced sensitivity across a wider range of DM masses compared to materials with higher thresholds. Nevertheless, as previously stated, this work will focus on characterizing the WIMP interaction with target nuclei.

\subsubsection{Expected dark matter signal}

The most likely interaction channel in the majority of WIMP models is the elastic scattering between the
WIMP and the nucleus within a detector, producing nuclear recoils~(NRs) of the target \cite{Lewin:1995rx,Baudis:2012ig,Schumann:2019eaa}. An inelastic scattering is also possible, but less probable, and requires
a minimum energy of the particle to excite the nucleus, being strongly dependent on the nuclear structure of the target nucleus.

The expected differential interaction rate of WIMPs in a target consisting of one type of nucleus of mass m$_N$ (being the total detector mass M$_{det}$) can be expressed as:

\begin{equation}
    \frac{dR}{dE_{\textnormal{NR}}} = \frac{M_{\text{det}} \rho_\chi}{m_N m_\chi} \int_{v_{\text{min}}}^{v_{max}} \frac{d\sigma_{WN}}{dE_{\textnormal{NR}}} \, v f_{det}(\vec{v}) \, d\vec{v},
    \label{rateTotalEq}
\end{equation}

where E\textsubscript{NR} is the energy of the recoiling nucleus and $v$ the WIMP velocity in the detector's  frame. The interaction rate depends on WIMP model parameters, such as the WIMP mass, $m_\chi$, and the differential cross section for WIMP–nucleus scattering,  \( \frac{d\sigma_{WN}}{dE_{\textnormal{NR}}} \), both of which remain unknown and will be discussed in detail in Section~\ref{sigmasection}. It also relies on astrophysical parameters related to the assumed halo model, such as the WIMP velocity distribution in the halo in the detector's reference frame, $f_{det}(\vec{v})$, and the local WIMP density, \( \rho_\chi \), for which estimates exist, though they are affected by significant uncertainties (see Section~\ref{fvsection}). In Equation \ref{rateTotalEq}, $v_{\text{max}}$ represents the maximum WIMP velocity considered in the integral, related to the escape velocity in the galactic rest frame, above which WIMPs are no longer gravitationally bound to the Milky Way.

WIMPs gravitationally bound to the Milky Way are expected to move at non-relativistic velocities on the order of hundreds of km/s. In the case of elastic scattering, the nuclear recoil energy \( E_{\textnormal{NR}} \) depends on the center of mass scattering angle $\theta^*$ as:

\begin{equation}
    E_{\textnormal{NR}} = \frac{\mu_{\chi N}^2}{m_N}v^2 (1-cos(\theta^*),
\end{equation}

\vspace{0.4cm}
being \( \mu_{\chi N} = \frac{m_\chi m_N}{m_\chi + m_N} \) the WIMP–nucleus reduced mass.

A simple kinematic argument shows that the maximum energy deposition occurs when the scattering angle in the center-of-mass frame is equal to 180$^\circ$:

\begin{equation}
    E_{\textnormal{NR}}^{max} = \frac{2\mu_{\chi N}^2}{m_N}v^2 .
\end{equation}

\begin{figure}[t!]
\begin{center}
\includegraphics[width=0.49\textwidth]{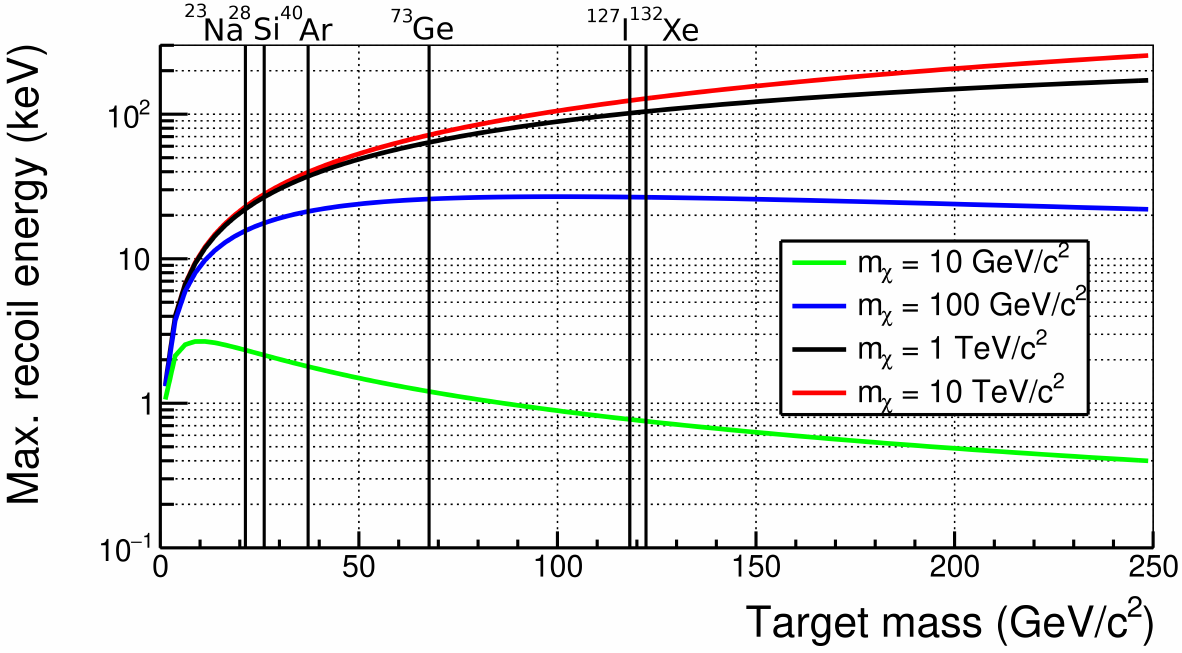}
\includegraphics[width=0.49\textwidth]{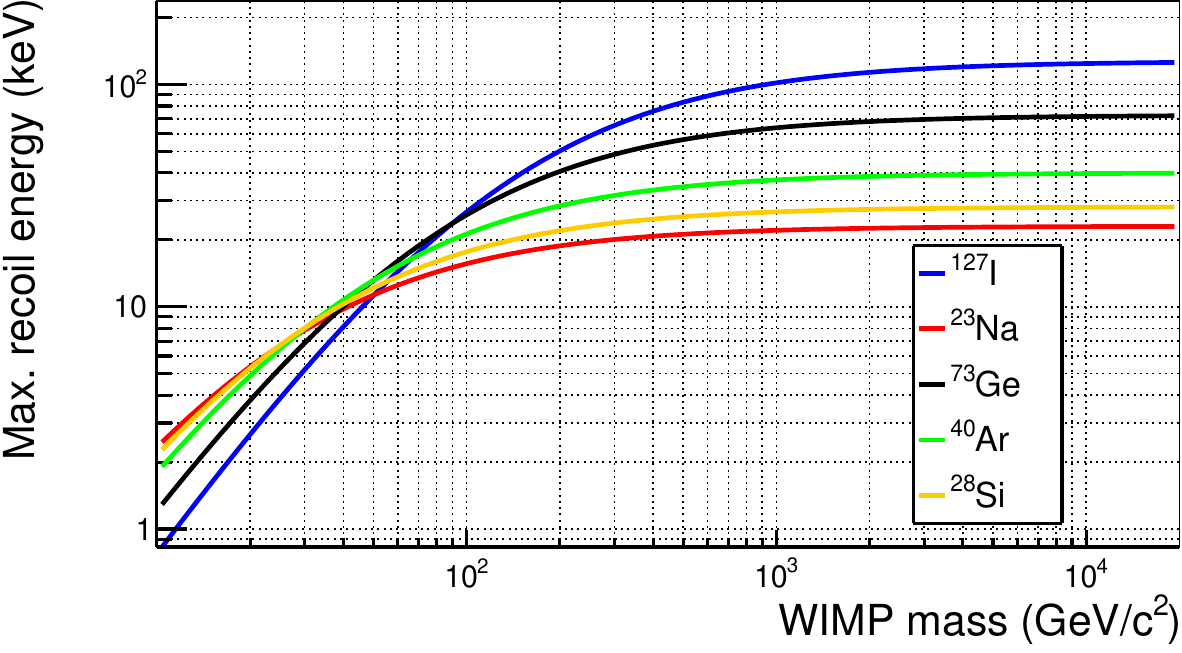}

\caption{\label{maximumEnergyDeposited} Maximum energy deposited by a WIMP through an elastic scattering
for a WIMP velocity of 220 km/s for some typical
target nuclei used in direct DM search experiments. \textbf{Left panel:} As a function of the target mass for different
WIMP masses. \textbf{Right panel:} As a function of the WIMP mass for different targets. } 

\end{center}
\end{figure}

The left panel of Figure \ref{maximumEnergyDeposited} shows the maximum recoil energy deposited by a WIMP at a velocity of 220 km/s through elastic scattering as a function of the target mass. These energies are further depicted in the right panel as a function of WIMP mass for different target nuclei. These figures indicate that the
elastic scattering of WIMPs with masses in the range of 10–100~GeV would produce typically 
nuclear recoil energies below $\sim$ 100 keV. In addition, it is inferred that the maximum recoil energy for low WIMP masses is
higher for light nuclei than for massive nuclei, and viceversa.

As it will be explained later, an important requirement for DDD experiments is the energy threshold, or the minimum energy deposition that the
detector is able to measure. Energy thresholds around 10 keV in nuclear recoil energy are common in DDD experiments, although they strongly depend on the specific detection technique and target material used. The minimum WIMP velocity needed to induce a nuclear recoil above the experimental energy threshold, $E_{th}$ is given by: \\

\begin{equation}
    v_{min} = \sqrt{\frac{E_{th} m_N}{2 \mu_{\chi N}}}.
\end{equation}
    \vspace{0.5cm}


The WIMP-nucleus cross-section $\sigma_{WN}$ and its mass $m_\chi$ are said to be the two observables of a DM experiment, parameters that are today loosely constrained. To interpret the results of DM experiments, additional assumptions must be made about the specific particle-physics model and the nuclear-physics processes involved.\\


\textbullet \quad  \textbf{WIMP-nucleus cross-section and and nuclear physics}\label{sigmasection}\\

The WIMP-nucleus differential cross section \( \frac{d\sigma_{WN}}{dE_{\textnormal{NR}}} \) is calculated based on the Lagrangian describing the interaction between a specific WIMP and ordinary matter, and encodes the input from particle physics. Additional inputs from hadron physics and nuclear physics are required in order to build the WIMP-nucleus cross-section from the WIMP-quark.

Direct detection experiments have traditionally relied on simplified assumptions regarding the type of WIMP–nucleon interactions, typically considering only two leading cases: the spin-independent (SI) interaction, in which the WIMP couples to the mass of the target nucleus via scalar couplings, and the spin-dependent (SD) interaction, where the coupling occurs with the spin of the nucleus.

However, a comprehensive Effective Field Theory (EFT) framework has been developed to describe WIMP–nucleus scattering in a more general way \cite{Fan:2010gt,fitzpatrick2013effective}. This approach systematically incorporates all leading- and next-to-leading-order operators permitted in the effective Lagrangian, allowing the WIMP–nucleus interaction to be parametrized in terms of 14 independent EFT operators. The inclusion of these operators introduces significant modifications, both in the spectral shape of the expected recoil energy and in the relative interaction rates across different detector targets. Therefore, a robust DM direct detection program employing diverse target materials becomes essential in order to disentangle the specific operator or combination of operators responsible for any potential signal.

Due to the large number of possible operators and their arbitrary combinations, a comprehensive treatment is computationally demanding and practically unfeasible. It is important to emphasize that combinations of operators cannot be ruled out, which significantly complicates the comparison of results across different experiments. Nevertheless, it is generally expected that the interaction is dominated by operators $\mathcal{O}_1$ (corresponding to SI interaction) and $\mathcal{O}_4$ (associated with the SD). Consequently, as commonly adopted in the DM search community, the WIMP-nucleus differential cross-section in the following analysis will be considered as a combination of these two dominant contributions,

\begin{equation}
    \frac{d\sigma_{WN}}{dE_{\textnormal{NR}}} = (\frac{d\sigma_{WN}}{dE_{\textnormal{NR}}})_{SI} + (\frac{d\sigma_{WN}}{dE_{\textnormal{NR}}})_{SD} = \frac{m_N}{2 \mu^2_{\chi N} v^2} \left[(\sigma^0_{WN})_{SI} F^2_{SI} (E_{\textnormal{NR}})+(\sigma^0_{WN})_{SD} F^2_{SD} (E_{\textnormal{NR}})\right].
    \label{eqsigma}
\end{equation}

Equation~\ref{eqsigma} relies on the assumption that the scattering cross section can be approximated as isotropic and independent of the recoil energy. Here, $(\sigma^0_{WN})_{SI}$ and $(\sigma^0_{WN})_{SD}$ are the SI and SD components of the total WIMP-nucleus cross sections in the limit of zero momentum transfer, and $F_{SI}$ and $F_{SD}$ are the corresponding nuclear form factors which encodes
the dependence on the
momentum transfer which is associated with finite-size effects, as the nuclei are not point-like.


For SI interactions, the WIMP–nucleus cross section in the limit of zero momentum transfer, $(\sigma^0_{WN})_{SI}$, can be expressed in a general form as:

\begin{equation}
    (\sigma^0_{WN})_{SI} = \left( Z+(A-Z)\frac{f_n}{f_p}\right)^2\frac{\mu^2_{\chi N}}{\mu^2_{\chi n}}(\sigma_{Wn})_{SI},
\end{equation}

where \( Z \) denotes the number of protons, \( (A - Z) \) the number of neutrons, \( f_p \) and \( f_n \) represent the effective WIMP couplings to protons and neutrons, respectively, $\mu_{\chi n}$ is the WIMP-nucleon reduced mass, \( \mu_{\chi n} = \frac{m_\chi m_n}{m_\chi + m_n} \), and $(\sigma_{Wn})_{SI}$ is the WIMP-nucleon SI
cross-section.

A common simplification applied to facilitate comparison among results from experiments employing different target nuclei is to assume \( f_p = f_n \) \cite{bertone2010particle}, thereby resulting:

\begin{equation}
    (\sigma^0_{WN})_{SI} =A^2 \frac{\mu^2_{\chi N}}{\mu^2_{\chi  n}} (\sigma_{Wn})_{SI}. 
    \label{sigmaSI}
\end{equation}

From an experimental perspective, a noteworthy feature of the SI cross section is its scaling with the square of the mass number of the target nucleus, $\propto A^2$. This enhancement arises because, in a SI interaction, DM
couples coherently to all the nucleons inside the nucleus, so its cross-section adds coherently for protons and neutrons. Thus, detectors using more massive nuclides are generally sensitive to smaller SI cross sections with the nucleon, which explains the popularity of liquid xenon-based DM searches.

On the other hand, the SD cross section depends on the average spin contributions of the protons
and neutrons in the target nucleus and can be written as: 

\begin{equation}
   (\sigma^0_{WN})_{SD} = \frac{\mu^2_{\chi N}}{\mu_{\chi n}^2} \frac{4}{3}\dfrac{J+1}{J} (\sigma^0_{Wn})_{SD}  \frac{1}{\bar a^2} \left[ a_p \langle
 S_p \rangle
 + a_n 
\langle S_n \rangle \right]^2 , 
\label{sdinteraction}
\end{equation}

where $(\sigma^0_{Wn})_{SD}$ is the SD cross section per nucleon, $J$ is the nuclear spin, $a_p$ and $a_n$ are the effective WIMP couplings to protons and neutrons, respectively, such that $\bar a^2 = a^2_p + a^2_n$, and $ \langle S_{p,n  }\rangle = \langle N | S_{p,n} | N \rangle$ are the expectation values of the spin content of the proton and neutron groups in the nucleus~(N).

Unlike the SI case, SD scattering does not benefit from the use of heavier target nuclide because its cross
sections depend only on the unpaired nucleons rather than the number of nucleons. Thus, it scales with the the spin of the nucleus and is only possible for J different from 0 nuclei. Depending on the material used, the detector will be mostly sensitive to DM-neutron interactions or to DM-proton interactions. In general, analyses focus on nuclei whose spin is dominated by either the proton or neutron component, allowing the contribution of the other component to be neglected. Detectors using $^{19}\mathrm{F}$, such as the PICO experiment \cite{PhysRevD.108.062003}, are particularly sensitive to proton interactions ($\langle S_{p}\rangle$ = 0.478 vs $\langle S_{n}\rangle$ = -0.002) \cite{klos2013large}. $^{23}\mathrm{Na}$ and $^{127}\mathrm{I}$, the target nuceli of ANAIS-112, are mostly sensitive to DM-proton interactions ($\langle S_{p}\rangle$ = 0.248 vs $\langle S_{n}\rangle$ = 0.020 for \textsuperscript{23}Na, and $\langle S_{p}\rangle$ = 0.264 vs $\langle~S_{n}~\rangle$~=~0.066 for \textsuperscript{127}I). Conversely, detectors based on xenon primarily probe neutron-coupled SD interactions. 

Accordingly, Equation \ref{eqsigma} can be written as:

\begin{equation}
    \frac{d\sigma_{WN}}{dE_{\textnormal{NR}}} = \dfrac{1}{E_{\textnormal{NR}}^{max}}  \frac{\mu^2_{\chi N}}{\mu^2_{\chi n}}\left[ A^2 (\sigma^0_{Wn})_{SI}F^2_{SI} (E_{\textnormal{NR}}) + \frac{4}{3} \frac{(J + 1)}{J} (\sigma^0_{Wn})_{SD}  \frac{1}{\bar a^2} \left[ a_p \langle
 S_p \rangle
 + a_n 
\langle S_n \rangle \right]^2F^2_{SD}(E_{\textnormal{NR}}) \right]
    \label{eqsigmadesarrollada}
\end{equation}

A complete analysis for targets with non-zero spin would require exploring a four- dimensional parameter space, $\sigma_{\text{SI}}$, $\sigma_{\text{SD},p}$, $\sigma_{\text{SD},n}$, and $m_\chi$, which is computationally demanding. As a result, exclusion limits are typically derived under the assumption that the interaction is dominated by a single operator at a time for simplicity. Experiments typically examine two-dimensional projections, ($\sigma_{\text{SI}}$ vs $m_\chi$), ($\sigma_{\text{SD},p}$ vs $m_\chi$), or ($\sigma_{\text{SD},n}$ vs $m_\chi$), while setting the remaining two couplings to zero. In this way, experiments provide constraints on $\sigma_{WN}$ for a given halo model, but meaningful comparisons between different results should be made using parameters independent of the target material, $\sigma_{Wn}$ for a given DM particle model. Nonetheless, any such comparison will inevitably be model-dependent.\\ 

\textbullet \quad  \textbf{Distribution of dark matter in the Milky Way}\label{fvsection}\\

The DM density in the Milky way at the position of the Earth and its velocity
distribution are astrophysical input parameters required for the estimation of the expected DM interaction rates and then, for the interpretation of the experimental results in terms of DM. 
Consequently, any uncertainty in these parameters will directly affect the measurement of the WIMP-induced scattering rate, and, ultimately, the constraints placed on the WIMP properties. 

With the aim of comparing their results, DDD  experiments typically assume a local WIMP density in the galactic halo of $\rho_\chi$~=~0.3~GeV/cm$^3$, which
results from mass modelling of the Milky Way, using parameters in agreement with
observational data~\cite{baxter2021recommended}. Although most recent measurements point to a local DM density of $\rho~=~0.4$~GeV/cm$^3$~\cite{de2019estimation,lim2025mapping}, it is recommended to continue using the traditional value of 0.3 GeV/cm$^3$ in order to ensure consistency and comparability with previously published limits. Since this parameter acts as a multiplicative factor in the total event rate, its impact on exclusion curves can be straightforwardly rescaled if required.


Regarding the WIMP velocity distribution in the galactic frame, $f(\vec{v_{gal}})$, a Maxwell- Boltzmann distribution truncated at
the Milky Way escape velocity, $v_{esc}$, is the simplest and most common choice used to describe the DM profile in the galactic reference system. This is the so-called
Standard Halo Model (SHM) \cite{Freese:1987wu}, in which DM particles are assumed to be distributed
according to a thermal distribution in the galaxy, and forming an isothermal
and spherical halo: 

\begin{equation}
    f(\vec{v_{gal}}) d^3\vec{v_{gal}}= \frac{1}{v_o^3 \pi^{\frac{3}{2}}} \exp({-\frac{\vec{v_{gal}}^2}{v_o^2})} d^3\vec{v_{gal}},
\end{equation}

where $v_0$ is the most probable speed value, related with the dispersion velocity of the DM particles in the
halo. This parameterization of the distribution function has the advantage of being fully defined with just two parameters, \(v_0\) and \(v_{\text{esc}}\), with standard values of 220~km/s \cite{mihalas1981galactic} and 544~km/s \cite{smith2007rave}, respectively.

However, the calculation of the interaction rate (Equation~\ref{rateTotalEq}) involves the velocity distribution of WIMPs in the detector's reference frame, $\vec{v}$. Since WIMPs are non-relativistic, their velocity distribution in the Earth's frame can be obtained by applying a Galilean transformation to the WIMP velocities in the Galactic frame using the Earth's velocity, $\vec{v}_E$, such that $\vec{v}_{\text{gal}} = \vec{v} + \vec{v}_E$. Due to the Earth's orbital motion around the Sun, $\vec{v_E}$ varies over the course of the year, making the DM flux on Earth a periodic function of time with a period of one year. This gives rise to the annual modulation signature (see Section \ref{AnnualModSec}). As a result, the differential rate of WIMP interactions depends not only on the WIMP velocity distribution but also on time,

\begin{equation}
    \frac{dR}{dE_{\textnormal{NR}}} = \frac{M_{\text{det}} \rho_\chi}{2 m_\chi \mu^2_{\chi N}} \left[ \sigma^0_{SI} F^2_{SI} (E_{\textnormal{NR}})+\sigma^0_{SD} F^2_{SD} (E_{\textnormal{NR}})\right]\int_{v_{\text{min}}}^{v_{max}} \dfrac{f_{det}(\vec{v},t)}{v} \, d\vec{v}.
    \label{rateTotalEqcontime}
\end{equation}

\begin{figure}[b!]
\begin{center}
\includegraphics[width=0.6\textwidth]{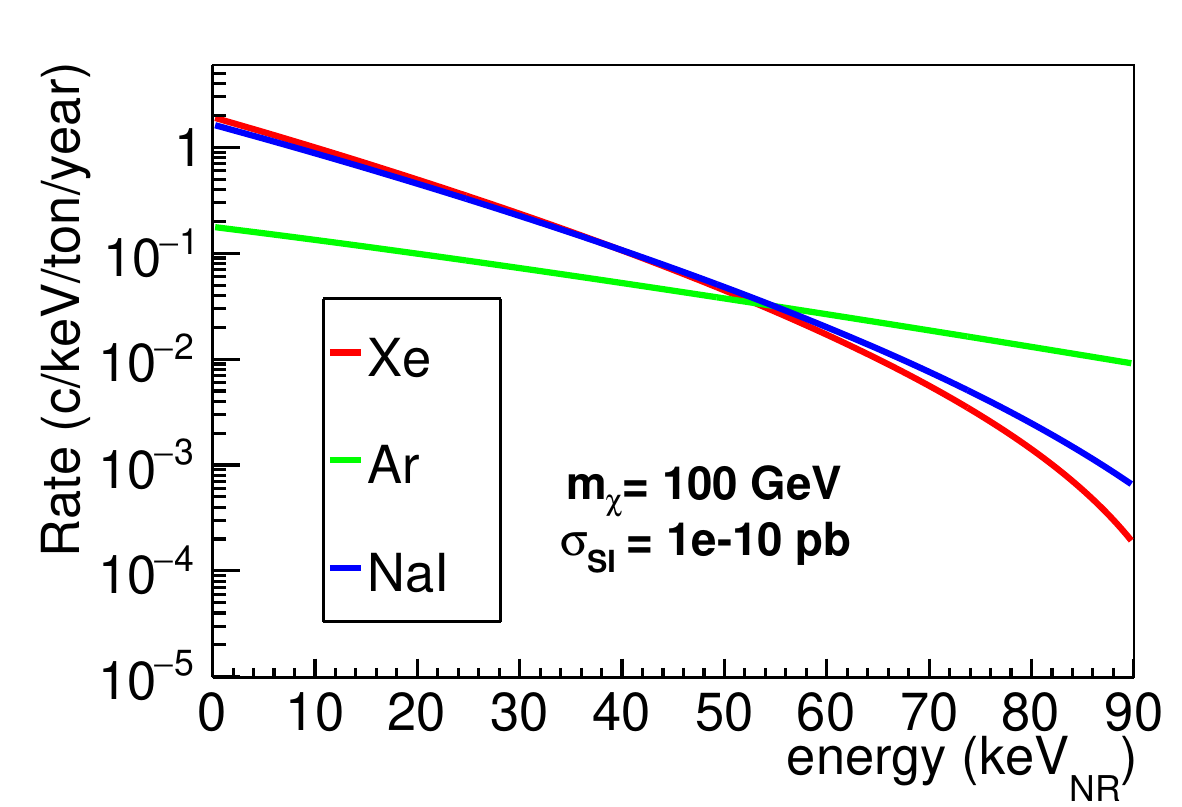}
\caption{\label{RateForTargets} Differential event rate as a function of the nuclear recoil energy induced by a $m_\chi$ = 100~GeV WIMP for
several target materials considering $\sigma_{\textnormal{SI}}$ = 10$^{-10}$ pb.} 

\end{center}
\end{figure}

\begin{figure}[t!]
\begin{center}
\includegraphics[width=0.49\textwidth]{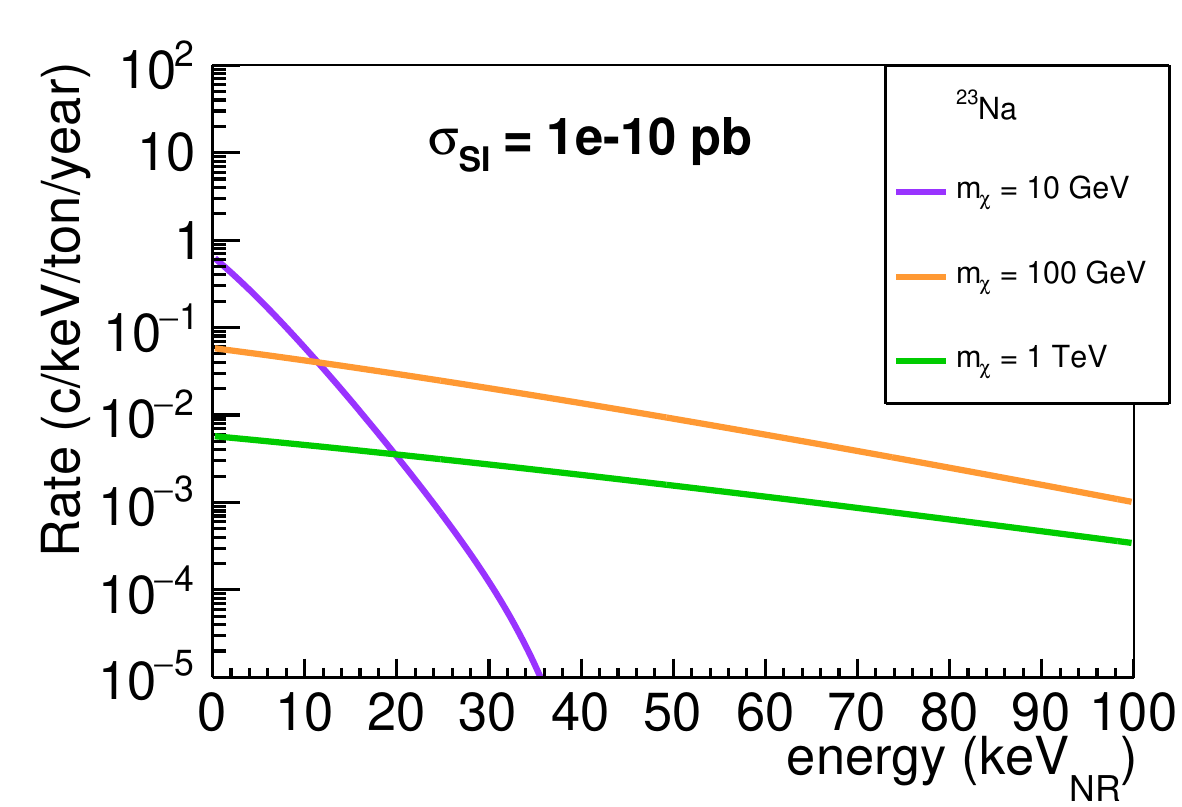}
\includegraphics[width=0.49\textwidth]{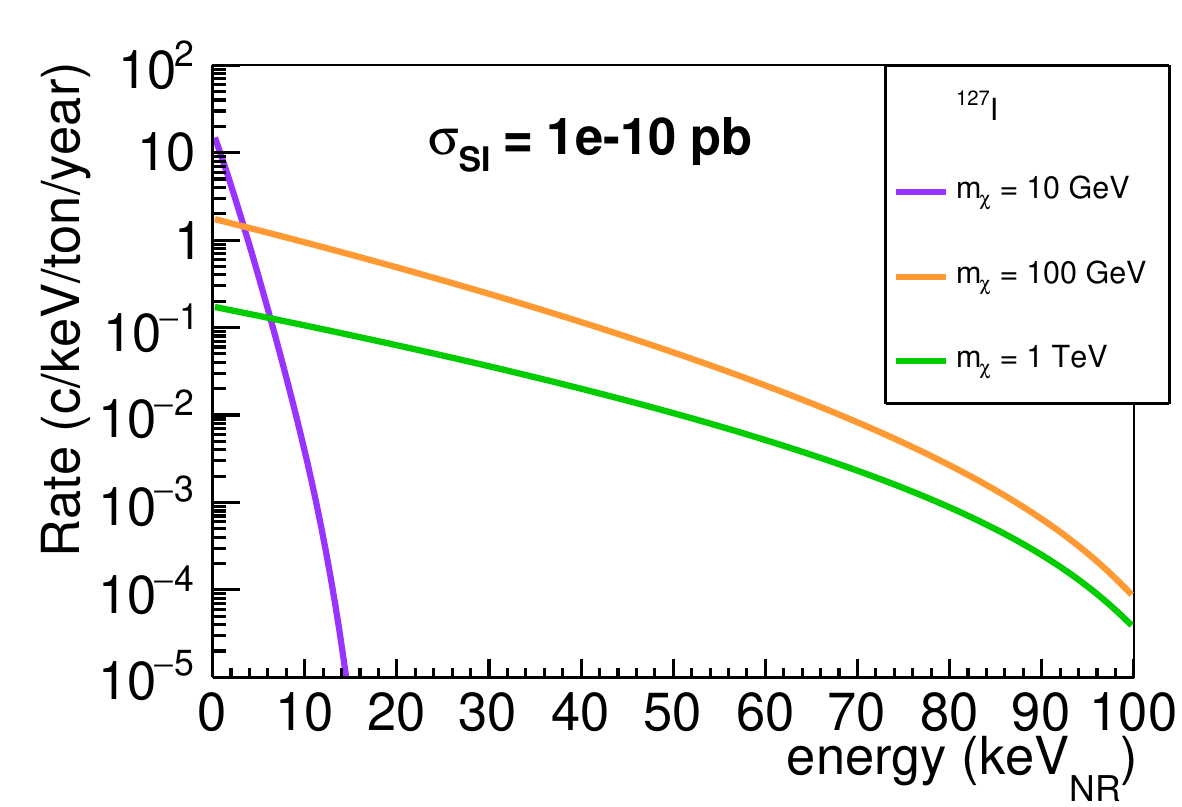}

\caption{\label{RateForTargetsNaI} Differential event rate as a function of the nuclear recoil energy induced by three different WIMP masses considering considering $\sigma_{\textnormal{SI}}$ = 10$^{-10}$ pb. \textbf{Left panel:} For $^{23}$Na as target nuclei. \textbf{Right panel:} For $^{127}$I as target nuclei.} 
\end{center}
\end{figure}

Figure~\ref{RateForTargets} shows the differential recoil rate as a function of the nuclear recoil energy for several typical target nuclei used in direct DM detection experiments considering SI interaction. For composite
materials, like NaI, the recoil rate is calculated by summing
the contributions from the individual nuclei scaled to their molecular fraction. The calculation is based on Equation~\ref{rateTotalEqcontime}, taking its average value, and assuming a WIMP mass of $m_\chi = 100~\text{GeV}$ and a SI WIMP–nucleon cross-section of $\sigma_{\textnormal{SI}} = 10^{-10}~\text{pb}$. It can be observed that the expected rate increases for heavy nuclei at lower energies. However, although the rate for SI interactions increases with \(A^2\), the rate for heavy nuclei (such as Xe) decreases at higher energies owing to the suppression imposed by the nuclear form factor.

The shape of the expected recoil spectrum also depends on the mass ratio between the WIMP and the target nucleus. Figure~\ref{RateForTargetsNaI} illustrates how the differential recoil rate varies with different WIMP masses for two typical target nuclei, $^{23}$Na and $^{127}$I, assuming a SI WIMP–nucleon cross-section of $10^{-10}$ pb. The spectra have been again computed using Equation~\ref{rateTotalEqcontime} taking its average value. As previously observed, the differential interaction rate exhibits an exponential decrease with increasing recoil energy. The figure suggests that different target materials can aid in exploring a broader WIMP mass range with greater sensitivity.

\subsubsection{Detector requirements}

Depending on the WIMP-nucleon interaction cross-section, the estimated rate ranges from 1~event/kg/day to 1 event/ton/year. This low expected rate, coupled with the fact that the spectral shape does not show distinctive features for the standard couplings (SI, SD), but an approximately continuous exponential shape (see Figures \ref{RateForTargets} and \ref{RateForTargetsNaI}), confirm that detecting DM is a highly challenging task. 

To maximize sensitivity and guarantee the discrimination of a potential DM signal against background sources, low-background techniques are applied. In particular, the reduction and
suppression of all the environmental radiation contributions to the background (from cosmic rays, natural and mand-made radioactive isotopes) is a key requirement for DDD experiments.\\

\textbullet \quad  \textbf{External backgrounds} 

Cosmic radiation is one of the main sources of background in rare event searches. Cosmic radiation reaching Earth atmosphere (primary cosmic rays) is predominantly composed of protons (about 90\%), with the remainder consisting of $\alpha$-particles, electrons, positrons, and a minor proportion of light nuclei like lithium, beryllium, and boron. 

High-energy collisions in the upper atmosphere produce cascades of particles that decay into muons, electrons, positrons, photons, and neutrinos. At sea level, the hadronic component (protons and neutrons) remains relevant but is easily suppressed by modest overburden. In contrast, muons, the most abundant and penetrating component, are harder to attenuate and contribute to detector backgrounds through various processes. 

On the one hand, muons can directly interact within the detector and the surrounding materials. In addition to their substantial energy deposition as they traverse matter, even when such events fall outside the region of interest (ROI) of the experiment, they can also induce nuclear reactions and generate fast neutrons, either promptly or with a delay. This represents the most
dangerous background for DM searches \cite{Wang:2001fq}, as these cosmogenic neutrons can lead to keV-level NRs through elastic scattering with the target nuclei in the detector \cite{Mei:2005gm}, in a similar way as WIMPs are expected to do, thereby posing an irreducible background. On the other hand, they can also produce secondary electrons, gammas
and spallation products in surrounding materials that may reach the detector.

For these reasons, it is essential to minimize the cosmic ray flux in the vicinity of the detectors by situating the experiment in a deep underground facility. Several underground laboratories have been established worldwide. Notable examples include LNGS in Italy under 3.800 meters of water equivalent (m.w.e.), the Yangyang Underground Laboratory (Y2L) in South Korea under 2.400 m.w.e., and the LSC in Spain, where the ANAIS-112 experiment is taking data under 2.450 m.w.e. 

The measured integrated muon flux in hall A of LSC is (5.26 $\pm$ 0.21)~$\times$~10$^{-3}$~m$^{-2}$s$^{-1}$, whereas in hall B, where ANAIS-112 is located, it is (4.29 $\pm$ 0.17)~$\times$~10$^{-3}$~m$^{-2}$s$^{-1}$~\cite{trzaska2019cosmic}. For comparison, the muon flux at LNGS is (3.35 $\pm$ 0.030)~$\times$~10$^{-4}$~m$^{-2}$s$^{-1}$ \cite{PhysRevD.100.062002}, which is one order of magnitude lower because it is located at a greater depth. The residual muon flux can be further reduced by using active veto detectors to identify and tag high-energy deposits from the original muon or its associated cascade as it crosses the experimental set-up.

Besides the cosmogenic neutrons, radiogenic neutrons may arise from ($\alpha$,n) reactions triggered in the rocks or building materials within the laboratory \cite{Mei:2008ir}, as well as from spontaneous fission in isotopes like $^{238}$U or $^{235}$U. The neutron flux measured by HENSA collaboration in the hall B of LSC (a neutron spectrometer installed in 2019 very near to the position of the ANAIS-12 experiment) is (20.90 $\pm$ 0.02)~x~10$^{-6}$~cm$^{-2}$s$^{-1}$ up to 20 MeV \cite{nil}, which has been found to be larger than what was previously measured in hall A. Once again, for comparison, the neutron flux at LNGS is measured to be (3.29~$\pm$~0.09)~x~10$^{-6}$~cm$^{-2}$s$^{-1}$ \cite{belli}. To shield against these neutrons and moderate them, a passive shielding based on materials with high hydrogen content, such as water or polyethylene, is typically employed to ultimately reduce its direct contribution. Moreover, most current experiments make use of active shielding systems, which enable a much more efficient rejection of neutrons through the identification of multiple scattering interactions.

Another source of background comes from natural
uranium and thorium decay chains, along with decays from common isotopes like $^{40}$K, $^{60}$Co, or $^{137}$Cs present in the surrounding materials. The integral gamma flux in the hall A of the LSC is 1.23~$\pm$~0.17~cm$^{-2}$s$^{-1}$~\cite{bettini2012canfranc}, being found higher in the hall B, with a  measured gamma flux of about $\mathcal{O}$(2 cm$^{-2}$s$^{-1}$). To reduce the contribution of this environmental background, passive shieldings made of high-Z materials with low radioactivity content, typically lead and copper, are employed to enclose completely the experimental space to house the detector. Such materials provide a good attentuation for gamma radiation.

Moreover, a dangerous background source for all rare event searches is the airborne radioactivity
coming from radon.$^{222}$Rn is part of the $^{238}$U decay chain and undergoes $\alpha$-decay to $^{218}$Po, with a half-life of 3.825 days. It is present not only in dwellings, but
also in underground facilities as it can be washed or diffused out from surrounding
rock and accumulates easily because of poor ventilation. $^{222}$Rn and its progeny emit a range of $\alpha$- and $\beta$-particles that can produce gamma radiation through bremsstrahlung or nuclear reactions, in addition to the photons intrinsically emitted throughout the decay chain. 

The annual $^{222}$Rn average activity concentration at hall B of LSC is 85.2~Bq~m$^{-3}$~\cite{amare2022long}. This value is notably influenced by seasonal variations and by the internal ventilation conditions of the laboratory. To establish a radon-free zone around the detectors, it is crucial to tightly isolate the inner part of the detector shielding from the laboratory air, keeping an inner atmosphere of clean nitrogen gas or radon-free air. In this context, the \(^{222}\text{Rn}\) content inside the ANAIS-112 shielding has been continuously flushed with radon-free nitrogen gas, achieving a best upper limit for the $^{222}$Rn content of 0.06 Bq m\(^{-3}\) (95\% C.L.), with its contribution being negligible in the ROI. 

The ultimate limitation on background is imposed by the neutrino flux \cite{arnaud2020first}, with the Sun or the atmosphere as the most relevant sources. Coherent elastic neutrino-nucleus scattering (CE$\nu$NS) in the detector target nuclei is an inherent, irreducible background for DDD experiments, sharing many characteristic features with the WIMP scattering. In 2024, the XENONnT collaboration and PANDAX-4T reported the first indication of $^8$B solar neutrino signal detection via CE$\nu$NS \cite{PandaX:2024muv,aprile2024first}. Although CE$\nu$NS represents a significant background, subject to substantial systematic uncertainties in DM searches, there are strong scientific motivations to study low-energy solar neutrinos as an interesting objective by themselves. Large-scale DDD experimental efforts could greatly contribute to advancing this research.

Sensitivity to the WIMP–nucleon cross section improves with increased target mass and reduced background. However, in presence of some background, any improvement in the explored WIMP-nucleus cross-section requires a significantly larger exposure compared to a background-free case. 
The neutrino background then implies a much reduced sensitivity to DM and because of this is typically called neutrino floor. To overcome this limitation  and probe below the so-called neutrino floor, characteristic WIMP signatures such as the directionality of the recoil are being explored. In particular, applying directional cuts, e.g., selecting events opposite to the Sun’s direction, can help suppress the solar neutrino background.\\

\textbullet \quad \textbf{Internal backgrounds}

Internal contamination, found in the materials of the detector, stands as the most significant background for the majority of experiments, as it is impossible to shield. The isotopes that typically contribute to this background are long-lived
natural radioisotopes, including those in the $^{232}$Th, $^{238}$U, $^{235}$U chains, and $^{40}$K, as well as cosmogenic activation isotopes like $^{3}$H and $^{39}$Ar, and anthropogenic isotopes such as $^{60}$Co, $^{85}$Kr, and $^{137}$Cs.

Working with the most radiopure materials is imperative to minimize internal radioactivity contamination and reach high sensitivities. For that purpose, meticulous material selection is employed while designing the detector and target materials, ensuring low radioactivity through various analytical methods. For instance, material screening is frequently conducted using high-purity germanium (HPGe) gamma spectrometry, a non-destructive technique capable of analyzing bulk samples to detect with high sensitivity gamma-emitting isotopes. In contrast, Inductively Coupled Plasma Mass Spectrometry (ICP-MS) requires the dissolution of the sample into a solution, making it a destructive technique. However, it allows for the detection of trace levels of specific isotopes with exceptional sensitivity, even in sub-milligram sample masses. Additional complementary techniques include Neutron Activation Analysis (NAA), which involves irradiating the sample with neutrons to induce radioactivity and subsequently measuring the resulting gamma emissions. The latter technique is not generic, but very sensitive to specific isotopes.

Regarding isotopes generated through cosmogenic activation resulting from the prior exposure of detector materials to cosmic radiation, their presence can be reduced by minimizing surface exposure time and avoiding air transport. Additionally, special consideration is necessary for the removal of $^{222}$Rn (and its daughters) deposits on surfaces, which can be achieved through processes like acid treatment or electropolishing. 

Even with the implementation of the low-background techniques outlined above, DDD experiments have not reached a zero-background regime, and the measured background still exceeds the level expected from DM interactions. Most background events in DM detectors produce electronic recoils, whereas WIMPs are expected to induce NRs. This fundamental difference motivates the development of discrimination techniques aimed at distinguishing signal from background.

While some level of background rejection can be achieved by using pulse shape discrimination (PSD) techniques in several detection approaches, the most effective background suppression to date is obtained in experiments capable of measuring two distinct energy conversion channels, such as scintillation and heat, providing event-by-event discrimination at the energies of interest. Another approach to background discrimination relies on the detector's ability to reconstruct the position of the interaction, as is the case in time projection chambers (TPCs). Event position reconstruction enables the implementation of volume fiducialisation, which allows for the rejection of events originating in regions of the detector with elevated background levels, such as surfaces or edges.

In addition to minimizing the background level, DDD experiments must meet other crucial criteria, including a very low energy threshold, large amount of target material, extended data-taking periods, and control over the stability of environmental and operational conditions of the detector. Even with the fulfillment of all these essentials, distinguishing the hypothetical signal generated by a DM particle remains challenging.

\subsubsection{Direct detection techniques: status and future}

As discussed previously, DDD experiments aim to measure the  scattering of WIMPs from the Milky
Way galactic halo with atomic nuclei of a convenient detector based on Earth. Focusing on the elastic scattering, such kinetic energy transfer to the nucleus can manifest itself in mainly three different channels: atomic motion (heat), ionisation (charge) or excitation of radiative decaying levels (light). 

Most of the deposited energy is converted into heat for all the detection approaches, while smaller fractions are released through ionization and scintillation. The ionization channel, which involves the liberation of charge carriers, offers a good compromise between signal size and readout complexity, and is widely used in detectors with electric field-based charge collection. The scintillation (light) channel, though typically the least probable, is comparatively easy to detect due to the straightforward photon readout. In contrast, the heat channel, which carries the largest fraction of the energy released in the interaction, requires cryogenic operation in the mK range to achieve the sensitivity needed for for single particle detection.

\begin{figure}[t!]
\begin{center}
\includegraphics[width=1\textwidth]{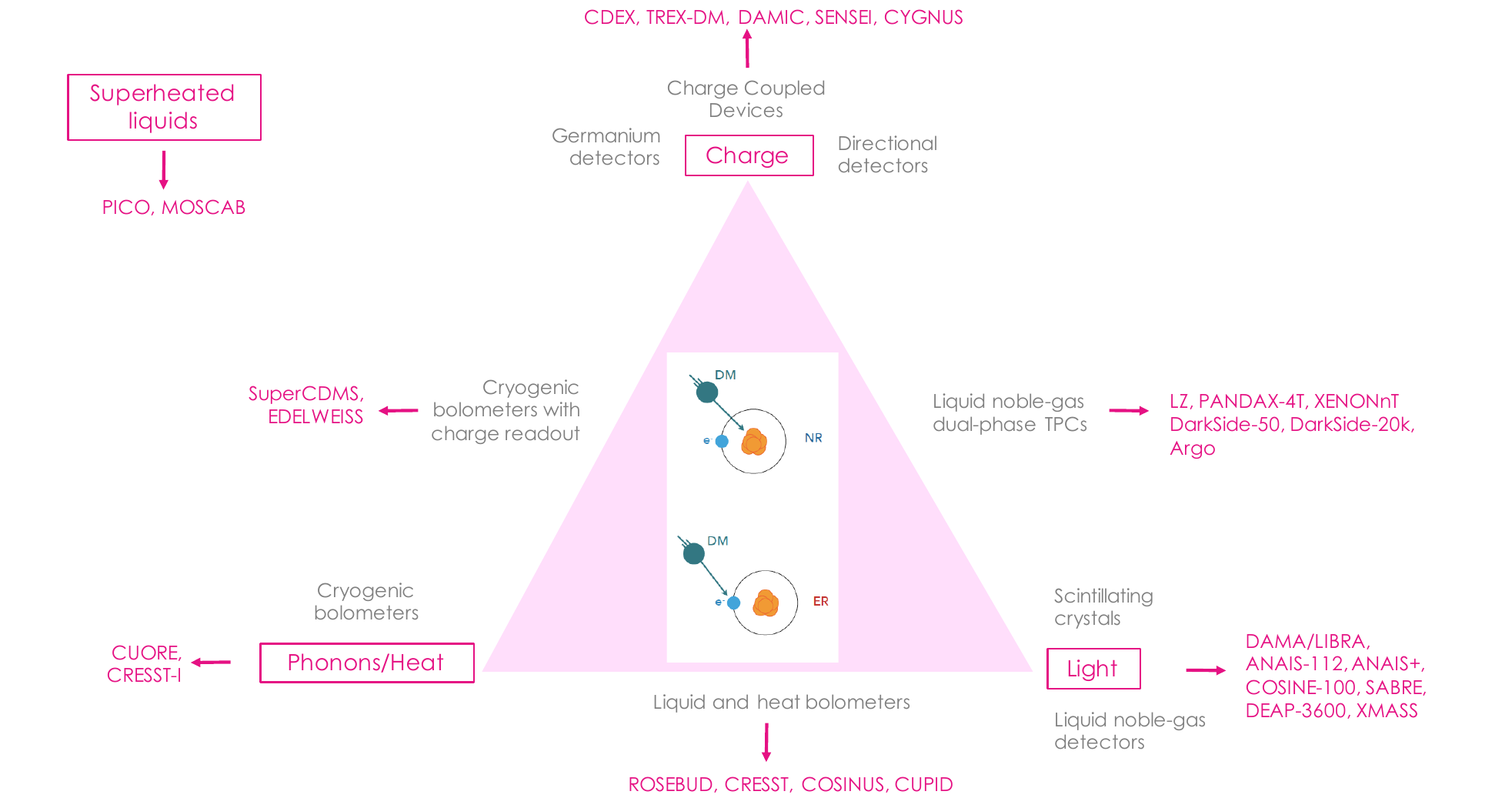}
\caption{\label{PossibleSignalsDirectDetection} Schematic representation of the three signal detection channels that can be measured in DDD experiments, together with the predominant technologies and main experiments associated with each detection mechanism.} 

\end{center}
\end{figure}

Figure \ref{PossibleSignalsDirectDetection} illustrates the different outcomes of a WIMP interaction and lists
some of the most common technologies designed to measure each type of signal. Detection strategies focus either on one of the three, or on the combination of two of these signals. The latter approach allows a powerful discrimination of NRs from
electronic recoils, considerably reducing the background of the experiment profiting from the different energy sharing between channels for different interacting particles. To date, there is no experiment that records all three signals
simultaneously.

Several experiments have reported rate excesses over the years \cite{boddy2022snowmass2021,Leane:2022bfm}. Some of these have been subsequently attributed to previously unaccounted-for background sources. Notable examples include the CRESST experiment, which observed an excess later explained by mechanical stress relaxation; the signal disappeared after modifying the crystal mounting system \cite{abdelhameed2019first}. The SuperCDMS experiment reported an excess in the silicon bolometer data, which, to date, remains unexplained \cite{agnese2018first}. XENON1T observed a low-energy excess that was later interpreted as being consistent with background from $^3$H contamination \cite{aprile2020excess}. These cases highlight the importance of comprehensive background modelling and continuous refinement of experimental techniques.

Most experiments conducted thus far have not claimed a positive DM signal and they have succesfully excluded candidates that would have otherwise produced an interaction rate higher than the measured value over known backgrounds. This allows to create exclusion plots in the WIMP-nucleus interaction cross-section as a function of the WIMP mass. In order to compare experiments, it is necessary to derive limits on the WIMP–nucleon cross-section within the framework of a chosen DM particle model.
\begin{figure}[t!]
\begin{center}
\includegraphics[width=0.7\textwidth]{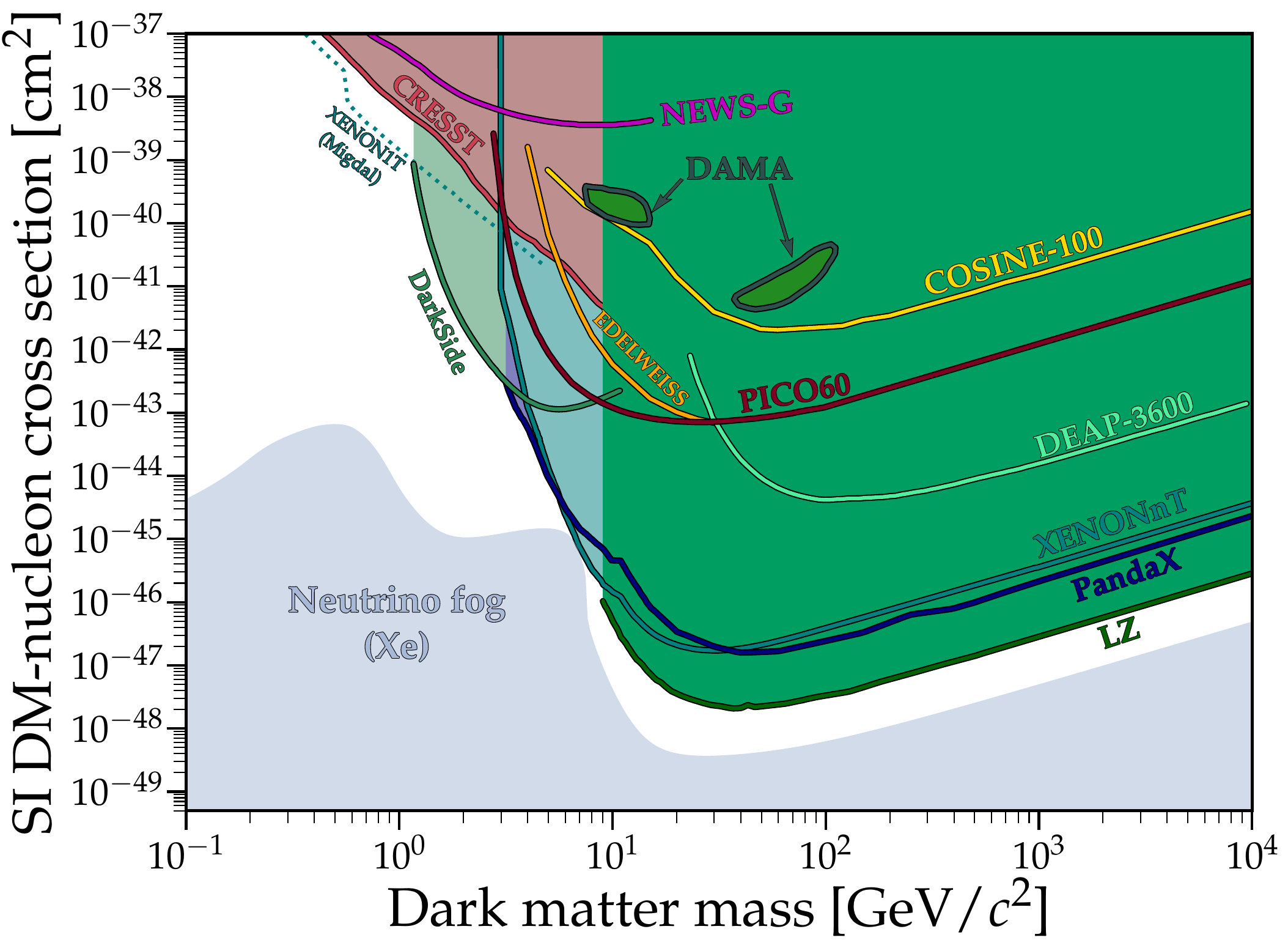}
\caption{\label{ExclusionPlot} Spin-independent WIMP-nucleon scattering exclusion plot with current
limits at 90\% C.L. considering the standard halo model. The green islands illustrates the
DAMA/LIBRA positive signal in the [2-6] keV range \cite{Bernabei:2020mon}, whereas the gray lower region represent
the neutrino floor calculated with a xenon target \cite{exclusionplot}.
} 
\vspace{-0.5cm}
\end{center}
\end{figure}

\begin{figure}[t!]
\begin{center}
\includegraphics[width=0.58\textwidth]{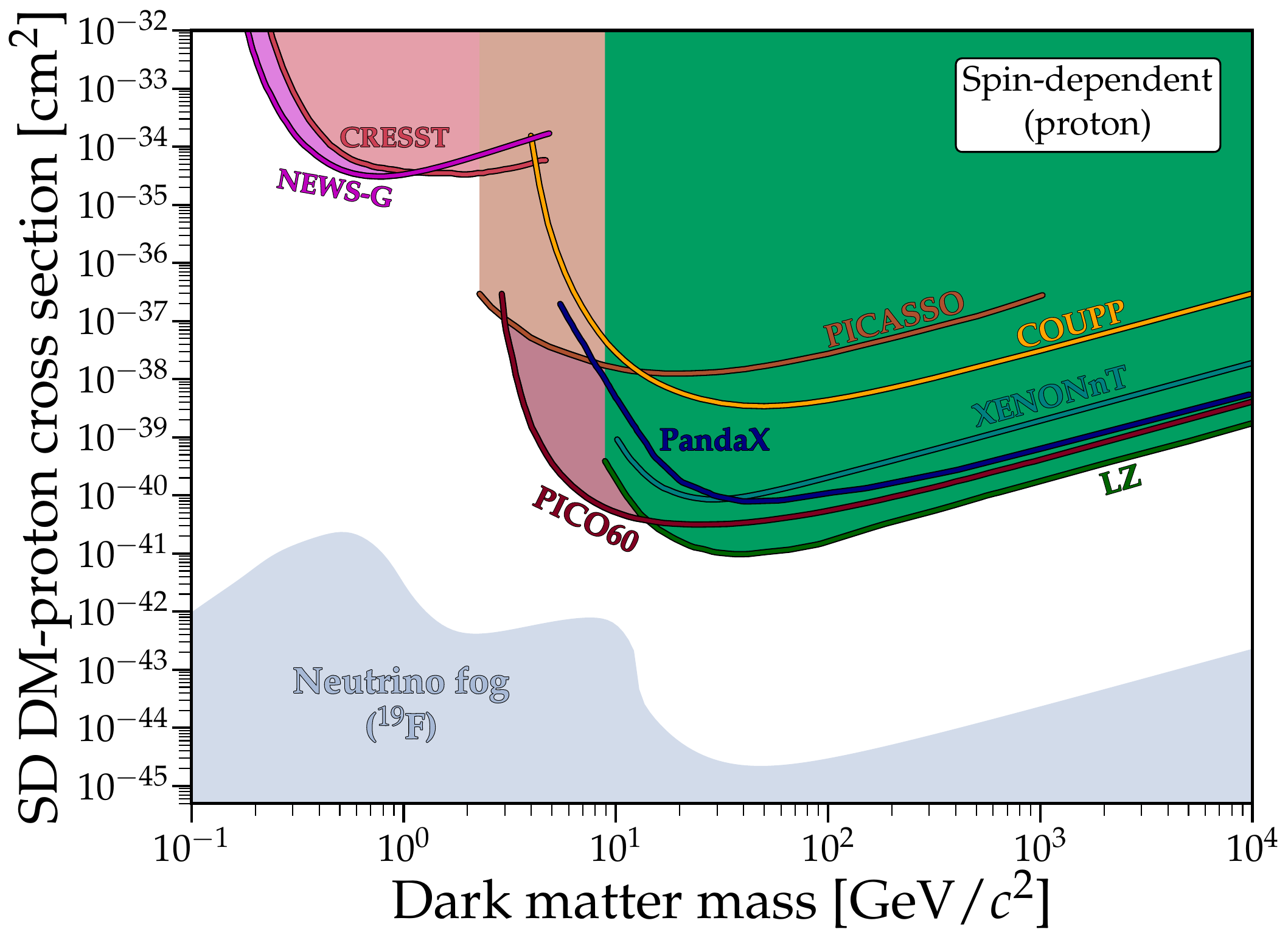}
\includegraphics[width=0.58\textwidth]{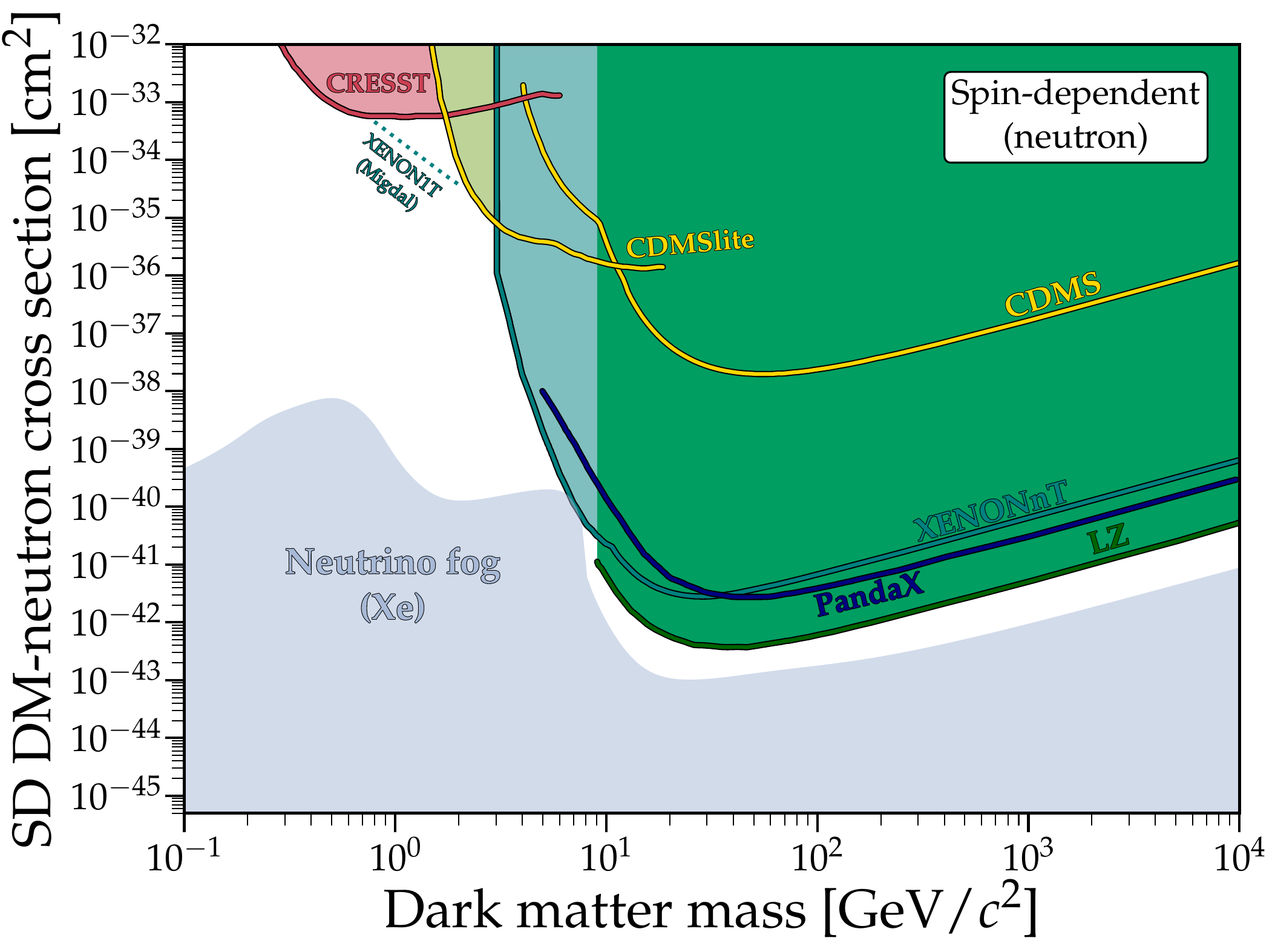}
\caption{\label{ExclusionPlotSD} Spin-dependent WIMP-nucleon scattering exclusion plot with current
limits at 90\% C.L. considering the standard halo model \cite{exclusionplot}. \textbf{Top panel:} proton coupling, with the neutrino floor calculated with a \textsuperscript{19}Fe target. \textbf{Bottom panel:} neutron coupling, with the neutrino floor calculated with a xenon target.
} 

\end{center}
\end{figure}

Figure \ref{ExclusionPlot} shows the main recent experimental results for
SI interaction in the ($\sigma_{\textnormal{SI}}$,$m_\chi$) plane, including the DAMA/LIBRA experiment compatible signal regions \cite{Bernabei:2020mon} (see Section~\ref{DAMAsec}). The curves in this figure represent the upper limit cross-sections for each WIMP mass that an experiment can exclude at a given C.L., typically 90\%. The excluded region reported by each experiment has been derived under a common set of simplified assumptions: SI interactions with equal couplings to protons and neutrons ($f_p = f_n$), and SHM with identical astrophysical parameters. Figure \ref{ExclusionPlotSD} shows the analogous plots for the main recent experimental results for
SD interaction for protons and neutrons.

The parameter space for WIMP-nucleus interactions is becoming increasingly constrained by DD experiments, as illustrated in Figures \ref{ExclusionPlot} and \ref{ExclusionPlotSD}. Moreover, in the coming decade, many experiments are expected to reach the neutrino floor, where CE$\nu$NS becomes an irreducible background for WIMP searches. In addition to the already mentioned DM–electron interactions aimed at probing sub-GeV candidates, another proposed approach to extend sensitivity into the MeV mass regime for DM–nucleus interactions involves exploiting the so-called Migdal effect \cite{ibe2018migdal,knapen2021migdal}. This effect predicts that a NR from a low-mass DM interaction could be accompanied by a secondary ER. This is because the electron cloud
does not follow the recoiling nucleus instantaneously. Such an ER may produce an ionization or excitation signal above threshold, making detection possible. However, this effect has not yet been experimentally confirmed in the context of nuclear scattering.

In most DM models, WIMPs are expected to interact primarily with the nuclei of the detector target material. To properly interpret these data in terms of DM it is essential to characterize the energy scale associated with NRs. However, calibrations in DM search experiments are typically performed using gamma-ray sources, which induce electron recoils (ERs). The energy scale derived from such calibrations is referred to as electron-equivalent energy (E\(_{\textnormal{ee}}\)) and is expressed in units of keV\(_{\textnormal{ee}}\)\footnote{Unless otherwise stated, all energies reported in the following will correspond to electron-equivalent energies (keV$_{\textnormal{ee}}$ will be abbreviated as keV).}. To relate the expected NR energy signal produce by WIMPs (E\(_{\textnormal{NR}}\), expressed in keV\(_{\textnormal{NR}}\)) with the corresponding measured (visible) signal (E\(_{\textnormal{ee}}\)) an scaling factor is required. This is known as the quenching factor (QF), given by QF(E) = E\(_{\textnormal{NR}}\)/E\(_{\textnormal{ER}}\). In scintillation and ionization detectors, this factor is less than unity.



The field of direct WIMP searches is very active and numerous experiments are running all over the world. The following is a brief overview of the most relevant techniques used in ongoing DDD searches, together with its most important results. A detailed overview on the current experimental landscape can be found in \cite{Billard:2021uyg,navas2024review}. \\

\textbullet \quad \textbf{Scintillating crystals.}

Scintillating crystals such as NaI(Tl) and CsI(Tl) are widely used in particle detection, and are also deployed as a target
for DM searches. Introducing a dopant in inorganic scintillators (tipically $\sim$0.1\% of the
molar mass), such as thallium in NaI(Tl), creates states between the valence and conduction bands that work as luminescent centers. When a particle interacts in the crystal, part of the deposited energy is devoted to excite electrons from the valence
to the conduction band. If these electrons reach a luminescent center, the subsequent de-excitation releases scintillation light as photons in the visible-UV range following an exponential decay time-distribution according to the lifetime of the luminiscent excited state. The light can be easily read out by low
background photomultipliers (PMTs). 

NaI(Tl) has been employed since the early stages of DM searches due to its suitability as a target material for several reasons. It offers sensitivity to both low-mass WIMPs, via interactions with sodium, and high-mass WIMPs, through interactions with iodine. Moreover, both isotopes contribute to SD interactions, enhancing its relevance for a wide range of DM models. NaI(Tl) also provides a high LY, approximately 40 photons/keV, enabling low energy thresholds essential for detecting faint signals.

As a mature, cost-effective, and operationally simple detector technology, NaI(Tl) scintillators facilitate the deployment of large target masses and ensure stable long-term operation. While they offer PSD capabilities to distinguish NRs from ERs, this discrimination is limited and does not allow for event-by-event classification in the ROI. As a result, achieving competitive sensitivities necessitates the use of ultra-pure crystals to minimize intrinsic background levels. 

Characteristic WIMP signatures, such as the annual modulation expected from DM interactions (see Section~\ref{AnnualModSec}), can be investigated using scintillating crystals. The primary motivation for exploiting this approach stems from the long-standing observation reported by the DAMA/LIBRA collaboration, which remains the only experiment to claim a statistically significant signal compatible with DM, supported by over two decades of data acquisition \cite{bernabei2020dama} (see Section \ref{DAMAsec}). Although DAMA/LIBRA results have been excluded by other experiments with superior sensitivity, these null results were obtained using different target materials, introducing model dependencies that prevent a conclusive confirmation or refutation of the DAMA/LIBRA claim. This has underscored the need for independent verification with the same target material, prompting the development of similar experiments. Further details on the DAMA/LIBRA experiment and ongoing international efforts, including ANAIS-112, the main focus of this thesis, are provided in Section~\ref{DAMAsec}.\\

\textbullet \quad\textbf{Cryogenic detectors.} 

Cryogenic detectors are solid state detectors operated at ultra-low temperatures. They are very sensitive devices providing a low energy threshold and
an excellent energy resolution and thus, are very suitable for the direct detection of
WIMPs. In particular, cryogenic detectors lead the search for
light DM, below few GeV/c$^2$, where expected recoils are more frequent but less
energetic. However, they cannot compete with other approaches for higher WIMP masses because they typically employ detectors with very low target masses, limiting the exposure and consequently the sensitivity to the rarer, high-energy recoil events expected from heavier DM particles.

Most of the recoil energy of the detector target nuclei induced by the elastic scattering of DM particles is dissipated via phonons, which result in a small,
but measurable, temperature increase in the material, $\Delta$T($\mathcal{O}$($\mu$K)). This temperature rise is detected by a highly sensitive thermometer located on the detector, such as a transition-edge sensor (TES). To measure this phonon signal,
detectors have to be cooled down to mK temperatures, which is only possible in
dedicated cryostats. 

A huge advantage of cryogenic detectors is their versatility in using different combinations of the three available readout channels. Thus, as hybrid detectors, it is possible to choose as target material either a scintillating material or a semiconductor, and simultaneously read either phonon plus light, or phonon plus charge signals, respectively. The event-by-event measurement of light and charge yield allows to identify the type of
interacting particle and guarantees the
discrimination of beta/gamma radioactive backgrounds from potential DM signal events. This high discrimination power is a huge advantage in the search for rare events. There is a wide choice of target materials available. Germanium and silicon semiconductor detectors are being used by SuperCDMS  \cite{SuperCDMS:2014cds,SuperCDMS:2017mbc}, located at SNOLAB, and EDELWEISS~\cite{EDELWEISS:2016nzl,EDELWEISS:2017lvq}, at the Laboratoire Souterrain de Modane (LSM). 

By contrast, the CRESST experiment \cite{CRESST:1999ynq}, running at the LNGS, uses scintillating crystals, mainly CaWO$_4$, as target material. The CRESST experiment stands out as a pioneer in the employment of scintillating crystals as cryogenic detectors for the search for DM. In this context, it is also important to mention the ROSEBUD experiment (Rare Objects SEarch with Bolometers Underground), established in 1999 at the LSC, which developed the first underground light-heat measurement \cite{cebrian2003first}. The collaboration, in which the University of Zaragoza participated, was also at the forefront of developing scintillating bolometers, making significant contributions to the advancement of cryogenic detection techniques for rare event searches.

CRESST-III has successfully reached a 30 eV a nuclear recoil energy threshold, achieving
the leading limits for SI interacting WIMPs to date in the lowest mass range down to 160~MeV~\cite{CRESST:2019jnq}. Based on the detector technology developed within CRESST over the past decades, the COSINUS (Cryogenic Observatory for SIgnatures seen in Next-generation Underground
Searches) \cite{angloher2016cosinus} experiment aims to cross-check the long-standing DAMA/LIBRA result (see Section \ref{DAMAsec}) using cryogenic NaI detectors. The two-channel readout on a NaI target allows COSINUS to verify the DAMA claim with significantly less exposure due to its lower recoil energy threshold. Moreover, this technology would allow for the identification of the recoil type in the case of a signal being observed above background levels.\\

\textbullet \quad \textbf{Noble liquid detectors.} 

Another approach to detect DM is to use liquid noble gases as detector materials. Noble gases are excellent scintillators, particularly argon
and xenon. In addition, particles interacting in their liquid phase (LAr, LXe) produce ionization, which can be measured independently allowing particle discrimination. Liquid noble gas experiments offer the clear advantages of an inherent low background level and a high exposure, since the scaling up of the target mass to the order of tons is relatively straightforward. In order to operate these experiments, two different detection technologies are followed: single-phase and dual-phase.

Single-phase detectors exclusively measure the scintillation signal released in the liquid by the scattering
particles while typically using 4$\pi$ PMT coverage in order to
increase the light collection efficiency. Background discrimination in these detectors is achieved either through PSD in LAr, which is particularly useful for distinguishing between NRs and ERs based on their different light time profiles, or through target fiducialization in LXe, which isolates a specific ROI within a large target volume. The DEAP–3600 experiment \cite{DEAP-3600:2017ker}, located at SNOLAB, has demonstrated an impressive background rejection capability through a single-phase LAr detector, establishing the best limit on the SI WIMP-nucleon interaction on a LAr target \cite{DEAP:2019yzn}.

Dual-phase detectors work as TPCs, combining a liquid and a gaseous phase in the same volume, measuring the primary scintillation signal (S1) in the liquid phase, and the  secondary
scintillation (S2) produced by electrons drift by an applied electric field into the gas phase, which is proportional to the initial
ionization or charge signal. Dual-phase detectors enable not only the discrimination of the interacting particle type based on the ratio of S2 to S1 signals, but also 3D event reconstruction with spatial
resolution at the order of mm by using the drift time of the electrons. 

Noble liquid detectors provide the most stringent exclusion limits in the high mass range, from a few GeV to
TeV. Yet, its energy threshold, of the order of $\mathcal{O}$(1~keV), is higher than the one achieved by cryogenic detectors. 

The most stringent limit above 10 GeV/c$^2$ has been provided by the LUX-ZEPLIN (LZ) experiment \cite{aalbers2023search}, at the Sanford Underground Research Facility (SURF), in the United States. LUX-ZEPLIN is the fusion of the LUX \cite{LUX:2016ggv} and ZEPLIN \cite{ZEPLIN} experiments, and runs with an active mass of 7 ton of LXe. The next stronger limit is that provided by The PandaX-4T experiment, operating with an active mass of 4 tons at the China Jinping Underground Laboratory (CPJL) \cite{PandaX:2022osq,bo2025dark}, serving as an upgrade to PandaX-II \cite{PandaX-II:2020oim}.

From 3.6 - 9.0 GeV/c$^2$, the XENON experiment, at LNGS, leads the searches with XENON1T \cite{XENON:2018voc}
and XENONnT~\cite{XENON:2020kmp}, the latest upgrade of this dual-phase experiment with an active mass of 5.9 tons. The region 1.2 - 3.6 GeV/c$^2$ is primarly explored by DarkSide-50 \cite{DarkSide-50:2022qzh}, with their LAr TPC experiment. Its next phase, DarkSide-20k \cite{DarkSide-20k:2020qfz} is currently under construction at LNGS and will have an active mass of 23 tons. 

The first detections of solar $^8\mathrm{B}$ neutrinos via CE$\nu$NS have recently been reported by XENONnT and PandaX-4T experiments \cite{PandaX:2024muv,aprile2024first}. However, these neutrinos dominate at low recoil energies. At higher recoil energies, where atmospheric neutrinos become the main background, there still remains a region of parameter space to be explored before reaching the neutrino fog, where neutrino backgrounds limit the sensitivity to WIMP signals. Next-generation xenon- and argon-based experiments, such as DARWIN/XLZD~\cite{BAUDIS2024116473} and Argo \cite{McDonald:2024osu}, have proposed detector concepts specifically designed to probe this challenging region.\\

\textbullet \quad \textbf{Semiconductor detectors.} 

Detectors that rely solely on ionization, such as semiconductors, are yielding highly interesting results in the quest to detect low-mass DM. Germanium-based experiments have a long history in direct detection. They take advantage from their extremely low level of intrinsic backgrounds and high energy resolution. Today,  point-contact Ge detectors can reach sub-keV thresholds, such as the CoGeNT \cite{cogent} experiment at Soudan, or CDEX \cite{CDEX:2018lau}, operated at CJPL.

Silicon charge-coupled devices (CCDs) are silicon-based detectors that consist
of a grid of pixels each acting as an individual detector.  Silicon CCD
detectors do
enable lower DM masses searches by offering high charge resolution, a lighter target nucleus, and imaging possibilities for background identification. The DAMIC experiment \cite{DAMIC:2011khz,DAMIC:2020cut}, consisting of 7 CCDs with a total mass
of 40 g, has been operating at SNOLAB since 2017. The subsequent phase, DAMIC-M \cite{DAMIC-M:2022aks}, is planned to be placed in LSM, and will consist of an array of 50 CCDs with a mass of 1 kg and a detection threshold of sub-electron. DAMIC-M is expected to explore SI WIMP-nucleon scattering at the GeV range with an unprecedented
sensitivity and to probe WIMP-electron interactions with masses down to
1 MeV/c$^2$. In the long term, the OSCURA experiment \cite{oscura} (DAMIC+SENSEI \cite{adari2025first}) is likely to confirm or dismiss DM-electron scattering for masses as low as 500 keV. \\

\textbullet \quad \textbf{Bubble chambers.}

A different type of excitation is exploited by bubble chambers: a liquid is kept above their boiling point
in such a metastable state that the interaction of a particle is able to trigger
the phase transition producing a gas bubble. Optical detection and counting of bubbles provide a measurement of particle interactions. Particle discrimination is performed based on the change of the sonic pressure in
piezoelectric materials detected during bubble formation. One advantage of bubble chambers is that gamma rays rarely cause bubble formation, which significantly reduces the primary background. The PICO experiment \cite{PICO:2019vsc} (the fusion of PICASSO \cite{PICASSO} and COUPP \cite{COUPP}
experiments), which uses C$_3$F$_8$ in a gel matrix, is leading the sensitivity for WIMP candidates with SD
coupling to protons. \\

\textbullet \quad \textbf{Directional detectors.}

Due to the Earth motion around the Sun and the solar system motion around the galactic center, an Earth-bound detector is expected to be exposed to a wind-like flux of DM particles, coming preferentially from the direction of the Earth's velocity. Consequently, an annual modulation in the WIMP interaction rate is expected, with a maximum around June and a minimum around December. This annual modulation signature constitutes the central focus of this thesis and will be addressed in detail in Section~\ref{AnnualModSec}.

A potentially more distinctive DM signature is encoded in the directionality of WIMP- induced NRs. Specifically, the orientation of the expected DM wind shifts by approximately 90 degrees every 12 sidereal hours due to the tilt of the Earth rotational axis with respect to the direction of the solar system motion. Measuring the incoming direction of WIMPs offers an unambiguous identification of a DM signal, as no known background exhibits such directional correlation. Moreover, this observable may gain increasing relevance as experiments approach the sensitivity required to detect elastic nuclear scattering from solar and other neutrinos, providing a method to discriminate between neutrino and DM events.

It is worth noting that only gaseous targets can preserve the directional information of NRs, which inherently limits the achievable exposure due to low target densities. Moreover, the direction of the recoil is not, in general, aligned with that of the incoming WIMP, although the two are correlated. Measuring the recoil direction would thus not provide the exact trajectory of the WIMP, but it would still offer a powerful, unambiguous signature of DM if a directional signal were observed.

Currently, directional detectors face significant technological challenges and are not yet competitive in the broader DM search landscape. One of the pioneering efforts in this domain is the DRIFT experiment \cite{drift}. The CYGNUS collaboration \cite{cygnus}, which brings together most research groups focused on directional DM detection, aims to enhance the sensitivity of this approach in the near future.

\subsection{The annual modulation signature in the DM signal}\label{AnnualModSec}

As previously discussed, the expected DM spectrum does not show distinctive features that would unequivocally set it apart from the aforementioned backgrounds (see Figures~\ref{RateForTargets} and \ref{RateForTargetsNaI}), assuming that the operators describing WIMP-nucleus scattering do not depend on the velocity. Hence, it becomes essential to  identify unique characteristics associated with the DM signal, providing a basis for a reliable detection. In this context, a time-dependent, DM-induced annual modulation in the WIMP detection rate is one of the most clear dependences which does not require a directional detector to be measured. 

\begin{figure}[b!]
\begin{center}
\includegraphics[width=0.6\textwidth]{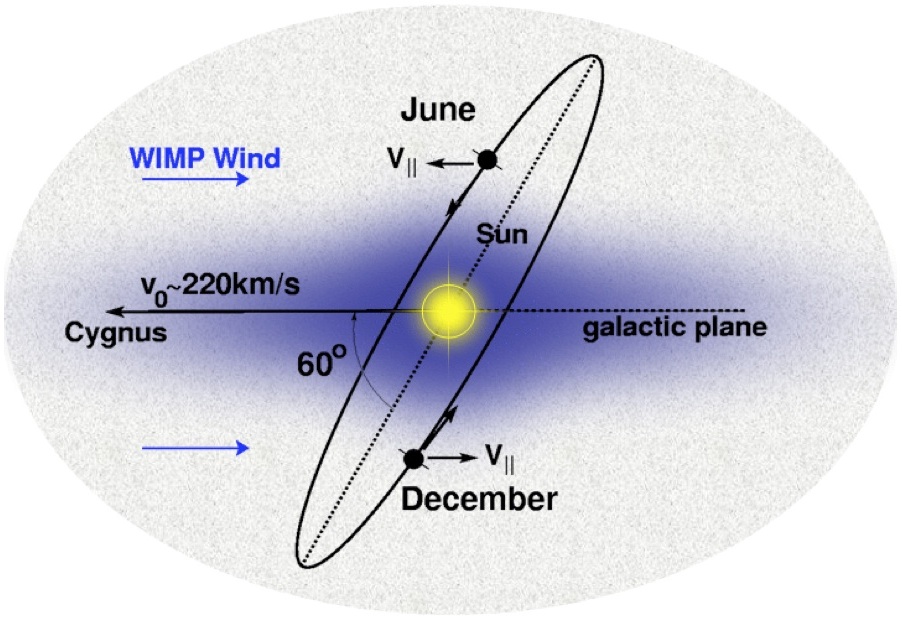}
\caption{\label{annualModulation} Illustration of the expected annual modulation effect. The Earth orbit around the Sun, combined with the Sun motion around the galactic center, induces a predictable annual modulation in the velocity of WIMPs reaching the Earth \cite{annualModulation}.   } 

\end{center}
\end{figure}

This signature is produced by the change in the relative velocity of the DM particles with respect to the detector nuclei over the course of the year, due to the Earth's motion around the Sun \cite{Drukier:1986tm,Freese:1987wu}, and is depicted in Figure \ref{annualModulation}. At the beginning of June, the velocity of the Sun and the projection of the Earth's velocity onto it are parallel, whereas in December they are antiparallel. Thus, when the relative velocity WIMP-nucleus is higher, WIMPs are
more energetic on average and leave more energy in the detector. This modifies the shape of the differential recoil spectrum, leading to annually modulated detection rates whose amplitude depends on the WIMP mass and the deposited energy, such that the event rate in June can be either higher or lower than in December, depending on the energy range considered.

The projection of the Earth velocity in the direction of motion of the Sun onto the Galactic plane, $v_T$, changes with the time of the year following:

\begin{equation}
    v_T = v_0 +  v_E \cos(\gamma) \cos (\frac{2\pi}{T} (t-t_0)).
    \label{vteq}
\end{equation}

Here, $v_0 \sim$ 220 km/s  is the velocity of the Sun respect to the galactic center; $v_E~\sim$~30~km/s is the average Earth orbital velocity around the Sun on a plane with inclination $\gamma$~=~\SI{60}{\degree} relative to the Galactic plane, such that $v_\parallel = v_E \cos(\gamma)$, as shown in Figure~\ref{annualModulation}. In addition, $T$ = 1 year is the period of the orbit of the Earth around the Sun, and $t_0$ corresponds to the 2$^{\textnormal{nd}}$ of June, when the velocities of the Sun and Earth are aligned, maximizing the sum of the two.


As shown in Equation \ref{rateTotalEqcontime}, the differential rate of WIMP interactions in the detector's rest frame depends on the WIMP velocity distribution, which depends on the time of the year. The velocity $v$ in Equation \ref{rateTotalEqcontime} can be approximated by $v_T$ in Equation \ref{vteq} since the perpendicular component of the velocity projection is very small. As the projection of Earth orbital velocity in the galactic plane is significantly smaller than the Sun velocity with respect to the Galactic halo (15/220 $\simeq$ 0.07), the modulation amplitude is expected to be small, which allows to perform a first-order Taylor expansion of the expected DM rate for the SHM model \cite{annualModulation}:


\begin{equation}
    \frac{dR}{dE_{\textnormal{NR}}} \approx S_0(E_{\textnormal{NR}}) + S_m(E_{\textnormal{NR}}) \cos(\frac{2\pi}{T}(t-t_0),
\end{equation}

where $S_0(E_{\textnormal{NR}})$ is the time-averaged differential rate and $S_m(E_{\textnormal{NR}})$ is the modulation amplitude. Contribution from higher-order terms is negligible (<0.1\%).\\

The DM annual modulation signature is searched for by requiring simultaneously: 

\begin{itemize}
    \item The rate of events is expected to modulate around the average value, following a cosinoidal behaviour with a
period of 1 year and reaching its maximum around the 2$^{\textnormal{nd}}$ of June. 

\item The modulation must only be found in a precisely defined low-energy range, where events induced by DM particles are expected to occur.

\item The modulation amplitude is required to be a weak effect, from 1 to 10\% of the total DM average rate depending on the WIMP mass and energy range, smaller than this when considering that the average of background is also contributing to the non-modulated rate.

\item A phase reversal which can produce a negative modulation amplitude is expected at very low energies.

\item  In a multi-detector configuration, only events that produce a signal in a single detector, the so-called single-hit events, should be considered, as long as the probability that a DM particle interacts with more than one detector is negligible.

\end{itemize}

Experiments searching for a DM annual modulation need to mitigate any potential detector-induced modulation signal. Between the sources that might mimic this signature, one notable example are muons, since it has been observed an annual modulation of the muon flux underground \cite{agafonova2019measurement}. This signal could lead to a modulation in the detection rate caused by muon-induced neutrons or other muon-related events. Furthermore, it is known that the radon concentration in the air of underground laboratories, such as in the LSC, shows an annual modulation that correlates with other environmental conditions, such as humidity~\cite{Amare:2022dgr}. In addition to carefully monitoring all those experimental parameters which could prompt a time-variable behaviour, a robust background understanding is essential, as certain background components may have a time-dependent contribution to the rate that could affect the search for the annual modulation.

\subsubsection{The DAMA/LIBRA experiment} \label{DAMAsec}

The DAMA/LIBRA (DM/Large sodium Iodide Bulk
for RAre processes) experiment, located at LNGS, in Italy, is the only experiment
reporting a positive annual modulation signal compatible with that expected for DM particles in the SHM with high statistical significance, accumulating more than two decades of data \cite{bernabei1999further,bernabei2000search,bernabei2008first,bernabei2020dama,bernabei2018first,bernabei2023dark}. The DAMA/LIBRA operation has been divided into three phases: DAMA/NaI (1996–2002), which used 100 kg of NaI(Tl) scintillators with a 2 keV threshold over 7 annual cycles; DAMA/LIBRA-phase~1 (2003–2010), with 250 kg of NaI(Tl), a 2 keV threshold, and 7~annual cycles; and DAMA/LIBRA-phase~2 (2010–2024), using the same 250~kg target mass but with a reduced threshold of 1~keV and 14 annual cycles (although only 10 have been published so far). Notably, in DAMA/LIBRA-phase~2, the energy threshold was lowered from 2 keV to 1 keV thanks to the replacement of the PMTs with new ones featuring higher quantum efficiency~(QE).

The DAMA/LIBRA experiment used $\sim$ 250 kg of highly radiopure NaI(Tl) scintillators produced by Saint
Gobain company distributed in 25 modules of 9.7 kg each in a 5×5 matrix configuration. Despite being manufactured two decades ago, its detectors remain the most radiopure NaI(Tl) that have ever
been deployed in a DM search. Each NaI(Tl) crystal was coupled to two PMTs on each side and the module was surrounded by
low-radioactive OFHC (Oxygen Free High Conductivity) freshly electrolyzed copper
shields. The detectors were housed within an anti-radon box
made of OFHC copper which was continuously flushed with high-purity nitrogen gas. The passive shield was made up of $\sim$ 15 cm copper, 15 cm of low-radioactivity lead, 1.5\  mm of cadmium foil, about 10-40 cm of polyethylene, and roughly 1 m of special concrete
made of the Gran Sasso rock.

\begin{figure}[t!]
    \centering
    \begin{subfigure}[b]{1.\textwidth}
       \hspace{-0.5cm} \includegraphics[width=\textwidth]{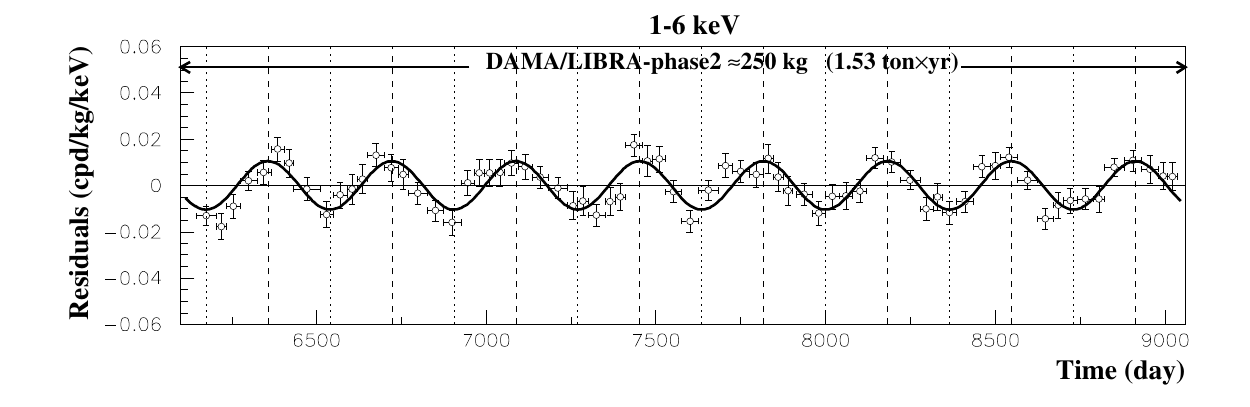}
        \label{DAMALIBRA2_6}
    \end{subfigure}
    \hfill
    \begin{subfigure}[b]{1.\textwidth}
        \includegraphics[width=\textwidth]{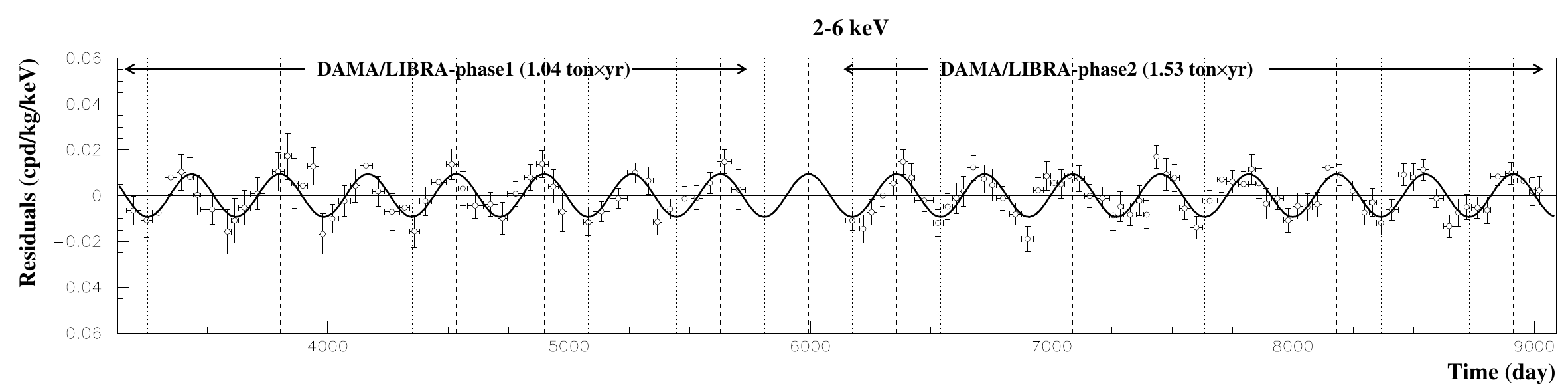}
        \label{DAMALIBRA1_6}
    \end{subfigure}

    \caption{Experimental residual rate of the single-hit scintillation events measured
in the [1-6] keV energy range by DAMA/LIBRA-phase2 \textbf{(top panel)}, and by DAMA/LIBRA-phase1 and DAMA/LIBRA-phase2 in the [2-6] keV energy region \textbf{(bottom panel)}\cite{bernabei2023dark}.}
\label{DAMALIBRA_results}

\end{figure}

DAMA/LIBRA data favor the presence of a modulation with proper features at 13.7$\sigma$~C.L. in the [2-6] keV energy region (with the full exposure, 2.86 ton$\times$year), and 11.8$\sigma$ C.L. in the [1-6] keV energy region (with data from DAMA/LIBRA-phase 2, 1.53~ton$\times$year) when fitting with free period and phase \cite{bernabei2020dama}. The DAMA/LIBRA final results with the full dataset are still pending. Figure \ref{DAMALIBRA_results} shows the residual
rate of the single-hit scintillation events measured by DAMA/LIBRA-phase2 in the [1-6] keV region (top pannel), and in the [2-6]~keV energy region by DAMA/LIBRA-phase1 and DAMA/LIBRA-phase2 (bottom pannel). It is worth highlighting that their total event rate has not been ever publicly disclosed. 

DAMA/LIBRA analysis strategy is based on subtracting every year the average measured rate in both energy regions considered. The corresponding residual rates are fitted to the function $S_m$ $cos(\frac{2 \pi}{T} (t-t_0)$ (black curve in the figures),  with period $T$ and phase $t_0$. DAMA/LIBRA has performed the modulation fit both with fixed period and phase (one year and the 2$^{\textnormal{nd}}$ of June, respectively) and allowing them to vary freely, obtaining results consistent with a DM interpretation in both cases. For ANAIS-112, the period and phase are fixed to the same values in order to allow a direct comparison with DAMA/LIBRA, as presented in \cite{Bernabei:2020mon}.



Several proposals of alternative origins of DAMA annual modulation signal have arised throughout the years. Many explanations claim that muons or muon-induced
and solar neutrino-induced neutrons could give rise to the observed modulation signal. In addition, upper limits on possible contributions to the measured modulation amplitude from sources such as radon, temperature variations, energy calibration or other background sources have been estimated. However, the DAMA collaboration discards any proposed systematic effects or backgrounds contributing to such a modulated signal.

If the DAMA signal ([2-6] keV region) is interpreted as resulting from the elastic interactions of WIMPs with the detector target materials, two WIMP candidates emerge with a good fit
for masses between 10–15 GeV for scattering off sodium, and 60–100~GeV for scattering off iodine considering the standard SI interaction scenario \cite{savage2009compatibility}. However, the energy depence of the modulation amplitude in the low energy region of DAMA/LIBRA-phase~2 ([1-6]~keV region) is not well fitted by standard WIMP models \cite{Baum:2018ekm}, which poses a significant caveat in interpreting the DAMA/LIBRA signal as a DM signature. Nonetheless, the DAMA/LIBRA collaboration argues that the interpretation strongly depends on the assumed QFs \cite{bernabei2019improved}.

Some theoretical models, although somewhat ad-hoc in nature, have been proposed to reconcile the DAMA/LIBRA signal with the null results obtained by other experiments~\cite{PhysRevD.99.023017,PhysRevD.98.123007,PhysRevD.99.103019,adhikari2021strong}. However, most reasonable theoretical models predict that a particle creating that kind of signal would be also observable by other experiments using different targets or techniques. Nevertheless, the DM implications of DAMA annual modulation seem to directly
conflict with the null results of other more sensitive experiments. As shown in Figure~\ref{ExclusionPlot}, many collaborations have excluded the parameter space region favored by DAMA/LIBRA by several orders of magnitude. Even so, all such constraints have been established by DM experiments that do not employ NaI(Tl) as target material. As a result, they cannot conclusively confirm or refute the DAMA/LIBRA claim due to many model-related uncertainties when comparing results obtained with different detector target. 


Therefore, a model-independent test of the DAMA/LIBRA signal using NaI detectors is essential to definitively clarify the nature of this long-standing anomaly. 
\vspace{-0.3cm}
\subsubsection{Experiments testing DAMA/LIBRA}\label{experimentstestingDAMA}

ANAIS–112 and COSINE-100 experiments have conducted dedicated efforts to clarify the origin of the DAMA/LIBRA signal using the same target material and detection techniques. This thesis is framed within the ANAIS-112 experiment. In particular, Section \ref{set-up} will provide a comprehensive description of the ANAIS-112 experimental set-up, data analysis procedures, and the most recent results obtained.

The COSINE-100 experiment was a direct detection DM experiment that aimed to test the DAMA/LIBRA positive annual modulation signal with low-background NaI(Tl) crystals at the Y2L in South Korea \cite{adhikari2019search}. It started the data taking on September,~2016 and was decommissioned in March of 2023.

The eight NaI(Tl) modules, produced by Alpha Spectra Inc., were hermetically
housed in 1.5 mm-thick Oxygen-Free Electronic (OFE) copper tubes with a calibration
window to allow for low energy calibrations, and were optically coupled with
quartz windows to two Hamamatsu PMTs. The 4x2 matrix of detectors was housed in an acrylic box immersed in a wall reflecting tank of
2200 L filled with a liquid scintillator which allows to strongly reduce the $^{40}$K emissions
from the crystals. The tank was encased in an OFC box and lead, with additional panels made of plastic scintillator for muon veto purposes. It is worth highlighting that, out of the total 106 kg of NaI(Tl) crystals (from the same provider as ANAIS-112), three of the COSINE-100 detectors were not suitable for DM searches due to lower LY, higher energy thresholds, or elevated noise rates. As a result, COSINE-100 operated with only 61.3 kg of active mass.

COSINE-100 reported recently the analysis of the annual modulation signal using the full data set, comprising 6.4 years of data and yielding a total exposure of 358~kg~x~yr~\cite{carlin2024cosine}. The best fit for the modulation amplitude of single-hit events, with the period and phase fixed to the expected values for the SHM, is S\textsubscript{m} = 5.3 ± 3.1 c/keV/ton/day   (1.7~±~2.9~c/keV/ton/day) for the [2–6] keV ([1–6] keV) energy region. These results are compatible with the absence of modulation within 1.5(2.8)$\sigma$ and incompatible with DAMA/LIBRA at 3.3(3.6)$\sigma$ C.L. for [1–6] ([2–6]) keV.

The COSINE collaboration is currently commissioning its second phase, COSINE-100U, at the Yemilab underground facility in Korea. This new stage employs the same NaI(Tl) crystals as COSINE-100 (99.1 kg), but uses a revised encapsulation design without a quartz window to enhance light collection by approximately 40\%. The data-taking campaign was scheduled to start in May 2025, and is expected to achieve world-competitive sensitivity to low-mas DM candidates. This upgrade aims to optimize detector performance in preparation for the future COSINE-200 experiment, which will deploy 200 kg of ultra-radiopure NaI(Tl) crystals with significantly reduced levels of $^{210}$Pb and $^{\textnormal{nat}}$K contamination \cite{park2020development}.

In the longer term, several experiments are pursuing the goal of a model-independent test of the DAMA/LIBRA signal. These include the SABRE project \cite{SABRE:2018lfp}, which aims to operate twin detectors in the northern (Italy) and southern (Australia) hemispheres to disentangle potential seasonal background contributions; the COSINUS experiment \cite{angloher2020cosinus}, based on cryogenic calorimetry with dual readout; and the PICOLON project \cite{fushimi2021development}, which focuses on developing ultra-radiopure NaI(Tl) crystals.

Additionally, the ANAIS+ project represents the next phase of the ANAIS-112 experiment. ANAIS+ aims to enhance detector performance by replacing conventional PMTs with silicon photomultipliers (SiPMs) operated at cryogenic temperatures ($\sim$100~K). This configuration is expected to suppress PMT-induced light noise, increase the QE of the light readout system, and exploit the improved scintillation yield of NaI(Tl) at low temperatures. With these improvements, an energy threshold as low as $\sim$100 eV could be achieved, enabling sensitivity to both SI and SD interactions from sub-GeV WIMPs, while also eliminating systematic uncertainties associated with QF differences.

The results from ANAIS-112 \cite{amare2025towards} and COSINE-100 \cite{carlin2024cosine}, each based on six years of data, strongly disfavor a DM interpretation of the DAMA/LIBRA modulation signal, even when conservative assumptions on QF are taken into account. This conclusion is further supported by the combined three-year analysis conducted jointly by both collaborations \cite{carlin2025combined}. Nonetheless, the new generation of NaI(Tl)-based experiments is expected to shed more light on this persistent discrepancy. In this context, reproducibility, transparency, and community-wide validation of experimental results, together with the adoption of open data policies, constitutes an essential pillar of science.

It must be acknowledged, however, that fully clarifying the nature of the DAMA/LIBRA signal may remain out of reach without direct access to their data and analysis protocols. In this context, a formal expression of intent has been submitted to INFN-LNGS to evaluate the feasibility of merging the expertise, operational know-how, and instrumentation of the ANAIS, COSINE, and SABRE collaborations for the potential reoperation of the DAMA/LIBRA detectors. Such a unified effort would enable a direct annual modulation search using the original detectors in a novel and independently controlled experimental environment. 

\subsubsection{NaI(Tl) response in the ROI: the QF role}

The uncertainties in the knowledge of the Na and I QFs in NaI poses a significant challenge for the model-independent interpretation of the DAMA/LIBRA signal. If the DM signal is attributed to NRs, discrepancies in the detector response may arise between DAMA/LIBRA and other NaI(Tl)-based experiments if the QF is not an intrinsic property of the NaI(Tl) scintillator.

The QF can be determined either via direct measurements using neutron scattering experiments or by comparing experimental NR spectra from neutron calibration sources with those produced by Monte Carlo simulations. Both approaches have been pursued in the context of ANAIS-112 as part of its
dedicated neutron calibration program, aimed at determining the QF of Na and I
to address this key uncertainty which significantly affects the comparison with the
DAMA/LIBRA result.  Regarding the first method, five different NaI(Tl) crystals were measured at the Triangle Universities Nuclear
Laboratory (TUNL) by ANAIS (in collaboration with COSINE researchers from Yale University) \cite{cintas2024measurement}, yielding consistent QF values across the crystals. However, notable variations were observed depending on the energy calibration procedure, attributed to the intrinsic non-linearity of NaI. Overall, the measured QF values were lower than those reported by the DAMA/LIBRA collaboration. In contrast, the second methodology is the subject of the present thesis and will be discussed in detail in Chapter \ref{Chapter:QF}, where the onsite neutron calibration campaign conducted in the ANAIS-112 experiment is presented.

\begin{figure}[t!]
\begin{center}
\includegraphics[width=0.49\textwidth]{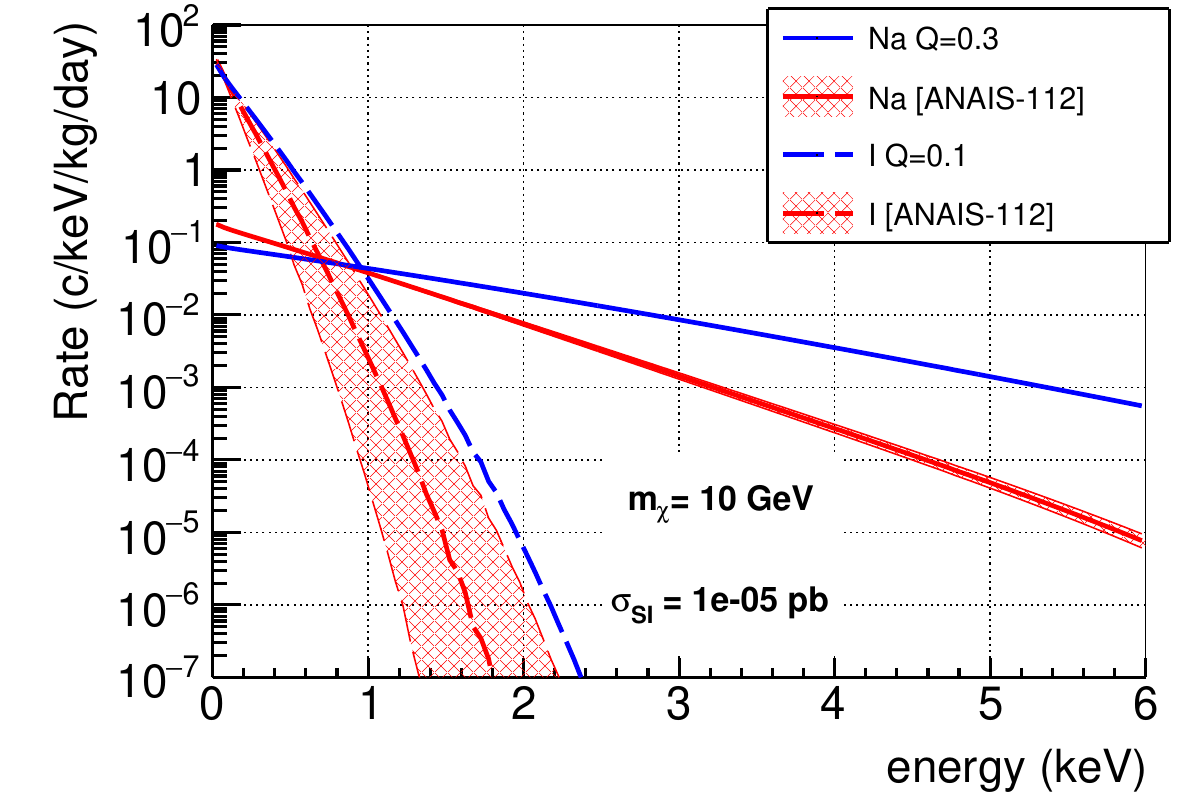}
\includegraphics[width=0.49\textwidth]{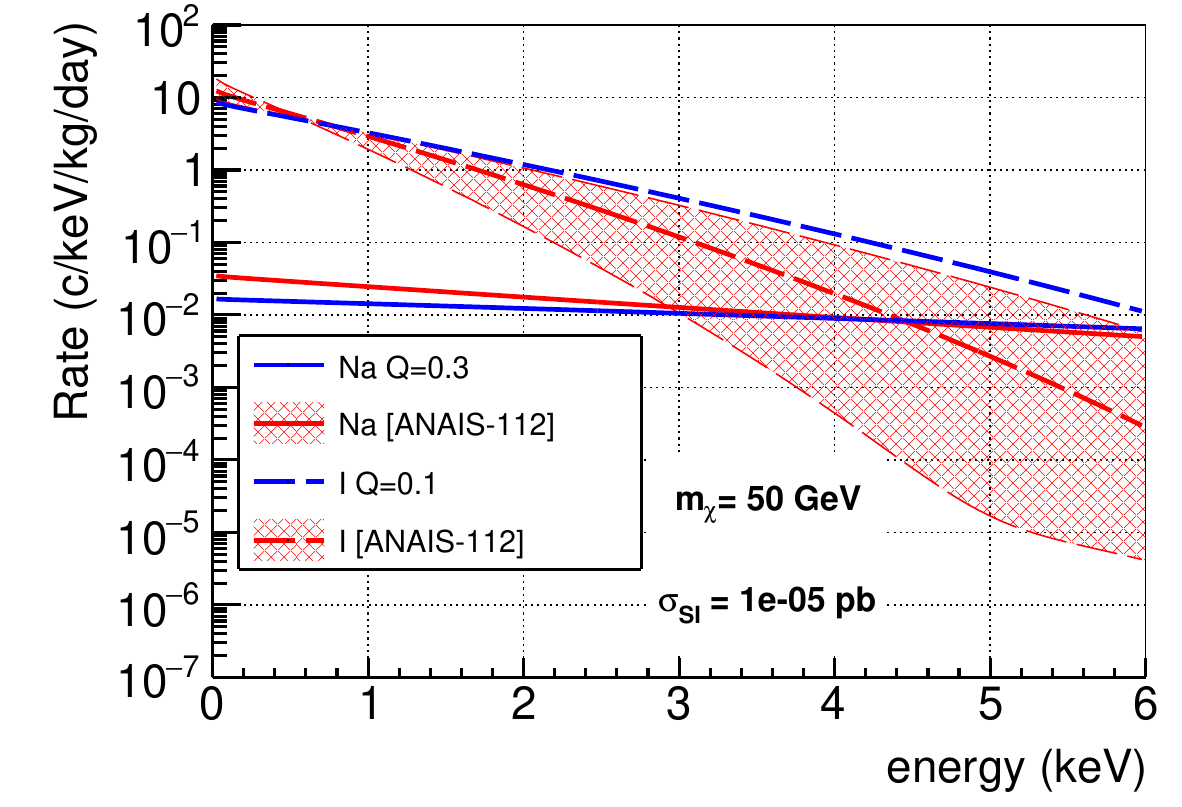}

\caption{\label{rateconqf} Expected DM interaction rates in NaI as a function of energy, expressed in electron-equivalent units. The spectra correspond to WIMPs with masses of 10 GeV (\textbf{left panel}) and 50 GeV  (\textbf{right panel}), assuming a spin-independent cross section \(\sigma_{\text{SI}} = 10^{-5}\)~pb. Different curves are shown for the DAMA/LIBRA QFs \cite{bernabei1996new}, as well as for those used by ANAIS \cite{cintas2024measurement,phddavid}, specifically ANAIS(1), an energy-dependent sodium QF, and a constant 6\% value for iodine (see Chapter \ref{Chapter:QF} for further details).} 
\vspace{-0.5cm}
\end{center}
\end{figure}

The expected DM signal in a NaI(Tl) target is shown in Figure \ref{rateconqf} for WIMP masses of 10 and 50 GeV, assuming a SI WIMP–nucleon cross section of \(\sigma_{\text{SI}} = 10^{-5}\) pb. The figure compares the predicted energy spectra assuming the QFs for sodium and iodine NRs as measured by DAMA/LIBRA \cite{bernabei1996new} and ANAIS-112 \cite{cintas2024measurement}. The critical dependence of the expected signal on the choice and uncertainty of the QF is clearly illustrated. Accurate knowledge of this parameter is essential for the interpretation of results from NaI-based experiments such as ANAIS-112 and for ensuring meaningful comparisons with other experiments using different detector materials. Misestimations in the QF can lead to significant displacements in the region of the WIMP parameter space accessible to the experiment, thereby compromising the robustness of any inferred DM properties. 

For this reason, experiments such as ANAIS-112 place great emphasis on an accurate determination of the QF in order to ensure a precise characterization of their detector response to NRs. In contrast, the status of the QF measurements by the DAMA/LIBRA collaboration remains uncertain. Their QF values, derived from measurements conducted in the 1990s, show significant disagreement with more recent experimental determinations, both in terms of their absolute values, tending to be lower in recent measurements, and in their energy dependence, as DAMA/LIBRA assumes constant QF values. Although the collaboration had previously indicated an intention to reassess their QF determinations, no definitive updates have been made publicly available to date.

Another aspect of the NaI(Tl) detector response that may impact the comparison with the DAMA/LIBRA signal is the well-known non-proportionality in the LY \cite{PhysRev.122.815,Rooney:1997}. This effect causes variations in the light output at the few-percent level in the range of up to 20~keV. In the course of this thesis, this behavior has also been characterized for the ANAIS-112 detectors, as will be presented in Section \ref{nonpropSec}.

\setcounter{chapter}{1} 

\chapter{The ANAIS-112 experiment}\label{Chapter:ANAIS}

\vspace{-0.2cm}
\vspace{0.2cm}

The ANAIS-112 (Annual modulation with NaI Scintillators) experiment \cite{Amare:2018sxx} is a direct DM experiment whose goal is to confirm or refute in a model independent way the positive annual modulation signal reported by the DAMA/LIBRA experiment \cite{Bernabei:2020mon}. ANAIS-112 is taking data at the LSC in Spain since the 3$^{\textnormal{rd}}$ of August 2017. 

In this chapter, the ANAIS-112 experimental set-up is described (Section \ref{set-up}), including the crystals, photomultipliers, electronics and data acquisition, muon veto system and blank module. Section \ref{energyCal} addresses the electron-equivalent energy calibration of the experiment. Section~\ref{Filtering} reviews the filtering protocols used in the ANAIS-112 experiment aimed at rejecting non-bulk scintillation events in the very low-energy region. The background model of ANAIS–112 is presented in Section \ref{BkgModel}. Finally, Section \ref{annualMod} describes the analysis strategy designed for ANAIS-112, followed by the annual modulation results of six years of data.\\

\section{ANAIS-112 experimental set-up}\label{set-up}
\vspace{-0.1cm}

\begin{figure}[b!]
\begin{center}
 \centering
    \begin{minipage}{0.48\textwidth}
        \centering
        \includegraphics[width=\textwidth]{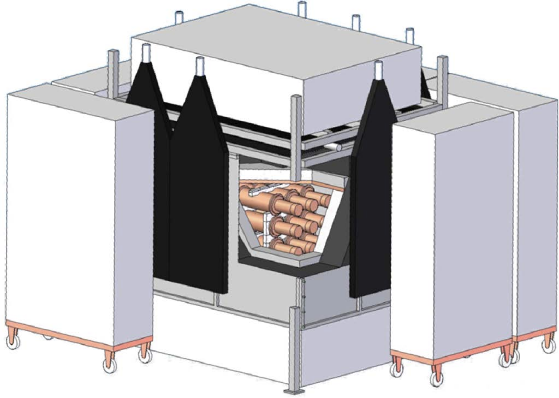}
    \end{minipage}
    \begin{minipage}{0.48\textwidth}
        \centering
        \includegraphics[width=\textwidth]{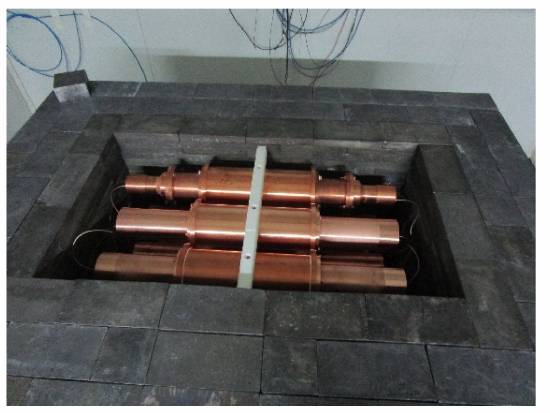}
    \end{minipage}
\caption{\label{ANAISset-up} \textbf{Left panel:} Artistic representation of the ANAIS-112 experimental set-up. \textbf{Right panel:} Image of the ANAIS-112 modules within the shielding.}

\end{center}
\end{figure}

The ANAIS-112 experiment consists of 112.5 kg of NaI(Tl) distributed in nine modules, each with a mass of 12.5 kg. These cylindrical modules, with dimensions of 6.03 cm in radius and 29.85 cm in length, were manufactured by Alpha Spectra Inc.~\cite{AlphaSpectra} and arranged in a 3~$\times$~3 configuration (see Figure~\ref{ANAISset-up}).

The crystals are coated with teflon film and housed in OFE copper. Each end of the crystals is optically coupled to a highly efficient Hamamatsu R12669SEL2 PMT via quartz windows. To monitor stability, the ANAIS-112 modules are designed with a mylar window in the lateral face of each module, allowing for low-energy calibration with external gamma sources.

Before the commissioning of ANAIS–112, several detectors were tested in smaller but functionally equivalent set-ups \cite{Amare:2013lca,Amare:2015uxa,Olivan:2017akd}, according to their respective arrival times at the LSC. Table \ref{caracteristicasANAIScrystals} summarizes this information, together with the powder used for the crystal growth of each ANAIS–112 module. As shown in the table, modules D0 and D1 were grown from the same powder and arrived simultaneously at the LSC in December~2012. Subsequently, module D2, grown from a different powder, was delivered in March 2015. From that point onwards, all remaining modules were grown using the same powder, though in different batches: D3 arrived in March 2016, followed by D4 and D5 in November 2016, and finally D6, D7, and D8, which were delivered together in March~2017.

\begin{table}[t!]
\begin{center}
 \centering
\begin{tabular}{ccc}
\hline
Detector & Powder name & Date of arrival at LSC \\
\hline
\hline
       D0  &   \multirow{2}{*}{<90 ppb K}          &         \multirow{2}{*}{December 2012}               \\
       D1  &             &                        \\
       \hline
       D2  &      WIMPScint-II    &          March 2015              \\
       \hline
       D3  &      \multirow{7}{*}{WIMPScint-III}       &   March 2016                     \\
       \cline{3-3} 
       D4  &             &              \multirow{2}{*}{November 2016}           \\
       D5  &             &                        \\
       \cline{3-3} 
       D6  &             &        \multirow{3}{*}{March 2017}                 \\
       D7  &             &                        \\
       D8  &             &                       \\
       \hline
\end{tabular}
\caption{Powder and date of arrival at LSC for the nine NaI(Tl) detectors of ANAIS-112 manufactured by Alpha Spectra.}
\label{caracteristicasANAIScrystals}
\vspace{-0.5cm}
\end{center}
\end{table}

The ANAIS-112 modules exhibit a remarkable light collection efficiency of approximately 15 photoelectrons/keV. This efficiency is higher and more homogeneous than that of the DAMA/LIBRA modules (5.5-7.5 photoelectrons/keV in phase 1 and 6–10 photoelectrons/keV in phase 2), and comparable to that of the COSINE-100 modules. This efficient light collection is attributed to the outstanding optical properties of the ANAIS crystals, the high quantum efficiency of the Hamamatsu PMTs, and the effective optical coupling between the crystals, quartz windows, and PMTs. Each end of the PMT is coupled to a layer of optical gel, which is followed by a quartz window and a layer of silicone.

 As depicted in Figure \ref{ANAISset-up}, detectors are shielded by a combination of 10 cm of archaeological lead, 20 cm of low-activity lead and an anti-radon box continuously flushed with radon-free nitrogen gas to prevent the entrance of airborne radon inside the
lead shielding. Additionally, there is a muon veto system composed of 16 plastic scintillators, followed by 40 cm of polyethylene bricks and water tanks serving as a neutron moderator.

\subsection{NaI(Tl) crystals} \label{NaIcrystals}

Sodium iodide thallium doped is a well known scintillator, widely used since the 1940s, due to its excellent scintillation properties and cost-effectivenesss, the latter largely due to the ability to grow large-volume crystals. In spite of its hygroscopic nature, it remains a crystal of choice for low-background applications.

Scintillation in inorganic materials like NaI(Tl) arises from the band structure of the crystal lattice \cite{leo1994techniques, Knoll:2000fj}. When charged particles deposit energy in the crystal, electrons are excited from the valence band to the conduction band. In pure insulating materials, photon emission is rare and too energetic to be visible. However, doping with small amounts of impurities, such as thallium, introduces discrete energy levels within the band gap, allowing electrons to de-excite to the valence band through these luminescent centers. The emission decays exponentially in time, with a rate determined by the lifetime of the excited states. The thallium concentration in NaI(Tl) is crucial for optimizing its LY. If the concentration is too low, energy is not efficiently converted into photons, while high concentrations increase self-absorption, reducing the LY. The optimal thallium concentration is $\sim$ 0.1\% of the molar mass at room temperature.

NaI(Tl) has a high LY, producing $\sim$40 photons/keV. Its primary scintillation component has a decay time of 230 ns \cite{Birks:1964zz}, with additional slower components around 1.5 $\mu$s and 0.15~s~\cite{Cuesta:2013vpa}. The LY increases slightly as the temperature decreases up to 240 K, after which it significantly declines \cite{Lee:2021aoi}. Scintillation time constants also change with temperature, with different rates for slow and fast components \cite{Lee:2021jfx, Schweitzer1983}. The high LY of NaI(Tl) and the emission peak  at 420 nm allow for building very high light collection detectors, profiting from the good matching with the spectral sensitivity of bialkali PMTs. Moreover, NaI(Tl) exhibits high photoelectric cross sections over a broad energy range, which makes it particularly suitable for spectroscopic applications. These properties have facilitated their widespread use as scintillation detectors for various types of ionizing radiation across a broad range of applications, such as medical imaging and nuclear spectroscopy.

NaI(Tl) crystals are an excellent target for DM searches, as they are sensitive to both low-mass WIMPs via sodium interactions and massive WIMPs via iodine interactions. Both elements are fully sensitive to SD interactions. The high LY of NaI(Tl) enables the achievement of low energy thresholds. Despite their versatility and availability in various sizes and shapes, NaI(Tl) crystals are fragile and hygroscopic, requiring careful handling and storage in airtight containers to prevent degradation. 

As previously mentioned, NaI(Tl) scintillators are characterized by an excellent LY, particularly when the excitation process involves photons transferring their energy to electrons. However, it is well established that ionizing particles with different stopping power produce different amounts of scintillation light for the same deposited energy. This variation in light output depending on the particle type is mainly attributed to the saturation of color centers induced by highly ionizing particles, which deposit energy with high spatial density.

As a result, a precise understanding of the detector response to NRs is essential when using NaI(Tl) crystals for WIMP-nucleus scattering searches. This is especially important in scintillators, since NRs generate significantly less light than ERs of the same energy. This reduction in LY is quantified by the QF, a critical parameter which, to date, constitutes the most important systematic uncertainty in the comparison of ANAIS-112 data with DAMA/LIBRA. The QF estimation for the ANAIS-112 crystals will be addressed in detail in Chapter \ref{Chapter:QF} of this thesis using onsite neutron calibrations within the same ANAIS-112 experimental set-up.

Additionally, non-linearities in the response of NaI(Tl) crystals, a common behavior in scintillators, have been reported \cite{Rooney:1997, Choong:2008, Payne:2009, Khodyuk:2010ydw, Payne:2011}. Ideally, the number of scintillation photons generated should be proportional to the deposited energy. However, in practice, a non-linear relationship between LY and deposited energy, referred to as non-proportionality, is observed. This phenomenon will be discussed in Section \ref{nonpropSec}, where the non-proportionality characteristics of NaI(Tl) scintillators will be examined using ANAIS-112 data from different isotopes and populations. 

\subsection{Photomultiplier tubes}

\begin{figure}[b!]
\begin{center}
 \centering
    \begin{minipage}{0.56\textwidth}
        \centering
        \includegraphics[width=\textwidth]{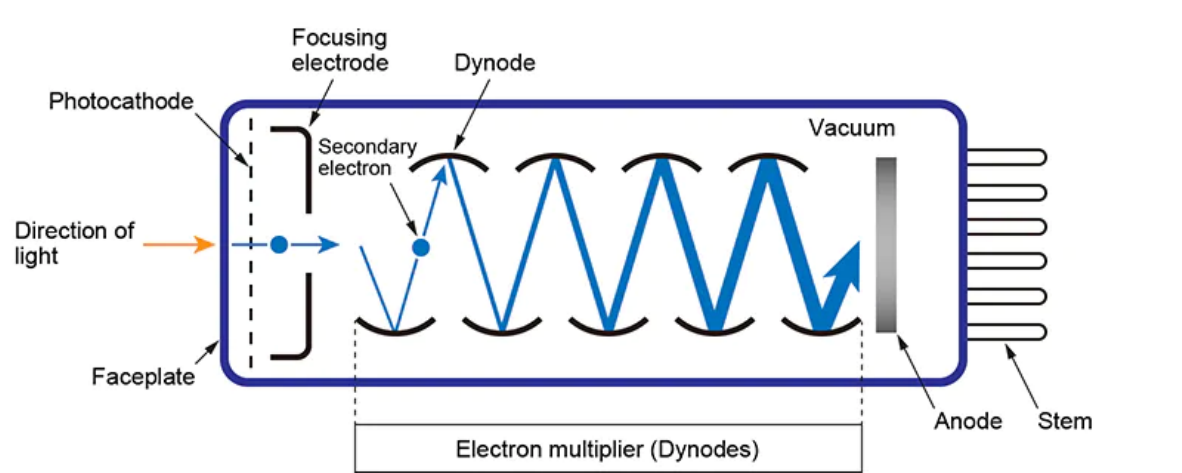}
    \end{minipage}
    \hfill
    \begin{minipage}{0.4\textwidth}
        \centering
        \includegraphics[width=\textwidth]{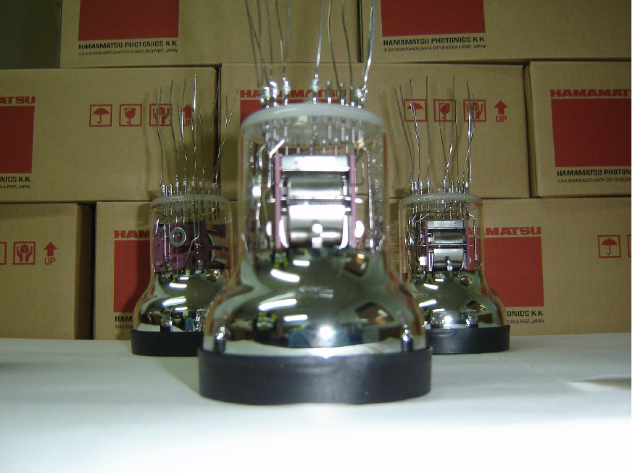}
    \end{minipage}
\caption{\label{ANAISPMTs} \textbf{Left panel:} Schematic view of a photomultiplier tube \cite{matsusada}. \textbf{Right panel:} Hamamatsu R12669SEL2 photomultipliers used in the ANAIS-112 experiment. }

\end{center}
\end{figure}
PMTs are highly sensitive light detectors that play an essential role in numerous fields, including scientific research, medical diagnostics and industrial applications. 

A PMT consists of a tube operating under vacuum conditions with a window made of quartz or borosilicate, onto which a thin layer of photosensitive material, known as the photocathode, is deposited. Most photocathodes are composed of alkali metals with a low work function, defined as the energy required to remove an electron from a surface. Inside the tube, there are electron multipliers (dynodes) and a collecting anode \cite{Leo:1987kd,Knoll:2000fj}. The number of amplification stages (or number of dynodes) and the different dynode designs available, besides the PMT dimensions, allow an offer of PMTs with very different gain regimes and time responses. A scheme of a PMT is shown in Figure \ref{ANAISPMTs} (left panel).

PMTs convert the incident light into a measurable electric current at the anode. PMTs sensitivity allows applications for the detection of single photons, although in the readout of scintillation detectors, typically they detect hundred-thousands of photons.  When a photon hit the photocathode, an electron is emitted via the photoelectric effect with a probability determined by the QE. This electron is then accelerated and focused onto the first dynode due to the voltage gradient. At the first dynode, the electron's kinetic energy is converted into several secondary electrons. These secondary electrons are subsequently accelerated to the next dynode, generating additional secondary electrons. This cascading process continues through the dynode chain. Ultimately, the electrons emitted by the final dynode are collected at the anode after suffering a multiplication process corresponding to a gain of the order o 10$^6$-10$^7$ depending on the PMT model and the high voltage (HV) operation point. 

\begin{figure}[b!]
\begin{center}
 \centering
    
        \includegraphics[width=0.49\textwidth]{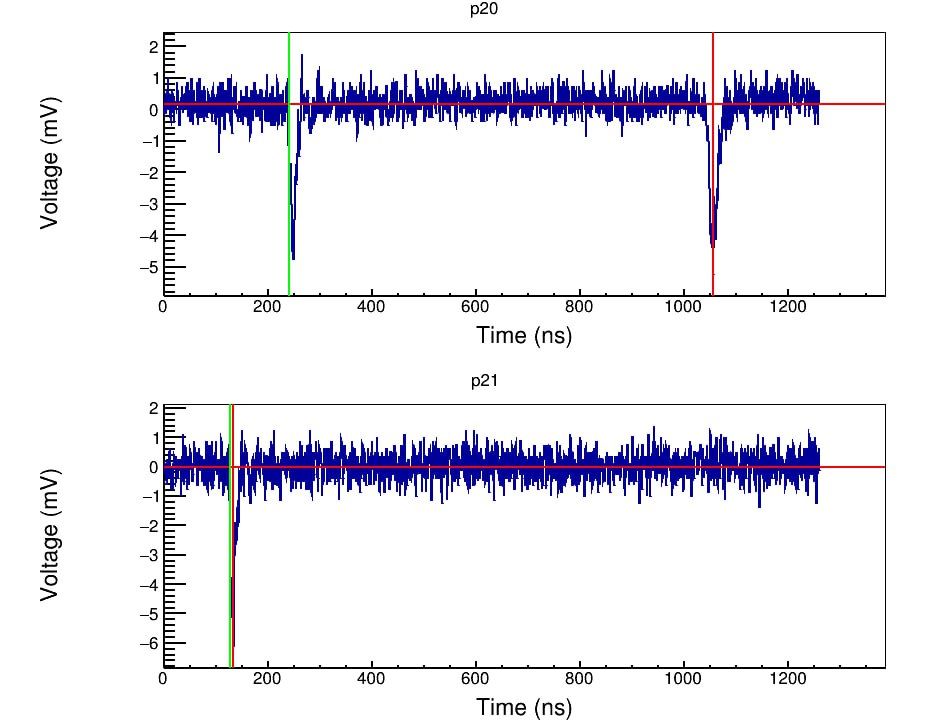}
        \includegraphics[width=0.49\textwidth]{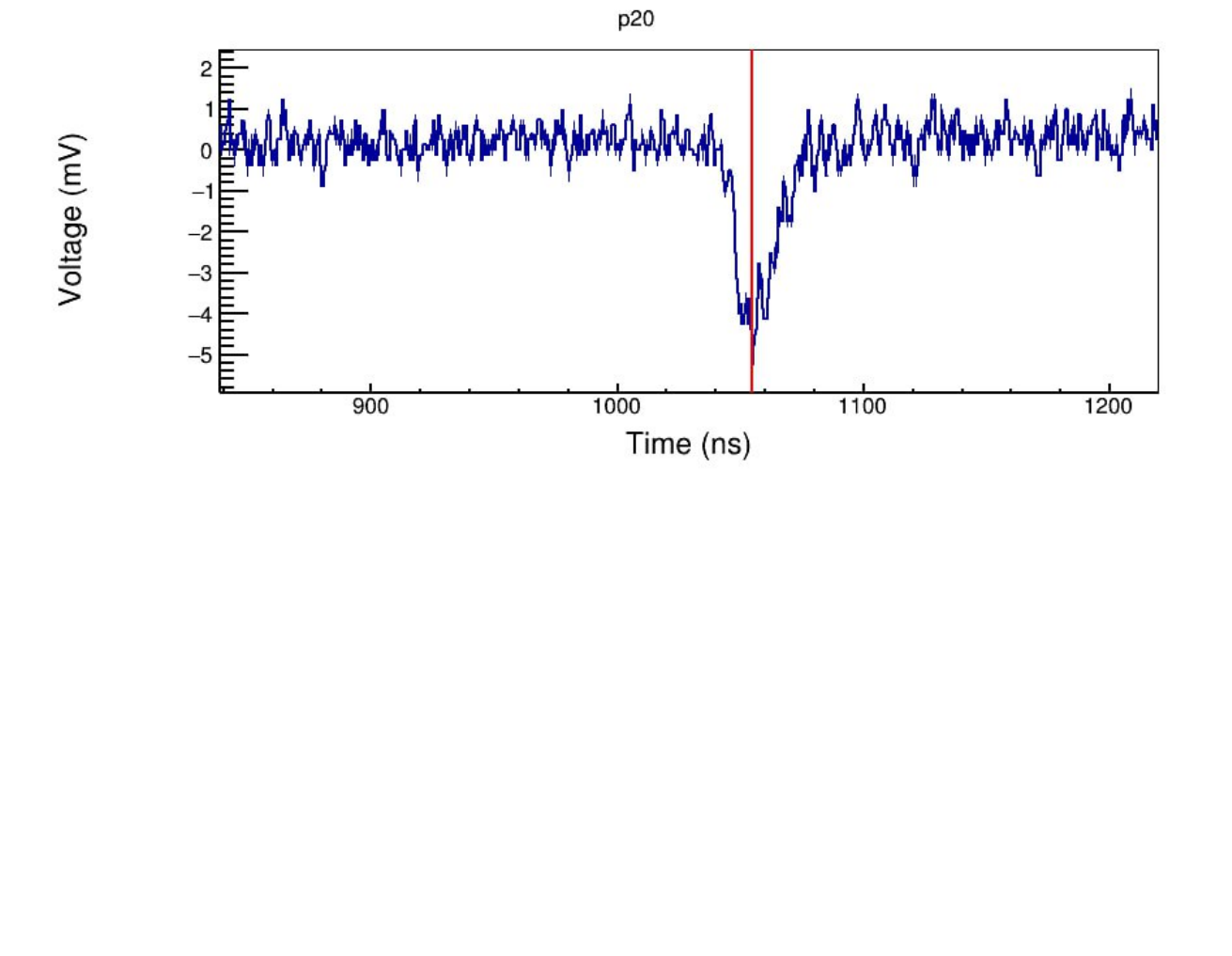}

\caption{\label{SERDavid} \textbf{Left panel:}  Example of a low-energy pulse with a low number of photoelectrons, where a single phe from the tail is selected for the SER estimation. The two traces correspond to the two
PMT signals. \textbf{Right panel:} Zoomed view of the single photoelectron. The green vertical lines indicate the first phe identified in the pulse, while the red vertical lines mark all detected peaks. The horizontal lines represent the baseline level.}

\end{center}
\end{figure}

Since their development in the 1930s, PMTs have achieved significant success and are used in a wide range of contexts and applications. This success is attributed to their high gain, ultra-fast response, low noise, excellent linearity, broad dynamic range, long lifespan, and large collection area. High QE models are available, with peak QE around or even above 50\% covering blue and near-UV regions.

In order to understand the scintillation signal in the very low-energy region, a good understanding of the PMT response to low light intensities is essential. A photoelectron (phe) is defined as the current pulse at the output of the PMT that corresponds to a single photon that hits the photocathode
and ejects an electron, which is then accelerated by the electric field toward the first dynode, where it triggers the generation of an electron cascade. The most basic PMT signal can be characterized by the so-called Single Electron Response (SER) distribution, which corresponds to a single phe emitted in the photocathode. The right panel of Figure \ref{SERDavid} shows the corresponding waveform.

The SER represents the average area of a phe and is obtained by averaging many such events. Two effects must be carefully considered to obtain a representative SER distribution. First, trigger bias must be avoided; this is done by not using the population of photoelectrons that triggers the readout electronics. Instead, isolated photoelectrons from the tail of higher-energy events are selected (see left panel of Figure~\ref{SERDavid}). Second, pile-up must be prevented, which is achieved by selecting low-energy events with a low number of~phe.


PMTs have some limitations that may affect their suitability for DM search experiments. One limitation is the dark rate, defined as the rate of events occurring without light excitation of the photocathode. Thermoionic noise caused by the thermal excitation of electrons in the photocathode is the dominant source of dark current at room temperature. However, its contribution is reduced in ANAIS-112 by measuring with two PMTs in coincidence. Moreover, the large size and complex material structure of PMTs pose challenges for experiments requiring high radiopurity. For instance, radioactive isotopes such as \(^{40}\text{K}\) and those in the natural decay chains distributed in the transparent glass of PMT housings are major contributors to the radioactive background in experiments requiring low backgrounds (see Section \ref{tablaPMT}). These contributions can arise from Cherenkov radiation, photons from radioactive decays, or electrons from \(\beta\)-decays hitting the dynodes and generating a signal. High trigger rates can result in higher dead times and a more complicated selection of real bulk scintillation events.


The Hamamatsu R12669SEL2 model (see Figure \ref{ANAISPMTs}, right panel) is selected for ANAIS-112 due to its favorable characteristics. This model features a bialkali photocathode, composed of two alkali metals and antimony (Sb-Rb-Cs), which ensures high QE (>33\%) with a peak response at 420 nm. It has a dark current below 500~Hz and ten dynode stages for signal amplification, achieving a gain factor of $\sim$10$^6$ at the nominal voltage. This PMT model is similar to the PMTs used in the DAMA/LIBRA-phase2 experiment \cite{Bernabei:2020mon,Bernabei:2012zzb}, and was also chosen by the KIMS/COSINE collaboration \cite{Adhikari:2017esn,Kim:2014toa}. The levels of radioactive contamination in the PMTs used in ANAIS-112, derived from measurements using high purity germanium (HPGe) spectrometry at LSC \cite{amare2019analysis}, will be discussed in Section \ref{BkgModel}. 

Silicon photomultipliers (SiPMs) offer an alternative with the potential to overcome some of the intrinsic limitations of photomultiplier tubes (PMTs). Their application for reading out the scintillation signals of NaI and NaI(Tl) crystals is currently under investigation in the forthcoming ANAIS+ project.


\subsection{Data acquisition system}\label{DAQsec}

The ANAIS electronics system is a robust and scalable design adapted to VME electronics~\cite{maolivan}. Key considerations include a stable trigger at the phe level, minimal electronic noise, complete digital processing of the two signals from the two PMTs of each module, and redundant information (high and low energy paths with different gains).

\begin{figure}[b!]
\begin{center}
 \centering
   \includegraphics[width=0.9\textwidth]{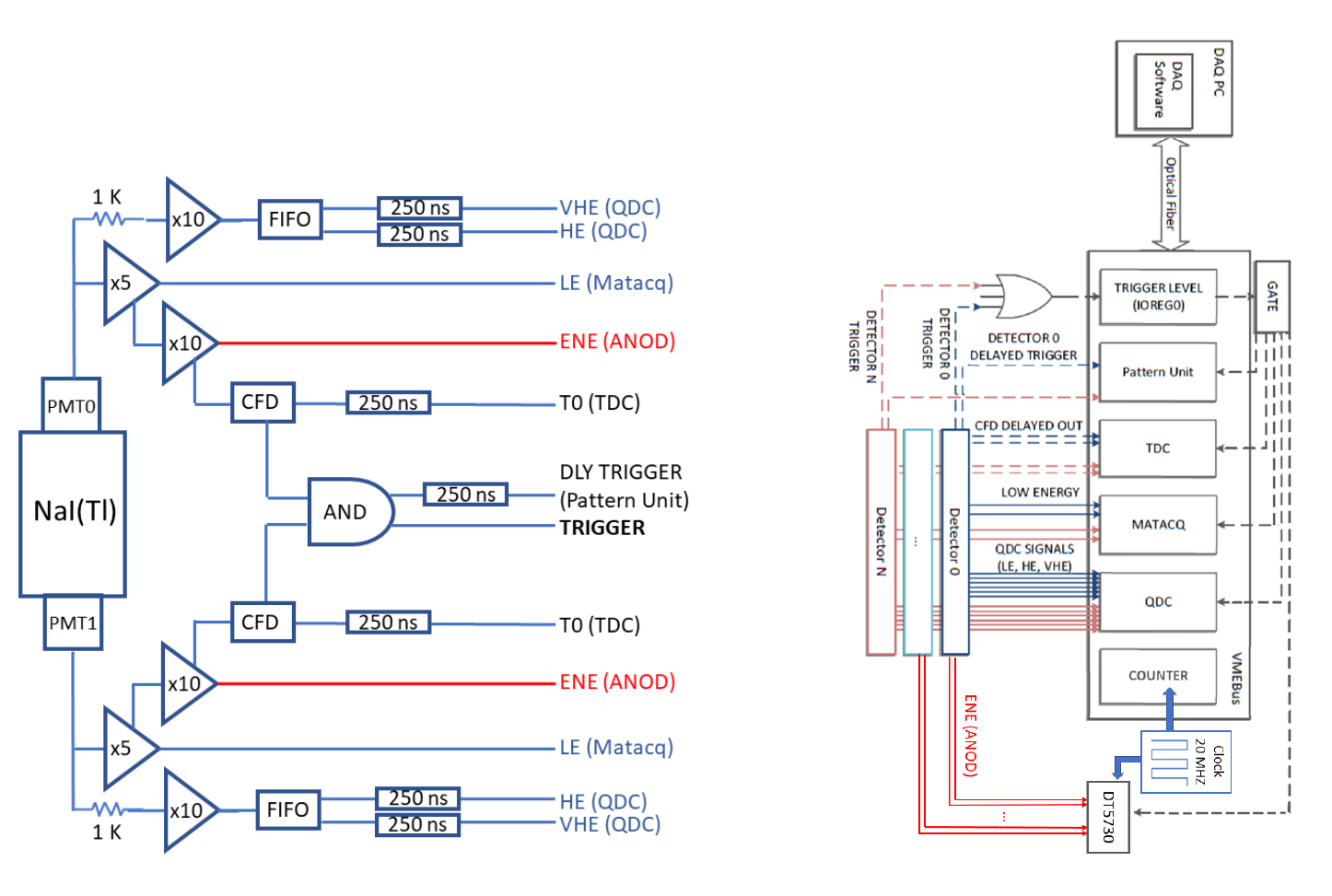}
    
\caption{\label{ANAISDAQ} Scheme of the ANAIS DAQ system, with the ANOD DAQ highlighted in light red, shown for a single detector \textbf{(left panel)} and for the complete set-up \textbf{(right panel)}. }

\end{center}
\end{figure}

Figure \ref{ANAISDAQ} shows an scheme of the ANAIS-112 signal processing chain. The charge signal from each PMT is processed and recorded separately, divided into a trigger signal performed by a CAEN N843 constant fraction discriminator (CFD), a low energy signal sent to a MATACQ-CAEN V1729A digitizer, and a high energy signal attenuated  and sent to CAEN V792 charge-to-digital converter (QDC). Triggering each module is achieved by the coincidence (logical AND) of the two PMT trigger signals within a 200~ns window, while the main acquisition trigger is the logical OR of individual detectors in 1 $\mu$s window. Low-energy signals are recorded with 2520 samples at a 2~GS/s sampling rate, providing a 1260 ns time acquisition window (pretrigger around 250 ns) with a time resolution of 0.5~ns/point. The dynamic range is 1 V with a 14-bit resolution, resulting in a vertical resolution of 0.06~mV/point. The transfer of the digitized pulse to the main PC, the readout of the other electronic modules and the rearm of the acquisition systems implies a dead time per event of about 4 ms.

For each event, the waveforms of the two PMT signals for each module triggering, besides the corresponding QDC signals, the time of the event and a variable including the information of the modules triggering are saved in a ROOT file.  Energy spectra in the low energy range are built offline by integrating the PMT waveforms.

The ANAIS-112 DAQ has operated smoothly for almost eight years of data acquisition. However, it has limitations that hinder the understanding and removal of anomalous low-energy events. These events are believed to originate at or near the PMTs. They are characterized by an asymmetric light-sharing among the two PMTs and hard to distinguish from bulk NaI(Tl) scintillation, as will be discussed in Section~\ref{ANODfiltering}. The 1260 ns window of the ANAIS DAQ system prevents the recording of pulse tails, which could aid in removing these anomalous scintillation events. 

\begin{table}[t!]
\centering

\begin{tabular}{ccc}
\hline
                                   & ANAIS & ANOD \\
 \hline \hline       
 Time acquisition window (ns)         & 1260  & 8000 \\
Number of samples                  & 2520  & 4000 \\
Sampling rate (GS/s)               &  2 & 0.5  \\
Temporal resolution (ns/point)               &  0.5 & 2  \\
Pretrigger (ns) & 250   & 715  \\
Time integration window (ns)         & 1000  & 1000, 2000 ... \\
Dead time (ms)                 & 4.5   & -   \\
\hline
\end{tabular}
\caption{\label{ANAISANODdiff} Comparison between ANAIS and ANOD systems, highlighting the main differences between both DAQ systems currently operating in parallel in ANAIS-112.}
\vspace{-0.3cm}
\end{table}

With the aim of reducing the background at low energy by allowing a better non-bulk scintillation event rejection and eliminating the dead time, a new DAQ system called ANOD (\underline{A}NAIS \underline{No} \underline{D}ead-time) has been implemented and operated in parallel with the ANAIS DAQ. ANOD was installed at the LSC at the beginning in Winter 2023. The first version of the system used a CAEN DT5730 digitizer with 8 channels, which enabled data acquisition from the two PMTs of four modules (modules 0, 2, 5, and~8). 

In Winter 2024, the system was upgraded with the installation of a CAEN VX2730 digitizer featuring 32 channels. This upgrade allows for the digitization of data from all ANAIS modules, including the blank module. The digitizer offers a configurable acquisition window, a sampling rate of 500 MS/s, 14-bit vertical resolution, and an internal buffer of 83.886 MS per channel. Data is transferred asynchronously from the digitizer to the main PC via USB 3.0, achieving a transfer rate of 280 MB/s, which ensures operation with no dead time. It is worth noting that ANOD operates using the same trigger as ANAIS, and that the synchronization between both systems is based on a common clock. The ANOD DAQ scheme is also shown in Figure~\ref{ANAISDAQ}, highlighted on the general ANAIS system. Table~\ref{ANAISANODdiff} summarizes the main differences between the ANAIS and ANOD DAQs.


\begin{figure}[t!]
\begin{center}
 \centering
   \includegraphics[width=1.\textwidth]{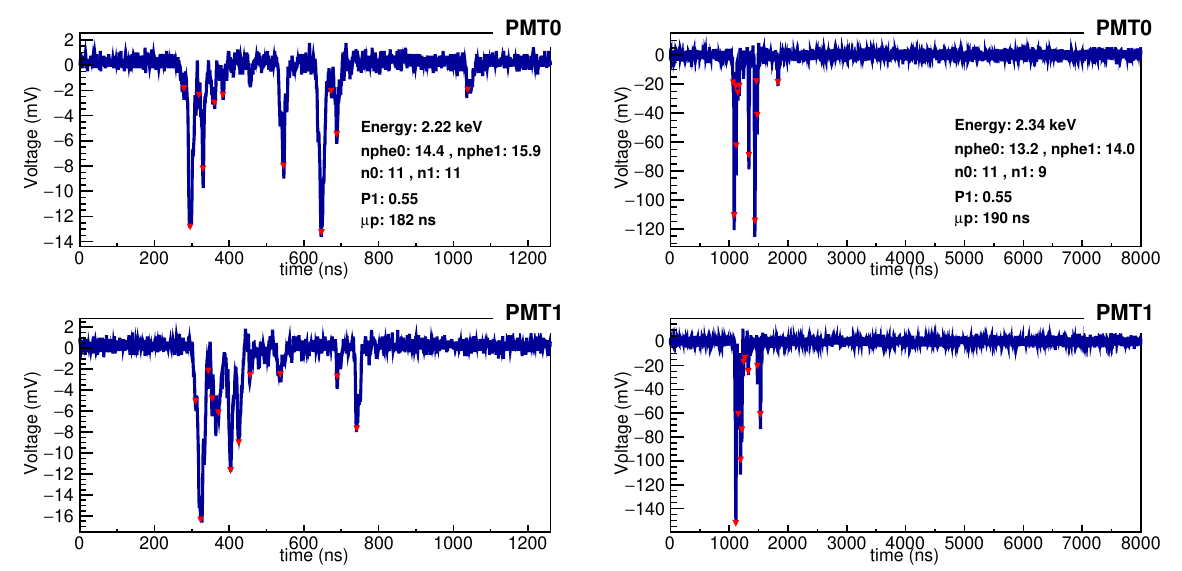}
    
\caption{\label{examplePulseANAISAnod} Example of the same bulk scintillation event acquired with ANAIS (\textbf{left panel}) and ANOD (\textbf{right panel}). The two traces in each panel correspond to the two PMT signals, while the legend displays the event energy, the number of photoelectrons detected in each trace (nphe0 and nphe1), the number of peaks detected by the peak identification algorithm at every trace (n0 and n1, shown as red triangles) and the PSA parameters P1 and $\mu$p (see Section \ref{Filtering}).  }

\end{center}
\end{figure}

ANOD acquires 4000 points at a frequency of 500 MS/s, resulting in an acquisition window of 8000 ns (pretrigger around 715 ns) with a time resolution of 2 ns/point. Figure~\ref{examplePulseANAISAnod} shows an example of the same bulk scintillation event using both DAQ systems, where the difference in the size of the acquisition window between the two systems, 1260 ns in ANAIS and 8000 ns in ANOD, becomes clear. The red triangles in the figure represent the number of peaks detected by a peak identification algorithm. The interpretation of the pulse shape parameters shown in the figure and an introduction to new filtering developments based on ANOD data, are presented in Section~\ref{Filtering}.

\subsection{Muon veto system}\label{muonsec}

Muons interacting directly with NaI(Tl) detectors are expected to produce high energy depositions far exceeding the ROI for DM analysis. Yet, these particles can also interact with other parts of the experimental set-up. Such interactions might produce events that fall within the ROI of ANAIS-112, potentially threatening the experiment goals. For instance, muons can generate neutrons within the ANAIS-112 shielding, or interact with components like PMT glass, quartz windows, and other materials, creating Cherenkov light or weak scintillation. Moreover, given that muon energy deposits can create long-lived phosphorescence within NaI(Tl) detectors lasting around 0.15 s \cite{TesisClara}, photons arriving in the tail of a muon pulse can lead to multiple detection events in the ANAIS DAQ \cite{Amare:2018sxx} that increases the total acquisition rate and can mimic low-energy DM signals.

By operating the ANAIS–112 experiment at the LSC under 2450 m.w.e., the cosmic ray muon flux is reduced by a factor of $\sim$10$^5$ compared to measurements taken at sea level. While this level of flux is considerably reduced, it is still enough to compromise the ANAIS–112 detectors performance. 

\begin{figure}[t!]
\begin{center}
\includegraphics[width=1.\textwidth]{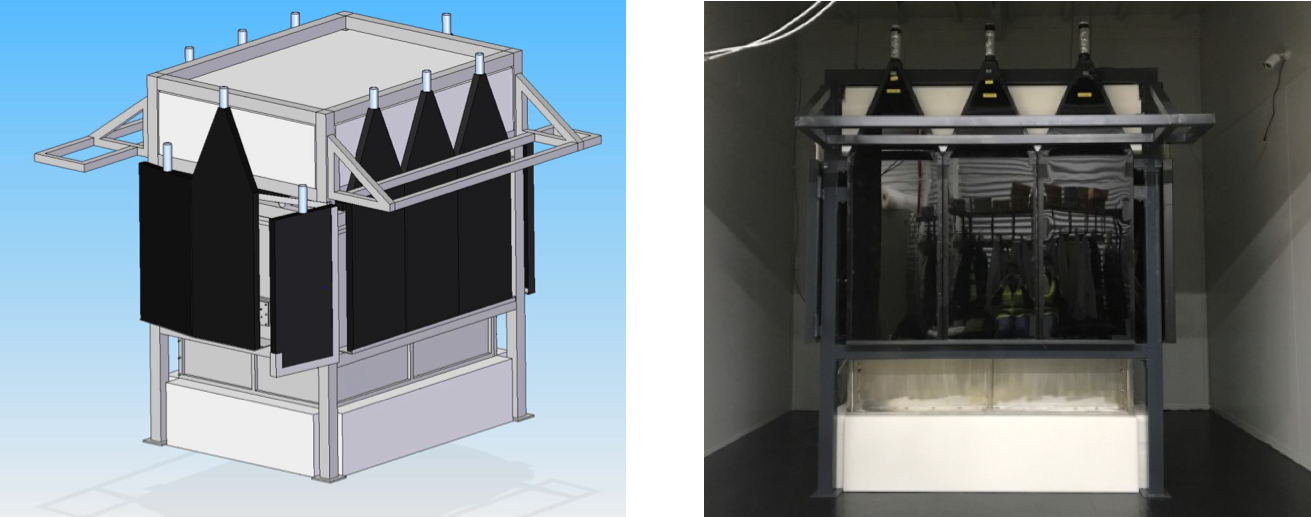}

\caption{\label{muonveto} \textbf{Left panel:} Artistic view of the ANAIS–112 experiment with the veto system. \textbf{Right panel:} Picture taken during the commissioning of ANAIS–112. }
\end{center}
\end{figure}

Therefore, in order to identify and reject muon-related background events, the ANAIS-112 experiment is equipped with a muon veto system consisting of 16 plastic scintillators covering the top and four sides of the anti-radon box (see Figure \ref{muonveto}). This system is designed to detect the residual muon flux that survive its passage through the earthen overburden and interact with the underground detector. By studying the correlation between muon interactions in the plastic scintillators and events in the NaI(Tl) crystals, a cut has been successfully designed to remove events induced by muons, by selecting a time window of 1 s after a veto signal for this purpose.

Cosmic ray muons have been extensively considered as a possible cause for the annual modulation observed by DAMA/LIBRA \cite{nygren2011testable,blum2011dama}, given that muon flux varies seasonally with upper atmospheric temperature changes \cite{agafonova2019measurement}. It is worth highlighting that the DAMA/LIBRA experiment does not have a muon detection system. 

Although the muon flux at LNGS \cite{Bernabei:2020mon} is about an order of magnitude lower than at LSC, residual muons could still interact directly with the DAMA/LIBRA crystals or produce muon-related events that might mimic DM signals. However, DAMA/LIBRA has largely rejected this explanation because the muon flux peaks in early July, approximately one month after the expected peak of DM flux on June 2\textsuperscript{nd} \cite{Bernabei:2012wp,Bernabei:2014tqa}. Data from the ANAIS-112 muon veto system have been used to analyze the characteristics of the muon flux at LSC. Preliminary results indicate a maximum on June 26 ± 6 days, with a relative modulation amplitude of 1.3\% \cite{tfgraul}.


\begin{figure}[t!]
\begin{center}
\includegraphics[width=0.5\textwidth]{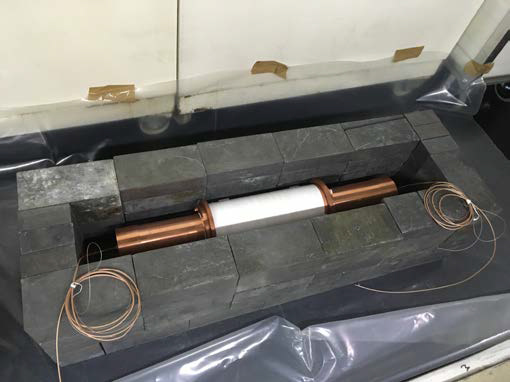}

\caption{\label{blankmodule} Picture of the ANAIS Blank module in the ANAIS hut enclosed in a specific 10~cm
thick lead shielding.}
\end{center}

\end{figure}

\vspace{-0.3cm}
\subsection{Blank module} \label{BlankModSec}

The Blank module is similar to the ANAIS-112 modules, except that it lacks a NaI(Tl) crystal. It consits of two PMTs connected to quartz optical windows within a copper housing whose interior surface is coated with a teflon diffusor film (see Figure~\ref{blankmodule}). Integrated in the ANAIS hut at LSC in August 2018, it is enclosed in a 10~cm thick lead shielding, independent from the main ANAIS–112 set-up, that can be flushed
with Radon-Free Air~(RFA) or nitrogen gas. The readout electronics and data analysis employed for this module are identical to those used for the ANAIS–112 modules. 

The Blank module aims to improve the understanding of low-energy events associated with non-bulk NaI(Tl) scintillation originated
in the PMTs, and to determine the efficiency of the ANAIS-112 filtering protocols in rejecting such events.

\section{ANAIS-112 energy calibration}\label{energyCal}


Energy calibration is a critical aspect of any DM search experiment, particularly for ANAIS,
 which aims to explore the same low-energy region where the DAMA experiment reports a signal. However, 
scintillating crystals exhibit non-proportional light response, as discussed in Section \ref{NaIcrystals}. Therefore, 
special care has been taken in ANAIS to develop a robust and reliable calibration protocol.

The following section describes the electron-equivalent energy calibration procedures followed in ANAIS-112. However, since WIMPs are expected to interact predominantly with detector nuclei, a dedicated neutron calibration program has been implemented within the experiment to characterize the ANAIS detector response to NRs. This program will be detailed in Chapter~\ref{Chapter:QF}.

Section \ref{LEcalibration} describes the external low-energy calibration system specifically designed for ANAIS-112.
 This system enables continuous monitoring of the detector gain at low energies. Additionally, the section discusses two types of internal radioactive
 contaminations that can be identified through coincidence with other crystals, providing bulk scintillation events in the ROI
 that serve as internal low-energy calibration peaks.

 Section \ref{nonpropSec} presents a study of the non-proportionality behavior of the ANAIS crystals carried out in this thesis,
 highlighting the importance of incorporating this effect into the calibration strategy. In ANAIS-112, this is addressed by dividing the calibration procedure into distinct energy ranges. Finally, the implementation of the energy calibration is described for three distinct energy ranges: very low energy ([1–20] keV) and low energy ([20–150]~keV) in Section~\ref{implementationenergycal}, and high energy ([150–1600] keV) in Section~\ref{HEcal}.

\subsection{External system and internal lines for low-energy calibration}\label{LEcalibration}

\begin{figure}[b!]
\begin{center}
\includegraphics[width=1.\textwidth]{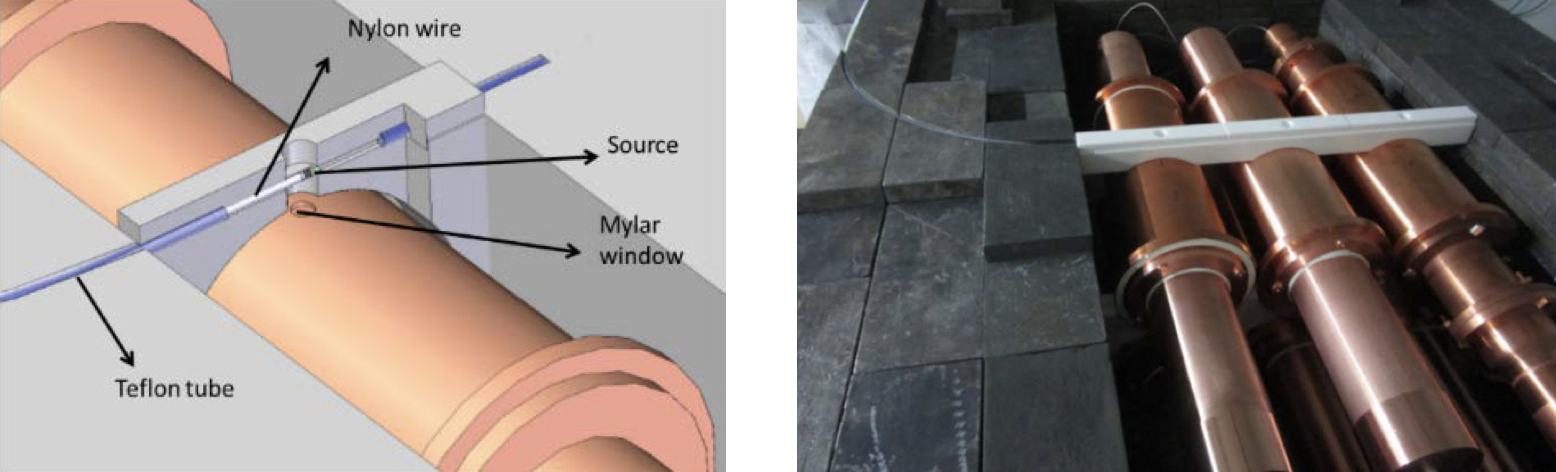}

\caption{\label{calibration} \textbf{Left panel:} Schematic diagram of the ANAIS–112 calibration set-up. \textbf{Right panel:} Picture of ANAIS–112 calibration system. }

\end{center}
\end{figure}

The ANAIS–112 modules have a mylar window on the lateral face, allowing for low-energy calibration with external gamma sources. The calibration is performed every two weeks using \(^{109}\)Cd sources. To minimize downtime, all nine modules are calibrated simultaneously with a multi-source system on flexible wires. The sources, introduced into the shielding during calibration, face the mylar windows while keeping the system sealed and radon-free. The calibration system is illustrated in Figure~\ref{calibration}.

\begin{figure}[t!]
    \centering
    \begin{subfigure}{0.49\textwidth}
        \centering
        \includegraphics[width=\linewidth]{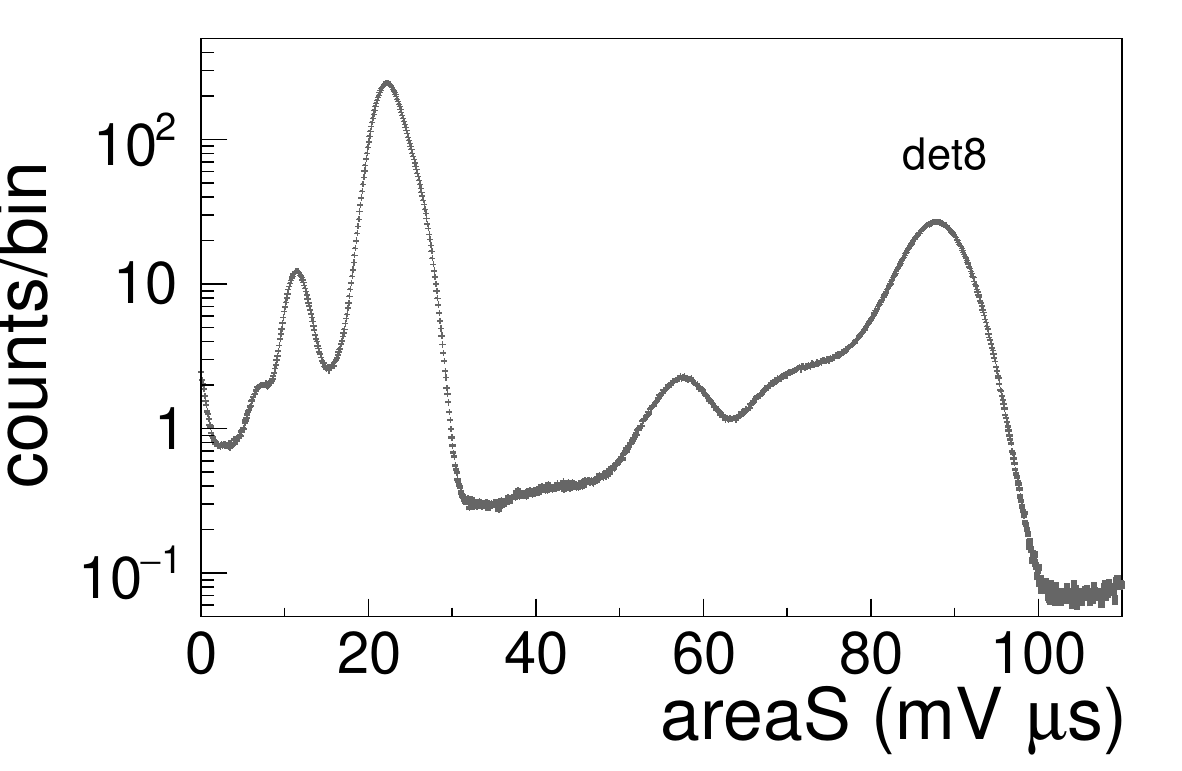}
    \end{subfigure}
    \begin{subfigure}{0.49\textwidth}
        \centering
        \includegraphics[width=\linewidth]{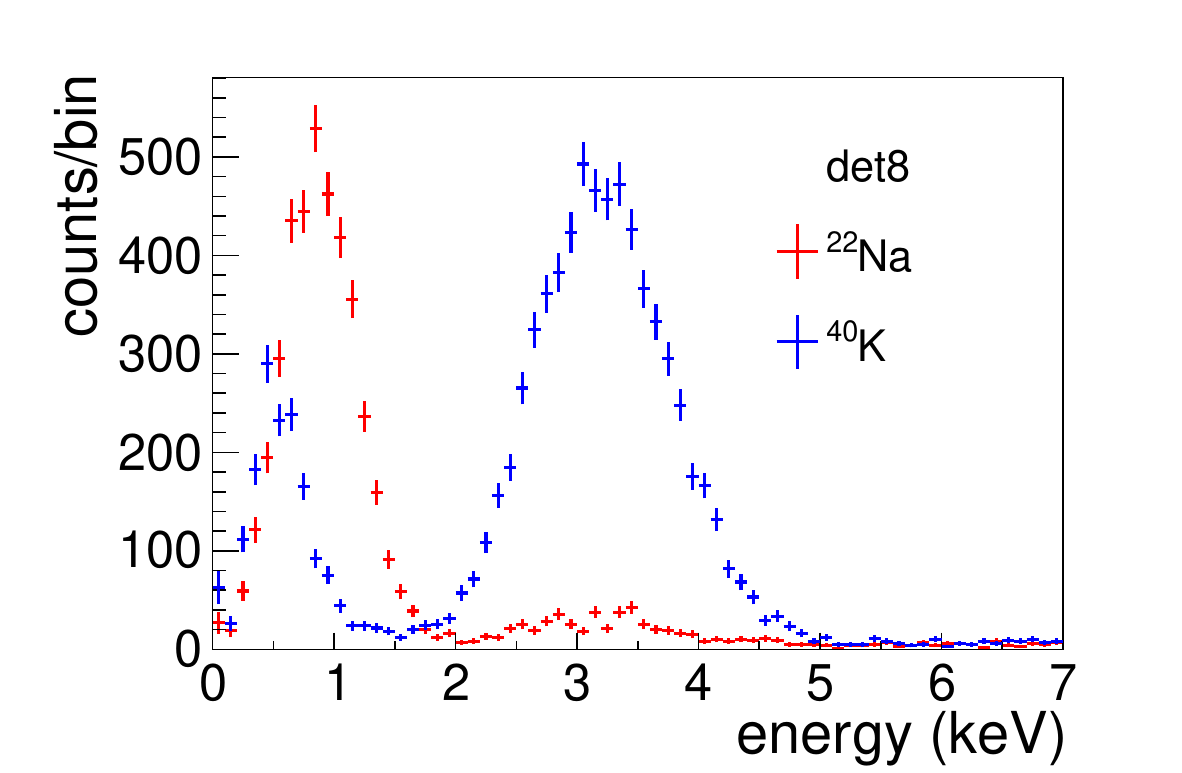}
    \end{subfigure}

    \caption{\textbf{Left panel}: Measured spectrum for the D8 module from one of the periodic
$^{109}$Cd calibration runs. \textbf{Right panel}: 0.9 and 3.2 keV \textsuperscript{22}Na and \textsuperscript{40}K deposits detected in the D8 module selected in coincidence
with a high energy gamma in another module during the first five
years of operation.  }
    \label{calSpectrum}
\end{figure}

The left panel of Figure~\ref{calSpectrum} shows the measured spectrum from one of the periodic \textsuperscript{109}Cd calibration runs. The $^{109}$Cd calibration produces a peak at 88.0~keV due to the decay to $^{109}$Ag via electron capture (EC). The peak around 60 keV results from the escape of iodine K-shell x-rays following the absorption of the 88.0 keV photon. K$_{\alpha}$ and K$_{\beta}$ Ag X-rays are also emitted with average energies of 22.1 keV and 25.1~keV, respectively, merging into a single broader peak due to the detector resolution. In addition, the plastic housing of the source contains a certain unknown amount of bromine, which under $^{109}$Cd irradiation, produces a calibration line at an averaged energy of 12.1 keV from K$_{\alpha}$ and K$_{\beta}$ Br X-rays.

The right panel of Figure \ref{calSpectrum} shows the measured lines in D8 module produced by the internal contamination of \(^{40}\)K and \(^{22}\)Na within the bulk crystal. These isotopes decay via EC, creating a vacancy in the inner atomic shells and leaving the daughter atom in an excited state. The subsequent atomic de-excitation releases 0.9~keV for \(^{22}\)Na and 3.2 keV for \(^{40}\)K from K-shell EC. Simultaneously, high-energy gamma photons are emitted, with energies of 1274.5 keV for \(^{22}\)Na and 1460.8 keV for \(^{40}\)K. The binding energy from the atomic shell is fully absorbed in the same crystal where the decay occurs, while the gamma photon may escape and interact with a different detector, producing a clearly identifiable coincident event. As shown in the right panel of Figure~\ref{calSpectrum}, in the case of \(^{40}\)K the L-shell peak around 0.3 keV can also be clearly observed, demonstrating the high trigger efficiency of the experiment.

\vspace{0.5cm}

\subsection{ANAIS-112 non-proportionality} \label{nonpropSec}


In scintillators, the non-proportionality effect refers to the deviation of the light output from a proportional dependence on the deposited energy. Despite extensive experimental and theoretical efforts spanning the past five decades, the underlying mechanisms of this phenomenon remain poorly understood, thus preventing the development of a theoretical model. Moreover, since non-proportionality manifests as a slight deviation from the linearity of the energy calibration, potential systematics associated with detector calibration, which is in turn affected by specific characteristics of the DAQ related to light-to-signal conversion, must be considered. Consequently, calibration should be performed in close proximity to the ROI of the experiment, as is done in ANAIS-112.

Several techniques have been developed to measure the non-proportional response of scintillators, which can be categorized into photon response and electron response measurements. The photon response is typically measured by analyzing the LY from a set of gamma rays and x-rays of various energies. While the electron response is more challenging to measure, it provides information over an energy continuum rather than at discrete photon energies. Most electron response measurements rely on the Compton scattering, which, unlike photon response, can be assumed to take place in the bulk of the crystal volume. In addition, microscopic semi-empirical models of LY non-linearity can be applied to simulated energy spectra in order to account for non-linearity in the measured data, as in \cite{saldanha2023cosmogenic}.

\begin{figure}[t!]
\begin{center}
\includegraphics[width=0.6\textwidth]{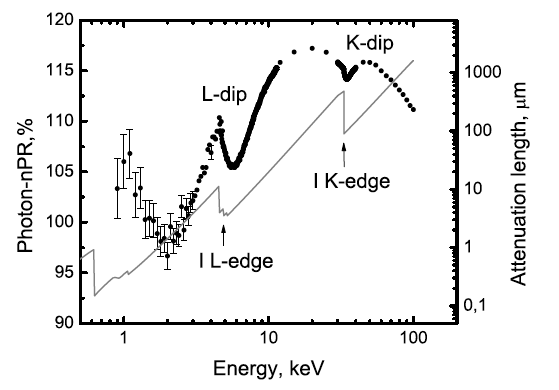}

\caption{\label{nonprop} Non-proportional response of NaI(Tl) as a function of x-ray or gamma photon energy, normalized to the \textsuperscript{137}Cs 662 keV photopeak \cite{khodyuk2010nonproportional}. The solid line represents the attenuation length for X-rays and gamma rays in NaI. Arrows indicate the energies corresponding to the iodine L-shell and K-shell, at 33.2 keV and $\sim$5 keV, respectively. 
}
\vspace{-0.5cm}
\end{center}
\end{figure}

In the specific case of NaI(Tl), a non-proportional behaviour has been widely reported for photons. The non-linearity becomes more pronounced at lower energies, with reported variations in LY of 10–15\% between approximately 3 and 20 keV \cite{Payne:2011}. Figure \ref{nonprop} shows the non-proportional response of NaI(Tl) as a function of incident X-ray or gamma-ray energy \cite{khodyuk2010nonproportional}. A pronounced non-linear feature is observed at the K-shell binding energy of iodine (33.2 keV), known as the K-dip, where photoelectrons produced via K-shell photoabsorption exhibit reduced kinetic energy, resulting in lower ionization density. A similar, though less well-characterized, feature, referred to as the L-dip, is expected near the L-shell binding energy of iodine ($\sim$ 5 keV), although empirical data in this low-energy region remain limited. Non-proportionality curves derived from the study of the scintillation response to photons and electrons are expected to differ due to the distinct mechanisms by which these particles transfer energy. While photons transfer their energy to lower-energy electrons, electrons deposit energy continuously through ionization. For example, the characteristic discontinuities at the iodine K and L dips, associated with the photoelectric effect, are not expected to appear when examining the response to electrons.

COSINE-100 recently presented a study of the non-proportional response of NaI(Tl) scintillators by combining internal COSINE-100 data with external $\gamma$-ray spectroscopy measurements \cite{cosine2024nonproportionality}. The resulting non-proportionality curve was parametrized using an empirical function, successfully capturing the characteristic K-dip. Based on this modelling, the L-dip was somehow predicted. However, the absence of data in the region around 10~keV limited the accuracy of the fit in that region, resulting in a broader uncertainty band.

In this context, the ANAIS-112 experiment presents a valuable opportunity to investigate the non-proportional behavior of NaI(Tl) scintillators. In particular, ANAIS-112 can provide additional data in the energy region between the L-shell and K-shell dip. Accordingly, in this thesis the response of the ANAIS-112 crystals has been characterized by analyzing the LY from a set of gamma rays, X-rays, and electron binding energies identified in the ANAIS-112 data.


\begin{figure}[b!]
\begin{center}
 \centering
   
        \includegraphics[width=0.9\textwidth]{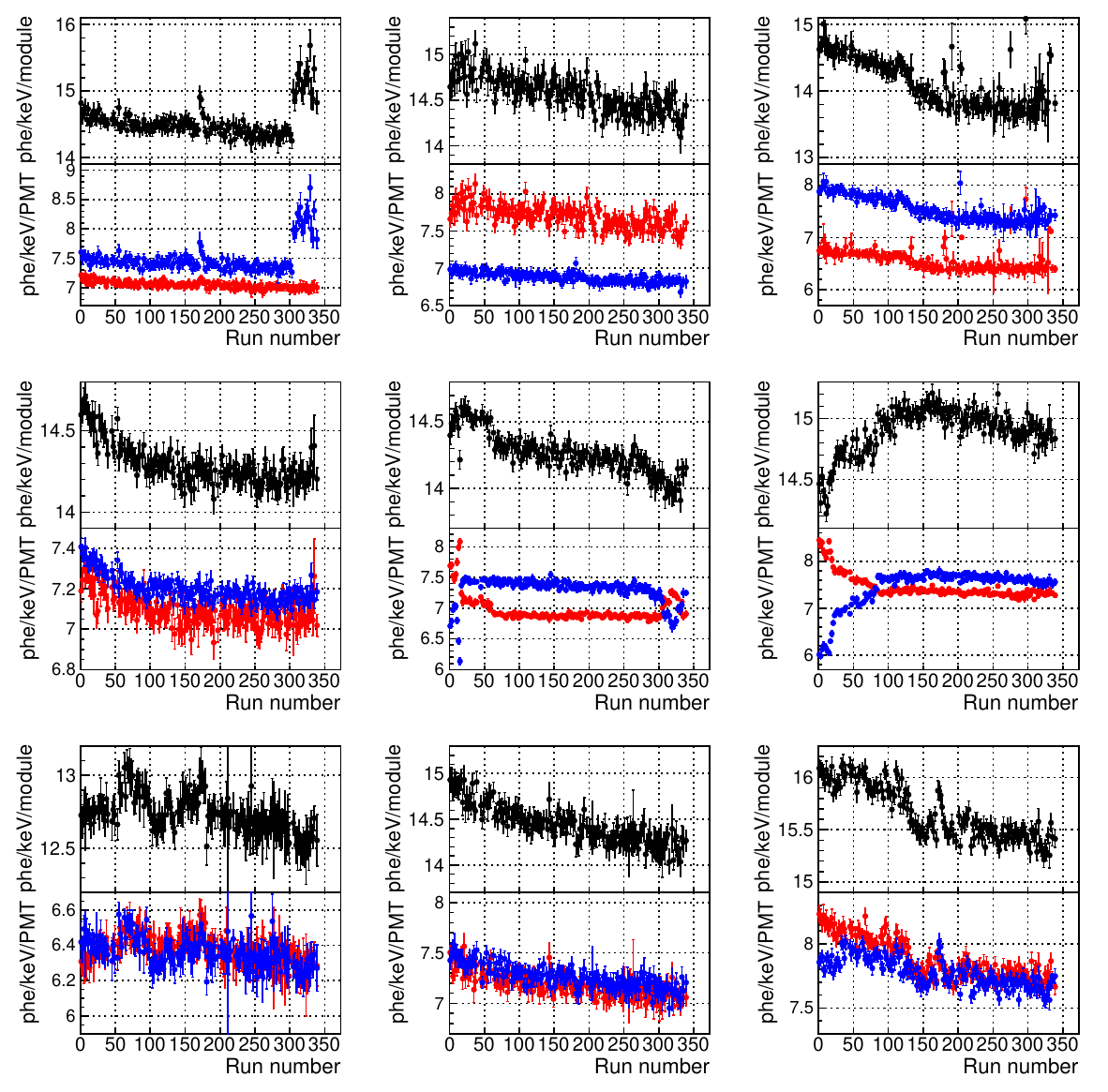}
   
\caption{\label{LYevolution} Evolution of the light collected per unit of energy deposited for the nine NaI(Tl) modules in ANAIS-112 over six years of data taking. Black dots indicate the total light collected per module, while red and blue dots represent the contributions from PMT0 and PMT1, respectively. }

\end{center}
\end{figure}

This analysis will be performed in terms of the number of photoelectrons (nphe), defined as the area of the PMT waveform divided by the average area of the SER. Working with the nphe variable provides the most direct and bias-free estimate of the LY, as it aligns with its physical definition: the number of detected photons per unit of deposited energy. The observable nphe is naturally corrected for PMT gain variations, since the SER is calculated every two weeks. Therefore, within the adopted methodology, nphe effectively measures the variation in~LY.


Since the LY will be studied across different years, it is crucial to apply corrections to account for temporal variations in LY throughout the measurement period. Figure~\ref{LYevolution} displays the temporal evolution of the LY. As expected, the LY exhibits a gradual decrease over time due to degradation of optical couplings, PMTs, and related components. Some fluctuations are observed, such as in detectors 4 and 5 during the first year of data taking, where an anti-correlation between the LY of PMT0 and PMT1 is clear. This is related to the HV applied to the PMTs, which was corrected by lowering the voltage. In detector 0, however, the issue is not the light collection but rather an incorrect determination of the SER due to the reduced gain of one PMT. This requires appropriate correction, which will be addressed in future work.

The LY evolution can be roughly modeled by a second-order polynomial to derive a run-by-run correction factor, thereby mitigating statistical fluctuations in LY estimates on a per-run basis. This correction will be systematically applied throughout this section whenever energy spectra are presented in units of nphe.

Table \ref{listaisotopos} presents all the isotopes identified in the ANAIS-112 data that are included in the non-proportionality study conducted in this thesis. The type of signal, as well as its determination method, are also listed in the table. As shown, the analysis incorporates all available data, including background events, calibration lines from a \(^{109}\)Cd source, neutron calibration runs, and cosmogenically induced isotopes identified during the first months of data-taking. Furthermore, it is important to note that, as reflected in the table, except in cases where the emission is gamma and the signal is produced by a primary photon depositing its full energy (as in the case of \textsuperscript{109}Cd calibration), the isotopes selected for this study typically result in more complex or distributed energy depositions. This is especially true when the isotope decay occurs via EC, leading to the sum of many Auger electrons and X-rays.

\begin{table}[t!]
    \centering
    \resizebox{0.9\textwidth}{!}{%
    \begin{tabular}{cccc}
    \hline
    Isotope  & Energy (keV) & Type & Determination \\
    \hline \hline
   $^{22}$Na  & 0.9  &  eBE$_\textnormal{K}$ &  Coincidence with 1274.5 keV \\
   $^{40}$K &  3.2  & eBE$_\textnormal{K}$ &  Coincidence with 1460.8 keV \\
   $^{109}$Cd & 3.8   & eBE$_\textnormal{L}$ &  y3 - y6  \\
   $^{113}$Sn & 4.2   & eBE$_\textnormal{L}$ &  y1m1\_4 - \textnormal{y1m10\_12}  \\
   $^{121}$Te & 4.7   & eBE$_\textnormal{L}$ &  Coincidence with 573.1 keV  \\
   Cu (D8)  & 8.0 & X-rays & $^{109}$Cd calibration\\
   Br & 12.1 & X-rays & $^{109}$Cd calibration \\
   $^{109}$Cd & 22.1 & X-rays  & $^{109}$Cd calibration\\
   $^{109}$Cd & 25.5 &  eBE$_\textnormal{K}$ & y3 - y6  \\
   $^{113}$Sn & 27.9 &  eBE$_\textnormal{K}$ & \textnormal{y1m1\_4} - \textnormal{y1m10\_12} \\
   $^{121}$Te & 30.5 &  eBE$_\textnormal{K}$ & Coincidence with 573.1 keV \\
   $^{121}$Sb & 30.5 &  eBE$_\textnormal{K}$ & Coincidence with 212.2 keV ($^{\textnormal{121m}}$Te)\\
   $^{127}$I & 31.8 &  eBE$_\textnormal{K}$ & neutron calibration\\
   $^{125}$I & 35.5 + 4.9 & $\gamma$ +eBE$_\textnormal{L}$ & \textnormal{y1m1\_4} - \textnormal{y1m10\_12}  \\
   $^{127}$I & 58.8 & $\gamma$ (+NR) & neutron calibration \\
   $^{125}$I & 35.5 + 31.8 & $\gamma$ +eBE$_\textnormal{K}$  & \textnormal{y1m1\_4} - \textnormal{y1m10\_12}  \\
   $^{\textnormal{121m}}$Te & 81.8 & $\gamma$ & Coincidence with 212.2 keV \\
   $^{109}$Cd & 88.0 & $\gamma$ & $^{109}$Cd calibration\\
   $^{\textnormal{127m}}$Te & 88.3 & $\gamma$ & \textnormal{y1m1\_4} - \textnormal{y1m10\_12} \\

    \hline
    \end{tabular}%
    }
    \caption{Main isotopes identified in ANAIS-112 data, included in the non-proportionality study. The table lists the energy of the analyzed signal, the type of emission (\(\gamma\) emissions, X-rays, and binding energies for EC decays (eBE\textsubscript{K}, eBE\textsubscript{L}, according to the shell where EC occur)), and the determination mechanisms. These mechanisms include coincidence with a higher-energy gamma in another module, time differences across years, background, or \(^{109}\)Cd/neutron calibration data. y3(6) refer to the third (sixth) year of ANAIS-112 data, while \textnormal{y1m1\_4} (\textnormal{y1m10\_12}) denotes the first four months (last three months) of the first year of data-taking.}
    \label{listaisotopos}
\end{table}

The procedure to be followed in this study consists of obtaining the spectra corresponding to the populations listed in Table \ref{listaisotopos} for each case. An energy spectrum in nphe units, proportional to the deposited energy, will be constructed, from which the signatures of interest can be identified. Each peak will be fitted with a gaussian function superimposed on a second-order polynomial to account for the background, yielding the mean nphe value for that peak. This value will then be divided by the nominal energy of the signal to determine the LY at that point.

For background events, \(^{109}\)Cd, and neutron calibration events, the full six-year dataset of ANAIS-112 will be employed, applying the standard ANAIS selection cuts. The $\sim$8~keV signature of copper X-rays, also identified during \textsuperscript{109}Cd calibration, is only visible in detector~8 because it exhibits the best energy resolution among the ANAIS-112 modules. Given the straightforward nature of the fitting procedure in these cases, it will not be elaborated further in this section, and only the results will be presented in terms of LY.

However, particularly relevant are the peaks arising from the decay of short-lived isotopes, as they correspond to energy depositions homogeneously distributed throughout the crystal volume. Although their extraction is hindered by limited statistics and potential overlap with other spectral components, these features provide a unique probe of the non-proportional response. Their successful identification is achieved through the analysis of the spectral shape of subtraction spectra across different years or specific data-taking periods, in combination with coincidence analysis between modules. It is important to emphasize that the study of cosmogenically induced isotopes will be restricted to detectors D6, D7, and D8, as the remaining modules arrived earlier at the LSC, and by the time data taking began, the signatures of these isotopes had already significantly decayed. 

\begin{figure}[b!]
\begin{center}
 \centering
    \begin{minipage}{0.48\textwidth}
        \centering
        \includegraphics[width=\textwidth]{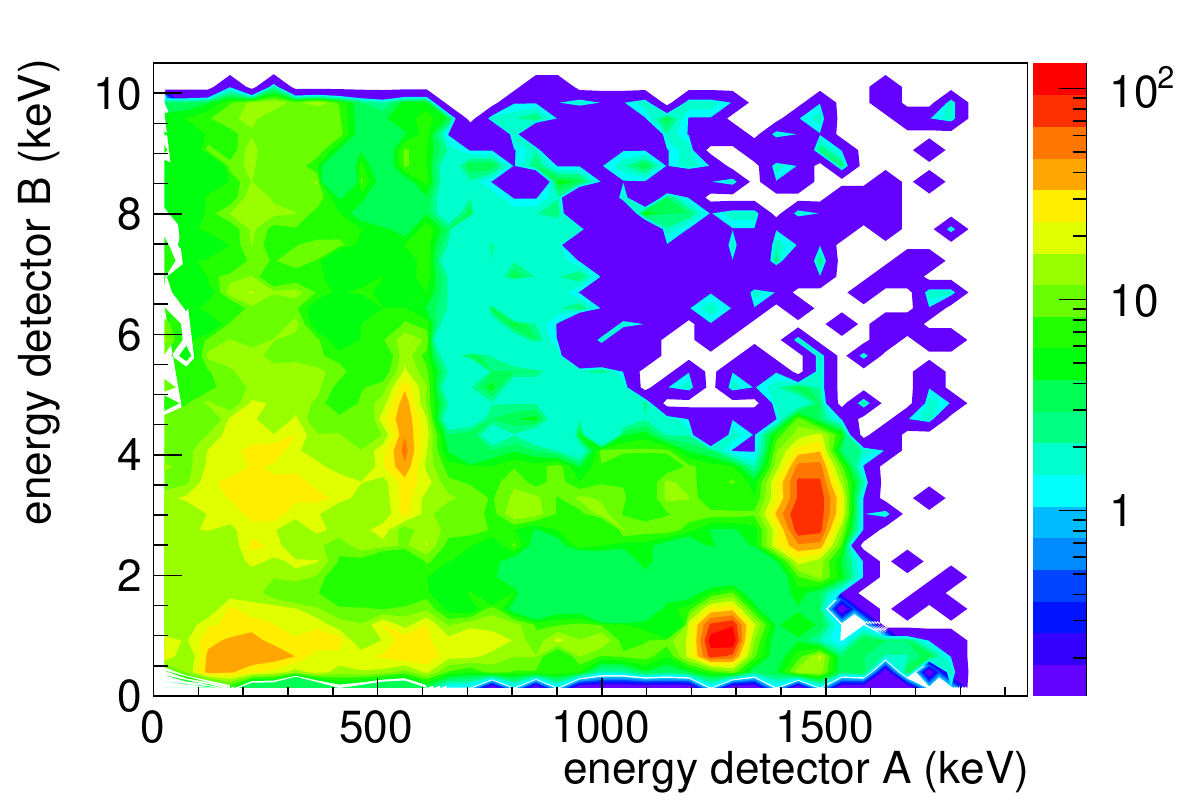}
    \end{minipage}
    \hfill
    \begin{minipage}{0.48\textwidth}
        \centering 
        \includegraphics[width=\textwidth]{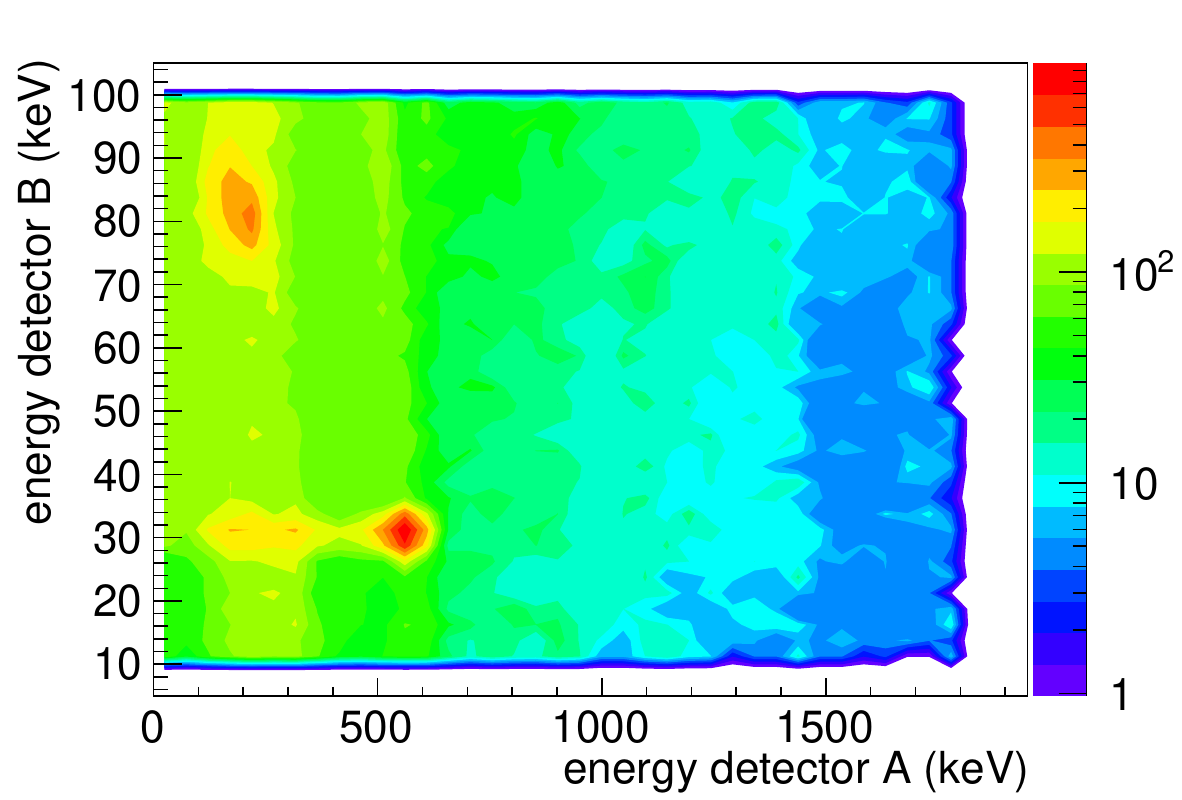}
    \end{minipage}
\caption{\label{plotcoin2d} Plot of the energies recorded in two detectors triggered in coincidence, with the low-energy data corresponding to modules D6, D7, and D8. Data correspond to the first four months of ANAIS-112. \textbf{Left panel:} detector B has an energy deposition below 10 keV. \textbf{Right panel:} detector B has an energy deposition from 10-100~keV.  }
\vspace{-0.5cm}

\end{center}
\end{figure}

The study of cosmogenically induced isotopes begins with the analysis of coincident signals. Figure \ref{plotcoin2d} shows the energies registered in two detectors triggered in coincidence during the first four months of ANAIS-112 data. These types of plots aim to identify the EC decay of isotopes in one module, accompanied by the emission of a higher gamma ray from the de-excitation of the daughter nucleus in another module. Several spots emerge in the plots, together with Compton-scattered gamma rays, which are also clearly visible.

The left panel of Figure \ref{plotcoin2d} displays coincident events where one of the energy depositions is below 10 keV. Two prominent features are visible in the figure, which can be attributed to bulk contamination of the NaI crystals by \(^{22}\)Na and \(^{40}\)K. The atomic de-excitation energy, 0.9 keV for \(^{22}\)Na and 3.2 keV for \(^{40}\)K in the case of K-shell EC, is fully absorbed in the crystal where the decay occurs. In contrast, the accompanying high-energy gamma rays (1274.5 keV for \(^{22}\)Na and 1460.8 keV for \(^{40}\)K) may escape and be absorbed in another module, thus producing a coincidence event. Additionally, a less intense spot can be identified in the same figure, corresponding to coincidences involving a high-energy gamma from \(^{121}\)Te (573.1 keV) absorbed in one detector and an energy deposition corresponding to the L-shell binding energy of Sb (4.7~keV) in another module.



The right panel of Figure \ref{plotcoin2d} shows the coincident event when one of the detectors has an energy between 10-100 keV. The same coincidence from the high-energy gamma of \(^{121}\)Te (573.1 keV) is observed, but in this case, the coincidence is with the K-shell binding energies of Sb, 30.5 keV. Additionally, the 212.2 keV emission from \(^{121\text{m}}\)Te also shows coincidence with the 81.9~keV cascade emission from the same isotope.

\begin{figure}[t!]
\begin{center}
 \centering
    \begin{minipage}{0.48\textwidth}
        \centering
        \includegraphics[width=\textwidth]{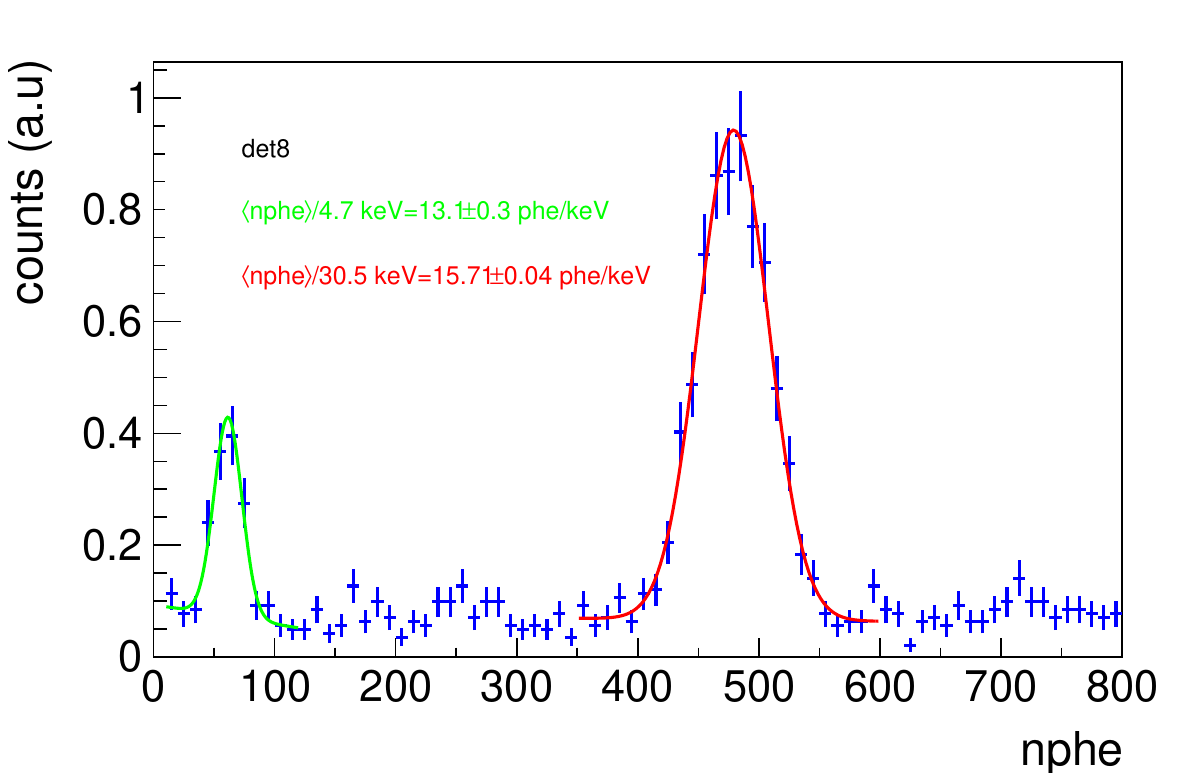}
    \end{minipage}
    \hfill
    \begin{minipage}{0.48\textwidth}
        \centering
        \includegraphics[width=\textwidth]{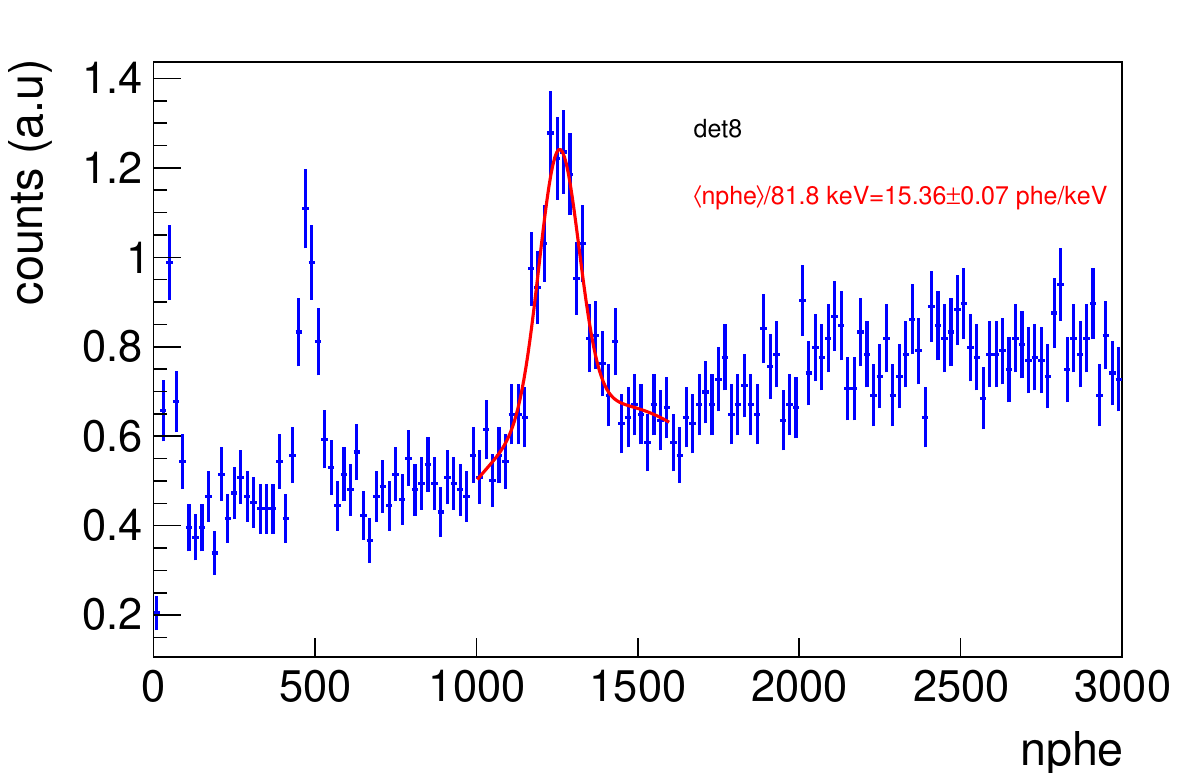}
    \end{minipage}
\caption{\label{LEcoincident} Spectrum of the low-energy D8 data (nphe units) in coincidence with high-energy data selected from another module. The fitting of the identified signals is shown, along with their corresponding LY estimation, obtained by dividing the mean nphe of the peak by its nominal energy. \textbf{Left panel:} Coincidence around the 573.1 keV emission from \(^{121}\)Te. Peaks corresponding to the L-shell and K-shell binding energies of Sb (4.7 and 30.5~keV, respectively) are fitten to a gaussian shape. \textbf{Right panel:} Coincidence around the 212.2 keV gamma emission from \(^{121\text{m}}\)Te. The peak corresponding to the 81.9 keV emission from the cascade decay of $^{121\text{m}}$Te is fitted. }

\end{center}
\vspace{-0.5cm}
\end{figure}

By selecting the 573.1 keV emission from \(^{121}\)Te in a window of [520-620] keV, the low-energy D8 coincident spectrum shown in the left panel of Figure \ref{LEcoincident} is obtained. Peaks corresponding to the L-shell and K-shell binding energies of Sb (4.7 and 30.5 keV, respectively) are clearly observed. These peaks are used to determine the mean nphe associated with that emission. The results of all fits from the non-proportionality study for detector 8 are summarized in Table \ref{LYresultsD8}. Similarly, the right panel of Figure \ref{LEcoincident} shows the low-energy D8 coincident spectrum obtained by selecting the 212.2 keV emission from \(^{121\text{m}}\)Te within a window of [100-270] keV. As shown in the two-dimensional plots, the 81.9~keV emission from the cascade decay of $^{121\text{m}}$Te is visible and then fitted. Two additional lower energy peaks are also shown. These are not fitted here because they do not originate from $^{121\text{m}}$Te emissions. Instead, they result from the Compton scattering of the 573.1 keV emission of $^{121}$Te that leak into  the selection window and correspond to the L-shell and K-shell binding energies of Sb (4.7 and 30.5 keV, respectively).

\begin{figure}[t!]
\begin{center}
 \centering
    \begin{minipage}{0.48\textwidth}
        \centering
        \includegraphics[width=\textwidth]{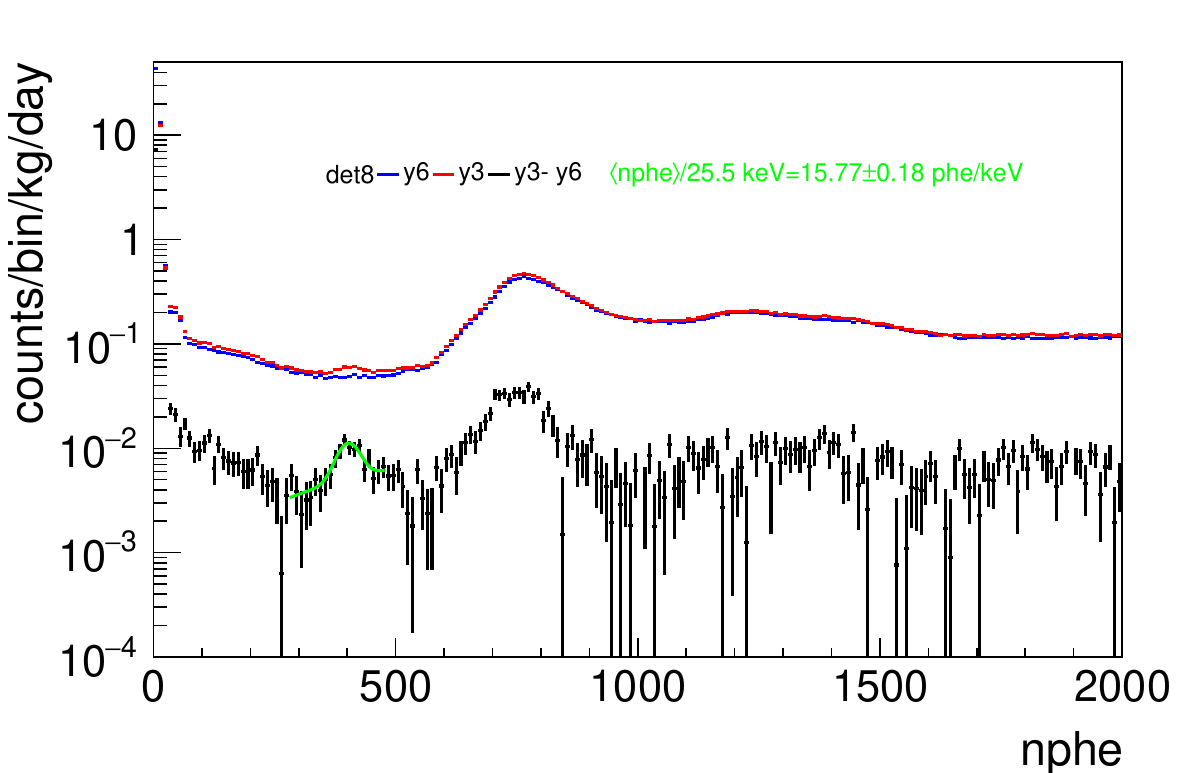}
    \end{minipage}
    \hfill
    \begin{minipage}{0.48\textwidth}
        \centering
        \includegraphics[width=\textwidth]{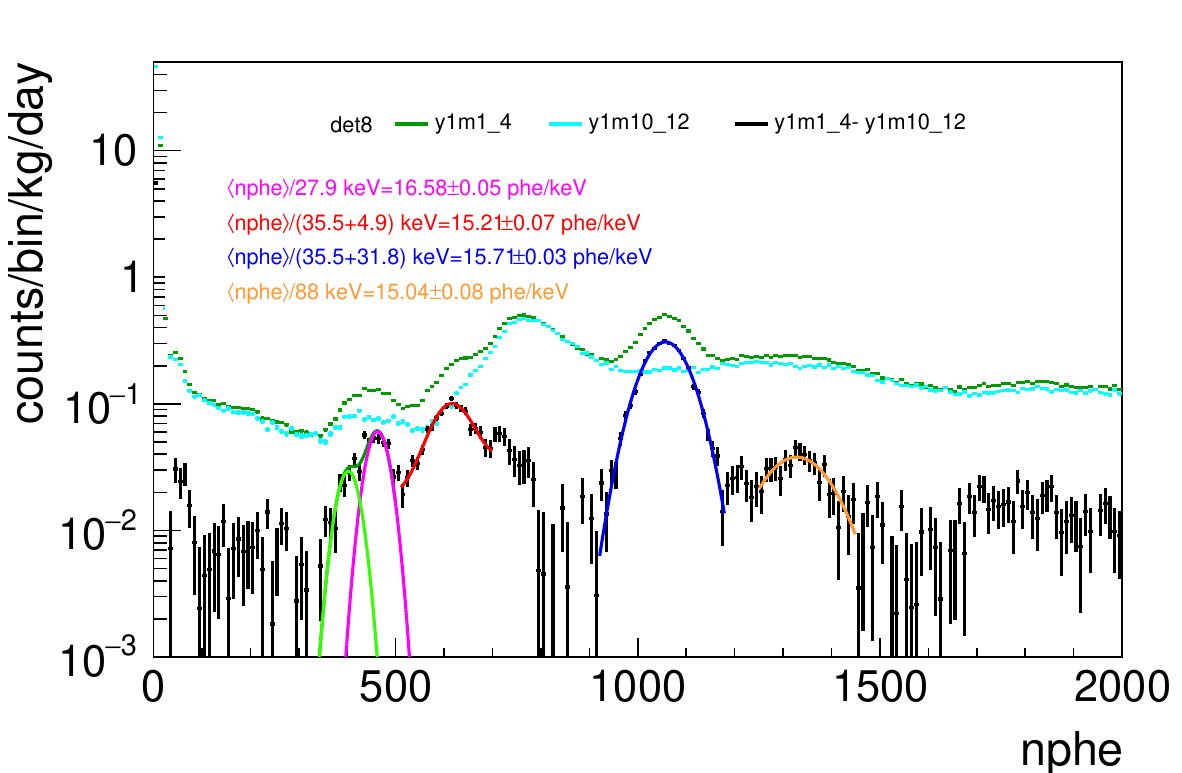}
    \end{minipage}
\caption{\label{restaDatos} \textbf{Left panel:} Difference spectrum in D8 obtained by subtracting year~6 data (blue) from year 3 data (red). The resulting difference (black) isolates the contribution from cosmogenic \(^{109}\)Cd. \textbf{Right panel:} Difference spectrum in D8 obtained by subtracting the last three months of year 1 (\textnormal{y1m10\_12}, cyan) from the first four months of year 1 (\textnormal{y1m1\_4}, green). The resulting spectrum (black) reveals the presence of cosmogenic \(^{109}\)Cd, \(^{113}\)Sn, and the two composite peaks of \(^{125}\)I. The fitting of the identified signals is shown, along with their corresponding LY estimation, obtained by dividing the mean nphe of the peak by its nominal energy. }

\end{center}
\vspace{-0.5cm}
\end{figure}

An alternative method to clearly identify cosmogenically induced isotopes, now that several years of data taking are available in ANAIS-112, is to analyze the differences observed between data sets corresponding to different time intervals. The time periods selected for subtraction are chosen to maximize the signal resulting from the difference between the datasets. Figure \ref{restaDatos} presents such differences for detector D8, allowing for the identification of the isotopes that will be discussed in the following. As evidenced in the figure, the correction for LY loss described previously has been properly accounted for, as illustrated by the consistent position of the \(^{210}\)Pb peak around 49 keV, without requiring any additional ad-hoc scaling factors.

The left panel of Figure \ref{restaDatos} displays the difference between D8 data from year~3 and year 6, specifically aimed at isolating the contribution from cosmogenic \(^{109}\)Cd. This isotope decays via EC to the 88.0 keV isomeric state of \(^{109}\)Ag, with a half-life of 461.9~days, producing a peak at the K-shell binding energy of Ag at 25.5 keV. In ANAIS-112, this isotope is generally difficult to resolve due to its spectral overlap with the peak from \(^{113}\)Sn, which decays via EC predominantly to a 391.7 keV isomeric state of the daughter nucleus, with a half-life of 115.1 days. Consequently, \(^{113}\)Sn produces a peak at the K-shell binding energy of In at 27.9 keV. Given the significant difference in half-lives, by approximately a factor of four, the approach adopted in this work involves first isolating the contribution of \(^{109}\)Cd by analyzing data from later years, thereby exploiting its longer half-life. 

Through the subtraction of year 3 data from year 6 data, the resulting spectrum isolates the contribution from $^{109}$Cd, as the expected activity of $^{113}$Sn at these late times is negligible and no detectable signature from this isotope is anticipated. The 88.0~keV $\gamma$-ray emission (T$_{1/2}$ = 39.7 s), corresponding to the first excited state of $^{109}$Ag, should in principle also be observed, offering a means to compare bulk and surface emissions from the same isotope. However, this line is not clearly visible. While a slight accumulation around 1400 nphe can be perceived in the left panel of Figure~\ref{restaDatos}, it is not sufficiently well-defined to allow for a reliable fit or quantitative analysis.

Regarding the low-energy region, this subtraction also allows for the identification of the L-shell binding energy of $^{109}$Cd at 3.8 keV, in addition to the K-shell line. In this case, however, the feature does not manifest as a well-defined peak. Since fitting a gaussian function is not feasible due to the dispersion of the distribution, the energy position is estimated by calculating the mean of the distribution, taking into account the detector-specific energy resolution at this energy in ANAIS-112. While this method is less precise than a peak fit, it provides an indicative value of the LY associated with this energy deposition. Furthermore, the associated uncertainty, derived from the standard deviation of the distribution, is accordingly larger.

With the contribution of \(^{109}\)Cd determined from the previous step, the analysis proceeds with the data subtraction shown in the right panel of Figure \ref{restaDatos}, where the spectrum from the first four months of year 1 (\textnormal{y1m1\_4}) is subtracted from that of the last three months of the same year (\textnormal{y1m10\_12}). The underlying reason behind this choice is that, having already determined the response of \(^{109}\)Cd, its contribution can be introduced as a gaussian with fixed mean and sigma. An additional gaussian component is included to model the \(^{113}\)Sn contribution, from which the corresponding mean nphe associated with this isotope is extracted.

\begin{table}[b!]
\begin{center}
 \centering
\begin{tabular}{cccc}
\hline
Isotope & Energy (keV) & Relative LY & $\sigma$ (\%) \\
\hline
\hline

  $^{22}$Na  & 0.9  & 0.805 ± 0.007 & 38.3 ± 0.9 \\
   $^{40}$K &  3.2  & 0.901 ± 0.00 & 18.1 ± 0.4\\
   $^{109}$Cd & 3.8   & \textcolor{green!50!black}{0.909 ± 0.024} & \textcolor{green!50!black}{43.6 ± 2.2}\\
   $^{113}$Sn & 4.2   & \textcolor{green!50!black}{0.850 ± 0.030} &   \textcolor{green!50!black}{26 ± 4}  \\
   $^{121}$Te & 4.7   &  0.617 ± 0.034 &  18.7 ± 2.0  \\
   Cu (D8)  & 8.0 & 0.9948 ± 0.0024 & 6.41 ± 0.27\\
   Br & 12.1 &  0.9994 ± 0.0004 &  9.80 ± 0.06 \\
   $^{109}$Cd & 22.1 & 1.0104 ± 0.0004  & 6.609 ± 0.004\\
   $^{109}$Cd & 25.5 &  1.064 ± 0.008 & 12.8 ± 0.6  \\
   $^{113}$Sn & 27.9 &  1.038 ± 0.003 & 5.706 ± 0.003 \\
   $^{121}$Te & 30.5 &  0.984 ± 0.007 & 6.03 ± 0.22 \\
   $^{121}$Sb & 30.5 &  0.983 ± 0.002 & 5.6 ± 0.7\\
   $^{127}$I & 31.8 &  1.005 ± 0.004 & 7.4 ± 0.5\\
   $^{125}$I & 35.5 + 4.9 & 0.951 ± 0.004 & 4.3 ± 0.6  \\
   $^{127}$I & 58.8 & 1.000 ± 0.002 & 4.99 ± 0.26 \\
   $^{125}$I & 35.5 + 31.8 & 0.983 ± 0.000  & 4.61 ± 0.05  \\
   $^{\textnormal{121m}}$Te & 81.8 & 0.960 ± 0.004  & 4.9 ± 0.7 \\
   $^{109}$Cd & 88.0 & 0.9680 ± 0.0004 & 3.750 ± 0.006\\
   $^{\textnormal{127m}}$Te & 88.3 & 0.941 ± 0.005 & 5.4 ± 0.7\\
       
       \hline
\end{tabular}
\caption{Results of the non-proportionality fits conducted in this study for detector~8, showing the isotope, the energy of the analyzed signal, the corresponding relative LY normalized to the LY of the neutron inelastic peak, and the energy resolution expressed as a percentage. Highlighted in green are point results obtained not from a gaussian fit but from the mean and standard deviation of the corresponding distribution. }
\label{LYresultsD8}
\end{center}
\end{table}

This subtraction also enables the identification of another isotope, \(^{125}\)I, which decays with 100\% probability by EC to an excited state of the Te daughter nucleus. This decay produces two observable peaks: one at 40.4 keV, resulting from the sum of the $\gamma$ excitation energy (35.5 keV) and the L-shell binding energy of Te (4.9~keV), and another at 67.3 keV, from the same excited state combined with the K-shell binding energy of Te (31.8 keV). These two composite peaks are clearly visible in the difference spectrum shown in the right panel of Figure \ref{restaDatos}, and their corresponding mean nphe values are extracted. From this subtraction, the LY associated with two additional signatures can also be estimated. First, the peak at 88.3 keV from $^{127\textnormal{m}}$Te is observed, corresponding to the transition from its metastable state. Second, analogous to the determination of the L-shell binding energy of $^{109}$Cd, the L-shell binding energy of $^{113}$Sn at 4.2 keV is identified. As in the previous case, its LY is estimated by computing the mean and standard deviation of the corresponding distribution, given that no clear peak allows for a standard gaussian fit.

\begin{figure}[t!]
\begin{center}
 \centering
    
        \includegraphics[width=0.65\textwidth]{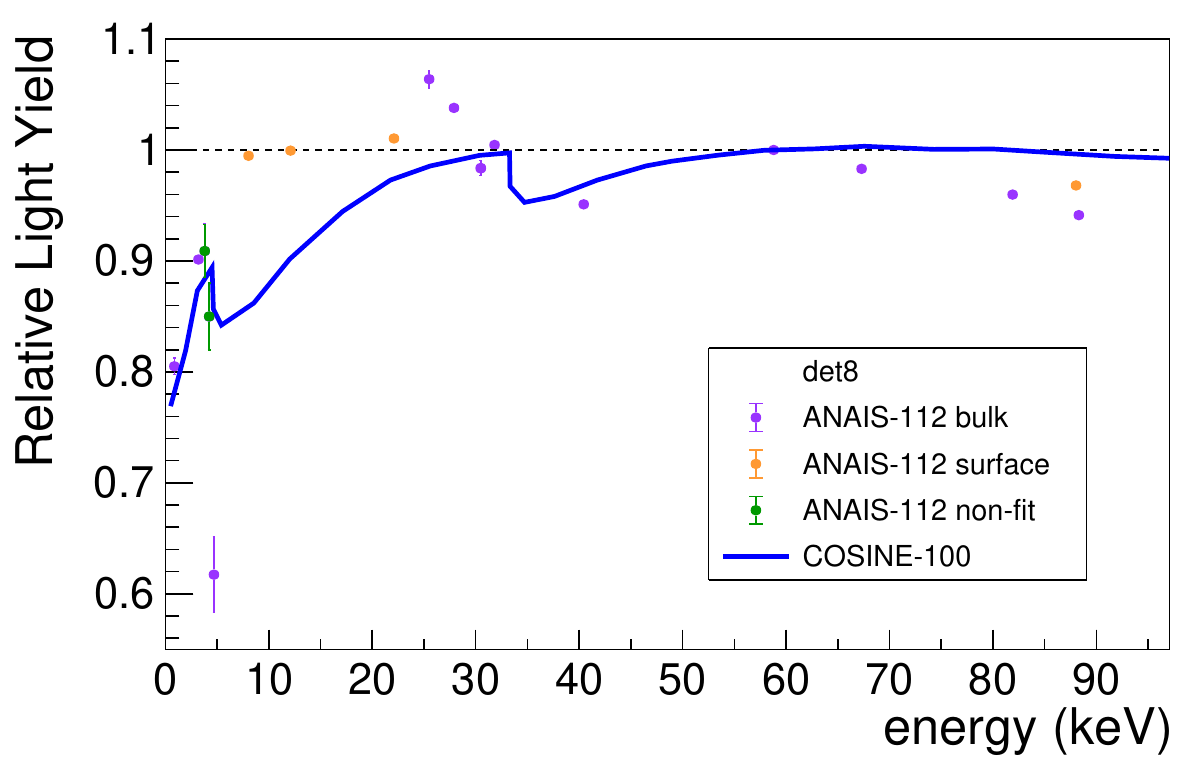}
                \includegraphics[width=0.65\textwidth]{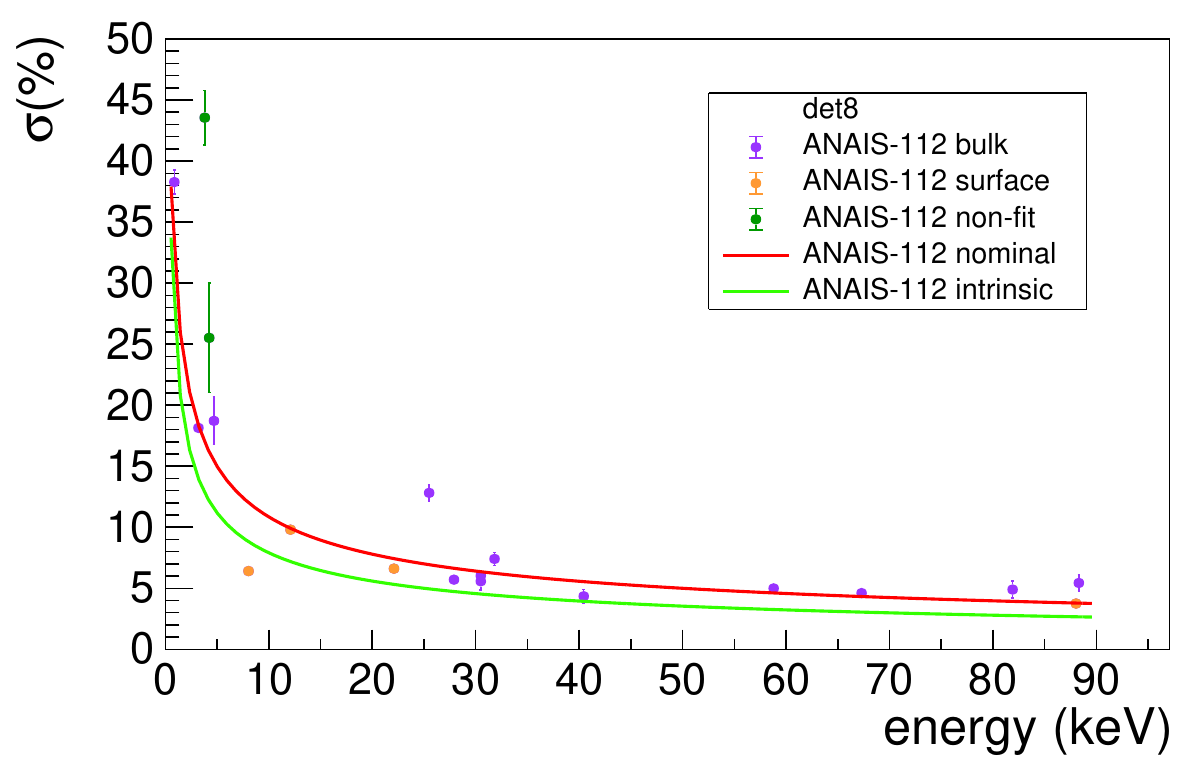}

\caption{\label{LY_soloD8} \textbf{Top panel:}
Relative LY as a function of energy. For comparison, the non-proportionality curve reported by COSINE-100 is shown \cite{cosine2024nonproportionality}. \textbf{Bottom panel:} Energy resolution as a function of energy. The nominal resolution of the module as revisited in this work (red line), and the intrinsic resolution calculated from the measured light collection (light green line) \cite{Amare:2018sxx} are shown. Data points corresponds to detector~8 and are normalized to the LY of the neutron inelastic peak (58.8 keV). Bulk (violet) and surface (orange) events in ANAIS-112 are distinguished. Highlighted in green are point results obtained not from a gaussian fit but from the mean and standard deviation of the corresponding distribution. }

\end{center}
\end{figure}


The top panel of Figure \ref{LY_soloD8} displays the relative non-proportionality curve for detector~8, where bulk and surface contributions are explicitly separated. Surface peaks primarily originate from periodic $^{109}$Cd calibration runs, whereas bulk peaks arise from internal and cosmogenic isotopes in background and neutron calibration runs. To emphasize the deviation from linearity, the LY values have been normalized by a constant factor for each crystal, defining the so-called relative LY. The normalization is fixed at 58.8 keV, corresponding to the inelastic scattering peak of $^{127}$I, as it is a bulk event located in the energy region where LY is expected to be approximately constant. Table \ref{LYresultsD8} presents the results for the relative LY and energy resolution as obtained from the fits for detector 8.

As shown, ANAIS-112 data reveal a clear reduction in LY at both the K- and L-shell binding energies, below 40 and 5~keV respectively, with a particularly pronounced drop at the L-dip, where the LY falls to approximately 60\% of the normalized value. According to Table \ref{LYresultsD8}, this point corresponds to the L shell of Sb and was determined from coincidences with the high-energy gamma line of $^{121}$Te, as shown in Figure \ref{LEcoincident}. The fit does not exhibit poor performance, thus the pronounced drop is not attributable to a misfit. 

However, the points at 3.8 and 4.2 keV, corresponding to the L-shell of \textsuperscript{109}Cd and \textsuperscript{113}Sn, were not obtained from a gaussian fit due to the difficulty in identifying a clear peak for these populations. Therefore, conclusions drawn from these points must be interpreted with caution. Moreover, the K-shell peaks of these same isotopes, at 25.5 and 27.9 keV, exhibit an elevated LY, suggesting the presence of a systematic issue in peak determination based on the year-subtraction method. For comparison, the non-proportionality curve reported by the COSINE-100 experiment is also included \cite{cosine2024nonproportionality}. Notably, surface events in ANAIS-112 produce significantly more light in the 10–30 keV range than expected from the COSINE-100 model.


The bottom panel of Figure \ref{LY_soloD8} presents the energy resolution, expressed as a percentage, as obtained from the non-proportionality fits. Again, bulk and surface peaks are distinguished, and the results are compared with both the nominal resolution of ANAIS-112 (revisited in this thesis, see Figure \ref{LEres}) and the intrinsic resolution expected from the light collection derived from the first year of ANAIS-112 data \cite{Amare:2018sxx}. 

As shown, the resolution derived from this analysis agrees generally well with the nominal resolution. The peak corresponding to the L-shell of cosmogenic $^{109}$Cd, where the L-dip is observed, indicates a poorer resolution than the nominal trend. However, the determination of this peak is subject to large uncertainties due to methodological limitations, as no actual fit was performed. Furthermore, the 25~keV peak from cosmogenic $^{109}$Cd exhibits a broader resolution compared not only to neighboring points but also to the nominal resolution, which is well determined in that energy range, further confirming the presence of systematic effects in the peak determination.

When compared with the intrinsic resolution, it is observed that the energy resolution derived from the fits is larger than the purely poissonian contribution. This discrepancy may be attributed to a non-uniform LY throughout the detector volume and to possible energy-dependent variations. Furthermore, above approximately 20 keV, the resolution does not continue to decrease with energy, but instead remains constant, suggesting the presence of additional contributions beyond the intrinsic poissonian in the number of photoelectrons produced at the PMTs.

\begin{figure}[t!]
\begin{center}
 \centering
    
        \includegraphics[width=0.7\textwidth]{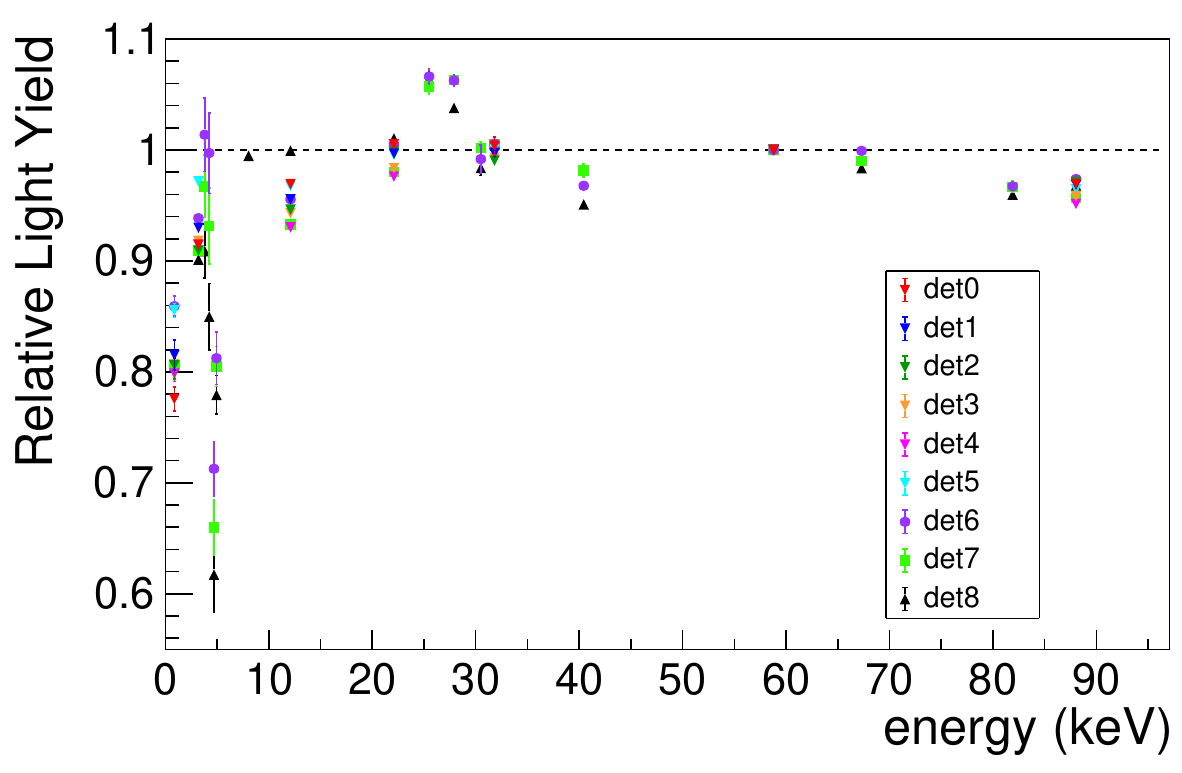}
 
\caption{\label{LY_todos} Relative LY as a function of energy, normalized to the LY of the neutron inelastic peak (58.8 keV), for the nine modules of ANAIS-112.  }

\end{center}

\end{figure}

Figure~\ref{LY_todos} shows the relative LY curves for the nine ANAIS-112 detectors. The results are consistent across modules and systematically exhibit a decrease in LY around the L-shell and K-shell binding energies. Detector 8, previously shown, stands out compared to the other detectors in terms of the LY of the 11 keV calibration peaks. No explanation has been found for this distinct behavior, apart from the fact that this module exhibits the best energy resolution.

However, as previously discussed, the non-proportionality plots presented in this section may be somewhat misleading in certain cases. In this work, all available information from different
event populations at various energies in ANAIS-112 has been exploited. Consequently, the
conclusions drawn from this study inherently involve the combined response to both
electrons and photons, reflecting the nature of the available data. Specifically, when analyzing signatures from isotopes that decay via EC, the measured energy does not correspond to a single gamma photon but rather to a sum of multiple, lower-energy deposits arising from emitted X-rays and Auger electrons. This implies that the energy deposition mechanisms involved are heterogeneous and dominated by numerous low-energy components, complicating the interpretation of the LY values as if they were associated with single, monoenergetic events.

Nevertheless, this study provides valuable insight into the light response of the ANAIS-112 detectors, confirming the expected decrease in relative LY at specific energy regions and revealing a homogeneous behavior across the nine ANAIS-112 detectors.

A precise characterization of the non-proportionality effect is essential for accurate modelling of background interactions involving photons and electrons, as well as for the interpretation of potential DM signals coupled to these particles. Having characterized in this work the light response of the ANAIS-112 crystals, future efforts should aim to consistently incorporate the non-proportional behavior of NaI(Tl) into the data analysis or Monte Carlo simulations. One possible approach involves deriving, from these data, a microscopic model that can be incorporated into simulations and accurately reproduce the underlying physical processes.

\subsection{Low energy calibration procedure} \label{implementationenergycal}

\begin{figure}[t!]
    \centering
    
    \begin{subfigure}{0.3\textwidth}
        \centering
        \includegraphics[width=\linewidth]{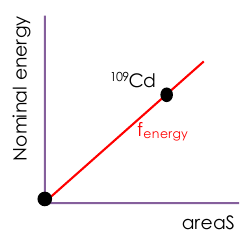}
    \end{subfigure}
    \begin{subfigure}{0.3\textwidth}
        \centering
        \includegraphics[width=\linewidth]{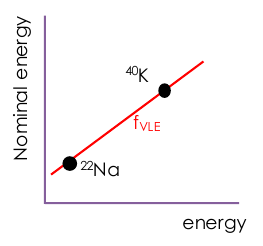}
    \end{subfigure}
    \begin{subfigure}{0.3\textwidth}
        \centering
        \includegraphics[width=\linewidth]{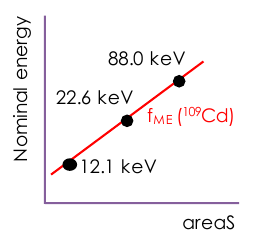}
    \end{subfigure}

    \caption{\textbf{Left panel}: First step of the very low-energy calibration procedure, which involves a proportional calibration using the $^{109}$Cd 22~keV line.  \textbf{Central panel}: Second step of the very low-energy calibration procedure, involving recalibration with the combined full exposure of internal background lines from $^{22}$Na and $^{40}$K.  \textbf{Right panel}: Medium-energy calibration procedure, based on three peaks from \(^{109}\)Cd at 12.1, 22.6, and 88.0 keV.}
    \label{implementationcal}
\end{figure}

The current two-step low-energy calibration of ANAIS-112 is illustrated in Figure~\ref{implementationcal}. 

The first step of the very low-energy calibration procedure includes a proportional calibration using the \(^{109}\)Cd 22.6 keV line to correct gain drifts (left panel), followed by a recalibration combining the full exposure of \(^{22}\)Na and \(^{40}\)K background lines (central panel) to enhance the reliability of the energy calibration in the ROI. This is the so-called VLE (Very Low Energy) energy variable, used for the lowest energy range in the the ROI. Its validity is established up to $\sim$20~keV.

The low-energy peaks from \(^{22}\)Na and \(^{40}\)K are ideal for calibration at very low energies, as they lie within or near the ROI. However, due to their low detection rates and limited coincidence efficiency, data from long-term operation over the first five years are required for each detector. In particular, for \(^{22}\)Na, its short half-life (T$_{1/2}$ = 2.6 yr) results in a significantly reduced contribution to the peak in later years compared to the initial ones, while for \(^{40}\)K the rate remains constant over time.

In the intermediate region, from 20 keV up to $\sim$150 keV, a different energy variable is used, referred to as ME (Medium Energy), whose calibration scheme is shown in the bottom panel of Figure \ref{calSpectrum}. The ME calibration is based on three peaks from \(^{109}\)Cd at 12.1, 22.6, and 88.0 keV.

\subsection{High energy calibration procedure}\label{HEcal}

The ANAIS-112 digitization scale is optimized for low-energy events, making events above $\sim$500 keV exceed the MATACQ digitizer dynamic range and causing saturation in the pulse area, affecting the energy estimation which is based on this parameter. To handle high-energy events, the ANAIS–112 DAQ system uses a second signal line conveniently attenuated before entering into the QDC modules for integration in 1 $\mu$s window (see Figure~\ref{ANAISDAQ}). However, the QDCs have poor resolution, so to reduce the quantization error in the energy signal, the QDC readout is not used directly as the energy estimator. Instead, it is used to linearize the sum of the truncated area of the digitized pulses.

For energies above $\sim$150 keV, the previously used method of linearizing the pulse area energy estimator with a modified logistic function described in \cite{Amare:2018sxx} led to deviations of up to 4\%. This has been improved by using a new linearization method based on a combination of a first-degree polynomial, a 12$^{\textnormal{th}}$-order Chebyshev polynomial, and a second-degree polynomial, which has successfully reduced the deviation to below 2\% \cite{nature}. 

The double readout system discriminates \(\alpha\) events from \(\beta\)/\(\gamma\) events by exploiting the different pulse shapes (\(\alpha\) pulses are faster). As a result, high-energy \(\alpha\) events exhibit smaller integrals in the microsecond window compared to \(\beta\)/\(\gamma\) events with the same QDC value.

Since external high-energy sources are not available for calibrating the ANAIS–112 high-energy regime, after the linearization,  events above $\sim$150 keV are calibrated independently for each detector and background run. This calibration uses several easily identifiable peaks present in the background (e.g., 238.6 keV from $^{212}$Pb, 609.3 keV from $^{214}$Bi), each fitted with a gaussian function and a second-degree polynomial. Calibration is then performed by applying a linear regression between the nominal energies of these peaks and the means of the gaussian fits.

\section{ANAIS-112 filtering protocols}\label{Filtering}

The sensitivity of ANAIS is limited by anomalous non-bulk scintillation events dominating the detection rate below 10~keV. Therefore, a robust analysis is required to select events associated with NaI(Tl) scintillation in the crystal bulk, as expected from DM particles. 

The standard developed filtering protocols of ANAIS–112 were introduced in \cite{Amare:2018sxx} and will be summarized in Section \ref{previosufiltering}. This former procedure worked very well above 2~keV, however in the region from 1 to
2~keV it exhibited some rather significant weakness. In particular, it showed a low acceptance efficiency while it resulted in a clear background excess over the background model estimates below 2~keV (see Section \ref{BkgModel}), likely due to non-scintillation events leaking the filtering. This issue motivated the development of more effective  low-energy filtering protocols based on machine learning (ML) techniques (Section~\ref{BDTnew}). In addition, new filtering developments carried out during this thesis based on the recent implementation of the new DAQ, ANOD, will be presented in Section \ref{ANODfiltering}.

\subsection{Standard filtering protocol} \label{previosufiltering}

The former ANAIS-112 filtering protocol employed multiparametric cuts based on the shape of the pulse and the asymmetry in light distribution between PMTs of the same module. Specifically, the previous ANAIS-112 filtering protocols, whose application to the ANAIS-112 data can be seen in Figure \ref{cuts}, comprised the following cuts:

\begin{itemize}
    \item \textbf{Single-hit events.} 
    
    WIMPs are expected to interact with the nuclei of detectors through processes with extremely low cross sections, ranging from \(10^{-50}\) to \(10^{-40}\)~cm\(^2\) per nucleon \cite{undagoitia2015dark}. As a result, the probability of DM interacting simultaneously in multiple modules or coinciding with background events is  negligible, allowing such coincidences to be attributed to background and be rejected.\\

    \item \textbf{Non-correlation with muons.} 
    
    Muon interactions in the NaI(Tl) result in large energy deposits that can cause long-lived phosphorescence in NaI(Tl) detectors~\cite{Cuesta:2013vpa}, leading to a high number of photons arriving in the tail of a muon pulse that potentially can trigger the ANAIS DAQ multiple times, thus increasing the total acquisition rate. To mitigate this issue, ANAIS employs a veto scintillator system that enables the correlation of muon interactions in the plastic scintillators with events in the NaI(Tl) crystals. Most of the increased trigger rates are removed by rejecting events arriving less than one second after a veto scintillation trigger, resulting in only a 3\% reduction in live time due to the filtering. \\

    \item \textbf{NaI scintillation time behaviour.} 

       \begin{figure}[b!]
\begin{center}
\includegraphics[width=0.6\textwidth]{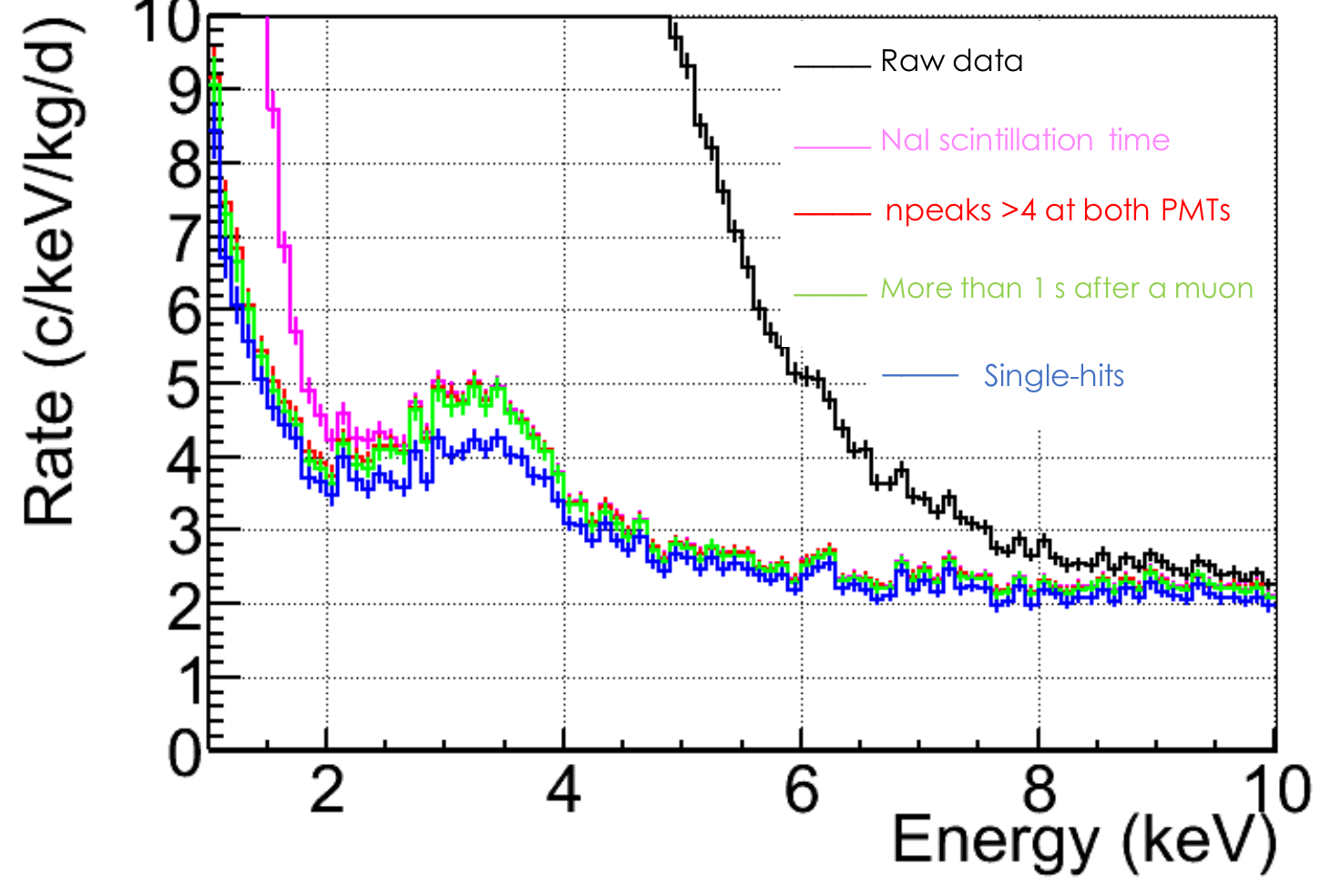}

\caption{\label{cuts} Application of the standard ANAIS-112 filtering protocol, which employs multiparametric cuts based on the shape of the pulse and asymmetry in light distribution between PMTs. }
\end{center}
\end{figure}

    Different light-emitting mechanisms in the PMTs or nearby materials contribute significantly to non-bulk scintillation events in the low-energy region. Cherenkov emission, often caused by radioactive contamination within the PMT or its surroundings, is a common source of light \cite{Amare:2014eea}. These events are typically very fast and thus can be discriminated from NaI(Tl) scintillation events, which last hundreds of nanoseconds, using the following PSA parameter \cite{DAMA:2008bis}:

    \begin{equation}
    P_1 = \frac{\sum_{t=100ns}^{t=600ns} V(t)}{\sum_{t=0}^{t=600ns} V(t)},
\end{equation}

    where \( V(t) \) represents the pulse amplitude of the waveforms from the two PMTs within the same module at time \( t \) after the trigger position. The P1 parameter also aids in removing random coincidences from dark current photoelectrons. For NaI(Tl) scintillation pulses, P1 is approximately 0.65. 

    \begin{figure}[b!]
\begin{center}
\includegraphics[width=0.5\textwidth]{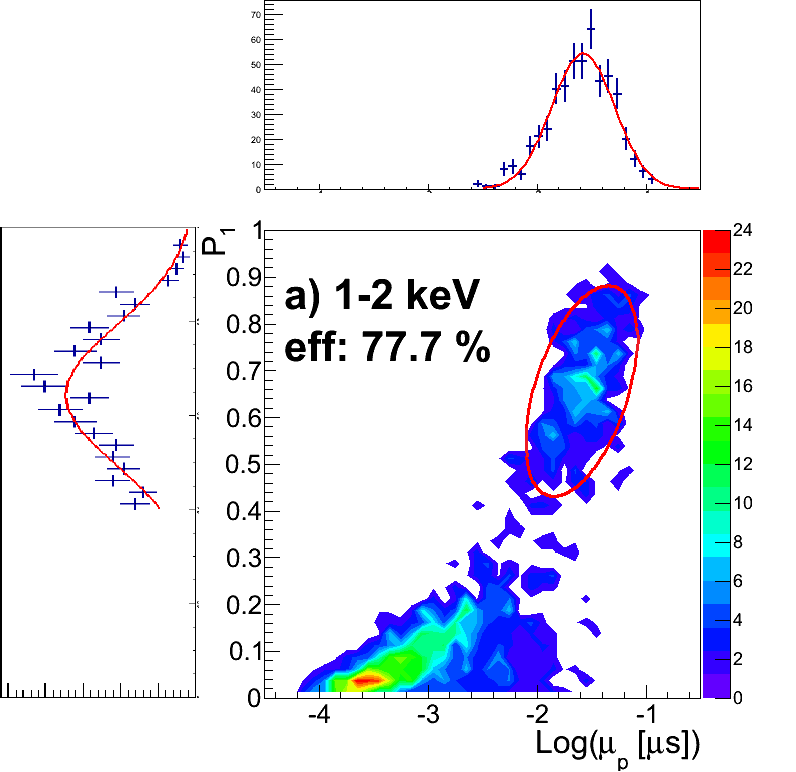}

\caption{\label{P1mu12} Distribution of \(^{22}\)Na/\(^{40}\)K events in the [1–2] keV energy range, selected by coincidence with a 1274.5 keV or 1460.8 keV $\gamma$-ray detected in a different module, shown in the (P1, Log($\mu_p$)) parameter space. The red line indicates the PSV = 3. The PSV variable is defined within the [1–2] keV interval, with a cut corresponding to 77.7\% efficiency in that range. Projections of the P1 and Log($\mu_p$) variables are shown in the left and top panels, respectively, together with gaussian fits (for P1 > 0.4 events), supporting the bivariate analysis performed \cite{Amare:2018sxx}. }
\vspace{-0.5cm}
\end{center}
\end{figure}

    However, P1 is not effective in discriminating between genuine low-energy bulk scintillation events and those caused by long phosphorescence or pulse tails. To address this, a new parameter based on the logarithm of the mean time of individual phe arrival times within the digitized window, $\mu$, is introduced \cite{kim2019limits}:

    \begin{equation}
    \mu = \frac{\sum_{i} A_i t_i}{\sum_{i} A_i},
    \end{equation}

    where \( A_i \) and \( t_i \) represent the amplitude and time of the \( i \)-th peak identified in the waveform by the peak-finding algorithm \cite{maolivan}, respectively. Bulk scintillation events are located around log($\mu$)= - 1.65.

    Since P$_1$ and $\mu$ are correlated, event selection was based on a bi-parametric
cut through the Pulse Shape Variable (PSV) parameter. Figure \ref{P1mu12} shows the distribution of these pulse shape parameters for \(^{22}\)Na/\(^{40}\)K events in the [1–2]~keV energy range, selected by coincidence with a high energy gamma in another module.
 
In ANAIS-112, an event was classified as bulk NaI(Tl) scintillation if its PSV was smaller than 3 (red line in the main panel of Figure \ref{P1mu12}). The PSV cut value was set to achieve a 77.7\% efficiency for accepting events in the [1–2]~keV energy range, specifically for $^{40}$K and $^{22}$Na events that are identified by coincidence with a high-energy gamma in another module \cite{Amare:2018sxx}. The PSV cut eliminates fast Cherenkov-like events originating in the PMTs, random coincidences between individual dark current photoelectrons, and long phosphorescence or pulse tails in the crystal.\\

    \item \textbf{Symmetric light sharing.} 
    
    Below 2 keV, a specific population of background events is observed. Unlike pulse tails, these events are characterized by an asymmetry in the light collected by the two PMTs within the same module~\cite{Akimov:2015cta,Li:2015qhq}. These asymmetric events, also observed by the COSINE collaboration \cite{Adhikari:2017esn}, cannot be discriminated from bulk scintillation using only the PSV parameter. 
    
    To reject this type of events, a cut based on the number of peaks detected in each PMT (n$_0$ and n$_1$) by the peak-finding
algorithm is applied. In particular, an event must have more than four peaks in each PMT to be selected (n$_0$>4 and n$_1$>4). These events, which are only observed in background runs below 2~keV, are not present neither in $^{109}$Cd calibration runs nor in the population from $^{22}$Na selected
by coincidence with a high energy deposition in another crystal. This suggests that they are likely caused by light emission near or within the PMTs.

\end{itemize}

The total efficiency for selecting bulk NaI(Tl) scintillation events derived from the previous ANAIS-112 filtering protocol is calculated before unblinding with the full statistics independently for each detector. This efficiency, depicted in Figure \ref{efficiencyBDT} in black, is estimated using \(^{109}\)Cd, \(^{22}\)Na, and \(^{40}\)K scintillation events by combining the trigger efficiency, PSV cut efficiency, and asymmetry cut efficiency. The detailed procedure for obtaining these efficiencies can be found in \cite{Amare:2018sxx}. 

The overall efficiency is nearly 100\% down to 2 keV, but then drops sharply to around 15\% at 1 keV. The asymmetry cut based on the number of peaks has been found to be the most restrictive cut, significantly reducing the efficiency at 1 keV. Additionally, as seen in Figure \ref{LEdatasim}, the former filtering protocol resulted in a background excess below 2~keV, likely due to non-scintillation events leaking the filtering. 

\vspace{0.5cm}

\subsection{New filtering protocol based on machine learning techniques}\label{BDTnew}

To improve the acceptance efficiency of bulk scintillation events and enhance the rejection of asymmetric or anomalous events below 2 keV, a supervised ML approach based on multivariate analysis has been implemented. Specifically, a Boosted Decision Tree (BDT) algorithm is employed, which combines several pulse-shape variables into a single powerful discriminator for more effective noise rejection. A complete description of the new low-energy filtering procedure is provided in \cite{coarasa2022improving,Coarasa_2023,phdivan}.

The most challenging aspect of BDT training is obtaining pure event samples to model signal and noise. To achieve this, specific selection criteria, called preselection cuts, are applied to the training data. These cuts ensure that the events used for training match the energy of the ROI  and meet certain quality standards in terms of their signal shape.

\begin{itemize}
    \item \textbf{Signal events.} Taking advantage of the dedicated neutron calibration program conducted within the ANAIS-112 experiment (see Chapter \ref{Chapter:QF}), signal events are selected as scintillation events caused by neutron interactions (mainly elastic NRs). The neutron training population includes [1-2] keV events with P1>0.35 and a trigger time within 400 ns of the main acquisition trigger. The latter cut removes neutron events with delayed software triggers, which may lead to inaccurate energy and pulse shape parameter estimations.

    \item \textbf{Noise events.} Noise events are obtained from the blank module (see Section~\ref{BlankModSec}). The selection criteria for building the noise training population are: pulse area equivalent to a number of phe between 10 and 28 (equivalent to the energy range between [1-2] keV in the modules with scintillator), baseline root mean square below 0.60 mV in each PMT to remove electrical noise, and, in 20\% of the selected events, P1<0.35 is required to boost the training to reject the populations more difficult to disentangle from bulk scintillation (asymmetric events).
\end{itemize}

\begin{figure}[b!]
\begin{center}
\includegraphics[width=0.6\textwidth]{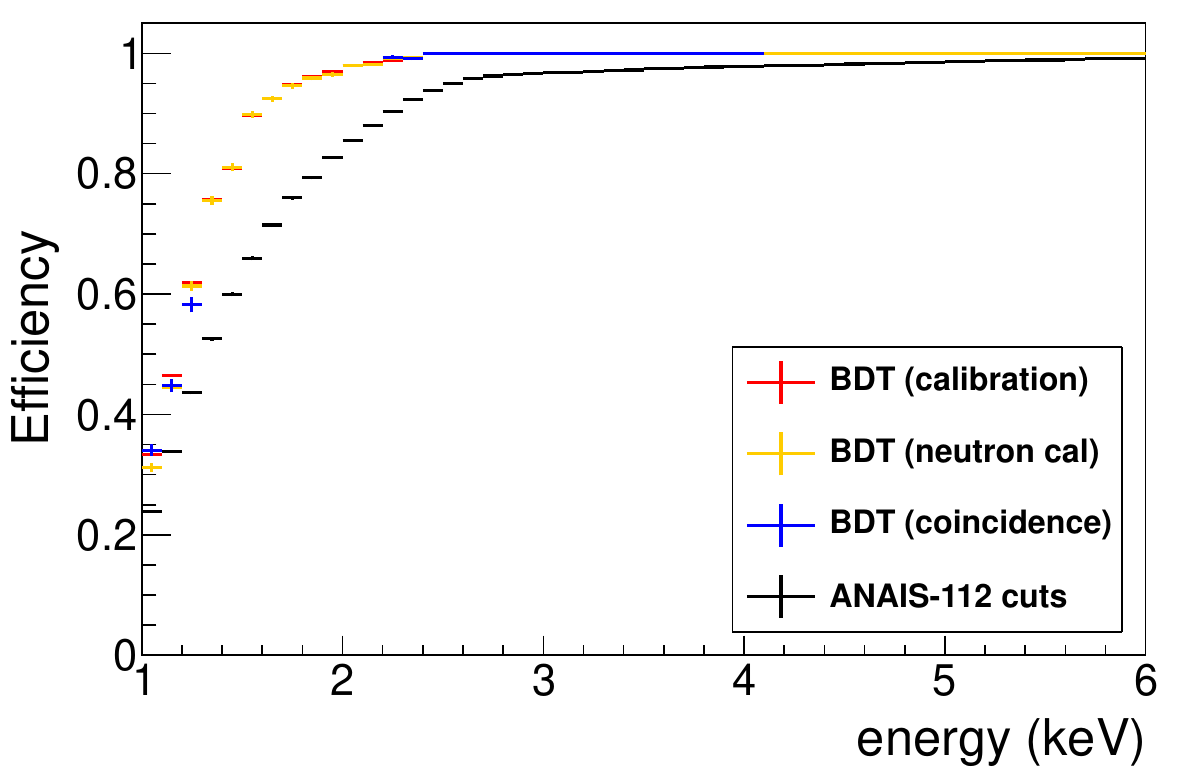}

\caption{\label{efficiencyBDT} Average total efficiency in all the ANAIS–112 modules as a function of energy, obtained using the BDT cut estimated from neutron
calibration (in orange), $^{109}$Cd calibration (in red) and $^{22}$Na/$^{40}$K coincidences (in blue). For comparison, the efficiency with the previous ANAIS–112 filtering protocols is shown in black \cite{coarasa2022improving}. }
\end{center}
\end{figure}

Training populations of over 30.000 events each for signal and noise, independent of background data, were selected. From these events, only 70\% of each population was used for training, with the remaining 30\% used for checking overtraining. 

For event selection, an energy-dependent BDT cut was defined  for every detector. Efficiencies were estimated independently for each detector using different populations of events (neutrons, \(^{109}\)Cd, \(^{40}\)K, and \(^{22}\)Na), with all estimations found to be fully compatible, as shown in Figure \ref{efficiencyBDT}. The BDT cut significantly improves the efficiency by about 30\% in the [1-2] keV range, reaching nearly 100\% at 1.8~keV. 

Furtuhermore, as will be depicted in Figure \ref{LEdatasim}, the BDT filtering procedure achieves a 18\% reduction in the background level below 2 keV compared to the previous ANAIS-112 filtering procedure. However, the excess of events below 2 keV with respect to the background model still persists, suggesting the presence of an unaccounted background contribution in this energy range. In parallel, alternative approaches to identify anomalous scintillation are being explored, in line with the objectives of the new DAQ system, ANOD (see Section~\ref{ANODfiltering}). A reanalysis of the background model will be presented in Chapter~\ref{Chapter:bkg}.


\subsection{New filtering developments using ANOD data} \label{ANODfiltering}

The same discrimination variables defined in ANAIS, which has been presented in Section~\ref{Filtering}), have also been implemented in this thesis in ANOD, enabling the application of the same filtering strategy with the new acquisition system. As will be discussed later in this chapter (see Figure \ref{LEdatasim}), a comparison between the background model and ANAIS-112 data reveals discrepancies below 3 keV, suggesting the presence of anomalous scintillation events that survive the standard cuts, such as asymmetric pulses with more than four peaks in each PMT and pulses with slow decays or long tails.

To select these anomalous scintillation events, the variables P1 and $\mu$ will be used. The definition of P1 remains the same as in ANAIS; however, due to the different temporal resolution, the resulting values might differ. For this reason, the notation P$_{1,\text{ANOD}}$ will be used to refer specifically to the P1 parameter computed in ANOD. In contrast, the first moment in ANOD, $\mu_{\text{ANOD}}$ is redefined within the full 8 $\mu$s DAQ window. Thus, events in ANOD appear in regions that were unpopulated in ANAIS, such as $\log(\mu_{\text{ANOD}}) > 0$.


\begin{figure}[b!]
\begin{center}
\includegraphics[width=0.8\textwidth]{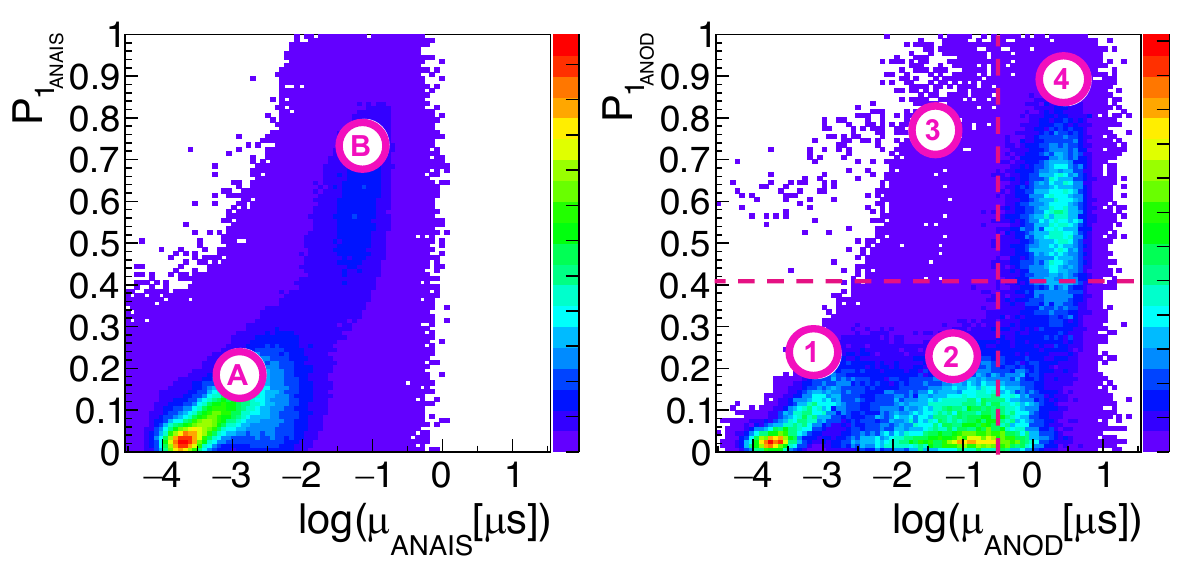}

\caption{\label{ANODfondo12} (P1, log($\mu_p$)) plane for background events in the [1-2] keV range. \textbf{Left panel:} ANAIS. \textbf{Right panel:} ANOD. ANOD is capable of further separating Region A and Region~B identified in ANAIS into two distinct populations. Region A is split into Regions~1 and 2, and Region B is separated into Regions 3 and 4. The selection cuts applied in this work to identify anomalous scintillation events in ANOD are indicated in the figure by dashed magenta lines. }
\end{center}
\end{figure}




Figure \ref{ANODfondo12} illustrates the distribution of background events in the [1-2] keV range for ANAIS and ANOD corresponding to the same run. It can be observed that in ANAIS, scintillation events concentrate in Region B, describing an ellipse centered at P$_{1,\text{ANAIS}}\sim0.65$ and \( \log(\mu_{\text{ANAIS}})\sim\)-1.5. Additionally, fast pulses are concentrated in Region A, where P$_{1,\text{ANAIS}}$<0.4 and long \( \log(\mu_{\text{ANAIS}}) \) <-2.

However, unlike ANAIS, ANOD is capable of further separating each of these two regions into two distinct populations. Region A in the left panel of Figure~\ref{ANODfondo12} is split into Regions~1 and 2, the latter characterized by P$_{1,\text{ANOD}}$<0.4 and $\log(\mu_{\text{ANOD}}) \sim -1$. Region B is separated into Regions 3 and 4, the latter exhibiting P$_{1,\text{ANOD}}$ $\sim$ 0.65  and $\log(\mu_{\text{ANOD}}) \sim 0.5$. Notably, Region 4 can be associated with non-bulk NaI scintillation events contaminating the ROI.

Region 2 comprises pulses not compatible with NaI(Tl) bulk scintillation. These events exhibit fast scintillation characteristics, compatible with Cherenkov events, and therefore do not present a significant challenge for rejection.

Photoelectrons contributing beyond the ANAIS window could be attributed to accidental coincidence of scintillation events or Cherenkov light emission with a low probability. In some cases, a second real correlated scintillation event could be expected when a short lifetime isotope decays after the parent nucleus does. The longer acquisition window, on the other hand, allows to discriminate slower scintillation time constants that could be originated by different scintillation mechanisms, and to identify the possible contribution from afterpulses (a PMT signal which correlates with a previous one, in general producing a large charge cloud within the PMT that ionizes the residual gas and  results in positive ions hitting the photocathode a few microseconds after the original pulse).

\begin{figure}[b!]
\begin{center}
 \centering
   \includegraphics[width=1.\textwidth]{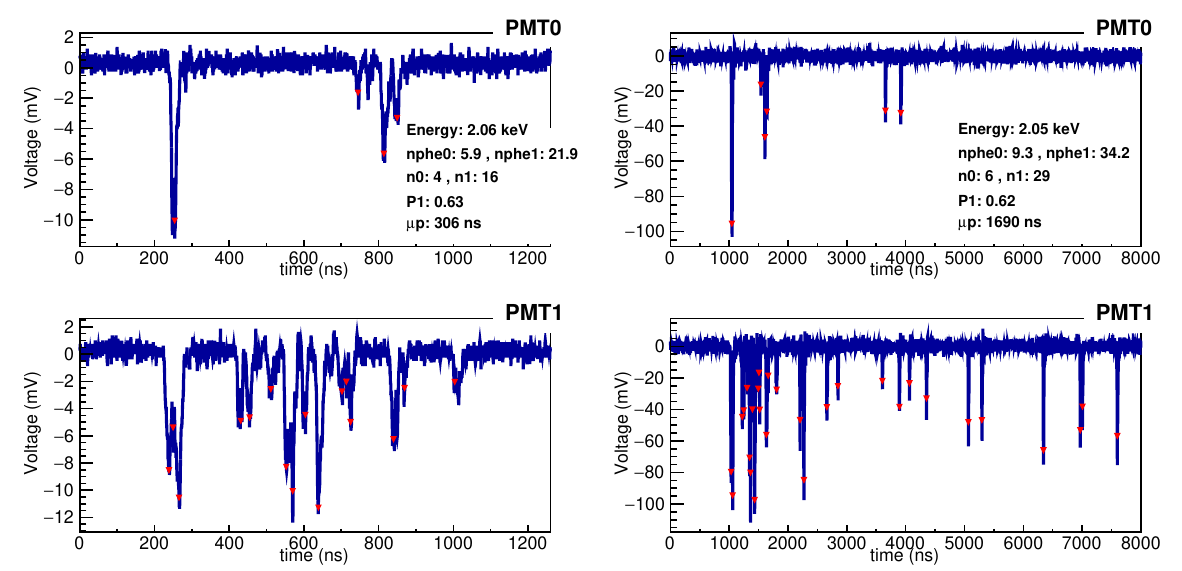}
    
\caption{\label{examplePulseANAISAnodAnomalous} Analogous to Figure \ref{examplePulseANAISAnod}, but for an anomalous scintillation event corresponding to Region 4 (see Figure \ref{ANODfondo12}). }
\end{center}
\end{figure}

\begin{figure}[b!]
\begin{center}
 \centering
    \includegraphics[width=0.49\textwidth]{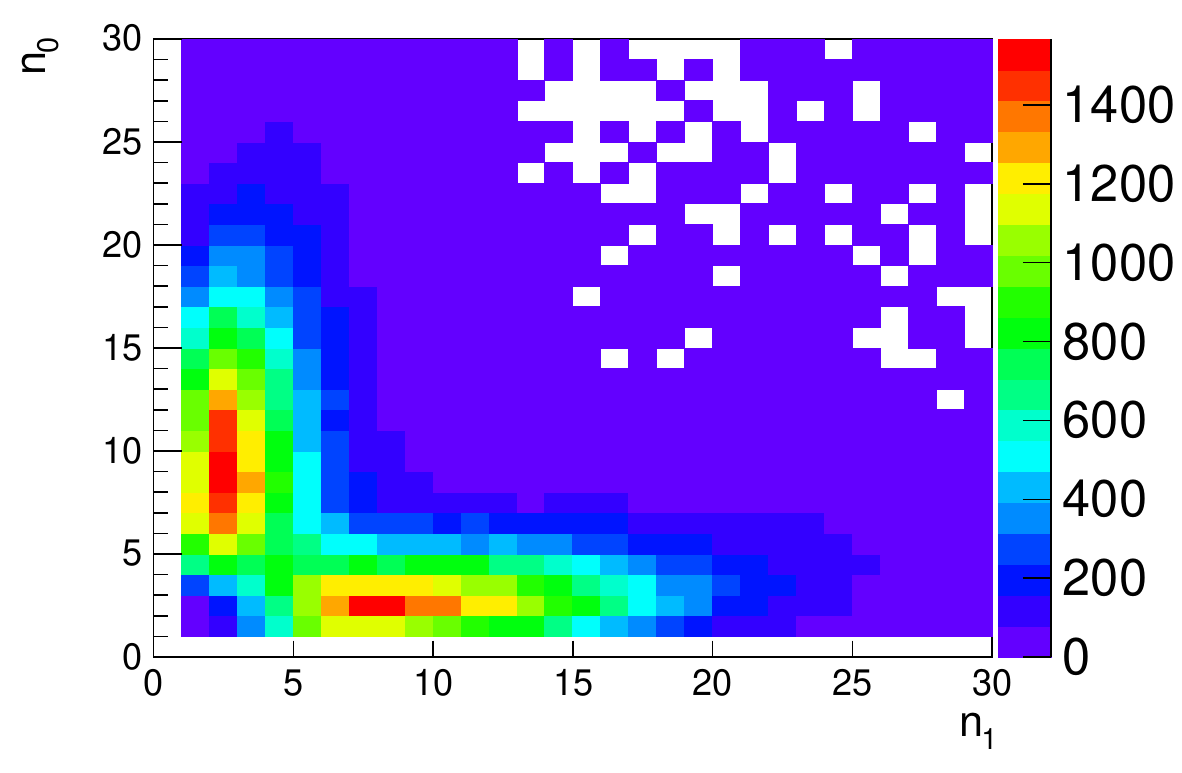}   \hfill \includegraphics[width=0.41\textwidth]{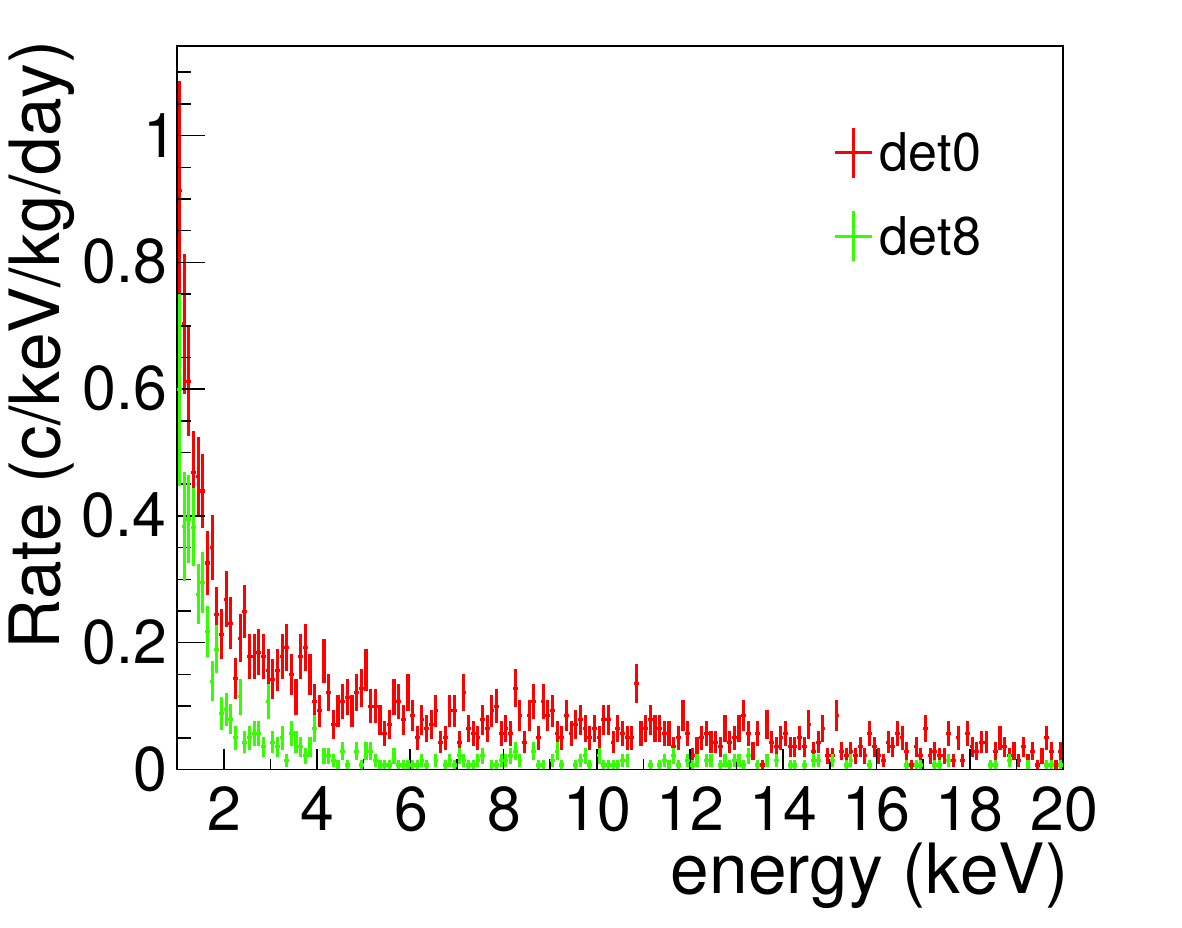}
    
\caption{\label{forman0n1} \textbf{Left panel:} Number of peaks detected in each PMT in ANAIS (n$_0$ and n$_1$) for background events in the [1–20] keV energy range that fall within Region 4 in ANOD, i.e., satisfying P$_{1,\text{ANOD}} >0.4$ and $\log(\mu_{\text{ANOD}}) > -0.5$.
\textbf{Right panel:} Energy deposited in ANAIS by background events belonging to Region 4 selected in ANOD, shown for detector~0 (red) and detector 8 (green).}
\end{center}
\end{figure}

More interesting is the separation observed in ANOD between Region 3 and Region~4 in the right panel of Figure~\ref{ANODfondo12}. In ANAIS, these events overlap within a single ellipse that includes both genuine scintillation events and others with asymmetric light sharing between the two PMTs, and then, the latter are uncorrectly tagged as bulk scintillation events. In contrast, ANOD data show that scintillation events cluster around Region 1, while events in Region 4 are anomalous scintillation events. Most of these Region 4 events are concentrated between 1 and 2 keV.

Figure \ref{examplePulseANAISAnodAnomalous} shows an example of such an anomalous (asymmetric) scintillation event that passes the ANAIS-112 filtering protocols but could be effectively rejected using ANOD. It is evident from the ANOD acquisition window that this event does not correspond to NaI scintillation, whereas in the ANAIS acquisition window it appears compatible.

It is therefore of interest for the experiment to identify and select this type of events in order to achieve a more detailed characterization and, ultimately, to reject them. This selection will be performed in this work using the synchronized ANAIS-ANOD event tree. Temporal synchronization between both DAQs is made possible by their shared clock. Using this synchronized dataset, anomalous events can be selected based on ANOD variables, while the corresponding energy deposited in ANAIS can be studied.

In this work, a conservative cut is applied to select this population of events. The cut is shown in the right panel of Figure \ref{ANODfondo12}. Specifically, events are selected that fulfill the following conditions. First, P$_{1,\text{ANOD}} > 0.4$ is imposed to eliminate fast pulses. Subsequently, due to the clear separation observed in the variable $\log(\mu_{\text{ANOD}})$, this parameter is used to define an anomalous first moment: $\log(\mu_{\text{ANOD}}) > -0.5$. It should be noted that this is a conservative cut. At low energies, a more stringent condition on $\log(\mu^{\mathrm{ANOD}}_p)$ would be required to effectively reject more events corresponding to anomalous scintillation. Nevertheless, for the scope of the present work, the applied selection is deemed sufficient. This selection can be applied to both background events and events from \textsuperscript{109}Cd and neutron calibration runs.

The left panel of Figure~\ref{forman0n1} shows the number of peaks detected by each ANAIS PMT (n\textsubscript{0} and n\textsubscript{1}) for the background population of anomalous events located around Region 4. As observed, these events are characterized by a strong asymmetry in the light sharing. The right panel of Figure~\ref{forman0n1} displays the energy deposited in ANAIS by background events when applying the same selection criteria for two different detectors. The distribution exhibits an exponential behavior at low energies.





\begin{figure}[t!]
    \centering
    {\includegraphics[width=0.9\textwidth]{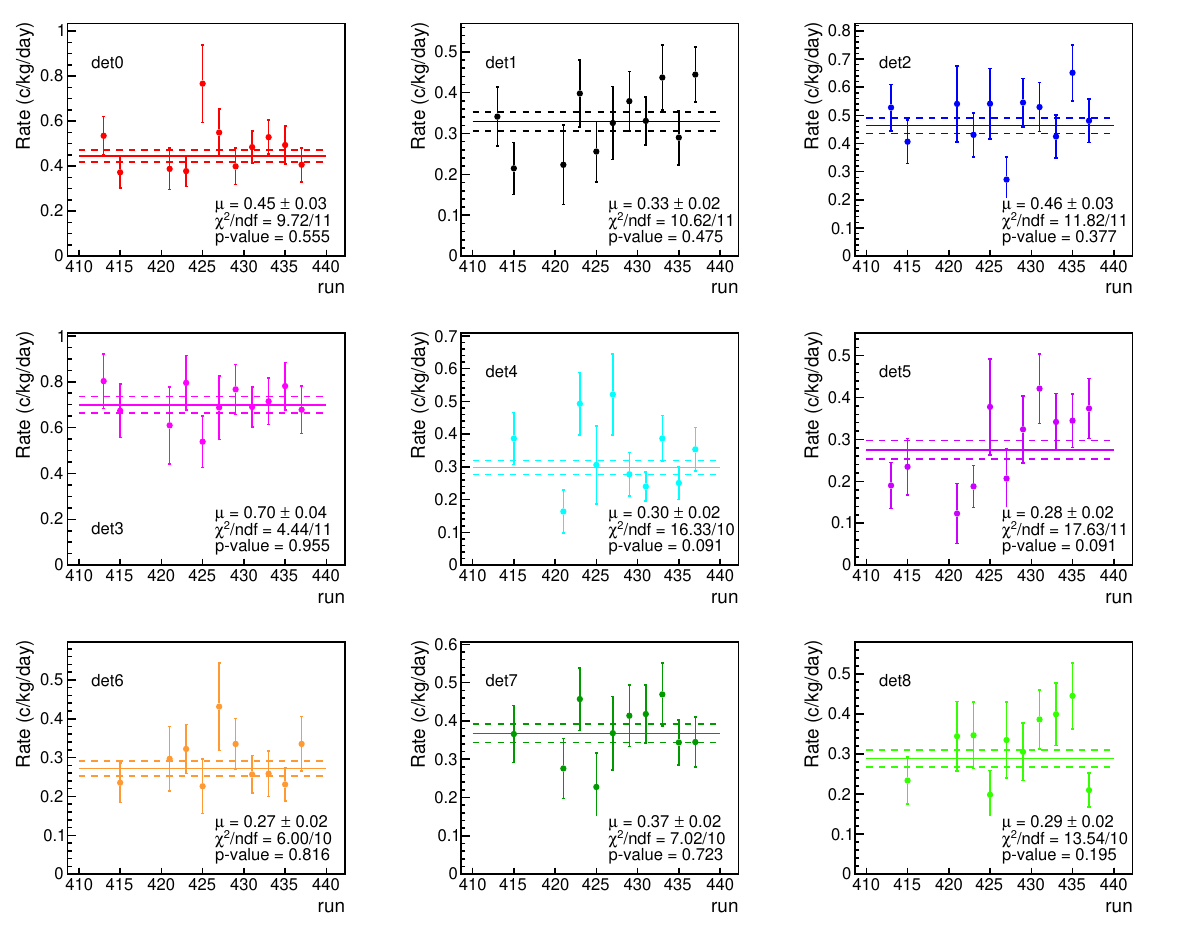}}
    
    \caption{\label{estabilidad} Evolution of the rate of anomalous events in Region 4 of background data within the [1–2] keV energy range, passing the BDT selection cut for the nine ANAIS–112 modules, since the integration of the ANOD DAQ into the ANAIS–112 setup in Winter~2024. The weighted mean rate and corresponding standard deviation for each detector are indicated in the panels. The results of a chi-squared test of the data points with respect to that mean, taking into account the associated uncertainties, are also shown, together with the corresponding p-value.}
\end{figure}

Figure~\ref{estabilidad} shows the temporal stability of the rate of these events in the [1–2]~keV energy interval across detectors since the integration of the ANOD DAQ into the ANAIS-112 setup in Winter~2024. It is important to note that the displayed rates correspond to anomalous asymmetric events that pass the BDT selection cut and are inherently present in the ANAIS-112 background data.

A $\chi^2$-squared test of the data points with respect to the mean, accounting for the associated uncertainties, has been performed. In all cases, the resulting p-values exceed 5\%, supporting the hypothesis of the temporal stability of this population. Nevertheless, additional data are required to confirm this hypothesis. Moreover, the mean rate is not uniform among detectors; for instance, detector 3 exhibits a substantially higher rate. This behavior is attributed to the specific characteristics of the PMTs and the applied HV. In addition, it is worth noting that the rate of anomalous events in the multiple-hit population is negligible.

This population of highly asymmetric anomalous events will be employed in Chapter~\ref{Chapter:QF} and Chapter~\ref{Chapter:bkg} for the study of the QFs of the ANAIS crystals and the improvement of the ANAIS-112 background model, respectively.

\section{ANAIS-112 background model}\label{BkgModel}

The sensitivity of a DM experiment is ultimately limited by its radioactive background levels. Consequently, substantial effort has been invested by experiments in enhancing detector radiopurity. For instance, ANAIS-112 collaborated with Alpha Spectra Inc. to develop low-background NaI(Tl) detectors. Since completely eliminating all the radioactive contributions is not feasible, understanding the background of an experiment is crucial for effective DM searches. A reliable background model aids in the experiment design, systematics analysis and accurate sensitivity estimation. 

The background model of the ANAIS–112 experiment was developed in 2019 with the Geant4 package \cite{amare2019analysis}. The ANAIS-112 set-up was thoroughly implemented, incorporating key experimental features. However, during the course of this thesis, several improvements have been introduced to the 2019 model. These include an updated Geant4 geometry of the ANAIS-112 set-up with a more accurate representation, a revision of the Geant4 physics list, and a post-processing stage of the simulation to account for the response of the ANAIS-112 detectors. These improvements will be described in detail in Chapter \ref{Chapter:Geant4}.

Within the ANAIS-112 background model the contributions are classified into three main groups: external, internal and cosmogenic backgrounds. \\

\begin{table}[t!]
\centering
\begin{tabular}{ccccc} 
\hline
\multicolumn{5}{c}{\multirow{2}{*}{Activity (mBq/PMT)}} \\
\multicolumn{5}{c}{}                                     \\

 \hline
Component                                                         & $^{40}$K & $^{232}$Th & $^{238}$U & $^{226}$Ra \\
\hline
 \hline
\multirow{2}{*}{PMTs D0}  &  97 $\pm$ 19 &  20 $\pm$ 2 &  128 $\pm$ 38 &  84 $\pm$ 3\\
 &   133 $\pm$  13  & 20 $\pm$  2  & 150 $\pm$  34  & 88 $\pm$  3          \\

 \multirow{2}{*}{PMTs D1}   &  105 $\pm$ 15 & 18 $\pm$ 2 & 159 $\pm$ 29 & 79 $\pm$ 3\\
 &     105 $\pm$ 21 & 22 $\pm$ 2 & 259 $\pm$ 59 & 59 $\pm$ 3\\
  \multirow{2}{*}{PMTs D2}   &  155 ± 36 & 20 ± 3 & 144 ± 33 & 89 ± 5 \\
 &     136 ± 26 & 18 ± 2 & 187 ± 58 & 59 ± 3 \\
 \multirow{2}{*}{PMTs D3}  &  108 ± 29 & 21 ± 3 & 161 ± 58 & 79 ± 5\\
 &     95 ± 24 & 22 ± 2 & 145 ± 29 & 88 ± 4 \\
 \multirow{2}{*}{PMTs D4}  &  98 ± 24 & 21 ± 2 & 162 ± 31 & 87 ± 4\\
 &     137 ± 19 &  26 ± 2 &  241 ± 46 &  64 ± 2 \\
 \multirow{2}{*}{PMTs D5}  &  90 ± 15 & 21 ± 1 & 244 ± 49 & 60 ± 2 \\
 &     128 ± 16 & 21 ± 1 & 198 ± 39 & 65 ± 2 \\
  \multirow{2}{*}{PMTs D6} &  83 ± 26 & 23 ± 2 & 238 ± 70 & 53 ± 3 \\
 &     139 ± 21  & 24 ± 2  & 228 ± 52  & 67 ± 3 \\
\multirow{2}{*}{PMTs D7}  &  104 ± 25 & 19 ± 2 & 300 ± 70 & 59 ± 3\\
 &     103 ± 19 & 26 ± 2 & 243 ± 57 & 63 ± 3\\
 \multirow{2}{*}{PMTs D8}  &  127 ± 19 & 23 ± 1 & 207 ± 47 & 63 ± 2\\
 &     124 ± 18 & 21 ± 2 & 199 ± 44 & 61 ± 2\\
 \hline
 Weighted mean   &  114.9 ± 4.6 & 21.6 ± 0.4 & 180.2 ± 9.8 & 66.7 ± 0.6\\
 \hline
\end{tabular}
\caption{\label{tablaPMT} Activity of the PMTs of the ANAIS-112 experiment. The values obtained for each Hamamatsu R12669SEL2 unit are reported together with the weighted mean \cite{amare2019analysis}.}

\end{table}

\begin{itemize}
    \item \textbf{External backgrounds.} 
    
    External background in the ANAIS NaI(Tl) crystals arise from surrounding sources such as PMTs, copper housings, lead shielding and $^{222}$Rn in the inner atmosphere. Among these, PMTs are the most important contribution. While external backgrounds are dominant at high energies, their effect on the DM search region is expected to be low. The main sources of external radiation are four primordial radioactive isotopes: $^{40}$K, $^{232}$Th, $^{238}$U and $^{226}$Ra. 
    
    The activities of these isotopes were measured using HPGe spectrometry at LSC before ANAIS commissioning. As shown in Table~\ref{tablaPMT}, compatible levels of activity among the different PMT units (eighteen) were found, with $^{40}$K and $^{238}$U the main contributions. 

     Since direct measurements of the activities from copper housing, lead shielding and $^{222}$Rn were not possible, upper limits were derived instead (see Table~\ref{tablaOthers}). For copper and quartz windows values are the same as was reported for ANAIS–0 prototype \cite{cebrian2012background}. Moreover, a value for the
radon content in the inner volume air of about one hundredth
of the external air radon content has been assumed in the 
background model (0.6~Bq/m$^3$)~\cite{amare2025towards}.
    
    Additional contributions from external neutrons, gammas and muons were considered negligible in the background model. 
\end{itemize}

   \begin{table}[t!]
    \centering
\begin{tabular}{ccccccc}
\hline
\multicolumn{6}{c}{\multirow{2}{*}{Activity (mBq/kg)}} \\
\multicolumn{6}{c}{}    \\
 \hline
Component & $^{40}$K & $^{232}$Th & $^{238}$U & $^{226}$Ra & Other \\
\hline
 \hline
Copper housing & < 4.9 & < 1.8 & < 62 & < 0.9 & $^{60}$Co: < 0.4\\
 Quartz windows & < 12 & < 2.2 & < 100 & < 1.9\\
 Silicone pads & < 181 & < 34 & & 51$\pm$7 \\
 Lead shielding & & < 0.3 & < 0.2 & & $^{210}$Pb: < 20 \\
Inner atmosphere & & & & & $^{222}$Rn: < 0.6 Bq/m$^3$\\

 \hline
\end{tabular}
\caption{\label{tablaOthers} Upper limits (95\% C.L.) of the external components (outside the crystal and excluding PMTs) of the ANAIS-112 detectors \cite{amare2019analysis}. All components are expressed in mBq/kg, except for the \textsuperscript{222}Rn content, which is given in Bq/m\textsuperscript{3}.}
\end{table}

\vspace{0.5cm}

\begin{itemize}

    \item \textbf{Internal backgrounds.}  
    
    Internal background in a NaI(Tl)-based experiment comes from isotopes either incorporated in the crystal growth or not fully removed from the raw powder used in its production. As their decays occur within the volume of the crystal, internal contamination is the dominant background source because of the high efficiency for the detection of the energy released. Key internal backgrounds include $^{40}$K and isotopes from the decay chains.

    \begin{itemize}
        \item \textbf{$^{40}$K.} 
        
        The 3.2 keV deposit (following the K-shell EC decay of $^{40}$K) lies in the ROI of ANAIS-112, making it a primary background concern in the experiment. The bulk \(^{40}\text{K}\) content of ANAIS–112 crystals was accurately estimated by detecting coincidences between the low-energy deposit in one detector and the 1460.8 keV gamma ray absorbed in another module, with efficiencies evaluated through Monte Carlo simulations. Table~\ref{tablaInternal} shows the measured \(^{40}\text{K}\) activity in ANAIS-112 crystals, with an average value of 0.96~mBq/kg. \\

        \begin{table}[b!]
\centering
\begin{tabular}{ccccc} 
\hline
\multicolumn{5}{c}{\multirow{2}{*}{Activity (mBq/kg)}} \\

\multicolumn{5}{c}{}               \\

\hline
Module                                                         & $^{40}$K & $^{232}$Th & $^{238}$U &  $^{22}$Na \\
\hline
 \hline
D0 & 1.33 $\pm$ 0.04 & (4.0 $\pm$ 1.0)$\times$10$^{-3}$ & (10.0 $\pm$ 2.0)$\times$10$^{-3}$ & 1.79 $\pm$ 0.13 \\
D1 & 1.21 $\pm$ 0.04 & & & 1.94 $\pm$ 0.13 \\
D2 & 1.07 $\pm$ 0.03 & (0.7 $\pm$ 0.1)$\times$10$^{-3}$ & (2.7 $\pm$ 0.2)$\times$10$^{-3}$ & 0.51 $\pm$ 0.07 \\
D3 & 0.07 $\pm$ 0.03 & & & 0.79 $\pm$ 0.05 \\
D4 & 0.54 $\pm$ 0.04 & & & 0.72 $\pm$ 0.04 \\
D5 & 1.11 $\pm$ 0.02 & & & 0.51 $\pm$ 0.03 \\
D6 & 0.95 $\pm$ 0.03 & (1.3 $\pm$ 0.1)$\times$10$^{-3}$ & & 0.62 $\pm$ 0.03 \\
D7 & 0.96 $\pm$ 0.03 & (1.0 $\pm$ 0.1)$\times$10$^{-3}$ & & 0.64 $\pm$ 0.03 \\
D8 & 0.76 $\pm$ 0.02 & (0.4 $\pm$ 0.1)$\times$10$^{-3}$ & & 0.65 $\pm$ 0.03 \\

 \hline

\end{tabular}

\caption{\label{tablaInternal} Measured internal contamination levels for the nine NaI(Tl) detectors in ANAIS–112 produced by Alpha Spectra Inc. at the time of detector installation underground~\cite{amare2019analysis}. }
\vspace{0.5cm}

\end{table} 
        
        \item \textbf{$^{238}$U and $^{232}$Th.} 
        
        Both \(^{238}\text{U}\) and \(^{232}\text{Th}\) are naturally occurring parent isotopes with complex decay chains, each comprising over ten isotopes. Many of these isotopes contribute to the radioactive background in ANAIS-112. Due to their long half-lives, most decay steps of these chains are expected to be in secular equilibrium, where the activity of any isotope in the chain is directly proportional to that of the parent. 
        
        The activities were measured by analyzing alpha rates using pulse shape analysis (PSA) and delayed coincidences (Bi/Po sequences). For both \(^{232}\text{Th}\) and \(^{238}\text{U}\) chains, very low activity levels, on the order of a few \text{$\mu$}Bq/kg, were found across all detectors, as shown in Table \ref{tablaInternal}. However, PSA revealed a significant presence of \(^{210}\text{Pb}\) out of equilibrium.\\

        \item \textbf{$^{210}$Pb.} 

 \begin{table}[b!]
\centering

\begin{tabular}{ccccc} 
\hline
\multicolumn{5}{c}{\multirow{2}{*}{Activity (mBq/kg)}} \\
\multicolumn{5}{c}{}               \\
\hline

Module                                                         & $^{210}$Pb & \% bulk & \% surface & depth (\( \mu m \)) \\
\hline
\hline

D0 & 3.15 $\pm$ 0.10 & 50 & 50 & \multirow{2}{*}{100} \\
D1 & 3.15 $\pm$ 0.10 & 50 & 50\\
D2 & 0.70 $\pm$ 0.10 & 25 & 75 & \multirow{6}{*}{30}\\
D3 & 1.80 $\pm$ 0.10 & 79 & 30\\
D4 & 1.80 $\pm$ 0.10 & 75 & 25\\
D5 & 0.78 $\pm$ 0.01 & 25 & 75\\
D6 & 0.81 $\pm$ 0.01 & 50 & 50\\
D7 & 0.80 $\pm$ 0.01 & 50 & 50\\
D8 & 0.74 $\pm$ 0.01 & 75 & 25\\

 \hline
 
\end{tabular}

\caption{\label{tablaPlomo} Internal contamination levels of $^{210}$Pb for the nine NaI(Tl) detectors in ANAIS–112, together with the fixed surface emission fraction and depth for each module, at the time of detector installation underground \cite{amare2019analysis}. }
\vspace{0.5cm}

\end{table}  

        $^{210}$Pb is the main background contributor in the ROI of the ANAIS-112 experiment. It is likely introduced in the detectors through gaseous $^{222}$Rn exposure during the final crystal growth and encapsulation procedures. Therefore, in addition to bulk contamination, a surface $^{210}$Pb activity is also expected. This surface contamination is necessary not only to accurately reproduce the shape of the low-energy region in the measured data but also was proposed as a hypothesis to explain the complex alpha peak structure arising from \(^{210}\)Po, a decay product in the \(^{210}\)Pb sequence. A more detailed discussion regarding this alpha peak structure will be presented in Section~\ref{alphasec}.
        
        The contamination levels of $^{210}$Pb for each ANAIS–112 module are detailed in Table \ref{tablaPlomo}. It is worth highlighting that Alpha Spectra Inc. implemented changes in the production protocols during the manufacturing of the detectors to reduce $^{210}$Pb activity in the crystals, as demonstrated by the lower $^{210}$Pb levels in the most recent modules. As presented in Table \ref{tablaPlomo}, the fraction and depth of $^{210}$Pb surface emission have been adjusted in the background model for each ANAIS-112 module to match the observed low-energy data, with values ranging from 25\% to 75\% and 30 to 100 $\mu$m, respectively. 
        
        In addition, reported in this category although not located directly in the crystal but in close proximity, a $^{210}$Pb activity in the teflon diffuser surrounding the crystals was also considered for certain modules (D3 and D4), based on the hypothesis of $^{222}$Rn-induced contamination.
 A value of 3 mBq/detector has been chosen to reproduce the structure observed in the low-energy background around 12 keV in certain detectors that can be explained by the X-rays from Bi emitted following the \textsuperscript{210}Pb decay.

    \end{itemize}

\end{itemize}

\begin{itemize}
    \item \textbf{Cosmogenic backgrounds.} 
    
    Cosmogenic isotopes are generated by interactions of the cosmic rays with the nuclei in the NaI(Tl) and other detector components when located above ground (fabrication, transport, storage of components...), mainly through secondary neutrons. Proper modelling of these time-dependent backgrounds is crucial for annual modulation searches due to the fact that half-lives of common cosmogenic components are of the order of ten years or less. In particular, most cosmogenic isotopes in NaI(Tl) detectors have half-lives ranging from 19 days (\(^{121}\)Te) to 12 years (\(^{3}\)H), with \(^{129}\)I being an exception due to its much longer half-life of \(1.57 \times 10^7\) years. 
    
    Thanks to the prompt data taking after placing the detectors underground, a dedicated study of cosmogenic radionuclide production in ANAIS-112 detectors was conducted. In ANAIS-112, the most significant cosmogenic isotopes are \(^{22}\)Na and \(^{3}\)H. Several iodine and tellurium isotopes, as well as \(^{109}\)Cd and \(^{113}\)Sn are also of greater concern.\\

    
    \begin{itemize}
        \item \textbf{$^{22}$Na.} 
        
        \(^{22}\)Na poses a significant challenge for DM searches due to the 0.9~keV K-shell binding energy of its daughter, \(^{22}\)Ne, which falls within the ROI of ANAIS-112. Its half-life (T$_{1/2}$ = 2.6 yr) is long enough to compromise the initial years of data taking. 
        
        Similar to \(^{40}\)K determination, to estimate the \(^{22}\)Na activity, coincidences of these low energy depositions in one detector (following the EC decay of \(^{22}\)Na) with 1274.5 keV gamma line absorbed in another module were analyzed. The activity levels of \(^{22}\)Na in the different modules at the moment of being moved underground are summarized in Table \ref{tablaInternal}. \\

        \item \textbf{$^{3}$H.} 
        
        \(^3\)H is a pure beta emitter with a Q-value of 18.6 keV and a long T$_{1/2}$ of 12.3 years. In DM experiments like ANAIS-112 and COSINE-100 \cite{adhikari2021background}, \(^3\)H is a significant background concern given the fact that 57\% of its $\beta$-decay electrons falls within the 1–7 keV energy range, which overlaps with the ROI of these experiments. 
        
\begin{figure}[b!]
\begin{center}
\includegraphics[width=0.7\textwidth]{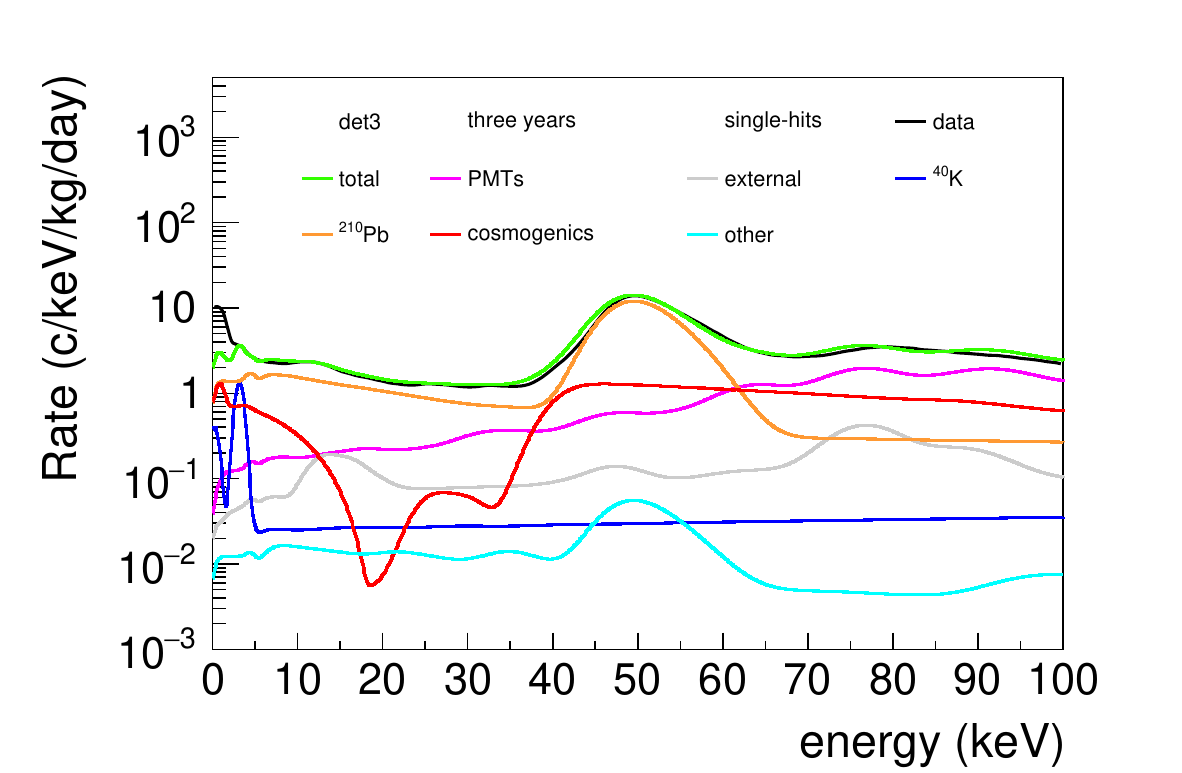}
\includegraphics[width=0.7\textwidth]{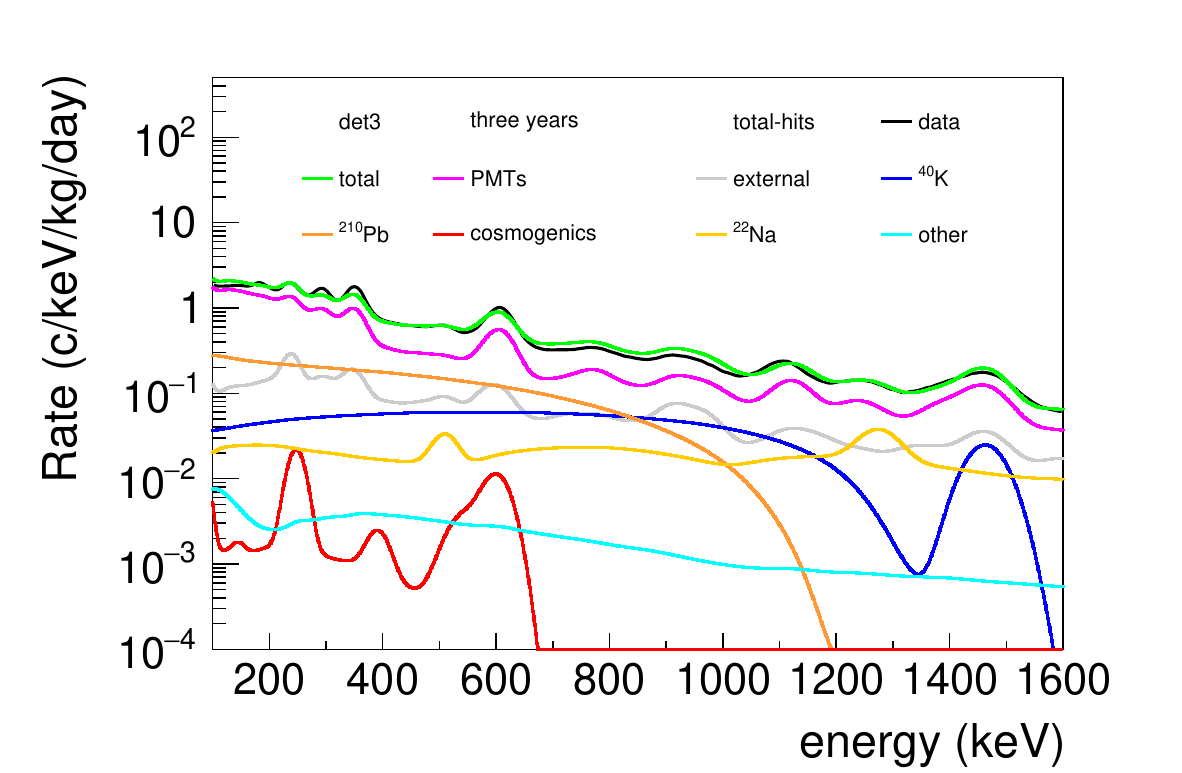}

\caption{\label{LEMEcomp} Energy spectra expected from individual background sources, with the sum of all the contributions (green) and the measured background (black). The results are shown for the D3 module for the first three years of data taking, with the anticoincidence spectrum in the low-energy region \textbf{(top panel)} and the total spectrum in the high-energy region \textbf{(bottom~panel)}. }
\vspace{-0.5cm}
\end{center}

\end{figure}

        Although direct measurement of \(^3\)H in the crystals of ANAIS-112 was not possible due to the difficulty of disentangling its continuous $\beta$-spectrum from other low-energy background components, including its contribution in the background model was required to explain the low-energy data. The initial activities needed to reproduce the data were 0.20 mBq/kg for the D0 and D1 modules and 0.09 mBq/kg for the D2-D8 modules.
        
        Recently, COSINE-100 conducted the first experimental measurement of cosmogenic neutron activation rate for \(^{3}\)H by irradiating NaI(Tl) detectors with a high-energy neutron beam that mimics the energy profile of cosmic ray neutrons at sea level \cite{saldanha2023cosmogenic}. The result agrees exceptionally well with the \(^{3}\)H production rate of \((83 \pm 27) \text{ kg}^{-1}\text{day}^{-1}\) determined by ANAIS-112 \cite{amare2018cosmogenic} with analytical calculations based on various cross-section models and the same Gordon cosmogenic neutron spectrum as COSINE-100 uses.

    \end{itemize}

\end{itemize}

The reliability of the background model is subject to continuous validation against experimental data across different populations (single- and multiple-hits) and energy ranges (low- and high-energy). Figure \ref{LEMEcomp} shows the agreement between data and simulation for the D3 module over the first three years of data taking, showing the anticoincidence spectrum in the top panel and the total spectrum in the bottom panel. 

Below 100 keV (upper plot), the single-hit event rates are mainly dominated by the intrinsic contamination of the NaI(Tl) crystals. According to the background model, the primary contributors to the measured rate in the [1–6] keV energy range during the first year of ANAIS-112 data taking are \(^{210}\)Pb (32.5\%), \(^{3}\)H (26.5\%), \(^{40}\)K (12.0\%), and \(^{22}\)Na (2.0\%). In the high-energy range (lower plot), the background model matches the total rate with a 5.6\% accuracy averaging all the ANAIS–112 detectors. As alredy stated, PMT activity is dominant in the high energy range. The regions that are most poorly described by the model are those between [150,400] keV and between [600,1100] keV, where the model underestimates and overestimates the measured data, respectively. 

\begin{figure}[b!]
\begin{center}
\includegraphics[width=0.6\textwidth]{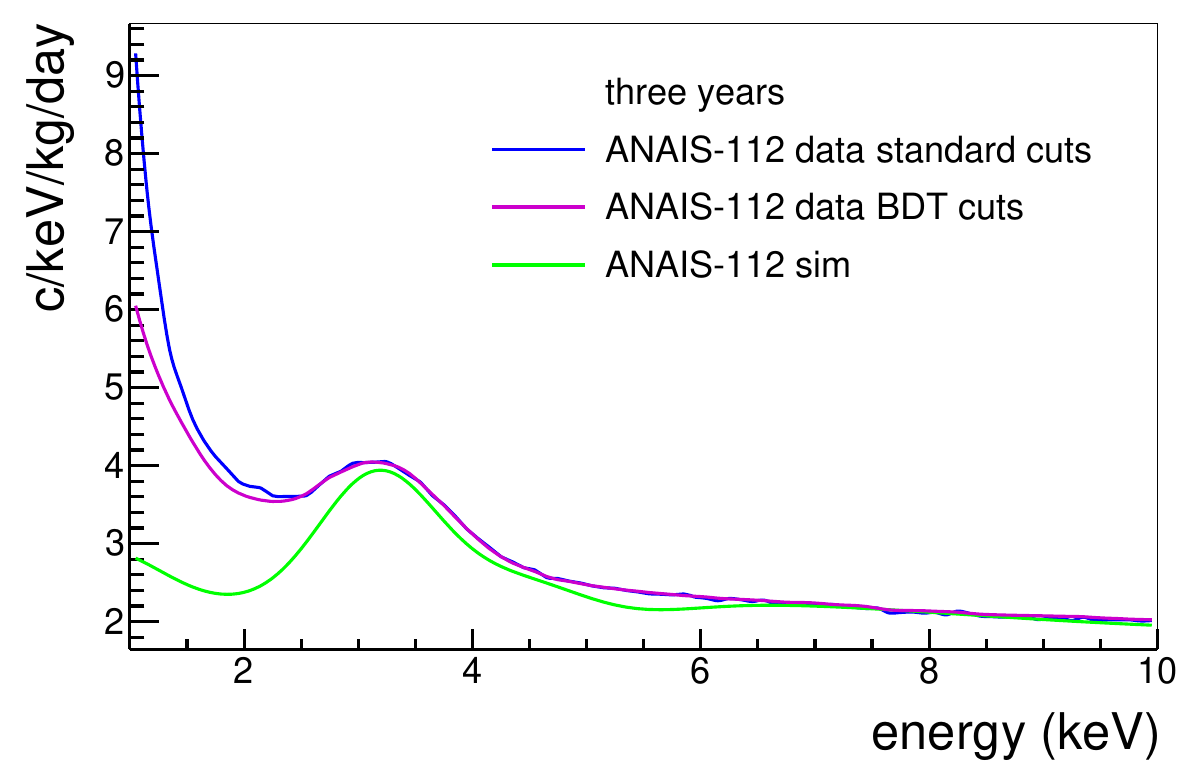}

\caption{\label{LEdatasim} Comparison of the total anticoincidence energy spectrum within the ROI~(blue) against the corresponding background model estimate (green), which includes all simulated contributions. The data presented corresponds to the first three years of ANAIS-112 data, following standard filtering protocols (blue) and the filtering protocols based on machine learning techniques (magenta).}
\end{center}
\end{figure}

The ANAIS-112 background model shows a robust agreement with data above 3~keV. However, as shown in Figure \ref{LEdatasim}, there is a significant discrepancy at lower energies, with a deviation of 54\%, as already mentioned in Section \ref{BDTnew}. The unexplained events below 2 keV could potentially be attributed to non-bulk scintillation events that have not been rejected by the filtering protocols or to background sources not accounted for in the current model. As can be derived from the figure, the application of filtering protocols based on ML techniques results in a significant reduction of the background level below 2 keV. This thesis includes a comprehensive revision of the ANAIS-112 background model, which will be presented in Chapter \ref{Chapter:bkg}.

\section{ANAIS-112 annual modulation search} \label{annualMod}

The following section will review the strategy for the annual modulation analysis, followed by a discussion of the latest ANAIS-112 annual modulation results corresponding to 6 years of data presented in \cite{amare2025towards}.

ANAIS–112 began data acquisition in DM mode on August 3, 2017, and is expected to complete operation by the end of 2025, accumulating more than eight years of data. In the six years of data-taking, ANAIS has maintained over 95.1\% live time across all nine modules. The dead time represents 2.2\%, while 2.7\% is attributed to down time primarily due
to bi-weekly $^{109}$Cd calibrations and nine $^{252}$Cf calibrations
carried out in the referred period. This remarkable duty cycle ensures continuous year-round coverage, which is essential for the annual modulation analysis. 

The energy and time distributions of single-hit events in the ROI ([1-6] keV) were kept blinded from the beginning of data taking. After 1.5, 2, 3 and 6 years
of data acquisition, the corresponding ROI events were unblinded for the annual
modulation analysis. The results, corresponding to exposures of 157.55, 220.69, 
322.82 and 625.75~kg$\times$yr \cite{PhysRevLett.123.031301,Amare_2020,Amare:2021yyu,amare2025towards}, respectively, are consistent with the absence of modulation and
confirm the sensitivity projections.

With data corresponding to 3 years of operation, ANAIS sensitivity to DAMA/LIBRA result is $\sim$ 2.5$\sigma$ in the energy regions between [1-6] keV and [2-6] keV \cite{Amare:2021yyu}. In addition, a reanalysis of the very same first 3 years of data-taking was performed using ML techniques, as previously described in Section~\ref{Filtering}. The improved analysis chain enhances the experiment sensitivity, increasing it from 2.5$\sigma$ to 2.8$\sigma$ \cite{nature}. 

Furthermore, given that ANAIS-112 and COSINE-100 are complementary NaI(Tl) experiments aiming to solve the DAMA/LIBRA puzzle, a combined annual modulation search has been carried out through a direct combination of their residuals \cite{carlin2025combined}. This joint analysis achieves an enhanced sensitivity of 3.7$\sigma$ ([1-6] keV) and 2.6$\sigma$ ([2-6] keV) to the DAMA/LIBRA result for 3 years of data. Very recently, the results for 6 years of ANAIS-112 data have been released \cite{amare2025towards}. Results show no modulation for a sensitivity of  $\sim$ 4$\sigma$  for both the [1–6] and [2–6]~keV energy regions.
In the following, the strategy for the annual modulation analysis,
followed by a discussion of  the last data release, will be reviewed.

\subsection{Annual modulation analysis strategy}\label{annualstrategy}
In the DAMA/LIBRA annual modulation analysis, the residual rate of anticoincidence events as a function of time is calculated by subtracting the annual average from the total rate year by year. These residuals are then fitted to a cosine function of the form \( A \cos(\omega (t - t_0)) \). While it has been suggested that this method may introduce a bias in the fit for slowly varying backgrounds, such a bias is unlikely to explain the DAMA signal. The phase reported by DAMA would correspond to a slightly increasing background, which is difficult to justify, and no significant bias is observed above the energy range where the DM signal is expected.

A different model-independent analysis is performed in ANAIS-112, directly searching for modulation in the overall event rate as a function of time to avoid any potential systematic effects. To enable a more direct comparison with the DAMA/LIBRA result, the same energy regions are considered, specifically [1-6] keV and [2-6] keV, the period is fixed at 1 year, and the phase set to June 2$^{\textnormal{nd}}$. 

Subsequently, a simultaneous fit of the 9 detectors is performed using 45-day time bins, followed by a chi-square minimization, where the $\chi^2$ function is defined as follows:

\begin{equation}
    \chi^2 = \sum_{i,d} \dfrac{(n_{i,d}-\mu_{i,d})^2}{\sigma^2_{i,d}}.
\end{equation}

Here, $n_{i,d}$ denotes the number of events within the ROI for a given time bin $t_i$ and detector $d$, where the measured event count is corrected for both the live time associated with the specific time bin and detector, as well as the corresponding acceptance efficiency. The Poisson uncertainty, denoted as $\sigma_{i,d}$, represents the statistical error associated with the event count, corrected by the same live time and efficiency factors. Furthermore, $\mu_{i,d}$ represents the expected number of events in the specified time bin and detector, including the contribution from a hypothetical DM signal. 

Over time, $\mu_{i,d}$ is expected to decrease due to background contributions from radioactive isotopes with half-lives on the order of a few years, such as $^{210}$Pb (T$_{1/2}$~=~22.3~y), $^{3}$H (T$_{1/2}$~=~12.3~y), and $^{22}$Na (T$_{1/2}$~=~2.6 y). Shorter-lived cosmogenically produced isotopes in the lastly arrived detectors (D6-D8) also contribute to this diminishing rate. 

The time evolution of the background rate  for every detector is modeled by the ANAIS-112 Monte Carlo background simulation,  allowing the event rate to be formulated as:

\begin{equation}
    \mu_{i,d} = [ R_{0,d}(f_d\phi^{MC}_{bkg,d}(t_i)+(1-f_d)\phi_{flat}(t_i))+S_m\cos(\omega(t_i-t_0))]M_d\Delta E\Delta t ,
    \label{rateevents}
\end{equation}
    \vspace{1cm}

wher  $R_{0,d}$ corresponds to the average
rate in the considered energy region, $M_d$ denotes the mass of each module, and $\Delta E$ and $\Delta t$ represent the energy and time intervals, respectively. Three distinct contributions can be clearly identified in Equation~\ref{rateevents}:

\begin{itemize}

    \item A decaying background, modeled by Monte Carlo simulations, where $\phi^{\text{MC}}_{\text{bkg},d}$ is the probability distribution function sampled from the model describing the expected background rate in time bin $t_i$ for detector $d$.

        \item A constant probability distribution function, $\phi_{flat}$, that accounts for noise not explained by the background model (linked to the observed excess below 3 keV) and found at a constant rate in the data, as well as for the average component of a hypothetical DM signal.
        
    \item A modulation signal, where $S_m$ represents the DM annual modulation amplitude. It is set to 0 in order to test the null hypothesis, and allowed to vary freely when testing for modulation. $\omega$ and $t_0$ denote the angular frequency and phase of the modulation searched for in the data.
\end{itemize}

In the fit, the modulation period is fixed to one year and the phase is set to June~2\textsuperscript{nd} according to the SHM, allowing for a direct comparison with the results reported by the DAMA/LIBRA experiment in \cite{Bernabei:2020mon}. $R_{0,d}$ and $f_d$ are free parameters for each detector, while $S_m$ represents a global DM annual modulation amplitude. This results in a total of 19 free parameters to be determined in the fit.

\subsection{Annual modulation results with 6 years}

Before the 6-year unblinding, the stability of event selection efficiencies derived from the ML-based filtering protocol was evaluated. Efficiencies from $^{252}$Cf and $^{109}$Cd calibrations were found to be consistent. Analysis of the deviations in $^{109}$Cd efficiency, averaged across all detectors, relative to their mean value in the [1–8] keV energy range, yielded a standard deviation of 0.13\%. For individual detectors, this value increased to 0.3\%. These results are comparable to or better than those reported by DAMA/LIBRA-phase2 in the same energy range (0.3\%) \cite{Bernabei:2020mon}, demonstrating the excellent stability of the filtering procedure over the entire six-year period.

\begin{figure}[t!]
\begin{center}
\includegraphics[width=0.6\textwidth]{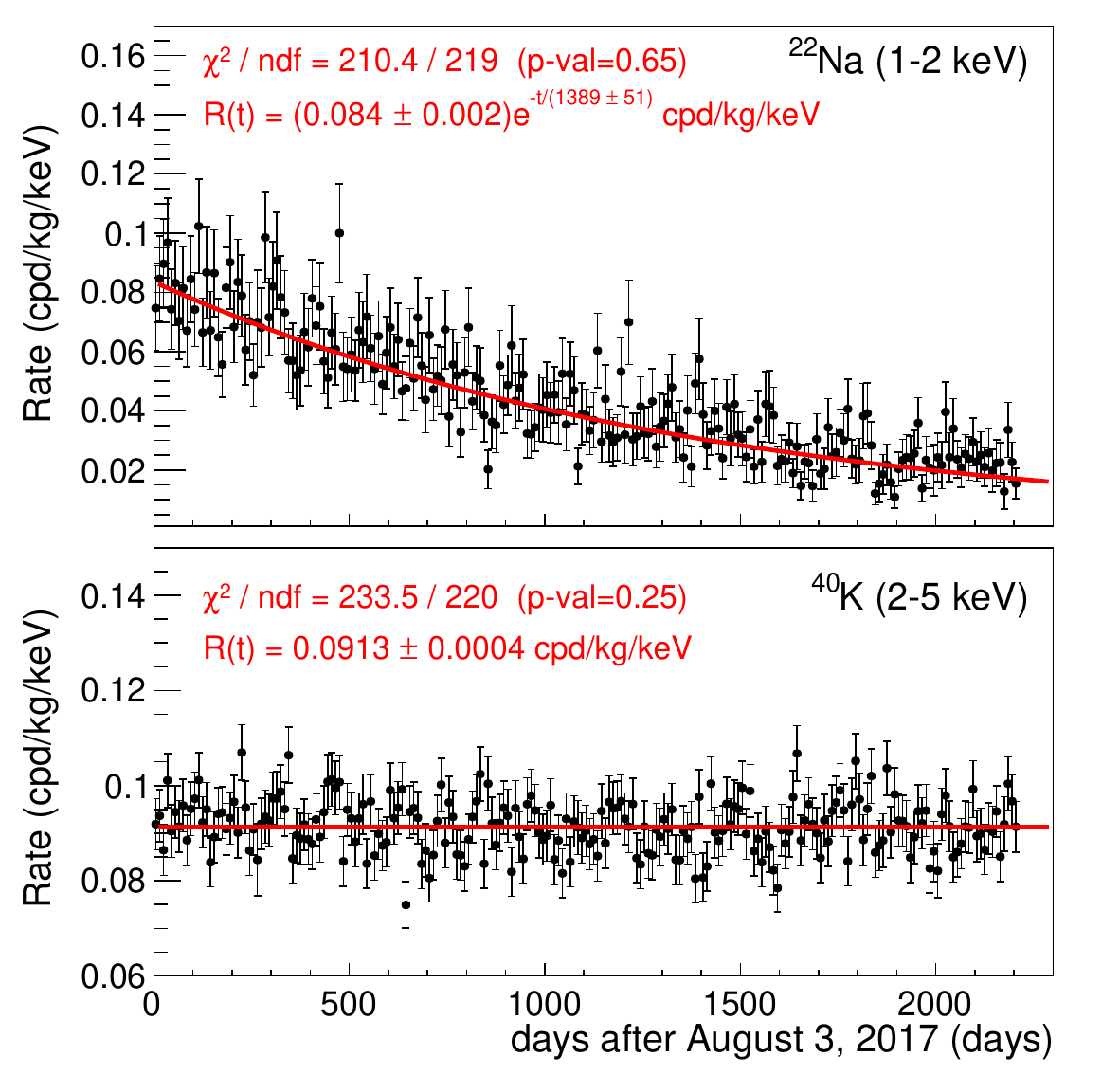}

\caption{\label{controlpopulations} Time evolution of the event rate for $^{22}$Na (upper panel) and $^{40}$K (lower panel) at low energies, identified by coincidence with the corresponding high-energy gamma in a second module \cite{amare2025towards}.}
\vspace{-0.5cm}
\end{center}
\end{figure}

\begin{table}[b!]
\centering
    \resizebox{\textwidth}{!}{\Large

\begin{tabular}{ccccc} 
\hline
Energy region & \begin{tabular}[c]{@{}c@{}}Experiment \\ and exposure\end{tabular}   & \begin{tabular}[c]{@{}c@{}}S$_m$\\ (cpd/ton/keV)\end{tabular}   & \begin{tabular}[c]{@{}c@{}}Incompatibility\\ with DAMA\end{tabular}  & \begin{tabular}[c]{@{}c@{}}Sensitivity\\ to DAMA\end{tabular}  \\
\hline
\hline

\multirow{6}{*}{ [1-6] keV} & \begin{tabular}[c]{@{}c@{}}ANAIS-112\\ (625.75 kg x year)\end{tabular}  & -0.4±2.5 & 4.0$\sigma$ & (4.2$\pm$0.4)$\sigma$\\
 & \begin{tabular}[c]{@{}c@{}}COSINE-100\\ (358.00 kg x year)\end{tabular}  & 1.7±2.9 & 2.8$\sigma$ & (3.6±0.4)$\sigma$\\
  & \begin{tabular}[c]{@{}c@{}}DAMA/LIBRA-phase2\\ (1126.40 kg x year)\end{tabular}  & 10.5±1.1 & - & - \\
\hline
  \multirow{7}{*}{ [2-6] keV} & \begin{tabular}[c]{@{}c@{}}ANAIS-112\\ (625.75 kg x year)\end{tabular}  & 1.1±2.5 & 3.5$\sigma$ & (4.1$\pm$0.3)$\sigma$\\
 & \begin{tabular}[c]{@{}c@{}}COSINE-100\\ (358.00 kg x year)\end{tabular}  & 5.3±3.1 & 1.5$\sigma$ & (3.3$\pm$0.3)$\sigma$\\
  & \begin{tabular}[c]{@{}c@{}}DAMA/NaI+ \\ DAMA/LIBRA\\ (2462.09 kg x year)\end{tabular}  & 10.2±0.8 & - & - \\

\hline

& & (cpd/ton/3.3 keV\textsubscript{NR}) & & \\
\hline
\multirow{7}{*}{ [6.7–20] keV\textsubscript{NR}}  & \begin{tabular}[c]{@{}c@{}}ANAIS-112\\ (625.75 kg x year)\end{tabular}  & 0.0±2.3 & 4.2$\sigma$ & (4.4$\pm$0.3)$\sigma$\\
 & \begin{tabular}[c]{@{}c@{}}COSINE-100\\ (358.00 kg x year)\end{tabular}  &  1.3±2.7 & 3.2$\sigma$ & (3.8$\pm$0.3)$\sigma$\\
  & \begin{tabular}[c]{@{}c@{}}DAMA/NaI+ \\ DAMA/LIBRA\\ (2462.09 kg x year)\end{tabular}  & 10.2±0.8 & - & - \\
\hline
\end{tabular}}
\caption{\label{tablaresults} Summary of the fits performed to search for an annual modulation in the [1–6]~keV, [2–6] keV and [6.7–20] keV\textsubscript{NR} sodium nuclear recoil energy region (corresponding to
[2–6] keV for DAMA/LIBRA): ANAIS-112 six-year exposure  \cite{amare2025towards}, full COSINE-100 dataset \cite{carlin2024cosine}, and DAMA/LIBRA \cite{Bernabei:2020mon}. The incompatibility with the DAMA/LIBRA signal is quantified as \( | S_m - S_m^{\text{DAMA}}| / \sqrt{\sigma^2(S_m)+\sigma^2_{\textnormal{DAMA}}(S_m)} \), while the sensitivity to the DAMA/LIBRA result is quoted as the ratio \( S_m^{\text{DAMA}} / \sigma(S_m) \),  where the uncertainty in the sensitivity corresponds
to the 68\% C.L. DAMA/LIBRA result uncertainty.}
\end{table}

To ensure the robustness of the analysis, it is equally crucial to monitor independent control populations for which no annual modulation is expected, such as the $^{22}$Na and $^{40}$K internal contaminants homogeneously distributed within the NaI(Tl) crystal bulk. The time evolution of the event rates of these populations over six years is shown in Figure~\ref{controlpopulations}. For $^{22}$Na,  the rate is expected to exhibit an exponential decay. The fit to the $^{22}$Na decay rate yields a mean lifetime of $1389 \pm 51$~days, consistent with the nominal lifetime of 1369 days within $1.7\sigma$. The $^{40}$K events, on the other hand, are expected to remain constant due to the long half-life of $^{40}$K (T$_{1/2} = 1.25 \times 10^9$~y). The fit to the $^{40}$K rates confirms compatibility with a constant rate. The results from $^{22}$Na and $^{40}$K further reinforce the long-term stability of the threshold and the calibration in the ROI over the six years of data.

\begin{figure}[t!]
\begin{center}
\includegraphics[width=0.7\textwidth]{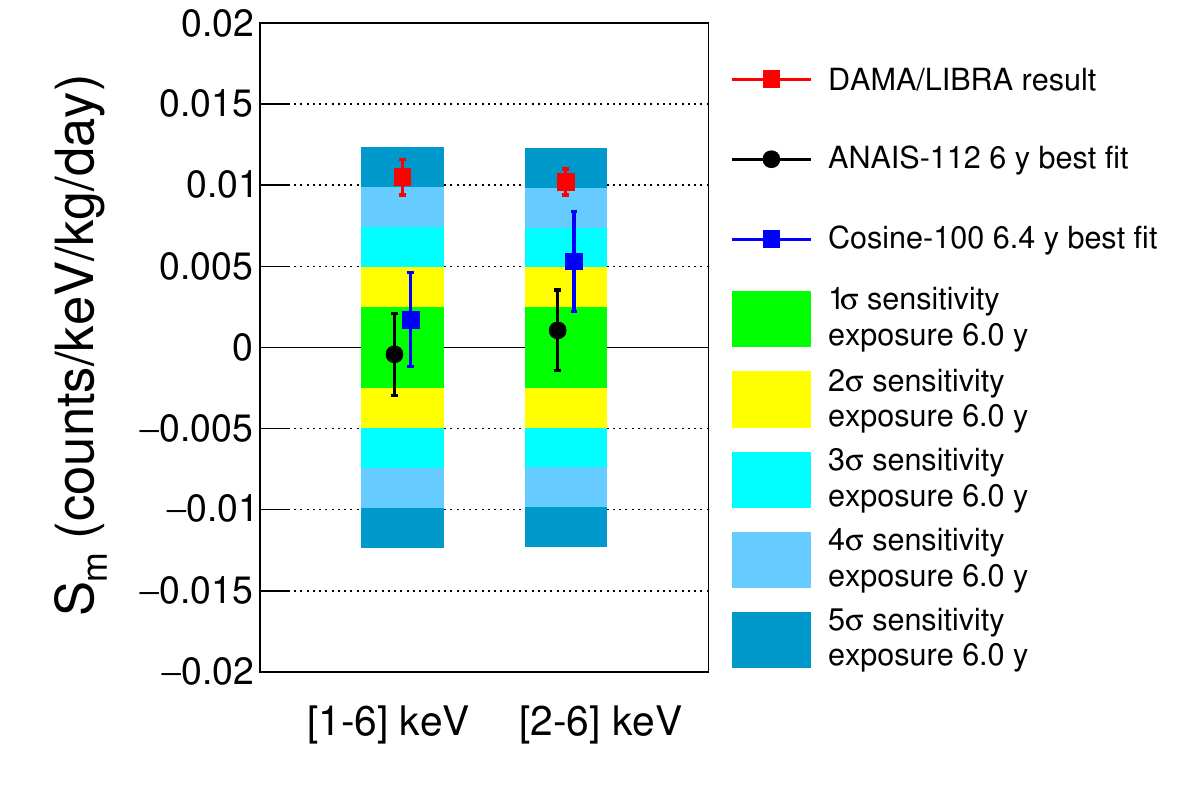}

\caption{\label{results} Comparison between results on annual modulation from
ANAIS-112 six-year exposure \cite{amare2025towards} (black), COSINE-100 full dataset  \cite{carlin2024cosine} (blue) and DAMA/LIBRA \cite{Bernabei:2020mon} (red). Estimated sensitivity of ANAIS-
112 is shown at different C.L. as the colored bands.}
\end{center}
\end{figure}

The results of the fit of the event rate following Equation \ref{rateevents} are presented in Table \ref{tablaresults} for the ANAIS-112 six-year exposure \cite{amare2025towards}, together with those from the full COSINE-100 dataset \cite{carlin2024cosine}, and DAMA/LIBRA \cite{Bernabei:2020mon}. Figure \ref{results} shows a graphical summary of these results. Both ANAIS-112 and COSINE-100 annual modulation searches are performed with a fixed phase in the [1–6]~keV and [2–6] keV energy regions. The first range is suitable for comparison with the DAMA/LIBRA-phase2 results, while the second allows for direct comparison with the total accumulated exposure of DAMA.

In addition, the annual modulation analysis carried out in \cite{amare2025towards} with the 6-year exposure data also considered the [1.3-4] keV energy interval. This region was selected by converting DAMA/LIBRA signal region [2-6] keV into NR-energy scale for sodium and iodine recoils. This conversion uses the QF values reported by the DAMA/LIBRA collaboration for their NaI(Tl) crystals \cite{bernabei1996new} (QF\textsubscript{Na} = 0.3 and QF\textsubscript{I} = 0.09) and those determined for crystals grown in the same batch as those used in ANAIS–112, assuming constant QF values (QF\textsubscript{Na} = 0.2 and QF\textsubscript{I} = 0.06)~\cite{cintas2024measurement,phddavid}. Then, the [1.3-4]~keV energy region in ANAIS corresponds to [6.7-20]~keV\textsubscript{NR} for sodium and [22.2-66.7]~keV\textsubscript{NR} for iodine recoils. Corresponding results are collected also in Table~\ref{tablaresults}, showing incompatibility and sensitivity levels to the DAMA/LIBRA signal.

Summarizing, the ANAIS-112 results for six years of data are consistent with the absence of modulation within one standard deviation and show an incompatibility with the DAMA/LIBRA signal of 4$\sigma$ and 3.5$\sigma$ C.L. for the [1–6]~keV and [2–6] keV energy regions, respectively. The sensitivity to the DAMA/LIBRA signal is quantified as the ratio \( S_m^{\text{DAMA}} / \sigma(S_m) \), which directly provides in $\sigma$ units the C.L. for testing the DAMA/LIBRA result. These correspond to sensitivities of \( (4.2\pm0.4)\sigma \) and \( (4.1\pm0.3)\sigma \) for both energy regions at 68\% C.L. These results stand out as the strongest model independent test
of the DAMA/LIBRA result to date using the same target material, achieving for the first time a statistical significance
beyond 4$\sigma$.

\begin{figure}[b!]
\begin{center}
\includegraphics[width=0.6\textwidth]{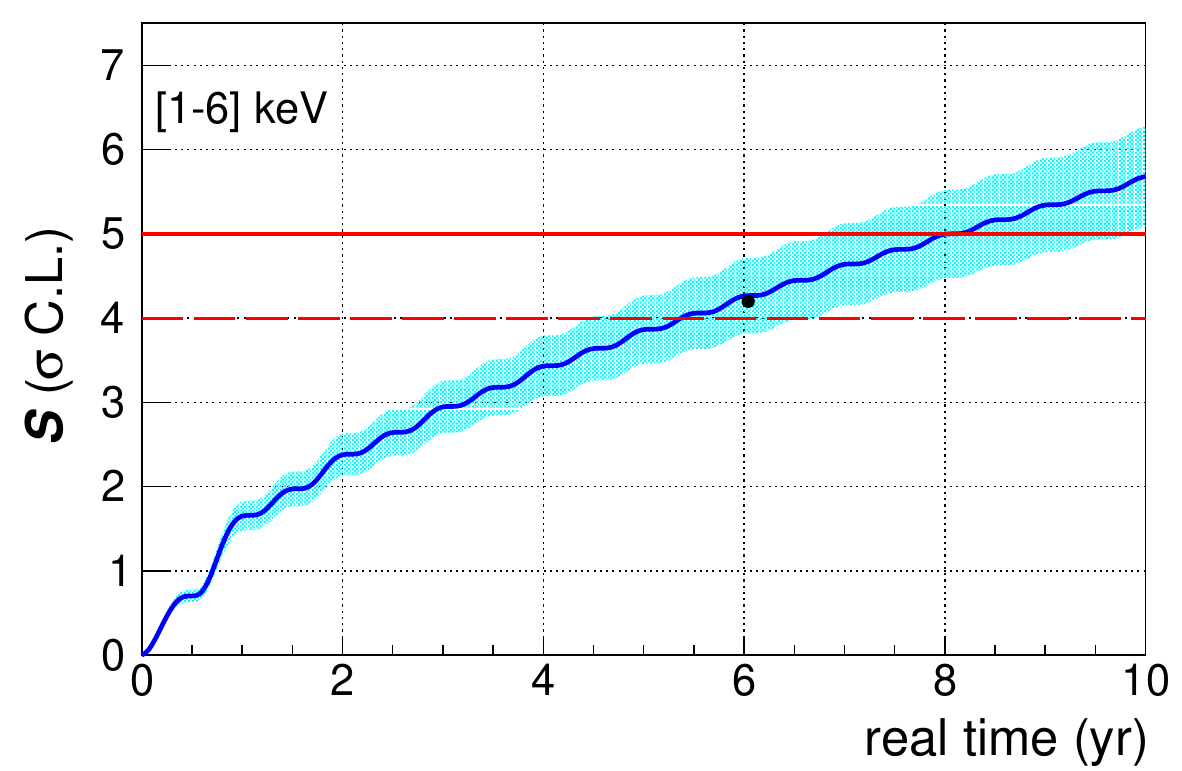}

\caption{\label{sproj} Sensitivity of ANAIS-112 to the DAMA/LIBRA signal in $\sigma$ C.L. units as a function of real time in the [1–6] keV energy region. The black dot represents the experimentally measured sensitivity for a 6-year exposure, while the cyan band indicates the 68\% C.L. uncertainty from DAMA/LIBRA \cite{amare2025towards}.}
\end{center}
\end{figure}

Figure \ref{sproj} shows the ANAIS-112 sensitivity projection in the [1–6] keV energy region, represented by the dark blue line, based on \cite{ANAISsproj}. Similar sensitivities are obtained for the [2–6] keV energy region. The cyan band represents the 68\% uncertainty in \( S_m^{\text{DAMA}} \). The black dot indicates the sensitivity derived from the 6-year result, which is in good agreement with prior estimates. This result reinforces the expectation of achieving a 5$\sigma$ sensitivity to the DAMA/LIBRA result by the end of 2025.

Among the systematics that could affect the direct comparison between ANAIS-112 and DAMA/LIBRA results, the most relevant one is the QF. While ANAIS-112 inconsistency with the DAMA/LIBRA signal is undeniable in the case of DM particles causing ER in the NaI(Tl) crystals, DM particles are expected to induce NR in the target nuclei in most of the DM models, as previously discussed. Therefore, the comparison should be made on the NR energy scale rather than the ER energy scale, which requires precise knowledge of the QF.

In this thesis, the QF of the ANAIS-112 crystals has been estimated through dedicated onsite calibrations using a \(^{252}\)Cf neutron source. A detailed discussion of this analysis is presented in Chapter \ref{Chapter:QF}. Additionally, the ANAIS-112 background model has been revised to achieve a more accurate description of the evolution in time of the measured event rate (see Chapter~\ref{Chapter:bkg}). The impact of incorporating different QF values, including those derived in this work, as well as the improved background modelling, on both the annual modulation analysis and the comparison between ANAIS-112 and DAMA/LIBRA results will be explored in Chapter \ref{Chapter:annual} of this thesis.

\setcounter{chapter}{2} 

\chapter{ANAIS simulation software}\label{Chapter:Geant4}

\vspace{-0.2cm}
\minitoc

Accurately understanding the background sources of ANAIS-112, especially those affecting its ROI, and properly modelling the detector response are key for a reliable DM analysis. One common approach for studying particle-induced backgrounds involves the use of general-purpose Monte Carlo simulation software. Several established toolkits are available for this goal, such as Geant4 \cite{GEANT4:2002zbu}, FLUKA \cite{fluka} and MCNPX \cite{MCNPX}.

Geant4 was chosen to simulate the response of ANAIS-112 to the various interactions occurring within the detector. Specifically, it has been employed to test different quenching factor models by comparing simulations against onsite neutron calibration data (Chapter~\ref{Chapter:QF}), and to revisit the ANAIS-112 background model (Chapter \ref{Chapter:bkg}). 

In this chapter, the characteristics of the simulation software used in ANAIS-112 are reviewed, with a particular focus on the new features implemented throughout this thesis. Special attention is given then to the geometry configuration and the selection of physics lists (Section \ref{ANAISG4}). To validate the model predictions against experimental data, it is necessary to replicate the detector response. Therefore, Section~\ref{ANAISresponse} summarizes the code efforts developed to mimic the detector response and DAQ characteristics.

\section{ANAIS Geant4 simulations}\label{ANAISG4}
\vspace{-0.4cm}
Geant4 (GEometry ANd Tracking) is a C++ toolkit originally developed at CERN used to simulate particles interaction with matter \cite{geantuserdocu}. Initially developed for high-energy physics, it has become a valuable tool in fields like particle physics, nuclear physics, and medical physics due to its versatility. Geant4 allows users to define complex geometries, access comprehensive and highly detailed material databases, and select appropriate physics models tailored to the particles and energy ranges of interest. These features make simulations with Geant4 highly customizable. Moreover, Geant4 is actively maintained and developed with strong support from CERN, ensuring regular updates and improvements. For a comprehensive overview of the toolkit, readers are referred to the official Geant4 references and user documentation \cite{GEANT4:2002zbu,Allison:2006ve,allison2016recent}.

In Geant4, a run is the basic unit of simulation, comprising a defined number of events, each statistically independent. A single event includes all the physical processes undergone by primary particles and any secondary particles they produce. The event continues until all particles are either absorbed, escape the outermost volume of the simulation, or are no longer tracked due to production or range cuts that prevent the generation of further secondaries.

Events are further divided into tracks, which represent the sequence of steps taken by a particle as it propagates through the simulation. Each track consists of one or more steps, where each step corresponds to a well-defined state of the particle, including its position, energy, and other dynamic properties.

Geant4 determines the step length by computing the minimum distance allowed by all relevant physics processes, geometry boundaries, and user-imposed limits. Each process (ionization, scattering, etc.) proposes a maximum step based on its probability; the shortest of these proposed steps is taken. Geant4 then updates the particle's state accordingly and repeats the process until the particle stops or leaves the simulation volume. 

Geant4 also allows for the definition of sensitive volumes, so that the energy deposited by a particle in them during a step is recorded. This includes energy lost through continuous processes, such as ionization and excitation. The information is stored in a G4Hit structure, which can record other relevant analysis observables, such as position, time, and particle type.

For each event, the user must specify the initial conditions of the primary particle(s), including its energy, position and direction (either a fixed value or following a distribution). Particles can be uniformly distributed across a particular volume or surface, such as a parallelepiped, sphere, cylinder, or specific detector components. In the simulations conducted in this work, the Geant4 G4GeneralParticleSource (GPS) class has been employed to implement the random distribution of the radioactive contaminants within the different detector components (volume or surface).

ANAIS used a Geant4 model based on version v9.4.p01 to develop its background model \cite{Amare:2018ndh}. During this thesis, however, an in-depth revision of both the ANAIS geometry and the physics list was performed. The updated simulation results presented in this work regarding the background model (see Chapter \ref{Chapter:bkg}) are based on Geant4 version v10.7.0. In the context of neutron calibration simulations (see Chapter \ref{Chapter:QF}), these versions together with version v11.1.1, will be used and compared. In particular, Section~\ref{datalibrary} discusses their differences and evaluates which version best aligns with the experimental~data.

\subsection{ANAIS geometry}

In Geant4, geometries are organized in a hierarchical structure, where volumes are placed within each other. The outermost volume is called the world volume. Subsequent volumes are placed inside the world using three main steps. First, a solid defines the shape of the volume. Second, a logical volume, based on the solid, is created, where the material is assigned. Third, a physical volume is defined by placing a copy of the logical volume into the simulated geometry. These steps can be repeated to build detailed set-ups with many different components and complex geometries. Sensitive volumes are especified at the logical volume level. In the case of ANAIS, the sensitive detector is defined as the NaI(Tl) crystals.

The experimental set-up of ANAIS-112 is detailed in Section \ref{set-up}. The former ANAIS Geant4 model included a simplified but sufficient description of the geometry of the experiment. While this was adequate for generating a solid background model \cite{Amare:2018ndh}, throughout the course of this thesis it was decided to refine the ANAIS geometry to enhance the simulation accuracy. Specifically, the following improvements have been implemented regarding the ANAIS geometry:

\begin{itemize}
    \item Integration of key components into the ANAIS-112 shielding set-up, including the anti-radon box, water and polyethylene blocks, muon veto system, and the mechanical structures supporting both detectors and calibration sources. The inclusion of water and polyethylene, along with the muon veto panels, is especially relevant for neutron simulations due to their high hydrogen content. These low-Z materials efficiently moderate neutrons down to thermal energies via elastic scattering, making their accurate modelling essential for obtaining reliable simulation results.
    
    \item  Improvement in the modelling of the mylar window and the $^{109}$Cd sources used for periodical calibration.

    \item Enhanced description of the PMT geometry, including a more accurate modelling of the borosilicate glass and the dynode structure. The previous background model of ANAIS \cite{amare2019analysis}, which serves as a reference for comparison in this thesis, did not include these internal PMT components. In an earlier study, the dynode system was approximated using a simplified model consisting of a compact steel parallelepiped \cite{phddavid}. In contrast, the present work replaces this approximation with a more detailed geometry explicitly modelling the 10 individual dynodes.
    
    The improvements in the PMT geometry are particularly relevant for the optical simulations of ANAIS-112, rather than for the background model simulations. Although such optical studies are not included in this dissertation, significant efforts were made during this thesis in refining the optical properties of the materials used in the ANAIS modules, the treatment of the involved surfaces, and the modelling of the PMT QE.

\end{itemize}

\begin{figure}[b!]
\begin{center}
\includegraphics[width=1.\textwidth]{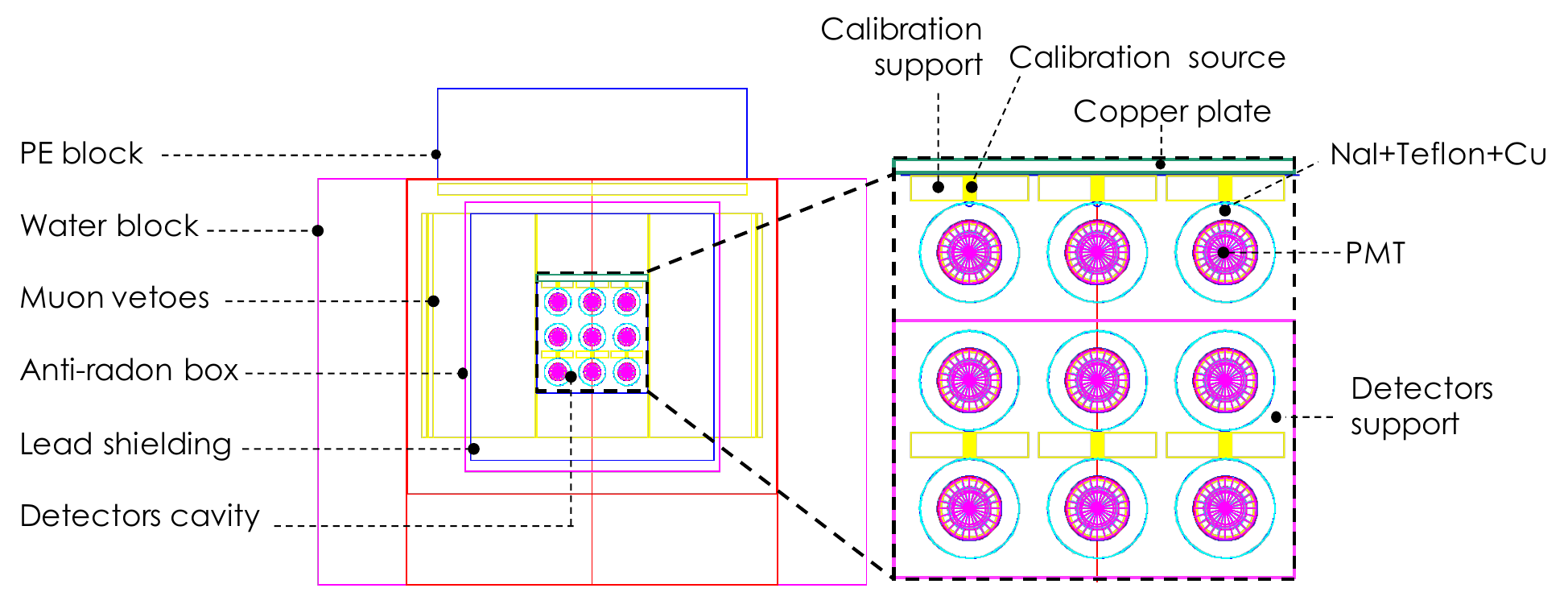}

\caption{\label{esquemaANAIS1} Frontal view of the complete ANAIS-112 geometry used in the simulations presented in this study, with key structural components highlighted. From the outside in, the water and polyethylene blocks are followed by the muon veto system, anti-radon box, and lead shielding. An inset shows a view of the \(9 \times 9\) NaI(Tl) detector array within the detector cavity.}
\end{center}
\end{figure} 

The new volumes were created using dimensions extracted from the CAD design of ANAIS-112. To avoid potential overlaps, a built-in Geant4 overlap test was performed subsequently. Overlapping volumes occur when a region of space is assigned to more than one volume, which can lead to incorrect simulation results. 

Figure \ref{esquemaANAIS1} shows the frontal view of the full ANAIS-112 geometry implemented in the simulations presented in this study, including a detailed view of the detector cavity with the arrangement of the nine NaI(Tl) modules. 

The structure is organized in layers. Starting from the outermost components and going inwards, there are two polyethylene blocks at the top and bottom of the set-up with dimensions of 2800$\times$400$\times$3000 mm$^3$ and 1640$\times$400$\times$3000 mm$^3$, respectively. The lateral blocks are modelled as water-filled, with dimensions of 400$\times$181.2$\times$2670~mm$^3$ and 820x181.2x400 mm$^3$. 

\begin{figure}[b!]
\begin{center}
\includegraphics[width=1.\textwidth]{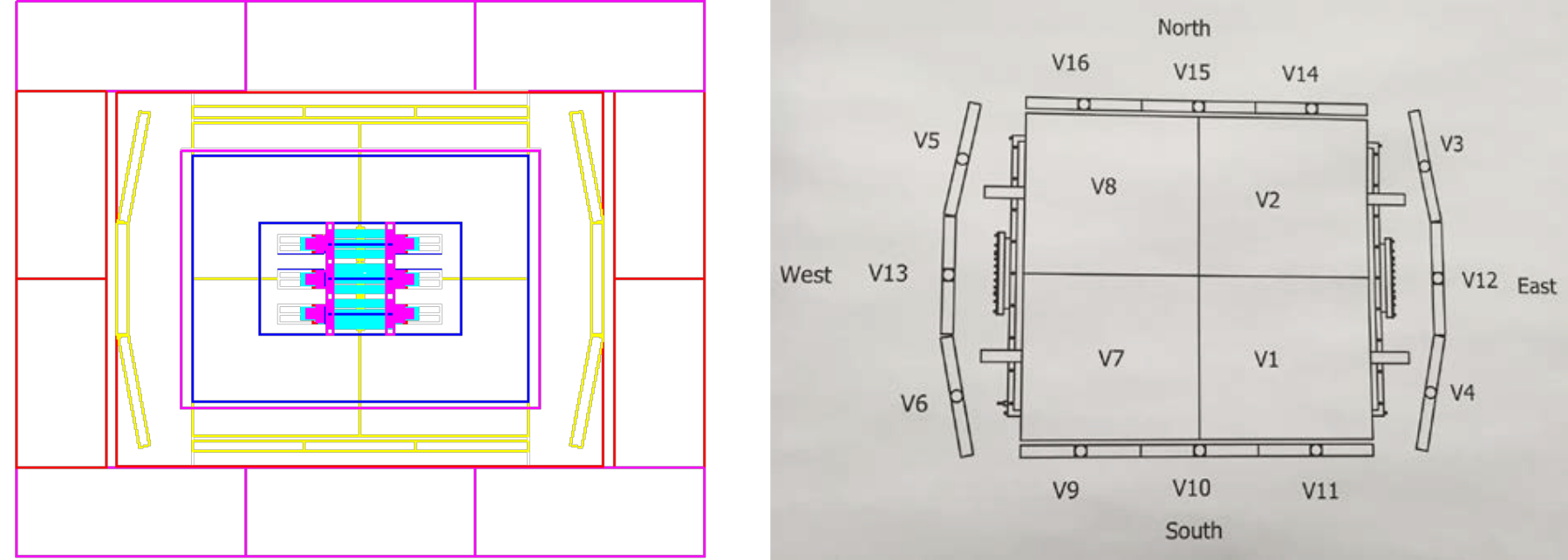}

\caption{\label{esquemavetos} \textbf{Left panel:} Top view of the complete ANAIS-112 geometry used in the simulations presented in this study. From the outside inwards, the water and PE blocks are followed by the muon veto system, anti-radon enclosure and lead shielding. \textbf{Right panel:} Top view of the experimental layout of the ANAIS-112 veto system.}
\end{center}
\end{figure}

The geometry also includes 16 muon veto plastic scintillators. Figure \ref{esquemavetos} shows the top-view layout of the veto system. The veto panels on the top, south, and north faces, together with panels nº12 on the east face and nº13 on the west face, measure 1000$\times$500$\times$50~mm$^3$. The remaining panels have dimensions of 750$\times$700$\times$50 mm$^3$. To accurately replicate the experimental set-up, the lateral panels on the east and west sides were positioned with an inclination angle of 11 degrees. The PMTs and light guides attached to these veto panels were excluded from the simulation, given that its contribution is expected to be non-relevant.

Surrounding the detector assembly is an anti-radon box of stainless steel continuously flushed with
radon-free nitrogen gas to prevent the entrance of airborne radon inside the
lead shielding. Its dimensions are 576$\times$601$\times$801~mm$^3$. Beyond this enclosure, a lead shielding measuring 550$\times$550$\times$750~mm$^3$ can be found. Following the experimental design, instead of lead directly covering the detectors from above, a 2.5~cm thick copper plate was installed as the first layer of the upper shielding. Thus, as shown in Figure~\ref{esquemaANAIS1} and Figure~\ref{esquemavetos}, the ANAIS set-up is symmetric along north-south and west-east horizontal axes, but not along the vertical axis, because of the copper plate.

At the core of this multi-layered set-up lies the detector cavity, housing the nine NaI(Tl) modules. Each module consists of a cylindrical NaI(Tl) crystal, 29.85 cm in length and 6.03~cm in radius, encased in a 0.5 mm-thick layer of teflon. Each end of the tube is coupled to a 1 cm-thick quartz window through a 3 mm-thick layer of silicone, both having a radius of 3.8~cm. This entire assembly is enclosed in a 1.5 mm-thick OFE copper housing, which protects the hygroscopic NaI(Tl) crystal from moisture damage. The detectors are arranged in a 3x3 configuration and held in place by a teflon structure (500$\times$500$\times$30~mm$^3$) at both ends that securely encases the modules, ensuring proper alignment and stability (see Figure~\ref{esquemaANAIS1}, "Detectors support").

\begin{figure}[t!]
\begin{center}
\includegraphics[width=1.\textwidth]{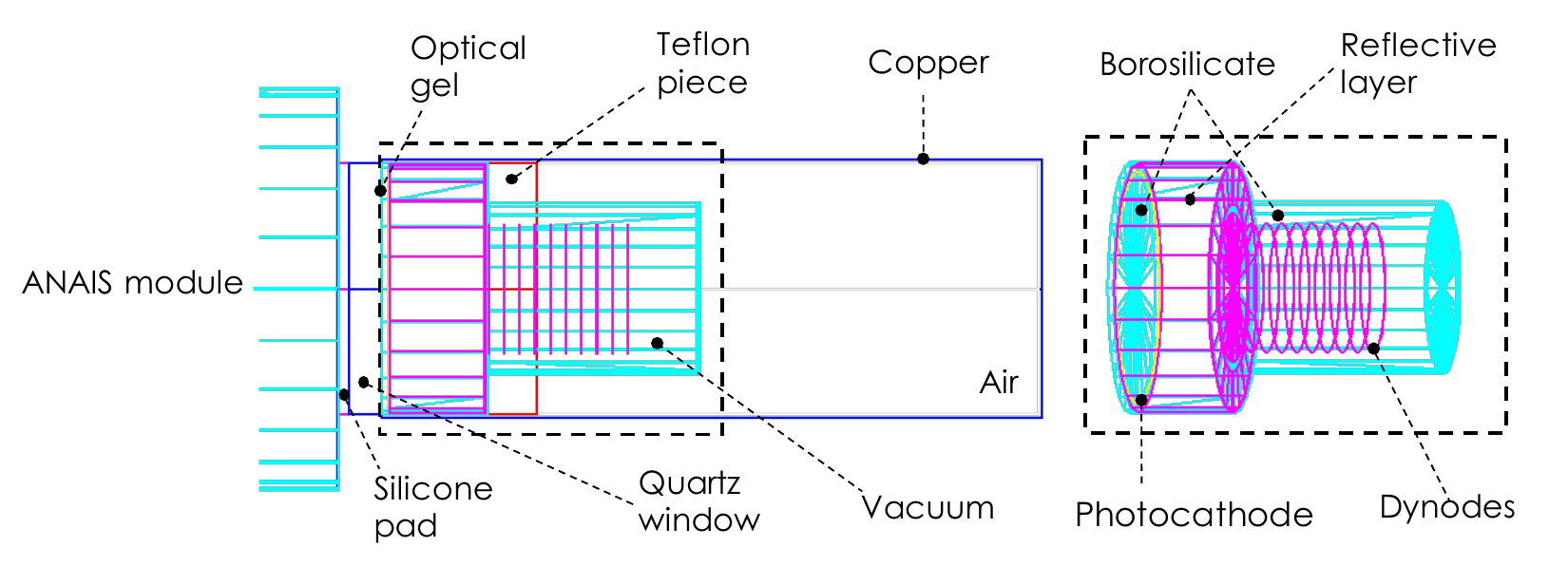}

\caption{\label{esquemaPMT} Geometry of the PMT used in the simulation, showing the optical coupling region between the PMT and NaI(Tl) crystal. Key components of the PMT, including the borosilicate, reflective layer, photocathode, and dynodes, are highlighted. }
\end{center}
\end{figure}

Each quartz window is coupled to a PMT via a 0.1 mm-thick layer of optical gel. Both the silicone and optical gel serve to match the refractive indices and ease light propagation towards the PMT. The PMTs in this simulation are based on the R12669SEL2 model used in the ANAIS-112 modules (see Figure \ref{ANAISPMTs}). Figure \ref{esquemaPMT} shows the geometry of the PMT considered in the simulation, as well as the optical coupling region between the PMT and the NaI(Tl) crystal.

Each PMT is enclosed in a copper volume filled with air, and is modelled as two concentric borosilicate cylinders, 0.5 mm thick, representing the head and body of the PMT. The head has a radius of 38.1 mm and a length of 29.4 mm, while the body has a radius of 26 mm and a length of 63.6 mm. These cylinders are connected in a single volume filled with vacuum. The borosilicate head features a 0.1 mm-thick copper reflective layer, while the body contains the dynodes. 

This simulation models the dynodes as a set of 10 thin, copper circular foils, each 0.1~mm thick and 9.4 mm in radius. These foils correspond to the ten amplification stages of the PMTs used in the ANAIS modules. They are arranged in parallel within the borosilicate body, providing a more accurate representation of light propagation within the PMT. Finally, the photocathode, a bialkali (Sb-Rb-Cs) disk with a radius of 35 mm and a thickness of 20 nm, is located within the head of the borosilicate volume, near the base closest to the NaI(Tl) crystal. A teflon ring of 15 mm in length is placed around the PMT to secure its position. In this thesis, when simulations are performed with contamination located in the photocathode of the PMT, the term frontal contamination will be used to encompass contributions from both the photocathode and the quartz window, as these components cannot be disentangled.

\begin{figure}[t!]
\begin{center}
\includegraphics[width=1.\textwidth]{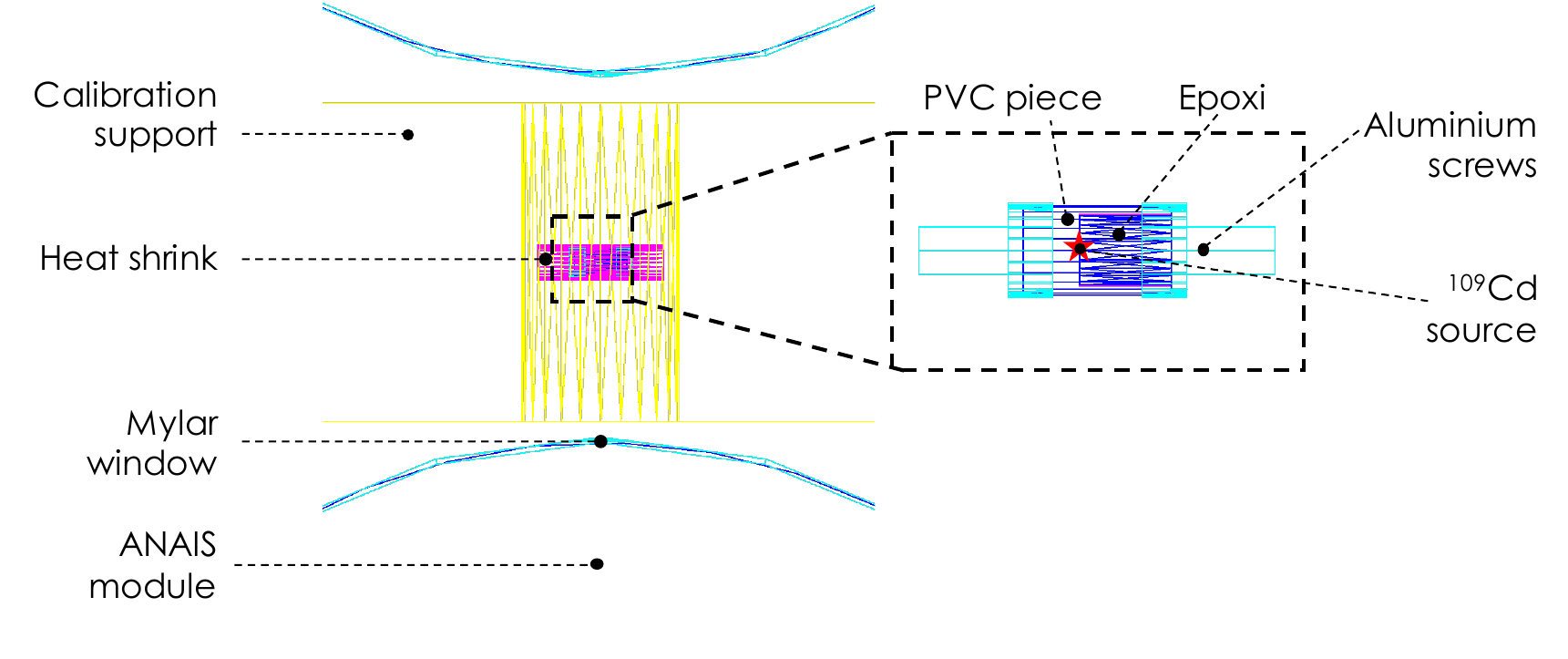}

\caption{\label{cd109source} Sketch of the calibration set-up, showing two modules from the first and second rows of the ANAIS matrix (starting from the bottom), both calibrated using the same $^{109}$Cd source. The mylar windows of both modules are aligned with the source. Included is a zoomed-in view of the calibration piece, highlighting its key components.}
\vspace{-0.5cm}
\end{center}
\end{figure}

A key feature of the module design is the aluminized mylar window, which has a radius of 5 mm and a thickness of 20 \textnormal{$\mu$}m. It is located on one side of each module to allow for low-energy gamma calibration using external sources (see Section \ref{LEcalibration}). In this simulation, the mylar window is modelled as two distinct layers, in line with the experimental design: a 0.1-\textnormal{$\mu$}m-thick aluminum layer and a mylar layer that accounts for the remaining thickness of the window. To accurately model the mylar window, the thickness of the copper and teflon layers surronding the NaI(Tl) crystal was subtracted in the region corresponding to the window. For the copper housing, the entire thickness was subtracted. In contrast, the exact residual teflon thickness after module manufacturing was uncertain. After conducting dedicated simulations of the $^{109}$Cd calibration and comparing them with the available experimental data, it was determined that leaving a 0.1 cm-thick layer of teflon (out of its original 0.5 cm thickness) provided the most compatible results. However, this thickness may actually vary between detectors.

Another important aspect of the $^{109}$Cd calibration modelling is the geometry of the piece housing the radioactive isotope. The calibration piece is modelled as three concentric cylinders. From the outside in, the first component is a heat-shrink tube made of PVC with a high but not well quantified bromine content. It has a length of 6 mm, a radius of 1.7 mm, and a thickness of 0.1 mm. Inside this tube is a second PVC cylinder, 5 mm long, with a radius of 1.5 mm and a thickness of 0.3~mm. The innermost cylinder is made of epoxy and contains the $^{109}$Cd source, modelled as a point-like emitter located at the center. The entire assembly is held together by an aluminum screw at both ends and is connected to a nylon wire that positions the source in front of the mylar windows of the ANAIS detectors during calibration runs. However, the wire itself is not included in the simulation. The calibration piece is supported by teflon holders.

Figure \ref{cd109source} illustrates the calibration set-up, showing a sketch of two modules from the first and second rows of the ANAIS matrix (starting from the bottom), both calibrated using the same $^{109}$Cd source. The topmost row of the matrix (third from the bottom) is calibrated independently. The figure also includes their respective calibration holders and a zoomed-in view of the source structure, highlighting the key components of the calibration piece.

\begin{figure}[b!]
\begin{center}
\includegraphics[width=0.6\textwidth]{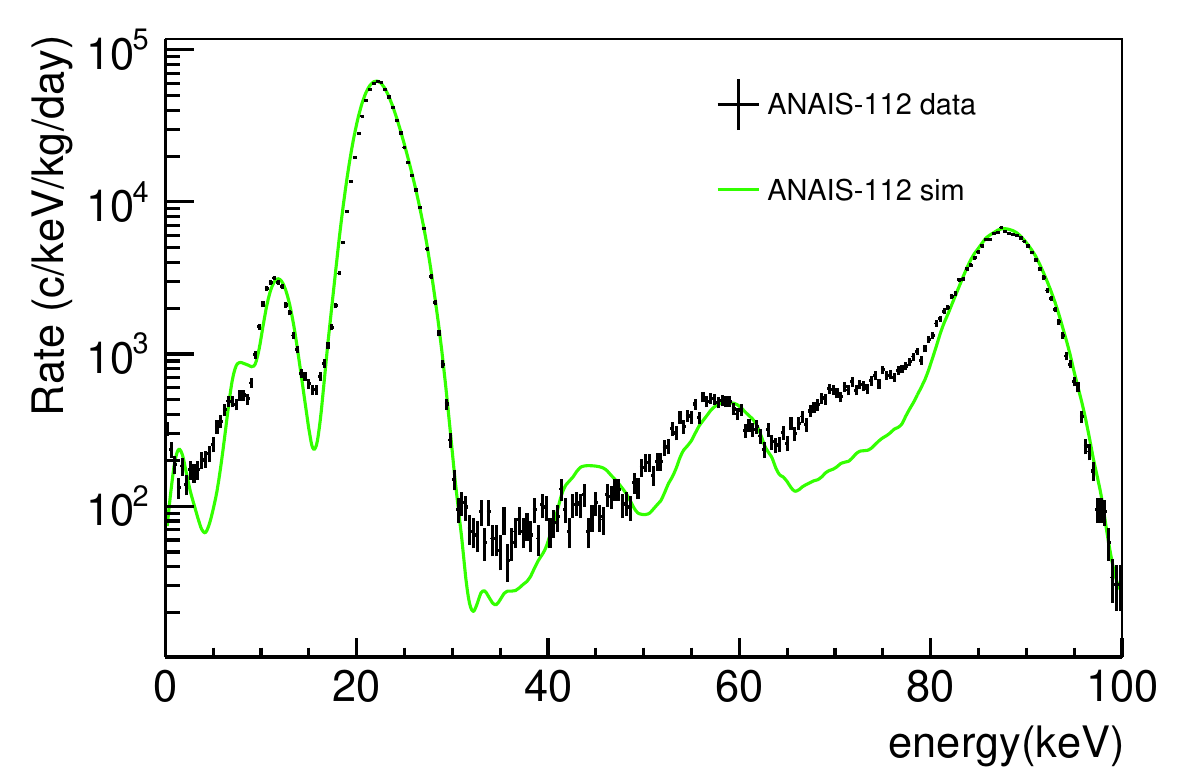}

\caption{\label{comparisoncd109} Comparison between experimental (black) and simulated (green) spectra for the $^{109}$Cd calibration of the ANAIS-112 experiment, after the improvements developed in this thesis in the modelling of the mylar window and the calibration piece.}
\end{center}
\end{figure}

Code validation and analysis of systematic effects related to the ANAIS geometry can be assessed by comparing the simulated and experimental energy spectra from the $^{109}$Cd calibration. This comparison is presented in Figure \ref{comparisoncd109}, where the simulated spectrum has been scaled according to the exposure time, considering both the nominal activity of the $^{109}$Cd source (1.5 \textnormal{$\mu$}Ci) and its decay over time. Additionally, the simulated data have been convolved with the updated energy resolution obtained in this work (see Figure \ref{LEres}), resulting in a good agreement with the experimental spectrum. The spectra clearly show the characteristic peaks of $^{109}$Cd decay at 22.6 keV and 88.1 keV, as well as the bromine x-ray line at 12.1 keV present in ANAIS-112. Additionally, the peak observed around 60 keV corresponds to iodine K-shell x-ray escape following the absorption of the 88 keV photon. Copper X-ray lines around 8 keV are also visible in the detector 8 spectrum.

Accurately reproducing the $^{109}$Cd calibration in simulation requires careful consideration of several factors. One key aspect is the bromine content in the PVC heat-shrink tube surrounding the source, which strongly influences the relative intensity of the 22.6 keV and 12.1 keV peaks. A higher bromine content leads to increased production of the 12.1~keV x-rays from bromine fluorescence and simultaneously attenuates more of the 22.6 keV photons before they reach the detector. The thickness of the PVC also plays a role, as it can absorb low-energy gammas. Additionally, the alignment of the source relative to the mylar window is crucial. Misalignments or small rotations of the detectors can prevent some low-energy photons from entering the crystal.

Previous studies suggested that a bromine mass content of 30\% would best match the experimental 22.6/12.1 keV peak ratio \cite{phddavid}. However, under that assumption, the overall intensity of the peaks in the simulation did not match experimental observations. In this thesis, improvements in the modelling of the source encapsulation, including geometry, material thicknesses, and positioning relative to the mylar window, have led to a remarkable agreement between simulation and data when assuming a bromine content of 15\%, as shown. Remaining discrepancies, such as the excess observed to the left of the 88 keV peak, are likely due to surface effects not fully accounted for in the simulation of energy depositions of this work.


\subsection{ANAIS physics list}

In Geant4 simulations, the version of the toolkit defines the state of the codebase, including bug fixes, updates, and newly introduced functionalities. Each Geant4 version is distributed with a specific set of physics lists and model classes that are developed and validated for that version. Running a simulation with physics lists from a different Geant4 version can lead to errors or incompatibilities due to changes in process definitions or functionalities that were not present in the previous version.

Physics lists specify the particles, interaction cross-sections, and energy thresholds for secondary particle production required in a given simulation. They are implemented via dedicated model classes, which handle specific physical processes such as electromagnetic interactions, decays, or hadronic processes. These classes are selected and combined, thereby determining particle behavior and interactions. The selection of an appropriate physics list is essential, as it directly affects the accuracy and reliability of simulation outcomes \cite{geantuserdocu}. For example, for high-energy applications, preferred physics lists such as FTFP\_BERT are commonly used, which include optimized production thresholds and cross-sections tailored to high-energy interactions, which are not optimized for low-energy processes.

Geant4 provides several reference physics lists, which are regularly maintained and validated across different applications. Additional physics lists found in user-contributed examples often serve highly specific use cases. In the previous Geant4-based model of the ANAIS experiment (version 9.4.p01) \cite{Amare:2018ndh}, the simulation followed an example tailored for underground DM experiments. Radioactive decays were simulated using the Geant4 Radioactive Decay Module (G4RadioactiveDecay), and low-energy electromagnetic interactions of $\alpha$ particles, electrons, and photons were based on the Livermore libraries.

For this thesis, the ANAIS simulations have been upgraded to Geant4 version 10.7.0. Unlike the earlier approach based on a predefined example, a custom physics list has been here implemented. This custom list incorporates not only electromagnetic processes, but also neutron and optical physics, to provide a more realistic description of the detector response to different background contributions. A detailed summary of the model 
                classes included in this custom physics list is presented below.

\begin{itemize}
    \item \textbf{G4EmStandardPhysics\_option4.}

 It is one of the standard electromagnetic physics lists provided by Geant4, designed to deliver high-precision simulations across a broad energy range, particularly below 100 TeV. Among the four standard options available, Option 4 is recommended for applications requiring enhanced accuracy at low energies, such as DM detection experiments. It employs a combination of theoretical and data-driven models optimized for precision in processes involving photons, electrons, and positrons. The minimum energy threshold for this physics list is 100 eV, although values below 250 eV are not typically advised due to increasing uncertainties. 
 
 In the energy range relevant for ANAIS-112, G4EmStandardPhysics\_option4 and the Livermore physics list rely on the same models for gamma and electron interactions. However, Option 4 is generally favored due to its superior accuracy and broader validation, despite its higher computational cost \cite{arce2021report}. \\

  \item \textbf{G4RadioactiveDecayPhysics.} 
 
 It handles the radioactive decay processes for a broad range of standard isotopes. However, certain isotopes such as $^3$H, which is especially relevant for ANAIS-112, are treated as stable by default in Geant4. In the ANAIS framework, $^3$H decay has been specifically incorporated.\\

 \item \textbf{G4DecayPhysics.} 
 
 It manages the decay of all long-lived hadrons and leptons. \\

 \item \textbf{NeutronHP physics.} 
 
 Low-energy neutrons (< 20 MeV) are modelled in Geant4 by the Neutron High-Precision (Neutron-HP) package. Neutron-HP is a data-driven model that relies on the Geant4 Neutron Data Library (G4NDL), which provides cross-section data for neutron interactions together with information on final states. This physics accurately describes processes such as elastic and inelastic neutron scattering, capture, and fission through dedicated classes like G4ParticleHPElastic, G4ParticleHPInelastic, G4ParticleHPCapture and G4ParticleHPFission, respectively.

At thermal energies (< 4 eV), additional factors such as atomic translational motion, vibrations, and rotational states of chemically bound atoms impact neutron scattering cross-sections and the distribution of secondary neutrons. In this energy regime, Geant4 refines its elastic neutron scattering models by replacing the standard G4ParticleHPElastic class by the G4NeutronHPThermalScattering one. The necessary thermal data are derived from specialized experiments or solid-state physics models. Thermal properties depend on the nuclei in the material. The Geant4 10.7.0 release includes thermal neutron data for only 19 elements, such as hydrogen in water, hydrogen in polyethylene and aluminium, which are included in the ANAIS-112 simulations conducted in this work.

The Geant4 version 10.7.0 employs the neutron library G4NDL4.6. In this thesis, discrepancies were observed when comparing experimental data from the onsite neutron calibration of the ANAIS-112 experiment with dedicated Geant4 simulations. Specifically, certain energy lines appeared in the simulation that were not present in the experimental data. Some of these lines correspond to excited states of $^{127}$I, which, according to nuclear data tables, should not be generated by inelastic neutron processes. 

As a result of this issue, it was decided to test the neutron library G4NDL3.14, corresponding to Geant4 version 9.4.p01, which was used in the previous background model. When using this older database, some of the lines disappear, while other differences were observed in the simulation outcome. A detailed study of the discrepancies between these neutron libraries, focusing on cross-sections and their better agreement with experimental data, will be conducted in Section \ref{datalibrary}.\\

 \item \textbf{G4HadronElasticPhysics.} 
 
 It models the elastic scattering of hadrons within a range from 0 to 10 TeV, particularly important for modelling $\alpha$ and high energy neutrons interactions.\\

\item \textbf{G4OpticalPhysics.} 

It enables the simulation of optical photons, their transport and interactions for detailed light simulations in scintillator-based detectors like ANAIS-112. This class is available for optical studies but deactivated by default.\\
 
\end{itemize}

The physics models registered in the ANAIS physics list each have a characteristic energy range within which their predictions are validated and considered reliable. Moreover, in Geant4, a production cut defines the minimum energy a secondary particle must have to be explicitly generated in the simulation. The default production threshold is 990 eV; particles with lower kinetic energy are not produced as individual tracks but their energy is instead added to the total
deposited energy of the interaction. In the ANAIS simulation framework, this threshold is reduced to 250 eV to improve accuracy at low energies. Although the ANAIS detection threshold is 1 keV, lower energy depositions can be observed in coincidence with high-energy gamma interactions, and reproducing these effects in simulations is considered valuable.

In addition to the production cut, Geant4 also determines a range cut, which defines the minimum distance a secondary particle must travel to be produced. This range cut is internally converted into an energy threshold based on the particle type and the material where the interaction is taking place. The purpose of this method is to optimize computational efficiency without distorting the simulation output. If the calculated energy threshold is lower than the production cut, this minimum value is applied. Range cuts apply to electrons, positrons, photons, and protons. 

The default range cut in Geant4 is 0.7 mm, but in ANAIS, it is reduced to 0.1~mm for gammas and 0.01~mm for electrons, positrons, and $\alpha$ particles in order to improve precision. The range cut for protons has a different meaning compared to other particles. It specifically applies to recoiling ions (including protons) in elastic scattering processes. For accurate simulation of low-energy NRs, ions with kinetic energies as low as 0 eV are tracked. Same range cut values were implemented in the previous ANAIS Geant4 model.

\section{ANAIS detector response and DAQ characteristics}\label{ANAISresponse}

After the Geant4 simulations are completed, a ROOT file \cite{Brun:1997pa} with a user-defined name is automatically generated. Variables corresponding to physical quantities of each simulation hit taking place in the sensitive volume, such as deposited energy, volume, position, particle type, time, and event~ID, are stored at the first level of analysis.

After that, a second-level analysis is applied to sum the energy depositions occurring within the detector's integration time and to identify coincidences. At this stage, the characteristics of the DAQ system must be incorporated to accurately account for event reconstruction. The following steps, listed in the order they are applied, are used to include the most relevant features of both the ANAIS and ANOD DAQ systems in the simulated signal construction (see Section \ref{DAQsec} for a description of their main characteristics):


\begin{itemize}
    \item \textbf{Integration window and dead time.} 
    
    The ANAIS electronics do not feature a continuous readout. Instead, it incorporates a 1 \textnormal{$\mu$}s integration window followed by a approximately 4.5 ms dead time, during which the DAQ system is busy in data processing and transfer, and therefore unavailable for acquiring new events. To accurately replicate the detector behavior in simulations, the second-level analysis combines the energy of all first-level Geant4 hits occurring within 1 \textnormal{$\mu$}s of the first hit for each simulated event to create a corresponding second-level single experimental event. Hits occurring outside the integration window and within the subsequent 4.5 ms are discarded at this second level of analysis. This process requires prior sorting by arrival time of all hits associated with an event within the ROOT tree.

In practice, once the data transfer is completed, the DAQ system is ready again and can record new events if additional photoelectrons arrive. In the simulation, this means that hits associated to the same simulated event occurring after the initial integration window and subsequent dead time, such as those from delayed isomeric decays or neutrons undergoing multiple interactions (e.g., bouncing back from the lead shielding), can trigger a new second-level event.

It is important to note that in the simulations of the ANOD DAQ system, the non-dead-time DAQ of ANAIS, the process differs slightly. As stated in Section~\ref{DAQsec}, the acquisition window of ANOD is 8 \textnormal{$\mu$}s, and since there is no dead time, a new acquisition window of 8 \textnormal{$\mu$}s, and therefore a new second-level event, can begin immediately after the previous one ends. 

In ANOD, the signal integration is performed over configurable sub-windows of shorter duration within the 8 $\mu$s acquisition window (see Section \ref{DAQsec}). Different energy observables can be defined depending on the integration time. In this work, both the energy integrated over a 1 \textnormal{$\mu$}s window (as used in ANAIS) and a 2 \textnormal{$\mu$}s integration window have been tested in the ANOD simulations. If the characteristics of the ANOD DAQ were to be considered in optical simulations, something not yet done in the experiment but potentially of interest in the future, it would be necessary to define pulse shape variables (see Section~\ref{previosufiltering}) in the same way as they are defined experimentally. In particular, P1 and P2 would be computed analogously to how they are defined in ANAIS. However, the first moment calculation, $\mu$, would be defined as the mean of all individual photoelectron arrival times within the acquisition window.\\

\vspace{-0.2cm}
    \item \textbf{Area loss correction in the integration window. } 
    
NaI(Tl) has a scintillation decay constant of 230 ns. The energy deposited in each hit is converted into optical photons, which reach the PMTs following an exponentially decaying time distribution. If photons arrive after the digitization window has closed, part of the energy from the hit will not be recorded.

This effect is not considered in the first level of the simulation, as optical transport is not modelled. It must, therefore, be implemented at the second level of analysis. This becomes particularly relevant in the case of multiple-hit events, where the integration window is opened by the first hit, but other hits may reach the same or different detectors with a certain delay. To account for this effect, the time between the hit arrival and the end of the digitization window is calculated, and only the corresponding portion of the scintillation pulse is integrated.\\

\item \textbf{Trigger efficiency.}

The ANAIS hardware threshold is set low enough to efficiently trigger at the photoelectron level for each PMT. However, at very low energies, the requirement for coincidence between the two PMT within a 200 ns window reduces the trigger efficiency. This is because fewer photoelectrons are collected, increasing the mean time interval between them. The ANAIS trigger efficiency has been evaluated through a Monte Carlo simulation and has been updated relative to the results reported in \cite{Amare:2018sxx}. As shown in Figure \ref{triggereff}, the efficiency remains above 98\% down to 1 keV, as expected.

Incorporating the trigger efficiency into the simulation is crucial to produce accurate energy spectra. Including very low-energy events that the electronics cannot detect would otherwise distort the simulated results, especially the multiplicity distributions.

\begin{figure}[b!]
\begin{center}
\includegraphics[width=0.7\textwidth]{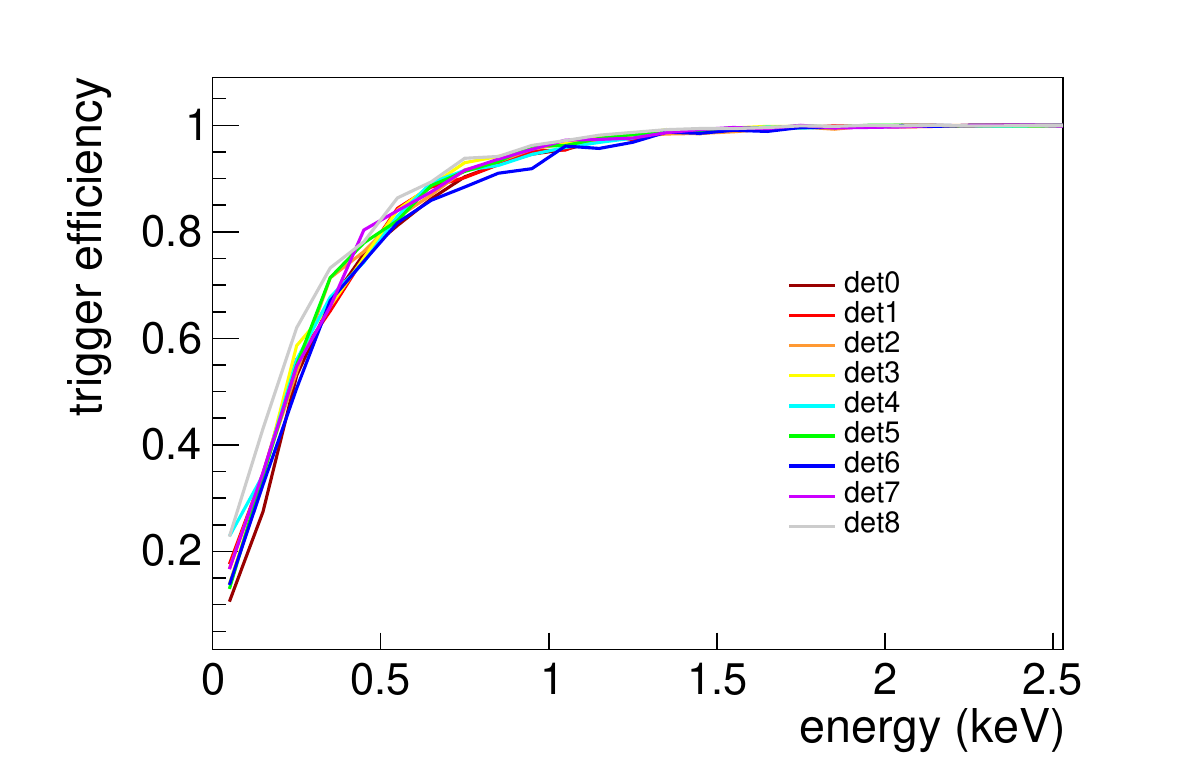}

\caption{\label{triggereff} Trigger efficiency for every detector, calculated by a Monte Carlo technique with the respective measured light collection, the NaI scintillation time and the PMT coincidence window used in ANAIS-112. }
\end{center}
\end{figure}

To account for this, the trigger efficiency was implemented in the simulation as follows. For each event, a random floating-point number between 0 and 1 is generated. This number is compared to the trigger efficiency value for each detector evaluated at the total energy deposited in that detector within the integration window (see Figure \ref{triggereff}). If the random number exceeds the efficiency value, the event fails the trigger condition and is excluded from the second-level analysis. \\

        \item \textbf{Coincidence definition.}

        Single-hit events are those acquired in anticoincidence. In contrast, multiple-hit events involve coincidences detected across modules. To analyze these populations effectively, the simulation must be able to tag them appropriately.

For this purpose, after recording events for each module within the integration window, a second analysis step examines all modules. Note that, as reflected in the order in which the processes are listed, the trigger efficiency is applied before the definition of multiplicity. Therefore, events that do not pass the trigger are excluded, both as events and as coincident events.

For a given detector, it is checked whether any event occurred in other detectors within the coincidence window. In the case of ANAIS, the integration and coincidence windows coincide (1 \textnormal{$\mu$}s in both cases), but this is not the case in ANOD, where the integration window is 1(2) \textnormal{$\mu$}s and the coincidence window is 8~\textnormal{$\mu$}s. This results in a difference in the definition of multiplicity between ANAIS and ANOD simulations, as coincidences between modules are searched within a broader time window in the latter.

Then, an event multiplicity variable is introduced: it takes a value of 1 for single-hit events and the number of modules having a hit within the coincidence window for multiple-hit events. In the second-level analysis, each event has a set of associated variables that correspond to its coincident event(s). Therefore, for a given event, the following data is associated: event multiplicity and arrays containing information on coincident events (ID, energy, event time, etc.). \\

    \item \textbf{Quenching factor.} 
    
    In the simulation, energy must be expressed in the electron-equivalent energy scale because ANAIS is calibrated using gamma sources (e.g. \(^{109}\)Cd) that produce ERs. 
    
    This approach is straightforward for energy deposited by electrons or gamma particles. However, special consideration is required when constructing the visible energy of NRs or alpha particles, as ERs generate significantly much more light than NRs of the same energy. This difference is accounted for by the quenching factor (QF), which relates the light yield produced by ERs to that of NRs or alpha particles with the same deposited energy. A detailed discussion about the QF for Na and I nuclei is presented in Chapter \ref{Chapter:QF}.

To determine the visible energy of NRs and alpha particles, the deposited energy is multiplied by an appropriate QF. In the initial analysis stage, particle identification is available, allowing the QF to be applied based on the particle type:

\begin{itemize}
    \item For alpha particles, a QF value of 0.6 is applied based on \cite{Amare:2018ndh}.
    \item For \(^{23}\)Na or \(^{127}\)I nuclei, as discussed in Chapter \ref{Chapter:QF}, different QF models for Na and I can be tested.
    \item  For other nuclei, although the expected effect in ANAIS simulations is minimal, the following approach based on the atomic number of the particle has been adopted due to the lack of information on the QF values. In particular, for particles with atomic numbers between 1 and 8, a QF of 0.6 is applied, similar to alpha particles. For atomic numbers between 9 and 20, the QF of sodium is used, while for those between 21 and 60, the QF of iodine is considered. For atomic numbers greater than 60, a QF of 0.04 is assumed, serving as a reasonable upper bound based on \(^{127}\)I data.  \\
\end{itemize}

\end{itemize}

\begin{itemize}

    \item \textbf{Energy resolution.}

Energy resolution in ANAIS-112 is primarily driven by poissonian fluctuations in the number of scintillation photons producing photoelectrons in the PMTs, and then a visible signal. Although previously determined \cite{Amare:2018sxx}, changes in the energy calibration strategy required its reevaluation employing the energy variables used in the updated calibration. Energy calibration in ANAIS-112 is performed in two ranges: low energy and high energy, with the boundary between these regions typically set $\sim$ 150 keV (see Section \ref{implementationenergycal}). The energy resolution is chosen accordingly, depending on the energy range being compared in the simulation.

\begin{figure}[t!]
\begin{center}
\includegraphics[width=0.9\textwidth]{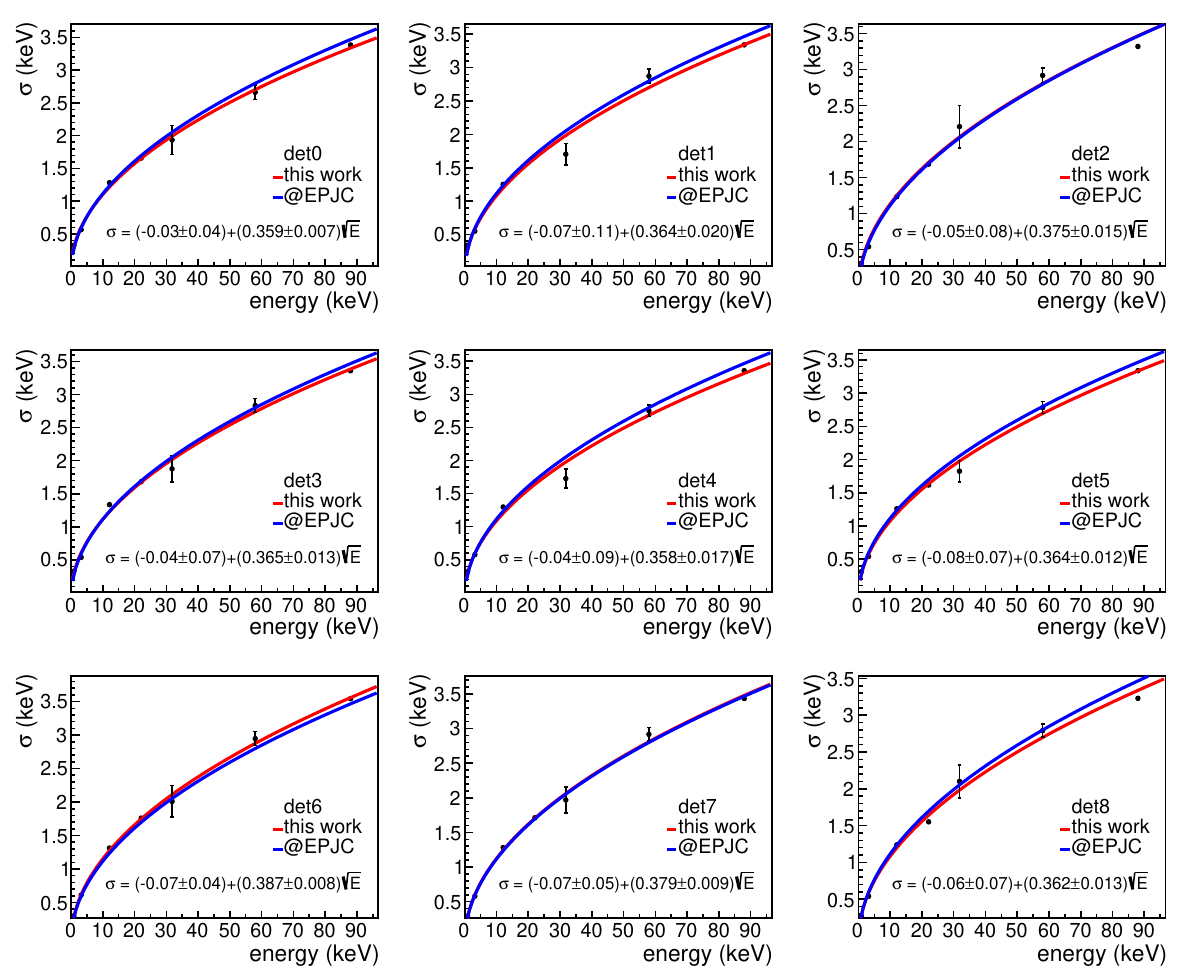}

\caption{\label{LEres} Energy resolution in the low-energy range as a function of energy. The estimation is based on efficiency-corrected coincidence background peaks from \(^{22}\)Na and \(^{40}\)K, the 12.1, 22.6, and 88 keV external \(^{109}\)Cd calibration peaks, and the 31.8 and inelastic peaks from neutron calibrations. The red line represents the fit to the \(\alpha + \beta \sqrt{E}\) function, with the fitted parameter values displayed in the plot. For comparison, the previous resolution estimation from \cite{Amare:2018sxx} is also shown in blue.}
\end{center}
\end{figure}

Figure \ref{LEres} shows the modelling of the energy resolution of ANAIS-112 in the low-energy region besides the experimental available information. Unlike previous estimations, the energy resolution is now modelled separately for each detector to account for their distinct behaviors, ensuring better accuracy in the simulated detector response. 

For determining the energy resolution with experimental data, the \(^{22}\)Na and \(^{40}\)K peaks, as well as the 12.1 keV and the 88 keV line from \(^{109}\)Cd, are fitted with a gaussian function plus a linear background. For the 22 keV line from \(^{109}\)Cd, a double gaussian fit with the same width is applied, accounting for the K\(_\alpha\) and K\(_\beta\) X-rays, emitted at average energies of 22.1 and 25.0 keV, respectively, plus a linear background.

Additionally, the two peaks originating from neutron calibrations are included in the resolution estimation: the 31.8 keV peak from the decay of \(^{128}\)I, produced by neutron capture in \(^{127}\)I, and the inelastic\footnote{According to the simulation, while the 31.8 keV peak is identified as purely electronic, the inelastic peak arises from a combination of electron and NRs. Specifically, it results from the superposition of the 57.6~keV de-excitation gamma with a tail of NR events. After convolution with the ANAIS-112 energy resolution, the fit to the simulated spectrum predicts the mean of this composite peak at (58.77 $\pm$ 0.03)~keV. For simplicity, it will hereafter be referred to as the inelastic peak.} peak of \(^{127}\)I.

Thus, for the 31.8 keV peak, a simple fit using a gaussian function plus a linear background, as in the case of \(^{109}\)Cd calibration, was sufficient. However, applying the same procedure to the inelastic peak systematically resulted in a resolution significantly higher than expected based on the trend dictated by the other peaks. To address this, a fit was performed in RooFit, incorporating the simulated peak shape, which accounts for the tail on the right side of the peak due to NRs, convoluted with a gaussian with the width as a free parameter. By applying this approach, the value obtained for the resolution follows the expected trend (see Figure \ref{LEres}).

The resolution obtained for the seven peaks is fitted to an \(\alpha + \beta \sqrt{E}\) parametrization, a functional dependence commonly used for scintillation detectors, where \(\beta\) represents the statistical fluctuations dominated by the number of detected photons. \(\beta\) is then a parameter related to LY, defined as the amount of light produced per unit of deposited energy. Specifically, \(\beta = \frac{1}{\sqrt{LY}}\).

As shown in Figure \ref{LEres}, the resolution estimate follows the expected trend. When compared to the previous modelling from \cite{Amare:2018sxx}, which was performed with the spectral information built by adding the nine modules signals, it can be observed that the updated resolution is better than previous estimates, except for, as expected, detector~6, which is known to exhibit a worse resolution than the average because of its worse light collection.

Table \ref{tablaLEres} presents the fit values obtained from the low-energy resolution estimation. Values of the \(1/\beta^2\) are compared with the LY values of the nine modules of ANAIS-112. These LY values were derived from the first year of ANAIS-112 data \cite{Amare:2018sxx}, and are of the order of 15 photoelectrons/keV. The table also includes the ratio between both quantities. As expected, the values obtained for the nine detectors deviate from those expected for the ANAIS-112 modules. This suggests the presence of additional contributions to the resolution, beyond the statistical fluctuations in the number of photoelectrons generated during the detection process, that play a significant role. Nevertheless, as shown in Figure \ref{LEres}, the energy dependence of the resolution exhibits the expected behavior. One possible explanation for this result is that the LY may not be uniform throughout the detector volume and that an energy-dependent variation, unaccounted for in the current analysis, could exist. 

Module D6 exhibits the worst energy resolution, as evidenced by both the resolution fit and the amount of light collected in ANAIS-112, although overall the results across detectors remain compatible. Furthermore, it is worth noting that for several detectors (D0, D1, D2, D3, D4, and D8), the independent term is compatible with zero within the uncertainty of the fit, which supports the conclusion that additional systematic contributions independent of energy, such as electronic noise, are not significant.

\begin{table}[t!]
\centering
\begin{tabular}{ccccc}
\hline
Detector & $\alpha$ (keV) & 1/$\beta^2$ (1/keV) & LY (photoelectrons/keV)  &   (1/$\beta^2$)/LY     \\
\hline
\hline

   D0  &  -0.03 ± 0.04  & 7.76 ± 0.15  & 14.532 ± 0.102 & 0.534 ± 0.010 \\
D1 & -0.07 ± 0.11  & 7.54 ± 0.41  & 14.745 ± 0.169 & 0.511 ± 0.030 \\
D2 & -0.05 ± 0.08 & 7.85 ± 0.33  & 14.506 ± 0.104 & 0.541 ± 0.025 \\
D3  & -0.04 ± 0.07  & 7.51 ± 0.27  & 14.453 ± 0.109 & 0.520 ± 0.020 \\
D4  & -0.04 ± 0.09 & 7.81 ± 0.37  & 14.483 ± 0.090 & 0.539 ± 0.026 \\
D5  & -0.08 ± 0.07 & 7.54 ± 0.25  & 14.572 ± 0.158 & 0.518 ± 0.019 \\
D6  & -0.07 ± 0.04 & 6.67 ± 0.14  & 12.707 ± 0.104 & 0.525 ± 0.012 \\
D7  & -0.07 ± 0.05 & 6.96 ± 0.17  & 14.743 ± 0.137 & 0.472 ± 0.013 \\
D8  & -0.06 ± 0.07 & 7.63 ± 0.27  & 15.994 ± 0.076 & 0.477 ± 0.017 \\
         \hline
\end{tabular}
\caption{\label{tablaLEres} Fit values obtained from the low-energy resolution study performed in this work, modelled as a \(\alpha + \beta \sqrt{E}\) function. The ratio \(1/\beta^2\) is compared with the light collected per unit of deposited energy by the nine NaI(Tl) modules of the ANAIS-112 set-up, derived from the first year of ANAIS-112 data \cite{Amare:2018sxx}. The table also includes the ratio between both quantities.}
\end{table}

\begin{figure}[t!]
\begin{center}
\includegraphics[width=1.\textwidth]{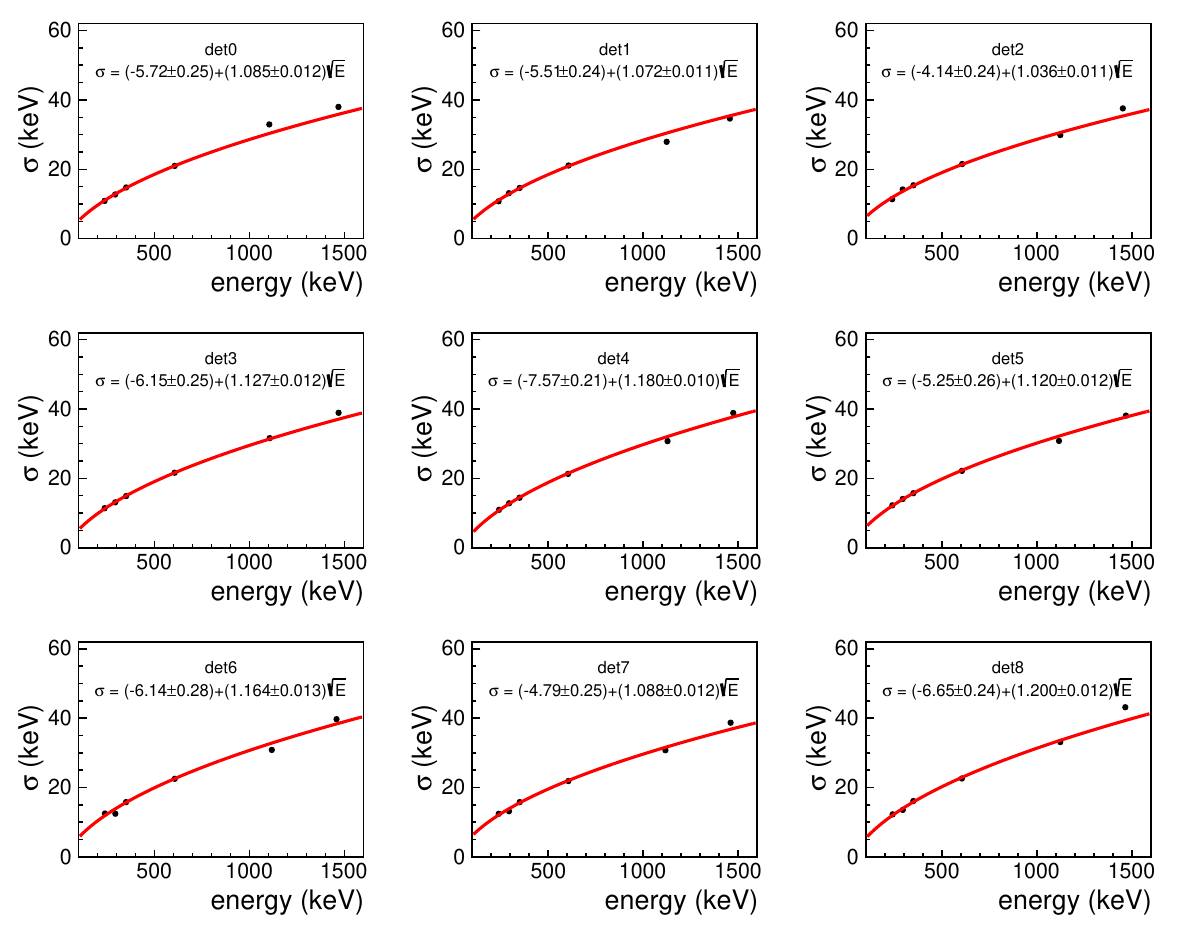}

\caption{\label{HEres} Energy resolution in the high-energy range as a function of energy. The estimation is based on identifiable peaks on background measurements. The red line represents the fit to the \(\alpha + \beta \sqrt{E}\) function, with the fitted parameter values displayed in the plot.}
\vspace{-0.5cm}
\end{center}
\end{figure}

In the high-energy range, calibration is done using background measurements, as no external high-energy sources are available. Calibration is performed independently for each background run using easily identifiable peaks (e.g., 238.6 keV from $^{212}$Pb, and 609.3 keV from $^{214}$Bi). Each peak is fitted with a gaussian, and calibration is achieved through a linear regression between the nominal peak energies and the gaussian means, using a second-degree polynomial. The energy resolution of ANAIS-112 in the high-energy range is shown in Figure \ref{HEres}.

To apply energy resolution in the simulation, an energy smearing is applied for each energy deposition within the time integration window of ANAIS-112, based on the detector resolution shown in Figures \ref{LEres} and \ref{HEres} for each detector. The observed energy is randomly sampled from a gaussian distribution centered on the Geant4-simulated deposited energy, with a width corresponding to the energy range under consideration. As previously mentioned, it is worth highlighting that the resolution is applied after correcting the deposited energy using the QF, ensuring it corresponds to the visible energy after all of the previously explained corrections.

\end{itemize}


\section{Conclusions}

This chapter summarizes the work undertaken in this thesis regarding simulation software. An update of the entire simulation framework has been developed, including a revision of the geometry, the Geant version, the physics lists, and the implementation of the ANAIS detector response and the most relevant features of the DAQ system for signal building. The goal of these efforts is to generate simulated data that can be directly compared with experimentally acquired data. Using this framework, the response of the ANAIS-112 detectors to NRs will be evaluated to assess the QF of Na and I nuclei (see Chapter \ref{Chapter:QF}), and the background model will be revisited (see Chapter~\ref{Chapter:bkg}).

\setcounter{chapter}{3} 
\chapter{Onsite neutron calibrations in ANAIS-112}\label{Chapter:QF}

\minitoc
\vspace{-1cm}
In DM search experiments, discriminating between nuclear recoils (NRs) and electron recoils (ERs) is crucial, as NRs are expected to be the signature of DM interactions in most of the WIMP models, while ERs represent most of the background events. 
Therefore, a precise understanding of the NaI(Tl) detector response to energy deposited by NRs is necessary for the proper calibration of the ROI in DM searches. However, practical constraints related to the availability of calibration sources often prevent this procedure. In practice, experiments such as ANAIS-112, DAMA/LIBRA and COSINE-100 use X-rays and gamma rays for calibration. In particular, ANAIS-112 relies on external $^{109}$Cd sources and internal radioactive contaminants, $^{22}$Na and $^{40}$K, for calibration purposes, as detailed in Section~\ref{energyCal}, while DAMA/LIBRA used $^{241}$Am sources, and COSINE relied on a combination of internal radioisotopes and external sources ($^{241}$Am, $^{57}$Co, $^{60}$Co, and $^{137}$Cs) for calibration. It is worth highlighting that calibration campaigns in COSINE-100 were conducted approximately every two years, in contrast to the biweekly calibration runs periodically performed by ANAIS-112.

The energy calibrated using this approach is referred to as electron equivalent energy (E$_{\textnormal{ee}}$, measured in keV$_{\textnormal{ee}}$). In order to convert E$_{\textnormal{ee}}$ to nuclear recoil energy (E$_{\textnormal{NR}}$, measured in keV$_{\textnormal{NR}}$), it is crucial to know the scaling factor between these two energy scales. This scaling factor is known as the quenching factor (QF), and measures the relative efficiency for the production of scintillation between NR\footnote{Different QF values correspond to ionizing particles with different stopping power, such as $\alpha$-particles, deuterons, protons and NR from heavier nuclei. However, throughout this chapter, the term will solely refer to NRs from Na and I nuclei. } and ER of the same energy. The objective of this chapter is to estimate the QFs for sodium (QF\textsubscript{Na}) and iodine (QF\textsubscript{I}) recoils in the ANAIS-112 crystals by comparing experimental data with dedicated simulations of the onsite neutron calibrations conducted within the experiment. Since the values of QF\textsubscript{Na} and QF\textsubscript{I} differ, the NR energy scale consequently depends on whether the recoil occurs in a sodium or iodine nucleus. Given the importance of energy scales in this chapter, it is clarified that, unless otherwise specified, all energies reported here are expressed in electron-equivalent units (keV$_{\textnormal{ee}}$ or simply keV). The notation keV$_{\textnormal{NR}}$ is used exclusively when referring to NR energies.

Section \ref{QFcaveat} introduces the concept of the QF, outlines the models available to describe this parameter, and presents an experimental overview of existing measurements of QF\textsubscript{Na} and QF\textsubscript{I} in NaI(Tl) crystals, focusing specifically on those conducted at TUNL for the ANAIS-112 crystals (Section \ref{TUNL}). Section \ref{neutronData} then presents the onsite neutron calibration program developed at ANAIS-112, outlining its objectives, measurement schedule and procedure, the behavior of neutron data in terms of PSA, the applied data selection cuts, and the adopted calibration strategy. Subsequently, the simulation used to compare with the data will be described (Section \ref{neutronsim}), with particular attention to the comparison of Geant4 neutron physics libraries, the adjustment of the $^{128}$I capture cross-section, which was found to be necessary when comparing with the data, and the different QF models introduced in the simulation. Finally, Section \ref{comparison} will present the comparison between the data and simulations for the different QF models under evaluation.

\section{Overview of QF\textsubscript{Na} and QF\textsubscript{I} measurements in NaI(Tl)} \label{QFcaveat}

A comparison between experimental results obtained with different target nuclei
in DM searches is affected by model-dependencies in the DM particle and halo models, as introduced in Chapter \ref{Chapter:Intro}. Using the same target should remove
these dependencies completely, but only under the condition that the detector response is fully accounted for in the comparison. This is particularly relevant for the ANAIS-112 test of the DAMA/LIBRA signal, because in scintillators only a portion of the energy  released by NRs is transferred to electrons and converted into scintillation, while the remainder can become trapped in long-lived excited states of the crystals or end up in lattice vibrations. This is generally driven by the high stopping power, resulting in saturation of scintillation centers and de-excitation via non-radiative decays. As a result, NRs yield less light compared to ERs of the same energy. 

The parameter describing this behavior is the QF, defined as the ratio of scintillation light produced by a NR ($L_{\text{NR}}$) to the
scintillation light produced by a ER of the same energy ($L_{\text{ER}}$):

\begin{equation}
    \textnormal{QF}  = \dfrac{L_{\text{NR}}}{L_{\text{ER}}}.
    \label{mainQFeq}
\end{equation}

The QF is
usually < 1 for scintillation and ionization measurements, while it could be larger than one for heat, and typically depends on energy. It is important to note that ANAIS detectors are not able to identify if an energy deposition corresponds to a NR or ER. Therefore, even if the detectors were calibrated using NR events, the reconstructed energy for all ER events would be incorrect. This underscores the importance of accurately determining the QF parameter.

Despite extensive research on QFs for various particles and scintillating materials, no fundamental theory can precisely predict the QF. The Lindhard theory and the Birks model only offer semi-empirical estimates. Both of these theoretical models will be tested against the ANAIS-112 results obtained at TUNL to evaluate their viability in Section \ref{QFmodels}.

\begin{figure}[t!]
\begin{center}
\includegraphics[width=0.49\textwidth]{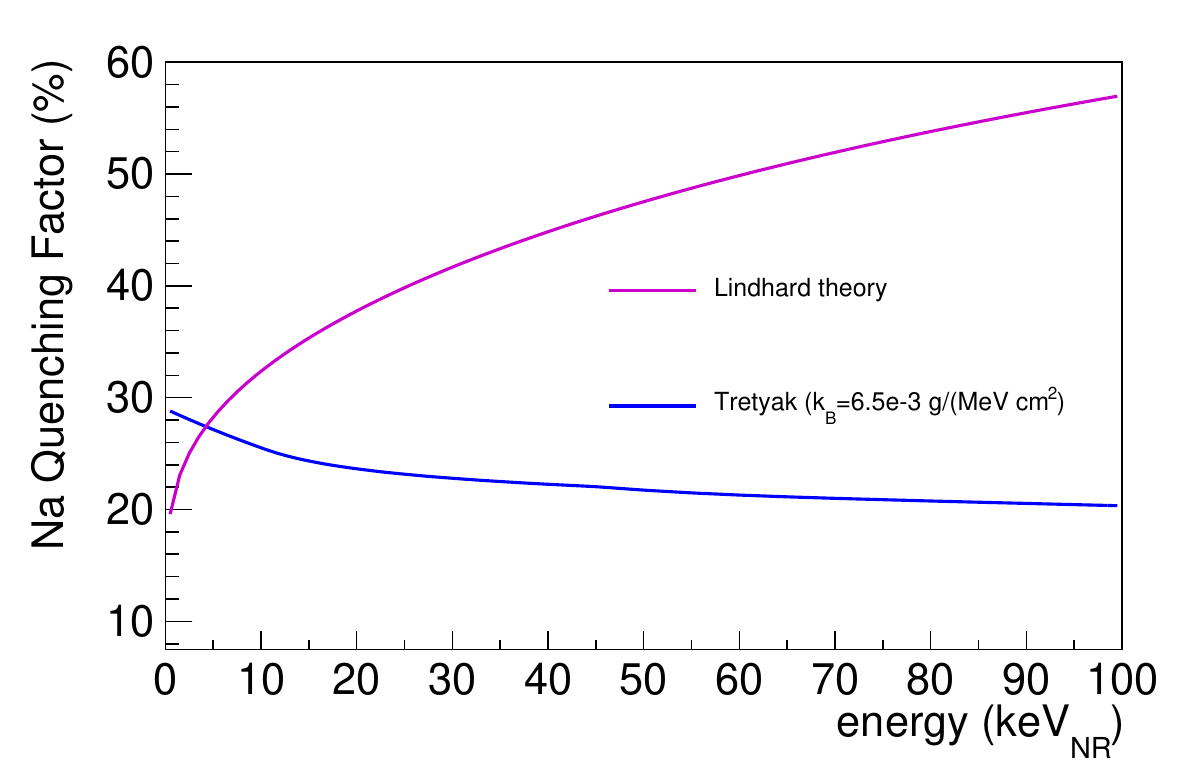}
\includegraphics[width=0.49\textwidth]{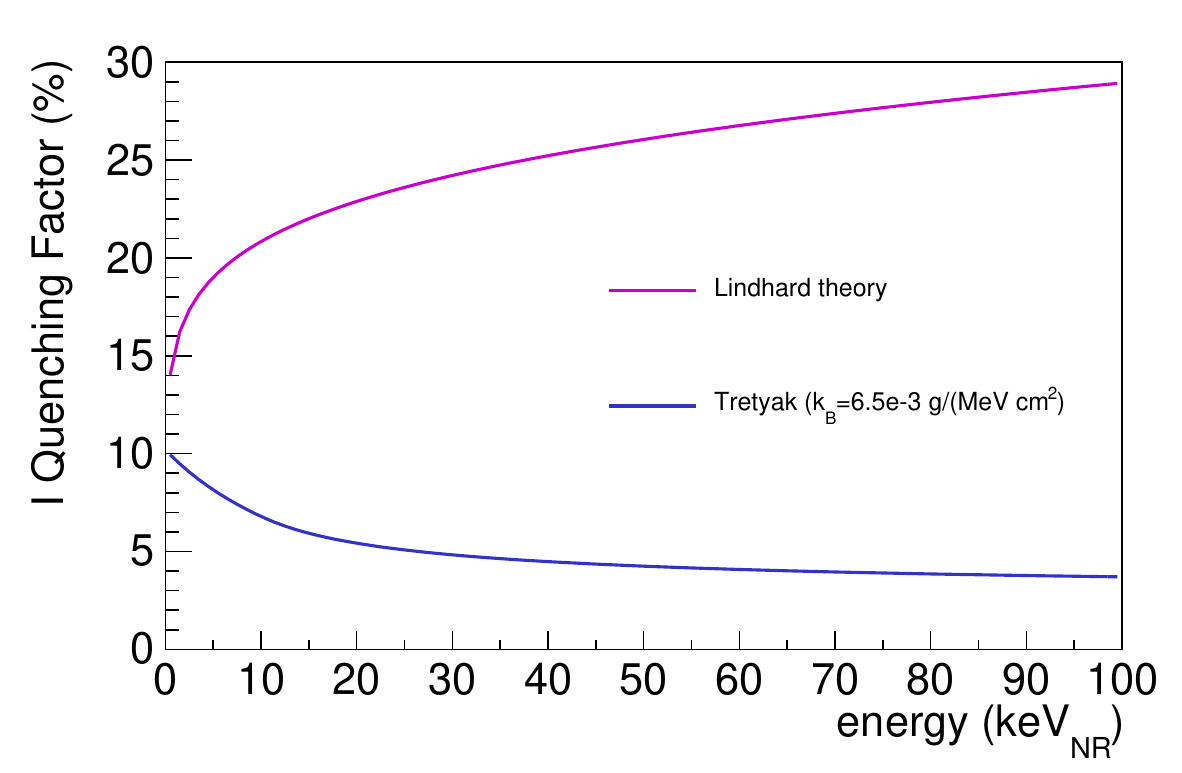}

\caption{\label{QFtheory} Predictions from the Lindhard theory (magenta), as described by Equation~\ref{eqlind}, and from the Birks model using the approximation proposed by Tretyak et al. \cite{tretyak2010semi}, as expressed in Equation \ref{approxTetry} (blue). \textbf{Left panel:} QF\textsubscript{Na}. \textbf{Right panel:} QF\textsubscript{I}.}

\end{center}
\end{figure}

The Lindhard theory \cite{lindhard1963range,LINDHARD} aims to quantify energy loss from first principles to theoretically describe the QF. In the Lindhard model, following a series of approximations, the QF is described as:

\begin{equation}
\textnormal{QF}(E_{\textnormal{NR}}) = \frac{k g(\epsilon)}{1 + k g(\epsilon)},
\label{eqlind}
\end{equation}

where $k$ is proportional to the ratio between the electronic stopping power and the velocity of the recoiling nucleus, with $k = 0.133 Z^{2/3}A^{-1/2}$, where $Z$ and $A$ are the atomic and mass numbers of the nucleus, respectively. The function $g(\epsilon)$ is proportional to the ratio of electronic to nuclear stopping power \cite{lenardo2015global}, and depends on the dimensionless energy variable $\epsilon = \frac{11.5}{\textnormal{keV}} E_{\textnormal{NR}}\textnormal{(keV)} Z^{-7/3}$. In this context, $g(\epsilon)$ is given by $g(\epsilon) = 3\epsilon^{0.15} + 0.7\epsilon^{0.6} + \epsilon$ \cite{Lewin:1995rx}. For sodium, Lindhard’s calculation yields $k$ = 0.137, while for iodine $k$~=~0.167. Figure \ref{QFtheory} shows the Lindhard model predictions for QF\textsubscript{Na} and QF\textsubscript{I}.

\begin{figure}[t!]
\begin{center}
\includegraphics[width=0.75\textwidth]{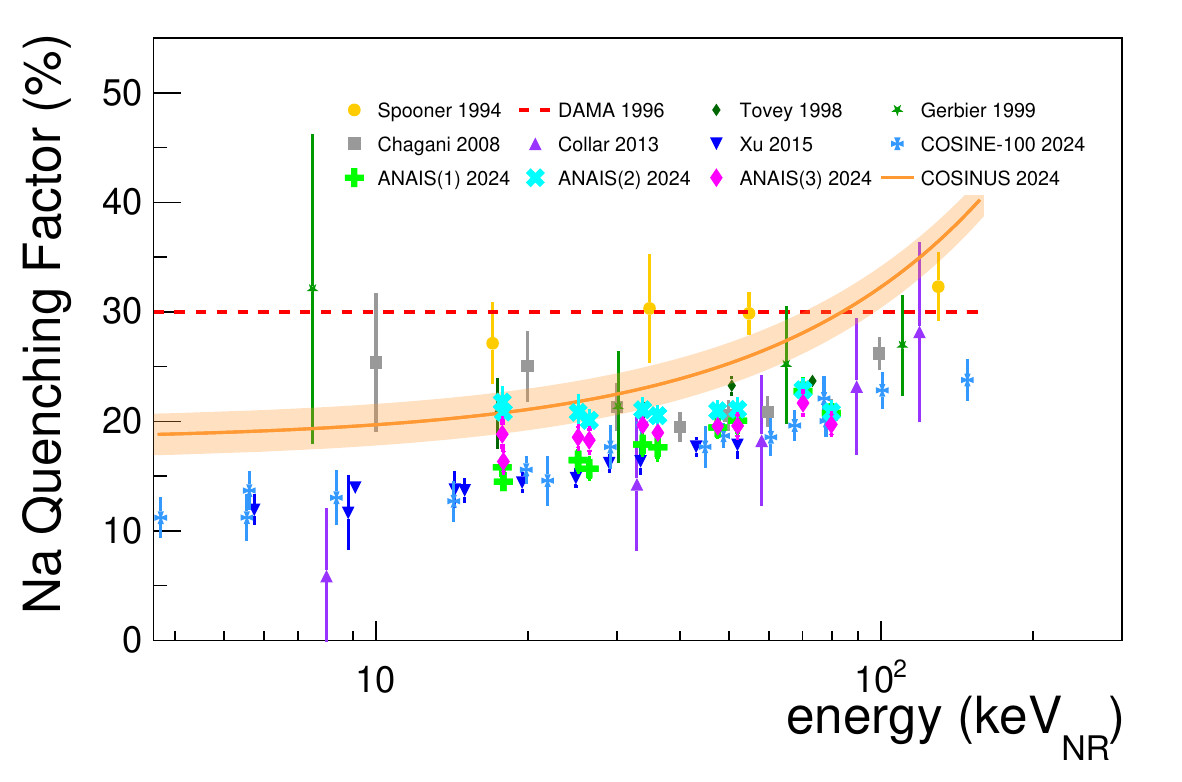}

\caption{\label{NaQFcomparison} Measurements of the QF for sodium nuclei in NaI(Tl) crystals \cite{Bernabei:1996vj,Spooner,Tovey:1998ex,Gerbier:1998dm,Chagani,Collar,Xu,Joo:2018hom,cintas2024measurement,PhysRevD.110.043010}.}

\end{center}
\end{figure}

Although the Lindhard model effectively describes the relative ionization efficiencies in semiconductors, its direct application to NaI(Tl) crystals does not accurately reproduce recent QF\textsubscript{Na} and QF\textsubscript{I} measurements, as will be illustrated in Figure \ref{NaQFcomparison}. In particular, the model predicts larger values for QF\textsubscript{Na} and QF\textsubscript{I} than what is observed experimentally. Moreover, it does not show the expected saturation around 100 keV\textsubscript{NR}. This requires modifications to the model for better alignment with data, as discussed in Section \ref{TUNL}. The Lindhard theory also fails to adequately describe the QF behavior in xenon and argon, for which dedicated models have also been developed \cite{szydagis2025review}. 

On the other hand, the Birks model \cite{Birks:1964zz} offers a semi-empirical alternative more specific for scintillation QF than the Lindhard model, suggesting that the light yield in scintillating materials is a function of both the particle energy and the total stopping power. Following the Birks model, the scintillation light $L$
produced by an ionizing particle  of energy $E$ is given by:



\begin{equation}
L(E) = \int_0^E dL = \int_0^E\frac{S \, dE}{1 + k_B \, \frac{dE}{dr}},
\end{equation}

where $S$ is the absolute scintillation efficiency, $\frac{dE}{dr}$ represents the stopping power of the ionizing particle considered in the material, and $k_B$ is the Birks factor, a parameter that depends on the material. Therefore, analogous to Equation \ref{mainQFeq}, the QF can be written as:


\begin{equation}
\textnormal{QF}(E) = \frac{L_{\textnormal{NR}}(E)}{L_\textnormal{ER}(E)} = 
\frac{\displaystyle \int_0^{E} \frac{S dE}{1 + k_B \left( \frac{dE}{dr} \right)_\textnormal{NR}}}{\displaystyle \int_0^E \frac{S dE}{1 + k_B \left( \frac{dE}{dr} \right)_\textnormal{ER}}}.
\label{eqbirks}
\end{equation}

To estimate the QF, Tetryak et al. \cite{tretyak2010semi} adopt the following simplifications of the Birks model. On the one hand, at low ionization density, such as that produced by electrons, there is no significant suppression due to stopping power, and thus the term \(k_B(\frac{dE}{dr})_{ER} << 1\).
On the other hand, at high ionization density, typical of heavy charged particles like recoiling nuclei or \(\alpha\)-particles, the term \(k_B(\frac{dE}{dr})_{NR} >>1\). Consequently, Equation \ref{eqbirks} can be approximated as:

\begin{equation}
    \textnormal{QF}\sim \dfrac{1}{k_B (\frac{dE}{dr})_{\textnormal{NR}}},
    \label{approxTetry}
\end{equation}

which implies that the QF is minimal when  \((\frac{dE}{dr})_{NR}\) is maximal.

Figure \ref{QFtheory} also presents the predictions of the Birks model following the Tretyak approximation for QF\textsubscript{Na} and QF\textsubscript{I}, employing a value of $k_B = 6.5 \times 10^{-3}$ g/(MeV cm$^2$) as proposed in \cite{tretyak2010semi} to reproduce the experimental results reported in \cite{chagani2008measurement}. As can be seen from the figure, the Tretyak model predicts an increase in QF with decreasing energy, which is not consistent with most recent measurements.

Since there is no universal description of the scintillation response due to the complex physics involved in ion interactions and scintillation light production, experiments using scintillators typically determine the QF of their crystal through dedicated calibration measurements under fast neutron irradiation. Neutrons deposit energy in the detector through various interaction mechanisms with nuclei, including elastic scattering, which produces NRs similar to those expected from WIMP interactions in most favored models. However, neutrons can undergo other processes such as inelastic scattering, neutron capture, and even spallation or other nuclear reactions if their energies are sufficiently high. Depending on the target nucleus and the neutron energy range, different interaction mechanisms may dominate.


Most measurements employ quasi-monoenergetic neutron sources and determine the NR energy released within small-sized targets by measuring the angle of the scattered neutron, with the scattered neutrons detected by an array of surrounding detectors \cite{Spooner,Tovey:1998ex,Gerbier:1998dm,Chagani,Collar,Xu,Joo:2018hom,cintas2024measurement}. These neutron detectors, typically made from liquid scintillators, enable discrimination between neutron and gamma events through PSA. By combining the measured scattering angle with the known energy of the incoming neutron, the total NR transferred by a single-scatter event on sodium or iodine nuclei can be directly calculated. The scintillation signal corresponding to the NR energy deposition is calibrated using X-ray or gamma sources (i.e. in the ER scale), allowing the QF to be determined as the ratio of the energy inferred from scintillation light to the NR energy.

However, DAMA/LIBRA QF measurements differ from this approach by relying on  irradiation with a non-monoenergetic neutron source following the decay of \(^{252}\)Cf \cite{Bernabei:1996vj}. This measurement, lacking from the knowledge of the NR energy deposited, produces an exponential-shape distribution of energies, combining recoil energies from sodium and iodine nuclei, which cannot be disentangled. Furthermore, depending on the target size, multiple scattering can be dominant, making more difficult to extract the QF from the measured spectral shape. DAMA/LIBRA collaboration first subtracts the background from the measured low-energy spectrum and then fits the resulting data using a general empirical exponential function, with the parameters estimated through Monte Carlo simulations, and assuming that the QFs for both Na and I nuclei are constant. DAMA/LIBRA reports values of 30\% and 9\%, respectively \cite{Bernabei:1996vj}.


Nevertheless, the analysis followed by DAMA/LIBRA is difficult to replicate. There is no information about the size of the crystal used, for instance, to evaluate the relevance of multiple scattering. In addition, the fitting relies on a complex and phenomenological exponential function with many free parameters, whose constraints were neither clearly defined nor physically motivated \cite{fushimi1993application}. Attempts were made to replicate a similar fitting using data from ANAIS-112 crystals under onsite irradiation with neutrons from a \textsuperscript{252}Cf source without success. 

As outlined, two strategies can be pursued for determining the QF: using monochromatic neutron sources at dedicated facilities or performing in-situ calibrations with neutron sources, such as $^{252}$Cf, at the experiment's site. Each approach presents specific advantages and limitations. The monochromatic neutron source approach enables a precise determination of the QF for each nuclei species at specific NR energies,  but requires small crystals, preventing from measuring those used in the experiment. This fact could be relevant if the QF depends on specific crystal properties in the case of not being an intrinsic property of the material. In contrast, the $^{252}$Cf approach allows for the calibration of the actual large-mass crystals used in the experiment, which is a significant advantage. Nevertheless, the extraction of QF values in this case relies more heavily on Monte Carlo simulations, which allow to take into account consistently the dominant multiple scattering contribution and the different nuclear species involved. Despite their respective advantages and disadvantages, both approaches are complementary and are expected to provide consistent results.


\begin{figure}[b!]
\begin{center}
\includegraphics[width=0.75\textwidth]{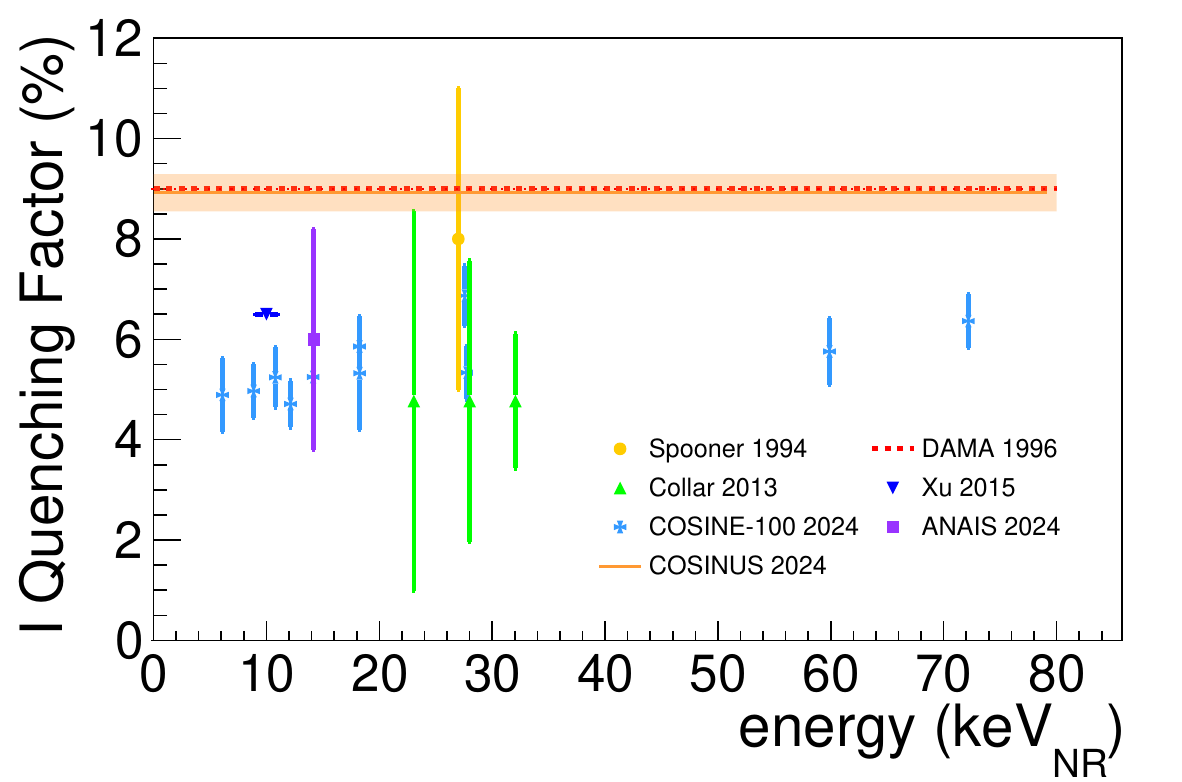}

\caption{\label{IQFcomparison} Measurements of the QF for iodine nuclei in NaI(Tl) crystals \cite{Bernabei:1996vj,Spooner,Collar,Xu,Joo:2018hom,cintas2024measurement,PhysRevD.110.043010}.}

\end{center}
\end{figure}

Given its crucial role in determining the energy scale of WIMP interactions in NaI(Tl)-based DM detectors, it is unsurprising that extensive research has been dedicated to measure the QFs of NaI(Tl) detectors. Figure \ref{NaQFcomparison} presents QF$_{\textnormal{Na}}$ results from various mesurements, including those carried out by ANAIS using the monochromatic neutron source approach, revealing significant discrepancies. Analogously, Figure \ref{IQFcomparison} shows the current experimental status of QF\textsubscript{I} measurements.

Early measurements reported relatively high QF$_{\textnormal{Na}}$ values~\cite{Bernabei:1996vj,Spooner,Chagani}. However, most recent measurements agree on a decreasing QF$_{\textnormal{Na}}$ at lower nuclear recoil energies, though there is some disagreement in the energy dependence
among the different measurements~\cite{Collar,Xu,Joo:2018hom,cintas2024measurement} and information is scarce at low energies. Figure \ref{NaQFcomparison} also shows the QF$_{\textnormal{Na}}$ reported by COSINUS \cite{PhysRevD.110.043010}, which constitutes the first measurement of the QF at cryogenic temperatures performed in situ using the dual-channel readout of phonon and light signals. The COSINUS results exhibit the same general energy dependence, but their values are slightly higher than those reported by other measurements. The QF values from COSINUS shown in Figures \ref{NaQFcomparison} and \ref{IQFcomparison} are displayed as lines, as also done by the collaboration, since they do not correspond to direct measurements. Instead, they are derived from the modelling and subsequent fitting of the LY–energy bands for sodium and iodine recoils (see Chapter~\ref{Chapter:COSINUS} for further details on this modelling).

Determining the QF$_{\textnormal{I}}$ in NaI(Tl) is particularly challenging due to iodine being a heavier nucleus, which leads to lower-energy recoils that are difficult to distinguish from the background, as QF$_{\textnormal{I}}$ are expected also to be lower (see Figure \ref{QFtheory}, for instance). This challenge has resulted in a limited number of precise measurements \cite{Bernabei:1996vj,Spooner,Collar,Xu,Joo:2018hom,cintas2024measurement,PhysRevD.110.043010}. It is worth highlighting that the COSINUS results for QF$_{\textnormal{I}}$ are compatible with that of DAMA/LIBRA within 1$\sigma$.

The discrepancies in the reported QF values can be attributed to two main reasons. 

\begin{itemize}
    \item One possibility is that the QF value is consistent across NaI(Tl) detectors, but differences in experimental procedures have introduced unaccounted systematic errors, leading to different results. This aligns with the common assumption that the QF is an inherent material property that remains constant across different detectors. 

    \item Alternatively, the observed differences could be genuine, suggesting that the QF varies between individual NaI(Tl) detectors, as proposed by DAMA/LIBRA~\cite{Bernabei:1996vj,bernabei2013dark}. If the QF is not an intrinsic property of NaI(Tl), it would be the only remaining uncertainty when comparing DM results obtained with DAMA/LIBRA and other NaI(Tl) experiments. This would complicate testing the DAMA/LIBRA claim, as it would require dedicated NR energy calibrations for each NaI(Tl) detector. Different QF\textsubscript{Na} and QF\textsubscript{I} would impact strongly the energy region where DAMA/LIBRA finds the modulation in the NR energy scale. For instance, the DAMA/LIBRA region from [2-6] keV would correspond to the ANAIS-112 region from [1.3-4] keV when assuming for QF\textsubscript{Na} and QF\textsubscript{I} values of 0.3 and 0.09 for DAMA/LIBRA and 0.2 and 0.06 for ANAIS.

\end{itemize}




To determine the QF of the ANAIS-112 crystals, a dedicated neutron calibration program was developed. This program followed two distinct approaches in parallel, corresponding to the different methodologies previously mentioned. On the one hand, QF measurements for small NaI(Tl) crystals, similar to those used in ANAIS-112, were carried out using a monoenergetic neutron beam at TUNL in North Carolina. The findings from this approach will be presented in the next subsection. On the other hand, onsite neutron calibrations have been conducted since 2021 using low-activity \(^{252}\)Cf sources at the LSC. This chapter focuses on the latter approach, with its methodology thoroughly explained in the following sections.

\subsection{ANAIS-112 QF measurements at TUNL}\label{TUNL}

The QF of five NaI(Tl) crystals has been measured through a collaborative effort initiated in 2018 by members of COSINE-100, COHERENT and ANAIS-112. All five crystals were measured in the same experimental set-up (see Figure \ref{QFTunl}), and were produced by Alpha Spectra Inc. using the same growth procedure, with one of them (Crystal No. 5) from the same batch as some of the ANAIS-112 crystals. One of the goals of the measurement was to check the consistency of the QF values across different crystals. The experimental set-up, measurement procedure and event analysis are explained in detail in \cite{cintas2024measurement,phddavid}.

In this study, outside the scope of this thesis, a comprehensive identification of potential systematic effects during the measurement and analysis protocols was conducted, highlighting the electron equivalent energy calibration method as the most critical. For that purpose, three calibration strategies were considered. Custom names are assigned and will be used throughout the chapter to facilitate the identification of each calibration approach:

\begin{figure}[b!]
\begin{center}
\includegraphics[width=0.6\textwidth]{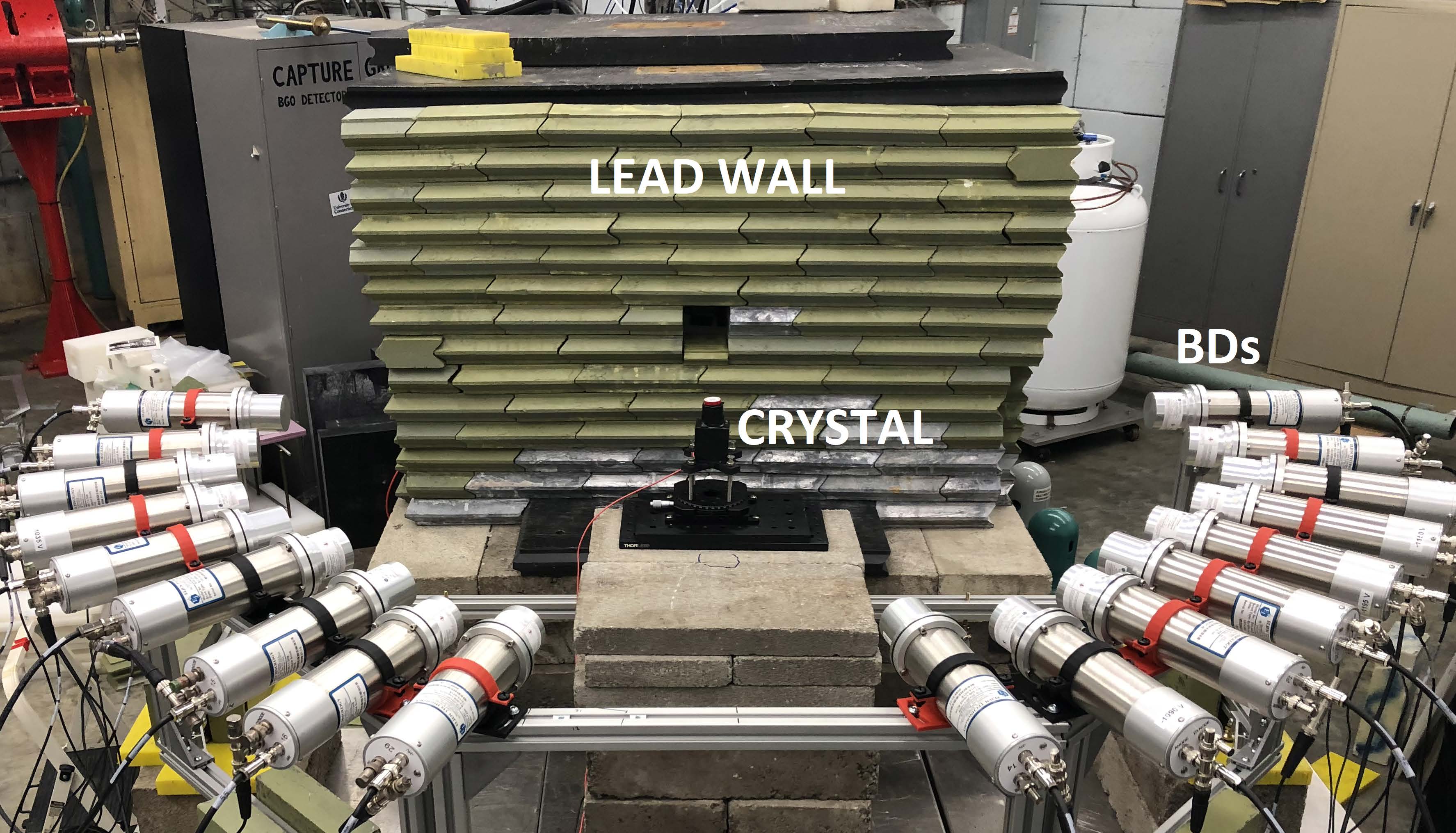}

\caption{\label{QFTunl} Picture of the experimental set-up using a monoenergetic neutron beam at the Triangle Universities Nuclear Laboratory (TUNL) in North Carolina \cite{cintas2024measurement,phddavid}. }
\end{center}
\end{figure}

\begin{enumerate}
    \item  \textbf{ANAIS(1):} A proportional calibration using only the inelastic peak from the $^{127}$I(n,n'$\gamma$) inelastic process. This approach is the most commonly used in other QF\textsubscript{Na} measurements.

    \item \textbf{ANAIS(2):} A linear calibration in the ROI using gamma and x-ray emissions from an external $^{133}$Ba source (6.6, 30.9 and 35.1 keV).
    
    \item \textbf{ANAIS(3):} A combination of both calibration methods (linear calibration above 6~keV using the lines coming from the $^{133}$Ba source, and a
proportional response below this energy) to improve calibration accuracy.

\end{enumerate}

\begin{figure}[b!]
\begin{center}
\includegraphics[width=0.8\textwidth]{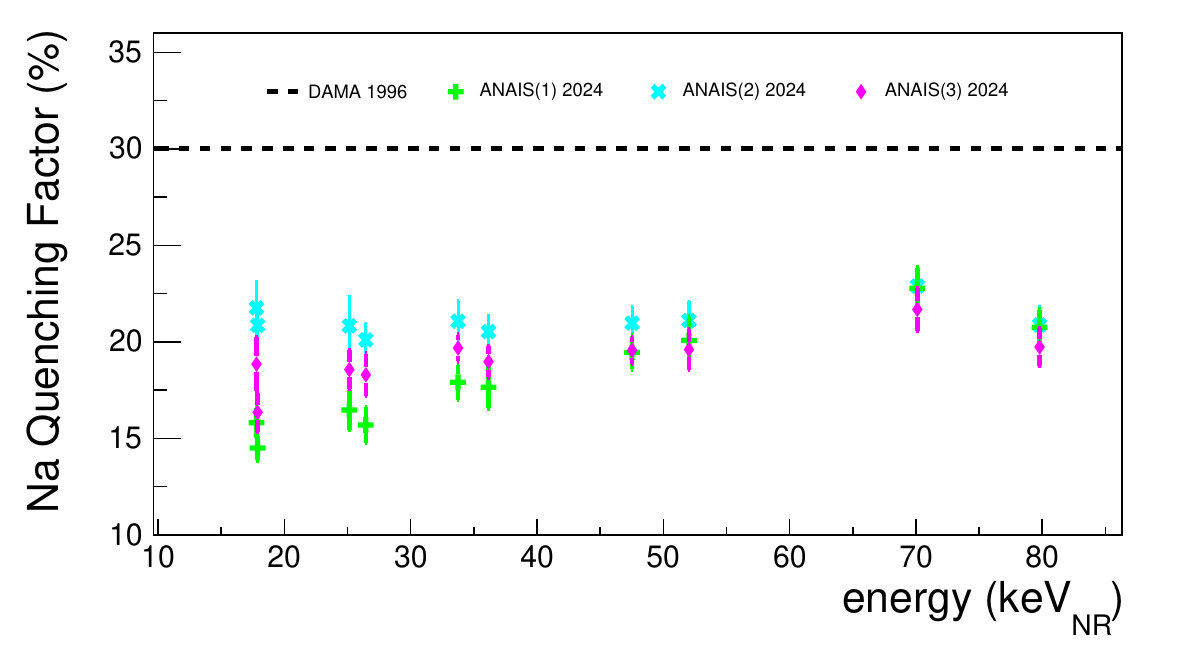}

\caption{\label{NaQFcomparisonANAIS} QF$_{\textnormal{Na}}$ results for Crystal No. 5 using the three calibration methods described in \cite{cintas2024measurement,phddavid}: ANAIS(1) (green); ANAIS(2) (cyan), and ANAIS(3) (magenta). The DAMA/LIBRA result is also shown for comparison.}
\end{center}
\end{figure}

The results obtained for QF$_{\textnormal{Na}}$ for all the five crystals are consistent with
each other, yet very
important differences can be found when applying the different energy calibrations. Figure \ref{NaQFcomparisonANAIS} shows the comparison between the QF$_{\textnormal{Na}}$ values for Crystal~No.~5, the only crystal coming from the same batch than ANAIS-112, using the three calibration
methods, and the corresponding result from DAMA/LIBRA. It is worth highlighting that a lower QF\(_{\text{Na}}\) value  than the one claimed by DAMA/LIBRA (30\%) is measured under all the calibration approaches considered in this analysis.

A decrease in QF$_{\textnormal{Na}}$ at lower energies is observed with both ANAIS(1) (proportional calibration with the inelastic line) and ANAIS(3) (linear calibration above 6 keV using the $^{133}$Ba lines but proportional below this energy) calibration methods, consistent with previous studies \cite{Collar,Xu,Joo:2018hom,cintas2024measurement}. In contrast, ANAIS(2) (linear calibration with the $^{133}$Ba lines) does not show a clear energy dependence of QF$_{\textnormal{Na}}$, which aligns with previous studies \cite{Bernabei:1996vj,Spooner,Chagani}, with a weighted mean for all crystals of (21.0 $\pm$ 0.3)\%.

Regarding the ANAIS-112 QF$_{\textnormal{I}}$ measurement at TUNL, iodine recoils  could not be separated from the background in any channel. Therefore, QF$_{\textnormal{I}}$ was analyzed instead by studying the position of the inelastic scattering peak on \(^{127}\text{I}\) (sum of the light produced by the gamma deposition and the iodine recoil) at different scattering angles, resulting in a combined weighted mean of (6.0~$\pm$~2.2)\% for crystals No.~2 and No.~3. Hence, a lower QF$_\textnormal{I}$ value than that reported by DAMA/LIBRA (9\%) has again been obtained.

This work suggests that the discrepancies in QFs reported by previous measurements might arise from differences in the electron-recoil energy calibration method used, rather than from actual variations in the QFs among NaI(Tl) detectors. This systematic effect, which is associated with the well-known non-proportional behavior of the NaI(Tl) light yield (see Section \ref{nonpropSec}), may also be present in most of earlier measurements. Onsite neutron calibrations of ANAIS-112, which is the primary goal of this thesis, serve as an important cross-check for the results obtained at TUNL. 

\section{Onsite ANAIS-112 neutron calibration}\label{neutronData}

In the second half of its operation, the ANAIS-112 experiment has undertaken a dedicated neutron calibration program to produce a NR-rich population to better evaluate the response of the detectors to WIMPs. The results from the monochromatic neutron source approach have been detailed in Section \ref{TUNL} and will be used as benchmarks for the subsequent analysis. From this point onwards, this chapter will focus on the onsite neutron calibration of ANAIS-112.


The section is organized as follows: first, the goals of the onsite neutron calibrations are revisited in Section \ref{goals}. Then, the decay of \(^{252}\)Cf and the details of the source used for the calibrations are described in Section \ref{cf}, followed by an overview of the measurement schedule and procedure in Section \ref{schedule}. Section \ref{features} addresses the data analysis of the neutron calibrations, and Section \ref{pulses} presents the average pulses corresponding to neutron and gamma events for different energy regions and analyze their temporal behaviour.

\subsection{Goals of the onsite neutron calibrations}\label{goals}


Neutron calibrations provide a clean population of bulk scintillation events, arising predominantly from nuclear elastic scattering. One of the main goals of this calibration program, and the primary focus of this chapter, is to improve the understanding of the QF of the ANAIS-112 crystals. However, neutron calibrations also fulfill two additional objectives within the ANAIS-112 experiment.

On the one hand, neutron calibrations are used to generate WIMP-like signal events, which serve as the signal sample in the ML–based filtering protocol developed in ANAIS to improve the identification and rejection of anomalous events below 2 keV. Although the BDT method, already introduced in Section \ref{Filtering}, cannot determine the origin of a given energy deposit in the ANAIS-112 modules since NaI(Tl) scintillators cannot distinguish between NRs and ERs on an event-by-event basis at low energies, dedicated neutron simulations (see Section \ref{neutronsim}) show that events below 2 keV in $^{252}$Cf calibrations are dominated by elastic scattering. In particular, 98.7\% of multiple-hit events and 97.1\% of single-hit events in the [1–20] keV energy range result from elastic neutron interactions. This enables the selection of a neutron-rich population composed almost exclusively of NRs.

\begin{figure}[b!]
\begin{center}
\includegraphics[width=0.9\textwidth]{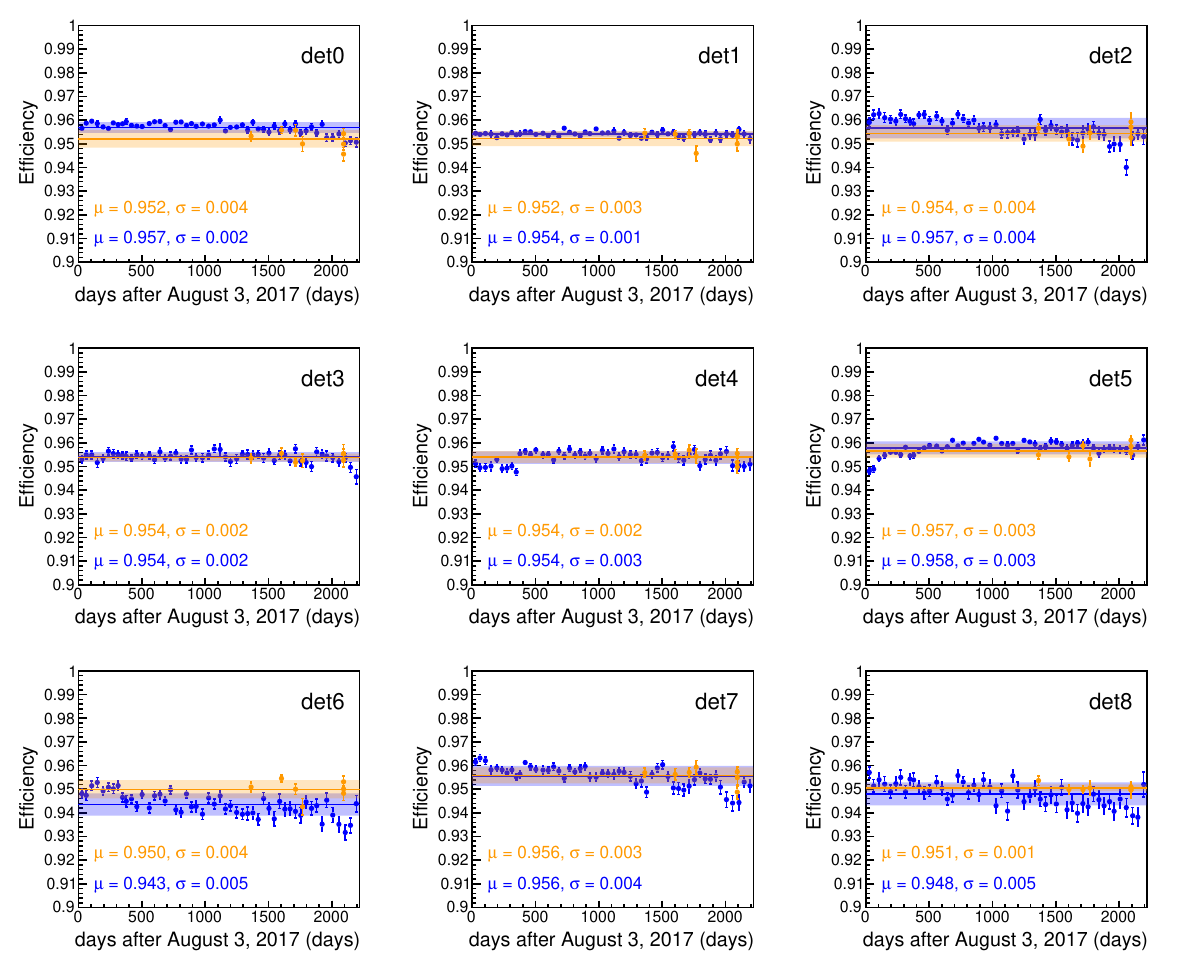}

\caption{\label{efficiencyneutrons} Evolution of the total detection efficiency in the [1–6] keV range, derived from neutron calibration (orange) and \textsuperscript{109}Cd calibration (blue) across six years of data taking for the nine ANAIS–112 modules \cite{amare2025towards}. The mean efficiency values and standard deviations are shown for each detector in the panels. }
\end{center}
\end{figure}

Once the training is completed, filtering efficiencies are calculated, introducing the third goal of the neutron calibration program: evaluating ANAIS-112 event selection efficiency with a bulk scintillation population primarily originating from NRs. Figure~\ref{efficiencyneutrons} shows the evolution of the total ANAIS-112 event selection efficiency using ML techniques derived from neutron calibration and \textsuperscript{109}Cd calibration in the [1–6] keV range over six years of data collection for the nine ANAIS-112 modules. The efficiency remains consistent across the different neutron calibration runs, with mean values above 95\% for all modules. A slight decrease in efficiency is observed in some detectors (e.g., D2, D3, D7, and D8), which will need to be confirmed with the data from the following years. Furthermore, the efficiencies obtained from $^{252}$Cf and $^{109}$Cd calibrations are in good agreement, confirming the compatibility of the selection efficiencies for a bulk population dominated by NRs and a surface population of ERs.


\subsection{The $^{252}$Cf source}\label{cf}

\(^{252}\)Cf has a half-life of 2.645 years and undergoes two primary decay processes: \(\alpha\)-emission and spontaneous fission (SF), with probabilities of 96.91\% and 3.09\%, respectively. The \(\alpha\)-decay of \(^{252}\)Cf leads to the formation of \(^{248}\)Cm, starting a long decay chain. In contrast, SF results in a range of fission products and the emission of fast neutrons. With an isotropic neutron emission rate of 2.30 × 10\(^6\) neutrons per second per microgram of material \cite{zeynalov2009neutron}, and the ability to be produced in small sizes, \(^{252}\)Cf is easy to control, store, and transport. These characteristics make it the most widely accessible fission neutron source in laboratories.

\begin{figure}[b!]
\begin{center}
\includegraphics[width=0.49\textwidth]{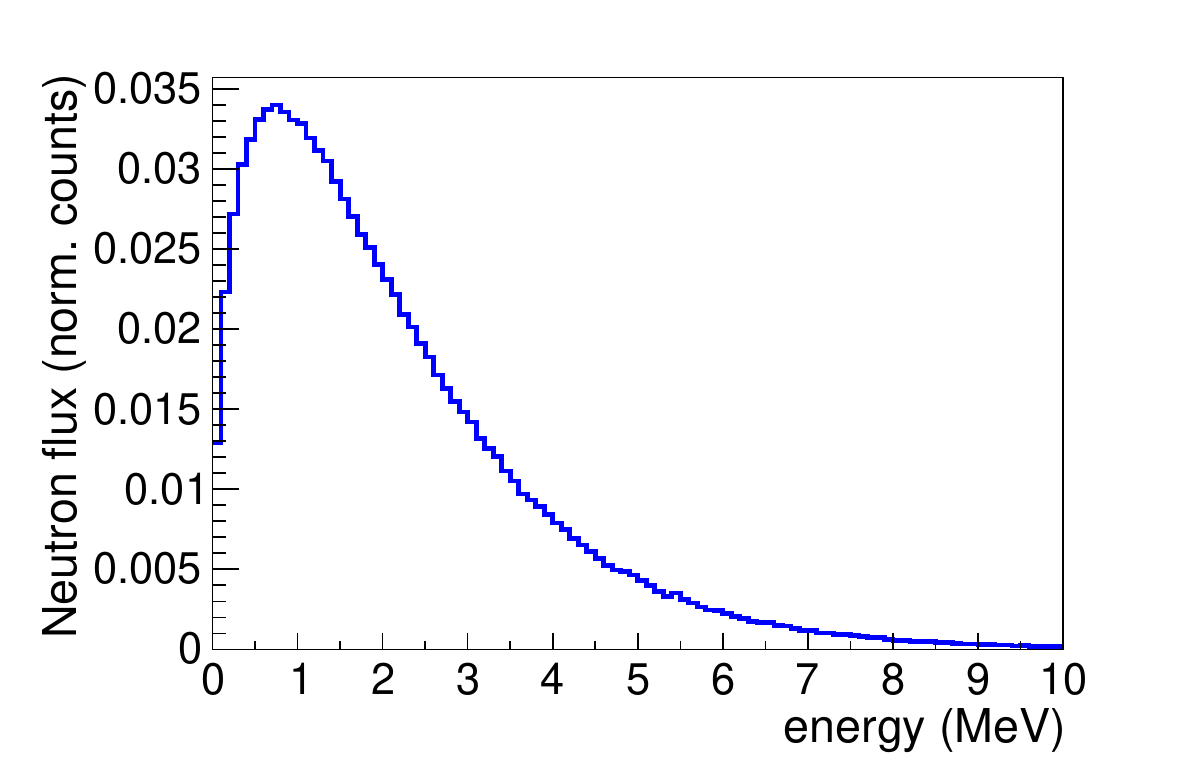}
\hfill
\includegraphics[width=0.49\textwidth]{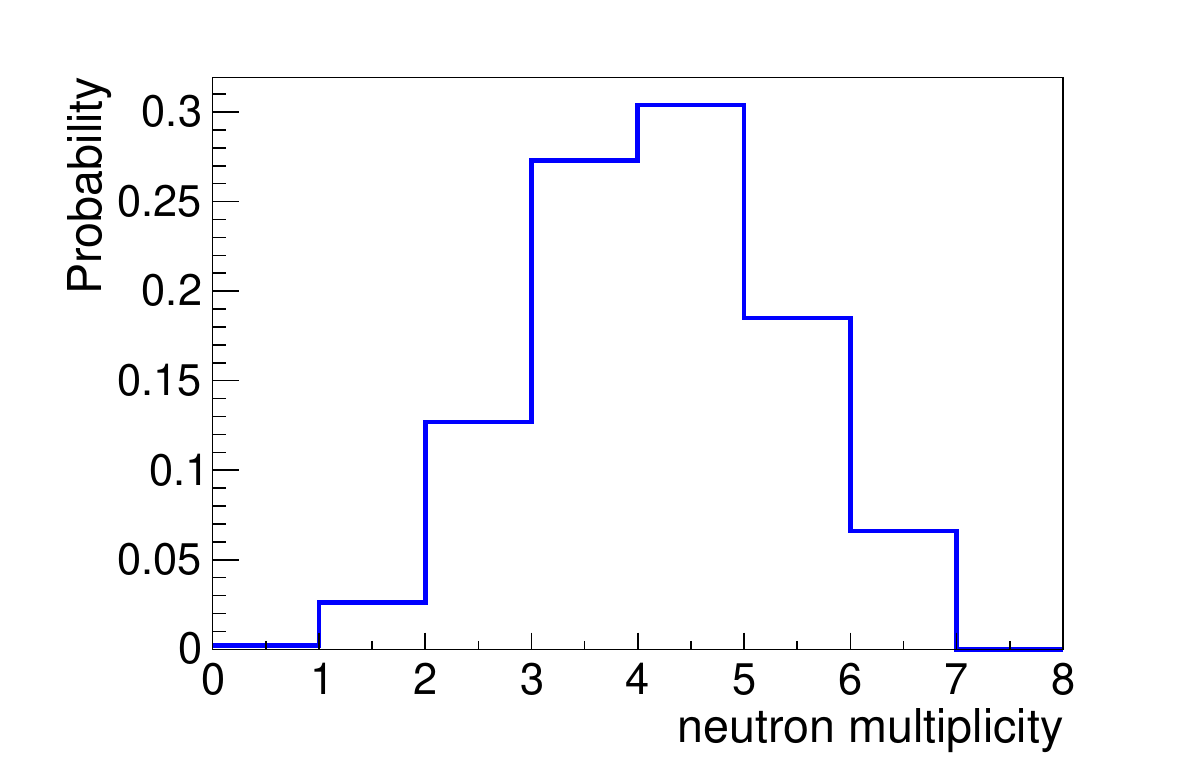}

\caption{\label{energymult} \textbf{Left panel:} Energy spectrum of \(^{252}\)Cf. \textbf{Right panel:} Neutron multiplicity distribution for \(^{252}\)Cf. Both distributions are obtained from Geant4 from the simulation of the neutron calibration conducted in this thesis. }
\end{center}
\end{figure}

Figure \ref{energymult} (left panel) shows the neutron energy distribution from the SF of \(^{252}\)Cf, based on the Mannhart evaluation \cite{mannhart1987evaluation}. As can be seen, it features a typical fission spectrum shape, with a most probable energy of 0.7 MeV and an average energy of 2.1~MeV \cite{langner1998application}. The right panel of Figure \ref{energymult} illustrates the neutron multiplicity distribution for \(^{252}\)Cf, where the number of neutrons emitted per fission event ranges from 0 to 8, depending on the fission kinematics, with an average multiplicity \(\bar{\nu}\) of 3.77 neutrons per fission.

There are two other spectral models commonly used to simulate the SF neutron energy distribution of \(^{252}\)Cf: the Madland-Nix model and the Froehner-Watt spectrum. The Madland-Nix model provides a theoretical description of the spectrum by considering the distribution of excitation energy between fission fragments \cite{marten1986description}, while the Watt distribution offers a simple, efficient macroscopic approximation of the \(^{252}\)Cf neutron spectrum \cite{frohner1990evaluation}, making it suitable for quick calculations. However, the Mannhart-corrected Maxwellian spectrum, an empirical model that adjusts the classical Maxwellian distribution by incorporating corrections for the high-energy to better align with experimental data, provides the highest fidelity to experimental measurements \cite{mannhart1987evaluation}. For this reason, the Mannhart spectrum is the default choice in Geant4 simulations involving \(^{252}\)Cf neutron sources and has been selected for this work.

Californium isotopes are produced in high thermal flux reactors through neutron capture on \(^{249}\)Bk, followed by subsequent $\beta$-minus decay to form \(^{250}\)Cf. Through additional neutron captures, \(^{251}\)Cf, \(^{252}\)Cf, and higher-A isotopes of californium are generated. Therefore, it is desirable to know the isotopic composition of the source, as different isotopes might also contribute to the total neutron emission.

In this thesis, two \(^{252}\)Cf sources have been used. One, referred to as the LSC source, is availabe at the LSC for the experiments and has an activity of 10 kBq as of August~2016. The other, the so-called HENSA source, has an activity of 10~kBq in December 2021 and was shared by the HENSA experiment. The technical data sheet for the HENSA source was available, unlike for the LSC source.

Both sources were produced by Eckert Ziegler Isotope Products and have a similar structural design, with each source in the form of a vial embedded in an aqueous solution. The HENSA source is in the chemical form of Cf(NO\(_3\))\(_3\) in 0.1~M~HNO\(_3\), with a volume of 1 mL, while the LSC source is larger, with a volume of 10 mL. A literature review was conducted to verify that the isotopic composition of the HENSA source aligns with typical commercial standards. Thus, it is reasonable to assume that the isotopic composition of the LSC source is the same as that of the HENSA source, and this assumption has been adopted for the purposes of this study.

\begin{table}[b!]
\centering
\begin{tabular}{cccccc}
\hline Nucleide & Mass (\%) & Activity (\%)  & T$_{1/2}$ (y) & SF BR (\%) & \(\bar{\nu}\)  \\
 \hline
 \hline
 $^{249}$Cf & 24.494 & 0.4881 & 351 & 5.2x10$^{-7}$ & 3.4   \\
 $^{250}$Cf & 30.799 & 16.323 & 13.20 & 0.079 & 3.53 \\
 $^{251}$Cf & 12.844 & 0.0991 & 898 &  9.0x10$^{-4}$ & 3.7\\ 
 $^{252}$Cf & 31.863 & 83.089 & 2.645 & 3.096 & 3.768 \\
 
 \hline
\end{tabular}
\caption{Isotopic composition of the \(^{252}\)Cf  HENSA source, provided by Oak Ridge National Laboratory, along with fundamental nuclear data for the Cf isotopes present. This includes the half-life (T\(_{\textnormal{1/2}}\)), the spontaneous fission branching ratio (SF BR), and the average number of neutrons emitted per fission event (\(\bar{\nu}\)).}
    \label{hensasourceinfo}
\end{table}

Table \ref{hensasourceinfo} presents the isotopic composition detailed in the technical datasheet, based on isotopic mass analysis of the HENSA source, certified by Oak Ridge National Laboratory. Basic nuclear data for the Cf isotopes are also provided. As can be derived from the table, \(^{249}\)Cf and \(^{251}\)Cf have long half-lives and low spontaneous fission branching ratios (SF BR), meaning they do not significantly contribute to the overall neutron emission of the source. This leaves just $^{250}$Cf and $^{252}$Cf as the only Cf neutron emitters that should
be taken into account during the source life. In addition, the alpha decay daughter of $^{252}$Cf, $^{248}$Cm, has a long half-life (3.48x10$^{5}$ y), but a relatively high SF probability (8.4\%). As a result, $^{248}$Cm may also contribute to the overall neutron emission rate and should be considered in the analysis of the source behavior.

\begin{figure}[t!]
\begin{center}
\includegraphics[width=0.6\textwidth]{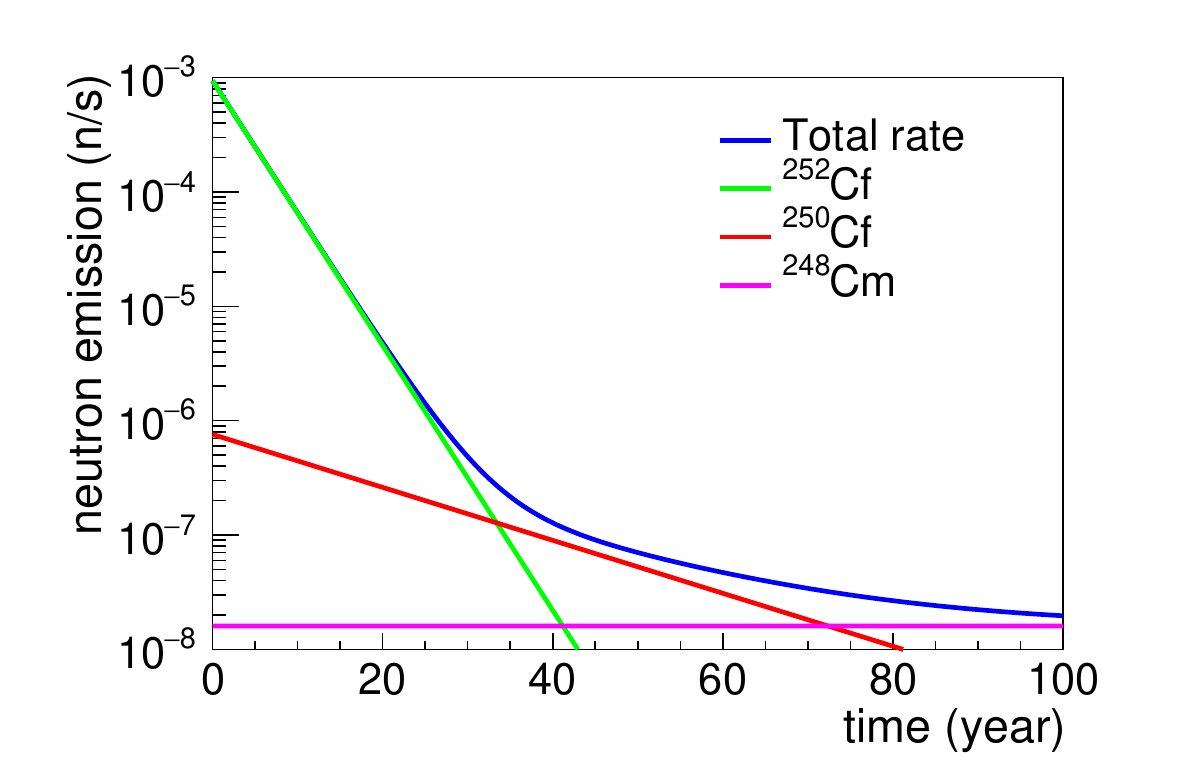}

\caption{\label{Cfsourceemission} Neutron emission evolution of the HENSA source with time. $^{252}$Cf, $^{250}$Cf and $^{248}$Cm contributions are presented. }
\end{center}
\end{figure}

The time evolution of the neutron emission rate of the HENSA source  with the initial composition shown in Table \ref{hensasourceinfo} considering $^{252}$Cf, $^{250}$Cf and $^{248}$Cm contributions is presented in Figure~\ref{Cfsourceemission}.

For the first few decades after the source is created, neutron emissions are primarily dominated by \(^{252}\)Cf. Approximately 35 years after the isotopic composition measurement, \(^{250}\)Cf will become the dominant neutron emitter. Around 65 years after the composition measurement, the neutron yield will be largely attributed to \(^{248}\)Cm. 

Therefore, it can be concluded that modelling the source as only $^{252}$Cf is a reasonable approximation for a newly fabricated source, particularly within the typical recommended working lifetime of 15~years for Cf sources. However, for older $^{252}$Cf  sources (>15~years), the neutron yield and multiplicity spectrum of the source will evolve differently with age. Regarding this work, the sources are well within the time range where 
other isotopic forms need not be considered, as they where fabricated in 2016 and 2021. Nevertheless, in this study the contribution of \(^{250}\)Cf will be accounted for to ensure completeness. Given the isotopic composition of the source shown in Table \ref{hensasourceinfo}, this corresponds to 10~kBq of $^{252}$Cf and 1.60 kBq of $^{250}$Cf as of August 2016 and December 2021 for the LSC and HENSA sources, respectively. It is worth noting that for each neutron calibration, the time evolution of the corresponding activities is properly taken into account in the simulation.


\vspace{-0.5cm}

\subsection{Measurement schedule and procedure}\label{schedule}

Throughout this thesis, nine neutron calibration runs were conducted over a period of four years, using two different \(^{252}\text{Cf}\) neutron sources placed at various positions in the ANAIS-112 set-up: the west, south, and top faces of ANAIS-112. Table \ref{infoneutroncalibration} summarizes key information about the onsite ANAIS-112 neutron calibrations, including the number of the calibration run, the face of the experiment where the source was placed, the \(^{252}\text{Cf}\) source used, the date of each calibration and its live time.

The neutron calibration runs performed with the LSC source typically lasted about 2–4~hours, as shown in Table \ref{infoneutroncalibration}. Notably, the calibrations using the HENSA source were shorter despite the same nominal activity (10 kBq), as the source was significantly younger (December 2021 vs. August 2016). Over the course of the neutron calibration campaign, the activity of the $^{252}\text{Cf}$ source decreased due to its 2.645-year half-life, necessitating longer runs to obtain comparable statistics, as evidenced by the extended live time of run 9014. In fact, the last neutron calibration run conducted within this thesis (run 9016) was left running for two full days, including overnight, to maximize the accumulated statistics.

\begin{table}[t!]
    \centering
    \begin{tabular}{ccccc}
    \hline
        Run  & Face & Source & Date & Live time (h) \\
        \hline \hline
        9000  & west & LSC & 27/04/2021 & 3.35 \\
9002  & south & LSC & 23/12/2021 & 2.57 \\
9004  & top & LSC & 13/04/2022 & 2.56 \\
9006  & west & LSC & 07/06/2022 & 2.80 \\
9008  & west & LSC & 27/04/2023 &  2.77 \\
9010  & west & HENSA & 27/04/2023 & 1.23 \\
9012  & south & HENSA & 27/04/2023 & 0.40 \\
9014   &  west & LSC & 02/04/2024 & 3.75 \\

9016  &  west & LSC & 11-13/02/2025 & 42.10   \\

         \hline
    \end{tabular}
    \caption{ Key information regarding the onsite ANAIS-112 neutron calibrations, including the calibration run number, the face of the experiment where the source was placed, the source used, the date of the calibration, and its live time. }
    \label{infoneutroncalibration}
\end{table}

\begin{figure}[t!]
\begin{center}
 \centering
    \begin{subfigure}{0.35\textwidth}
        \centering
        \includegraphics[width=\textwidth]{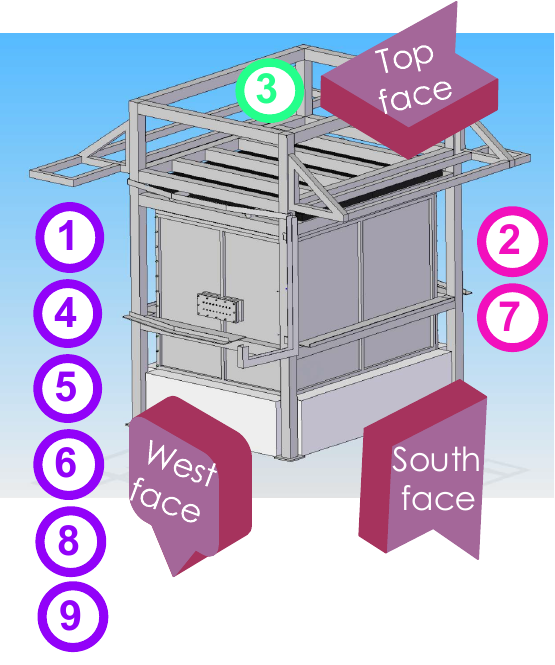}
        \caption{}
    \end{subfigure}
    \hspace{1cm}
    \begin{subfigure}{0.3\textwidth}
        \centering
        \includegraphics[width=\textwidth]{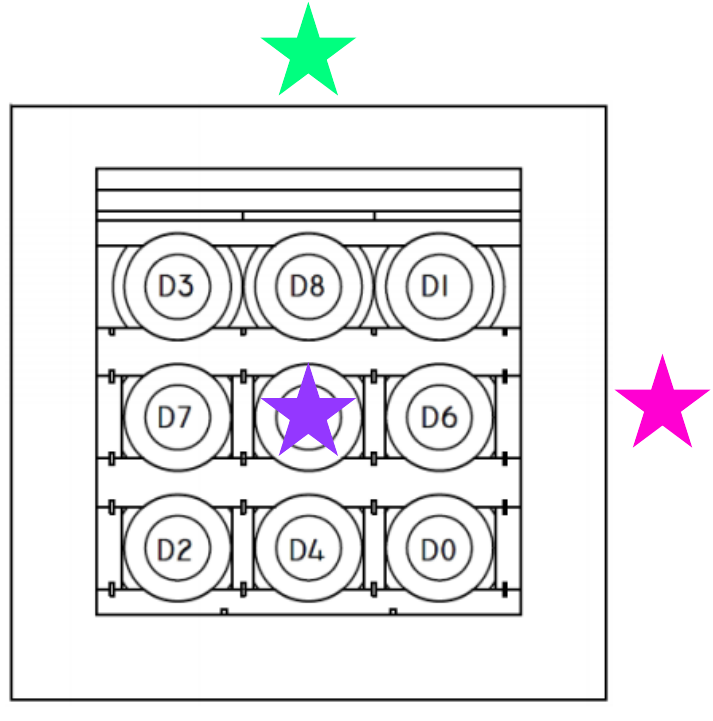}
        \caption{}
    \end{subfigure}
    \begin{subfigure}{0.6\textwidth}
        \centering
        \includegraphics[width=\textwidth]{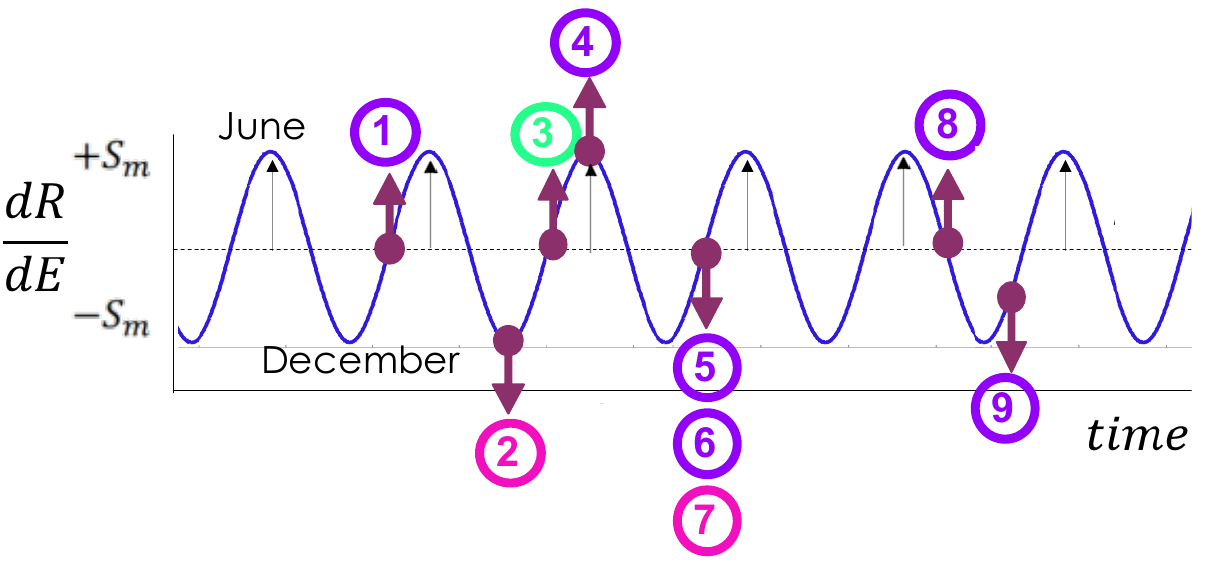}
        \caption{}
    \end{subfigure}
\caption{\label{distribution}\textbf{(a):} Position of the \(^{252}\text{Cf}\) neutron source during
the different calibration runs carried out: west side (1,4,5,6,8 and 9), south side (2 and 7) and top side (3). The source is placed outside
the anti-radon box and lead shielding, but inside the muon vetoes and neutron shielding. \textbf{(b):} West face view of detector positions in the 3×3 modules matrix, showing the positions of the \(^{252}\text{Cf}\) neutron source for calibrations conducted on different faces of the experiment. \textbf{(c):} Time distribution
of the neutron calibration runs carried out in the ANAIS-112 experiment. }
\vspace{-0.3cm}
\end{center}
\end{figure}

Panel (a) of Figure \ref{distribution} shows the position of the \(^{252}\text{Cf}\) neutron source during the calibration runs, highlighting the three faces of the ANAIS-112 experiment. Panel~(b) provides a west-facing view of the detector positions in the 3×3 module matrix, indicating the positions of the \(^{252}\text{Cf}\) neutron source for calibrations conducted on different faces of the experiment. Of the nine runs, six were carried out on the west face, two on the south face, and one on the top face. 

It is important to note that the \(^{252}\text{Cf}\) source was consistently placed outside the anti-radon box and the 30 cm lead shielding, but inside the muon vetoes and neutron shielding. This positioning is significant because, in addition to the fission neutrons described in Section \ref{cf}, gamma rays are also emitted during the SF of \(^{252}\text{Cf}\). The prompt gamma-ray multiplicity ranges from 0 to 20 per fission, with an average of 8.32 gamma rays per fission \cite{zeynalov2009neutron}. However, given the source placement and the shielding, neither the gamma emissions nor the alpha decay chain contribute to the detected events in ANAIS-112, as none of their emissions actually reach the detectors.

Moreover, it is worth highlighting that the background spectra before and after a neutron calibration run have been systematically compared. No significant differences in event rates or any pulse shape parameter were observed, suggesting that activation induced by the neutrons from the $^{252}$Cf source is negligible. This is consistent with expectations, given that the source has low activity and is located outside the lead shielding, though still within the neutron moderators. To confirm the lack of activation induced by $^{252}$Cf, an ad-hoc simulation was performed by setting all the volumes present in the ANAIS Geant4 set-up as sensitive volumes, meaning information was stored for all of them. This allowed checking for the generation of isotopes, and none were found that would pose any concern for the performance of ANAIS-112.

Panel (c) of Figure \ref{distribution} shows the temporal distribution of the neutron calibration runs plotted against the expected modulation corresponding to a DM signal, dR/dt~(t), integrated over a specific energy window. The modulation signal is expected to exhibit maxima in June and minima in December. The timing of the neutron calibrations was aligned with the experimental schedule, typically coinciding with periodic $^{109}$Cd calibrations in order to minimize the downtime of the experiment, a period during which ANAIS-112 does not collect background data due to regular calibration activities. However, these calibration runs were scheduled at different times of the year to assess the stability of the ANAIS detector efficiency over time. As derived from the figure, one calibration was conducted during the signal minimum, one during the maximum, and the remaining runs at intermediate points in the modulation cycle.

\begin{figure}[t!]
    \centering

    \begin{subfigure}{0.3\textwidth}
        \centering
        \includegraphics[width=\linewidth, angle=-90]{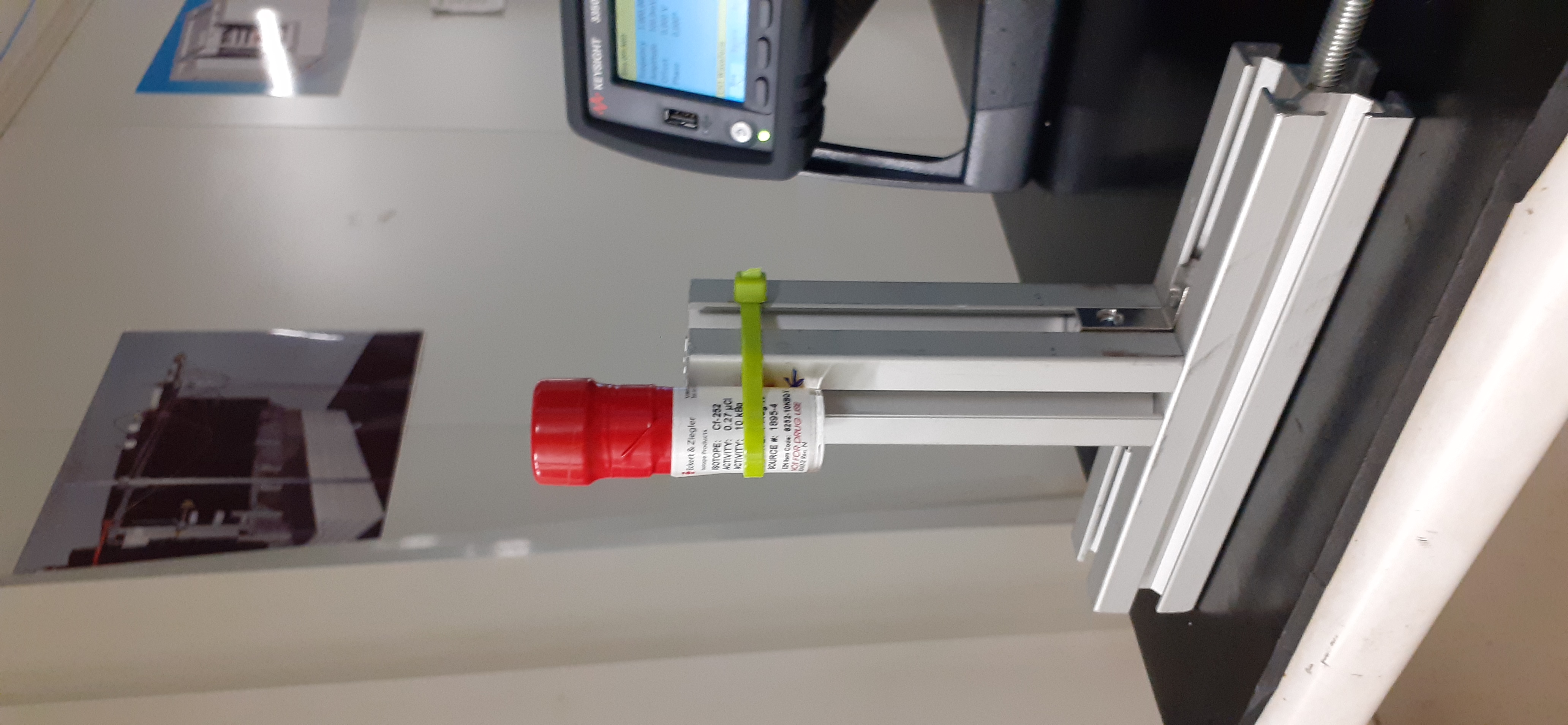}
        \caption{}
    \end{subfigure}
    \hfill
    \begin{subfigure}{0.3\textwidth}
        \centering
        \includegraphics[width=\linewidth, angle=-90]{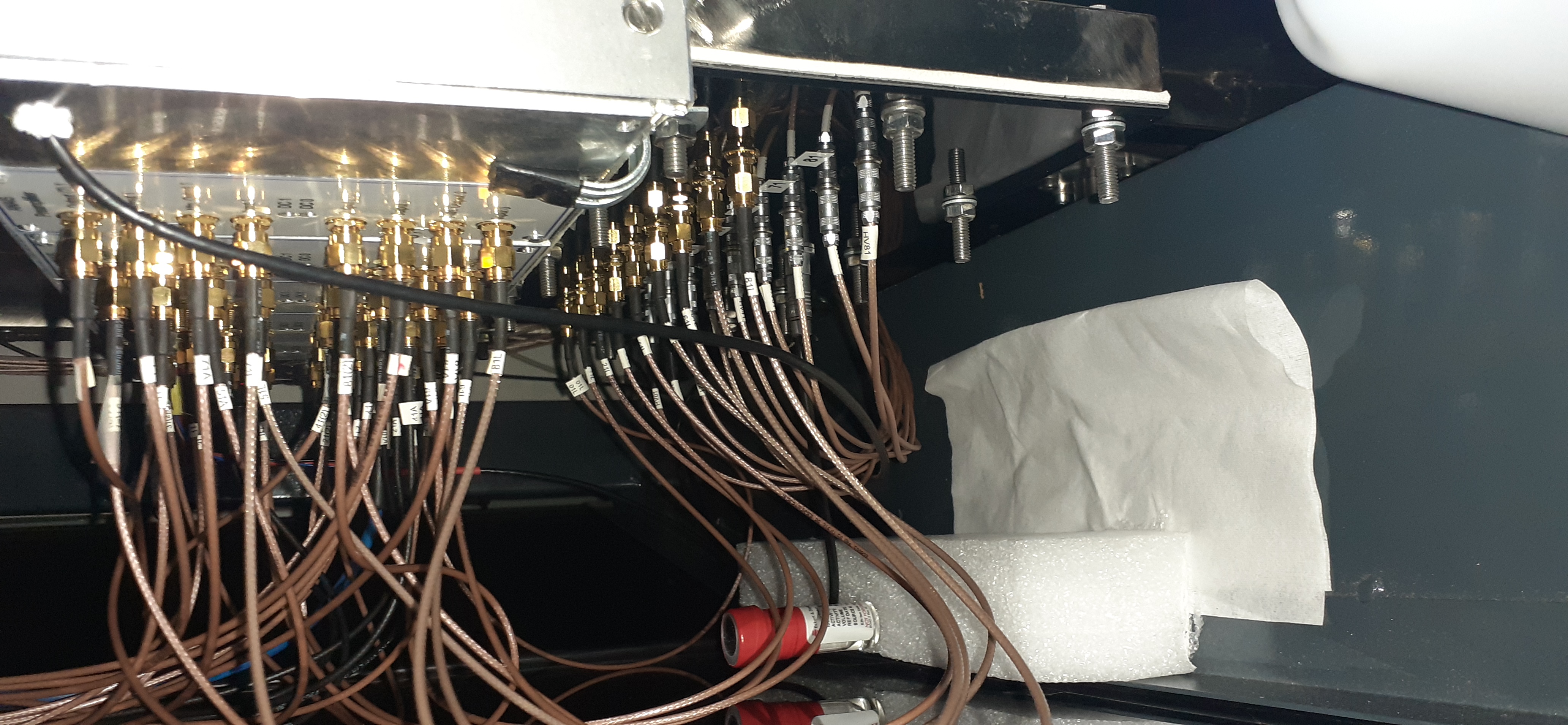}
        \caption{}
    \end{subfigure}
    \hfill
    \begin{subfigure}{0.3\textwidth}
        \centering
        \includegraphics[width=\linewidth, angle=-90]{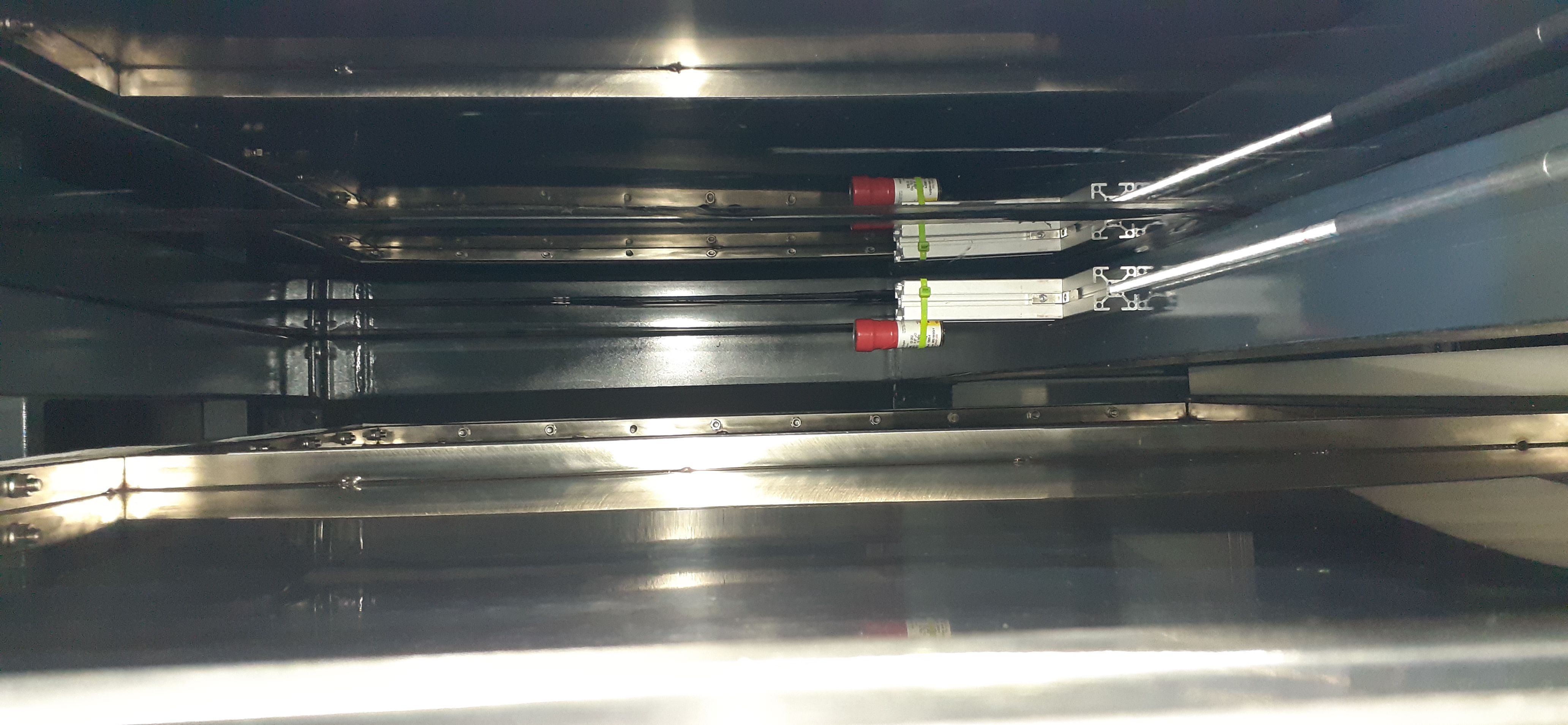}
        \caption{}
    \end{subfigure}

    \vspace{1em} 

    \begin{subfigure}{0.6\textwidth}
        \centering
        \includegraphics[width=\linewidth]{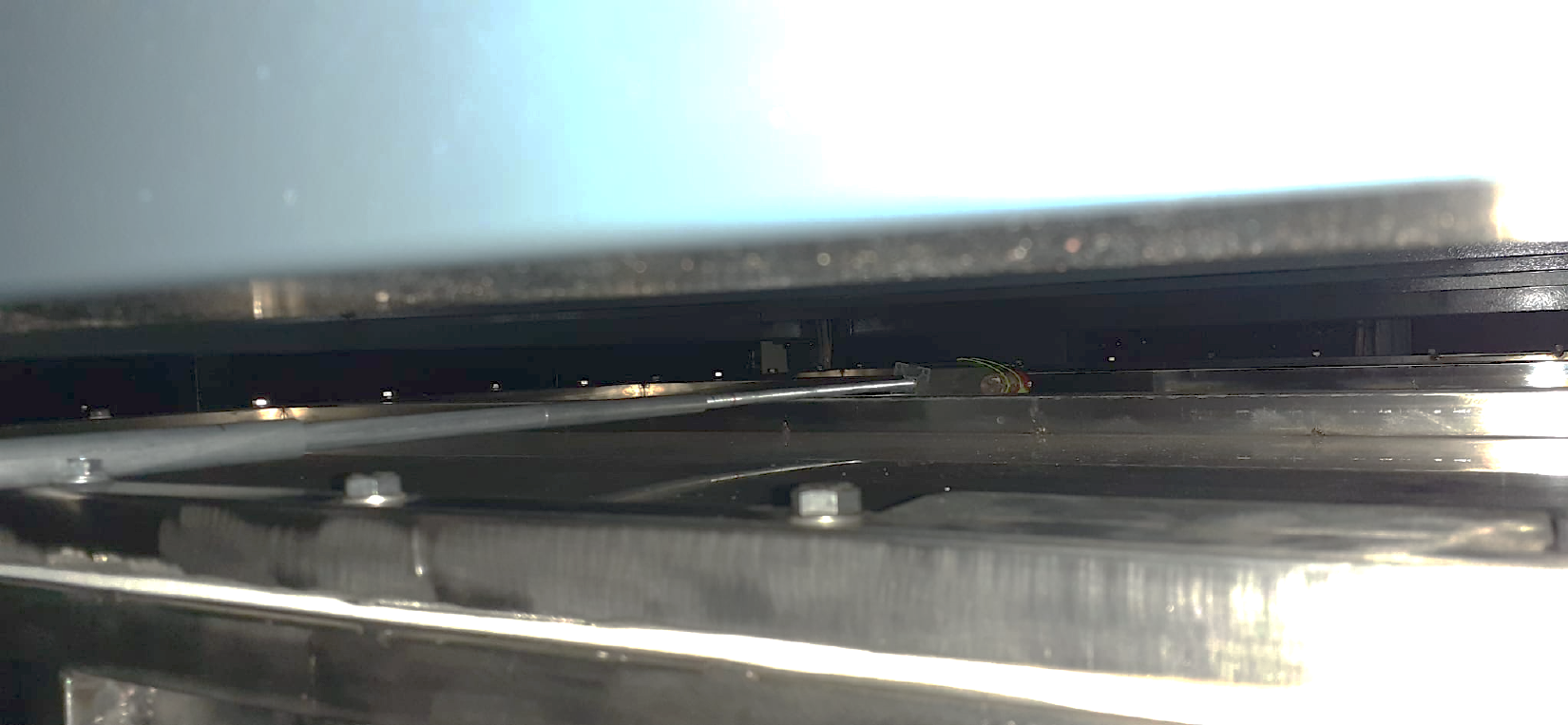}
        \caption{}
    \end{subfigure}

    \caption{ \textbf{(a):} $^{252}$Cf source attached to the end of an aluminum profile using a zip tie. The remaining panels show the position of the $^{252}$Cf source during different neutron calibration runs performed in ANAIS-112. \textbf{(b):} West calibration \textbf{(c):} South calibration. \textbf{(d):} Top calibration.}
    \label{position2}
    \vspace{0.2cm}
\end{figure}

The calibration procedure requires placing the \(^{252}\text{Cf}\) neutron source within the experimental set-up on the three different faces of the ANAIS-112 detector. For the west-face calibration, the source was positioned at the center of the anti-radon box along the x-axis, adjacent to the preamplifiers and signal extraction point, and supported by a polyethylene piece. A picture showing the source position during one of the west-face calibration runs is provided in Figure \ref{position2}, b.

In contrast, positioning the source for the south and top-face calibrations proved to be more challenging due to the lack of direct access to these faces, unlike the west face. To address this, the source was attached to the end of an aluminum profile using a zip tie, allowing it to be extended to the required position (see Figure \ref{position2},~a). Although this approach enabled calibration for the south (Figure \ref{position2}, c) and top (Figure~\ref{position2},~d) faces, the resulting placement was less reproducible compared to the west-face calibration, complicating efforts to replicate these positions in future campaigns.

The challenges in reproducing the source placement on the south and top faces, combined with the much easier access to the west face of ANAIS-112, justified conducting a greater number of runs on the west face from the experimental point of view.

\subsection{Neutron calibration data}\label{features}

Once the structure of the $^{252}$Cf source, the measurement schedule, and procedure have been explained, this section delves into the measured data obtained from the onsite neutron calibrations of ANAIS-112, with an initial focus on describing the population in terms of pulse shape behavior (Section \ref{psa}). Based on this, the cuts applied to this population to isolate bulk scintillation events will be detailed (Section \ref{criteria}), followed by the presentation of pulses from this and other populations to highlight differences in the scintillation time for NR and ER events (Section \ref{pulses}). Additionally, the energy calibration process, specifically developed for this neutron population, will be discussed in Section~\ref{energycal}, addressing the challenges in accurately calibrating energy for neutron events and the impact on the QF analysis.

\subsubsection{Neutron pulse shape behaviour}\label{psa}

Neutron calibration runs provided a well-defined population of NR events uniformly distributed throughout the crystal volume. To verify that neutrons predominantly result in a clean population of events corresponding to elastic scattering with sodium and iodine nuclei, the distribution of PSA parameters, as detailed in Section \ref{Filtering}, is analyzed for neutron calibration events. Comparing their behavior with other well-characterized populations, such as events from \(^{109}\text{Cd}\) calibration and background data, provides additional insight.

\begin{figure}[t!]
\begin{center}
\includegraphics[width=1.\textwidth]{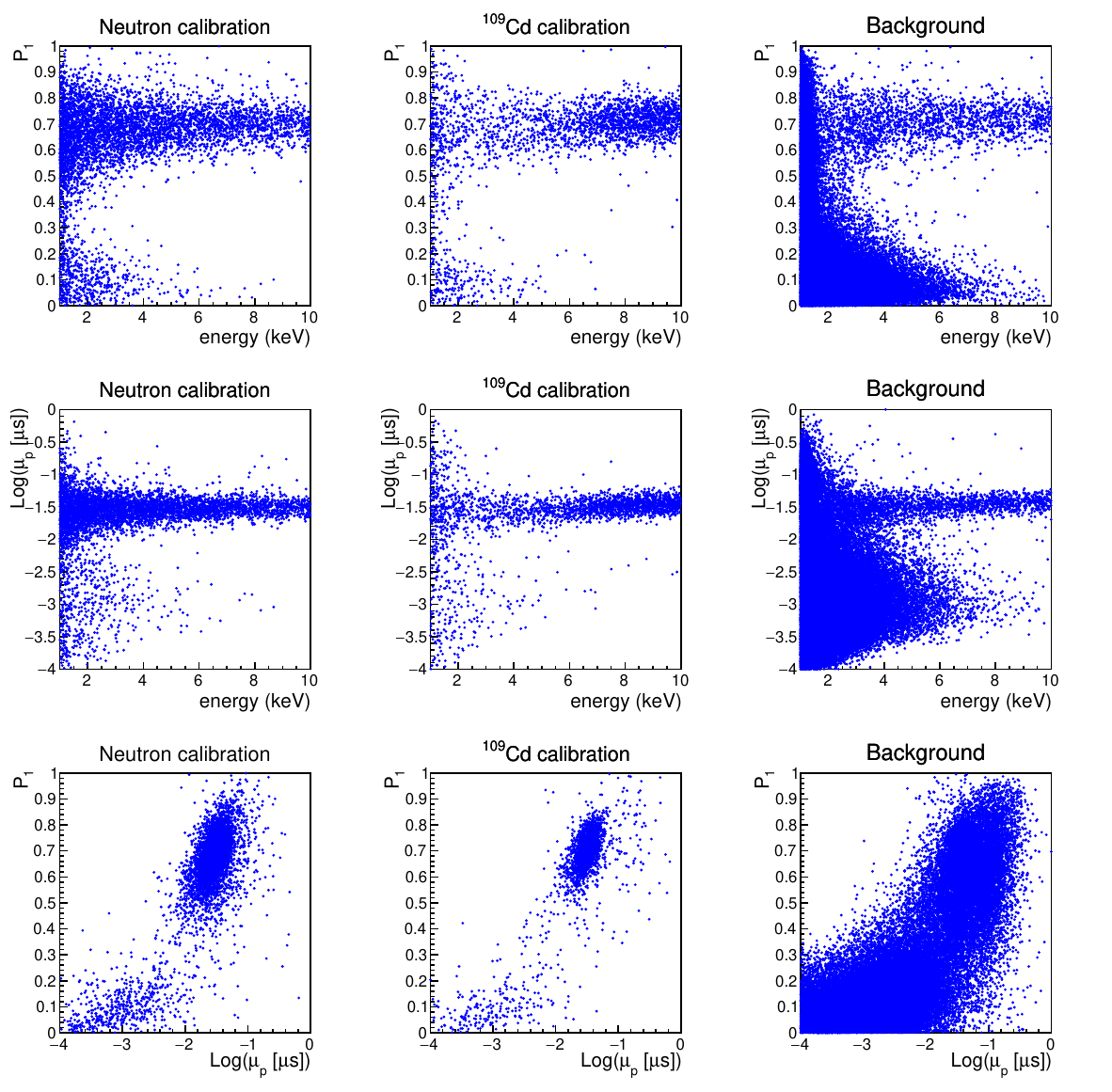}

\caption{\label{P1muPSVar} Distribution of two representative PSA parameters and their combination. \textbf{Upper row:} P1 as a function of energy; \textbf{Middle row:} Mean time, Log($\mu_p[\mu s]$), as a function of energy; and \textbf{Lower row:}  (P1-Log($\mu_p[\mu s]$)) plane in the [1-10] keV energy range. The first column shows the event distribution from the first neutron calibration run conducted in the ANAIS experiment (total-hits), the second column displays events from the associated \(^{109}\text{Cd}\) calibration run (single-hits), and the third column represents events from the corresponding background data (single-hits).  \vspace{-0.3cm}}
\end{center}
\end{figure}

Figure \ref{P1muPSVar} presents the distribution of three PSA parameters for neutron calibration, \(^{109}\text{Cd}\) calibration, and background data. The parameters displayed include the P1 parameter as a function of energy (upper panels), the mean time Log(\(\mu_p\) [\(\mu\)s]) as a function of energy (middle panels), and the P1-Log(\(\mu_p\) [\(\mu\)s]) plane (lower panels), for [1-10] keV events. The neutron data corresponds to the first neutron calibration run carried out in the ANAIS experiment (total-hits), whereas the remaining panels show data from the previous $^{109}\text{Cd}$ calibration and background runs (single-hits).

As derived from \(^{109}\text{Cd}\) events, dominated by NaI(Tl) scintillation, the P1 parameter exhibits values centered around 0.65 for bulk scintillation events. Both the \(^{109}\text{Cd}\) and neutron calibration populations are slightly polluted by fast events with P1 < 0.4, particularly at low energy, due to the intrinsic background present also during calibration. These fast events, predominantly produced by Cherenkov emission in the PMT glass, dominate in background data, forming a dense population with lower P1 values.

\begin{figure}[b!]
\begin{center}
\includegraphics[width=1.\textwidth]{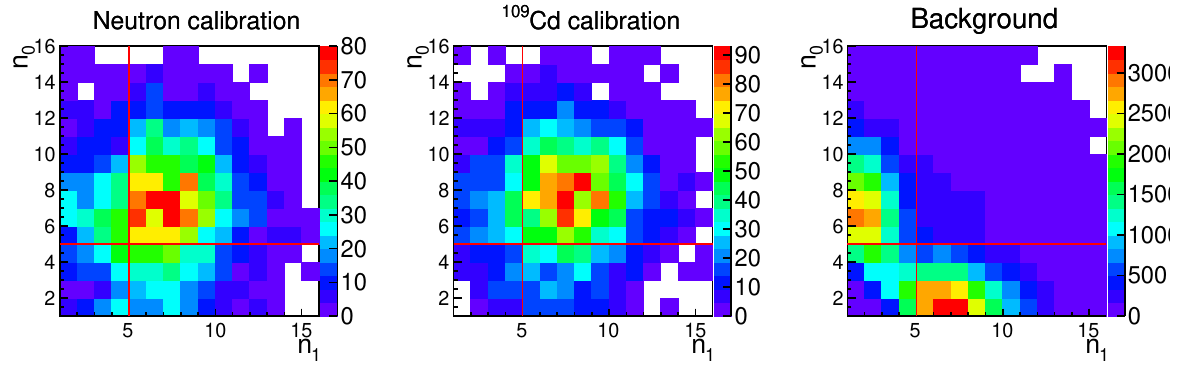}

\caption{\label{n0n1todos} Distribution in the (\(\textnormal{n}_0, \textnormal{n}_1\)) plane, representing the number of peaks detected by the algorithm for each of the two PMT signals from the same module, for: \textbf{Left panel:} neutron calibration events (total-hits); \textbf{Middle panel:} \(^{109}\text{Cd}\) calibration events (single-hits); \textbf{Right panel:} background events (single-hits). In all three populations, events were selected in the [1-2]~keV energy range after applying previous cuts in the pulse shape parameters.}
\end{center}
\end{figure}

The middle panels of Figure \ref{P1muPSVar} illustrate the distribution of Log(\(\mu_p\)) as a function of energy, a parameter used to identify and exclude spurious low-energy events caused by long phosphorescence in the crystal or pulse tails. Bulk scintillation events, again derived from \(^{109}\text{Cd}\) calibration, are located around Log(\(\mu_p\)) $\approx$ - 1.65. A significant difference in event density is evident when comparing neutron and \(^{109}\text{Cd}\) calibration events to background data. In the background data, a noticeable overlap between signal and noise events at low energies is observed, with a substantial fraction of events exhibiting mean times inconsistent with bulk scintillation.

Since P1 and Log(\(\mu_p\) [\(\mu\)s]) are correlated, it is useful to represent the distribution of events from the three populations on the P1-Log(\(\mu_p\)) plane for the [1-10] keV energy range (lower panels). Bulk scintillation events cluster around (P1 = 0.65, Log(\(\mu_p\))~=~-1.65), while most background events at these energies correspond to very fast events with P1 < 0.2.

As previously discussed in Chapter \ref{Chapter:ANAIS}, a different population of background events is observed below 2 keV, compatible with bulk scintillation. However, the main feature of these events is a pronounced asymmetry in the energy partition between the two PMTs of each detector. This asymmetry is easily identifiable through the difference in the number of peaks detected in the signals of the two PMTs (n$_\textnormal{0}$ and n$_\textnormal{1}$) using the peak-finding algorithm. 

Figure \ref{n0n1todos} (right panel) illustrates the distribution of n$_\textnormal{0}$ and n$_\textnormal{1}$ for background events in the [1-2] keV energy range. The majority of these events exhibit a highly asymmetric pattern, with one PMT registering only a few peaks (one or two), while the other records more than five. Notably, such events, which are only observed below 2~keV in DM runs, are not present in the \(^{109}\text{Cd}\) calibration runs (middle panel), nor in neutron calibration data (left panel). This strongly suggests that they are likely spurious light events originating at or near the PMTs, but not corresponding to bulk NaI scintillation.

\begin{figure}[t!]
\begin{center}
\includegraphics[width=.8\textwidth]{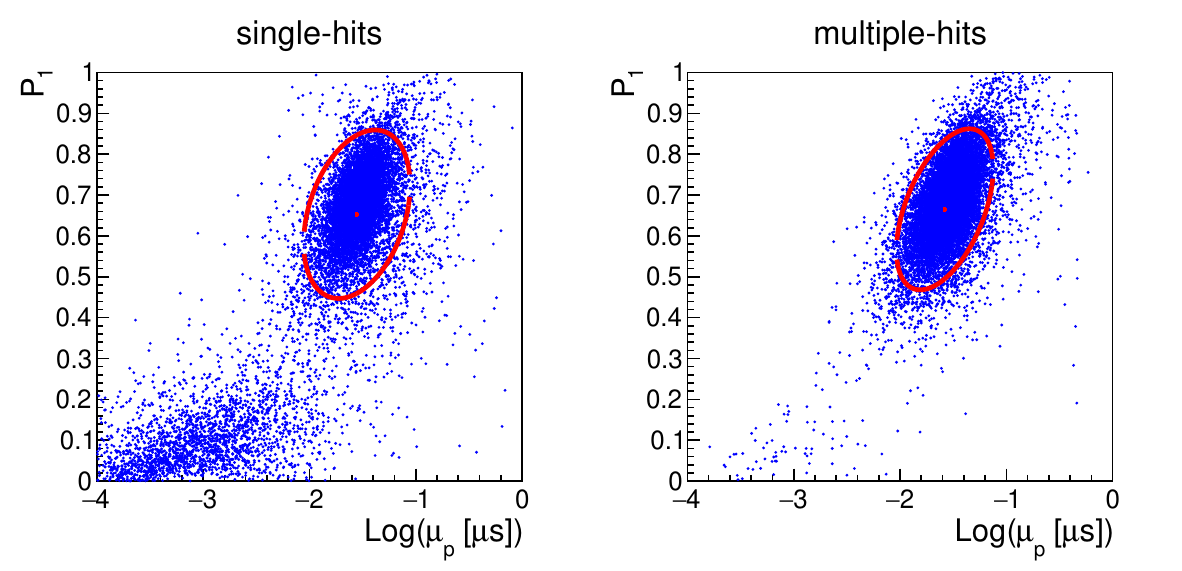}
\includegraphics[width=.8\textwidth]{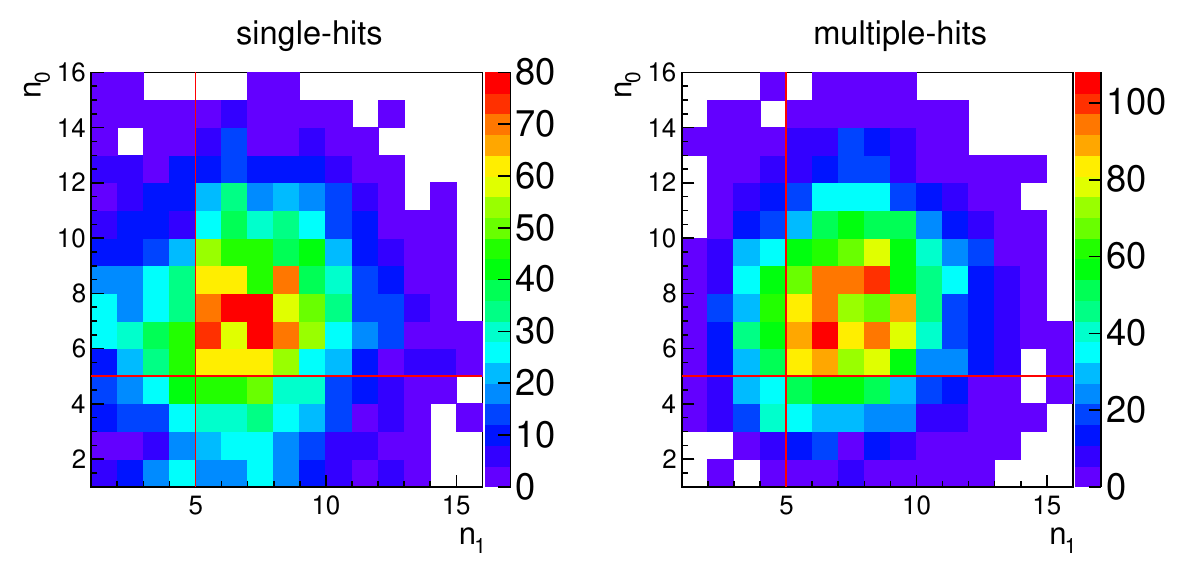}

\caption{\label{comparisonsinglemulti}\textbf{Top panel:}  (P1-Log($\mu_p[\mu s]$)) plane in the [1-10] keV energy range. The red ellipsoid in the figures represents the cut defined by PSV$_{\textnormal{cut}}$ = 3 for each population. The center of the ellipsoid is also shown. \textbf{Bottom panel:} Distribution of the (\(\textnormal{n}_0, \textnormal{n}_1\)) plane, representing the number of peaks detected by the algorithm for each of the two PMT signals from the same module, after applying previous cuts in the pulse shape parameters. Both panels shows the corresponding distributions for single-hits and multiple-hits neutron populations. }
\end{center}
\end{figure}

Figure \ref{comparisonsinglemulti} shows the (P1-Log($\mu_p[\mu s]$)) plane and the (\(\textnormal{n}_0, \textnormal{n}_1\)) peak distributions for single and multiple-hits populations from the neutron calibrations. The red ellipsoid in the figure represents the PSV cut. The center of the ellipsoid is also displayed. In this case, the ellipses are calculated for neutron events as performed for Na/K coincidences in \cite{Amare:2018sxx}, ensuring that the same percentage of events lies
outside the ellipsoid after applying the condition P1~>~0.4. Under this assumption, the PSV cut guarantees that 77.8\% of good events in the [1–2] keV energy range survive.

The (P1-Log($\mu_p[\mu s]$)) plane shows a clear reduction in the population of fast events when selecting multiple-hit events, as these contain a smaller fraction of anomalous scintillation components due to their lower background contribution. As a result, multiple-hit events offer a cleaner dataset, with fewer events falling outside the ellipsoid, which is particularly beneficial for applications such as training ML-based
filtering protocols in ANAIS-112, where a well-defined dataset for training is crucial. In addition, the difference between the ellipsoids of single-hit and multiple-hit events is minimal, indicating that the temporal behavior of both neutron populations is highly consistent. The temporal behavior of the various populations present in ANAIS-112 will be assessed in Section \ref{pulse} through the analysis of the average pulses. Regarding the number of peaks distribution, it can be observed that both single-hit and multiple-hit populations exhibit a rather symmetric pattern.

\subsubsection{Selection criteria}\label{criteria}

The following selection criteria are applied for identifying bulk scintillation events to be later used in the analysis of the QF in ANAIS-112 crystals.

\begin{itemize}
    \item \textbf{P1 > 0.4}

As shown in Figure \ref{comparisonsinglemulti}, applying the P1 > 0.4 cut effectively removes most fast pulses originating from the PMTs in neutron calibration data, likely due to Cherenkov emissions within the PMT glass. A comparison between the P1~>~0.4 cut and the bi-parametric pulse shape cut based on PSV cut reveals no significant differences in the resulting energy spectra for neutron populations. 

However, the P1 > 0.4 cut is assumed to have 100\% efficiency, while the PSV cut requires efficiency corrections. The efficiency of the PSV cut has been reliably characterized in ANAIS-112 using ER populations ($^{22}$Na and $^{40}$K coincidences), but not NRs. A Monte Carlo simulation of the NR population has been performed assuming a scintillation constant $\tau_{\text{NR}} \sim 0.9 \times \tau_{\text{ER}} \approx$ 205~ns~\cite{Amare:2018sxx}, and computing the efficiency of the PSV cut when applied under the assumption of ER behavior, indicating that the maximum efficiency discrepancy is 0.03 in the 1–2 keV energy range.

Given these considerations, the present analysis only employs the P1 > 0.4 criterion to reject fast PMT-related events, thereby avoiding the need for efficiency corrections and the associated assumptions.\\

    \item \textbf{Trigger requirements}

    As detailed in Section \ref{DAQsec}, the triggering of each detector module is done by the coincidence of signals
from the two PMTs at the photoelectron level within a 200~ns window. Later, in the analysis procedure, the software trigger position is searched for in each pulse by a threshold over baseline condition. 

In the case of coincident events, there is one main trigger and subsequent triggers in other modules within a coincidence window of 1 $\mu$s opened by the first module  triggering. Because the coincidence window in ANAIS has the same size than the integration window (1 $\mu$s), some of the coincident pulses can be strongly displaced within the acquisition window. 

        \begin{figure}[b!]
\begin{center}
\includegraphics[width=.49\textwidth]{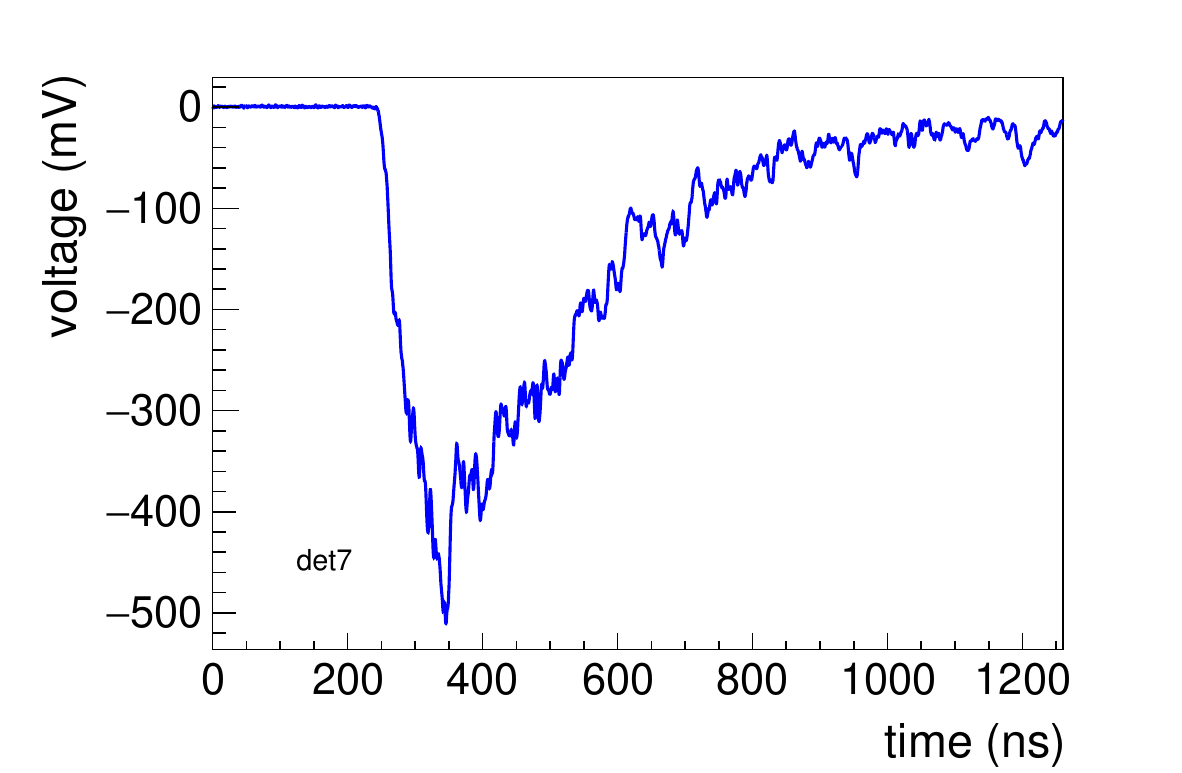}
\includegraphics[width=.49\textwidth]{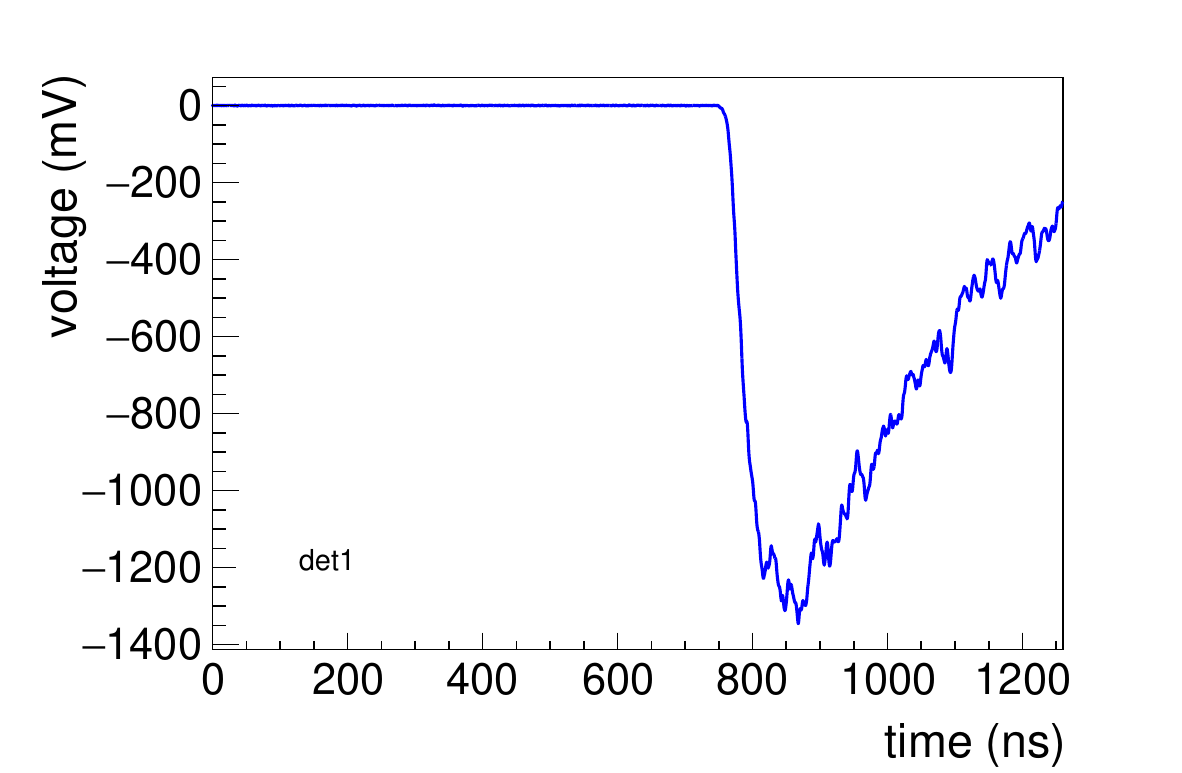}

\caption{\label{pulse} Example of two digitized pulses (sum of the two PMT signals each) in ANAIS from a neutron calibration run, corresponding to a multiple-hit event. \textbf{Left panel:} event containing the main acquisition trigger detected in detector 7. \textbf{Rigth panel:} corresponding coincident event with trigger position beyond 400 ns from the main trigger detected in detector 1. }
\end{center}
\end{figure}

       Figure \ref{pulse} illustrates an example of two pulses (sum of the two PMT signals each) digitized in ANAIS during a neutron calibration run, corresponding to a multiple-hit event. The event containing the main acquisition trigger, which opens the integration window, is shown in the left panel, while the right panel shows the coincident event with a trigger position occurring more than 400 ns after the main trigger. As evident from the figure, delayed triggers can cause incomplete pulse acquisition, resulting in an inaccurate energy and pulse shape parameter estimations. 

       Because of this, only events with a software trigger position within 400~ns of the main acquisition trigger are selected, as reported in \cite{coarasa2022improving}. This cut is not relevant in the case of the ANAIS-112 background runs, but it is in neutron calibrations because the propagation times of the neutrons within the set-up allow for multiple events within the coincidence window.
       
       To ensure reliable QF determination for the ANAIS crystals, such events are excluded from the analysis. In particular, this cut removes approximately 17\% of events in the 1–100~keV range for multiple-hit events. When the same selection cut is applied to the simulation, 19\% of the events are rejected, demonstrating that the DAQ system has been very well reproduced in the simulation. It is worth noting that this limitation does not apply to the ANOD DAQ, as it features an 8~$\mu$s acquisition window. In fact, when applying the same selection to the neutron calibration data acquired with ANOD, the fraction of excluded events under the same conditions is only 0.06\%.

\end{itemize}

It is important to note that, unlike in background data, events occurring within 1 s after a muon veto trigger are not rejected in neutron calibration data. This decision is motivated by the fact that applying such a cut would result in a substantial loss of live time, approximately 44\%, during neutron calibration runs. In contrast, the same cut during background data acquisition only results in a live time loss of about 3\%. The significantly higher rejection rate observed during neutron calibration runs is attributed to interactions between neutrons and the muon veto system. This interpretation aligns well with the experimental set-up: the $^{252}$Cf neutron source, which emits neutrons isotropically, is located between the muon veto panels and the lead shielding. Applying this cut would lead to the rejection of genuine neutron events due to multiple scattering coincidences registered in the veto system. Furthermore, muon-induced events are not a concern in this context, as the rate of neutron calibration events is approximately 25~Hz, while the identified muon rate during these runs is below 50 mHz.

%




In standard neutron simulations, the veto system is typically not implemented as a sensitive volume in order to reduce computational load and storage requirements. However, dedicated simulations were performed configured to register energy deposits in the veto detectors. The simulated fraction of multiplicity-2 events depositing energy in both the crystals and in one veto reaches 78\%, confirming the significant interaction of neutrons emitted from the $^{252}$Cf source or other secondaries produced by such neutrons with the veto system.

Based on the (\(\textnormal{n}_0, \textnormal{n}_1\)) distributions presented above, neutron events exhibit a highly symmetric signal distribution between the two PMTs. Consequently, no additional cuts based on this parameter are applied to the data in this study.

\begin{figure}[t!]
\begin{center}
 \centering
    \begin{minipage}{0.6\textwidth}
        \centering
        \includegraphics[width=\textwidth]{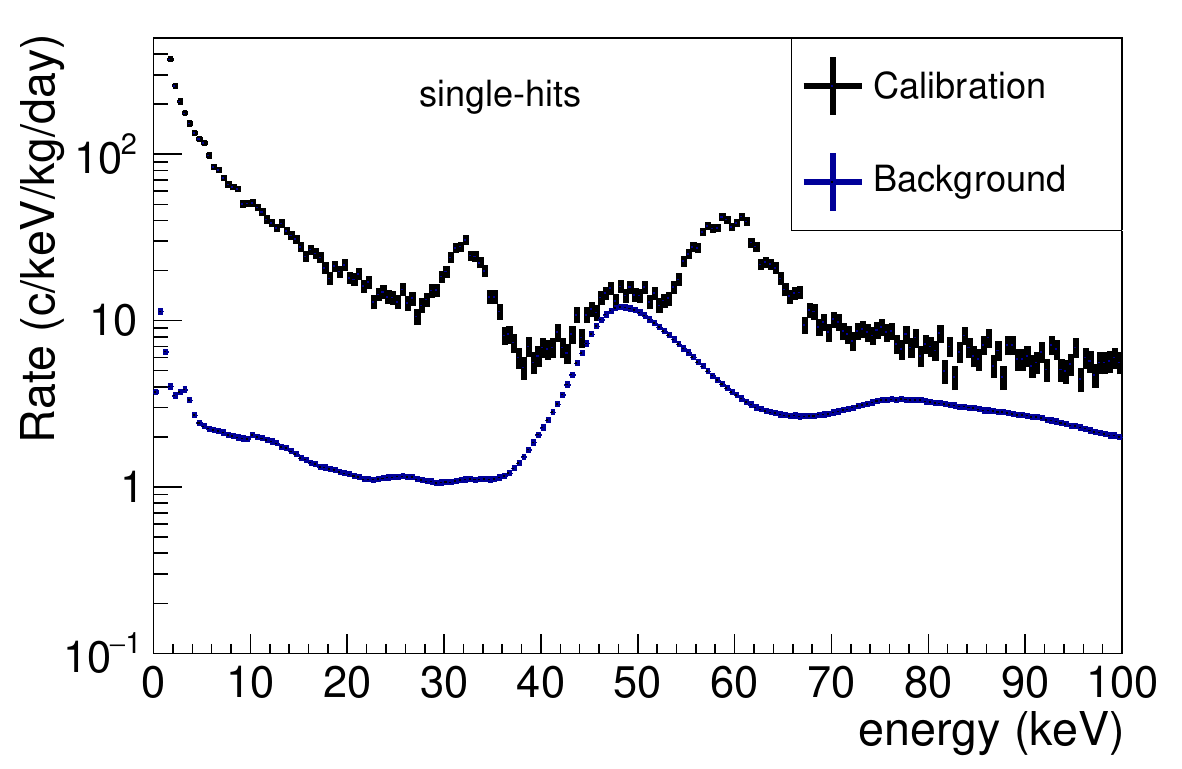}
    \end{minipage}
    \begin{minipage}{0.6\textwidth}
        \centering
        \includegraphics[width=\textwidth]{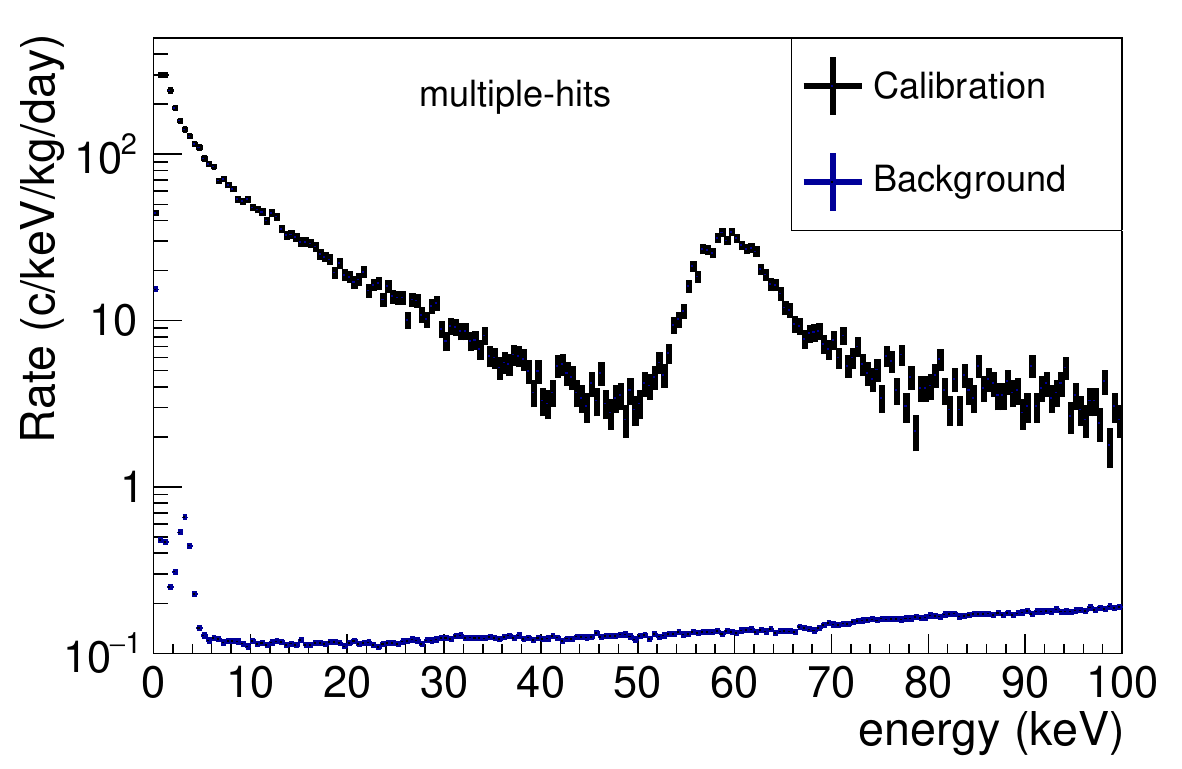}
    \end{minipage}
\caption{\label{comparesinglemulti} Energy spectrum from a \(^{252}\text{Cf}\) calibration on the west side of the ANAIS–112 experiment (black), with the background contribution shown as reference (blue). \textbf{Upper panel:} Single-hit events. \textbf{Lower panel:} Multiple-hit events.
}

\end{center}
\end{figure}

\begin{figure}[t!]
\begin{center}
\includegraphics[width=.9\textwidth]{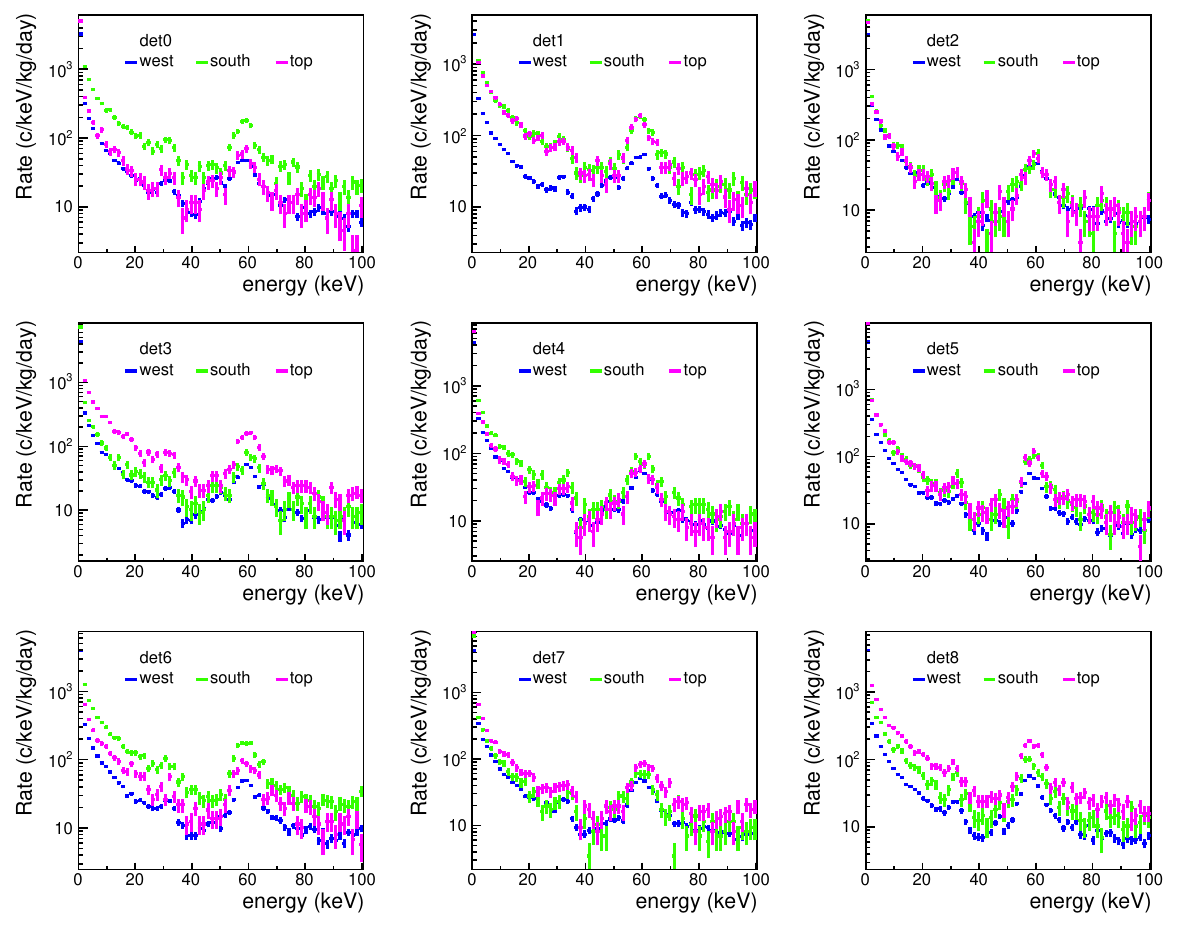}

\caption{\label{comparedatatotaldet} Total energy spectrum of the nine modules from \(^{252}\text{Cf}\) neutron  calibrations performed on the west face (blue), south face (green) and top face (magenta) of the ANAIS–112 experiment.}
\end{center}
\end{figure}

Figure \ref{comparesinglemulti} illustrates the low-energy spectrum from a \(^{252}\text{Cf}\) calibration conducted on the west side of the ANAIS–112 experiment, with the background contribution estimated according to the live time included for comparison. The background contribution is negligible, except in the $^{210}\text{Pb}$ region around 40–50 keV in the single-hit spectrum. In fact, the acquisition rate during neutron calibration is approximately five times higher than the ANAIS-112 background rate ($\sim$ 4.5~Hz) when using the LSC source, while it increases to about fifteen times higher when using the HENSA source. 


Significant contributions from electron/gamma events are observed at specific energies. At 31.8 keV, visible in the single-hit spectrum, the peak corresponds to the EC decay of \(^{128}\text{I}\) produced by neutron activation of \(^{127}\text{I}\). $^{128}$I decays via EC or $\beta^+$ emission (6.9\%) to $^{128}$Te. In the case of EC decay, a distinct peak at 31.8 keV, corresponding to the K-shell binding energy of Te, is expected (80\%). Similary (with 15.6\%) L-shell EC should be observed, corresponding to peaks at $\sim$ 4.3-4.9 keV, which are hidden within the large elastic scattering contribution. Another contribution appears at $\sim$ 58 keV, present in both single-hit and multiple-hit spectra, which originates from neutron inelastic scattering on \(^{127}\text{I}\). As will be shown in Figure \ref{simdistribucion}, this peak presents a more complex structure than that of a pure gamma event, corresponding to that gamma line combined with a NR recoil tail.

Below 20 keV in the single-hit spectrum and 40 keV in the multiple-hit spectrum, the rates show a quasi-exponential dependence and are predominantly due to multiple elastic scatterings on Na and I nuclei, as will be further discussed in this chapter in the analysis of the neutron calibration simulation (see Section \ref{neutronsim}). Because of the quasi-exponential low-energy dependence of the elastic scattering differential spectrum, a large number of events in the ROI can be collected in just a few calibration runs, each lasting 3–4 hours of real time.



Moreover, it is interesting to compare the energy spectrum measured  when the source is placed on each face of the ANAIS-112 experiment. Referring back to Panel (b) of Figure~\ref{distribution}, which illustrates the position of the neutron source for each calibration face, it can be observed that the detector axis is aligned in the east-west direction. Given that the \(^{252}\text{Cf}\) source emits isotropically, it is expected that the event rate during neutron calibration runs performed on the south and top faces will be higher than on the west face, because of the larger solid angle covered by the NaI crystals, as Figure~\ref{comparedatatotaldet} illustrates. 

The same spectral characteristics explained earlier are visible in this comparison. The west calibration configuration results in a more homogeneous exposure across the nine detectors, achieving comparable statistics for all modules. In contrast, in the south (top) calibration, detectors D0, D1, and D6 (D1, D3, and D8) are more directly exposed to the source. That is, the west calibration provides a more uniform exposure across the nine modules, which is desirable for the QF study in order to obtain sufficient statistics in all detectors, reduce possible bias and thus derive consistent results.

\subsubsection{Average pulses}\label{pulses}


This section focuses on investigating the differences in scintillation time constants between NRs and ERs, a behavior previously studied by other collaborations \cite{lee2015pulse,kim2019limits}.

\begin{figure}[t!]
    \centering
    \begin{subfigure}[b]{0.45\textwidth}
        \centering
        \includegraphics[width=\textwidth]{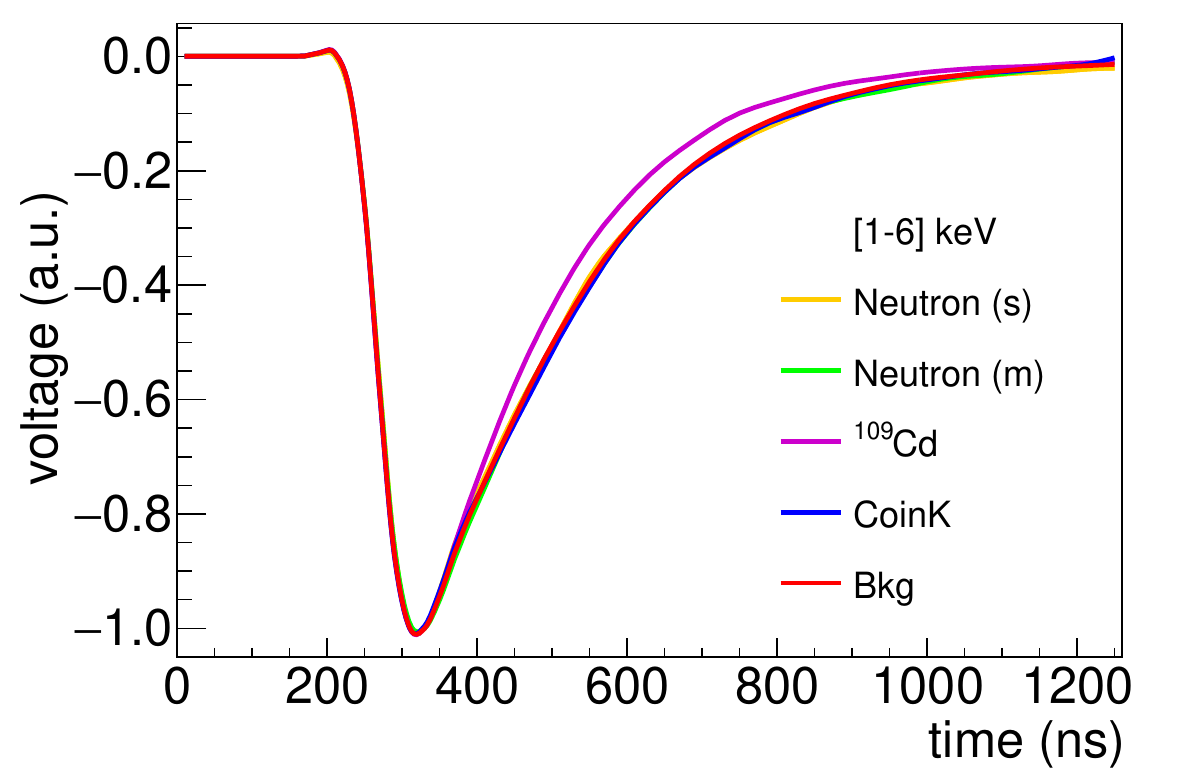}
        \caption{}
    \end{subfigure}
    \hspace{0.05\textwidth} 
    \begin{subfigure}[b]{0.45\textwidth}
        \centering
        \includegraphics[width=\textwidth]{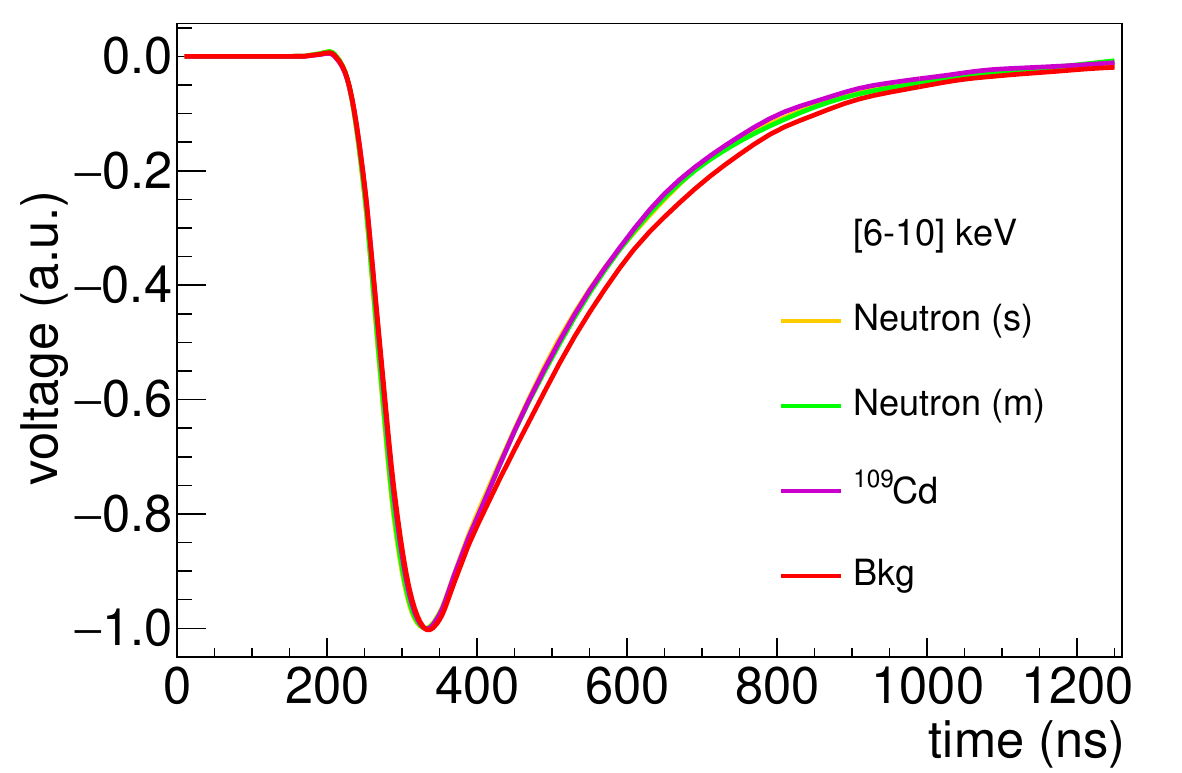}
        \caption{}
    \end{subfigure}
    \vspace{0.4cm} 
    \begin{subfigure}[b]{0.45\textwidth}
        \centering
        \includegraphics[width=\textwidth]{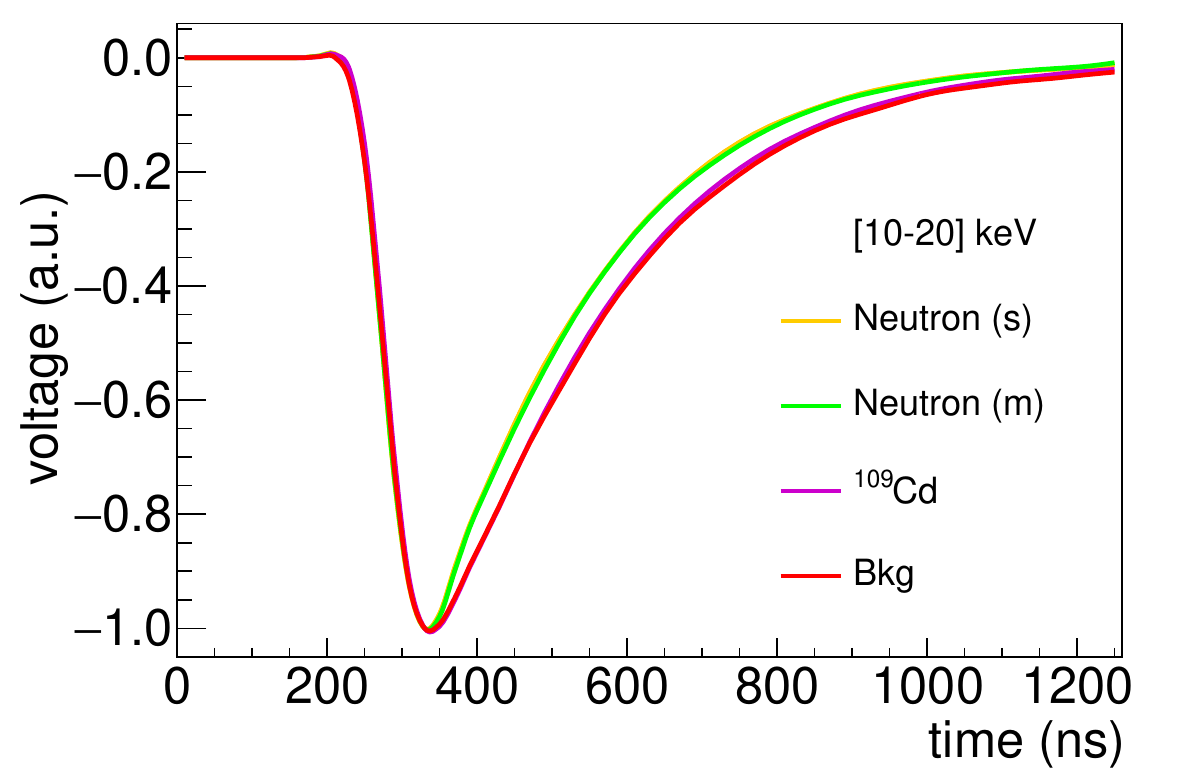}
        \caption{}
    \end{subfigure}
     \hspace{0.05\textwidth} 
    \begin{subfigure}[b]{0.45\textwidth}
        \centering
        \includegraphics[width=\textwidth]{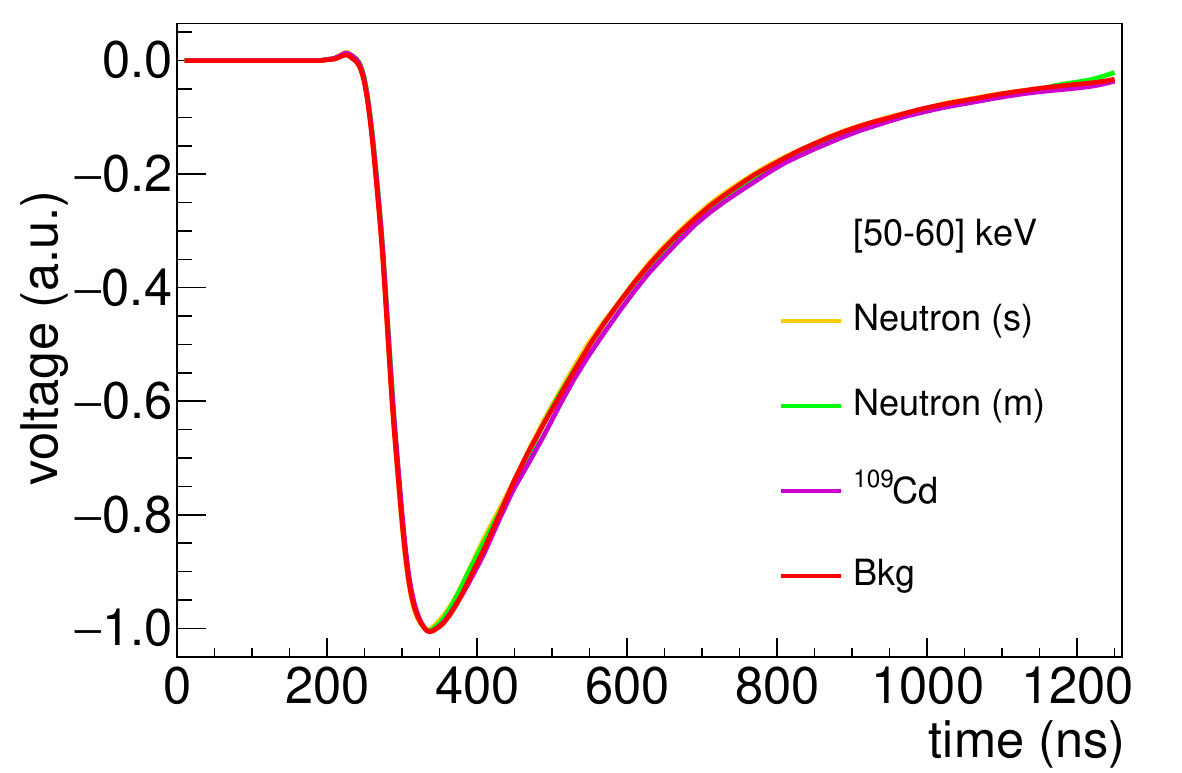}
        \caption{}
    \end{subfigure}

    \caption{ Average pulses corresponding to different populations: single-hit neutrons (yellow), multiple-hit neutrons (green), single-hit \textsuperscript{109}Cd events (magenta), single-hit background scintillation events (red), and gamma events selected in coincidence with a high-energy $\gamma$ from $^{40}$K (blue). The evolution of the pulses is analyzed across different energy ranges: \textbf{(a)} within the ANAIS ROI [1–6] keV, \textbf{(b)} [6–10] keV, \textbf{(c)} [10–20] keV, and \textbf{(f)} [50–60] keV, where the events labeled as neutron single (s) and neutron multiple (m) within this energy window are actually ER.}
    \label{pulsestodos}
\end{figure}

To this end, analyzing the evolution of average pulses across different event populations as a function of energy provides valuable insight, as illustrated in Figure~\ref{pulsestodos}. This figure compares the average pulse shapes for single-hit and multiple-hit neutron events with those from background and \textsuperscript{109}Cd calibration (single-hit) events, and $^{40}$K coincidence events. For events with energies below 10 keV, an average over 20000 pulses is performed, while for higher energies, 4000 pulses are averaged, followed by a smoothing of the averaged pulse.

For background events, standard selection criteria are applied: PSV<3, number of peaks detected in
each PMT (n$_0$ and n$_1$, separately) > 4, and coincidence of the two PMT triggers in a 200 ns window. Neutron events are selected using the bulk scintillation cuts described in Section \ref{criteria}. For \textsuperscript{109}Cd calibration events, a straightforward P1>0.4 cut is sufficient. In the case of $^{40}$K coincidences, the selection requires P1>0.4 along with a coincidence cut, selecting the high-energy gamma emitted in the associated decay and detected in a different module.



Panel (a) in Figure \ref{pulsestodos} presents a comparison of average pulse shapes within the ANAIS ROI, [1–6] keV. In this energy range, neutron calibration events (both single- and multiple-hit), background events, and $^{40}$K coincidence events exhibit nearly identical temporal behavior. However, the pulses from \textsuperscript{109}Cd calibration events appear noticeably faster. This discrepancy is likely due to the superficial nature of these energy depositions, which do not represent a homogeneously distributed bulk gamma population and may introduce surface-related effects not accounted for. Thus, with the exception of the \textsuperscript{109}Cd events, NR events, represented by neutron-induced single- and multiple-hit populations, show no discernible temporal differences from ER events, represented by background and $^{40}$K coincidences, within the ROI.

Above 6 keV, as shown in Panel (b), NR events begin to display slightly faster scintillation decay times compared to ER events. This behavior has been previously reported in the literature \cite{lee2015pulse,kim2019limits,coarasa2022improving}. Again, the \textsuperscript{109}Cd events exhibit faster pulse shapes than other ER populations, such as those from \textsuperscript{40}K coincidences, which are homogeneously distributed in the bulk of the crystal. This is consistent with the hypothesis of surface effects, particularly at low energies. This is one of the aspects that could be further addressed in future calibration campaigns after the stop of data taking of ANAIS-112. In particular, dedicated calibrations are needed to identify Compton events depositing energy in the bulk of the crystals, selected through coincidence techniques. This would require relatively intense sources placed appropriately within the set-up, along with accurate simulations to support the event selection. Beyond 10 keV, \textsuperscript{109}Cd events begin to display pulse shapes aligned with those of background ER events.


As the energy increases further, Panel (c) shows that in the [10–20] keV range, NRs continue to exhibit systematically faster pulse shapes than ERs. Beyond 50 keV, as seen in Panel (d), no differences are observed between neutron and background pulses. This is because the events dominating the neutron calibrations in this energy region are ERs. From this point onwards, although not shown, the pulse shape converges uniformly across background and neutron calibration events, as both correspond to ER populations.

\begin{figure}[t!]
\begin{center}

\includegraphics[width=0.49\textwidth]{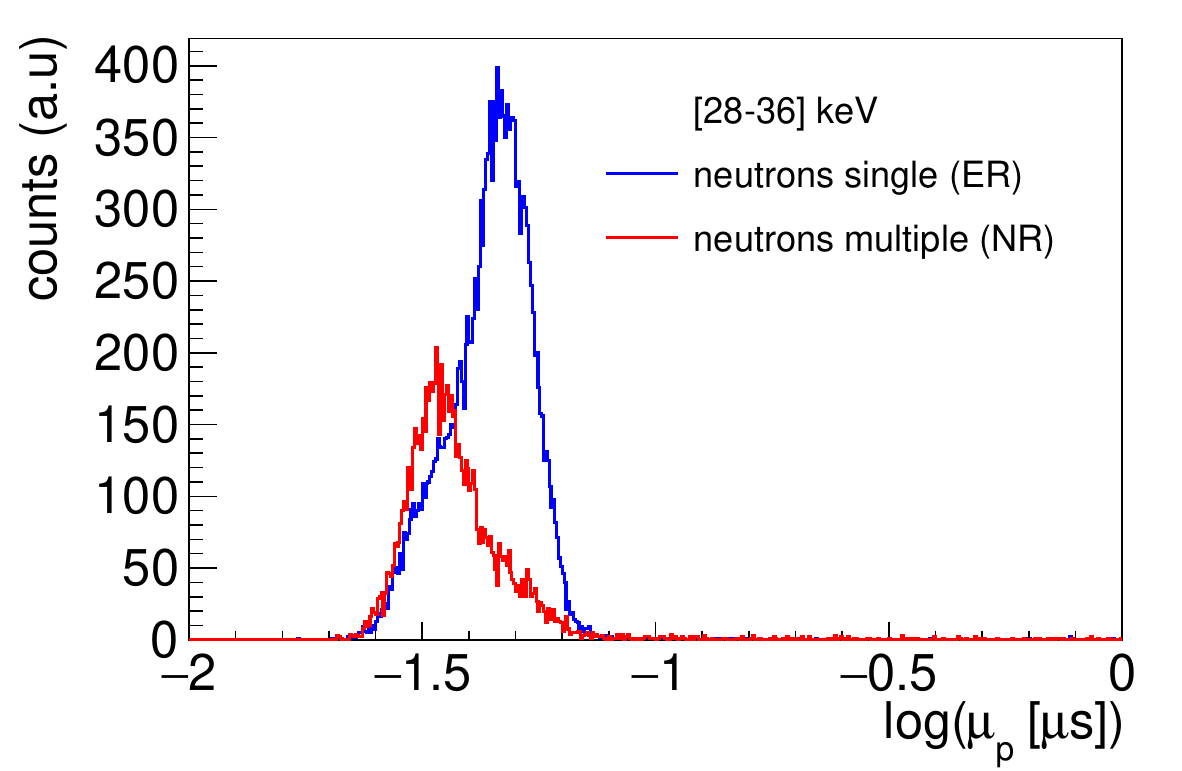}
\includegraphics[width=0.49\textwidth]{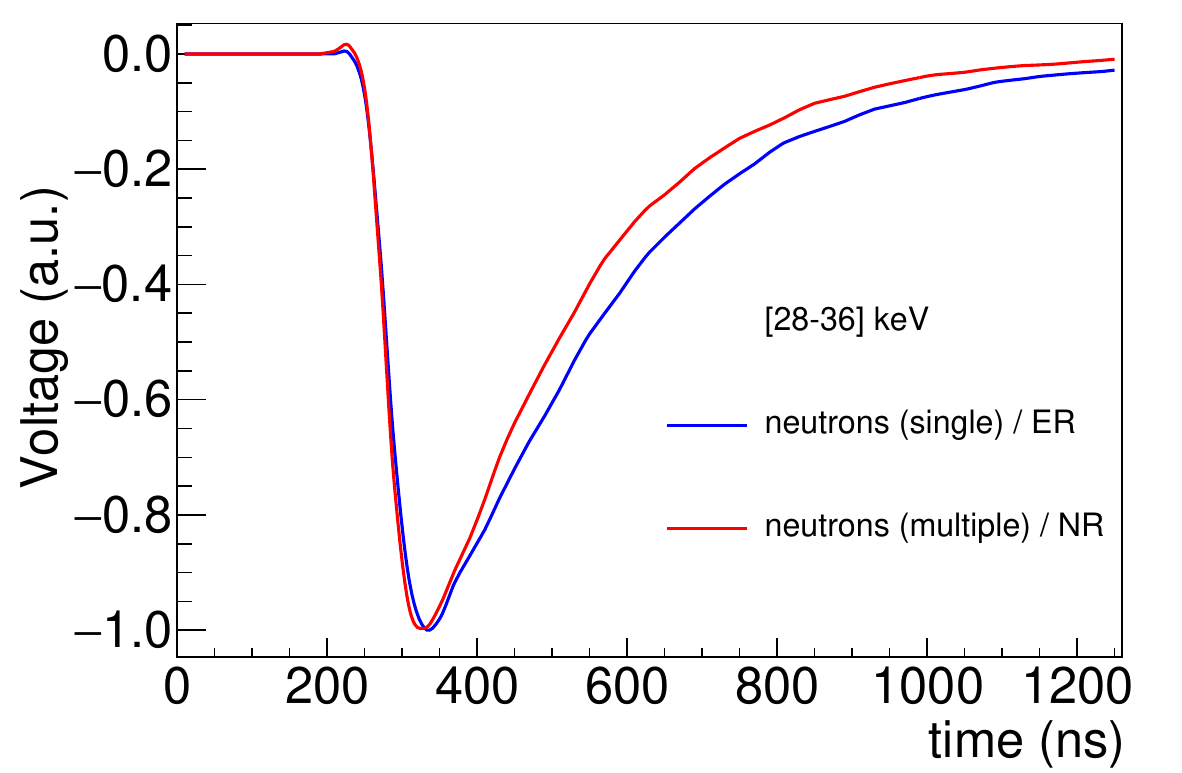}

\caption{\label{comparisonpulseneutronsinglemulti}  \textbf{Left panel:}  log($\mu_p$) distribution corresponding to single-hit neutrons (blue) and multiple-hit neutrons (red), selected around the 31.8 keV peak region ([28–36]~keV). \textbf{Right panel:} Average pulses for the same selected populations. This region is dominated by ERs in the single-hit population, whereas it is dominated by NRs in the multiple-hit population. NRs are clearly observed to be slightly faster than ERs.}
\end{center}
\end{figure}

The existence of different temporal profiles for ERs and NRs can be further confirmed by analyzing neutron data around the 31.8 keV peak. In single-hit events, the neutron population is dominated by ERs, as this feature originates from the X-ray de-excitation of Te following EC decay. In contrast, within the same energy window for multiple-hit neutron events, the population is dominated by NRs. The left panel of Figure \ref{comparisonpulseneutronsinglemulti} shows the distribution of log($\mu_p$) for neutron data around the 31.8 keV peak in both single-hit and multiple-hit events. Differences between the distributions for NRs and ERs are clearly visible: the peaks of both asymmetric gaussian-like distributions are shifted relative to each other. However, it is also evident that the populations are mixed, as both NR and ER events are present in the neutron single-hit and multiple-hit samples. For instance, in the neutron single-hit case, a bump on the left side of the distribution corresponds to NRs, while a similar feature on the right side of the multiple-hit distribution indicates the presence of~ERs.

The comparison of the average pulses is shown in the right panel of Figure \ref{comparisonpulseneutronsinglemulti}. Based on the features observed in the left panel of the same figure, and in an attempt to isolate NR and ER populations, an additional selection was applied on top of the standard cuts by averaging only those pulses within a log($\mu$\textsubscript{p}) window of $\pm$ 0.05 around the maximum of the distribution. As expected, NRs exhibit faster pulse shapes than ERs. However, in the single-hit case, the presence of a residual population of NRs remains visible even after applying the log($\mu$\textsubscript{p}) cut. This could be attributed to the inability of ANAIS-112 to discriminate between NRs and ERs on an event-by-event basis.


The fact that NRs exhibit faster pulse shapes than ERs originates from the different stopping powers of NRs, alphas, and ERs (or any other particle under consideration), which significantly affect the scintillation process. In NaI scintillators, $\alpha$ particles produce faster pulses, and NRs, having an even shorter mean free path, also exhibit this behavior. Just as they quench light production due to saturation of scintillation levels, they also modify the pulse shape by enabling alternative scintillation pathways, such as the excitation of different scintillation states.


To study this behavior across the full energy range, log($\mu_p$) distributions have been obtained in 1 keV energy bins from 1 to 100 keV for multiple-hit neutron events (which, as discussed earlier, feature reduced background contamination) and for single-hit background events. The mean of each distribution is then computed.

\begin{figure}[t!]
\begin{center}
\includegraphics[width=0.49\textwidth]{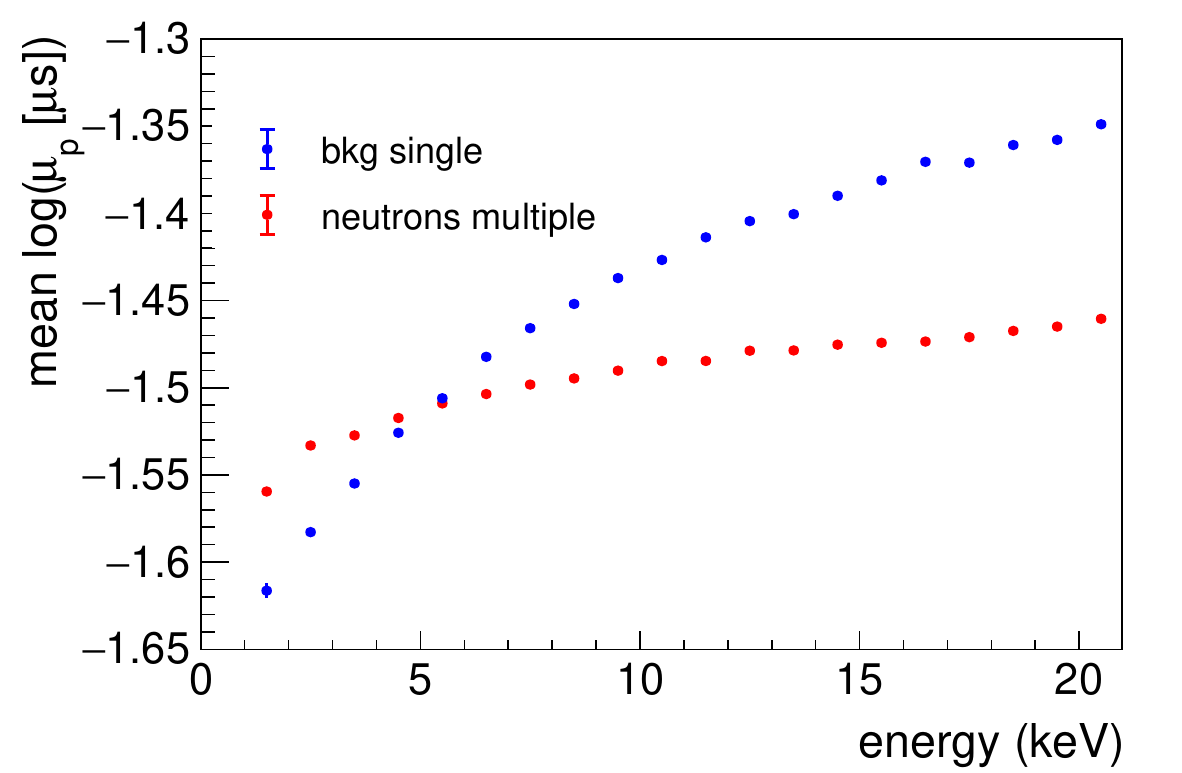}
\includegraphics[width=0.49\textwidth]{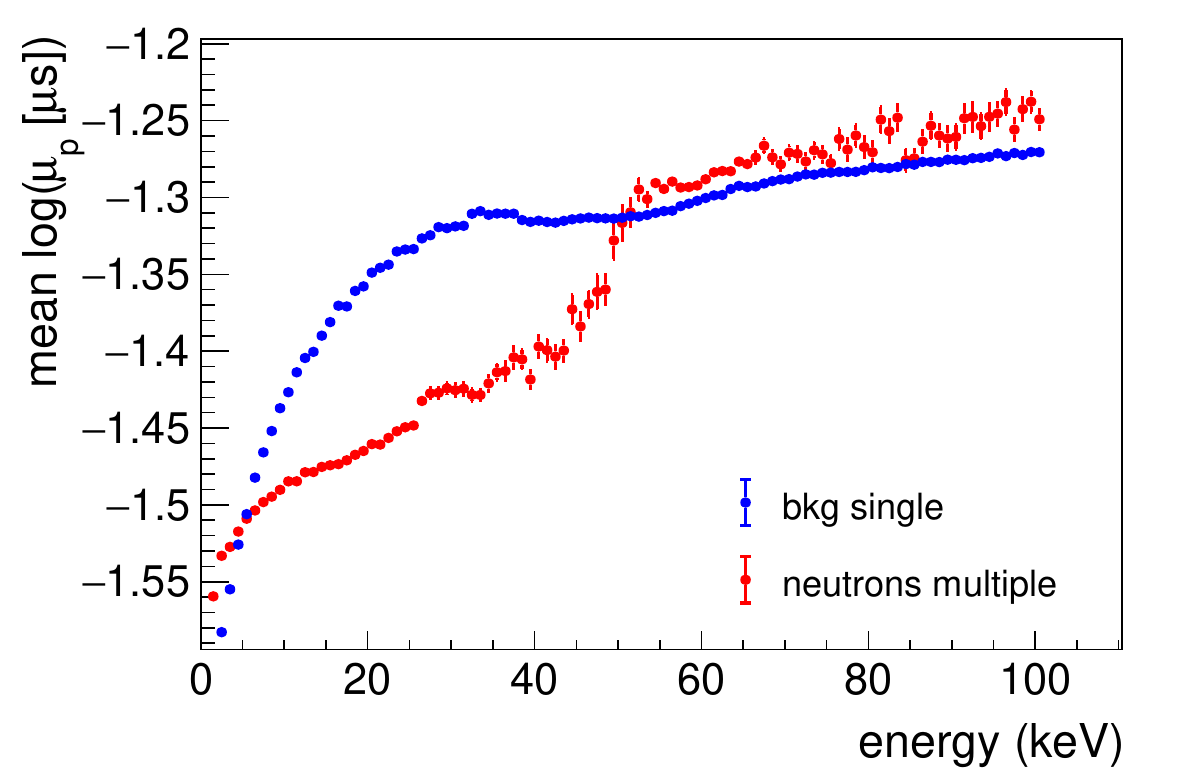}

\caption{\label{meanmu} Evolution of the mean of log($\mu_p$) with energy for multiple-hits neutron events (red) and single-hits background events (blue). \textbf{Left panel:} up to 20~keV. \textbf{Right panel:} same distribution extended up to 100 keV. 
}
\end{center}
\end{figure}

Figure \ref{meanmu} shows the evolution of the mean log($\mu_p$) as a function of energy. The left panel focuses on the low-energy region up to 20 keV, while the right panel extends up to 100 keV. As seen in the plots, the mean values for background and neutron events differ significantly below 50 keV. In particular, when the neutron data is dominated by NRs, the mean log($\mu_p$) remains relatively constant and increases more slowly than for ERs, with systematically lower values. In contrast, the mean log($\mu_p$) for ER-dominated background increases with energy and stabilizes around 40 keV. This behaviour has already been observed in \cite{miramonti2002study}.

Similar mean log($\mu_p$) values are obtained for neutron multiple-hit events above 50~keV, where the population becomes ER-dominated. Therefore, above this energy, the mean log($\mu_p$) values for both populations converge, as both consist of ERs. This analysis demonstrates that the ability to distinguish between ER and NR based on PSA decreases at low energies. As a result, average pulses in the [1–6] keV region (see Panel~(a) of Figure \ref{pulsestodos}) appear to have similar temporal behaviour for ERs and NRs. Thus, within the ROI, below 6–7 keV, clear discrimination between NR and ER events is not achievable.

From these distributions, the mean decay times of NaI(Tl) scintillation pulses are estimated for NRs ($\tau$\textsubscript{NR}) and ERs ($\tau$\textsubscript{ER}). The value of $\tau$\textsubscript{NR}, calculated by averaging mean log($\mu_p$) values for multiple-hit neutron events below 50 keV (where NR dominates, as supported by simulation), is 236.2 $\pm$ 1.7 ns. For ERs, the average $\tau$\textsubscript{ER} computed from background single-hit events between 60 and 100 keV is 277.4 $\pm$ 0.4 ns. This yields a ratio of $\tau$\textsubscript{NR}/$\tau$\textsubscript{ER} $\sim$ 0.85, consistent with previous estimations \cite{lee2015pulse}.

This result is slightly higher than the often-quoted 230 ns decay constant for NaI(Tl). However, literature values range from 200 to 320 ns \cite{kudryavtsev1999characteristics,murray2007energy}. Moreover, these values depend heavily on how the decay parameter is defined: NaI(Tl) scintillation includes both fast and slow components, with a long tail extending up to 1 $\mu$s \cite{Birks:1964zz}. Consequently, the average decay time inferred from pulse shape parameters is typically longer than the time constant extracted from fits to the primary scintillation exponential, which are themselves sensitive to the chosen fitting range and model.


\subsubsection{Energy calibration}\label{energycal}

A critical aspect in the analysis of the neutron calibration data is the energy calibration procedure. In ANAIS-112, energy calibration is performed in three different energy ranges  (see Section \ref{LEcalibration}): low (<20 keV), medium ([20-150] keV, and high energy ([150-1600]~keV), with an energy threshold established at 1 keV.

For the low-energy region, the same energy calibration applied in the ANAIS ROI will be used. This calibration is based on the full exposure of the \(^{22}\)Na and \(^{40}\)K lines present in the background after correcting possible gain drifts with the $^{109}$Cd calibration 22 keV line. Similarly, for the high-energy region, the method of linearizing the response of the pulse-area, as discussed in Section \ref{HEcal}, has proven to be successful.

However, in the medium-energy region, discrepancies and differing behaviors arise depending on the calibration procedure followed. Figure \ref{calibrations} illustrates the various calibrations considered in this energy range. Custom names are assigned to ease the identification of each calibration procedure. Two of these calibration methods correspond to pre-existing calibrations developed in the ANAIS experiment for different purposes, while the remaining two were specifically designed for the study of the QF of the ANAIS-112 crystals conducted in this work. The key points of each calibration method are summarized below:

\begin{figure}[b!]
\begin{center}
\includegraphics[width=.65\textwidth]{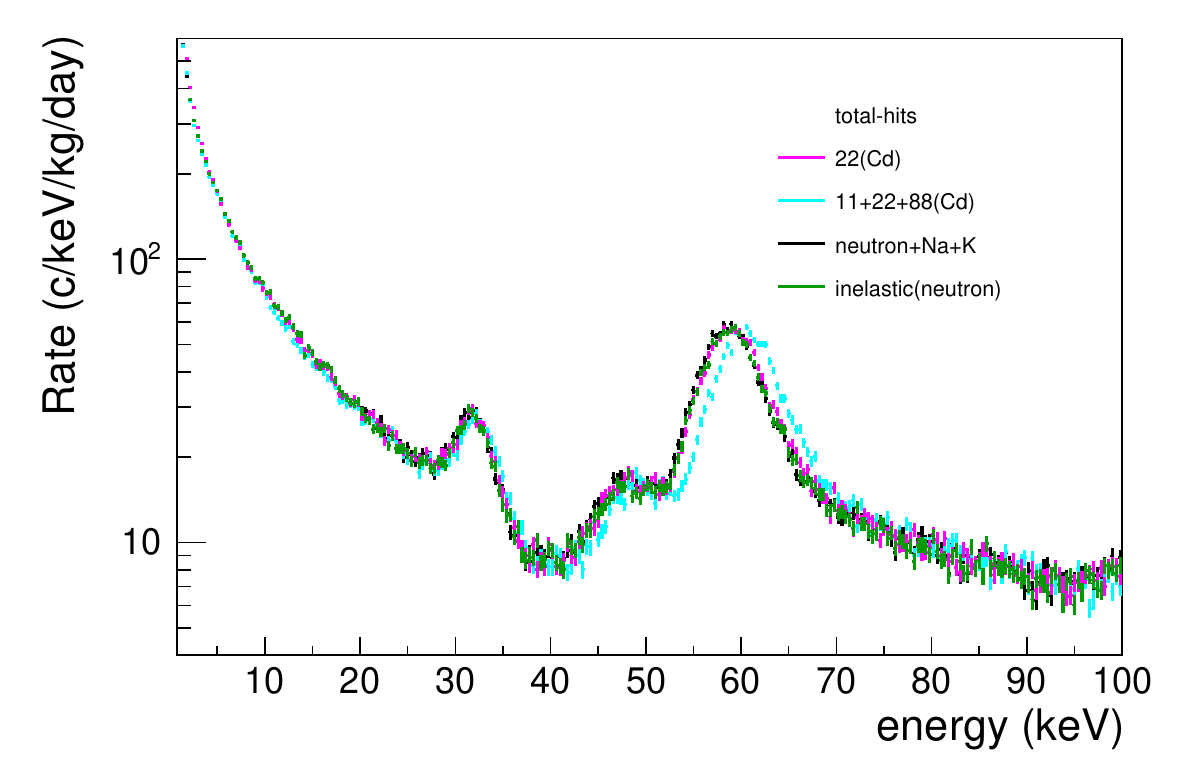}

\caption{\label{calibrations} Total energy spectrum from \(^{252}\text{Cf}\) ANAIS-112 neutron calibrations. Data is presented applying several energy calibrations considered for this study, using for that purpose the following information: the \(^{109}\)Cd 22 keV line (magenta); the three energy peaks of \(^{109}\)Cd (11.9, 22.6, and 88.0~keV) (cyan); the two peaks observed during the neutron calibration (31.8 keV and the inelastic peak), along with the \(^{22}\)Na and \(^{40}\)K internal background lines (black); and the inelastic peak derived from the neutron calibration alone (green). A deviation of $\sim$ 1.6 keV is observed at the inelastic peak using the the three energy peaks of \(^{109}\)Cd for calibration compared to the others, that are mostly compatible. }
\end{center}
\end{figure}

\begin{itemize}
    \item \textbf{22(Cd):} It uses a proportional calibration based on the \(^{109}\)Cd 22 keV line.
    \item \textbf{11+22+88(Cd):} It employs the three energy peaks of \(^{109}\)Cd (11.9, 22.6, and 88.0~keV) to calibrate the medium-energy region.
    \item \textbf{neutron+Na+K:} It includes the two peaks observed during the neutron calibration (31.8 keV and the inelastic peak), as well as the \(^{22}\)Na and \(^{40}\)K internal background lines.
    \vspace{0.3cm}
    \item \textbf{inelastic(neutron):} It relies solely on the inelastic peak observed from the neutron calibration. As previously stated, this peak arises from the superposition of the 57.6 keV gamma and the iodine recoil, with the potential addition of other recoil contributions. The energy used for calibration is determined by fitting the simulated spectrum. For the mean of this composite peak, the fit to the simulated data after convolution with the ANAIS-112 energy resolution yields a value of (58.77~$\pm$~0.03)~keV.

    It is worth noting that the proportional calibration using the 31 keV peak from neutrons also yields similar results. However, the calibration with the inelastic peak was preferred, as the latter lies at the center of the medium-energy range under study. This central position offers better calibration precision at both ends of the range. Additionally, non-linear effects are expected at the K-shell iodine binding energy (33.2~keV), further justifying the choice of the inelastic peak.
\end{itemize}

As shown in Figure \ref{calibrations}, all calibrations, except the one based on the three $^{109}$Cd peaks, exhibit consistent behavior. In particular, the energy spectrum remains in good agreement across all calibrations, except at the inelastic peak, where the mean value obtained using the three $^{109}$Cd peaks deviates by $\sim$ 1.6 keV from the others.

Later in this chapter, when comparing data to the neutron calibration simulation, it will be seen that the simulated position of the inelastic peak agrees with the calibrations that do not rely on the three $^{109}$Cd peaks. This observation supports the hypothesis that surface effects may be responsible for the discrepancy observed in calibrations using $^{109}$Cd sources, whereas bulk calibrations are unaffected. A full explanation of these surface-related effects lies beyond the scope of this work. However, it is important to highlight that such discrepancies do not affect the low-energy region, where the QF of the ANAIS-112 crystals is determined, since the same energy calibration applied in the ANAIS ROI is used for this purpose.

In order to employ a variable already integrated into the ANAIS-112 analysis pipeline and to avoid relying on additional ad-hoc calibrations, and given that all three approaches yielded compatible results, this work adopts the proportional calibration using the 22~keV line from the $^{109}$Cd source for the medium energy region.

\section{Neutron calibration simulation}\label{neutronsim}

A dedicated Geant4 simulation of the ANAIS-112 neutron calibration has been performed in this work to evaluate the QF of the ANAIS crystals for Na and I recoils through a precise quantitative comparison between simulation and measurement. A detailed description of the simulation configuration, including detector geometry, physics list, and the reproduction of the ANAIS detector response, has been provided in Chapter~\ref{Chapter:Geant4}.

\begin{figure}[b!]
\begin{center}
\includegraphics[width=0.7\textwidth]{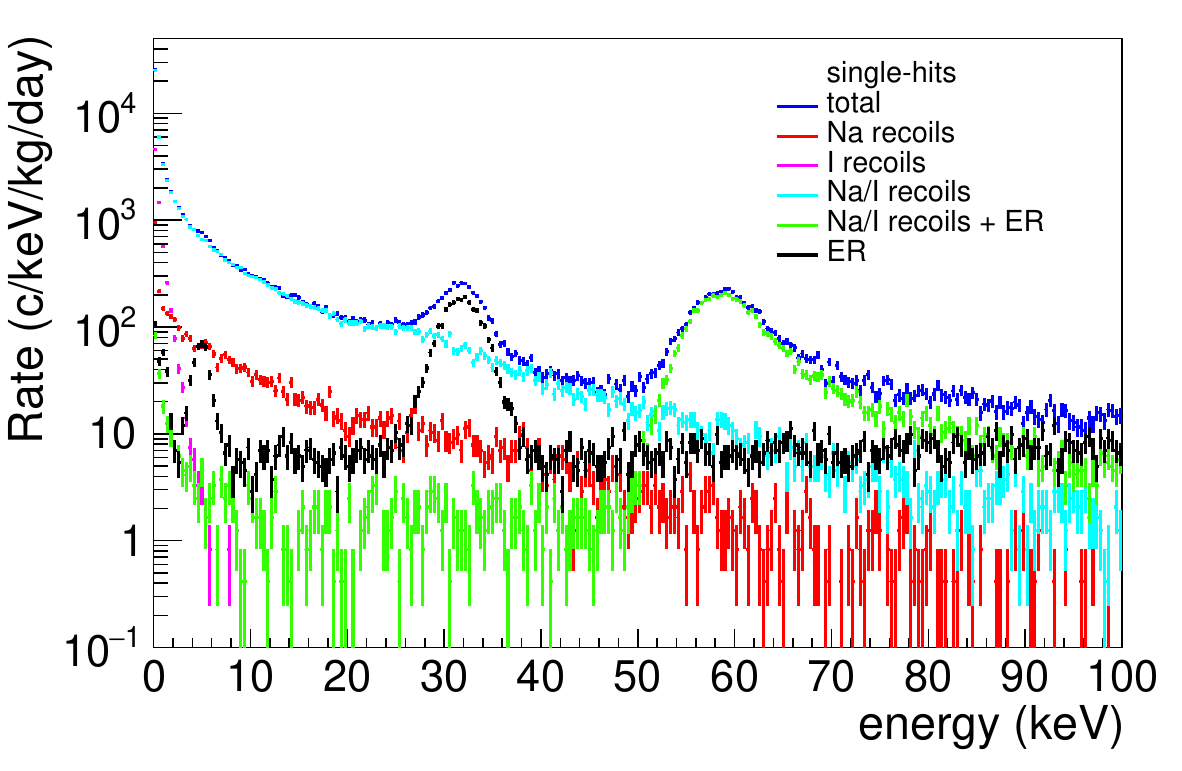}

\includegraphics[width=0.7\textwidth]{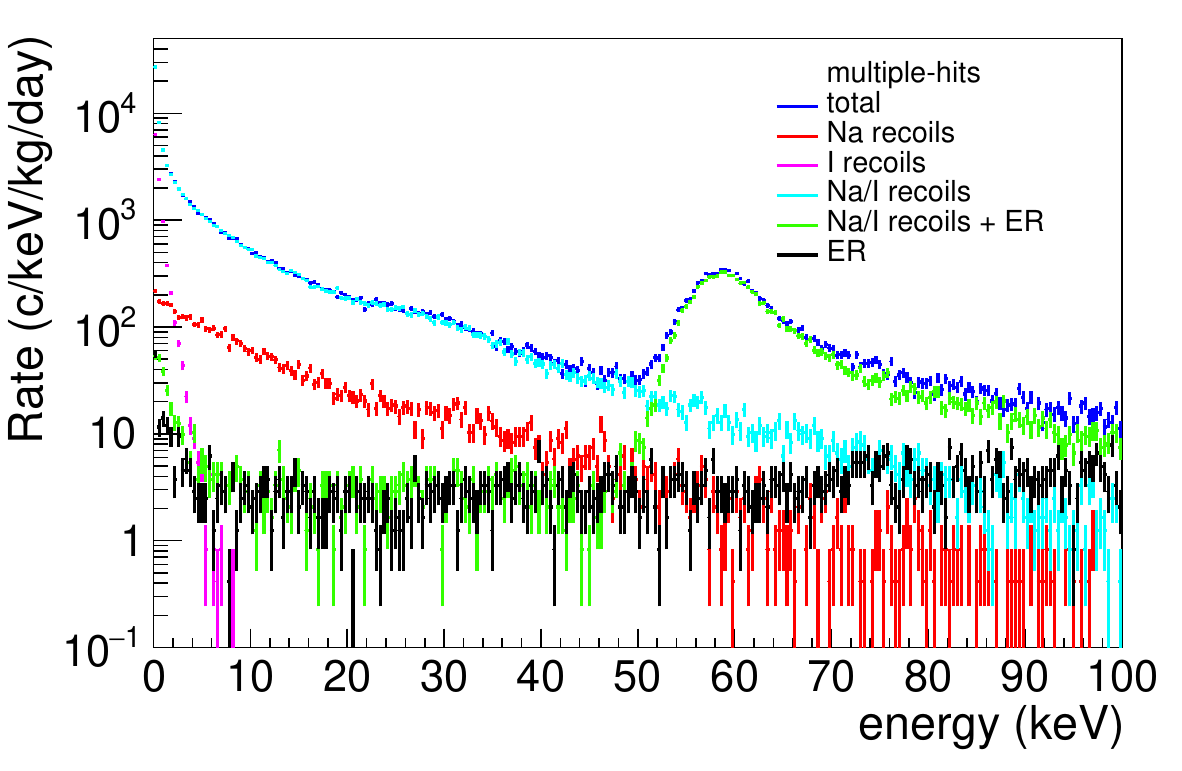}

\caption{\label{simdistribucion} Simulated neutron calibration spectra, showing the total rate (blue) and individual contributions: Na recoils (red), I recoils (magenta), combined Na and I recoils (cyan), Na or I recoils with ERs (green), and pure ERs (black). The low-energy region is primarily dominated by multiple scattering on Na and I nuclei, with distinct localized ER features. \textbf{Top panel:} single-hits. \textbf{Bottom panel:} multiple-hits. In both cases, the electron-equivalent energy is shown after applying ANAIS(1) QF${_\textnormal{Na}}$ and QF$_{\textnormal{I}}$=0.06. }
\end{center}
\end{figure}

As explained in that chapter, the Geant4 simulation enables particle identification. Leveraging this capability, a variable corresponding to the energy deposited by specific particle types —electrons/gammas, Na or I nuclei, and other ions— has been generated. With these definitions, various energy observables can be constructed to verify which particle population dominates each energy range. This allows selecting depositions from Na recoils, I recoils, combinations of Na and I recoils (with or without accompanying ERs), or pure ER events. By distinguishing the particle types involved, appropriate QF corrections can be applied even to mixed events containing ERs, sodium NRs, and iodine NRs. Such categorization facilitates a more accurate assessment of the sensitivity to QF variations across different energy regions.

\begin{figure}[b!]
\begin{center}
\includegraphics[width=0.7\textwidth]{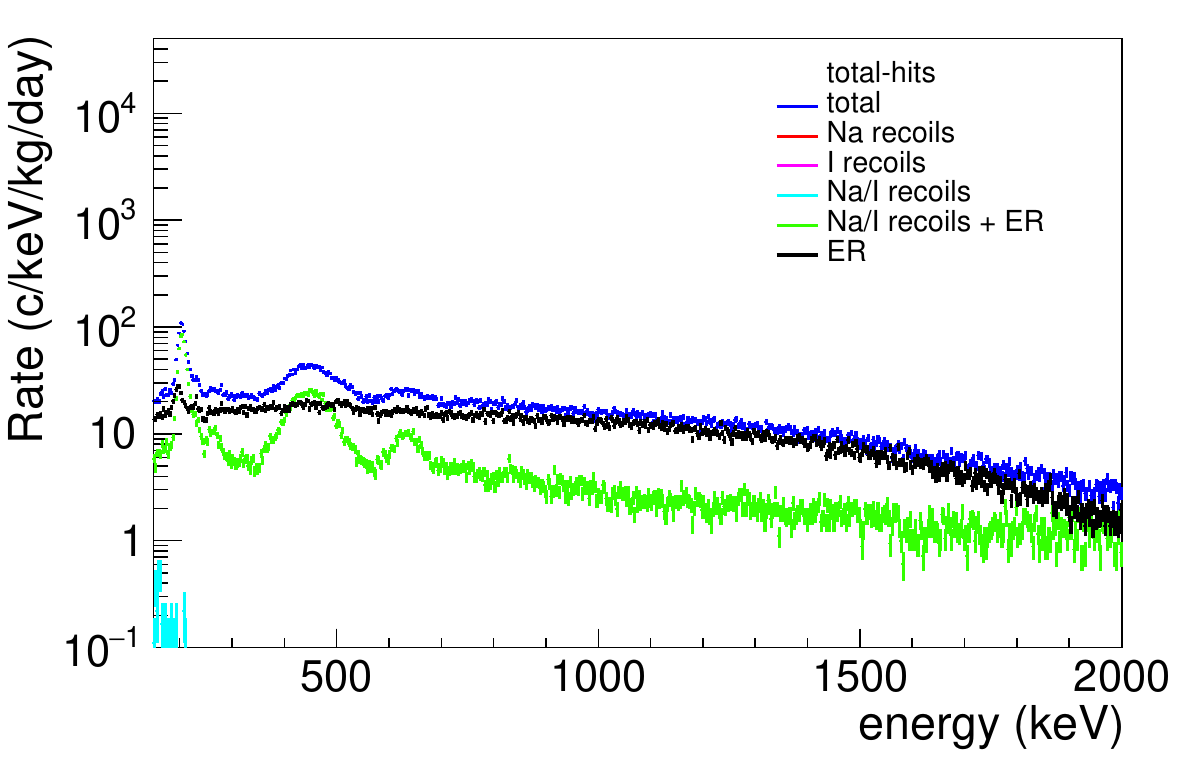}

\caption{\label{simdistribucionHE} Total high-energy simulated neutron calibration spectra, showing the total rate (blue) and individual contributions: Na recoils (red), I recoils (magenta), combined Na and I recoils (cyan), Na or I recoils with ERs (green), and pure ERs (black). The electron-equivalent energy is shown applying ANAIS(1) QF${_\textnormal{Na}}$ and QF$_{\textnormal{I}}$=0.06. 
}
\end{center}
\end{figure}

Figure~\ref{simdistribucion} presents the simulated energy spectra corresponding to neutron calibration for both the single-hit (top panel) and multiple-hit (bottom panel) detection configurations, incorporating the energy resolution function of the ANAIS-112 detectors. These spectra are in electron-equivalent energy, and therefore are based on a specific QF model. For the purpose of the figure, which aims to illustrate the different components of the spectrum, the choice of QF is not relevant. As derived from the figure, large ANAIS-112 crystals exposed to fast neutrons exhibit low-energy rates dominated by multiple scattering on both Na and I nuclei. Given the considerable size of the ANAIS crystals (29.85 cm in length and 6.03 cm in radius), according to the simulation neutrons should interact in average 4.4 times within a single module. The simulation shows that events below 20 keV are completely dominated by multiple elastic scattering (97.1\% for single-hits and 98.7\% for multiple-hit events).

NRs are the primary contribution up to approximately 50 keV, with iodine recoils being important at very low energies. The electron/gamma contribution corresponds to those features described in the previous section, where the 31.8 keV peak is identified as a pure ER peak in the single-hit spectrum, while the 58 keV peak combines ERs with NRs  from I, always present, and possibly also from Na, depending on the specific interaction conditions. Additionally, a Te L-shell ER peak analogous to the 31.8 keV peak is also generated, though it is orders of magnitude weaker than the NR signals and is not visible in either the data or simulation after resolution convolution. 

Figure \ref{simdistribucionHE} presents the high-energy simulated spectra for the total-hits case. In the high-energy range, no significant changes are observed regarding the dominant populations for single-hits and multiple-hits. It can be seen that the recoil events from Na or I reach up to approximately 200 keV, beyond which the ERs dominates, except for the inelastic interactions in I, such as the 202.86 keV transition. A broader peak encompasses the contributions from the 417.99, 473.4, as well as the 618.31, 628.69 and 650.92 keV lines, which also appear merged due to the finite energy resolution of the detectors. Other inelastic interactions in Na or I are not visible due to the same latter reason. Given the total dominance of ER, it is expected that changes in the QF in this specific region will not have significant effects, as has been confirmed through ad-hoc comparison between simulation and measurement.



\begin{figure}[b!]
\begin{center}
\includegraphics[width=0.65\textwidth]{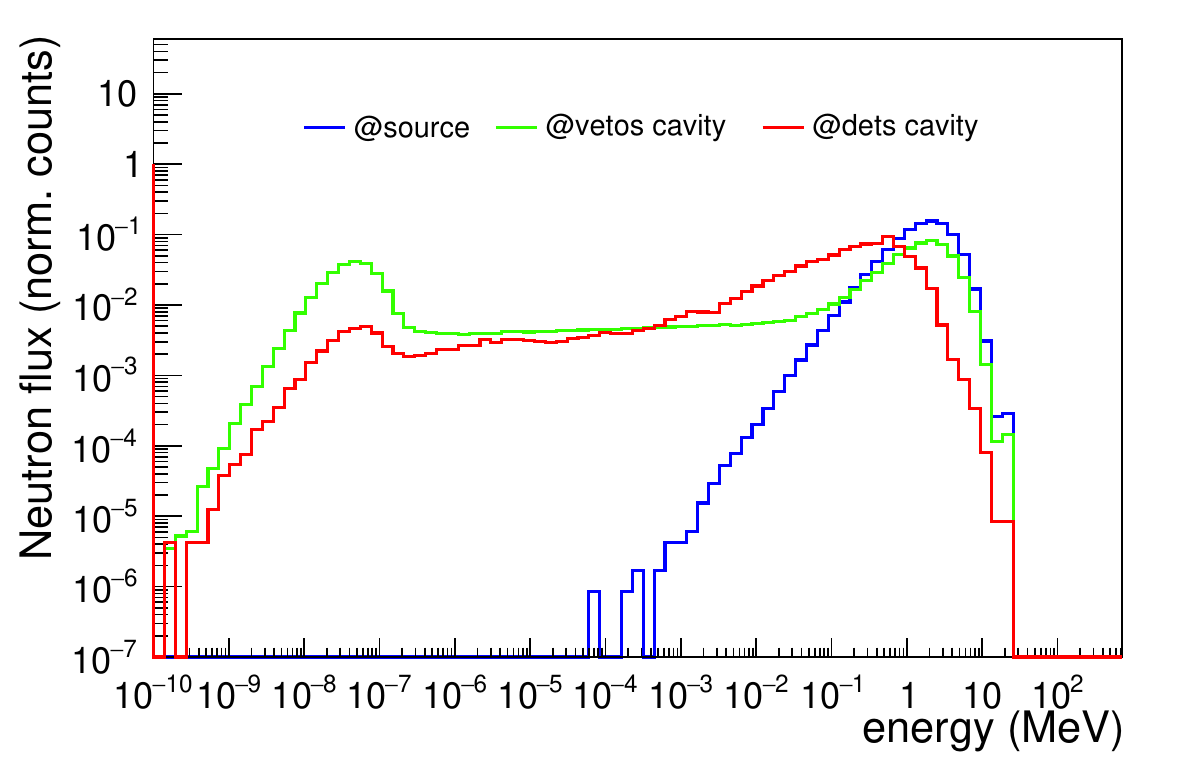}

\caption{\label{fluxdifferentcav} Neutron flux at different locations within the ANAIS-112 set-up as obtained from the neutron simulation, showing the flux emitted by the $^{252}$Cf source via SF (blue), the flux reaching the veto cavity (green), and the flux reaching the detector cavity (red).
}
\end{center}
\end{figure}

The simulation allows full tracking of neutrons passage through the ANAIS-112 set-up. Figure \ref{fluxdifferentcav} presents the neutron flux obtained from the simulation, showing both the flux emitted by the $^{252}$Cf source and the neutron flux reaching both the cavity that houses the vetoes and detectors. The detailed procedure used to derive these spectra is described in Section~\ref{NeutronHENSA}, where the environmental neutron flux incident on ANAIS-112 and its contribution to the experimental background is analyzed. As a brief introduction to the method, counters are positioned within the Geant4 geometry at different locations. These counters are defined as vacuum volumes with negligible thickness (1 $\mu$m). Within these regions, neutrons are tagged without undergoing interactions, enabling a precise characterization of the incident flux.

Figure \ref{fluxdifferentcav} shows that the energy spectrum of fast neutrons emitted by the $^{252}$Cf source peaks around 2 MeV, in agreement with the typical energy distribution of fast
fission neutrons (see Figure \ref{energymult}). The neutron population reaching the detector cavity exhibits a lower thermal neutron flux than that found in the veto cavity. This difference arises because neutrons in the veto cavity have, in most cases, already interacted with the polyethylene and water blocks, or have undergone scattering in the lead shielding and/or detectors. 

As neutrons interact with the polyethylene and water blocks, they undergo frequent elastic scattering with hydrogen nuclei. This process is highly effective at reducing neutron energy, leading to thermalization. As a result, the intensity of the fast neutron peak diminishes, and the energy distribution shifts towards lower values, producing a thermal neutron component. In contrast, neutron transport through lead is dominated by inelastic scattering and absorption. Due to its high atomic number and density, lead is efficient at absorbing thermal neutrons but plays a limited role in moderating them, as it lacks light nuclei like hydrogen. The behavior shown in Figure \ref{fluxdifferentcav} is especially relevant for understanding and improving the shielding configuration in experiments such as ANAIS-112, where minimizing the contribution of irreducible neutron backgrounds is essential.

The study of the QF of the ANAIS crystals will focus on the low-energy regime, with particular emphasis on neutron elastic scattering events releasing energies below 30 keV. Nonetheless, achieving a precise quantitative agreement between experimental data and simulation requires an accurate modelling of the various spectral features arising from the different interaction processes within the NaI(Tl) target and other detector components. 

In this context, neutron simulation can serve as a powerful tool to validate, by comparison with experimental data, the cross-sections implemented in Geant4 (see Section~\ref{testinggeant4}). Once it is established that the reliability of the simulation critically depends on the accuracy of these cross-sections, as variations among them can significantly impact the predicted outcomes, Section~\ref{QFmodels} will present the different QF models for Na and I recoils that will be used within the simulation.


\subsection{Testing Geant4 using ANAIS-112 neutron calibrations}\label{testinggeant4}

In recent years, the Geant4 Neutron-HP package, which enables accurate transport of low-energy neutrons (below 20 MeV) using evaluated cross-section libraries, has undergone significant development \cite{mendoza2014new,mendoza2018update,thulliez2022improvement}. As a result, Geant4 Neutron-HP is now considered on-par in terms of neutron physics with reference neutron transport codes such as Tripoli-4 \cite{tripolii} and MCNP6.2 \cite{mcnpp}. Despite this status, Geant4 remains subject to continuous validation by its user community, who regularly identify and report issues as part of its ongoing refinement. 

In this thesis, the simulation of neutron calibration data has revealed specific limitations within the Geant4 framework in two particular versions. On the one hand, a performance comparison between the two neutron cross-section libraries on which these versions rely has been conducted, highlighting discrepancies when compared to experimental data, which enabled the selection of one library over the other for the QF estimation study presented in this thesis. On the other hand, again by comparison with experimental results, an overestimation of the neutron capture cross section for $^{128}$I production has been identified in both versions.

\subsubsection{The data library caveat}\label{datalibrary}

The neutron simulations of ANAIS-112 conducted in this thesis have been carried out using several Geant4 versions, ranging from v9.4.p01, previously used in the development of the ANAIS-112 background model, to the latest stable release available in the ANAIS framework at the time of writing this thesis, v11.1.1. Intermediate versions, including v10.07 and v11.0.3, were also considered in test simulations without the full ANAIS-112 set-up.

For neutron-related processes, the relevant transport and cross-section information is provided through dedicated neutron libraries. Specifically, Geant4 v9.4.p01 relies on the G4NDL3.14 database, whereas v10.07 and v11.0.3 are based on G4NDL4.6, and v11.1.1 uses G4NDL4.7. A preliminary comparison of simulation outputs showed that versions v10.07, v11.0.3, and v11.1.1 yield identical results, which differ significantly from those of v9.4.p01. As a result, the present study focuses on comparing the performance of Geant4 v9.4.p01 and v11.1.1.

\begin{figure}[t!]
\begin{center}
\includegraphics[width=0.7\textwidth]{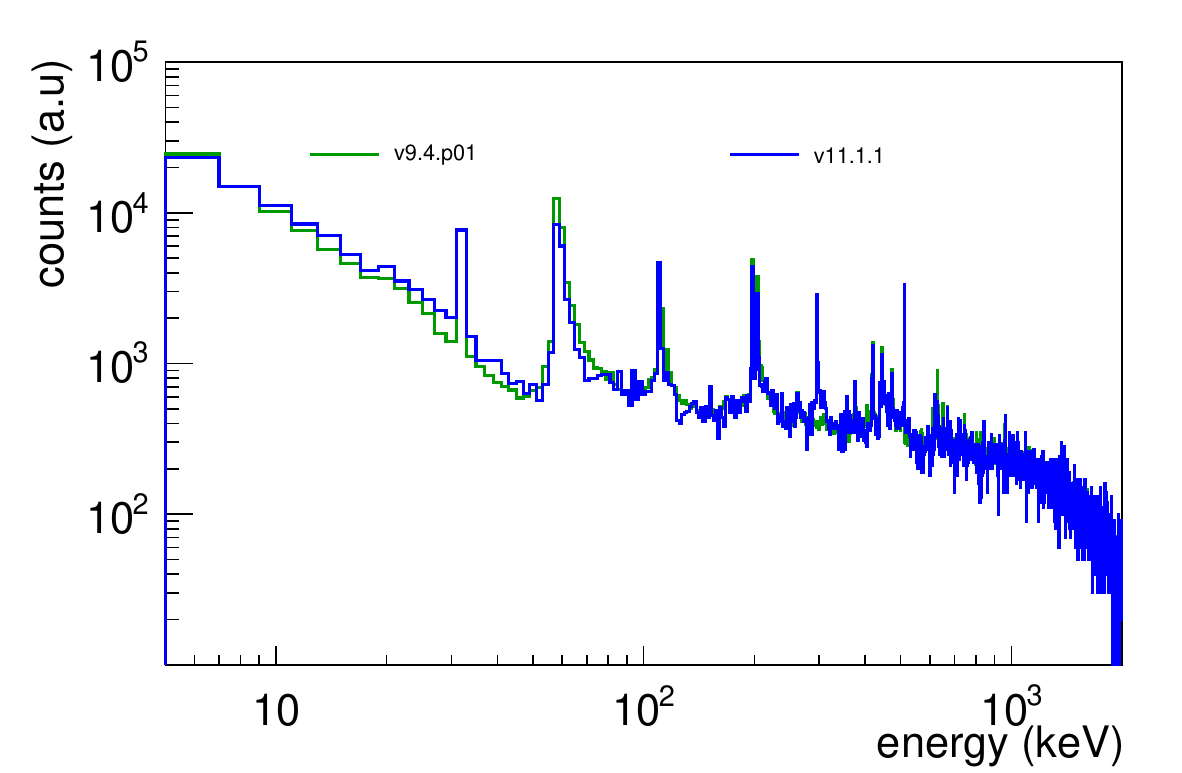}

\caption{\label{versioncomparion}  Comparison of the ANAIS-112 neutron simulation using Geant4 versions v9.4.p01 (blue) and v11.1.1 (green), with no resolution correction and the same QF\textsubscript{Na} and QF\textsubscript{I} applied in both cases, ANAIS(1) QF${_\textnormal{Na}}$ and QF${_\textnormal{I}}$=0.06. The 295 keV de-excitation line is absent in the v9.4.p01 simulation, but present in the v11.1.1 simulation. Additional discrepancies are observed in the elastic region and the intensities of several lines.  }
\end{center}
\end{figure}
\begin{figure}[b!]
\begin{center}
\includegraphics[width=0.49\textwidth]{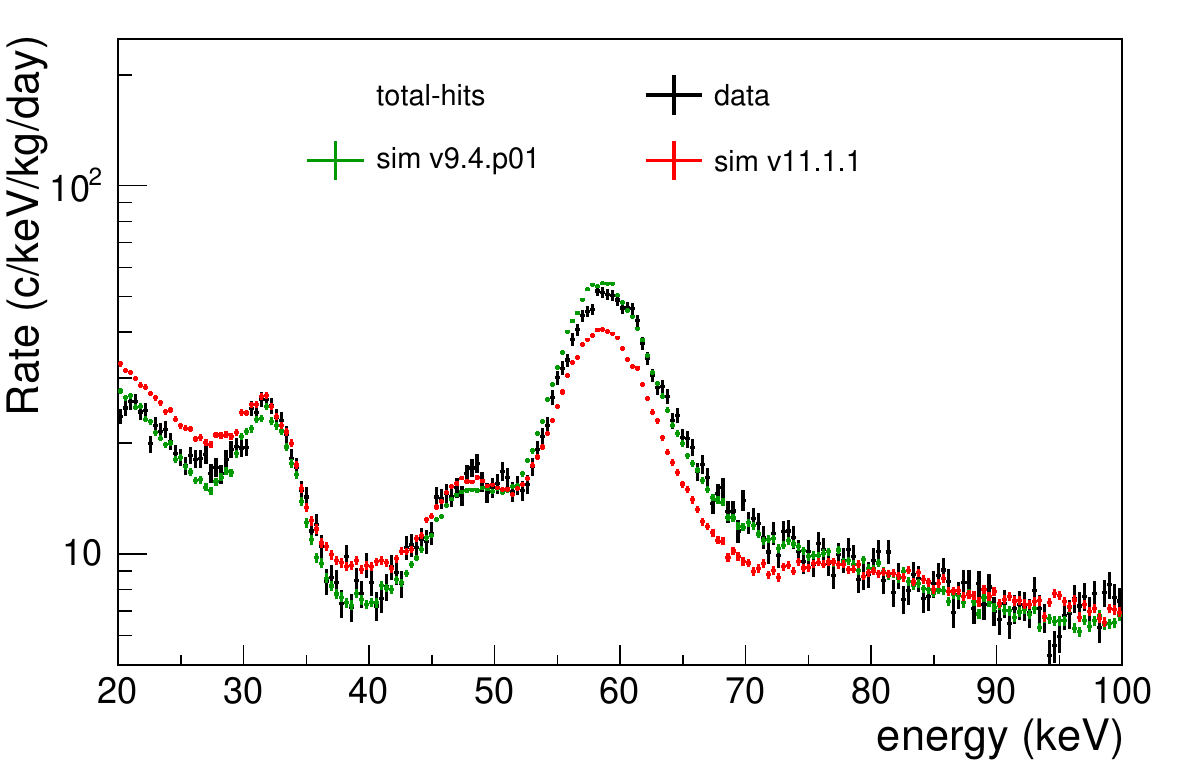}
\includegraphics[width=0.49\textwidth]{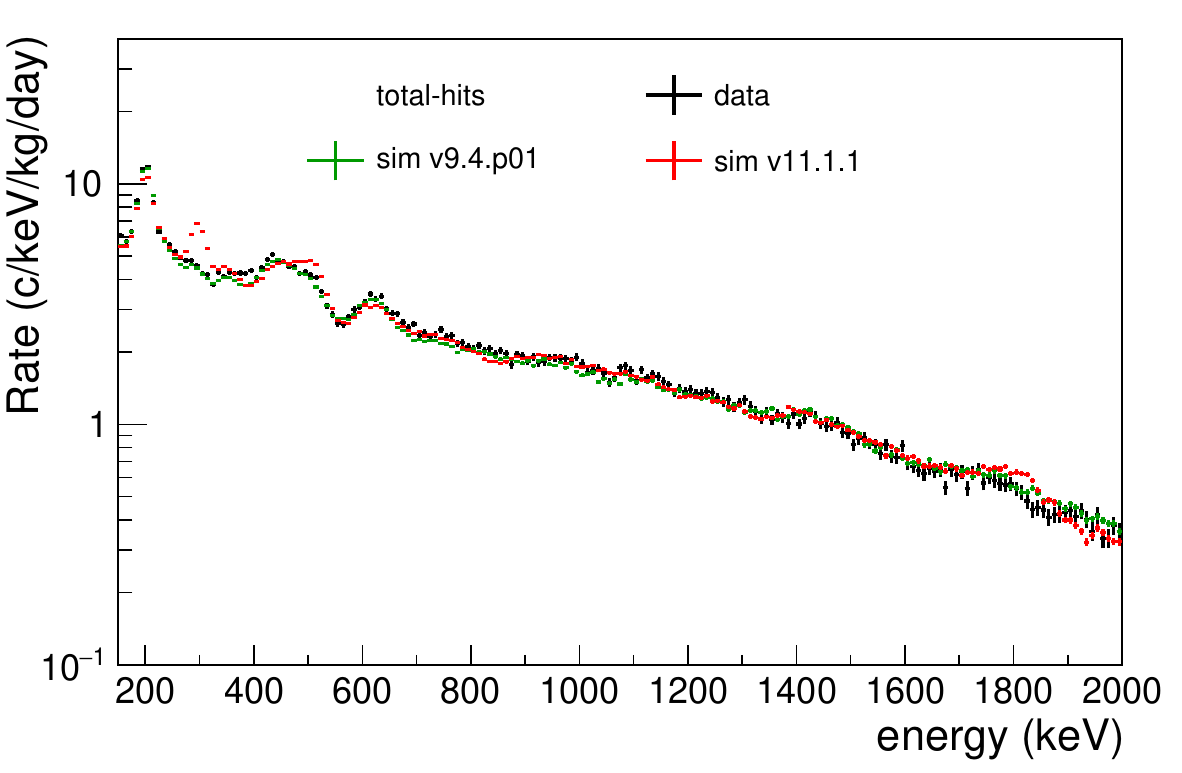}

\caption{\label{compareversioncondatos} Comparison between ANAIS-112 neutron calibration data (black) and simulation results using Geant4 v9.4.p01 (green) and v11.1.1 (red). The left panel shows the medium-energy region, while the right panel displays the high-energy region. ANAIS(1) QF${_\textnormal{Na}}$ and QF${_\textnormal{I}}$=0.06 have been applied in both simulations, although in these regions its impact their expected to be minimal. In these plots, the capture cross section for the production of $^{128}$I has already been corrected (see Section \ref{I128prodsec} for further details). }
\end{center}
\end{figure}

Figure \ref{versioncomparion} shows the comparison between neutron calibration simulations of ANAIS-112 using Geant4 versions v9.4.p01 and v11.1.1. No energy resolution correction has been applied, and the same QF\textsubscript{Na} and QF\textsubscript{I} are used in both simulations at the second level of analysis (see Chapter \ref{Chapter:Geant4}). The 295 keV de-excitation line is absent in the v9.4.p01 simulation, which uses the G4NDL3.14 library, and it is also not observed in the experimental data (as it will be shown in Figure \ref{compareversioncondatos}). However, this line appears in the simulation of the newer version. This line correspond to excited states of $^{127}$I, which, according to evaluated nuclear data, should not be populated through neutron inelastic scattering \cite{AUBLE197277}. Additionally, the intensities of several high-energy spectral features differ between versions. In the medium energy region, further discrepancies are evident, particularly in the elastic scattering region below the 31 keV peak and in the intensity of the inelastic peak, with the older version predicting a higher rate for the latter.

\begin{figure}[b!]
    \centering
    \begin{subfigure}[b]{0.45\textwidth}
        \centering
        \includegraphics[width=\textwidth]{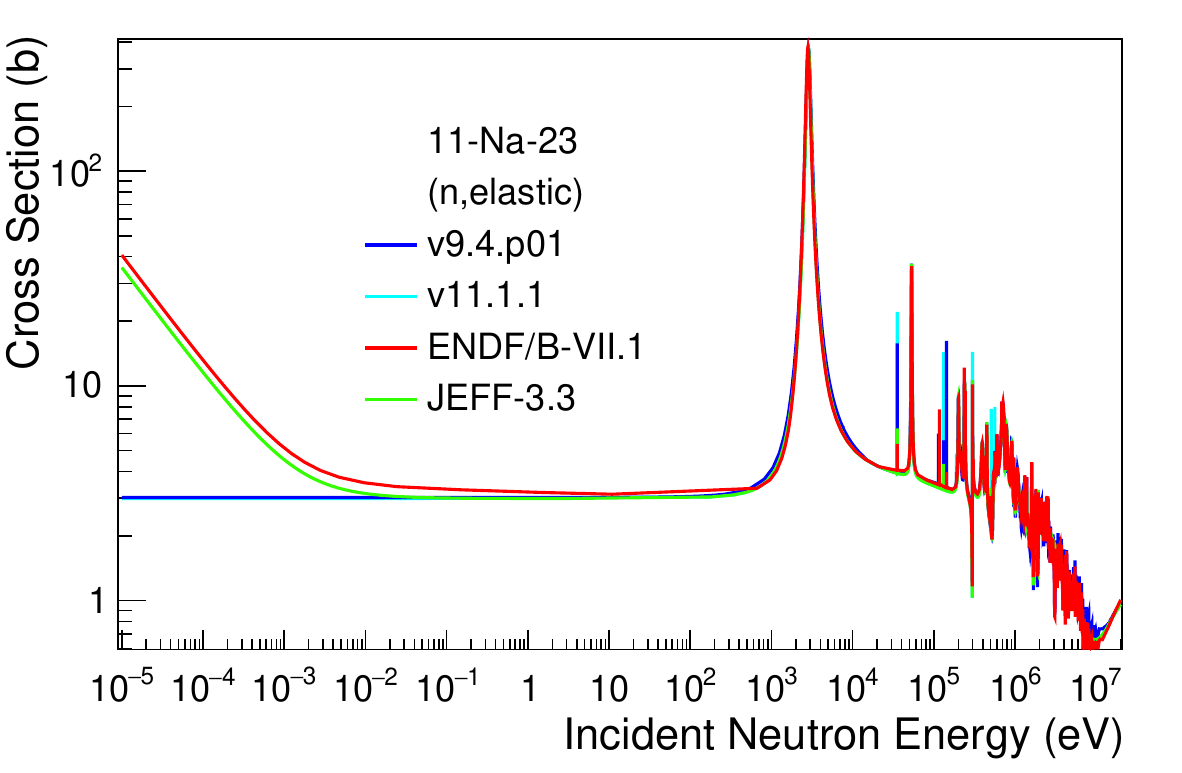}
        \caption{}
    \end{subfigure}
    \hspace{0.05\textwidth} 
    \begin{subfigure}[b]{0.45\textwidth}
        \centering
        \includegraphics[width=\textwidth]{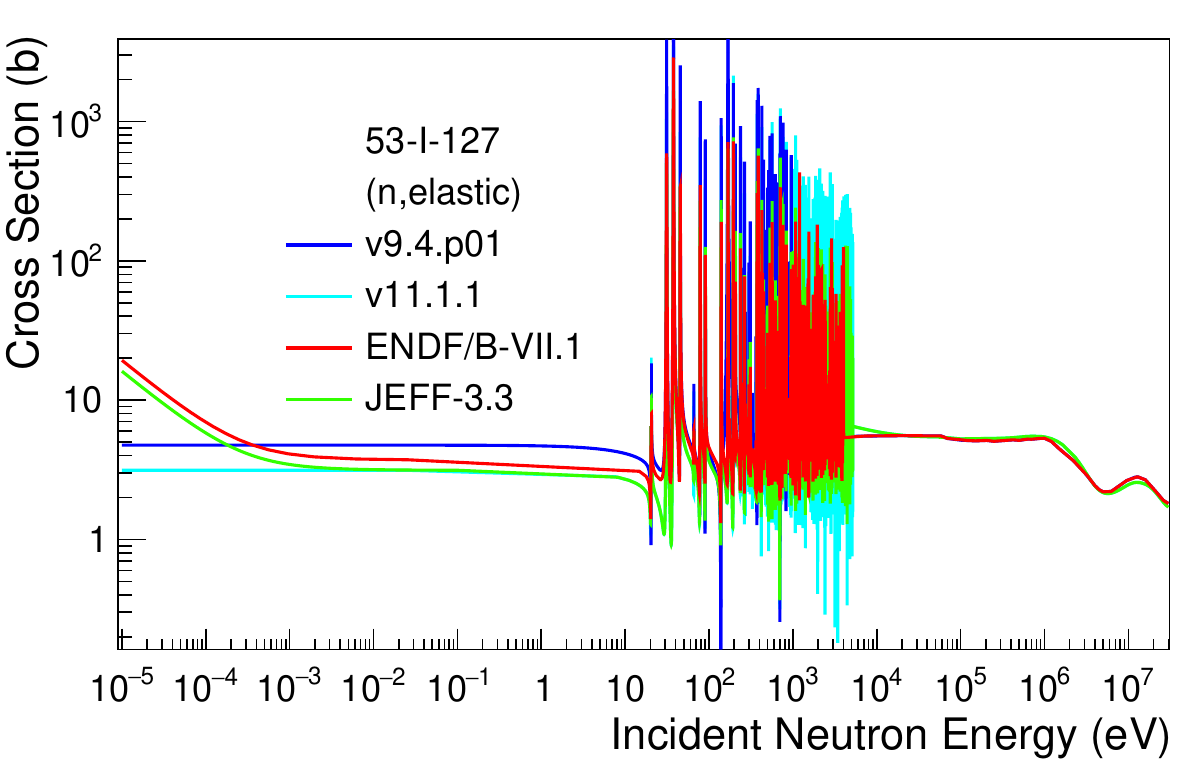}
        \caption{}
    \end{subfigure}
    \vspace{0.4cm} 
    \begin{subfigure}[b]{0.45\textwidth}
        \centering
        \includegraphics[width=\textwidth]{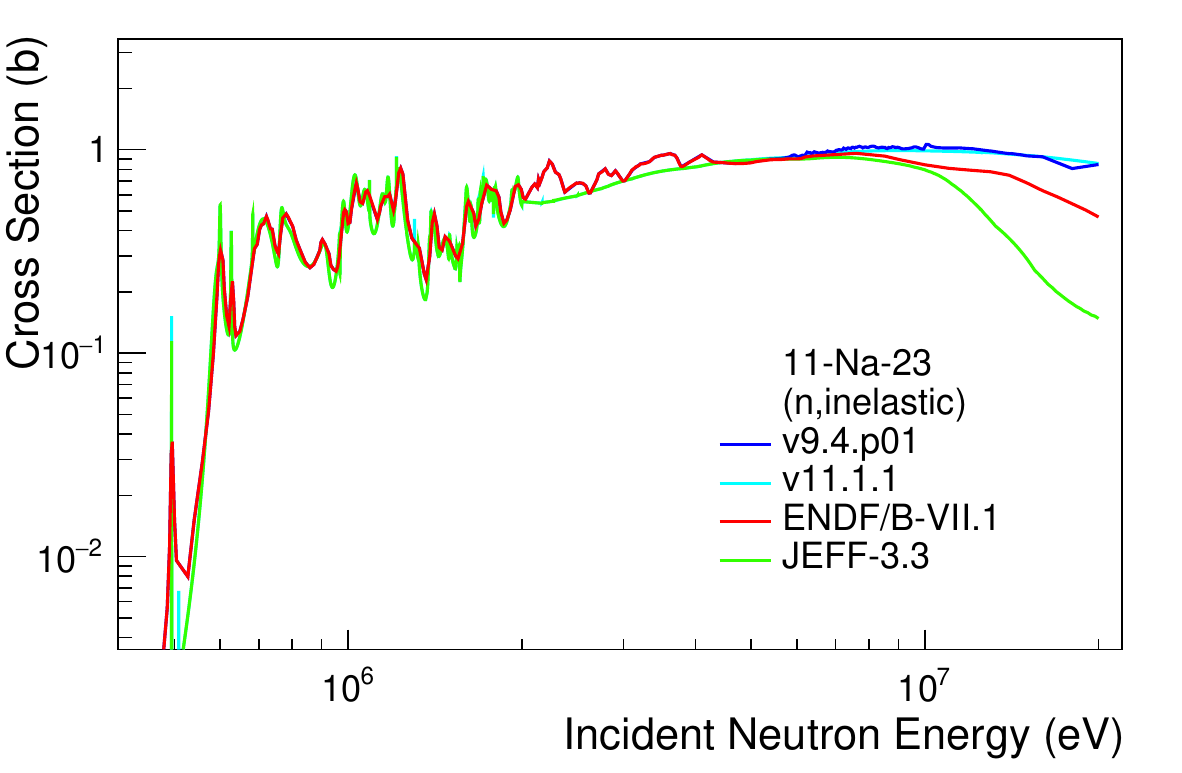}
        \caption{}
    \end{subfigure}
     \hspace{0.05\textwidth} 
    \begin{subfigure}[b]{0.45\textwidth}
        \centering
        \includegraphics[width=\textwidth]{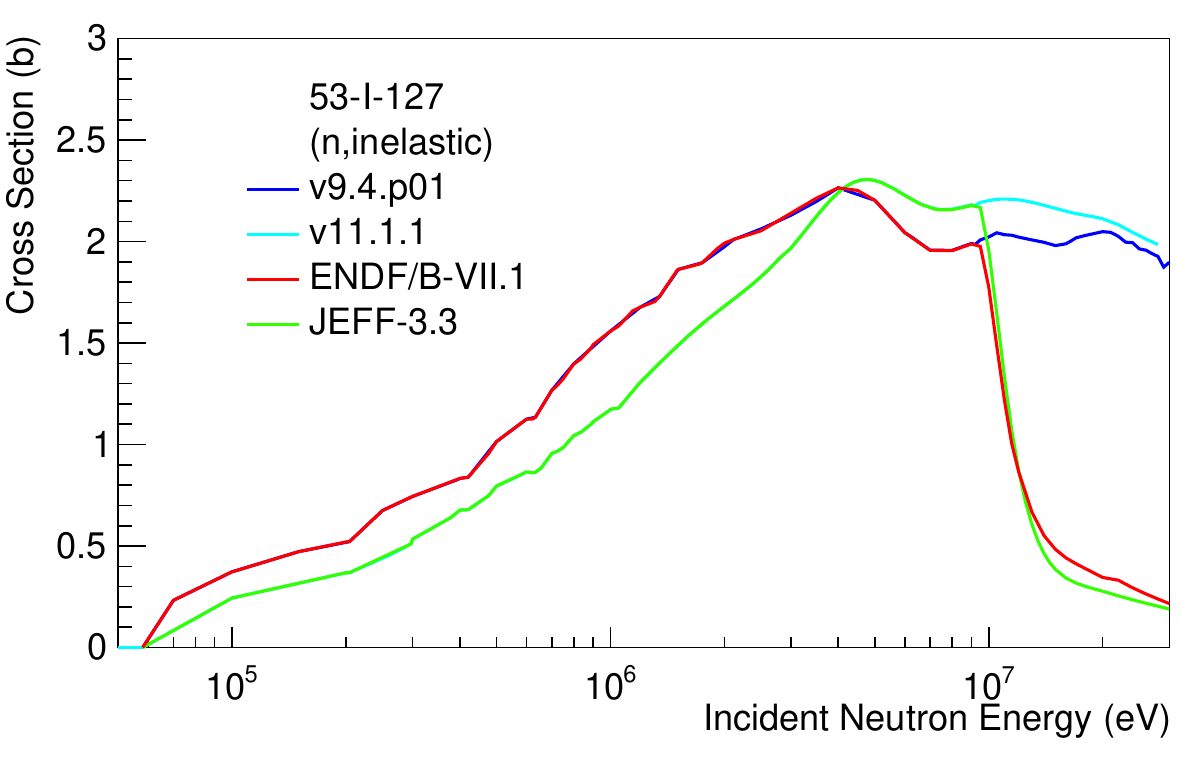}
        \caption{}
    \end{subfigure}
     \vspace{0.3cm} 
    \begin{subfigure}[b]{0.45\textwidth}
        \centering
        \includegraphics[width=\textwidth]{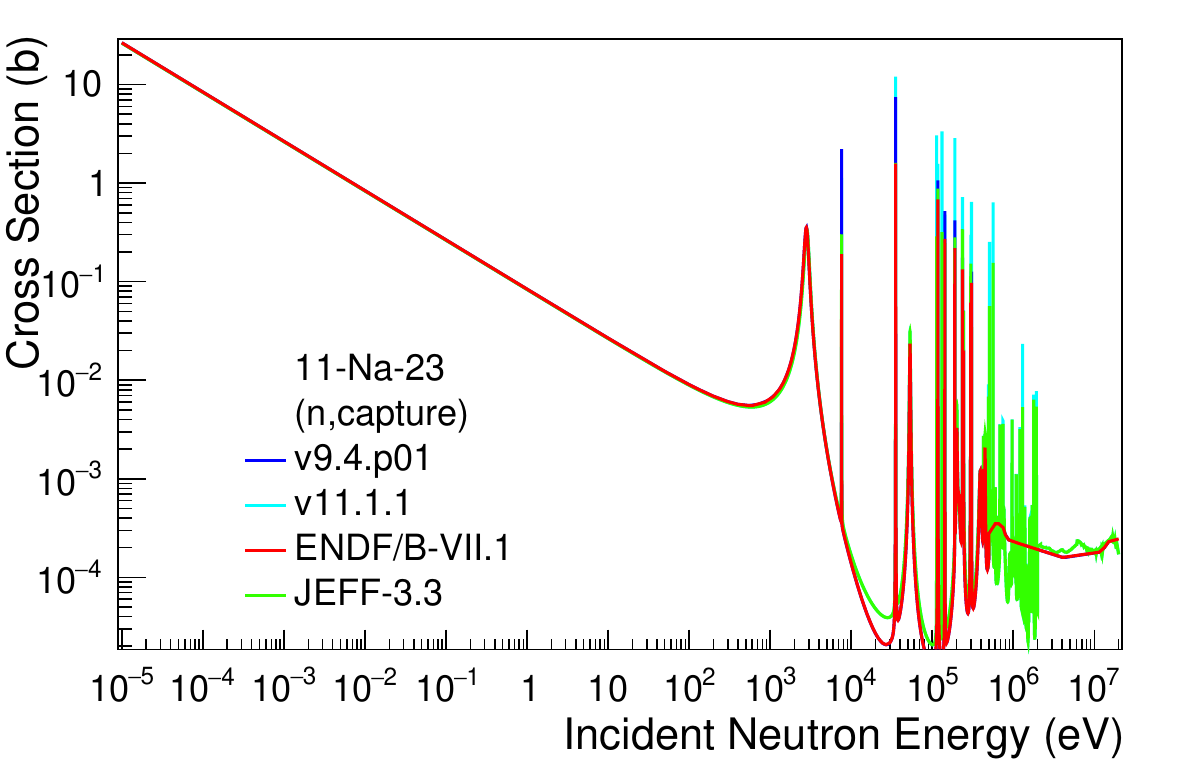}
        \caption{}
    \end{subfigure}
    \hspace{0.05\textwidth} 
    \begin{subfigure}[b]{0.45\textwidth}
        \centering
        \includegraphics[width=\textwidth]{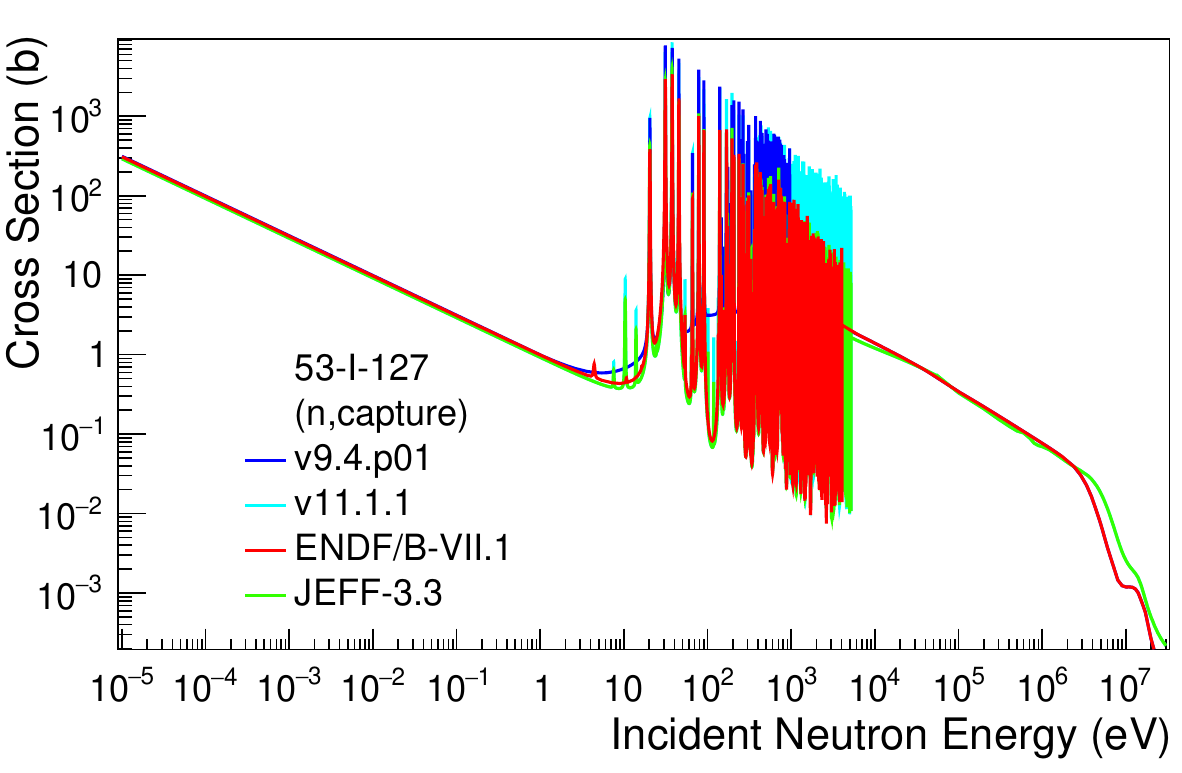}
        \caption{}
    \end{subfigure}

    \caption{Cross-sections for different neutron interactions in a NaI(Tl) target, derived from Geant4 v9.4.p01 (blue), Geant4 v11.1.1 (cyan), JEFF-3.3 (green), and ENDF/B-VII.1 (red). \textbf{(a and b)} Elastic scattering on Na and I nuclei; \textbf{(c~and~d)} Inelastic scattering on Na and I nuclei; \textbf{(e~and~f)} Neutron capture on Na and I nuclei.}
    \label{CSversiones}
\end{figure}

Since the treatment of elastic scattering directly impacts the evaluation of the QF, it is crucial to understand what has changed between the different nuclear data libraries used in the simulations, in order to assess which one best describes the ANAIS-112 data and whether there is a strong reason to favor one over the other. To address this, Figure \ref{CSversiones} presents a comparison of the elastic, inelastic, and capture cross-sections for both Na and I nuclei, using the two Geant4 versions employed in this study, along with the standard evaluated cross-section libraries ENDF/B-VII.1 and JEFF-3.3.

In general, the G4NDL3.14 data library relies on ENDF/B-VII.1, while G4NDL4.7 contains data primarily sourced from JEFF-3.3. Both ENDF/B-VII.1 and JEFF-3.3 are reference nuclear data libraries in the scientific community and are continuously validated across various applications. Nevertheless, as shown in Figure \ref{CSversiones}, the translation of the standard libraries into a format compatible with Geant4 in a linearly interpolable form is not a direct conversion but rather a custom evaluation carried out by the Geant4 collaboration. This is evident in the figure, where the cross-sections from version v9.4.p01 (v11.1.1) do not match exactly with those taken from the ENDF/B-VII.1 (JEFF-3.3) library, although they do follow a quite similar trend.

For elastic scattering, the cross-sections for sodium at 2 MeV show no significant variation between versions, with differences below 1\%. However, for iodine at the same energy, although in the figure the corresponding lines are not visible due to their exact overlap with those from standard evaluated cross-section libraries, the discrepancy is approximately 9\%, with reported values of 3.91 b and 4.32 b for versions v9.4.p01 and v11.1.1, respectively. This is consistent with the trend shown in Figure \ref{versioncomparion}, where the newer Geant4 version predicts a higher elastic scattering signal. Although this difference may appear small, it can lead to a noticeable variation in event rates due to the relatively large cross-sections.

It is worth highlighting that the Geant4 implementation of the versions under study does not fully account for the description of the 1/v region of the elastic cross sections, although recent releases are addressing the issue. However, thermal neutrons in the range of a few meV no longer produce a measurable signal in ANAIS. Therefore, the differences between versions and standard databases in this range are considered non-relevant for the purpose of analyzing energy deposition in the ANAIS-112 detectors.


Continuing towards higher energies in the energy spectrum, the 31.8 keV peak is primarily attributed to thermal neutron capture on iodine. The discrepancy in the capture cross-sections for iodine between versions is approximately 6\%, with reported values of 0.047~b and 0.044 b for versions v9.4.p01 and v11.1.1, respectively. Given the relatively small magnitude of these cross-sections compared to other processes involved, it is reasonable that no noticeable differences between versions are observed in Figure~\ref{versioncomparion}. Consequently, only the capture cross-section for \textsuperscript{128}I production is relevant for this peak, which requires additional corrections, as will be discussed in the following section.

On the other hand, the Geant4 v11.1.1 version results in a substantially lower rate for the inelastic peak, which arises from interactions involving both NR and ER (see Figure \ref{versioncomparion}). In this case, the inelastic cross-section for iodine at 2 MeV differs between libraries, with values of 1.97 b in v9.4.p01 and 1.68 b in v11.1.1, a difference of $\sim$ 14\%. This variation accounts for the higher amplitude of the inelastic peak observed in the older version.

Once the differences in cross-sections and their influence on the simulated spectrum have been analyzed and understood, the next step is to definitively select one version over the other, if possible. This requires validation through comparison with experimental data. Figure \ref{compareversioncondatos} shows the comparison between data and simulations using the two Geant4 versions in the medium- and high-energy regions. Data below 20 keV have been excluded from the plot to avoid a range where the QF plays a significant role. Above 20 keV, the QF variations do not affect significantly the spectral shape.

As can be seen, the inelastic peak is not compatible with ANAIS-112 data when using Geant4 v11.1.1, while it is properly reproduced by the simulation using the v9.4.p01 version. In the high-energy region, the 295 keV peak is clearly incompatible with ANAIS-112 data, although the rest of the spectrum is well described by both simulations. It is important to note that this agreement is achieved after applying the correction to the neutron capture cross-section for $^{128}$I production, which will be discussed in the following subsection.

While newer versions are typically expected to include updates to nuclear data, they do not necessarily guarantee improved accuracy, as is often observed. Moreover, NaI(Tl) is not a commonly studied target in neutron applications, which may explain why these changes have not been extensively validated by the scientific community. In light of the above, since the results from G4NDL3.14 show a better agreement with the measurements for a NaI(Tl) target in terms of inelastic de-excitation processes, this library has been used to describe the neutron physics and obtain the results presented in this chapter. However, the effect in the determination of the QF of the choice of the Geant4 version will be considered as a systematic uncertainty.



\subsubsection{Adjusting the $^{128}$I neutron capture cross section}\label{I128prodsec}

According to the ANAIS-112 simulations, detection rate in the NaI(Tl) detectors
resulting from the irradiation of the $^{252}$Cf source placed outside the lead
shielding, but inside the neutron shielding, is dominated by the interaction of
neutrons, while contributions from electrons and gammas only become relevant
in specific energy ranges (see Figure \ref{simdistribucion}). Below 100 keV, in particular, two interesting spectral features are the peak at 31 keV from
the decay of $^{128}$I produced by neutron capture in $^{127}$I and the $^{127}$I inelastic peak. These features can aid in disentangling the uncertainties associated with both the QFs and the cross-sections implemented in the simulation; however, they do not provide direct information about the elastic scattering cross-sections, which are more relevant for QF determination.

\begin{figure}[b!]
\begin{center}
\includegraphics[width=0.65\textwidth]{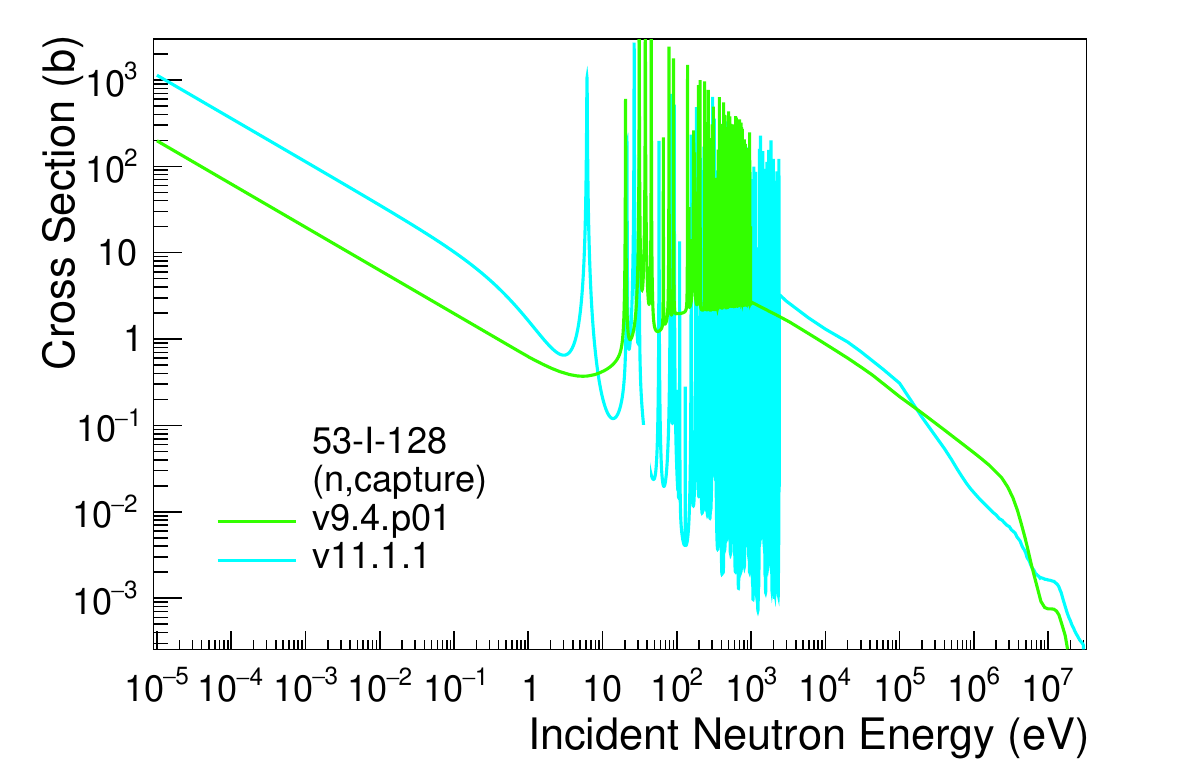}

\caption{\label{i128versioncs} Neutron capture cross sections for $^{128}$I production as implemented in Geant4 versions v9.4.p01 (green) and v11.1.1 (cyan).}
\end{center}
\end{figure}

The comparison between data and simulation revealed a significant excess in the latter contribution, not only around the 31 keV region but also above 1 MeV, energy ranges that are no longer affected by the choice of the QF model, neither for Na nor for I. This excess
was consistently observed across all detectors, on all faces, and in both Geant4
versions, but only in the single-hit spectrum, where the
decay of $^{128}$I leaves a signature. 

$^{128}$I decays via $\beta^-$ emission (93.1\%) to $^{128}$Xe and via EC or $\beta^+$ emission (6.9\%) to $^{128}$Te. The total decay energy for the $\beta^-$ channel is 2.14 MeV, resulting in a characteristic continuous $\beta$ spectrum. In the case of EC decay, as already stated, a distinct peak at 31.8~keV, corresponding to the K-shell binding energy of Te, is expected due to atomic de-excitation following the EC process.

The observed discrepancy indicates an overestimation of
the neutron capture cross section for the production of $^{128}$I. While elastic cross
sections are extensively validated, neutron capture measurements and their verification
remain challenging. Recent experimental results for this cross-section
in particular suggest lower cross-section values than previous estimates \cite{gandhi2021neutron}.

Figure \ref{i128versioncs} shows the neutron capture cross sections for $^{128}$I production as implemented in Geant4 versions v9.4.p01 and v11.1.1. As can be seen, there is again a clear discrepancy between the cross sections of both versions, as alredy highlighted in the previous section. In particular, at 2 MeV, the cross section in version v9.4.p01 is 0.030 b, whereas in v11.1.1 it is 0.009 b, corresponding to a factor of $\sim$ 70\% between them.


To accurately reproduce the ANAIS-112 data, the neutron capture cross-section for the production of $^{128}$I has been adjusted for both Geant4 versions v9.4.p01 and v11.1.1. The adjustment was based on the spectral
shape of the energy deposited following the decay of $^{128}$I, which, due to its
half-life, 24.99 min, does not overlap with elastic or inelastic neutron scattering events. A
fit restricted to the high-energy region ([200-1600] keV) was performed using the RooFit tool \cite{verkerke2006roofit}, introducing a single free-floating parameter to account for the excess of $^{128}$I in
the simulation. This parameter was determined in a simultaneous fit to all nine detectors
of ANAIS-112, as any misinterpretation of the capture cross section should result
in a consistent correction across all evaluated data.

According to the adjustment of the observed contribution from
$^{128}$I in ANAIS data, the Geant4 cross-section is overestimated by (37.0~$\pm$~0.3)\% in v9.4.p01 and (20.9~$\pm$~0.4)\% in v11.1.1. This correction factor does not provide a direct measurement of the cross-section at a given energy, but a global scaling of the convolution of the capture cross-section and the $^{252}$Cf neutron spectrum.

\begin{figure}[t!]
\begin{center}
\includegraphics[width=0.49\textwidth]{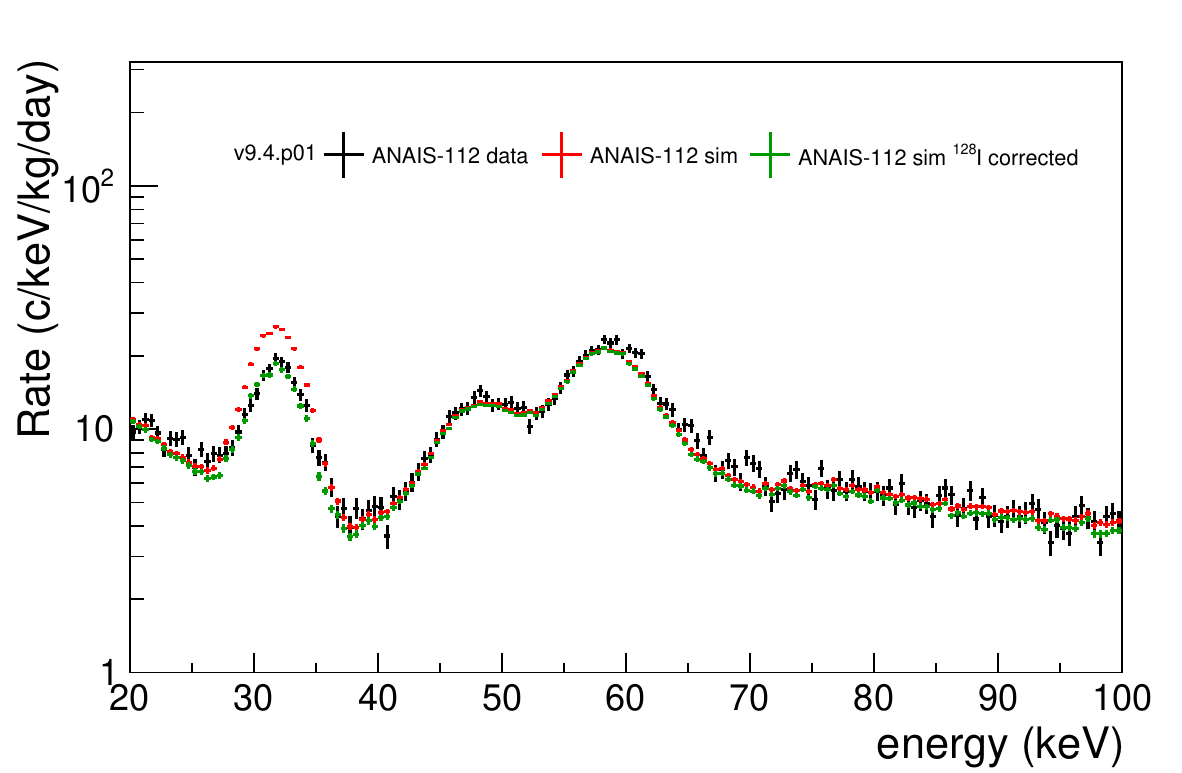}
\includegraphics[width=0.49\textwidth]{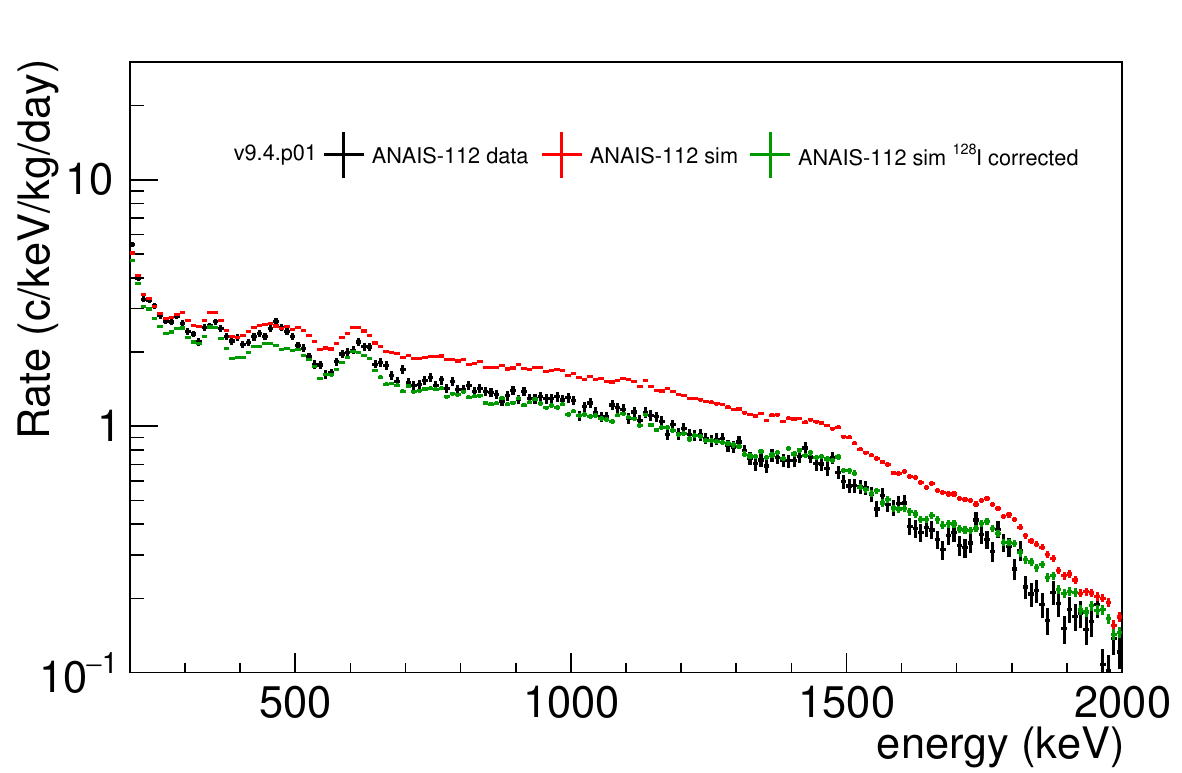}
\includegraphics[width=0.49\textwidth]{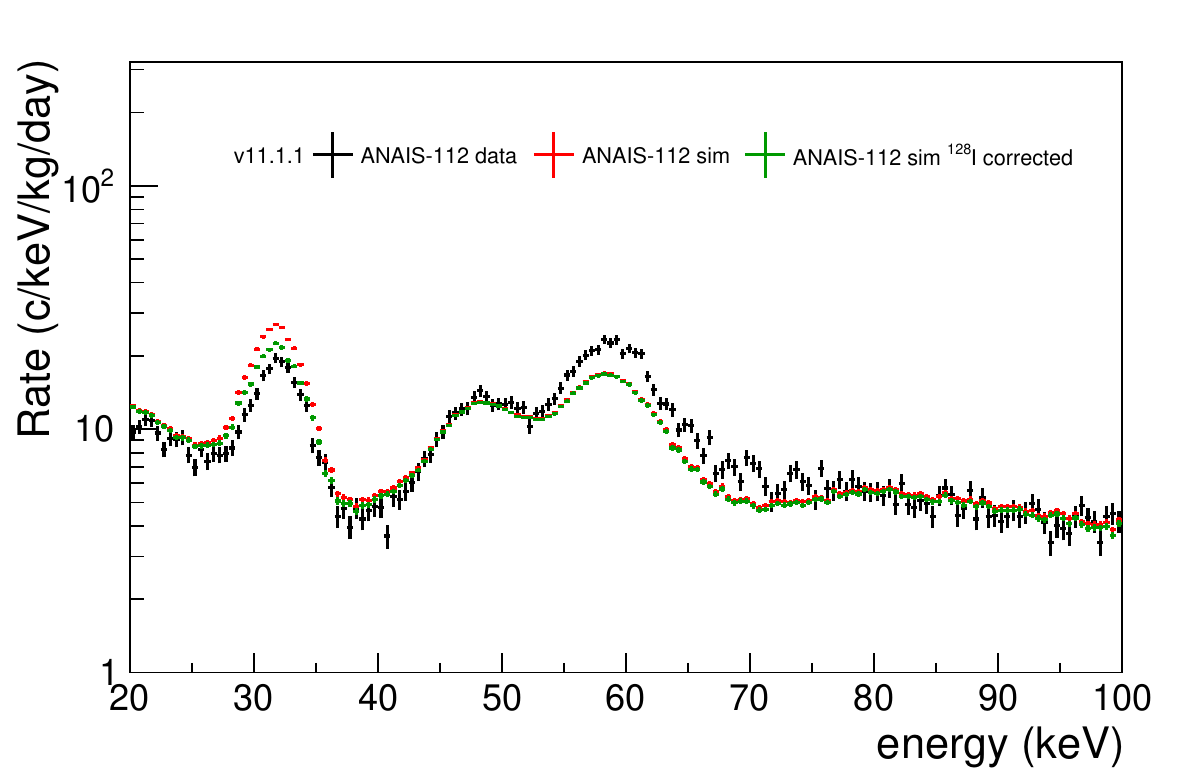}
\includegraphics[width=0.49\textwidth]{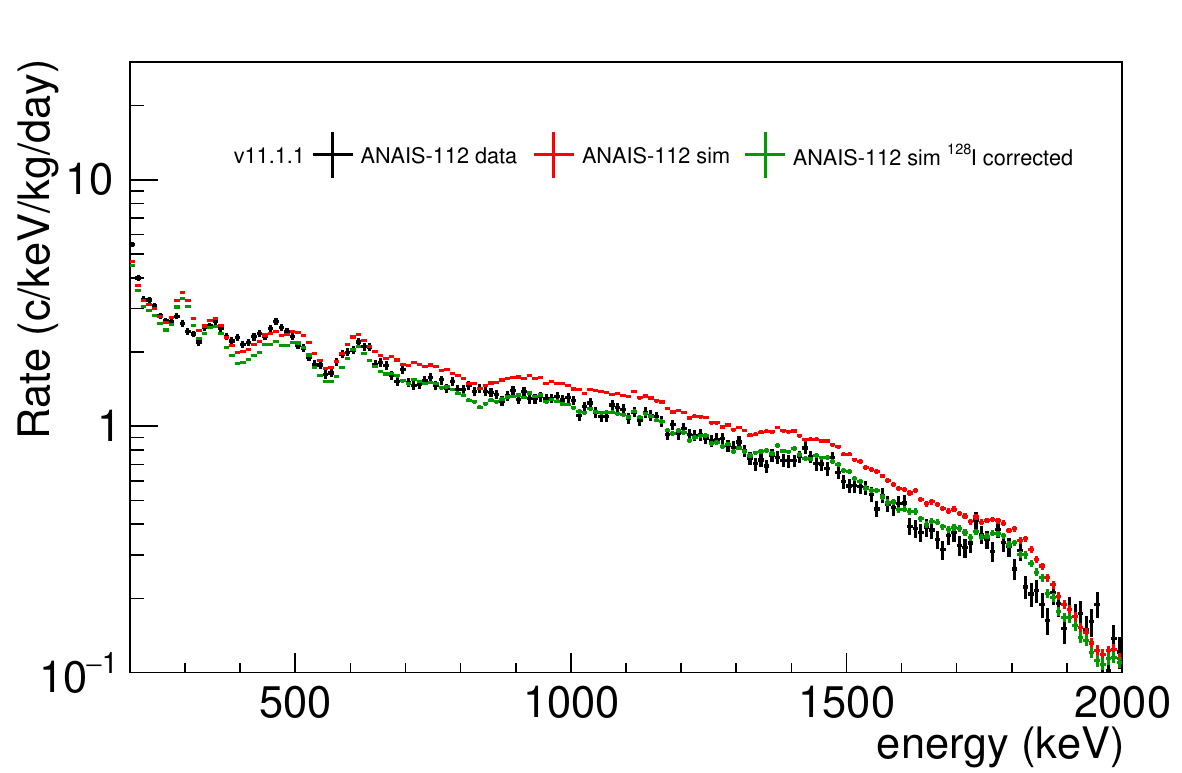}

\caption{\label{resultadosmodificari128} Comparison between neutron calibration data (black) and simulation results before (red) and after (green) applying the $^{128}$I cross-section adjustment. Data and simulation correspond to single-hits spectra. The top panels show the simulation using Geant4 v9.4.p01, while the bottom panels correspond to v11.1.1. ANAIS(1) QF${_\textnormal{Na}}$ and QF${_\textnormal{I}}$=0.06 have been applied in both simulations. The left panels display the medium-energy region, and the right panels show the high-energy region.}
\end{center}
\end{figure}

The agreement between data and simulation before and after applying the $^{128}$I correction to the simulation is shown in Figure \ref{resultadosmodificari128}, with the results for the medium and high-energy regions displayed for v9.4.p01 (upper panel) and v11.1.1 (lower panel). As can be seen, after the correction, a good agreement is achieved between the experimental data and the simulation for v9.4.p01, not only in the high-energy region where the fit was performed, but also in the 31 keV peak, supporting the validity of the hypothesis.

The application of the analogous adjustment for version 11.1.1 is shown in the bottom panels of the same figure. As can be observed, after the correction, a good agreement between data and simulation is achieved in the high-energy region where the fit is performed. However, when the same correction factor for $^{128}$I, obtained using the same protocol as for the previous version, is applied to the low-energy region, the $^{128}$I peak is not reproduced with same accuracy. While the decay spectrum of $^{128}$I itself has been verified to be consistent between the two Geant4 versions, the observed discrepancy might be explained by the higher elastic scattering cross section for iodine, as demonstrated in the previous section, which causes the exponential contribution to add up on the left side of the 31 keV peak, resulting in a slight distortion. Again, as displayed in the bottom left panel, the amplitude of the inelastic peak is not reproduced at all with the newer version.


\begin{figure}[t!]
\begin{center}
\includegraphics[width=0.7\textwidth]{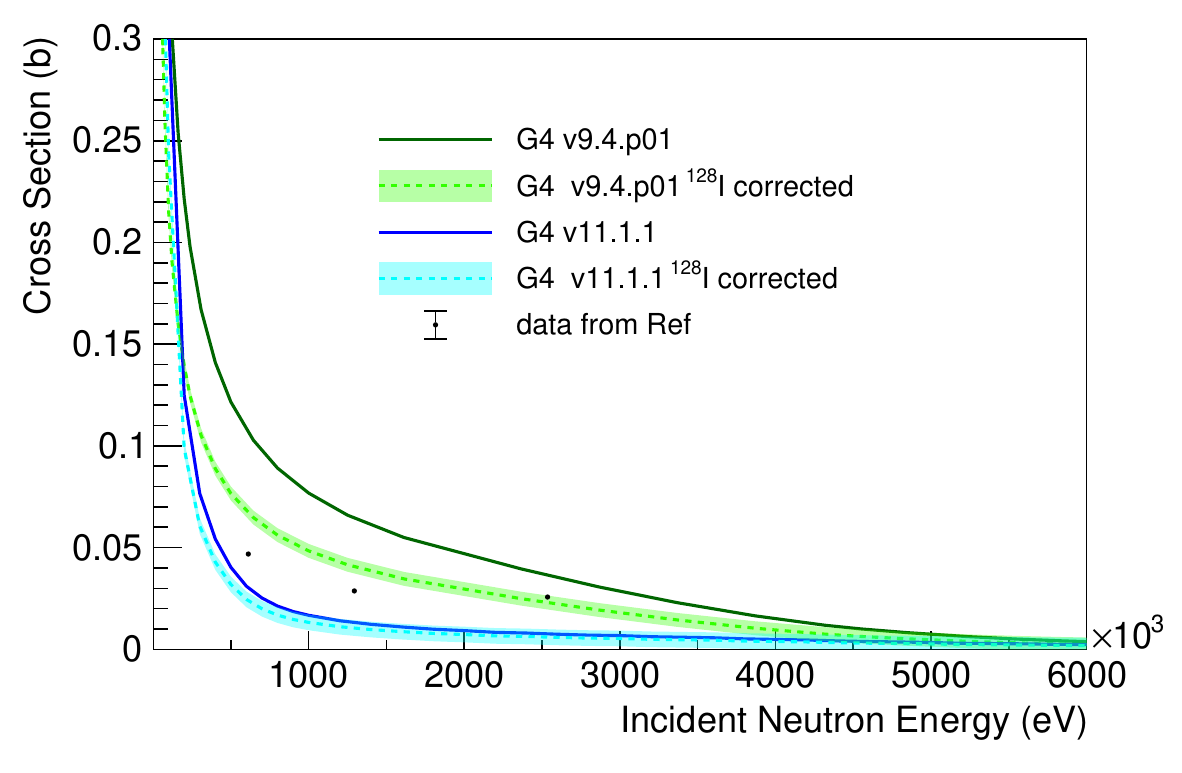}

\caption{\label{comparisoncs}   Comparison of the neutron capture cross section for the production of $^{128}$I. Shown are the cross sections implemented in Geant4 version 9.4.p01, the same corrected by the scaling factor determined in this study through the fitting procedure, and analogously, the cross section from Geant4 version 11.1.1 together with its correction. For comparison, the experimentally measured cross sections from \cite{gandhi2021neutron} are also included.
}
\end{center}
\end{figure}

Figure \ref{comparisoncs} presents the comparison of the neutron capture cross section for the production of $^{128}$I, including the cross section implemented in Geant4 version v9.4.p01, the same corrected by the scaling factor determined in this study through the fitting procedure, and analogously, the cross section from Geant4 version v11.1.1 together with its correction. For comparison, the experimentally measured cross sections from \cite{gandhi2021neutron} are also included. 


Following the measurements reported in \cite{gandhi2021neutron}, and taking their second experimental point at 1.3 MeV as a reference, the cross-section for the production of \textsuperscript{128}I in Geant4 is found to be 2.24 times larger in version v9.4.p01, and 0.45 times smaller in version v11.1.1, relative to the measured value. Interestingly, despite the cross-section in v11.1.1 being lower than experimental data at that energy, the neutron capture cross-section in this version appears to be overestimated by about 21\%, as inferred from the fitting procedure carried out in this study.

This seemingly contradictory result can be explained by the distinct energy dependence of the cross-sections in each Geant4 version, as illustrated in Figure \ref{i128versioncs}. The curves intersect around a few hundred keV, indicating that version v11.1.1 exhibits a higher cross-section at lower neutron energies. Consequently, it is the integral of the cross-section over the full energy spectrum that results in the overestimation reported in this study for v11.1.1. Because the analysis presented here does not constrain the shape of the cross-section, but only determines the total excess needed to reproduce the experimental data of ANAIS, it is not possible to directly relate the observed overestimation to a specific value of the cross-section at a given energy. Nonetheless, when considering both the experimental measurements and the general behavior of the cross-section with energy, the shape given by version v9.4 appears to be more compatible with measurements.

In summary, this work has identified a clear overestimation of the capture cross section for the production of $^{128}$I. However, the precise energy dependence of this cross section has not been investigated. In the adopted approach, the 
dependences considered in each version of Geant4 have been kept scaled by a global factor. The better agreement with version v9.4.p01 after correction, as well as the consistency of the correction factor across both high and low energies, constitutes an additional argument in favor of version v9.4.p01 over v11.1.1, in addition to the improved description of inelastic processes shown in the previous section.

\subsection{QF models evaluated in this study}\label{QFmodels}

The neutron simulation provides the energy deposited in the ANAIS detectors in terms of NR energy, as long as neutrons produce NRs. However, for comparison with ANAIS-112 data, this energy must be converted into the electron-equivalent energy scale. This step is crucial because ANAIS is calibrated using gamma/electron populations, and its data are therefore expressed in that scale.



Figures \ref{NaQFcomparison} and \ref{IQFcomparison} show the current experimental status of QF measurements for Na and I, respectively. As already mentioned, for Na, multiple measurements of the QF are available \cite{Bernabei:1996vj,Spooner,Tovey:1998ex,Gerbier:1998dm,Chagani,Collar,Xu,Joo:2018hom,cintas2024measurement,PhysRevD.110.043010}. However, there is some dispersion, especially at low energies, where the available data are limited and associated with large uncertainties. In the case of I recoils, the measurements are even more limited. Thus, the ANAIS-112 neutron simulation can represent a valuable tool for testing the consistency of different QF models or measurements, complementing the results provided by monochromatic source measurements.



\subsubsection{QF\textsubscript{Na} models}

\begin{figure}[b!]
\begin{center}
\includegraphics[width=0.8\textwidth]{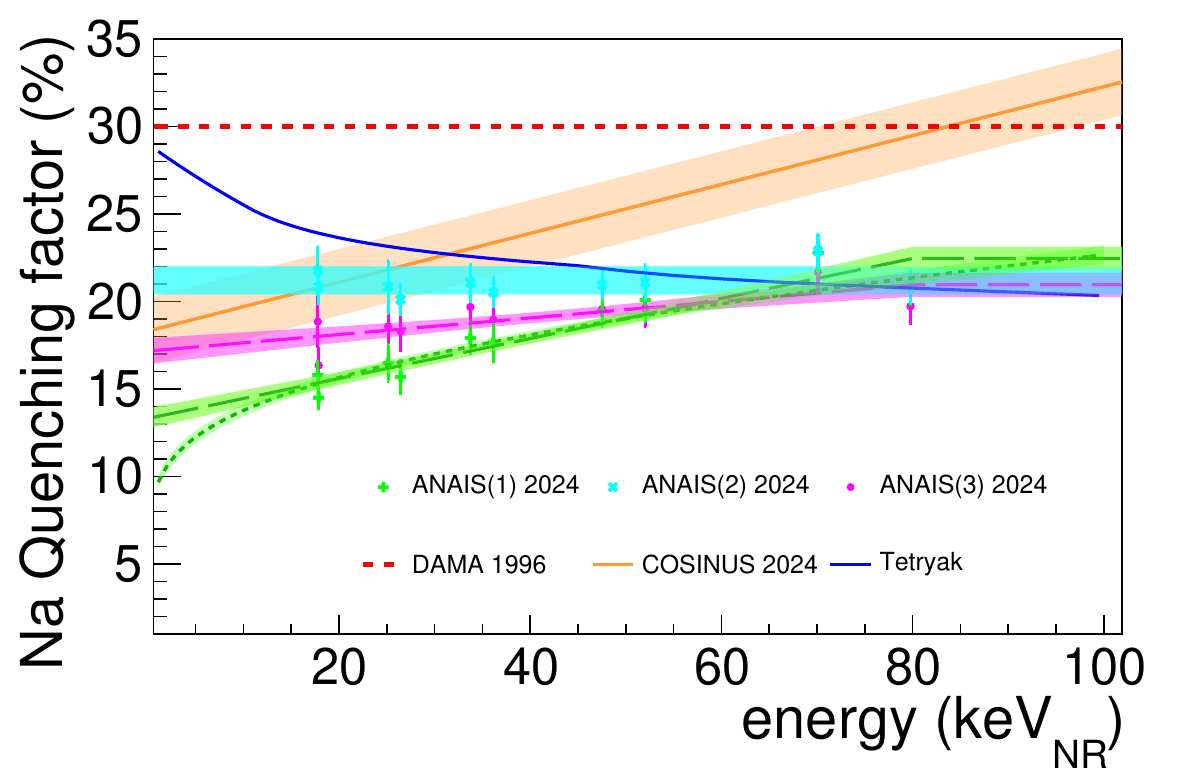}

\caption{\label{NaQFmodels} QF$_{\textnormal{Na}}$ models evaluated in this thesis. The DAMA/LIBRA result \cite{Bernabei:1996vj} (red), together with the three results for the ANAIS crystals following \cite{cintas2024measurement,phddavid}: ANAIS(1) (green), ANAIS(2) (cyan), and ANAIS(3) (magenta). For ANAIS(1) QF\textsubscript{Na}, two models will be considered: a linear fit (represented by long-dashed lines) and the energy dependence predicted by the modified Lindhard model (represented by short-dashed lines). For ANAIS(3) QF\textsubscript{Na}, only the linear fit will be taken into account. Additionally, the COSINUS result \cite{PhysRevD.110.043010} (orange), and the Tetryak model \cite{tretyak2010semi} (blue), are also considered. The figure displays 1$\sigma$ uncertainty bands for the experimental measurements and their corresponding models, except for the DAMA/LIBRA result, which was reported without associated uncertainties, and the Tretyak model, which is a semi-empirical approximation.}
\end{center}
\end{figure}

Figure \ref{NaQFmodels} illustrates the QF\textsubscript{Na} models considered in this study.

A primary output of the neutron simulation is to confirm or refute the QF estimation performed by DAMA/LIBRA \cite{Bernabei:1996vj}, which would be an interesting result, as DAMA/LIBRA determined the QFs of their crystals using the same methodology as in this study, that is, exposing them to $^{252}$Cf sources. 

Furthermore, this study aims to validate the QF$_{\textnormal{Na}}$ results obtained for crystals of the same batch that some of the ANAIS-112 crystals using a monochromatic source \cite{cintas2024measurement,phddavid}. Based on those measurements, three QF$_\text{Na}$ input models, ANAIS(1), ANAIS(2), and ANAIS(3), will be considered in this analysis. The reader is referred to Section \ref{TUNL} for a detailed description of the calibration strategy on which these results are based.

\begin{figure}[b!]
\begin{center}
\includegraphics[width=0.49\textwidth]{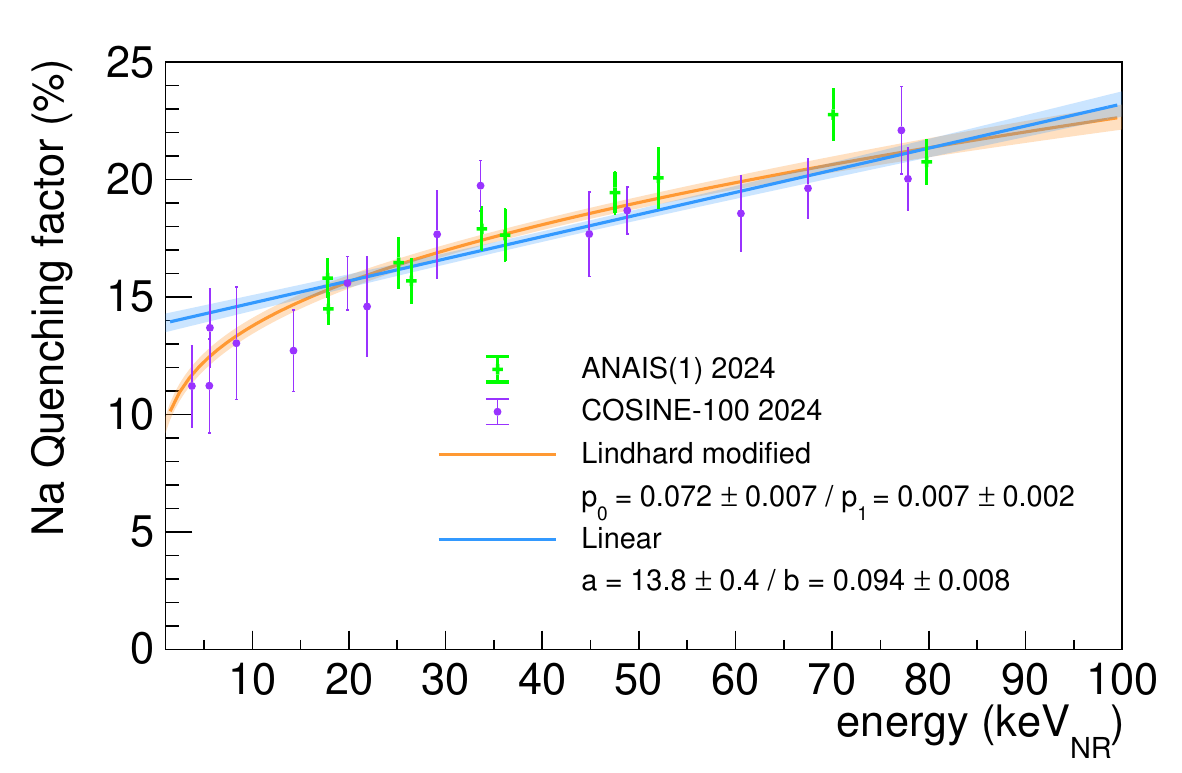}
\includegraphics[width=0.49\textwidth]{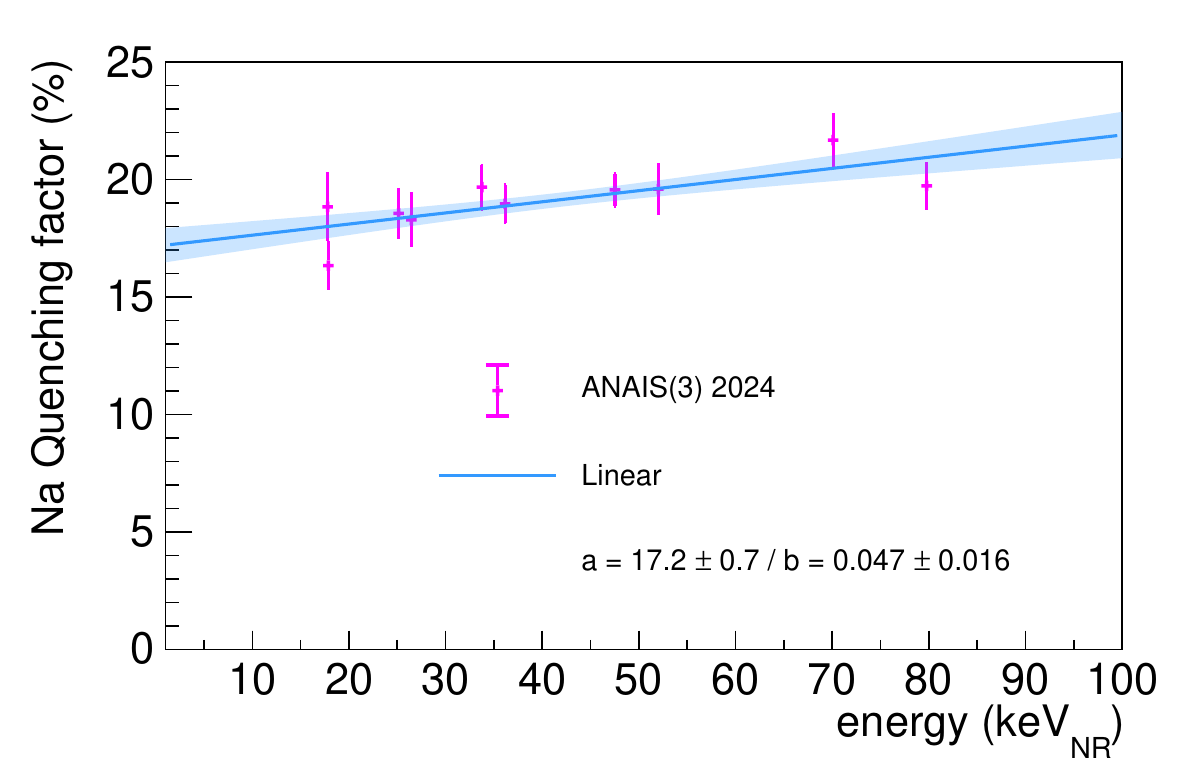}

\caption{\label{QFfitLind} Energy-dependent QF$_{\textnormal{Na}}$. \textbf{Left panel:} ANAIS(1) (green) and COSINE-100 (violet) QF\textsubscript{Na}, together with the linear fit and the fit based on the modified Lindhard model \cite{chagani2008measurement}. \textbf{Right panel:} ANAIS(3) QF\textsubscript{Na}, along with the linear fit. The plots show the fit results, with shaded bands representing the 1$\sigma$ uncertainty derived from the fits.}
\vspace{-0.5cm}
\end{center}
\end{figure}

To model the energy dependence of QF\textsubscript{Na} in ANAIS-112, two approaches are considered. The first is a linear fit to the experimental data points, shown in blue in Figure \ref{QFfitLind}. As represented by long-dashed lines in Figure~\ref{NaQFmodels}, the linear fit is constrained above 80 keV\textsubscript{NR} to avoid extrapolations leading to QF\textsubscript{Na} values greater than 0.4, which are not supported by any existing measurements, including those from DAMA/LIBRA. Since the ANAIS data, as well as most other measurements, do not extend beyond that energy, a flat extrapolation is imposed above 80 keV\textsubscript{NR}, assuming a constant QF value. Several boundary values were tested, and the effect on the results was found to be minimal.

The second approach involves using a modified Lindhard model, corresponding to the short-dashed lines in Figure~\ref{NaQFmodels}. As seen in Figure \ref{QFtheory}, the direct application of the original Lindhard theory fails to reproduce the measured QF\textsubscript{Na} and QF\textsubscript{I} values in NaI(Tl) scintillators, pointing towards higher QF\textsubscript{Na} values compared to available measurements. To address this discrepancy, a modified version of the Lindhard model is here used. Following \cite{chagani2008measurement}, the modified model maintains the original functional form of the Lindhard theory (see Equation~\ref{eqlind}), but introduces two free parameters $p_0$ and $p_1$: \(k = p_0\), and a rescaled reduced energy \(\epsilon = p_1 E_{\text{NR}}\).

A key aspect of the modified Lindhard modelling is its behavior at low energies. However, the ANAIS data points extend only down to 18 keV\textsubscript{NR}, and therefore, attempting the fit using only ANAIS data would yield arbitrary and unreliable results in the low-energy region. Referring back to Figure \ref{NaQFcomparison}, it is clear that the results of ANAIS(1) QF\textsubscript{Na} are fully compatible with those reported by COSINE-100 \cite{Joo:2018hom}. The latter data set is particularly relevant, as it not only includes several measurements at low energies but also corresponds to crystals from the same manufacturer of the ANAIS detectors, Alpha Spectra. Given the observed compatibility, the ANAIS(1) and COSINE-100 data are combined into a single dataset to perform the fit using the modified Lindhard model.

The left panel of Figure \ref{QFfitLind} shows the fit to the ANAIS(1)+COSINE-100 QF\textsubscript{Na} data. The fit performs well, and the resulting parameters are consistent with those reported in \cite{ko2019comparison}. Both the linear and modified Lindhard models provide comparable results for ANAIS(1) data overall, though significant differences emerge at low energies. In particular, the linear model predicts a QF\textsubscript{Na} of approximately 14\% at low energies, while the modified Lindhard model yields values around 10\% in the same region. The impact of this steeper decline at low recoil energies will be further examined in this work through detailed data–simulation comparison results using both QF\textsubscript{Na} modellings.

An analogously modified Lindhard fit was attempted for the ANAIS(3) QF\textsubscript{Na}. In this case, the COSINE-100 \cite{Joo:2018hom} dataset is not combined due to a lack of sufficient compatibility, which would not justify a joint analysis. When fitting the ANAIS(3) data alone, the results obtained for the modified Lindhard parameter $p_1$ were not physically meaningful, with values compatible with zero or negative. This is due to the fact that, in the absence of data at lower energies, attempting this fit is not meaningful. As a result, the modified Lindhard model is not applicable in this case, and only the linear modelling approach is considered for ANAIS(3).

Furthermore, the neutron simulation can be used to study the viability of other results or models. In this context, referring back to Figure \ref{NaQFcomparison}, the entire measurement region using monochromatic sources is generally consistent with the ANAIS-112 measurements. Nonetheless, it would be insightful to compare the results from COSINUS \cite{PhysRevD.110.043010}. Although these measurements are conducted at temperatures on the order of mK, the methodology closely resembles that of the ANAIS onsite neutron calibrations: it uses neutrons from a continuous source, not monoenergetic, but COSINUS measurement includes NR discrimination. It can be observed that its QF$_{\textnormal{Na}}$, although energy-dependent, is slightly higher than that of other measurements. 


On the other hand, semi-empirical models such as the one proposed by Tretyak \cite{tretyak2010semi} can also be considered. This model was selected because it does not agree with recent measurements from monochromatic neutron sources, including those obtained by ANAIS, as it predicts a decreasing QF\textsubscript{Na} with energy. Therefore, it is of particular interest to investigate whether neutron simulations employing this model for QF\textsubscript{Na} can be excluded when compared to the ANAIS data. When a fit was attempted using the Tretyak functional form on the energy-dependent ANAIS QF\textsubscript{Na} results, the fit failed to converge and is thus not presented here. As a result, the Tretyak model tested in this work employs a value of $k_B = 6.5 \times 10^{-3}$~g/(MeV cm$^2$), as suggested in \cite{tretyak2010semi}, which reproduces the experimental results reported in \cite{chagani2008measurement}.




\subsubsection{QF\textsubscript{I} models}
\begin{figure}[t!]
\begin{center}
\includegraphics[width=0.7\textwidth]{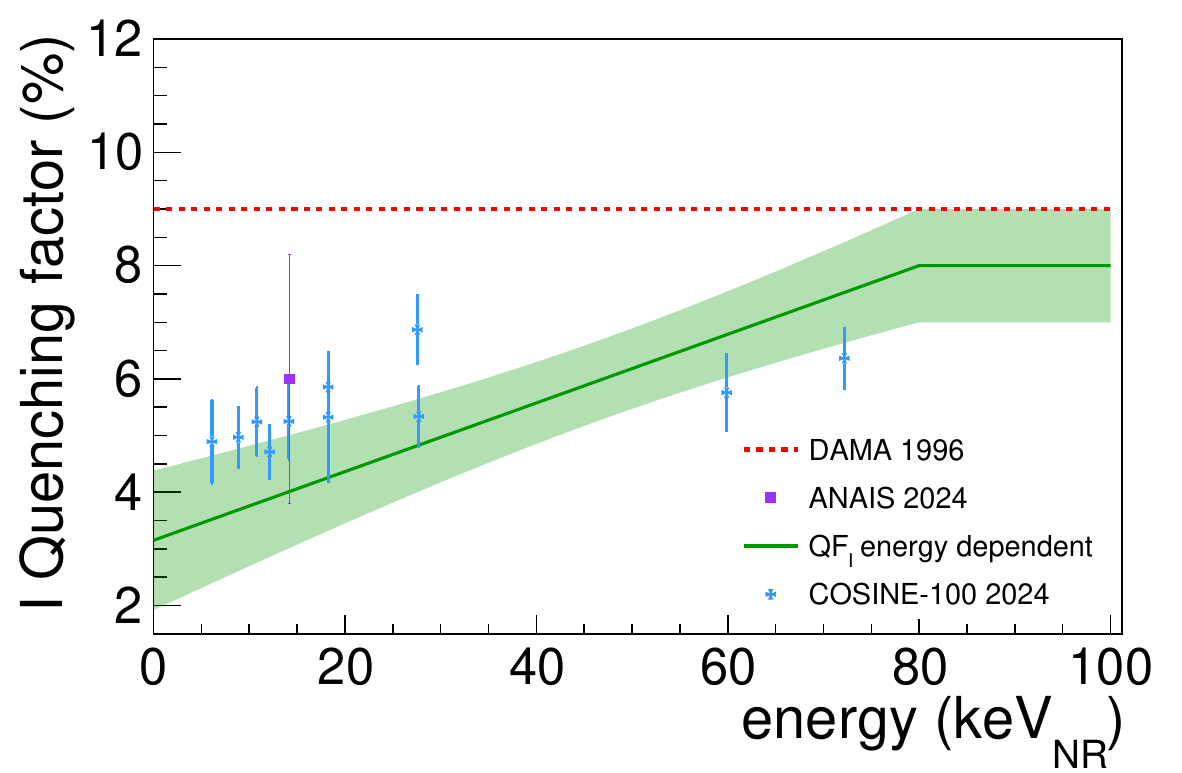}

\caption{\label{IQFmodels}QF$_{\textnormal{I}}$ models evaluated in this thesis. The DAMA/LIBRA result \cite{Bernabei:1996vj} (red), together with the result obtained using the neutron monochromatic source following \cite{cintas2024measurement,phddavid} (violet). Additionally, an energy-dependent QF$_{\textnormal{I}}$ model proposed in this work, compatible with the ANAIS-112 measurements, will be also evaluated. The QF\textsubscript{I} measurements from COSINE-100 (blue) are also presented \cite{Joo:2018hom}, demonstrating their compatibility with the considered energy dependence. The figure displays 1$\sigma$ uncertainty bands for the experimental measurements and the energy-dependent QF$_{\textnormal{I}}$ model, except for the DAMA/LIBRA result, which was reported without associated uncertainties.}
\end{center}
\end{figure}

Figure \ref{IQFmodels} displays the QF\textsubscript{I} models considered in this study.

This study considers the QF$_{\textnormal{I}}$ estimation of DAMA/LIBRA~\cite{Bernabei:1996vj}, the measured at TUNL for ANAIS-like crystals \cite{cintas2024measurement,phddavid}, and an energy-dependent QF$_{\textnormal{I}}$ model proposed in this work, which is compatible with the ANAIS results and follows a trend similar to the QF$_{\textnormal{Na}}$. Given the large uncertainty in the QF$_{\textnormal{I}}$ estimates, an energy dependence cannot be excluded, especially considering the experimental difficulty of measuring such low-energy deposits due to threshold limitations. Therefore, an energy-dependent QF$_{\textnormal{I}}$ model has been introduced, assumed to decrease from 8\% to 4\% over the energy range of 80 keV$_{\textnormal{NR}}$ to 14 keV$_{\textnormal{NR}}$. As done for QF\textsubscript{Na}, the QF\textsubscript{I} is constrained above 80 keV\textsubscript{NR} to avoid extrapolations leading to QF\textsubscript{I} values greater than 0.09.


As shown in Figure~\ref{IQFmodels}, the QF\textsubscript{I} values from COSINE-100 are fully compatible with the QF\textsubscript{I} measured for the ANAIS crystals. Thus, it is also worth noting that a modified Lindhard model was considered for the QF\textsubscript{I}, attempting a fit to the QF\textsubscript{I} values reported by COSINE-100 \cite{Joo:2018hom}.  However, the fit again yielded values of $p_1$ compatible with zero, including unphysical negative values. Therefore, introducing an energy-dependent Lindhard-like behavior for QF\textsubscript{I} was discarded. Nonetheless, the COSINE-100 QF\textsubscript{I} measurements are consistent with the energy dependence considered in this study.

Moreover, the energy dependence of COSINE-100 QF\textsubscript{I} is much less pronounced than that observed for QF\textsubscript{Na}. In the case of QF\textsubscript{I}, it is clear that the choice between a constant or Lindhard-like model has a non-relevant impact; in any case, any residual effect can be effectively accounted for within the linear energy dependence framework introduced in this study.



\subsection{Building simulated electron-equivalent spectra}\label{building}
The simulated spectra are converted into electron-equivalent energies in a second-level analysis by applying the QF correction corresponding to the selected model for Na and I nuclei, as presented in the previous section. In this step, since each energy deposition is tagged with the type of recoiling nucleus responsible, the corresponding QF can be accurately applied (see Chapter \ref{Chapter:Geant4} for a detailed description of the procedure to generate ANAIS-like data). 

It is important to note that the QF correction, i.e., multiplying the energy deposited in NRs by the QF to obtain electron-equivalent energies, is applied to each individual energy deposit, not to the total event energy within the ANAIS integration window. In Geant4 simulations, energy deposition by NRs occurs in a single step, in contrast to electrons, which lose energy gradually through multiple interactions. In the case of multiple scattering within a crystal, the QF correction is applied accurately to each individual interaction. While this distinction is not relevant for constant QF models, it becomes crucial when applying energy-dependent QFs, as neglecting it can result in inaccurate energy reconstruction.

Moreover, the simulated spectrum is scaled to match the measurement time exposure, accounting for the nominal activity of the $^{252}$Cf source on the date of the neutron calibration and considering its decay since production, as detailed in Table \ref{infoneutroncalibration}. It should be emphasized that the spectra presented in this chapter have not been scaled or corrected on the y-axis using ad-hoc normalization factors to match the data; the results are solely determined by the simulation output and the source activity normalization. This will underscore, in the following section, the sensitivity of the ANAIS-112 neutron simulation not only to the QF, but also to the accurate description of the main experimental features.

If the simulation were to be directly compared (after corrections of energy resolution and QF) with experimental data, the agreement would not be as satisfactory as expected. This is because two additional corrections need to be applied, which consist of adding data inputs that the neutron simulation cannot predict on its own. These corrections, as will be shown in Section \ref{comparison}, lead to a better agreement between data and simulation.

\begin{figure}[b!]
\begin{center}
\includegraphics[width=0.49\textwidth]{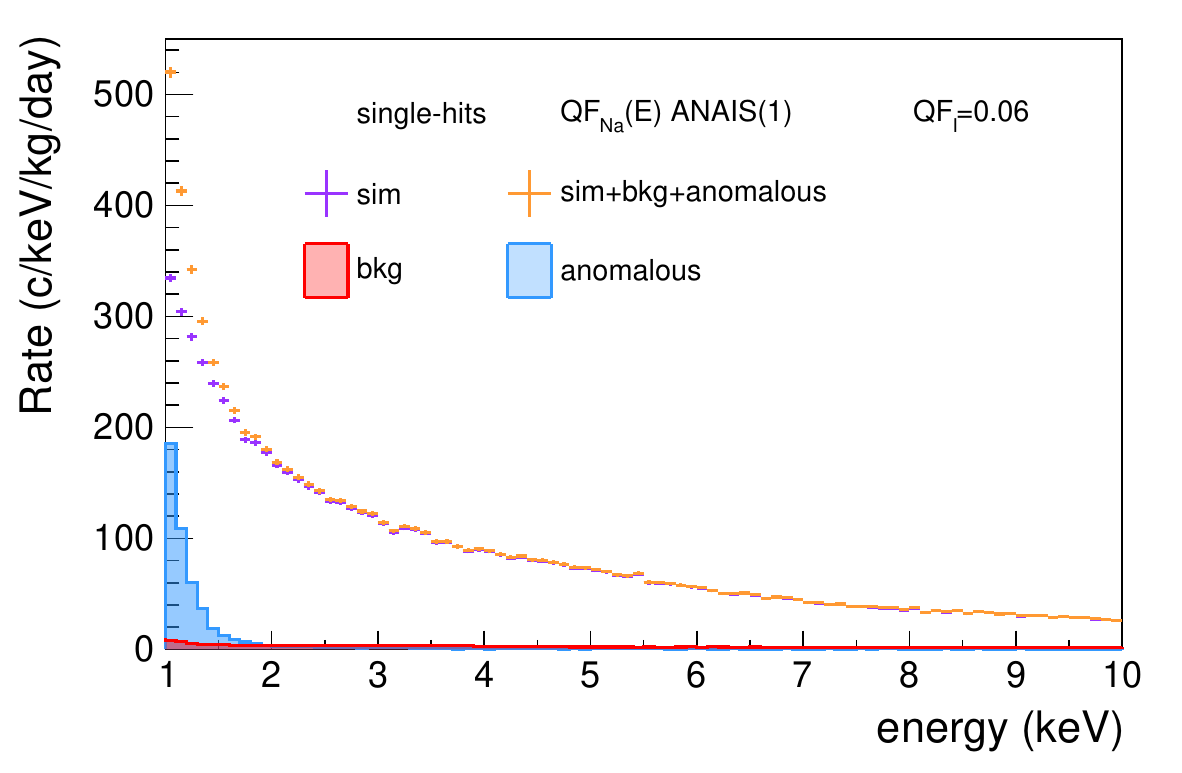}
\includegraphics[width=0.49\textwidth]{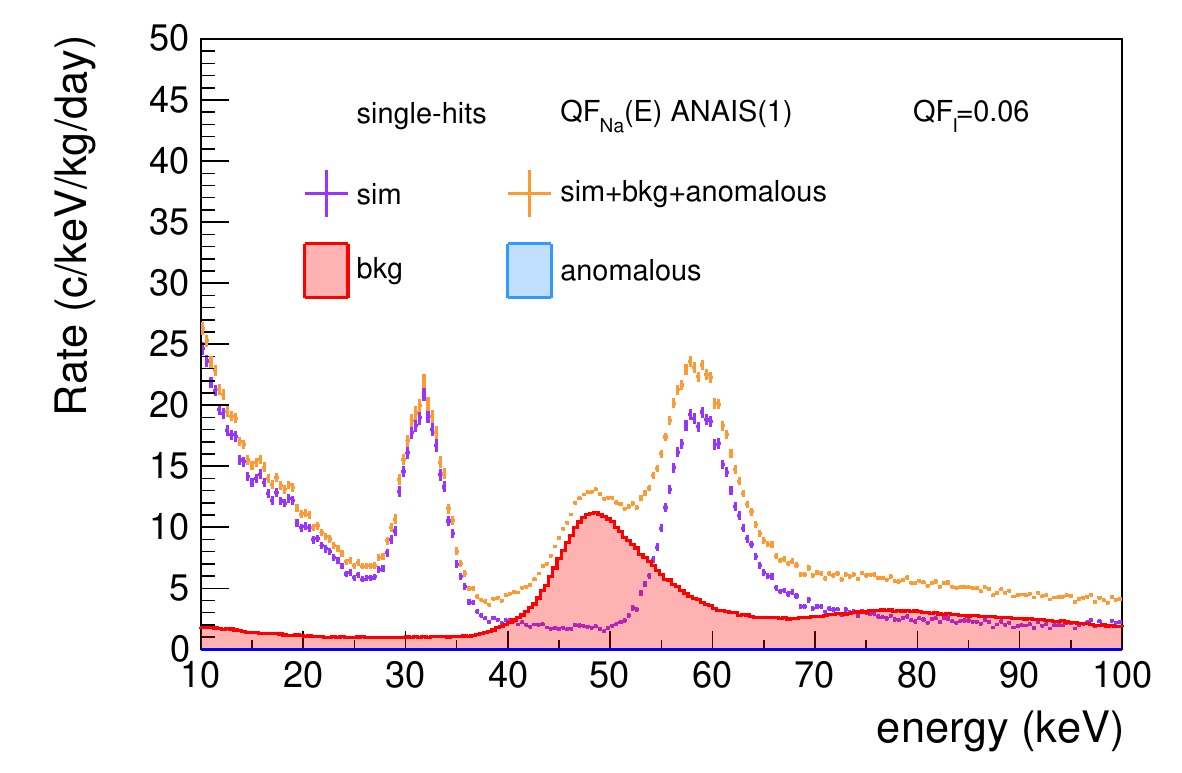}
\includegraphics[width=0.49\textwidth]{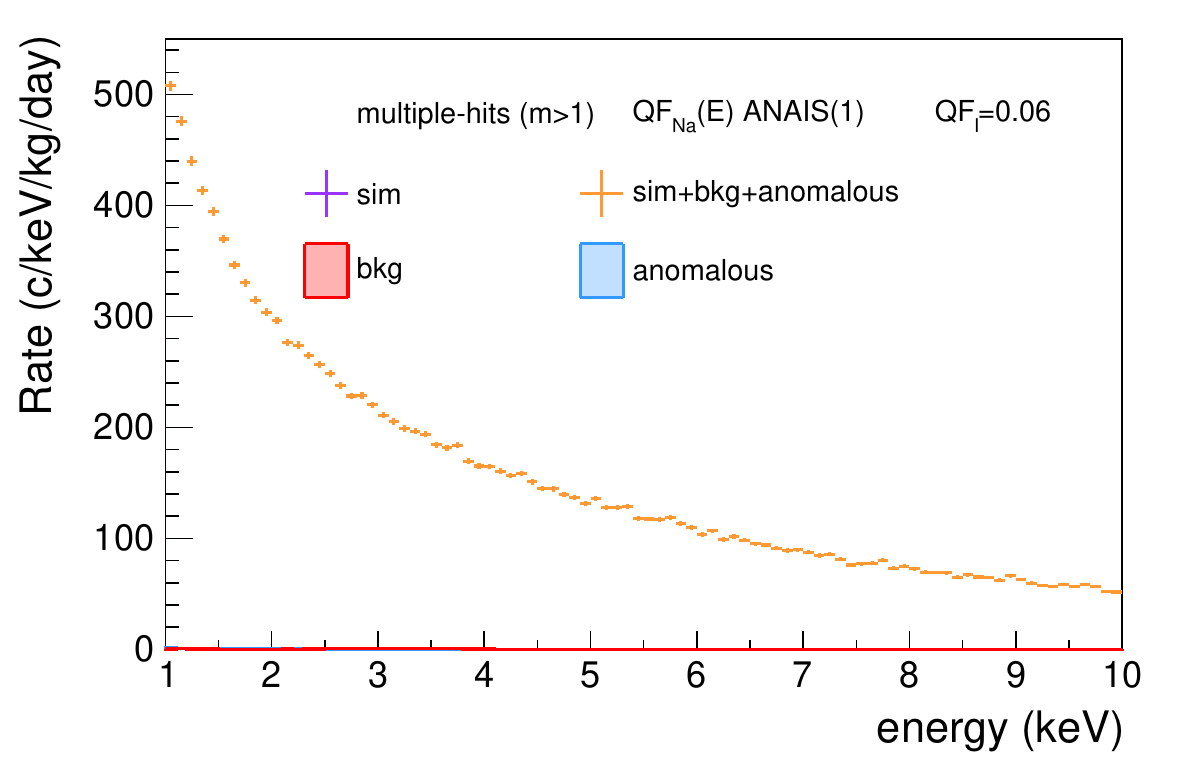}
\includegraphics[width=0.49\textwidth]{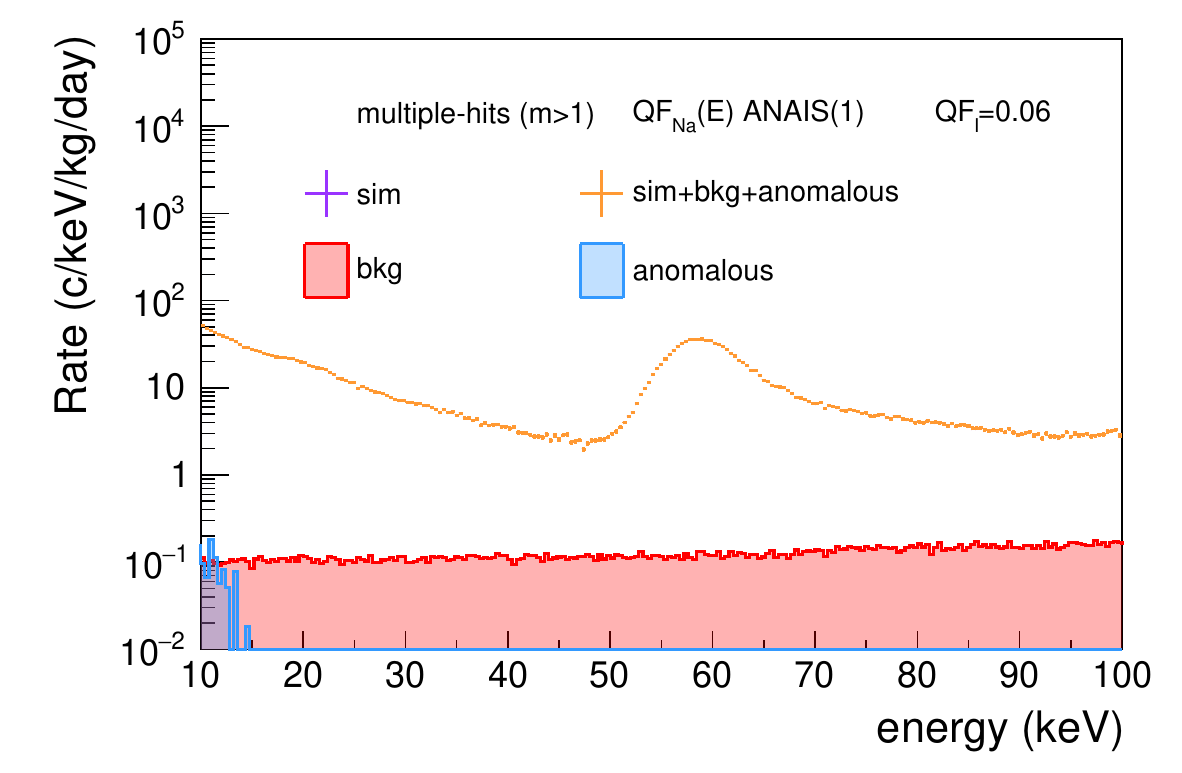}

\caption{\label{añadecosas} Experimental inputs added to the neutron simulation, which include the measured background in ANAIS (red) and the anomalous event population identified with the ANOD DAQ (blue). The figure shows the behavior of the simulation before (violet) and after (orange) incorporating these contributions. The ANAIS(1) QF$_\textnormal{Na}$ model and a constant QF$_\textnormal{I}$=0.06 are considered. The left panel displays the low-energy region, while the right panel covers the medium-energy range. \textbf{Top panels:} single-hits. \textbf{Bottom panels:} multiple-hits.}
\vspace{-0.2cm}

\end{center}
\end{figure}

On one hand, as shown in the presentation of the experimental spectrum (see Figure~\ref{comparesinglemulti}), the neutron calibration data exhibit a spectral feature around 49 keV corresponding to the $^{210}$Pb background present in ANAIS-112. This peak is not predicted by the simulation, as only the $^{252}$Cf source emission is simulated. Therefore, the background spectrum acquired prior to the neutron calibration, conveniently normalized to the effective exposure of each calibration run, is added to the simulation to account for this component. Figure \ref{añadecosas} shows the effect of including this contribution in the simulation for both single- and multiple-hit spectra. The results indicate that this background component becomes relevant only in the 40–50 keV region of the single-hit spectrum.

On the other hand, the contribution of anomalous events, leaking the filtering protocols, and considered to be responsible of the background excess in the 1-2 keV energy region, needs to be added. With the incorporation of the ANOD DAQ, it is now possible to discriminate these worrisome population thanks to its new features: a larger 8-$\mu$s integration window and no dead time. It should be noted that this population cannot be selected in ANAIS because, given its characteristics, it overlaps with bulk scintillation events. 

The process of selecting this anomalous event population was already explained in Section \ref{ANODfiltering}, when the new DAQ was introduced. In this case, the energy spectrum of these events (relevant only in the [1-2] keV range) is considered, taking into account the rate of synchronized events in ANAIS and ANOD during run 9016, the neutron calibration conducted with the ANOD DAQ. By selecting anomalous events in ANOD, it is possible to determine what ANAIS detects, even though ANAIS cannot discriminate them on its own, using the synchronized tree. 

Figure \ref{añadecosas} also illustrates the impact of including this contribution in the simulation, showing that the effect is only significant in the single-hit spectrum below 2 keV. As will be shown in Section \ref{comparison}, the inclusion of these anomalous events allows for much better agreement between data and simulation in the low-energy range of single-hits. This is particularly important when drawing conclusions about the QF$_{\textnormal{I}}$, as QF$_{\textnormal{I}}$ plays a significant role precisely in the [1–2] keV region, which is strongly affected by the presence of these anomalous events.

The two aforementioned corrections will be applied to all event populations and energy ranges, although they are only relevant for single-hits.


In the study of the QF estimation, a distinction will be made between the simulations of the west, south, and top faces. The west face calibration is the one with the most accumulated statistics and the most homogeneous irradiation of the detectors, making it the reference benchmark. However, as will be shown in the next section, where the comparison between simulation and data will be carried out, the neutron simulation is able to satisfactorily reproduce the experimental data of all faces, and the conclusions derived from one face are compatible with those that could be drawn from the others. As seen in Table \ref{infoneutroncalibration}, the west face accumulates a total of 6 calibration runs, the south face has 2, and the top face has 1. 

To combine multiple runs in the simulation, each run is post-processed in physical units (c/keV/kg/day), considering the age of the source and including the corresponding background spectrum for each case. Then, the runs for each face are summed, taking into account the live time of each run, to create a simulation histogram that can be compared with the total exposure of the west face, for instance.\\



Having discussed the details of the neutron simulation, including the comparison between Geant4 versions, the correction for the neutron capture cross-section for the production of $^{128}$I, the QF models for Na and I considered in this study, and the process of converting absolute counts to rates, the neutron data presented in the previous section will now be compared with the neutron simulation. This comparison will allow to derive conclusions about the QF of Na and I for the ANAIS crystals.


\section{Comparison between data and simulation}\label{comparison}

In this section, a comparison between the experimental data and the simulations described in Sections~\ref{neutronData} and~\ref{neutronsim} is presented. The analysis begins with the high-energy region (Section~\ref{highenergyrange}). Subsequently, the QF values reported by DAMA/LIBRA are evaluated (Section~\ref{DAMAQF}), followed by an assessment of the QF results obtained for ANAIS-like crystals in \cite{cintas2024measurement,phddavid} (Section~\ref{ANAISQF}). Other QF\textsubscript{Na} models are then examined, in particular the COSINUS QF\textsubscript{Na} measurement and the Tetryak model (Section~\ref{otherNa}). Finally, after assessing the impact of QF\textsubscript{Na}, different models for QF\textsubscript{I} are explored (Section~\ref{QFImodels}).

It is worth noting that error bands are omitted in the initial comparisons to highlight the overall behavior; they will be included in the final analysis when presenting the QF models most compatible with the ANAIS-112 neutron calibration data. On the other hand, the $\chi^2$/ndf statistic has been manually computed as a criterion to evaluate which model better describes the ANAIS-112 data. It is important to note that this parameter is not the result of a fit, since no such fit has been performed here; it is merely intended to provide an indication of the goodness of the comparison in an attempt to shed light on the QF model compatibility.

With the exception of the comparison performed in the high-energy range, which is not affected by variations in the QF, this study will consistently present the total spectra, as well as the single-hit, multiple-hit (m>1), and m2-hit populations. The distinction between multiple-hits and m2-hits is relevant because, according to the data, 71\% of neutron events are single-hits, while the remaining 29\% are multiple-hits. Among the multiple-hit population, 16\% correspond to m2-hits, 8\% to m3-hits, 3\% to m4-hits, and 1\% to m5-hits, with higher multiplicities being increasingly rare. Therefore, it is particularly informative to evaluate the agreement between simulation and data for the general multiple-hit population, as well as for the dominant m2-hit class.

\vspace{-0.1cm}
\subsection{High-energy range} \label{highenergyrange}

\begin{figure}[b!]
    \centering
    {\includegraphics[width=1.\textwidth]{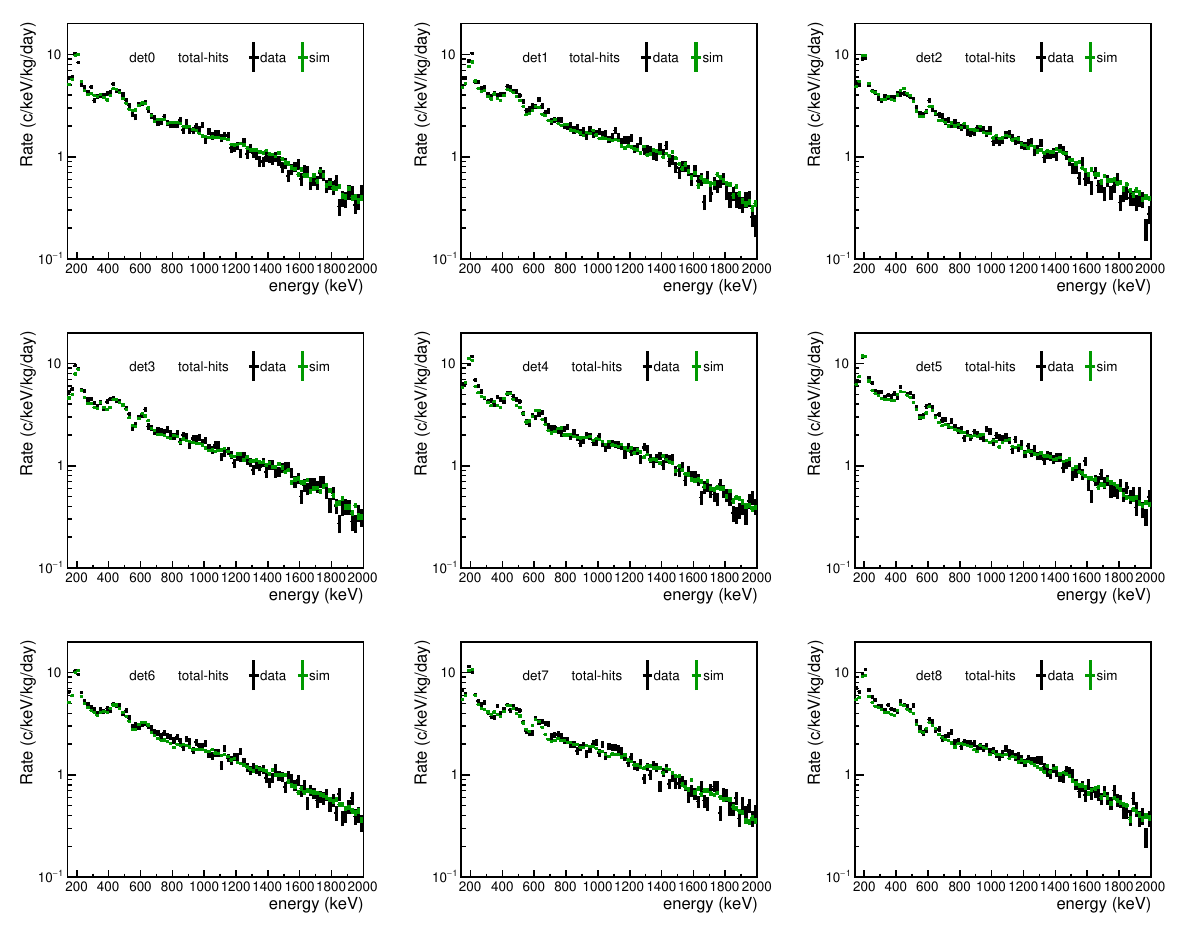}}
  
    \caption{\label{HEtotalperdet} 
Comparison between the total-hits high-energy spectra measured in the west-face neutron calibration for each ANAIS-112 detector (black) and the corresponding simulation (green), assuming ANAIS(1) QF$_\textnormal{Na}$ and a constant QF$_\textnormal{I}$=0.06. 
}
\end{figure}

\begin{figure}[t!]
    \centering
    {\includegraphics[width=0.49\textwidth]{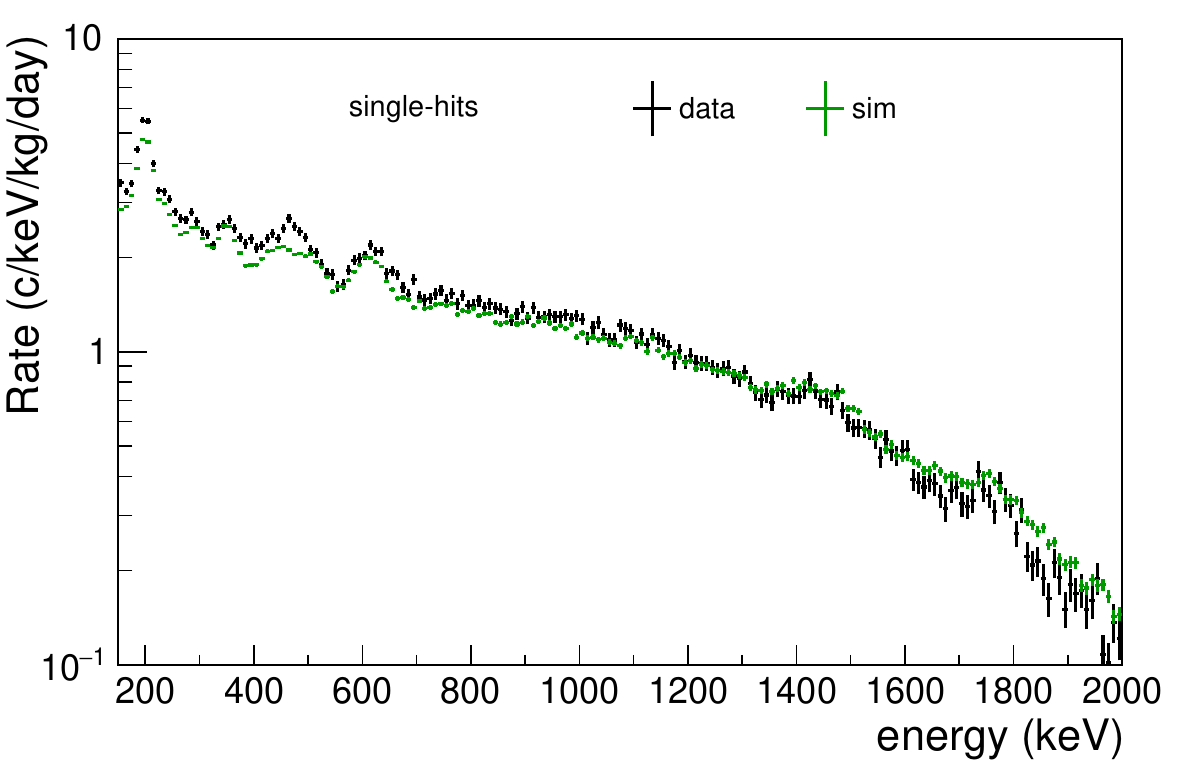}}
    {\includegraphics[width=0.49\textwidth]{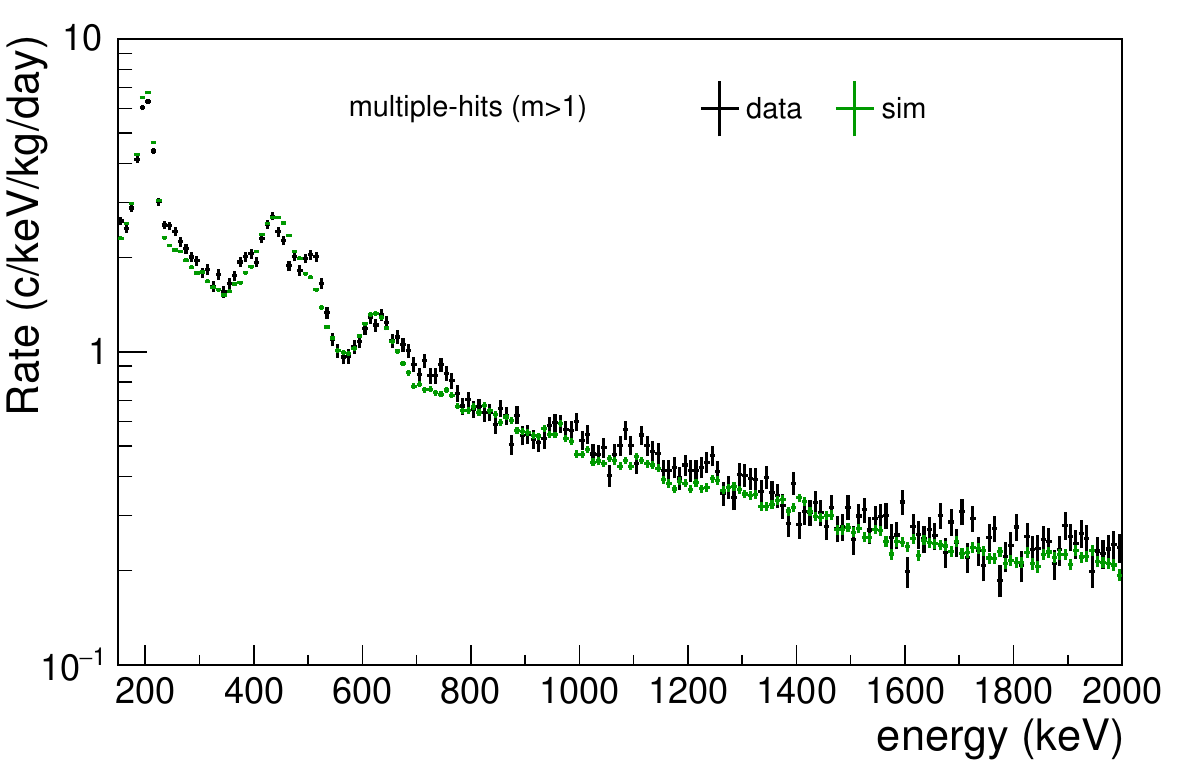}}

    \caption{\label{HEsinglemulti} 
Comparison between the high energy spectra measured in the west-face neutron calibration for the sum of the nine ANAIS-112 detectors (black) and the corresponding simulation (green), assuming ANAIS(1) QF$_\textnormal{Na}$ and a constant QF$_\textnormal{I}$=0.06. \textbf{Left panel:} single-hits. \textbf{Right panel:} multiple-hits.
}
\end{figure}

As shown in Figure \ref{simdistribucionHE}, the high-energy range is dominated by ER. Accordingly, no changes have been observed in the simulation when introducing different models for either QF$_{\textnormal{Na}}$ or QF$_{\textnormal{I}}$. Nevertheless, this population provides an important cross-check to verify that the entire electromagnetic component is being properly modelled by the simulation. Figure~\ref{HEtotalperdet} shows the comparison between data and the simulation (total-hits) in the high-energy range, analogous to what is presented in Figure~\ref{compareversioncondatos}, but shown detector by detector. Figure~\ref{HEsinglemulti} presents the corresponding comparison for single-hit and multiple-hit events, summed over all detectors. 



The agreement between data and simulation, both at the individual detector level and in the sum over all nine detectors, is highly satisfactory. Although certain features, such as the structure around 400-500 keV in the single-hit population, are not fully captured, the overall description is very good. 

Additionally, in the multiple-hit spectrum, two peaks are observed around 500 keV and 750 keV that are not reproduced by the simulation. Since the measured background in ANAIS has already been incorporated into the simulation and these features remain unexplained, this suggests that the isotopes responsible for these peaks must originate from neutron-induced activity. However, the exact origin of these peaks has not been identified.

Overall, these results confirm the reliability of the simulation in reproducing the high-energy spectra across all event populations, supporting its use as a robust tool for studying the QF in ANAIS-112 crystals.

\subsection{DAMA/LIBRA QFs} \label{DAMAQF}

Once the high-energy description has been verified, the validity of the QF reported by DAMA/LIBRA, i.e., QF$_\textnormal{Na}$=0.3 and QF$_\textnormal{I}$=0.09, is assessed. As previously mentioned, most recent measurements point to an energy-dependent QF$_\textnormal{Na}$ that increases with energy, whereas DAMA/LIBRA reports a constant value higher than most of the others \cite{Bernabei:1996vj} .

\begin{figure}[t!]
    \centering
    {\includegraphics[width=1\textwidth]{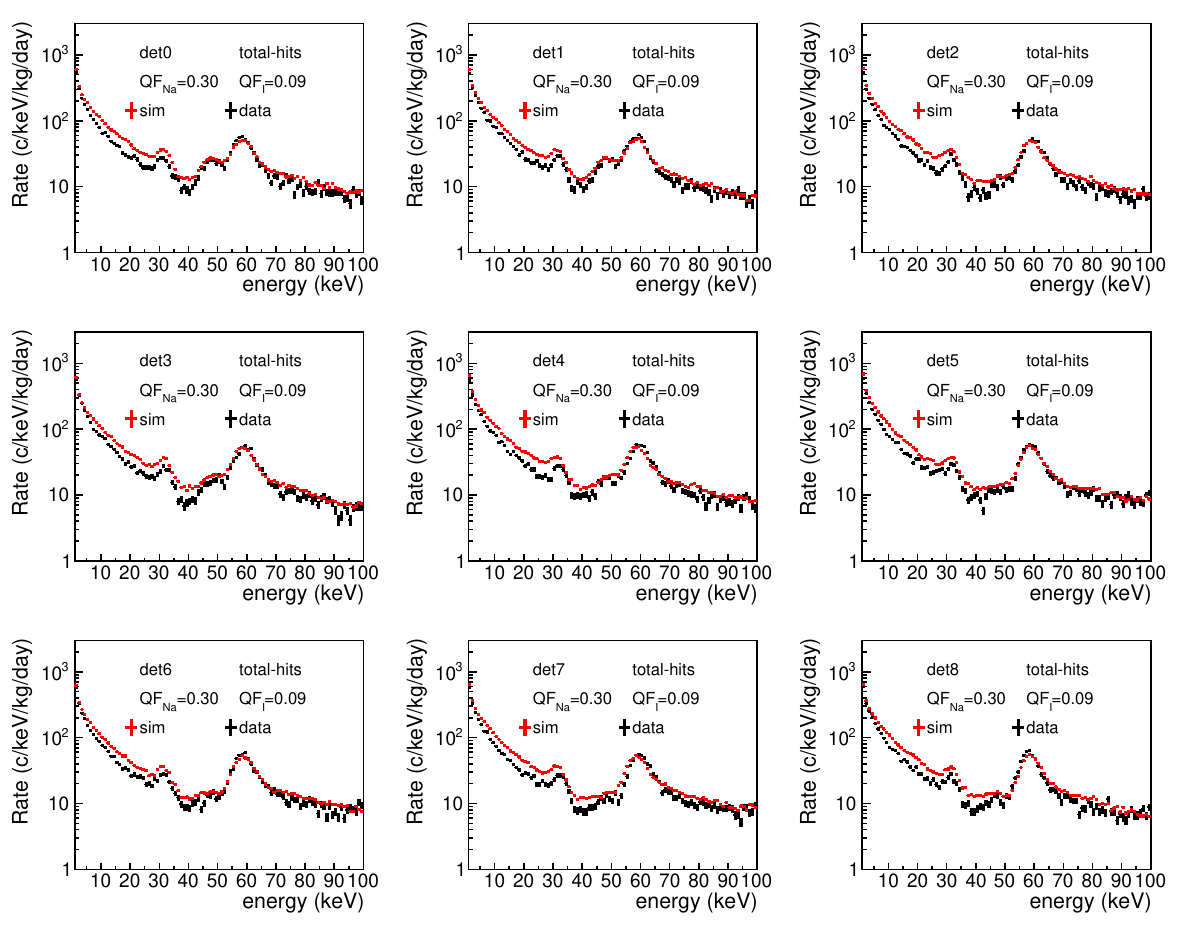}}

    \caption{\label{DAMAperdetTotal} Comparison between the total-hits medium-energy spectra measured in the west-face neutron calibration for each ANAIS-112 detector (black) and the simulation (red), assuming DAMA/LIBRA QFs, QF${_\textnormal{Na}}$=0.3 and QF${_\textnormal{I}}$=0.09.}
\end{figure}

\begin{figure}[t!]
    \centering
    {\includegraphics[width=0.45\textwidth]{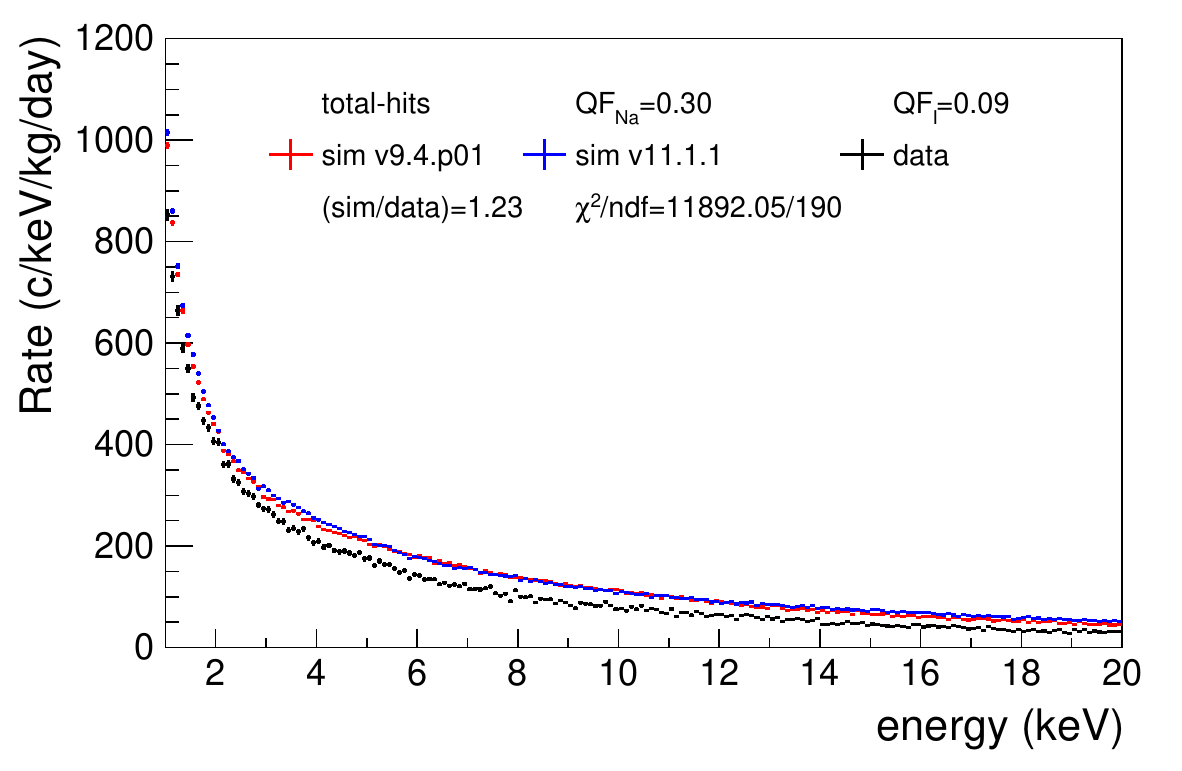}}
    {\includegraphics[width=0.45\textwidth]{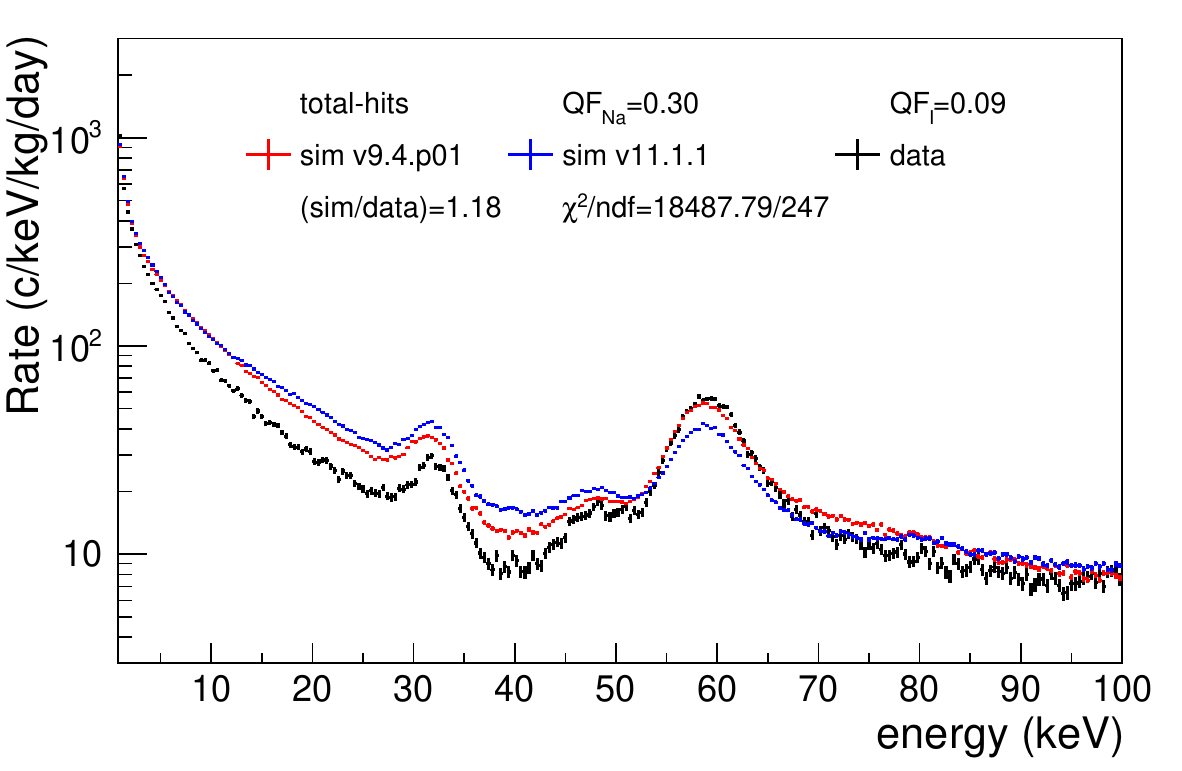}}
    {\includegraphics[width=0.45\textwidth]{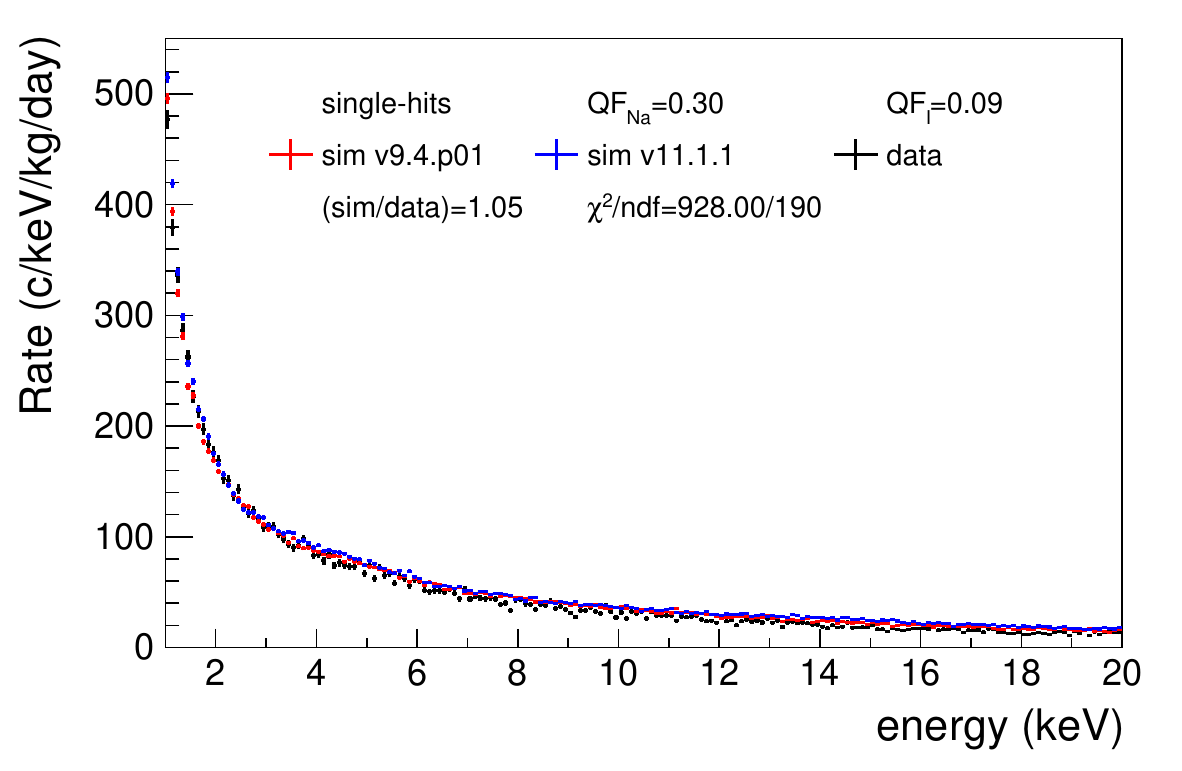}}
    {\includegraphics[width=0.45\textwidth]{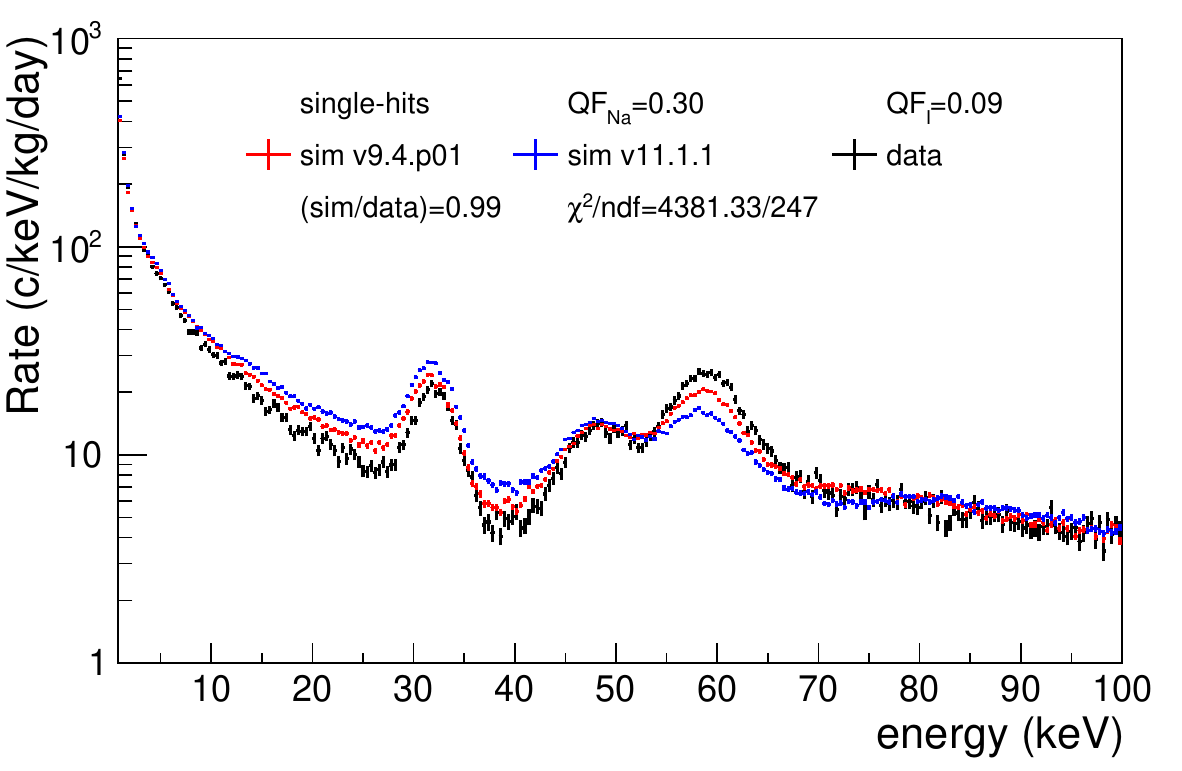}}
    {\includegraphics[width=0.45\textwidth]{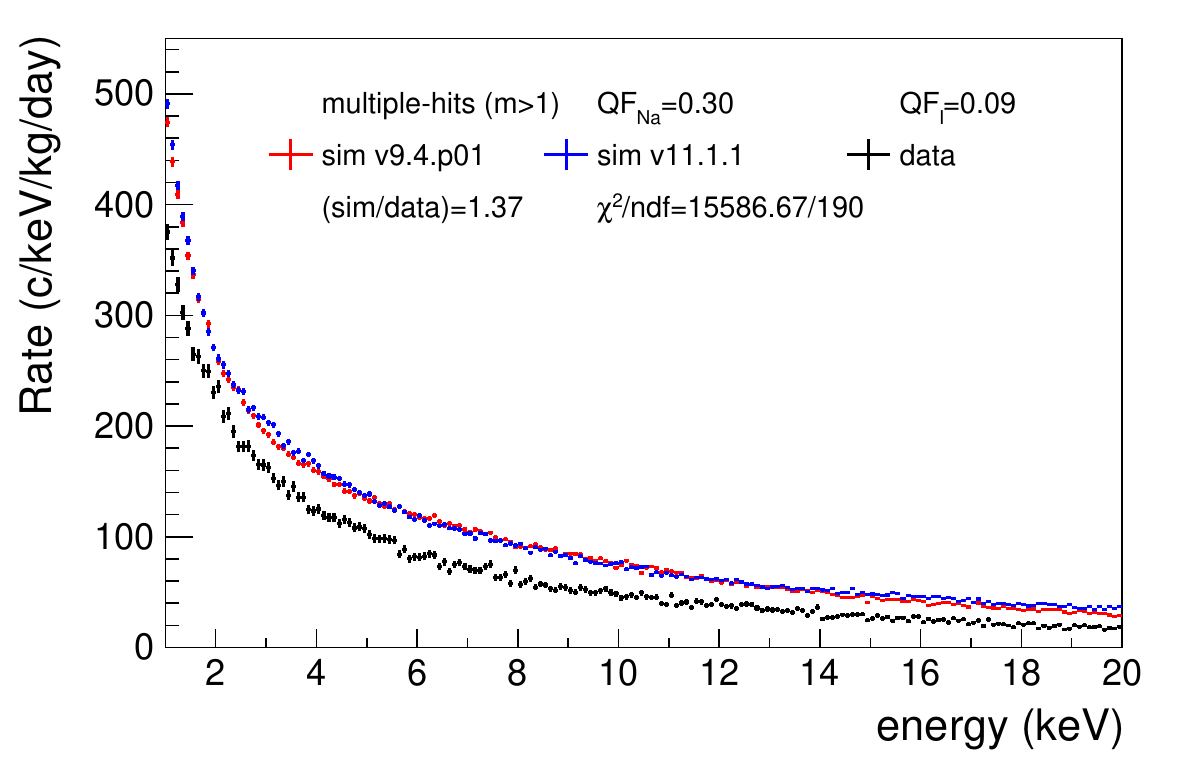}}
    {\includegraphics[width=0.45\textwidth]{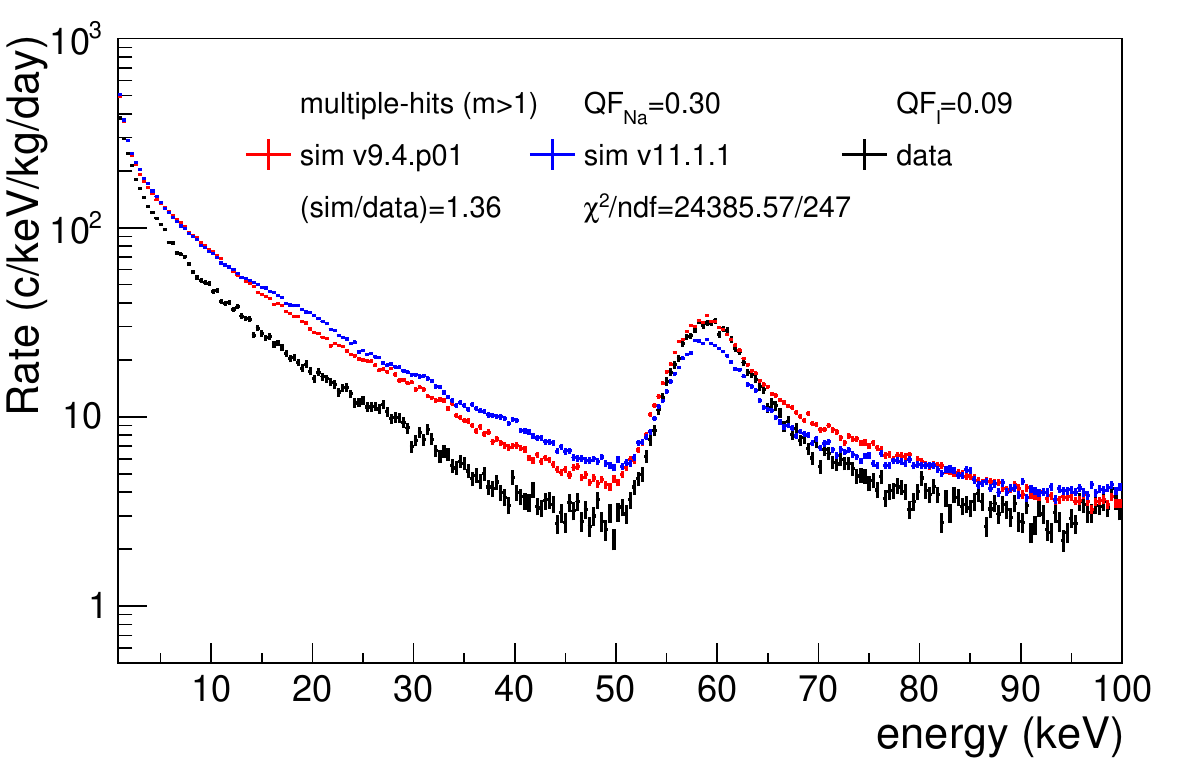}}
    {\includegraphics[width=0.45\textwidth]{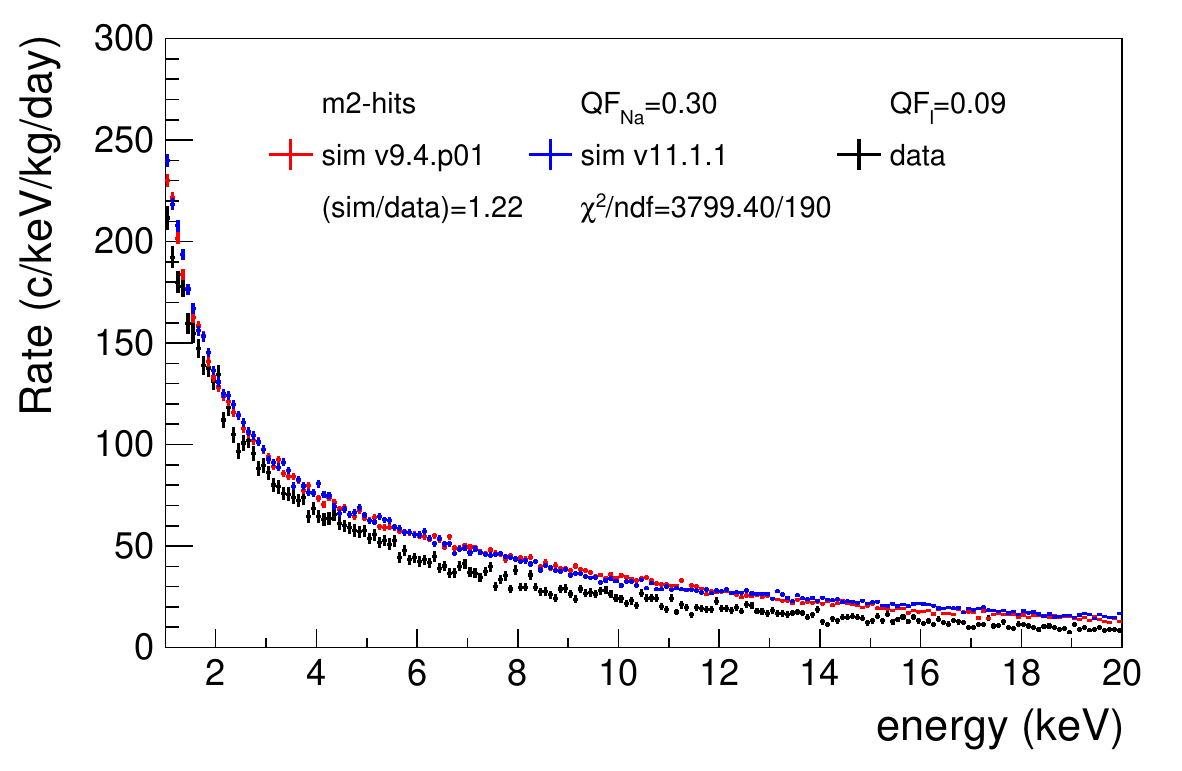}}
    {\includegraphics[width=0.45\textwidth]{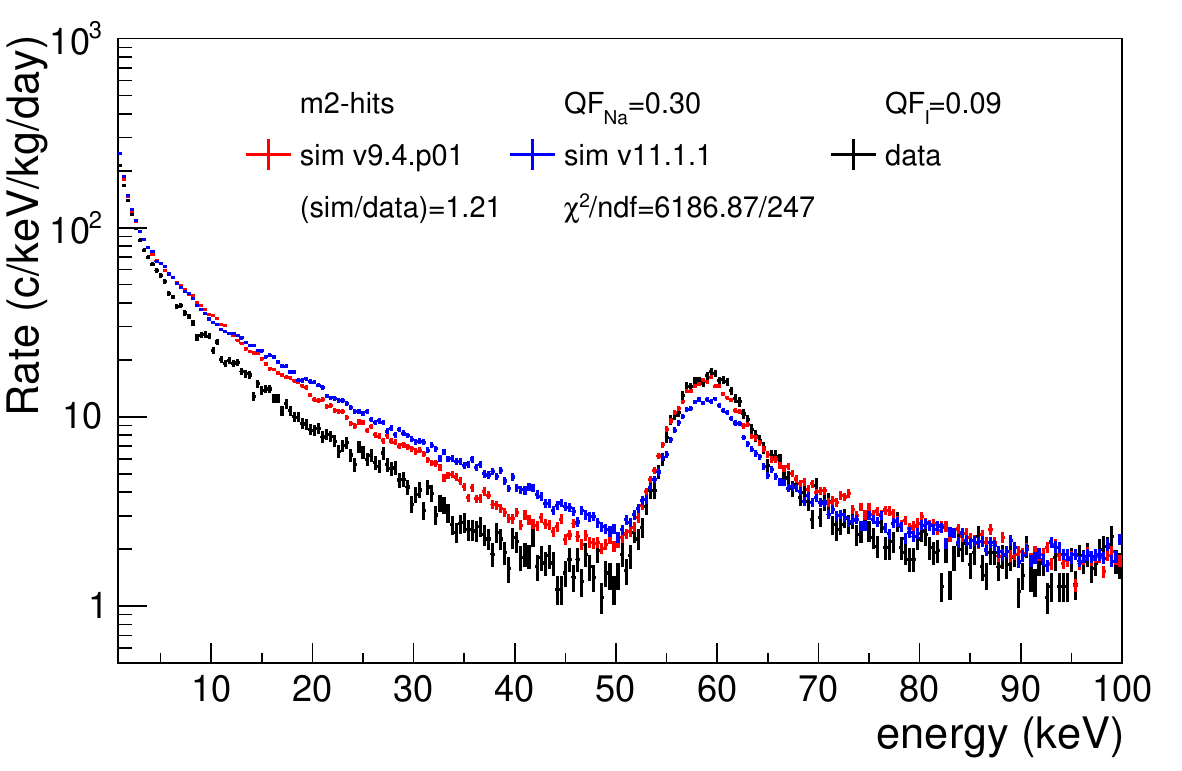}}

    \caption{\label{DAMATotal} 
Comparison between the energy spectra measured in the west-face neutron calibration for the sum of the nine ANAIS-112 detectors (black) and the corresponding simulations, assuming DAMA/LIBRA QFs, QF${_\textnormal{Na}}$=0.3 and QF${_\textnormal{I}}$=0.09. Simulated spectra obtained using Geant4 versions v9.4.p01 (red) and v11.1.1 (blue) are shown. Left panels show the ratio between simulation (v9.4.p01) and experimental data, as well as the goodness of the comparison in the [1–20] keV range; these parameters are displayed in the right panels for the [1–100] keV range. The left column displays the low-energy region, while the right column shows the medium-energy range. \textbf{First row:} total-hits. \textbf{Second row:} single-hits. \textbf{Third row:} multiple-hits (m>1). \textbf{Fourth row:} m2-hits.
}
\end{figure}

\begin{figure}[t!]
    \centering
    {\includegraphics[width=0.49\textwidth]{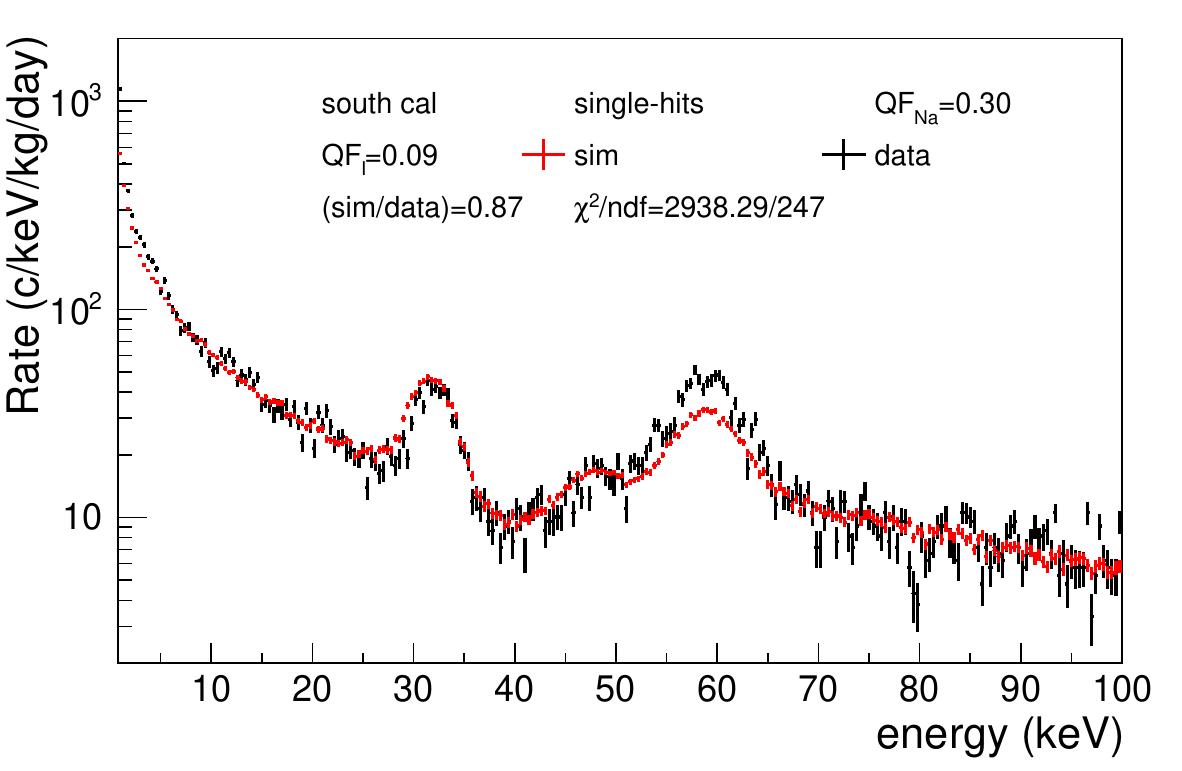}}
    {\includegraphics[width=0.49\textwidth]{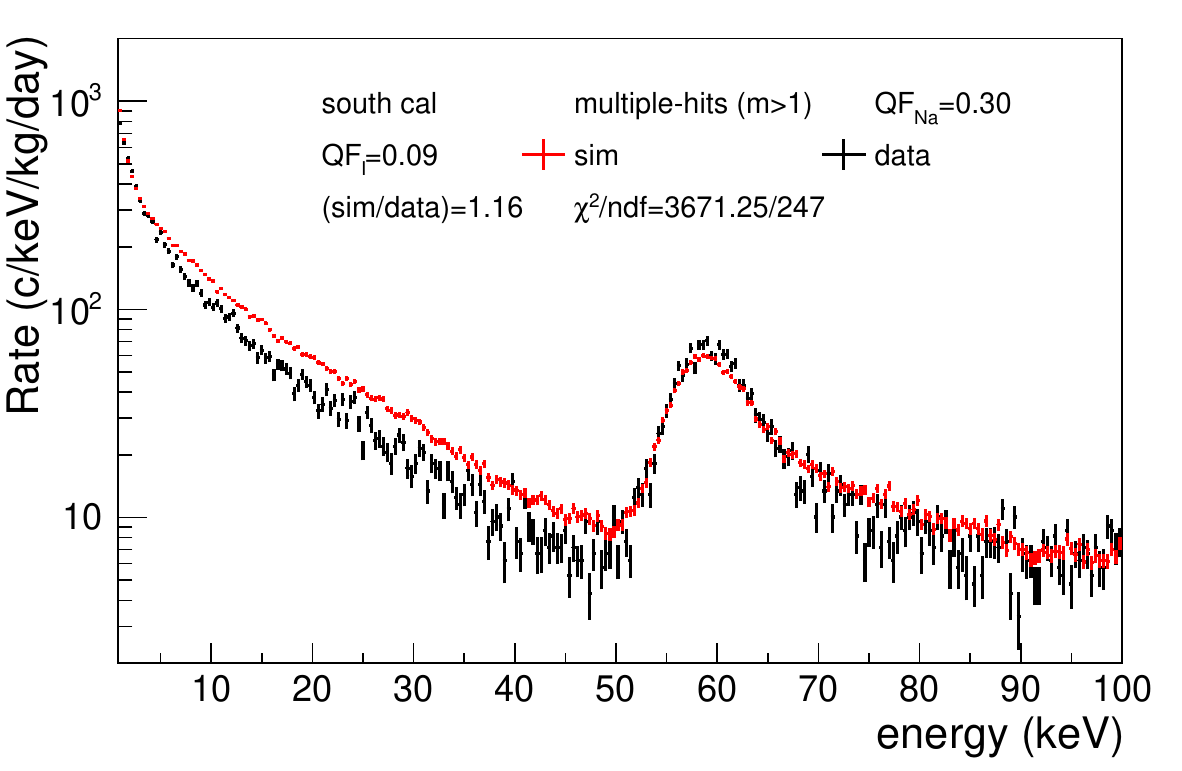}}
    {\includegraphics[width=0.49\textwidth]{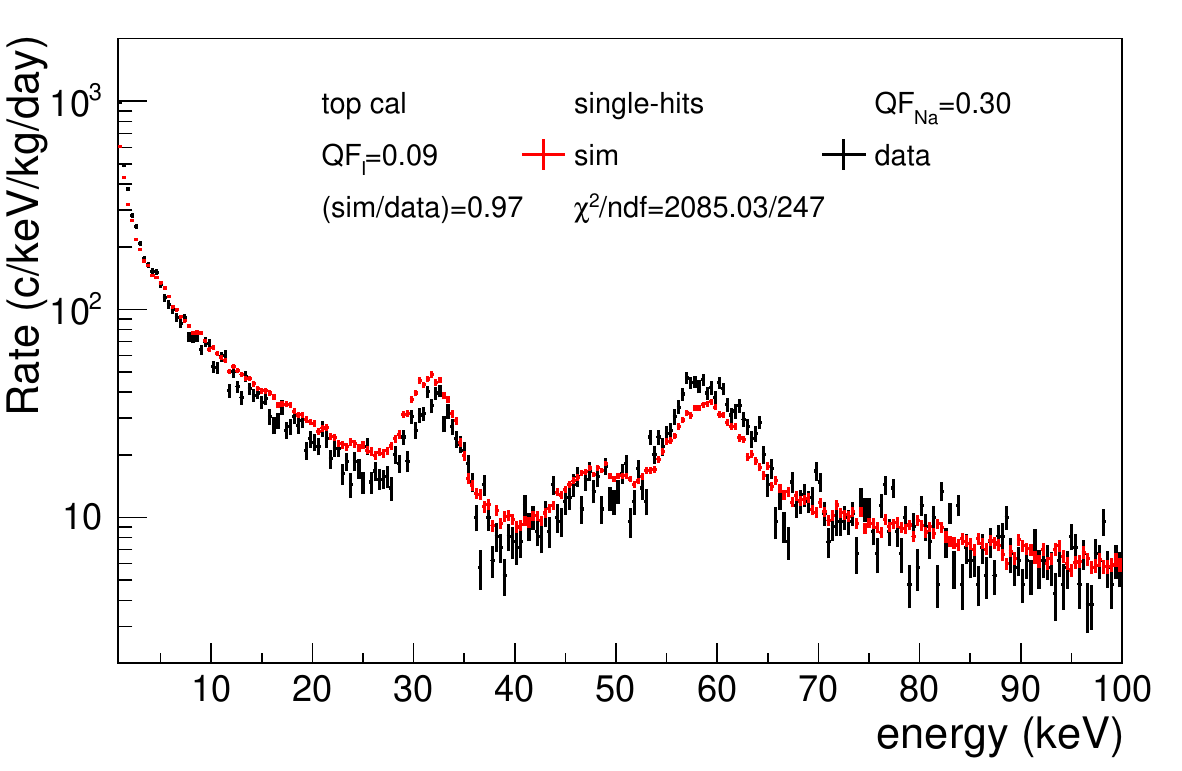}}
    {\includegraphics[width=0.49\textwidth]{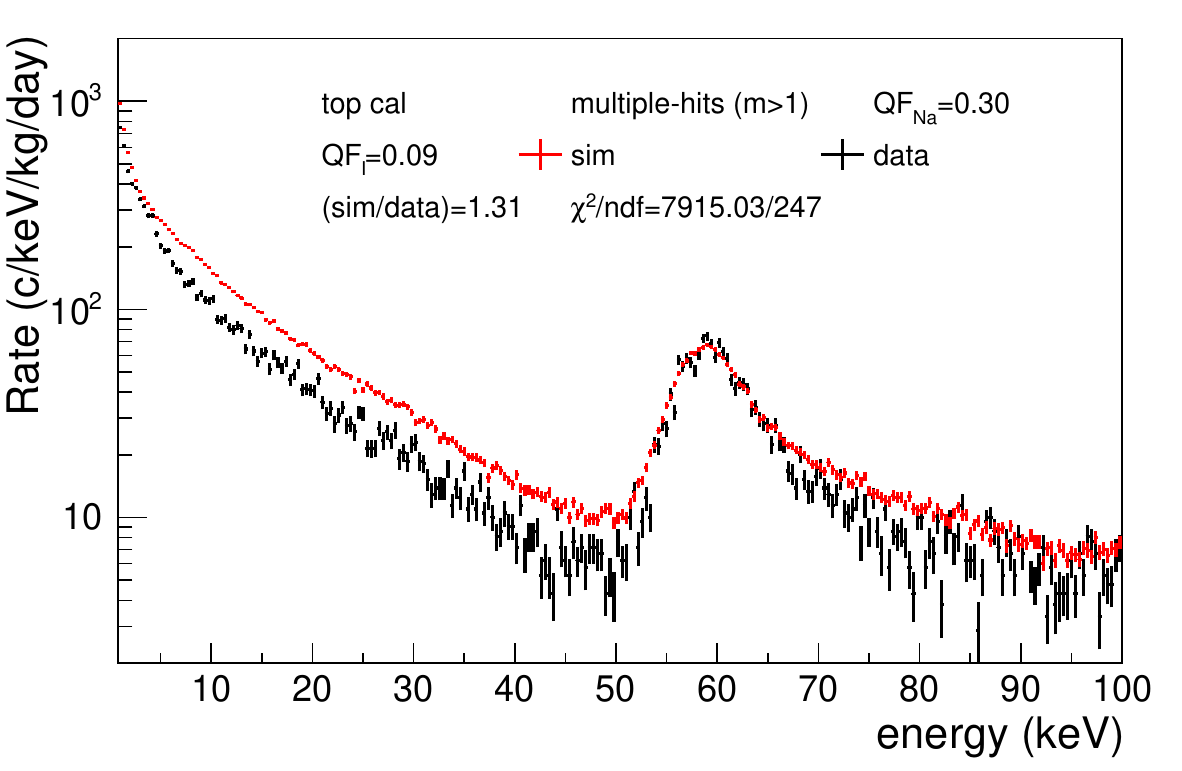}}

    \caption{\label{DAMAsurtop} Comparison between the medium-energy spectra measured in the neutron calibrations for the sum of the nine ANAIS-112 detectors (black) and the corresponding simulation using Geant4 v9.4.p01 (red), assuming DAMA/LIBRA QFs, QF${_\textnormal{Na}}$=0.3 and QF${_\textnormal{I}}$=0.09. The left column displays the single-hits spectra, while the right column shows the multiple-hits (m>1) spectra. The panels show the ratio between simulation and experimental data, as well as the goodness of the comparison in the [1–100] keV range. \textbf{Top row:} south neutron calibration. \textbf{Bottom row:} top neutron calibration.  }
\end{figure}
Figure \ref{DAMAperdetTotal} presents the comparison between data and simulation (total-hits) in the medium-energy range, analyzed detector by detector. First row of Figure \ref{DAMATotal} shows the corresponding energy spectra obtained by summing the nine ANAIS-112 detectors, in both the low- and medium-energy regions. The latter figure includes results from the version v11.1.1 of the simulation to illustrate that the agreement with the DAMA/LIBRA QFs does not depend on the Geant4 version employed. As shown in Figure \ref{DAMAperdetTotal}, the response is consistent across all detectors. While the inelastic peak is well reproduced (with v9.4.p01), the NR region is clearly overestimated. 


Figure \ref{DAMATotal} also shows the compatibility across different event populations, single-hits (second row), multiple-hits (m>1) (third row) and m2-hits (fourth row), for both Geant4 versions. Single-hit events are nicely reproduced at low energies using the DAMA/LIBRA QFs; however, the agreement becomes less satisfactory as the energy increases. It is within the multiple-hit population that the discrepancy between data and simulation becomes most evident. Multiple-hit events are clearly overestimated, with the ratio between simulation and data in the [1–20] keV region reaching 1.37. This overestimation is not uniform; for instance, the inelastic peak is accurately reproduced, indicating that the discrepancy cannot be attributed to a global scaling factor across the spectrum. The same behavior is observed in the m2-hit population. Altogether, these results indicate that the QFs reported by DAMA/LIBRA result in clear discrepancies with the ANAIS-112 neutron calibration data.

Figure \ref{DAMAsurtop} presents the analogous comparison for the neutron calibrations conducted on the south and top faces of the ANAIS set-up using Geant4 v9.4.p01. It is observed that the statistics in these two faces is more limited compared to the west face. Yet, similar behavior is observed, further reinforcing the incompatibility with the DAMA/LIBRA QFs. Specifically, single-hit events are well reproduced, except for the underestimation of the inelastic peak, while the NR region in the multiple-hit population is clearly overestimated, resulting in a total spectrum that significantly overestimates the measured rates.

\subsection{QFs of the ANAIS crystals} \label{ANAISQF}

After assessing the compatibility with the DAMA/LIBRA QFs results, the evaluation proceeds with the QFs of the ANAIS crystals as determined in \cite{cintas2024measurement,phddavid}. Specifically, the three QF\textsubscript{Na} results obtained using the monochromatic source are considered: ANAIS(1), ANAIS(2) and ANAIS(3), previously introduced in Section \ref{QFmodels}. For iodine, a QF\textsubscript{I} value of 0.06 is assumed.

\begin{figure}[t!]
    \centering
    {\includegraphics[width=0.43\textwidth]{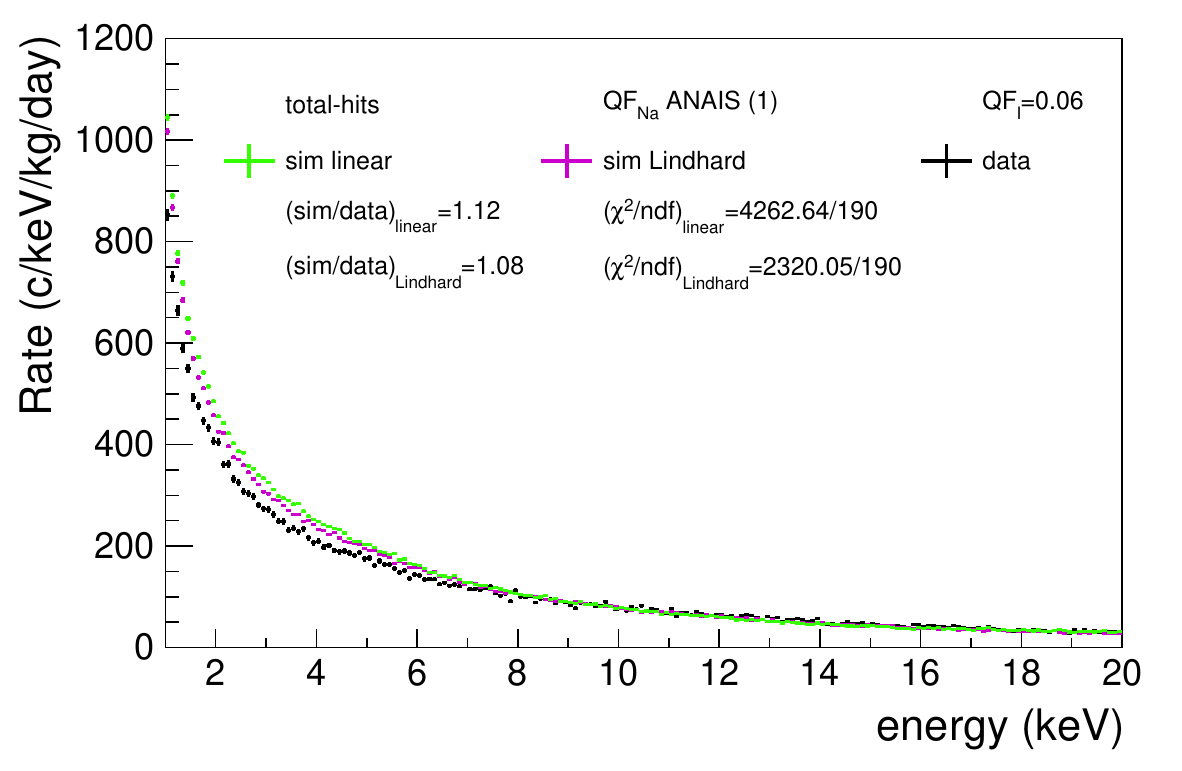}}
    {\includegraphics[width=0.43\textwidth]{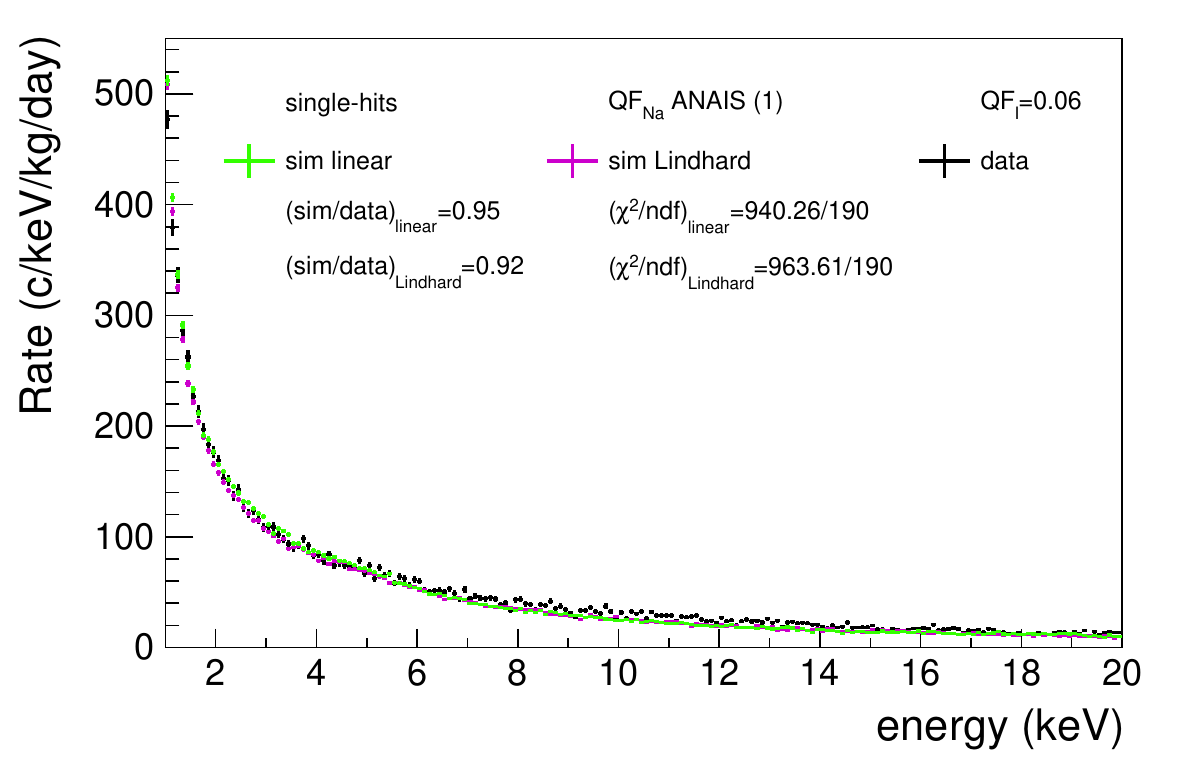}}
    {\includegraphics[width=0.43\textwidth]{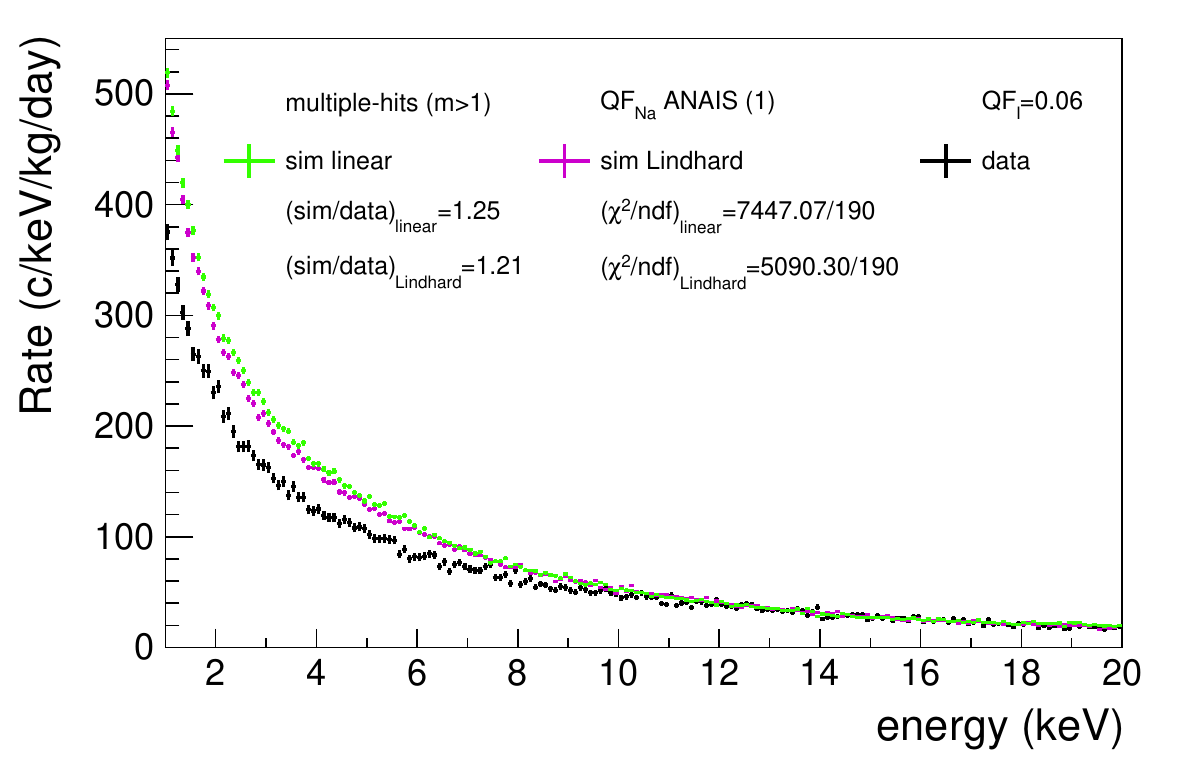}}
    {\includegraphics[width=0.43\textwidth]{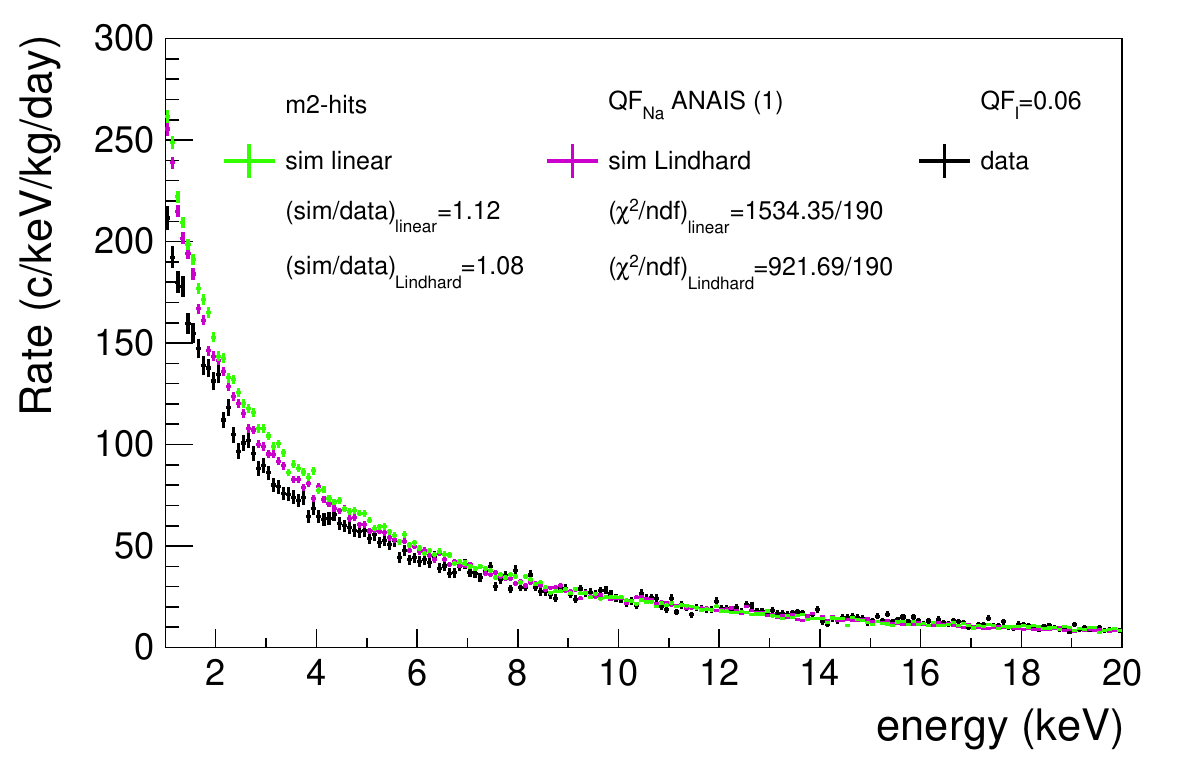}}

    \caption{\label{QFLinhardlineal} Comparison between the low-energy spectra measured in the west-face neutron calibration for the sum of the nine ANAIS-112 detectors (black) and the corresponding simulations, assuming ANAIS(1) QF\textsubscript{Na} and a constant QF\textsubscript{I} of 0.06. The simulation results assuming a linear dependence (green) and a modified Lindhard model (magenta) for modelling the QF\textsubscript{Na} are compared.  Each panel includes the ratio between simulation and experimental data for each case, as well as the goodness of the comparison, both computed in the [1–20]~keV energy range.  \textbf{Top left panel:} total-hits. \textbf{Top right panel:} single-hits. \textbf{Bottom left panel:} multiple-hits (m>1). \textbf{Bottom right panel:} m2-hits. }
\end{figure}

Two models will be used to describe the energy dependence of ANAIS(1) QF\textsubscript{Na}, as discussed in Section \ref{NaQFmodels}: one obtained from a linear fit to the experimental data, and another derived from a fit to the modified Lindhard model. For ANAIS(3) QF\textsubscript{Na}, it should be noted that only the linear modelling approach will be considered. This comparison is shown in Figure~\ref{QFLinhardlineal}, which presents the simulation results for ANAIS(1) QF\textsubscript{Na} modelled using both dependencies. As observed, these two approaches differ only in the region below approximately 5 keV. The difference between both modelling approaches is very subtle; however, it is true that the Lindhard-like dependence provides a slightly better agreement with the data. Referring back to Figure \ref{QFfitLind}, this improvement arises from the steeper decrease of QF\textsubscript{Na} at very low NR energies.


These results suggest that the NR signals in ANAIS primarily originate from very low energies, and the simulation is sensitive enough to reflect the effect of QF variations in the definition of coincidences and electron-equivalent energy spectral shape. It is inferred, and has been verified, that a sharper decrease in the QF\textsubscript{Na} could yield a better description of the data at low energies. However, no physically motivated model has been identified to support such a parametrization. Other energy dependencies cannot be ruled out, although a stronger reduction of the QF with energy appears to be favored over a linear dependence. Nevertheless, the overall agreement remains good, considering the uncertainties.

It is interesting to observe that a worse agreement between data and simulation is found in the general multiple-hit spectra. In fact, the analysis indicates that the discrepancy is more pronounced for events with higher multiplicity, as will be shown in Figure \ref{repartomulti}. While the underlying causes remain unclear, different possible explanations were evaluated. One hypothesis was that this mismatch could be related to the trigger efficiency. Variations in trigger efficiency can affect the balance distribution of single- and multiple-hit events, whereas no significant change is expected in the total-hit rate. Specifically, a slightly reduced trigger efficiency would decrease the number of multiple-hit events and increase the number of single-hit events, which could be beneficial given the agreement observed between data and simulation. To test this possibility, several scenarios with modified trigger efficiencies were simulated, showing that reproducing the observed number of single-hit events would require a significant reduction in the trigger efficiency at low energies. Therefore, considering the reasonable agreement observed in the total-hit rate and the lack of any other experimental evidence supporting a significantly degraded efficiency, the nominal ANAIS-112 trigger efficiency has been used throughout this work. 

\begin{figure}[b!]
    \centering
    {\includegraphics[width=1\textwidth]{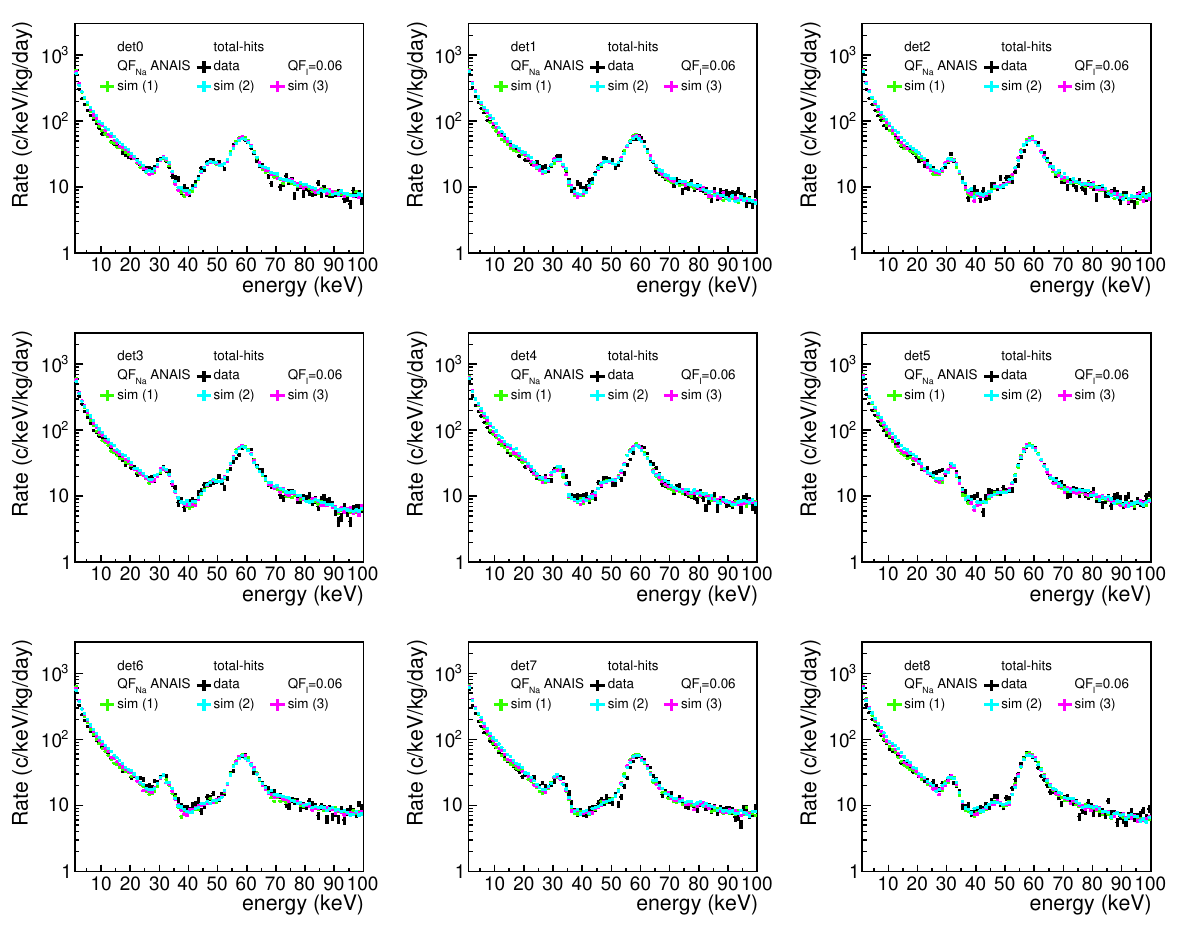}}

    \caption{\label{QFanaisperdet}Comparison between the total-hits medium-energy spectra measured during the west-face neutron calibration for each ANAIS-112 detector (black) and the simulation, assuming the QFs for the ANAIS crystals. This includes ANAIS(1) (green), ANAIS(2) (cyan), and ANAIS(3) (magenta). For iodine, a constant QF\textsubscript{I} of 0.06 is assumed.}
\end{figure}

\begin{figure}[t!]
    \centering
    {\includegraphics[width=0.46\textwidth]{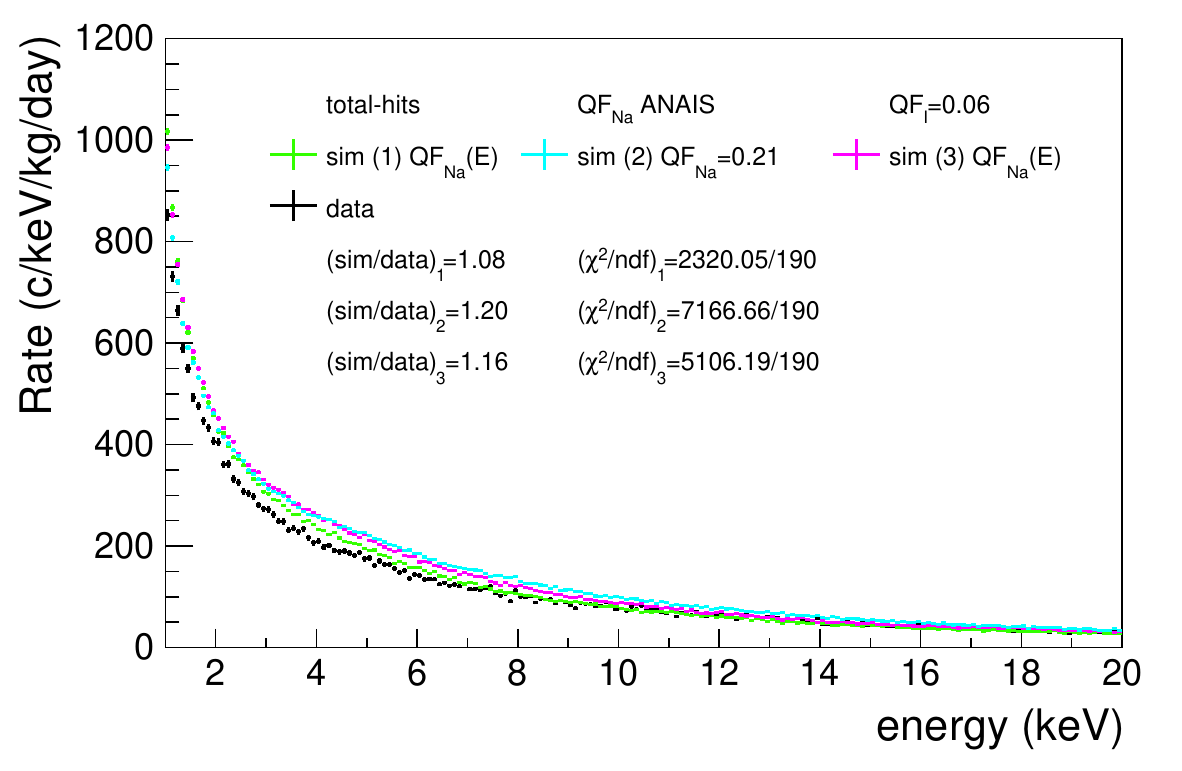}}
    {\includegraphics[width=0.46\textwidth]{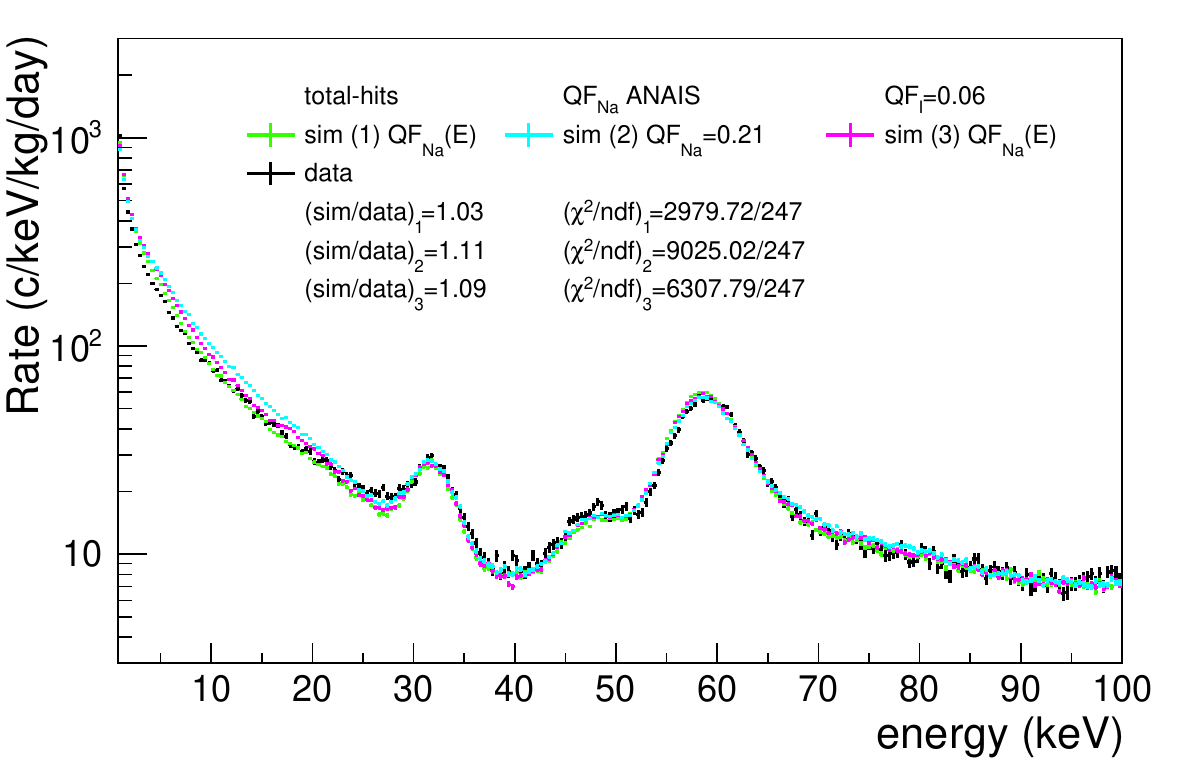}}
    {\includegraphics[width=0.46\textwidth]{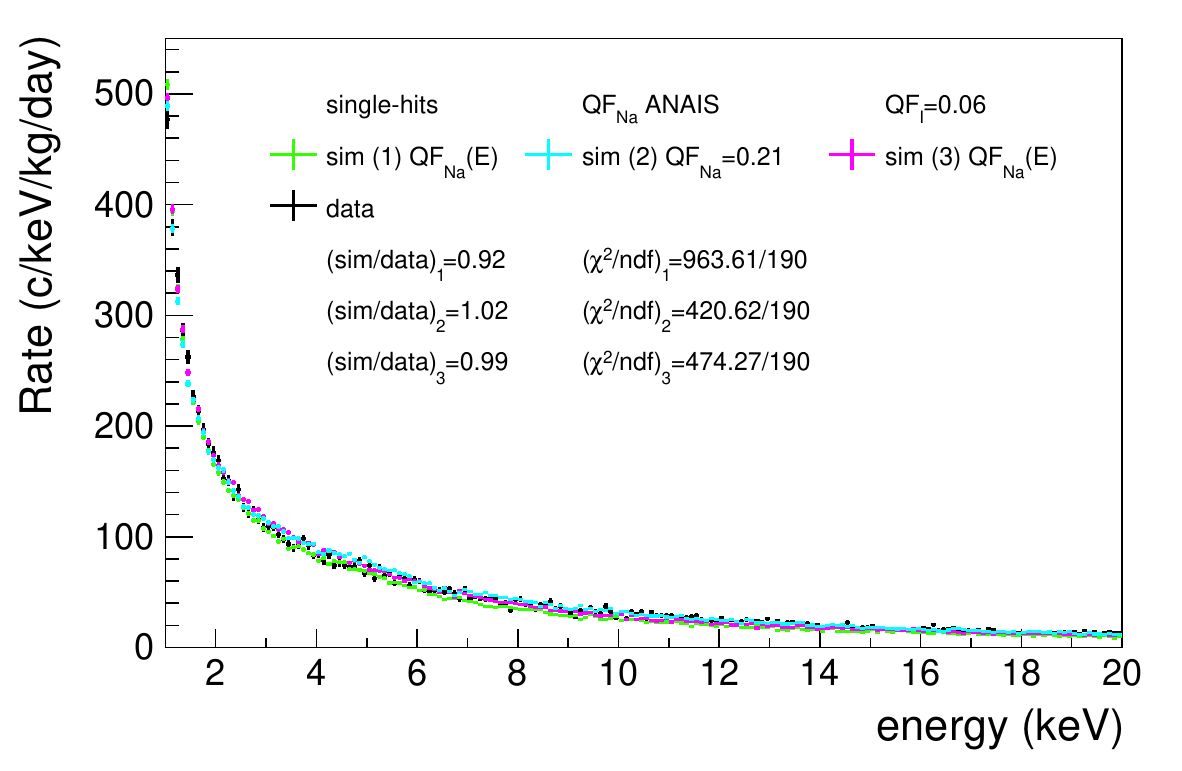}}
    {\includegraphics[width=0.46\textwidth]{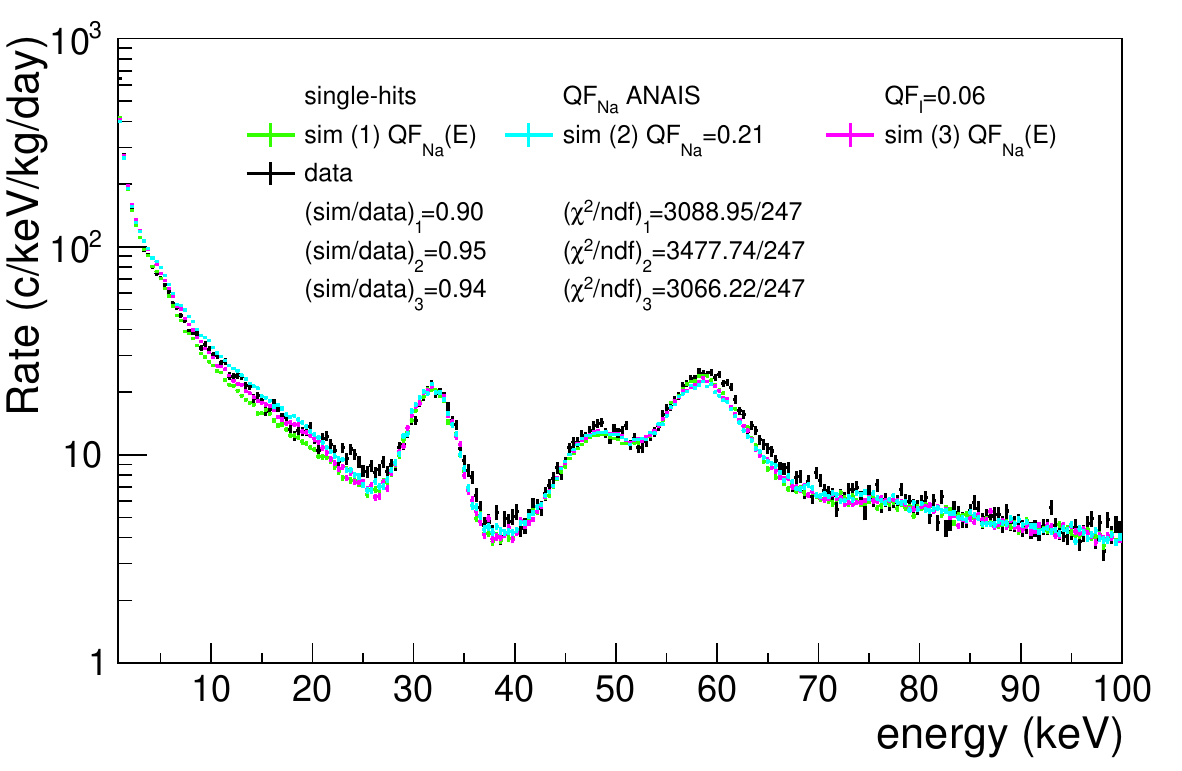}}
    {\includegraphics[width=0.46\textwidth]{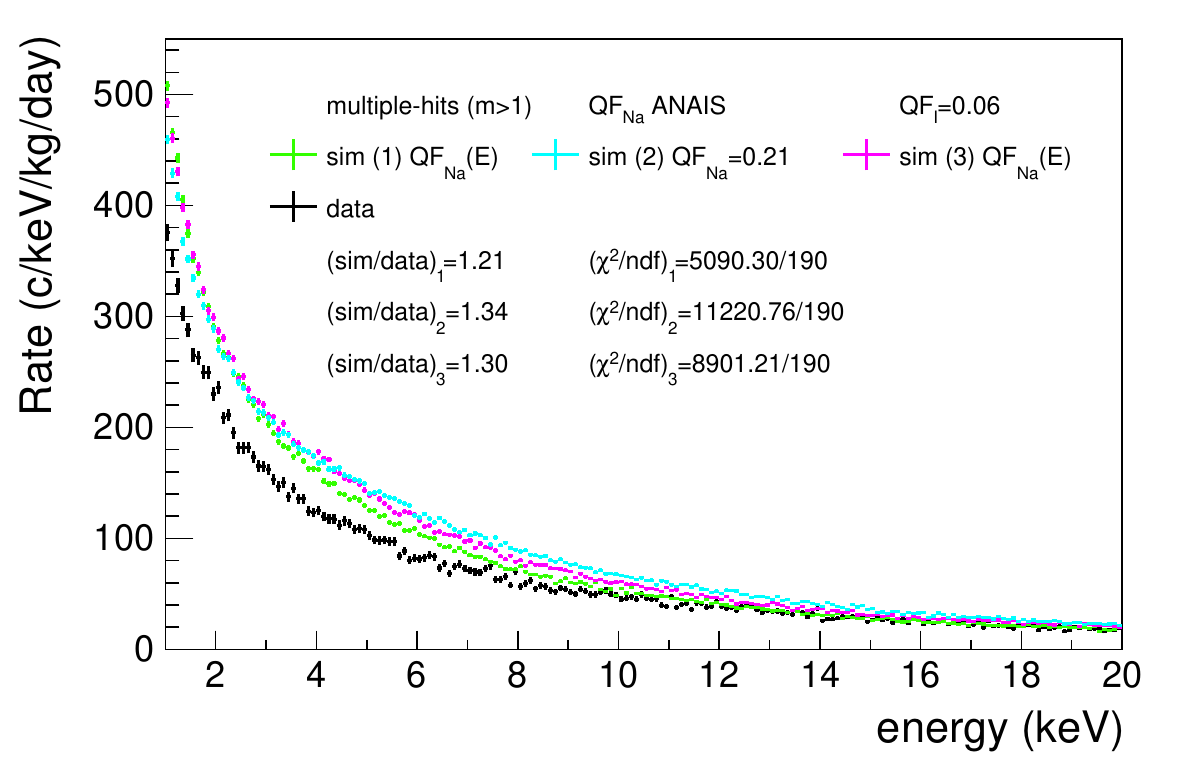}}
    {\includegraphics[width=0.46\textwidth]{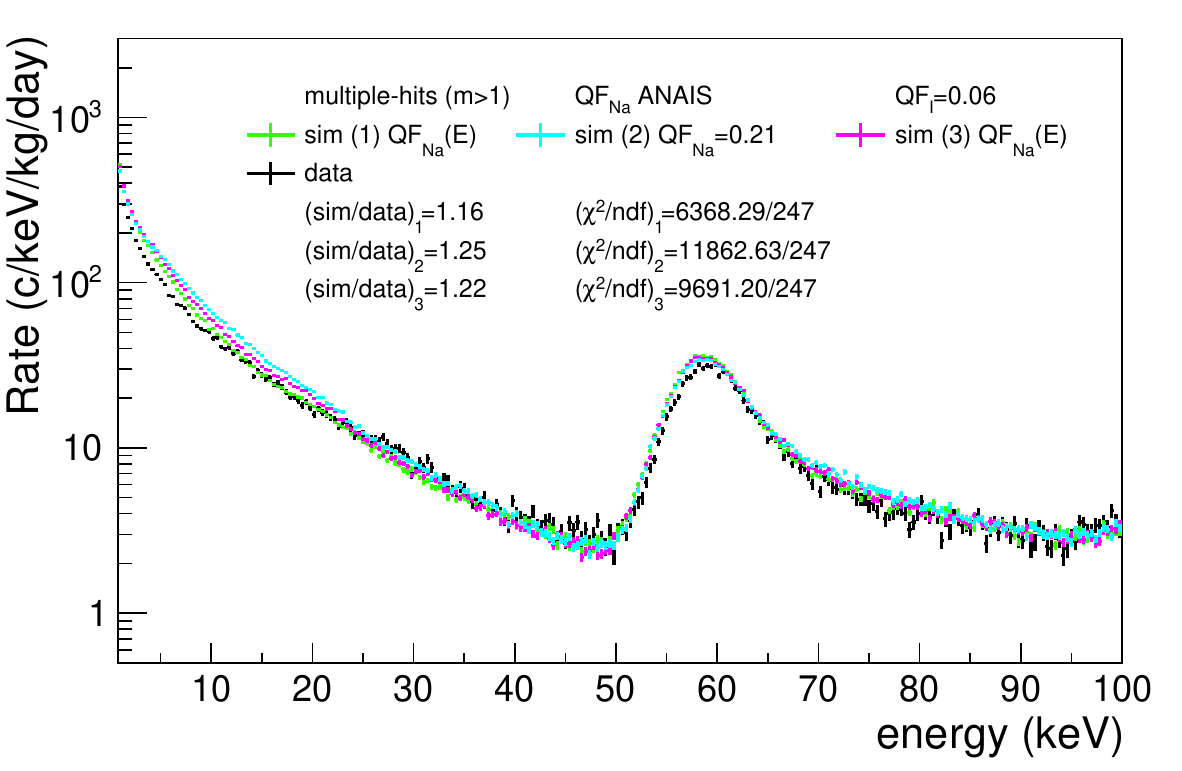}}
    {\includegraphics[width=0.46\textwidth]{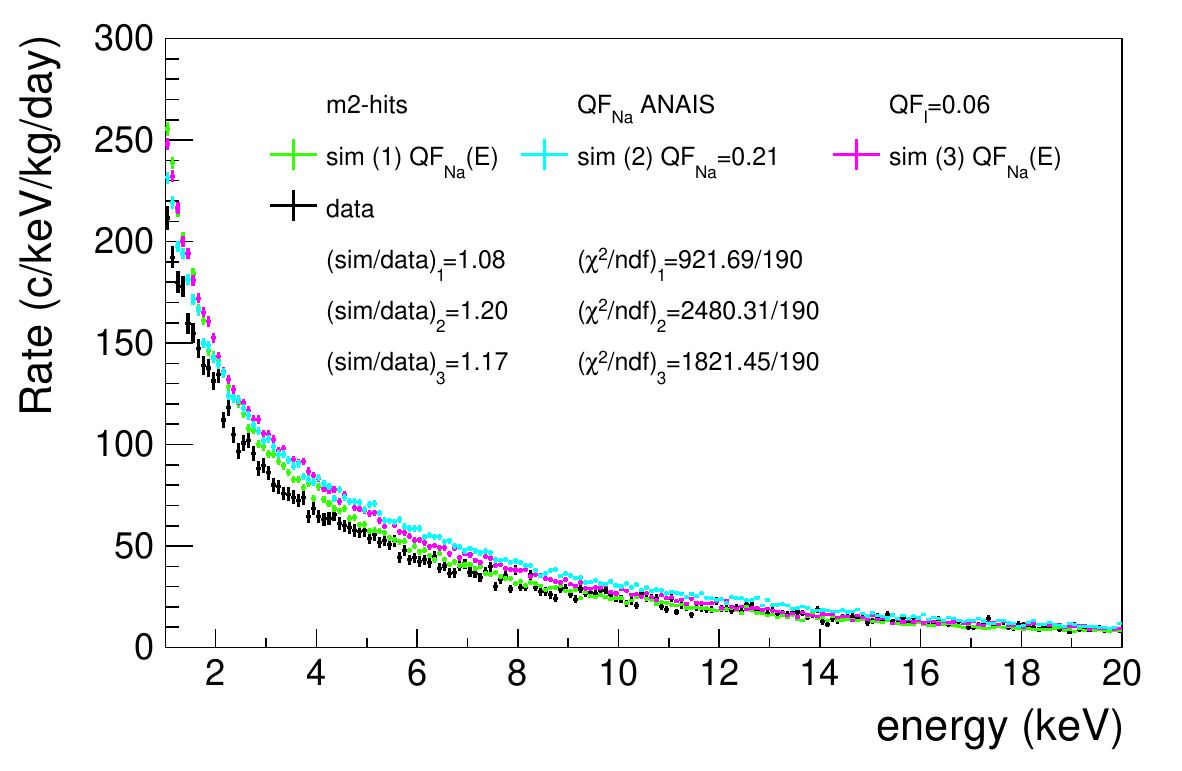}}
    {\includegraphics[width=0.46\textwidth]{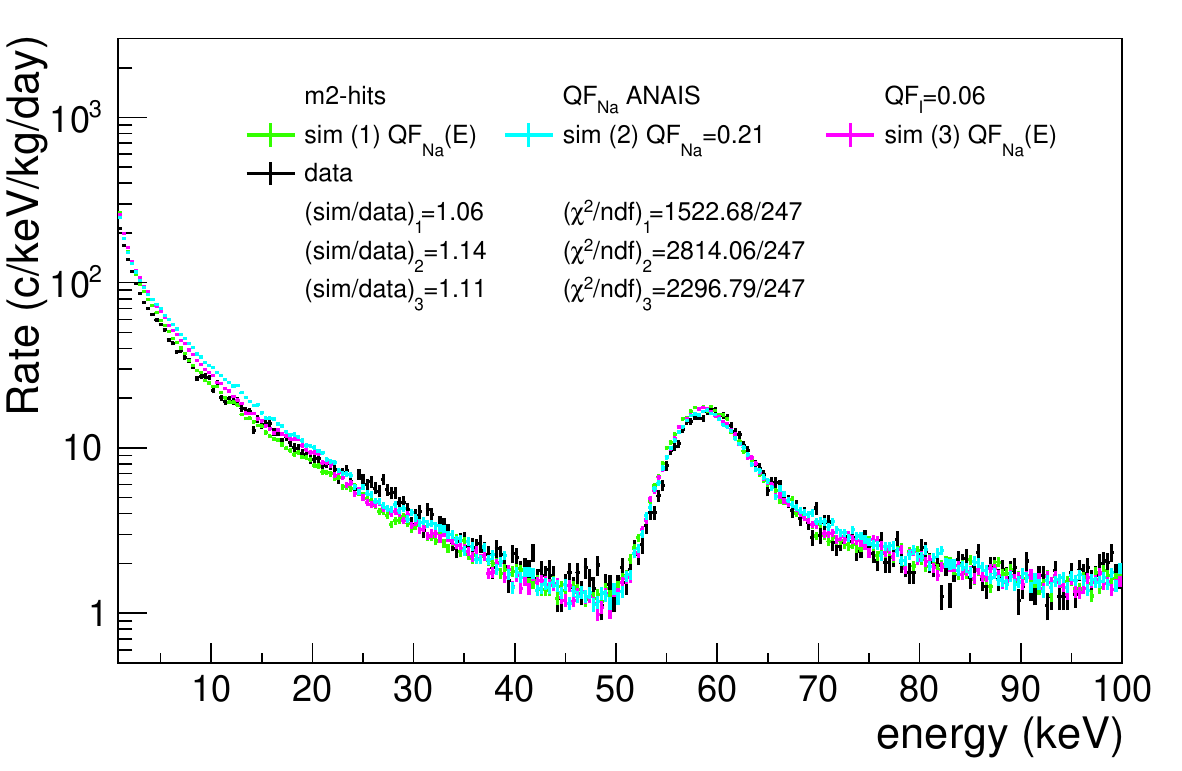}}

    \caption{\label{QFanaistodos} Comparison between the energy spectra measured in the west-face neutron calibration for the sum of the nine ANAIS-112 detectors (black) and the corresponding simulations, assuming ANAIS-112 QFs. This includes ANAIS(1) (green), ANAIS(2) (cyan), and ANAIS(3) (magenta). For iodine, a constant QF\textsubscript{I} of 0.06 is assumed. Panels show the ratio between simulation and experimental data, as well as the goodness of the comparison. The left column displays the low-energy region, while the right column shows the medium-energy range. \textbf{First row:} total-hits. \textbf{Second row:} single-hits. \textbf{Third row:} multiple-hits (m>1). \textbf{Fourth row:} m2-hits. }
\end{figure}

Another potential explanation is an incorrect implementation of the multiplicity distribution of $^{252}$Cf neutrons in Geant4, which cannot be fully excluded. This possibility could be tested in future measurements with other neutron sources, once the ANAIS-112 DM search is completed, for example by using monoenergetic sources or by modifying the neutron energy range with alternative sources. Nonetheless, the agreement observed for the m2-hit population, expected to be less sensitive to such uncertainties, is very good.

Because the Lindhard dependence is more physically motivated than a simple linear fit, this dependence will be used from this point onwards to present the ANAIS(1) QF\textsubscript{Na} results reported by ANAIS.

Figure \ref{QFanaisperdet} shows the detector-by-detector agreement for total-hit events for the three QF\textsubscript{Na} models of the ANAIS crsytals. The overall behavior and spectral shape are consistent across all nine detectors, with a very satisfactory level of agreement for the three of them. The differences between the simulated spectra obtained with the various QF\textsubscript{Na} models become relevant below 30 keV, where the response is dominated by NRs. This effect is more clearly illustrated in Figure \ref{QFanaistodos}, which presents the summed spectra from all nine detectors for the total-, single-, multiple-, and m2-hit populations, in both the low- and medium-energy ranges.



The simulation is sensitive enough to show preference among the three  QF\textsubscript{Na} models. The panels of Figure \ref{QFanaistodos}, as well as those throughout this chapter, display the simulation-to-data ratios along with the reconstructed $\chi^2/\mathrm{ndf}$ in the [1–20] keV and [1–100] keV energy regions, depending on the range shown in each figure. The ANAIS(2) model (constant model) generally yields a poorer agreement, both in spectral shape, simulation-to-data ratios and $\chi^2$ statistics.

Regarding the energy-dependent QF\textsubscript{Na} models of ANAIS, ANAIS(1) and ANAIS(3), it is not straightforward to favor one over the other. The $\chi^2/\mathrm{ndf}$ values are significantly better for ANAIS(1) when using the Lindhard model in the multiple-hit and total-hit distributions, though not in the single-hit case. However, in terms of spectral shape, particularly for multiple-hit events, ANAIS(1) provides a better agreement. This improved description in the multiple-hit population translates into a better match in the total-hit spectrum when using this QF\textsubscript{Na}. Nonetheless, ANAIS(3) also provides a good description, and it can be concluded that the ANAIS-112 neutron calibration data are more compatible with energy-dependent QF\textsubscript{Na} models.

When compared to DAMA/LIBRA QFs in Figure \ref{DAMATotal}, the agreement between data and simulation is significantly better when using the ANAIS QF models.

\subsection{Other QF\textsubscript{Na} models}\label{otherNa}

The validity of alternative QF\textsubscript{Na} models is now explored. 

\begin{figure}[t!]
    \centering
    {\includegraphics[width=0.46\textwidth]{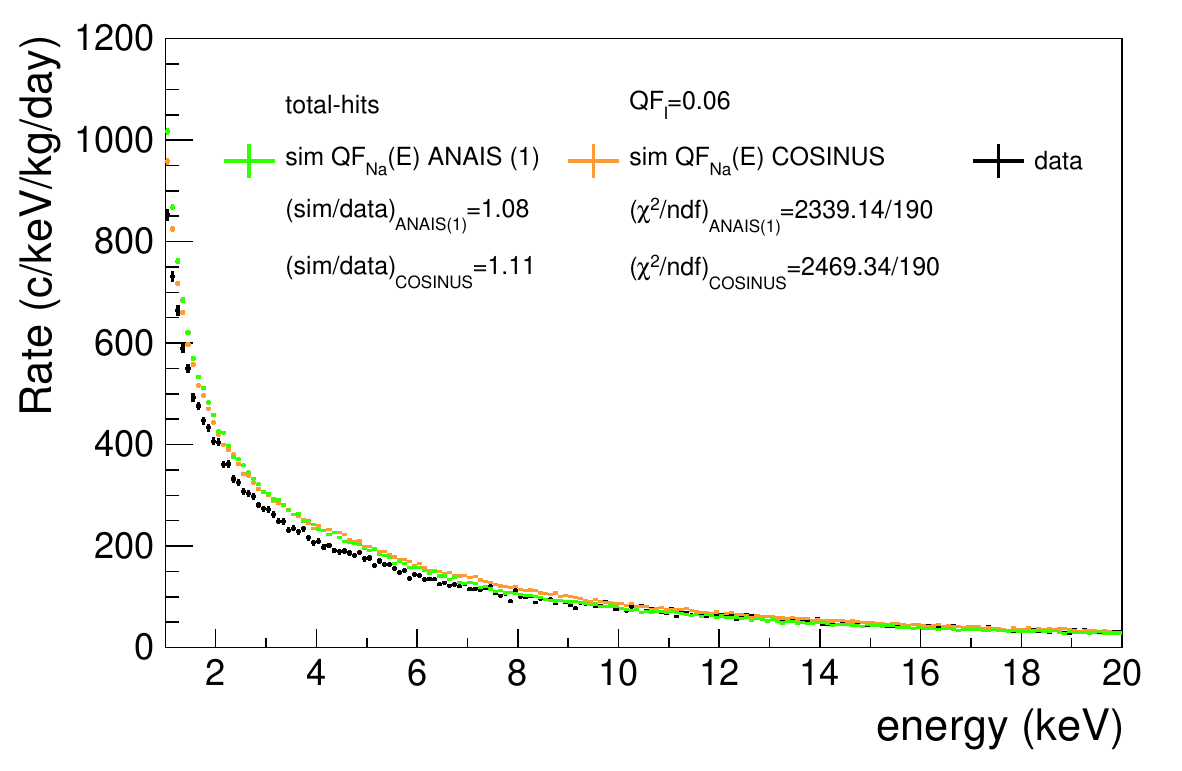}}
    {\includegraphics[width=0.46\textwidth]{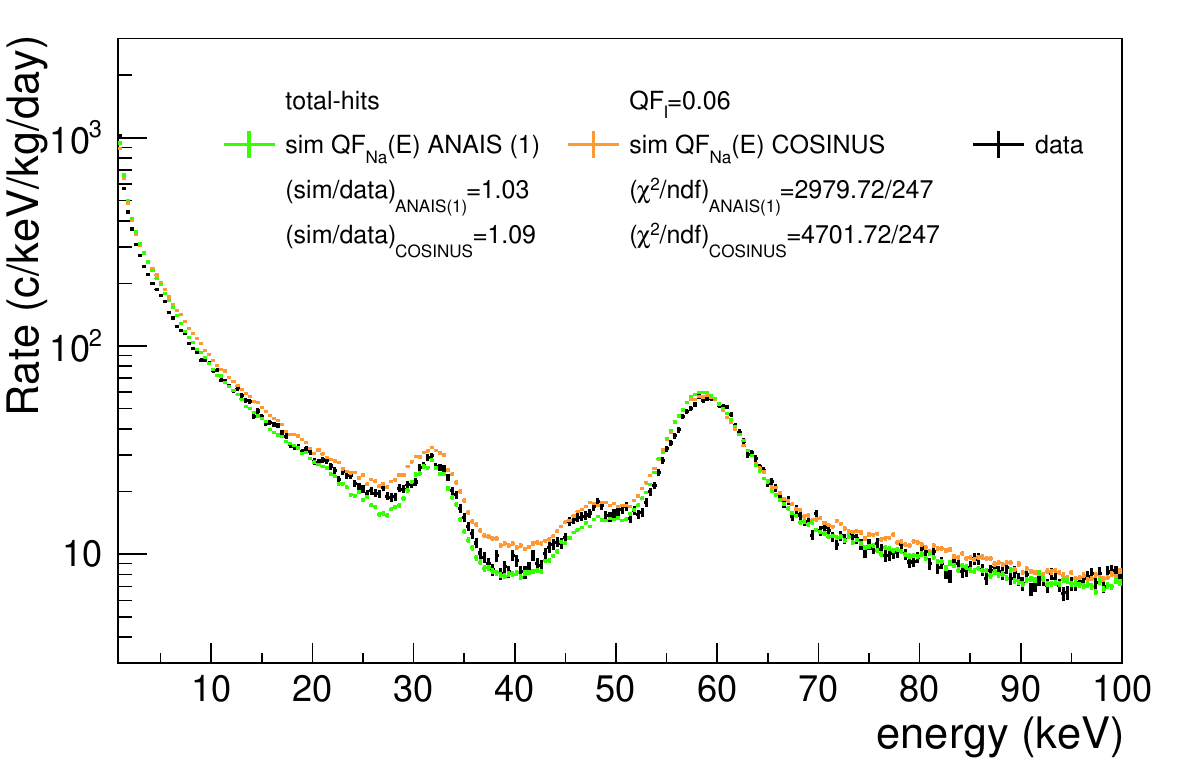}}
    {\includegraphics[width=0.46\textwidth]{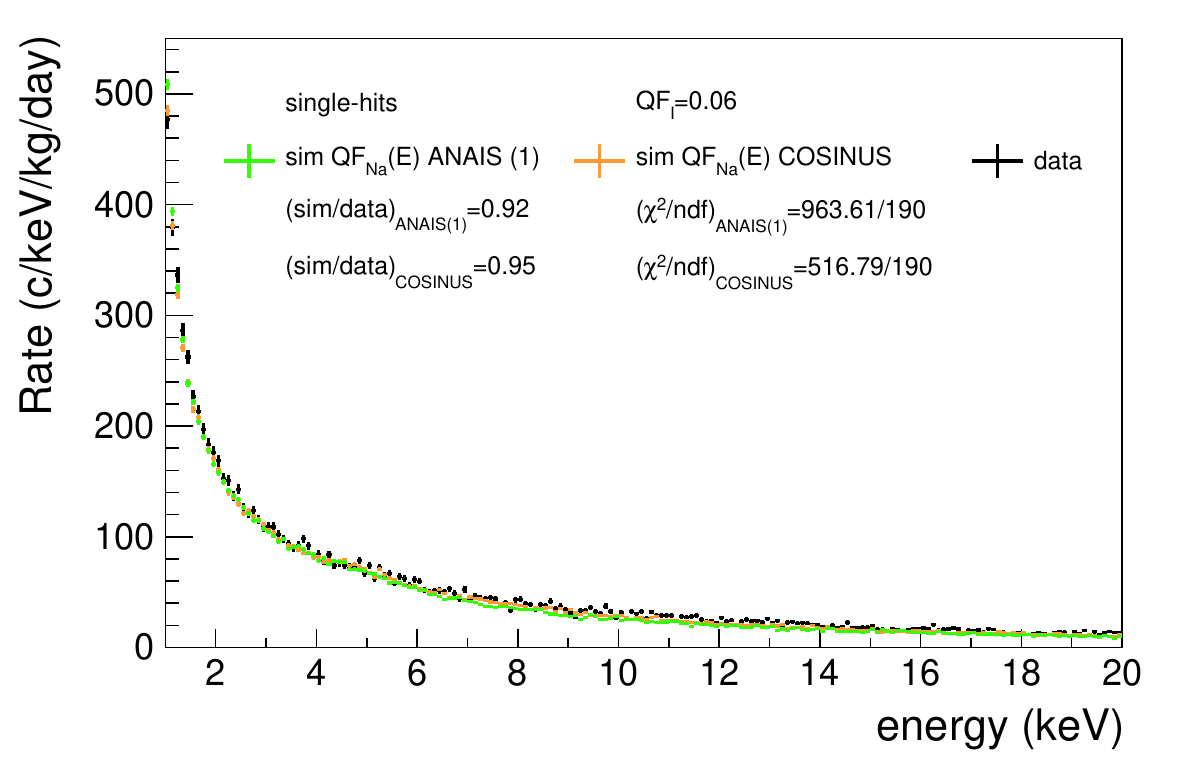}}
    {\includegraphics[width=0.46\textwidth]{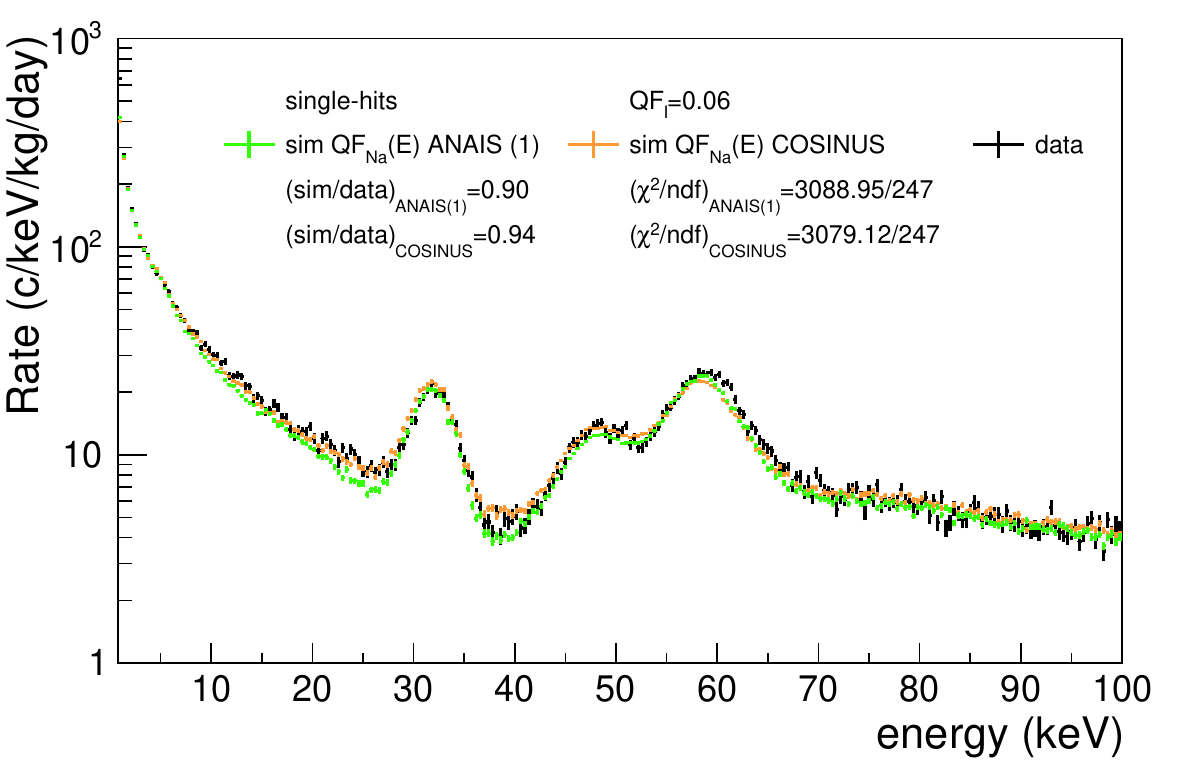}}
    {\includegraphics[width=0.46\textwidth]{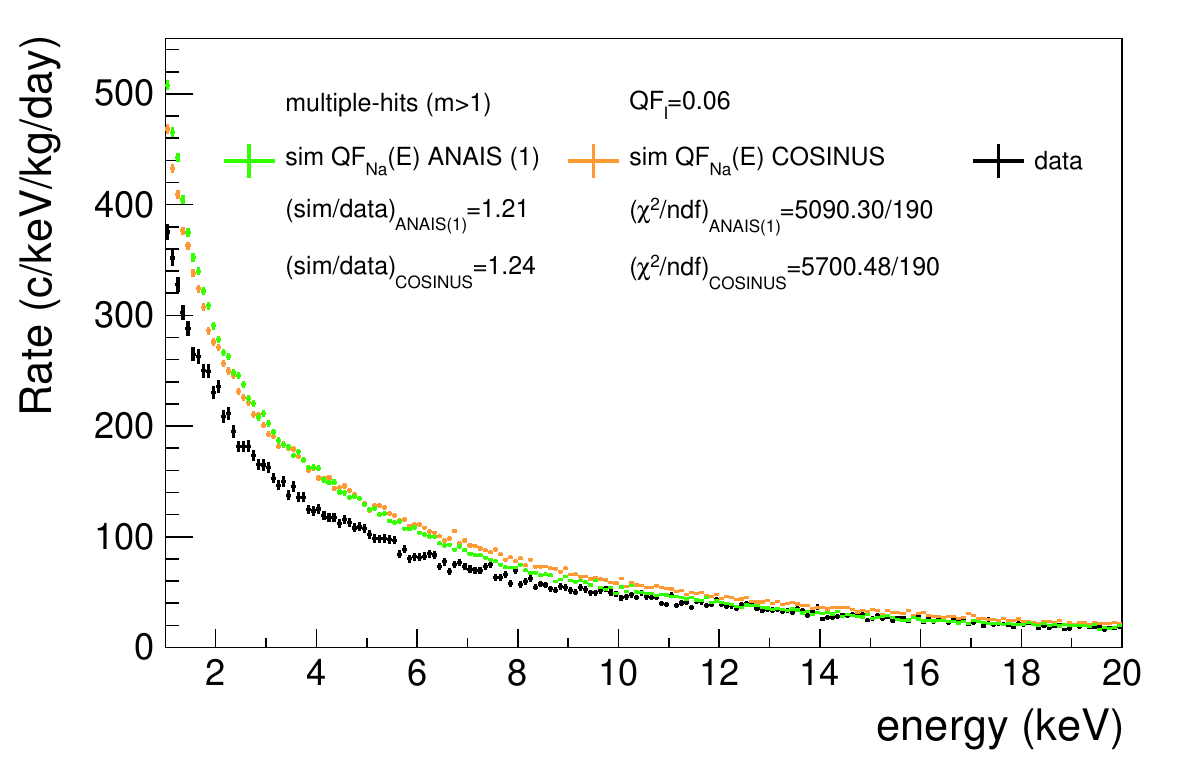}}
    {\includegraphics[width=0.46\textwidth]{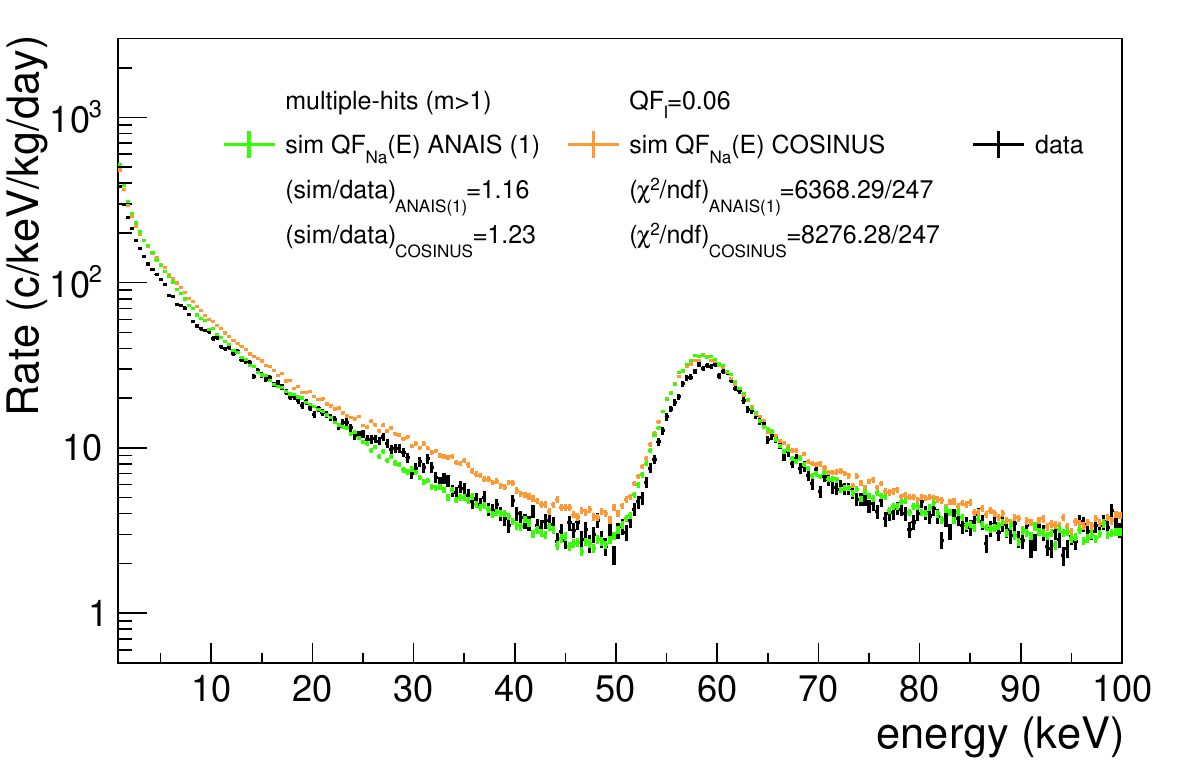}}
    {\includegraphics[width=0.46\textwidth]{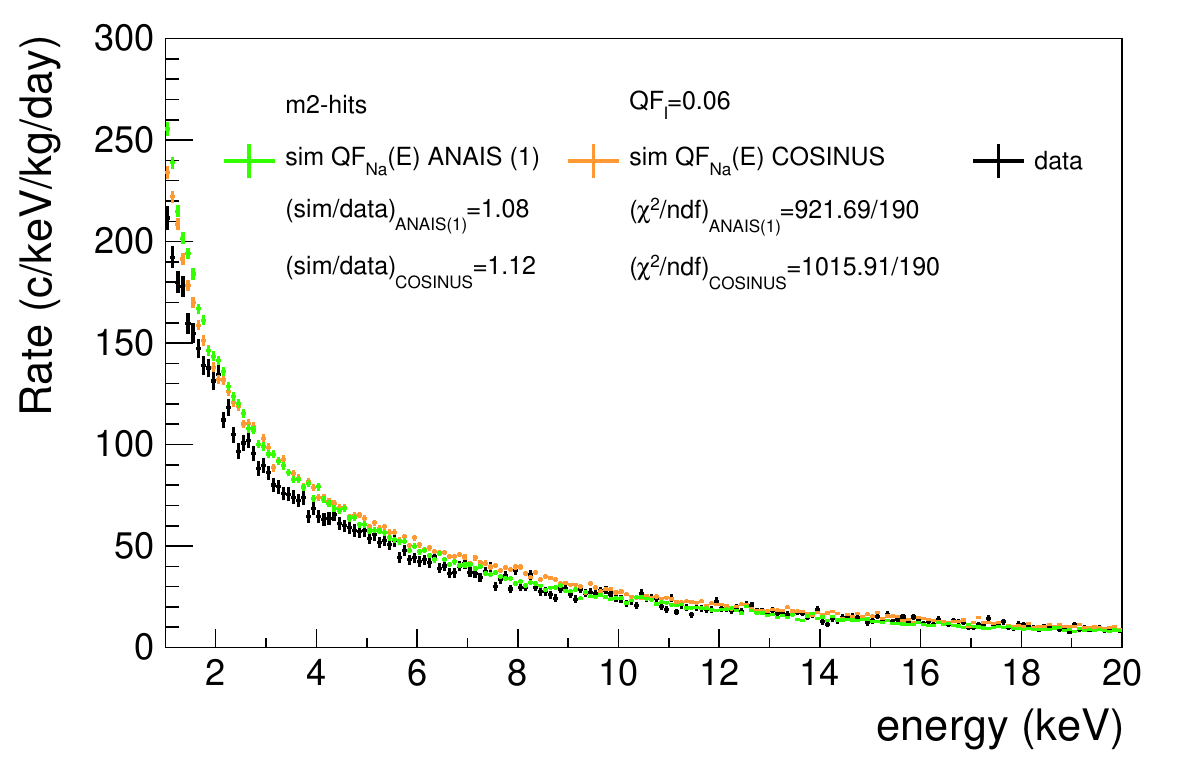}}
    {\includegraphics[width=0.46\textwidth]{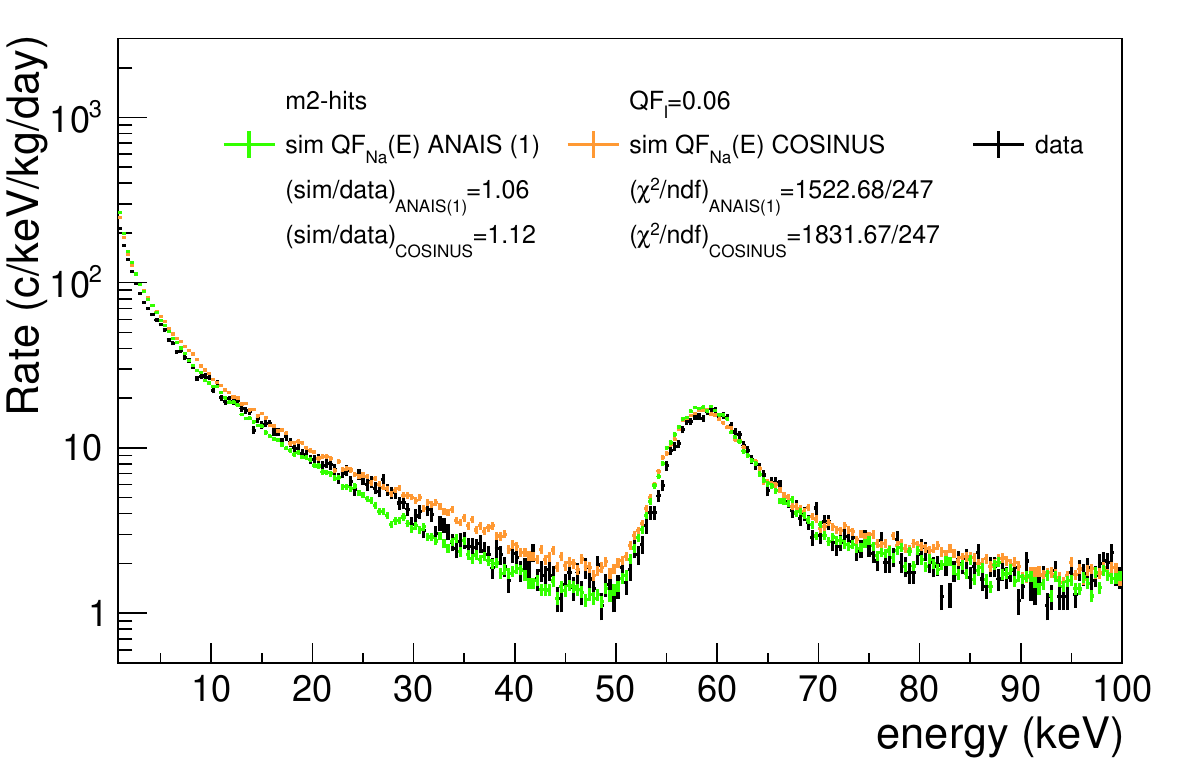}}

    \caption{\label{COSINUSQF} Comparison between the energy spectra measured in the west-face neutron calibration for the sum of the nine ANAIS-112 detectors (black) and the corresponding simulations, assuming: ANAIS(1) QF\textsubscript{Na} (green) and COSINUS QF\textsubscript{Na} (orange). For iodine, a constant QF\textsubscript{I} of 0.06 is assumed. Panels show the ratio between simulation and experimental data, as well as the goodness of the comparison. The left column displays the low-energy region, while the right column shows the medium-energy range. \textbf{First row:} total-hits. \textbf{Second row:} single-hits. \textbf{Third row:} multiple-hits (m>1). \textbf{Fourth row:} m2-hits. }
\end{figure}

\begin{figure}[t!]
    \centering
    {\includegraphics[width=0.46\textwidth]{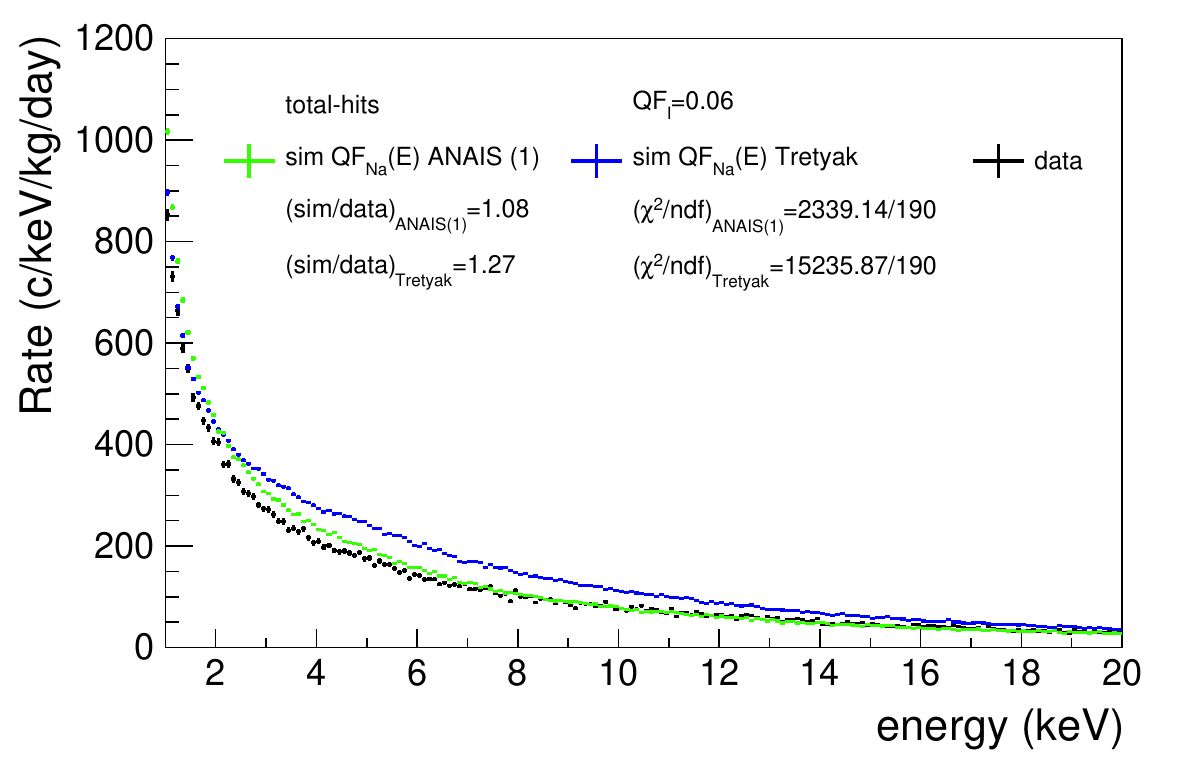}}
    {\includegraphics[width=0.46\textwidth]{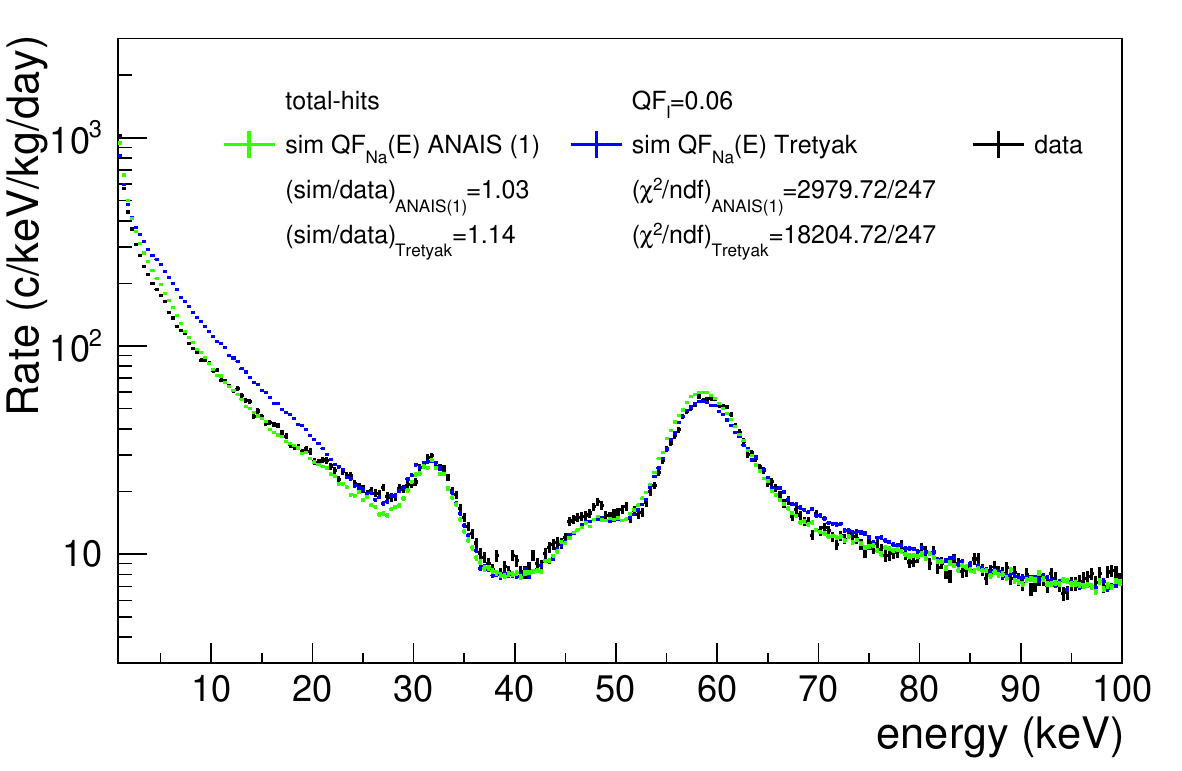}}
    {\includegraphics[width=0.46\textwidth]{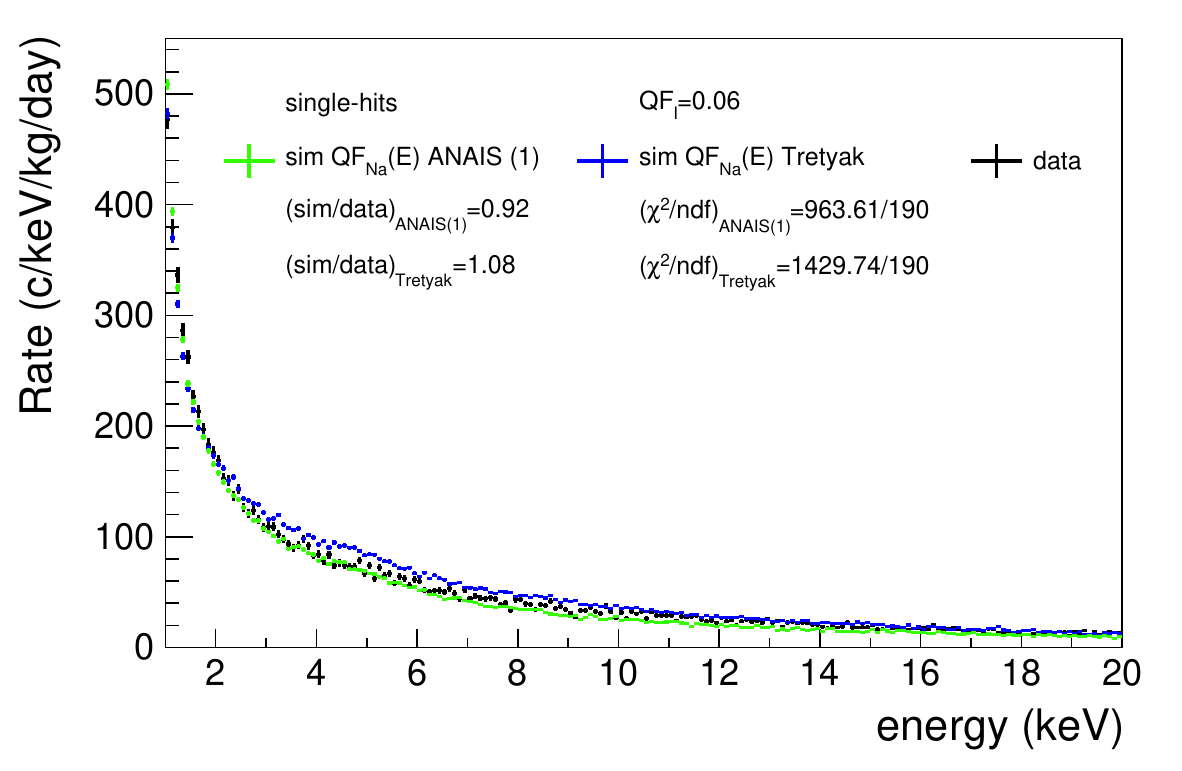}}
    {\includegraphics[width=0.46\textwidth]{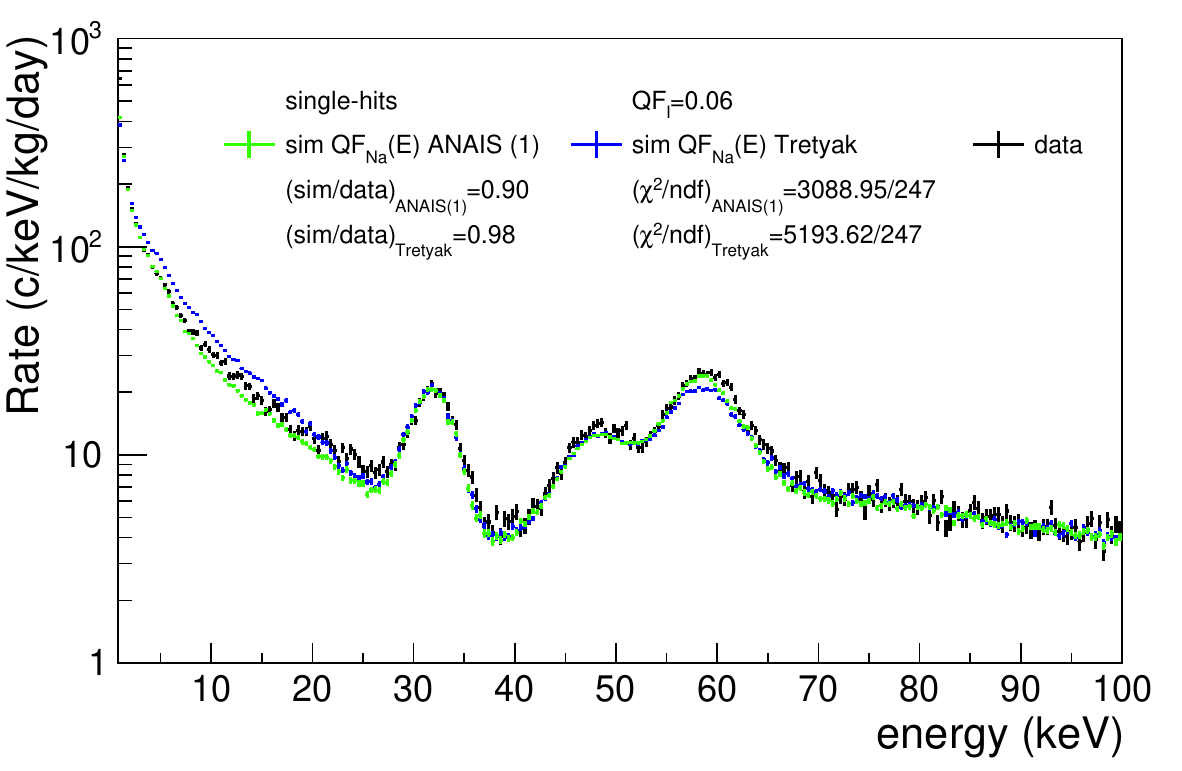}}
    {\includegraphics[width=0.46\textwidth]{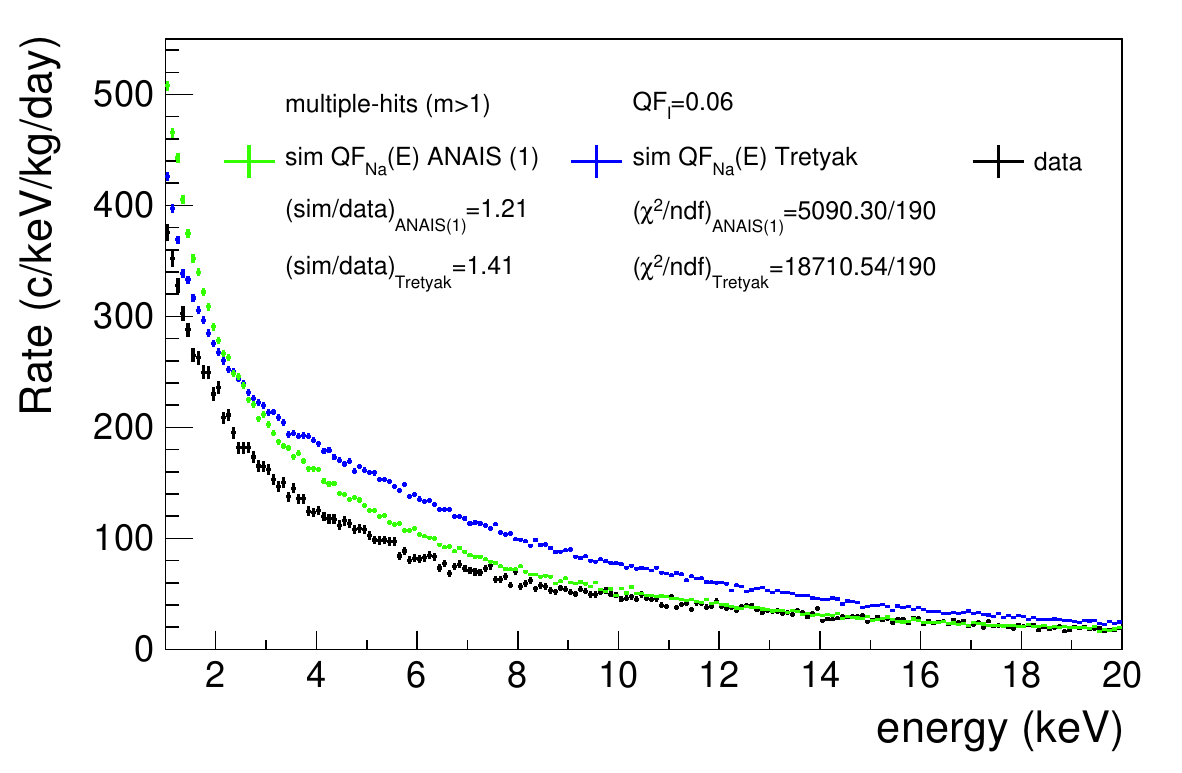}}
    {\includegraphics[width=0.46\textwidth]{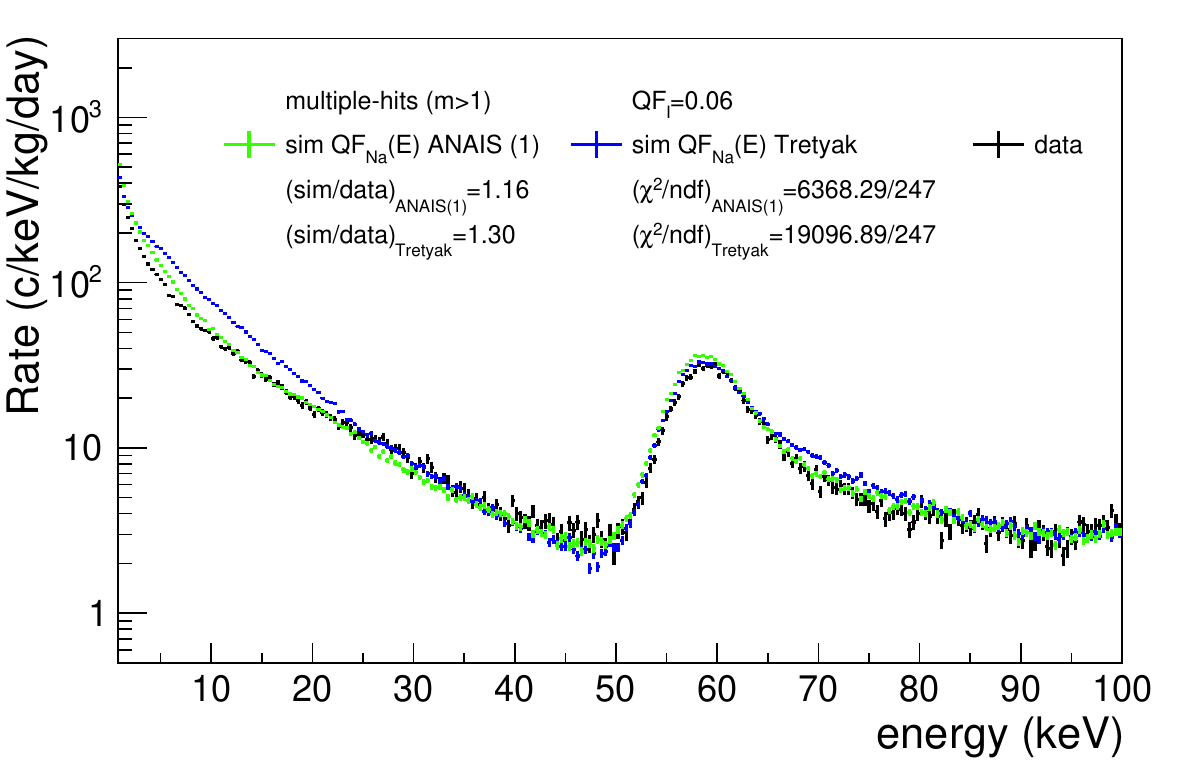}}
    {\includegraphics[width=0.46\textwidth]{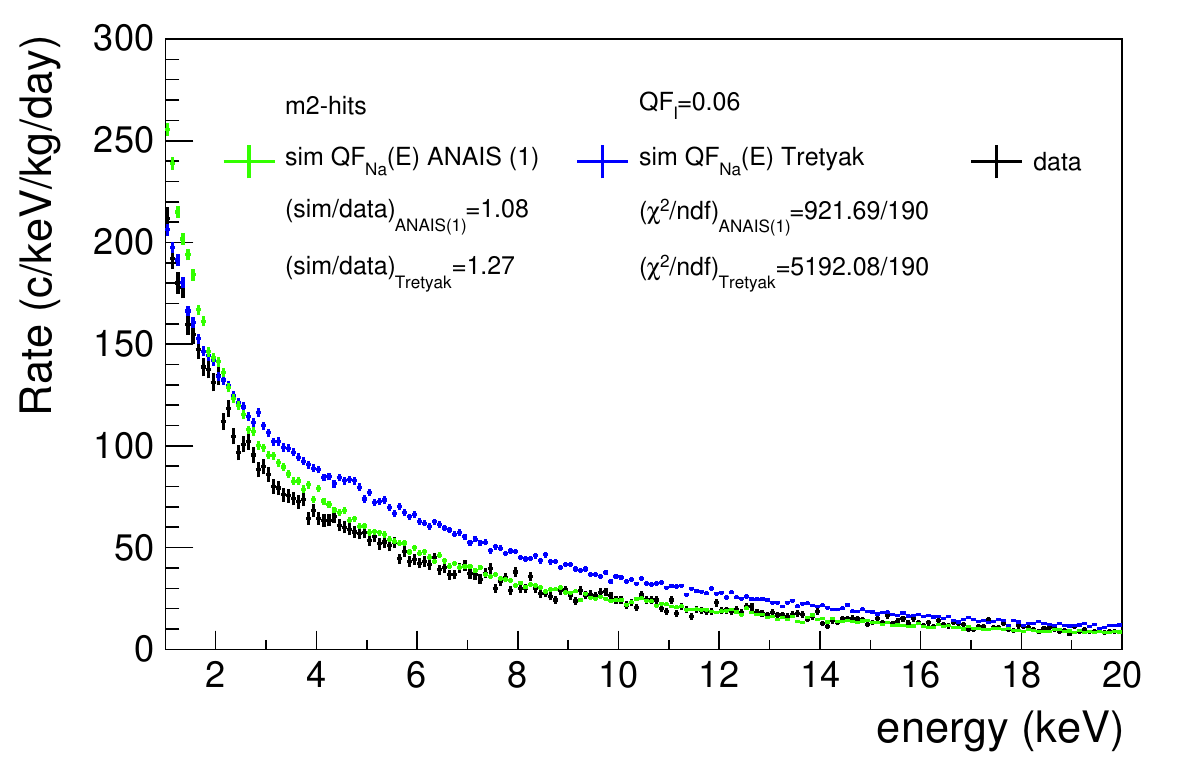}}
    {\includegraphics[width=0.46\textwidth]{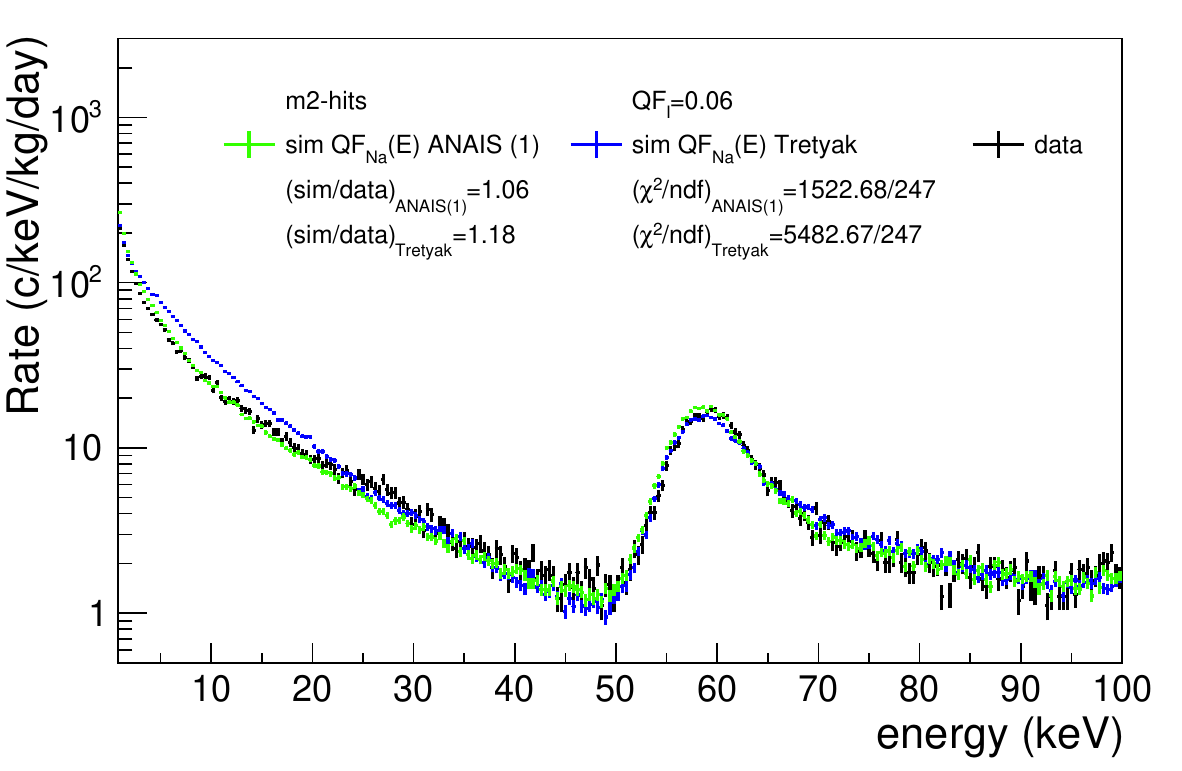}}

    \caption{\label{tetrytodos} Comparison between the energy spectra measured in the west-face neutron calibration for the sum of the nine ANAIS-112 detectors (black) and the corresponding simulations, assuming: ANAIS(1) QF\textsubscript{Na} (green) and the Tretyak et al. QF\textsubscript{Na} (blue). For iodine, a constant QF\textsubscript{I} of 0.06 is assumed. Panels show the ratio between simulation and experimental data, as well as the goodness of the comparison. The left column displays the low-energy region, while the right column shows the medium-energy range. \textbf{First row:} total-hits. \textbf{Second row:} single-hits. \textbf{Third row:} multiple-hits (m>1). \textbf{Fourth row:} m2-hits. }
\end{figure}

\begin{figure}[b!]
    \centering
    {\includegraphics[width=0.49\textwidth]{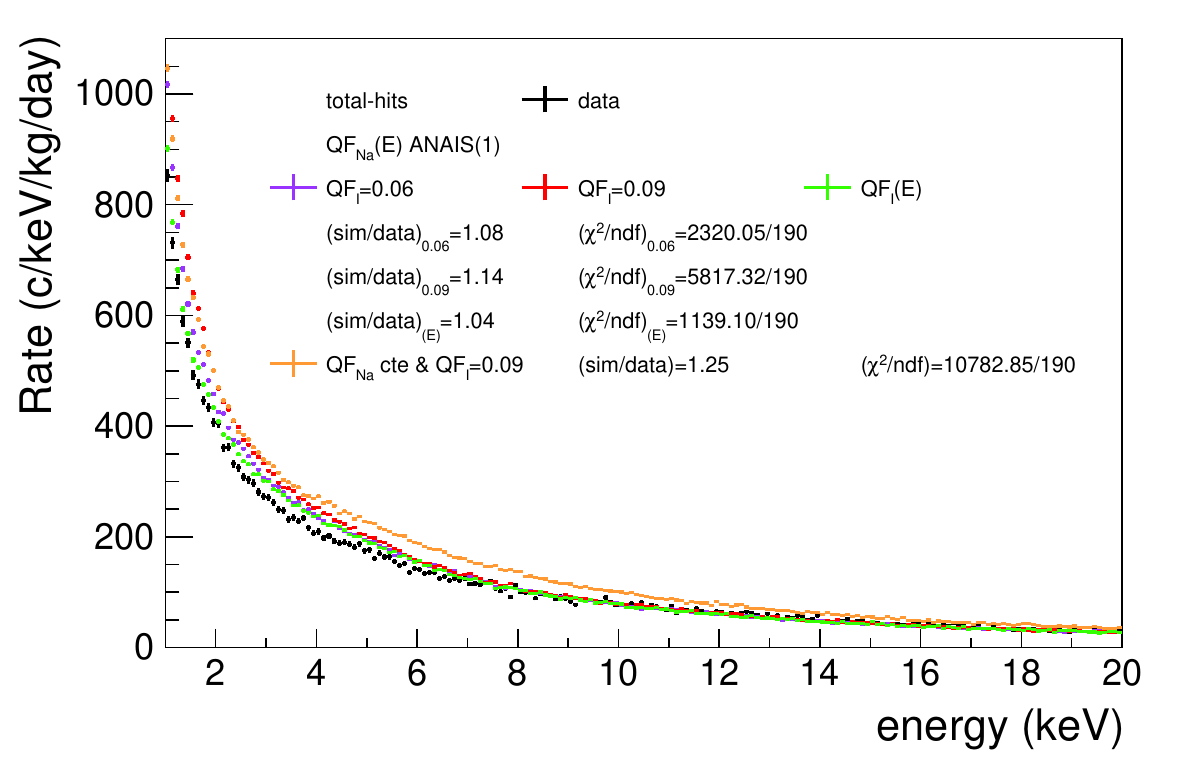}}
    {\includegraphics[width=0.49\textwidth]{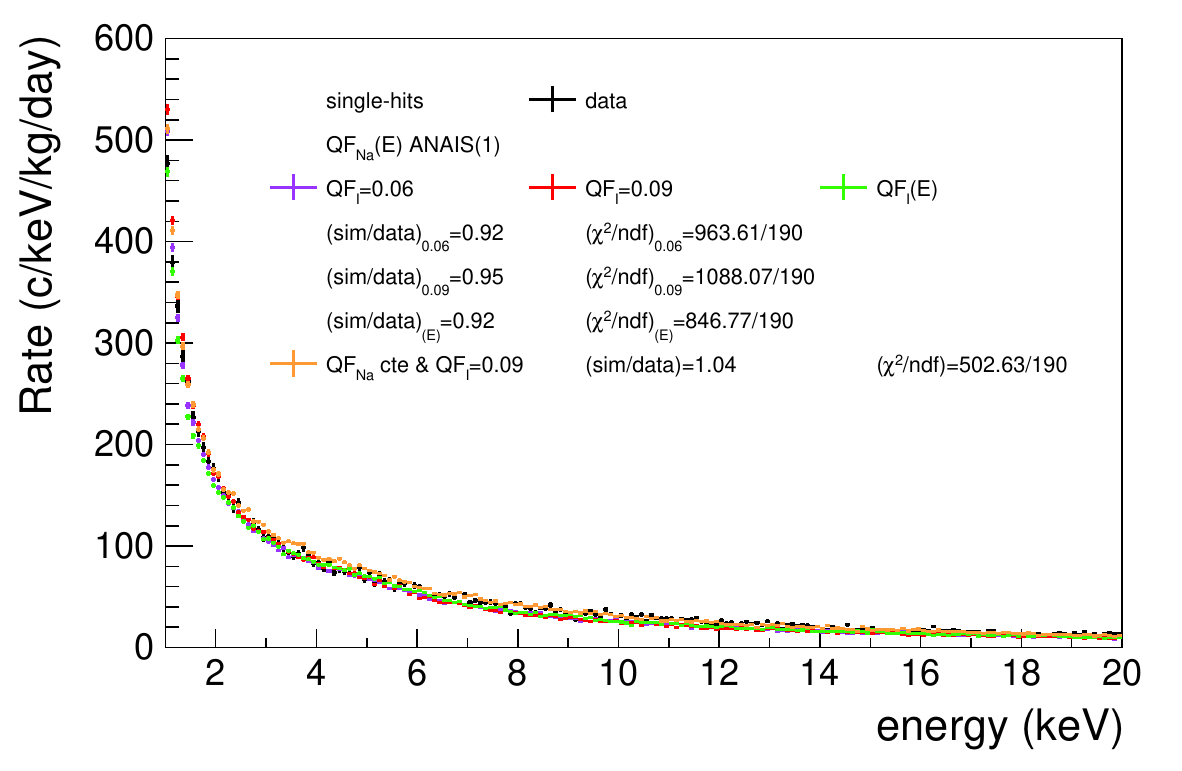}}
    {\includegraphics[width=0.49\textwidth]{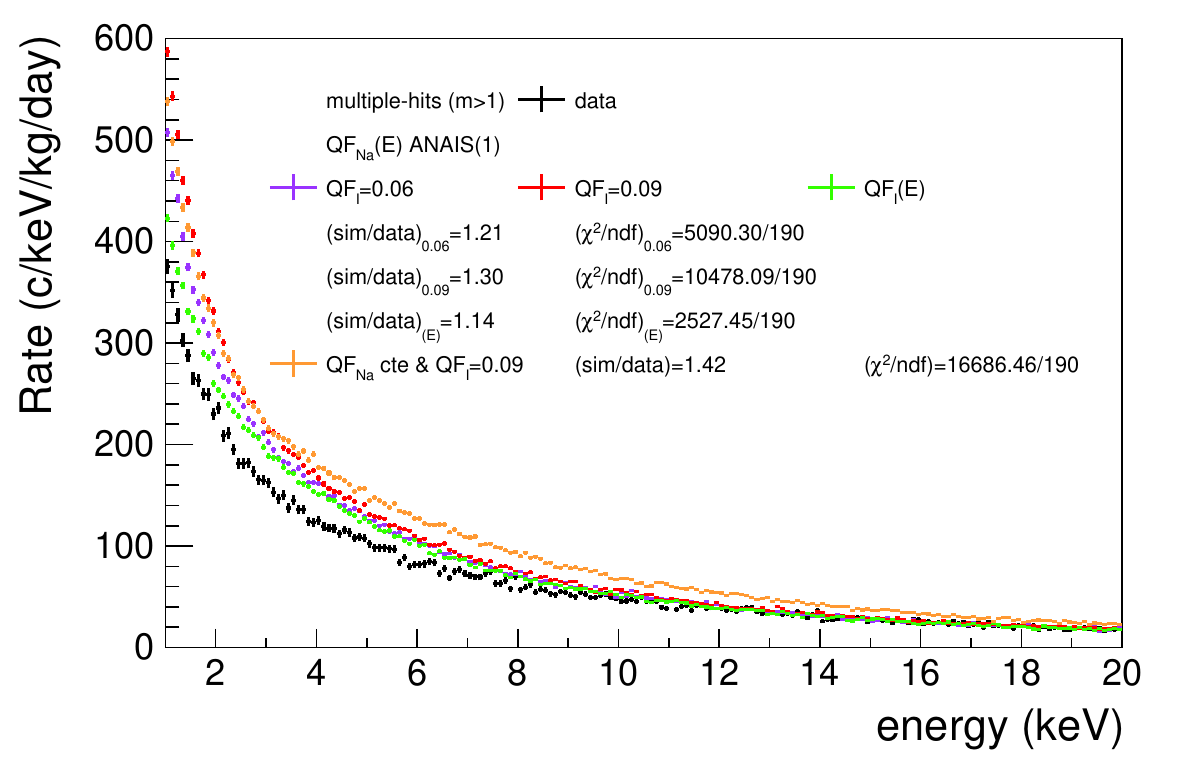}}
     {\includegraphics[width=0.49\textwidth]{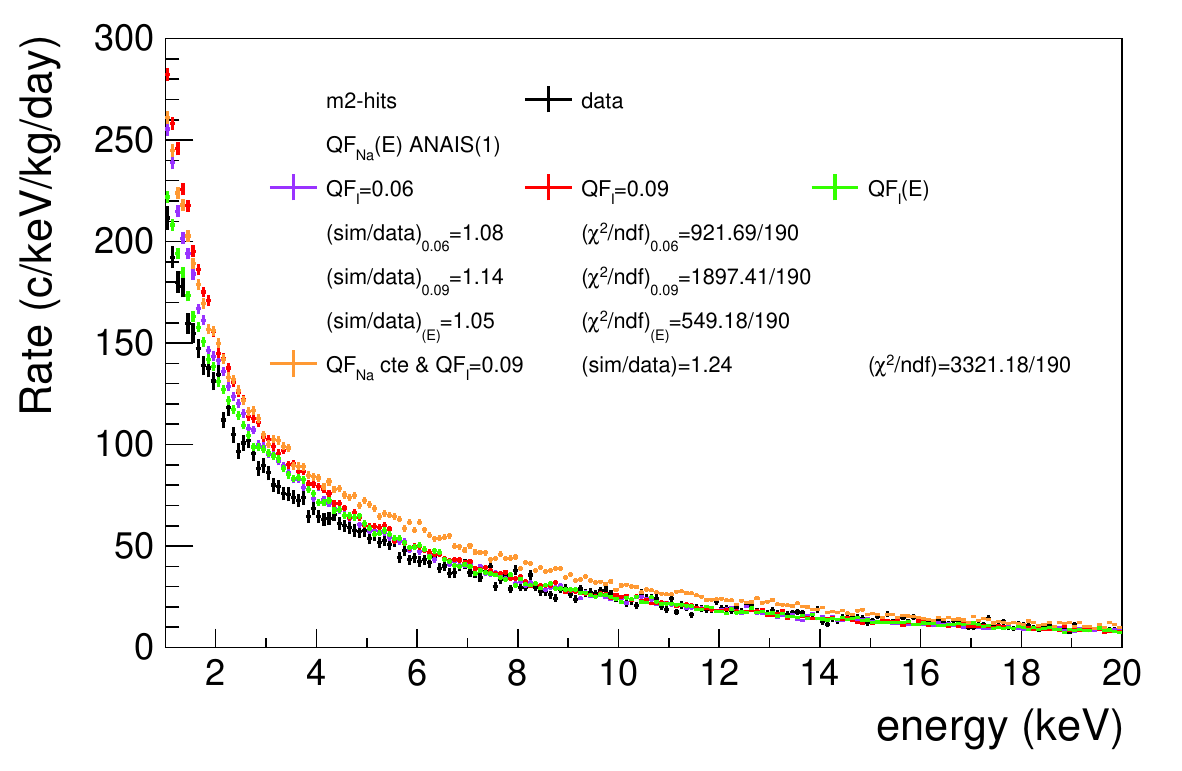}}

    \caption{\label{compareIodine} Comparison between the low-energy spectra measured in the west-face neutron calibration for the sum of the nine ANAIS-112 detectors (black) and the corresponding simulations, assuming ANAIS(1) QF\textsubscript{Na} and various iodine models, specifically a constant QF\textsubscript{I} of 0.09 following DAMA/LIBRA (red), a constant QF\textsubscript{I} of 0.06 following ANAIS crystals value (violet), and an energy-dependent QF\textsubscript{I} compatible with ANAIS-112 value (green). Results are also shown for the ANAIS(2) QF\textsubscript{Na} model, with QF\textsubscript{Na}~=~0.21, asumming a constant QF\textsubscript{I} of 0.09 (orange). Each panel includes the ratio between simulation and experimental data, as well as the goodness of the comparison, in the [1–20] keV energy range. \textbf{Top left panel:} total-hits. \textbf{Top right panel:} single-hits. \textbf{Bottom left:} multiple-hits (m>1). \textbf{Bottom right:} m2-hits. }
\end{figure}

Figure \ref{COSINUSQF} shows the comparison for the sum over all detectors and all event populations considered in this study, evaluating the ANAIS(1) QF\textsubscript{Na} model against the COSINUS QF\textsubscript{Na}. The COSINUS QF yields a good description of the single-hit spectrum; however, its limitations become apparent in the multiple-hit population. In particular, the NR exponential rise up to 50 keV is clearly overestimated and does not match the data as well as the ANAIS QF does. This discrepancy arise from the COSINUS QF\textsubscript{Na} values being higher, leading to a less compressed Na recoil contribution spectrum.

Figure \ref{tetrytodos} presents the analogous comparison when testing the Tetryak QF\textsubscript{Na} model~\cite{tretyak2010semi}. As clearly observed, consistent with the conclusions drawn from monochromatic source measurements, this study disfavors a decreasing QF\textsubscript{Na} with energy, as it is incompatible with the ANAIS-112 data across all event populations. In particular, it leads to a NR population that is clearly overestimated.

\subsection{QF$_{\textnormal{I}}$ models} \label{QFImodels}

After assessing the impact of variations in the QF\textsubscript{Na}, attention is now turned to the study of variations in the QF\textsubscript{I} model. Assuming ANAIS(1) QF\textsubscript{Na}, three different QF\textsubscript{I} models are considered: two constant values, 0.09 and 0.06, as measured by DAMA/LIBRA and ANAIS-112 respectively, and an energy-dependent model compatible with the ANAIS-112 results as presented in Section \ref{TUNL}. In addition, results are also shown for the ANAIS(2) QF\textsubscript{Na} model, with QF\textsubscript{Na}~=~0.21, assuming QF\textsubscript{I} = 0.09.

Figure \ref{compareIodine} presents the comparison of the summed spectra from the nine detectors. It is observed that variations in QF\textsubscript{I} significantly affect the spectra up to approximately 10~keV, in agreement with the behavior shown in Figure \ref{simdistribucion}, where iodine-induced events are dominant only at low energies. Notably, neutron simulations are clearly sensitive to variations in QF\textsubscript{I}, with the most pronounced effects observed in multiple-hit events. The DAMA/LIBRA iodine QF (QF\textsubscript{I} = 0.09) fails to reproduce the data in any population, resulting in a significant excess of events at very low energies. When the ANAIS(2) QF\textsubscript{Na} value of 0.21 is combined with QF\textsubscript{I} = 0.09, the disagreement persists, with multiple-hit events clearly overestimated. This confirms that a QF\textsubscript{I} value of 0.09 does not improve the agreement with data in any scenario.

In contrast, the use of QF\textsubscript{I} = 0.06, as measured by ANAIS, yields better agreement overall, although an excess remains in the multiple-hit spectrum. Part of the discrepancy is effectively corrected when using an energy-dependent QF\textsubscript{I}, resulting in a highly satisfactory agreement between data and simulation.

\subsection{QFs best-describing ANAIS \textsuperscript{252}Cf calibrations}

\begin{figure}[t!]
    \centering
    {\includegraphics[width=0.45\textwidth]{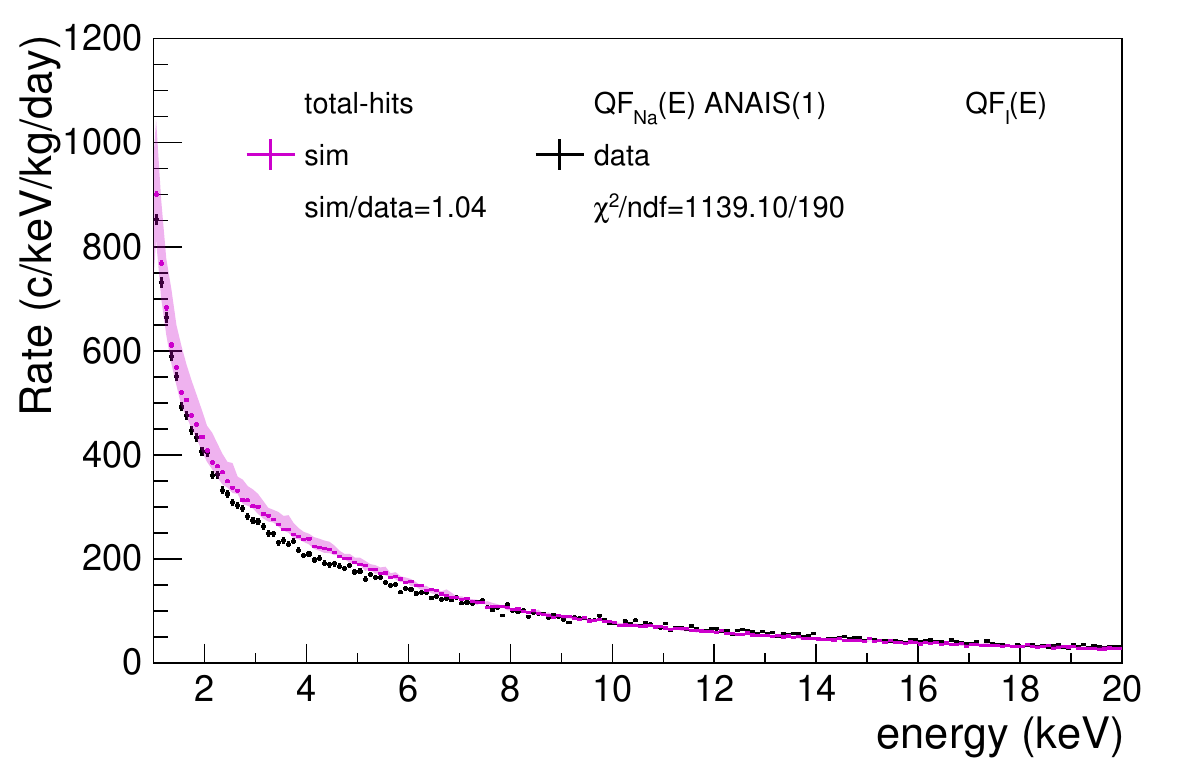}}
    {\includegraphics[width=0.45\textwidth]{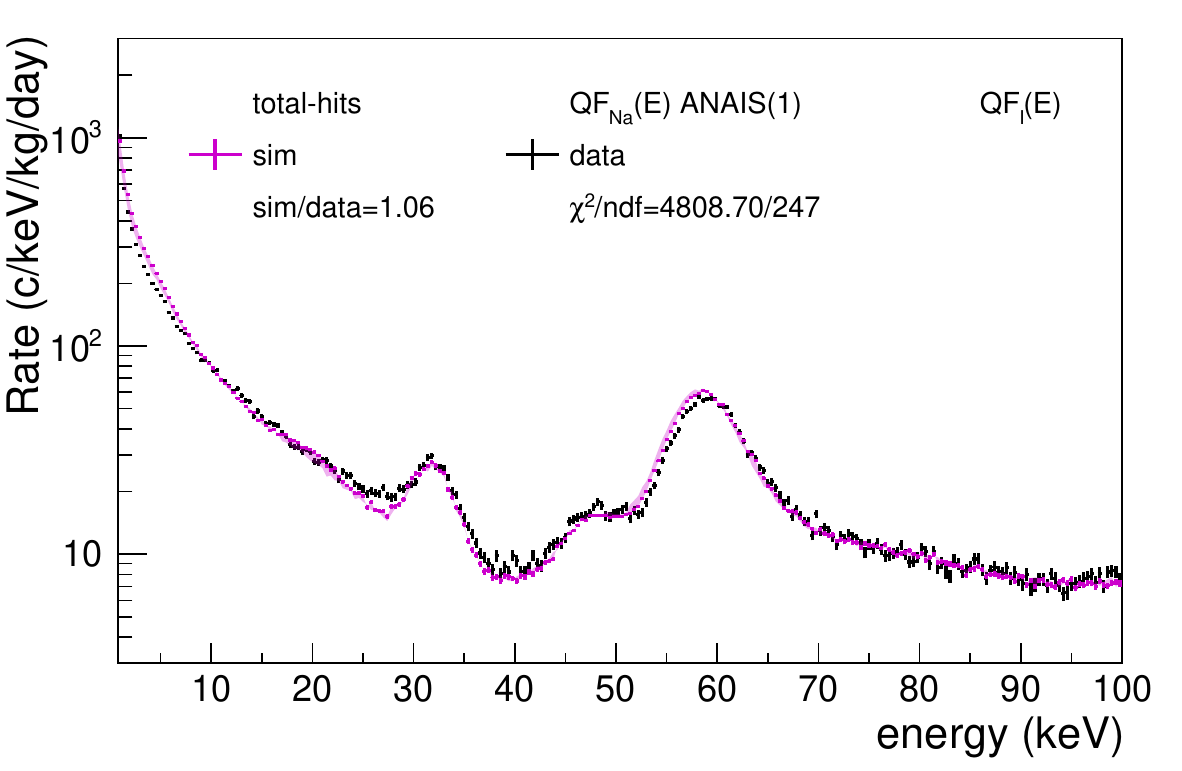}}
    {\includegraphics[width=0.45\textwidth]{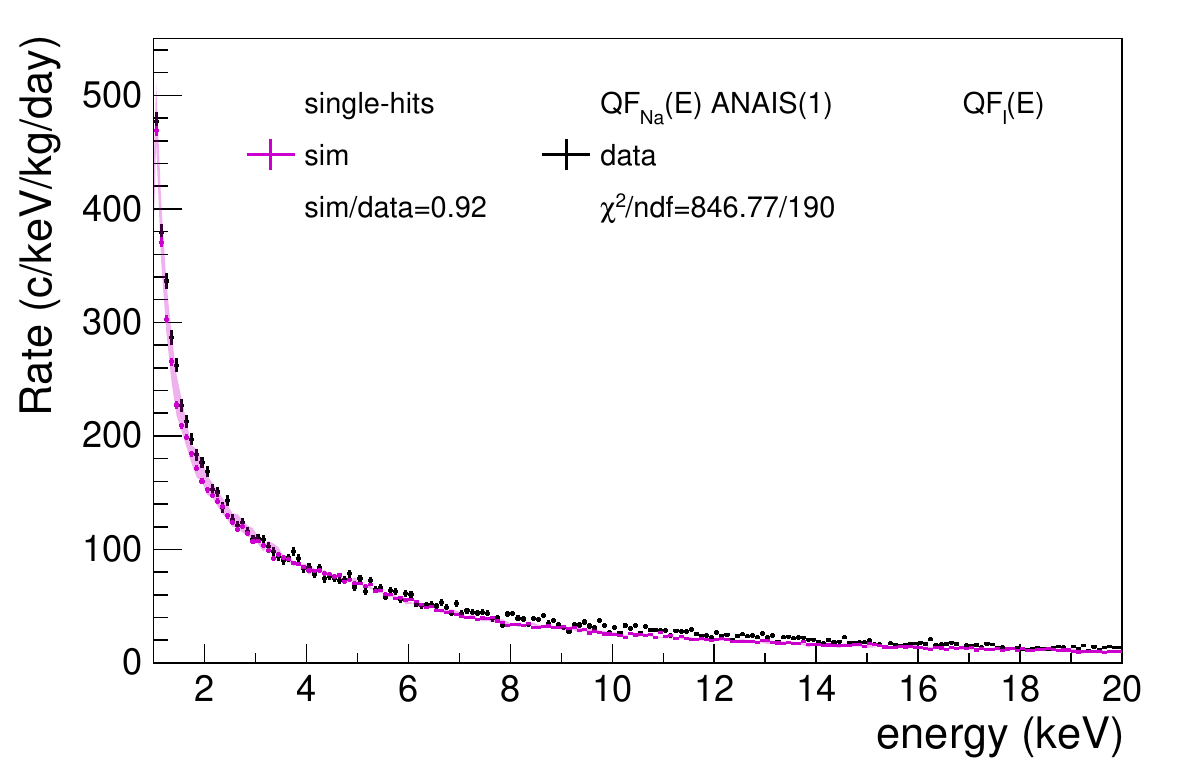}}
    {\includegraphics[width=0.45\textwidth]{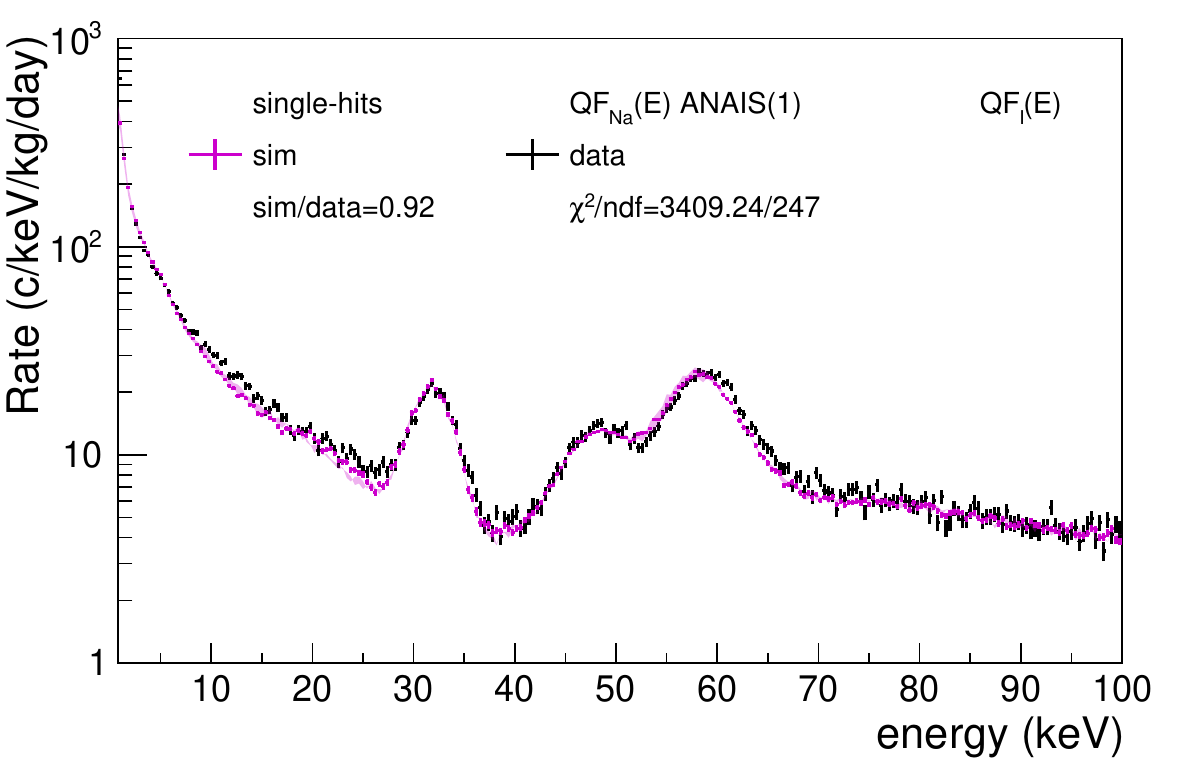}}
    {\includegraphics[width=0.45\textwidth]{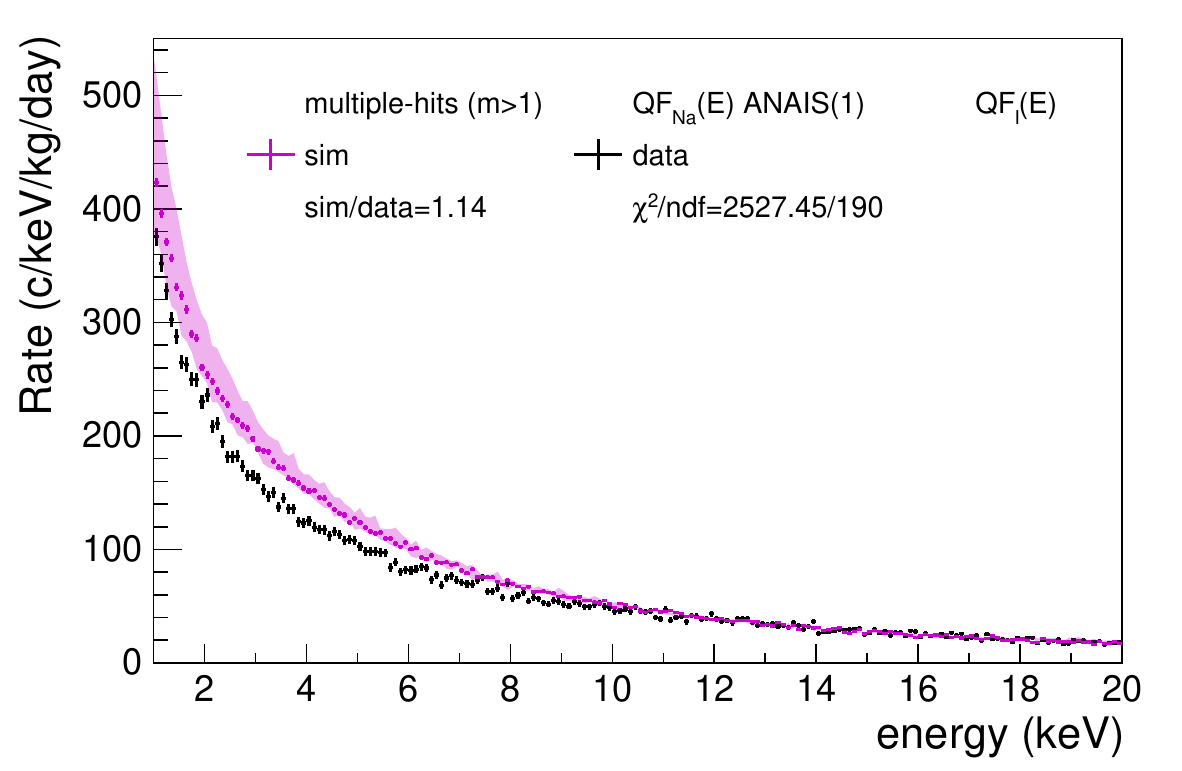}}
    {\includegraphics[width=0.45\textwidth]{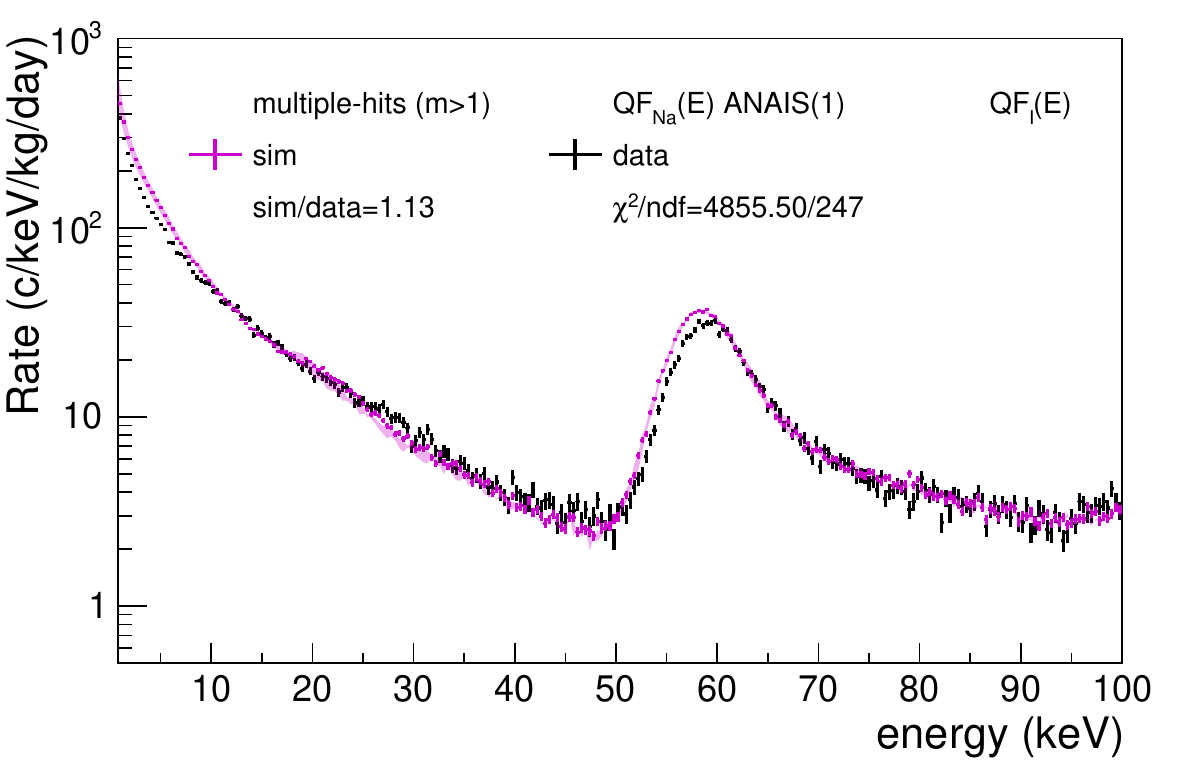}}
    {\includegraphics[width=0.45\textwidth]{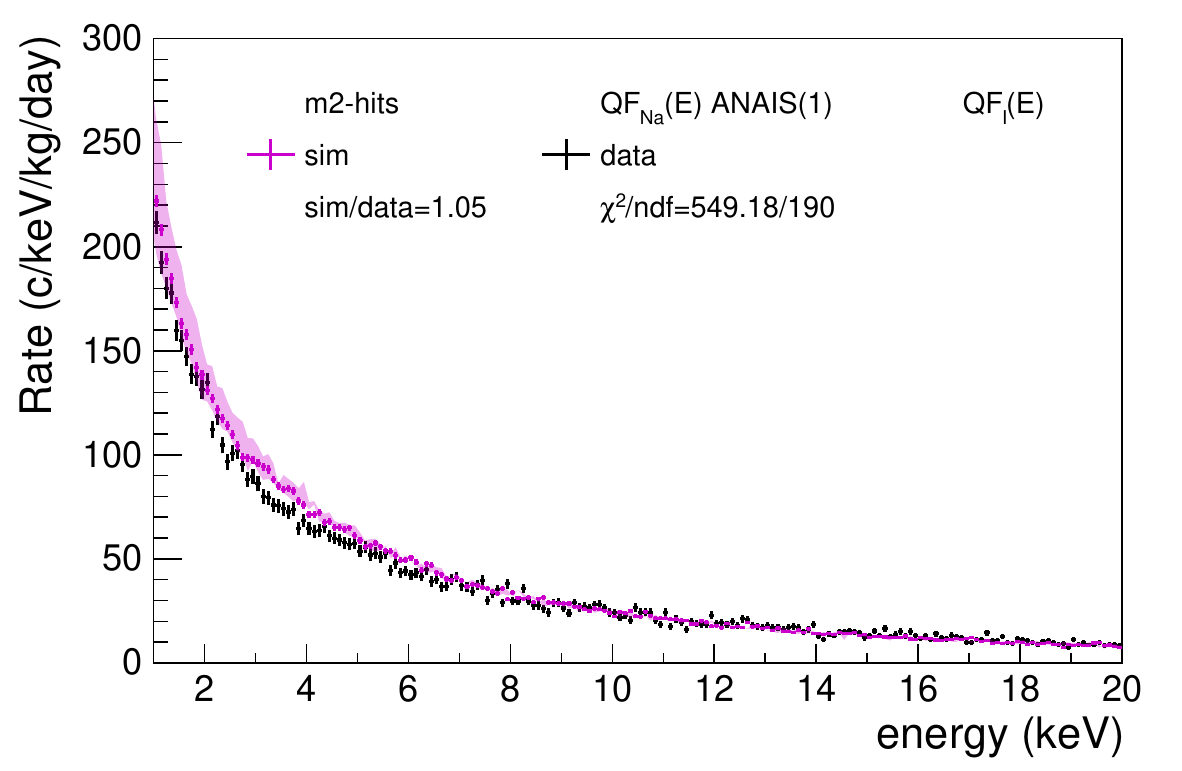}}
    {\includegraphics[width=0.45\textwidth]{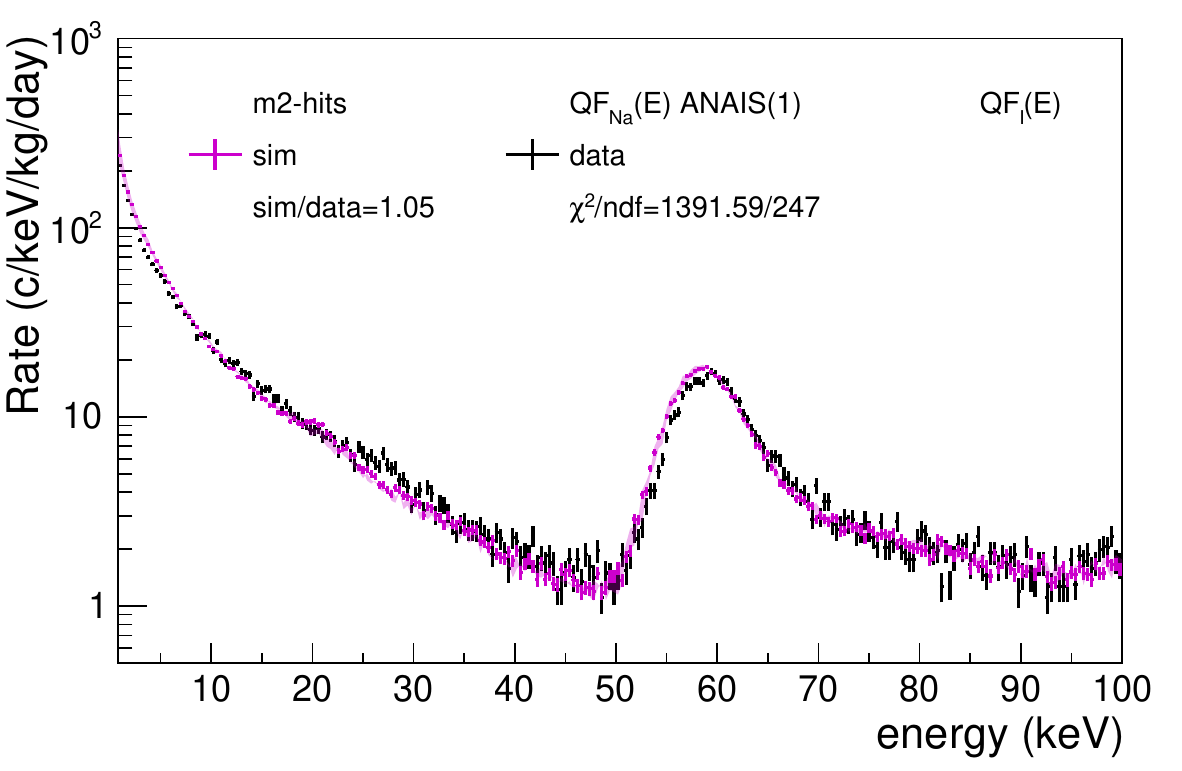}}

    \caption{\label{figurafinal} Comparison between the energy spectra measured in the west-face neutron calibration for the sum of the nine ANAIS-112 detectors (black) and the corresponding simulations (magenta), assuming ANAIS(1) QF\textsubscript{Na} and an energy-dependent QF\textsubscript{I} compatible with ANAIS-112 values. 1$\sigma$ uncertainty bands are plotted. The left column displays the low-energy region, while the right column shows the medium-energy range. Panels show the ratio between simulation and experimental data, as well as the goodness of the comparison. \textbf{First row:} total-hits. \textbf{Second row:} single-hits. \textbf{Third row:} multiple-hits (m>1). \textbf{Fourth row:} m2-hits.}
\end{figure}

\begin{figure}[t!]
    \centering
    {\includegraphics[width=0.78\textwidth]{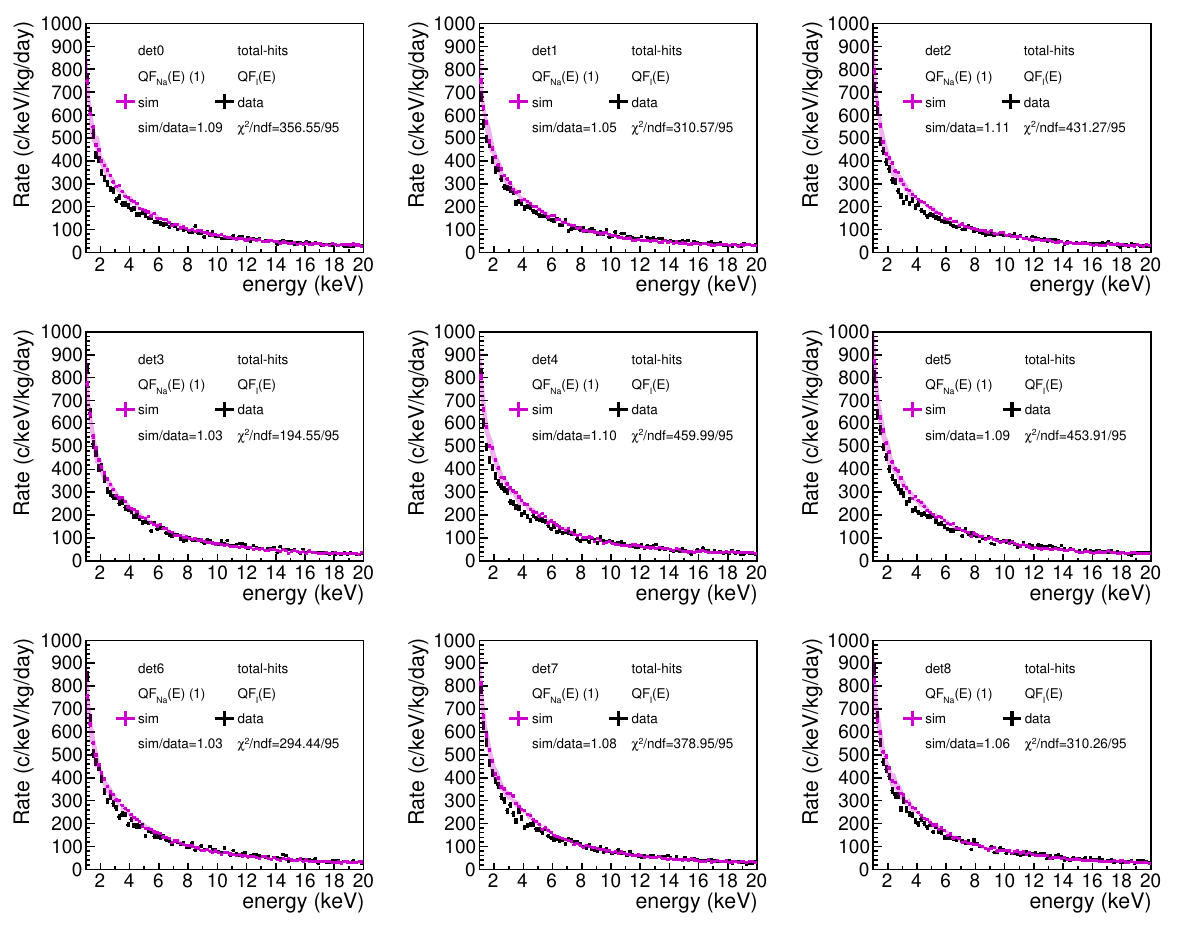}}

    \caption{\label{totalfinalperdet} Comparison between the total-hits low-energy spectra measured during the west-face neutron calibration for each ANAIS-112 detector (black) and the simulation (magenta) assuming ANAIS(1) QF\textsubscript{Na} and an energy-dependent QF\textsubscript{I} compatible with ANAIS-112 values. 1$\sigma$ uncertainty bands are plotted. Panels show the ratio between simulation and experimental data, as well as the goodness of the comparison.}
\end{figure}

\begin{figure}[b!]
    \centering
    {\includegraphics[width=0.78\textwidth]{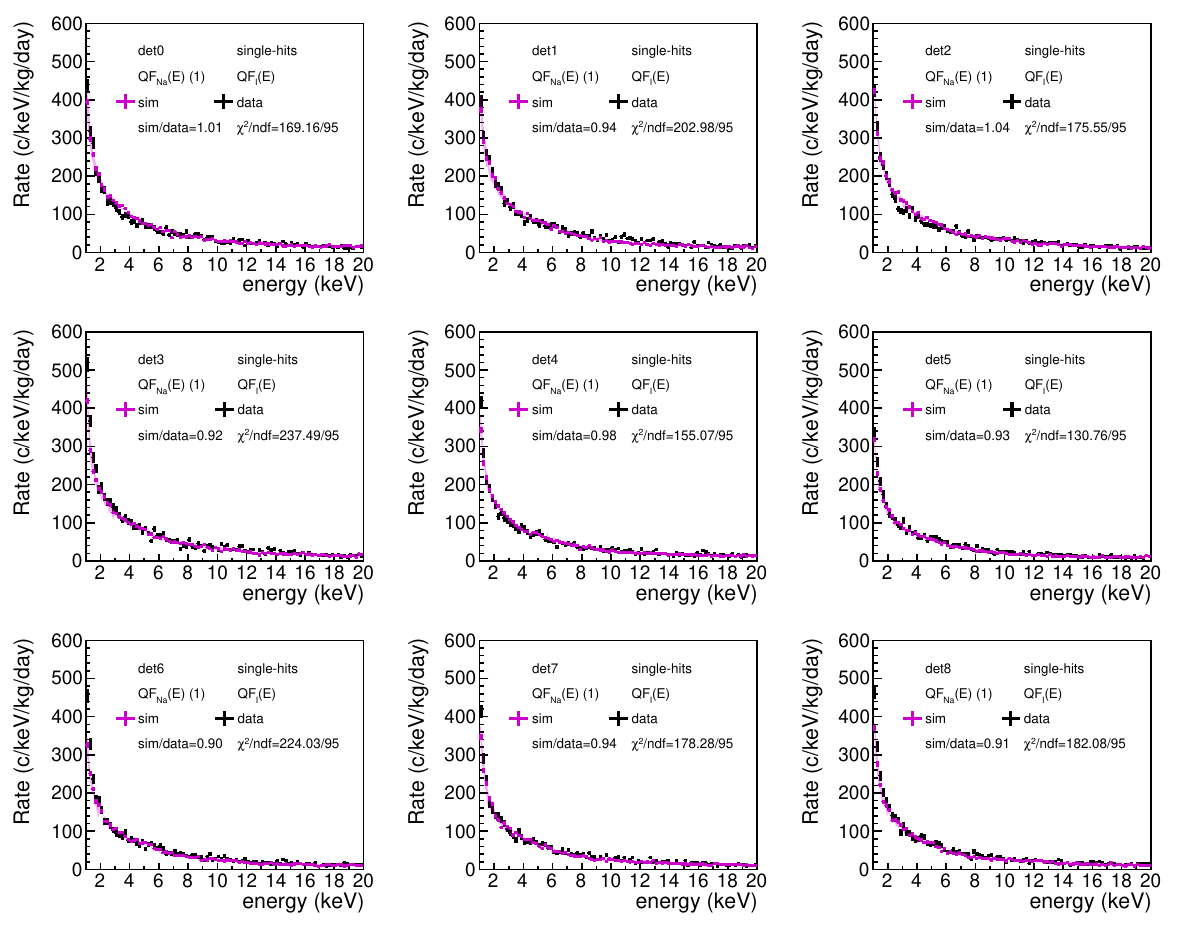}}

    \caption{\label{singlefinalperdet} Analogous to Figure \ref{totalfinalperdet}, but corresponding to single-hits. }
\end{figure}

\begin{figure}[t!]
    \centering
    {\includegraphics[width=0.85\textwidth]{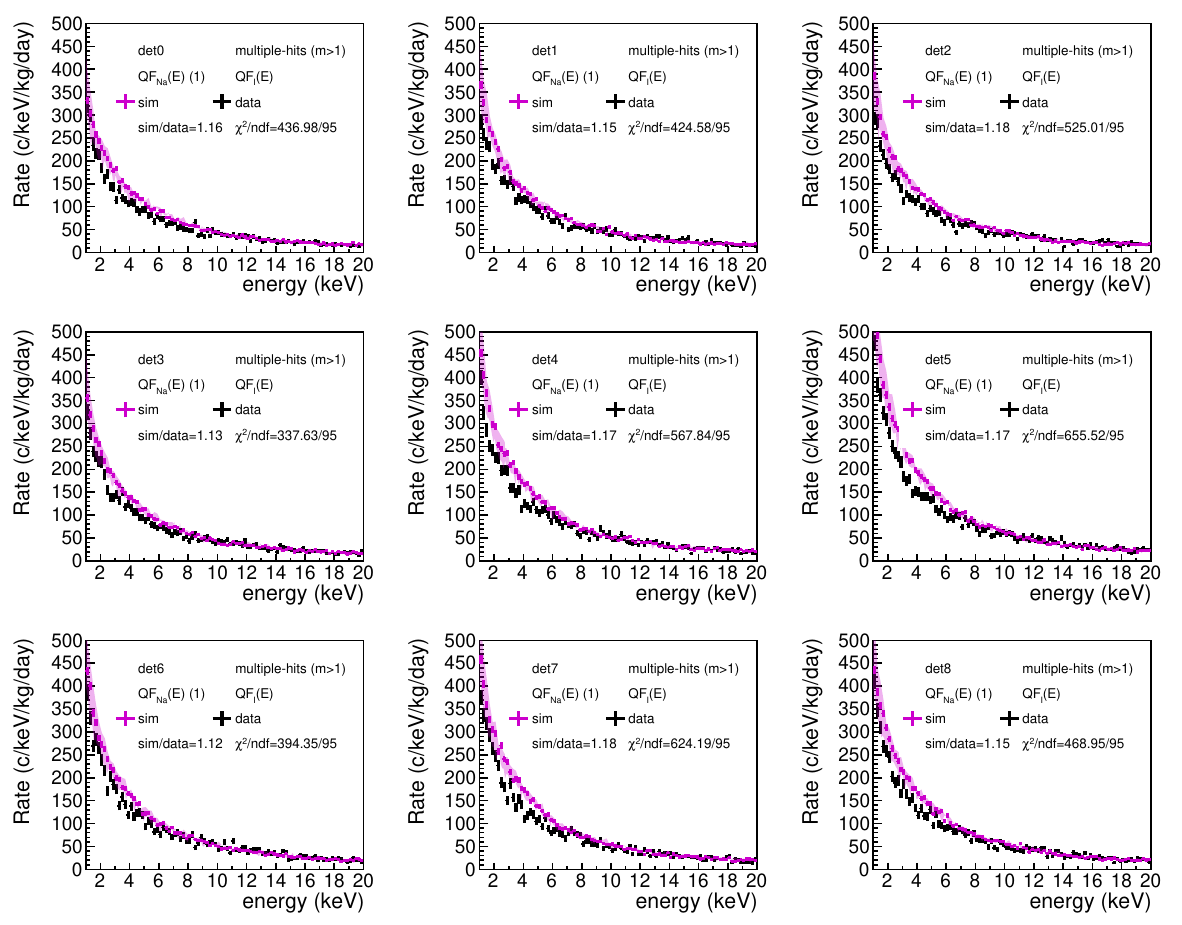}}

    \caption{\label{multifinalperdet} Analogous to Figure \ref{totalfinalperdet}, but corresponding to multiple-hits. }
\end{figure}

\begin{figure}[b!]
    \centering
    {\includegraphics[width=0.85\textwidth]{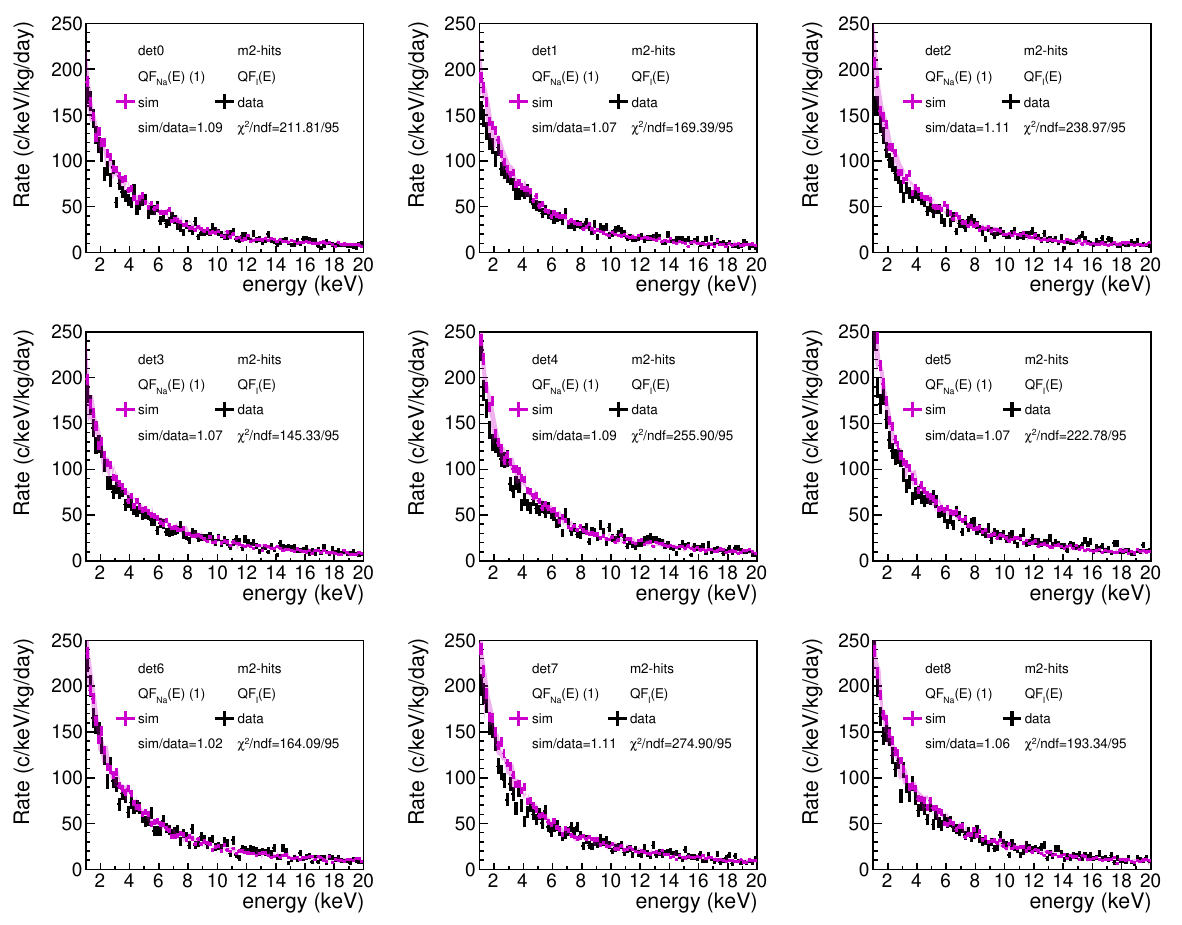}}

    \caption{\label{m2finalperdet}  Analogous to Figure \ref{totalfinalperdet}, but corresponding to m2-hits. }
\end{figure}

\begin{figure}[t!]
    \centering
    {\includegraphics[width=0.7\textwidth]{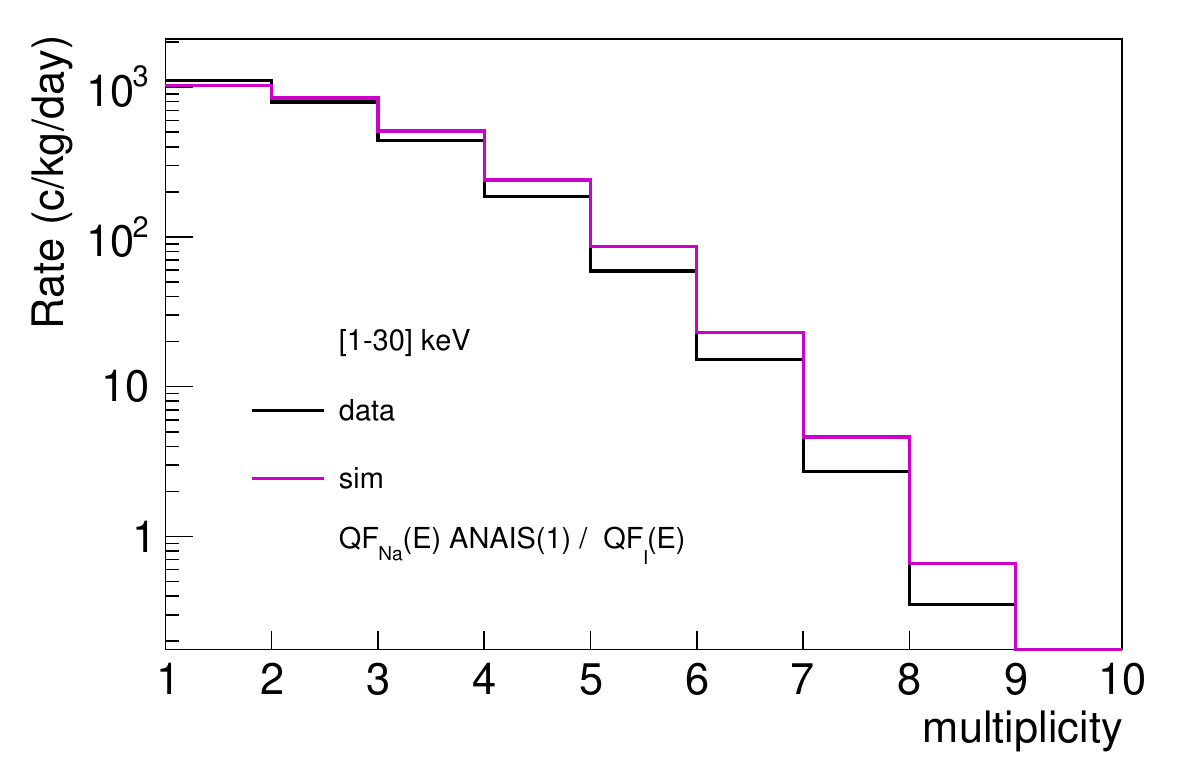}}

    \caption{\label{repartomulti} Rate in the [1-30] keV energy region for single-hit, m2-hit, and m>2-hit events, for both data (black) and simulation (magenta) using the preferred QFs selected in this work, ANAIS(1) QF\textsubscript{Na} and an energy-dependent QF\textsubscript{I} compatible with ANAIS-112 values.}
\end{figure}

Based on the analysis, the combination that best reproduces the neutron calibration data of ANAIS-112 consists of an energy-dependent QF\textsubscript{Na}, following the proportional calibration ANAIS(1) and fitted to a modified Lindhard model, and an energy-dependent QF\textsubscript{I} compatible with the values reported in \cite{cintas2024measurement,phddavid}. Figure~\ref{figurafinal} shows the agreement between data and simulation for the selected model, including 1$\sigma$ uncertainty bands. These uncertainty bands were derived from the fitting errors associated with the modified Lindhard model for QF\textsubscript{Na} and the energy-dependent modelling of  QF\textsubscript{I}. As shown, the associated uncertainty is small and becomes relevant only at low energies, where the uncertainty in QF\textsubscript{I} dominates.

The level of agreement between data and simulation is very good. This agreement is consistently observed across all detectors, as illustrated in Figures \ref{totalfinalperdet}, \ref{singlefinalperdet}, \ref{multifinalperdet}, and \ref{m2finalperdet}, which show detector-by-detector comparisons for total-hits, single-hits, multiple-hits, and m2-hits, respectively. It can be verified that all the ANAIS-112 crystals exhibit the same
NR spectrum in response to \textsuperscript{252}Cf sources, despite originating from different ingots, though grown following
similar protocols, 
confirming compatible results of the measurements performed at TUNL across several NaI(Tl) crystals built at Alpha Spectra. These results provide strong support for the selected QF models and confirm the robustness of the simulation in improving the understanding of QF\textsubscript{Na} and QF\textsubscript{I} in NaI(Tl) scintillators.

Figure \ref{repartomulti} presents the rate in the [1-30] keV energy region for single-hit, m2-hit, and m>2-hit events, for both data and simulation using the preferred QFs selected in this work. As already observed, there is a slight deficit in single-hit events, with the data-to-simulation ratio being 0.92. Furthermore, as previously noted, the agreement between data and simulation worsens with increasing multiplicity. However, it is from m4- or m5-hits onwards that the agreement notably deteriorates, although these represent only a small fraction of the total neutron calibration events. Specifically, as mentioned earlier, m4-hits correspond to 3\% of the total events, m5-hits to 1\%, with higher multiplicities having even lower probabilities. The majority of events, specifically single-hit and m2-hit populations, are nicely reproduced by the simulation.

\begin{figure}[h!]
    \centering
    {\includegraphics[width=0.45\textwidth]{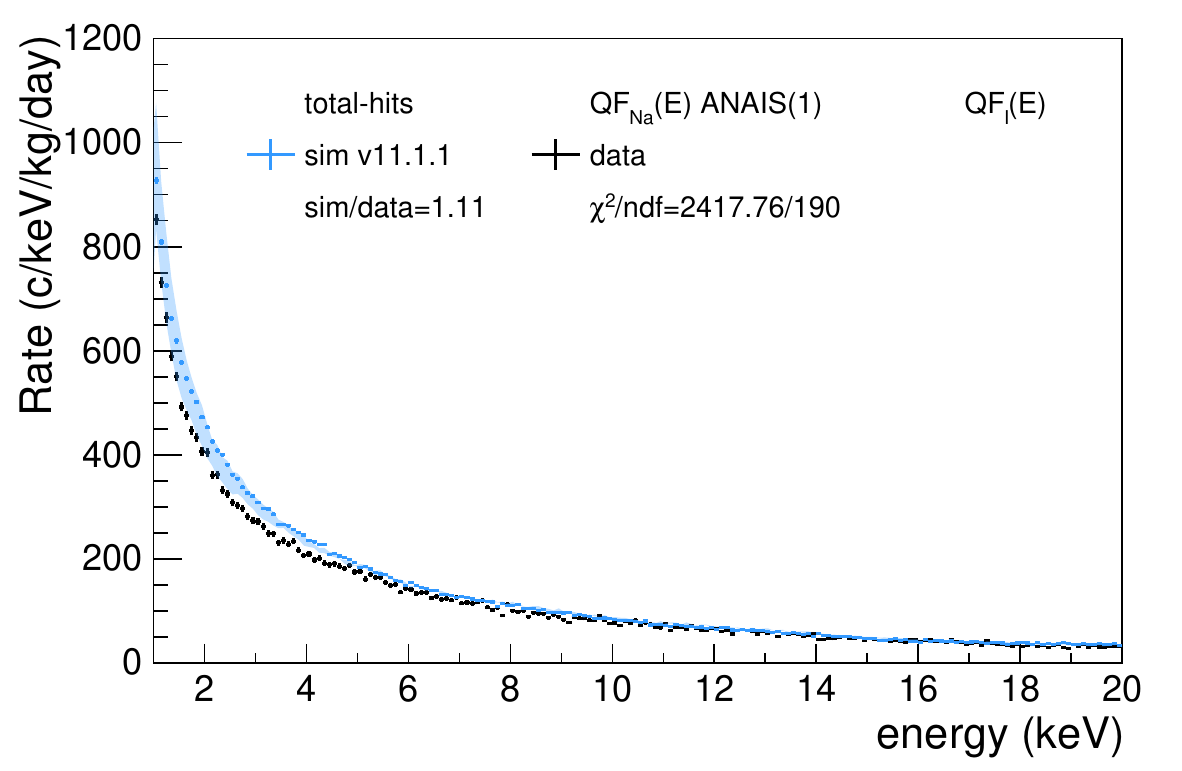}}
    {\includegraphics[width=0.45\textwidth]{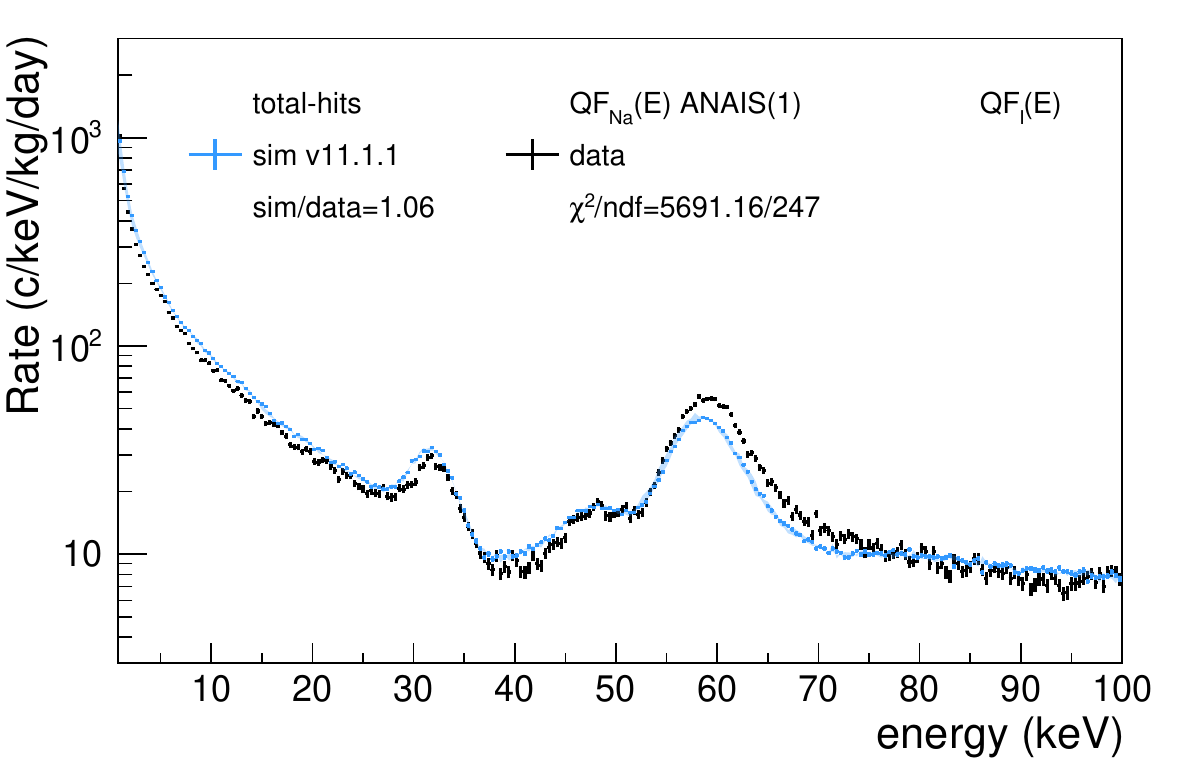}}
    {\includegraphics[width=0.45\textwidth]{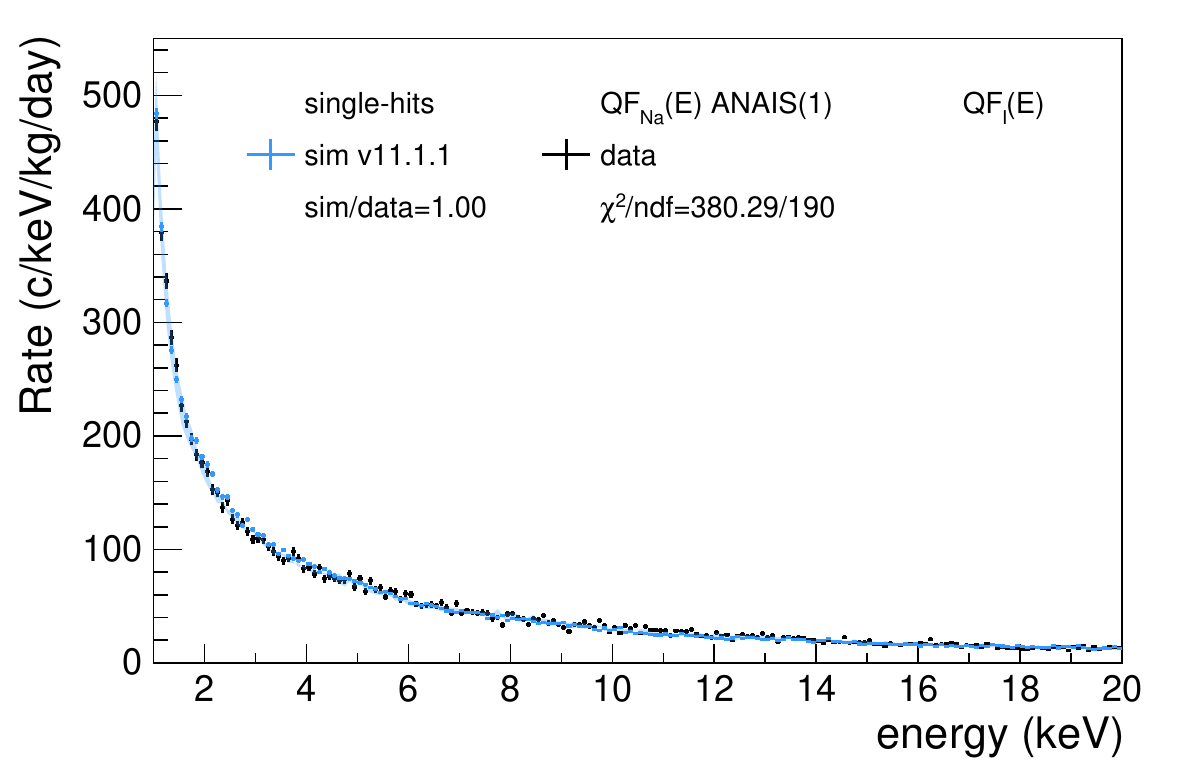}}
    {\includegraphics[width=0.45\textwidth]{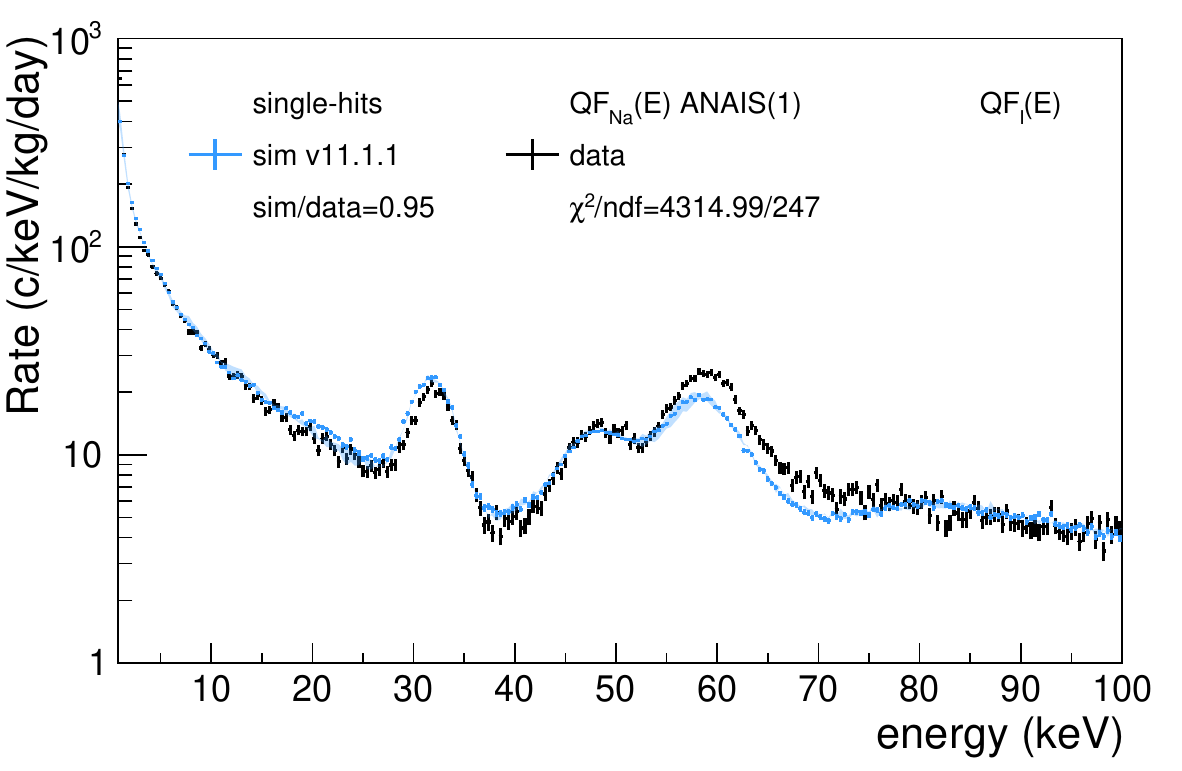}}
    {\includegraphics[width=0.45\textwidth]{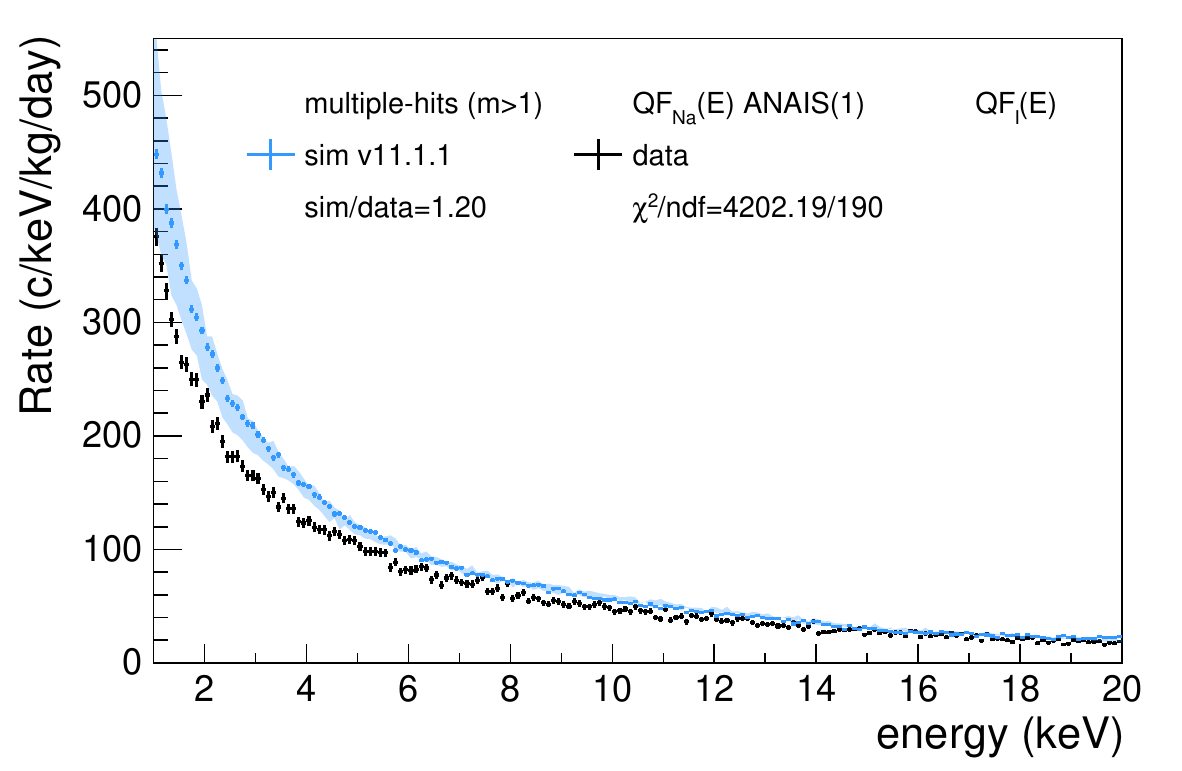}}
    {\includegraphics[width=0.45\textwidth]{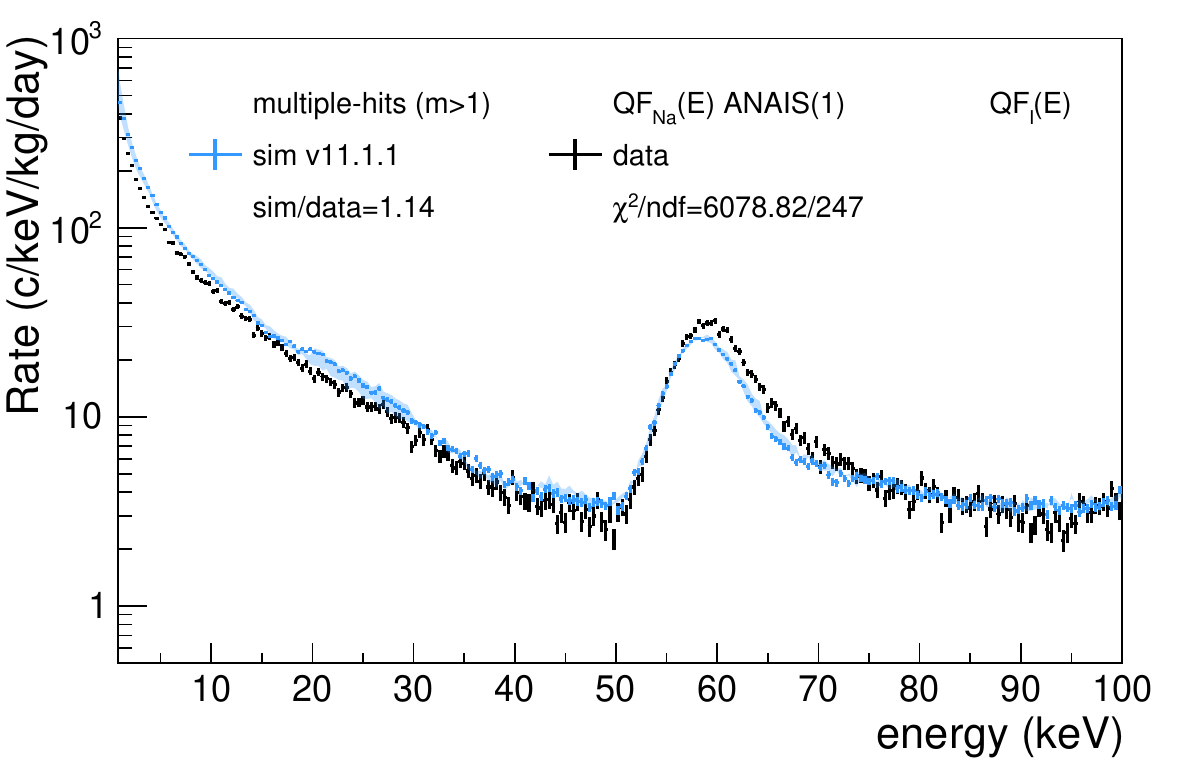}}
    {\includegraphics[width=0.45\textwidth]{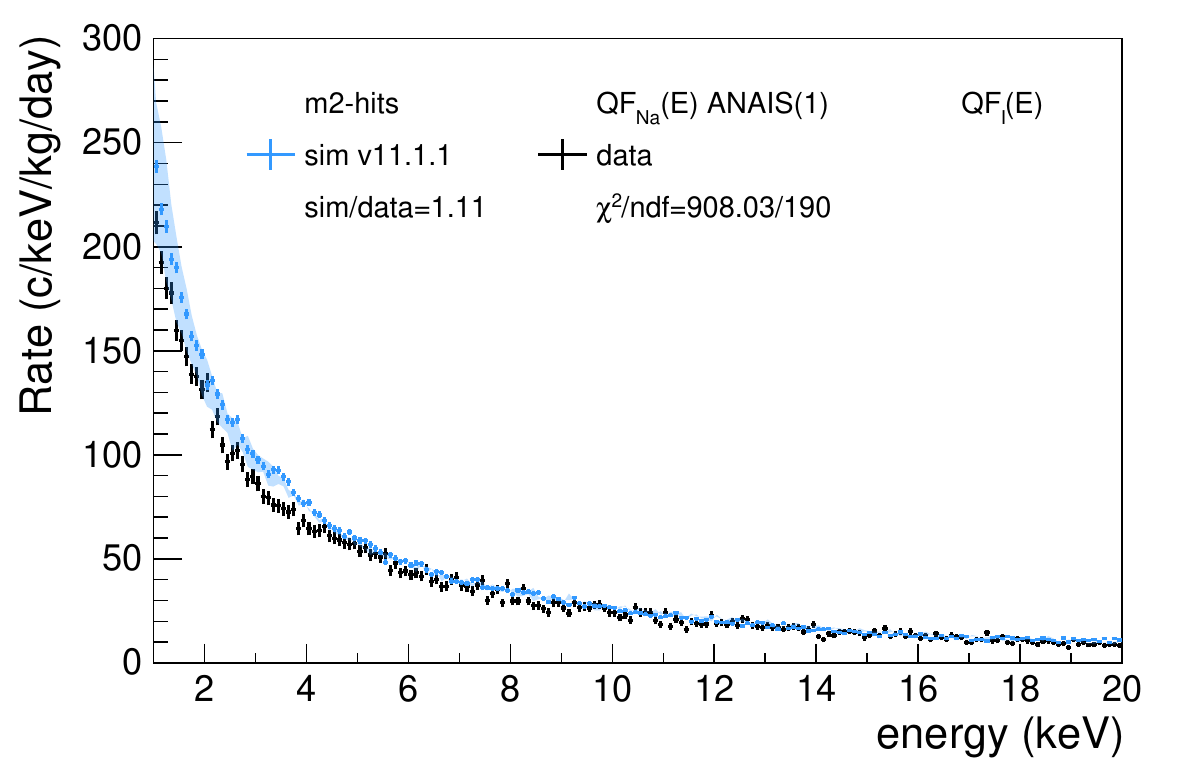}}
    {\includegraphics[width=0.45\textwidth]{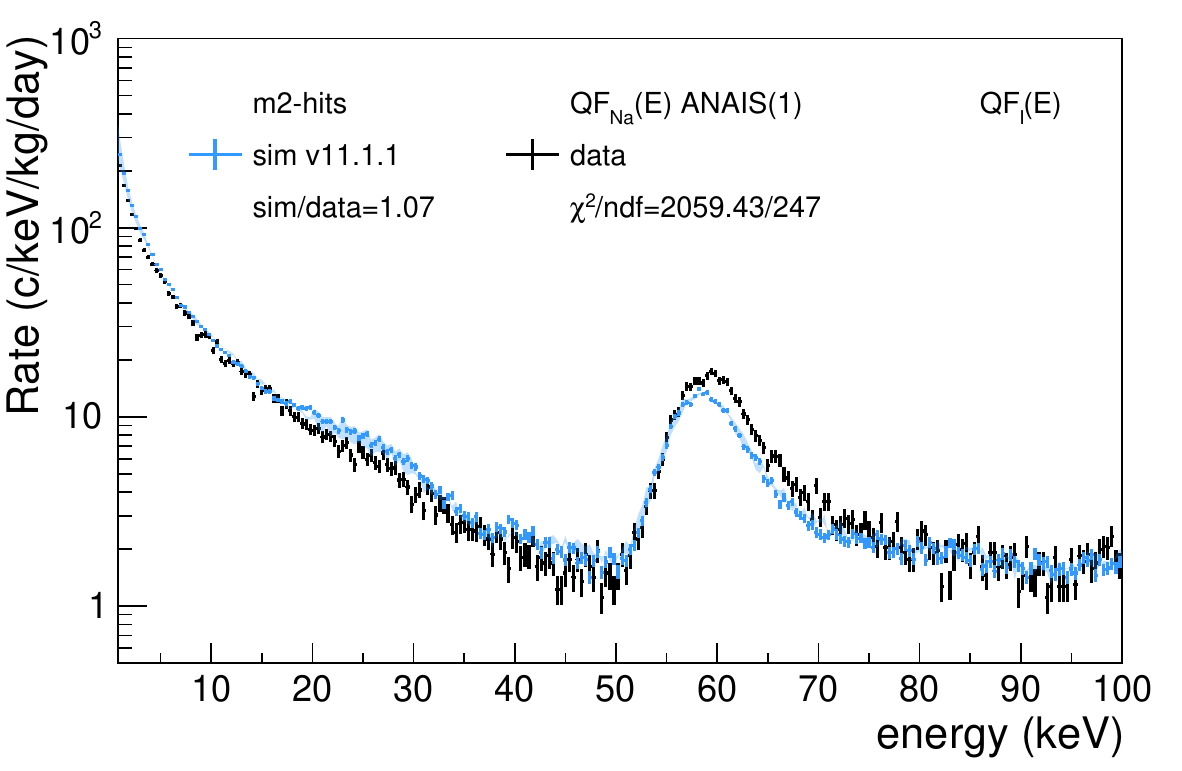}}

    \caption{\label{figurafinalv11} Comparison between the energy spectra measured in the west-face neutron calibration for the sum of the nine ANAIS-112 detectors (black) and the corresponding simulations using Geant v11.1.1 version (blue), assuming ANAIS(1) QF\textsubscript{Na} and an energy-dependent QF\textsubscript{I} compatible with ANAIS-112 values. 1$\sigma$ uncertainty bands are plotted. Panels show the ratio between simulation and experimental data, as well as the goodness of the comparison. \textbf{First row:} total-hits. \textbf{Second row:} single-hits. \textbf{Third row:} multiple-hits (m>1). \textbf{Fourth row:} m2-hits. \vspace{0.5cm} }
\end{figure}

The agreement between data and simulation is also evaluated employing Geant4 version v11.1.1 for the simulation using the QFs selected in this work. The objective is to determine whether the selection of QFs depends heavily on the elastic scattering cross section implemented in each Geant4 version, and whether this constitutes a limiting factor for the conclusions that can be drawn from this study. This comparison is shown in Figure \ref{figurafinalv11}. As previously discussed, the inelastic peak is not correctly reproduced with v11.1.1, and the 31.8 keV peak is slightly overestimated even after adjusting the capture cross section for the production of \textsuperscript{128}I. Nevertheless, beyond these specific features, the agreement below 30 keV remains very similar to that obtained with the reference version used in this work, Geant4 v9.4.p01 (see Figure \ref{figurafinal}). In particular, the single-hit rate is slightly better reproduced, with the data-to-simulation ratio improving from 0.92 to 1.00 between versions v9.4.p01 and v11.1.1. A similar increase is observed in multiple-hit events: for m > 1, the ratio rises from 1.14 to 1.20, and for m2, from 1.05 to 1.11. This overall trend results in a minor worsening of the total-hit agreement, with the ratio increasing from 1.04 to 1.11 across the same versions.

These results confirm that, although the elastic scattering processes are indeed modified in the more recent Geant4 version, the agreement with the ANAIS-112 neutron calibration data remains very reasonable when using the QFs selected in this work, particularly in the NR-dominated region. Furthermore, just as Geant4 v11.1.1 is compatible with the QFs proposed in this work, it is clearly incompatible with the DAMA/LIBRA QFs, as shown in Figure \ref{DAMATotal}. Therefore, neither of these conclusions is affected by the choice of the Geant4 version. That said, future work involving neutron calibrations with monoenergetic neutron sources should aim to further investigate the uncertainties associated with Geant4’s elastic scattering models.

\subsection{QF validity with ANOD data}

\begin{figure}[t!]
    \centering
    {\includegraphics[width=0.45\textwidth]{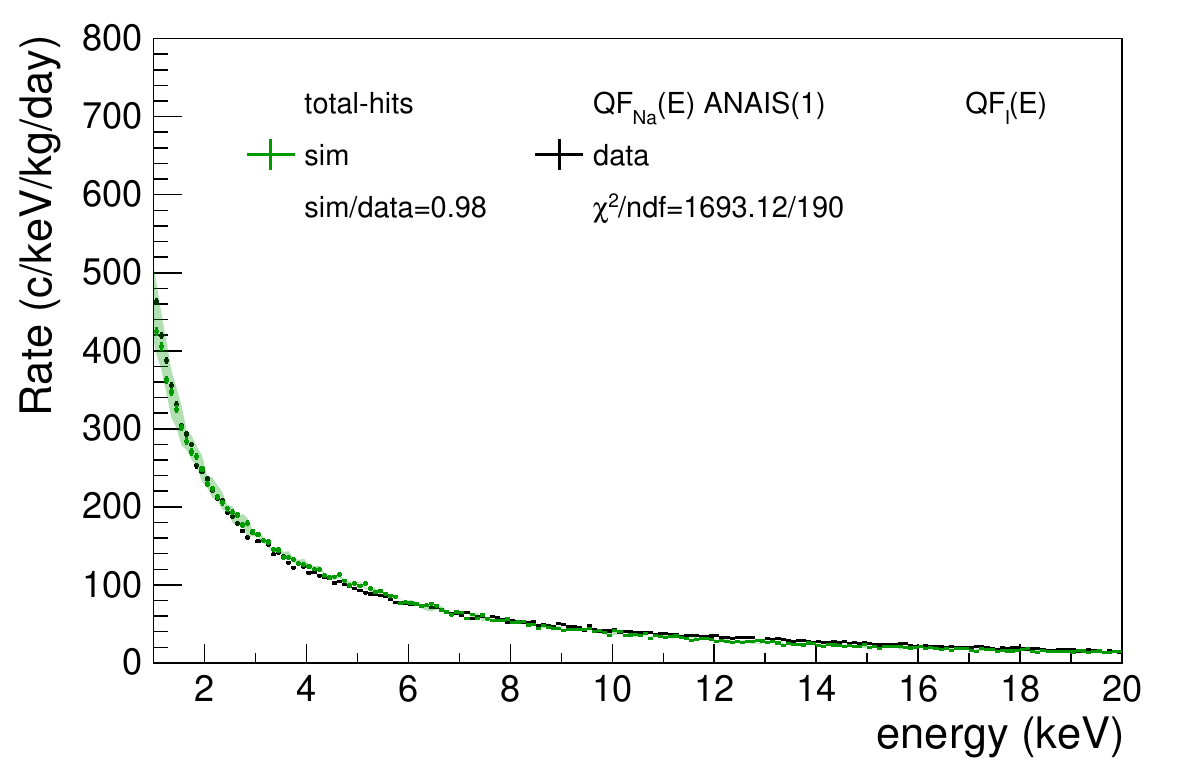}}
    {\includegraphics[width=0.45\textwidth]{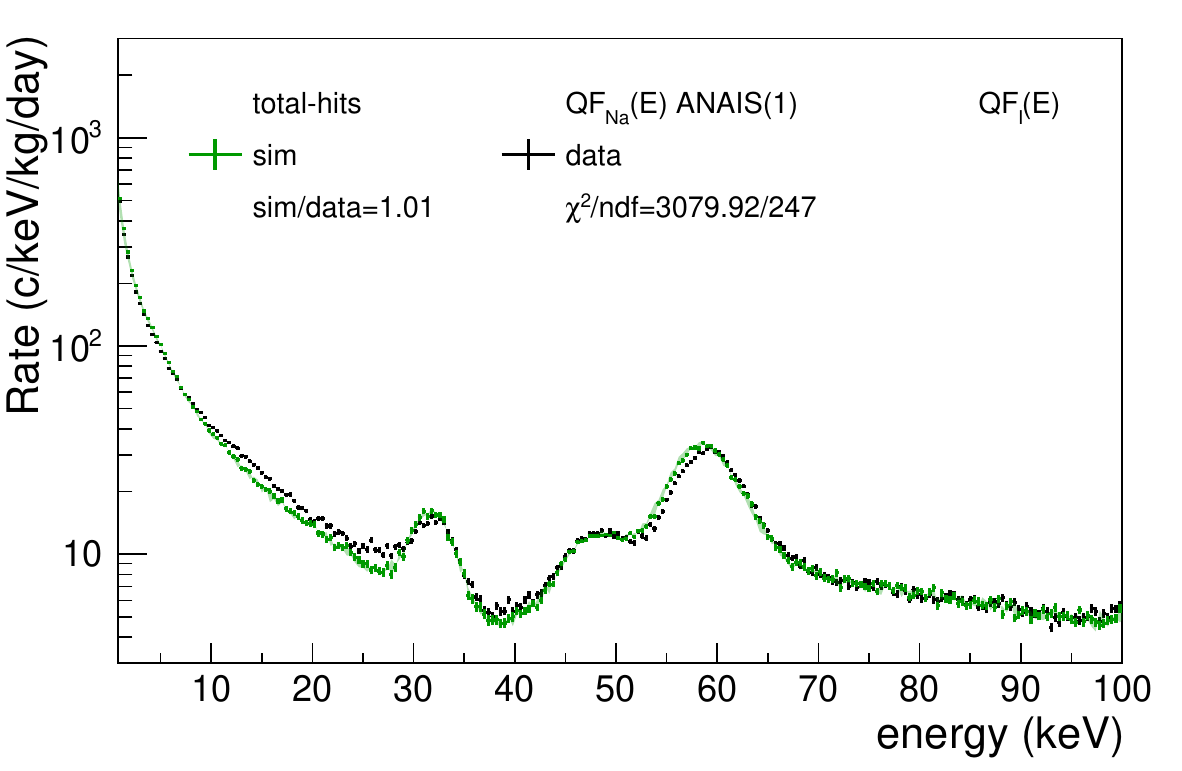}}
    {\includegraphics[width=0.45\textwidth]{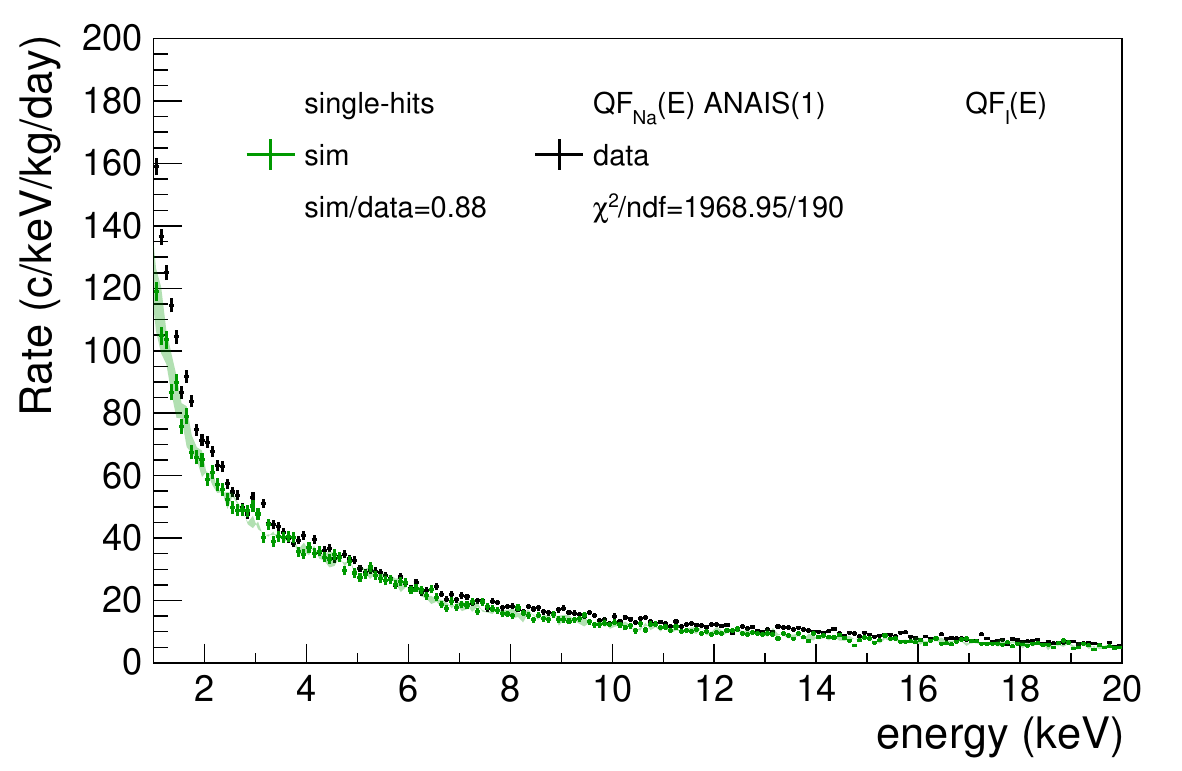}}
    {\includegraphics[width=0.45\textwidth]{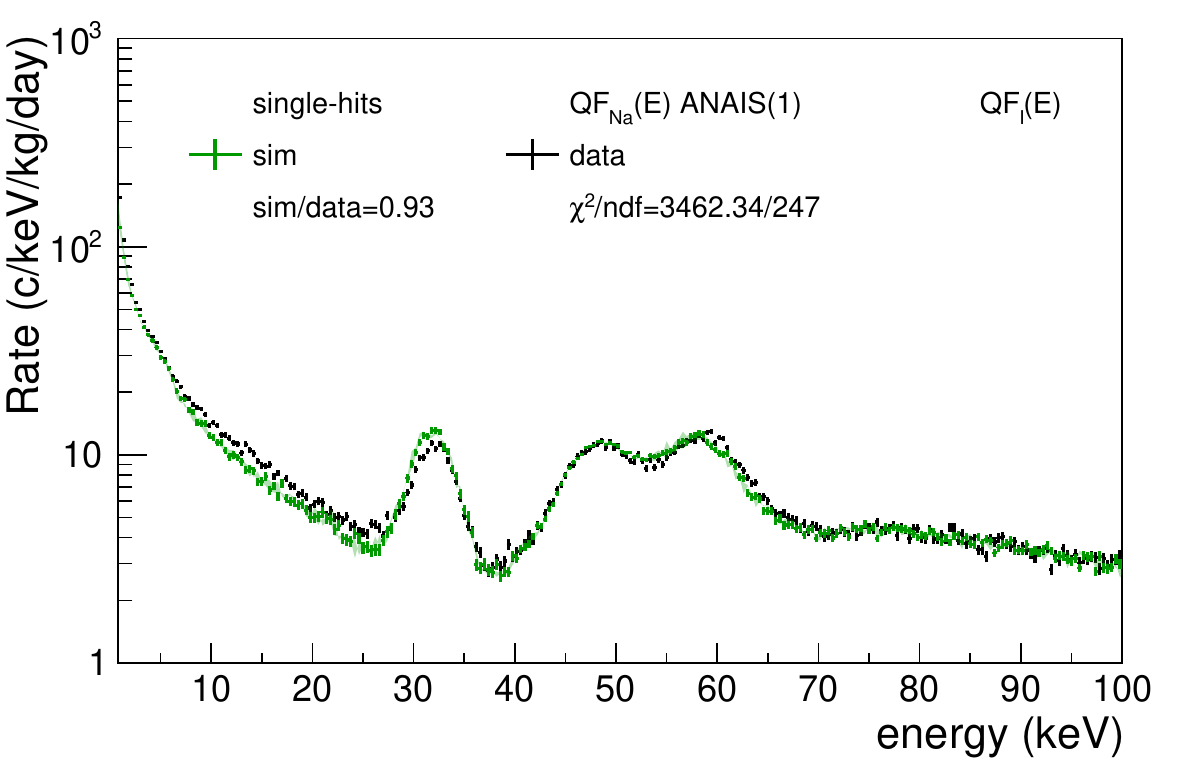}}
    {\includegraphics[width=0.45\textwidth]{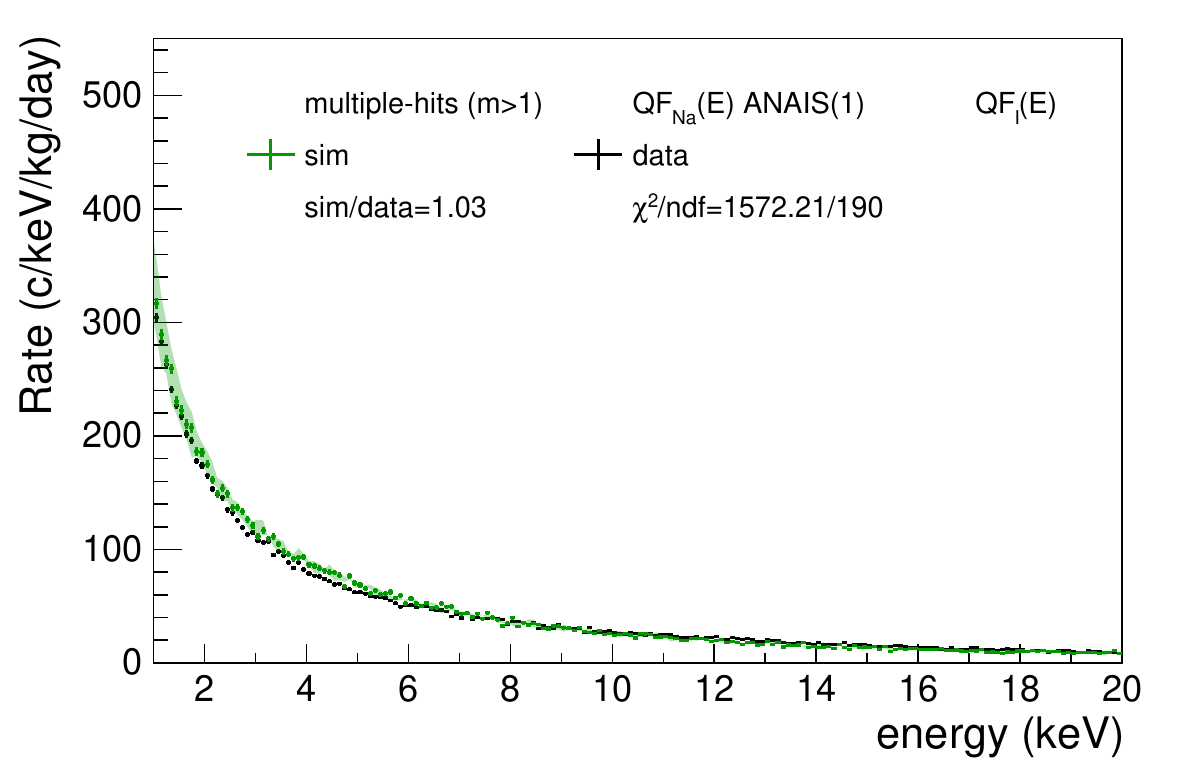}}
    {\includegraphics[width=0.45\textwidth]{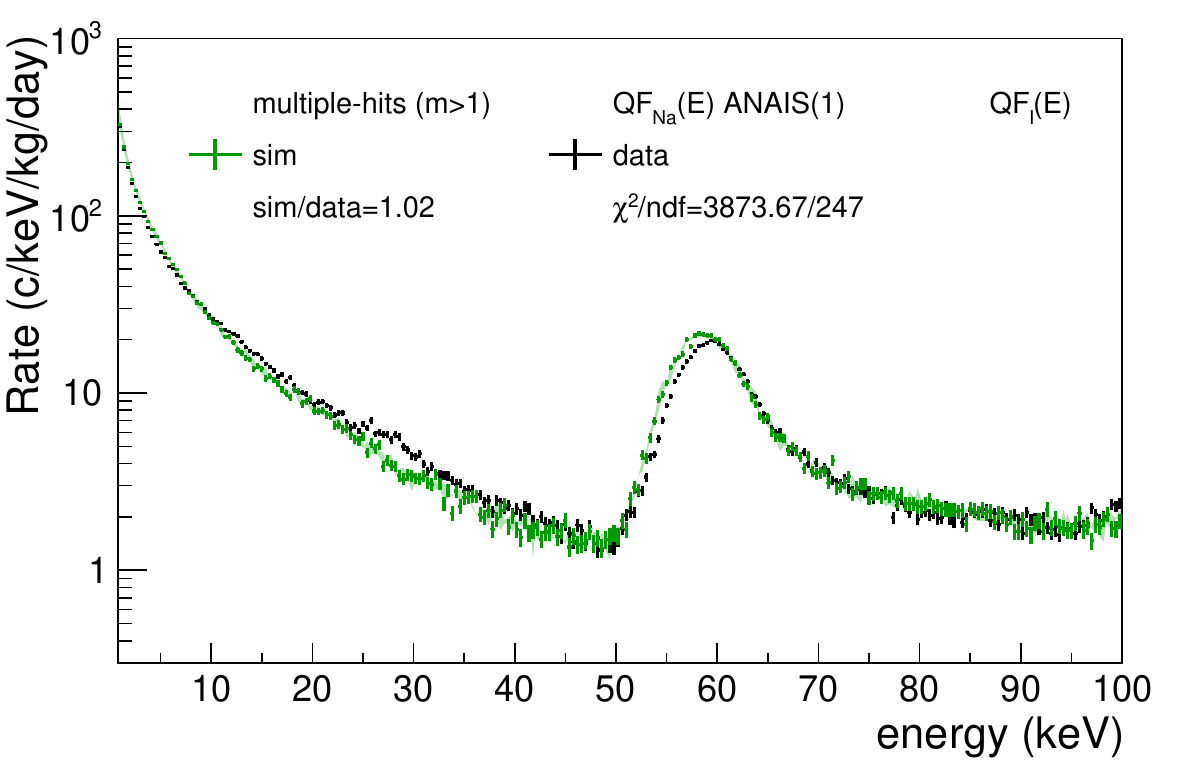}}
    {\includegraphics[width=0.45\textwidth]{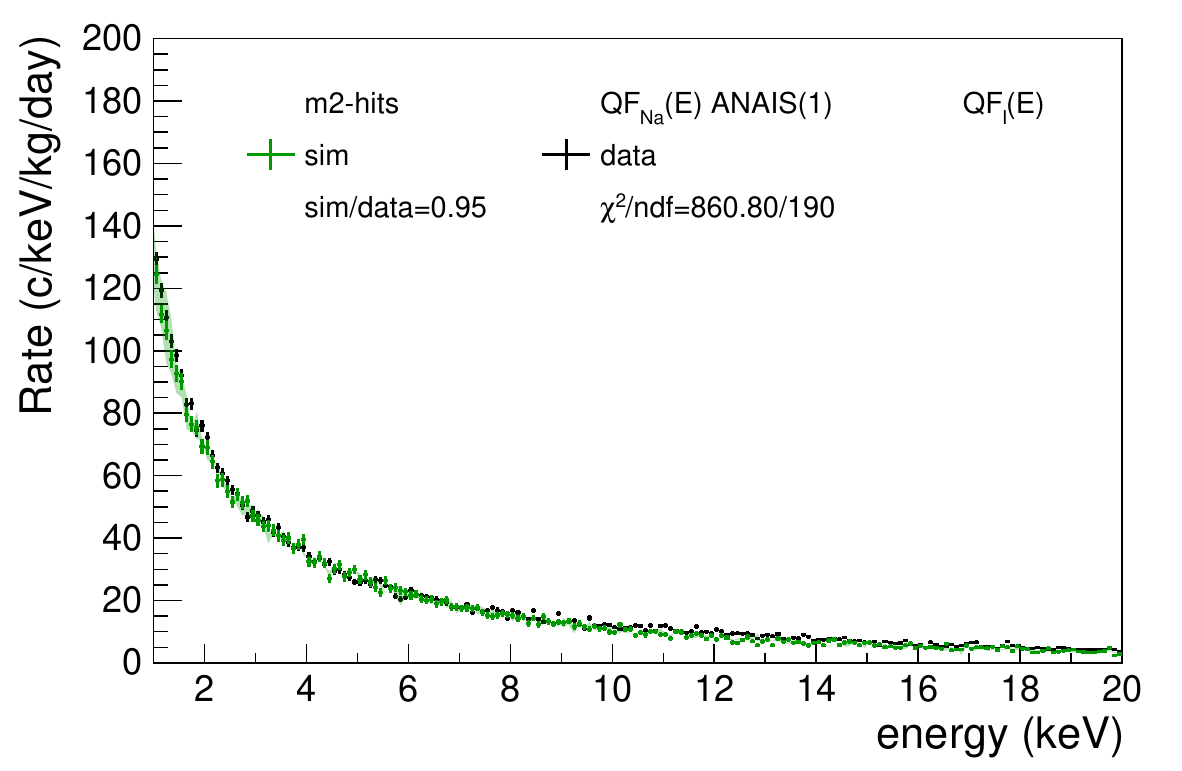}}
    {\includegraphics[width=0.45\textwidth]{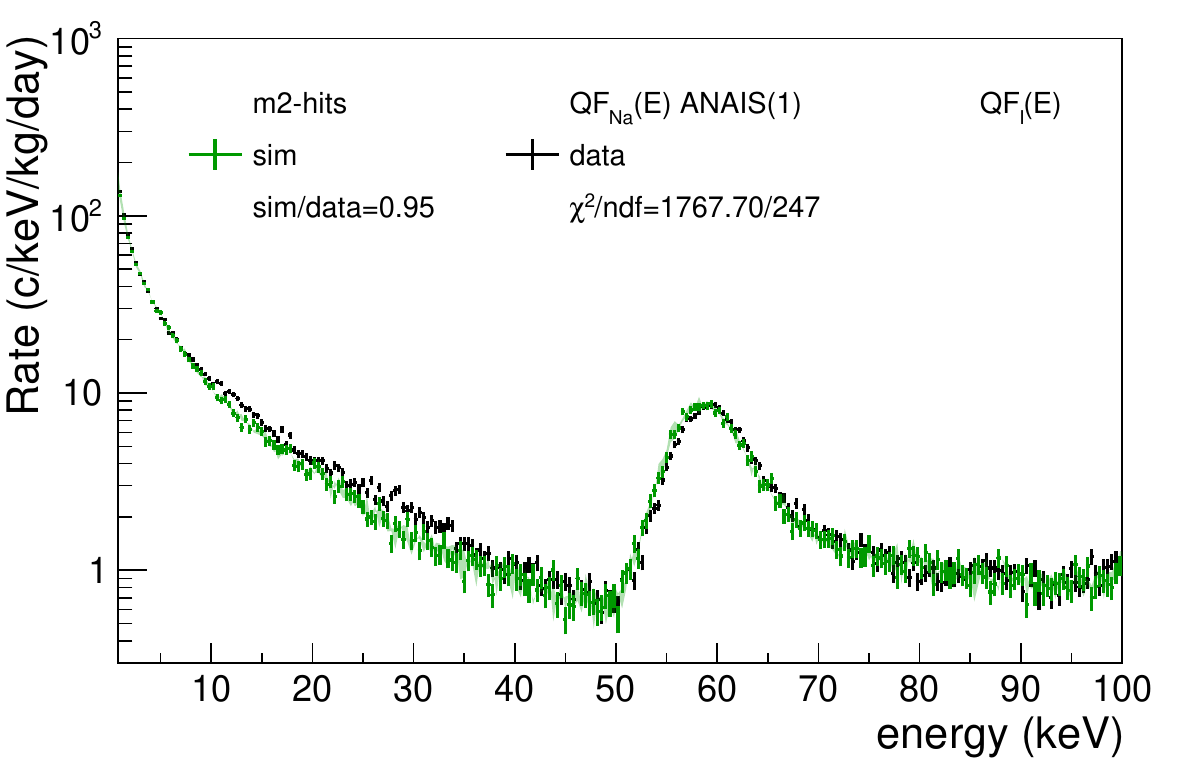}}

    \caption{\label{figurafinalANOD} Comparison between the energy spectra measured with the ANOD DAQ in the west-face neutron calibration for the sum of the nine ANAIS-112 detectors (black) and the corresponding ANOD simulations (green), assuming ANAIS(1) QF\textsubscript{Na} and an energy-dependent QF\textsubscript{I} compatible with ANAIS-112 values. 1$\sigma$ uncertainty bands are plotted. The left column displays the low-energy region, while the right column shows the medium-energy range. Panels show the ratio between simulation and experimental data, as well as the goodness of the comparison. \textbf{First row:} total-hits. \textbf{Second row:} single-hits. \textbf{Third row:} multiple-hits (m>1). \textbf{Fourth row:} m2-hits.}
\end{figure}

Finally, the results regarding the QF derived in this thesis have been validated using the data acquired with the ANOD DAQ.  This analysis has been applied to the only neutron run using ANOD DAQ for the nine detectors readout (run 9016, see Table \ref{infoneutroncalibration}). This run was recorded overnight, thus providing sufficient statistics.

The analysis approach adopted for the comparison between ANOD data and simulations is different from that conducted with ANAIS data, explained in previous sections. For ANOD data, asymmetric filtering of events below 2 keV is applied directly (using ANOD pulse shape variables, $\log(\mu_{\mathrm{ANOD}})$~>~-0.5) instead of adding to the simulation the corresponding contribution, as explained in Section~\ref{building} for ANAIS data. Regarding the simulation, the processing of simulated events has been adapted to the specific features of the ANOD DAQ, namely the 8~$\mu$s coincidence window and the absence of dead time, while the integration window remains fixed at 1~$\mu$s, as in ANAIS.


Figure~\ref{figurafinalANOD} shows the comparison between data and simulation assuming the QFs that best describe the ANAIS neutron calibrations, i.e. ANAIS(1) QF\textsubscript{Na} and an energy-dependent QF\textsubscript{I} consistent with the ANAIS-112 values. The agreement is remarkably good, reinforcing the validity of the selected QF. Of particular relevance is the agreement in the multiple-hit population, both globally and in the m2-hit subset. When comparing these populations for ANOD data (Figure~\ref{figurafinalANOD}) with those obtained for ANAIS (Figure~\ref{figurafinal}), the improvement at low energy is evident. Specifically, the data-to-simulation ratio for general multiple-hits improves from 1.14 in ANAIS to 1.03 in ANOD, leading to an excellent agreement in the total-hit population for ANOD. This improvement arises because, although extrapolation of ANAIS data processing has been carefully applied to the simulation, associated systematic effects that hinder satisfactory agreement in the multiple-hit population can be present, particularly at low energies. In contrast, for ANOD the comparison is more straightforward, given the simpler DAQ constraints.


For the single-hit population, in the energy region below 3 keV the ANOD simulation underestimates the data. This behavior is attributed to the cut applied to ANOD data to reject asymmetric events, which was implemented as a conservative constant cut rather than optimized as a function of energy. While this approach is valid for energies above 2.5~keV, it is conservative below this value, as already discussed in Section~\ref{ANODfiltering}. Furthermore, no efficiency correction associated with this cut has been applied in the present study, since such corrections are still under study. Establishing an optimized, energy-dependent cut is therefore identified as a line of future improvement, beyond the scope of this thesis. Overall, the results demonstrate that the selected QF is compatible with the ANOD data, providing further support for its validity.

\section{Discussion and conclusions}

This chapter has presented the neutron calibrations conducted onsite in ANAIS-112. These calibrations serve three primary objectives within the experiment: evaluating the ANAIS-112 event selection efficiency with a bulk scintillation population dominated by NRs, generating signal-like events for ML training, and improving the understanding of the QF. The present thesis has focused specifically on this last objective, by comparing the spectra measured during neutron calibrations with those obtained from dedicated Geant4-based neutron simulations.

To date, the QF of NaI(Tl) remains poorly constrained, with significant experimental uncertainties affecting both its absolute value and its energy dependence. It constitutes a particularly critical parameter for ANAIS-112, as it represents the dominant source of systematic uncertainty in the comparison with DAMA/LIBRA. The inconsistency between the six-year ANAIS-112 results and those reported by DAMA/LIBRA is well established in the scenario where DM particles interact via ERs in NaI(Tl) crystals \cite{amare2025towards}. However, in the case of DM particles producing NRs, an accurate knowledge of the QF is essential for converting electron-equivalent energy scales, commonly used for calibration in DM detectors, into the corresponding NR energy scale of the interaction channel. If the QF values assumed by DAMA/LIBRA and ANAIS-112 differ, the energy regions selected for the annual modulation search must also be conveniently adapted.

With the aim of shedding light on this parameter, this chapter has presented the on-site neutron calibration carried out in ANAIS-112. The calibration is based on the exposure of the full detector array to \textsuperscript{252}Cf neutron sources, placed outside the anti-radon box and the lead shielding, but inside the muon veto system and the neutron moderator. The characteristics of the source, the measurement schedule and procedure, as well as the specific features of the acquired data in comparison with other event populations have been described. This analysis has enabled the investigation of the distinct responses of NR and ER by comparing the average pulse shapes corresponding to each population.

Subsequently, the neutron simulations conducted to compare with the calibration data have been presented, which have enabled a critical evaluation of the Geant4 physics databases. Specifically, the ANAIS-112 neutron calibration data have revealed differences between various Geant4 versions regarding cross sections and the presence and intensity of certain inelastic scattering lines, which impact the agreement between simulation and experimental data. Consequently, Geant4 version v9.4.p01 has been selected for this work due to its better description of the measured data. Additionally, an overestimation of the neutron capture cross section for the production of \textsuperscript{128}I has been identified in all the versions. The process of constructing simulated spectra for comparison with the experimental data has been detailed, incorporating several experimental inputs such as the intrinsic background present during neutron calibrations and anomalous events identified in the ANOD DAQ.

The comparison between data and simulation demonstrates that ANAIS has a very good capability to reproduce the response of the detectors, validating the modelling of the experimental set-up and the DAQ. Although this study does not allow for the extraction of an explicit energy dependence of the QF as performed in monochromatic source studies, it enables the comparison of different QF models, favoring some of them over the others, or even allowing to exclude them, thus providing complementary insights to those obtained from such dedicated measurements. In particular, the DAMA/LIBRA QF models are disfavored due to poorer agreement with ANAIS-112 data, as are QF\textsubscript{Na} models exhibiting a decreasing dependence on energy. 

Regarding QF\textsubscript{Na} of the ANAIS crystals, the comparison suggests a preference for increasing with energy QF\textsubscript{Na} models rather than constant ones. It has been verified that the modified Lindhard model provides a better description than a linear fit to the data, further reinforcing this empirical model for estimating the QF. This implies a QF\textsubscript{Na} that decreases with energy in regions lacking extensive experimental data. Concerning iodine, the data-simulation agreement improves significantly when employing an energy-dependent QF\textsubscript{I} compatible with ANAIS-112 results, although the constant 6\% value reported by ANAIS also yields reasonable agreement.

Moreover, the systematic associated with the choice of Geant4 version in the determination of the QFs has been taken into account. The agreement between data and simulation using Geant4 version v11.1.1 supports the preferred QFs derived from this study. This indicates that, while the Geant4 version does affect the elastic scattering cross-section used in the simulation, it does not limit the QFs selection; rather, it reinforces it. Furthermore, the newest version also yields a poorer agreement with the DAMA/LIBRA QFs, thereby disfavoring their compatibility with ANAIS-112 neutron data.

On the one hand, the ANAIS experiment has recently implemented a new DAQ system, ANOD, which provides a longer acquisition window and operates without dead time (see Section~\ref{DAQsec}). The data acquired with this system are particularly relevant for neutron studies, since neutrons are expected to produce multiple interactions within the ANAIS crystals. At the time of writing this thesis, a neutron calibration run using the ANOD DAQ system in parallel with the standard ANAIS DAQ system was performed. The QF selected using ANAIS data has been revisited with the information provided by ANOD. For this purpose, the data selection procedure in ANOD has been redefined, allowing the direct rejection of the asymmetric-event population, rather than selecting them through the synchronized tree and subsequently adding them to the simulation. In addition, the simulation has been adapted to the specific characteristics of the ANOD DAQ system. The comparison between data and simulation using ANOD shows very good agreement, with a notably improved consistency in the multiple-hit populations, thereby further reinforcing the QF-related conclusions of this chapter.




Taking these results into account, Chapter \ref{Chapter:annual} will present the study of the annual modulation signal over six years of exposure. This study considers both the QF values reported by the DAMA/LIBRA collaboration for their crystals \cite{Bernabei:1996vj}, and the preferred QF model for the ANAIS-112 crystals derived from this work: an energy-dependent QF\textsubscript{Na} based on the ANAIS(1) model and an energy-dependent QF\textsubscript{I} consistent with the values obtained for ANAIS crystals \cite{cintas2024measurement,phddavid}. Since this study indicates that the QFs of DAMA/LIBRA and ANAIS crystals differ, the ROI for the annual modulation search must be redefined to correspond to the same NR energy range in both experiments. This reanalysis will also incorporate the updated background model, which is presented in the following chapter.

\vspace{0.5cm}

Based on the findings presented in this chapter, future work is proposed to further confirm the QF of the ANAIS-112 crystals.

The conclusions drawn in this work have been based on calibrations performed with $^{252}$Cf neutron sources. However, combining the data obtained from $^{252}$Cf calibrations with those from additional neutron sources could provide valuable complementary insights into the QF. In this regard, the use of ($\alpha$,n) sources such as Am-Be or Po-Be may be considered. However, these sources also emit neutrons with a broad energy spectrum. 

In contrast, the use of monoenergetic neutron sources may yield a more accurate determination of the QF by enabling a well-defined correlation between neutron energy and NR energy in the detector and because the results would not be affected by a wrong modelling of the neutron multiplicity in the simulation. A promising class of such sources is based on ($\gamma$,n) reactions, such as those induced in Y-Be sources. When an appropriate photon source is selected, these systems can produce monoenergetic neutrons. Nevertheless, they present several challenges, including low neutron production rates and the management of strong accompanying gamma radiation, which may interfere with neutron detection in ANAIS-112. Some feasibility studies have already explored the use of Y-Be sources in DM experiments \cite{collar2013applications}. However, a key limitation of $^{88}$Y lies in its short half-life (106.65~days), which prevents its use in long-term calibration campaigns, although periodic calibrations using specially prepared sources remain feasible.

An alternative approach involves the use of high-energy particle accelerators, which can generate neutrons by exceeding the energy threshold for neutron production in suitable targets. Under controlled conditions, such set-ups can provide monoenergetic neutrons. Accelerators offer the advantage of precise control over exposure time and can be deactivated when not in use. However, their implementation is limited by logistical and technical constraints, such as their large size, the difficulty of installing them in underground facilities near detectors, and the risk of electromagnetic interference with sensitive electronics.

\begin{figure}[t!]
    \centering
    {\includegraphics[width=0.73\textwidth]{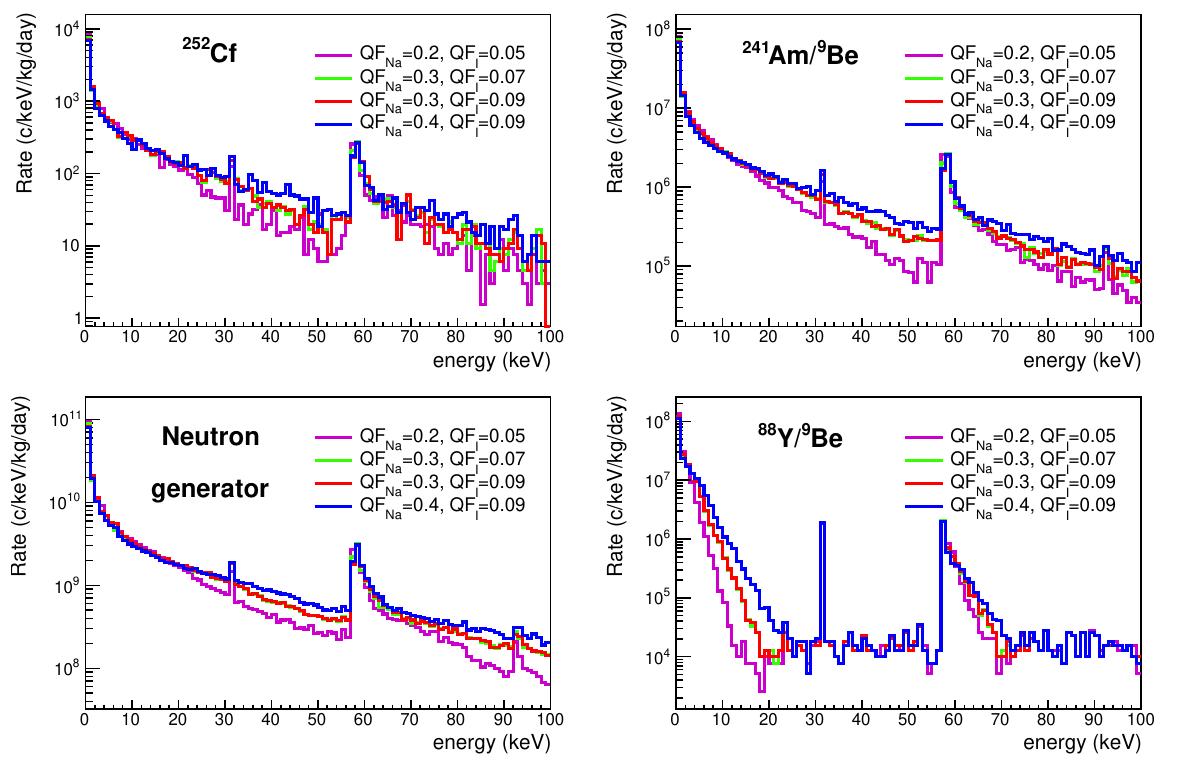}}

    \caption{\label{othersources} Simulated spectra of energy deposited in the ANAIS-112 detectors for different neutron sources under various QF combinations \cite{tfgainara}: QF\textsubscript{Na} = 0.2, QF\textsubscript{I}~=~0.05 (magenta); QF\textsubscript{Na} = 0.3, QF\textsubscript{I} = 0.07 (green); QF\textsubscript{Na} = 0.3, QF\textsubscript{I}~=~0.09 (red); and QF\textsubscript{Na}~=~0.4, QF\textsubscript{I}~=~0.09 (blue). No energy resolution correction has been applied. \textbf{Top left panel:} $^{252}$Cf source. \textbf{Top right panel:}  $^{241}$Am/$^9$Be source. \textbf{Bottom left panel:} Neutron generator emitting monoenergetic 2.45 MeV neutrons. \textbf{Bottom right panel:} $^{88}$Y/$^9$Be source. }
\end{figure}

Figure \ref{othersources} shows the comparison of simulated spectra of energy deposited in the ANAIS-112 detectors for different neutron sources under various QF combinations. The figure indicates that mono-energetic sources produce deposited energy spectra that are more sensitive to variations of the QFs values. In particular, at low energies, the $^{88}$Y/$^9$Be source exhibits the highest sensitivity. Despite their inherent limitations, further investigation into the use of neutron generators or Y-Be sources for onsite calibrations of ANAIS-112 would be valuable. A preliminary feasibility study on these options within the ANAIS experiment has already been conducted~\cite{tfgainara}. Should such developments be pursued, the support of the LSC will be essential in assessing the feasibility of integrating these sources into the experimental set-up.  There is a convenient window to do this in the first semester of 2026, after ending the scheduled ANAIS-112 data taking. A thorough calibration campaign will be planned within the next months, including additional neutron calibrations.

Future neutron calibrations using monoenergetic sources or sources covering different neutron energy ranges may also help further investigate uncertainties related to Geant4 modelling of the elastic scattering. Although this has been shown not to significantly affect the choice of QF, it still represents a systematic uncertainty. On the other hand, the neutron multiplicity spectrum from the \textsuperscript{252}Cf could also be analyzed in more detail with future neutron calibration campaigns to shed light on the observed discrepancies in the multiple-hit spectrum, where agreement with data worsens as multiplicity increases. Moreover, future simulation efforts could focus on manually modifying the neutron multiplicity spectrum to evaluate its impact on the results.

Moreover, comparison of simulations of neutrons using Geant4 and FLUKA frameworks have shown relevant differences in the propagation of fast neutrons through lead (as will be discussed in Section \ref{NeutronHENSA}), which could also introduce an additional systematic uncertainty affecting the conclusions on QFs of this chapter. Further work is required to better understand and quantify such uncertainties. 

Finally, it is worth highlighting a potential future scenario in which one or more crystals from the DAMA/LIBRA experiment could be distributed among the ANAIS, COSINE, and SABRE collaborations for further operation. Operating a DAMA/LIBRA crystal would not only allow for the measurement and characterization of its background, but also enable an independent determination of its QFs following the same methodology applied to the ANAIS-112 crystals in this work, either using $^{252}$Cf sources or alternative neutron sources. This would allow the QF of the DAMA/LIBRA detectors to be measured directly under the same experimental conditions and analysis strategies, thus minimizing systematic differences and resolving one of the most critical uncertainties in the comparison between DAMA/LIBRA and other experiments. The DAMA/LIBRA collaboration has repeatedly stated its intention to revisit its QF estimation. However, as of the time of writing this thesis, no new results have been published in this regard. Clarification may come with the final analysis of their full data exposure, which is still pending.

\setcounter{chapter}{4} 

\chapter{The improved ANAIS-112 background model}\label{Chapter:bkg}

\vspace{-1.cm}

\minitoc
\vspace{-1cm}
The sensitivity of a direct detection DM experiment is inherently constrained by its radioactive background levels. As a result, significant research and development efforts have been dedicated to improving detector radiopurity, shielding and background rejection strategies. However, the complete suppression of all radioactive backgrounds remains unfeasible. Thus, developing a precise background model is crucial for identifying and understanding the various background sources that contribute to the data. In the case of ANAIS-112, rigorous background characterization is particularly critical, as the time-dependent evolution of the background is directly incorporated into the annual modulation search strategy.

This chapter provides a comprehensive revision of the ANAIS-112 background model, initially developed and implemented in 2019 \cite{amare2019analysis}, by a multiparametric fit of the different background components. To begin with, Section \ref{previousmodel} examines the estimates of the previous background model in comparison with the measured background corresponding to six years of ANAIS data, identifying its strengths and limitations, and motivating the need for its refinement. Thereafter, Section \ref{keyprior} discusses prior considerations relevant to the fitting procedure. Following this, Section \ref{fittingprocedure} details the selection of data and simulated contributions involved in the fit, its sequential structure, and the minimization function employed. Next, Section \ref{backgroundfitting} presents the background fitting results and introduces the improved background model derived from the fitted activities. In Section~\ref{validation}, the new background model is validated across different event populations and energy ranges, and is compared to the previous ANAIS-112 background model.

In addition, Section \ref{additionalcontributions} introduces two additional contributions under consideration: the flux of environmental neutrons reaching the ANAIS detectors, and the possible contribution from events identified by the ANOD DAQ system but not by the ANAIS DAQ, that would be associated to an anomalous non-bulk scintillation population. Finally, in Section \ref{rateevolution}, a time-dependent background model is constructed. This updated model will be then employed in Chapter \ref{Chapter:annual} to reevaluate the annual modulation results over the full six-year exposure, assessing the degree of improvement achieved with respect to those previously published by the ANAIS collaboration.

\section{The previous ANAIS-112 background model} \label{previousmodel}

The background sources considered in the model of ANAIS-112 include both internal activity from the NaI(Tl) crystals and contributions from external components \cite{amare2016assessment,villar2018study,amare2018cosmogenic,amare2019analysis}, as described in Section \ref{BkgModel}. External background components were screened using HPGe spectrometry at the LSC, while internal activity was directly assessed, including contributions from \(^{40}\)K and \(^{210}\)Pb. Additionally, cosmogenic activation was evaluated, considering short-lived isotopes of Te and I, as well as \(^{3}\)H, \(^{22}\)Na, \(^{109}\)Cd, and \(^{113}\)Sn. 

A detailed background model incorporating these quantified contributions was successfully implemented in 2019 \cite{amare2019analysis}. The model was validated against experimental data from the $\sim$10\% first year of unblinded data of data taking, demonstrating robust performance across various experimental conditions (anticoincidence and coincidence) and energy ranges.

In the high-energy range (100 keV to 2 MeV) and in the medium-energy range (10 to 100 keV), the deviation between simulated and measured spectra remained below 10\%. In the low-energy region, the background model also provided a satisfactory description when compared to the first year of data. Specifically, the average deviation between simulation and data across all detectors was only 10.7\% in the [2–6] keV energy range. Moreover, the background model revealed that the region of interest for DM searches ([1–6] keV) is predominantly influenced by crystal emissions, with \(^{210}\)Pb, \(^{3}\)H, and \(^{40}\)K contributing 32.5\%, 26.5\%, and 12.0\% of the measured rate in ANAIS-112, respectively.  Nevertheless, significant discrepancies arise below 2 keV. In particular, the model underestimates the measured rate by 54\% in the [1–2] keV region. 

ANAIS work subsequently progressed along two directions: on the one hand, improving the background model to account for the observed excess, and on the other, better understanding those non-bulk scintillation events which have not been rejected by the standard filtering
protocols. Regarding the latter, to enhance the rejection of non-bulk scintillation events in the low-energy region, ANAIS implemented a supervised ML filtering protocol based on a BDT \cite{coarasa2022improving,Coarasa_2023}. This strategy resulted in an approximate 20\% reduction in the background level in the [1-2] keV energy range  with a very important improvement of the efficiency for the selection of bulk scintillation. (see Section \ref{BDTnew}) In this way, the discrepancy between the background model and the data after applying the BDT filtering method was reduced to 37\% and 7.0\% in the [1–2] keV and [2–6] keV regions, respectively.  At present, this line of work continues by using a new filtering procedure based on ANOD data (see Section \ref{ANODfiltering}).

In this section, an updated comparison between the data and the background model is performed for an accumulated exposure of six years of data taking. 

\begin{figure}[b!]
\begin{center}
\includegraphics[width=0.7\textwidth]{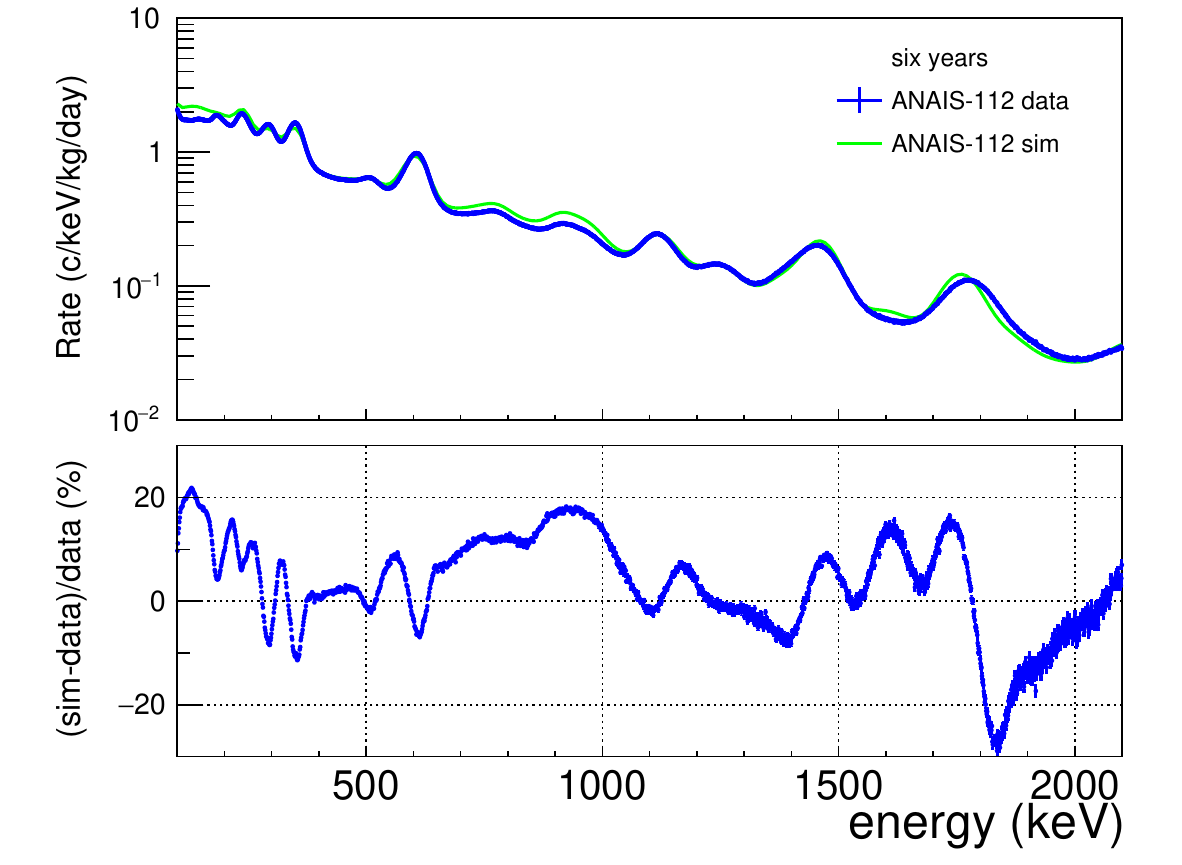}
\includegraphics[width=0.7\textwidth]{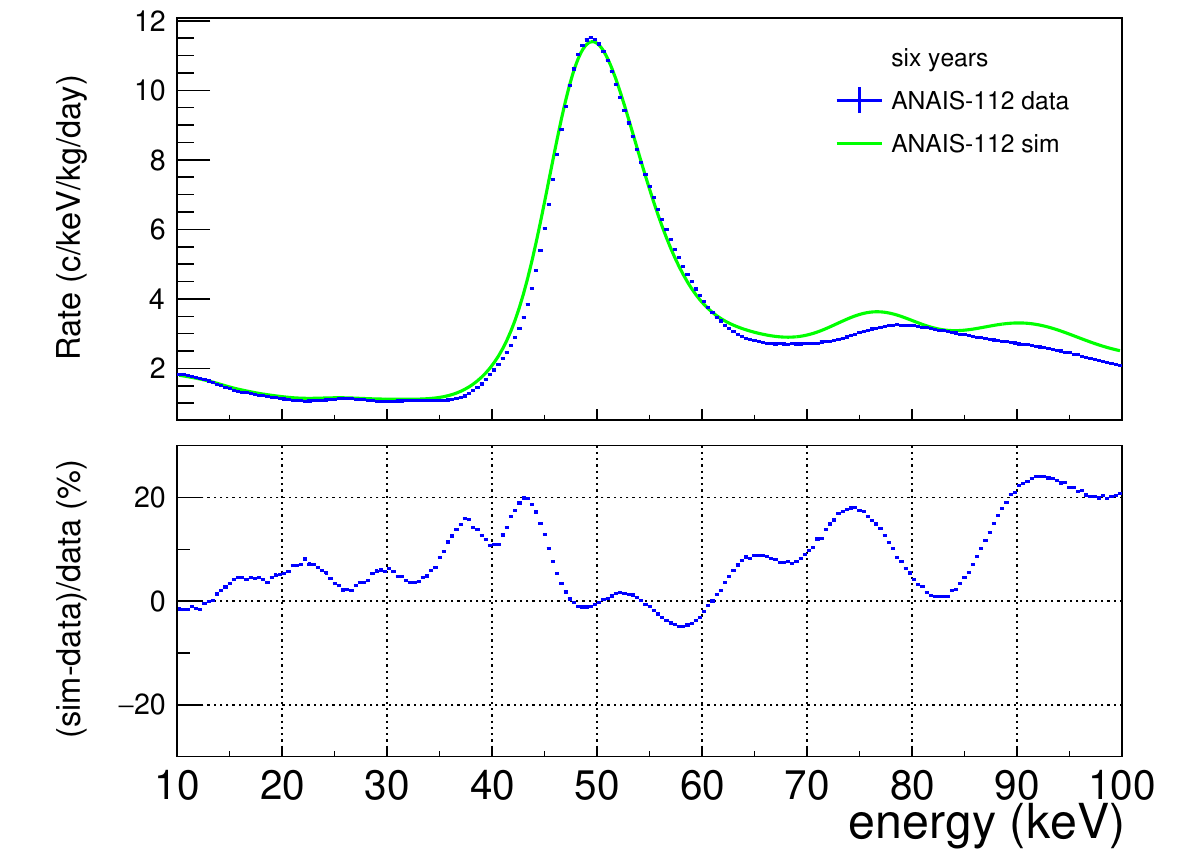}

\caption{\label{HEandMEold} Simulated background spectrum (green) compared to experimental data (blue) in the high-energy range (\textbf{top panel}) and medium-energy range (\textbf{bottom panel}). The high-energy spectrum shows the total spectrum, while the medium-energy spectrum shows the anticoincidence spectrum. Data correspond to the summed spectra of the nine detectors over a six-year exposure. The relative residuals between simulation and data are shown in each panel.}
\end{center}
\end{figure}

Figure \ref{HEandMEold} presents this comparison, displaying the high-energy range in the top panel and the medium-energy range in the bottom panel, along with the corresponding residuals. The high-energy shows the total spectrum, while the medium-energy shows the anticoincidence spectrum. Despite some localized structure at specific energies observed in the residuals, the overall agreement reinforces the robustness of the model. Among the discrepancies, two main issues stand out in the high-energy region. First, in the range up to $\sim$ 300 keV, a continuum excess leads to an overestimation in the model, while the peaks are not well reproduced. Second, the excess in the region [600-1000]~keV remains unexplained. Regarding the medium-energy region, a notable mismatch is observed above 60 keV in the spectral shape of the simulation. Such discrepancies will be thoroughly discussed later in this chapter (see Section \ref{keyprior}). 

\begin{figure}[t!]
\begin{center}
\includegraphics[width=0.65\textwidth]{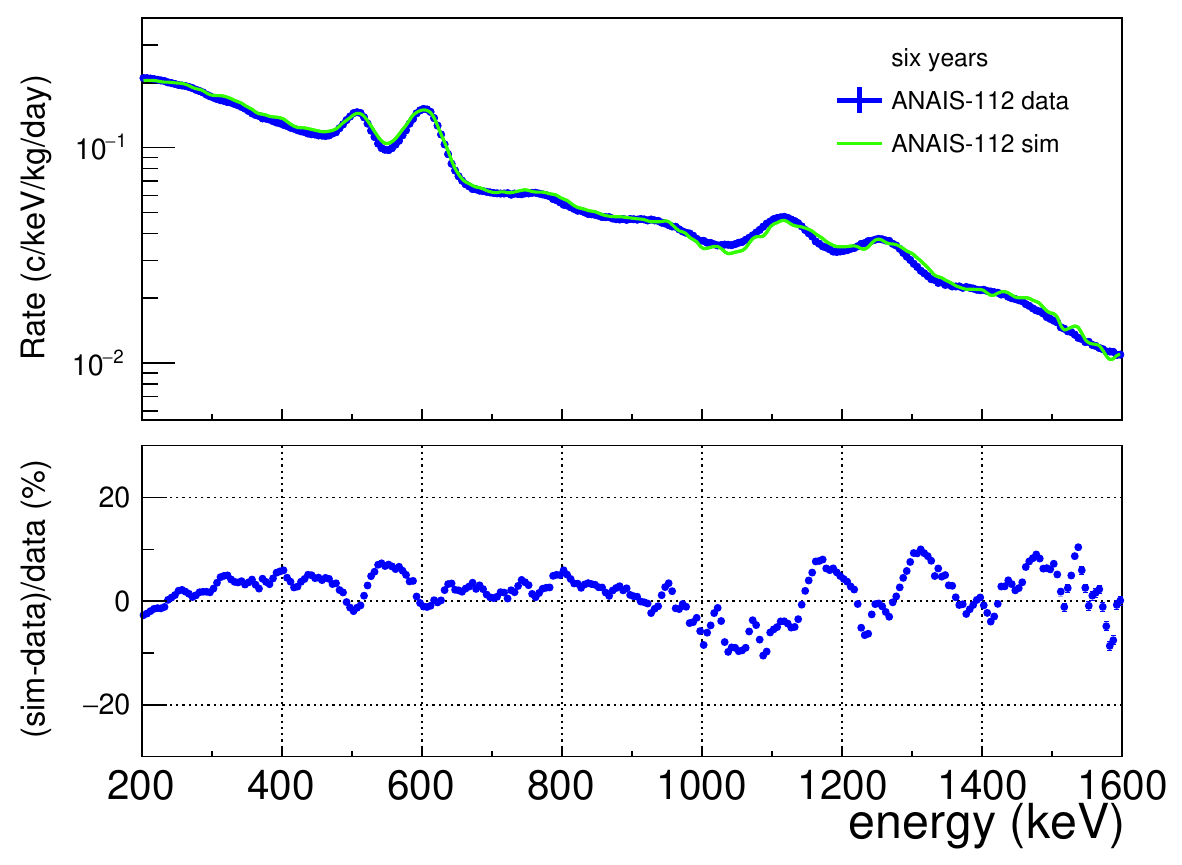}
\includegraphics[width=0.65\textwidth]{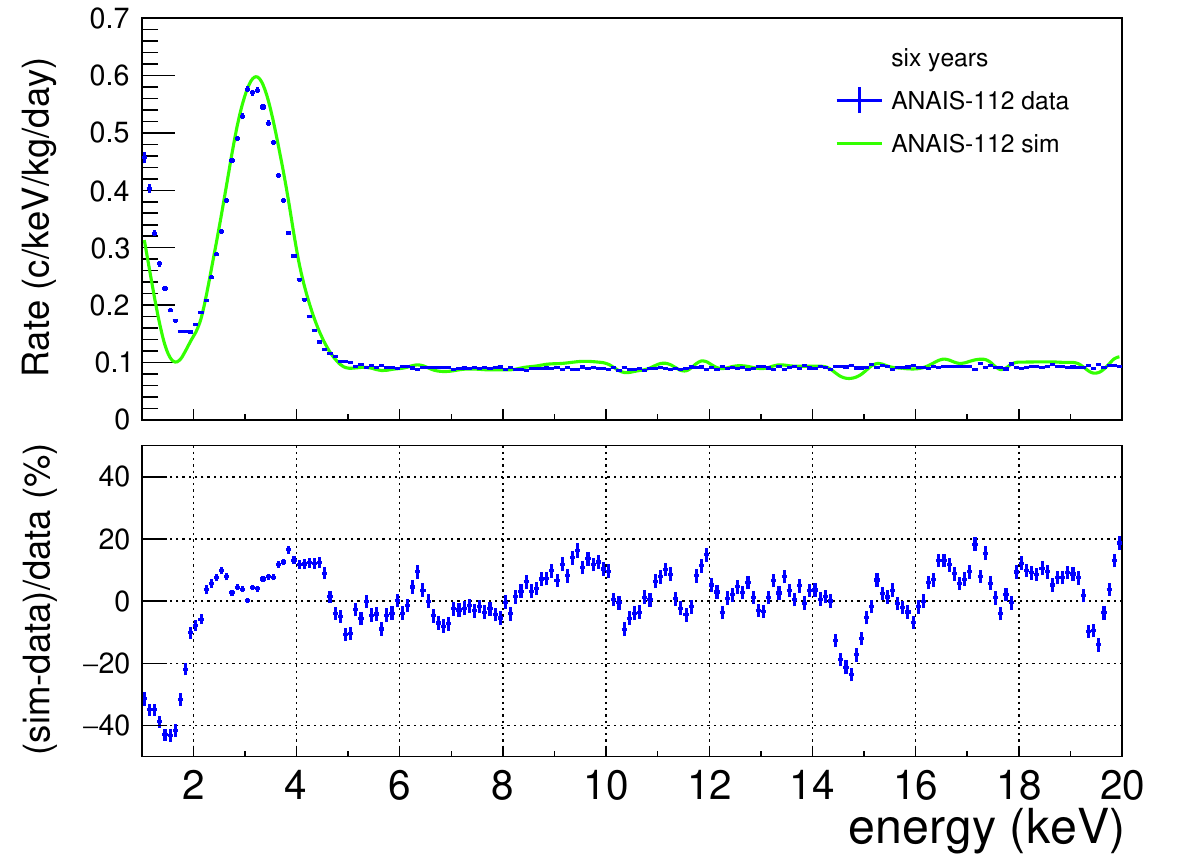}

\caption{\label{m2oldmodelo} Simulated reconstructed background spectrum (green) compared to experimental data (blue) in the high-energy range (\textbf{top panel}) and low-energy range (\textbf{bottom panel}). Both panels shows the m2-hits spectrum, corresponding to events having energy deposited at two detectors. Data correspond to the summed spectra of the nine detectors over a six-year exposure. The relative residuals between simulation and data are shown in each panel.}
\end{center}
\vspace{-0.7cm}
\end{figure}

The coincidences population is also well described by the previous background model. Figure \ref{m2oldmodelo} presents this comparison, showing the high-energy range in the upper panel and the low-energy range in the lower panel, along with the corresponding residuals, for events with energy deposition in two detectors, hereafter referred to as m2-hits\footnote{It is important to emphasize that the background model employed for the coincidence population does not strictly correspond to the previous background model, as the coincidence dataset was generated exclusively for the first year of data-taking. Instead, the background model presented here for the coincidences is derived from simulations performed within this thesis, using the activity values from the earlier background model as input. In principle, the simulation output is not expected to exhibit significant deviations, except for possible unaccounted changes introduced in subsequent Geant4 versions. Throughout this chapter, this configuration, i.e., simulations performed in the context of this work using the activity values from the former background model, will be referred to as the reconstructed background model.}. As observed, the agreement between data and simulation is satisfactory.

\begin{figure}[t!]
\begin{center}
\includegraphics[width=0.8\textwidth]{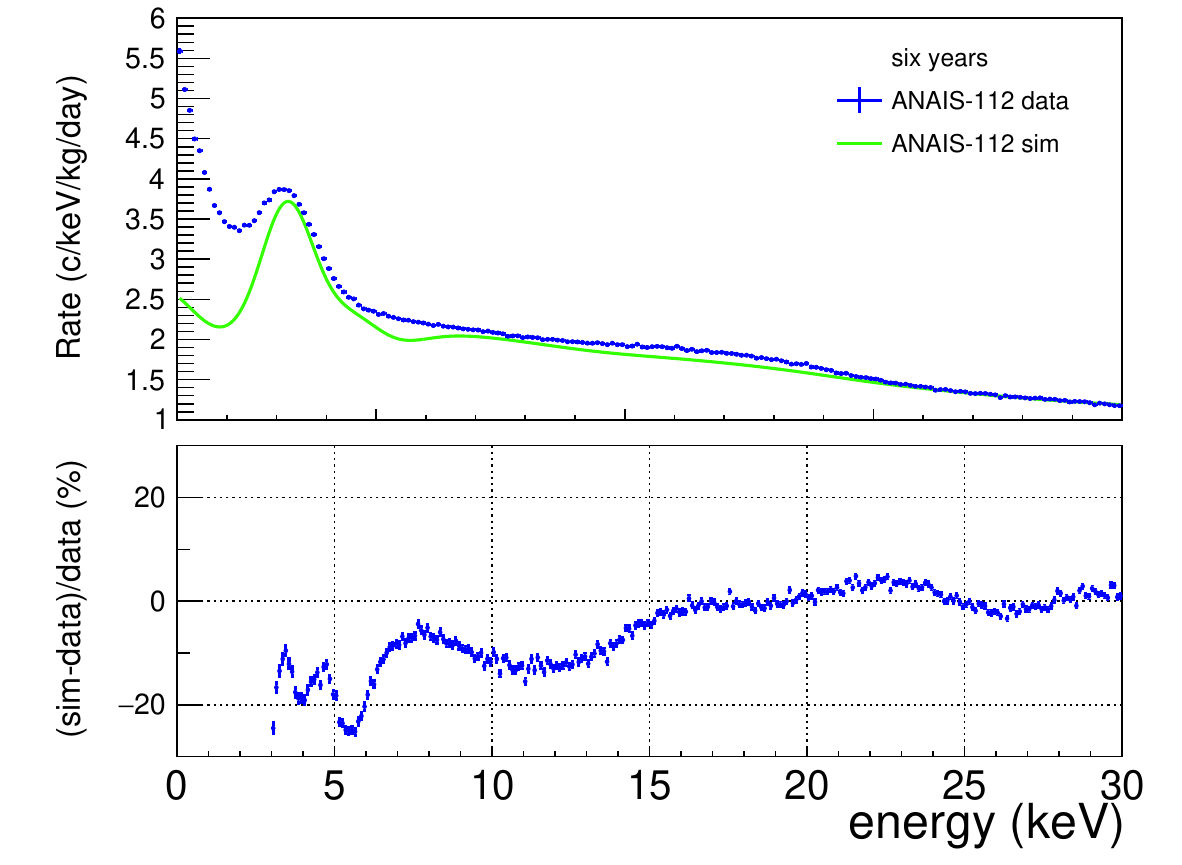}

\caption{\label{LEold} Simulated anticoincidence background spectrum (green) compared to experimental data (blue) in the low-energy range. Data correspond to the summed spectra of the nine detectors over a six-year exposure. The relative residuals between simulation and data are shown.}
\vspace{-0.5cm}
\end{center}
\end{figure}

However, in the low-energy region, certain limitations of the background model became evident as more data were accumulated and calibration strategies improved. Figure \ref{LEold} presents the comparison between the six-year ANAIS-112 dataset and the corresponding background model in the region below 30 keV, revealing a clear discrepancy between the measured and simulated spectra. Despite the implementation of enhanced filtering protocols, a significant mismatch persists in the [1–2] keV region. Furthermore, residual analysis beyond 6 keV suggests that the model does not account for the presence of \(^{210}\)Pb in some detectors, originating from the teflon diffuser film surrounding the crystals, which would produce a characteristic X-ray peak at 12 keV. 

This comparison highlights the necessity to reassess the background model for the full nine-detector set-up. In particular, for the full six-year dataset, the deviation between data and simulation has been quantified as 20\% in the [1–6] keV range, 11\% in the [2–6]~keV range, and 46\% in the [1–2]~keV region. In particular, within the [1–6] keV range, the overall background level measured in ANAIS-112 over the six years of data taking is \(3.2834~\pm~0.0018\)~keV\(^{-1}\)~kg\(^{-1}\)~day\(^{-1}\), whereas the background model predicts only \(2.611 \pm 0.004\)~keV\(^{-1}\)~kg\(^{-1}\)~day\(^{-1}\).

Another key issue concerns the energy resolution: the previous background model relied on the resolution measured from the ANAIS prototypes and assumed a common resolution for all modules. A better modelling of the resolution could help to improve the agreement and would be also pursued in the update of the background model (see next sections). Consequently, the goal of this chapter is to revisit the ANAIS-112 background model, focusing on reviewing the dominant contributions that compromise the general agreement between data and simulation. In the previous background model, no direct fitting of the individual components was performed; instead, the model was constructed based on activities determined through various methods. In this study, however, a multiparametric fit of the different background components in ANAIS-112 is proposed for a better estimation of the background of the experiment.

For this purpose, the background contributions identified in the previous ANAIS-112 background model (see Section \ref{BkgModel}) have been re-simulated using Geant4 version 10.7.0. The previous background model was based on Geant4 version v9.4.p01. As discussed in Section \ref{datalibrary}, the neutron interaction libraries differ between versions v9.4.p01 and v11.1.1. The former was selected to obtain the results presented in Chapter \ref{Chapter:QF} regarding the QF of the ANAIS crystals, as it showed better agreement with measurements for a NaI(Tl) target, particularly regarding inelastic de-excitation processes. However, no relevant differences were found between versions v9.4.p01 and v10.7.0 in the modelling of electromagnetic interactions. The latter is the version used for the simulations performed throughout this chapter.


In addition to the primary reason for re-simulating these contributions - namely, the need to update the geometry to better reflect the full experimental set-up, as described in Chapter \ref{Chapter:Geant4} - there is one further underlying reason.


The previous background model stored only partial information (specifically energy spectra), which is now expanded in order to use a broader set of observables for comparison between simulation and data. Moreover, in the previous background model, histograms were stored with fixed binning and resolution. The plan now is to store much more detailed information in ROOT structures, including all energy deposits un-convoluted with resolution and un-binned, categorized by interacting particle type, along with their position, time, and various other observables. Furthermore, the new model will allow for a better reconstruction of the visible energy by using appropriate QFs depending on the particle type, which were not implemented in the previous model, as only $\beta/\gamma$ events were stored. In addition, effects related with the DAQ system used for the signal readout can be also introduced in the analysis of the simulation.



The production of the simulated spectra follows a two-step process. 

In the first step, simulations for each source within the ANAIS-112 set-up are generated using Geant4 version v10.7.0. The contaminants to be simulated are primarily identified from the previous ANAIS-112 background model, with additional contaminants detected in this study based on distinctive features observed in the full six-year exposure. 

The second step involves post-processing the raw Monte Carlo simulation to account for the detector response (see Chapter \ref{Chapter:Geant4}), generating simulated events that closely match real ANAIS-112 data. For this purpose, the updated resolution determined in this thesis will be employed (see Figure \ref{LEres}). The simulated data undergo the same selection cuts as the real data. In particular, the new simulation accounts for dead time, which was not previously considered, and allows for the selection of integration or coincidence windows to reconstruct event multiplicity. This means that the simulation can be adapted to different DAQs (as has been done for ANOD in Chapter \ref{Chapter:Geant4}), cut efficiencies, trigger settings, resolution, and QFs. In general, once all these characteristics are incorporated into the generation of ANAIS-like events from the simulation, events depositing energy in a single detector are labeled as multiplicity-1 (m1 or single-hit) events. Events depositing energy in several crystals are grouped into higher multiplicities (m2, m3, ..., m8-hit events).

In the background fitting procedure that will be conducted in this chapter, each component is modelled as a probability density function (PDF), directly drawn from these simulations, with an independent floating parameter representing its fractional activity. The simulated spectra at different multiplicities are then employed in the background fitting when compared to the ANAIS-112 data, combining information from different populations and years to optimize the machinery of the fitting algorithm, as will be discussed later in this chapter. The following section outlines key considerations that must be addressed prior to the implementation of the background fitting procedure.


\section{Key considerations prior to the fitting}\label{keyprior}

Before proceeding with the fit, this section revisits key aspects of the ANAIS-112 background that justify the restriction of the fit to specific energy regions and underscore the need for particular care in treating certain contaminations, such as that from \(^{210}\)Pb or those originating in the PMTs.

\subsection{Asymmetry in light sharing} \label{asymmetry}

When comparing the single-hit simulated spectrum with the experimental data in the medium-energy range (see bottom panel of Figure \ref{HEandMEold}), a significant deviation was observed above 60 keV, marking the region with the poorest agreement in terms of spectral shape. According to the previous background model, events in this energy range are primarily attributed to background contributions from the PMTs coupled to the crystals, particularly X-ray deposits from the decay of the \(^{238}\)U chain.  

During the course of this thesis, the COSINE collaboration reported similar discrepancies between data and the background model in the very same energy range~\cite{cosine2025improved}. This inadequate background modelling has been linked by the collaboration to the asymmetry in the spatial distribution of background sources. Specifically, COSINE-100 suggests that contamination from the PMTs leads to asymmetric energy deposition, where background events tend to deposit more energy near the contaminated PMT than in the opposite one, causing an asymmetry in light sharing. Other external background sources have been ruled out by the experiment, as they primarily contribute to this energy range in multiple-hit events, and the background model shows good agreement with data for this population.

The crystals and PMTs used in COSINE-100 and ANAIS-112 are very similar in terms of radioactive contamination, as they correspond to the same model and manufacturer. Therefore, such asymmetry in light sharing and its possible origin have been investigated in this thesis to evaluate their impact on the discrepancies observed in ANAIS-112. To quantify this effect, the number of photoelectrons (nphe) in the pulse is analyzed. The asymmetry in nphe, \(\textnormal{nphe}_{\textnormal{Asy}}\), is then computed as the ratio between the difference in the number of photoelectrons collected by the two PMTs and the total number of photoelectrons, as follows:

\begin{equation}
\textnormal{nphe}_{\textnormal{Asy}} = \frac{\textnormal{nphe}_0 - \textnormal{nphe}_1}{\textnormal{nphe}_0 + \textnormal{nphe}_1}
\label{npheasy}
\end{equation}

Here, \(\textnormal{nphe}_0\) and \(\textnormal{nphe}_1\) represent the number of photoelectrons in the pulses from PMT0 and PMT1, respectively. While pulse area asymmetry could also be used to characterize this effect, \(\textnormal{nphe}_{\textnormal{Asy}}\) serves as a more direct and reliable variable, as it inherently accounts for PMT-specific gain variations. This choice ensures that the asymmetry is measured intrinsically, minimizing biases and improving the accuracy of light-sharing studies. It is worth noting that COSINE-100 employs an analogous definition to investigate this effect, but instead of using the number of photoelectrons, they quantify the charge asymmetry. In the case of ANAIS-112, the SER is calibrated periodically, allowing for time-dependent gain variations to be corrected for each PMT individually.
l

\begin{figure}[t!]
\begin{center}
\includegraphics[width=1.\textwidth]{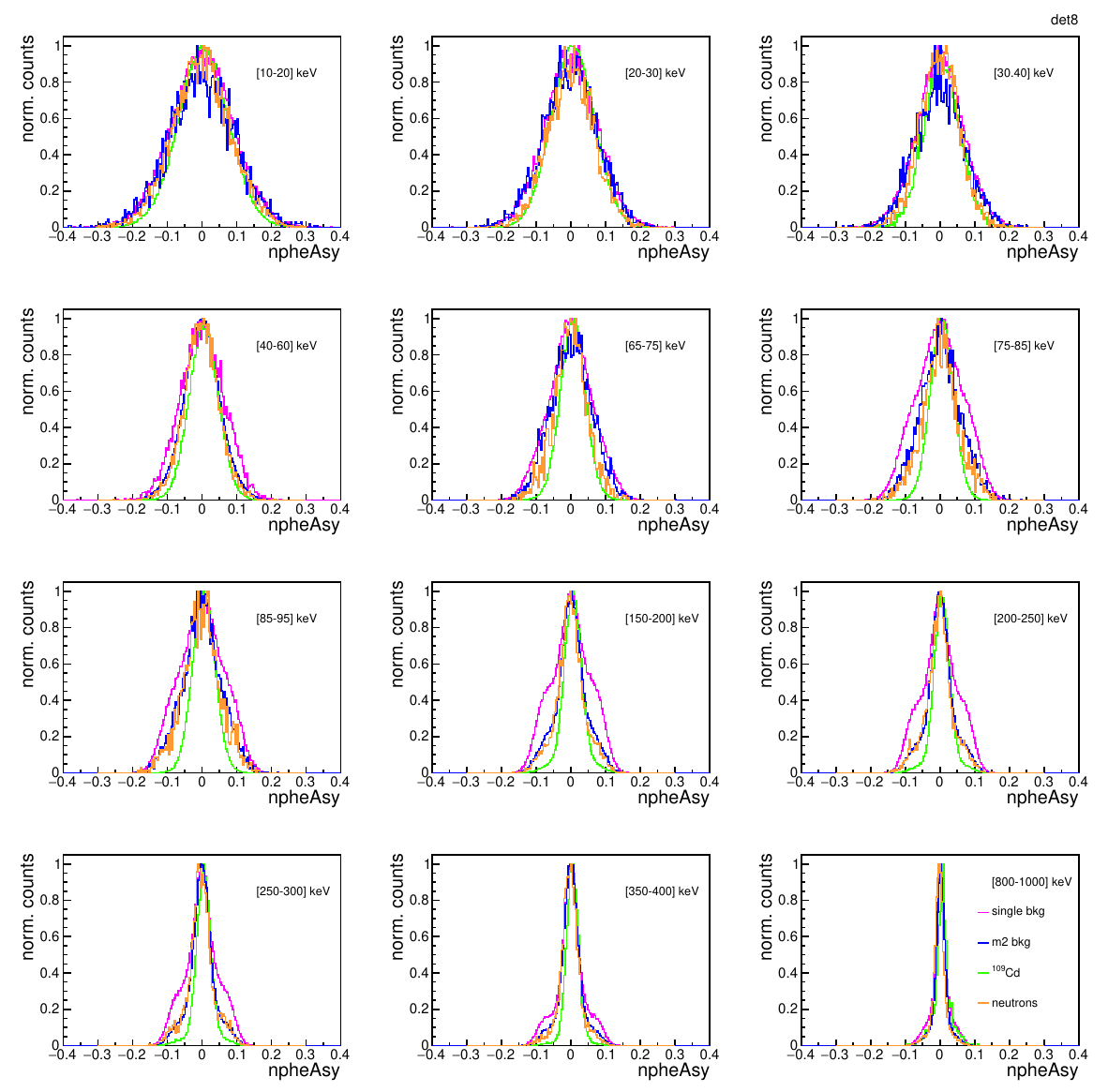}

\caption{\label{asy} Light-sharing asymmetry observed in detector 8 during year 6 across different energy intervals for various event populations measured in ANAIS-112: background single-hit events (magenta), background m2-hit events (blue), single-hit events from \(^{109}\)Cd calibration (green), and total-hit events from neutron calibrations (orange). Noticeable differences in the distribution are observed only in the [40-60] keV region and above 75 keV in the single-hit event population, gradually disappearing above 350 keV.}
\end{center}
\end{figure}

Figure \ref{asy} illustrates the light-sharing asymmetry observed in detector 8 during year 6 across different energy intervals. The distribution is analyzed for various event populations measured in ANAIS-112, specifically background single-hit events, background m2-hit events, single-hit events from \(^{109}\)Cd calibration, and total-hit events from neutron calibrations.  

At low energies, up to approximately 40 keV, all distributions exhibit a similar shape. Specifically, it can be observed that, as expected, the distributions remain centered around zero and become narrower with increasing energy. In the energy region between [40–60]~keV, which is dominated by the $^{210}$Pb contribution, the single-hit distribution exhibits a broader shape compared to other populations. A possible explanation for this feature will be proposed later in the text. In the [65–75] keV range, the single-hit distribution becomes again compatible with that of the multiple-hit population. However, above 75 keV, a clear deviation emerges. Notably, the single-hit distribution exhibits two distinct secondary bumps, deviating markedly from the expected gaussian profile. In particular, the single-hit background distribution begins to diverge significantly from the other event classes, with a clear separation emerging above 75~keV. This divergence does not diminish at higher energies; on the contrary, it becomes more pronounced, gradually disappearing above 350~keV. Above 350~keV, a certain deviation of the single-hit population is indeed observed; however, it is significantly less pronounced than in the previously highlighted energy regions.

These bumps are symmetrically located with respect to the central maximum, suggesting that the underlying asymmetric background component is present on both sides of the crystal. Consequently, the overall distribution could be effectively modelled by a central gaussian function, complemented by two additional gaussian components symmetrically positioned on either side. Moreover, although the \(^{109}\)Cd calibration distribution is narrower and more well-defined due to its localized surface but centered energy deposition, the neutron and m2-hit background distributions follow a nearly identical pattern without significant asymmetry. This reinforces the interpretation that neutron events represent a homogeneous and bulk-like population. 

\begin{figure}[t!]
    \centering
    {\includegraphics[width=0.49\textwidth]{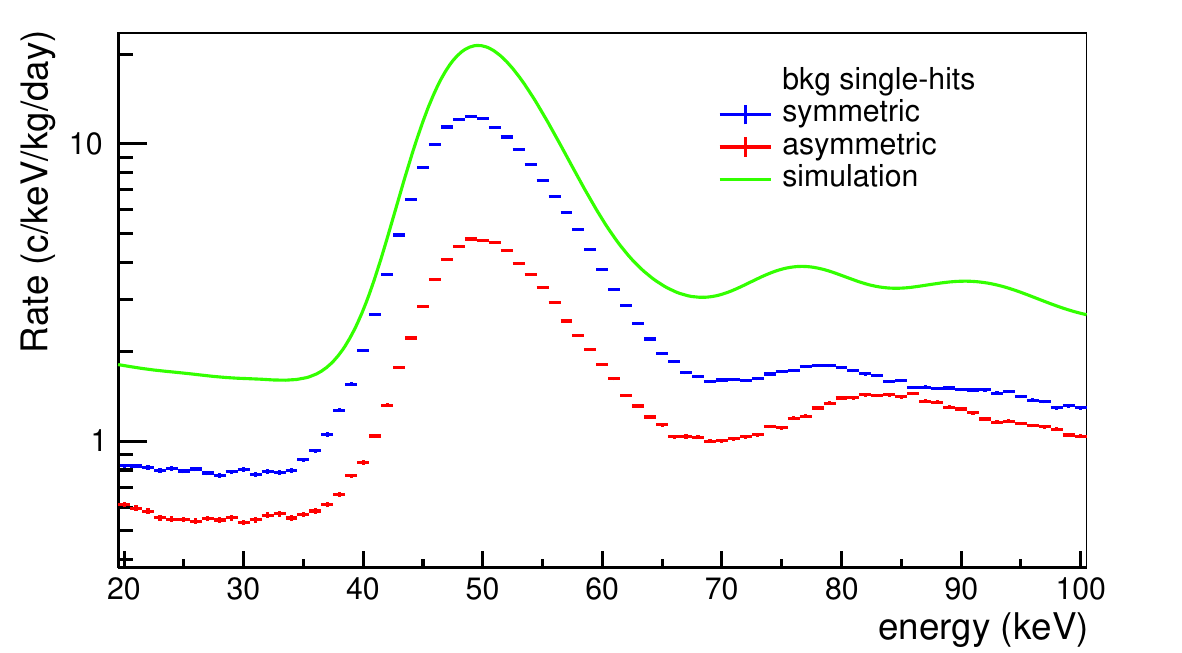}}
    \hfill
    {\includegraphics[width=0.49\textwidth]{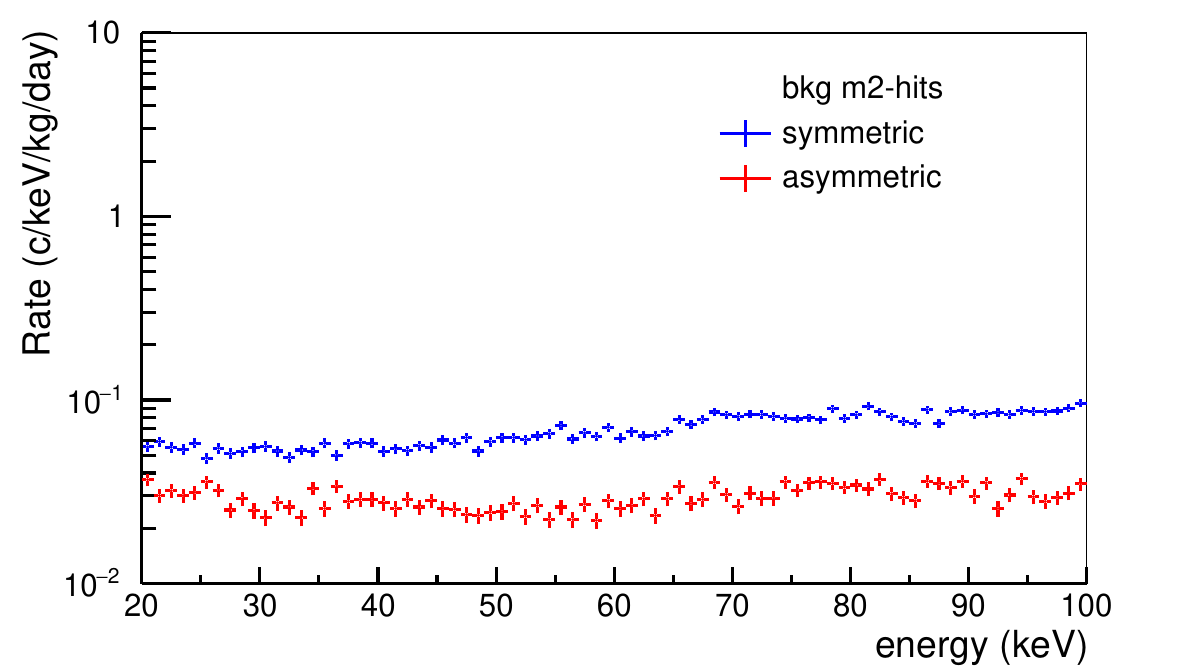}}
     {\includegraphics[width=0.49\textwidth]{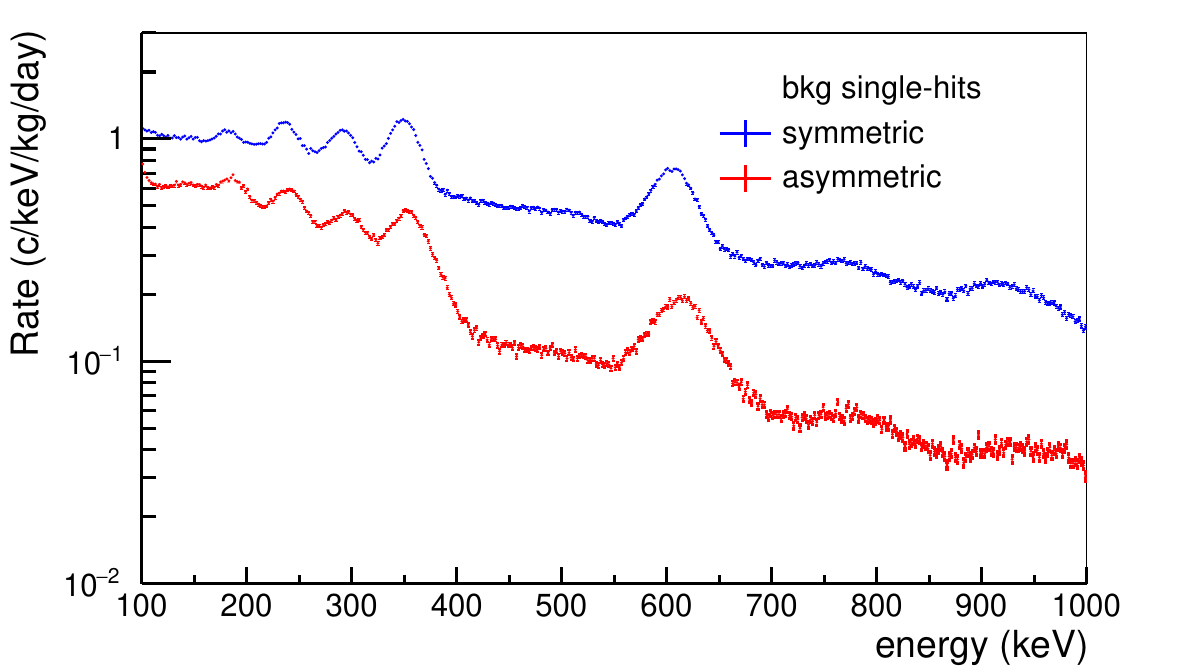}}
    \hfill
    {\includegraphics[width=0.49\textwidth]{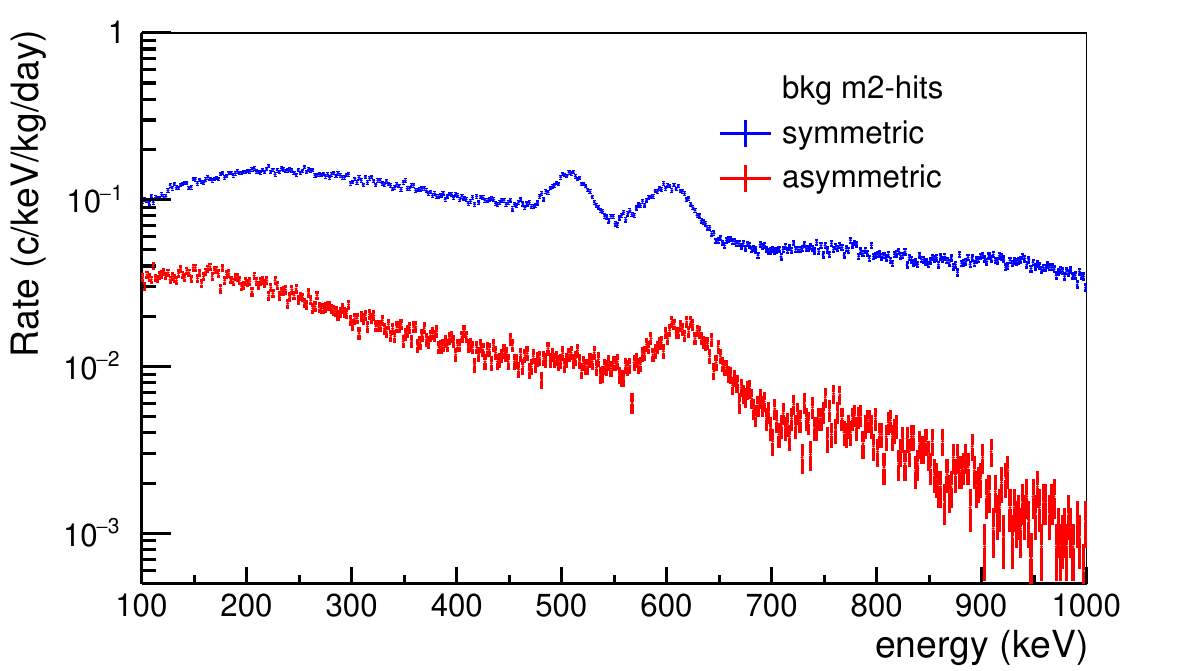}}
    \caption{Comparison of the background energy spectra of symmetric (blue) and asymmetric (red) events corresponding to detector 8. Symmetric events are defined as those lying within two standard deviations from the \textsuperscript{109}Cd $\textnormal{nphe}_{\textnormal{Asy}}$ distribution’s mean, while asymmetric events correspond to values beyond this range. \textbf{Top left panel:} medium-energy, single-hit event. The previous background model (green) is also shown for comparison. \textbf{Top right panel:} medium-energy, m2-hit events. \textbf{Bottom left panel:} high-energy, single-hit events. \textbf{Bottom right panel:} high-energy, m2-hit events. }
    \label{datalowhighasy}
\end{figure}

The energy distributions of symmetric and asymmetric events are compared in Figure~\ref{datalowhighasy}. As shown in Figure \ref{asy}, the distribution broadens significantly at low energies, preventing the application of a uniform selection criterion across the full energy range. Therefore, the classification of events is based on the width of the $\textnormal{nphe}_{\textnormal{Asy}}$ distribution derived from the \textsuperscript{109}Cd calibration source, which serves as the reference population for symmetric events: symmetric events are defined as those lying within two standard deviations from the \textsuperscript{109}Cd distribution’s mean, while asymmetric events correspond to values beyond this range. 

The comparison presented in Figure \ref{datalowhighasy} includes the single-hit and m2-hit background distributions at both medium and high energies. 


In the medium-energy single-hit spectrum (top left panel), a distinct change in the spectral shape is observed above approximately 75 keV between the symmetric and asymmetric event distributions. Notably, the symmetric distribution closely resembles the background model predictions, which are also shown in the figure for comparison. Specifically, both the symmetric distribution and the background model exhibit a double-peak structure with features around 75 keV and 95 keV, whereas the asymmetric distribution presents a smoother profile with maximum near 85 keV. In contrast, no such spectral difference is observed in the m2-hit background distribution.  In addition, a comparative study of the pulse-shape parameters between the symmetric and asymmetric populations, specifically the P$_1$ and FM distributions, has been performed for events with energies above 75 keV. No significant differences were observed, indicating the absence of anomalous contributions in this energy range, such as those arising from mixed Cherenkov and scintillation signals.

At higher energies, no significant shape differences between the symmetric and asymmetric components are observed either. Notably, the 511 keV peak associated with \textsuperscript{22}Na contamination appears fully symmetric in the m2-hit spectrum, as expected from a bulk contamination. Conversely, the 609~keV peak, primarily originating from PMT contamination, exhibits both symmetric and asymmetric components.

The hypothesis that bulk events are symmetric, supported by observations from the \textsuperscript{22}Na contribution, provides an alternative means to assess whether the \textsuperscript{210}Pb contamination of ANAIS-112 is entirely bulk. Based on this assumption, such contamination should appear exclusively in the symmetric component; however, this is not what is observed in the top left panel of Figure \ref{datalowhighasy}. Symmetric events are defined as those lying within two standard deviations of the \textsuperscript{109}Cd $\textnormal{nphe}_{\textnormal{Asy}}$ distribution’s mean, which implies that only 4.45\% of symmetric events would be expected to leak into the asymmetric selection.

To test this, the integral of the energy range [40–60] keV, where the \textsuperscript{210}Pb feature is located, is computed for single-hit events in both the symmetric and asymmetric components. The results for detector 8 show that $\sim$ 71\% of the events in this energy window belong to the symmetric component, while $\sim$ 29\% belong to the asymmetric one. This observation is further supported by the $\mathrm{nphe}_{\text{Asy}}$ distribution shown in Figure~\ref{asy}, where the single-hit population deviates from the distribution of other populations, exhibiting a clear asymmetry. This significant excess in the asymmetric population strongly indicates that not all \textsuperscript{210}Pb contamination of ANAIS detectors originates from a bulk contamination.

Beyond the possibility of a uniformly distributed surface contamination, either on the crystal surface or the teflon reflector, this result suggests the presence of \textsuperscript{210}Pb contamination on the NaI endcaps. Such contamination would naturally lead to an asymmetric event distribution. This is plausible given that the endcaps undergone a different surface treatment compared to the lateral faces of the crystal because they were polished under non-disclosed protocols. This symmetric/asymmetric event analysis could be extended to alpha events to investigate whether they also exhibit symmetric or asymmetric components, potentially shedding light on the origin of the \textsuperscript{210}Pb contamination. However, this is currently not feasible due to the saturation of alpha signals above 1 MeV. In the future, this approach could be explored with data from the ANOD DAQ or with dedicated calibrations
after finishing the data taking and before decommissioning the ANAIS-112 set-up.

\begin{figure}[t!]
    \centering
    {\includegraphics[width=0.49\textwidth]{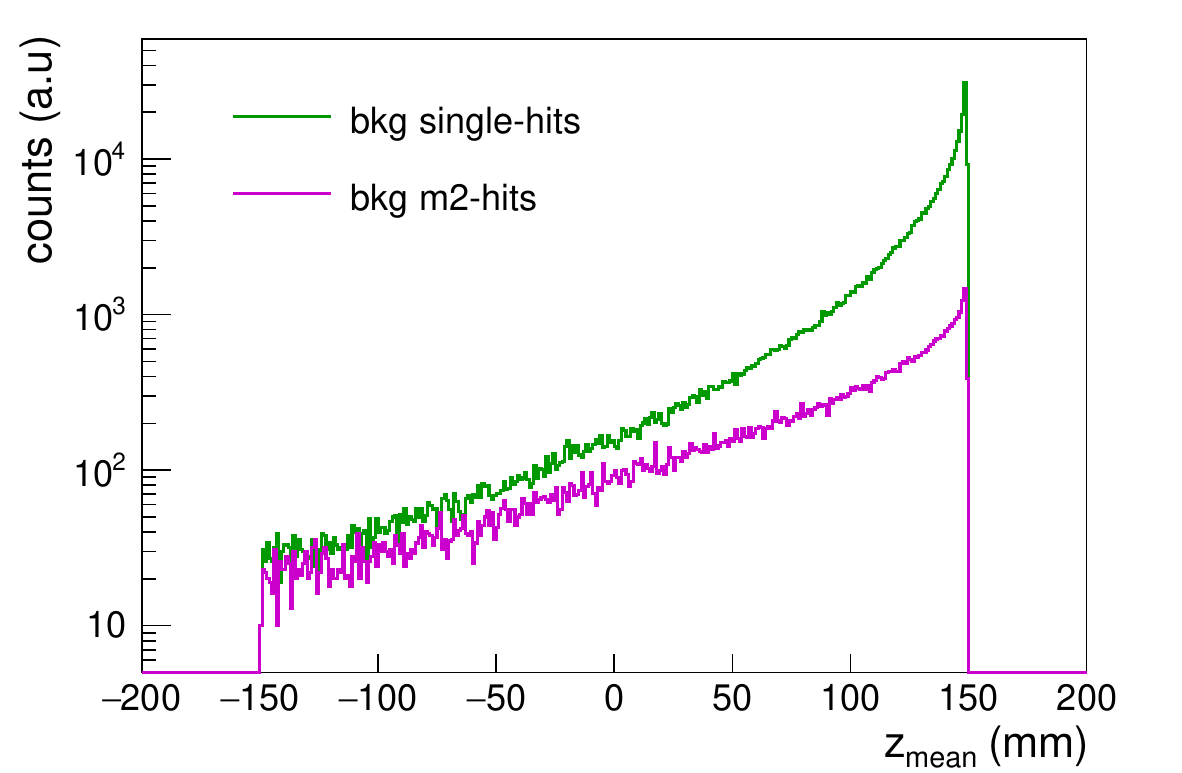}}
    \hfill
    {\includegraphics[width=0.49\textwidth]{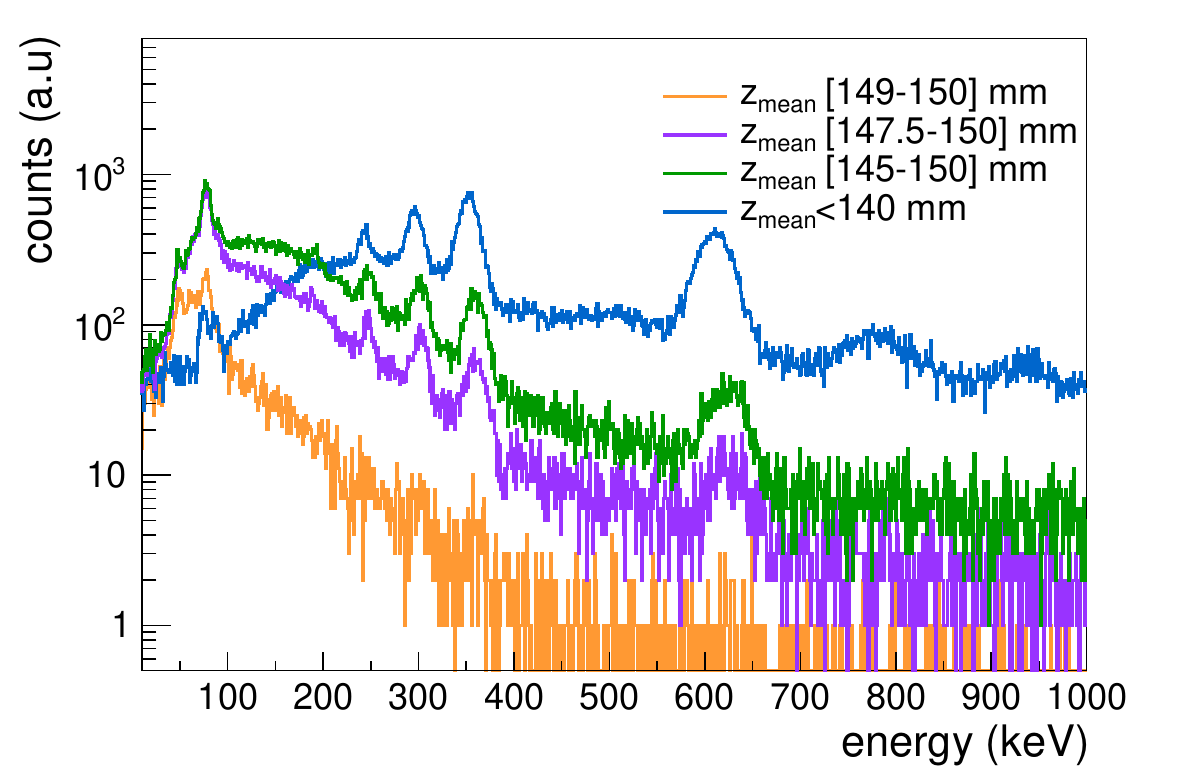}}
    
    \caption{\textbf{Left panel:} Average interaction position along the longitudinal axis (z-axis) of the ANAIS crystals for single-hit (green) and multiple-hit events (magenta) originating from a simulated \textsuperscript{226}Ra contamination located in one of the PMTs. \textbf{Right panel:} Simulated energy spectra for events occurring within the first 1 mm of the crystal (orange), the first 2.5 mm (violet), the first 5 mm (green), and for those occurring beyond 10 mm from the contamination source (blue).}
    \label{compareasysim}
\end{figure}

The hypothesis that PMTs may be responsible for the observed asymmetry in light sharing continues to be explored, in this case by exploiting the information obtained from simulations. The left panel of Figure~\ref{compareasysim} shows the average interaction position along the longitudinal axis (z-axis) of the ANAIS crystals for single-hit and multiple-hit events originating from a simulated \textsuperscript{226}Ra contamination located in one of the PMTs. The distributions differ noticeably: single-hit events exhibit a higher density of interactions occurring within the first few centimeters from the contamination source, supporting a higher degree of asymmetry in energy deposition along the crystal with respect to multiple hits.

The right panel of the same figure presents the simulated energy spectra for events occurring within the first mm (as well as within 2.5 mm and 5 mm) of the crystal, regions presumably associated with asymmetric light sharing, as well as for events occurring beyond 10 mm from the contamination source. A clear difference in the spectral shape is observed, particularly in the critical region above 65 keV. Specifically, the spectra corresponding to interactions within the first few mm show a noticeably enhanced contribution in the 60–100~keV energy range. It is found that 39\% of the events occurring in the first millimeter fall within the 60-100
keV energy range, while this percentage decreases to 32\% when considering depths up
to 2.5 mm, and to 22\% at 5 mm. By contrast, only 2\% of the events at depths beyond 10
mm fall in that energy region. When all events in the 60–100 keV energy range are
considered, 73\% are located within the first 2.5 mm, which evidences the significant
contribution of asymmetric light sharing expected in this population.


It is important to note that the threshold used to define proximity to the contamination source has been chosen in a somewhat arbitrary manner. A more physically motivated approach would involve modelling the differential light collection efficiency as a function of the interaction position, taking into account the solid angle subtended by each PMT. This function could be related to the ratio of solid angles, which, considering the dimensions of the ANAIS crystal and PMTs, suggests that within the first ~5 cm, the solid angle diminishes by approximately one order of magnitude. Consequently, interactions occurring closer to a PMT would result in higher detected signals compared to those farther away, effectively shifting the energy distribution of asymmetric events toward higher energies. Nonetheless, accurately converting this type of asymmetric event into energy and reproducing the correct spectral shape is challenging, as the differing gains of the PMTs play a significant role, an effect that is not explicitly considered for symmetric events. For this reason, the implementation of a position-dependent light collection model lies beyond the scope of the present thesis but constitutes a promising approach to incorporate this effect into the simulation, with the potential to enhance its agreement with experimental data.

COSINE-100 observes the same trend for single-hit events, attributing it to asymmetric events resulting from the PMT contaminations. Other alternatives has been explored in ANAIS, involving alpha contaminations in the quartz windows and silicon pads used for the optical coupling between crystals and PMTs.


An $\alpha$-decay occurring in a layer near the surface of the NaI(Tl) crystal would produce scintillation light and generate a peak in the energy spectrum, but at an equivalent energy significantly lower than that expected from the crystal, due to the reduced light yield in weak scintillating media (as previously observed in \cite{AMARE20141408}). This scintillation would be highly asymmetric and it could be difficult to distinguish from the NaI scintillation. However, it does not explain the asymmetric contributions identified in the different energy regions. 

To accurately account for spectral shifts arising from light-sharing asymmetry, COSINE-100 applied a correction to the simulated background spectrum from PMT contaminations, although the precise details of this correction are not fully clarified in their publication \cite{cosine2025improved}. However, since neither the origin of this event population is well understood nor an established physical model has been found to correct its effect, in this work the [60–100]~keV region is excluded from the fit. This restriction does not pose a limitation in terms of fixing background components, as PMT-related contributions are constrained by the information from the high energy spectra, while other external components are determined at lower energies. Once the fitting process is completed, the integral of the data in this region will be analyzed and compared to that of the total simulated spectrum. The goal is to determine whether the latter underestimates the observed counts and, if confirmed, to quantify the additional contribution required in terms of total event counts. This would support the hypothesis of $\alpha$-particles originating in the quartz or silicon pads. Such analysis will be conducted in Section \ref{validation}.

\subsection{PMT contribution modelling}\label{PMTsim}

The updated geometry of the PMT was presented in Chapter \ref{Chapter:Geant4}, where the main components of the detection system were discussed, namely the borosilicate, the dynodes, and the photocathode (PK), along with the corresponding updates conducted in this study. Compared to the previous background model, the geometry of the borosilicate has been updated and revised, and the internal dynode system is now explicitly included. Although the latter component was partially incorporated in the geometry described in \cite{phddavid}, its contribution to the background had not been previously investigated.

Regarding the distribution of the contamination, in the previous background model, the contaminations from the PMTs were simulated by uniformly sampling the volume occupied by the entire PMT (including the vacuum), giving equal weight to all parts of the PMT, implying that all parts were equally contaminated. However, in this work, in order to better control the contamination and more accurately reproduce the data, it was decided to simulate the decay of \(^{226}\text{Ra}\) and \(^{232}\text{Th}\) in the different parts of the PMT. Other contaminations of the PMTs were only simulated in the borosilicate since they are not dominant, and no significant differences in the contribution to the ANAIS background are expected.

\begin{figure}[b!]
    \centering
    {\includegraphics[width=0.49\textwidth]{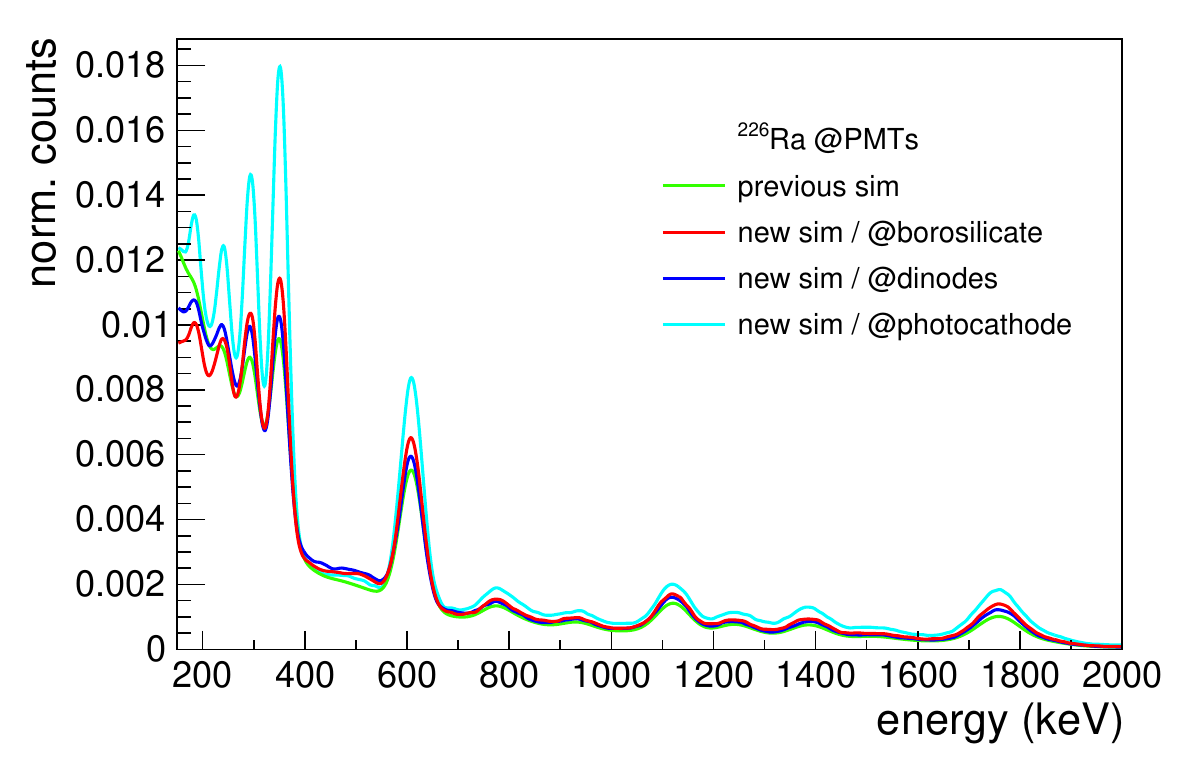}}
    \hfill
    {\includegraphics[width=0.49\textwidth]{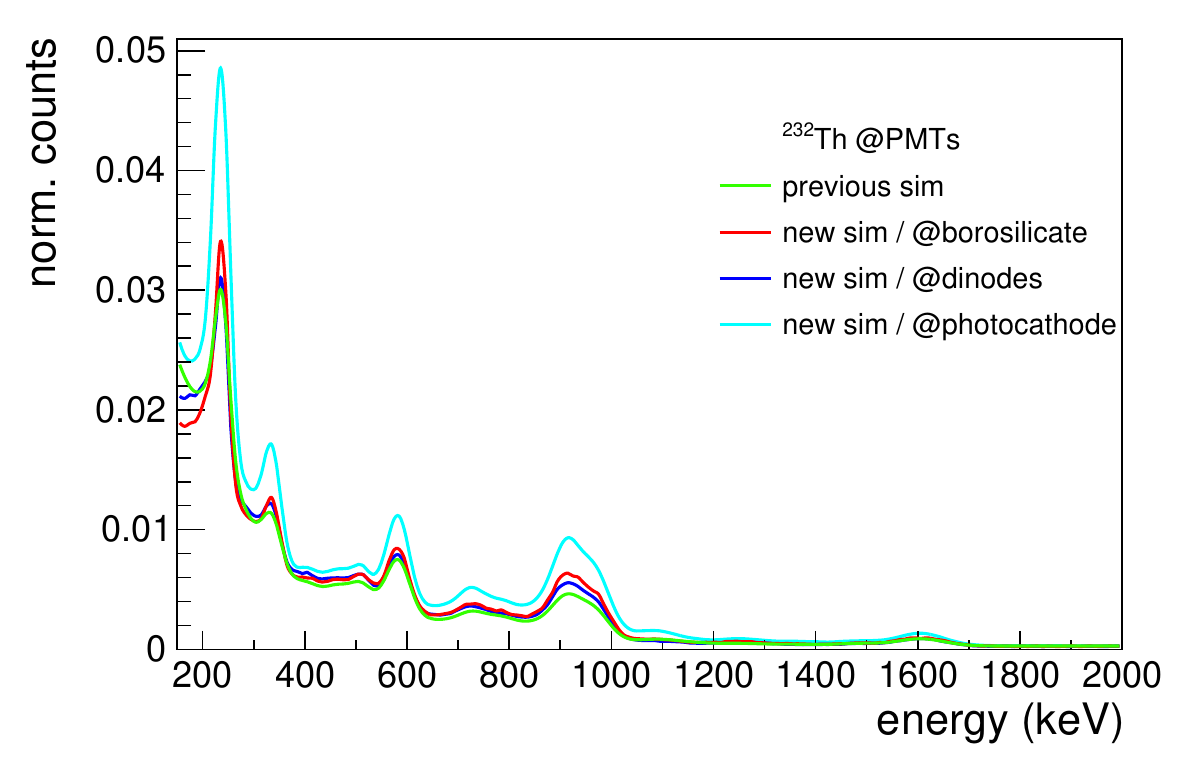}}
    \caption{Comparison of the total simulated spectra for \(^{226}\)Ra (\textbf{left panel}) and \(^{232}\)Th (\textbf{right panel}) in different PMT components: borosilicate (red), dynodes (dark blue), and photocathode (cyan). The simulation from the previous background model, where contamination was uniformly distributed across the entire PMT volume, is shown in green for comparison. All spectra correspond to the same number of initial decays of the parent. }
    \label{comparePMTsimulation}
\end{figure}

Figure \ref{comparePMTsimulation} shows the difference in the deposited energy in the high-energy range of the total spectrum, depending on whether the contamination occurs in the borosilicate, in the dynodes (specifically the first dynode), or in the photocathode, for both \(^{226}\text{Ra}\) and \(^{232}\text{Th}\). In both cases, the contamination incorporated in the previous background model is also shown for comparison.

As it can be observed, there is a significant difference when the contamination occurs in the photocathode compared to the other two components, which exhibit a rather similar behavior. This consistency aligns with the solid angle calculations, which supports that the expected number of events is compatible regardless of whether the contamination is uniformly distributed throughout the volume or concentrated at the center of the PMT, where the first dynode is located. However, as expected, the detection efficiency is significantly higher when the contamination is located in the photocathode, resulting in much larger peak amplitudes and a reduced continuum for both \(^{226}\text{Ra}\) and \(^{232}\text{Th}\).

Given the significant differences in the spectral shape, this work incorporates in the background fitting contamination from two PMT components,  the borosilicate and the photocathode, for \(^{226}\text{Ra}\) and \(^{232}\text{Th}\). Contamination from the dynodes is disregarded due to the similarity of the deposited energy spectrum to that originating from the borosilicate. Specifically, the two free parameters in the fit are the total activity and the fraction of contamination assigned to the borosilicate. 


This modification in the modelling of not only the PMT geometry but also the distribution of PMT contaminations implies that the PMT activity values measured by HPGe detectors in the commissioning of ANAIS-112 (see Table \ref{tablaPMT}) cannot be directly adopted or compared. Previously, they were compatible because the simulation procedure of these contributions was extrapolatable to the activity extraction process from HPGe measurements. In a germanium detector, the required efficiency is usually estimated using Monte Carlo simulations of an extended source. Consequently, the results obtained for the PMTs  with the fitting procedure will not be directly comparable with the activity values measured by the HPGe detectors. Instead, these measurements will be used solely as reference values and initial inputs for the fitting minimization process. As will be shown below, if the fitting procedure assigns a larger fraction of the activity to the photocathode, the derived activities are expected to be lower than the HPGe estimates due to the higher detection efficiency for photocathode contaminants.



Due to the discrepancies observed in the simulation depending on the location of the contamination within the PMT components, a dedicated measurement on a HPGe detector located at the LSC was performed on a spare PMT from the set-up, previously characterized during the comissioning of ANAIS-112. All PMTs acquired for ANAIS-112 came from the same production batch and consequently exhibited similar contamination levels according to earlier measurements \cite{amare2019analysis} (see Table \ref{tablaPMT}), making the results obtained from a single unit representative. To avoid disassembling and destroying the PMT to isolate the borosilicate and photocathode contributions, an alternative method was adopted: comparing the number of counts in the characteristic gamma peaks from the natural decay chains when placing the PMT in two different orientations with respect to the HPGe detector. Specifically, the PMT was measured first with the photocathode facing the detector, and then rotated such that the photocathode was far from it.

This configuration enables a qualitative assessment of the contamination distribution. Since borosilicate contamination is assumed to be uniformly distributed, it contributes similarly in both configurations. Therefore, the ratio of counts observed in each peak can provide insight into the relative contamination of the photocathode. The measurement durations were comparable, with 11.6 days for the configuration with the photocathode away from the HPGe and 13.5~days for the configuration with the photocathode facing the detector.

\begin{figure}[t!]
\begin{center}
\includegraphics[width=0.49\textwidth]{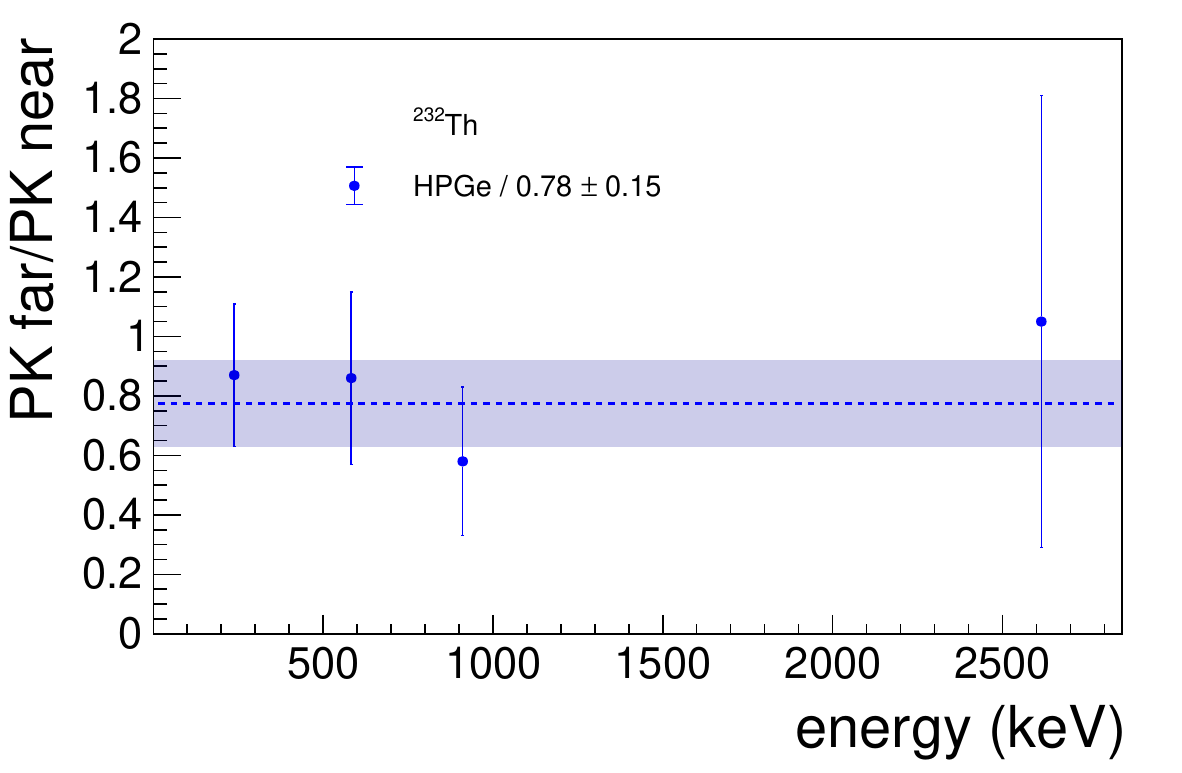}
\includegraphics[width=0.49\textwidth]{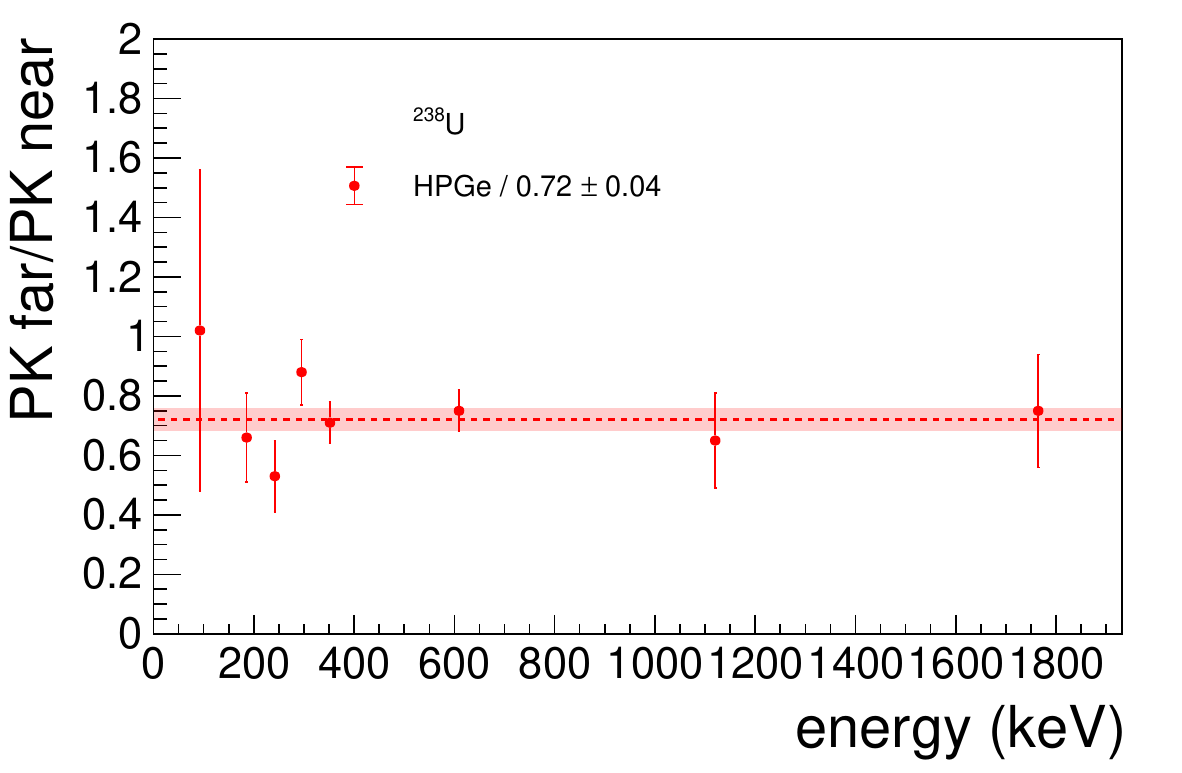}

\caption{\label{hpgepmt} Ratio of the number of counts identified in the peaks corresponding to the main gamma emissions from the natural decay chains, measured at two different PMT positions: photocathode far from the HPGe detector divided by photocathode close to the HPGe detector.
\textbf{Left panel:} \textsuperscript{232}Th decay chain. \textbf{Right panel:} \textsuperscript{238}U decay chain. In each case, the weighted average of the ratios is also shown.}
    \vspace{-0.5cm}

\end{center}
\end{figure}

The ratio of counts for the identified gamma peaks from the \textsuperscript{232}Th (238.6, 583.1, 911.1, and 2614.6 keV) and \textsuperscript{238}U (92.6, 185.7+186.0, 241.9, 295.2, 351.9, 609.3, 1120.3, and 1764.5 keV) decay chains is shown in Figure~\ref{hpgepmt}. The weighted mean of these ratios is 0.78 $\pm$ 0.15 for \textsuperscript{232}Th and 0.73 $\pm$ 0.04 for \textsuperscript{238}U. These values indicate that the front part of the PMT (photocathode) contains in fact more radioactive contamination than the borosilicate, in agreement with the modelling approach adopted in this work, where separate contamination components are introduced for the borosilicate bulk and the photocathode.

It is worth noting that no activity values were calculated from this measurement in order to avoid introducing additional assumptions. Instead, a straightforward comparison of count rates was performed. While the resulting ratios cannot be directly compared to the parameters extracted from the global fit (which estimates the fraction of activity assigned to each PMT component), they qualitatively support the adopted methodology. In particular, these results confirm the presence of a non-negligible contamination component in the frontal region of the PMT, which was neglected in the previous model. Indeed, as will be shown, introducing a contamination in the photocathode improves the overall fit quality significantly, reducing the previously observed excess in the continuum and better describing the identified peaks.

In parallel with these new measurements performed using the HPGe detector, the measurements conducted during the ANAIS-112 commissioning phase \cite{amare2019analysis} for the PMT units were re-evaluated. This review revealed that the uncertainties associated with the \textsuperscript{238}U activities had been underestimated, and that for most PMT units, the activity of this isotope is actually compatible with zero.

\subsection{$^{210}$Pb contribution modelling}\label{210Pbmodelling}

\(^{210}\)Pb is one of the main contaminants in ANAIS-112 and is part of the natural radioactive decay chain of \(^{238}\)U. \(^{238}\)U is naturally present in the Earth's crust due to its long half-life, making it a common component of many materials. One of its decay products is \(^{222}\)Rn, a gaseous isotope that easily escapes from rocks and construction materials, spreading and accumulating in poorly ventilated spaces. Another key isotope in this chain is precisely \(^{210}\)Pb, which has the longest half-life among the progeny of \(^{222}\)Rn, 22.3 years. As a result, any \(^{222}\)Rn contamination in a material will transform into \(^{210}\)Pb contamination within days.

\begin{figure}[b!]
\begin{center}
\includegraphics[width=0.8\textwidth]{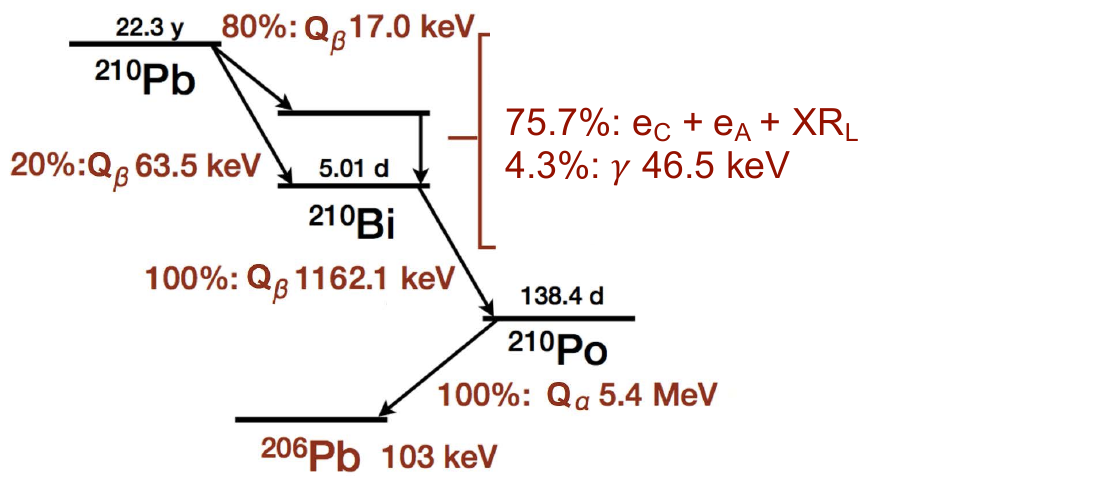}

\caption{\label{esquemades} Decay scheme of \(^{210}\)Pb leading to the stable isotope \(^{206}\)Pb. Figure adapted from \cite{redl2014accurate}.  }

\end{center}
\end{figure}

Figure \ref{esquemades} shows the decay scheme of \(^{210}\)Pb to the stable isotope \(^{206}\)Pb. The \(^{210}\)Pb undergoes \(\beta\)-decay to \(^{210}\)Bi, emitting an electron and an electron antineutrino, sharing an excess energy of 63.5 keV when the decay occurs to the ground state of \(^{210}\)Bi (20\%). When the decay occurs to the first excited state (80\%), 46.5 keV of energy is emitted in the form of gamma radiation or conversion electrons (accompanied by X-rays or Auger electrons), while the electron and the antineutrino share 17 keV. \(^{210}\)Bi also undergoes \(\beta\)-decay to \(^{210}\)Po, but in this case, the available energy is much higher, 1.16 MeV. Finally, \(^{210}\)Po undergoes $\alpha$-decay, emitting an $\alpha$ with an energy of 5.304 MeV. The recoiling $^{206}$Pb nucleus carries the rest of the Q-value available energy, approximately 103 keV, and moves in the opposite direction to the $\alpha$-particle.

The alpha rate following PSA was measured in ANAIS-112, revealing that the observed rate ($\sim$ mBq/kg level) exceeded the values expected under the assumption of secular equilibrium of the \(^{232}\)Th and \(^{238}\)U decay chains  ($\sim$ $\mu$Bq/kg level). Specifically, an analysis of the detector background low energy spectrum indicated an excess of events around 50~keV, as \textsuperscript{210}Pb decay should produce, which pointed to the fact that the lower part of the \(^{238}\)U chain was found out of equilibrium.

Figure \ref{alphadecay} presents the time evolution of the alpha activity for the nine ANAIS-112 detectors over six years of data. The noticeable increase in activity observed in certain detectors at the begining of the data taking is consistent with \(^{210}\)Pb contamination introduced by \(^{222}\)Rn occurring at the final stages of purification, crystal growth or encapsulation of the detectors, followed by the subsequent production of \(^{210}\)Po until equilibrium in the chain was restored. 

Due to its short half-life, \(^{210}\)Bi rapidly reaches secular equilibrium with \(^{210}\)Pb (after approximately 50 days). The secular equilibrium between \(^{210}\)Pb and \(^{210}\)Po, on the other hand, takes longer and then, the alpha activity measured is observed to increase with time, being described by the following equation:

\begin{figure}[b!]
\begin{center}
\includegraphics[width=1.\textwidth]{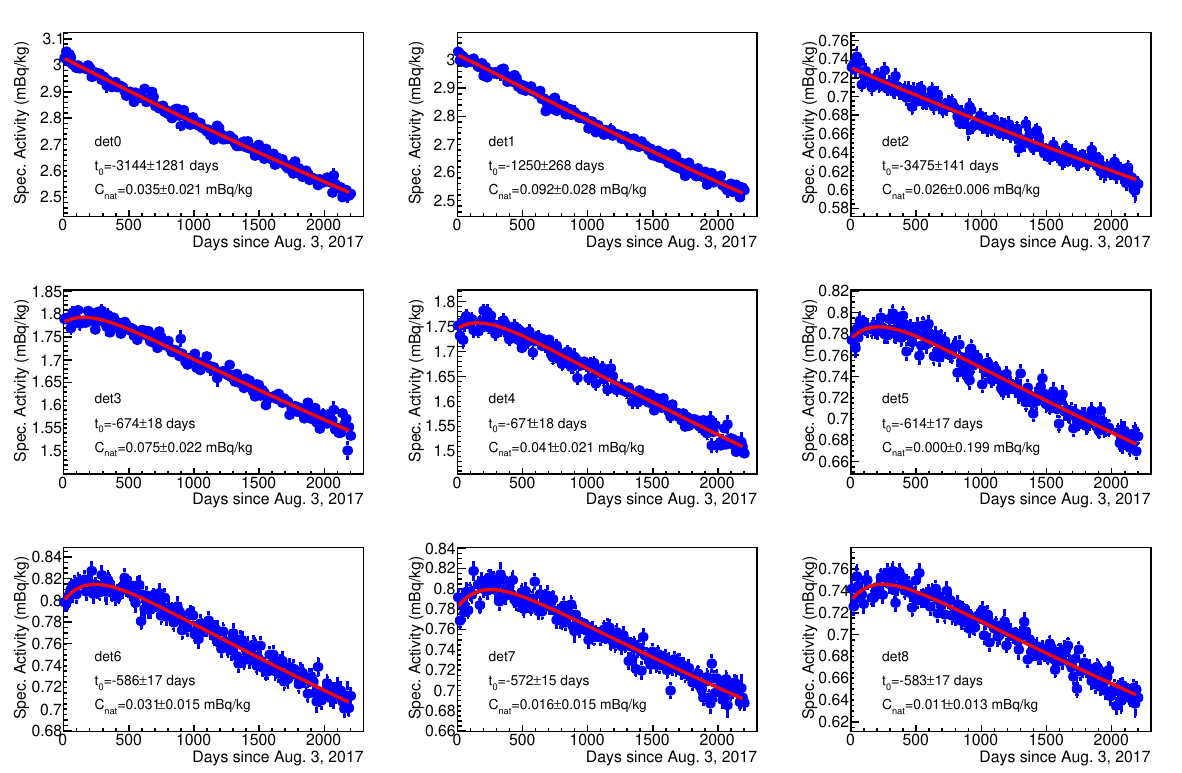}

\caption{\label{alphadecay} Measured alpha specific activity determined via PSA as a function of time for the ANAIS-112 detectors over six years of data taking. The red line represents the fitted model from Equation \ref{eqEq}. Each panel displays the inferred exposure date to \(^{222}\)Rn and the constant background contribution from \(^{238}\)U and \(^{232}\)Th contamination in the crystals.}
\vspace{-0.5cm}
\end{center}
\end{figure}

\begin{equation}
    A(t) = A_0 \frac{\lambda_{\text{Po}}}{\lambda_{\text{Po}} - \lambda_{\text{Pb}}} 
    \left( e^{-\lambda_{\text{Pb}}(t - t_0)} - e^{-\lambda_{\text{Po}}(t - t_0)} \right) 
    + C_{\text{nat}}
    \label{eqEq}
\end{equation}

where \( A(t) \) represents the specific $\alpha$ activity  of the system at a given time \( t \), \( A_0 \) is the initial activity of \(^{210}\)Pb at time \( t_0 \), and \( \lambda_{\text{Pb}} \) and \( \lambda_{\text{Po}} \) are the decay constants of \(^{210}\)Pb and \(^{210}\)Po, respectively. Note that here $t_0$ represents the expected day of exposure to $^{222}$Rn. Additionally, \( C_{\text{nat}} \) represents the constant background contribution of long-lived $^{238}$U and $^{232}$Th contamination of the ANAIS-112 crystals.

\begin{table}[t!]
\centering
       \begin{minipage}{0.45\textwidth}
    
    \centering
    \resizebox{\textwidth}{!}{\Large
    \begin{tabular}{c|c}
        \hline
       detector &  measured $\alpha$-rate (mBq/kg) \\
       \hline
       0 & 3.15 ± 0.10\\
       1 & 3.15 ± 0.10\\
       2 & 0.70 ± 0.10\\
       3 & 1.80 ± 0.10\\
       4 & 1.80 ± 0.10\\
       5 & 0.78 ± 0.01\\
       6 & 0.81 ± 0.01\\
       7 & 0.80 ± 0.01\\
       8 & 0.74 ± 0.01\\
        \hline
    \end{tabular}}

    \end{minipage}
    \caption{Measured $\alpha$-rate for each detector at the start of data taking on August~3,~2017 (corresponding to the origin of the x-axis in Figure \ref{alphadecay}).\label{ritmoalphaexp}}
\end{table}

The time evolution of the $\alpha$ activity for each detector is fitted to Equation~\ref{eqEq}, as shown by the red line in Figure \ref{alphadecay}. As shown in the figure, the fitting procedure yielded satisfactory results for all the detectors. Table \ref{ritmoalphaexp} shows the alpha rate for each detector at the start of data taking on August 3, 2017 (corresponding to the origin of the x-axis in Figure \ref{alphadecay}).


From the fit, the inferred date of exposure to \(^{222}\)Rn was determined and subsequently compared to the known arrival date of the NaI(Tl) detectors at the LSC and their expected construction date. In most cases, the inferred contamination date was consistent with a $^{222}$Rn exposure occurring at the end of the purification process and/or during the storage of the sodium iodide powder, crystal growth, or detector assembly stages. In particular, D0 and D1 were the first to arrive at Canfranc in December~2012, followed by D2 in March 2015, D3 in March 2016, D4 and D5 together in November 2016, and finally D6, D7, and D8 in March 2017. Notably, for the last three detectors, the inferred exposure date to \(^{222}\)Rn was found to be the same in the fit, consistent with their simultaneous fabrication. 

Regarding the results for $t_0$, as shown in Figure \ref{alphadecay}, the detectors that were the first to arrive at the LSC had already reached secular equilibrium by the time data taking began. As a result, $^{210}$Po is already in equilibrium, making the extraction of $t_0$ challenging and subject to large uncertainties, particularly for detector 0. However, it can be concluded that detectors D3, D4, and D5 followed similar growth and contamination protocols, despite their arrival at LSC at different times. Similarly, detectors D6, D7, and D8 must also have been constructed concurrently. These $t_0$ values will be used in Section \ref{mediumenergyfit} when comparing the alpha rate derived from the \textsuperscript{210}Pb activity values obtained from the fit, using Equation \ref{eqEq}, with the experimentally measured alpha rate.

Additionally, the fit allowed for the extraction of contamination levels from the \(^{238}\)U and \(^{232}\)Th decay chains within the crystals. The average activities of \(^{238}\)U and \(^{232}\)Th, determined through PSA methods, were found to be \(1.90 \times 10^{-3}\) mBq/kg and \(7.56 \times 10^{-3}\)~mBq/kg, respectively \cite{amare2019analysis}. Each decay chain contributes eight and six alpha decays, respectively (excluding highly improbable branches), yielding a total average $\alpha$ activity of 0.061 mBq/kg.A preliminary analysis using data from the ANOD DAQ, considering only the alpha activity from $^{232}$Th and $^{238}$U under the assumption of secular equilibrium within their decay chains, indicates a lower average $\alpha$ activity of 0.027 mBq/kg, which is more consistent with the results of the present fit, where the average long-lived $\alpha$ activity is found to be 0.036~mBq/kg.




\subsubsection{Distribution of $^{210}$Pb contamination in ANAIS-112}

As supported by the previous section, bulk contamination in ANAIS-112 crystals is expected from $^{210}$Pb either during the storage of the powder before crystal growth, or due to the exposure to normal air during the growth process. However, surface contamination is also motivated if \(^{222}\)Rn was deposited after the crystal growth. Additionally, \(^{210}\)Pb contamination could have also occurred in the teflon diffuser surrounding the crystals and while the polishing of the laterall faces of the crystals. Nevertheless, the exact distribution of \(^{210}\)Pb contamination within the ANAIS modules remains unknown. As becomes clear from the measurements, not all detectors could have been exposed in the same manner, and therefore, they may exhibit different contamination levels.




Moreover, as will be shown in Section \ref{alphasec}, the $\alpha$-spectrum of the ANAIS-112 detectors show a double peak structure. This observation motivated the consideration of a bulk and a surface $^{210}$Pb contamination of the crystals. In the previous background model, surface contamination of \(^{210}\)Pb from a fixed depth of 100 \(\mu\)m or 30 \(\mu\)m, depending on the detector, was considered \cite{amare2019analysis}. The fraction and depth of the \(^{210}\)Pb surface emission,  summarized in Table \ref{tablaPlomo}, were fixed for each crystal to accurately reproduce the measured low-energy data and the alpha energy distribution. In addition, \(^{210}\)Pb activity in the teflon diffuser surrounding the crystals was considered for some modules (D3 and D4). A value of 3~mBq/detector was chosen to reproduce the structure observed in the low-energy background at 12 keV in those detectors.


In this study, regarding the surface contamination, exponentially decaying density profiles have been considered instead to account for the diffusion of the gas at different depths. The choice of an exponential profile is based on the general solution of a diffusion equation, as in \cite{ribeiro1996general} applied to diffusion of a radioactive isotope in a solid medium, and is preferred over models with fixed contamination depths. Such profiles have already been considered by other DM search experiments such as the COSINE-100 collaboration, which reported exponential profiles for surface \(^{210}\)Pb contamination with mean depths of 0.107 $\mu$m and 1.39~$\mu$m~\cite{yu2021depth}. 

\begin{figure}[b!]
    \centering
    {\includegraphics[width=0.49\textwidth]{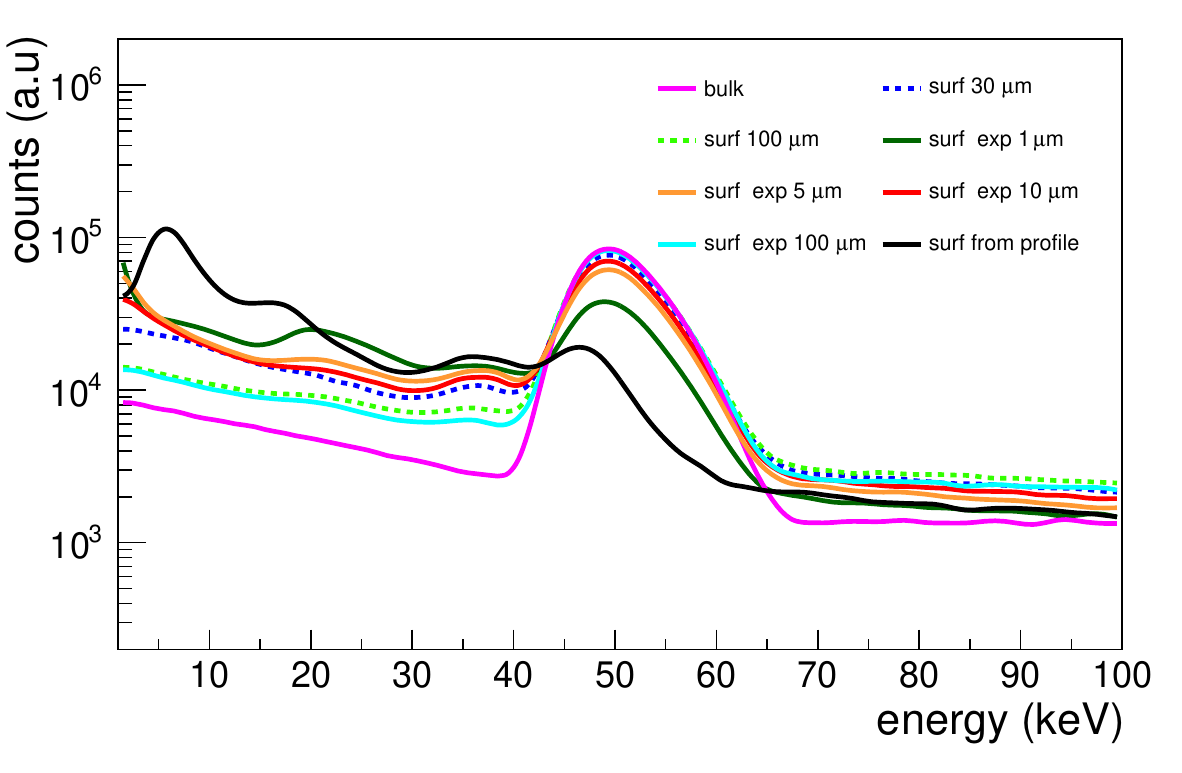}}
    \hfill
    {\includegraphics[width=0.49\textwidth]{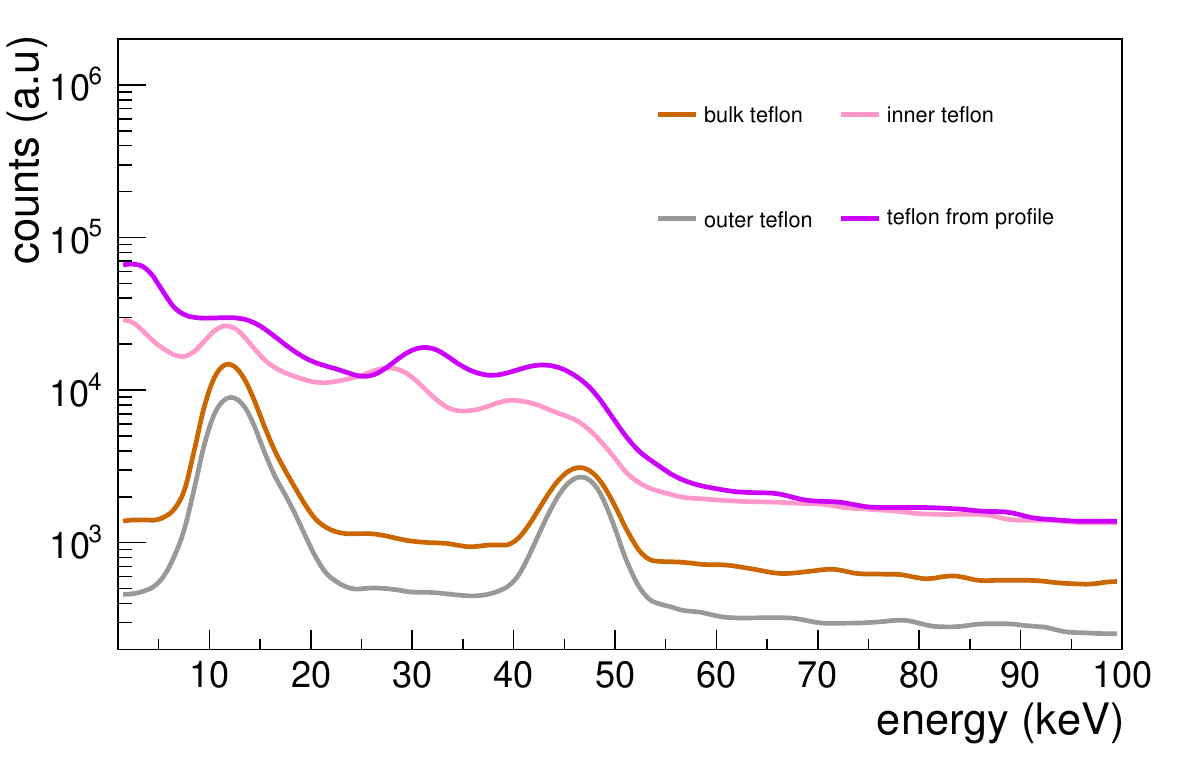}}
    \caption{Low-energy simulated spectra assuming \(^{210}\)Pb is distributed either in the bulk or at the surface. \textbf{Left panel:} Contamination in the NaI crystal, considering different mean exponential depths (1, 5, 10 and 100 $\mu$m). \textbf{Right panel:} Contamination in the teflon diffuser film surrounding the crystal, either in the bulk or on its inner or outer face considering an exponential mean depth of the contamination
layer of 2~$\mu$m. In the left panel, the surface distributions considered in the former background model are also shown in dashed lines, together with the energy spectra derived from a previous ANAIS-112 study (black), where the deposited energy spectra were obtained from the \(^{210}\)Pb implantation profiles in NaI and teflon using Geant4 \cite{tfgcarmen}. All spectra correspond to the same number of initial decays of
the parent.}
    \label{compare210PbLE}
\end{figure}

The left panel of Figure \ref{compare210PbLE} presents a comparison of the low-energy simulated spectra under the assumptions of \(^{210}\)Pb being distributed either in the bulk or at the surface of the crystal. In this work, several exponential mean depths of the contamination layer (1, 5, 10, and 100 $\mu$m) are considered, with the contamination density profile assumed to decrease exponentially with depth from the surface to the bulk of the crystal. In this work, unlike the approach adopted by COSINE-100, no actual fit of the mean deposition depth has been performed. Instead, contamination depths of 1, 10, and 100~$\mu$m have been considered as benchmark scenarios for a qualitative assessment of their impact on the background estimation. The figure also displays in dashed lines the energy spectrum assumed in the previous background model, with constant depths of 30 and 100 $\mu$m.

Additionally, the figure shows in black the energy spectra derived from the \(^{210}\)Pb implantation profile estimated using Geant4. This study was part of an undergraduate thesis co-supervised by the doctoral student, in which the penetration depth of \(^{210}\)Pb ions in NaI and teflon targets was compared using SRIM and Geant4 \cite{tfgcarmen}. SRIM is a widely used tool for simulating ion stopping in matter, particularly in ion implantation, sputtering, or transmission \cite{ziegler1985stopping}. Significant discrepancies were observed in the estimates for penetration depths and stopping powers between the two methods, as SRIM and Geant4 differ in their approaches to simulating ion transport. 

For instance, the mean penetration depth for \(^{210}\)Pb ions in NaI obtained with Geant4 was 67.8 Å, whereas SRIM yielded 567.5 Å, a discrepancy of nearly an order of magnitude. Similar discrepancies have been reported in other studies \cite{gruner2024enhancing}, although their origin remains unknown. Furthermore, it is observed that the low-energy spectral features obtained from the penetration profile analysis (see left panel of Figure \ref{compare210PbLE}) are not compatible with the ANAIS-112 data. Consequently, in the present work, a background modelling approach more closely aligned with the previous model is adopted, assuming exponential profiles with mean penetration depths of 1, 10, and 100 $\mu$m.



In the left panel of Figure \ref{compare210PbLE}, it can be seen that for more superficial distributions, the escape effects clearly dominate, with X-rays and gamma rays escaping most notably, and some energy from electrons, whether beta, Auger, or conversion, possibly escaping as well. Additionally, it is observed that the deposited energy spectrum derived from a 100 $\mu$m exponential depth profile is very similar to that from a constant 100 $\mu$m profile, as used in the previous background model.

The right panel of Figure \ref{compare210PbLE} shows the simulated energy spectra for energy deposited in the NaI crystal for \(^{210}\)Pb contamination on the teflon coating of the crystals.  X-rays, gamma rays, and higher-energy electrons emerging from the teflon are observed. However, they are detected with degraded energy due to partial energy loss within the teflon. According to the literature, the diffusion length of \(^{210}\)Pb in teflon is on the order of 100 nm \cite{pattavina2011radon}. In ANAIS-112, the thickness of the teflon layer is 0.5 mm, so bulk contamination of teflon, given this diffusion value, is not plausible. This was the \textsuperscript{210}Pb contamination in the teflon coating assumed in the previous background model. 

In this thesis, different mean penetration depths have been evaluated. In the previous study based on \textsuperscript{210}Pb implantation profiles using Geant4 simulations \cite{tfgcarmen}, a mean penetration depth of 190.5 Å was estimated for a teflon target. The corresponding energy spectrum is shown in the right panel of Figure \ref{compare210PbLE}. As can be observed, when the contamination depth is significantly below the micron scale, the low-energy region of the spectrum is largely overestimated, a feature not observed in the ANAIS-112 data. Moreover, as previously mentioned, the teflon film could have been exposed to \textsuperscript{222}Rn in Alpha Spectra facilities, either in one or the two faces. Therefore, in this work it is assumed that the \textsuperscript{210}Pb contamination is located either on the inner or the outer surface of the teflon layer, both modeled with an exponential depth profile with a mean penetration depth of 2 $\mu$m. For the background fitting procedure, both contaminations are considered with equal weight, directly summing the spectra, since there is no evidence suggesting that contamination occurred preferentially on one face. As derived from Figure \ref{compare210PbLE}, the energy spectrum for contamination in the bulk of teflon or its outer face follows a similar trend, showing the 46.5 keV gamma transition without the accompanying $\beta$-spectrum and X-rays around 12 keV.

\begin{figure}[t!]
    \centering
    {\includegraphics[width=0.49\textwidth]{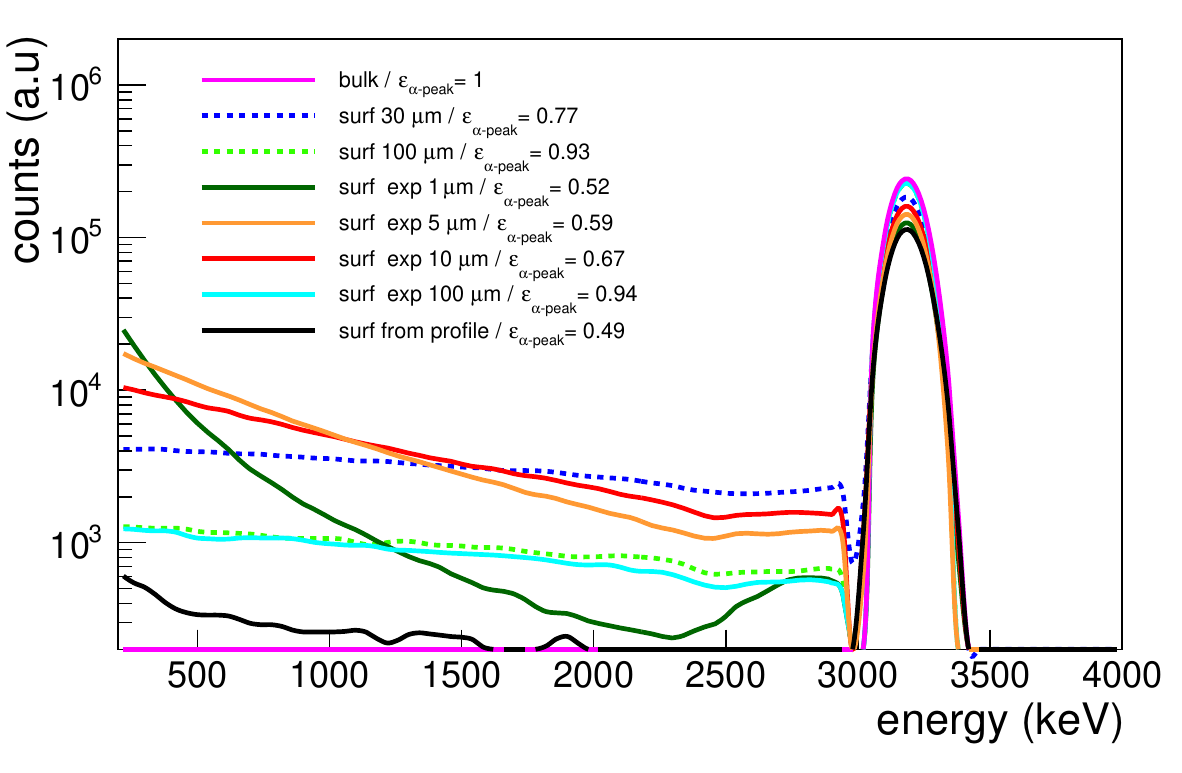}}
    \hfill
    {\includegraphics[width=0.49\textwidth]{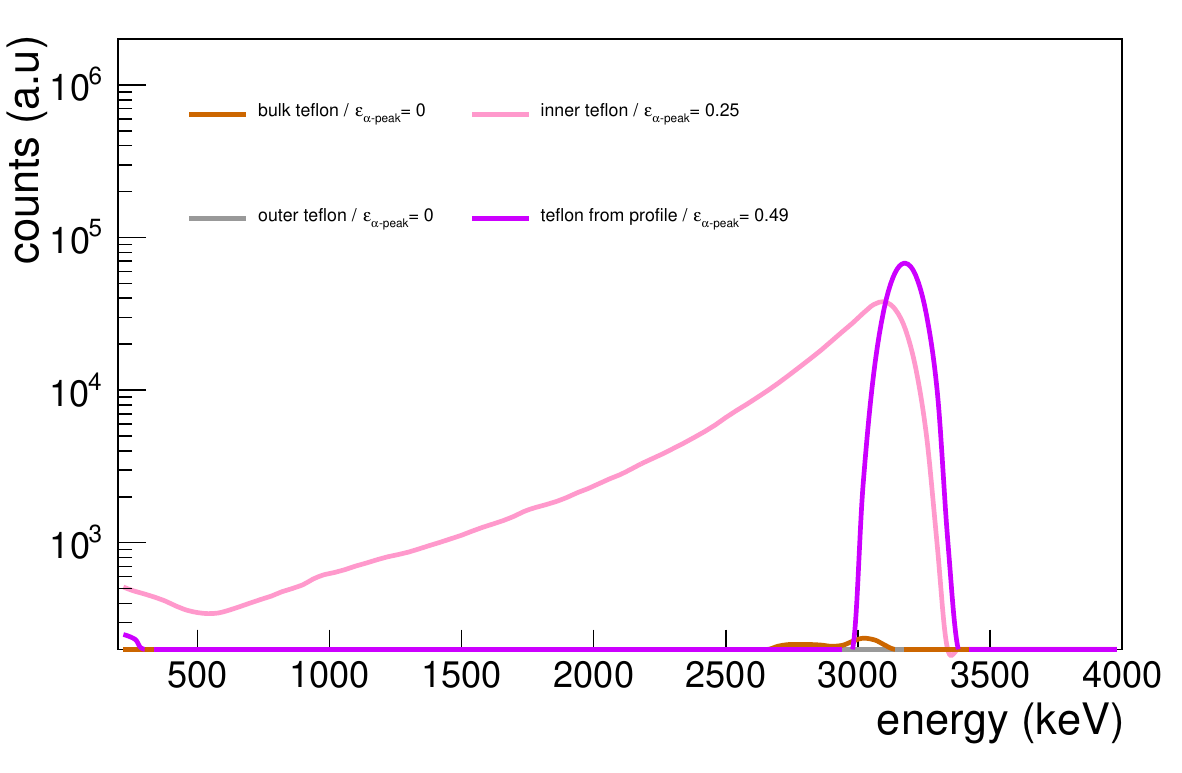}}
    \caption{High-energy simulated spectra assuming \(^{210}\)Pb is distributed either in the bulk or at the surface. Only $\alpha$-particles are considered. \textbf{Left panel:} Contamination in the NaI crystal, considering different mean exponential depths (1, 5, 10 and 100~$\mu$m). \textbf{Right panel:} Contamination in the teflon diffuser film surrounding the crystal, either in the bulk or on its inner or outer face considering an exponential mean depth of the contamination
layer of 2~$\mu$m. In the left panel, the distributions considered in the former background model are also shown in a dashed line, together with the energy spectra derived from a previous ANAIS-112 study (black), where the deposited energy spectra were obtained from the \(^{210}\)Pb implantation profiles in NaI and teflon using Geant4 \cite{tfgcarmen}. The $\alpha$-peak efficiency, $\varepsilon_{\alpha\text{-peak}}$, is displayed in the panels (see text for further details). All spectra correspond to the same number of initial decays of
the parent.}
    \label{compare210PbHE}
\end{figure}

\(^{210}\)Po is an \(\alpha\)-emitter and decays to stable \(^{206}\)Pb via the emission of an \(\alpha\)-particle with an energy of 5.304 MeV. Therefore, in the higher energy region, a peak from this decay is observed. Figure \ref{compare210PbHE} compares the $\alpha$-simulated spectra under the assumptions of \(^{210}\)Pb being distributed either in the bulk or at the surface of the crystal (left panel) or in the teflon diffuser film (right panel), analogous to Figure \ref{compare210PbLE}. In this figure, only \(\alpha\)-particles are considered, and a constant QF$_\alpha$ of 0.6 has been adopted for the $\alpha$-particles in the construction of the electron-equivalent energy spectra \cite{amare2019analysis}.

It can be observed in both panels of Figure \ref{compare210PbHE} that the area of the $\alpha$-peak differs between bulk and surface contaminations, while no significant energy shift is present. Additionally, the continuous component to the left of the $\alpha$-peak differs between the two cases. The $\alpha$-peak efficiency, $\varepsilon_{\alpha\text{-peak}}$, is defined as the ratio between the total number of counts under the $\alpha$-peak ([2900–4000] keV) and the number of initial decays of the parent. The value of $\varepsilon_{\alpha\text{-peak}}$ is shown in the panels of Figure~\ref{compare210PbHE} for each $^{210}$Pb contamination distribution. As observed, $\varepsilon_{\alpha\text{-peak}}$ reaches 100\% for a bulk contamination, since the entire $\alpha$-signal is contained within the crystal volume, and decreases progressively as the contamination becomes more superficial. When the contamination occurs in the bulk of the teflon coating or on its outer face, no \(\alpha\)-peak is observed, as $\alpha$ particles are absorbed within the teflon. This is because the mean free path of an \(\alpha\)-particle in teflon is approximately 23 $\mu$m, while the thickness of the teflon film in ANAIS-112 is 0.5 mm.


\subsubsection{$^{210}$Bi $\beta$-spectrum shape}\label{betashape}

As observed in Figure \ref{HEandMEold}, the region between [600-1000] keV has been systematically overestimated by the background model. This behavior was present in all detectors, being more pronounced in those with higher levels of \(^{210}\)Pb contamination. 

\begin{figure}[b!]
\begin{center}
\includegraphics[width=0.7\textwidth]{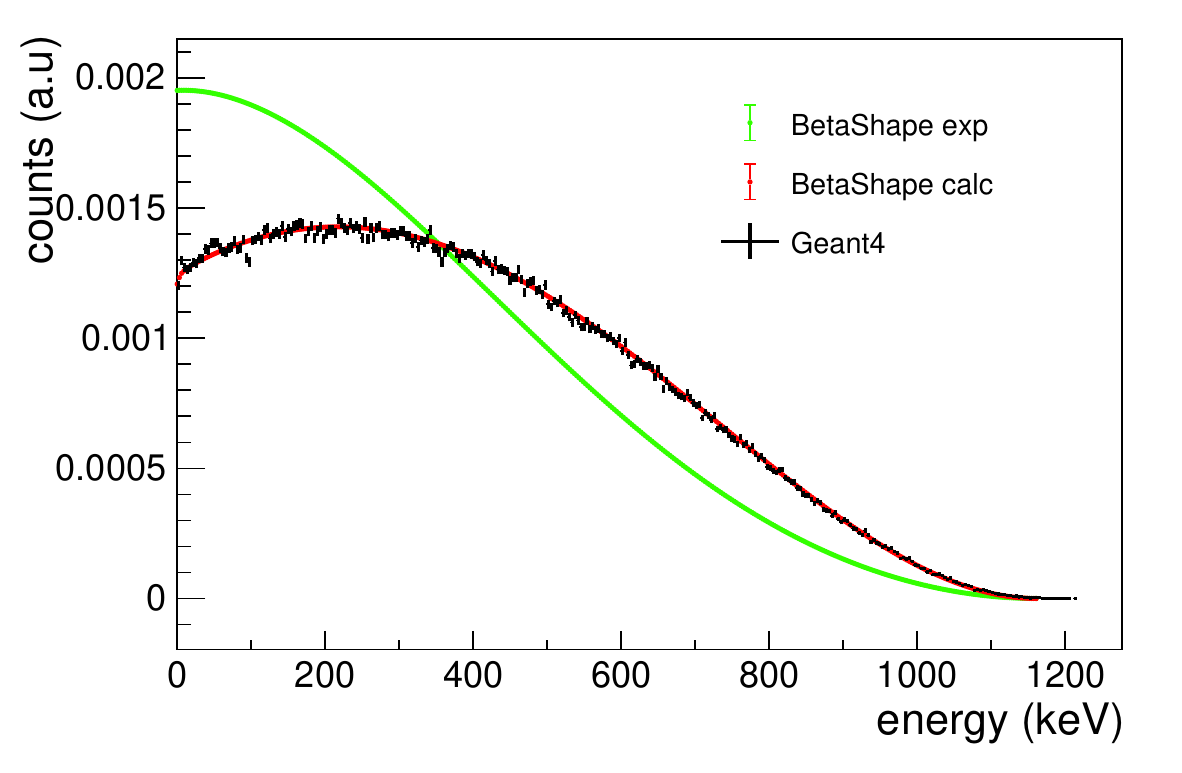}

\caption{\label{betashapebulk} Spectrum of the \(\beta\)-decay of \(^{210}\)Bi, a first-forbidden non-unique transition. The figure presents the experimental spectrum (green) and an analytically calculated spectrum (red), both from the BetaShape code, along with the spectrum used by Geant4 (black) for comparison, which adopts the analytical calculations for modelling this \(\beta\)-decay. The significant discrepancy in terms of spectral shape between the experimental and analytical estimations is evident.}
\end{center}
\end{figure}

In this energy range, the dominant contributions to the single-hit spectrum come primarily from PMT contamination, as well as from the \(\beta\)$^-$ decays of \(^{40}\)K and \(^{210}\)Bi, as will be seen in the next section (Figure \ref{HEdatasim}). The PMTs, specifically \(^{226}\)Ra, exhibited good agreement with the data in other regions of the spectrum, with the nearby 609 keV peak, for instance, being well reproduced. Consequently, it was hypothesized that the observed discrepancy between data and simulation could be attributed to an incorrect modelling of the \(\beta\)$^-$ decay spectrum of \(^{40}\)K and/or \(^{210}\)Pb. Such a kind of solution had already been proposed to address similar inconsistencies in the decays of certain isotopes in the first steps of the ANAIS background modelling. 

To this end, the BetaShape database was consulted. BetaShape is an improved code developed to enhance nuclear data related to \(\beta\)$^-$ decays which has become the standard reference code for modelling \(\beta\)$^-$ spectra \cite{mougeot2017betashape}. BetaShape provides two sets of results, one derived from analytical calculations and another based on experimental measurements.  

The \(\beta^-\) spectrum of \(^{40}\)K was analyzed using three different estimations: the spectrum used by Geant4, the calculated spectrum from BetaShape, and the experimental spectrum provided by BetaShape. All three predictions were found to be in good agreement. As reported in the BetaShape database, the discrepancy between the experimental and calculated spectra is only 1.21\%, indicating that this transition is well understood.

However, when the same comparison was performed for the \(\beta^-\) decay of \(^{210}\)Bi, an isotope undergoing a first-forbidden non-unique \(\beta^-\) decay, significant discrepancies were observed, as shown in Figure \ref{betashapebulk}. As can be derived from the figure, a clear mismatch is evident between the spectral shapes of the experimental and analytically calculated $\beta^-$~spectra, with Geant4 v10.7.0 following the analytical model. The analytically computed spectrum and the experimental measurement differ by 60.49\% in the region of interest for ANAIS-112, from 600 keV to the end-point, demonstrating that this transition is poorly characterized \cite{carles2005beta}. Notably, the experimental $\beta^-$ spectrum seems to align more closely with ANAIS-112 measured data, contributing significantly less in the [600–1000] keV energy region which has been always overestimated by the simulation. Consequently, in this work, the experimental \(\beta^-\) spectrum of $^{210}$Bi has been incorporated instead of the analytically calculated one.


\begin{figure}[b!]
\begin{center}
\includegraphics[width=1.\textwidth]{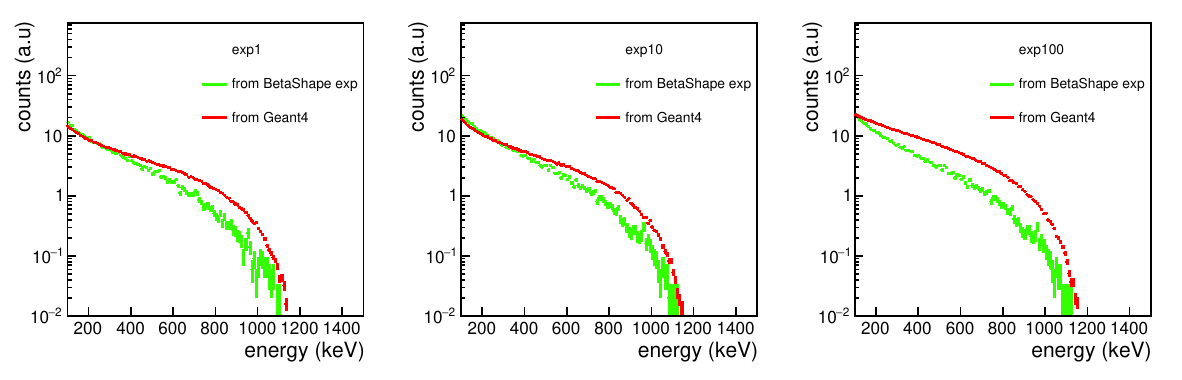}

\caption{\label{betashapesurf} Spectrum of the \(\beta\)$^-$ decay of \(^{210}\)Bi for a surface contamination in the NaI crystal, considering exponential mean depths of 1 $\mu$m (left panel), 10 $\mu$m (center panel), and 100 $\mu$m (right panel). The figure compares the \(\beta\)-spectrum obtained from Geant4 (red) with an alternative approach in which electrons are simulated at the surface with an energy directly sampled from the experimentally measured $\beta^-$ spectrum taken from BetaShape. At higher energies, the spectrum obtained using the experimental $\beta^-$ shape shows a significantly lower contribution.}
\end{center}
\end{figure}

This replacement is pretty straightforward for the simulation of \(^{210}\)Pb in the bulk of the crystal, as in this region there are no gamma rays or any other contributions except for the \(\beta^-\) decay of \(^{210}\)Bi. However, replacing the \(\beta\)$^-$spectrum for the surface contamination is not as direct and requires an ad-hoc simulation, as the energy deposited in the crystal does not directly correspond to the experimental \(\beta^-\) shape. 

To achieve this, electrons were simulated at the surface using exponentially decaying density profiles, directly sampling the experimental \(\beta\)$^-$ spectrum to obtain the particle energy, without using the isotope decay process. The results of the simulation are shown in Figure \ref{betashapesurf} for the three exponential profiles considered in this work. As seen, the shape of the \(\beta\)$^-$ spectrum differs from the one obtained with Geant4, contributing significantly less at high energies. The same procedure has been applied in this work to obtain the \(\beta\)$^-$ spectrum for \(^{210}\)Bi contamination in the external or internal layers of the teflon coating, where the behavior is analogous to that observed in the case of surface contamination in the NaI crystal.

In this way, during the fitting process carried out in this chapter, the experimental \(\beta\)-spectrum of \(^{210}\)Bi will be used, achieving, as will be shown later, a much better agreement between data and simulation in the [600-1000] keV region by incorporating this correction.

\subsection{Alpha backgrounds in ANAIS-112}\label{alphasec}

At the \(\alpha\)-region of the energy spectra measured in ANAIS-112, the prominent peak due to the $\alpha$-decay of \(^{210}\)Po does not exhibit the pure structure expected from either a bulk or a surface \(^{210}\)Pb contamination in the crystal, as it is shown in the simulation (see Figure~\ref{compare210PbHE}). 

\begin{figure}[b!]
\begin{center}
\includegraphics[width=1.\textwidth]{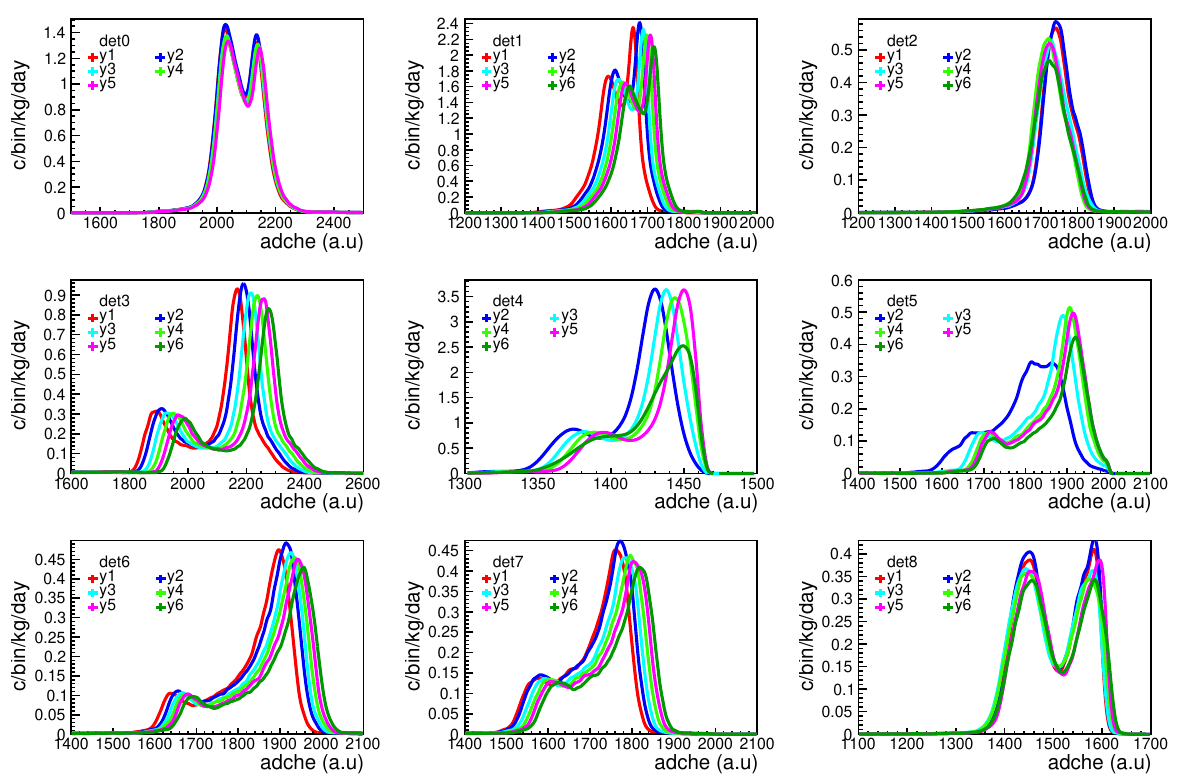}

\caption{\label{alphaspectrum} Alpha spectrum for the ANAIS-112 detectors over six years of data-taking. Some spectra, such as the one corresponding to year 6 of detector 0, are not plotted due to gain issues, which do not affect the annual modulation analysis. The x-axis is not calibrated in electron equivalent energy due to signal saturation in ANAIS-112, but is instead shown in arbitrary ADC units of the high-energy line. The spectra are also not corrected for gain drifts. The y-axis is scaled per bin, which implies that the bin size must be taken into account when comparing the rates of different detectors. }
\end{center}
\vspace{-0.5cm}
\end{figure}

Figure \ref{alphaspectrum} shows the measured \(\alpha\)-spectra of the ANAIS-112 detectors over six years of data-taking. As derived from the figure, most detectors clearly show two distinct peaks, with the exception of D2, where only one peak is observed, although it has a bump on the right side. However, the area of each peak and their relative separation varies from detector to detector without an underlying reason, at least not one that has been fully understood by ANAIS.

The two peaks must correspond to $^{210}$Po, based on the observed lifetime. Regarding the origin of the two peaks, one hypothesis considered by ANAIS is that the lower-energy peak arises from surface contamination, while the higher-energy peak originates from bulk contamination. In this thesis, the background fitting results will be used to evaluate the hypothesis that the lower-energy peak could be explained by surface deposition (see Section~\ref{mediumenergyfit}). Moreover, the presence of a scintillation dead layer or a low-light collection region in the surface of the detectors was also proposed as a possible explanation. In fact, the undergraduate thesis mentioned earlier explored for the first time in ANAIS-112 simulations the implementation of a partial light collection model \cite{tfgcarmen}. This study found that, in the high-energy region, the \(\alpha\)-peak for surface events would shift to lower energies, on the order of hundreds of keV, compared to the bulk distribution, roughly reproducing the double-peak structure observed in the experimental data. 

On the other hand, it is worth noting that a significant number of degraded $\alpha$ events is not observed, as the peaks do not exhibit a pronounced low-energy tail. However, it should also be noted that the $\alpha$/$\beta^-$/$\gamma$ discrimination power in ANAIS is not useful below $\sim$700 keV. The absence of a substantial population of degraded $\alpha$-events does not support the hypothesis that \textsuperscript{210}Pb contamination is found at lower depths. On the other hand, this hypothesis cannot be completely ruled out. As illustrated in Figure \ref{compare210PbHE}, if the \textsuperscript{210}Pb contamination of the crystal is very superficial, if the $\alpha$-particle is emitted towards the detector, its full energy is recorded. Conversely, if the $\alpha$-particle escapes in the opposite direction, it deposits only a small fraction of its energy. As a result, regions near the full-energy $\alpha$-peak receive very little contribution.





However, COSINE-100 explored both hypotheses regarding the origin of the observed $\alpha$-peak structure. Nevertheless, both scenarios failed to account for the large energy separation between the $\alpha$ peak-like structures observed in their data. Instead, motivated by a study conducted by the COSINUS experiment \cite{raghunath2023quenching}, which provides evidence for a QF dependence on the thallium doping concentration, COSINE-100 proposed that the double-peak structure could be explained by a spatial variation in Tl concentration, resulting in two distinct QF values for $\alpha$ particles \cite{adhikari2024alpha}. However, this interpretation raises several questions. If the QF for $\alpha$ particles varies with the local Tl concentration, a similar effect might be expected for other types of interactions, such as neutron-induced NRs and $\beta^-$/$\gamma$ events. Given that the Tl concentration influences the overall scintillation yield, a spatial dependence in its distribution would likely affect the light output for all interaction types, although not necessarily to the same extent.

\section{The fitting procedure}\label{fittingprocedure}
This section summarizes the data and background source selection performed in preparation for the fit (see Section~\ref{simulselection}). The sequence in which the fit is executed will also be outlined prior to introducing the background fitting function and the expression used to derive the initial activity of each isotope from the fit results (see Section \ref{function}).

\subsection{Data and background source selection}\label{simulselection}
To avoid over-parameterization of the model, only a selected subset of background sources is considered. This is due to the inherent limitations of the fitter, which may struggle to uniquely distinguish between simulated spectra with similar spectral shapes. Consequently, a prior selection of both data and simulated contributions, as well as careful consideration of the fitting strategy, including the choice of energy ranges and constraints applied to the model, are required.

The background model fit is performed over separate energy ranges to isolate as much as possible the contribution of each background source. Alternative approaches would be to fit the entire energy range at once or to use separate fit ranges in a simultaneous fit. However, in the first case, the complexity of the fit led to an incorrect modelling of the spectrum, as the fitter adjusted the contributions of certain sources in a compensatory manner, artificially increasing or decreasing them. Additionally, minor mismatches between spectral regions requiring different calibration procedures, although expected and not significant, introduced energy discontinuities that further complicated a global fit. Regarding the second option, the fitting was already considerably complex, involving 93 parameters (as detailed in Table \ref{componentesdelfit}), and extending it to a simultaneous fit over multiple energy ranges would further increase its complexity. As a result, the final approach adopted for the background fitting involved dividing the fit into three energy ranges, following the calibration energy ranges defined in ANAIS-112: low (<20~keV), medium ([20-150] keV, and high energy ([150-1600] keV). It is worth highlighting that the entire procedure is carried out on a detector-by-detector basis.

\begin{sidewaysfigure}
    \centering
    \includegraphics[width=\textheight]{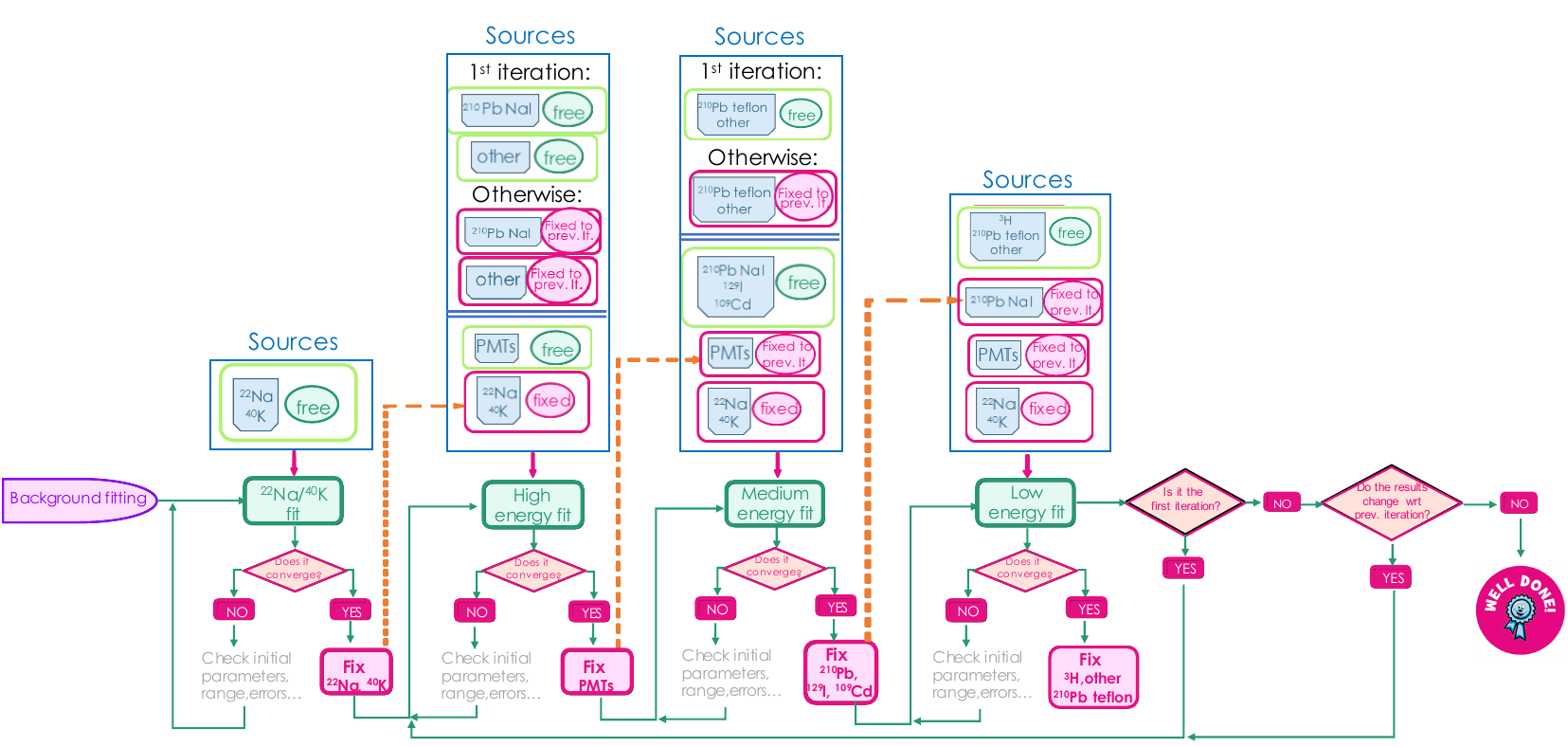}
    \caption{Flowchart illustrating the strategy followed in the background fitting procedure. Each step displays the considered background sources and the parameter status at the first iteration and in subsequent iterations (either free, fixed, or fixed to the value from the previous iteration). In the initial iteration under the desired conditions, the $^{22}$Na/$^{40}$K coincidence population is fitted first and subsequently fixed for the remainder of the procedure. The high-energy, medium-energy, and low-energy fits are then performed sequentially, with free parameters initialized to the values from the previous background model. After this initial iteration, all fitted parameters are fixed to the values obtained in the previous iteration, and the sequence of high-, medium-, and low-energy fits is repeated iteratively, updating the fit values at each step until convergence is reached, that is, when the parameters do not change across successive iterations.}
    \label{esquemaflow}
\end{sidewaysfigure}

Figure \ref{esquemaflow} presents the flowchart of the strategy followed in the background fitting procedure. As shown in the figure, if it is the first iteration under the selected conditions, such as the energy range or the set of included PDFs, the procedure begins with fitting the $^{22}$Na/$^{40}$K coincidence population, after which these contributions are fixed for subsequent steps. Next, in the first iteration, the remaining parameters are treated as free parameters, initialized to the values obtained from the previous background model. 

The high-energy fit is performed first, and the PMT-related parameters obtained are then fixed for the medium-energy fit, which includes $^{210}$Pb in the crystal bulk, $^{129}$I, and $^{109}$Cd as free parameters. Subsequently, they are fixed and used in the low-energy fit, which incorporates $^{210}$Pb in the teflon, $^{3}$H, and a generic external component encompassing all sources external to the crystals as free parameters. After this initial iteration, all parameters, except those intended to be adjusted within each energy region, are fixed to the values obtained in the previous iteration and the full sequence—high-, medium-, and low-energy fits—is repeated iteratively, updating parameter values at each step. The results are checked for consistency, and if they are not compatible, the fitting machinery is repeated until the fit results do not change across successive iterations.

 Table \ref{componentesdelfit} summarizes the contaminations considered in each volume across the different energy ranges, which include the contributions that dominate in each background component and energy region, together with their status in the fit, specifying whether the parameter is fixed or floating. Specifically, the background fit incorporates the following contributions:


\begin{enumerate}
    \item PMTs: \(^{40}\text{K}\), \(^{226}\text{Ra}\) and \(^{232}\text{Th}\). 
    
    The primary signature of \(^{40}\text{K}\) in the PMTs, a high-energy gamma emission at 1461~keV, lies outside the high-energy fitting range ([200-700] keV, see Table~\ref{componentesdelfit}) and its activity is therefore taken from the previous background model. However, its value will be adjusted if necessary during the post-fitting stage for each detector. Concerning the contamination from $^{226}\text{Ra}$ and $^{232}\text{Th}$, as previously mentioned, a uniform contamination is considered in the borosilicate, together with an additional contamination in the photocathode.

    Regarding \(^{238}\text{U}\), its activity was initially set as another free parameter, which also described the contribution of \(^{235}\text{U}\) (according to their natural activity ratio and half-lives). However, it was observed that during the fitting process, \(^{238}\text{U}\) activities were consistently driven to zero, reaching the limit. Therefore, it was decided to disregard its activity. This agrees with the re-evaluation of HPGe measurements of the PMT units during ANAIS-112 commissioning \cite{amare2019analysis}, indicating that most PMT units have \textsuperscript{238}U activity compatible with zero.

    \item Crystal: \(^{40}\text{K}\) and \(^{210}\text{Pb}\), both uniformly distributed within the bulk of the crystal, along with \(^{210}\text{Pb}\) distributed on its surface with different depth profiles, and \(^{210}\text{Pb}\) distributed on the inner and outer faces of the teflon coating (see Section \ref{210Pbmodelling}).

    \item Cosmogenic: \(^{22}\text{Na}\), \(^{129}\text{I}\), \(^{109}\text{Cd}\) and $^3$H from cosmogenic activation, uniformly distributed within the bulk of the crystal.
    
    \item Other external sources: To avoid degeneracies, it is unfeasible to include each external component individually in the fit and ensure its convergence. This category encompasses a wide range of contributions (see Table \ref{tablaOthers}), including the copper encapsulation, quartz windows, silicone pads, archaeological lead, and the inner volume atmosphere of the detector cavity. For each source, natural decay chains and specific isotopes, such as \(^{60}\text{Co}\) in copper, were simulated. No direct measurements were derived for these components, but upper limits on their activities.

    In this study, instead of fitting each contribution separately, all external sources are combined into a single histogram and a single free parameter representing  this component is included in the fit. After the fitting, the upper limit on the combined contribution is revisited to determine whether a more stringent constraint is achieved, but for the total contribution of these external sources. This method of estimating the contribution from this contamination is not optimal, as it constrains the possibility that some external components may dominate while others do not, by assigning equal weight to all of them. Nevertheless, it is the approach adopted in this work due to the large number of contributing sources involved.

\end{enumerate}

\begin{table}[t!]
\centering
\renewcommand{\arraystretch}{1.1} 
 \resizebox{\textwidth}{!}{\Large
\begin{tabular}{ccccccc}
\hline
\begin{tabular}[c]{@{}c@{}}Energy\\range\end{tabular} & Population & Volume & Contaminant & \begin{tabular}[c]{@{}c@{}}Status\\ 1\textsuperscript{st} iteration \end{tabular} & \begin{tabular}[c]{@{}c@{}}Status\\ otherwise \end{tabular}  & \begin{tabular}[c]{@{}c@{}}Free\\parameters\end{tabular} \\ \hline \hline

\multirow{3}{*}{\begin{tabular}[c]{@{}c@{}}[0.3--5] \\ keV\end{tabular}} 
& \multirow{3}{*}{\begin{tabular}[c]{@{}c@{}}$^{22}$Na coin \\ + $^{40}$K coin \\ y1 \end{tabular}} 
& \multirow{3}{*}{\begin{tabular}[c]{@{}c@{}}crystal \\ cosmogenic\end{tabular}}

& \multirow{3}{*}{\begin{tabular}[c]{@{}c@{}}$^{40}$K \\ $^{22}$Na\end{tabular}} &  \multirow{3}{*}{free} &  \multirow{3}{*}{-}
& \multirow{3}{*}{\begin{tabular}[c]{@{}c@{}}18 \\ 9 × A$_{^{40}\mathrm{K}}$ \\ 9 × A$_{^{22}\mathrm{Na}}$\end{tabular}} \\
& & &  &  & \\
& & & & & \\
\hline


\multirow{8}{*}{\begin{tabular}[c]{@{}c@{}}[200-700]\\keV\end{tabular}} 
    & \multirow{8}{*}{\begin{tabular}[c]{@{}c@{}}single-hits\\ + m2-hits \\ + m3m4-hits \\ y6\end{tabular}} 
    & \multirow{3}{*}{PMTs} & $^{40}$K & fixed & fixed  & \multirow{8}{*}{\begin{tabular}[c]{@{}c@{}}12 \\ 9 x A$_{^{226}\mathrm{Ra},\text{Boro}}$ \\ 1 x A$_{^{232}\mathrm{Th},\text{Boro}}$ \\ 1 x f$_{^{226}\mathrm{Ra},\text{Boro}}$ \\ 1 x f$_{^{232}\mathrm{Th},\text{Boro}}$ \end{tabular}} \\ 
    &  &  &  $^{226}$Ra &  free & free \\ 
    &  &  &   $^{232}$Th & free  & free \\ \cline{3-6} 
    &  & \multirow{3}{*}{crystal} & $^{40}$K & fixed & fixed\\ 
    &  &  & $^{210}$Pb bulk NaI & free & fixed to prev. it. \\ 
    &  &  & $^{210}$Pb surf NaI & free & fixed to prev. it. \\ \cline{3-6} 
    &  & cosmogenic & $^{22}$Na & fixed & fixed \\ \cline{3-6} 
    &  & other & - & free & fixed to prev. it. 
    \\ \hline 

\multirow{11}{*}{\begin{tabular}[c]{@{}c@{}}[20-60]\\keV\end{tabular}}
    & \multirow{11}{*}{\begin{tabular}[c]{@{}c@{}}single-hits\\ y6 + (y3-y6)\end{tabular}}  
    & \multirow{3}{*}{PMTs} & $^{40}$K & fixed & fixed & \multirow{11}{*}{\begin{tabular}[c]{@{}c@{}}36 \\ 9 x A$_{^{210}\mathrm{Pb,NaI,bulk}}$ \\ 9 x A$_{^{210}\mathrm{Pb,NaI,surf}}$ \\ 9 x A$_{^{129}\mathrm{I}}$ \\ 9 x A$_{^{109}\mathrm{Cd}}$ \end{tabular}} \\ 
    &  &  & $^{226}$Ra & fixed to prev. it. & fixed to prev. it. \\
    &  &  & $^{232}$Th & fixed to prev. it. & fixed to prev. it. \\ \cline{3-6} 
    &  & \multirow{4}{*}{crystal} & $^{40}$K & fixed & fixed \\
    &  &  &   $^{210}$Pb bulk NaI & free & free \\
    &  &  &   $^{210}$Pb surf NaI & free & free \\
    &  &  & $^{210}$Pb surf teflon & free & fixed to prev. it. \\ \cline{3-6} 
    &  & \multirow{3}{*}{cosmogenic} & $^{22}$Na & fixed & fixed \\
    &  &  &   $^{129}$I &  free & free \\
    &  &  &   $^{109}$Cd & free & free \\ \cline{3-6} 
    &  & other & - & free  & fixed to prev. it. \\ \hline

\multirow{10}{*}{\begin{tabular}[c]{@{}c@{}}[6-20]\\keV\end{tabular}}
    & \multirow{10}{*}{\begin{tabular}[c]{@{}c@{}}single-hits\\ y6\end{tabular}}  
    & \multirow{3}{*}{PMTs} & $^{40}$K & fixed & fixed & \multirow{10}{*}{\begin{tabular}[c]{@{}c@{}}27 \\ 9 x A$_{^{210}\mathrm{Pb,teflon,surf}}$ \\ 9 x A$_{^{3}\mathrm{H}}$ \\ 9 x A$_{\mathrm{other}}$  \end{tabular}} \\ 
    &  &  & $^{226}$Ra & fixed to prev. it. & fixed to prev. it. \\
    &  &  & $^{232}$Th & fixed to prev. it. & fixed to prev. it. \\ \cline{3-6} 
    &  & \multirow{4}{*}{crystal} & $^{40}$K & fixed & fixed \\
    &  &  & $^{210}$Pb bulk NaI & fixed to prev. it. & fixed to prev. it. \\
    &  &  & $^{210}$Pb surf NaI & fixed to prev. it. & fixed to prev. it. \\
    &  &  & $^{210}$Pb surf teflon & free & free \\ \cline{3-6} 
    &  & \multirow{2}{*}{cosmogenic} & $^{22}$Na & fixed & fixed \\
    &  &  &   $^{3}$H &  free & free \\ \cline{3-6} 
    &  &   other & - & free & free \\ \hline
    & & & & & & Total\\
    \cline{7-7} 
     & & & & & & 93\\
     \cline{7-7}
\end{tabular}}
\caption{Summary of the contaminations included at each step of the fitting procedure. For each step, the table reports the fitted energy range, the fit population, the considered contamination and its associated volume, the parameter status at the first iteration and in subsequent iterations (either free, fixed, or fixed to the value from the previous iteration), the number of free parameters introduced at that step and the total free parameters of the fit.}
\label{componentesdelfit}
\end{table}

For an accurate assessment of background sources, it is crucial to emphasize that all nine modules act as potential sources of contamination for the other detectors. This consideration is particularly important when contamination in a specific crystal or module component, such as PMTs, can produce a signal in a different module. Consequently, the high-energy fit is performed simultaneously across all nine detectors, as each detector can contribute to the background observed in the others. It is worth noting that this approach does not apply to the medium- and low-energy ranges, where the fit is performed individually for each detector, as it is verified that the signal observed by one detector is not correlated with other sources beyond itself. 

First, the fit of the $^{22}$Na and $^{40}$K coincidence population from the first year of data-taking is performed, involving a total of 18 free parameters, one activity value for $^{22}$Na and one for $^{40}$K for each detector.

The high-energy fit is then conducted by simultaneously fitting the spectra of single-hit, m2-hit, and m3m4-hit events within the energy range of [200-700] keV. Note that single-hit, m2-hit, and m3m4-hit refer to events where energy is deposited in one, two, or three to four detectors, respectively. The fit is carried out using only the sixth year of data to exclude the contribution from cosmogenic events. The outcomes corresponding to this stage of the fit are presented in Section \ref{highenergyfit}.

\begin{figure}[t!]
\begin{center}
\includegraphics[width=0.52\textwidth]{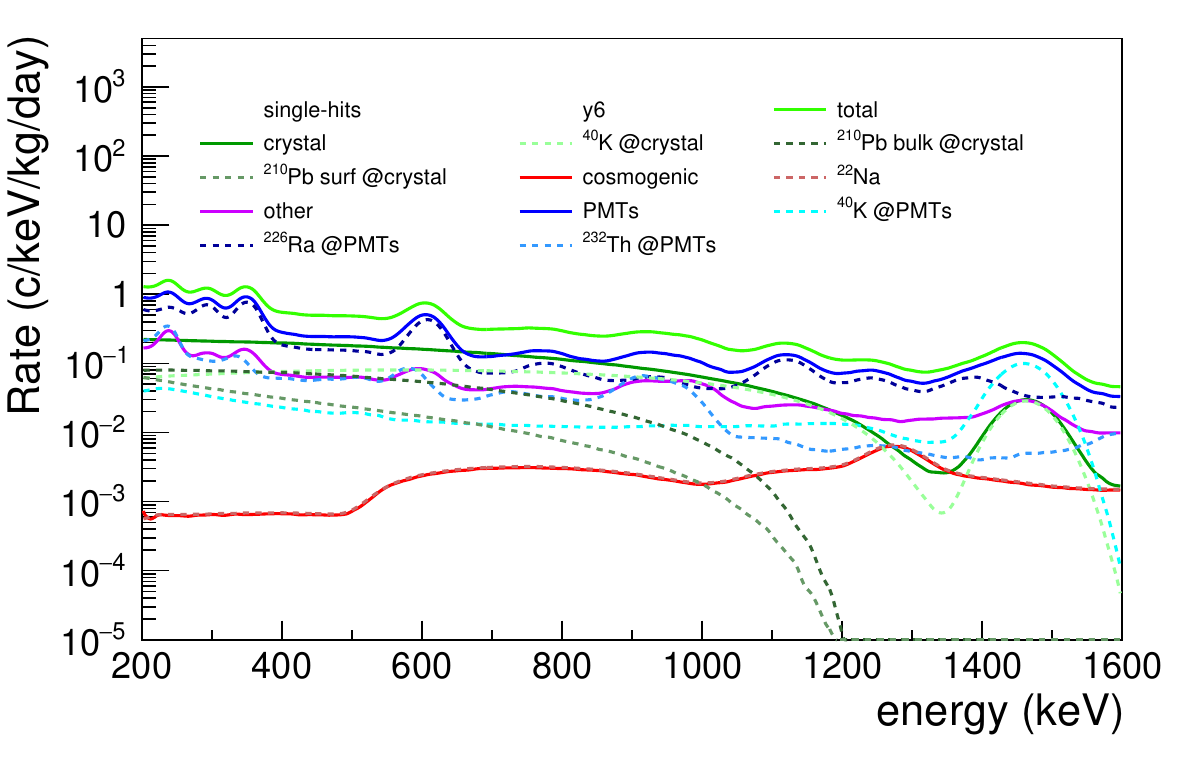}
\includegraphics[width=0.49\textwidth]{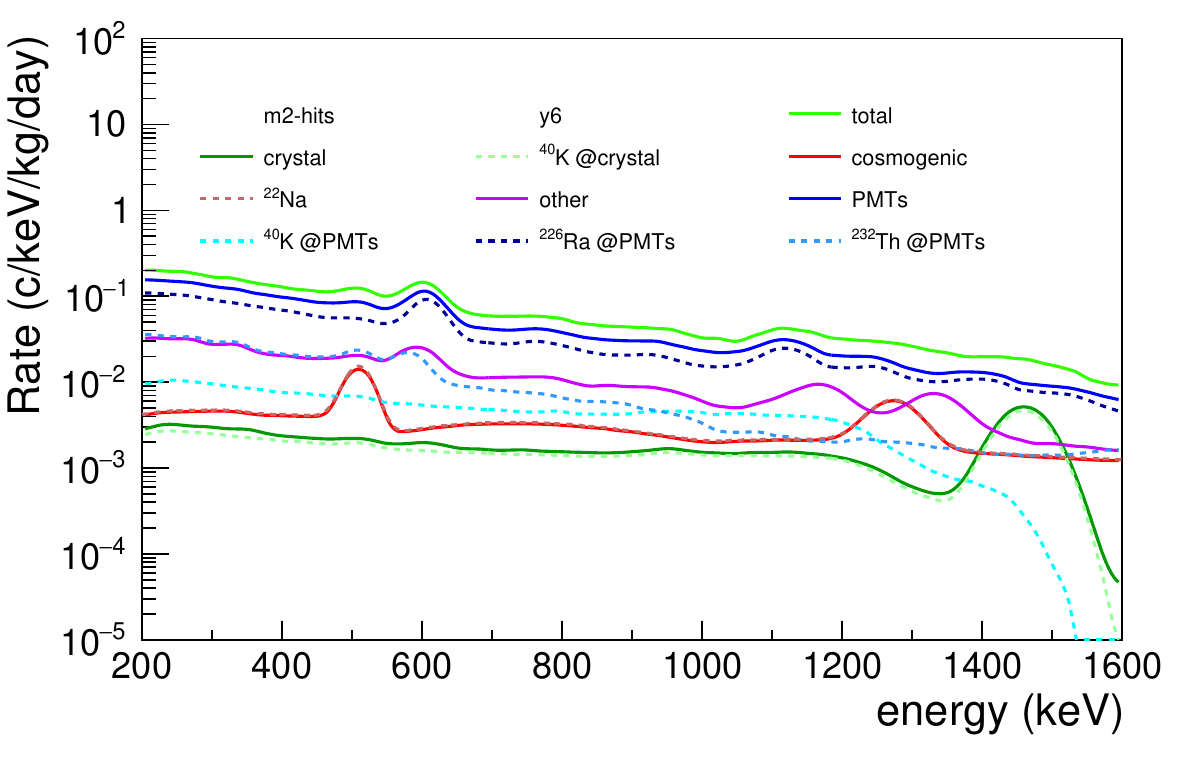}
\includegraphics[width=0.49\textwidth]{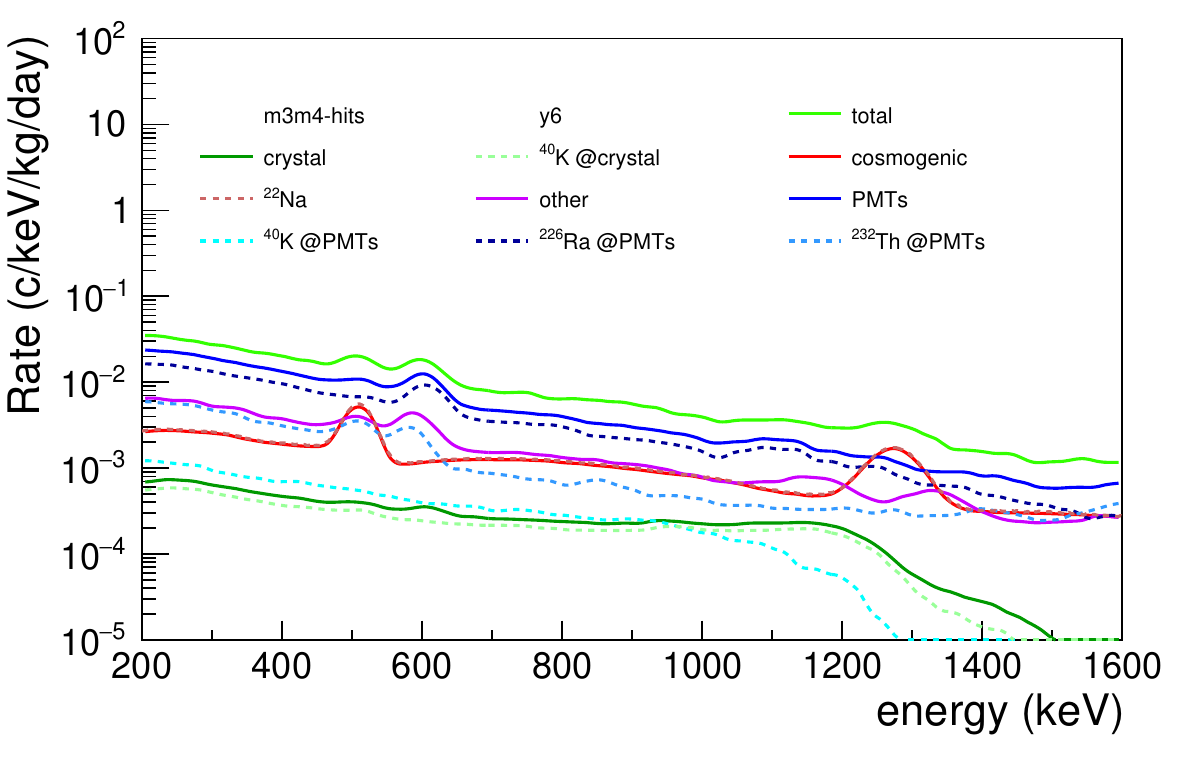}
\caption{\label{HEdatasim} High-energy spectra expected from individual background sources as dictated by the reconstructed background model, together with the sum of all contributions, for the sum of the nine detectors in the single-hits (\textbf{top panel}), m2-hits (\textbf{bottom, left panel}), and m3m4-hits (\textbf{bottom, right panel}) populations. Solid lines represent the four main groups of background contributions, while dashed lines indicate the most significant individual sources within each group. }
\end{center}
\end{figure}

Regarding the event-selection cuts applied to the high energy data, events arriving more than 1 s after the last muon veto trigger and $\beta/\gamma$ events, identified from the double readout system of the high-energy line, are selected. Concerning the fitting range, energies below 200 keV are excluded from the fitting process because the region between [100-200]~keV corresponds to the transition between two calibration procedures, which might be affected by some energy discontinuities or artifacts. The region above 700~keV is also excluded from the fit because, beyond this energy, the $\beta$$^-$ spectrum of \(^{210}\)Bi from the decay of \(^{210}\)Pb becomes dominant (see Figure \ref{HEdatasim}). As a result, the fitting procedure, in its minimization process, prevents an increase in the contributions from the PMTs to avoid distorting the [700-1000] keV region, leaving the peaks in the [200-400]~keV range clearly underfitted. To mitigate this systematic effect, the fit is first performed on the peaks region, prioritizing its description, and the behavior of the resulting fit is later evaluated in the energy region above 700 keV.

Figure \ref{HEdatasim} show the breakdown of the reconstructed background model contributions in each population. As can be seen, in the high-energy range, the contamination from the PMTs completely dominates, allowing this component to be fitted in this region and then introduced as a fixed parameter when performing the fit in the medium and low-energy ranges. The number of free parameters in the high-energy range increases to 12 in this second step of the fit. These parameters correspond to nine independent PMT \(^{226}\text{Ra}\) activities (one for each detector), a single PMT \(^{232}\text{Th}\) activity (as prior HPGe measurements indicated similar contamination levels across all detectors), and two additional free parameters representing the fractional activities of \(^{226}\text{Ra}\) and \(^{232}\text{Th}\) assigned to the borosilicate and the photocathode of the PMT (see Section \ref{PMTsim}). Note that this fractional activity is assumed to be the same across all detectors, based on the premise that all the Hamamatsu PMTs originated from the same production batch.

\begin{figure}[b!]
\begin{center}
\includegraphics[width=0.7\textwidth]{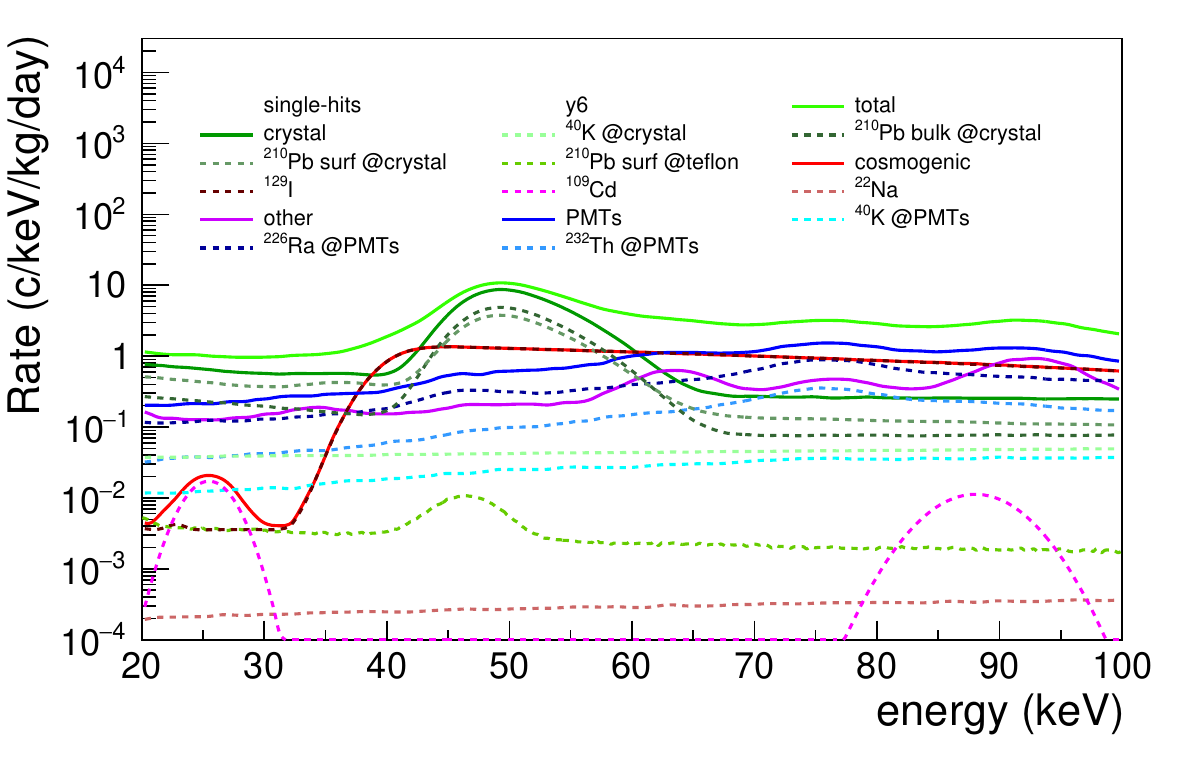}

\caption{\label{MEdatasim} Medium-energy spectra expected from individual background sources as dictated by the reconstructed background model, together with the sum of all contributions, for the sum of the nine detectors, in the single-hits population. Solid lines represent the four main groups of background contributions, while dashed lines indicate the most significant individual sources within each group.
}
\end{center}
\end{figure}

The medium-energy fit is carried out by simultaneously fitting the single-hit spectrum from the sixth year and the differential single-hit spectrum between the third and sixth years. The inclusion of this time difference improves the fit convergence by incorporating the decay of \(^{210}\text{Pb}\) over this period, providing an additional constraint to guide the fitting of this particular isotope. Year 3 is chosen trying to optimize the ratio signal/background, as the first two years have a large contribution from short life cosmogenic isotopes that will disturb the $^{210}$Pb fitting and the 22-year half-life of this isotope requires an interval of several years to observe a significant reduction of the isotope activity. 


In the medium-energy spectrum presented in this chapter, it was necessary to apply a correction shift to the data to align the measured \(^{210}\text{Pb}\) structure feature at about 50 keV, which dominates this region, with the background model prediction. This adjustment was required to better reproduce the observations, as the misalignment between the simulated and measured peak positions otherwise prevented the fit from converging.


The correction was determined by identifying the maximum of the \textsuperscript{210}Pb feature in both the simulation and the data, and then, shifting the data to match the simulated spectra. Given that a shift is performed, the shape of the peak remains unchanged. This correction is consistent across detectors, although not identical, and ranges between 0.2 and 1 keV depending on the detector. The origin of this discrepancy is not fully understood, but it is likely related to minor calibration issues. Non-proportionality in the LY of NaI(Tl) crystals (see Section \ref{nonpropSec}) could also contribute; however, according to the non-proportionality curve (see Figure), the light output in this energy range decreases by approximately 5\% in ANAIS crsytals, which would systematically shift the peak to lower energies with respect to the simulation. This behavior is not observed, but the contrary, making difficult to account for the discrepancy by the non-proportionality in the LY.


Regarding the fitting range, the region [20-60] keV was chosen for the medium-energy range. The lower limit was set at 20 keV to decouple the effects of \(^{210}\text{Pb}\) and \(^{3}\text{H}\), the latter having an endpoint at 18.6 keV. When attempting a fit from [6-60] keV, it was found that these two components clearly correlated. On the other hand, the upper limit is set at 60~keV due to the asymmetry in light sharing discussed in Section \ref{asymmetry}. Above 65 keV, highly asymmetric events begin to appear in the single-hit spectrum, which the simulation fails to accurately reproduce. To avoid biasing the fit, this energy range is excluded from the fitting range.

\begin{figure}[b!]
\begin{center}
\includegraphics[width=0.7\textwidth]{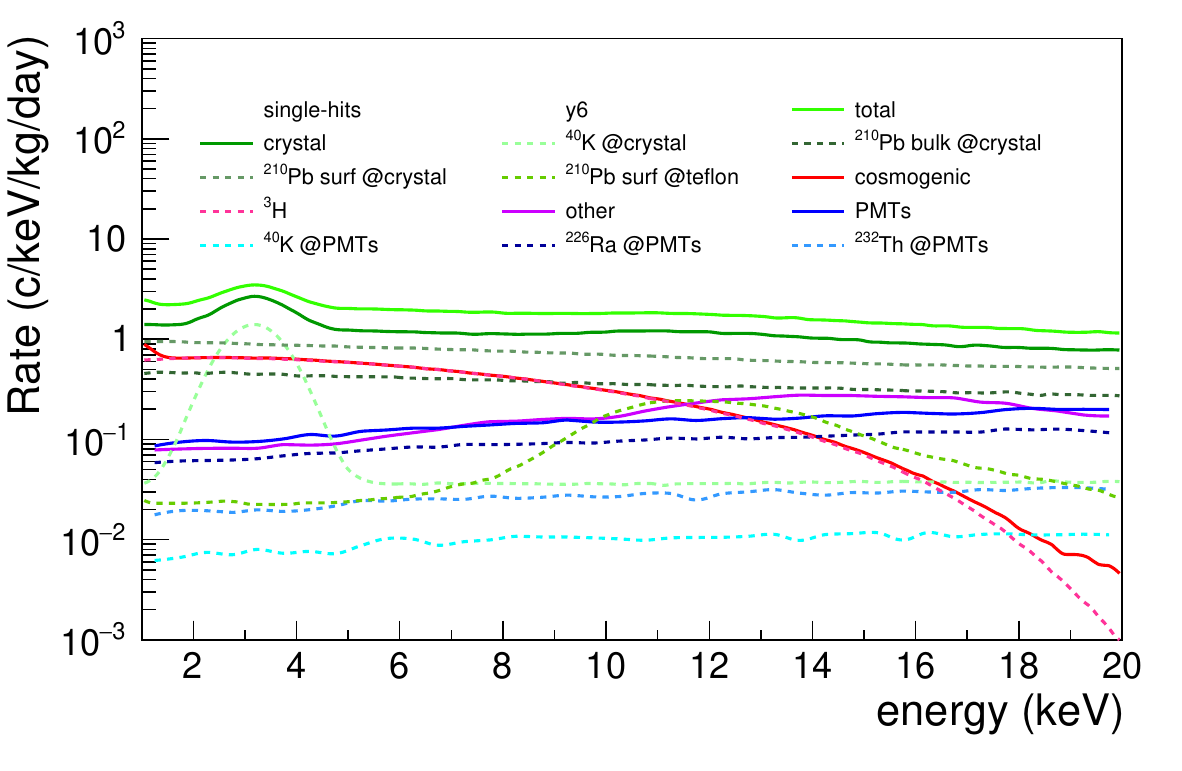}

\caption{\label{LEdatasimnew} Low-energy spectra expected from individual background sources as dictated by the reconstructed background model, together with the sum of all contributions, for the sum of the nine detectors, in the single-hits populations. Solid lines represent the four main groups of background contributions, while dashed lines indicate the most significant individual sources within each group.
}
\end{center}
\end{figure}

Figure \ref{MEdatasim}, together with Table \ref{componentesdelfit}, displays the components included at this stage of the fit. Since the fit is performed independently for each detector, the number of free parameters per detector is 4, resulting in a total of 36 free parameters for the nine detectors.

Finally, the low-energy fit is performed by fitting the single-hit spectrum of the sixth year for each detector individually. Preliminary fits were made where the difference between the third and sixth year was also fitted, as was done in the medium-energy fit. The objective is to distinguish $^3$H from $^{210}$Pb, which have different half-lives, though these are not vastly different, and three years is relatively short compared to the half-lives of both. However, no improvement in the convergence process was found.


The energy estimator used in this region is the one specifically calibrated in ANAIS for the low-energy range, to which filtering techniques based on machine learning are applied. The fitting range is set from [6-20] keV. The fitting starts at 6 keV to avoid mainly interferences with the annual modulation analysis (done within [1-6] keV), and other possible axion (see Chapter \ref{Chapter:annual}, Section \ref{axions}) and WIMP-model dependent searches. That is, the background in the ROI will not be fitted but instead extrapolated from the results of the fit in higher energy regions. 

Figure \ref{LEdatasimnew} shows the components that dominate this energy region, and Table \ref{componentesdelfit} lists those included in the fit for this region. In the low-energy region, 3~free parameters are used per detector, resulting in a total of 27 free parameters for the 9 detectors. Note that the PMT contribution is almost negligible at both low and medium energies.

Thus, the total number of free parameters to be determined in this work is 93, which will be sequentially determined across the different spectral regions.

\vspace{-0.2cm}

\subsection{The background fitting function}\label{function}

In this thesis, the ANAIS-112 background is modelled as a linear combination of the spectra induced by each background source detailed in the previous section, as obtained from Geant4 simulations. The relative contribution of each component is determined by fitting their weights to the measured background data, incorporating information from single-hit and multiple-hit event populations, as well as temporal variations observed in the spectra across different data-taking years.

The fitting procedure is carried out using the RooFit package, which is part of the ROOT framework \cite{antcheva2009root}. RooFit is a comprehensive toolkit for modelling the expected distribution of events in physics analyses and performing (un)binned maximum likelihood fits \cite{verkerke2006roofit}. In this work, a frequentist method using a $\chi^2$ minimization has been followed using the Minuit package of RooFit as the minimization algorithm. 

In RooFit, the basic ingredient is the PDF, $f^{PDF}$, that describes the distribution of data under certain hypothesis. RooFit can use both analytical or binned PDFs based on histograms, typically
obtained from MC simulations. This allows for the simulation output to be used directly in the fit, ensuring both accuracy and flexibility in the fitting procedure. The fit in this thesis uses RooFit's extended likelihood method (LM), which estimates the number of events. This differs from the standard LM, which returns fractions of each model component. The extended LM fits both the shape and the total number of events, whereas the standard LM considers only the shape of the distribution.


However, the extended LM fit must be performed in event (integer) counts rather than rates (counts/keV/kg/day), which is the standard unit for expressing measurements in ANAIS and other DM search experiments. Working in event counts requires introducing a normalization factor in the fit, $(P\epsilon)_{jl}$ which is related to the probability of producing a deposit in an energy interval  \([E_1, E_2]\) in detector $j$ for a given number of initial simulated decays of source $l$,

\begin{equation}
(P\epsilon)_{jl} = \frac{\text{number of events in } [E_1, E_2] \text{ in detector } j}{\text{total number of simulated initial decays of source } l}.
\label{pej}
\end{equation}

The fitting procedure allows the inclusion of as many PDFs as necessary, each weighted by a corresponding coefficient. For each energy interval \([E_1, E_2]\) and each population \(k\), the following function is defined:

\begin{equation}
F_{B}(E) = \sum_{ijl} N_{ijl} f^{PDF}_{jl}(E).
\end{equation}

In this formulation, $N_{ijl}$ are the free parameters in the fit associated with each PDF, $f^{PDF}_{jl}$, representing scaling factors for each detector $j$ and source $l$. Since the area under each PDF is normalized to 1 in the fitting range, $N_{ijl}$ directly corresponds to the expected number of counts deposited during the year $i$ in detector $j$ from each source $l$.





The goal is to express \(N_{ijl}\) in terms of the initial activities of each source \(l\) at the time when detector \(j\) was installed underground. Let \(t_j\) denote the time for each detector \(j\), such that \(t_j = 0\) marks the moment the detector moved underground, and \(A_{0l}\) denote the initial activity of source \(l\). The activity at time \(t_j\) evolves as:

\begin{equation}
A_l(t_j) = A_{0l} e^{-\lambda_l t_j}
\end{equation}

where \(\lambda_l\) is the decay constant of source \(l\).

The rate induced by source \(l\) in detector \(j\), at time \(t_j\), within the energy interval \([E_1, E_2]\), is given by:
\begin{equation}
    R_{jl}(t_j) = (P\epsilon)_{jl} \, A_l(t_j)
\end{equation}

To compute \(N_{ijl}\), the rate \(R_{jl}(t_j)\) must be integrated between \(t_j^{\text{ini-i}}\) and \(t_j^{\text{end-i}}\), corresponding to the start and end times of year \(i\):

\begin{equation}
N_{ijl} = \int_{t_j^{\text{ini-i}}}^{t_j^{\text{end-i}}} R_{jl}(t_j) \, dt_j = (P\epsilon)_{jl}  \frac{A_{0l}}{\lambda_l} \left( e^{-\lambda_l t_j^{\text{ini-i}}} - e^{-\lambda_l t_j^{\text{end-i}}} \right)
\label{EqN}
\end{equation}

Solving for \(A_{0l}\) yields:

\begin{equation}
A_{0l} = \frac{N_{ijl} \, \lambda_l}{(P\epsilon)_{jl} \left( e^{-\lambda_l t_j^{\text{ini-i}}} - e^{-\lambda_l t_j^{\text{end-i}}} \right)}
\label{EqA0}
\end{equation}

Equation \ref{EqA0} allows converting $N_{ijl}$, which is the free parameter of the fit, into the initial activity of the different isotopes when the detectors arrived underground. Note that this equation is only valid for isotopes not involved in decay chains. To account for the behavior of a decay chain, the Bateman equations must be employed, as shown in Equation~\ref{eqEq} to describe the secular equilibrium between $^{210}$Pb and $^{210}$Po. From the initial activities obtained in the fit conducted in this work, the total time-dependent background model for each crystal will be derived.

\section{The background fitting}\label{backgroundfitting}

The following section presents the results finales of the background fitting, following the iterative procedure detailed in Figure \ref{esquemaflow}. As previously mentioned, the fitting procedure is carried out in a stepwise manner, sequentially across distinct energy regions.

\subsection{$^{22}$Na and $^{40}$K fit}\label{NaK}

The first step of the fitting procedure is to determine the content of $^{22}$Na and $^{40}$K in the bulk of the crystal. While the former originates from cosmogenic activation, the latter is an intrinsic contamination of the crystal itself. These isotopes decay via EC to an excited state of the daughter nucleus, leading to the emission of a $\gamma$-ray from the de-excitation\footnote{\hfill $^{40}\text{K}$ has  also a $\beta^-$ decay channel. Specifically, its decay modes are $\beta^-$ (89.25\%), $\beta^+$ (0.001\%), and EC (10.75\%).}. The following (and simultaneous) atomic de-excitation energy emitted (0.87 keV for $^{22}$Na and 3.2 keV for $^{40}$K from K-shell EC) is fully absorbed within the crystal where the decay occurs. Meanwhile, the high-energy $\gamma$-ray (1274.5 keV and 1460.8 keV, respectively) may escape and interact within another detector, producing a coincidence event.

Data will be selected by searching for coincidences in the [1200,1340]~keV energy windows for $^{22}$Na and [1340,1560] keV for $^{40}$K, applying the same selection criteria to both data and simulation. The fit is then performed in the low-energy range of [0.3,5]~keV. Results are corrected by accounting for the number of sigmas corresponding to the selected coincidence energy windows in both the low- and high-energy ranges, which determine the selection efficiencies, because the $(P\epsilon)_{jl}$ (see Equation \ref{pej}) is calculated before convolution of the simulation with the energy resolution.

\begin{figure}[t!]
\begin{center}
\includegraphics[width=0.72\textwidth]{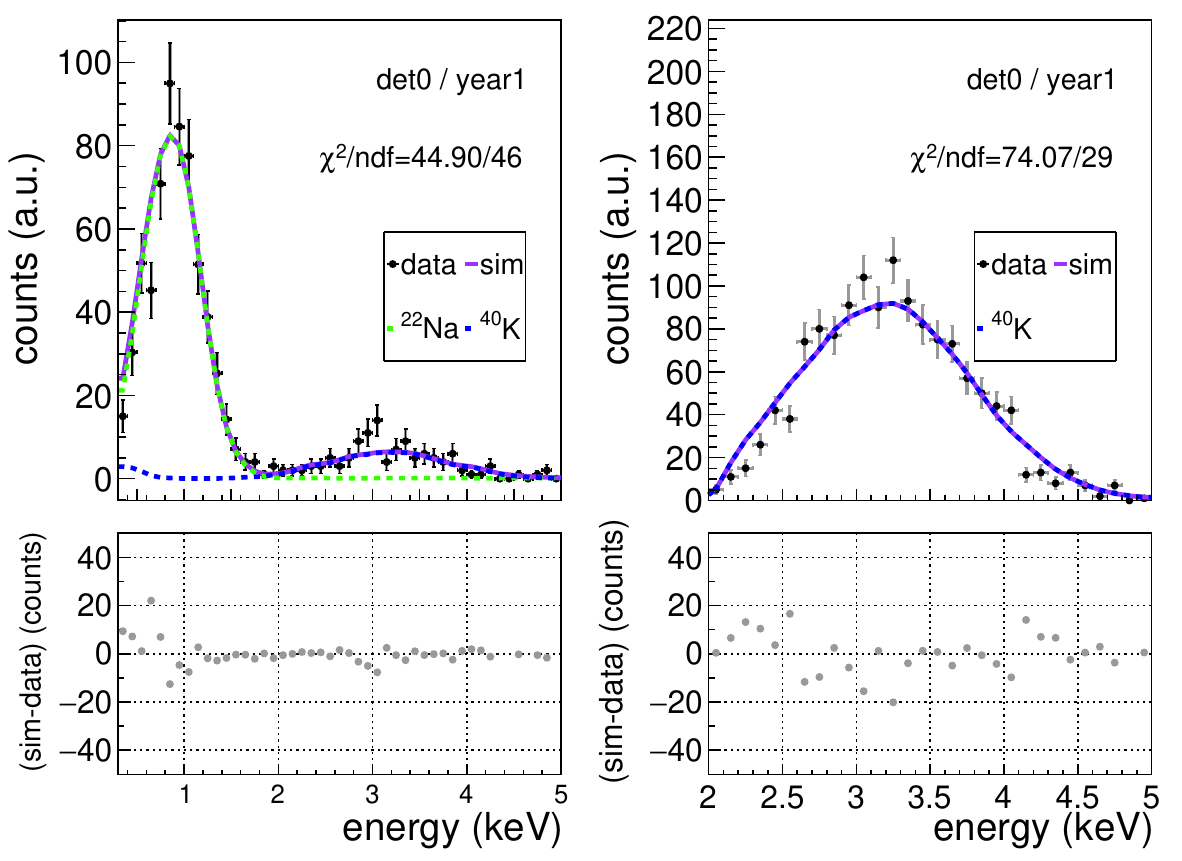}

\caption{\label{NaKfitD0} Low-energy spectra of the D0 crystal in coincidence with a high-energy gamma in the range [1200-1340] keV (\textbf{left panel}) and [1340-1560] keV (\textbf{right panel}) detected in a second module. The black points represent data from the full first year of ANAIS-112. Peaks at 0.87 and 3.2 keV, corresponding to the decays of $^{22}$Na and $^{40}$K in the NaI bulk, respectively, are clearly visible. A simultaneous fit of both populations is performed to determine the activity of $^{22}$Na (green, dashed line) and $^{40}$K (blue, dashed line). The total fit is shown in violet. The plot also displays the goodness of fit and the residuals (lower panels). }
\end{center}
\end{figure}

\begin{figure}[b!]
\begin{center}
\includegraphics[width=0.72\textwidth]{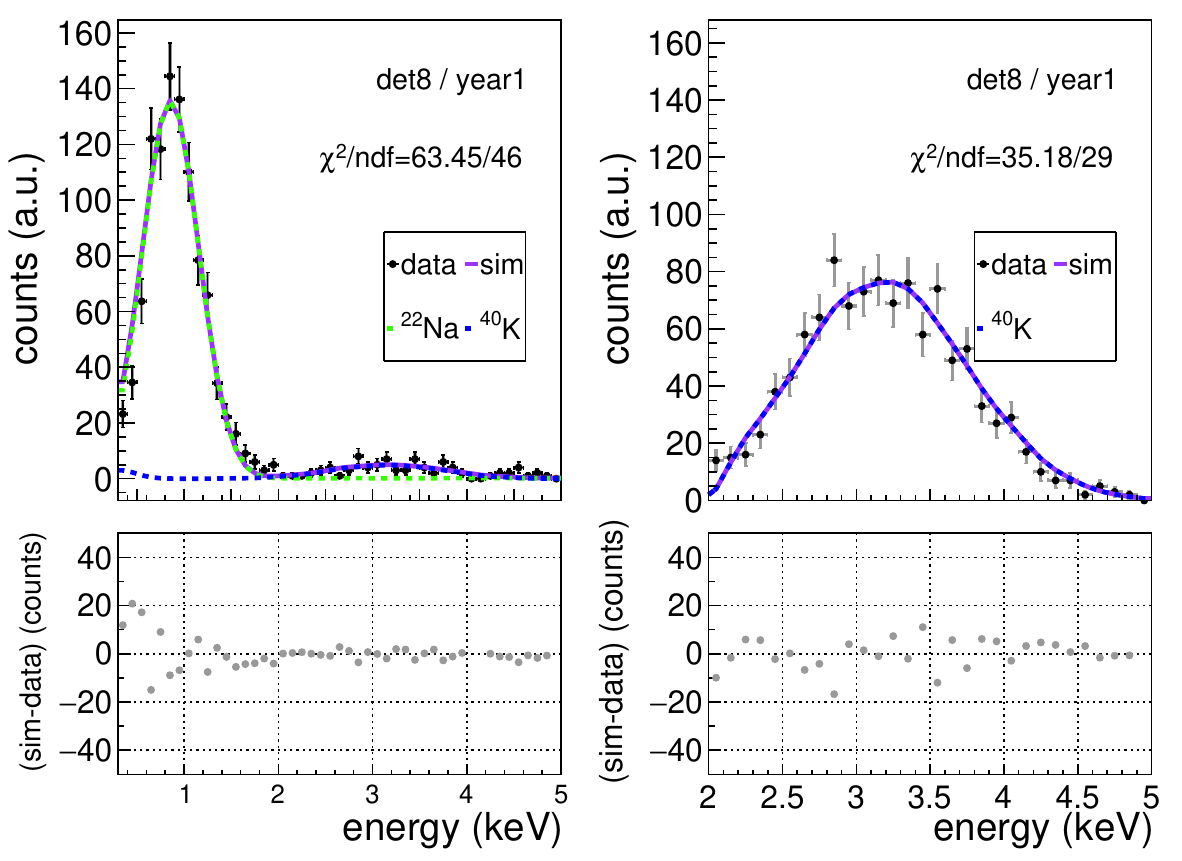}

\caption{\label{NaKfit} Analogous to Figure \ref{NaKfitD0}, but corresponding to crystal D8. }
\end{center}
\end{figure}

Figure \ref{NaKfitD0} shows the selected spectra for each case. Two peaks at 0.87 keV (left panel) and 3.2 keV (right panel) are clearly visible. Additionally, the selection criteria for the $^{22}$Na population also include Compton-scattered 1460.8 keV events, which generate the low-amplitude 3.2 keV peak observed in the left panel. Since a $^{40}$K signal is present in both coincidence windows, a simultaneous fit of both populations is required. In this case, the only contributions fitted are those from each individual crystal, as contaminations from other crystals do not contribute to the signal being searched for. This is because the characteristic considered corresponds to a very low-energy emission, which can only be attributed to contamination within the crystal itself.

\begin{table}[t!]
    \centering
    \begin{tabular}{ccc}
        \hline
        detector &  $^{22}$Na (mBq/kg) & $^{40}$K (mBq/kg) \\
        \hline
        0 & 2.18 $\pm$ 0.09 & 1.19 $\pm$ 0.03 \\
        1 & 2.04 $\pm$ 0.08 & 1.02 $\pm$ 0.03 \\
        2 & 0.87 $\pm$ 0.04 & 0.96 $\pm$ 0.03 \\
        3 & 0.95 $\pm$ 0.04 & 0.57 $\pm$ 0.02 \\
        4 & 0.86 $\pm$ 0.03 & 0.45 $\pm$ 0.02 \\
        5 & 0.75 $\pm$ 0.02 & 0.94 $\pm$ 0.02 \\
        6 & 0.74 $\pm$ 0.02 & 0.84 $\pm$ 0.02 \\
        7 & 0.67 $\pm$ 0.02 & 0.84 $\pm$ 0.02 \\
        8 & 0.69 $\pm$ 0.02 & 0.62 $\pm$ 0.02 \\
        \hline
\end{tabular}

\caption{\label{resultsNaK} Initial activities (at the time of detector installation underground) for cosmogenically produced $^{22}$Na and intrinsic contamination of $^{40}$K in the NaI crystals obtained from the fit. Values are expressed in mBq/kg for the nine ANAIS-112 detectors.}

\end{table}

\begin{figure}[b!]
    \centering
    {\includegraphics[width=0.49\textwidth]{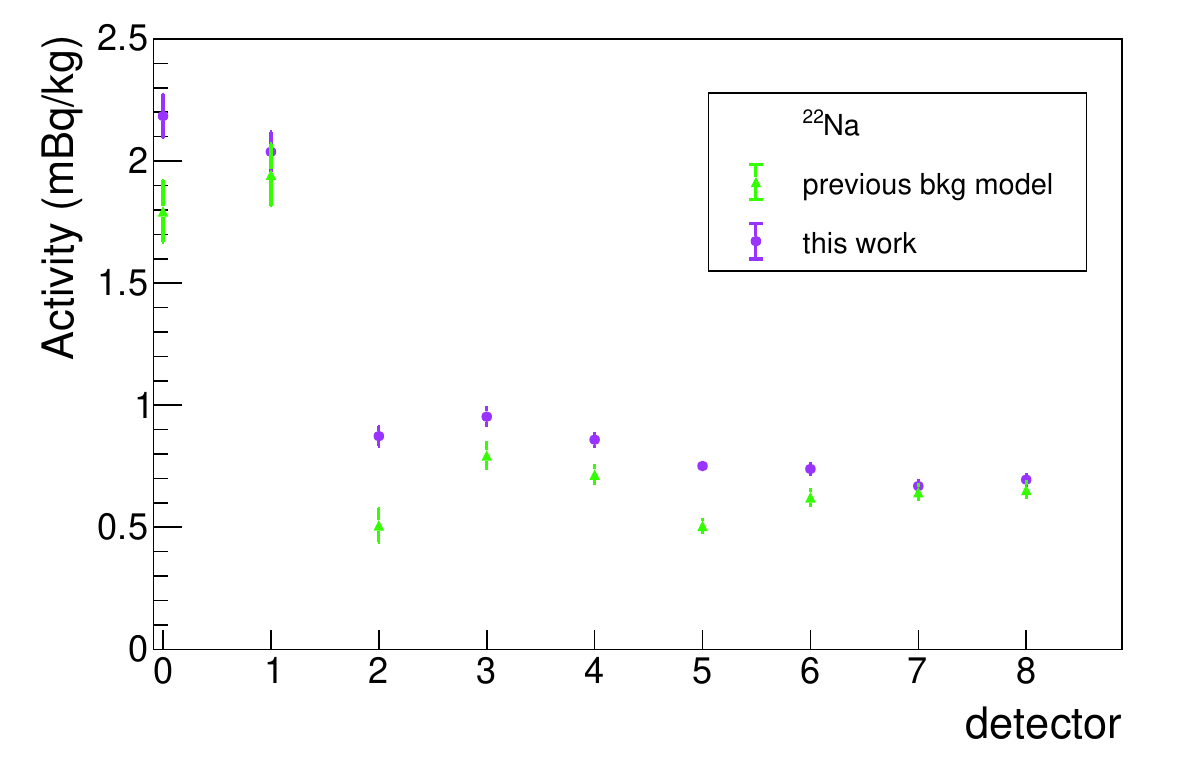}}
    \hfill
    {\includegraphics[width=0.49\textwidth]{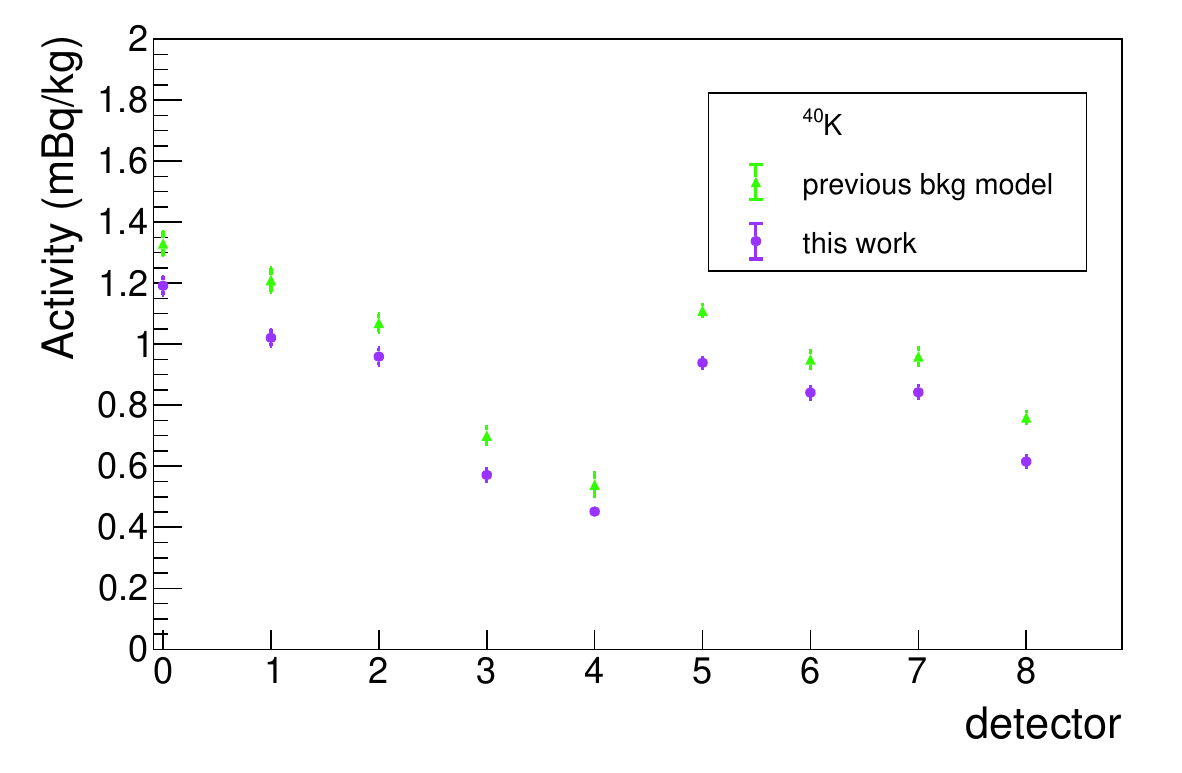}}
    
    \caption{\label{comparacionNaK}Comparison of the initial activities of cosmogenically produced $^{22}$Na (\textbf{left panel}) and intrinsic $^{40}$K contamination (\textbf{right panel}) at the time of detector storage underground. The values obtained from the fit in this study are compared with those measured and reported in the previous ANAIS-112 background model. Values are expressed in mBq/kg for the nine ANAIS-112 detectors.}
\end{figure}

Figure~\ref{NaKfitD0} presents the simultaneous fit for crystal D0, along with the corresponding residuals. Similarly, Figure~\ref{NaKfit} shows the fit results for crystal D8. A comparative analysis of the fit behavior of these two modules is particularly useful, as they correspond to the detectors with the highest and lowest contributions of $^{22}$Na and $^{40}$K, respectively, according to the previous background model. It is worth noting that, in the case of D0, the $^{22}$Na signal has significantly decayed due to the earlier arrival of this module at the LSC and the short half-life of the isotope. The fitting is conducted using data from the first year, when the \(^{22}\)Na content was at its highest, obtaining comparable values for the other years taking into account its radioactive decay over the corresponding time period. The fit exhibits satisfactory performance across all detectors, demonstrating a consistent agreement between the model and the observed data within the statistical uncertainties.

The results of the fit for $^{22}$Na and $^{40}$K are summarized in Table \ref{resultsNaK}.  Figure \ref{comparacionNaK} compares the initial activities derived from the fit conducted in this study with those measured and considered in the previous ANAIS-112 background model \cite{amare2019analysis}. 

As can be inferred from both the figure and the table, the fitted values follow the same trend as the previously measured ones. In particular, for $^{22}$Na, the initial activity in detectors D0 and D1 is significantly higher than in the other detectors produced later, suggesting a reduction in exposure time during crystal growth and detector assembly in Colorado. However, a slight increase in the fitted $^{22}$Na values and a decrease in $^{40}$K activity are observed compared to previous measurements, while this deviation is not uniform across detectors. These results are consistent with the signatures identified in the high-energy spectrum of some detectors.

These \(^{22}\)Na and \(^{40}\)K values are kept fixed in the subsequent steps of the fit.

\subsection{High-energy fit}\label{highenergyfit}
\vspace{-0.2cm}
The high-energy fitting procedure is carried out with $^{22}$Na and $^{40}$K fixed, involving a simultaneous fit of the nine detectors across single-hits, m2-hits, and m3/m4-hits in the energy range from 200 to 700 keV. Returning to Table \ref{componentesdelfit}, the free parameters in this fit are the \(^{226}\)Ra activity in the PMTs (one for each detector), the \(^{232}\)Th activity in the PMTs common to all nine detectors, and the contamination fraction in the borosilicate for each of these two components (see Section \ref{PMTsim}, where this distribution of contaminations is explained). It is worth highlighting that, in this case, the contributions from the PMTs of different modules affect the spectra measured by the others, making the combined fit necessary.

\begin{figure}[t!]
\begin{center}
\includegraphics[width=0.92\textwidth]{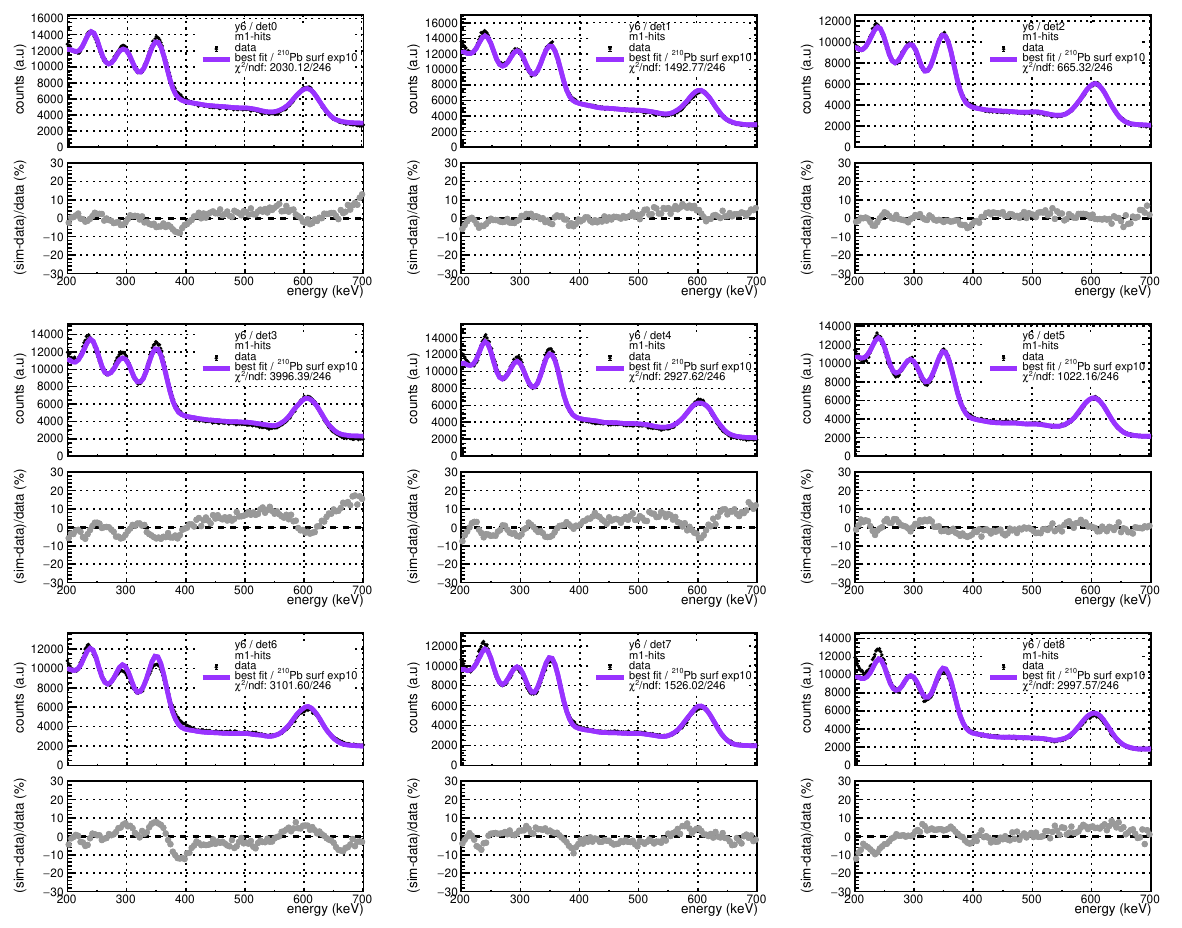}

\caption{\label{HEpop0} High-energy best fits of the single-hit spectra for the nine ANAIS-112 detectors. Black points correspond to data from the full sixth year of ANAIS-112, while the best fit is shown in violet. The exponential profile of 10 $\mu$m is considered for the superficial contamination of \(^{210}\)Pb. Residuals are displayed as grey points, and the goodness-of-fit is indicated in each panel. }
\end{center}
\end{figure}

In this work, three different depths of exponentially decaying profiles as benchmarks for the superficial contamination of \(^{210}\)Pb are considered: 1, 10, and 100 $\mu$m. The objective is to determine whether any particular distribution is preferred over the others. However, this is not a fitting of the depth of the superficial component itself, but rather an effort to identify which range depth provides the best fit across all energy ranges. This contamination is not necessarily uniform across all detectors, as each module may have been subjected to different contamination conditions. Regarding this contamination, it is worth highlighting that, for the single-hits in the high-energy spectra, this work incorporates the experimental shape obtained from BetaShape to describe the $\beta^-$ decay of \(^{210}\)Bi (see Section \ref{betashape}).

Figure \ref{HEpop0} shows the high-energy fit of the single-hits population for the nine ANAIS detectors using year 6. As can be seen, the fit performs well, with residuals below 10\%. However, the goodness of fit is found to be poor. This may be attributed to the energy resolution of the peaks, where small mismodelling can lead to a rapid increase in the $\chi^2$, as well as to minor energy miscalibrations. Overall, the reproduction of the peaks in the [200-400] keV region is generally satisfactory. The worst performance is actually observed for D8, where the first peak could not be reproduced more accurately.

\begin{figure}[t!]
    \centering
    {\includegraphics[width=0.45\textwidth]{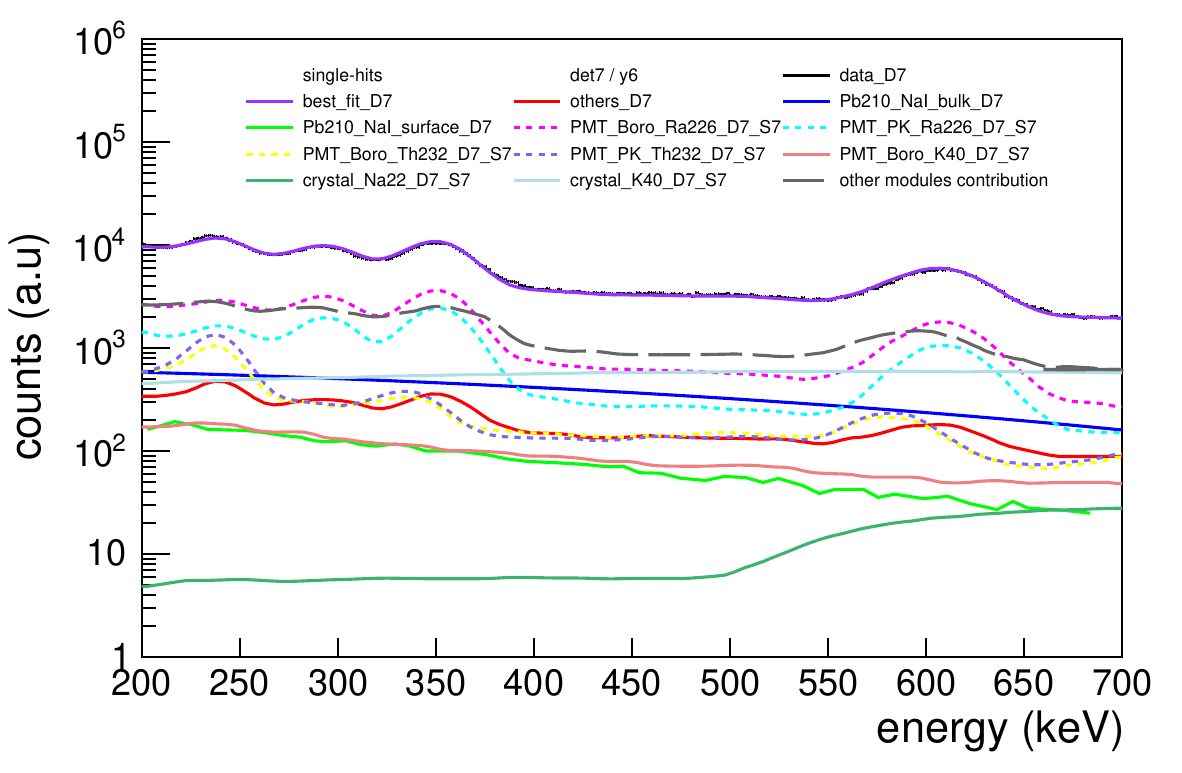}}
        \hfill
    {\includegraphics[width=0.45\textwidth]{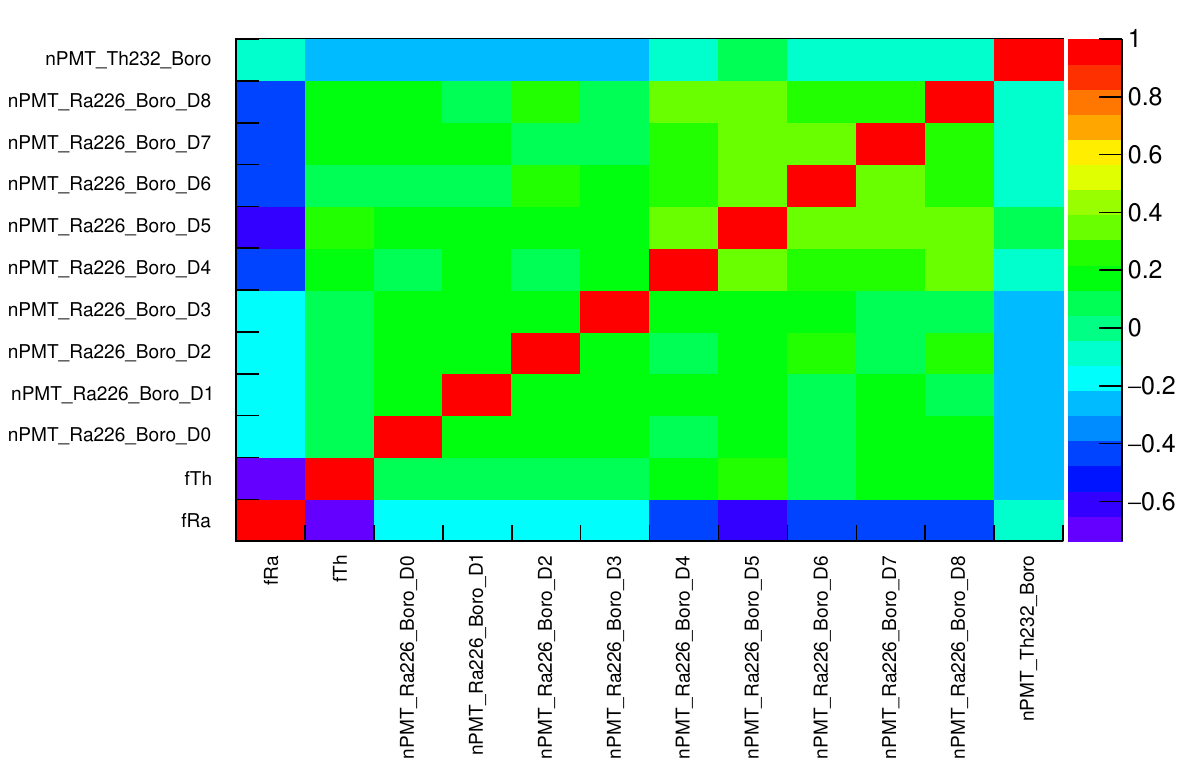}}

    \caption{\label{PMTspop0porcomponentes} \textbf{Left panel:} Breakdown of the single-hits high-energy fit components for D7 module. The exponential profile of 10 $\mu$m is considered for the superficial contamination of \(^{210}\)Pb. The contaminations from module D7 are shown separately, with dashed lines representing the components left free in the fit, and solid lines corresponding to the fixed components. The contributions from the remaining modules are grouped into a single component for clarity in the visualization, shown as a gray long dash-dotted line. The best fit is displayed in violet. \textbf{Right panel:} Correlation matrix from the simultaneous high-energy fit for the nine detectors, including single-hits, m2-hits, and m3m4-hits. }
\end{figure}

\begin{figure}[b!]
\begin{center}
\includegraphics[width=1\textwidth]{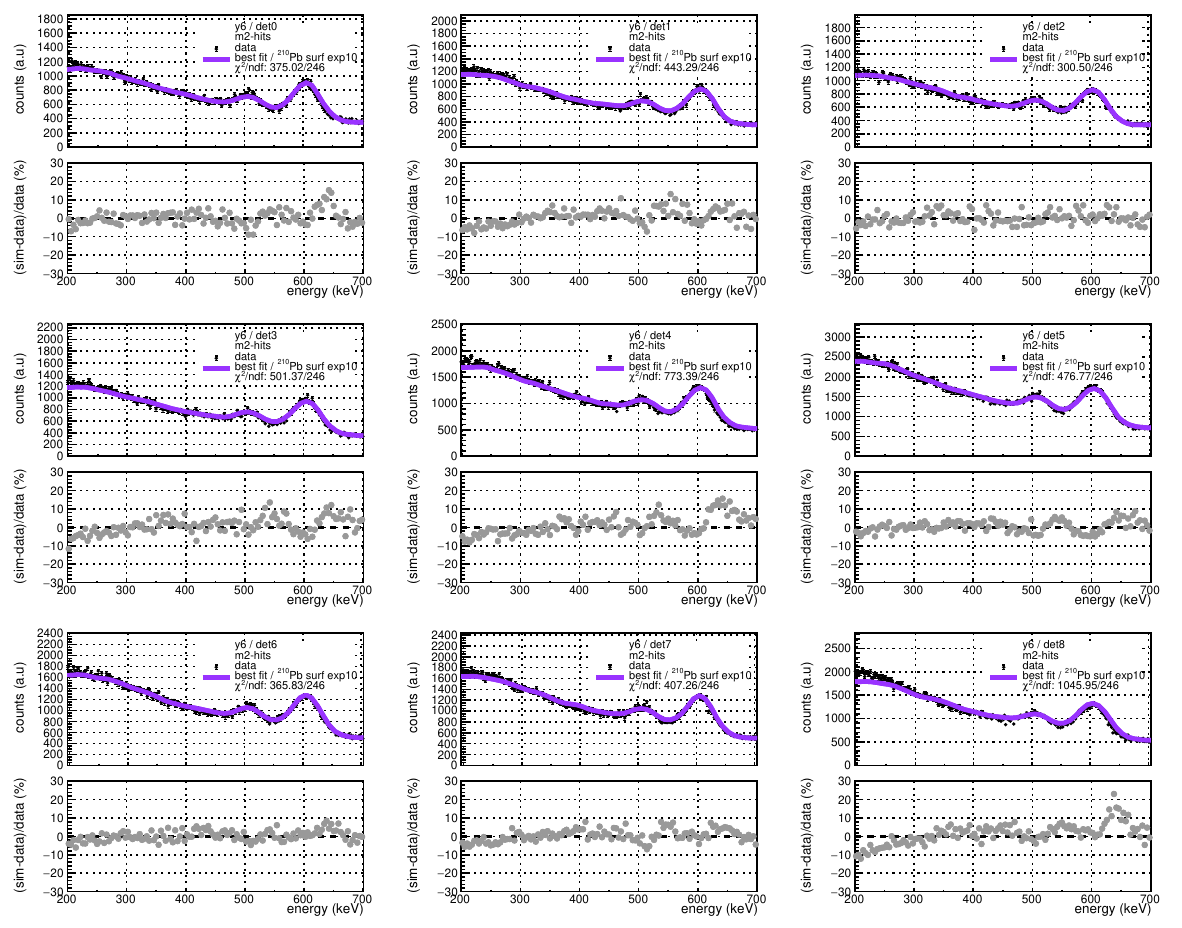}

\caption{\label{HEpop1} Analogous to Figure \ref{HEpop0}, but for m2-hits. }
\end{center}
\end{figure}

\begin{figure}[t!]
\begin{center}
\includegraphics[width=1\textwidth]{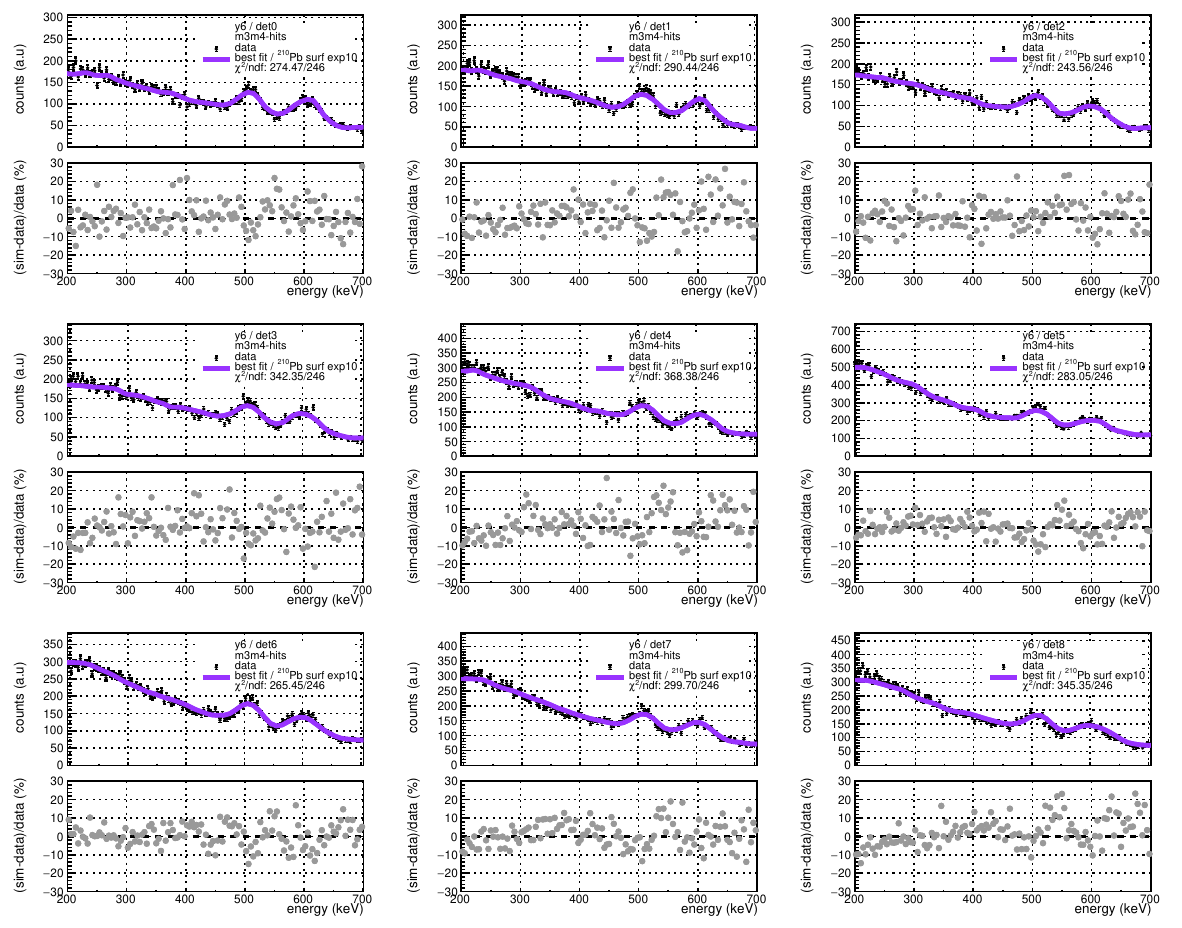}

\caption{\label{HEpop2} Analogous to Figure \ref{HEpop0}, but for m3m4-hits. }
\end{center}
\end{figure}

The left panel of Figure \ref{PMTspop0porcomponentes} presents the breakdown of the single-hit fit components for D7 module. The contaminations from module D7 are shown separately, while the contributions from the remaining modules are grouped into a single component for clarity in the visualization. In single-hit events, contaminations from the PMTs of the corresponding module dominate, although the contribution from others is significant and comparable when considered jointly. As shown in the figure, the dominant component is \(^{226}\)Ra, either in the borosilicate or the photocathode, followed by \(^{232}\)Th in both components. 

Moreover, the right panel of Figure \ref{PMTspop0porcomponentes} displays the correlation matrix corresponding to the high-energy fit. The off-diagonal elements are predominantly centered around zero, indicating weak correlations among the majority of the fit parameters. Certain parameter pairs, particularly within the subset of fractional activities and $^{226}$Ra activities, exhibit notable anticorrelations in the range –0.5 to –0.7. The observed block-diagonal structure highlights the robustness and competitiveness of the fit.

\begin{figure}[t!]
    \centering
    {\includegraphics[width=0.47\textwidth]{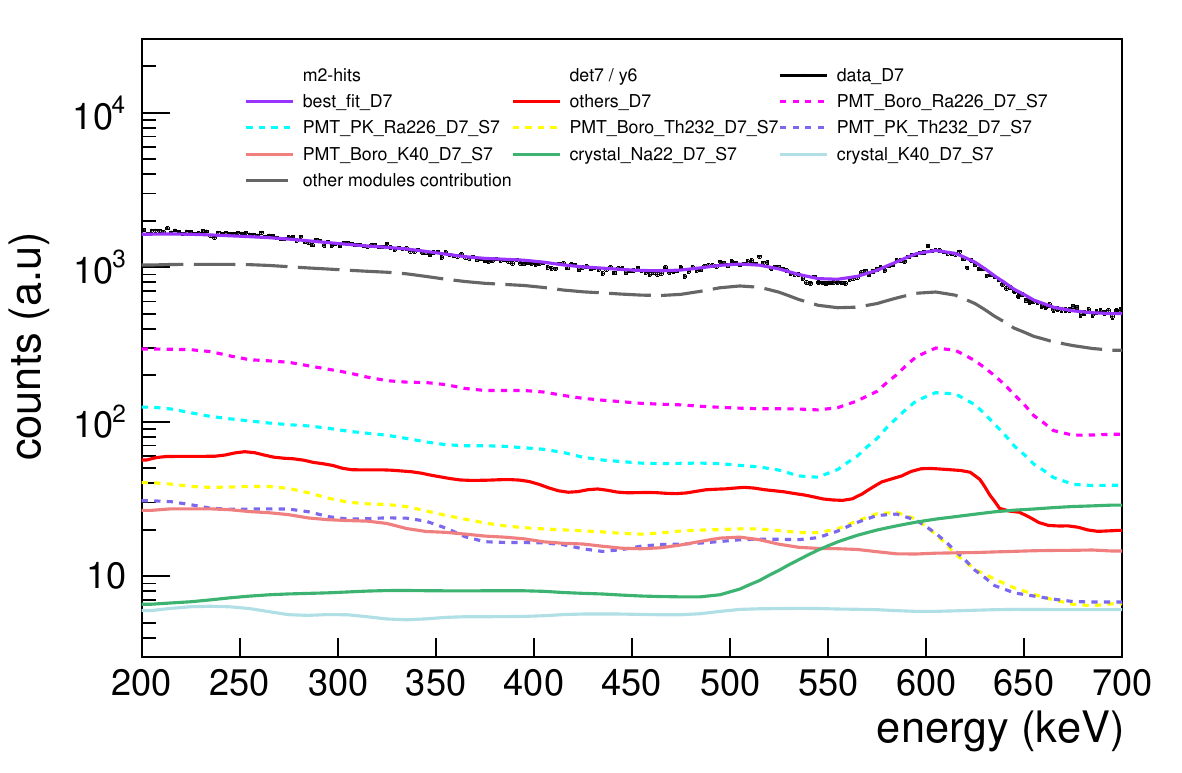}}
    \hfill
    {\includegraphics[width=0.47\textwidth]{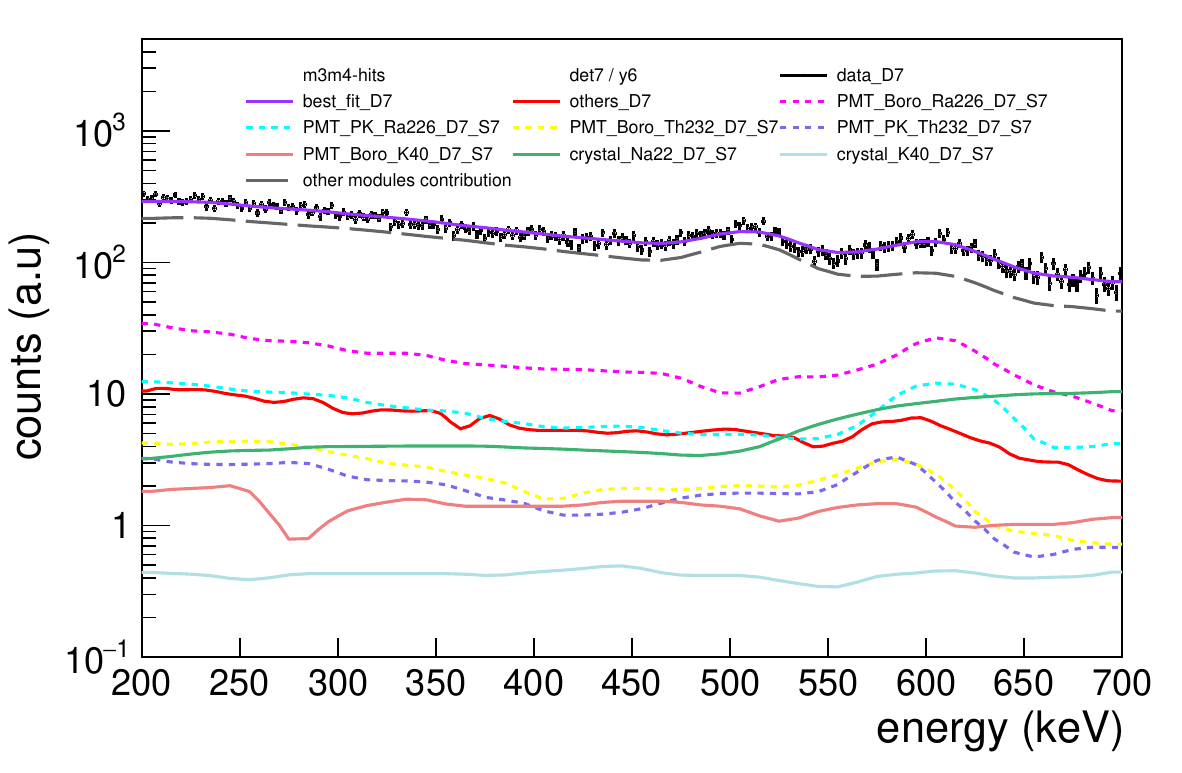}}

    \caption{\label{PMTspop12porcomponentes} Breakdown of the fit components for D7 module following the high-energy fit. The best fit is displayed in violet. \textbf{Left panel:} m2-hits. \textbf{Right panel:} m3m4-hits. The contaminations from module D7 are shown separately, with dashed lines representing the components left free in the fit, and solid lines corresponding to the fixed components. The contributions from the remaining modules are grouped into a single component for clarity in the visualization, shown as a gray long dash-dotted line. }
\end{figure}

Similarly, Figures \ref{HEpop1} and \ref{HEpop2} show the fit results for the m2-hits and m3m4-hits, respectively. Again, the fit performance in both populations and the residuals are adquate. The multiple-hit spectra are well reproduced, which ensures that the contamination in the PMTs is being properly determined. Figure \ref{PMTspop12porcomponentes} provides a breakdown of the fit components for these populations. In this case, unlike in the single-hit events (see Figure~\ref{PMTspop0porcomponentes}), the dominant component is clearly the contribution from modules other than the detector whose energy spectrum is being shown. Specifically, the largest contribution originates from the PMTs of the other modules. This makes a simultaneous fit of all nine detectors essential to accurately determine the PMT activities, as these are correlated across detectors.


From the fit, the activities of \(^{226}\)Ra and \(^{232}\)Th are determined according to Equation~\ref{EqA0}. The corresponding results for the different surface profiles of \(^{210}\)Pb contamination are presented in Table~\ref{resultadosdelpmt}.

As shown, the fraction of contamination attributed to the borosilicate varies significantly depending on the assumed depth profile of the superficial \(^{210}\)Pb. Based on the measurements performed on the PMT using the HPGe detector in the course of this thesis (see Figure~\ref{hpgepmt}), it is not expected that the total contamination from these isotopes resides exclusively in either the borosilicate or the photocathode; rather, a uniform distribution between both components can be reasonably assumed. In this regard, the results obtained for the 10~\(\mu\)m exponential profile appear to be more physically motivated, as the fractional activities derived for both \(^{226}\)Ra and \(^{232}\)Th fall within the intermediate range, avoiding the extremes of complete localization in either material. In addition, more homogeneous contamination for the different PMTs are obtained, compatible with the HPGe screening results \cite{amare2019analysis}.

\begin{table}[t!]
    \centering
    
    \begin{minipage}{\textwidth}

    \resizebox{\textwidth}{!}{\Large
    \begin{tabular}{c|ccc|ccc}
    \hline
  \multirow{2}{*}{detector} & \multicolumn{6}{c}{$^{210}$Pb surf exp1  } \\
  \cline{2-7}
        &  f$_{\text{Boro},^{226}\mathrm{Ra}}$ & \textsuperscript{226}Ra$_{\textnormal{Boro}}$ (mBq) & \textsuperscript{226}Ra$_{\textnormal{PK}}$ (mBq) &  f$_{\text{Boro},^{232}\mathrm{Th}}$ & \textsuperscript{232}Th$_{\textnormal{Boro}}$ (mBq) & \textsuperscript{232}Th$_{\textnormal{PK}}$ (mBq)   \\
        \hline
        0 & \multirow{9}{*}{0.9456 $\pm$ 0.0050} & 163.52 $\pm$ 0.43 & 3.93 $\pm$ 0.01 & \multirow{9}{*}{0.0000 $\pm$ 0.0026} & \multirow{9}{*}{0.0000 $\pm$ 0.0000} & \multirow{9}{*}{13.38 $\pm$ 0.07} \\
        1 & ~ & 304.96 $\pm$ 0.42 & 7.40 $\pm$ 0.01~ \\
        2 & ~ & 171.03 $\pm$ 0.39 & 4.11 $\pm$ 0.01~ \\
        3 & ~ & 181.73 $\pm$ 0.41 & 4.40 $\pm$ 0.01~ \\
        4 & ~ & 141.42 $\pm$ 0.48 & 3.38 $\pm$ 0.01~ \\
        5 & ~ & 101.18 $\pm$ 0.64 & 2.39 $\pm$ 0.02~ \\
        6 & ~ & 144.01 $\pm$ 0.48 & 3.44 $\pm$ 0.01~ \\
        7 & ~ & 149.84 $\pm$ 0.45 & 3.59 $\pm$ 0.01~ \\
        8 & ~ & 149.80 $\pm$ 0.47 & 3.60 $\pm$ 0.01~ \\
        \hline
    \end{tabular}}
    \subcaption{}
    \end{minipage}
    \vskip 0.5 cm

    \begin{minipage}{\textwidth}

    \resizebox{\textwidth}{!}{\Large
    \begin{tabular}{c|ccc|ccc}
    \hline
  \multirow{2}{*}{detector} & \multicolumn{6}{c}{$^{210}$Pb surf exp10  } \\
  \cline{2-7}
        &  f$_{\text{Boro},^{226}\mathrm{Ra}}$ & \textsuperscript{226}Ra$_{\textnormal{Boro}}$ (mBq) & \textsuperscript{226}Ra$_{\textnormal{PK}}$ (mBq) &  f$_{\text{Boro},^{232}\mathrm{Th}}$ & \textsuperscript{232}Th$_{\textnormal{Boro}}$ (mBq) & \textsuperscript{232}Th$_{\textnormal{PK}}$ (mBq)   \\
        \hline
        0 & \multirow{9}{*}{0.6383 $\pm$ 0.0069} & 116.01 $\pm$ 0.29 & 27.43 $\pm$ 0.07 & \multirow{9}{*}{0.4774 $\pm$ 0.0226} & \multirow{9}{*}{15.1094 $\pm$ 0.0843} & \multirow{9}{*}{6.77 $\pm$ 0.04} \\
        1 & ~ & 117.16 $\pm$ 0.28 & 28.00 $\pm$ 0.07~ \\
        2 & ~ & 118.07 $\pm$ 0.26 & 27.94 $\pm$ 0.06~ \\
        3 & ~ & 125.78 $\pm$ 0.28 & 30.01 $\pm$ 0.07~ \\
        4 & ~ & 105.85 $\pm$ 0.32 & 24.93 $\pm$ 0.08~ \\
        5 & ~ & 91.29 $\pm$ 0.40 & 21.21 $\pm$ 0.09~ \\
        6 & ~ & 117.79 $\pm$ 0.30 & 27.72 $\pm$ 0.07~ \\
        7 & ~ & 112.25 $\pm$ 0.30 & 26.50 $\pm$ 0.07~ \\
        8 & ~ & 120.44 $\pm$ 0.29 & 28.51 $\pm$ 0.07~ \\
        \hline
    \end{tabular}}
    \subcaption{}
    \end{minipage}
    \vskip 0.5 cm

    \begin{minipage}{\textwidth}

    \resizebox{\textwidth}{!}{\Large
    \begin{tabular}{c|ccc|ccc}
    \hline
  \multirow{2}{*}{detector} & \multicolumn{6}{c}{$^{210}$Pb surf exp100 } \\
  \cline{2-7}
        &  f$_{\text{Boro},^{226}\mathrm{Ra}}$ & \textsuperscript{226}Ra$_{\textnormal{Boro}}$ (mBq) & \textsuperscript{226}Ra$_{\textnormal{PK}}$ (mBq) &  f$_{\text{Boro},^{232}\mathrm{Th}}$ & \textsuperscript{232}Th$_{\textnormal{Boro}}$ (mBq) & \textsuperscript{232}Th$_{\textnormal{PK}}$ (mBq)   \\
        \hline
        0 & \multirow{9}{*}{0.9526 $\pm$ 0.0054} & 228.68 $\pm$ 0.43 & 4.74 $\pm$ 0.01 & \multirow{9}{*}{1.0000 $\pm$ 0.0014} & \multirow{9}{*}{29.9401 $\pm$ 0.1650} & \multirow{9}{*}{0.00 $\pm$ 0.00} \\
        1 & ~ & 236.09 $\pm$ 0.43 & 4.95 $\pm$ 0.01~ \\
        2 & ~ & 193.29 $\pm$ 0.39 & 4.01 $\pm$ 0.01~ \\
        3 & ~ & 244.07 $\pm$ 0.42 & 5.11 $\pm$ 0.01~ \\
        4 & ~ & 189.73 $\pm$ 0.47 & 3.92 $\pm$ 0.01~ \\
        5 & ~ & 65.61 $\pm$ 0.70 & 1.34 $\pm$ 0.01~ \\
        6 & ~ & 129.57 $\pm$ 0.49 & 2.68 $\pm$ 0.01~ \\
        7 & ~ & 127.58 $\pm$ 0.48 & 2.64 $\pm$ 0.01~ \\
        8 & ~ & 158.33 $\pm$ 0.45 & 3.29 $\pm$ 0.01~ \\
        \hline
    \end{tabular}}
    \subcaption{}
    \end{minipage}
    \caption{\label{resultadosdelpmt}Activity of the ANAIS-112 PMTs obtained from the fit. Values are expressed in mBq and refer to the contamination of the two PMTs of each module. The fraction of contamination attributed to the borosilicate by the fit, together with the corresponding activity in the borosilicate and the photocathode (here abbreviated as 'Boro' and 'PK', respectively), are displayed. Results are provided for the three exponential $^{210}$ Pb profiles considered in this work.
\textbf{(a)} 1 $\mu$m. \textbf{(b)} 10 $\mu$m. \textbf{(c)} 100 $\mu$m.}
\end{table}

\begin{figure}[t!]
    \centering
    {\includegraphics[width=0.45\textwidth]{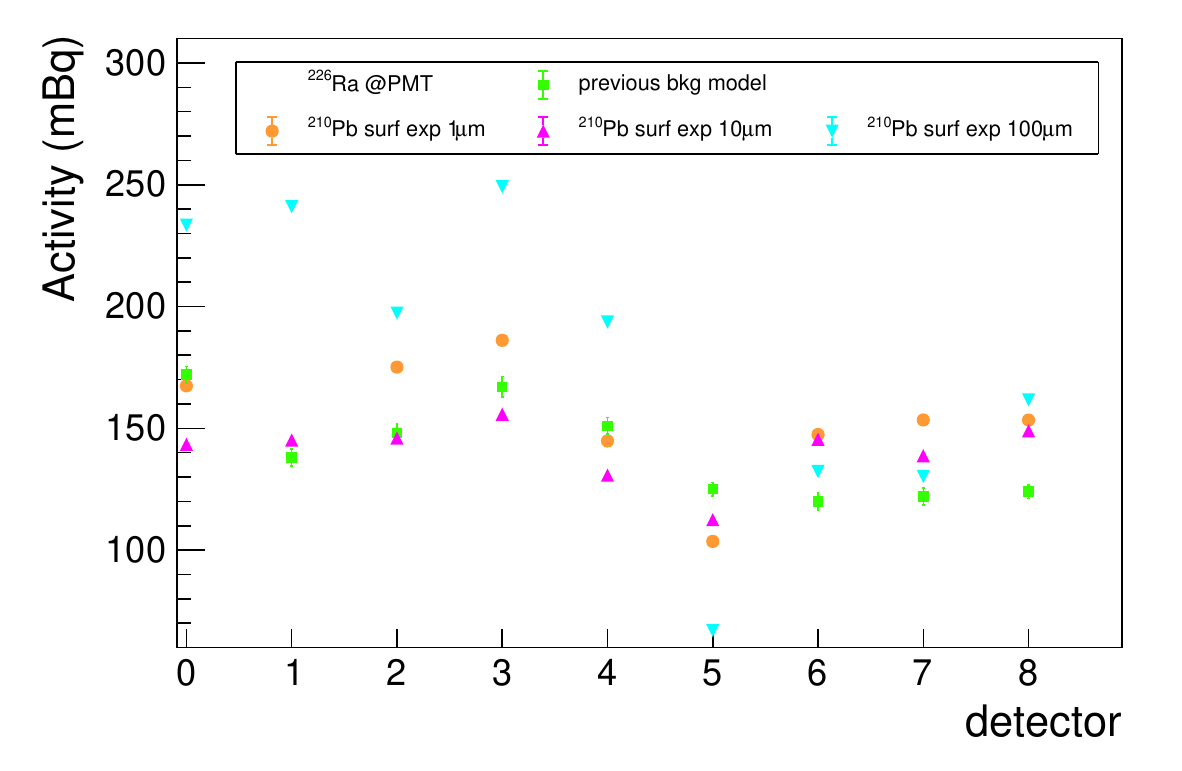}}
    \hfill
    {\includegraphics[width=0.45\textwidth]{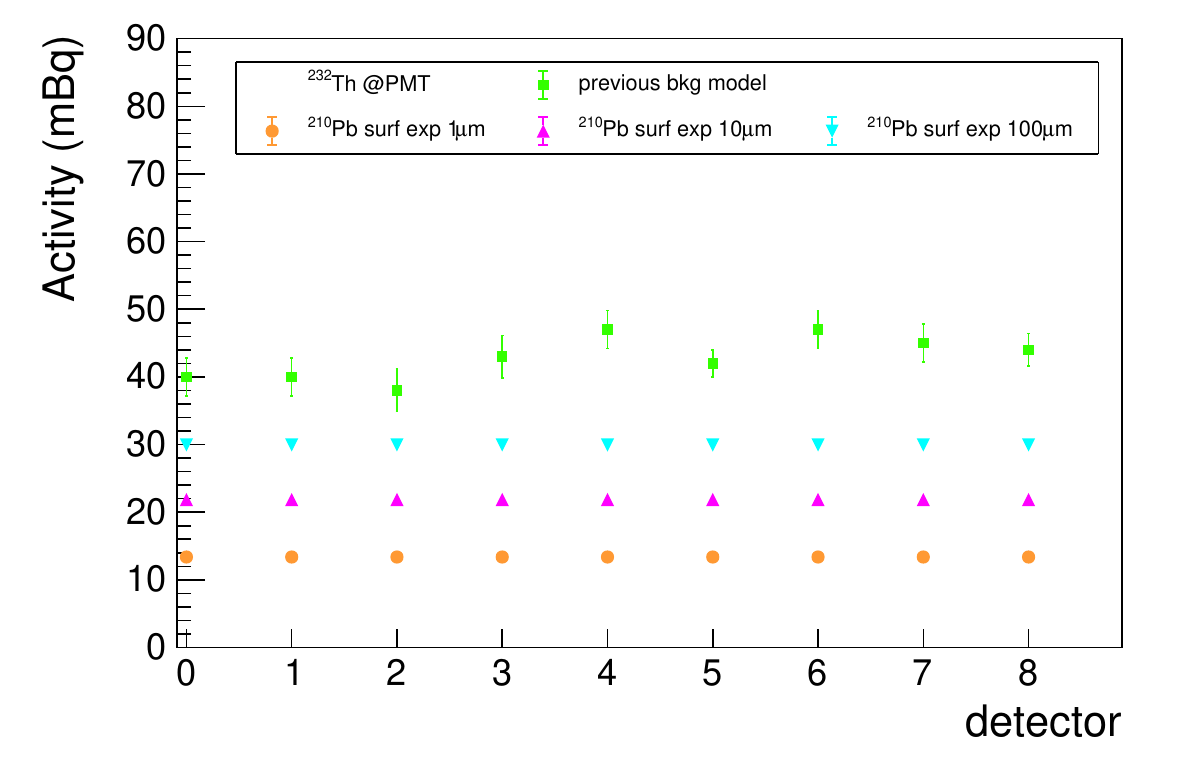}}
    
    \caption{\label{resultadosPMTsFigure} Comparison of the activity of the ANAIS-112 PMTs for $^{226}$Ra \textbf{(left panel)} and $^{232}$Th \textbf{(right panel)} obtained in this work, assuming three different exponential depth profiles for the $^{210}$Pb surface contamination (1, 10, and 100 $\mu$m, shown in orange, magenta, and cyan, respectively), with those reported in the previous ANAIS-112 background model (shown in green). All values are expressed in mBq and refer to the total contamination of the two PMTs coupled to each detector module considering both, photocathode and borosilicate components. }
    
\end{figure}

Figure \ref{resultadosPMTsFigure} shows the fit results obtained in this work for both \(^{226}\)Ra and \(^{232}\)Th, in comparison with the activities measured using the HPGe detector during the commissioning of ANAIS-112, and adopted in the previous model. The higher efficiency of photocathode contamination, as discussed in Section~\ref{comparePMTsimulation}, is particularly evident for \(^{232}\)Th (right panel). For the 1 \(\mu\)m exponential profile, the fit places the entire \(^{232}\)Th contamination in the photocathode (see Table \ref{resultadosdelpmt}), resulting in lower activity values than those measured. In contrast, when the 100 \(\mu\)m exponential profile is assumed, placing the contamination entirely in the borosilicate, the fitted values approach the measured ones, consistent with the hypothesis of a homogeneous distribution. It should be emphasized that the measured values are only indicative of the order of magnitude, as they cannot be directly compared with the fitted activities due to the differing assumptions regarding the spatial distribution of the contamination. In the case of \(^{226}\)Ra (left panel), it is worth noting that for the 100~\(\mu\)m profile, the activity associated with the first-arrived detectors is significantly higher than both the measured value and the fitted results obtained with the other exponential profiles, which is a hint of an incorrect modelling of the background, given that the PMTs from all modules are similar and were chosen randomly.

\begin{figure}[t!]
    \centering
    {\includegraphics[width=0.85\textwidth]{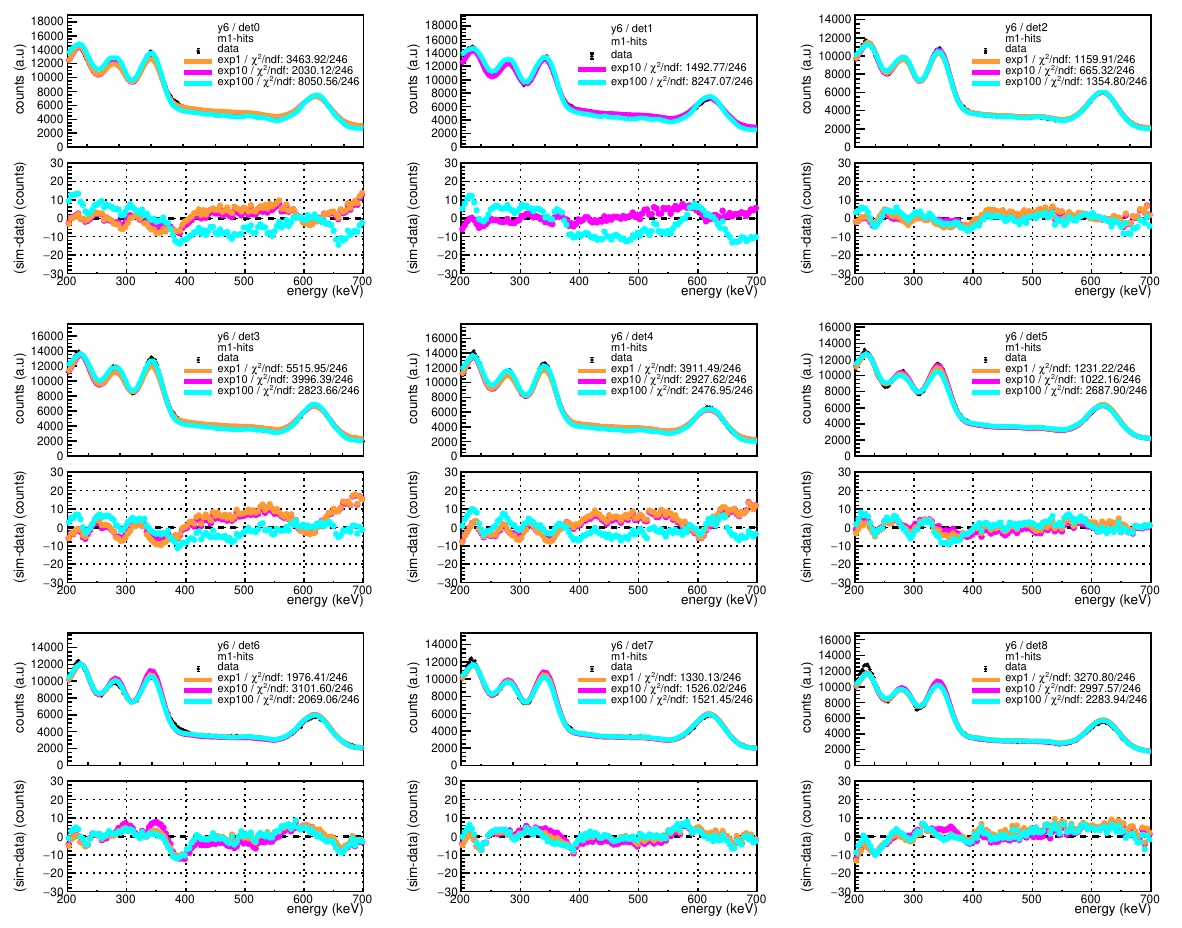}}    \caption{\label{comapraexpsHE} Comparison of the high-energy best fits for the single-hit population across the nine ANAIS-112 detectors, considering different exponential mean depth profiles for the $^{210}$Pb surface contamination. The profiles for 1 $\mu$m, 10 $\mu$m, and 100 $\mu$m contamination depths are represented in orange, magenta, and cyan, respectively. Data points are shown in black. The goodness of the fit and the corresponding residuals are also presented in the panels. }
\end{figure}

Figure~\ref{comapraexpsHE} shows the best fit results obtained for the different exponential depth profiles. In the specific case of the 1 \(\mu\)m exponential for detector D1, the best fit is not displayed in the plot, as the fit fails to converge. For the remaining detectors, all tested profiles yield similarly good fits, and no definitive preference can be established at this stage based solely on fit quality. However, it is important to note that the distribution of contamination between the borosilicate and the photocathode is more physically reasonable when assuming the 10 \(\mu\)m exponential profile, in contrast to the other two cases. Nevertheless, the subsequent steps of the fit will be performed using each of the three proposed benchmarks for the surface contamination of $^{210}$Pb, 1, 10, and 100 $\mu$m, in order to assess whether any particular mean depth improves the overall agreement across all energy regions.

As previously discussed and justified, the high-energy fit has been performed within the [200–700] keV range, with the objective of accurately reproducing the most prominent peaks that are easily identifiable within this region. Consequently, at this stage, no conclusions can be drawn regarding the model performance above the upper limit of the fit range.

It is important to recall that this fit incorporates the energy spectrum derived from the \(\beta^-\) decay of \(^{210}\)Bi, based on the experimental BetaShape parametrization. As will be shown in Figure~\ref{totalmodelHE}, the new \(\beta^-\) spectral shape exhibits significantly better agreement with the measured data. This not only improves the description of the high-energy region but also effectively constrains the spectral shape of this poorly characterized $\beta$ emission using ANAIS-112 data.

\subsection{Medium-energy fitting}\label{mediumenergyfit}

After completing the high-energy fit, the resulting activities of $^{226}$Ra and $^{232}$Th in the PMTs are fixed, the fitting procedure continues with the analysis of the medium-energy region. Referring back to Table~\ref{componentesdelfit}, this stage of the fit aims to determine the $^{210}$Pb contamination in the NaI(Tl) crystals, both in the bulk and at the surface, as well as the activity of $^{129}$I and $^{109}$Cd. A simultaneous fit is performed for each detector, combining the single-hit energy spectrum with the spectrum resulting from the subtraction of the sixth-year data from that of the third year.

\begin{figure}[t!]
\begin{center}
\includegraphics[width=1.\textwidth]{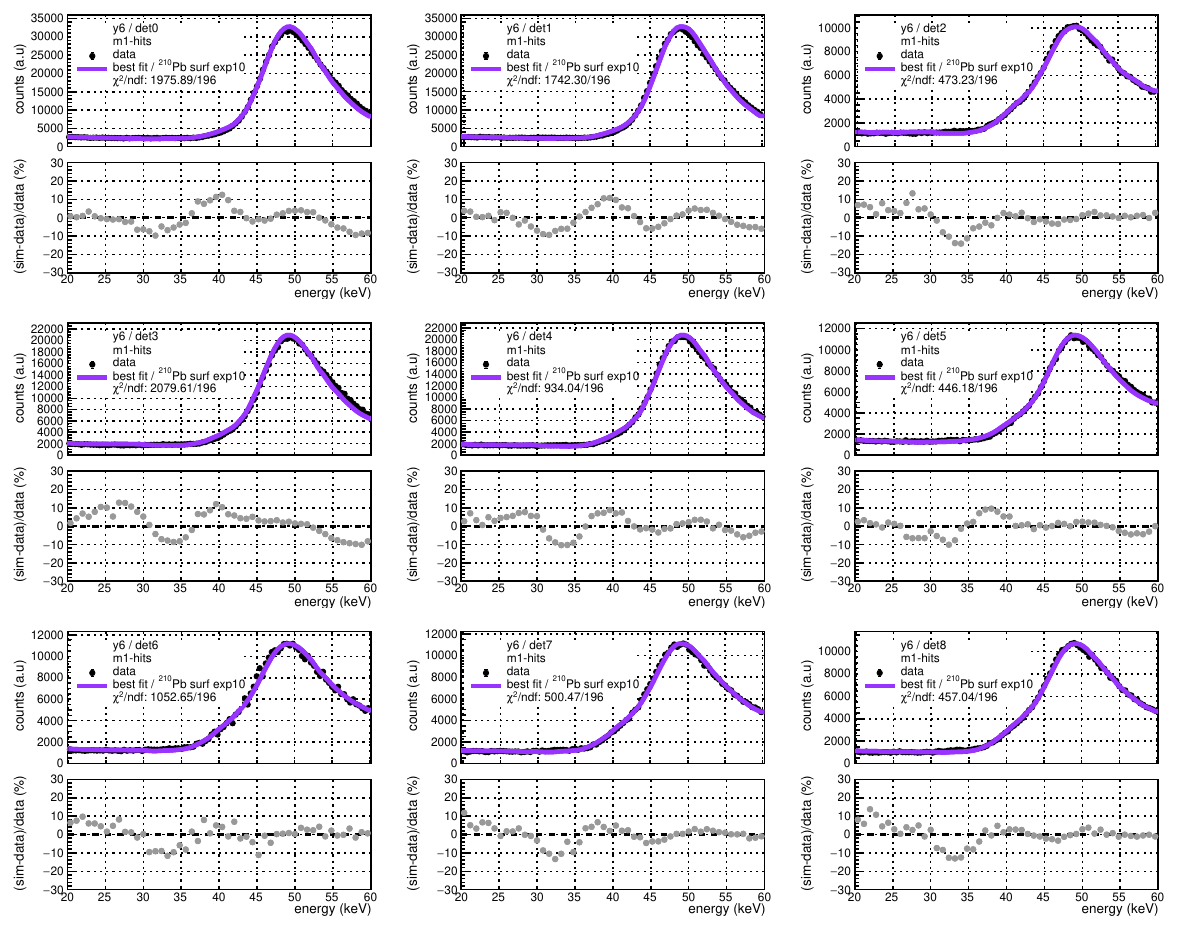}

\caption{\label{MEpop0} Medium-energy best fits of the single-hit spectra for the nine ANAIS-112 detectors. Black points correspond to data from the full sixth year of ANAIS-112, while the best fit is shown in violet. The exponential profile of 10 $\mu$m is considered for the superficial contamination of \(^{210}\)Pb. Residuals are displayed as grey points, and the goodness-of-fit is indicated in each panel. }
\end{center}
\end{figure}
\begin{figure}[b!]
    \centering
    {\includegraphics[width=0.6\textwidth]{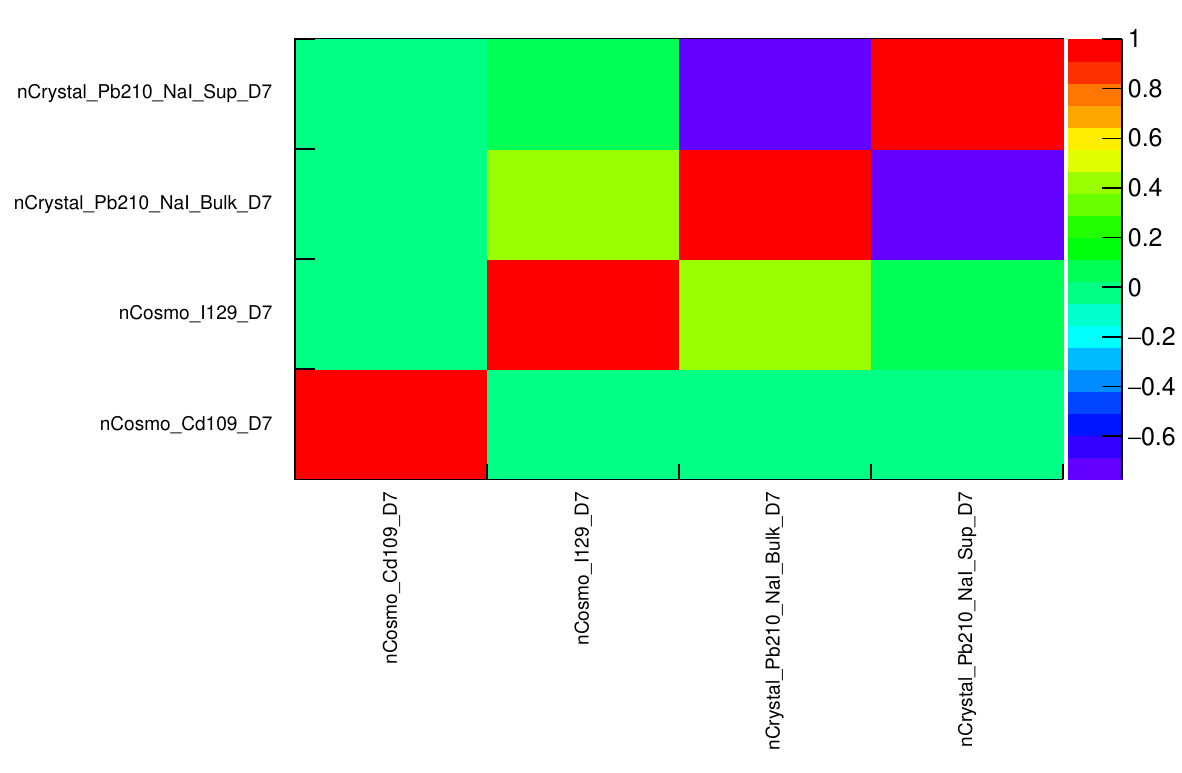}}
    \caption{\label{corrmatrixme} Correlation matrix from the simultaneous medium-energy fit for D7 module. }
\end{figure}

Figure~\ref{MEpop0} presents the results of the medium-energy fit of the single-hit population for the nine ANAIS-112 detectors. The overall agreement between data and the best fit is satisfactory, with residuals remaining below 20\% in all cases. Nevertheless, a structured deviation is observed in the residuals around 30–35 keV, where the experimental spectral shape is not fully reproduced. The discrepancy between data and simulation is more pronounced in detectors D0, D1, D3, and D4, which exhibit a higher $^{210}$Pb content. In these cases, the fit also fails to accurately reproduce the tail of this complex peak (below $\sim$ 50 keV), whereas in the other detectors, characterized by lower lead contamination, a better agreement is generally achieved in this energy region. This deviation is indicative of a systematic effect related to the $^{210}$Pb contamination affecting this energy region that is not yet fully understood. Figure~\ref{corrmatrixme} shows the correlation matrix for D7, demonstrating that, despite a certain correlation between the $^{210}$Pb contamination in the different components, the fit remains competitive.

\begin{figure}[t!]
\begin{center}
\includegraphics[width=0.55\textwidth]{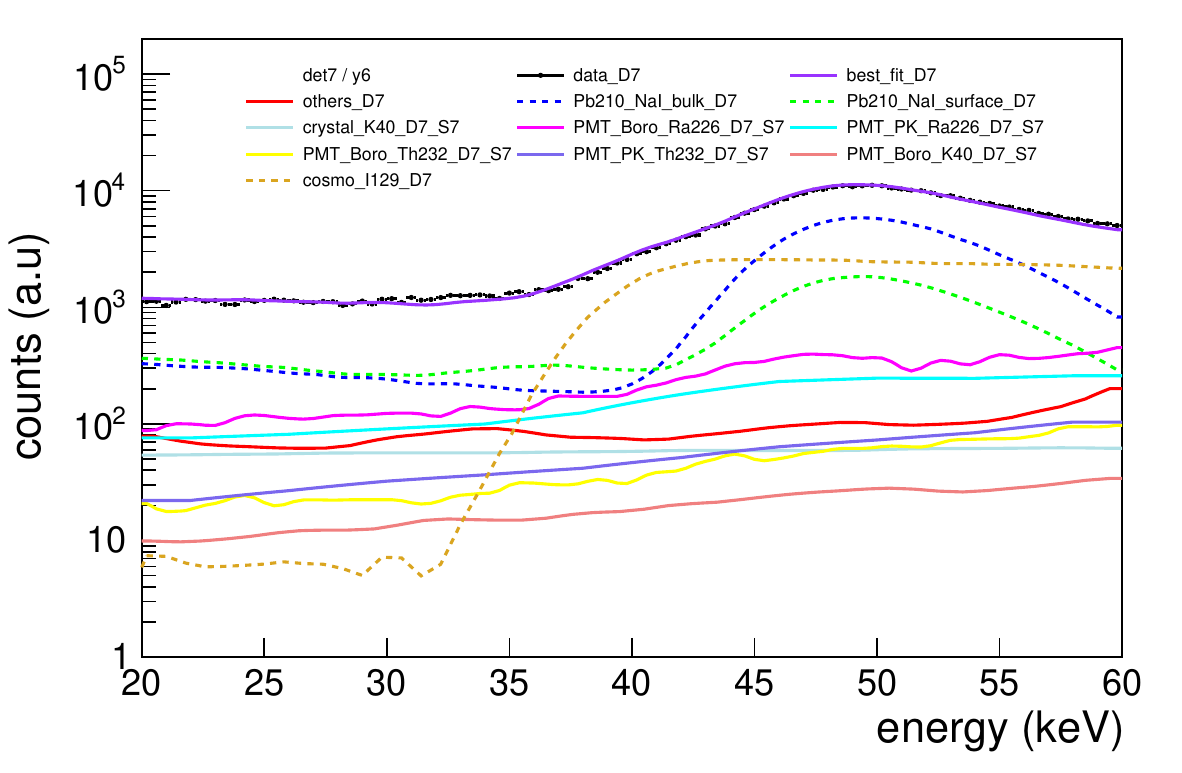}

\caption{\label{desgloseMEpop0} Breakdown of the individual components contributing to the medium-energy fit for the D7 detector. The figure displays the single-hit event population, with dashed lines representing the components left free in the fit, and solid lines corresponding to the fixed components. The best fit is displayed in violet.   }
\end{center}
\end{figure}

Figure~\ref{desgloseMEpop0} shows the breakdown of the fit components for detector D7. As previously discussed, the model fails to reproduce the spectral shape in the 30–35 keV region, which is visible in this figure. Moreover, the figure highlights the contribution of the surface contamination of $^{210}$Pb which increases towards the low-energy region.

\begin{figure}[t!]
\begin{center}
\includegraphics[width=0.95\textwidth]{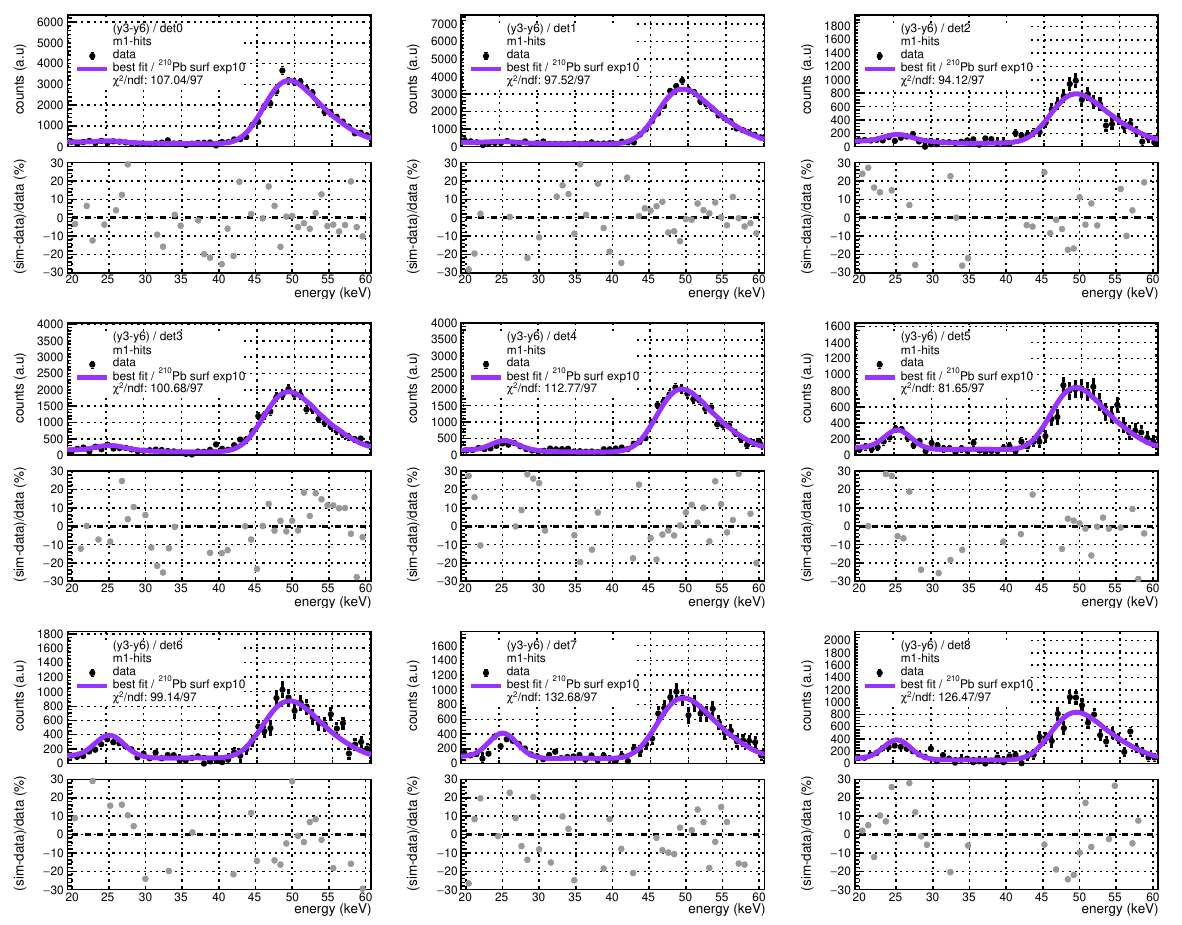}
\caption{\label{MEpop1} Medium-energy best fits of the single-hit spectra corresponding to the subtraction of sixth-year data from third-year data for the nine ANAIS-112 detectors. Black points correspond to data, while the best fit is shown in violet. The exponential profile of 10 $\mu$m is considered for the superficial contamination of \(^{210}\)Pb. Residuals are displayed as grey points, and the goodness-of-fit is indicated in each panel.  }
\end{center}
\end{figure}
 
The robustness of the fit is evident in Figure~\ref{MEpop1}, which presents the results of the fit to the spectrum obtained from the difference between the third and sixth years of data. In this figure, the residuals are shown for completeness. However, the data resulting from the temporal difference have larger statistical uncertainties due to reduced statistics. As a result, even small deviations between the model and the data can appear large in relative terms, giving the impression of highly scattered residuals.

The decay of $^{210}$Pb is well captured, with the exception of detector D8, where it appears slightly underestimated. The $^{109}$Cd peak around 25 keV is also satisfactorily described by the model. In general, the $^{210}$Pb contamination in the ANAIS data is accurately reproduced, which is crucial, as this isotope is one of the most significant contributors to the experiment background at both low and high energies.

\begin{table}[h!]

\centering

    \begin{minipage}{0.76\textwidth}
    
    \centering
    \resizebox{\textwidth}{!}{\Large
    \begin{tabular}{c|ccc}
        \hline
        \multirow{2}{*}{detector} & \multicolumn{3}{c}{$^{210}$Pb surf exp1 }  \\
        \cline{2-4}
        & $^{210}$Pb bulk (mBq/kg) & $^{210}$Pb surf (mBq/kg) & $\alpha$-rate (mBq/kg)  \\
        \hline
        0 & 3.047 $\pm$ 0.007 & 0.221 $\pm$ 0.004 & - \\
        1 & 0.000 $\pm$ 0.000 & 1.851 $\pm$ 0.004 & - \\
        2 & 0.667 $\pm$ 0.004 & 0.000 $\pm$ 0.000 & - \\
        3 & 1.708 $\pm$ 0.005 & 0.000 $\pm$ 0.000 &  1.581 \\
        4 & 1.628 $\pm$ 0.004 & 0.000 $\pm$ 0.000 & 1.507 \\
        5 & 0.665 $\pm$ 0.004 & 0.206 $\pm$ 0.003 & 0.800 \\
        6 & 0.655 $\pm$ 0.004 & 0.143 $\pm$ 0.003 & 0.729 \\
        7 & 0.653 $\pm$ 0.004 & 0.140 $\pm$ 0.003 & 0.722 \\
        8 & 0.634 $\pm$ 0.004 & 0.091 $\pm$ 0.003 & 0.662 \\
        \hline
    \end{tabular}}

    \label{tab:exp_data}\subcaption{}
    \end{minipage}
    \vskip 0.5cm

        \begin{minipage}{0.76\textwidth}
    
    \centering
    \resizebox{\textwidth}{!}{\Large
    \begin{tabular}{c|ccc}
        \hline
        \multirow{2}{*}{detector} & \multicolumn{3}{c}{$^{210}$Pb surf exp10 }  \\
        \cline{2-4}
        & $^{210}$Pb bulk (mBq/kg) & $^{210}$Pb surf (mBq/kg) & $\alpha$-rate (mBq/kg)  \\
        \hline
        0 & 2.851 $\pm$ 0.009 & 0.374 $\pm$ 0.008 & - \\
        1 & 2.837 $\pm$ 0.009 & 0.399 $\pm$ 0.008 & - \\
        2 & 0.664 $\pm$ 0.004 & 0.000 $\pm$ 0.000 & - \\
        3 & 1.706 $\pm$ 0.005 & 0.000 $\pm$ 0.000 & 1.579 \\
        4 & 1.624 $\pm$ 0.004 & 0.000 $\pm$ 0.000 & 1.503 \\
        5 & 0.485 $\pm$ 0.006 & 0.342 $\pm$ 0.005 & 0.759\\
        6 & 0.511 $\pm$ 0.006 & 0.260 $\pm$ 0.005 & 0.705 \\
        7 & 0.578 $\pm$ 0.006 & 0.221 $\pm$ 0.005 & 0.728 \\
        8 & 0.538 $\pm$ 0.006 & 0.168 $\pm$ 0.005 & 0.644\\
        \hline
    \end{tabular}}

    \label{tab:exp_data}\subcaption{}
    \end{minipage}
    \vskip 0.5cm

        \begin{minipage}{0.76\textwidth}
    
    \centering
    \resizebox{\textwidth}{!}{\Large
    \begin{tabular}{c|ccc}
        \hline
        \multirow{2}{*}{detector} & \multicolumn{3}{c}{$^{210}$Pb surf exp100 }  \\
        \cline{2-4}
        & $^{210}$Pb bulk (mBq/kg) & $^{210}$Pb surf (mBq/kg) & $\alpha$-rate (mBq/kg)  \\
        \hline
        0 & 0.000 $\pm$ 0.000 & 3.132 $\pm$ 0.006 & - \\
        1 & 0.000 $\pm$ 0.000 & 3.159 $\pm$ 0.006 & - \\
        2 & 0.000 $\pm$ 0.000 & 0.668 $\pm$ 0.004 & - \\
        3 & 0.000 $\pm$ 0.000 & 1.681 $\pm$ 0.005 & 1.556 \\
        4 & 0.000 $\pm$ 0.000 & 1.610 $\pm$ 0.004 & 1.490 \\
        5 & 0.000 $\pm$ 0.001 & 0.811 $\pm$ 0.004 & 1.059 \\
        6 & 0.000 $\pm$ 0.002 & 0.759 $\pm$ 0.004 & 0.932 \\
        7 & 0.000 $\pm$ 0.000 & 0.754 $\pm$ 0.004 & 0.888 \\
        8 & 0.000 $\pm$ 0.000 & 0.697 $\pm$ 0.003 & 0.789 \\
        \hline
    \end{tabular}}

    \label{tab:exp_data}\subcaption{}
    \end{minipage}

    \caption{\label{resultadosdeME} Initial $^{210}$Pb activities in the crystal, both in the bulk and at the surface, at the time of detector installation underground, as determined from the fit. The $\alpha$-rate is also reported, calculated following Equation~\ref{eqEq} and accounting for both bulk and surface \textsuperscript{210}Pb contributions, at the start of ANAIS-112 data taking on August 3, 2017. Values are expressed in mBq/kg for the nine ANAIS-112 detectors. Results are provided for the three exponential $^{210}$ Pb profiles considered in this work.
\textbf{(a)} 1 $\mu$m. \textbf{(b)} 10 $\mu$m. \textbf{(c)} 100 $\mu$m. }
    
\end{table}

Following Equation~\ref{EqA0}, the results of the medium-energy fit are obtained, as shown in Table~\ref{resultadosdeME}, which presents the bulk and surface \textsuperscript{210}Pb activities, as well as the $\alpha$-rate derived from the fit, accounting for both contributions according to Equation~\ref{eqEq}. Results for the $\alpha$-rate derived from the fit are not reported for detectors 0, 1, and 2, since, as discussed in Figure~\ref{alphadecay}, these detectors had already reached secular equilibrium between \textsuperscript{210}Pb and \textsuperscript{210}Po by the time data taking began, making it difficult to determine the time of the \textsuperscript{210}Pb contamination. For comparison, the experimental $\alpha$-rate is reported in Table~\ref{ritmoalphaexp}.

As can be inferred from Table \ref{resultadosdeME}, for the $^{210}$Pb surface component with a mean depth of 100 $\mu$m, the bulk component is zero for all detectors, and all of the contamination from this isotope is assigned to the surface. This is due to the fact that, as shown in Figure~\ref{compare210PbLE}, the 100 $\mu$m surface component exhibits a rather similar behaviour to the bulk component. Therefore, the fitting machinery cannot distinguish between the contributions of the bulk and surface components and arbitrarily assigns all the weight to one of them. Regarding the other two surface components, the results appear to be consistent, except for detector D1 in the case of the 1 $\mu$m surface component, where the fit does not converge.


Moreover, the $\alpha$-rate derived from the fit is compatible across the different exponential profiles of \textsuperscript{210}Pb surface contamination, and it is comparable with the measured $\alpha$-rate reported in Table~\ref{ritmoalphaexp}. Although it may be slightly lower than the experimental value, it approaches the measured rate more closely when the contribution from decay chains shown in Figure~\ref{alphadecay} is considered. Nevertheless, significant uncertainties remain in this comparison, both from the fit and from the experimental determination of the $\alpha$-rate due to the selection of this population, and overall, the measured $\alpha$-rate is reasonably well reproduced by the fitting.

\begin{table}[b!]

\centering
    \begin{minipage}{0.4\textwidth}
    
    \centering
    \resizebox{\textwidth}{!}{\Large
    \begin{tabular}{c|cc}
        \hline
        \multirow{2}{*}{detector} & \multicolumn{2}{c}{$^{210}$Pb surf exp1 }  \\
        \cline{2-3}
        & $^{109}$Cd (mBq/kg) & $^{129}$I (mBq/kg)   \\
        \hline
        0 & 1.389 $\pm$ 0.011 & 0.143 $\pm$ 0.065 \\
        1 & 1.410 $\pm$ 0.000 & 0.010 $\pm$ 0.005 \\
        2 & 1.133 $\pm$ 0.008 & 0.057 $\pm$ 0.013 \\
        3 & 1.142 $\pm$ 0.010 & 0.050 $\pm$ 0.009 \\
        4 & 1.296 $\pm$ 0.010 & 0.064 $\pm$ 0.006 \\
        5 & 1.195 $\pm$ 0.008 & 0.053 $\pm$ 0.006 \\
        6 & 1.324 $\pm$ 0.009 & 0.057 $\pm$ 0.005 \\
        7 & 1.287 $\pm$ 0.009 & 0.061 $\pm$ 0.004 \\
        8 & 1.255 $\pm$ 0.008 & 0.056 $\pm$ 0.004 \\
        \hline
    \end{tabular}}
   
    \label{tab:exp_data}\subcaption{}
    \end{minipage}
    \vskip 0.5cm

    \begin{minipage}{0.4\textwidth}
    
    \centering
    \resizebox{\textwidth}{!}{\Large
    \begin{tabular}{c|cc}
        \hline
        \multirow{2}{*}{detector} & \multicolumn{2}{c}{$^{210}$Pb surf exp10 }  \\
        \cline{2-3}
        & $^{109}$Cd (mBq/kg) & $^{129}$I (mBq/kg)   \\
        \hline
        0 & 1.320 $\pm$ 0.011 & 0.165 $\pm$ 0.065 \\
        1 & 1.357 $\pm$ 0.011 & 0.160 $\pm$ 0.066 \\
        2 & 1.120 $\pm$ 0.008 & 0.058 $\pm$ 0.013 \\
        3 & 1.128 $\pm$ 0.010 & 0.051 $\pm$ 0.009 \\
        4 & 1.274 $\pm$ 0.010 & 0.065 $\pm$ 0.006 \\
        5 & 1.096 $\pm$ 0.008 & 0.057 $\pm$ 0.006 \\
        6 & 1.249 $\pm$ 0.009 & 0.059 $\pm$ 0.005 \\
        7 & 1.108 $\pm$ 0.008 & 0.063 $\pm$ 0.004 \\
        8 & 1.200 $\pm$ 0.008 & 0.058 $\pm$ 0.004 \\
        \hline
    \end{tabular}}
   
    \label{tab:exp_data}\subcaption{}
    \end{minipage}
    \vskip 0.5cm

    \begin{minipage}{0.4\textwidth}
    
    \centering
    \resizebox{\textwidth}{!}{\Large
    \begin{tabular}{c|cc}
        \hline
        \multirow{2}{*}{detector} & \multicolumn{2}{c}{$^{210}$Pb surf exp100 }  \\
        \cline{2-3}
        & $^{109}$Cd (mBq/kg) & $^{129}$I (mBq/kg)   \\
        \hline
        0 & 1.342 $\pm$ 0.011 & 0.010 $\pm$ 0.080 \\
        1 & 1.377 $\pm$ 0.011 & 0.010 $\pm$ 0.622 \\
        2 & 1.172 $\pm$ 0.008 & 0.038 $\pm$ 0.013 \\
        3 & 1.177 $\pm$ 0.010 & 0.028 $\pm$ 0.009 \\
        4 & 1.344 $\pm$ 0.010 & 0.050 $\pm$ 0.006 \\
        5 & 1.144 $\pm$ 0.008 & 0.056 $\pm$ 0.006 \\
        6 & 1.275 $\pm$ 0.009 & 0.058 $\pm$ 0.005 \\
        7 & 1.249 $\pm$ 0.009 & 0.061 $\pm$ 0.004 \\
        8 & 1.239 $\pm$ 0.008 & 0.055 $\pm$ 0.004 \\
        \hline
    \end{tabular}}
   
    \label{tab:exp_data}\subcaption{}
    \end{minipage}

   \caption{ \label{resultadoscosmodeME} Initial \(^{109}\)Cd and \(^{129}\)I activities at the time of detector movement underground, determined from the fit. Values are given in mBq/kg for the nine ANAIS-112 detectors. Results are provided for the three exponential $^{210}$ Pb profiles considered in this work.
\textbf{(a)}~1~$\mu$m. \textbf{(b)} 10 $\mu$m. \textbf{(c)} 100 $\mu$m.  }
    
\end{table}

Moreover, as discussed in Section~\ref{alphasec}, the $\alpha$-region of the ANAIS-112 spectrum does not exhibit a single peak associated with the $^{210}$Po $\alpha$-decay, but rather two distinct peaks, with a detector-dependent structure. Therefore, one of the objectives of this stage of the fit was to assess whether the $\alpha$-spectrum observed in ANAIS-112 can be fully accounted for by the combination of bulk and surface $^{210}$Pb contamination inferred from the fit. If the experimental area of each of the peaks shown in Figure \ref{alphaspectrum} is computed and compared with the activity ratio of bulk to surface $^{210}$Pb reported in Table~\ref{resultadosdeME}, it is found that the fitted results do not support the hypothesis that attributes the leftmost peak to surface contamination and the rightmost to bulk activity, as assumed in the previous background model. Alternative explanations should therefore be considered, such as a possible spatially inhomogeneous distribution of thallium concentration, which could lead to variations in the QF\textsubscript{$\alpha$} along the crystal.

\begin{figure}[b!]
    \centering
    {\includegraphics[width=0.49\textwidth]{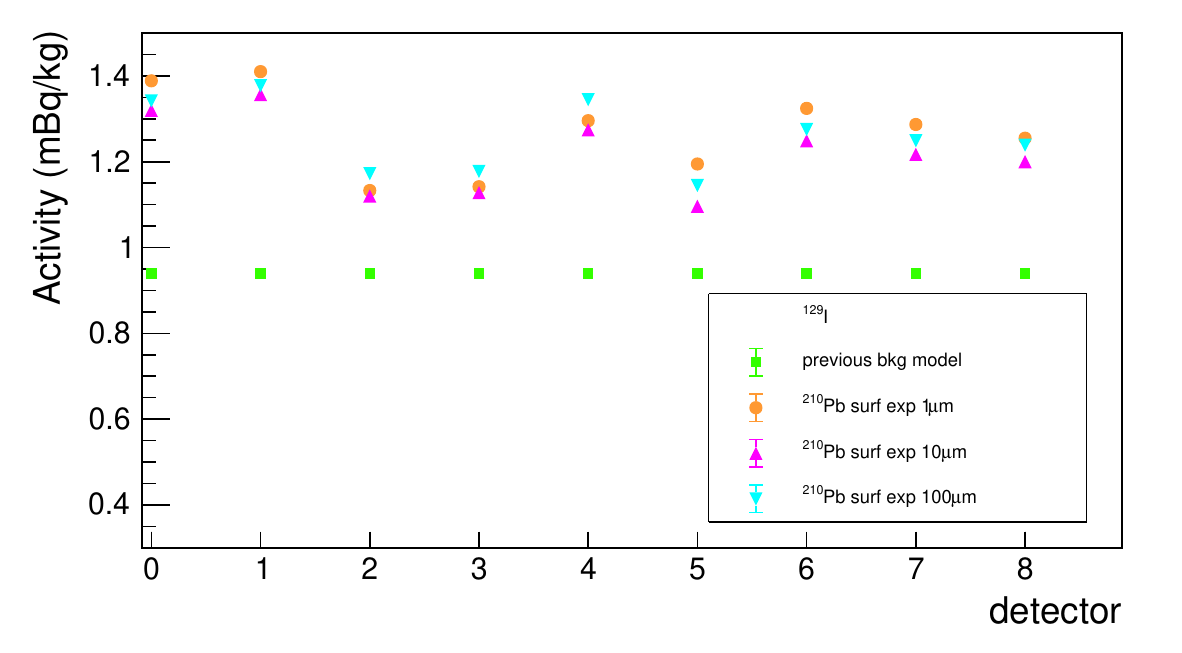}}
    \hfill
    {\includegraphics[width=0.49\textwidth]{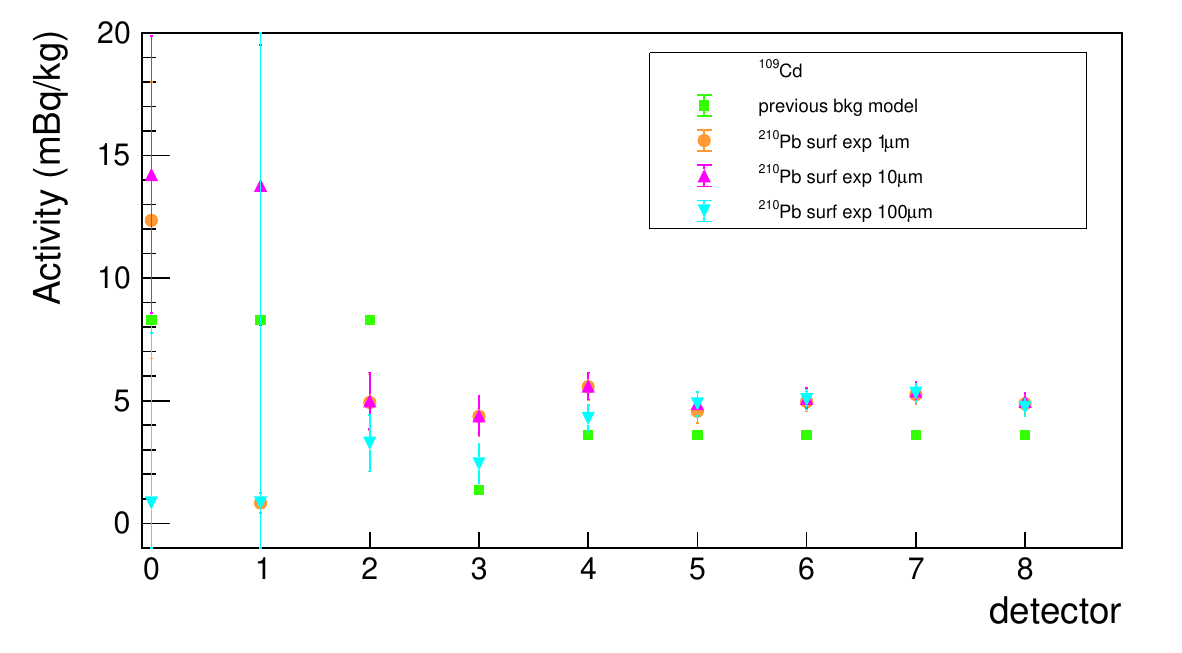}}
    
    \caption{\label{cdIplot} Comparison of the initial \(^{129}\)I \textbf{(left panel)} and \(^{109}\)Cd \textbf{(right panel)} activities at the time of detector movement underground, as determined from the fit. Three different exponential depth profiles for the \(^{210}\)Pb surface contamination (1, 10, and 100 $\mu$m, shown in orange, magenta, and cyan, respectively) are considered. The results are compared with those reported in the previous ANAIS-112 background model (shown in green). Activity values are expressed in mBq/kg for the nine ANAIS-112 detectors.}
\end{figure}

From the medium-energy fit, the activities of \(^{109}\)Cd and \(^{129}\)I are also obtained. Table~\ref{resultadoscosmodeME} shows these activities at the time of detector movement underground. Figure~\ref{cdIplot} compares the results with the values adopted in the previous background model. The $^{109}\mathrm{Cd}$ values of the previous model used for comparison include the corresponding correction based on the respective half-lives and surface activation exposure, as detailed in \cite{Amare:2021yyu}. Regarding \(^{129}\)I (left panel), the previous background model assumed its concentration to be the same as estimated by DAMA/LIBRA, corresponding to an activity of 0.94 mBq/kg. The results obtained from the fit described in this thesis are slightly higher, but similar with each other for the different exponential profiles. For \(^{109}\)Cd (right panel), in the cases of detectors D0 and D1, being the first to arrive at the LSC, the data show virtually no evidence of this isotope (see Figure~\ref{MEpop1}), which justifies the larger associated uncertainties. For the remaining detectors, the fitted activities are consistent across modules and slightly higher than those assumed in the previous background model for detectors D4-D8.

\begin{figure}[t!]
    \centering
    {\includegraphics[width=1\textwidth]{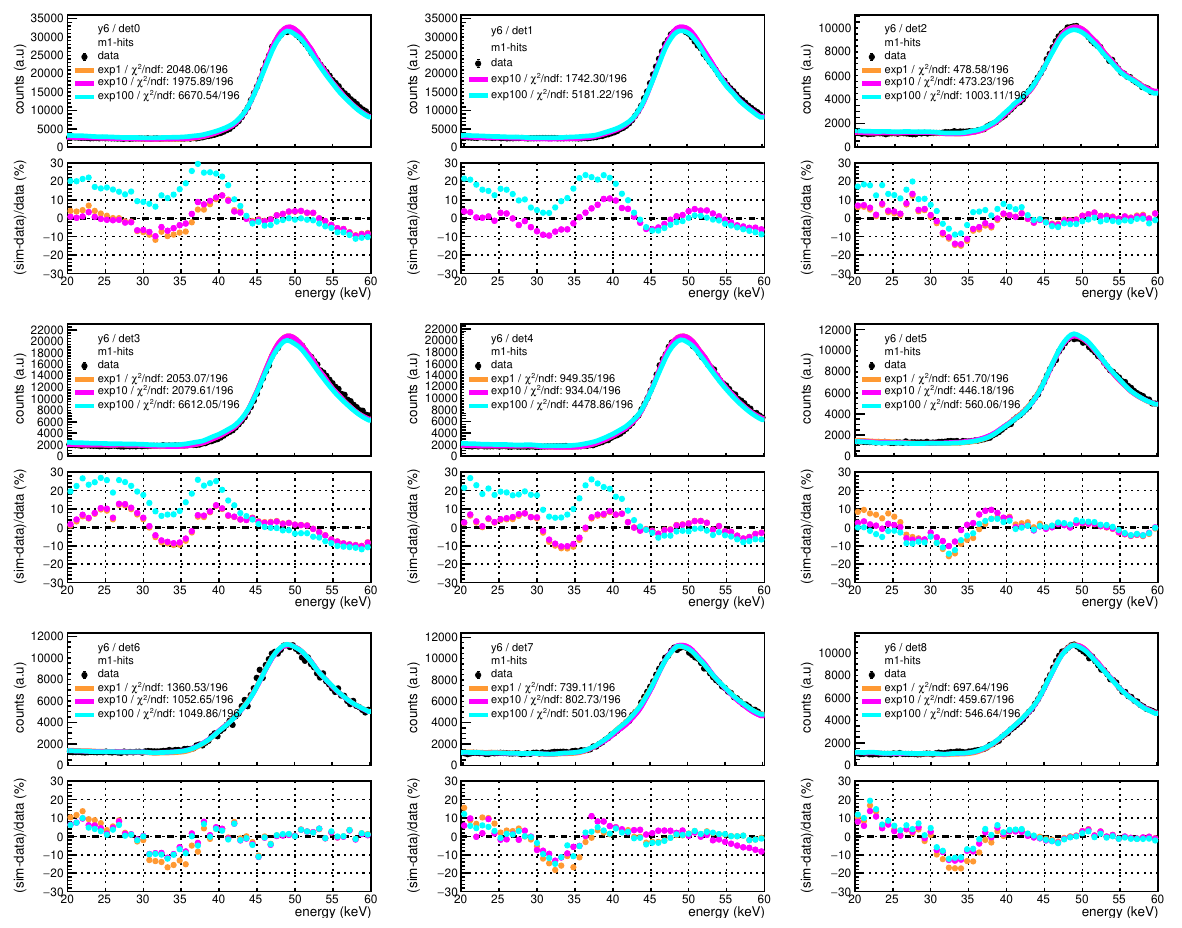}}

    \caption{\label{comapraexpsME} Comparison of the medium-energy best fits for the single-hit population across the nine ANAIS-112 detectors, considering different exponential mean depth profiles for the $^{210}$Pb surface contamination. The profiles for 1, 10 and 100 $\mu$m contamination depths are represented in orange, magenta, and cyan, respectively. Measured data points are shown in black. The goodness of the fit and the corresponding residuals are also presented in the panels.  }
\end{figure}

Figure \ref{comapraexpsME} presents the best fit for various exponential profiles in the medium-energy range. In the case of the 1$\mu$m exponential for D1, the best fit is again not shown, as the fit fails to converge, consistent with the poor convergence observed at high energies. For detectors D0, D2, D3, and D4, the \(^{210}\)Pb surface contamination at a depth of 100~$\mu$m appears to provide a poorer description of the measured data. For the remaining detectors, all profiles yield comparable fits.

Nevertheless, it is important to note that the detectors with higher $^{210}$Pb activity are fitted less accurately, suggesting that this component is still not properly modelled. Additionally, it is worth mentioning that the region between 30 and 35 keV remains unexplained by the current background model.

\subsection{Low-energy fitting}\label{lowenergyfit}

With the PMT contamination and both bulk and surface \(^{210}\)Pb activities in the crystals fixed, the low-energy fit is performed for each detector using the single-hit spectrum from the sixth year of data-taking.

\begin{figure}[b!]
    \centering
    {\includegraphics[width=0.95\textwidth]{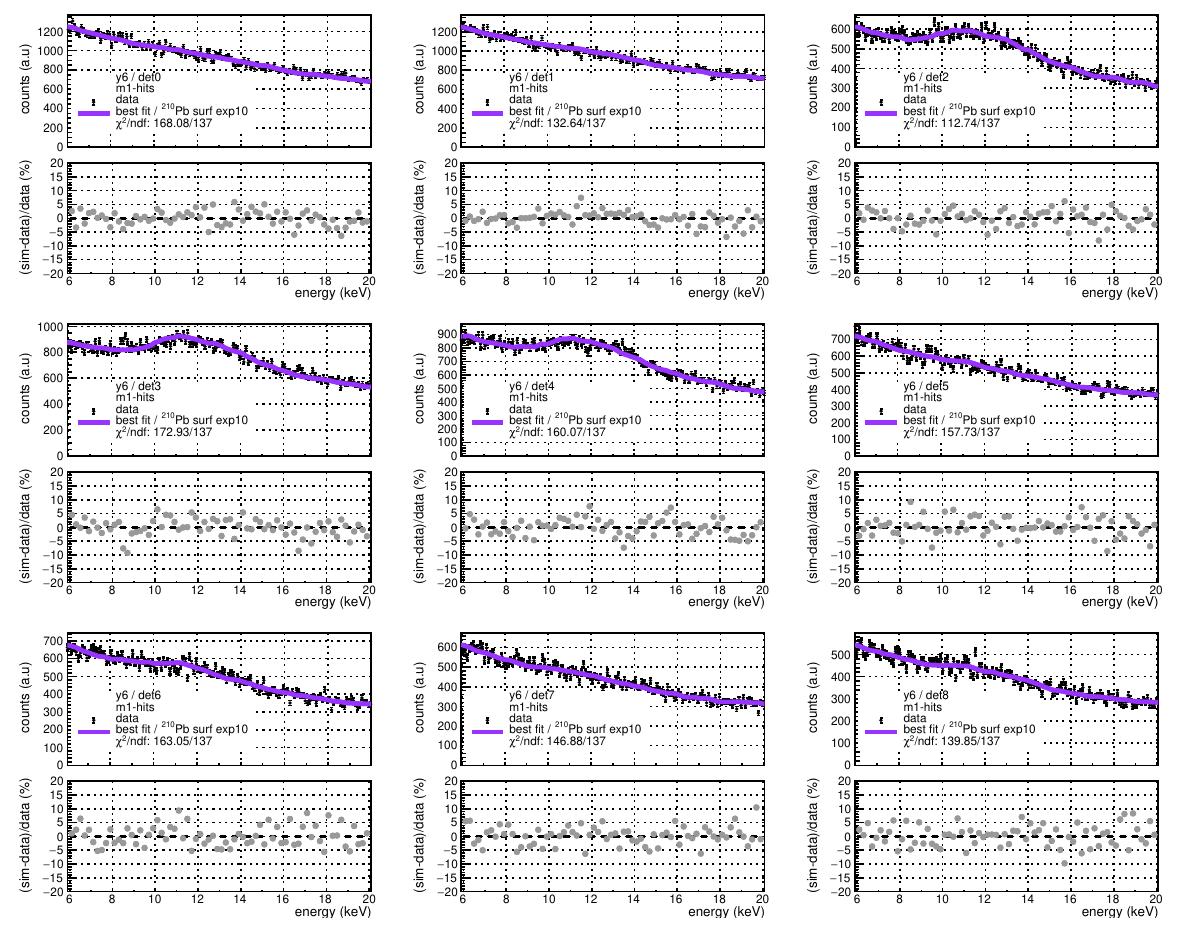}}

    \caption{\label{BEfitpop0} Low-energy best fits of the single-hit spectra for the nine ANAIS-112 detectors. Black points correspond to data from the full sixth year of ANAIS-112, while the best fit is shown in violet. The exponential profile of 10 $\mu$m is considered for the superficial contamination of \(^{210}\)Pb. Residuals are displayed as grey points, and the goodness-of-fit is indicated in each panel.}
\end{figure}

\begin{figure}[t!]
    \centering
    {\includegraphics[width=0.47\textwidth]{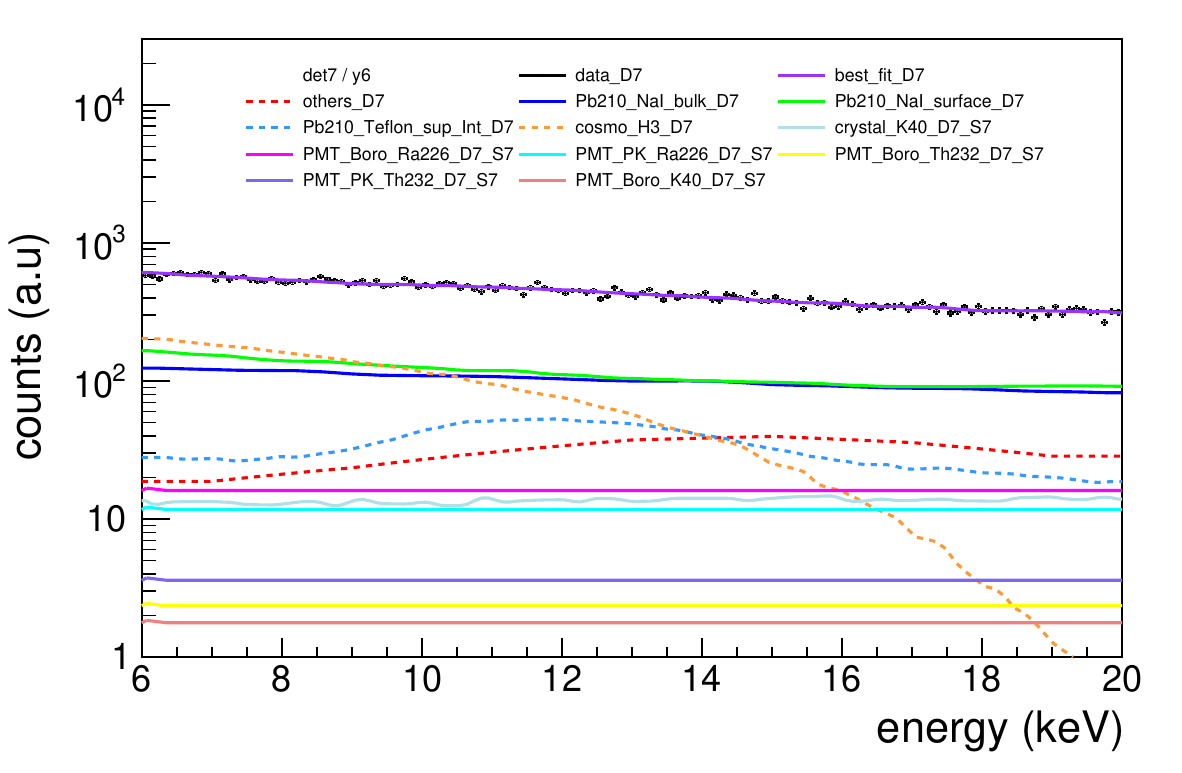}}
   \hfill
    {\includegraphics[width=0.47\textwidth]{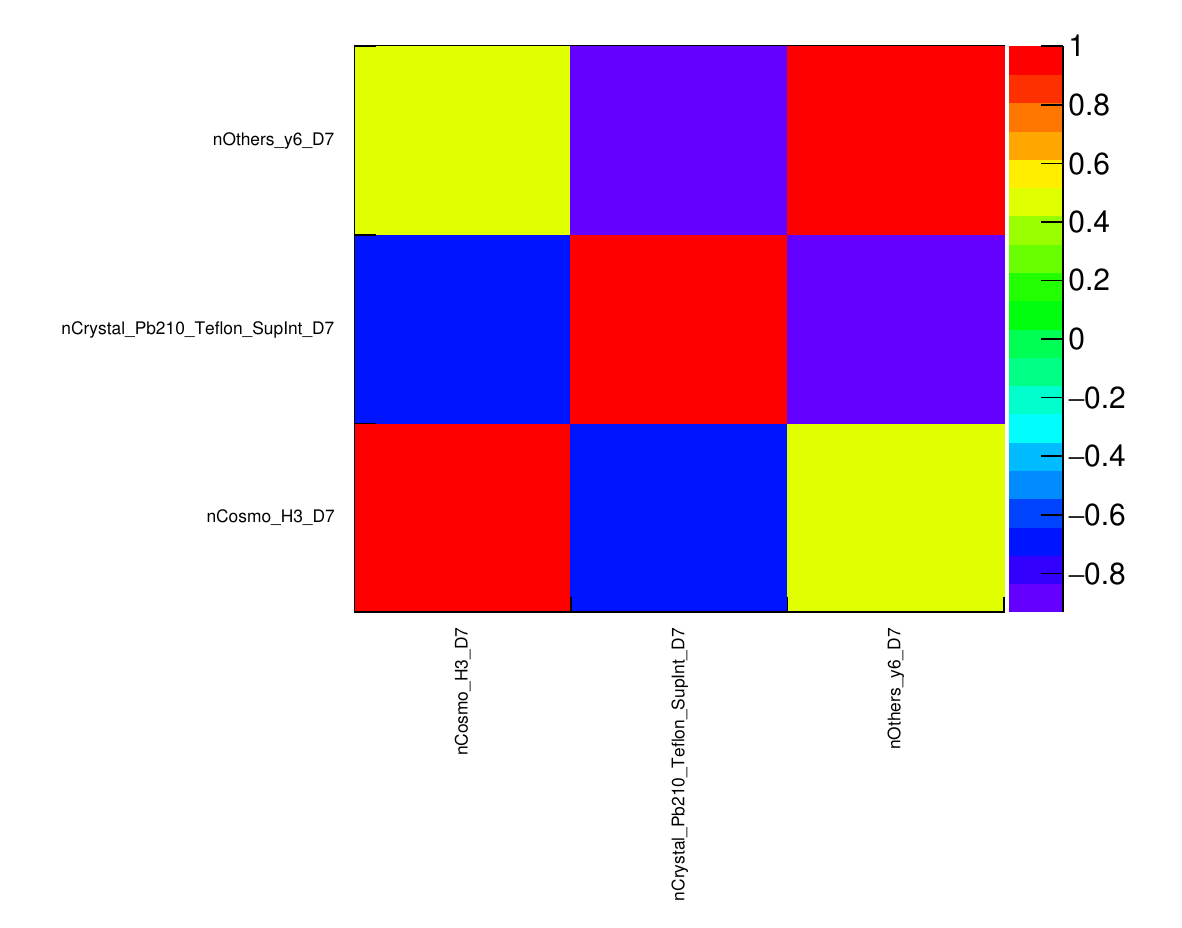}}
    \caption{\label{desgloseBE} \textbf{Left panel:} Breakdown of the individual components contributing to the low-energy fit for the D7 detector. The figure displays the single-hit event population, with dashed lines representing the components left free in the fit, and solid lines corresponding to the fixed components.  The best fit is displayed in violet. The exponential profile of 10 $\mu$m is considered for the superficial contamination of \(^{210}\)Pb. \textbf{Right panel:} Correlation matrix from the low-energy
fit for D7 module.}
\end{figure}

Figure \ref{BEfitpop0} displays the best fit results in the low-energy region. As shown, the fit adequately reproduces the data, with residuals remaining flat and within 10\%. Analogously to the analysis in previous energy regions, the left panel of Figure \ref{desgloseBE} presents the breakdown of the fit components for detector D7, highlighting the dominant contributions of \(^{3}\)H and \(^{210}\)Pb in this energy range. 

The right panel of Figure~\ref{desgloseBE} shows the correlation matrix for detector D7, which reveals significant dependencies among the fitted background components. In particular, a strong anticorrelation is observed between the contributions from \(^{210}\)Pb in the teflon and the cosmogenic \(^{3}\)H component, indicating spectral degeneracy between these sources. Additionally, both the \(^{3}\)H and the external background components exhibit a notable positive correlation. Despite these correlations, the fit remains stable and yields a consistent description of the background.

\begin{figure}[t!]
    \centering
    {\includegraphics[width=0.49\textwidth]{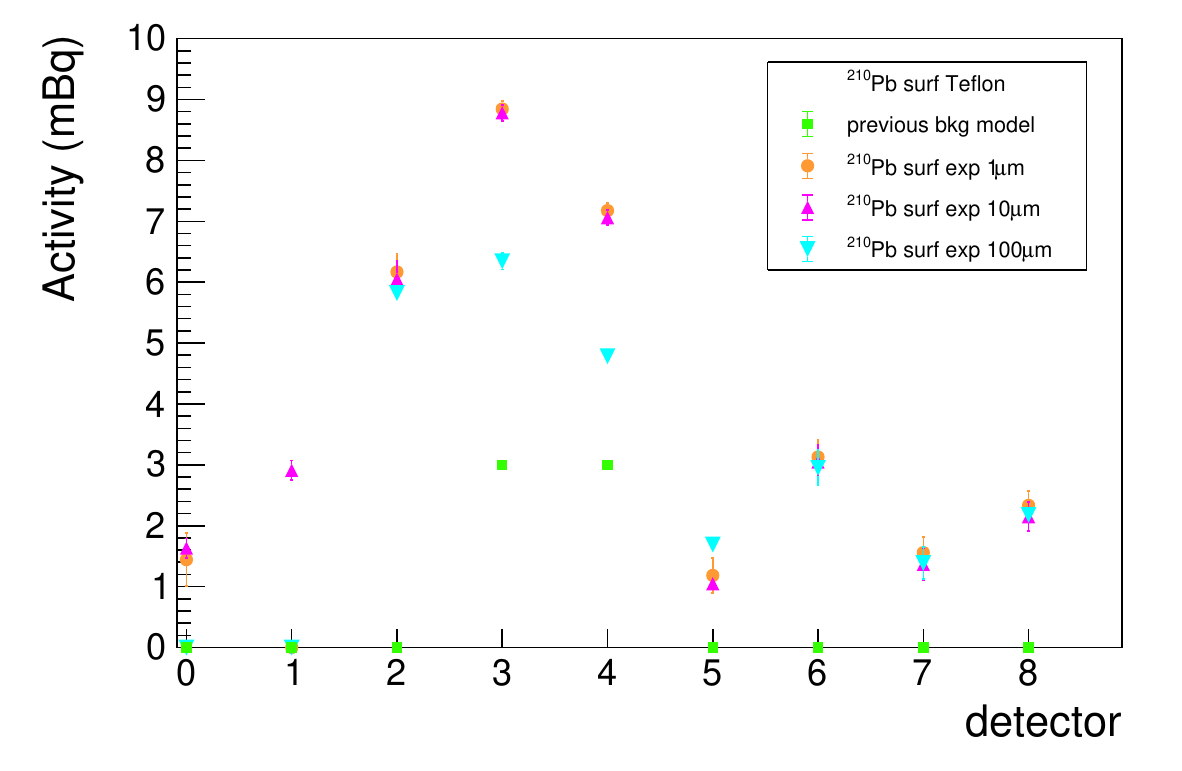}}
    \hfill
    {\includegraphics[width=0.49\textwidth]{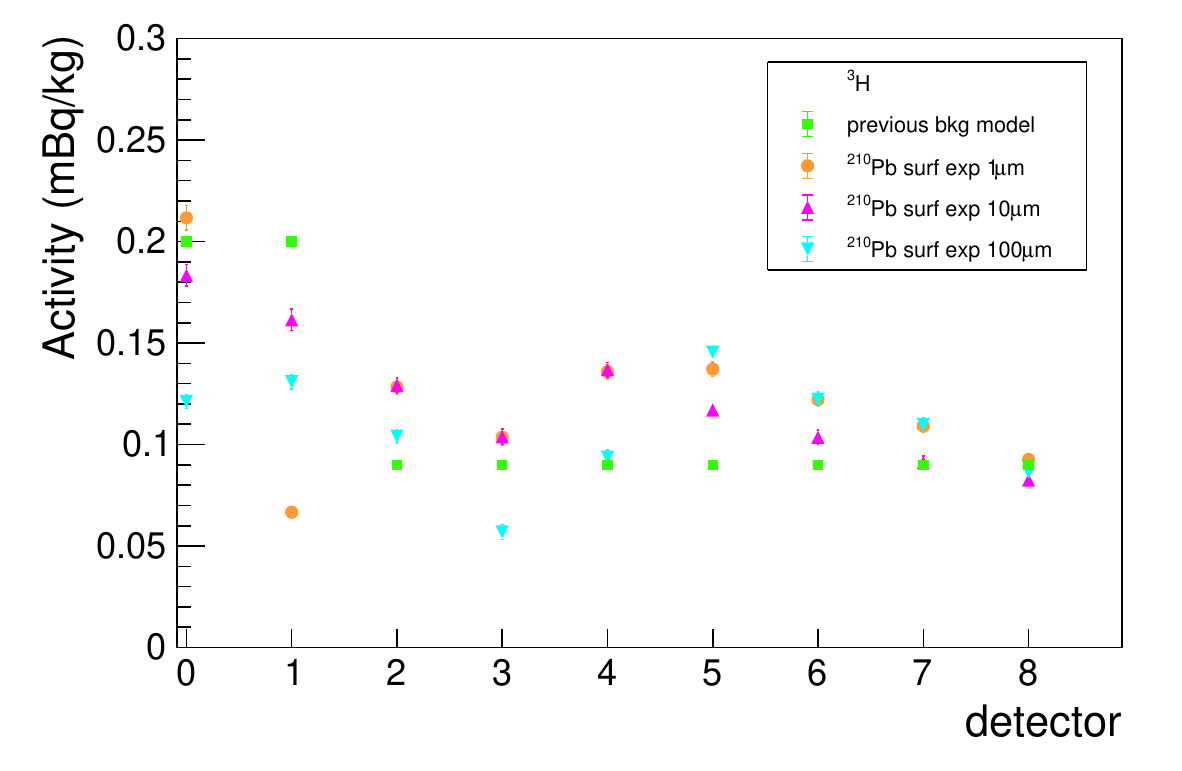}}
    
    \caption{\label{teflon3hplot} Comparison of the initial \(^{210}\)Pb contamination in the surface of the teflon film surrounding the crystals (considering deposits on both sides of the teflon) \textbf{(left panel)} and \(^{3}\)H cosmogenic \textbf{(right panel)} activities at the time of detector installation underground, as determined from the fit. Three different exponential depth profiles for the \(^{210}\)Pb surface contamination (1, 10, and 100~$\mu$m, shown in orange, magenta, and cyan, respectively) are considered. The results are compared with those reported in the previous ANAIS-112 background model (shown in green). Activity values are expressed in mBq and mBq/kg, respectively, for the nine ANAIS-112 detectors.}
\end{figure}

Following Equation \ref{EqA0}, the initial activities at the time of underground deployment are extracted for \(^{210}\)Pb on the teflon surface and for \(^{3}\)H within the crystal. These values are reported in Table \ref{valoresdelfitBE} for the three surface contamination profiles of \(^{210}\)Pb in the crystal considered in this work. A comparison with the activities assumed in the previous background model is shown in Figure \ref{teflon3hplot}.

Regarding \(^{210}\)Pb on the teflon surfaces, the previous background model only included an activity of 3 mBq per detector for D3 and D4, as no indication of such contamination was present in the first years of data for the other modules. However, with six years of exposure, a smaller yet significant contribution is observed across all modules, most notably in D2.

Concerning \(^{3}\)H, although its presence could not be directly identified in the previous background model, an activity of approximately 0.20 mBq/kg for D0 and D1 and 0.09~mBq/kg for D2–D8 was assumed to reproduce the data. The fit results obtained in this analysis also indicate a higher \(^{3}\)H activity for the first two detectors, although the difference between these and the rest of detectors is less pronounced, yielding comparable values across all modules.

\begin{figure}[b!]
    \centering
    {\includegraphics[width=0.49\textwidth]{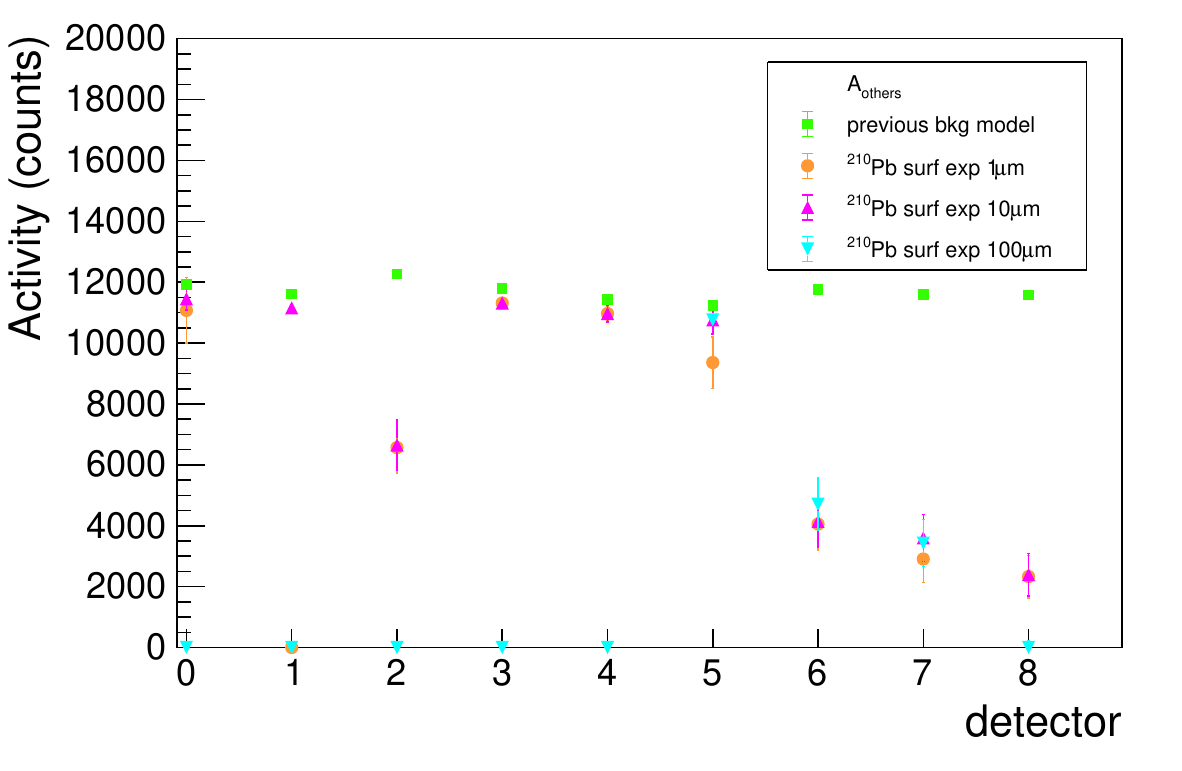}}

    \caption{\label{componenteothers} Comparison of the external contamination (excluding PMT contributions) as determined from the fit. Three exponential depth profiles for the \(^{210}\)Pb surface contamination (1, 10, and 100 $\mu$m, shown in orange, magenta, and cyan, respectively) are considered. The results are compared with those assumed in the previous ANAIS-112 background model (shown in green). Values are expressed in counts and correspond to the number of events expected within the fitting range for each of the nine ANAIS-112 detectors. }
\end{figure}

The fit in the low-energy region also includes the determination of the external component, which encompasses contributions from radioactive contamination in the copper housing surrounding the detectors, as well as from the quartz and silicon pads used in the PMT coupling, among others. Figure \ref{componenteothers} shows the fit results of this component, compared with the values assumed in the previous background model.

\begin{table}[t!]

\centering
    \begin{minipage}{0.5\textwidth}
    
    \centering
    \resizebox{\textwidth}{!}{\Large
    \begin{tabular}{c|cc}
        \hline
        \multirow{2}{*}{detector} & \multicolumn{2}{c}{$^{210}$Pb surf exp1 }  \\
        \cline{2-3}
        & $^{210}$Pb teflon surf (mBq) & $^3$H (mBq/kg)    \\
        \hline
        0 & 1.44 $\pm$ 0.43 & 0.212 $\pm$ 0.006 \\
        1 & 0.00 $\pm$ 0.00 & 0.067 $\pm$ 0.000 \\
        2 & 6.17 $\pm$ 0.31 & 0.128 $\pm$ 0.004 \\
        3 & 8.84 $\pm$ 0.13 & 0.104 $\pm$ 0.004 \\
        4 & 7.18 $\pm$ 0.12 & 0.136 $\pm$ 0.004 \\
        5 & 1.18 $\pm$ 0.29 & 0.137 $\pm$ 0.004 \\
        6 & 3.13 $\pm$ 0.29 & 0.122 $\pm$ 0.004 \\
        7 & 1.56 $\pm$ 0.26 & 0.109 $\pm$ 0.003 \\
        8 & 2.33 $\pm$ 0.24 & 0.093 $\pm$ 0.003 \\
        \hline
    \end{tabular}}
   
    \label{tab:exp_data}\subcaption{}
    \end{minipage}
    \vskip 0.5cm

    \begin{minipage}{0.5\textwidth}
    
    \centering
    \resizebox{\textwidth}{!}{\Large
    \begin{tabular}{c|cc}
        \hline
        \multirow{2}{*}{detector} & \multicolumn{2}{c}{$^{210}$Pb surf exp10 }  \\
        \cline{2-3}
        & $^{210}$Pb teflon surf (mBq) & $^3$H (mBq/kg)    \\
        \hline
        0 & 1.63 $\pm$ 0.16 & 0.183 $\pm$ 0.005 \\
        1 & 2.91 $\pm$ 0.16 & 0.161 $\pm$ 0.005 \\
        2 & 6.05 $\pm$ 0.31 & 0.129 $\pm$ 0.004 \\
        3 & 8.78 $\pm$ 0.13 & 0.104 $\pm$ 0.004 \\
        4 & 7.06 $\pm$ 0.12 & 0.137 $\pm$ 0.004 \\
        5 & 1.05 $\pm$ 0.10 & 0.117 $\pm$ 0.003 \\
        6 & 3.05 $\pm$ 0.29 & 0.104 $\pm$ 0.004 \\
        7 & 1.28 $\pm$ 0.26 & 0.096 $\pm$ 0.003 \\
        8 & 2.15 $\pm$ 0.24 & 0.082 $\pm$ 0.003 \\
        \hline
    \end{tabular}}
   
    \label{tab:exp_data}\subcaption{}
    \end{minipage}
    \vskip 0.5cm

    \begin{minipage}{0.5\textwidth}
    
    \centering
    \resizebox{\textwidth}{!}{\Large
    \begin{tabular}{c|cc}
        \hline
        \multirow{2}{*}{detector} & \multicolumn{2}{c}{$^{210}$Pb surf exp100 }  \\
        \cline{2-3}
        & $^{210}$Pb teflon surf (mBq) & $^3$H (mBq/kg)    \\
        \hline
        0 & 0.00 $\pm$ 0.01 & 0.121 $\pm$ 0.004 \\
        1 & 0.00 $\pm$ 0.03 & 0.131 $\pm$ 0.004 \\
        2 & 5.82 $\pm$ 0.11 & 0.104 $\pm$ 0.003 \\
        3 & 6.34 $\pm$ 0.13 & 0.057 $\pm$ 0.004 \\
        4 & 4.78 $\pm$ 0.12 & 0.094 $\pm$ 0.004 \\
        5 & 1.69 $\pm$ 0.10 & 0.146 $\pm$ 0.003 \\
        6 & 2.95 $\pm$ 0.29 & 0.122 $\pm$ 0.004 \\
        7 & 1.39 $\pm$ 0.26 & 0.110 $\pm$ 0.003 \\
        8 & 2.18 $\pm$ 0.09 & 0.086 $\pm$ 0.003 \\
        \hline
    \end{tabular}}
   
    \label{tab:exp_data}\subcaption{}
    \end{minipage}

   \caption{\label{valoresdelfitBE} Initial \(^{210}\)Pb in the surface of the teflon film surrounding the crystals and \(^{3}\)H activities at the time of detector movement underground, determined from the fit. Values are given in mBq and mBq/kg, respectively, for the nine ANAIS-112 detectors.  Results are provided for the three exponential $^{210}$ Pb profiles considered in this work.
\textbf{(a)} 1 $\mu$m. \textbf{(b)}~10~$\mu$m. \textbf{(c)} 100~$\mu$m.}
    
\end{table}

No initial activity is derived for this component, as all contributions falling within this category are merged into a single PDF. The goal is not to extract individual activities, but rather to evaluate whether the global contribution of this external component may be bounded more stringently because only upper limits for these contributions could be placed by the screening of materials before the experiment commissioning. Accordingly, Figure \ref{componenteothers} presents the number of counts attributed to this component for each of the three \(^{210}\)Pb surface contamination profiles considered in this work, compared to the expectations from the previous background model, where the upper limit of activity was adopted for each contribution.

Initially, it was hypothesized that this contamination could be common across all detectors. However, extensive attempts in this direction failed to yield convergent fits. Consequently, the model was adapted to include individual external components for each detector, which led to successful convergence. From this, it can be inferred that, for detectors D0, D1, D3, D4, and D5, the fitted upper limits are consistent with those previously assumed. In contrast, for detectors D2 and D6–D8, the contribution of the external component is significantly lower and mutually compatible. Interestingly, D2 was delivered to the LSC separately from the other detectors, while D6, D7, and D8 were produced at Alpha Spectra, received and installed at LSC simultaneously. The compatibility among the latter may suggest the possibility of a shared characteristic in their construction, potentially related to differences in the copper housing or other external elements compared to the rest of detectors with higher derived activity. Note that for the $^{210}$Pb surface contamination of 100~$\mu$m mean depth the result of this component from the fit is zero counts in most of detectors which is not a motivated scenario since some contamination coming from outside the crystal is indeed expected. 

At this stage, having completed the fit for all components and energy regions, it is necessary to determine which exponential depth profile of $^{210}$Pb surface contamination best describes the ANAIS-112 data. Except for the 1$\mu$m profile in the D1 module, which can be excluded due to lack of fit convergence, and some disfavoring of the 100 $\mu$m profile in the medium-energy range, no profile appears to be clearly preferred based solely on the fit quality. Nevertheless, as already observed in the high-energy fit, the distribution of \(^{226}\)Ra and \(^{232}\)Th contamination between the borosilicate and the photocathode appears more physically reasonable when assuming the 10\(\mu\)m exponential profile, as welll as the homogeneity in contamination values derived for the different modules (in particular for the PMTs) is better with this depth profile.

\begin{figure}[b!]
    \centering
    {\includegraphics[width=1\textwidth]{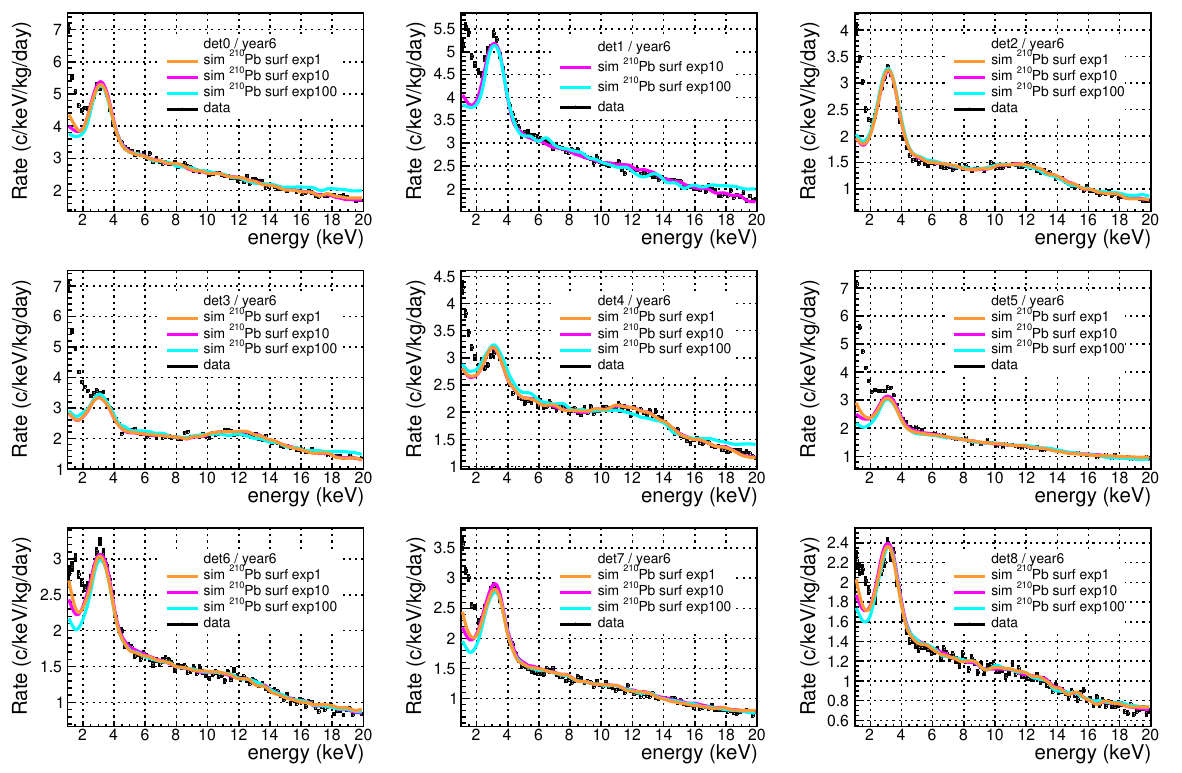}}

    \caption{\label{modeloBEvariasexp} Comparison of the total model for the single-hit population across the nine ANAIS-112 detectors in the low-energy region, considering different exponential mean depth profiles for the $^{210}$Pb surface contamination. The profiles for 1 $\mu$m, 10~$\mu$m, and 100 $\mu$m contamination depths are represented in orange, magenta, and cyan, respectively. Data points are shown in black. }
\end{figure}

\begin{table}[b!]
    \centering
    \resizebox{0.95\textwidth}{!}{%
    \begin{tabular}{c|cc|cc|cc}
        \hline
        \multirow{2}{*}{detector} & \multicolumn{2}{c}{$^{210}$Pb surf exp1} & \multicolumn{2}{c}{$^{210}$Pb surf exp10} & \multicolumn{2}{c}{$^{210}$Pb surf exp100} \\
        \cline{2-7}
        & $\chi^2$/ndf & p-value & $\chi^2$/ndf & p-value & $\chi^2$/ndf & p-value \\
        \hline
        0 & 154.94/137 & 0.14 & 168.08/137 & 0.04 & 458.03/137 & 0.00 \\
        1 & 2206.92/137 & 0.00 & 132.64/137 & 0.59 & 318.49/137 & 0.00 \\
        2 & 116.43/137 & 0.90 & 112.74/137 & 0.94 & 163.91/137 & 0.06 \\
        3 & 175.84/137 & 0.01 & 172.93/137 & 0.02 & 371.41/137 & 0.00 \\
        4 & 165.75/137 & 0.05 & 160.07/137 & 0.09 & 388.63/137 & 0.00 \\
        5 & 171.97/137 & 0.02 & 157.73/137 & 0.11 & 189.11/137 & 0.00 \\
        6 & 178.04/137 & 0.01 & 163.05/137 & 0.06 & 160.81/137 & 0.08 \\
        7 & 161.84/137 & 0.07 & 145.55/137 & 0.29 & 151.62/137 & 0.19 \\
        8 & 148.20/137 & 0.24 & 139.91/137 & 0.41 & 154.02/137 & 0.15 \\
        \hline
    \end{tabular}%
    }
    \caption{\label{chi2fitBE} Performance of the low-energy fit conducted in this study, conducted in the [6-20] keV range. The $\chi^2$/ndf and the corresponding p-value for the three exponential depth profiles assumed for the surface contamination of \(^{210}\)Pb in this analysis (mean depths of 1, 10, and 100 $\mu$m) are shown.}
    
\end{table}

Figure \ref{modeloBEvariasexp} shows the total model obtained from the fit parameters for each of the three exponential profiles below 20 keV. As can be seen, the 100 $\mu$m exponential profile significantly overestimates the data above 16 keV for detectors D0 through D4, and is therefore disfavored. Between the remaining two profiles (1 and 10 $\mu$m), the overall model shape is very similar across most detectors, except for D1, where the 1~$\mu$m profile is already excluded and therefore not displayed in the figure. The most superficial profile, 1 $\mu$m, does show a slightly higher contribution in the [1–2] keV region due to its shallow depth, but in general both profiles yield comparable results for the fit components and overall model. This behavior is reflected in Table \ref{chi2fitBE}, which lists the $\chi^2$/ndf values obtained for the fit for each surface distribution of \(^{210}\)Pb, along with their corresponding p-values. Much worse values are obtained for the exponential profile with a 100 $\mu$m depth, while somewhat more competitive, yet still similar, values are found for the 10 $\mu$m profile, with results comparable to the 1 $\mu$m profile.

Nevertheless, a persistent discrepancy remains in the [1–2] keV region across all superficial models considered, indicating that the excess events in this energy range likely originate from a source other than $^{210}$Pb contamination in the crystal or its immediate surroundings. Other surface contaminations have not been explored, as no reasonable alternatives have been identified. In view of this, and with the underlying reason of adopting a unified and consistent background description for all detectors, the ANAIS-112 background model will incorporate the 10~$\mu$m exponential depth profile for the \(^{210}\)Pb surface contamination. This choice does not imply, however, that such a profile is uniquely determined. The 1 $\mu$m profile also yields acceptable fits for all detectors except D1. Within this context, it is conservative to conclude that the actual depth of the \(^{210}\)Pb surface contamination in ANAIS-112 crystals likely lies between 1 and 10 $\mu$m.

\section{Validation of the improved background model}\label{validation}

In this section, the background model developed in this work, based on the fitted activities of each component as derived in the previous section, is presented for the full six-year exposure of ANAIS-112 across the different energy regions. First, a detailed comparison is carried out with respect to the background model previously adopted by the experiment. The spectra are shown individually for each detector, as well as for the combined response of the nine modules. The results presented for the revised ANAIS-112 background model will incorporate the 10 $\mu$m exponential spatial distribution for the superficial \(^{210}\)Pb contamination. Subsequently, the comparison is carried out for coincidence events and year differences across different energy ranges in order to further support the model.

\begin{figure}[t!]
    \centering
    {\includegraphics[width=0.95\textwidth]{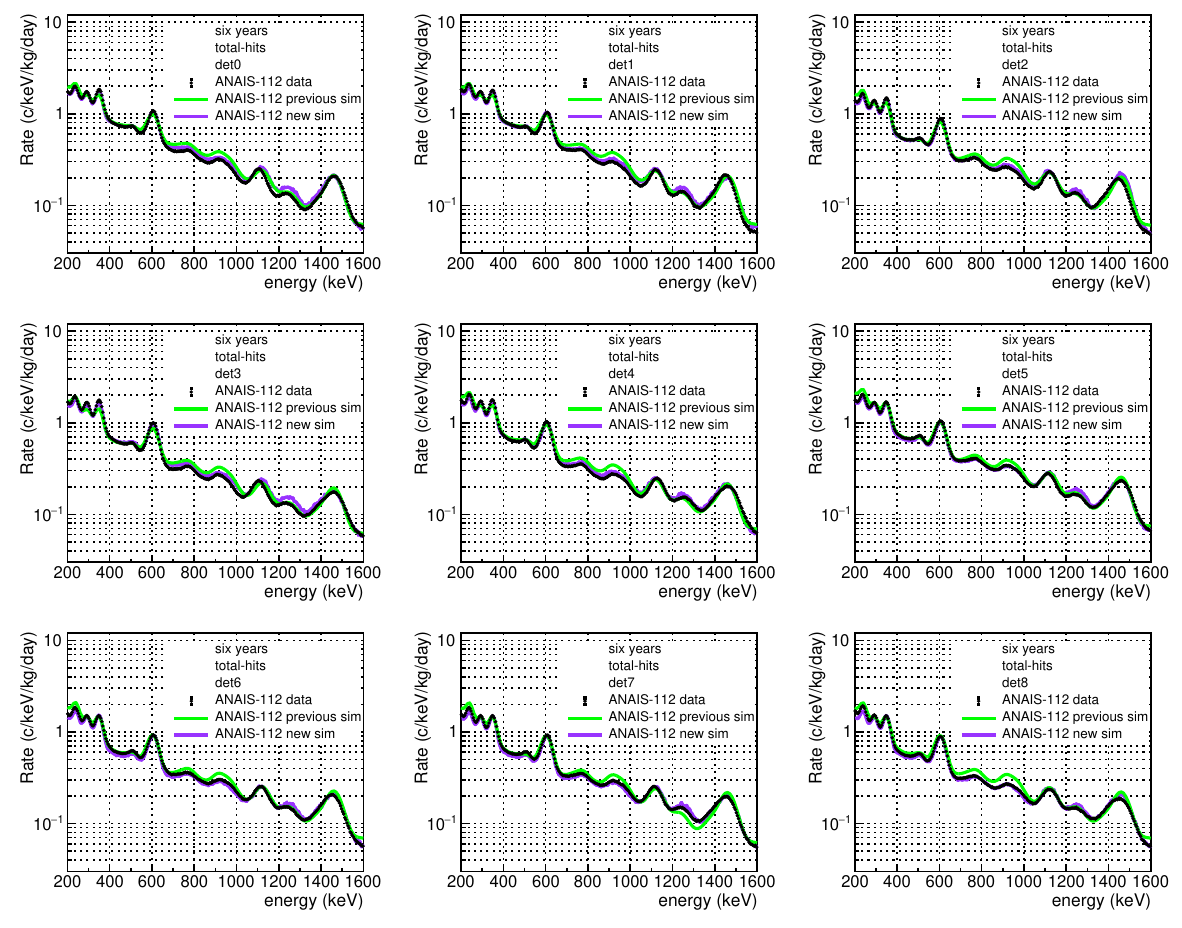}}

    \caption{\label{totalmodelHE_perdet} Comparison of the total high-energy spectra measured over six years for each ANAIS-112 detector (black) with the total background model developed in this thesis (violet) and the previous ANAIS-112 model (green). The model assumes a 10~$\mu$m exponential profile for the \(^{210}\)Pb surface contamination.  }
\end{figure}

\begin{figure}[b!]
    \centering
    {\includegraphics[width=0.6\textwidth]
    {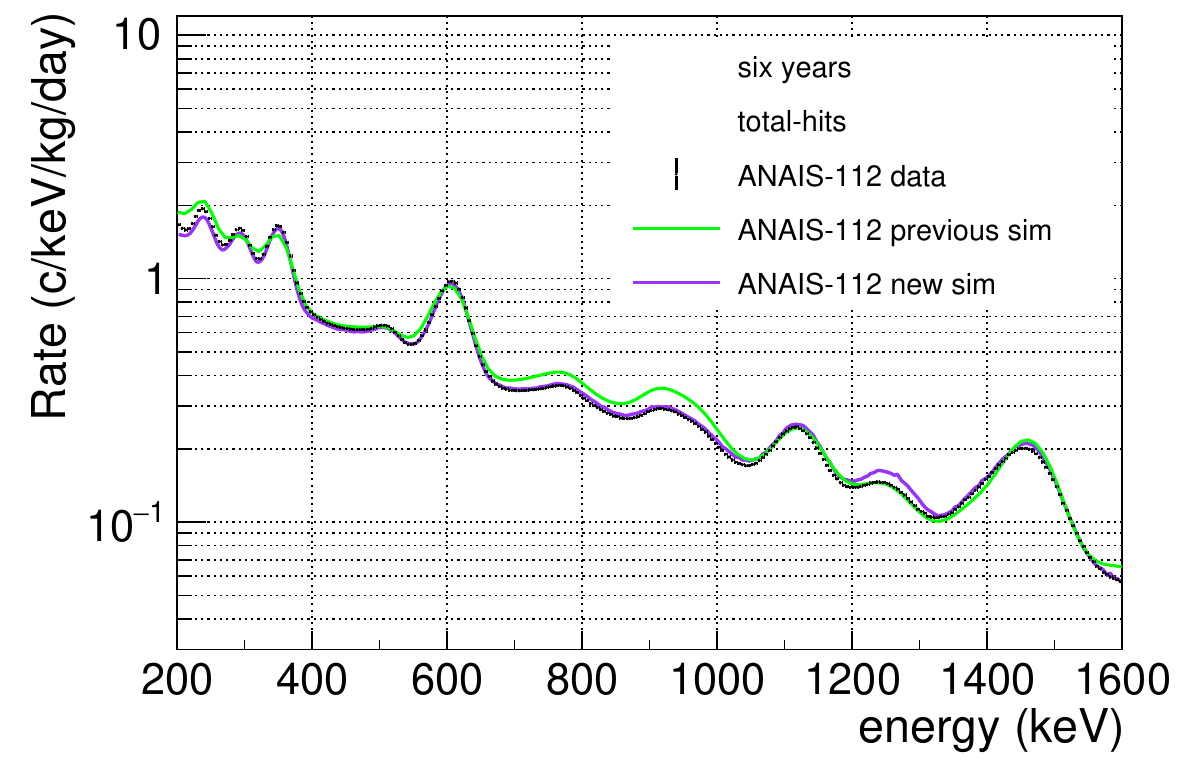}}

    \caption{\label{totalmodelHE}  Comparison of the total high-energy spectra measured over six years for the summed spectra of the nine ANAIS-112 detectors (black) with the total background model developed in this thesis (violet) and the previous ANAIS-112 model (green). The model assumes a 10~$\mu$m exponential profile for the \(^{210}\)Pb surface contamination. }
\end{figure}

Figure~\ref{totalmodelHE_perdet} presents the comparison between the measured total high-energy spectra over six years for each ANAIS-112 detector, the background model developed in this work, and the previous ANAIS-112 model. Complementarily, Figure~\ref{totalmodelHE} shows the summed spectrum of the nine detectors. It should be noted that spectra are now shown in regions that were not included in the fit, since, as previously described, the high-energy fit was performed in the [200–700] keV range. A substantially improved agreement between data and simulation is observed with the new model, both in the peak regions and in the continuum. It is worth highlighting that, to achieve this level of consistency, the contribution from \(^{232}\)Th in the copper housing surrounding the crystals was set to zero in all the energy regions. This assumption does not pose any inconsistency, as the previous estimation of this component was based solely on an upper limit. In the [200–700] keV range, the characteristic peaks are more accurately reproduced, and the modelling of the energy resolution in the high-energy dataset appears to have been improved.

Of particular relevance is the modelling of the continuum in the [600–1000]~keV region. For the first time in the ANAIS-112 background modelling, this continuum is successfully reproduced. This improvement results from replacing the analytically computed \(\beta^-\) spectrum of \(^{210}\)Bi implemented in Geant4 with the experimentally determined shape available in the BetaShape database \cite{carles2005beta}. This substitution indicates that ANAIS-112 data clearly favor the measured spectral shape over the theoretical one for this poorly characterized transition.

Beyond the continuum, a slight overestimation is observed around 1300 keV in the new model. The origin of this discrepancy was investigated, but no dominant contribution was identified from external sources or from \textsuperscript{226}Ra contamination, as in the latter case the preceding peak around 1100 keV is well reproduced and simulations predict similar amplitudes for both peaks. The most plausible candidate is \textsuperscript{40}K contamination in the PMTs, which dominates this energy region. In the current model, this contamination is assumed to be homogeneously distributed throughout the borosilicate glass. However, it is possible that the actual distribution is more concentrated near the base of the PMTs, particularly around the pins. Such a configuration would likely result in a reduced Compton contribution while preserving the full-energy peak, potentially explaining the observed excess.

\begin{table}[b!]
    \centering
    \resizebox{0.9\textwidth}{!}{\Large
    \begin{tabular}{c|c|cc|cc}
        \hline
       \multirow{3}{*}{detector} & \multicolumn{5}{c}{total-hits [100,1600] keV}  \\
       \cline{2-6}
       & \makecell{data \\ ($\text{kg}^{-1} \text{day}^{-1}$)} 
 & \makecell{previous sim \\ ($\text{kg}^{-1} \text{day}^{-1}$)} 
 & \makecell{previous desv \\ (\%)} 
 & \makecell{new sim \\ ($\text{kg}^{-1} \text{day}^{-1}$)} 
 & \makecell{new desv \\ (\%)} \\
        
        \hline
        0 & 893.13 $\pm$ 0.19 & 967.33 & 8.31 & 899.94 & 0.76 \\
        1 & 904.38 $\pm$ 0.19 & 957.99 & 5.93 & 892.75 & -1.29 \\
        2 & 720.68 $\pm$ 0.17 & 780.72 & 8.33 & 718.10 & -0.36 \\
        3 & 818.51 $\pm$ 0.18 & 833.76 & 1.86 & 810.73 & -0.95 \\
        4 & 855.52 $\pm$ 0.18 & 915.73 & 7.04 & 841.84 & -1.60 \\
        5 & 898.02 $\pm$ 0.19 & 991.22 & 10.38 & 878.41 & -2.18 \\
        6 & 804.43 $\pm$ 0.18 & 881.39 & 9.57 & 773.68 & -3.82 \\
        7 & 787.71 $\pm$ 0.18 & 857.39 & 8.85 & 752.67 & -4.45 \\
        8 & 796.51 $\pm$ 0.18 & 868.11 & 8.99 & 754.37 & -5.29 \\
        \hline
         ANAIS-112  & 830.99 $\pm$ 0.06 & 894.85 & 7.68 & 813.61 & -2.09 \\
        \hline
    \end{tabular}}
    \caption{\label{tablaDesvHE} Measured total rates in the [100–1600] keV range for each ANAIS-112 detector and their average over the full array during the six years of data taking. The corresponding simulated rates expected from both the previous background model and the revised model developed in this thesis are provided for each case, together with their respective deviations from the measured values. The associated statistical uncertainty in the simulations, not shown in the table, is $\sim$0.1\%.}
    
\end{table}

On the other hand, the peak near 250 keV tends to be slightly underestimated in certain detectors, a deficit of counts that is also observed in the summed spectrum. This may indicate the presence of other unaccounted background contributions that would add to the identified components in the lower part of the spectrum. There are indications of a possible contribution from \textsuperscript{235}U, and further investigations will be carried out to improve the characterization of the PMTs.

\begin{figure}[t!]
    \centering
    {\includegraphics[width=0.93\textwidth]{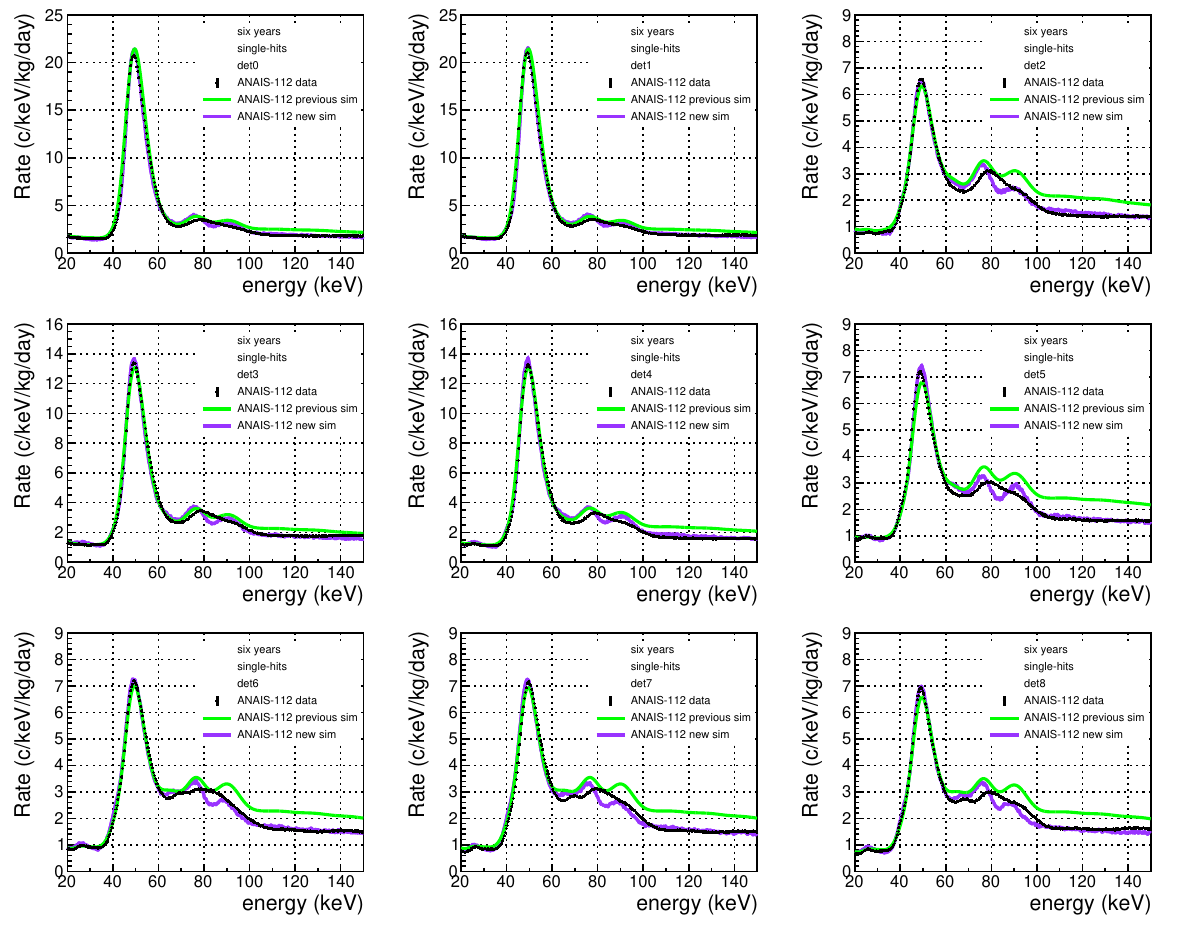}}

    \caption{\label{totalmodelME_perdet} Comparison of the single-hits medium-energy spectra measured over six years for each ANAIS-112 detector (black) with the single-hits background model developed in this thesis (violet) and the previous ANAIS-112 model (green). The model assumes a 10~$\mu$m exponential profile for the \(^{210}\)Pb surface contamination.  }
\end{figure}

\begin{figure}[b!]
    \centering
    {\includegraphics[width=0.6\textwidth]{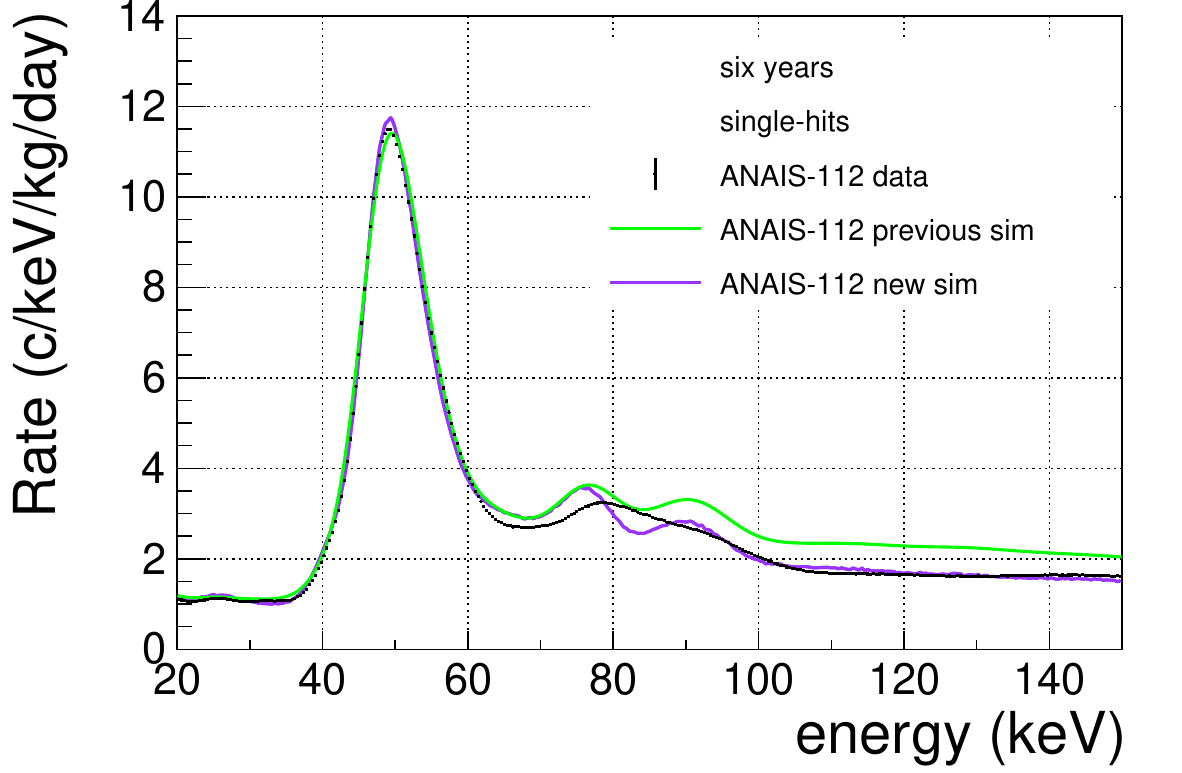}}

    \caption{\label{totalmodelME}  Comparison of the singe-hits medium-energy spectra measured over six years for the summed spectra of the nine ANAIS-112 detectors (black) with the total background model developed in this thesis (violet) and the previous ANAIS-112 model (green). The model assumes a 10~$\mu$m exponential profile for the \(^{210}\)Pb surface contamination.}
\end{figure}

However, the overall agreement between data and simulation is notably improved with the updated model. This is quantitatively supported by Table~\ref{tablaDesvHE}, which presents the integrated measured rates in the total high-energy ([100-1600] keV) for each individual detector and for the full detector array, together with the corresponding rates from both the previous and current simulations, as well as the relative deviations. The statistical uncertainty of the simulated rates remains below 0.1\% and is therefore omitted. The improvement achieved with the new simulation is consistent across all detectors, yielding deviations below 5\%, which clearly reinforces the improved reliability of the updated background model, although a slight global underestimation is observed.

Proceeding with the medium-energy region, Figure~\ref{totalmodelME_perdet} shows the comparison in the medium-energy region between the measured single-hit spectra over six years for each ANAIS-112 detector, the background model developed in this work, and the previous ANAIS-112 model. Figure~\ref{totalmodelME} displays the corresponding summed spectrum for the nine detectors. It should be noted that the representation of these spectra includes the previously discussed correction consisting of a shift in the x-axis, applied to properly visualize the consistency between data and simulation.

\begin{table}[b!]
    \centering
    \resizebox{0.9\textwidth}{!}{\Large
    \begin{tabular}{c|c|cc|cc}
        \hline
       \multirow{3}{*}{detector} & \multicolumn{4}{c}{single-hits [60-100] keV}  \\
       \cline{2-6}
        
        & \makecell{data \\ ($\text{kg}^{-1} \text{day}^{-1}$)} 
 & \makecell{previous sim \\ ($\text{kg}^{-1} \text{day}^{-1}$)} 
 & \makecell{previous desv \\ (\%)} 
 & \makecell{new sim \\ ($\text{kg}^{-1} \text{day}^{-1}$)} 
 & \makecell{new desv \\ (\%)} \\
 
        \hline
        0 & 3.13 $\pm$ 0.07 & 3.45 & 10.03 & 3.28  & 4.90 \\
        1 & 3.17 $\pm$ 0.07 & 3.44 & 8.39 & 3.32  & 4.65 \\
        2 & 2.56 $\pm$ 0.06 & 2.95 & 15.07 & 2.57  & 0.19 \\
        3 & 2.98 $\pm$ 0.07 & 3.11 & 4.51 & 2.99  & 0.39 \\
        4 & 2.88 $\pm$ 0.07 & 3.20 & 10.91 & 3.05  & 5.69 \\
        5 & 2.66 $\pm$ 0.07 & 3.12 & 17.13 & 2.72  & 2.12 \\
        6 & 2.80 $\pm$ 0.07 & 3.13 & 11.69 & 2.75  & -1.71 \\
        7 & 2.76 $\pm$ 0.07 & 3.12 & 13.04 & 2.67  & -3.46 \\
        8 & 2.65 $\pm$ 0.07 & 3.08 & 16.09 & 2.63  & -1.06 \\
        \cline{1-6}
         ANAIS-112  & 2.84 $\pm$ 0.02 & 3.18 & 11.66 & 2.89  & 1.43 \\
       \hline
    \end{tabular}}
    \caption{\label{tablaresultadosMEtotal} Measured single-hits rates in the [60–100] keV range for each ANAIS-112 detector and their average over the full array during the six years of data taking. The corresponding simulated rates expected from both the previous background model and the revised model developed in this thesis are provided for each case, along with their respective deviations from the measured values. The associated statistical uncertainty in the simulations, not shown in the table, is $\sim$0.1\%.}
    
\end{table}

The overall agreement is satisfactory. The compatibility was already reasonably good up to 60 keV with the previous background model, and in the present work, the measured \(^{210}\text{Pb}\) structure amplitude for the six years exposure is better reproduced in some detectors, notably D5 and D8. However, as discussed earlier, the region from 60 to 100 keV was not included in the fit, since this energy range is dominated by asymmetric single-hit events whose origin is not understood and the energy estimator is probably introducing some systematics that prevents this region from being well reproduced by the MC (see Section~\ref{asymmetry}). Therefore, the results shown in the figures reflect the extrapolation of the fitted activities obtained in other regions, principally influenced here by the PMT contributions, which were fitted in the high-energy region.

Table~\ref{tablaresultadosMEtotal} presents the integrated measured rates in the [60–100] keV interval for each detector and for the full array, along with the corresponding rates from both the previous and current simulations, as well as the relative deviations. Compared to the previous background model, the new model achieves a reduction in the predicted rate in this region. As can be observed, the total integral in this energy range is well reproduced in numerical terms, leaving no room for the introduction of an additional contamination component.


To further understand the spectral shape in the [60–100]~keV region, future simulations should incorporate a model describing the differential light collection efficiency as a function of the interaction position. Moreover, in the future, the impact of introducing a partial light production or collection at the detector surfaces could be investigated. Preliminary studies in this direction have already been conducted \cite{tfgcarmen}. This approach is motivated by the fact that while the
circular bases of the ANAIS scintillators are polished, the lateral surface remains rough,
reducing light emission and collection efficiency. Such roughness affects the probability
of decay products escaping the surface and depositing part of their energy, which can
significantly influence the detector response for events occurring in this region.

\begin{figure}[t!]
    \centering
    {\includegraphics[width=0.95\textwidth]{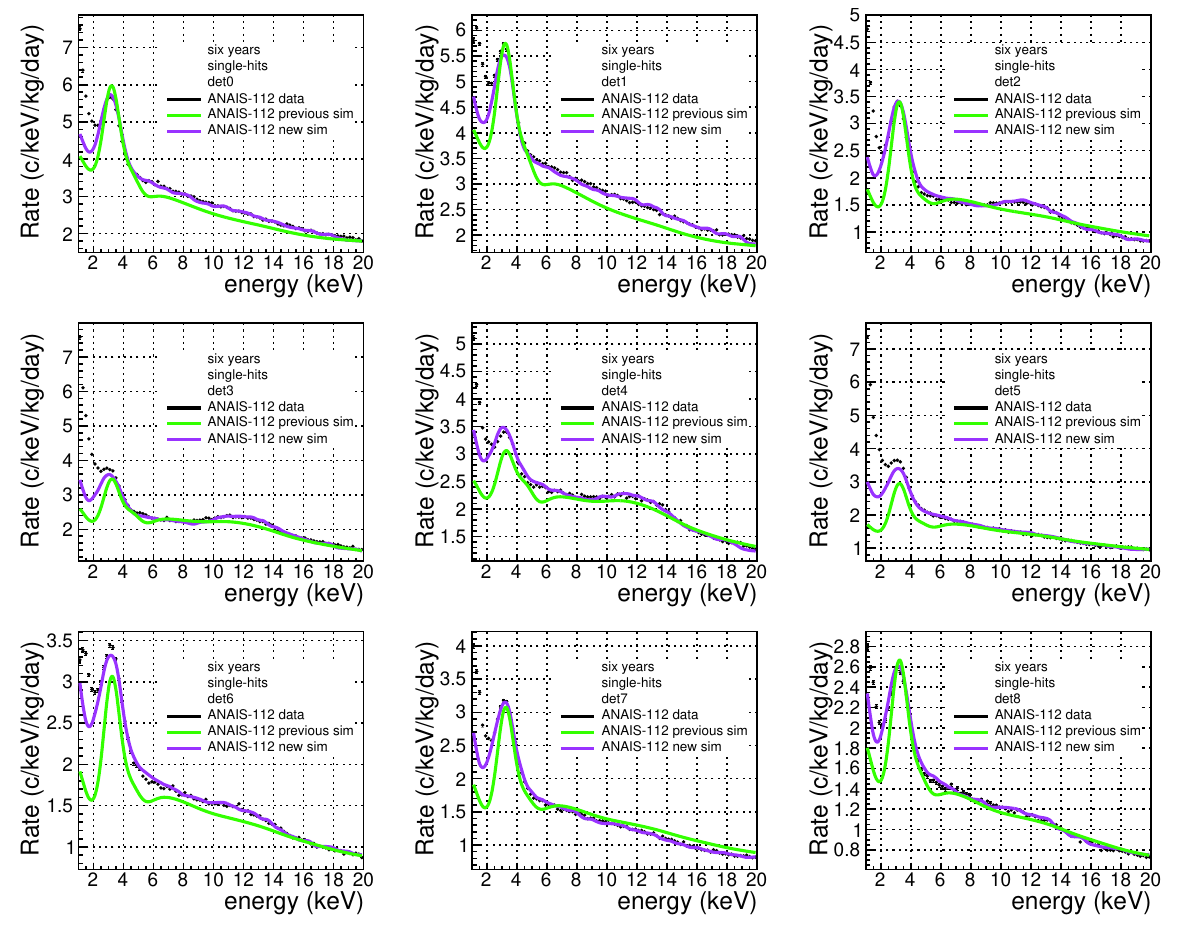}}

    \caption{\label{totalmodelLE_perdet} Comparison of the single-hits low-energy spectra measured over six years for each ANAIS-112 detector (black) with the background model developed in this thesis (violet) and the previous ANAIS-112 model (green). The model assumes a 10~$\mu$m exponential profile for the \(^{210}\)Pb surface contamination. }
\end{figure}

\begin{figure}[b!]
    \centering
    {\includegraphics[width=0.6\textwidth]{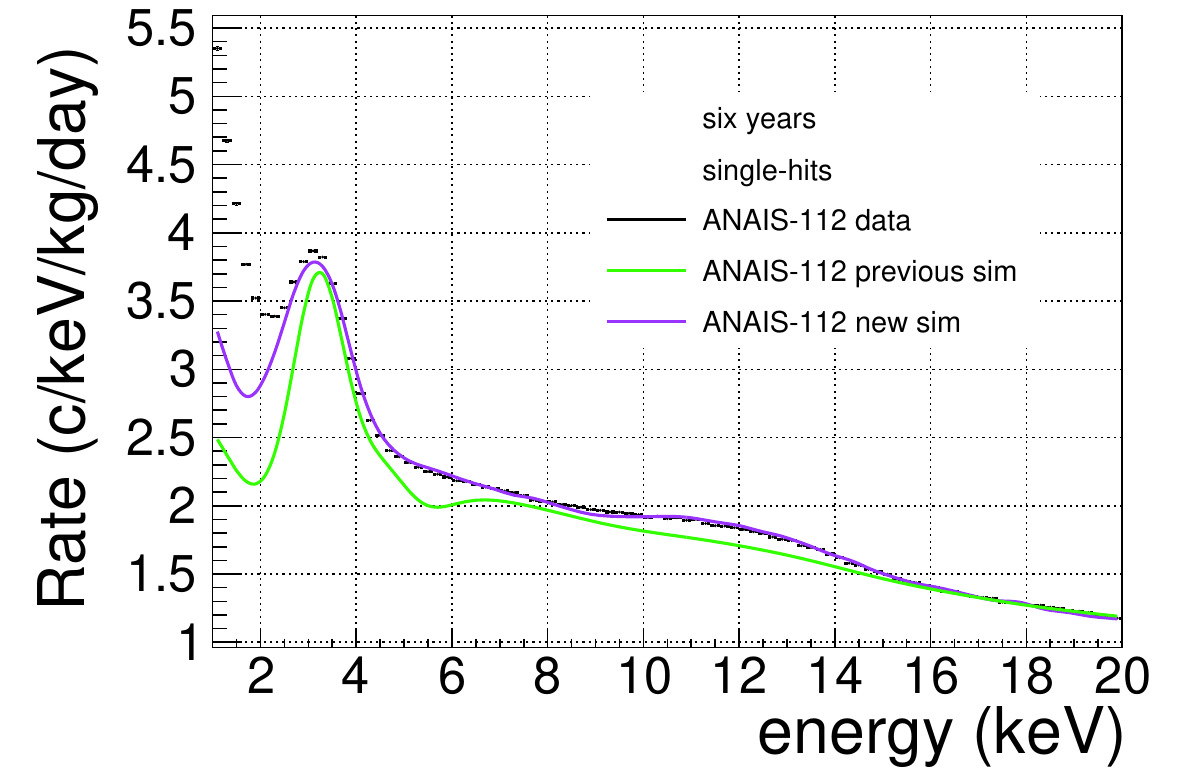}}

    \caption{\label{totalmodelLE} Comparison of the single-hits low-energy spectra measured over six years for the summed spectra of the nine ANAIS-112 detectors (black) with the background model developed in this thesis (violet) and the previous ANAIS-112 model (green). The model assumes a 10~$\mu$m exponential profile for the \(^{210}\)Pb surface contamination.}
\end{figure}

Focusing on the low-energy region, which is ultimately the most critical for the annual modulation analysis,  Figure~\ref{totalmodelLE_perdet} presents the comparison between the measured single-hit spectra over six years for each ANAIS-112 detector, the background model developed in this work, and the previous ANAIS-112 model. Figure~\ref{totalmodelLE} shows the corresponding summed spectrum over the nine modules.

As observed, the improvement over the previous background model is evident across all detectors. It is worth highlighting that this result has been obtained without including the ROI (single-hits, [1–6] keV) in the fit. All fits have been performed above 6~keV, except for the cases of the \textsuperscript{22}Na and \textsuperscript{40}K populations, for which the m2 spectrum was used. On one hand, the new model more accurately reproduces the continuum above 6~keV, correctly accounting for the \(^{210}\)Pb contamination in the teflon coating of the crystals. This improvement is particularly visible in detectors D2, D3, and D4, which exhibit higher levels of this isotope in that volume. On the other hand, the description of the 3.2 keV peak from \(^{40}\)K is significantly improved, as the resolution was previously underestimated in some detectors due to the assumption of a common resolution for all the detectors. Furthermore, the modelling of the [1–2] keV region is clearly enhanced. The overall agreement between the data and the simulation is generally very satisfactory.


\begin{table}[t!]
    \centering
    \resizebox{0.9\textwidth}{!}{\Large
    \begin{tabular}{c|c|cc|cc}
        \hline
       \multirow{3}{*}{detector} & \multicolumn{5}{c}{single-hits  [2,6] keV}  \\
       \cline{2-6}
        & \makecell{data \\ ($\text{kg}^{-1} \text{day}^{-1}$)} 
 & \makecell{previous sim \\ ($\text{kg}^{-1} \text{day}^{-1}$)} 
 & \makecell{previous desv \\ (\%)} 
 & \makecell{new sim \\ ($\text{kg}^{-1} \text{day}^{-1}$)} 
 & \makecell{new desv \\ (\%)} \\
        \hline
        0 & 4.39 $\pm$ 0.03 & 4.23 & -3.82 & 4.41 & 0.32 \\
        1 & 4.43 $\pm$ 0.03 & 4.15 & -6.42 & 4.30 & -2.97 \\
        2 & 2.33 $\pm$ 0.02 & 2.13 & -8.59 & 2.40 & 2.81 \\
        3 & 3.06 $\pm$ 0.02 & 2.67 & -12.81 & 2.90 & -5.21 \\
        4 & 2.87 $\pm$ 0.02 & 2.52 & -12.10 & 2.91 & 1.53 \\
        5 & 2.80 $\pm$ 0.02 & 2.06 & -26.35 & 2.61 & -6.87 \\
        6 & 2.57 $\pm$ 0.02 & 2.08 & -18.91 & 2.55 & -0.66 \\
        7 & 2.34 $\pm$ 0.02 & 2.08 & -10.89 & 2.33 & -0.12 \\
        8 & 1.95 $\pm$ 0.02 & 1.84 & -5.74 & 2.00 & 2.34 \\
        \hline
        ANAIS-112  & 2.97 $\pm$ 0.01 & 2.64 & -11.15 & 2.94 & -1.25 \\
        \hline
        
    \end{tabular}}
    \caption{\label{tablaDesvLE_26} Measured single-hits rates in the [2–6] keV range for each ANAIS-112 detector and their average over the full array during the six years of data taking. The corresponding simulated rates expected from both the previous background model and the revised model developed in this thesis are provided for each case, together with their respective deviations from the measured values. The associated statistical uncertainty in the simulations, not shown in the table, is $\sim$0.1\%.}
\end{table}

\begin{table}[b!]
    \centering
    \resizebox{0.9\textwidth}{!}{\Large
    \begin{tabular}{c|c|cc|cc}
        \hline
       \multirow{3}{*}{detector}  & \multicolumn{5}{c}{single-hits [1-2] keV}  \\
       \cline{2-6}
        & \makecell{data \\ ($\text{kg}^{-1} \text{day}^{-1}$)} 
 & \makecell{previous sim \\ ($\text{kg}^{-1} \text{day}^{-1}$)} 
 & \makecell{previous desv \\ (\%)} 
 & \makecell{new sim \\ ($\text{kg}^{-1} \text{day}^{-1}$)} 
 & \makecell{new desv \\ (\%)} \\
        \hline
        0 & 5.97 $\pm$ 0.02 & 3.85 & -35.40 & 4.37  & -26.75 \\
        1 & 5.61 $\pm$ 0.02 & 3.84 & -31.49 & 4.37  & -22.14 \\
        2 & 3.41 $\pm$ 0.02 & 1.59 & -53.48 & 2.16  & -36.60 \\
        3 & 5.56 $\pm$ 0.02 & 2.39 & -57.02 & 3.03  & -45.51 \\
        4 & 4.01 $\pm$ 0.02 & 2.34 & -41.61 & 3.06  & -23.67 \\
        5 & 5.32 $\pm$ 0.02 & 1.59 & -70.17 & 2.69  & -49.45 \\
        6 & 3.20 $\pm$ 0.01 & 1.71 & -46.61 & 2.62  & -18.08 \\
        7 & 3.27 $\pm$ 0.01 & 1.70 & -48.01 & 2.33  & -28.78 \\
        8 & 2.42 $\pm$ 0.01 & 1.61 & -33.41 & 2.02  & -16.30 \\
        \hline
        ANAIS-112  & 4.31 $\pm$ 0.01 & 2.29 & -46.80 & 2.96 & -31.24 \\
        \hline
    \end{tabular}}
    \caption{\label{tablaDesvLE_12} Measured single-hits rates in the [1-2] keV range for each ANAIS-112 detector and their average over the full array during the six years of data taking. The corresponding simulated rates expected from both the previous background model and the revised model developed in this thesis are provided for each case, together with their respective deviations from the measured values. The associated statistical uncertainty in the simulations, not shown in the table, is $\sim$0.1\%.}
    \vspace{-0.3cm}
\end{table}

Table~\ref{tablaDesvLE_26} presents the integrated measured rates in the [2–6] keV range for each individual detector and for the full array, together with the corresponding rates from both the previous and revisited background models, as well as the relative deviations. A clear improvement is observed across all detectors, with an average deviation of -1.25\% for this region, in contrast to -11.15\% with the previous background model.

As derived from Figure \ref{totalmodelLE_perdet} and Figure \ref{totalmodelLE}, the new model accounts for a substantially larger number of counts in the [1-2] keV range thanks to the better modelling of both bulk and surface components of \(^{210}\)Pb, a key isotope for ANAIS-112, as has been consistently highlighted. However, some discrepancies still persist below this energy. Table~\ref{tablaDesvLE_12} shows the analogous comparison in the [1–2] keV region. As can be seen, although the revisited model provides a substantially improved description across all energy ranges of ANAIS-112, it still fails to fully account for the observed rates in the [1–2] keV interval, with an average deviation of -31.24\% for this region, in contrast to -46.80\% obtained from the previous model. These results motivate the exploration of possible contributions beyond those arising from intrinsic radioactive contaminations in the various components of ANAIS-112, as will be discussed in the following section.

In summary, the revisited background model provides an unambiguously better description of the measured data. Nevertheless, the previous model should still be acknowledged for supporting the ANAIS-112 annual modulation analysis and for its important role as a starting point of this work, especially considering that it was developed with the limited data of only one year of exposure. 

After establishing the improvement over the previous background model, the following presents a comparison between the background model developed in this thesis and the data, for coincidences in different energy ranges. The predictions from the previous background model are not shown hereafter, as they were not generated for these observables.

\begin{figure}[t!]
    \centering
    {\includegraphics[width=0.68\textwidth]{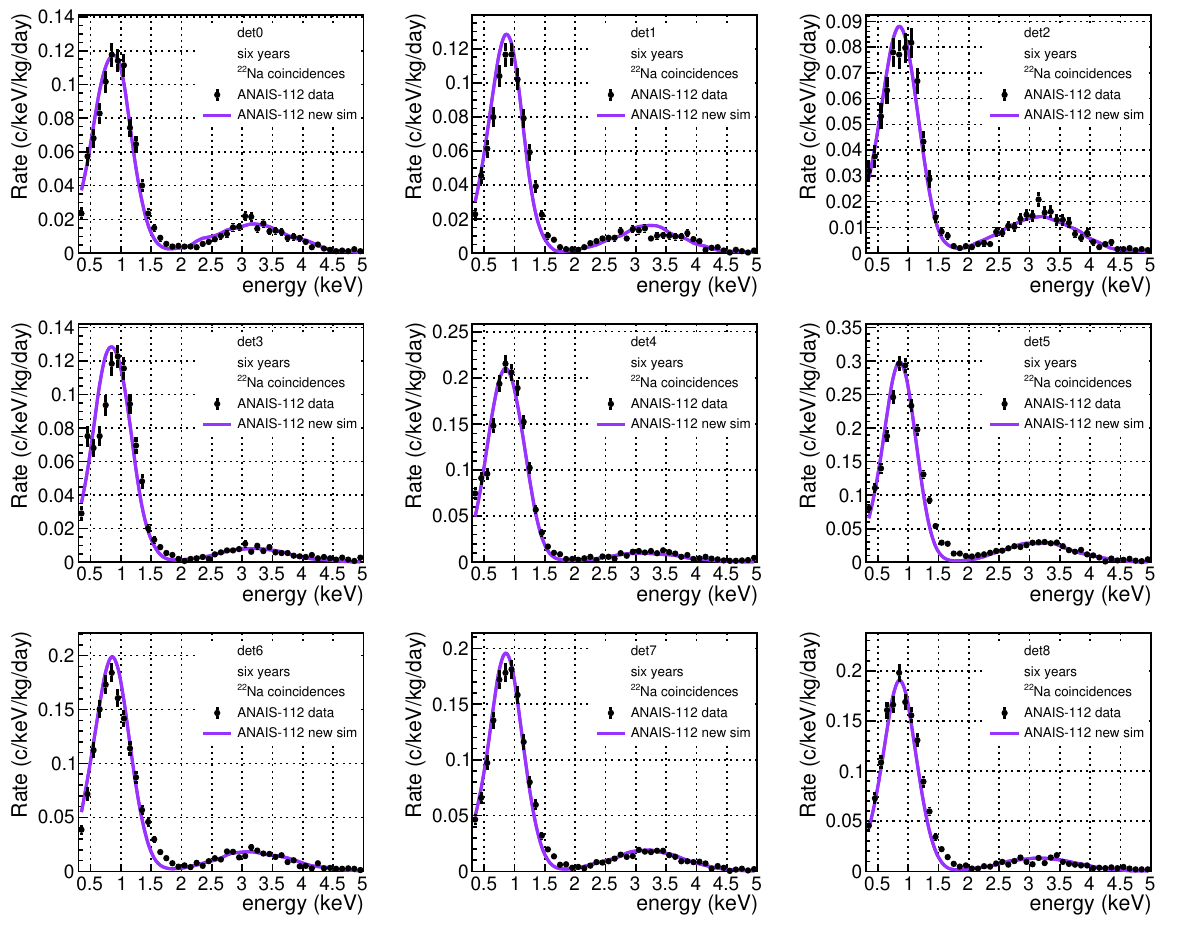}}

    \caption{\label{sixyearscoinna} Comparison of the low-energy spectra measured over six years in coincidence with a high-energy gamma at 1274.5 keV ([1200–1340] keV) detected in a second module for each ANAIS-112 detector (black), with the background model developed in this thesis (violet). A clear peak at 0.87 keV is attributed to the decay of $^{22}$Na in the NaI bulk. The low-amplitude peak at 3.2 keV corresponds to Compton-scattered 1460.8~keV gammas from $^{40}$K that satisfy the $^{22}$Na selection criteria.}

\end{figure}
\begin{figure}[b!]
    \centering
    {\includegraphics[width=0.68\textwidth]{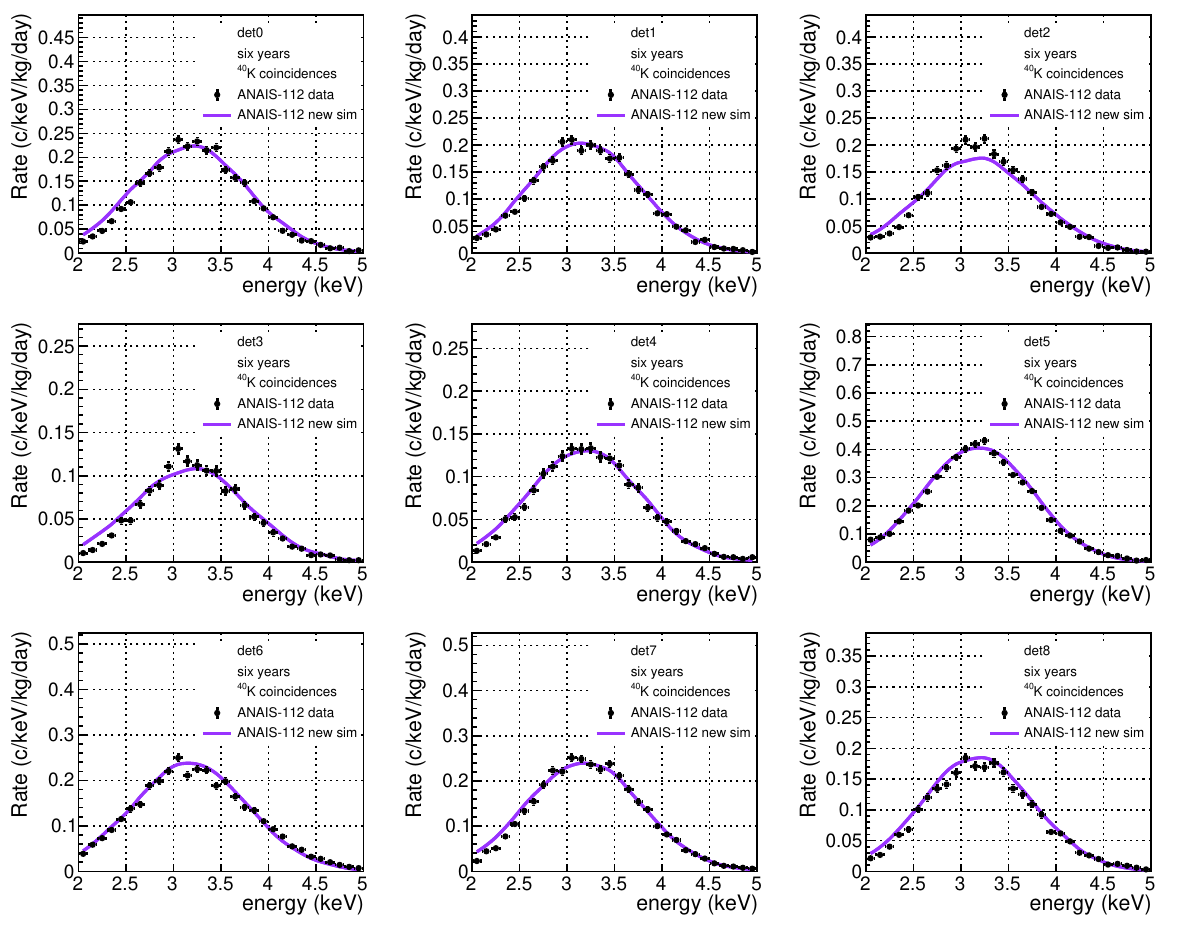}}

    \caption{\label{sixyearscoinK} Comparison of the low-energy spectra measured over six years in coincidence with a high-energy gamma at 1460.8 keV ([1340–1560] keV) detected in a second module for each ANAIS-112 detector (black), with the background model developed in this thesis (violet). A clear peak at 3.2 keV is attributed to the decay of $^{40}$K in the NaI bulk. }

\end{figure}

To start with, Figure \ref{sixyearscoinna} shows the comparison between data and the predictions of the improved background model for the six-year exposure, for the nine detectors, in the low-energy spectra corresponding to $^{22}$Na coincidences. Figure \ref{sixyearscoinK} presents
the analogous comparison for \textsuperscript{40}K coincidences. In some detectors, such as D2 and D3, the assumed energy resolution (see Figure \ref{LEres}) appears to be conservative at that energy range, suggesting that the actual resolution could be slightly better than modelled in these detectors. Nevertheless, the overall agreement is generally satisfactory.


\begin{figure}[t!]
    \centering
    {\includegraphics[width=0.82\textwidth]{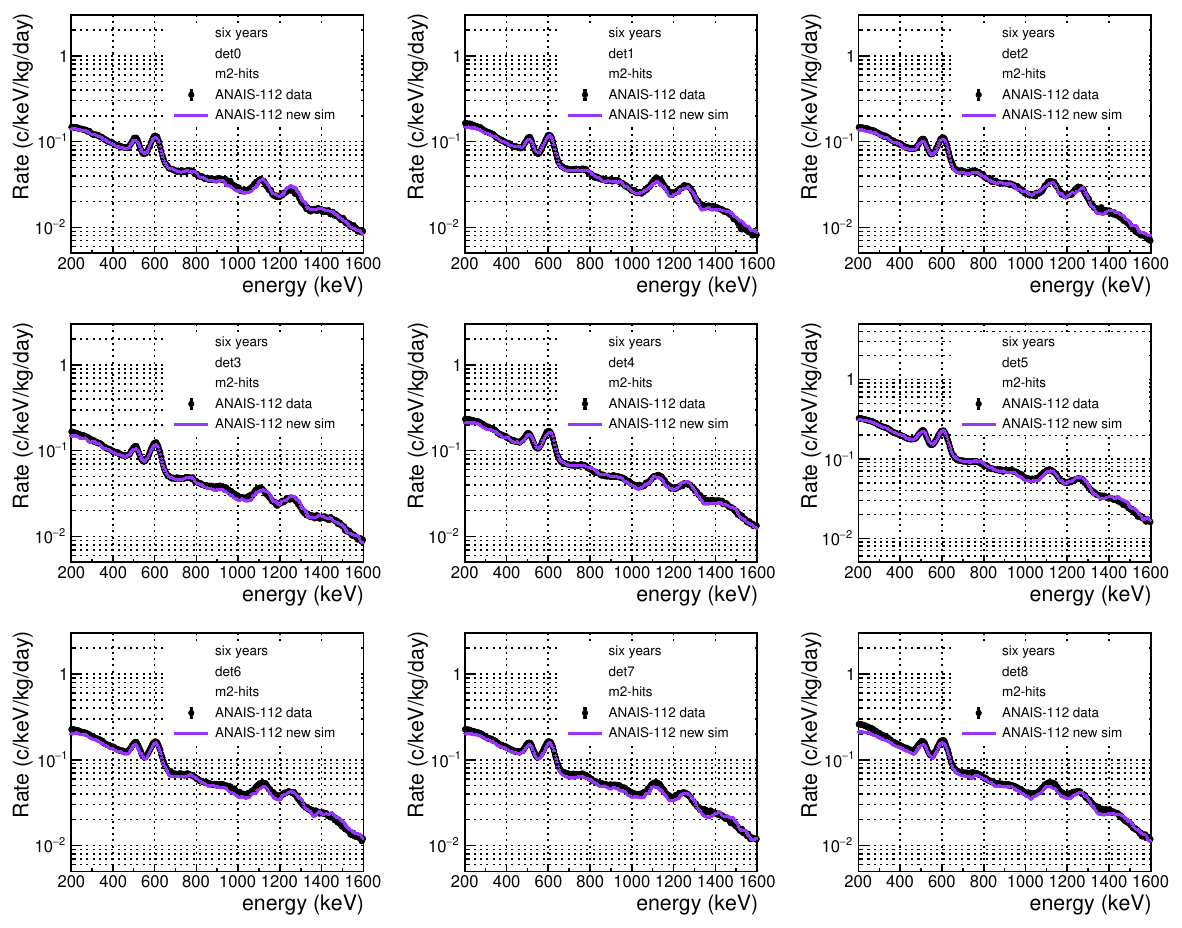}}

    \caption{\label{HEm2} Comparison of the m2-hits high-energy spectra measured over six years for each ANAIS-112 detector (black) with the background model developed in this thesis (violet).}

\end{figure}   

\begin{figure}[b!]
    \centering
    {\includegraphics[width=0.65\textwidth]{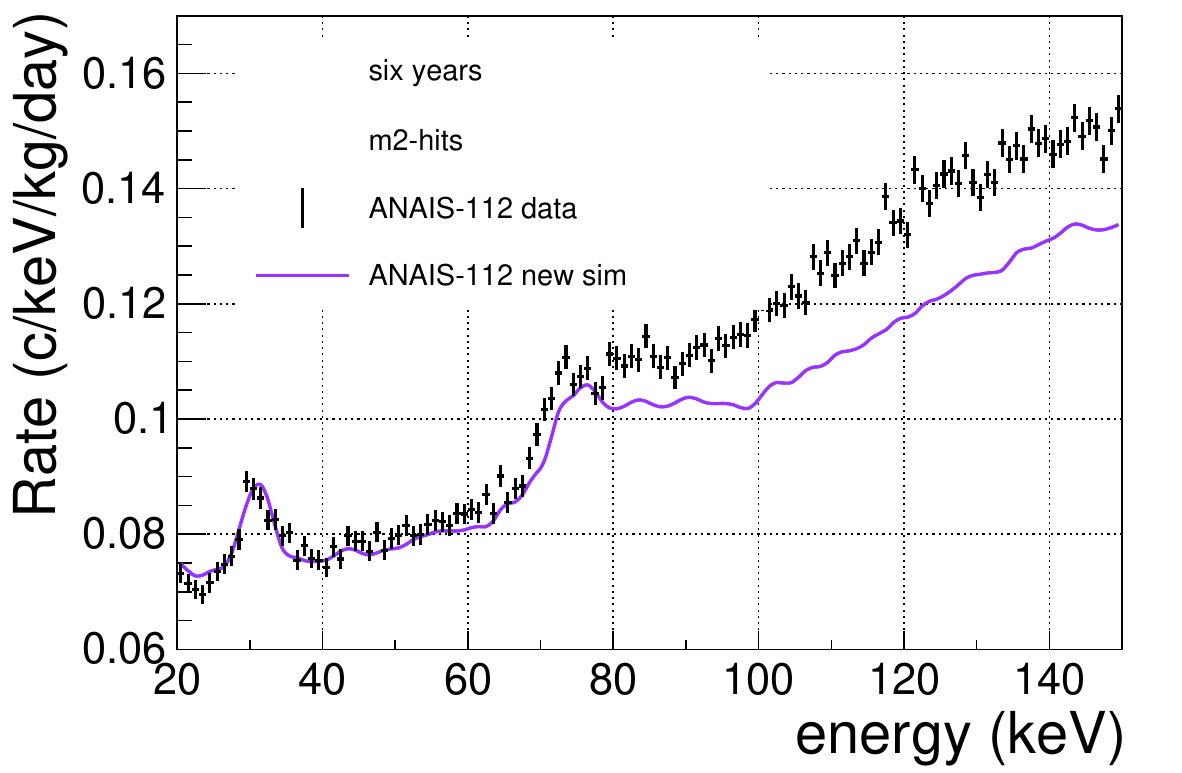}}

    \caption{\label{MEm2} Comparison of the m2-hits medium-energy spectra measured over six years for the summed spectra of the nine ANAIS-112 detectors (black) with the background model developed in this thesis (violet).}
    \vspace{-0.5cm}

\end{figure}
Moreover, the agreement between data and simulation in the m2-hit population for the six-year exposure is explored. Figure \ref{HEm2} shows the comparison between the new model and the data for this population in the high-energy region. All detectors exhibit strong consistency with the model, accurately reproducing the main data features. As was also the case in the single-hit comparison, detector 8 provides the poorest agreement below 400~keV. The origin of this discrepancy in this specific detector remains unidentified.

Figure \ref{MEm2} presents the corresponding comparison in the medium energy range. In this case, the comparison is shown for the sum over all detectors, since plotting detector-by-detector resulted in large statistical fluctuations that obscured any clear trends. The simulation correctly reproduces the 31.8 keV peak corresponding to the K-shell of Te, originating from the decay of Te isotopes. Interestingly, however, it fails to reproduce the behaviour observed above 80 keV, an energy region characterized by high-asymmetry events in single hits, which, notably, did not appear asymmetric in multiple-hit events (see Figure \ref{asy}).

\begin{figure}[t!]
    \centering
    {\includegraphics[width=0.9\textwidth]{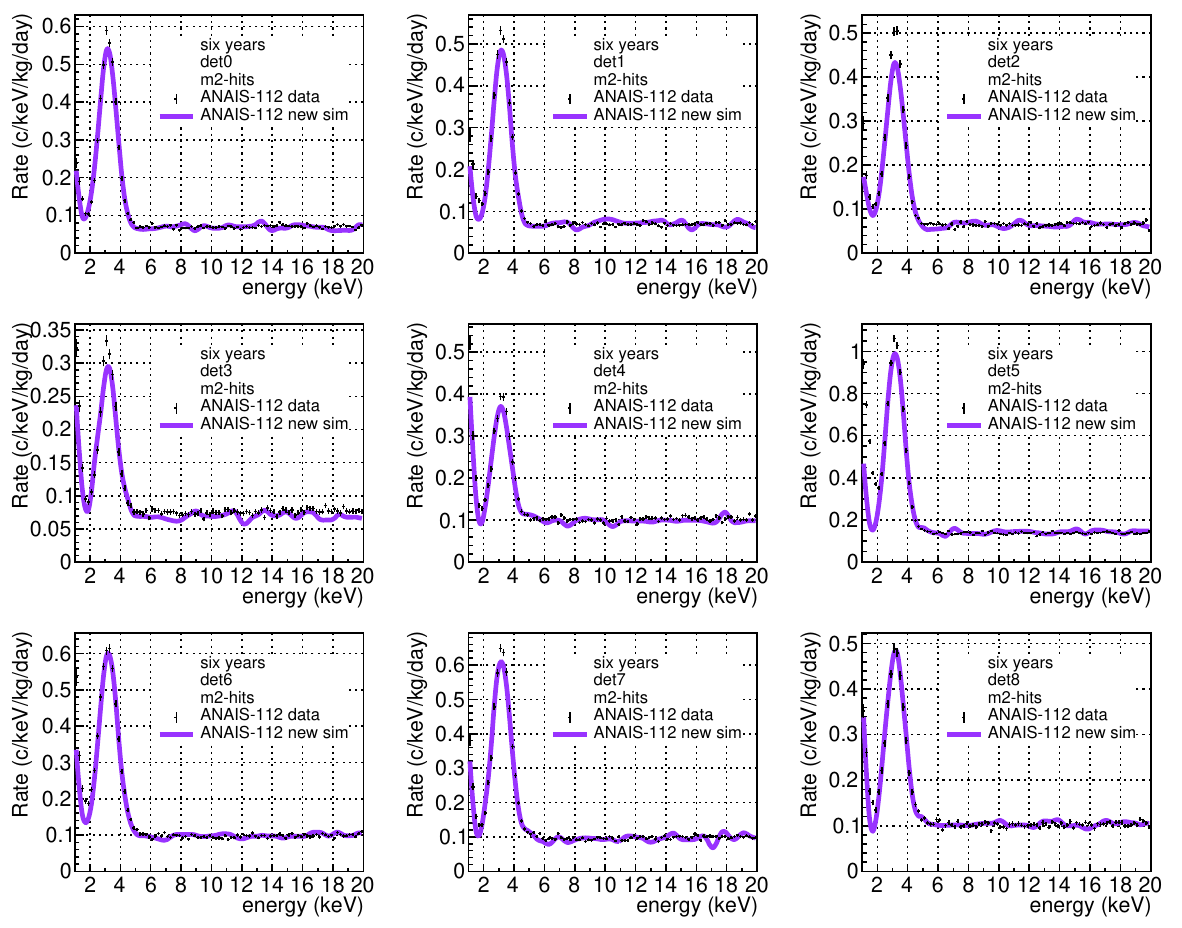}}

    \caption{\label{BEm2} Comparison of the m2-hits low-energy spectra measured over six years for each ANAIS-112 detector (black) with the background model developed in this thesis (violet).}
    
\end{figure}

In addition, Figure \ref{BEm2} shows the comparison of m2 hits between simulation and data in the low-energy region, correctly reproducing the $^{40}$K peak in this general m2-hit population. The largest discrepancies are perhaps found in detector D2, but overall, the description is satisfactory, and the deviations may stem from limitations in the resolution modelling.

On the other hand, it is particularly relevant to assess whether the new background model is capable of reproducing time differences, as its ultimate goal is to predict time evolution for its inclusion in the annual modulation analysis. Additionally, performing year-to-year subtractions allows to better visualize specific populations that may not be apparent when analyzing full-year data, as they are not dominant. Therefore, the comparison between data and simulation obtained by subtracting the sixth-year data from the first-year data is presented below, both for single-hit and multiple-hit populations.

\begin{figure}[t!]
    \centering
    {\includegraphics[width=0.72\textwidth]{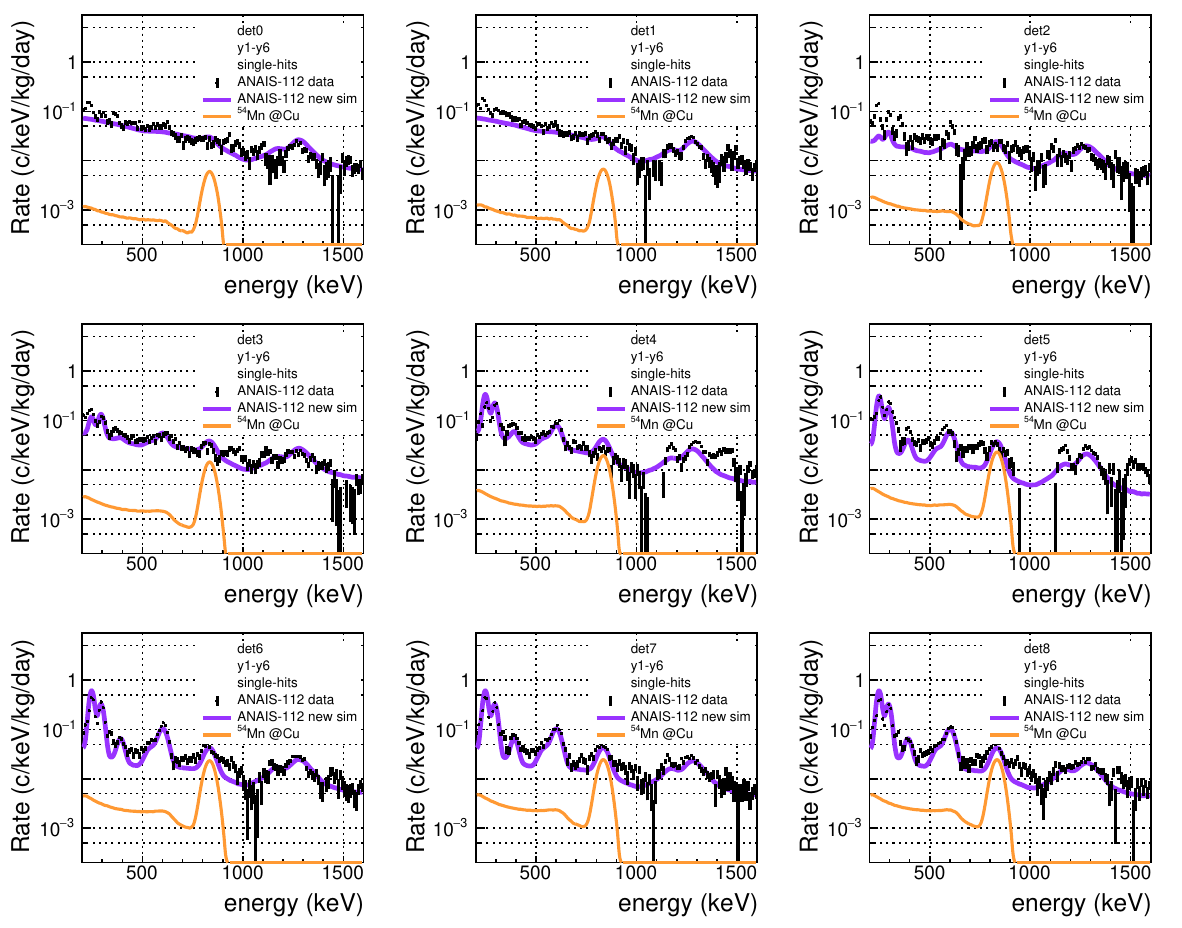}}

    \caption{\label{diffHEsingle} Comparison of the single-hit high-energy spectra obtained by subtracting the sixth-year data from the first-year data for the nine ANAIS-112 detectors (black), with the background model developed in this thesis (violet), which already includes the contribution from the newly identified $^{54}$Mn contamination in the copper housing surrounding the crystals. The $^{54}$Mn contribution separately is shown in orange.}
\end{figure}

\begin{figure}[b!]
    \centering
    {\includegraphics[width=0.72\textwidth]{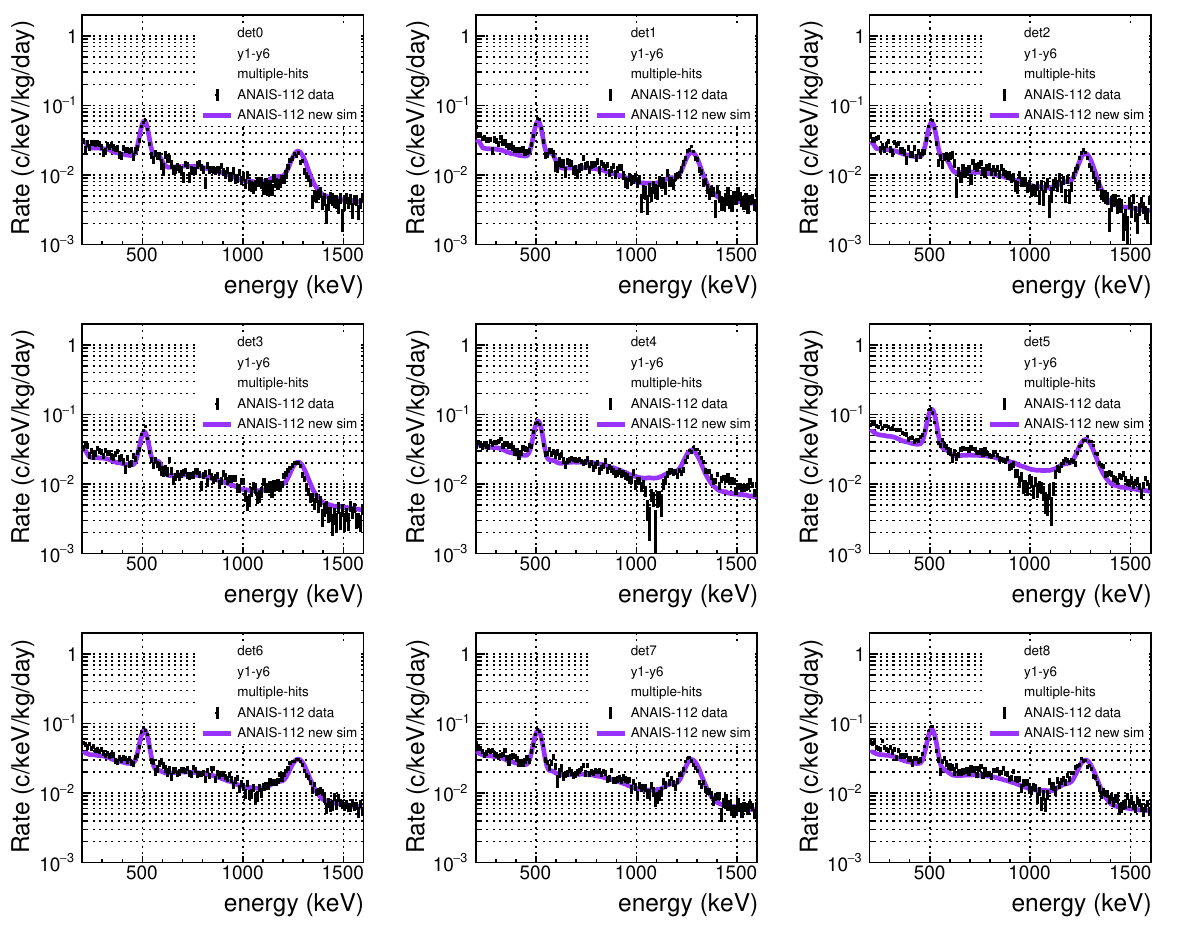}}

    \caption{\label{diffHEmulti} Comparison of the multiple-hits high-energy spectra obtained by subtracting the sixth-year data from the first-year data for the nine ANAIS-112 detectors (black), with the background model developed in this thesis (violet).  }
\end{figure}

\begin{figure}[t!]
    \centering
    {\includegraphics[width=0.76\textwidth]{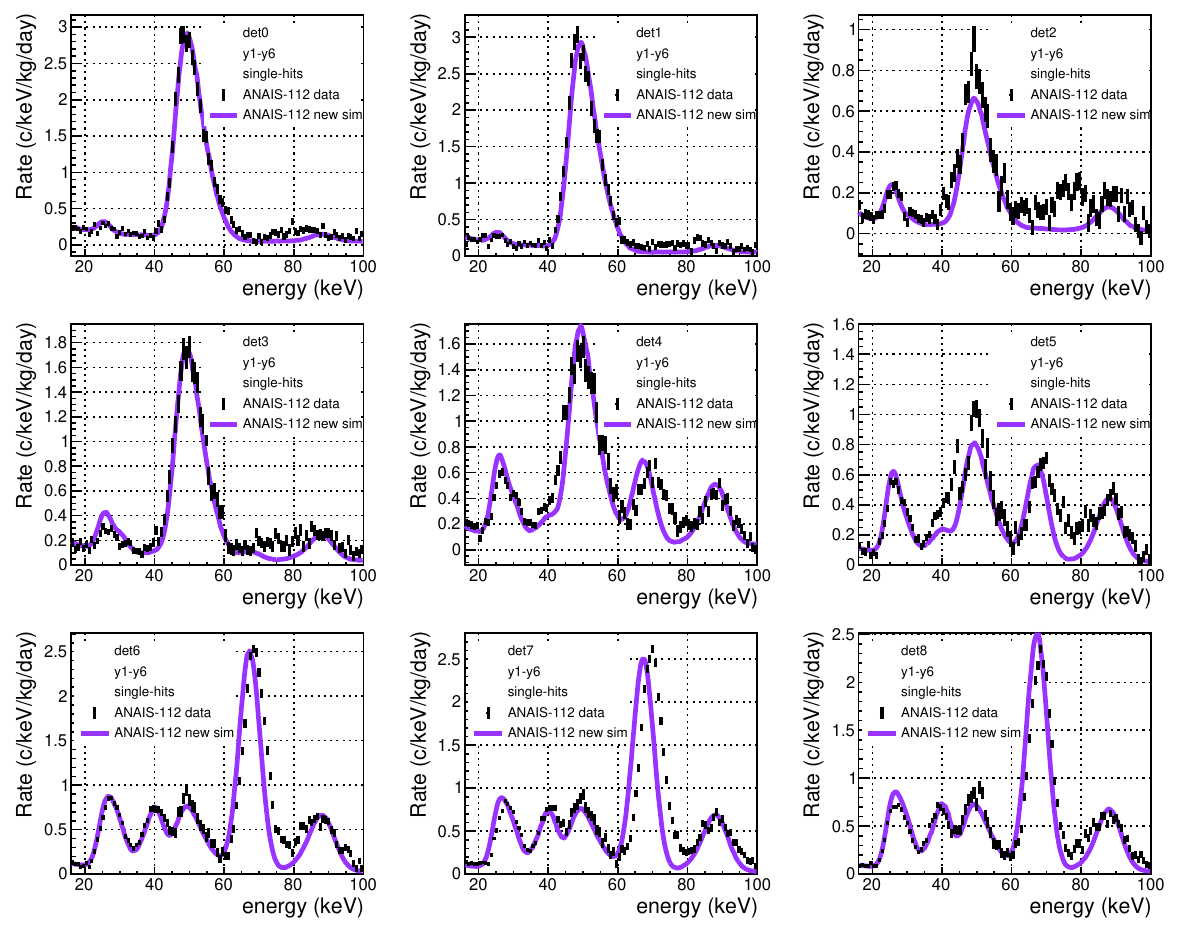}}

    \caption{\label{diffMEsingle} Comparison of the single-hits medium-energy spectra obtained by subtracting the sixth-year data from the first-year data for the nine ANAIS-112 detectors (black), with the background model developed in this thesis (violet). }
\end{figure}

\begin{figure}[b!]
    \centering
    {\includegraphics[width=0.76\textwidth]{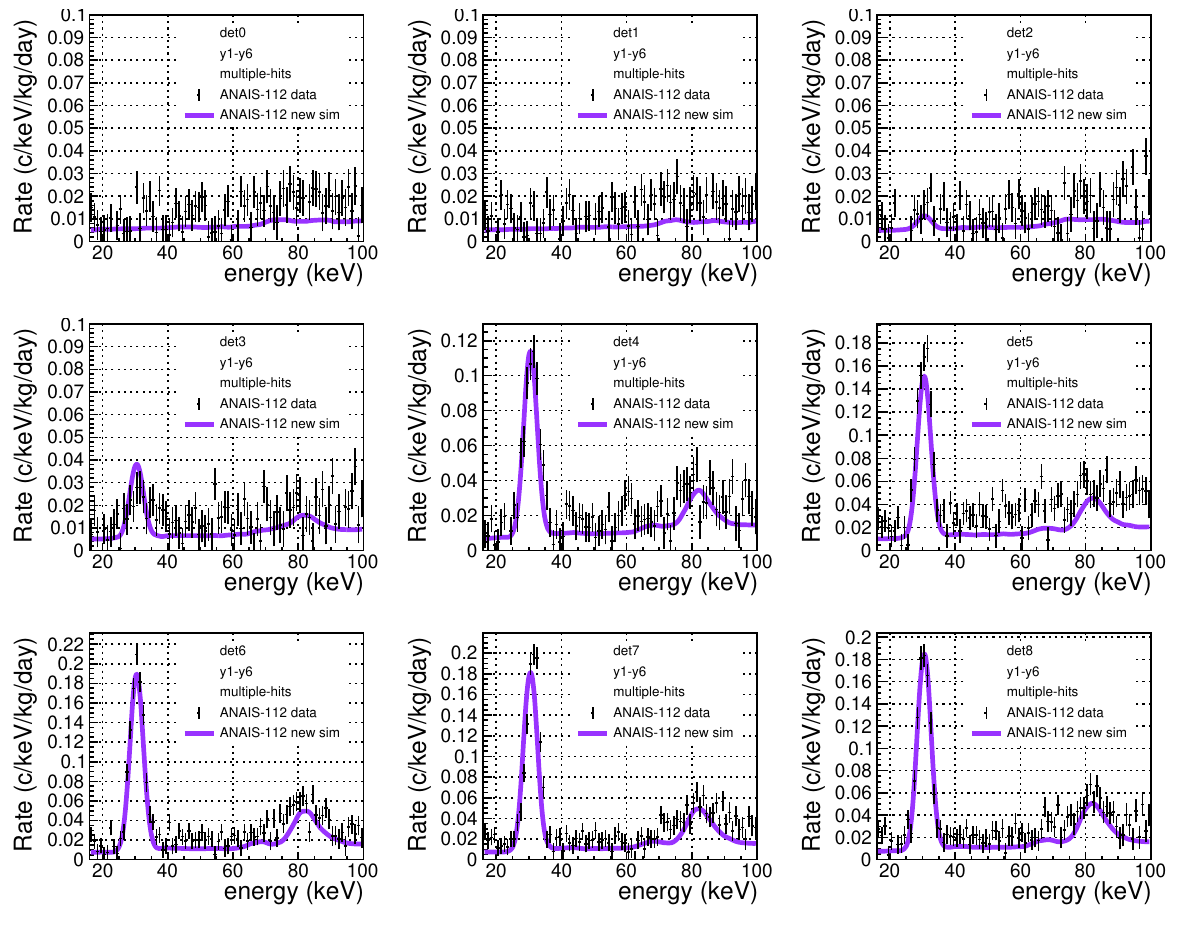}}

    \caption{\label{diffMEmulti} Comparison of the multiple-hits medium-energy spectra obtained by subtracting the sixth-year data from the first-year data for the nine ANAIS-112 detectors (black), with the background model developed in this thesis (violet). }
\end{figure}

Figure \ref{diffHEsingle} shows the comparison of these year-difference spectra for single-hit events. This approach revealed generally good agreement, although a distinct peak in the data suggested a missing contribution from an isotope. Specifically, this subtraction made it possible to identify the presence of an additional isotope, $^{54}$Mn, homogeneously distributed in the copper housing surrounding the crystals. $^{54}$Mn decays via electron capture (99.99\%) primarily to the 834.9 keV excited state of $^{54}$Cr, with minor EC and $\beta^+$ transitions to the ground state. The contribution from this isotope was simulated and manually included in the model, without being fitted, assuming a common activity for all detectors. This approach follows the procedure adopted for other external sources in \cite{amare2019analysis}. The best agreement was obtained with an activity of 2.71 mBq  homogeneously distributed in the copper housing. After incorporating this isotope into the background model, Figure \ref{diffHEsingle} shows that, with the exception of detector D2, whose behavior cannot be fully captured by the simulation for reasons not yet understood, the temporal difference is very well described across all detectors. In detectors D0 and D1, the new $\beta^-$ shape of $^{210}$Bi reproduces the data accurately, as these detectors are not significantly affected by other cosmogenic contributions, while in the rest, the temporal variation is also well modelled.

Figure \ref{diffHEmulti} presents the comparison between simulation and data for the same time difference in the multiple-hit population. The peaks at 511 keV and 1274 keV from $^{22}$Na are well described, further reinforcing the accuracy of the activity determination for this isotope.

Regarding the medium-energy region, Figure~\ref{diffMEsingle} shows the difference between year~1 and year 6 for single-hit events. The \textsuperscript{210}Pb feature around 50 keV is clearly visible and well captured by the simulation, although detectors D2 and D5 show a poorer description. In addition, the later detectors exhibit contributions from cosmogenic isotopes: \textsuperscript{113}Sn and \textsuperscript{109}Cd producing the peak around 25 keV, \textsuperscript{125}I with its composite peaks at 40.4 keV and 67.3~keV, and again the cosmogenic \textsuperscript{109}Cd giving rise to the 88~keV signal.

\begin{figure}[b!]
    \centering
    {\includegraphics[width=0.77\textwidth]{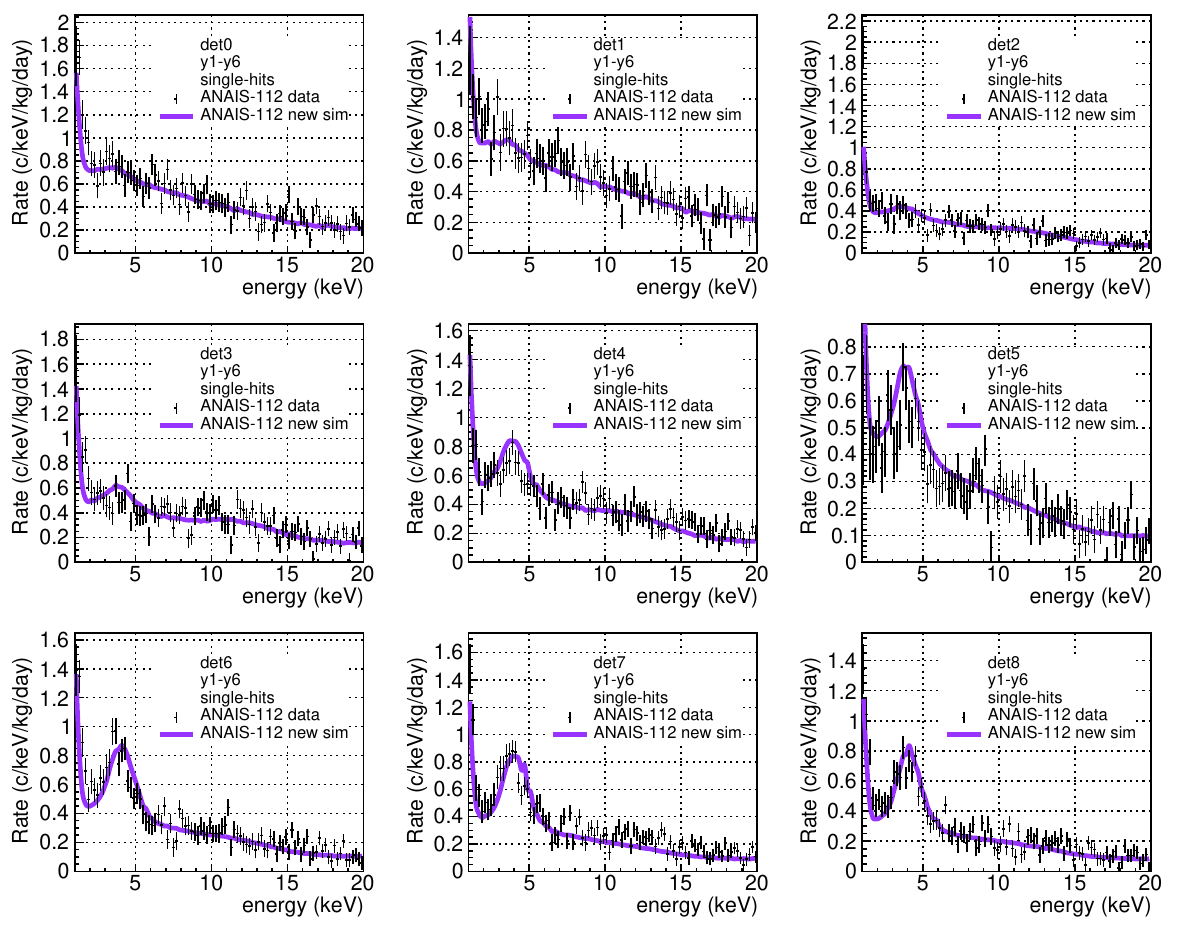}}

    \caption{\label{diffBEsingle} Comparison of the single-hits low-energy spectra obtained by subtracting the sixth-year data from the first-year data for the nine ANAIS-112 detectors (black), with the background model developed in this thesis (violet). }
\end{figure}

It is evident that the position of the \textsuperscript{125}I peak does not match between data and simulation, and this may be attributed to non-proportionality effects. The 67.3 keV feature results from the sum of a 35.5 keV gamma and 31.8 keV atomic de-excitation energy from K-shell EC. The latter component corresponds to a highly converted line involving both conversion and Auger electrons. According to the non-proportionality curve derived from ANAIS data (see Section \ref{nonpropSec}), both emissions lie near the K-shell dip, where a reduced light yield is expected for photons being absorbed. Therefore, based on this curve, the data points should systematically lie to the left of the simulation (i.e., exhibit lower light output). However, the opposite is observed: the data fall to the right of the simulation, similarly to what was commented about the shift applied to the \textsuperscript{210}Pb feature around 50~keV. As previously discussed, the light yield response differs for electrons and gammas, and the ANAIS-derived curve includes mixed contributions due to the inability to disentangle them for internal emissions. Thus, it is possible that the electron-specific curve near the K-dip might predict an increase in light yield, which cannot be identified with the currently available data.

Figure~\ref{diffMEmulti} shows the analogous comparison for multiple-hit events, where contributions from \textsuperscript{113}Sn and \textsuperscript{109}Cd can be identified. These are generally well described.

Finally, Figure~\ref{diffBEsingle} presents the year variation for single-hit events, which is well reproduced by the background model developed in this work. The multiple-hit counterpart is not included, as the limited statistics in this population hinder any robust comparison or conclusion.

Overall, the background model developed in this thesis provides an improved description compared to the previous ANAIS model across all event populations and energy ranges. Furthermore, its robustness is demonstrated consistently, including coincidence analyses and year variations.

\section{Additional contributions to the model}\label{additionalcontributions}

In the previous section, the new background model of ANAIS-112 has been constructed based on the multiparametric fit performed in this thesis of the different background components. As observed, the agreement between data and simulation systematically and substantially improves with the new model. However, a certain discrepancy still exists, particularly in the region between [1-2] keV. These events may originate from non-bulk scintillation signals that have not been effectively rejected by the filtering procedure or from background sources that have not
been considered in the current model. 

Therefore, in this section, additional contributions that could impact this energy region will be considered. These include, on the one hand, a background component not previously included in the model: the environmental neutron flux reaching the ANAIS-112 detectors and producing detectable signals. On the other hand, a contribution from noise events escaping the current selection cuts is also evaluated. The ANOD DAQ (see Section \ref{ANODfiltering}) has enabled the identification of a population of non-bulk scintillation events that is not being rejected by the selection protocols used in ANAIS so far, now allowing their contribution at the lowest energies to be estimated.


\subsection{Neutron background in ANAIS-112}\label{NeutronHENSA}

Rare event search experiments, such as ANAIS-112, require an extremely low background event rate. Underground laboratories provide a low-background environment due to the significant suppression of the muon flux from cosmic rays by the rock overburden. In particular, the measurement of the muon flux in the LSC indicates a reduction by a factor of \(10^5\) compared to surface levels \cite{trzaska2019cosmic}.  

While the component of the neutron flux produced by cosmic-ray muons is largely suppressed in underground laboratories, radiogenic neutrons are still generated in the surrounding rocks and cavity walls through ($\alpha$,n) reactions, induced by the $\alpha$-particles from the decay chains of \(^{238}\)U and \(^{232}\)Th on light elements, as well as through the spontaneous fission of \(^{238}\)U. These neutrons stand out as one of the most dangerous background sources in DM search experiments aiming to detect WIMPs, as the NRs produced by neutrons are indistinguishable from those induced by WIMPs. 

In the case of ANAIS-112, high-energy neutrons are particularly relevant, as they pose a significant challenge for DM detection since common shielding materials such as polyethylene and water cannot completely moderate them. Thermal neutrons are not as critical as they can no longer produce a measurable signal in the detector. Therefore, it is of paramount importance to measure and
fully characterize the neutron flux at the precise location
where the experiment is conducted, that is the hall B of the LSC.

 Although the ambient neutron background is presumed to be low, a precise characterization is essential since the effect of neutron background on the energy deposited in ANAIS-112 had not been previously explored. The improved understanding of the QFs for the ANAIS crystals also enables a more accurate estimation of their actual contribution to the measured spectrum. Throughout this thesis, a collaboration has been established with the HENSA experiment to assess the neutron background affecting ANAIS-112.

The High Efficiency Neutron Spectrometry Array (HENSA) \cite{hensa} is a state-of-the-art detection system based on the same principles as Bonner Sphere spectrometers \cite{thomas2002bonner}, which are among the most recognized and widely used techniques for neutron spectrometry. It consists of ten long \(^{3}\)He-filled neutron proportional counters embedded in blocks of high-density polyethylene of varying sizes, enabling sensitivity across a broad range of neutron energies. HENSA provides energy sensitivity ranging from thermal energies up to 10 GeV.

\begin{figure}[t!]
    \centering
    {\includegraphics[width=0.44\textwidth]{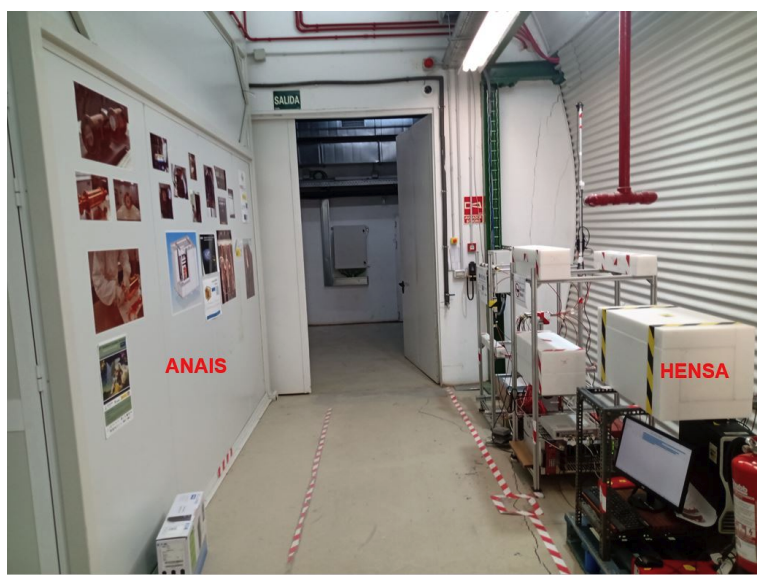}}
    {\includegraphics[width=0.55\textwidth]{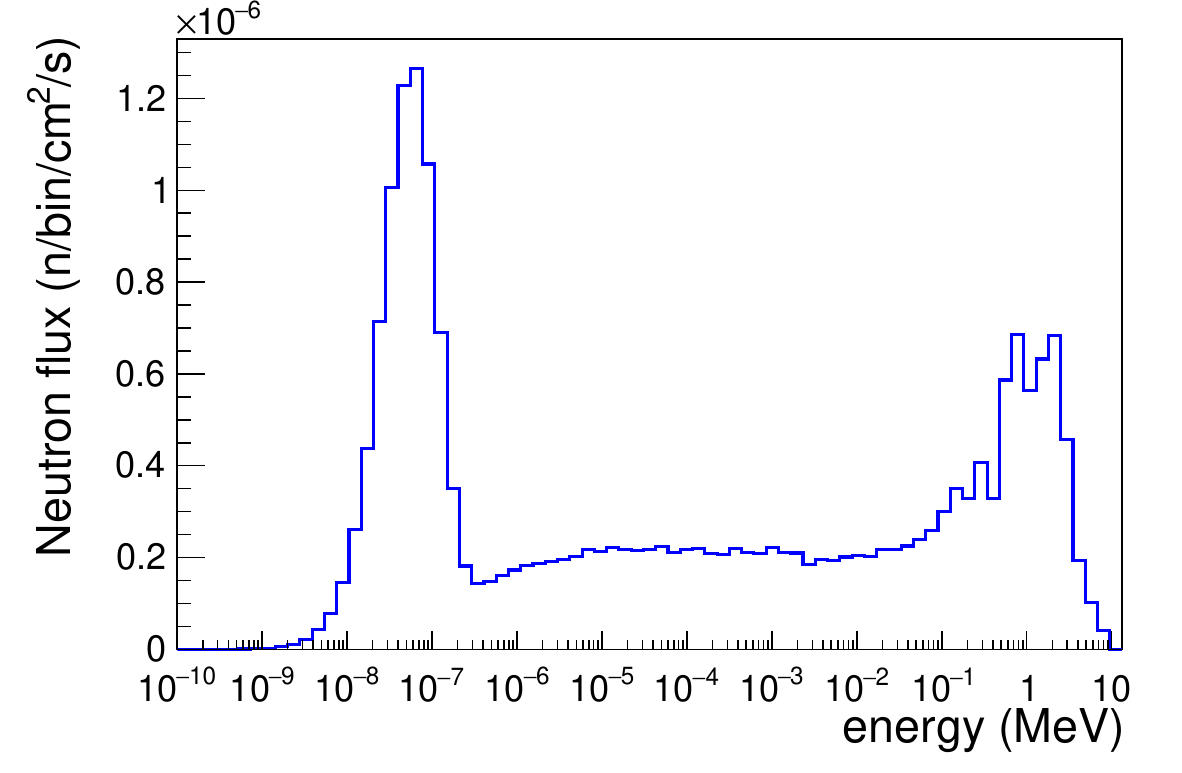}}
    
      \caption{\label{fondoneuInput} \textbf{Left panel:} HENSA set-up in hall B, indicating its relative position with respect to the ANAIS-112 experiment. \textbf{Rigth panel:} Neutron flux energy spectrum measured by HENSA in the hall B of the LSC, where the ANAIS-112 experiment is located. The integral neutron flux is \(\Phi_B = 20.41 \times 10^{-6} \text{ cm}^{-2} \text{ s}^{-1}\). The spectrum exhibits characteristic features, including the thermal neutron peak on the left, the fast neutron peak on the right, and a continuous component corresponding to intermediate-energy neutrons \cite{nil}.} 

\end{figure}

Since 2019, HENSA has carried out a long-term characterization of the neutron background at the LSC, including continuous measurements in both hall A and hall B. The left panel of Figure \ref{fondoneuInput} shows the HENSA set-up in hall B, indicating its relative position with respect to the ANAIS-112 experiment. The right panel of the same figure presents the neutron flux energy spectrum as measured by HENSA collaboration in hall B of the LSC, reconstructed from the counting rates using an unfolding algorithm \cite{nil}. The spectrum reveals distinct features, including the thermal neutron peak on the left, the fast neutron peak on the right, and a continuous component associated with intermediate-energy neutrons.  In the very high-energy region (above 20 MeV), muon-induced neutrons constitute the sole contribution and most of the prompt muon-induced neutron component would be vetoed by the plastic scintillators coincidence. However, this component is not considered in this work, as calculations indicate it is negligible relative to the rest of the total flux. The spectrum can typically be divided into three regions: thermal neutrons, from \( 1 \times 10^{-10} \) to \( 3.2 \times 10^{-7} \) MeV; intermediate neutrons, from \( 3.2 \times 10^{-7} \) to 0.1 MeV; and fast neutrons, from 0.1 to 100~MeV.


The integral of the neutron flux in hall B is found to be \(\Phi_B = 20.41 \times 10^{-6} \text{ n}/\text{cm}^{2}/\text{s}\), which exceeds that previously measured in hall A, \(\Phi_A = 16.22 \times 10^{-6} \text{ n}/\text{cm}^{2}/\text{s}\). This discrepancy is attributed to differences in the composition of the rock and concrete between the two locations, as well as variations in the mountain profiles above each hall.


FLUKA \cite{fluka1} is the primary simulation code employed by HENSA, an experiment entirely dedicated to the study of neutron backgrounds. Nevertheless, significant advancements have been made in recent years in the Geant4 Neutron-HP package. These developments have brought Geant4 to a level of accuracy comparable with reference neutron transport codes such as FLUKA \cite{fluka1,fluka2}, Tripoli-4 \cite{tripoli}, or MCNP6.2 \cite{mcnp} in terms of neutron physics.  

Consequently, the collaboration between ANAIS and HENSA arises in the context of cross-checking the ambient neutron flux reaching the ANAIS detector cavity, using the neutron flux measured by HENSA in hall B as the initial input (right panel of Figure~\ref{fondoneuInput}). To verify the neutron interaction and transport processes through the ANAIS-112 set-up, two simulation codes will be employed: Geant4 and FLUKA. While Geant4 will be used in this thesis, HENSA will rely on FLUKA. Once the compatibility of both codes in assessing the neutron flux reaching the ANAIS-112 detectors is evaluated, the energy deposited in the ANAIS modules will be estimated and compared between the two codes.

\subsubsection{Neutron flux assesment}

\begin{figure}[b!]
    \centering
    {\includegraphics[width=0.42\textwidth]{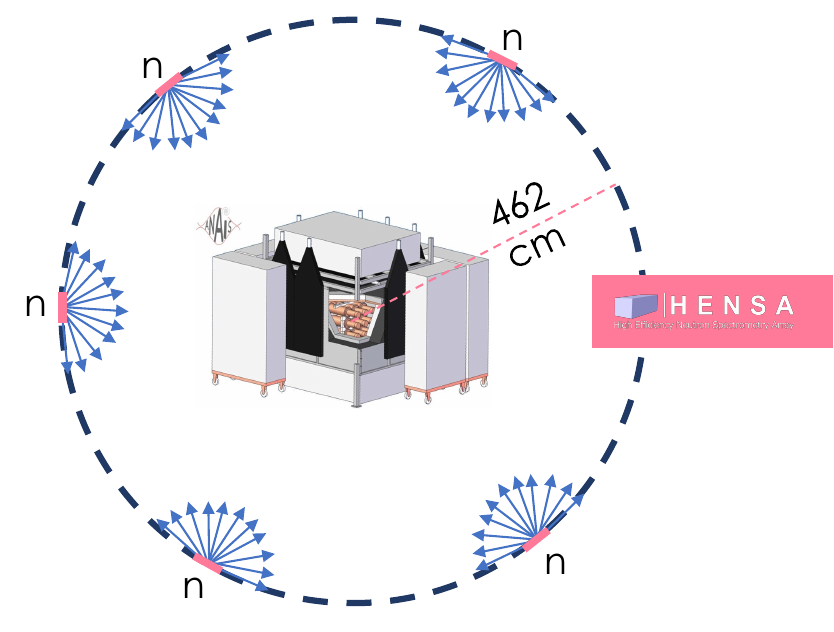}}
    \hfill
    {\includegraphics[width=0.49\textwidth]{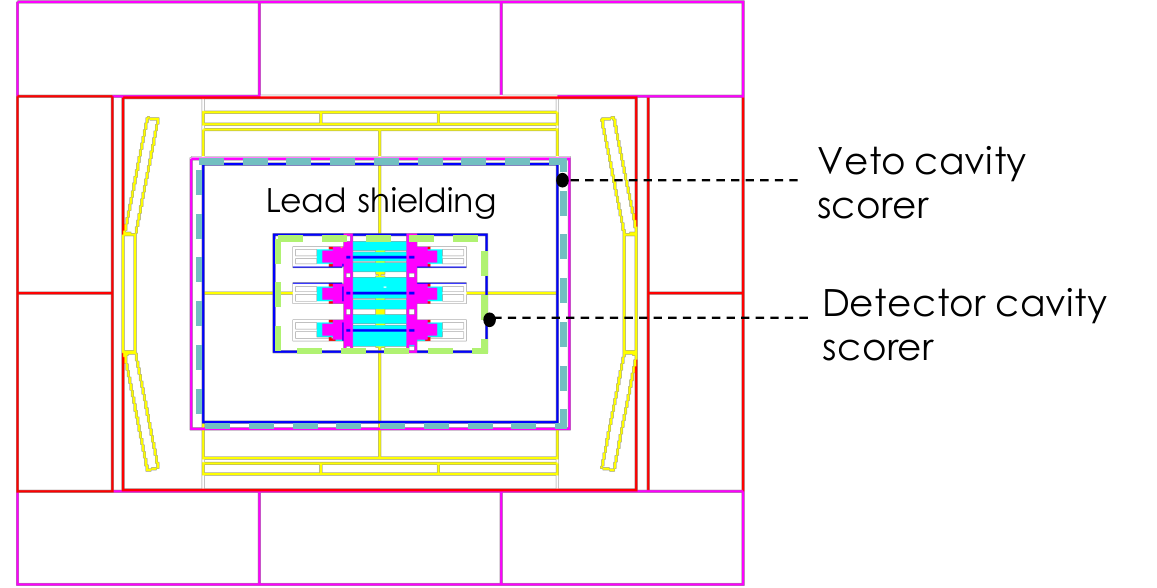}}
    \caption{\textbf{Left panel:} Scheme of the implementation of the environmental neutron source in Geant4, where neutrons are generated from the surface of a 462 cm-radius sphere following a cosine-law angular distribution. Note that, in reality, such a symmetric distribution would not be expected if neutrons originated from the laboratory walls; however, this represents the most reasonable and straightforward modelling approach adopted in Geant4. \textbf{Right panel:} Configuration of the scoring volumes: the veto cavity scorer, located outside the lead shielding, and the detector cavity scorer. }
    \label{esquemahensa}
\end{figure}

In order to estimate the neutron flux reaching the ANAIS detector cavity, an ad-hoc simulation has been performed in which the energy of the neutrons is directly sampled from the the neutron flux measured by HENSA in hall B (right panel of Figure \ref{fondoneuInput}). The initial position of the neutrons is correspondingly sampled from the surface of a sphere with a radius of 462~cm, which corresponds to the measured distance between ANAIS-112 and HENSA set-ups at LSC (see left panel of Figure \ref{fondoneuInput}). This represents the expected spatial distribution of neutrons in the laboratory as measured by HENSA. A schematic representation of the initial source implemented in Geant4 can be found in the left panel of Figure \ref{esquemahensa}.

To model an isotropic flux emanating from a spherical surface, as is the case in this work, the cosine law angular distribution from the General Particle Source module in Geant4 has been used. This approach ensures that the fluence in each direction is proportional to the cosine of the angle between the source direction and the local normal to the sphere surface. This angular distribution guarantees that neutrons are always emitted towards the interior of the sphere.     


FLUKA is generally considered less versatile than Geant4 in terms of the available output formats and the procedures required to extract them. For this reason, in the present work, the methodology used to compute the neutron flux in FLUKA was adapted and implemented within the Geant4 framework. According to the FLUKA documentation, the flux is computed as the track-length density within the scoring volume, i.e., the total track length travelled by particles (computed by summing the length of each step between two consecutive interactions) within a predefined geometric region, normalized by the volume of that region. In FLUKA, these scoring volumes are user-defined spatial regions in which specific physical quantities are tallied during the simulation, such as deposited energy, flux density, particle production rates, and other relevant observables.

In both the FLUKA simulations performed by HENSA and the Geant4 simulations carried out in this thesis, two scoring volumes are implemented within the ANAIS-112 set-up trying to evaluate the effect of the two main materials present in the experimental configuration: water and polyethylene, which are present in the neutron moderator blocks, and lead, which is found in the lead shielding. Right panel of Figure~\ref{esquemahensa} shows the distribution of these scorers within the ANAIS-112 set-up. One scorer is located in the veto cavity, specifically on the outer surface of the lead volume. The second scorer is placed in the detector cavity, adjacent to the lead. Both scorers are defined analogously to the FLUKA calculation, using a 1 $\mu$m depth vacuum volume to avoid the need for corrections related to the mean free path of neutrons in the material.

\begin{table}[b!]
\begin{tabular}{c|c|c|c|c|c}
\hline
\multicolumn{6}{c}{Neutrons ($\times$ 10$^{-6}$ n/cm$^2$/s) } \\

\hline
                       & Cavity        & Total & Thermal & Intermediate & Fast   \\
                       \hline \hline
                & outside &    20.96      &      7.64       &    8.09     &  5.66 \\
                \hline

                \multirow{3}{*}{HENSA} &      vetoes &    \begin{tabular}[c]{@{}c@{}} 2.335  \\ $\pm$ 0.002     \end{tabular}                                  &   \begin{tabular}[c]{@{}c@{}} 1.087 \\ $\pm$ 0.001    \end{tabular}  &   \begin{tabular}[c]{@{}c@{}}     0.5025 \\ $\pm$ 0.0009  \end{tabular}    &   \begin{tabular}[c]{@{}c@{}}    0.7825 \\ $\pm$ 0.001 \end{tabular}  \\
\cline{2-6}
                       &       detectors     &  \begin{tabular}[c]{@{}c@{}} 1.077 \\ $\pm$ 0.004  \end{tabular}      &     \begin{tabular}[c]{@{}c@{}} 0.150 \\ $\pm$ 0.001     \end{tabular}   & \begin{tabular}[c]{@{}c@{}}  0.457 \\ $\pm$ 0.002   \end{tabular}   &  \begin{tabular}[c]{@{}c@{}} 0.523 \\ $\pm$ 0.005 \end{tabular}\\    
                       \hline
                       

\multirow{3}{*}{\begin{tabular}[c]{@{}c@{}}ANAIS \\ Geant4 v9.4.p01  \end{tabular}} &  vetoes  & \begin{tabular}[c]{@{}c@{}}3.636 \\ $\pm$ 0.002 \end{tabular} & \begin{tabular}[c]{@{}c@{}}2.186 \\ $\pm$ 0.002 \end{tabular} &  \begin{tabular}[c]{@{}c@{}}0.978 \\ $\pm$ 0.004 \end{tabular} & \begin{tabular}[c]{@{}c@{}}0.4718 \\ $\pm$ 0.0005 \end{tabular} \\ 
\cline{2-6}
                       &     detectors  & \begin{tabular}[c]{@{}c@{}}0.716\\ $\pm$ 0.004\end{tabular} & \begin{tabular}[c]{@{}c@{}} 0.176 \\ $\pm$ 0.002 \end{tabular}& \begin{tabular}[c]{@{}c@{}} 0.381 \\ $\pm$ 0.004 \end{tabular} & \begin{tabular}[c]{@{}c@{}} 0.159 \\ $\pm$ 0.003  \end{tabular}\\
                       \hline


 \multirow{3}{*}{\begin{tabular}[c]{@{}c@{}}ANAIS \\ Geant4 v11.1.1 \end{tabular}} &  vetoes  & \begin{tabular}[c]{@{}c@{}}3.722 \\ $\pm$ 0.003 \end{tabular} & \begin{tabular}[c]{@{}c@{}}2.2725 \\ $\pm$ 0.0019 \end{tabular} &  \begin{tabular}[c]{@{}c@{}} 0.9803 \\ $\pm$ 0.0011 \end{tabular} & \begin{tabular}[c]{@{}c@{}} 0.4697 \\ $\pm$ 0.0006 \end{tabular} \\ 
\cline{2-6}
                       &     detectors  & \begin{tabular}[c]{@{}c@{}}0.709\\ $\pm$ 0.009 \end{tabular} & \begin{tabular}[c]{@{}c@{}} 0.184 \\ $\pm$ 0.004  \end{tabular}& \begin{tabular}[c]{@{}c@{}} 0.344 \\ $\pm$ 0.0004  \end{tabular} & \begin{tabular}[c]{@{}c@{}} 0.181 \\ $\pm$ 0.004 \end{tabular}\\
                       \hline

                   
                       


 \multirow{2}{*}{HENSA/ANAIS} & vetoes     & 0.64  & 0.50  & 0.51  & 1.66 \\
                   
                      \cline{2-6}
 & detectors & 1.50   & 0.85  & 1.20  & 3.29  \\
              
                      \hline
\end{tabular}
\caption{\label{neutronfluxresults} Neutron flux estimations obtained using Geant4 in this thesis and those reported by the HENSA collaboration using FLUKA in the two cavities within the ANAIS-112 set-up: the vetoes cavity and the detectors cavity. Units of the flux are $\times$~10$^{-6}$ n/cm$^2$/s. Both the total flux and the flux within three characteristic energy regions are shown: thermal neutrons ($1 \times 10^{-10}$ to $3.2 \times 10^{-7}$ MeV), intermediate neutrons ($3.2 \times 10^{-7}$ to 0.1 MeV), and fast neutrons (0.1 to 100 MeV). The table includes results from two Geant4 versions tested in this work, v9.4.p01 and v11.1.1, as well as the ratio between the neutron flux measured by HENSA and that obtained with ANAIS (Geant4 v9.4.p01). For reference, the integral of neutron flux outside the ANAIS-112 shielding measured in hall B is also included.
}
\end{table}

Table \ref{neutronfluxresults} presents the neutron flux estimations obtained using Geant4 in this work and those reported by the HENSA collaboration using FLUKA for the two cavities within the ANAIS-112 set-up. Both the total flux and the flux in three characteristic energy regions are reported. The table includes results for two Geant4 versions, v9.4.p01 and v11.1.1, analogous to the comparison performed in Section \ref{datalibrary} for the QF estimation of the ANAIS-112 crystals. Additionally, the ratio between the neutron flux measured by HENSA and that obtained with ANAIS (Geant4 v9.4.p01) is also provided.

It is observed that the neutron flux estimated by HENSA using FLUKA is lower than that obtained in this work with Geant4 in the veto cavity, but higher in the detector cavity, independently of the Geant4 version employed. When considering the total flux, the ratio between the HENSA and ANAIS results amounts to 0.64 in the veto cavity and 1.5 in the detector cavity. This means that the difference between both simulation codes corresponds to approximately a factor of 2. While this deviation is non-negligible and must be regarded as a systematic uncertainty in the extraction of results, it is nevertheless acceptable given the intrinsic differences between the simulation tools. Therefore, the overall agreement between both codes can be considered satisfactory.

However, it is useful to examine the flux distribution across the different energy regions. In the thermal and intermediate energy ranges, the discrepancies, although non-negligible, are notably smaller than those observed for fast neutrons. The largest differences between both simulation codes arise in the fast neutron component reaching the ANAIS cavities, with Geant4 predicting  $\sim$1.7~(3.3)~times fewer fast neutrons in the vetoes (detectors) cavity. Moreover, it is important to note that the results obtained with the two Geant4 versions are in general agreement, consistent with the observations made in Section \ref{datalibrary}, where differences between versions were present but not significant.

\begin{figure}[t!]
    \centering
    {\includegraphics[width=0.49\textwidth]{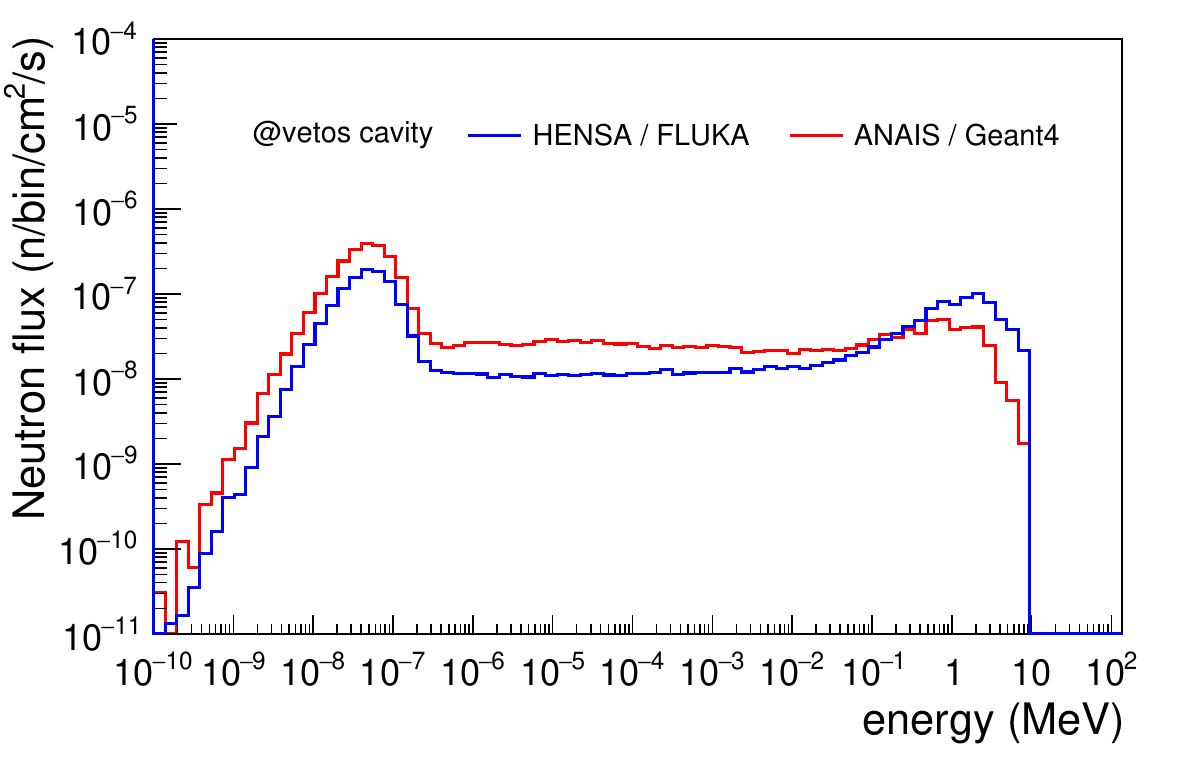}}
    {\includegraphics[width=0.49\textwidth]{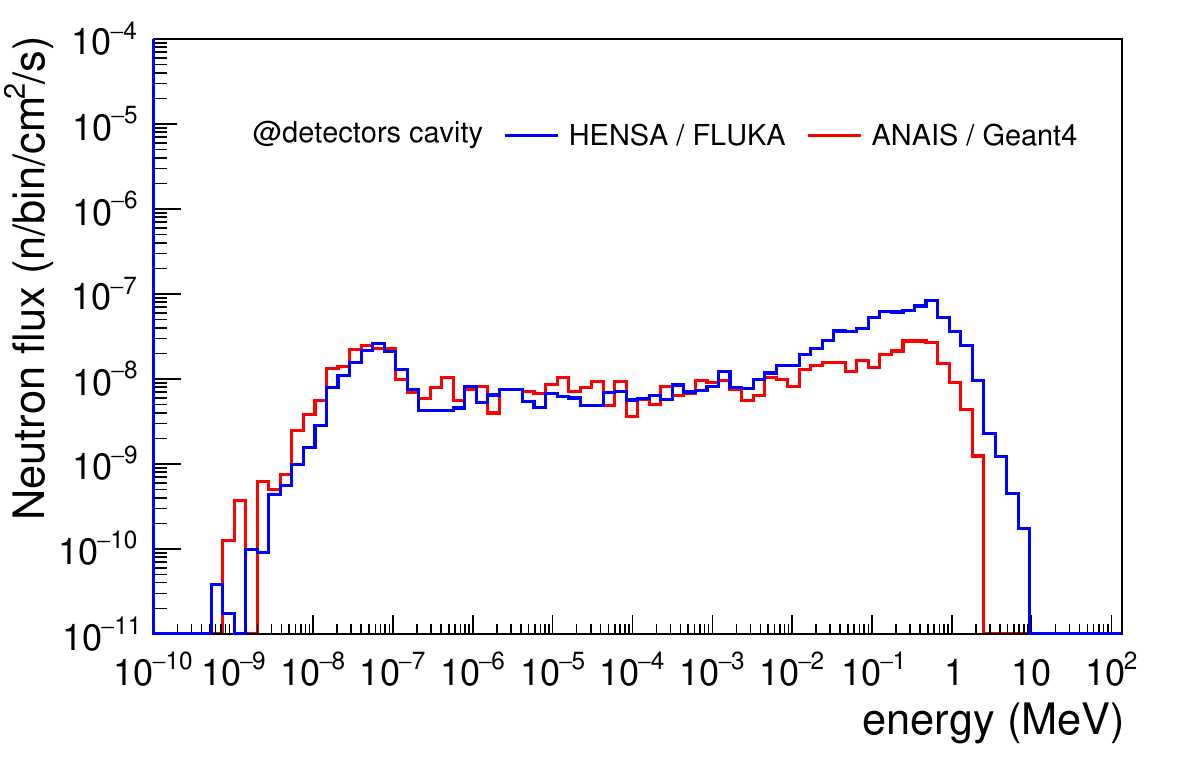}}
    \caption{\label{flujocavidadesHENSA} Ambient neutron flux reaching ANAIS-112 as estimated using FLUKA by HENSA (blue) and Geant4 by ANAIS (red). \textbf{Left panel:} Reaching the veto cavity. \textbf{Rigth panel:} Reaching the detector cavity.}
    
\end{figure}

Figure~\ref{flujocavidadesHENSA} shows the neutron flux spectra as estimated using FLUKA by HENSA and Geant4 by ANAIS. The moderation behavior of the flux as neutrons penetrate the different shielding regions is consistent with expectations. As neutrons traverse the polyethylene and water tanks surrounding the veto cavity, they undergo multiple elastic scatterings primarily on hydrogen nuclei, resulting in efficient thermalization. However, Geant4 predicts a significantly lower number of fast neutrons, as was already evident in Table~\ref{neutronfluxresults}. Further attenuation and moderation occur as neutrons reach the detector cavity, having passed not only through the polyethylene and water but also through the lead shielding and, in some cases, the detectors themselves. This additional shielding results in a further reduction in flux intensity. 

When comparing both spectra, the difference between the two simulation codes becomes evident, as not only the spectral shapes differ but also the relative weight of the neutron components. Specifically, in the veto cavity a smaller contribution of fast neutrons is predicted by Geant4, as already shown in Table~\ref{neutronfluxresults}. This reduced fast-neutron component is compensated by an increased contribution from intermediate and thermal neutrons, evidencing a redistribution among the spectral components. In the detector cavity, a lower contribution of fast neutrons is again observed, while the rest of the spectrum remains largely consistent between the two simulations.


The discrepancy between the simulation codes may indicate either differences in the geometrical modelling of the set-up in each framework, or in the neutron interaction cross-sections. Concerning the former, the hypothesis of a significantly different geometry in the HENSA simulation leading to a larger number of fast neutrons does not appear to be applicable. The ANAIS-112 set-up implementation in FLUKA was carried out with considerable detail and was based on the same source geometry used in the Geant4 simulations. Dimensions and material densities were thoroughly cross-checked between both implementations, with no discrepancies found that could account for the observed differences in the results. Nevertheless, it is acknowledged that the FLUKA geometry is slightly more simplified, due to the inherent complexity of building the full experimental set-up from scratch in FLUKA. For instance, FLUKA geometry does not consider the PMT structure.

Regarding the second hypothesis, the differences in cross-section data, it was verified that the FLUKA simulations performed by the HENSA collaboration employed the JEFF-3.3 nuclear data library, which is also the basis of the Geant4 v11.1.1 version, as discussed in Section \ref{datalibrary}. However, to further investigate this issue, and particularly to explore the possibility that the neutron flux may be computed differently within the Geant4 framework compared to FLUKA, additional tests were conducted using highly simplified geometries. In these tests, a monoenergetic 1 MeV neutron beam was directed onto water and lead blocks with thicknesses of 40 cm and 30 cm, respectively, corresponding to the same shielding dimensions present in the ANAIS set-up. The number of neutrons transmitted through each material was recorded. The results of this simplified study are presented in Table \ref{flujosimply}.

\begin{table}[t!]
\begin{tabular}{c|c|c|c|c|c}
\hline
\multicolumn{6}{c}{Neutrons (counts)} \\
\hline
                       & Material        & Total & Thermal & Intermediate & Fast   \\
                       \hline \hline

               \multirow{2}{*}{HENSA} &      Water &    361 $\pm$ 19                      &   244 $\pm$ 16 &   80 $\pm$ 9    &   40 $\pm$ 6  \\
\cline{2-6}
                       &       Lead     &    333944 $\pm$ 578                      &   0  &   28518 $\pm$ 169    &   314993 $\pm$ 561  \\ 
                       \hline
                       
                \multirow{2}{*}{ANAIS} &      Water &    424 $\pm$ 21                      &   312 $\pm$ 18 &   81 $\pm$ 9    &   31 $\pm$ 6  \\
\cline{2-6}
                       &       Lead     &    323597 $\pm$ 569                      &   0  &   38409 $\pm$ 196    &   285188 $\pm$ 534  \\ 
                       \hline

\end{tabular}
\caption{\label{flujosimply} Number of neutrons obtained by both collaborations after transmission through 40 cm of water or 30 cm of lead. Both the total numer of counts and the number of counts within three characteristic energy regions are shown: thermal, intermediate, and fast neutrons. The Geant4 results correspond to version v9.4.p01.
}
\end{table}


\begin{figure}[b!]
    \centering
    {\includegraphics[width=0.49\textwidth]{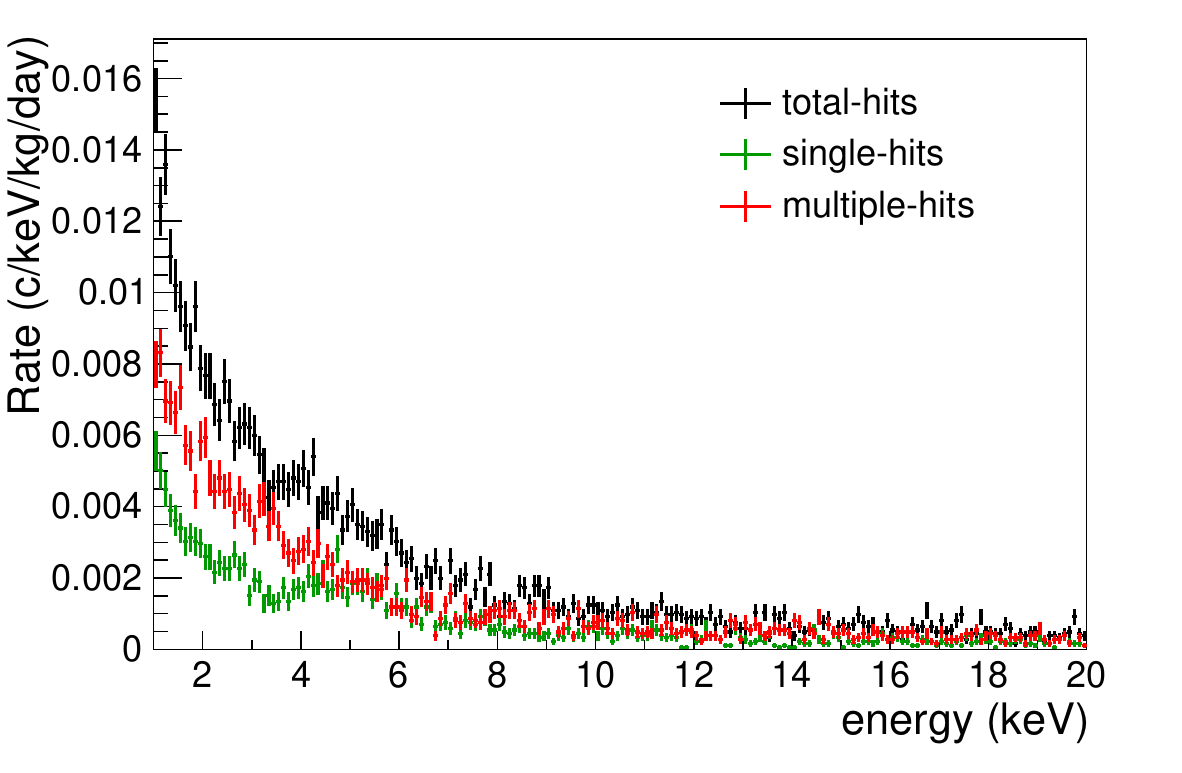}}
    {\includegraphics[width=0.49\textwidth]{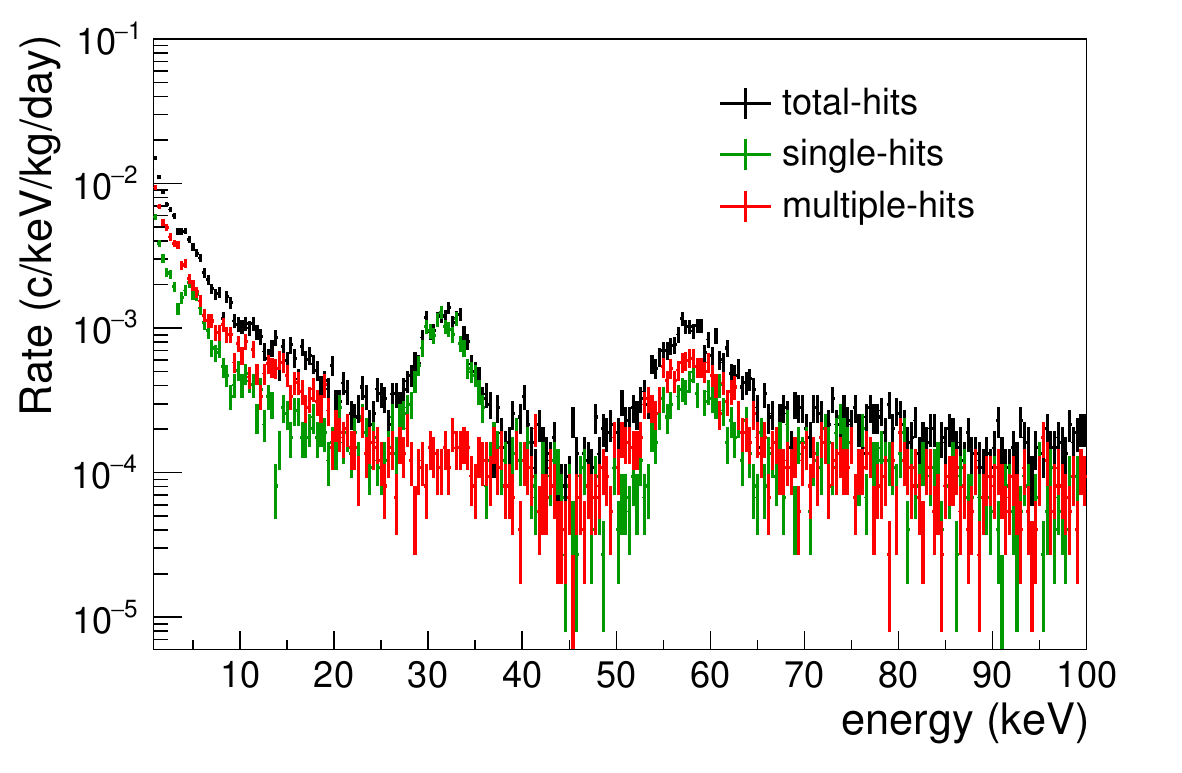}}
    \caption{ \label{depoenergyambientneu} Simulated deposited energy spectra from environmental neutron background, with separate contributions from total, single-hit, and multiple-hit events. \textbf{Left panel:} Low-energy. \textbf{Right panel:} Medium-energy. }
   
\end{figure}

After traversing the water block, the number of transmitted neutrons predicted by both ANAIS and HENSA simulations is in good agreement. However, the most pronounce differences emerge when neutrons pass through the lead block. FLUKA predicts a significantly higher number of neutrons transmitted through lead, particularly in the fast energy range, and the differences do not appear to be attributable to statistical fluctuations. In the Geant4 simulation, neutrons emerging from the lead block are more likely to be in the intermediate energy range, with a suppressed fast neutron component compared to the FLUKA predictions. This suggests that non-negligible differences in the neutron interaction cross sections between the two codes exist, although it remains uncertain whether these differences alone can fully account for the large discrepancies reported in Table \ref{flujocavidadesHENSA}.

These findings reinforce the relevance of ongoing efforts to benchmark and compare the performance of Geant4 and FLUKA in neutron transport simulations, in line with previous studies such as ~\cite{frosio2021calculation,zaman2022modeling}. These discrepancies may originate from differences in the nuclear data libraries employed as well as from the underlying physics models, including electromagnetic and neutron transport. However, considering the excellent agreement between measurement and simulation in the \textsuperscript{252}Cf calibrations, it is considered safe to rely in the Geant4 simulations of the ANAIS set-up. Nonetheless, possible systematics in the neutron simulations will be further investigated through dedicated neutron calibration campaigns.




\subsubsection{Deposited energy by the neutron background}

Once the neutron flux reaching the ANAIS-112 set-up has been analyzed, this section focuses on studying the energy deposited in the ANAIS detectors by the neutron background. The corresponding energy deposition spectra are shown in Figure \ref{depoenergyambientneu}, with separate contributions from total-hit, single-hit, and multiple-hit events.

As expected, the shape of the spectrum closely resembles that obtained in Chapter~\ref{Chapter:QF} from exposure to a $^{252}$Cf calibration source, with a prominent low-energy scattering component and characteristic gamma lines such as the 31.8 keV peak from neutron capture on $^{127}$I and the inelastic scattering peak. 

A key aspect at this stage is the comparison between the contribution from this neutron background and the total measured background, as well as the background model developed in the previous section. Table~\ref{tablaratesneu} summarizes this comparison in the [1–6] keV energy range across various event populations: total-hit, single-hit, multiple-hit and m3m4-hit events.

\begin{table}[b!]
\centering
\begin{tabular}{c|cccc}
\hline
\multicolumn{5}{c}{Rate (c/keV/kg/day) [1--6]~keV} \\
\hline
   & Total & Single & Multiple & m3m4 \\
\hline 
data & 3.54 & 3.24 & 0.30 & 0.04\\                       
background model & 3.23 & 2.95 & 0.27 & 0.03 \\
neutron background ANAIS & 0.007 & 0.0025 & 0.0041 & 0.0018\\
\hline              
\end{tabular}
\caption{\label{tablaratesneu} Rates in the [1--6]~keV energy range for the full ANAIS-112 detector array over six years of data taking. The table includes the corresponding simulated rates from the background model developed in this chapter, and the neutron background estimated in this section with Geant4. Results are shown for total-, single-, multiple-, and m3m4-hit events. }
\end{table}

In all cases, the contribution from ambient neutrons, as predicted by the ANAIS simulation, is negligible across all event populations. The table does not include the corresponding rates from the HENSA simulation, as it was not possible during the course of this work to implement the QF correction in FLUKA, an essential requirement for meaningful comparison with the measured ANAIS data. Moreover, event multiplicity was not available as an observable within the FLUKA framework. However, just by scaling with the fast neutron total flux the rates determined, an increase of about one order of magnitude in all the contributions could be expected.

According to the Geant4-based simulations performed in this study, the contribution of ambient neutrons to the background is negligible and is therefore not included in the background model. Considering the discrepancy by a factor of 3.3 in the fast neutron flux reported in Table \ref{neutronfluxresults}, a larger contribution could be expected in FLUKA. However, this would in no case represent a significant component of the experimental background.

\subsection{Non-bulk scintillation contribution}

The following section explores an alternative hypothesis that could account for the excess observed in the ANAIS-112 data. The ANOD DAQ system provides a promising approach to better characterize low-energy anomalous events exhibiting asymmetric light sharing and, ultimately, to suppress them.

The ANOD DAQ system has been described in detail in Section~\ref{DAQsec}, where its main features and capabilities were outlined, and it has already been used in the study of the QF of the ANAIS crystals (see Chapter~\ref{Chapter:QF}). While in the standard ANAIS DAQ scintillation events and low-energy asymmetric events below 2 keV are indistinguishable and appear as a single mixed population, the ANOD DAQ system provides a clear separation of these two classes of events based on their mean time distributions. The cuts and procedure used to select this highly asymmetric anomalous population, as well as its spectral shape, have been described in Section \ref{ANODfiltering}. The reader is referred to that section for further details.

\begin{figure}[b!]
    \centering
    {\includegraphics[width=0.9\textwidth]{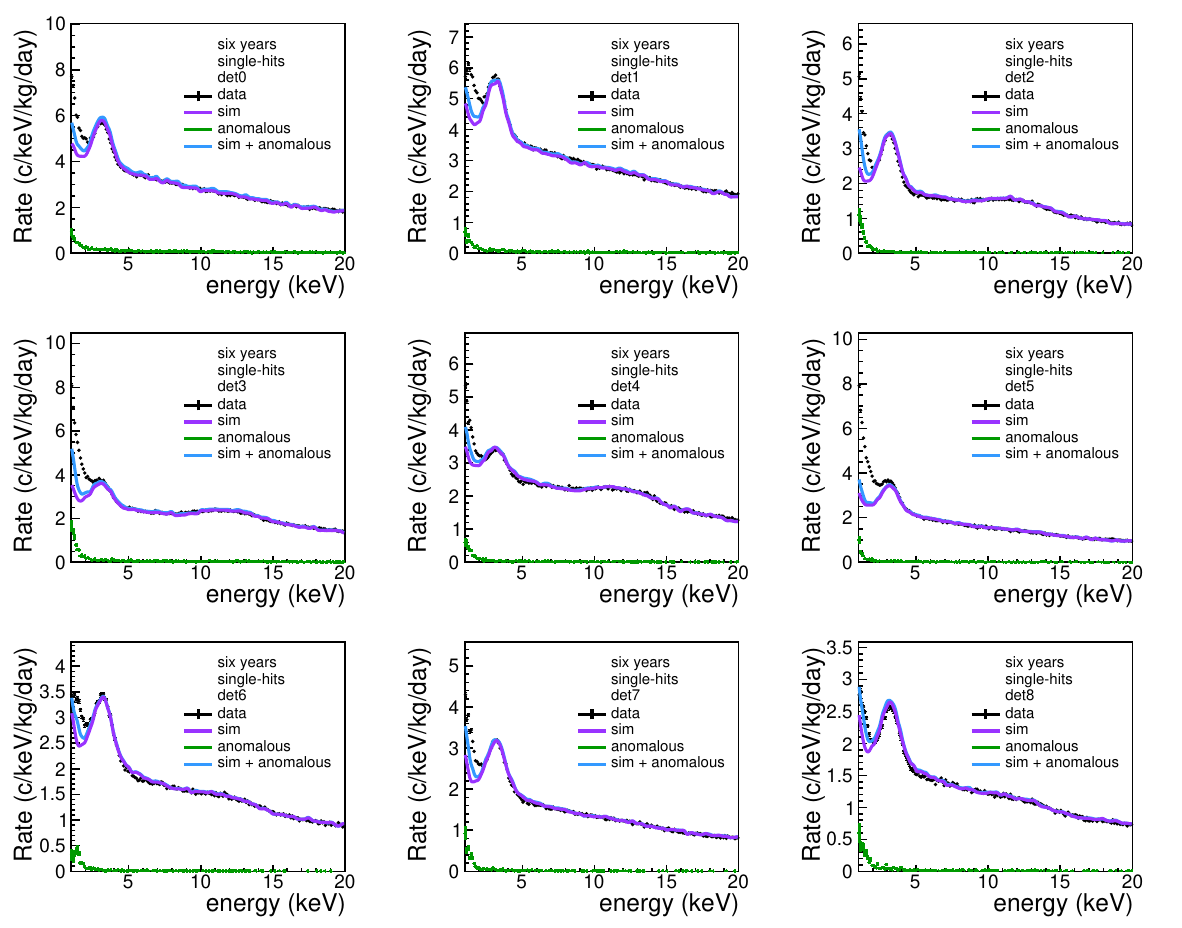}}
    
    \caption{\label{suma_malosbuenos} Comparison of the single-hits low-energy spectra measured over six years
for the nine ANAIS-112 detectors (black) with the improved background
model developed in this thesis. The figure shows the simulation before (violet) and after (blue) incorporating the anomalous population of events identified in ANOD corrected by the BDT efficiency (green). }
\vspace{-0.5cm}
\end{figure}

This section aims to assess whether the spectral shape of these asymmetric events matches the excess observed in ANAIS data following the background model improvement carried out in this chapter, and to what extent this anomalous population contributes to the measured data.

This population corresponds to the green spectrum in Figure~\ref{suma_malosbuenos}, which shows an exponential component plus a continuum above 2 keV. It is important to note that in the ANAIS-112 background data, the event
selection was performed using a cut in the BDT variable, and the
resulting spectrum was corrected for the efficiency of this cut (see Figure~\ref{efficiencyBDT}). Consequently, any residual asymmetric events that pass the BDT cut are
also artificially amplified by the efficiency correction. Therefore, to
ensure a consistent comparison, the selected anomalous population must
also be corrected for the BDT efficiency.

When this population is incorporated into the background model, the resulting spectrum exhibits qualitatively improved agreement with the observed excess, detector~8, for instance, shows particularly good consistency. Nevertheless, the spectral shape of these anomalous events appears more abrupt than that of the measured data. If this population were scaled to fully match the observed excess at low energy, it would result in a significant overestimation above 1.5 keV, particularly in the continuum region.

\begin{table}[b!]
    \centering
    \resizebox{\textwidth}{!}{\Large
    \begin{tabular}{c|c|cc|cc}
        \hline
       \multirow{3}{*}{detector}  & \multicolumn{5}{c}{single-hits [1-2] keV}  \\
       \cline{2-6}
        & \makecell{data \\ ($\text{kg}^{-1} \text{day}^{-1}$)} 
 & \makecell{new sim \\ ($\text{kg}^{-1} \text{day}^{-1}$)} 
 & \makecell{new desv \\ (\%)} 
 & \makecell{new sim + anomalous \\ ($\text{kg}^{-1} \text{day}^{-1}$)} 
 & \makecell{new + anomalous desv \\ (\%)} \\
        \hline
        0 & 5.97 $\pm$ 0.02 & 4.37  & -26.75 & 4.86 & -18.54 \\
        1 & 5.61 $\pm$ 0.02  & 4.37  & -22.14 & 4.71 & -15.94 \\
        2 & 3.41 $\pm$ 0.02  & 2.16  & -36.60 & 2.65 & -22.34 \\
        3 & 5.56 $\pm$ 0.02  & 3.03  & -45.51 & 3.76 & -32.41 \\
        4 & 4.01 $\pm$ 0.02 & 3.06  & -23.67 & 3.38 & -15.58\\
        5 & 5.32 $\pm$ 0.02 &  2.69  & -49.45  & 2.94 & -44.70 \\
        6 & 3.20 $\pm$ 0.01  & 2.62  & -18.08 & 2.88 & -9.82 \\
        7 & 3.27 $\pm$ 0.01 & 2.33  & -28.78 & 2.70 & -17.46\\
        8 & 2.42 $\pm$ 0.01 &  2.02  & -16.30  & 2.31 & -4.53\\
        \hline
        ANAIS-112  & 4.31 $\pm$ 0.01  & 2.96 & -31.24 & 3.36 & -22.08 \\
        \hline
    \end{tabular}}
    \caption{\label{tablaañadeanod} Measured single-hits rates in the [1-2] keV range for each ANAIS-112 detector and their average over the full array during the six years of data taking. The corresponding simulated rates expected from both the improved background model and the improved background model incorporating the anomalous events identified with ANOD are provided for each case, together with their respective deviations from the measured values. The associated statistical uncertainty in the simulations, not shown in the table, is $\sim$0.1\%.}
\end{table}

Furthermore, incorporating this population into the model still leaves an excess in the data. One possible explanation is that the BDT efficiency applied for the correction was derived from scintillation events, whereas the selected population consists of anomalous events. A lower BDT efficiency for this class of events could reconcile the observed discrepancy, which would support the BDT’s ability to discriminate such events by allowing fewer of them to pass the selection criteria. Additionally, the selection of anomalous events in ANOD was based on a constant cut in mean time, independent of energy (see Section \ref{ANODfiltering}). This cut was intentionally non-restrictive to avoid the complexity of calculating the corresponding efficiency, but it may result in the loss of anomalous events in the [1–2]~keV region. Efforts are currently underway to develop an energy-dependent selection of this population, which has proven effective in selecting additional events in this energy region \cite{tfgmartarojo}.

Overall, this study demonstrates the potential of the ANOD DAQ system to account for the population of anomalous events in the [1–2] keV region of the ANAIS background. Although this is only a preliminary study, the inclusion of this population in the background model is justified and will be incorporated into the time evolution of the improved background model as a constant contribution. In particular, Table~\ref{tablaañadeanod} shows the deviation between data and the improved background model before and after incorporating the contribution from these anomalous events. Including them reduces the average discrepancy from 31.24\% to 22.08\%. Further studies with increased ANOD statistics are already underway within the ANAIS collaboration and are expected to provide additional insights \cite{tfgmartarojo}.

\section{The time-dependent background model}\label{rateevolution}

A key feature of the ANAIS background model is its ability to predict the time evolution of event rates for different populations. The time-dependent background model includes radioisotopes from the time-integrated model with half-lives of a few years. This selection primarily consists of \(^{210}\)Pb (T\(_{1/2}\) = 22.3 yr), \(^{3}\)H (T\(_{1/2}\) = 12.3~yr), and \(^{22}\)Na (T\(_{1/2}\) =~2.6~yr). For detectors D6, D7, and D8, cosmogenic contributions from Te and I with shorter half-lives are also considered. Isotopes with constant activity over time are grouped into a single, time-independent constant background component for each detector.

The background event rate as a function of time induced by a radioactive background component follows the basic radioactive decay law, such that the number of events generated between an initial and final time is given by Equation \ref{EqN}. In the following, \( t = 0 \) denotes the start of the measurement on August 3, 2017. The background rate evolution as predicted from the model is computed for each detector in 15-day intervals.

For the background model, a 10 $\mu$m exponential depth profile is assumed for the $^{210}$Pb surface contamination. Moreover, it is important to emphasize that the figures shown below are not normalized by any ad-hoc scaling factor; they directly reflect the output of the background model based on the fitted activities obtained in this chapter. Neither has the previous model been scaled.

To begin with, the time evolution of event populations outside the ROI will be presented to assess how accurately the background model developed in this work is able to predict the temporal behavior of the data, without making any assumptions or implications in the region where the DM signal is searched for. 
\begin{figure}[t!]
    \centering
    {\includegraphics[width=0.73\textwidth]{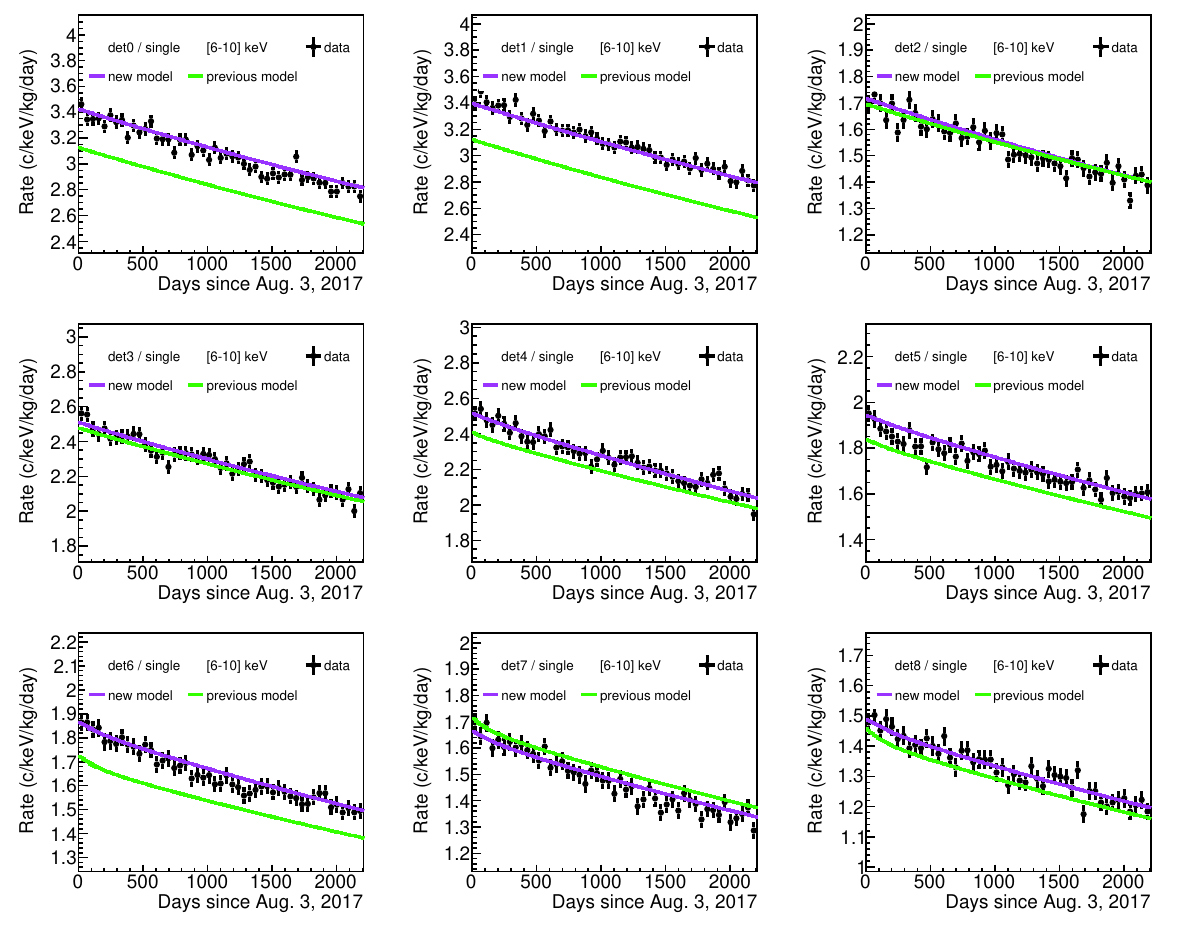}}
    
    \caption{\label{rateevol610} Time evolution of the rate of single-hit events above the ROI, [6–10]~keV, for each ANAIS-112 detector. The measured rates are shown in black, together with the time evolution predicted by the background model developed in this thesis (violet) and the expectations from the previous background model (green).  }
\end{figure}

\begin{figure}[t!]
    \centering
    {\includegraphics[width=0.73\textwidth]{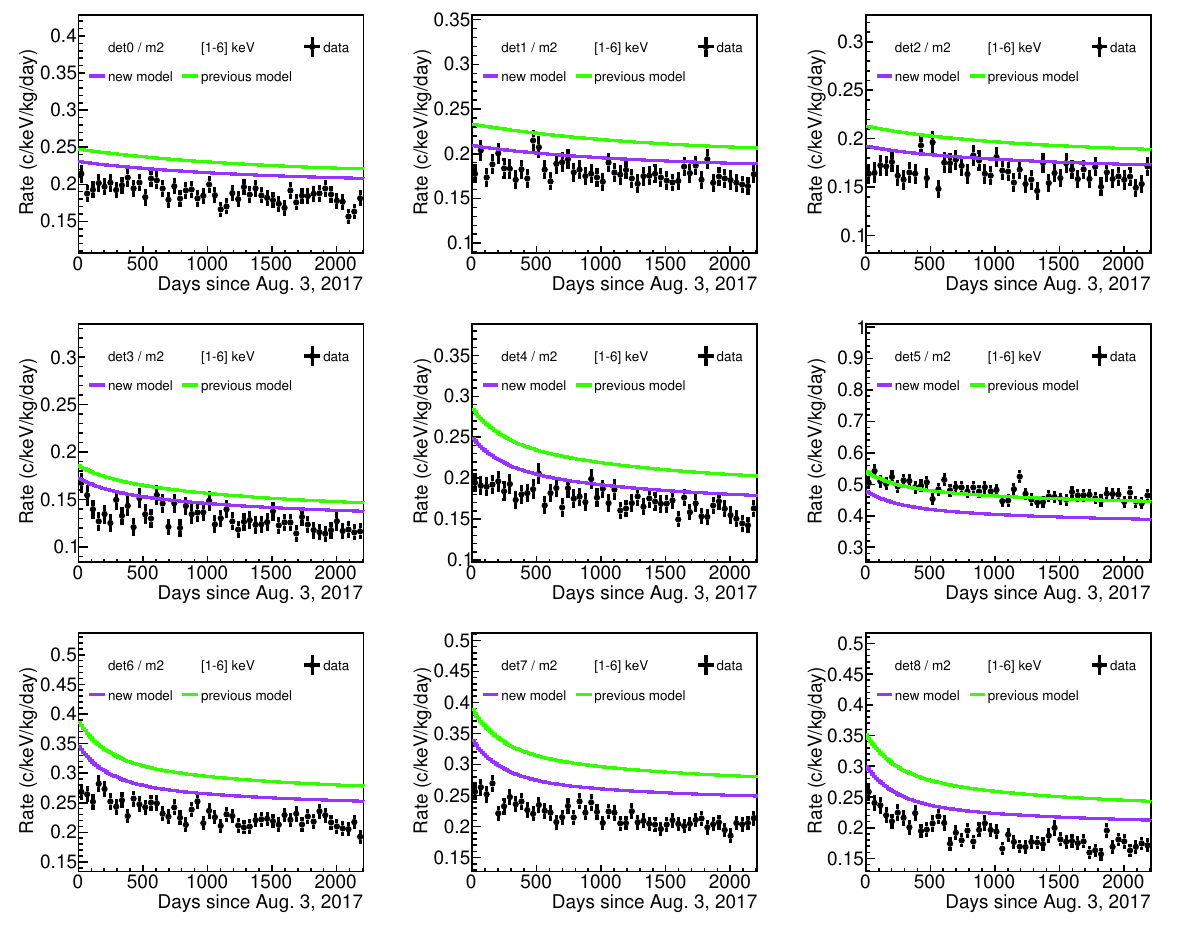}}
    
    \caption{\label{rateevol16M2} Time evolution of the rate of m2-hit events in the ROI, [1–6]~keV, for each ANAIS-112 detector. The measured rates are shown in black, together with the time evolution predicted by the background model developed in this thesis (violet) and the expectations from the previous background model (green). }
\end{figure}

\begin{figure}[b!]
    \centering
    {\includegraphics[width=0.72\textwidth]{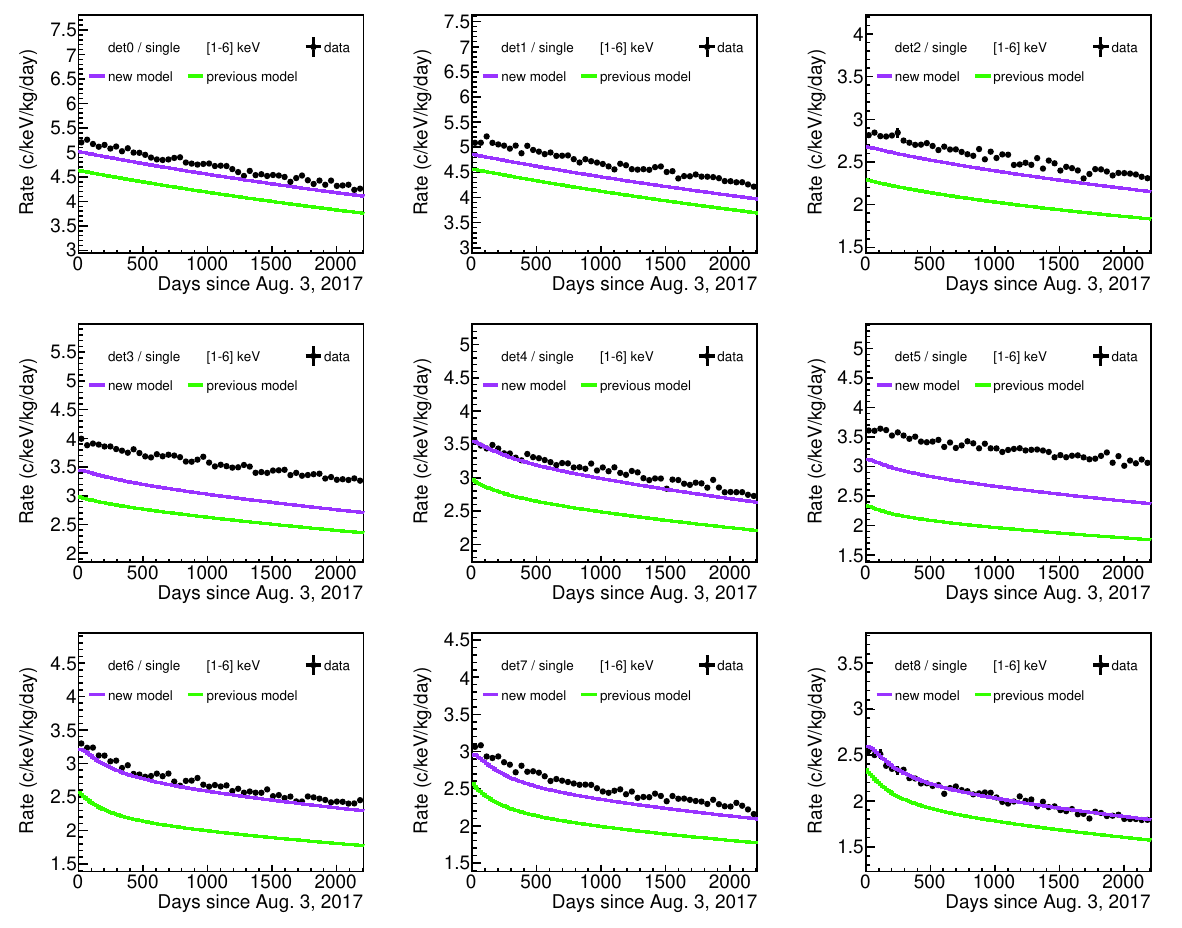}}
    
    \caption{\label{rateevol16}  Time evolution of the rate of single-hit events in the ROI, [1–6]~keV, for each ANAIS-112 detector. The measured rates are shown in black, together with the time evolution predicted by the background model developed in this thesis (violet) and the expectations from the previous background model (green). }
\end{figure}

\begin{figure}[t!]
    \centering
    {\includegraphics[width=0.73\textwidth]{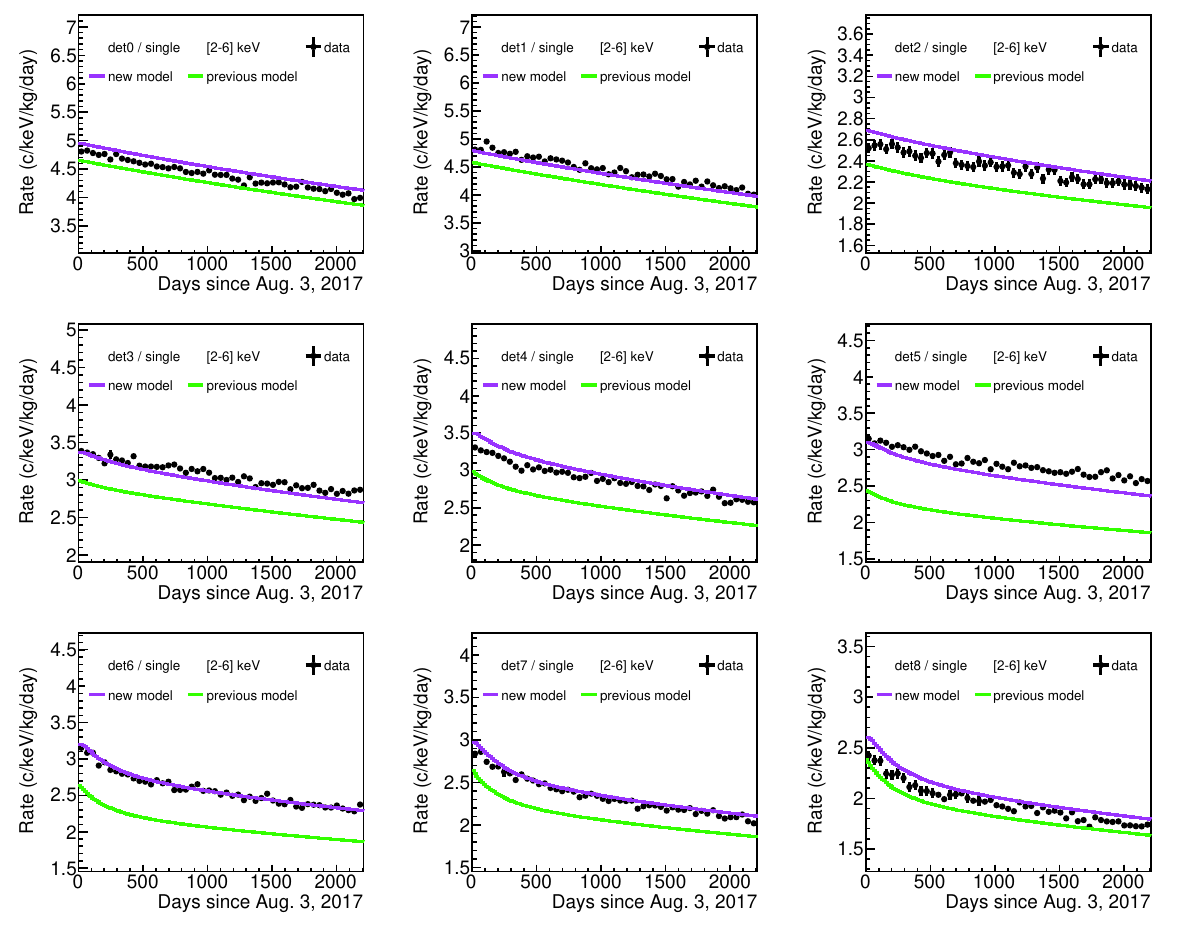}}
    
    \caption{\label{rateevol26}  Time evolution of the rate of single-hit events in the ROI, [2–6]~keV, for each ANAIS-112 detector. The measured rates are shown in black, together with the time evolution predicted by the background model developed in this thesis (violet) and the expectations from the previous background model (green). }
\end{figure}

\begin{figure}[b!]
    \centering
    {\includegraphics[width=0.73\textwidth]{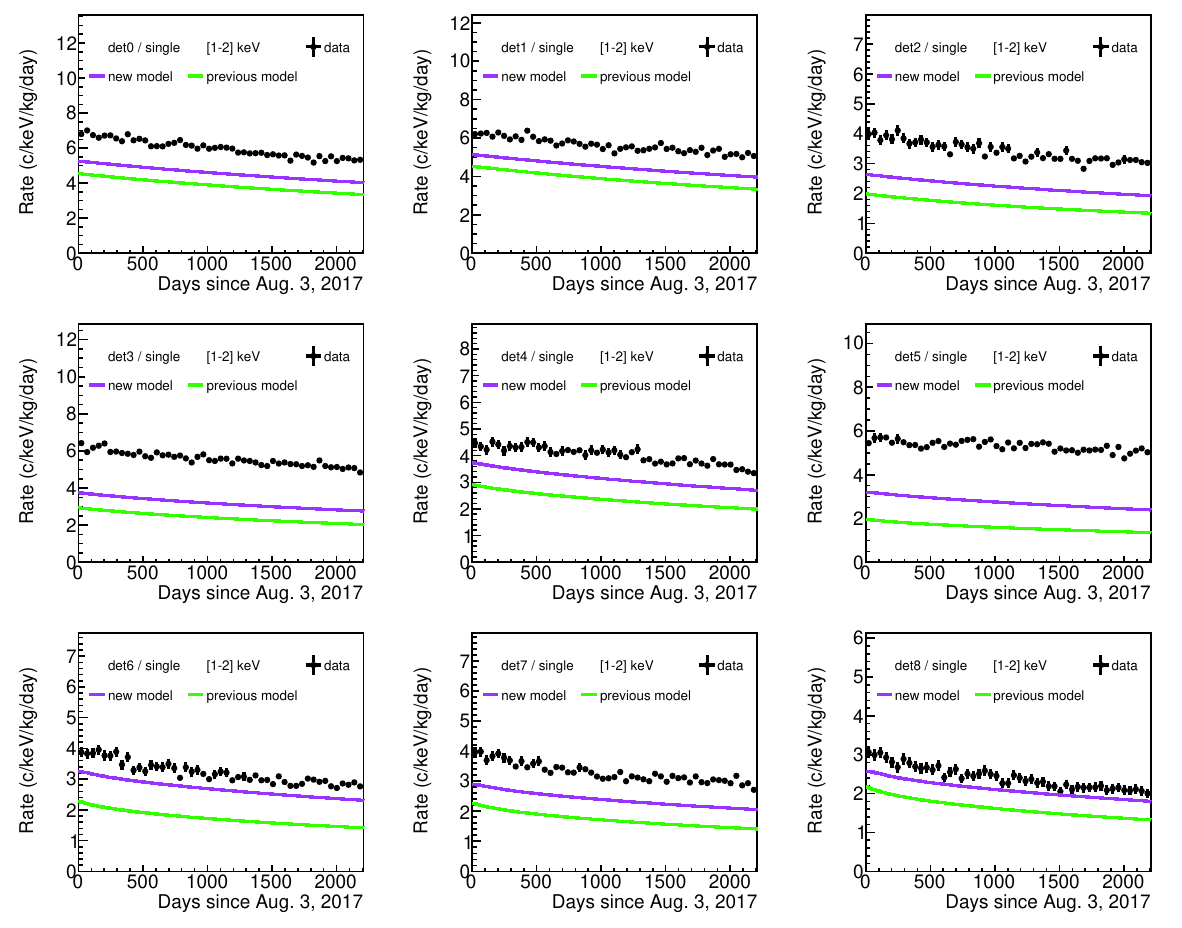}}
    
    \caption{\label{rateevol12}  Time evolution of the rate of single-hit events in the ROI, [1–2]~keV, for each ANAIS-112 detector. The measured rates are shown in black, together with the time evolution predicted by the background model developed in this thesis (violet) and the expectations from the previous background model (green). }
\end{figure}

Figure \ref{rateevol610} shows the time evolution of the rate of single-hit events above the ROI, [6–10]~keV, for each ANAIS-112 detector. The measured rates are shown, corrected for the corresponding live time and average detection efficiency in 45-day bins. These are compared with the predictions from the background model developed in this thesis and those from the previous background model. As can be observed, the agreement between data and the new model is satisfactory and improves upon the performance of the previous background model.

Coincident events serve as reliable tracers of radioactive backgrounds, as they are largely free from other populations that may leak into the lowest energy region of the ROI. Accordingly, Figure \ref{rateevol16M2} presents the time evolution of m2-hit events in the [1–6]~keV region. It can be observed that both models tend to overestimate the measured rate, with the new background model developed in this work showing closer agreement with the data, except for D5, which exhibits a higher coincidence rate due to its central position in the detector array.


Having addressed the behavior outside the ROI, the analysis proceeds with the time evolution within the ANAIS ROI. Figure~\ref{rateevol16} presents the rate evolution in the [1–6]~keV region, while Figure~\ref{rateevol26} shows it for [2–6] keV. A significantly improved agreement is observed across all detectors for both energy intervals, with the updated background model accounting for a larger fraction of the events compared to the previous version. It is true that in the [2–6] keV region, the new model slightly overestimates the rate in some detectors, such as D2, D4, and D8, which may explain why, for example in the latter, the [1–6] keV rate is well reproduced despite a residual excess being evident in the corresponding energy spectrum (see Figure~\ref{suma_malosbuenos}). Overall, the improvement over the previous background model is clear. 

It is important to emphasize that in both the [1–6] keV and [2–6] keV regions, the data-model discrepancy remains constant over time. Therefore, the background model can be used as a PDF to fit the measured background when searching for annual modulation, as will be done in Chapter \ref{Chapter:annual}. The previous background model also satisfied this condition and thus did not introduce any bias in the annual modulation analyses presented to date.

Analogously, Figure~\ref{rateevol12} displays the temporal evolution in the [1–2] keV region. Consistent with previous figures and tables, the new background model accounts for a larger fraction of events in this anomalous population region. Nevertheless, a non-negligible discrepancy between data and simulation remains. Further work is currently underway to address this.

\begin{figure}[t!]
    \centering
    {\includegraphics[width=0.6\textwidth]{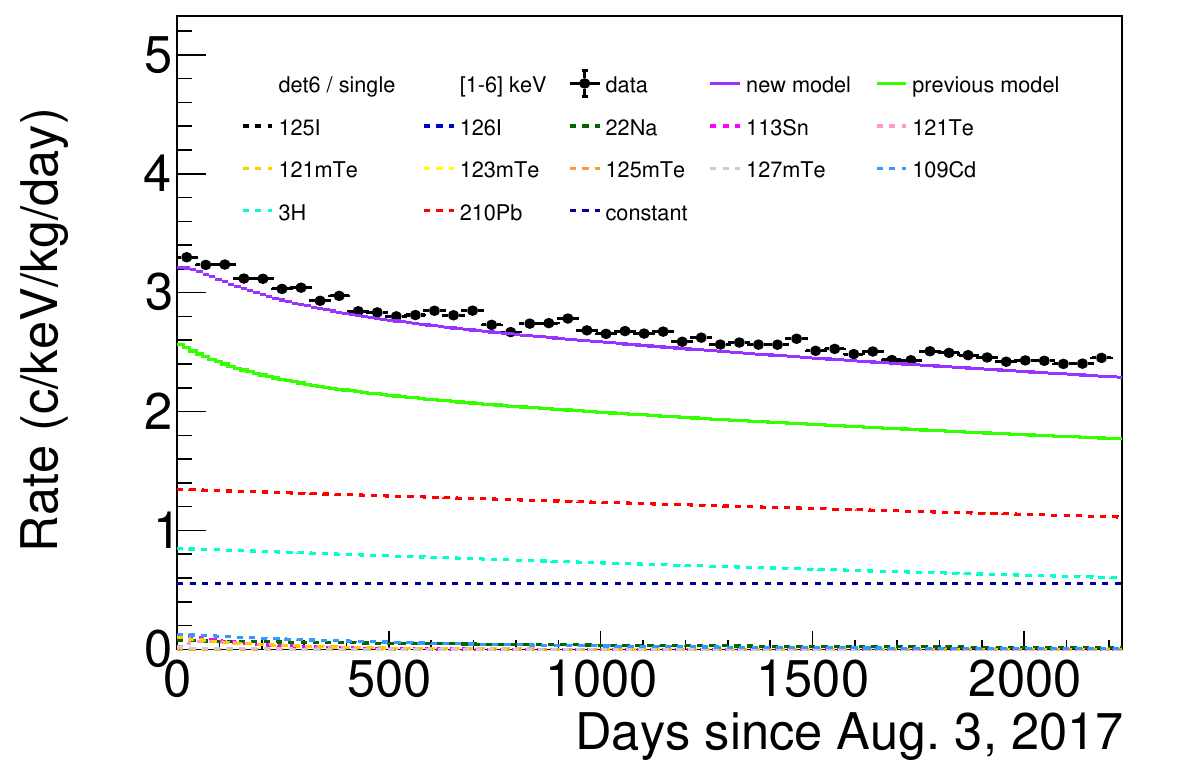}}
    
    \caption{\label{componentsrateevol16} Time evolution predicted by the background model developed in this thesis (violet) in the single-hit ROI, [1–6] keV energy region, for D6 detector. Predictions from the previous background model are shown in green, while the measured data are shown in blue. The individual components of the improved background model contributing to the time evolution are represented with dashed lines in different colors. }
\end{figure}

Ultimately, the time evolution of the individual background components for the D6 detector in the [1–6] keV ROI for single-hit in Figure \ref{componentsrateevol16}. As can be derived from the figure, the dominant contribution arises from $^{210}$Pb, which includes contamination in the crystal bulk, on the crystal surface, and on the teflon wrapping. This is followed by the contribution from \(^{3}\)H and the constant component, which encompasses all background sources in the model that do not vary with time.

\section{Conclusions}

This chapter has focused on the revision and improvement of the ANAIS-112 background model. A key difference with respect to the previous background model is that, in that case, the activity estimation was not performed through a multiparametric fit of the various contributions, an approach implemented for the first time in this thesis, but was instead based on the assessment of activities using the information available from one year of data across different event populations, along with empirical tuning of the activity values.

The chapter began by comparing the previous model with the data corresponding to the full six-year exposure, highlighting its robustness while also demonstrating the need for a revised model. Several key aspects were identified in the process of guiding the fitting strategy. The first was the asymmetry in light sharing, which revealed a population of strongly asymmetric single-hit events in the ANAIS-112 data above approximately 65 keV. The origin and nature of these events were investigated by distinguishing between symmetric and asymmetric populations, revealing a change in the spectral shape of the asymmetric single-hit events around 80~keV. The most plausible origin of these events is that contamination from the PMTs leads to asymmetric energy deposition. This hypothesis was further explored through dedicated PMT simulations, which predict spectral variations depending on the interaction position, laying the groundwork for implementing a position-dependent light collection model in future simulations. Furthermore, based on the assumption that bulk contaminations exhibit symmetry, additional evidence supports the presence of an asymmetric surface contamination of \textsuperscript{210}Pb, likely localized on the polished endcaps of the NaI crystals.

Additionally, the modelling of the PMT contribution has been revisited, proposing a contamination distribution not limited to the borosilicate volume, as assumed in previous background model, but also including a frontal contamination in the photocathode, which is more efficient in depositing energy within the crystal. This choice is supported by dedicated HPGe measurements performed during this thesis on a PMT unit of the same batch than the ones used in the ANAIS-112 set-up, which indicated that the contamination is not uniformly distributed throughout the borosilicate volume. Regarding \textsuperscript{210}Pb, both bulk and surface contamination hypotheses have been explored. To model the latter, three benchmark scenarios have been proposed, corresponding to surface contaminations of 1, 10, and 100 $\mu$m, assuming an exponential depth profile decreasing from the surface into the crystal bulk. Furthermore, discrepancies in the shape of the $\beta^-$ spectrum from \textsuperscript{210}Bi decay were identified, leading to the implementation of an improved spectral shape based on experimental measurements.

Subsequently, the data and simulation selection required to perform the fit has been described, including the procedure itself, detailing the classification of background components as fixed or free at each step, and the strategy for releasing these components during the process. The fitting range, the fit function, and the method to derive initial activities from the fit outcomes have also been detailed. The performance of the fit at each stage has been evaluated, and the fitted activity results have been compared with those from the previous background model. 

A surface contamination depth of 10 $\mu$m has been selected as the most plausible scenario for \textsuperscript{210}Pb, based on the overall agreement across all detectors and fitting intervals. This assumption has been adopted for the construction of the new background model. Its performance has been compared to the previous one, showing a clear improvement across all energy regions and a significant reduction in the residuals. This robustness is further supported by the good agreement between data and simulation for coincidence event populations and, most importantly, for the year-difference spectra, which the background model developed in this thesis successfully reproduces. The fit results do not support the hypothesis that the lower-energy alpha peak arises from surface \textsuperscript{210}Pb contamination while the higher-energy peak originates from a bulk contamination, based on the values of the fitted activities. However, further investigation is required to fully assess this interpretation. Additionally, the experimental \textsuperscript{210}Bi $\beta^-$ decay shape provides a significantly improved fit to the data, particularly in the [600–1000]~keV region, where a persistent excess had not been satisfactorily explained until now.

Two additional background contributions not included in the main fitting strategy have also been presented: the environmental neutron background in hall B, as measured by the HENSA experiment, and the population of anomalous events identified with the new ANOD DAQ. For the former, simulations of the neutron flux reaching the ANAIS experimental set-up, propagated from the external flux measured in hall B, have been performed using GEANT4 and compared with results obtained by HENSA using FLUKA. It has been shown that differences between simulation codes are non-negligible, especially in the treatment of fast neutrons. The resulting background contribution from environmental neutrons has been presented and found to be non-relevant compared to the measured background. 

Regarding the population of asymmetric events, those seen by ANAIS but discriminated only by ANOD due to its distinct acquisition characteristics, they have been incorporated into the model as a time-independent contribution. Their measured rates are compatible with part of the excess observed in ANAIS data in the [1–2]~keV region that remains unexplained by the background model. Efforts are currently focused on exploiting the enhanced discrimination capabilities of the ANOD system to understand and eventually eliminate this anomalous population of events.


Finally, the time evolution of the new background model has been developed and demonstrated to offer a significantly improved description compared to the previous version. In ANAIS-112, the annual modulation search strategy explicitly relies on the temporal evolution of the background, making accurate and robust backgorund modelling essential. The new model consistently provides a better description across all event populations and energy regions, reinforcing the validity of this work. Although it accounts for a greater number of events, the overall spectral shape remains largely consistent with that of the previous model, which was already in good agreement with the data. This improved model will be employed in Chapter~\ref{Chapter:annual} to reevaluate the annual modulation results over the full six-year exposure, allowing for a quantitative assessment of the performance gain. While the expected impact on the modulation analysis may be moderate given the similar spectral shapes, the refined modelling is anticipated to enhance the reliability of the results.

\setcounter{chapter}{5} 

\chapter{New physics searches with ANAIS-112 data}\label{Chapter:annual}

\vspace{-0.2cm}
\vspace{0.5cm}

\minitoc

In this thesis, an improved version of the previous ANAIS-112 background model has been developed. This updated model is used to describe the time evolution of the background components for each detector, which is required for the annual modulation analysis, but it also enables other searches for rare processes. In addition, the QF\textsubscript{Na} and QF\textsubscript{I} of the ANAIS-112 crystals have been evaluated in Chapter \ref{Chapter:QF} by comparing performed onsite neutron calibrations with dedicated Geant4 simulations. This approach has allowed the selection of an energy-dependent QF\textsubscript{Na} (ANAIS(1)), and an energy-dependent QF\textsubscript{I} as the QFs of the ANAIS crystals \cite{cintas2024measurement,phddavid}.  The QFs have to be taken into account for comparing DAMA/LIBRA and ANAIS-112 results on annual modulation in the same NR-energy scale.

This chapter first presents the annual modulation search based on the updated background model, using six years of exposure (Section \ref{withbkg}). The results are compared with those from the previous annual modulation analysis for the same dataset \cite{amare2025towards}. Furthermore, considering the QF\textsubscript{Na} and QF\textsubscript{I} estimations obtained in this work and their inconsistency with the values adopted by DAMA/LIBRA, the annual modulation search is also performed in the NR-energy range in ANAIS-112 corresponding to the [2-6]~keV energy range of DAMA/LIBRA, considering their measured QF (Section~\ref{annualwithQF}). In addition, the dependence of the modulation amplitude on the energy scale for both sodium and iodine NRs is examined. Subsequently, the solar axion search enabled by the improved background model developed in this thesis is presented (Section \ref{axions}).

\section{Annual modulation analysis}

The annual modulation analysis strategy has already been presented in Section \ref{annualstrategy} when reporting the ANAIS-112 annual modulation results corresponding to six-year exposure recently released \cite{amare2025towards}. It is briefly revisited here to highlight the key aspects of the analysis.

To independently test the DAMA/LIBRA annual modulation signal, ANAIS-112 employs a different analysis strategy. Unlike DAMA, which fits residual rates (total rate minus the average rate, calculated on a yearly basis) to a cosine function, ANAIS-112 performs a direct least-squares fit to the total event count over time, where $\chi^2$ is defined as:

\begin{equation}
\chi^2 = \sum_{i,d} \frac{\left(n_{i,d} - \mu_{i,d}\right)^2}{\sigma_{i,d}^2}
\end{equation}

The $\chi^2$ function quantifies the agreement between the observed number of events $n_{i,d}$ in each time bin $t_i$ and detector $d$, corrected for live time and detection efficiency, and the expected number of events $\mu_{i,d}$, which incorporates both the background and a potential modulated signal. The uncertainty $\sigma_{i,d}$ represents the statistical Poisson standard deviation of the observed counts, also corrected accordingly.

$\mu_{i,d}$ is expected to decrease over time due to background contributions from radioactive isotopes with half-lives of several years, mainly $\mathrm{^{210}Pb}$ (22.3~y), $\mathrm{^{3}H}$ (12.3~y), and $\mathrm{^{22}Na}$~(2.6~y). After selecting the low energy events using the BDT technique and applying the corresponding efficiency correction, the time
evolution of the resulting rate of events is modelled by:

\begin{equation}
    \mu_{i,d} = [ R_{0,d}(f_d\phi^{MC}_{bkg,d}(t_i)+(1-f_d)\phi_{flat}(t_i))+S_m\cos(\omega(t_i-t_0))]M_d\Delta E\Delta t ,
    \label{rateevents2}
\end{equation}

where $\phi^{\text{MC}}_{\text{bkg},d}$ is the PDF obtained from the improved MC background model, describing the expected background rate in time bin $t_i$ for detector $d$; and $\phi_{flat}$ is a constant probability distribution function that accounts for noise not explained by the background model (linked to the observed excess below 3 keV) and found at a constant rate in the data, as well as representing the average component of a hypothetical DM signal. $M_d$ denotes the mass of each detector module, while $\Delta E$ and $\Delta t$ represent the energy and time intervals, respectively. $R_{0,d}$ and $f_d$ are nuisance parameters in the fitting procedure specific to each detector: $R_{0,d}$ represents the average
rate in the considered energy region, while $f_d$ measures the relative weight of the MC time-dependent background with respect to the total rate. Moreover, $\textnormal{S}_\textnormal{m}$ is the amplitude of the annual modulation signal, and $\omega$ and $t_0$ denote the angular frequency and phase of the modulation searched for in the data, corresponding to a period of 1~year and June 2, respectively, according to the standard halo model. To test the null hypothesis, $\textnormal{S}_\textnormal{m}$ is fixed to zero, whereas it is treated as a free parameter under the modulation hypothesis.

\begin{figure}[b!]
    \centering
    {\includegraphics[width=0.49\textwidth]{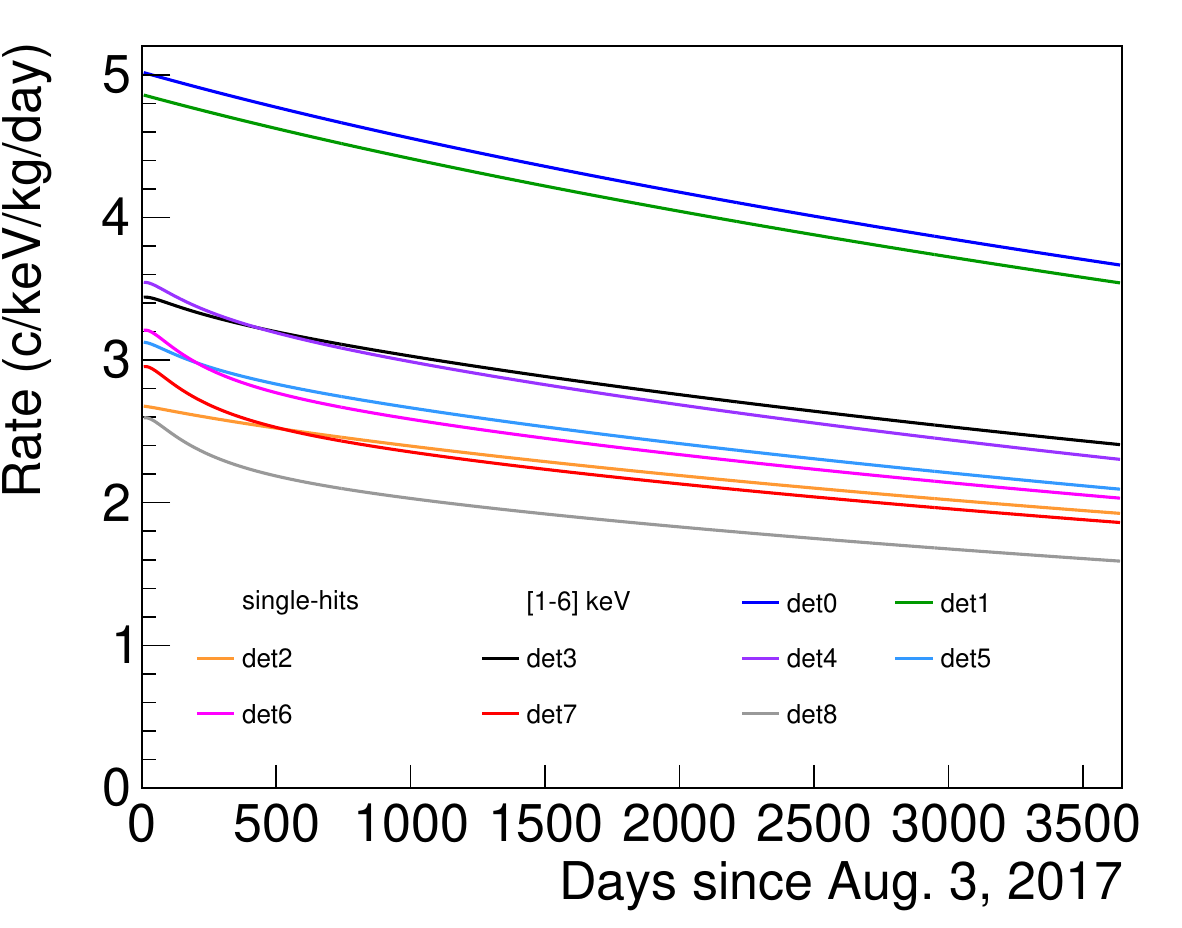}}{\includegraphics[width=0.49\textwidth]{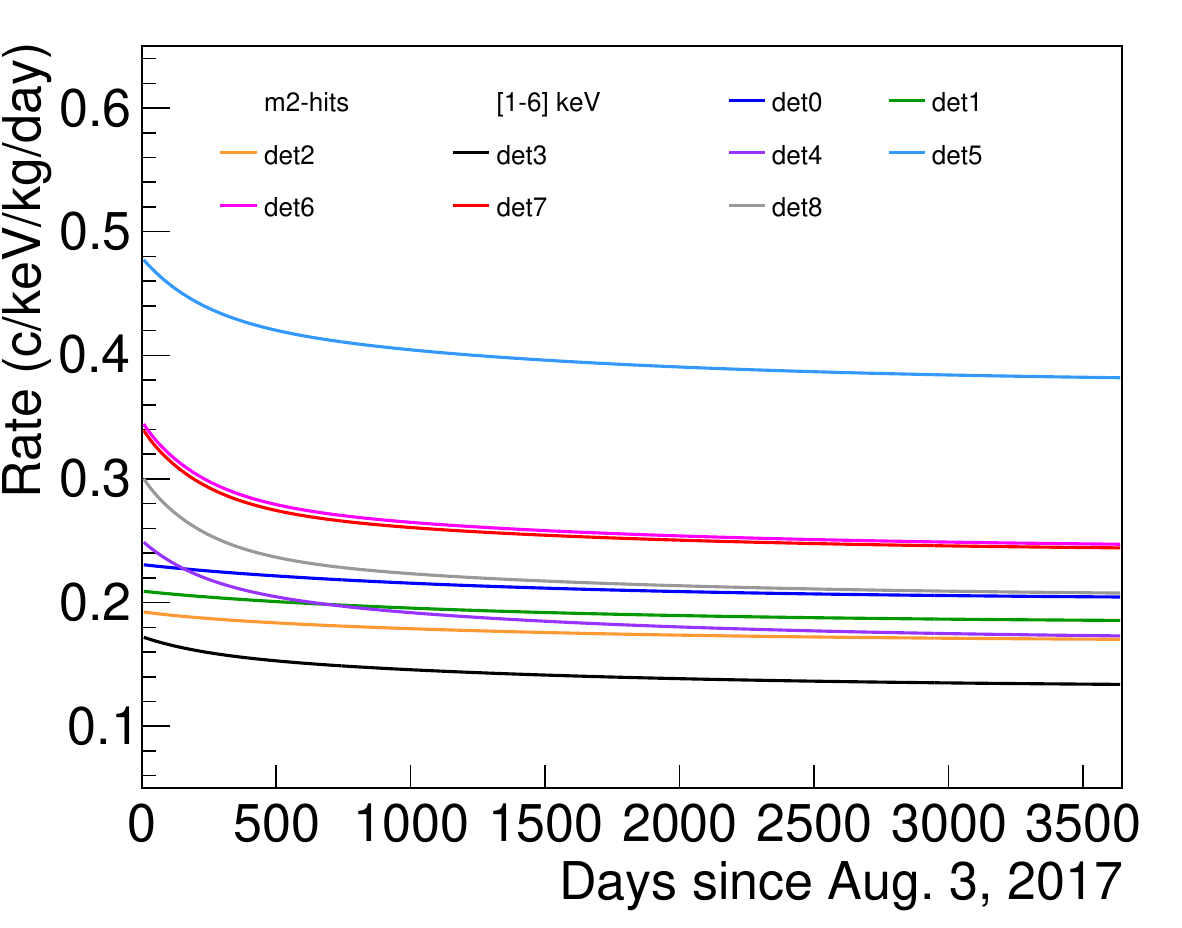}}

    \caption{\label{pdfbkg}  Rate evolution in time predicted by the improved background model in the [1–6]~keV energy region, for each ANAIS-112 module. \textbf{Left panel:} single-hits. \textbf{Right panel:} m2-hits. }
\end{figure}

It is clear that precise background modelling is critical for the annual modulation analysis, ensuring that the fit remains unbiased \cite{Amare:2021yyu,amare2025towards}. The time-dependent background model has been presented in Section \ref{rateevolution}. Figure \ref{pdfbkg} shows the time evolution predicted by the improved background model in the [1–6] keV energy region for each ANAIS-112 module, both in the single-hit and m2-hit spectra. As observed, the temporal behavior varies across detectors. As thoroughly discussed in Chapter \ref{Chapter:bkg}, modules D0 and D1 are the most contaminated, while modules D7, D6, and D8 still show signs of cosmogenic activity during the initial months of data taking. Furthermore, in the coincidence spectra, crystal D5 exhibits the highest coincidence rate, which is attributed to its central position within the ANAIS-112 set-up.

In the fit, the modulation period is fixed to one year and the phase is set to June~2\textsuperscript{nd}, allowing for a direct comparison with the results reported by the DAMA/LIBRA experiment in \cite{Bernabei:2020mon}. DAMA/LIBRA
best fits in the [1–6] keV ([2–6] keV) energy region results in a modulation amplitude of 10.5 $\pm$ 1.1 cpd/ton/keV
(10.2 $\pm$ 0.8 cpd/ton/keV), favouring the presence of a modulation with proper DM features.

\subsection{Using the improved background model}\label{withbkg}

\begin{table}[b!]
    \centering
    \resizebox{\textwidth}{!}{\Large
    \begin{tabular}{c|c|cc|cc|c}
        \hline
        
            \makecell{energy \\ region} & Results & \makecell{$\chi^2$/ndf \\ null hyp}& \makecell{p-value \\ null hyp} & \makecell{$\chi^2$/ndf \\ mod hyp} & \makecell{p-value \\ mod hyp} &  \makecell{S\textsubscript{m}\\  (cpd/ton/keV)}\\
           
            \hline
           \multirow{3}{*}{[1--6] keV} &  \makecell{ANAIS-112 \\ \cite{amare2025towards}}
 & 451.34 / 423 & 0.164 & 451.31 / 422 & 0.156 & -0.4 $\pm$ 2.5 \\

            \cline{2-7}
 & \makecell{ANAIS-112 \\ (this work)}
  &  435.39 / 423 & 0.328 & 435.39 / 422 & 0.316 & -0.2 $\pm$ 2.5 \\
        \hline
           \multirow{3}{*}{[2--6] keV} &  \makecell{ANAIS-112 \\ \cite{amare2025towards}}
  & 414.46 / 423 & 0.607 & 414.28 / 422 & 0.596   & 1.1 $\pm$ 2.5 \\
 \cline{2-7}
 & \makecell{ANAIS-112 \\ (this work)}
    &  399.78 / 423 & 0.785 & 399.55 / 422 & 0.778 & 1.2 $\pm$ 2.5 \\
        \hline
         
        & & & & & &  \textcolor{green}{(cpd/ton/3.3 keV\textsubscript{NR})}   \\
        & & & & & &  \textcolor{blue}{(cpd/ton/11.1 keV\textsubscript{NR})}   \\

        \hline
      
         \multirow{3}{*}{\shortstack{\textcolor{green}{[6.7–20] keV\textsubscript{NR}} \\ \textcolor{green}{(Na recoils)}  \\ \shortstack{\textcolor{blue}{[22.2-66.7] keV\textsubscript{NR}} \\ \textcolor{blue}{(I recoils)}}}}
 &  \makecell{ANAIS-112 \\ \cite{amare2025towards}}
 & 416.51 / 423 & 0.580 & 416.51 / 422 & 0.566   & 0.0 $\pm$ 2.3 \\
          \cline{2-7}
& \makecell{ANAIS-112 \\ (this work)}
   &  401.25 / 423 & 0.770 & 401.25 / 422 & 0.759 & 0.0 $\pm$ 2.3 \\
& & & & & & \\
   \hline

           & & & & & &  (cpd/ton/11.1 keV\textsubscript{NR})   \\
\hline
\makecell{[22.2-66.7] keV\textsubscript{NR} \\ (I recoils) }  & \makecell{ANAIS-112 \\ (this work)}
   &  427.34 / 423 & 0.432 & 427.29 / 422 & 0.419 & 0.7 $\pm$ 2.9 \\

        \hline
\end{tabular}}
\caption{\label{tablageneral}Summary of the fit results (goodness of the fit and best fit value for the modulation amplitude) for the annual modulation search with fixed period and phase in the [1–6] and [2–6] keV energy regions for ANAIS–112 six years of data.  Results in the [6.7–20] keV\textsubscript{NR} sodium and [22.2-66.7] keV\textsubscript{NR} iodine recoil energy region (corresponding to [2–6]~keV for DAMA/LIBRA when using QF\textsubscript{Na}=0.3, QF\textsubscript{I}=0.09) are also shown, using the constant values QF\textsubscript{Na}=0.2 and QF\textsubscript{I}=0.06, and the energy-dependent QF\textsubscript{I} selected in this thesis (see text for details). The results published by ANAIS-112 in \cite{amare2025towards} with the previous background model are compared with those obtained in this work, which incorporates the improved background model. }
\end{table} 



\begin{figure}[t!]
    \centering
    {\includegraphics[width=1.\textwidth]{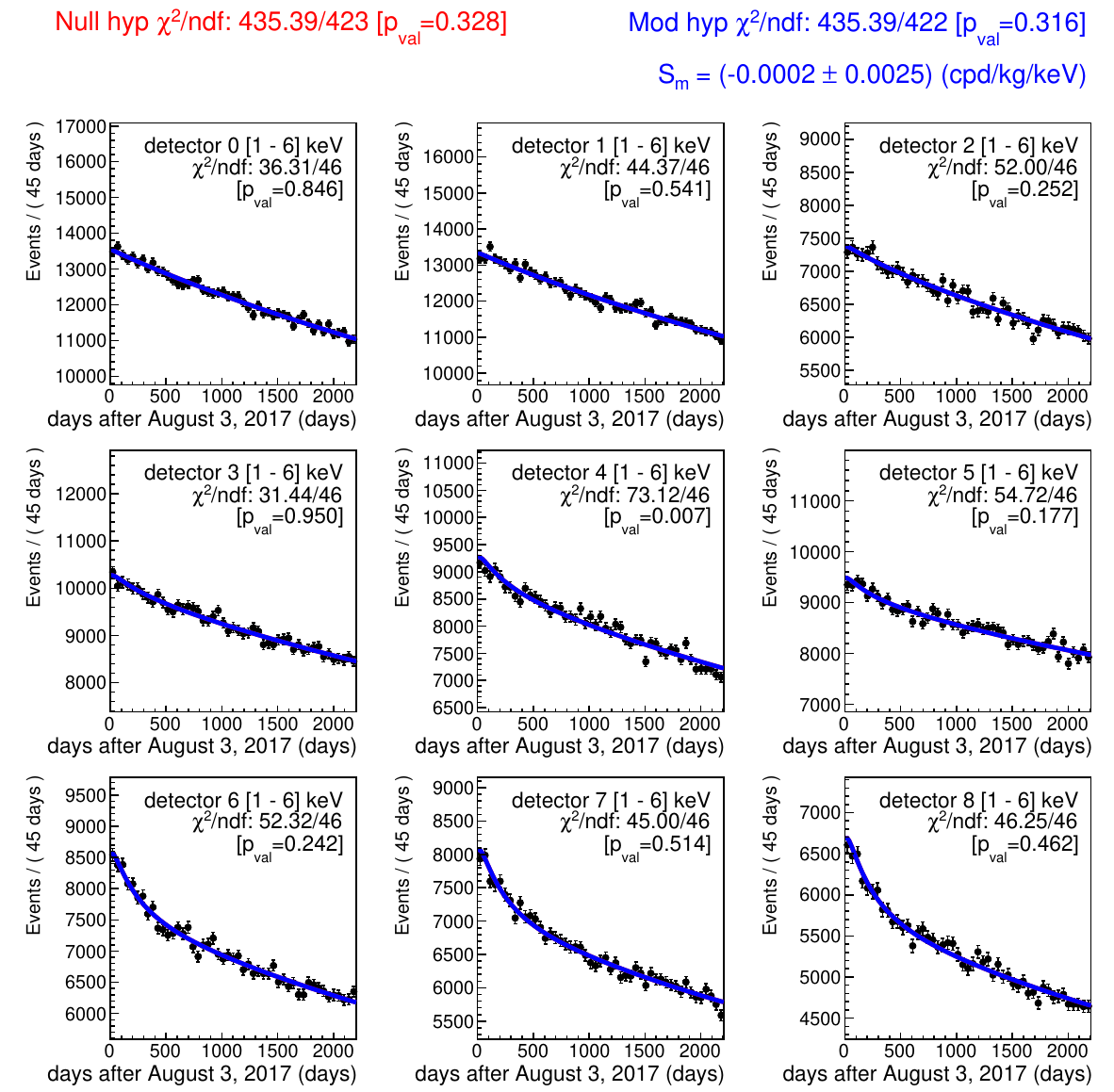}}

    \caption{\label{annual16}  Results of the fit for the data from the nine modules in the [1–6] keV energy region, under the modulation (blue)
and null hypotheses (red). In all the panels, the red line is masked by the blue one, as the fit obtained for the modulated
hypothesis is consistent with S\textsubscript{m} = 0. $\chi^2$/ndf and p-values under the modulation hypothesis are also individually displayed for
each module.}
\end{figure}

\begin{figure}[t!]
    \centering
    {\includegraphics[width=1.\textwidth]{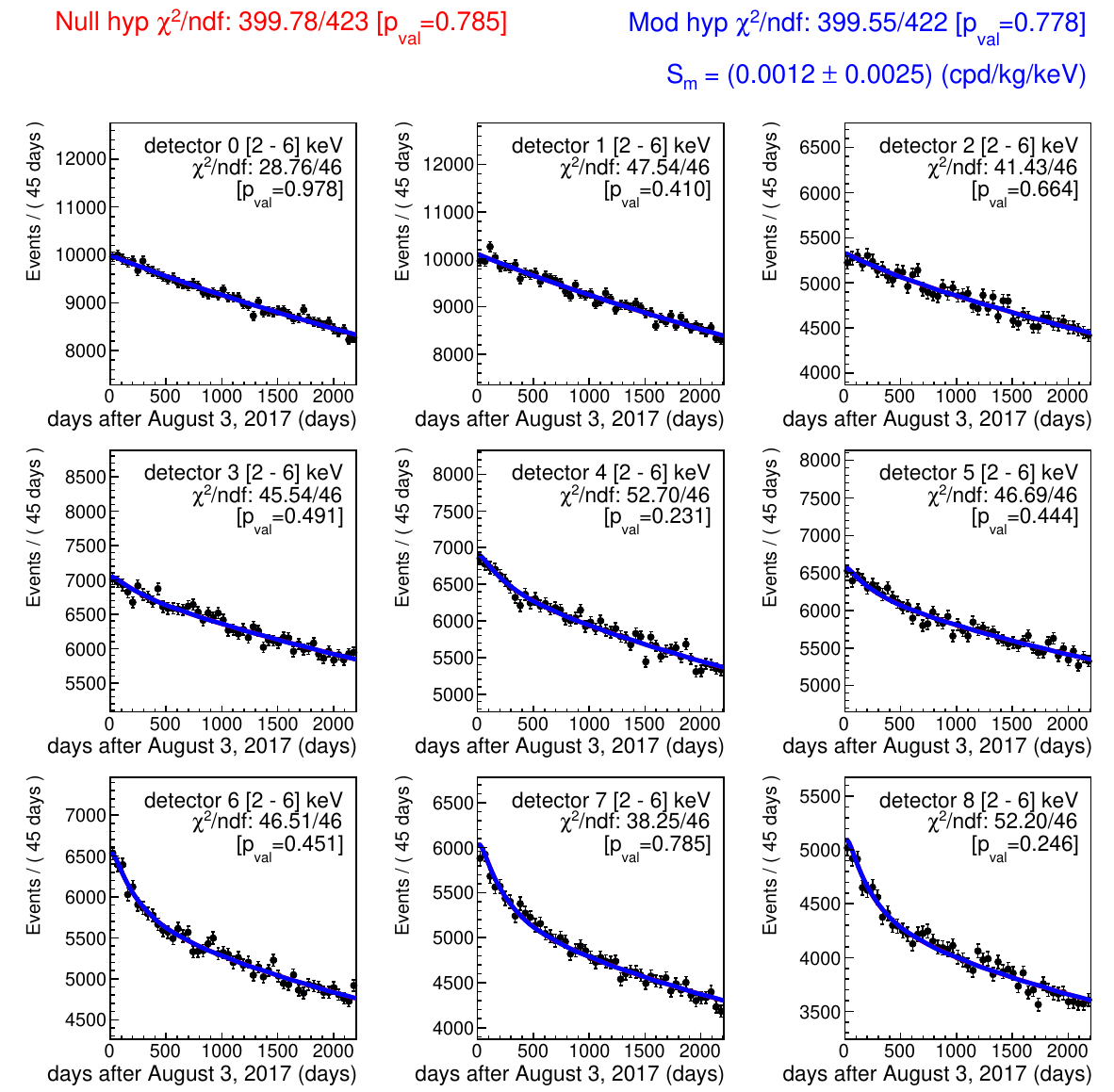}}

    \caption{\label{annual26}  Results of the fit for the data from the nine modules in the [2-6] keV energy region, under the modulation (blue)
and null hypotheses (red). In all the panels, the red line is masked by the blue one, as the fit obtained for the modulated
hypothesis is consistent with S\textsubscript{m} = 0. $\chi^2$/ndf and p-values under the modulation hypothesis are also individually displayed for
each module.}
\end{figure}

The $\chi^2$ minimization results for both the null and modulation hypotheses are shown in Figures~\ref{annual16} and \ref{annual26}, corresponding to the [1–6]~keV and [2–6]~keV energy intervals, respectively, with data grouped in 45-day bins. 

Table~\ref{tablageneral} summarizes the fit results (goodness of the fit and best fit value for the modulation amplitude) for both energy regions using the full six-year ANAIS–112 dataset as presented in \cite{amare2025towards} using the previous background model, and compares them with those obtained in this work using the improved background model. The $\chi^2/\mathrm{NDF}$ and corresponding p-values have been computed separately for each detector module and are reported in both the figure legends and Table~\ref{tabladetectoradetector}.

\begin{table}[t!]

\centering
    \begin{minipage}{0.52\textwidth}
    
    \centering
    \resizebox{\textwidth}{!}{\Large
   
    \begin{tabular}{c|cc|cc}
        \hline
        \multicolumn{5}{c}{[1-6] keV}\\
        \hline
        \multirow{2}{*}{detector}  & \multicolumn{2}{c|}{ANAIS-112 \cite{amare2025towards}} &  \multicolumn{2}{c}{ANAIS-112 (this work)}\\
        \cline{2-5}
         &  $\chi^2$/ndf & p-value & $\chi^2$/ndf & p-value \\
         
        \hline
               0 & 36.38 / 46 & 0.844 & 36.31 / 46 & 0.846 \\
        1 & 44.27 / 46 & 0.545 & 44.37 / 46 & 0.541 \\
        2 & 52.30 / 46 & 0.243 & 52.00 / 46 & 0.252 \\
        3 & 30.44 / 46 & 0.963 & 31.44 / 46 & 0.950 \\
        4 & 75.55 / 46 & 0.004 & 73.12 / 46 & 0.007 \\
        5 & 57.71 / 46 & 0.115 & 54.72 / 46 & 0.177 \\
        6 & 55.43 / 46 & 0.161 & 52.32 / 46 & 0.242 \\
        7 & 50.85 / 46 & 0.289 & 45.00 / 46 & 0.514 \\
        8 & 48.51 / 46 & 0.372 & 46.25 / 46 & 0.462 \\
        \hline
\end{tabular}}

    \label{tab:exp_data}\subcaption{}
    \end{minipage}
    \vskip 0.5cm

    \begin{minipage}{0.52\textwidth}
    
    \centering
    \resizebox{\textwidth}{!}{\Large
   
    \begin{tabular}{c|cc|cc}
        \hline
        \multicolumn{5}{c}{[2-6] keV}\\
        \hline
        \multirow{2}{*}{detector}  & \multicolumn{2}{c|}{ANAIS-112 \cite{amare2025towards}} &  \multicolumn{2}{c}{ANAIS-112 (this work)}\\
        \cline{2-5}
         &  $\chi^2$/ndf & p-value & $\chi^2$/ndf & p-value \\
         
        \hline
            0 & 28.97 / 46 & 0.977 & 28.76 / 46 & 0.978 \\
        1 & 47.53 / 46 & 0.410 & 47.54 / 46 & 0.410 \\
        2 & 43.27 / 46 & 0.587 & 41.43 / 46 & 0.664 \\
        3 & 43.72 / 46 & 0.568 & 45.54 / 46 & 0.491 \\
        4 & 53.12 / 46 & 0.219 & 52.70 / 46 & 0.231 \\
        5 & 48.65 / 46 & 0.367 & 46.69 / 46 & 0.444 \\
        6 & 49.17 / 46 & 0.347 & 46.51 / 46 & 0.451 \\
        7 & 45.72 / 46 & 0.484 & 38.25 / 46 & 0.785 \\
        8 & 54.19 / 46 & 0.190 & 52.20 / 46 & 0.246 \\
        \hline
\end{tabular}}

    \label{tab:exp_data}\subcaption{}
    \end{minipage}
    \vskip 0.5cm

    \begin{minipage}{0.52\textwidth}
    
    \centering
    \resizebox{\textwidth}{!}{\Large
   
    \begin{tabular}{c|cc|cc}
        \hline
        \multicolumn{5}{c}{[1.3-4] keV}\\
        \hline
        \multirow{2}{*}{detector}  & \multicolumn{2}{c|}{ANAIS-112 \cite{amare2025towards}} &  \multicolumn{2}{c}{ANAIS-112 (this work)}\\
        \cline{2-5}
         &  $\chi^2$/ndf & p-value & $\chi^2$/ndf & p-value \\
         
        \hline
        0 & 54.21 / 46 & 0.190 & 53.08 / 46 & 0.220 \\
        1 & 38.79 / 46 & 0.766 & 39.25 / 46 & 0.749 \\
        2 & 38.23 / 46 & 0.785 & 37.80 / 46 & 0.800 \\
        3 & 27.93 / 46 & 0.984 & 27.99 / 46 & 0.983 \\
        4 & 61.73 / 46 & 0.060 & 55.81 / 46 & 0.152 \\
        5 & 45.09 / 46 & 0.510 & 41.20 / 46 & 0.673 \\
        6 & 51.47 / 46 & 0.268 & 54.04 / 46 & 0.194 \\
        7 & 49.78 / 46 & 0.325 & 44.55 / 46 & 0.533 \\
        8 & 49.35 / 46 & 0.341 & 47.60 / 46 & 0.407 \\
        \hline
\end{tabular}}

    \label{tab:exp_data}\subcaption{}
    \end{minipage}  

   \caption{\label{tabladetectoradetector} Summary of the goodness-of-fit results for the annual modulation search with fixed period and phase, using six years of ANAIS–112 data for each detector. The results published by ANAIS-112 in \cite{amare2025towards} are compared with those obtained in this thesis, which incorporates an improved background model. \textbf{(a)} [1-6] keV. \textbf{(b)} [2-6] keV. \textbf{(c)}~[1.3-4]~keV (corresponding to [2–6]~keV for DAMA/LIBRA for both sodium and iodine under the hypothesis of QF\textsubscript{Na}=0.2, QF\textsubscript{I}=0.06).}
    
\end{table}

The ANAIS–112 results obtained in this work remain consistent with the null hypothesis, yielding p-values of 0.328 and 0.785 for the [1–6]~keV and [2–6]~keV regions, respectively. Comparable p-values are also found in the modulation hypothesis. As evidenced in Table~\ref{tablageneral}, the analysis with the improved background model leads to a clear enhancement of the fit behaviour in the [1–6] keV region, improving the global p-value from 0.164 (previous model) to 0.328 (revisited model).

In contrast, in the [2–6]~keV interval, both background models provide a comparable description of the data, indicating that the original model was already very adequate in this energy range. As discussed in the previous chapter (and shown in Figures \ref{rateevol610}, \ref{rateevol16}, \ref{rateevol26} and \ref{rateevol12}), the improved background model does explain a larger fraction of the observed events across all regions, demonstrating a better overall agreement with data. However, since RooFit normalizes each PDF to unity \cite{verkerke2006roofit}, variations in the predicted total event rate have little impact on the fit results unless the shape of the temporal evolution is significantly altered, which is not the case in this energy interval. 

As a result, the fitted parameters $R_{0,d}$ and $f_d$ remain fully consistent with those obtained using the previous background model and are therefore not displayed here. The same holds for the associated goodness-of-fit values, which also show no significant deviation. This consistency is observed both globally (Table~\ref{tablageneral}) and on a detector by detector basis (Table~\ref{tabladetectoradetector}).

The best-fit modulation amplitudes derived are $\textnormal{S}_\textnormal{m}~= (-0.2 \pm 2.5)$~cpd/ton/keV for the [1–6]~keV region and $\textnormal{S}_\textnormal{m} = (1.2 \pm 2.5)$~cpd/ton/keV for the [2–6]~keV region. These results are fully consistent with the previously published ANAIS–112 values~\cite{amare2025towards}, $\textnormal{S}_\textnormal{m} = (-0.4 \pm 2.5)$~cpd/ton/keV and $\textnormal{S}_\textnormal{m} = (1.1 \pm 2.5)$~cpd/ton/keV for the same energy intervals, respectively.


 \begin{figure}[b!]
    \centering
    {\includegraphics[width=1.\textwidth]{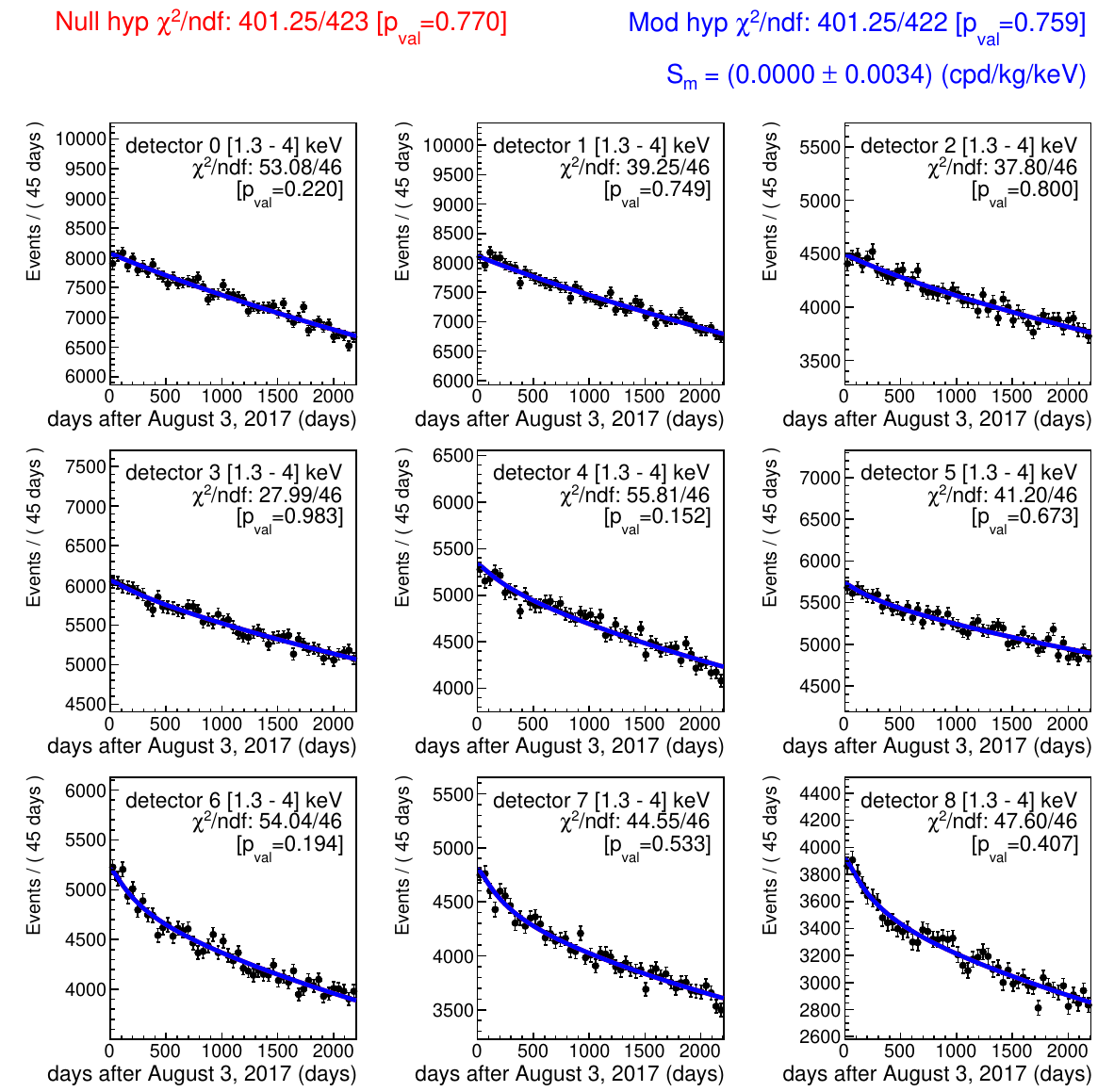}}

    \caption{\label{annual1p34}  Results of the fit for the data from the nine modules in the [1.3–4] keV energy region, corresponding to the sodium
nuclear recoil energy range of [6.7–20] keV\textsubscript{NR} under the modulation (blue)
and null hypotheses (red). In all the panels, the red line is masked by the blue one, as the fit obtained for the modulated
hypothesis is consistent with S\textsubscript{m}~=~0. $\chi^2$/ndf and p-values under the modulation hypothesis are also individually displayed for
each module.}
\end{figure}

The annual modulation analysis carried out in \cite{amare2025towards} with the 6-year exposure data also considered the possibility of DAMA/LIBRA and ANAIS-112 having different QF. This analysis was performed for the DAMA/LIBRA [2-6] keV result.
By converting this energy region into NR-energy scale for sodium and iodine assuming the QF values reported by DAMA/LIBRA for their NaI(Tl) crystals \cite{bernabei1996new} (QF\textsubscript{Na}=0.3, QF\textsubscript{I}=0.09), the resulting NR energy ranges are [6.7-20] keV\textsubscript{NR} for sodium and [22.2-66.7] keV\textsubscript{NR} for iodine. Given that the constant QFs for ANAIS-112 reported in \cite{cintas2024measurement,phddavid} (QF\textsubscript{Na}=0.2, QF\textsubscript{I}=0.06) are approximately 2/3 of those used by DAMA/LIBRA for both Na and I, both NR energy regions correspond to the same electron-equivalent energy region in ANAIS-112, i.e., [1.3–4] keV.

Figure~\ref{annual1p34} presents the $\chi^2$ minimization results under the null and modulation hypotheses for this energy region. The global fit parameters and the detector-by-detector goodness-of-fit are displayed in Tables~\ref{tablageneral} and \ref{tabladetectoradetector}, respectively, where they are compared to the results obtained using the previous background model. As in the [2–6]~keV region, both background models yield equivalent performance in terms of goodness-of-fit. In addition, the same result is obtained for the annual modulation amplitude as with the previous background model \cite{amare2025towards}, $\textnormal{S}_\textnormal{m}$ = (0.0 $\pm$ 2.3) cpd/ton/3.3 keV\textsubscript{NR}.

Thus, in the three energy regions that can be compared with the previously published ANAIS-112 results \cite{amare2025towards}, compatible values for the modulation amplitude are consistently obtained. This leads to same levels of incompatibility and sensitivity with respect to the DAMA/LIBRA result. There is no denying that, as shown in Chapter~\ref{Chapter:bkg}, the new background model provides a better description of the ANAIS-112 data across all event populations and energy ranges. Nevertheless, since the overall shape of the time evolution in the ROI does not change substantially, the fact that the final results remain very similar to those obtained with the previous model demonstrates the robustness of the ANAIS-112 annual modulation analysis.

Moreover, the energy dependence of the annual modulation amplitude, $\textnormal{S}_\textnormal{m}$(E), constitutes a powerful tool for testing the presence of a DM induced modulation signal. The positive result reported by the DAMA/LIBRA collaboration is only present below 6~keV, with modulation amplitudes compatible with zero above this threshold. 

 \begin{figure}[t!]
    \centering
    {\includegraphics[width=0.8\textwidth]{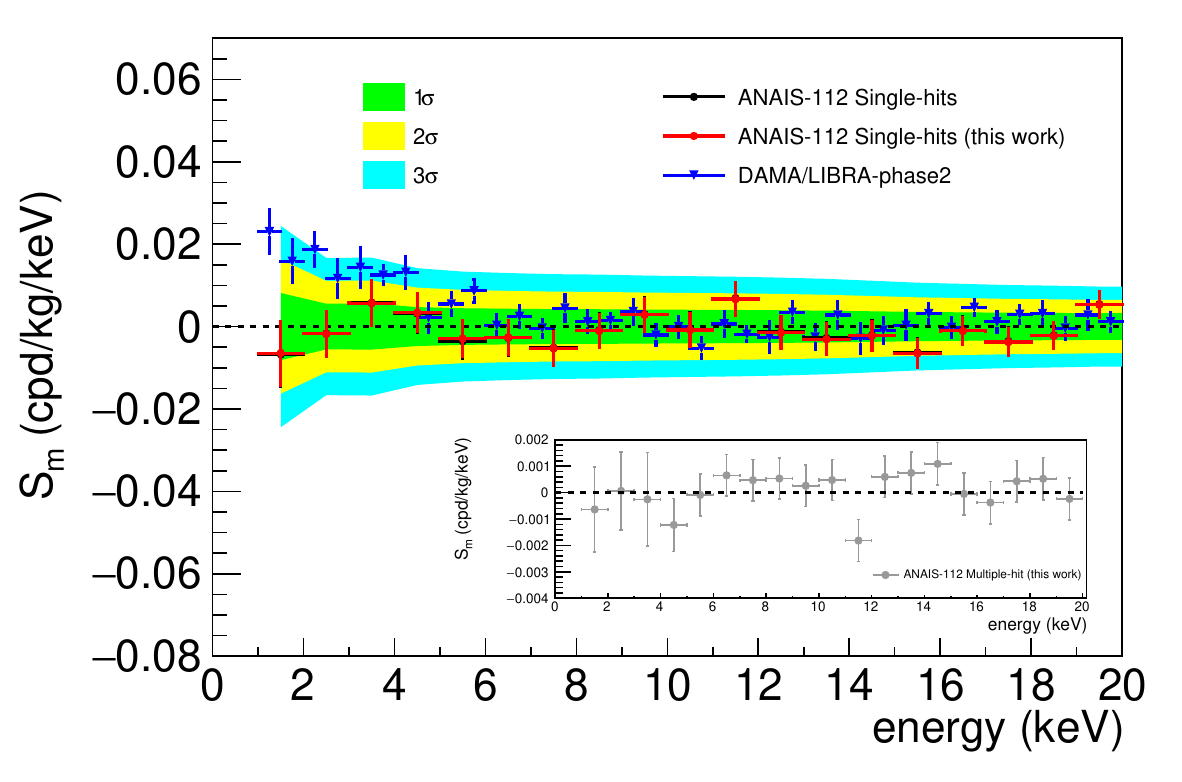}}

    \caption{\label{Smeesingle} Modulation amplitude obtained in this work in 1 keV energy bins for single-hit events (red dots), compared with the DAMA/LIBRA-phase2 result\cite{Bernabei:2020mon} (blue triangles), and the results from the previous background model \cite{amare2025towards} (black dots). The shaded bands represent the 1$\sigma$, 2$\sigma$, and 3$\sigma$ sensitivity intervals derived from ANAIS–112 data.  Inset: modulation amplitude obtained in this work in 1 keV energy bins for multiple-hit events. }
\end{figure}

In the present analysis, following the same approach as in \cite{amare2025towards}, $\textnormal{S}_\textnormal{m}$ has been evaluated in 1 keV energy bins from 1 to 20~keV, separately for single-hit and multiple-hit events, combining data from all nine ANAIS–112 modules. The results are displayed in Figure~\ref{Smeesingle}, together with those obtained using the previous background model~\cite{amare2025towards}, and the modulation amplitude reported by DAMA/LIBRA-phase2~\cite{Bernabei:2020mon}. Expected 1$\sigma$, 2$\sigma$, and 3$\sigma$ sensitivity bands, derived from ANAIS–112 data~\cite{ANAISsproj}, are also included. The corresponding distribution obtained in this work for multiple-hit events is shown in the same figure as an inset. As illustrated, the updated background model yields modulation amplitudes compatible with the previous analysis, further reinforcing the robustness of the annual modulation search strategy.

A $\chi^2$ test is performed by comparing the modulation amplitude values obtained with the improved background model to the DAMA/LIBRA modulation amplitude (DAMA/LIBRA hypothesis) and the null hypothesis ($\textnormal{S}_\textnormal{m}$=0). For the DAMA/LIBRA hypothesis in the [1–6]~keV region for single-hit events, a  $\chi^2/\text{ndf}$ = 23.22/5 (p-value = 3.06 $\times$ 10$^{-4}$) is obtained, taking into account the uncertainties in the DAMA/LIBRA result. In contrast, for the null hypothesis $\chi^2/\text{ndf}$ = 2.88/5 (p-value = 0.72), indicating compatibility with the absence of modulation. For events with multiplicity 2, a $\chi^2/\text{ndf}$ value of 1.69/5 (p-value = 0.89) is obtained, in agreement with the results for single-hit events, supporting the conclusion that no significant systematic effects are present in the analysis. Compatible values are also obtained using the previous background model \cite{amare2025towards}, and  similar results are also found in the [2–6] keV region.

\subsection{Using energy-dependent ANAIS-112 QF\textsubscript{Na} and QF\textsubscript{I}}\label{annualwithQF}

In the previous section, the annual modulation search was conducted in the [1.3–4]~keV energy region of ANAIS–112, which corresponds to the [2-6] keV region of DAMA/LIBRA region under the assumption that both experiments have QFs that fulfill a $\sim$ 3/2 proportionality. However, in this thesis, an energy-dependent QF for both nuclei has been proposed in Chapter \ref{Chapter:QF} as a more accurate description of the ANAIS-112 detector response, altering this correspondence between energy scales.


\begin{figure}[b!]
    \centering
    {\includegraphics[width=1.\textwidth]{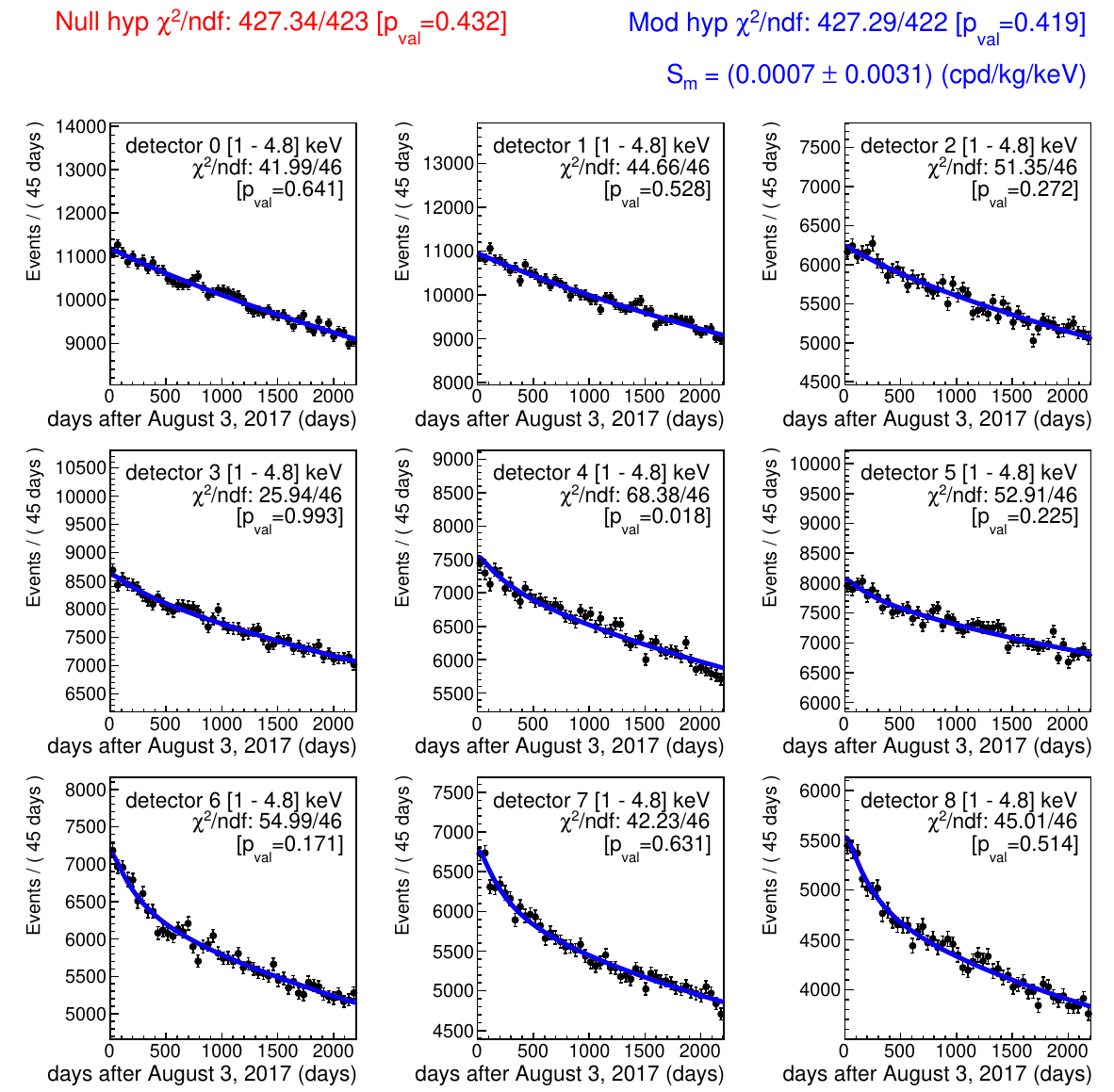}}

    \caption{\label{annual148}  Results of the fit for the data from the nine modules in the [1–4.8] keV energy region, under the modulation (blue)
and null hypotheses (red). In all the panels, the red line is masked by the blue one, as the fit obtained for the modulated
hypothesis is consistent with S\textsubscript{m} = 0. $\chi^2$/ndf and p-values under the modulation hypothesis are also individually displayed for
each module.}
\end{figure}

According to the energy-dependent QF models preferred in this work, the ANAIS(1) QF\textsubscript{Na} using the modified Lindhard model dependence varies from 0.129 at 6.7 keV\textsubscript{NR} to 0.157 at 20 keV\textsubscript{NR}, implying a corresponding electron-equivalent region of [0.87–2.6]~keV in ANAIS–112 data. However, the lower bound lies below the current experimental energy threshold (1 keV), making the [6.7-20]~keV\textsubscript{NR} sodium recoil energy region experimentally inaccessible. Nevertheless, there is a possibility of lowering the energy threshold to at least 0.6 keV by leveraging ML in upcoming data releases, which would make this region accessible in the near future.

In contrast, the energy-dependent QF\textsubscript{I} ranges from 0.045 at 22.2 keV\textsubscript{NR} to 0.072 at 66.7~keV\textsubscript{NR}, yielding an explorable electron-equivalent energy region of [1–4.8]~keV in ANAIS–112. Thus, Figure~\ref{annual148} presents the best fits for both the null and modulation hypotheses in this region, corresponding to the [22.2-66.7] keV\textsubscript{NR} iodine recoil energy region. Table~\ref{tablageneral} lists the corresponding results, in comparison with those obtained in other energy regions.

ANAIS–112 results remain compatible with the null hypothesis, with a p-value of 0.432. The best-fit modulation amplitude is $\textnormal{S}_\textnormal{m}$ = (0.7 $\pm$ 2.9)~cpd/ton/11.1 keV\textsubscript{NR}, fully consistent with the absence of modulation within one standard deviation. These results are incompatible with the DAMA/LIBRA signal at the 3.2$\sigma$ level, which correspond to a sensitivity to the DAMA/LIBRA signal of (3.5 $\pm$ 0.3)$\sigma$, where the uncertainty corresponds to the 68\% C.L. DAMA/LIBRA result uncertainty.


The incompatibility with the DAMA/LIBRA result and sensitivity are reduced compared to the search performed with a constant QF\textsubscript{I} of 0.06, 4.2$\sigma$ and 4.4$\sigma$ respectively. However, this effect may not be attributed to the use of the improved background model, given that all previously analyzed energy regions reported modulation amplitudes compatible with those from the previous model. This reduction in both the incompatibility and sensitivity may stem from the fact that the signal search window is significantly larger in terms of the electron equivalent energy but the same in NR energy. For the energy-dependent QF\textsubscript{I}, the window extends from [1-4.8] keV (3.8 keV wide); for the constant QF\textsubscript{I}, it spans from [1.3-4] keV (2.7 keV wide). Since the expected DM signal remains unchanged, the larger search window when using the energy-dependent QF\textsubscript{I} leads to a higher background, thereby reducing the statistical significance of a potential DM induced modulation signal. Overall, the fit performs well using the energy-dependent QF\textsubscript{I}, and the results indicate incompatibility with DAMA/LIBRA.



\begin{figure}[t!]
    \centering
    {\includegraphics[width=0.8\textwidth]{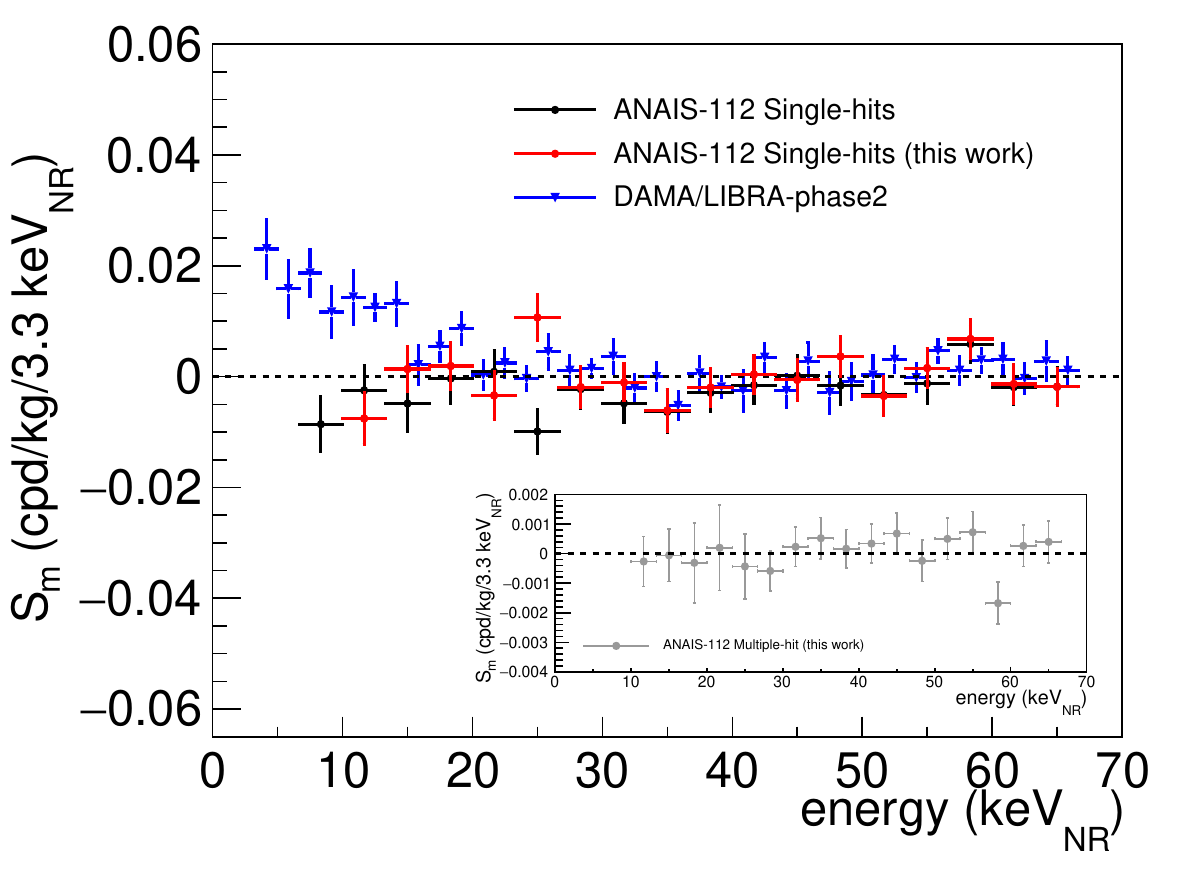}}

    \caption{\label{smNRNa} Modulation amplitude obtained in this work in 3.3 keV\textsubscript{NR} energy bins in the energy scale of sodium nuclear recoils for single-hit events (red dots), compared with the DAMA/LIBRA-phase2 result\cite{Bernabei:2020mon} (blue triangles), and the results from the previous background model \cite{amare2025towards} (black dots). Inset: modulation amplitude obtained in this work in 3.3 keV\textsubscript{NR} energy bins in the energy scale of sodium nuclear recoils for multiple-hit events. }
\end{figure}

Figure~\ref{smNRNa} shows the modulation amplitude on the energy scale
of sodium nuclear recoils, for 3.3~keV\textsubscript{NR} energy bins, for single-hit events, combining data from the nine ANAIS–112 modules. For comparison, results obtained with the previous background model of ANAIS \cite{amare2025towards}, and those from DAMA/LIBRA-phase 2 \cite{Bernabei:2020mon} are also shown. The corresponding distribution obtained in this work for multiple-hit events is presented in the same figure as an inset. These figures assume a constant QF\textsubscript{Na} of 0.3 for DAMA/LIBRA, a constant QF\textsubscript{Na} of 0.2 for the previous ANAIS-112 results, and the energy-dependent ANAIS(1) QF\textsubscript{Na} for the present analysis. For each DAMA/LIBRA NR energy bin, ANAIS(1) QF\textsubscript{Na} is evaluated to determine the electron-equivalent energy interval in ANAIS–112 data, in which the annual modulation search is then performed.


Due to the rapid decrease of ANAIS(1) QF\textsubscript{Na} at low recoil energies, the first 
bin from DAMA/LIBRA above the ANAIS-112 experimental threshold that can be reliably explored is [3–4] keV, which corresponds to [1.38–1.93]~keV in ANAIS–112 considering ANAIS(1) QF\textsubscript{Na}. This implies that using the energy-dependent QF\textsubscript{Na} from ANAIS(1) results in one fewer energy bin being explored compared to using a constant QF\textsubscript{Na} of~0.2.

The ANAIS fit results in the [10–20] keV\textsubscript{NR} region are compatible with zero modulation. Specifically, the modulation amplitudes yield a $\chi^2$/ndf = 2.95/3 (p-value = 0.400) for the null hypothesis, and $\chi^2$/ndf = 14.90/3 (p-value = 1.90 × 10$^{-3}$) for the DAMA/LIBRA modulation hypothesis, strongly supporting the statistical incompatibility between the two experimental results.


The only notable deviation between both results occurs at 25 keV\textsubscript{NR}, which corresponds to [4.7–5.3] keV for a QF\textsubscript{Na} of 0.2 and to [7–8] keV for the ANAIS(1) QF\textsubscript{Na}. In this energy bin, a positive amplitude incompatible with zero at the 1$\sigma$ level is observed in this work. The origin of this deviation remains unclear. Interestingly, in the previous ANAIS–112 results \cite{amare2025towards}, the same NR bin exhibited the most negative modulation amplitude, also incompatible with zero at the 1$\sigma$ level. Nonetheless, as previously mentioned, the energy regions explored in \cite{amare2025towards} and in this work are not the same due to differing QF\textsubscript{Na} assumptions, making differences in the results plausible.

 \begin{figure}[t!]
    \centering
    {\includegraphics[width=0.8\textwidth]{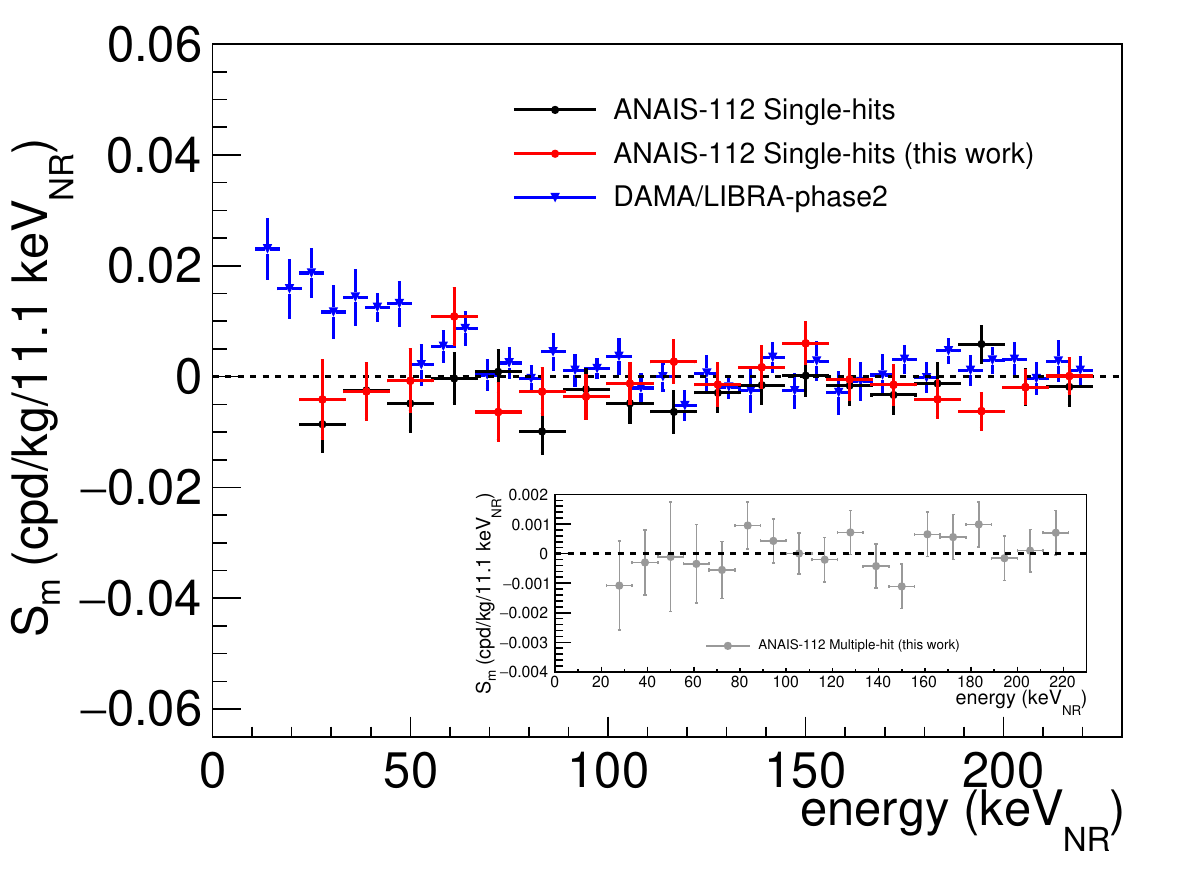}}
    \caption{\label{SmNRI}  Modulation amplitude obtained in this work in 11.1 keV\textsubscript{NR} energy bins, in the energy scale of iodine nuclear recoils for single-hit events (red dots), compared with the DAMA/LIBRA-phase2 result\cite{Bernabei:2020mon} (blue triangles), and the results from the previous background model \cite{amare2025towards} (black dots). Inset: modulation amplitude obtained in this work in 11.1 keV\textsubscript{NR} energy bins in the energy scale of iodine nuclear recoils for multiple-hit events.   }
\end{figure}

Analogously, the same analysis can be performed on the iodine nuclear recoil scale. Figure~\ref{SmNRI} shows the modulation amplitude on this scale, using 11.1 keV\textsubscript{NR} energy bins for single-hit events, combining data from the nine ANAIS–112 modules. In the same figure, the distribution corresponding to multiple-hit events from this work is displayed as an inset. In the case of DAMA/LIBRA, a constant QF\textsubscript{I} value of 0.09 is assumed. For the previous ANAIS-112 result based on the previous background model \cite{amare2025towards}, a constant QF\textsubscript{I} of 0.06 is adopted. In contrast, for the present analysis with the ANAIS-112 detectors, the energy-dependent QF\textsubscript{I} is used.


In the case of the iodine recoil scale, the first DAMA/LIBRA bin above the ANAIS-112 experimental threshold that can be explored when considering an energy-dependent QF\textsubscript{I} is [2–3] keV, which corresponds to [1-1.72]~keV in ANAIS–112. The modulation amplitudes in the [22.2–66.7]~keV\textsubscript{NR} region yield a $\chi^2$/ndf = 5.16/4 (p-value = 0.27) for the null hypothesis, and $\chi^2$/ndf = 15.35/4 (p-value = 0.004) for the DAMA/LIBRA modulation hypothesis, supporting compatibility with zero modulation. 


The annual modulation analysis using six years of ANAIS-112 data has been revisited. Although the background model developed in this thesis provides a significantly improved description of the ANAIS-112 background, its temporal evolution shape remains similar to that of the previous model, which already offered a reliable characterization. As a result, the fit outcomes are consistent with those reported in \cite{amare2025towards}, showing improved performance in the [1–6] keV region. The analysis, based on the energy-dependent QF\textsubscript{Na} and QF\textsubscript{I} values selected in Chapter \ref{Chapter:QF} that more accurately describe the ANAIS-112 detector response, remains compatible with the null hypothesis. However, due to current detector performance, the analysis is restricted to the [3–6] keV range, thus covering only part of the DAMA/LIBRA signal region. This limitation highlights the need for future experimental efforts, such as ANAIS+, which aims to lower the energy threshold below 0.5 keV in NaI detectors by replacing PMTs with SiPMs for light collection.

\section{Solar axion search in ANAIS-112} \label{axions}

The Sun is expected to be a potential source of axions, whose detection could be investigated on Earth. When these axions, referred to as solar axions, interact with a strong magnetic field on Earth, they can be converted into X-ray photons by the Primakoff effect, a process exploited by helioscope experiments such as CAST \cite{CAST:2017uph} and its successor IAXO \cite{IAXO:2019mpb}. However, solar axions can also be searched for using experiments originally designed for WIMP detection, such as ANAIS-112, offering additional insights. In fact, direct detection experiments dedicated to WIMP searches, such as XENONnT~\cite{aprile2022search}, PANDAX-4T~\cite{zeng2025exploring}, and LZ \cite{aalbers2023search}, have already reported highly competitive axion limits. 

This section is devoted to the search for solar axions with ANAIS-112, enabled by the improved background model developed in this thesis.


\subsection{Axions production and detection}


All axion and ALPs models (PQ, KSVZ, DFSZ,...) predict an axion-photon coupling, enabling the conversion of axions into photons in the presence of electromagnetic fields, and the decay into two photons, for instance. Thus, most of the experimental efforts devoted to axion search are focused on the specific coupling of the axion to photons \cite{irastorza2018new}. However, depending on the specific theoretical model, the axion may also exhibit couplings to other SM particles such as electrons or nucleons.

\begin{figure}[t!]
\begin{center}
\includegraphics[width=0.6\textwidth]{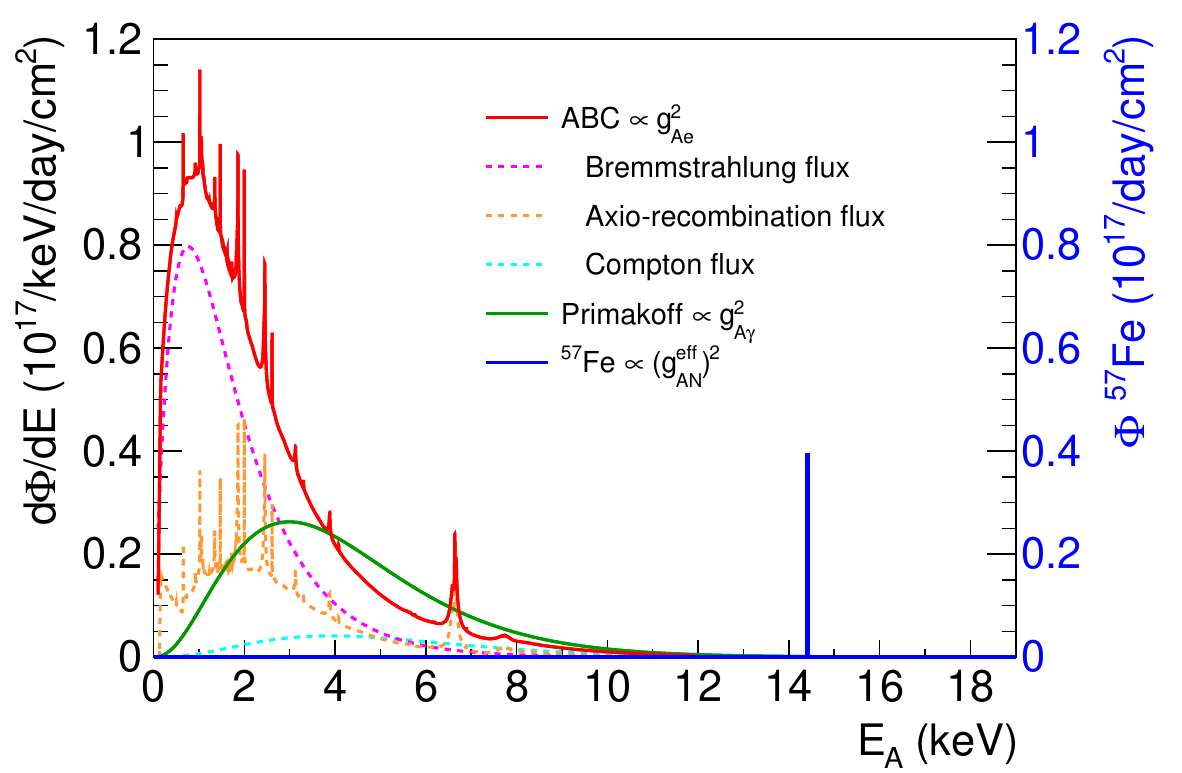}

\caption{\label{flux}Fluxes of solar axions from ABC processes (red), the Primakoff effect (green),  and the \textsuperscript{57}Fe transition (blue), assuming couplings of \( g_{\text{Ae}} = 3.5 \times 10^{-12} \),  \( g_{\text{A}\gamma} = 2 \times 10^{-10} \, \text{GeV}^{-1} \), and \( g_{\text{AN}}^{\text{eff}} = 1 \times 10^{-6}  \), respectively. The different contributions of the ABC flux are represented by dashed lines. The right axis applies exclusively to the contribution from the \textsuperscript{57}Fe transition. Each flux scales with the square of its corresponding coupling. }
\end{center}
\end{figure}

\begin{itemize}
    \item \textbf{Axion production via electron coupling (g$_{Ae}$)}: If axions couple to electrons, they can be generated through several electron-mediated processes, collectively known as ABC reactions~\cite{redondo2013solar}. These include atomic de-excitation and recombination, bremsstrahlung, and Compton scattering, where the role of the photon is assumed by the axion. 
    
    Among these mechanisms, bremsstrahlung is the primary contributor to the solar ABC axion flux. During an electron interaction with an ion or another atomic electron, an axion can be emitted instead of a photon. This phenomenon is particularly relevant given the high ionization state of hydrogen and helium in the Sun, which is a potential strong emitter of axions/ALPs. 

The ABC 
flux scales with the axion-electron
coupling g$_{Ae}$ as $\phi^{ABC}_A \propto g_{Ae}^2$. Its spectral shape, characterized by a complex energy dependence, has been adopted from \cite{redondo2013solar} and is presented in Figure~\ref{flux}, together with its different contributions.\\

    \item \textbf{Axion production via photon coupling (g$_{A\gamma}$)}: If axions couple to photons, they should be produced in the Sun via the Primakoff effect, whereby solar photons convert into axions in the presence of the solar magnetic field. Accordingly, major dedicated axion experiments seek to detect photons generated through the inverse axion-to-photon conversion, known as the inverse Primakoff effect, under strong magnetic fields. Although this objective is shared, the experimental techniques and detection strategies employed can differ significantly \cite{irastorza2018new}.

    The Primakoff axion solar flux is given by \cite{kuster2007axions}:

    \begin{equation}
    \frac{d\Phi_A^{\text{Prim}}}{dE_A} = \left( \frac{g_{A\gamma}}{\text{GeV}^{-1}} \right)^2 
    \left( \frac{E_A}{\text{keV}} \right)^{2.481} 
    e^{-E_A / (1.205 \text{ keV})} 
    \times 6 \times 10^{30} \text{ cm}^{-2} \text{s}^{-1} \text{keV}^{-1},
\end{equation}

where \( E_A \) denotes the axion energy and \( g_{A\gamma} \) is the axion–photon coupling constant.\\

    \item \textbf{Axion production via nuclear de-excitation (g$^{eff}_{AN}$)}: Axions may also be emitted from nuclear transitions instead of a photon/gamma ray. The 14.4 keV M1 transition of \textsuperscript{57}Fe is of particular interest due to the relatively high abundance of Fe in the Sun core, estimated at $(9.0 \pm 1.2) \times 10^{19} \text{ cm}^{-3}$, corresponding to a 2.12\% mass fraction \cite{andriamonje2009search}. 

    The \textsuperscript{57}Fe solar axion flux is given as:
    \begin{equation}
    \Phi_A^{57Fe} = \left( \frac{k_A}{k_\gamma} \right)^3 
    \times 4.56 \times 10^{23} \left( g_{AN}^{\text{eff}} \right)^2 
    \text{ cm}^{-2} \text{s}^{-1},
    \end{equation}
where \( k_A \) and \( k_\gamma \) represent the momenta of the produced axions and photons, respectively, given by \( k_A = \sqrt{E^2 - m_A^2 c^4}/c \) and \( k_\gamma = E/c \). The effective axion-nucleon coupling, \( g_{AN}^{\text{eff}} \), is defined as:  

\begin{equation}
    g_{AN}^{\text{eff}} = g_{AN}^{(3)} - 1.19 \, g_{AN}^{(0)}
\end{equation}  

where \( g_{AN}^{(0)} \) and \( g_{AN}^{(3)} \) denote the isoscalar and isovector coupling constants, respectively, and the coefficient 1.19 arises from the specific nuclear structure of \({}^{57}\)Fe. It is important to note that the definition of \( g_{AN}^{\text{eff}} \) is isotope-dependent and applies specifically to axions emitted by \({}^{57}\)Fe.

The \textsuperscript{57}Fe solar axion emission
spectrum is a gaussian centered at
the transition energy, 14.4 keV. Owing to Doppler broadening caused by the thermal motion of \textsuperscript{57}Fe nuclei in the hot solar interior, the standard deviation of the gaussian is taken to be $\sigma~\sim$~2~eV~\cite{andriamonje2009search}.

\end{itemize}

Thus, each component of the flux scales with the square of the corresponding coupling, $\Phi_A^{ABC} \propto g^2_{Ae}$; $\Phi_A^{Prim} \propto g^2_{A\gamma}$; $\Phi_A^{57Fe} \propto (g_{AN}^{\text{eff}})^2$. Figure~\ref{flux} displays the differential flux of solar axions originating from ABC processes and the Primakoff effect, computed assuming couplings of \( g_{\text{Ae}} = 3.5 \times 10^{-12} \), \( g_{\text{A}\gamma} = 2 \times 10^{-10} \, \text{GeV}^{-1} \) and \( g_{\text{AN}}^{\text{eff}} = 1 \times 10^{-6}  \), respectively. 


Besides dedicated solar axion experiments, DM direct detection experiments can also search for axions by the axio-electric effect. In particular, NaI-based detectors like ANAIS-112 are sensitive to this process, as will be shown in this section. The axio-electric effect is analogous to the photoelectric effect, with an axion replacing the photon and being absorbed by an atom, releasing a bound atomic electron. This absorption transfers the axion’s full energy to the electron, causing atomic ionization and resulting in a measurable energy deposition in the detector.

\begin{figure}[t!]
\begin{center}
\includegraphics[width=0.49\textwidth]{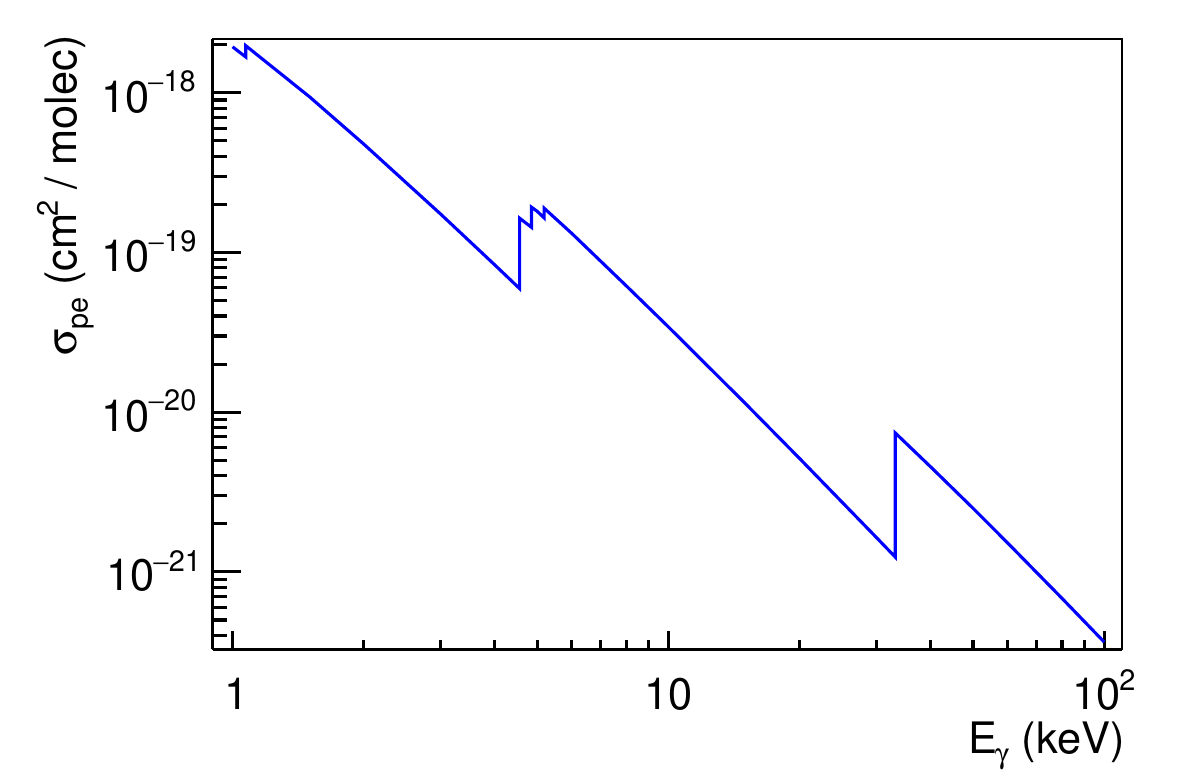}
\includegraphics[width=0.49\textwidth]{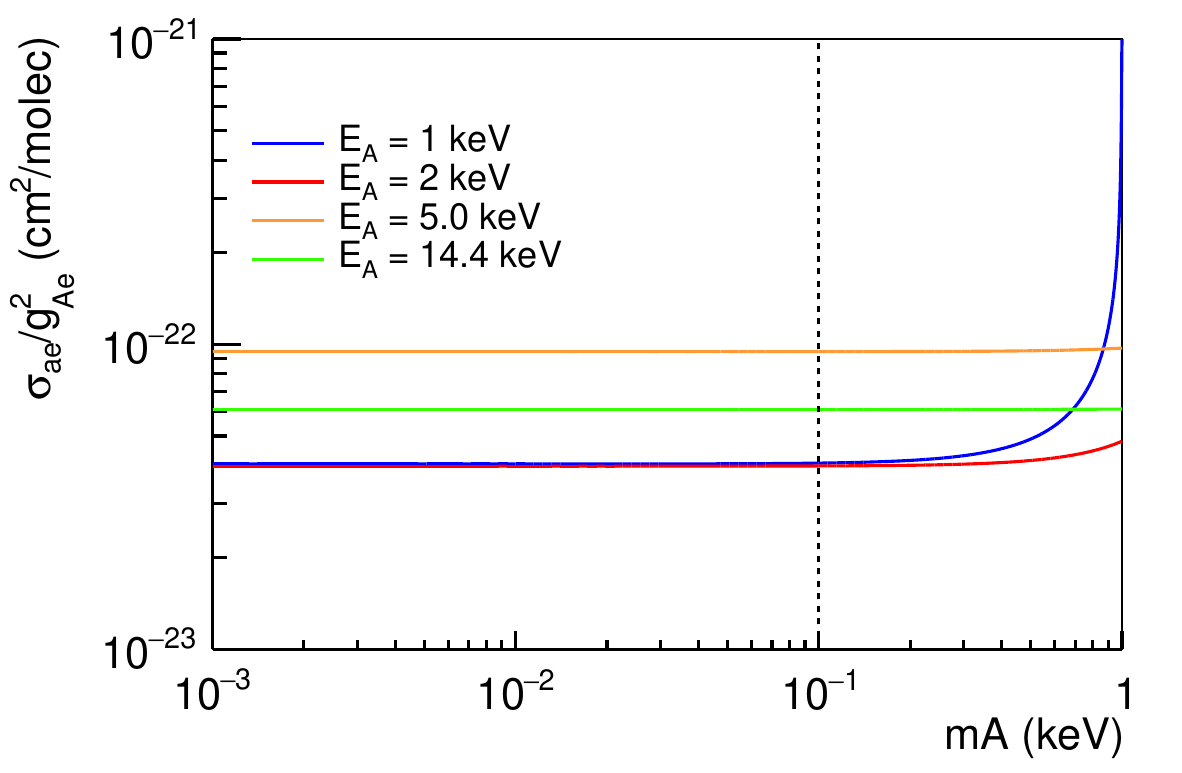}

\caption{\label{sigamae} \textbf{Left panel:} Photoelectric cross section $\sigma_{pe}$ for a NaI target. \textbf{Right panel:} Axio-electric cross section \( \sigma_{\text{Ae}}/g^2_{Ae} \) for a NaI target as a function of axion mass \( m_A \) for four representative energies for solar axions: 1 keV (blue), 2 keV (red), 5 keV (orange), and 14.4~keV (green), corresponding to typical solar axion energies. In all cases, \( \sigma_{\text{Ae}} \) remains constant for \( m_A \lesssim 100 \, \text{eV}/c^2 \).}
\end{center}
\end{figure}

Unlike WIMP-induced signals, which predominantly produce NRs, axions would generate ERs signals. The axio-electric cross-section \( \sigma_{\text{Ae}} \) is given by \cite{dimopoulos1986atomic,pospelov2008bosonic}:

\begin{equation}
    \sigma_{Ae}(E) = \sigma_{pe} \frac{g_{Ae}^2}{\beta} \frac{3E_A^2}{16\pi \alpha m_e^2} 
    \left( 1 - \frac{\beta^{2/3}}{3} \right) \text{ cm}^2/\text{molec}
    \label{eqsigmaae}
\end{equation}

where \( \sigma_{pe} \) is the photoelectric cross-section for a NaI target (see left panel of Figure~\ref{sigamae}), \( \beta = \frac{v_A}{c} = \sqrt{1 - \left(\frac{m_A c^2}{E_A}\right)^2} \) corresponds to the axion velocity in units of the speed of light, and \( E_A \) denotes the axion energy. Additionally, $\alpha=\frac{1}{137}$ represents the fine structure constant, and \( m_e \) is the electron mass. As can be derived from Equation~\ref{eqsigmaae}, \( \sigma_{\text{Ae}} \) exhibits a quadratic dependence on the axion-electron coupling, scaling as \( \sigma_{\text{Ae}} \propto g_{\text{Ae}}^2 \).

In addition, as shown in Equation \ref{eqsigmaae}, \( \sigma_{\text{Ae}} \) also depends on the axion mass \( m_A \) through the velocity parameter \( \beta \). In typical axion searches, the massless axion approximation is commonly adopted. To assess the validity of this assumption, right panel of Figure~\ref{sigamae} displays \( \sigma_{\text{Ae}}/g^2_{Ae} \) as a function of \( m_A \) for four representative axion energies for a NaI target. From this comparison, it is clear that \( \sigma_{Ae} \) remains effectively constant for \( m_A \lesssim 100 \, \text{eV}/c^2 \), making this a conservative upper limit for the validity of the massless approximation. 

The expected solar axion detection rate over a measurement period \( T \), and for a total detector mass \( M \), is given by:

\begin{equation}
    R(E) = \Phi_A (E) \times \sigma_{Ae}(E) \times MT \times F_{det}(E),
    \label{rateaxions}
\end{equation}

where \( F_{\text{det}}(E) \) is the detector response function, which accounts for both its energy resolution and efficiency.

\begin{figure}[t!]
\begin{center}
\includegraphics[width=0.5\textwidth]{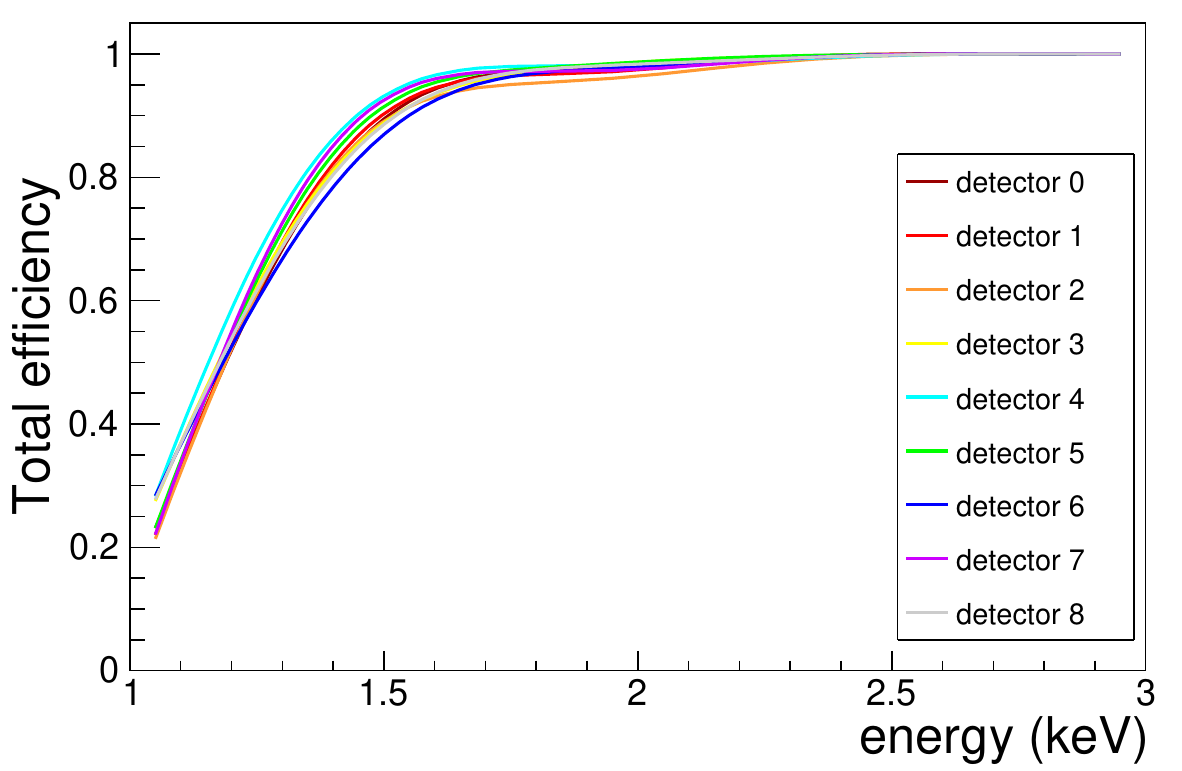}

\caption{\label{totaleff}Total detection efficiency as a function of energy for the ANAIS–112 modules. The efficiency curve results from the combination of the trigger efficiency and the efficiency of the BDT-based event selection.
}
\vspace{-0.5cm}
\end{center}
\end{figure}

Consequently, the solar axion signal predicted for an ideal NaI(Tl) detector can be transformed into the experimentally expected spectrum in ANAIS-112 by incorporating both energy resolution and detection efficiency effects. The ideal spectrum is first convolved with a gaussian function using the detector specific energy resolution shown in Figure~\ref{LEres}, accounting for intrinsic performance differences among the modules. Subsequently, the resulting spectrum is scaled by the total detection efficiency presented in Figure~\ref{totaleff}, which combines the trigger efficiency and the event selection efficiency obtained through BDT machine learning techniques.

As shown in Figure \ref{totaleff}, the total acceptance efficiency in ANAIS-112 decreases to $\sim$20\%–30\% at 1~keV, depending on the module, reaching around 80\% at 1.5 keV, and approaching $\sim$100\% above 2 keV. Other leading solar axion experiments such as XENONnT achieves an energy resolution of 9\% at 1 keV, an efficiency of $\sim$50\% at 1.5 keV, and maintains a constant $\sim$100\% efficiency above 3 keV \cite{aprile2022search}. Therefore, in the low-energy regime, where the axion flux is particularly relevant (see Figure \ref{flux}), ANAIS-112 demonstrates a notably superior detection efficiency.

\begin{figure}[b!]
\begin{center}
\includegraphics[width=0.6\textwidth]{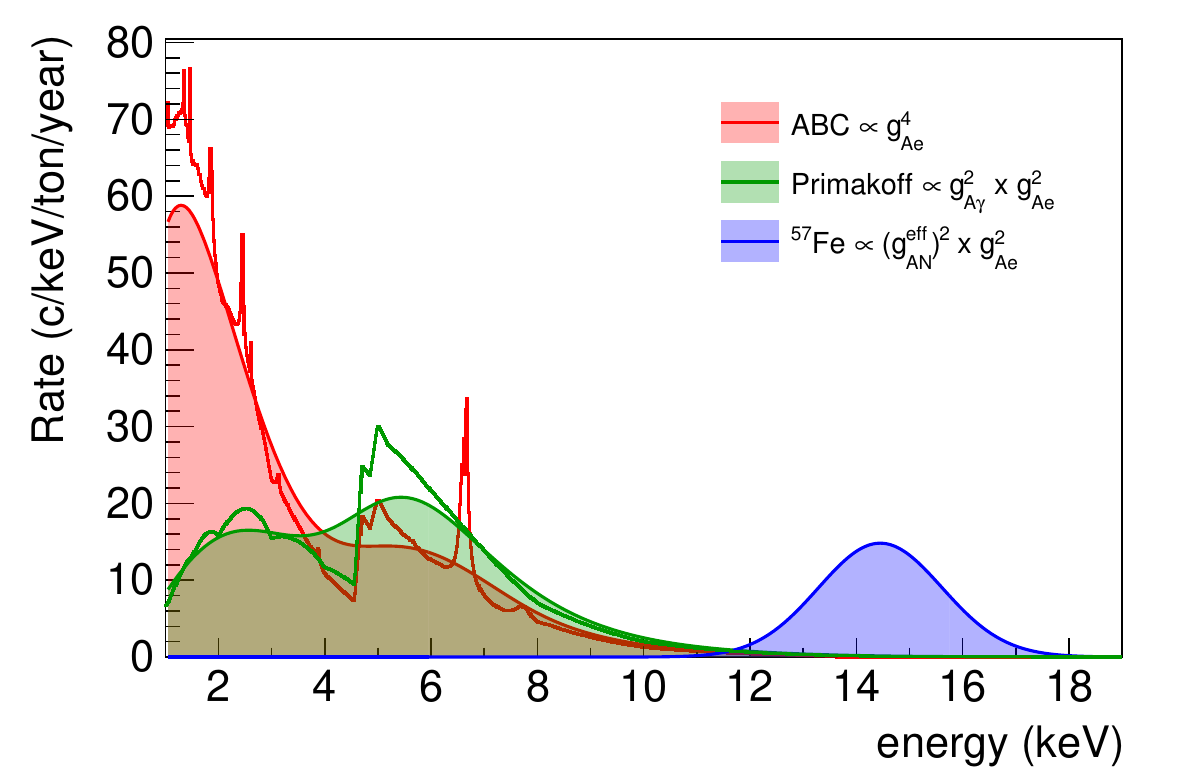}

\caption{\label{rateANAIS} Expected solar axion detection rates in ANAIS-112 before (unshaded) and after (shaded) accounting for the detector energy resolution and efficiency. The ABC spectrum is represented in red, the Primakoff component in green, and the \(^{57}\)Fe contribution in blue. The plots are generated assuming coupling constants of \( g_{\text{Ae}} = 3.5 \times 10^{-12} \), \( g_{\text{A}\gamma} = 2 \times 10^{-10} \, \text{GeV}^{-1} \) and  \( g_{\text{AN}}^{\text{eff}} = 1 \times 10^{-6}  \). The expected detection rates for ABC, Primakoff, and \(^{57}\)Fe axions scale as \( g_{\text{Ae}}^4 \), \( g_{\text{A}\gamma}^2 \times g_{\text{Ae}}^2 \), and \( (g_{\text{AN}}^{\text{eff}})^2 \times g_{\text{Ae}}^2 \), respectively.
}
\end{center}
\end{figure}

Figure~\ref{rateANAIS} displays the expected solar axion detection rates in ANAIS-112 for the three axion components, shown both before (unshaded) and after (shaded) applying the detector energy resolution and total efficiency. 

The sharp rise observed around 5~keV in both the ABC and Primakoff spectra arises from the increase observed in the NaI photoelectric cross-section (see right panel of Figure \ref{sigamae}), leading to a double peak structure, particularly pronounced in the Primakoff component. In the case of ABC axions, the sharp features present in the solar axion spectrum, originating from atomic de-excitation and recombination processes, are significantly smeared by the detector finite energy resolution. 


As displayed in the figure, although the expected event rate in the detector strongly depends on the assumed coupling constants, which are somewhat arbitrary and entirely model-dependent, the sensitivity to ABC axions is enhanced at low energies, particularly below $\sim$ 4 keV. The Primakoff signal is expected to be observable up to about 10 keV, while the \textsuperscript{57}Fe axion signal is expected to appear around 14.4~keV, following a gaussian energy distribution.



Since the flux of each axion component scales with the square of its respective coupling constant, and \( \sigma_{\text{Ae}} \) scales as \( g_{\text{Ae}}^2 \), the resulting detection rates in solar axion searches exhibit the following parametric dependencies: \( \Phi_A^{\text{ABC}} \propto g_{\text{Ae}}^4 \), \( \Phi_A^{\text{Prim}} \propto g_{A\gamma}^2 g_{\text{Ae}}^2 \), and \( \Phi_A^{^{57}\text{Fe}} \propto (g_{\text{AN}}^{\text{eff}})^2 g_{\text{Ae}}^2 \). Consequently, for the Primakoff and \(^{57}\)Fe contributions, experimental sensitivities are inherently limited to the product of the two couplings.


\subsection{Fitting procedure}

Given the coupling dependencies discussed above, it is important to emphasize that, in principle, axions arising from all three interaction channels may appear simultaneously, as their only interplay within ANAIS-112 is through their detection mechanism, the axio-electric effect common to all three. Consequently, the methodology adopted in this chapter consists of fitting the  ANAIS-112 single-hit accumulated rate, corresponding to six years of exposure (0.63 ton$\times$year), under different hypotheses.

The null hypothesis (considering only known background components) has already been addressed in Chapter \ref{Chapter:bkg}. Here, however, the axion hypothesis, considering the possibility of a hypothetical axion signal present in the ANAIS-112 data, will be explored. To avoid biasing the fit, not only the axion contribution, but also those background components fitted in Chapter \ref{Chapter:bkg} using data restricted to the same energy region of interest for the axion signal, must be treated as nuisance parameters in the fit. These background contributions include $^{3}$H, $^{210}$Pb in the teflon, and the other external component (see Section \ref{lowenergyfit}). 

Regarding the axion signal, the simultaneous contribution of the three axion flux components (ABC, Primakoff, and $^{57}$Fe), as well as each component contribution separately, will be considered in the fits to assess whether any of them yields a statistically significant best fit over the others and to identify correlations among them. It should be noted that the background-only fit was performed using only year 6 data, whereas this analysis, which includes axions, will be carried out using the full six-year exposure of ANAIS-112, being the latter more dependent on the modelling of the cosmogenic contributions.

The fitting region is defined in the energy interval [3–20] keV, extending the lower bound with respect to the [6–20] keV range employed in the background model developed earlier in this thesis. As discussed in the previous chapter, there is still some discrepancy between the data and the background model in the  [1-3] keV range, even after incorporating the anomalous population identified by the ANOD DAQ into the revised background model. Including this region in the fit introduces an excess in the data that could be misinterpreted as an axion-like signal, especially affecting the ABC component. To mitigate this risk and avoid biasing the results, the lower bound of the fitting region is conservatively set at 3 keV, at the cost of reduced sensitivity to low-energy axions.


In addition to the known background contributions, which are modelled independently for each detector and are uncorrelated, the axion signal, if present, is expected to be common across all detectors for each of the three production mechanisms. This condition imposes the need for a simultaneous fit across all nine ANAIS-112 detectors. Consequently, when considering the simultaneous presence of all three axion components (ABC, Primakoff, and \(^{57}\)Fe), the total number of free parameters in the fit rises to 30: 9 associated with \(^{3}\)H activity in the crystals, 9 with \(^{210}\)Pb in the teflon layer surronding the crystals, 9 with the other external backgrounds, and 3 corresponding to the axion signal, one for each axion component. If only one axion component is considered in a given hypothesis, the number of free parameters is reduced to 28 accordingly.

\begin{table}[b!]

\centering

    \begin{minipage}{0.9\textwidth}
    
    \centering
    \resizebox{\textwidth}{!}{\Large
   
    \begin{tabular}{c|c|c|c}
\hline        
\multicolumn{4}{c}{Fitted number of events in [3–20] keV (counts)} \\
\hline
Axion hypothesis & N\textsubscript{ABC} & N\textsubscript{Primakoff} & N\textsubscript{\textsuperscript{57}Fe} \\
\hline
Bkg + ABC + Primakoff + Fe & $8466 \pm 866$ & -$14962 \pm 1020$ & $723 \pm 194$ \\
Bkg + ABC & -$208 \pm 649$ & - & - \\
Bkg + Primakoff & - & -$7799 \pm 717$ & - \\
Bkg + Fe & - & - & $669 \pm 193$ \\
\hline
\end{tabular}}
       \end{minipage}

   \caption{\label{cuentashipotesis} Fitted number of events for the axion signals in the [3-20] keV region. The results correspond to the fitted number of signal counts for the ABC component (N$_\text{ABC}$), the Primakoff component (N$_\text{Primakoff}$), and the 14.4 keV M1 transition of \textsuperscript{57}Fe (N\textsubscript{$^{57}$Fe}), under different axion hypotheses: background plus all three axion components simultaneously, and background plus each axion component individually.}

\end{table}

It is important to note that the fit in this work is performed using the extended likelihood framework of RooFit \cite{verkerke2006roofit}, as done for the background model fit (Chapter~\ref{Chapter:bkg}), yielding results directly in terms of the number of signal counts. In the absence of an axion signal, the fit is expected to return values consistent with zero for each axion production channel. Any significant deviation could indicate either a possible axion detection or a correlation with background components, as discussed later. To prevent bias, the number of axion counts must be allowed to vary freely, including negative values. Constraining this parameter to be positive can lead to artificially stringent limits that are not physically meaningful given ANAIS-112 background levels.

In the absence of a statistically significant excess, upper limits on the signal, $N_{\text{up}}$, will be derived at 90\% C.L. To derive $N_{\text{up}}$, it is assumed that the number of axion-induced counts follows a gaussian distribution centered at the best-fit value $N_{\text{best}}$, with a standard deviation $\sigma_{N_{\text{best}}}$ given by its statistical uncertainty, as reported in Table~\ref{cuentashipotesis}. The upper limit is defined such that 90\% of the integral of this distribution, restricted to the physical (positive) domain, lies below $N_{\text{up}}$. Consequently, assuming the experiment were repeated many times, 90\% of the fitted values would be expected to fall below $N_{\text{up}}$ for each axion production channel This allows the derivation of upper limits on the axion-induced signal.

Corresponding constraints can be set on the relevant coupling constants (or their products) for each axion production mechanism, according to the following expression: 

\begin{equation}
    N_{\text{up}} \geqslant \Phi_A \; \sigma_{\text{Ae}} \; M \; T
\end{equation}

The product $\Phi_A \times \sigma_{\text{Ae}} \times M \times T$ is obtained by integrating the expected solar axion detection rates shown in Figure~\ref{rateANAIS} over the fitting range $[3, 20]$ keV\footnote{Note that the solar axion detection rates shown in Figure~\ref{rateANAIS} have been corrected for the event selection efficiency. When comparing with the ANAIS-112 data,  which are already corrected by the event selection efficiency, no additional efficiency correction should be applied, in order to avoid double correction and a reduction in the axion expected signal.}. In this calculation, the coupling constants assumed for generating the spectra are factored out explicitly, since upper limits on these parameters will be derived from the fit. Therefore, the resulting integrals represent the expected number of axion signal counts per unit of the relevant combination of coupling constants, allowing the equation to be solved for the couplings in order to derive corresponding limits. Accordingly, for the ABC component, the upper limit will be derived on the fourth root of g\textsubscript{Ae}; for the Primakoff component, on the square root of the product $(g_{\text{A}\gamma} \times g_{Ae})$; and for the \textsuperscript{57}Fe component, on the square root of the product $(g_{\text{AN}}^{\text{eff}} \times g_{Ae})$.

Table \ref{cuentashipotesis} summarizes the fitted number of events of axion-related parameters for each hypothesis considered in this work: background with all three axion channels simultaneously, and background with each axion channel independently. 

When all three axion signals are included simultaneously, a clear anticorrelation between the ABC and Primakoff components is observed: the ABC channel becomes strongly positive while the Primakoff becomes strongly negative. This behavior is reasonable, since, as seen in Figure \ref{rateANAIS}, both signals have very similar shapes in the [3–20] keV range, motivating the analysis of each component separately. 

When fitting only the ABC component, the result is a negative value compatible with zero. However, the Primakoff component continues to yield a negative value when analyzed independently, which may indicate that this component is anticorrelated with one of the background components that are left free in the fit. Consequently, no meaningful upper limit can be derived for the Primakoff channel in any scenario, neither from the global analysis nor from the independent fit. The \textsuperscript{57}Fe transition shows a positive value compatible with zero at 3.6$\sigma$.

\begin{figure}[b!]
\begin{center}
\includegraphics[width=1\textwidth]{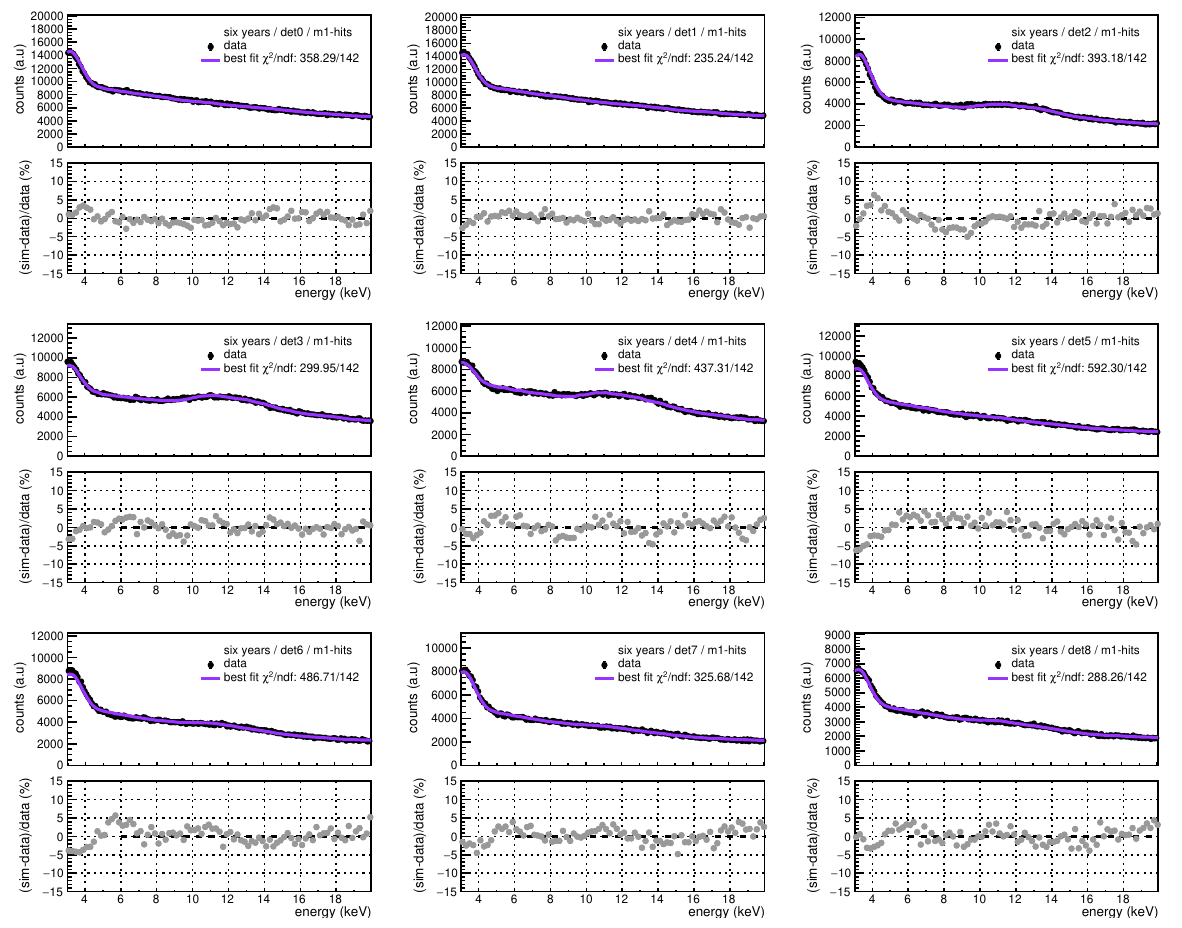}

\caption{\label{axionfit} Low-energy best fits of the single-hit spectra for the nine ANAIS-112 detectors. Black points represent data collected over the full six-year exposure of ANAIS-112 (0.63~ton$\times$year), while the best-fit, which includes both the background and the ABC axion signal contribution, is shown in violet. Residuals are displayed as grey points below each spectrum, and the goodness-of-fit is indicated in each panel.
}
\vspace{-0.5cm}
\end{center}
\end{figure}

As an example, Figure~\ref{axionfit} displays the best fit results under the background plus ABC axion hypothesis. The model provides an excellent description of the data, with residuals remaining below 5\%. Comparable fit quality and $\chi^2/\text{ndf}$ values are obtained for the other fit hypotheses considered in this study.


\begin{figure}[t!]
\begin{center}

\includegraphics[width=0.6\textwidth]{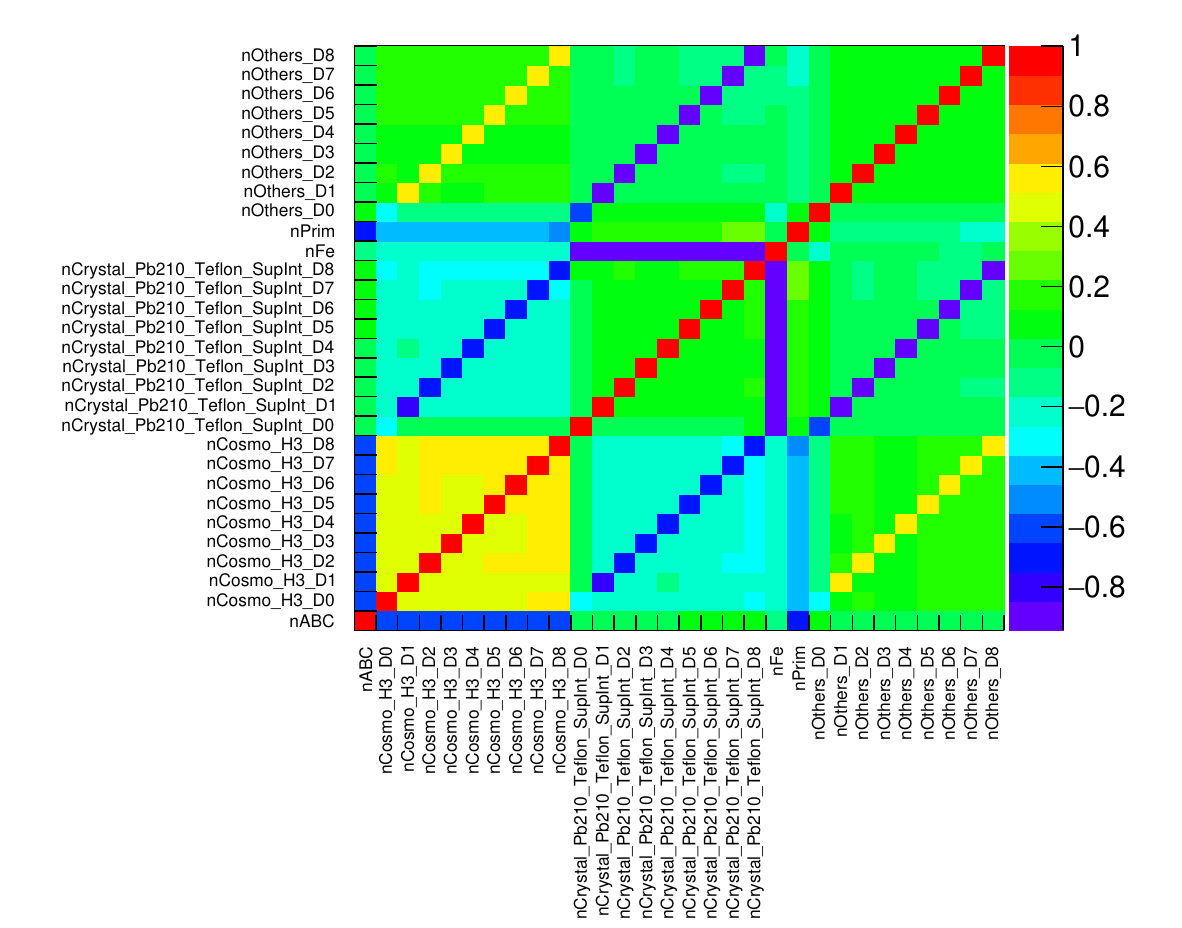}

\includegraphics[width=0.6\textwidth]{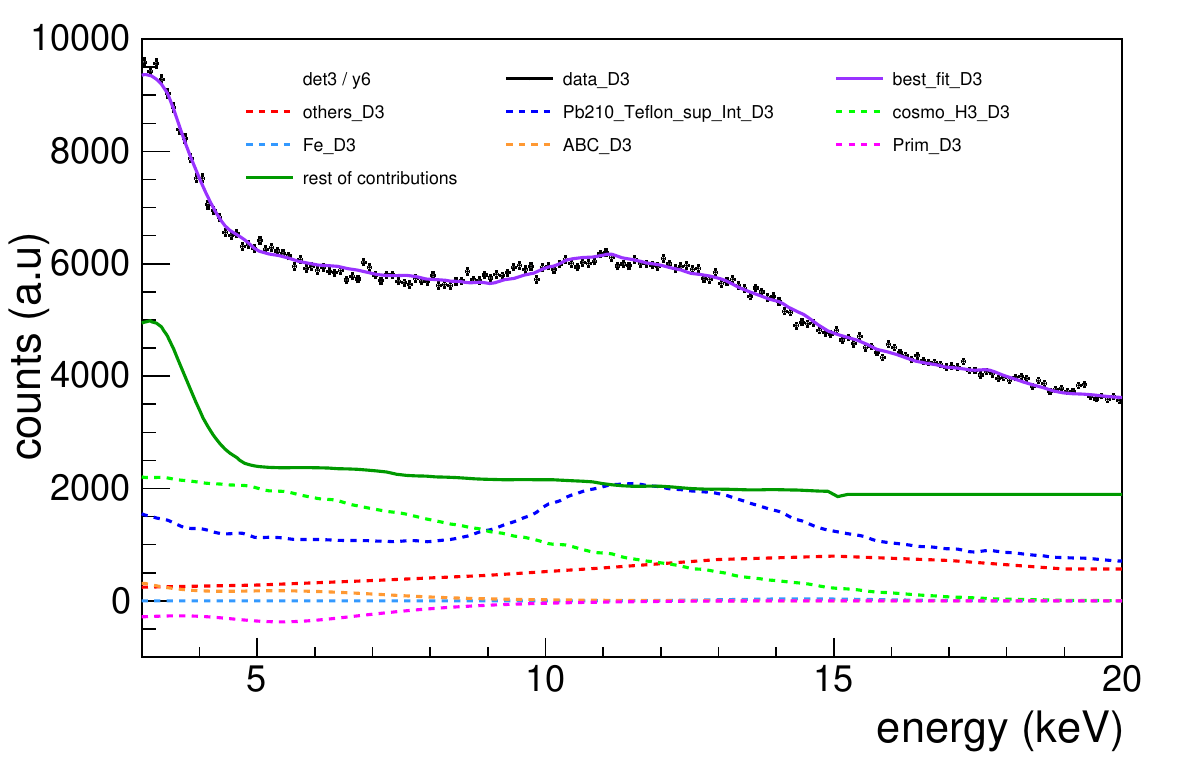}

\caption{\label{componentesaxions} Performance of the fit under the background-plus-axion hypothesis, including the three axion flux contributions. \textbf{Top panel:} Correlation matrix from the fit. \textbf{Bottom panel:} Comparison between data (black) and the best fit (violet) for detector D3. The figure displays the components fitted in this energy range with dashed-lines: $^{3}$H (light green), $^{210}$Pb in the teflon (blue), the other external component (red), ABC axions (orange), Primakoff axions (magenta), and the \textsuperscript{57}Fe M1 transition axions (light blue), together with the sum of rest of contributions of the background model (dark green), which are fixed in this fit as they are determined in other energy regions. }
\end{center}
\vspace{-0.7cm}
\end{figure}


The top panel of Figure~\ref{componentesaxions} shows the correlation matrix of the fit. The anticorrelations arising from the fit are clearly visible. Notably, the ABC component shows a strong anticorrelation with \textsuperscript{3}H, while external components appear to correlate with \textsuperscript{3}H. The observed correlations among the \textsuperscript{3}H contributions from all detectors are less intuitive but can be attributed to the simultaneous nature of the fit: adjusting the axion signal requires modifying the \textsuperscript{3}H content individually in each detector. In addition, the Primakoff component exhibits an anticorrelation with ABC and with \textsuperscript{3}H. Moreover, a particularly strong anticorrelation is observed between the \textsuperscript{57}Fe contribution and the \textsuperscript{210}Pb contamination in the teflon coating surrounding the crystals, which also exhibits an anticorrelation with \textsuperscript{3}H, although weaker than with other axion components.

This latter correlation is explicitly illustrated in the bottom panel of Figure~\ref{componentesaxions}, which shows the breakdown of the fitted contributions for detector D3 under the background plus three axion hypothesis. The figure highlights that the 14.4 keV M1 transition from \textsuperscript{57}Fe lies in a particularly challenging region of the ANAIS-112 background spectrum, near the $\sim$12~keV X-ray peak arising from $^{210}$Pb contamination in the teflon reflector. This important contamination complicates the determination of the \textsuperscript{57}Fe signal. Tests excluding detectors with the highest teflon $^{210}$Pb levels (specifically D2, D3, and D4) yielded similar results for this axion channel. This suggests that it is the spectral shape of the teflon $^{210}$Pb contribution, rather than its absolute contamination level, that introduces a systematic effect, biasing the estimation of the \textsuperscript{57}Fe contribution. The figure also supports the pronounced degeneracy among the axion components: the Primakoff contribution becomes strongly negative, while the ABC component shifts significantly positive, reflecting the spectral similarity of their expected signals within the fit range.

\begin{figure}[b!]
\begin{center}
\includegraphics[width=0.49\textwidth]{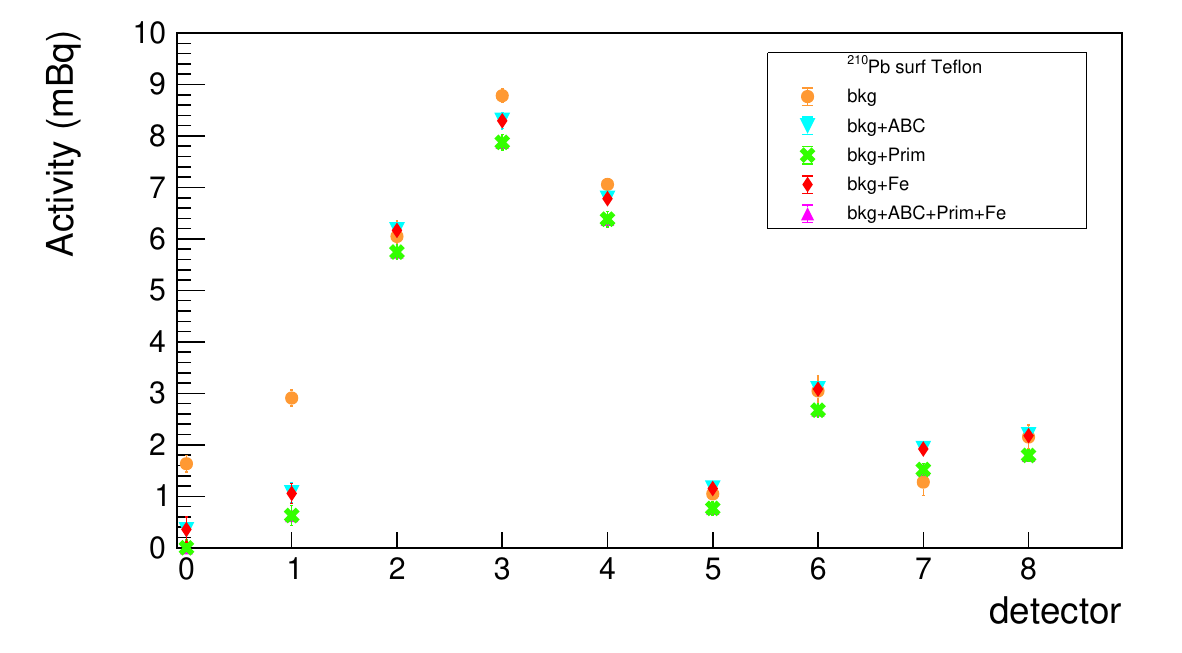}
\includegraphics[width=0.49\textwidth]{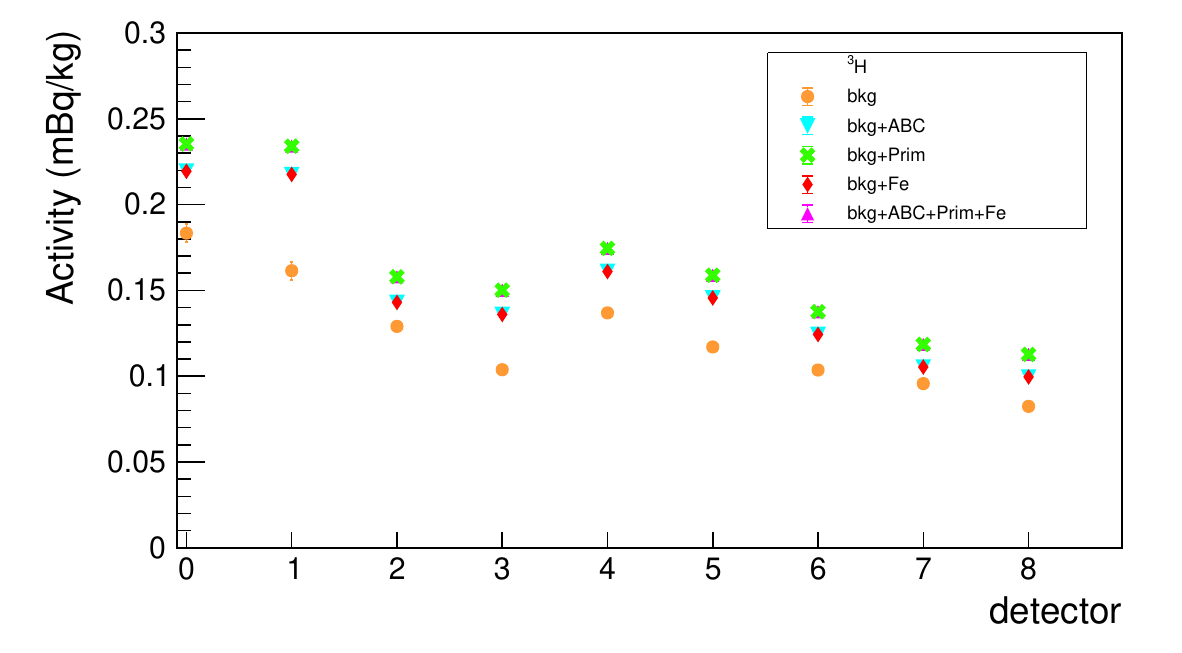}

\caption{\label{compareactivites} Comparison of the initial $^{210}$Pb contamination on the surface of the teflon film surrounding the crystals \textbf{(left panel)} and the cosmogenic $^{3}$H activities \textbf{(right panel)} at the time of detector movement underground. Results obtained considering the null hypothesis (background only) are shown in orange, and are compared to those obtained under the four axion hypothesis: three axion channels simultaneously (magenta), ABC (cyan), Primakoff (green), and \textsuperscript{57}Fe (red).}
\vspace{-0.5cm}
\end{center}
\end{figure}

To ensure the consistency of the fit results, it is also important to verify whether the fitted background contributions agree with those reported in Chapter \ref{Chapter:bkg} or there are strong discrepancies. This comparison is presented in Figure~\ref{compareactivites}, where the activities obtained under the null hypothesis from the previous chapter are compared to those from the four axion hypotheses considered in this section. Overall, the values are quite comparable and follow the same general trend. However, the $^{3}$H activity is found to be somewhat higher under the axion hypothesis, which is justified by the clear anticorrelation between $^{3}$H and all three axion components.



Given the fit performance under each axion hypothesis and the evident correlation between the axion components, this work derives exclusion limits by considering each axion component individually. Accordingly, the 90\% C.L. upper limits on the solar axion–electron coupling constant $g_{\text{Ae}}$ are reported below. In addition, the corresponding upper limit for the \textsuperscript{57}Fe axion component is also derived, assuming that the observed 3.6$\sigma$ excess remains statistically compatible with zero under the assumption that a physically meaningful positive signal is not supported.

This conclusion is further supported by the comparison of fit performance under the null and axion hypotheses, taking into account the contributions from both ABC and \textsuperscript{57}Fe axions. As shown in Table \ref{nullaxionhyp}, the goodness-of-fit is substantially better under the null hypothesis (i.e., background-only), which reinforces the argument that no axion signal is observed in the ANAIS-112 data. Moreover, neither the ABC component nor the \textsuperscript{57}Fe component appears to be preferred over the other, with both fits exhibiting similar performance behavior. 


\begin{table}[t!]
\centering
\begin{minipage}{0.7\textwidth}
\centering    
    \centering
    \resizebox{\textwidth}{!}{\Large
   
   \begin{tabular}{c|cc|cc}
   \hline
    \multirow{3}{*}{Detector} & \multicolumn{2}{c|}{\multirow{2}{*}{Null hypothesis}} & \multicolumn{2}{c}{Axion hypothesis} \\
    \cline{4-5}
    &   &  & ABC & \textsuperscript{57}Fe \\
    \cline{2-5}
    & $\chi^2$/ndf & p-value & $\chi^2$/ndf  & $\chi^2$/ndf  \\
    \hline
    0 & 168.08 / 137 & 0.037 & 358.29 / 142  & 370.92 / 142  \\
    1 & 132.64 / 137 & 0.589 & 235.24 / 142  & 235.96 / 142  \\
    2 & 112.74 / 137 & 0.936 & 393.18 / 142  & 399.56 / 142  \\
    3 & 172.93 / 137 & 0.020 & 299.95 / 142  & 304.07 / 142  \\
    4 & 160.07 / 137 & 0.087 & 437.31 / 142  & 434.68 / 142  \\
    5 & 157.73 / 137 & 0.109 & 592.30 / 142  & 585.15 / 142  \\
    6 & 163.05 / 137 & 0.064 & 486.71 / 142  & 471.08 / 142  \\
    7 & 145.55 / 137 & 0.292 & 325.68 / 142  & 308.05 / 142  \\
    8 & 139.91 / 137 & 0.415 & 288.26 / 142  & 274.99 / 142  \\
    \hline
\end{tabular}}
\end{minipage}
\caption{\label{nullaxionhyp} Performance of the low-energy fits conducted in this thesis. The $\chi^2$/ndf and the corresponding p-value for the null hypothesis (background only) and $\chi^2$/ndf for the axion hypothesis (background plus ABC axions, and background plus \textsuperscript{57}Fe axions). p-values for the axion hypotheses are omitted, as they are consistently zero.}
\end{table}
\subsection{Axion search results}

According to the background-plus-ABC-axion signal hypothesis, the upper limit on the axion–electron coupling constant $g_{\text{Ae}}$ derived from the six-year exposure of the ANAIS-112 experiment is found to be:

\begin{equation}
    g_{\textnormal{Ae}} < 7.40\times10^{-12}  \quad \text{at 90\% C.L.}
\end{equation}

\begin{figure}[t!]
\begin{center}
\includegraphics[width=0.65\textwidth]{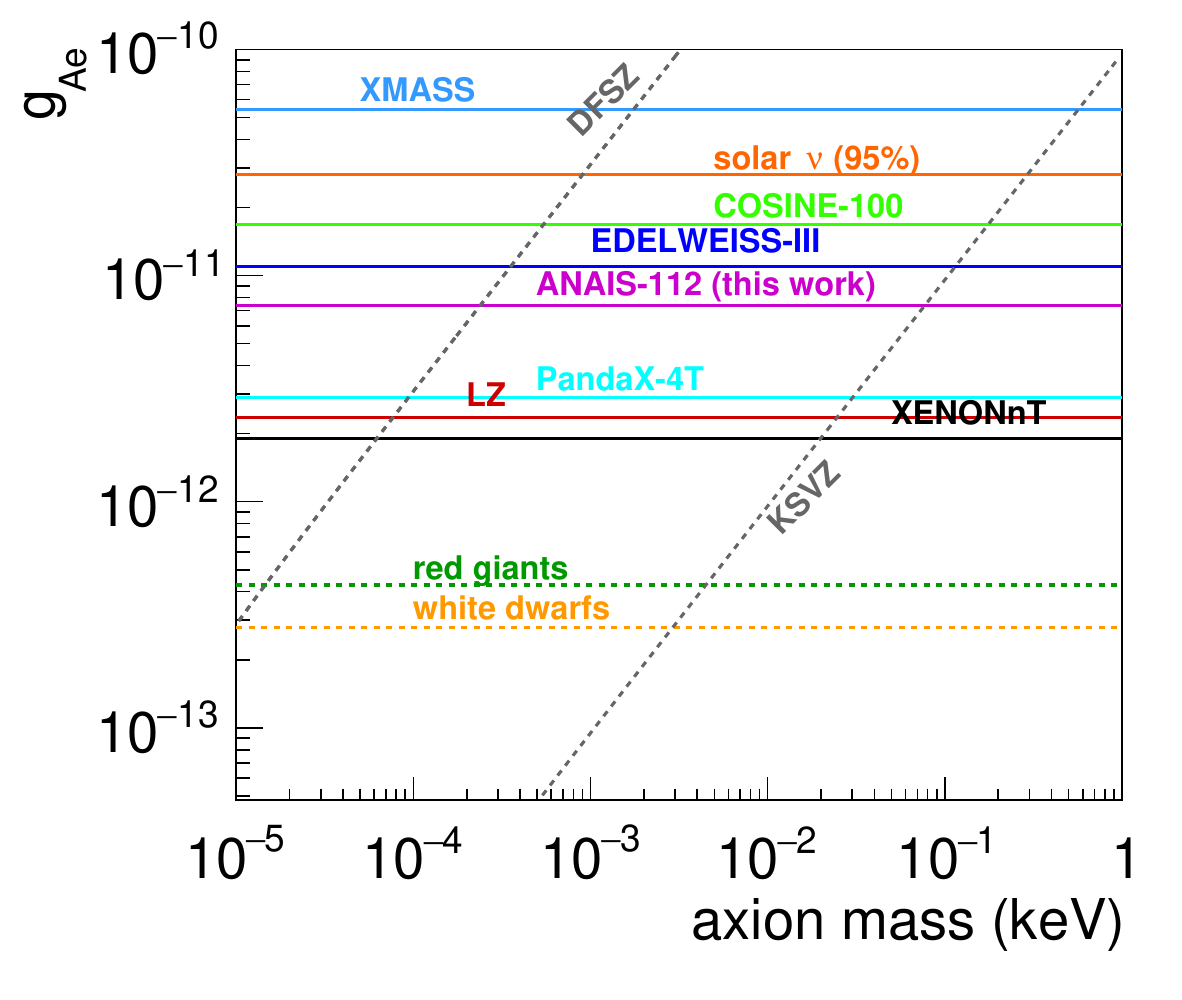}

\caption{\label{limitgae} 90\% C.L. upper limit on the axion–electron coupling constant $g_{\text{Ae}}$ derived from the six-year exposure of ANAIS-112 (magenta). For comparison, selected limits from other experiments and astrophysical observations are also shown \cite{aprile2022search,aalbers2023search,zeng2025exploring,adhikari2023search,abe2013search,armengaud2018searches,altherr1994axion,gondolo2009solar}. The dotted gray lines represent the benchmark predictions from QCD axion models DFSZ \cite{dine1981simple} and KSVZ \cite{kim1979weak}.
}
\end{center}
\end{figure}

Figure~\ref{limitgae} shows this limit together with constraints from other axion searches \cite{aprile2022search,aalbers2023search,zeng2025exploring,adhikari2023search,abe2013search,armengaud2018searches,altherr1994axion,gondolo2009solar}, as well as the predicted values from the benchmark QCD axion models DFSZ \cite{dine1981simple} and KSVZ \cite{kim1979weak}.

\begin{figure}[b!]
\begin{center}
\includegraphics[width=0.65\textwidth]{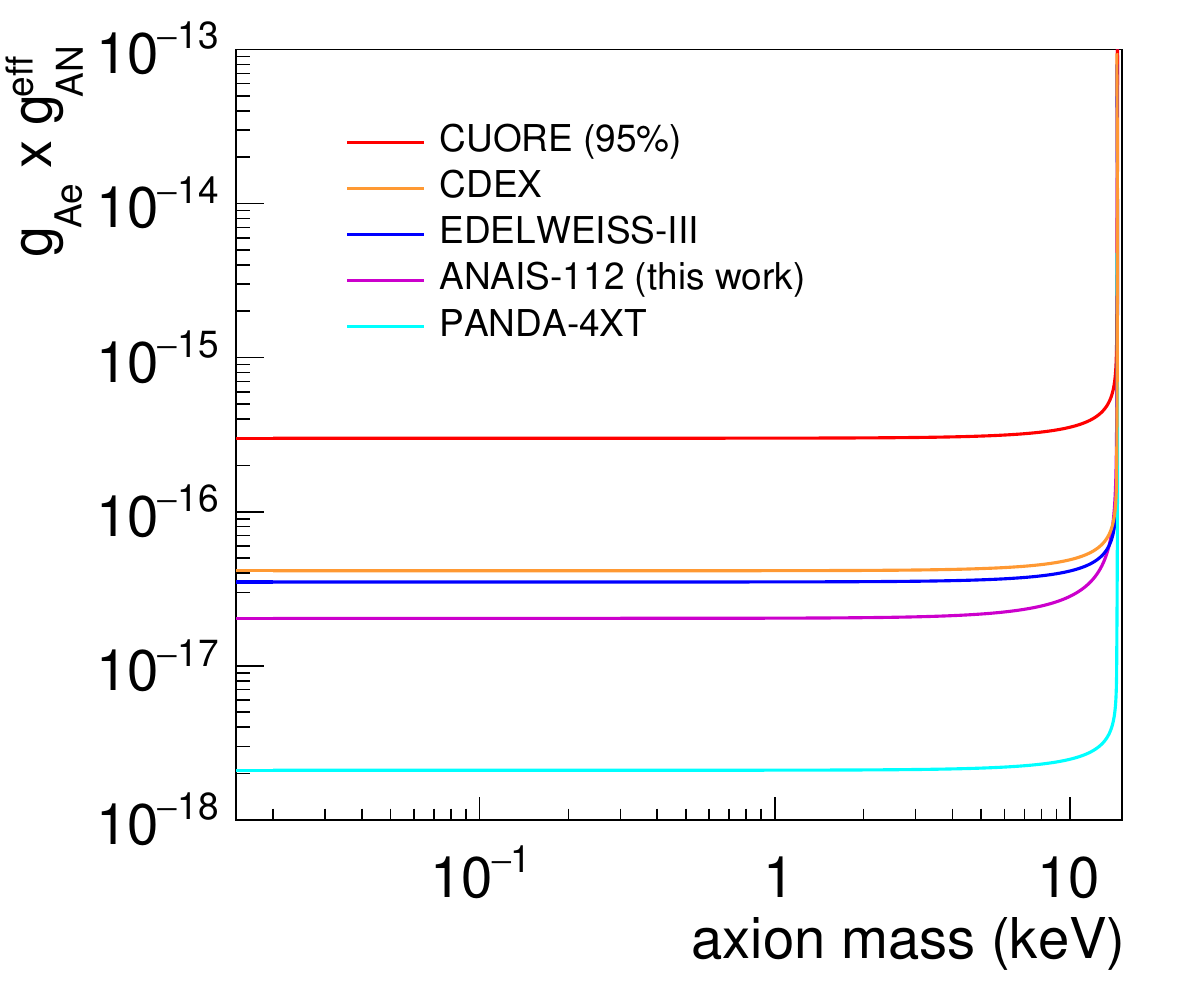}

\caption{\label{limitFe} 90\% C.L. upper limit on the solar axion coupling constant product \( g_{\text{AN}}^{\text{eff}} \times g_{\text{Ae}} \) derived from the six-year exposure of ANAIS-112 (magenta). For comparison, selected limits from other experiments are also shown \cite{alessandria2012search,wang2020improved,armengaud2018searches,zeng2025exploring}. 
}
\end{center}
\end{figure}

The ANAIS-112 search excludes QCD axion models with masses m$_A$ > 0.24 eV/c$^2$ (DFSZ) and m$_A$ > 77.23 eV/c$^2$ (KSVZ), representing the most stringent bound obtained to date using NaI(Tl) detectors. For comparison, the COSINE-100 experiment excludes QCD axions with masses above 0.59 eV/c$^2$ (DFSZ) and 168.1 eV/c$^2$ (KSVZ), implying an improvement by a factor of $\sim$ 2 relative to this other NaI(Tl)-based search. The exposure of ANAIS-112 in this search (0.63 ton$\times$year) is lower than that of leading experiments such as XENONnT (1.16 ton$\times$year) \cite{aprile2022search}, PANDA-4XT (1.54 ton$\times$year) \cite{zeng2025exploring}, and LZ (0.91~ton$\times$year)~\cite{aalbers2023search}. These experiments achieve superior sensitivities not only due to their larger exposures, but also to their significantly lower background levels.

Similarly, following the background-plus-\textsuperscript{57}Fe-axion signal hypothesis, the upper limit on the solar axion–electron coupling constant product $g_{\text{AN}}^{\text{eff}} \times g_{\text{Ae}}$ evaluated at m$_A$=0 is found to be:

\begin{equation}
    g^{\text{eff}}_{\text{AN}} \times g_{\textnormal{Ae}} < 2.03\times10^{-17}  \quad \text{at 90\% C.L.}
\end{equation}

Figure~\ref{limitFe} presents the ANAIS-112 upper limit together with results from other experiments \cite{alessandria2012search,wang2020improved,armengaud2018searches,zeng2025exploring}. As observed previously, experiments based on dual-phase xenon TPCs exhibit superior sensitivities; however, ANAIS-112 proves competitive with respect to the remaining experiments, providing a more stringent constraint than EDELWEISS-III, CDEX, or CUORE, whose limits on this coupling product evaluated at m$_A$=0 are $3.50 \times 10^{-17}$, $4.14 \times 10^{-17}$, and $3.00 \times 10^{-16}$ (95\%), respectively.


\section{Conclusions}

This chapter has presented new physics searches that can be conducted within the ANAIS-112 experiment using the improved background model and the QF estimations conducted in this thesis. Specifically, a reanalysis of the annual modulation signal has been performed, together with a dedicated search for solar axions, both using the six-year exposure of the experiment.

The annual modulation analysis has been performed in the same energy windows as those explored in previous ANAIS-112 analyses \cite{amare2025towards}. Compatible results have been obtained for the annual modulation amplitude, implying similar levels of incompatibility and sensitivity to the DAMA/LIBRA result. A clear improvement in fit performance has been observed in the [1–6] keV region, where the p-value for the null hypothesis has increased from 0.164 using the previous background model to 0.328 with the updated model. In the remaining energy regions, the performance of both models has remained comparable.

The result in the [6.7–20] sodium keV\textsubscript{NR} and [22.2–66.7] iodine keV\textsubscript{NR} energy region, corresponding to [2–6] keV in DAMA/LIBRA when assuming QF\textsubscript{Na}=0.3 and QF\textsubscript{I}=0.09, has been reexamined considering ANAIS constant values of QF\textsubscript{Na}=0.2 and QF\textsubscript{I}=0.06, yielding similar results than previous analysis. The energy-dependent QF\textsubscript{Na} selected in this work does not allow the exploration of this region, although this interval can still be probed on the iodine recoil energy scale. The use of the energy-dependent QF\textsubscript{I} results in reduced sensitivity and incompatibility with the DAMA/LIBRA signal. This is primarily due to the larger signal search window in terms of electron-equivalent energy, while covering the same NR energy range, associated with the energy-dependent QF\textsubscript{I} compared to the constant QF\textsubscript{I}. This results in a higher integrated background and, consequently, reduces the statistical significance of a potential DM–induced modulation.

Although the new background model explains more accurately the ANAIS background, the rate evolution shape, which is the key input for the analysis, has not changed significantly compared to the previous background model, which already performed well. Given that compatible results have been obtained, the robustness of the ANAIS-112 annual modulation analysis has been further reinforced.

Furthermore, the energy dependence of the modulation amplitude has been presented for both single-hit and multiple-hit events under two scenarios: one in which both experiments assume identical QFs (i.e., in electron-equivalent energy), and another in which the QFs differ (i.e., in NR energy for sodium and iodine recoils). In all cases, the results have been consistent and have favored the null hypothesis.

Nevertheless, when the search is conducted in the NR energy scale, part of the energy region in which DAMA/LIBRA observes modulation cannot be explored with the current performance of ANAIS-112, assuming the QF reported in this thesis. Specifically, given the current ANAIS-112 energy threshold of 1 keV, the analysis is restricted to probing sodium and iodine recoils above 3 keV and 2 keV, respectively, in the DAMA/LIBRA energy scale. This limitation justifies the need for future projects such as ANAIS+, which aims to lower the energy threshold below 0.5~keV in NaI detectors by replacing conventional PMTs with SiPMs for light detection. Such improvements would enable the exploration of the entire energy region where DAMA/LIBRA observes annual modulation, thereby allowing a more comprehensive test of the DAMA/LIBRA result overcoming systematic uncertainties related to the QFs. Moreover, ANAIS+ is expected to achieve good sensitivity to light WIMPs even with reasonable exposure and background levels. It also presents promising prospects for neutrino detection through CE$\nu$NS.

On the other hand, the improved background model has enabled the search for solar axions with ANAIS-112 data. A previous \textsuperscript{57}Fe solar axion search was performed in \cite{tfgmpellicer} using a simplified fitting approach. In contrast, the present analysis constitutes the first effort within the experiment to conduct such a search with a comprehensive treatment of background modelling and a detailed implementation of the expected spectral shapes for solar axion detection rates. In this work, four axion hypotheses have been considered, simultaneously accounting for the presence of the three axion channels as well as each independently, in addition to the background components adjusted in this low-energy region.

The fit has identified correlations between the background and axion components, as well as among the axion components themselves. In particular, correlations have been observed between $^{3}\mathrm{H}$ and the axion components, between the Primakoff and ABC channels, and between the $^{57}\mathrm{Fe}$ line and the $^{210}\mathrm{Pb}$ component in the teflon, which clearly overlap (peaks around 14.4 keV in the former case and around 12 keV in the latter). The performance of the null hypothesis and axion hypotheses has been compared, with the ANAIS-112 data favoring the absence of axions.

Due to these correlations, the decision was made to report results considering each axion contribution separately, and only for the ABC and $^{57}\mathrm{Fe}$ components, as the fit result for the Primakoff component was found to be highly incompatible with zero. The upper limits derived from the six-year exposure of the ANAIS-112 experiment are $g_{\text{Ae}} < 7.40 \times 10^{-12}$ and $g^{\text{eff}}_{\text{AN}} \times g_{\text{Ae}} < 2.03 \times 10^{-17}$ at 90\% C.L. While these limits are not as competitive as those obtained by dual-phase xenon TPCs, which lead the field due to their larger exposures and lower backgrounds, they nevertheless improve upon the results reported by other experiments in both cases.

Thus, this work highlights the potential of ANAIS-112 to probe axion parameters. However, it also identifies some current limitations of the analysis. Background contamination restricts the ability to report combined results for all three axion components, as clear correlations have been observed between them. In addition, the limit associated with the $^{57}\mathrm{Fe}$ line could be improved if the background shape in this region were smoother. Nevertheless, these challenges are common to all NaI-based detectors with similar contamination levels and are therefore not unique to ANAIS-112. Overall, this work establishes a solid foundation for future axion searches with increased statistics and enhanced detector performance.

\setcounter{chapter}{6} 
\chapter{Study of background for WIMP searches in COSINUS}\label{Chapter:COSINUS}

\minitoc

As part of this doctoral thesis, a research stay was undertaken at the Max Planck Institute for Physics in Munich, contributing to the COSINUS experiment, which is described in Section \ref{cosinussec}. This stay constitutes a first step towards a potential future collaboration between ANAIS-112 and COSINUS, two DM experiments which share clear synergies in objectives and target material, despite employing different detection techniques. 

This experience provided an opportunity to engage with cryogenic systems and gain valuable insight into alternative experimental techniques beyond only-scintillation detectors. Furthermore, the stay at COSINUS directly aligns with one fundamental objective of this doctoral thesis: the development of simulations for background modelling. Accordingly, during the research stay, simulation work was conducted to contribute to the development of the internal electromagnetic background model of COSINUS (Section \ref{internalbkg}). Additionally, the internal (or intrinsic) radiogenic neutron background simulation previously conducted by the collaboration will be presented (Section \ref{neutronsimulations}). With these two inputs, an estimation of the potential presence of backgrounds within the expected DM induced signal region (NR bands) has been carried out (Section~\ref{LYyleakage}), with particular focus on the region of interest for the DAMA/LIBRA signal. These events cannot be distinguished from a DM-induced signal and represent an irreducible background, limiting the experimental sensitivity. The goal of this study is to evaluate the effect of the detector performance on the sensitivity of COSINUS, by analyzing the influence of the energy resolution of the light and heat signals on the potential contamination of background events in the NR bands.

\section{The COSINUS experiment}\label{cosinussec}

COSINUS (Cryogenic Observatory for SIgnatures seen in Next-generation Underground Searches) \cite{angloher2020cosinus} is a DM experiment located  at the LNGS with the primary goal of providing a model-independent cross-check of the annual modulation signal observed by the DAMA/LIBRA experiment \cite{bernabei1999further,bernabei2000search,bernabei2008first,bernabei2020dama,bernabei2018first}. This is achieved by using the same target material, NaI, in low-temperature ($\sim$mK) calorimeters equipped with dual read-out. This configuration enables the simultaneous measurement of both light and heat, allowing for particle discrimination on an event-by-event basis because of the different energy sharing between both channels for NR and ERs. The measurement of the energy deposited in the heat channel removes the signal interpretation's dependence on the scintillation QFs for Na and I, affected by large uncertainties (as explained in Section \ref{bandsdef}). On the other hand, the QF could be determined by COSINUS in situ, providing a measurement for the QF of NaI at cryogenic temperatures.

A detailed description of the COSINUS experimental set-up can be found in \cite{angloher2016cosinus,angloher2022simulation}. It includes a water tank ($\sim$ 7x7 m$^2$) housing a dry dilution refrigerator equipped with a pulse tube cryocooler, enabling the operation of NaI at ultra-low temperatures of $\sim$10~mK. 

\begin{figure}[b!]
\begin{center}
 \centering
    \begin{minipage}{0.3\textwidth}
        \centering
        \includegraphics[width=\textwidth]{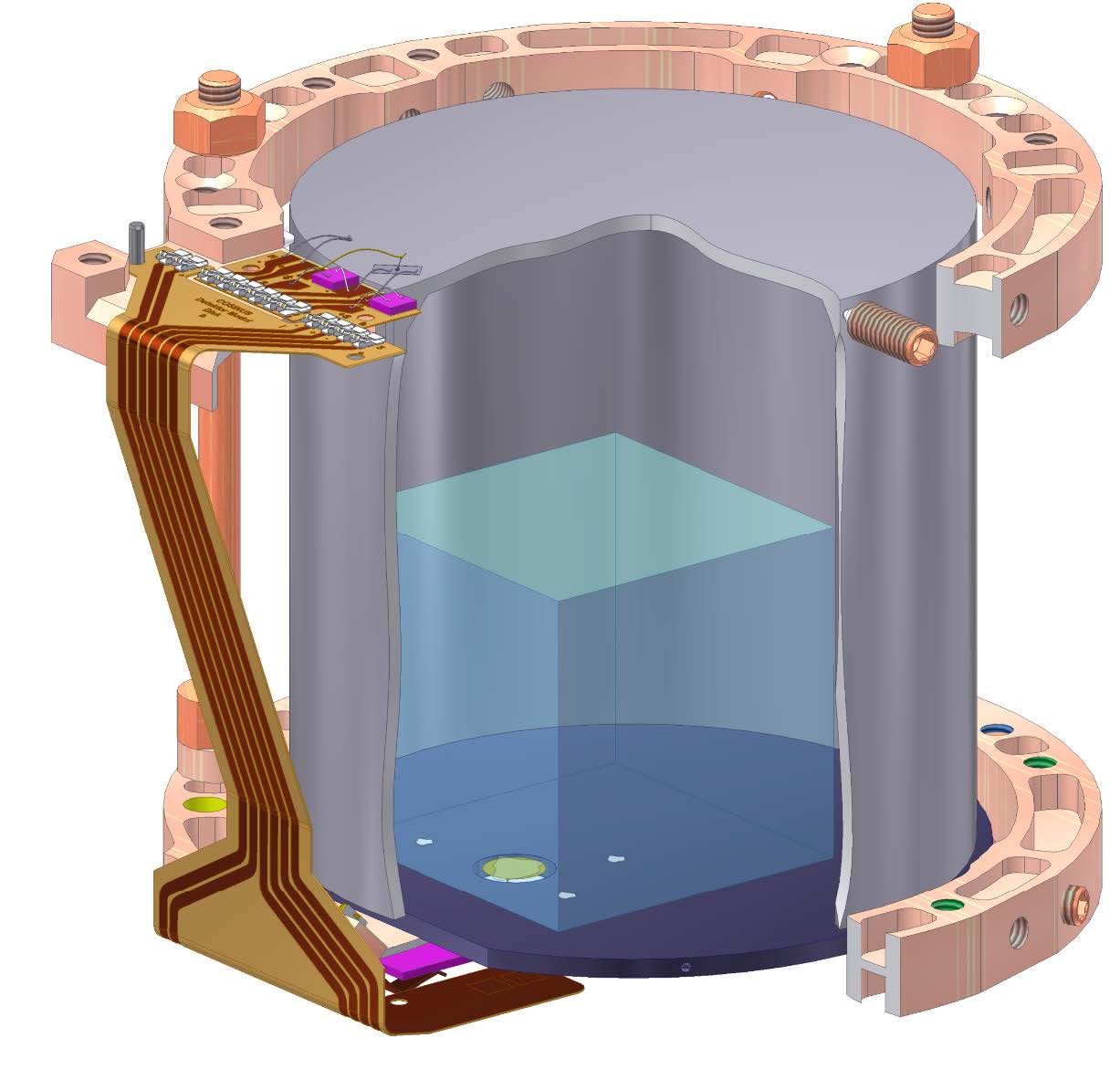}
    \end{minipage}
    \hspace{0.5cm}
    \begin{minipage}{0.58\textwidth}
        \centering
        \includegraphics[width=\textwidth]{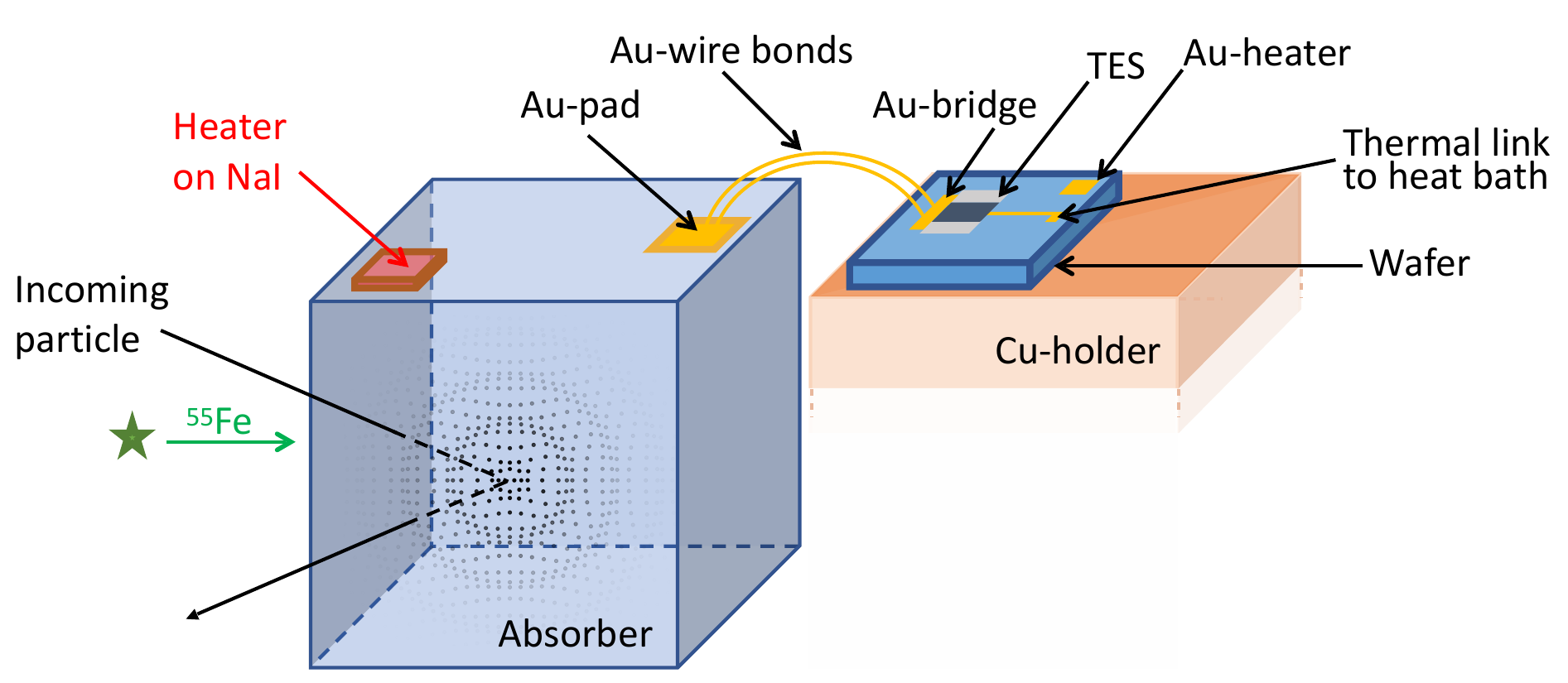}
    \end{minipage}
\caption{\label{COSINUSset-up} \textbf{Left panel:} Scheme of a fully assembled COSINUS detector module. \textbf{Right panel:} Illustration of the remoTES detector design, featuring a NaI crystal equipped with a gold pad thermally connected to a TES on a separate wafer via gold bonding \cite{angloher2024particle}. }

\end{center}
\end{figure}

The NaI crystal is a scintillating material,
which generates both lattice vibrations (phonons) and scintillation light upon interaction with
incident particles. COSINUS detectors are designed as modules consisting of an energy absorber (NaI crystal), a phonon detector and a light detector, both using cryogenic technologies originally developed within the CRESST collaboration \cite{angloher2016results}.  The left panel of Figure \ref{COSINUSset-up} shows a scheme of the fully assembled COSINUS 4$\pi$ detector module. This configuration enables the simultaneous measurement of phonon and light signals, which allows for particle discrimination.

When a particle deposits energy in the absorber crystal, it induces a temperature increase, $\Delta T$ (typically in the $\mu$K range), detected by a highly sensitive thermometer known as a Transition Edge Sensor (TES). A TES is a superconducting thin film operated in the transition region between its superconducting and normal-conducting states. This enables the detection of small temperature variations through the resulting resistance change, $\Delta R$. On the other hand, part of the deposited energy results in scintillation light that escapes the crystal and is collected by a 4$\pi$ beaker-shaped light detector made of high-purity silicon, which surrounds the NaI crystal (see left panel of Figure~\ref{COSINUSset-up}). When a particle interacts with the NaI crystal or the silicon detector, it causes a temperature rise that is read out via a TES as a resistance change. Consequently, both channels are effectively read out through a phonon detector. The phonon and light signals are read out by Superconducting Quantum Interference Devices (SQUIDs), which are capable of detecting extremely small changes in magnetic fields, induced by the resistance variations resulting from temperature changes in the TES. 

\begin{table}[b!]
\centering
\begin{tabular}{ccccc} 
\hline
\multicolumn{5}{c}{Contamination level (ppb)} \\
 \hline
Experiment & Sample & K & Th & U                                                         \\
\hline
\hline

ANAIS-112 & NaI(Tl) crystal & 27.3 $\pm$ 0.3 & (230 $\pm$ 16)x10$^{-6}$ & (188 $\pm$ 24)x10$^{-6}$                                                   \\

COSINUS & NaI powder & <15 & <0.01 & <0.005                                                         \\

\hline

\end{tabular}

\caption{\label{tabla} Intrinsic contamination levels of the NaI crystals used in ANAIS-112 and COSINUS. In COSINUS, material screening was performed using HR-ICP-MS on NaI astrograde powder, with the upper limit set at three times the standard deviation of a blank measurement. In ANAIS-112, the values shown in the table correspond to the average across the nine detectors. The K content has been revisited in this thesis  via coincidence detection of low-energy deposits and high-energy gamma rays (see Section~\ref{NaK}), while Th and U activities were measured through alpha rates using pulse shape analysis and delayed Bi/Po coincidences \cite{amare2019analysis}.}
\end{table}

In the light channel, the TES is directly deposited on the silicon beaker. However, direct TES fabrication on NaI crystals is impractical due to their hygroscopy and low melting point. To address this, a novel approach known as remoTES has been implemented \cite{angloher2023first} by COSINUS, in which the TES is fabricated on a separate substrate and thermally coupled to the absorber crystal via gold-wire bonding (see right panel of Figure \ref{COSINUSset-up}).

The target material consists of highly radiopure, undoped NaI crystals produced by the Shanghai Institute for Ceramics (SICCAS).  High-resolution Inductively Coupled Plasma Mass Spectrometry (HR-ICP-MS) analysis of the NaI powder was conducted at the LNGS Chemical Services facility, with the results presented in Table \ref{tabla}, together with the contamination values found in the ANAIS experiment (average of all detectors) for comparison. The \(^{40}\)K content in the ANAIS-112 crystals corresponds to the revised values obtained in this thesis, (0.844 $\pm$ 0.008) mBq/kg for the average of all detectors, and was evaluated through the identification of coincident events between low-energy deposits and high-energy gamma emissions in separate modules (see Section~\ref{NaK}). In contrast, thorium and uranium concentrations were determined by analyzing alpha-induced event rates using pulse shape discrimination techniques and by identifying delayed Bi/Po coincidences \cite{amare2019analysis}.

As shown in the table, the radiopurity levels of the NaI powder used for crystal growth in the COSINUS experiment could be comparable to those of the NaI(Tl) crystals employed in ANAIS-112, although there are only upper limits. In particular, the potassium content in the COSINUS powder is at least a factor of $\sim$2 lower, indicating a notable improvement. For thorium and uranium, only upper limits are available for COSINUS, which are less stringent than the values measured for ANAIS-112 crystals. 

However, since these values correspond to the raw NaI powder prior to crystal growth, the comparison remains only indicative. The actual radiopurity must be reassessed once the COSINUS crystals are grown and available at LNGS. Radiopurity could improve by the growth procedure, but potential contamination at growth or detector assembly could substantially modify the final radiopurity levels. Moreover, no information is yet available for the COSINUS crystals regarding the presence of \(^{210}\)Pb, which continues to be a significant challenge in the fabrication of NaI-based detectors \cite{amare2019analysis,adhikari2021background,barberio2023simulation}. If not carefully controlled, the presence of \(^{210}\)Pb could have a considerable impact on the overall background.

Regarding the future schedule, the COSINUS experiment will collect data in two runs. The first run is scheduled to begin in 2025, last for one year, and will employ eight detector modules of approximately 34 g each (2.1 $\times$ 2.1 $\times$ 2.1 cm$^3$ cubes), yielding a total absorber mass of 270 g. A future upgrade to 24 modules is planned, with an absorber mass between 30 and 100 g of NaI. The objective of the first phase is to collect 100 kg x days of data, which will be sufficient to rule out the DAMA/LIBRA signal in the context of DM scattering off sodium and/or iodine in the SI scenario \cite{angloher2016cosinus}. In the near future, COSINUS may determine whether the DAMA signal originates from interactions with nuclei or electrons. Notably, the first phase of COSINUS will not study modulation, distinguishing it from other experiments such as ANAIS-112 and COSINE-100. The second run is scheduled to collect 1000~kg~×~days over 3–4 years of data taking, aiming to achieve sensitivity to arbitrary recoil spectra.


\subsection{Event discrimination based on Light-Phonon signals}\label{bandsdef}

The COSINUS detector design enables simultaneous measurement of energy in the phonon channel (E\(_P\)) and the light channel (E\(_L\)). This dual-channel approach allows for effective particle discrimination and background rejection by using the light yield (LY) parameter defined as the ratio of light to phonon energy:

\begin{equation}
    LY = \dfrac{E_{L}}{E_{P}}.
    \label{LYeq}
\end{equation}

To avoid confusion, it is important to clarify that the LY parameter referenced in this chapter, following the definition from the COSINUS collaboration, differs from the LY parameter mentioned earlier in this thesis. The latter refers to a fundamental property of the scintillator, specifically the number of photons emitted per unit of deposited energy, while here it refers to a relative value chosen to be 1 for ERs at a given energy. In particular, the LY in COSINUS is defined such that gamma events at the calibration energy, 5.9 keV from $^{55}$Fe calibration sources, have a mean LY of 1 \cite{angloher2024particle}. A LY~<~1 means that less energy escapes the crystal as scintillation light, and more energy remains within the crystal contributing to the phonon signal, when compared to a $^{55}$Fe x-ray event.

Given that a $\gamma$-source is typically the predominant choice for calibration sources, as will be the case in COSINUS, this study describes both the E$_{P}$ and E$_{L}$ energy scales in terms of electron-equivalent energy. It should be noted that the electron-equivalent energy reported for the light detector does not correspond to energy deposited in the light detector itself, but rather to the energy deposited in the absorber, which produces the scintillation light subsequently detected.



The following equation describes the relationship of \(E_P\) and \(E_L\) with the total deposited energy (\(E\)):

\begin{equation}
    E = \eta E_L + (1-\eta) E_P,
    \label{relacionenes}
\end{equation}

where \(\eta\) is the scintillation efficiency, which quantifies the fraction of deposited energy converted into scintillation light in a \(\gamma\)-calibration event. It is worth highlighting that \(\eta\) is, in general, energy- and particle-dependent. However, in COSINUS, it is assumed to be constant across all energies and particle types, with this dependence being incorporated in the description of the electron recoil band based on the light collected (see Equations \ref{meane} and \ref{meanNa}). 

From Equation \ref{relacionenes}, it follows that \(E = E_p(1 - \eta(1 - LY))\), indicating that the phonon channel would provide a mostly particle-independent measurement of the deposited energy if \(\eta\) is small enough. In COSINUS, \(\eta\) is measured to be 9.1\% \cite{PhysRevD.110.043010} for NaI, while in other cryogenic materials such as CaWO\(_4\) used in CRESST, it typically reaches only about 1\% \cite{angloher2014results}. Given that in the light channel, the amount of scintillation light produced shows a much stronger dependence on the nature of the interacting particle, the dual readout system becomes a powerful tool for particle identification.

\begin{figure}[t!]
\begin{center}
 \centering
     \includegraphics[width=1\textwidth]{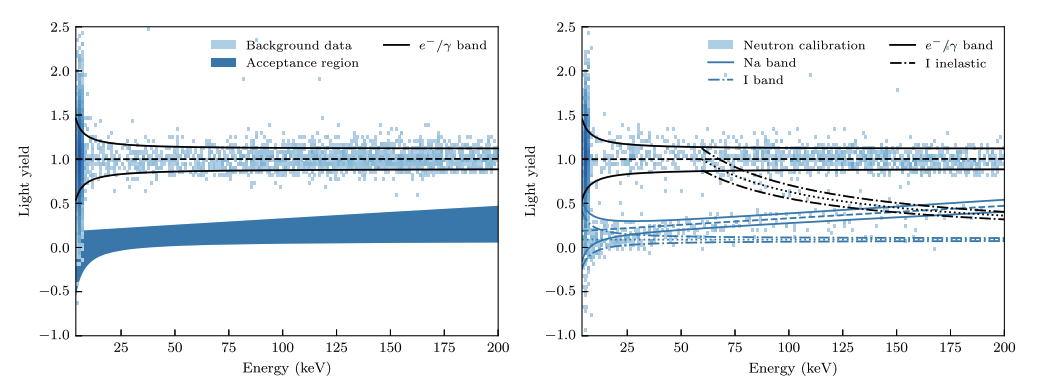}
    
\caption{\label{LYplane} LY-energy plane for background data (left panel) and neutron calibration data (right panel) \cite{PhysRevD.110.043010}. The bands corresponding to e$^-$/$\gamma$, Na recoils, I recoils, and inelastic recoils are shown, along with the acceptance region for DM analysis.}

\end{center}
\end{figure}

The most common representation of data using the dual detection technique is the LY~vs.~E plot, where E refers to the total deposited energy (see Equation~\ref{relacionenes}). Figure~\ref{LYplane} depicts the LY-E plane for both background (left panel) and neutron calibration data (right panel) \cite{PhysRevD.110.043010}. The densely populated band around a LY of approximately 1 corresponds to the $e^-/\gamma$ background. In contrast, NRs, arising from neutron interactions produce substantially less scintillation light than ERs of equivalent total energy, a behavior quantified by the QF. Due to the differing nuclear masses in NaI targets, two distinct NR bands emerge: one for sodium and another for iodine, as illustrated in Figure \ref{LYplane}. Under the assumption that nonrelativistic, WIMP-like DM particles primarily scatter off nuclei, the acceptance region for DM analysis is determined by the position of these NR bands.

The band positions are obtained from an unbinned likelihood fit to the full data set, including background and neutron calibration, in both phonon and light energy. Their parameterization consists of phenomenological descriptions of light output as a function of the total deposited energy. 

The modelling of the bands corresponding to different event classes is presented below. It should be emphasized that, although Figure \ref{LYplane} is represented in the LY–E plane, each band is rigorously defined by an energy-dependent mean in the light channel, \( E_L(E) \), and an
energy-dependent gaussian width \( \sigma(E_L(E)) \), both parametrized then in the \( E_L \)–\( E \) plane. A comprehensive parameterization of the recoil bands, including all relevant functional forms and associated parameters, can be found in \cite{angloher2024likelihood}.\\

\textbf{Mean of the bands.}

The mean energy in the light channel for the electron recoil band (e) and the elastic nuclear recoil bands (n, with n \(\in\) \{Na, I\}) is modelled according to the following expressions:

\begin{equation}
    E_{L,e} (E) = l_0E + l_1E^2,
    \label{meane}
\end{equation}

\begin{equation}
    E_{L,n} (E) = (l_0E + l_1E^2) k_{\text{QF},n} \left(1 - a_n e^{-E/d_n}\right); \, \text{n} \in \text{\{Na,I\}} \, .\\
    \label{meanNa}
\end{equation}

l$_0$, l$_1$, k$_{\textnormal{QF,n}}$, a$_\textnormal{n}$ and d$_\textnormal{n}$ are free parameters determined from the maximum likelihood bandfit, with their values provided in \cite{PhysRevD.110.043010}.

l$_0$ represents the proportional component of the light output, while l$_1$ is an empirical parameter that explains slight deviations from linearity generally improving the fit result. The exponentially decaying factor of Equation \ref{meanNa} accounts for the well-known non-proportionality of scintillators (see Section \ref{nonpropSec}), causing the band to bend towards lower LYs at very low energy. This exponential description of the NR band is purely phenomenological. In particular, \(a_n\) accounts for the magnitude of the non-proportionality effect, while \(d_n\) characterizes the curvature of the downward bending. On the other hand, $k_{\text{QF},n}$ is a parameter proportional to the QF that incorporates an overall scaling to account for crystal-specific light yield variations, such as those induced by impurities. 

The QF can be then described as QF$_n$(E)=E$_{L,n}$(E)/E$_{L,e}$(E) based on this modelling of the bands, enabling in-situ QF measurements based on neutron calibration data. In~\cite{PhysRevD.110.043010}, the QFs at 10 keV energy were determined as \( \text{QF}_{\text{Na}}(10 \, \text{keV}) = 0.197 \pm 0.019 \) and \( \text{QF}_{\text{I}}(10 \, \text{keV}) = 0.0892 \pm 0.0037 \). The energy dependence of the QF derived from this study is shown in Figure~\ref{NaQFcomparison}. Its viability as a QF model has been assessed by comparison with the measurements obtained from the onsite neutron calibrations conducted in ANAIS-112 (see Section \ref{otherNa}).

It is worth highlighting that in the analysis presented in this chapter, e$^-$ and $\gamma$ interactions are combined into a single band. However, the LY for $\gamma$ interactions is typically somewhat lower than for electrons. This is because $\gamma$ photons generate multiple low-energy secondary electrons, and because of the non-linearities found in the response to electrons, the $\gamma$-band experiences a slight quenching compared to the electron band. \\

\textbf{Width of the bands.}

The finite resolution of the phonon and light detectors, together with statistical fluctuations in the number of generated scintillation photons, causes the measured data
points to spread around the mean energy line. The width of the bands is determined by
the achievable resolution of both the light channel ($\sigma_L$) and the phonon channel ($\sigma_P$), to either an electron or a NR. The resolution can be modelled as \cite{angloher2024likelihood}:\\

\begin{equation}
    \sigma_L(E_{L,x})=\sqrt{\sigma^2_{L,0}+S_1E_{L,x}+S_2E^2_{L,x}}\\
    \label{sigmaL}
\end{equation}

\begin{equation}
    \sigma_P(E)=\sqrt{\sigma^2_{P,0}+\sigma^2_{P,1}(E^2-E^2_{thr})}\\
    \label{sigmaP}
\end{equation}

These dependencies do not follow a strict poissonian behavior, where \(\sigma \propto \sqrt{E}\), in order to account for other factors affecting the resolution. In particular, \( \sigma_L \) is primarily affected by the baseline noise (resolution at zero energy, \( \sigma_{\textnormal{L,0}} \)), poisson fluctuations in scintillation photon production (\( S1 \)), and position dependencies (\( S2 \)). For the phonon channel, \( \sigma_P \) is dominated by the baseline resolution (\( \sigma_{\textnormal{P,0}} \)), while \( \sigma_{P,1} \) accounts for a potential energy dependence.  These parameters, which model the resolution, are treated as free parameters and are determined from the first underground measurement of a COSINUS prototype by a maximum likelihood fit \cite{PhysRevD.110.043010}.

A common rule of thumb in cryogenic experiments for establishing the energy threshold \( E_{\text{thr}} \) is to set it at five times the baseline resolution in the phonon channel, such that \( E_{\text{thr}} = 5 \times \sigma_{P,0} \), resulting in very few noise triggers for an exposure on the order of one kg x day\cite{strauss2017prototype}. Consequently, this criterion will be adopted in this work. On the other hand, for the band description conducted in this thesis, the parameters shown in Equations \ref{meane}, \ref{meanNa}, \ref{sigmaL}, and \ref{sigmaP} will be assigned the values determined from the COSINUS prototype measurement \cite{PhysRevD.110.043010}, assuming they are representative of future COSINUS detectors. An exception is made for  \( \sigma_{\textnormal{L,0}} \) and  \( \sigma_{\textnormal{P,0}} \), which will be varied within reasonable limits around the value reported by this prototype, in order to evaluate the impact of detector performance on the sensitivity of the experiment.

Combining the resolution of both channels, the total width of the bands in the E$_L-E$ plane can be written as:

\begin{equation}
    \sigma_x(E_L)=\sqrt{\sigma^2_L(E_{L,x}(E))+(\frac{dE_{L,x}}{dE}(E))^2\sigma^2_P(E)}.
    \label{sigmatot}
\end{equation}

The phonon resolution must be scaled by the slope of the corresponding band, \( \frac{dE_{L,x}}{dE} \), at each point, which is calculated analytically. This accounts for the increased broadening of bands with higher light yield values.

As shown in Figure \ref{LYplane}, at high energies, the electronic and NR bands are well separated, highlighting the excellent discrimination power of COSINUS. The larger the separation
of the bands, the better the discrimination of background events. 

However, at low recoil energies, the NR and ER bands width increases and finally begin to overlap, with the NR band flaring upwards and the ER band flaring downwards. This overlap worsens the discrimination power between the two types of events, causing any ER that leaks into the NR band to represent a non-discriminable background that could mimic a WIMP candidate event. 

Owing to its capability for event-by-event particle discrimination, COSINUS will be uniquely positioned to determine whether the DAMA signal arises from  particles scattering off electrons
or nuclei in NaI(Tl). The COSINUS performance goal is to achieve an event rate below 0.1 c/kg/day within the DAMA/LIBRA ROI, defined as the energy interval from 1 to 6 keV, in electron-equivalent energy for DAMA. 

This sensitivity estimate is based on the fact that in a positive modulation signal, the modulation amplitude
cannot be larger than the (average) absolute rate \cite{kahlhoefer2018model}. In the case of electron interactions, the DAMA/LIBRA ROI directly corresponds to the same visible energy range.  However, for NRs, the scintillation signal of DAMA/LIBRA is quenched, and the equivalent ROI must be scaled according to the QF. Specifically, for Na recoils (QF\(_\text{Na,DAMA}\) = 0.3), the equivalent DAMA/LIBRA ROI is [3.33–20] keV\textsubscript{NR}, and for I recoils (QF\(_\text{I,DAMA}\)~=~0.09), it is [11.11–66.66] keV\textsubscript{NR}.


This study aims to evaluate the event rate bound by estimating the leakage of \(e^-/\gamma\) events into the DAMA/LIBRA ROI for each recoil band. In addition, COSINUS is also expected to be sensitive to low-mass WIMPs, which motivates the evaluation of \(e^-/\gamma\) leakage across the full nuclear recoil band starting from the energy threshold.

Unlike previous approaches based on empirical models \cite{angloher2024likelihood}, the analysis conducted in this thesis incorporates results from detailed background simulations. The dominant background in the COSINUS experiment, similarly to other NaI-based DM searches such as ANAIS \cite{amare2019analysis}, is expected to arise from intrinsic radioactive contamination within the NaI crystal itself. Section \ref{internalbkg} details the simulations developed in this thesis to model the internal electromagnetic background of COSINUS in order to quantify the resulting \(e^-/\gamma\) leakage into the NR bands. 

Additionally, radiogenic neutrons produced via \((\alpha\),n) reactions and SF in the surrounding materials or the detector itself can generate NRs indistinguishable from those expected from WIMP interactions, thereby constituting a particularly concerning irreducible background source. In this work, only the radiogenic neutron background arising from the NaI crystal itself, as provided by the COSINUS collaboration \cite{privcomMatt}, is considered, since simulations of the ambient radiogenic neutron background have not yet been developed by the experiment. A key aim of this study is to assess the contributions of internal radiogenic neutrons to the NR bands. Although these contributions are expected to be minor, these neutrons cannot be shielded due to their origin within the crystal.

\section{Internal electromagnetic background model}\label{internalbkg}

\vspace{-0.4cm}
As previously emphasized, the main source of background contamination in NaI DM search experiments comes from natural radioactivity embedded within the detectors themselves. The most prevalent isotopes are those of $^{238}$U, $^{235}$U, $^{232}$Th and their progeny,
which eventually decay down to a stable isotope of lead via the emission of multiple $\alpha$, $\beta$ and subsequent $\gamma$-rays. In addition, $^{40}$K contamination is typically a relevant radioactive contribution for
NaI detectors.

In order to develop the intrinsic electromagnetic background model for the COSINUS experiment, simulation studies have been conducted using a simple geometry consisting of a NaI cube. This cube accurately reproduces the dimensions and mass of the COSINUS crystals, with a volume of 9.3 cm$^3$ (2.1 $\times$ 2.1 $\times$ 2.1 cm$^3$ cubes) and a mass of $\sim$ 34~g. Monte Carlo simulations were carried out using Geant4 (v10.2.3) and ImpCRESST \cite{abdelhameed2019geant4}, a Geant4/ROOT-based simulation framework originally developed for CRESST and now also employed by COSINUS. ImpCRESST includes optimized low-energy physics, flexible geometry handling, and structured ROOT data output. The software is currently limited for internal use.

The background contribution from \(^{40}\)K and the \(^{238}\)U, \(^{235}\)U, and \(^{232}\)Th decay chains was simulated, assuming a uniform distribution of nuclides within the NaI crystal. Following the COSINUS procedure, each step of the decay chains was modelled separately: a radionuclide was placed in the crystal, and the simulation ran until it decayed into the next nuclide in the chain, repeating the process for subsequent decays. This approach allows for easy adjustment of specific nuclide activities by rescaling individual decay spectra once the actual COSINUS crystals become available and activity reassessment is performed.

\begin{figure}[t!]
    \centering
    \begin{subfigure}[b]{0.7\textwidth}
        \centering
        \includegraphics[width=\textwidth]{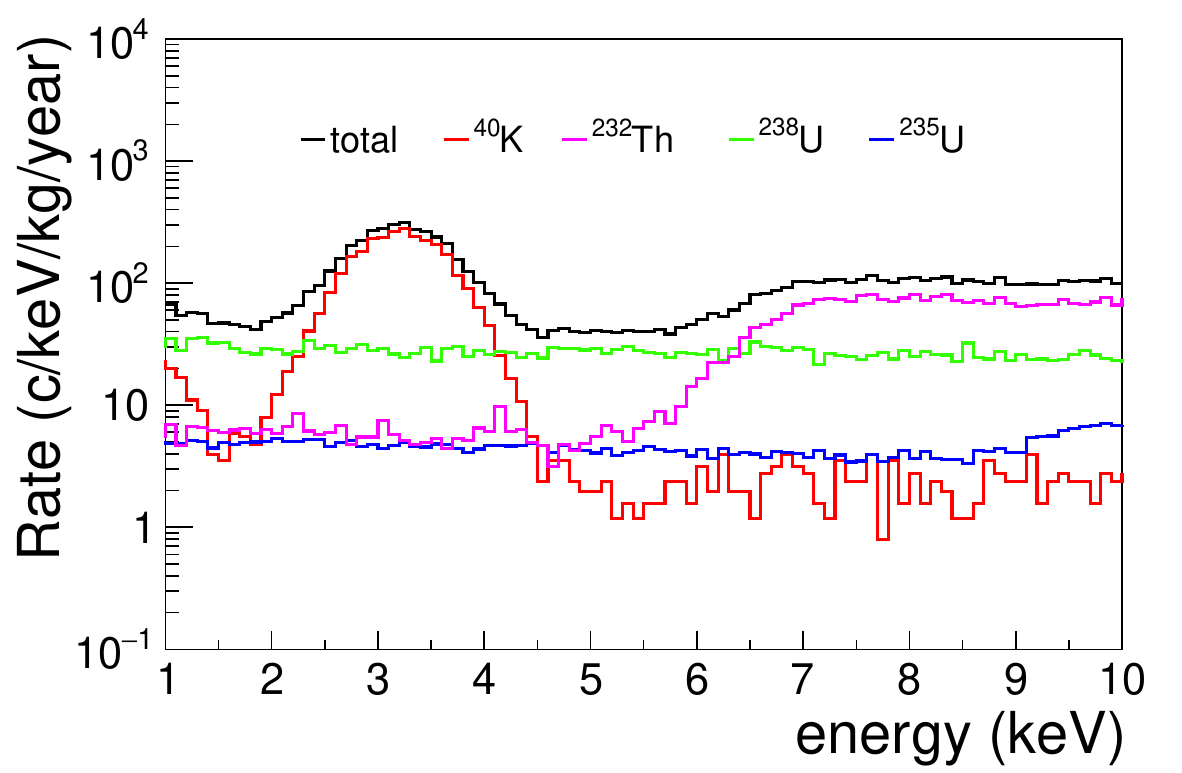}        \label{fig:imagen1}
    \end{subfigure}
       \hspace{-0.5cm}
    \begin{subfigure}[b]{0.7\textwidth}
        \centering
        \includegraphics[width=\textwidth]{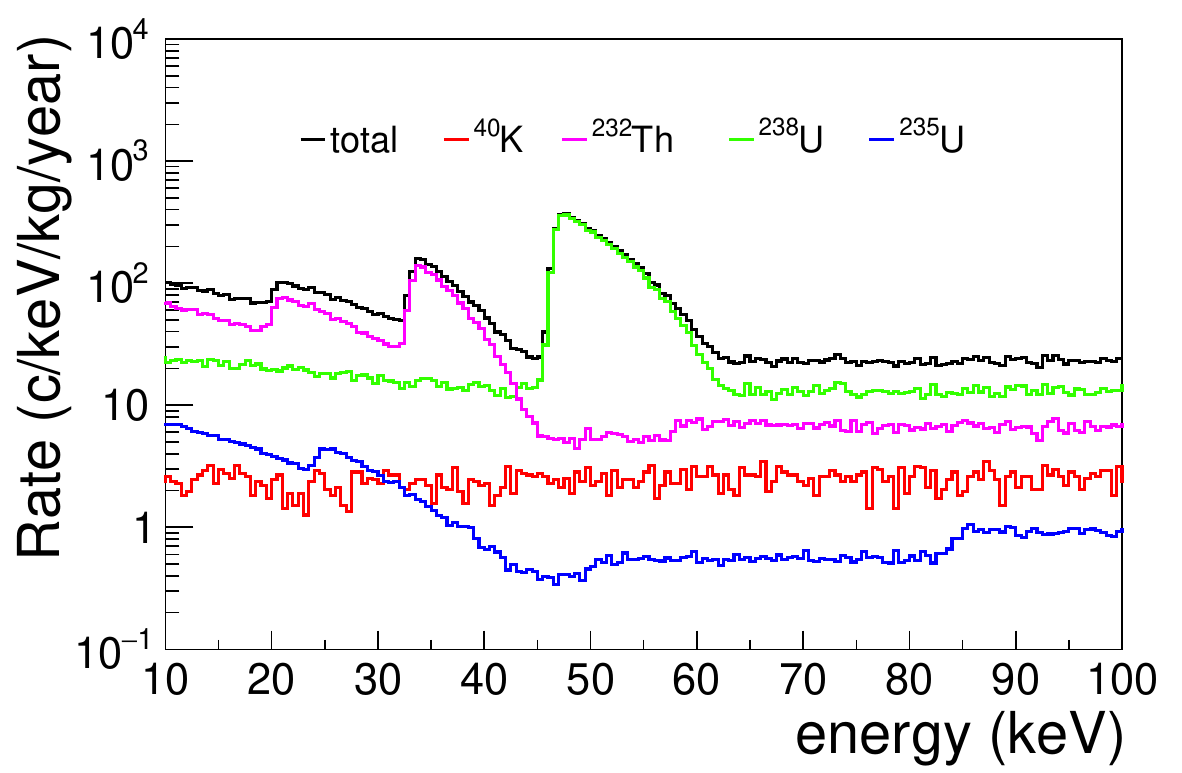}
        \label{fig:imagen2}
    \end{subfigure}
    \vspace{-0.3cm}
    \caption{Simulated energy spectra expected from internal background sources ($^{40}$K, $^{232}$Th,$^{238}$U and $^{235}$U) in a COSINUS NaI crystal. \textbf{Top panel:}
low energy range. \textbf{Bottom panel:} medium energy range.}
    \label{internalbkgplot}
\end{figure}

In the results presented here, secular equilibrium has been assumed in the radioactive chains taking as initial activity levels the upper limits listed in Table \ref{tabla} from the NaI powder screening by HR-ICP-MS. However, as previously noted, natural decay chains frequently present disruptions in secular equilibrium, necessitating a reassessment of radiopurity levels once the final COSINUS crystals become available.

The simulated spectrum is post-processed by applying the detector energy resolution. In this study, only the resolution of the phonon channel is considered, as it is the dominant channel used for energy reconstruction. For CaWO\(_4\), a resolution of 0.09~keV is achieved at 2.2 keV, as reported in \cite{angloher2014results}. The phonon channel resolution for COSINUS at 2.2 keV is 0.42 keV, as determined from Equation~\ref{sigmaP}. For comparison, the resolution at the same energy in ANAIS-112, based on the resolution revisit conducted in this work (see Figure~\ref{LEres}), is 0.49~keV.  Thus, the energy resolution of COSINUS prototypes is considerably worse than that of other bolometers, being comparable to that of ANAIS-112, although there is still room for improvement in the final set-up configuration. This highlights the challenges associated with operating NaI as a target material.

Figure \ref{internalbkgplot} presents the simulated internal electromagnetic background for the COSINUS experiment in both low- and medium-energy ranges. Although not shown, characteristic $\alpha$ lines from natural chains are clearly identifiable in the MeV region.

Table \ref{tablaexpectedrated} provides the expected rates in different energy windows, highlighting the relevance of \(^{40}\)K in the very low-energy region due to its 3.2 keV energy deposit following K-shell EC. In the medium- and high-energy ranges, \(^{238}\)U dominates the internal electromagnetic background. The counting rates per detector per year coming from the internal electromagnetic background of COSINUS are 17.22 ± 0.14 c/det/year in [1-6]~keV, 212.8~±~0.4~c/det/year in [10-100] keV, and 760.0 ± 0.8 c/det/year in [100-2000] keV. It should be recalled that this estimate is likely to be significantly overestimated, as only upper limits exist for the contamination in the NaI powder, which could in fact be substantially lower than assumed. Nevertheless, although these components may be overestimated, COSINUS final crystals could contain out-of-equilibrium \textsuperscript{210}Pb, as well as other cosmogenically activated isotopes, which can contribute to the background in the ROI.

\begin{table}[t!]
\centering
\resizebox{0.95\textwidth}{!}{
\begin{tabular}{c|cc|c|c}
 \hline
\multicolumn{5}{c}{Rate (c/kg/year)}                                                   \\
 \hline
  & \multicolumn{2}{c}{[1-6] keV} & \multicolumn{1}{|c|}{[10-100] keV} & \multicolumn{1}{c}{[100-2000] keV} \\
 \hline
Isotope & ANAIS-112 & COSINUS    & COSINUS    & COSINUS \\ 
\hline
\hline

$^{40}$K & 725.20 & 305 ± 3 & 219 ± 3 & 3324 ± 12\\
$^{238}$U & 22.95  & 145 ± 2 & 3621 ± 10 & 13069 ± 19 \\
$^{235}$U & 0.29 & 23.63 ± 0.18 & 0.4 ± 1 & 571.9 ± 0.9 \\
$^{232}$Th & 0.67 & 32.7 ± 0.8 & 2271 ± 7  & 5388  ± 25 \\
\hline
Total & 749.11 & 506 ± 4 & 6257 ± 12 &  22353 ± 25 \\
\hline

\end{tabular}
}
\caption{\label{tablaexpectedrated}Expected rates from the internal electromagnetic background sources for the
COSINUS experiment in different energy regions. For comparison, the simulated rate predicted by the ANAIS-112 background model (single-hit events) in the [1–6]~keV region is also shown. The associated statistical uncertainty in the ANAIS-112 simulation, not shown in the table, is $\sim$0.1\%.}
\end{table}

These results cannot be directly extrapolated to a final set-up involving multiple detectors, as the present simulation considers only a single crystal. In a multi-detector configuration, the background contribution from \(^{40}\)K in particular would need to be reevaluated, since anticoincidence techniques are expected to lead to a partial reduction in its contribution. Consequently, these results should be interpreted as indicative estimates of the background levels achievable under the assumption that no additional contamination is introduced during the crystal growth or detector assembly processes. In such a case, the $^{40}$K contributions should be regarded as upper limits. 



Table~\ref{tablaexpectedrated} also includes, for comparison, the simulated rates (single-hit events) of ANAIS-112 within the [1–6]~keV range, which defines the ROI for both experiments. Analogously to COSINUS, the value reported for ANAIS-112 does not correspond to a directly measured rate but rather to the background rate predicted by its background model. As observed in Table \ref{tablaexpectedrated}, comparable background levels in the ROI are obtained in this scenario for both experiments. ANAIS crystals exhibit higher $^{40}$K contamination but a lower contribution from natural chains, and includes $^{3}$H and other components such as high out-of-equilibrium $^{210}$Pb.

Unlike the ANAIS-112 simulations discussed in this thesis, the present study does not incorporate the detector time response, i.e., the integration window of the data acquisition system, nor does it include the modelling of threshold and the trigger efficiency during the event-building stage. The omission of the time response of the detector could be relevant in the case of decay chains, where temporal correlations between successive decays may influence the detector response. In the natural chains, there are several isotopes (e.g. $^{228}$Ac) that can produce different signals in different detectors, depending on their time response. For detectors with larger signal integration windows, all the energy will be summed into a single event, whereas this is not the case for detectors with shorter integration windows or dead time, such as ANAIS, where only part of the decay may be recorded. Consequently, including the detector response in the background modelling can have a clear impact for some isotopes, although, as it will be shown later in this chapter, the background estimates are well below the COSINUS design goal to be compromised by such a correction.

\section{Internal radiogenic neutron background}\label{neutronsimulations}

\begin{figure}[b!]
\begin{center}
 \centering
     \includegraphics[width=0.7\textwidth]{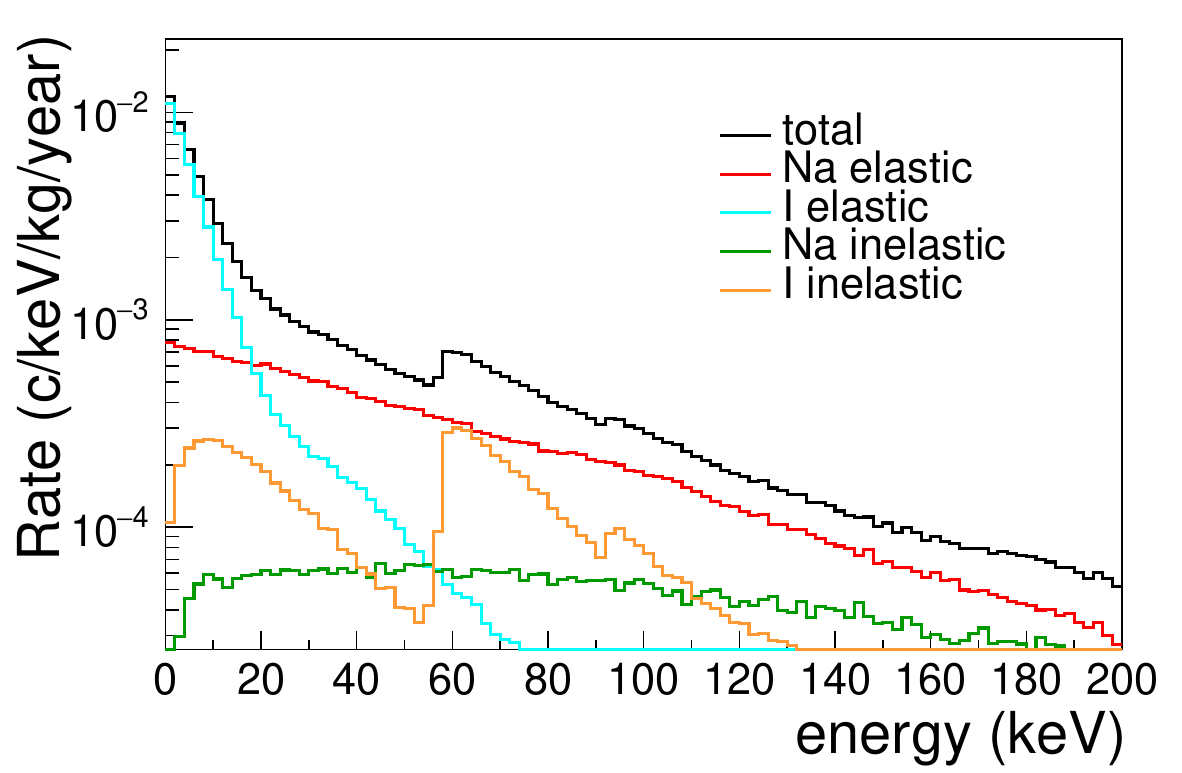}
    
\caption{\label{fondoneutrones} Simulated energy spectrum expected in a COSINUS crystal by internal radiogenic neutrons originating from (\(\alpha\),n) reactions and spontaneous fission, including both elastic and inelastic scattering on Na and I nuclei \cite{privcomMatt}. Natural decay chains of \(^{238}\text{U}\), \(^{235}\text{U}\), and \(^{232}\text{Th}\) were considered. Simulations were performed using the SOURCES-4C code with TENDL-2021 cross sections. }

\end{center}
\end{figure}

Figure \ref{fondoneutrones} shows the simulated internal radiogenic neutron background as modelled by the COSINUS collaboration \cite{privcomMatt}. This simulation includes both (\(\alpha,\textnormal{n})\) reactions and contributions from SF processes originating from naturally occurring decay chains within the NaI crystal itself.


Neutron production calculations were conducted using the SOURCES-4C code \cite{wilson2005sources}, a widely used tool for modelling neutron yields. SOURCES-4C does not directly provide interaction rates within the crystal, but rather calculates the energy spectrum and total flux of neutrons. These results are intended to be used for subsequent neutron transport simulations from the source to the detector using Monte Carlo codes. For this study, the radioactive decay chains of \(^{238}\text{U}\), \(^{235}\text{U}\), and \(^{232}\text{Th}\) were simulated in SOURCES-4C by the collaboration, utilizing cross sections from the TENDL-2021 (TALYS-based Evaluated Nuclear Data Library) library. The transport of neutrons within the detector was then modelled using Geant4. Additionally, the TENDL-2019 version, along with other databases, was compared, showing compatible results.

The NR background from the internal radiogenic neutron background is expected to contribute \((0.066~\pm~0.024)\)~c/kg/year in the [1–10]~keV energy range. For an exposure of 500~kg × year, which will be adopted in this study to estimate the neutron background contribution to the NR bands  (as detailed in the following subsection), this corresponds to an expectation of \(33 \pm 12\) neutron-induced events in the [1–10]~keV range. However, due to the exponential rise of the neutron energy spectrum at low energies (see Figure~\ref{fondoneutrones}), these events are more likely to lie below the energy threshold, depending on the specific detector performance. This estimate further reinforces the prediction that leakage of the internal \(e^-/\gamma\) background will constitute the dominant source of background events in the NR bands.

\vspace{0.8cm}

Once the internal electromagnetic background has been simulated and the radiogenic intrinsic (or internal) neutron background has been characterized, the resulting event distributions in the LY-energy plane are analyzed in the following section. The objective is to quantify the presence of background in the NR bands and to evaluate the detector sensitivity as a function of the energy resolution in both the light and phonon channels.

\section{Simulation of the LY-energy bands}\label{LYyleakage}

This section explores the sensitivity of COSINUS to the DAMA/LIBRA signal and to WIMPs, in general, with special focus on low-mass WIMPs (below 10 GeV/c$^2$), by quantifying the leakage of background-induced events into the NR bands. These false positives, which cannot be distinguished from a genuine DM signal, are expected to predominantly arise from internal $e^-/\gamma$ contaminations in the crystal, as well as from the internal radiogenic neutron background. The contribution of both sources to the NR bands is assessed in the following analysis.

Until this point in the chapter, only the total deposited energy
spectrum of the internal \(e^-/\gamma\) background (Figure \ref{internalbkg}) and the intrinsic neutron background (Figure \ref{fondoneutrones}) has been presented. However, to evaluate the particle discrimination power of COSINUS, it is essential to represent these contributions in the LY-energy plane, according to the corresponding bands defined in Section \ref{bandsdef}.

The process of obtaining the LY-E plane is as follows:

\begin{itemize}
    \item First, the mean of the bands is drawn as given by Equations \ref{meane} and \ref{meanNa}. All free parameters in these expressions are fixed to the values reported in \cite{PhysRevD.110.043010}, except for the baseline noise in both channels (and, consequently, the energy threshold), which will be varied in this study. Afterwards, the 90\% upper and lower boundaries, taking into account the total width of the bands as described in Equation \ref{sigmatot}, are plotted. 

    \item The spectra shown in Figures~\ref{internalbkgplot} and~\ref{fondoneutrones} are sampled a number of times corresponding to the product of their integral and the chosen exposure. It is worth noting that, in the case of the neutron background, only elastic scattering on Na and I nuclei is considered as interaction channel, as the inelastic recoil band has not been modelled in this study. These interactions are independently sampled and subsequently assigned to their corresponding recoil bands. 

In this analysis, results are reported for an exposure of 1 kg × year, which corresponds to a slightly larger exposure than that expected for the first COSINUS run. However, the simulation is performed with a much larger statistical sample. Specifically, an exposure equivalent to 50 kg × year is used for the $e^-/\gamma$ background, while for the neutron background, to ensure sufficient statistics, an exposure of 500~kg~×~year is considered.

 The number of false positives is first computed using this high-statistics dataset and then scaled down by the corresponding factor to estimate the expected leakage for the 1 kg × year scenario. This approach mitigates the statistical limitations inherent to small exposure datasets while preserving a realistic estimation of the leakage.

    \item The sampled energy from the simulation corresponds to the total deposited energy. From this, the energy of the light channel is obtained using Equations \ref{meane} and \ref{meanNa}. Subsequently, the energy of the phonon channel is determined according to Equation~\ref{relacionenes}. Then, the LY is defined as the ratio of the energy deposited in both channels (see Equation \ref{LYeq}).
    
    \item The total width of each band in the light channel is calculated according to Equation~\ref{sigmatot}. This is the width of the band in the E$_L$-E plane. Yet, to obtain the width in the LY-E plane, the obtained width is divided by the energy of the phonon channel following the LY definition.

    \item A LY smearing is applied for each LY value. Specifically, the LY value is randomly sampled from a gaussian distribution centered on that value, with a width corresponding to the total width calculated in the previous step.
    
\end{itemize}

\begin{figure}[t!]
\begin{center}
 \centering
    \begin{minipage}{0.7\textwidth}
        \centering
        \includegraphics[width=\textwidth]{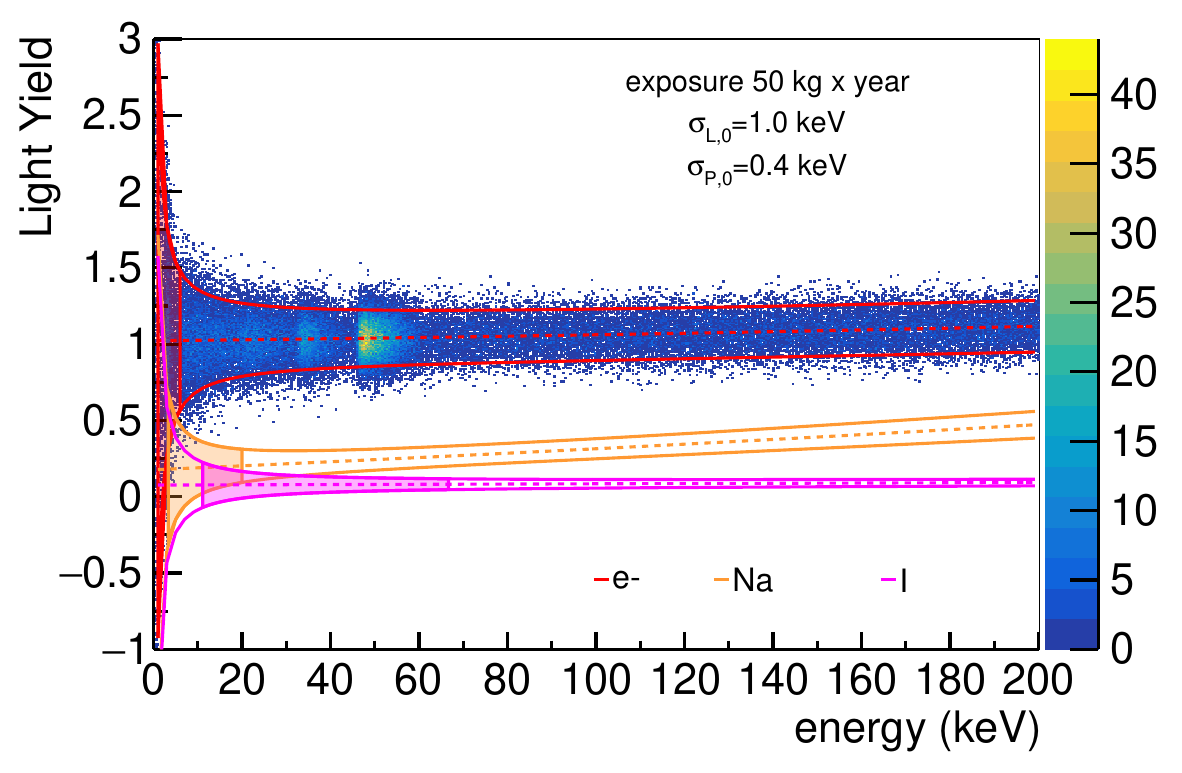}
    \end{minipage}
    \hspace{0.5cm}
    \begin{minipage}{0.7\textwidth}
        \centering
        \includegraphics[width=\textwidth]{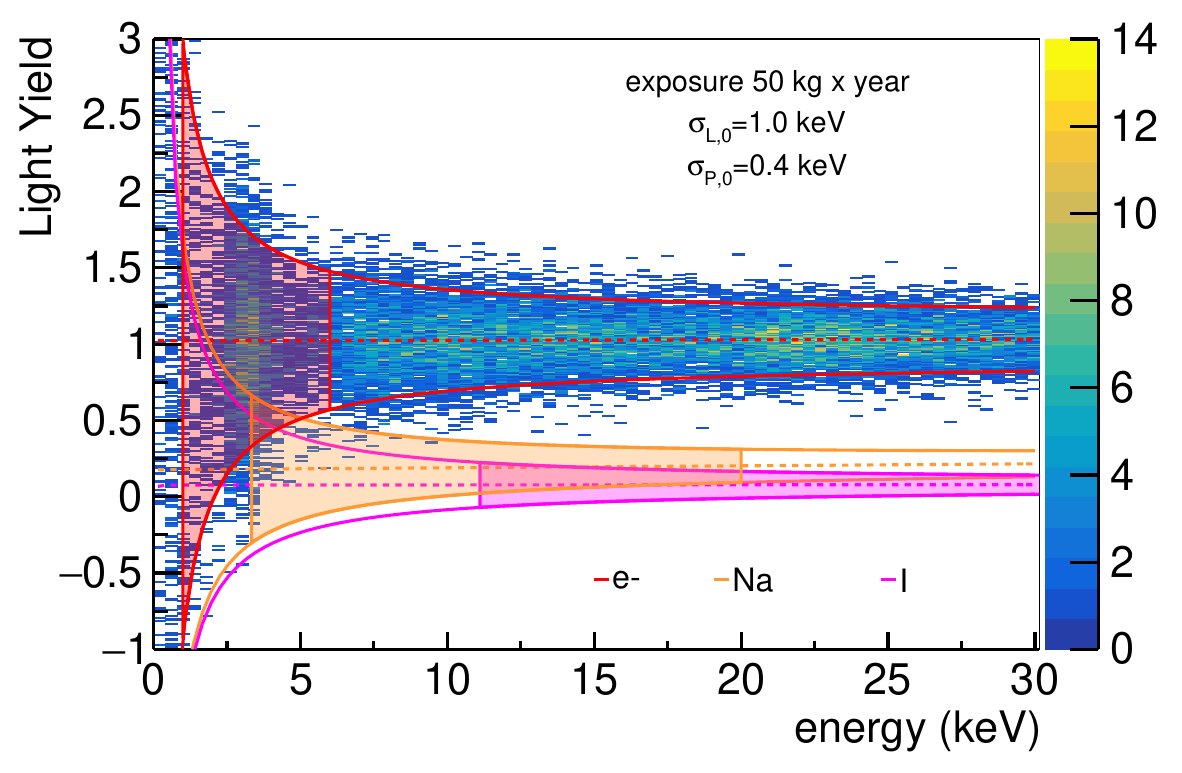}
    \end{minipage}
\caption{\label{LYoficial} Simulated data in the LY-energy plane for an exposure of 50 kg x year, considering the internal electromagnetic background sources expected in COSINUS. Values from the COSINUS module performance, as detailed in \cite{PhysRevD.110.043010}, have been considered for the free parameters in the band modelling. The \textbf{top panel} displays the full spectrum up to 200 keV, while the \textbf{bottom panel} focuses on the low-energy region relevant for DM searches. The red band corresponds to the \(e^-/\gamma\) population (centered around LY=1 by definition), while the orange and magenta bands represent NR off Na and I, respectively. The DAMA/LIBRA ROI ([1–6] keV) is shown in shaded red within the $e^-/\gamma$ band, and in shaded orange and magenta within the Na and I NR bands, respectively, after applying the QF correction.  }
\vspace{-0.3cm}

\end{center}
\end{figure}

Figure \ref{LYoficial} shows the simulated data in the LY-energy plane considering the internal electromagnetic background of COSINUS for an exposure of 50 kg x year. The red band, centered around a LY of 1, represents the \(e^-/\gamma\) band. The LY-energy plot reproduces the spectral shape of the simulated data, revealing, for instance, an accumulation around 50~keV corresponding to the $\gamma$-line at 46.5 keV from the decay of $^{210}$Pb, together with its corresponding $\beta$-spectrum, as well as the low-energy peak arising from \(^{40}\)K contamination. The bands corresponding to elastic scattering on Na and I are visible at lower LYs, shown in orange and magenta, respectively. The DAMA/LIBRA ROI ([1–6]~keV) is shown in shaded red within the $e^-/\gamma$ band, and in shaded orange and magenta within the Na and I NR bands, respectively, after applying the QF correction.




\begin{figure}[b!]
\begin{center}
 \centering
    \begin{minipage}{0.7\textwidth}
        \centering
        \includegraphics[width=\textwidth]{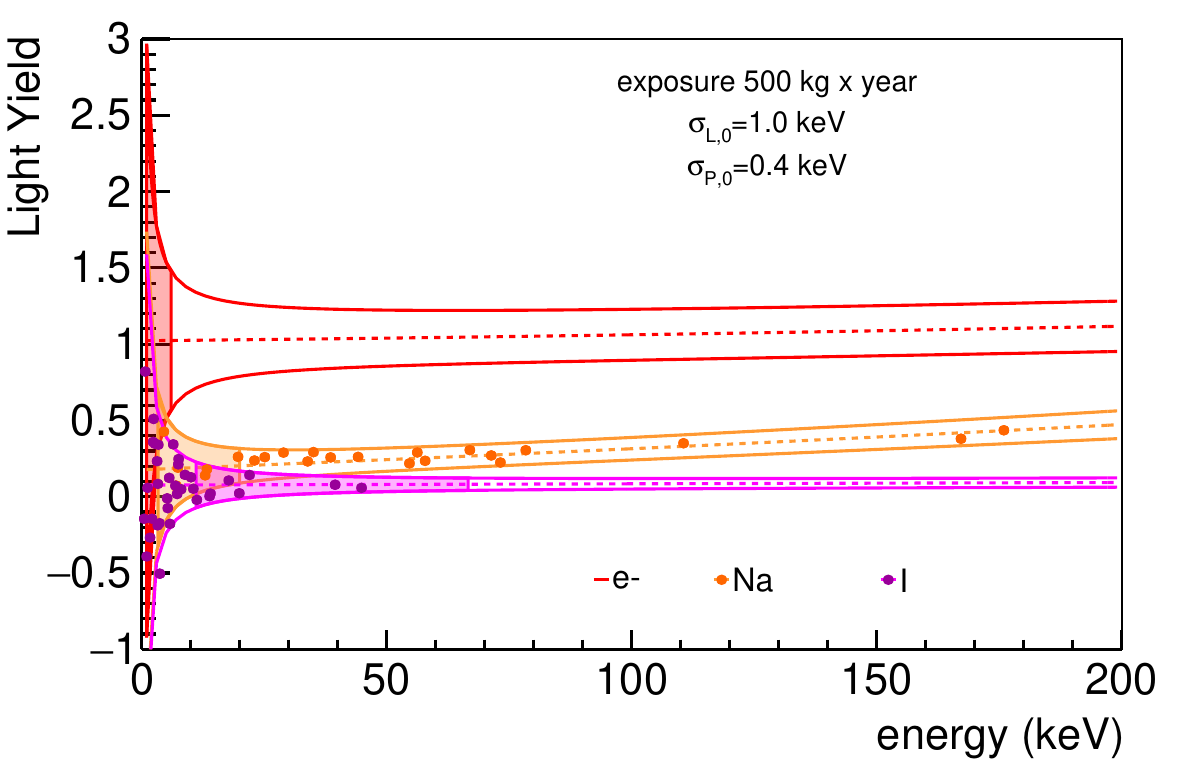}
    \end{minipage}
    \hspace{0.5cm}
    \begin{minipage}{0.7\textwidth}
        \centering
        \includegraphics[width=\textwidth]{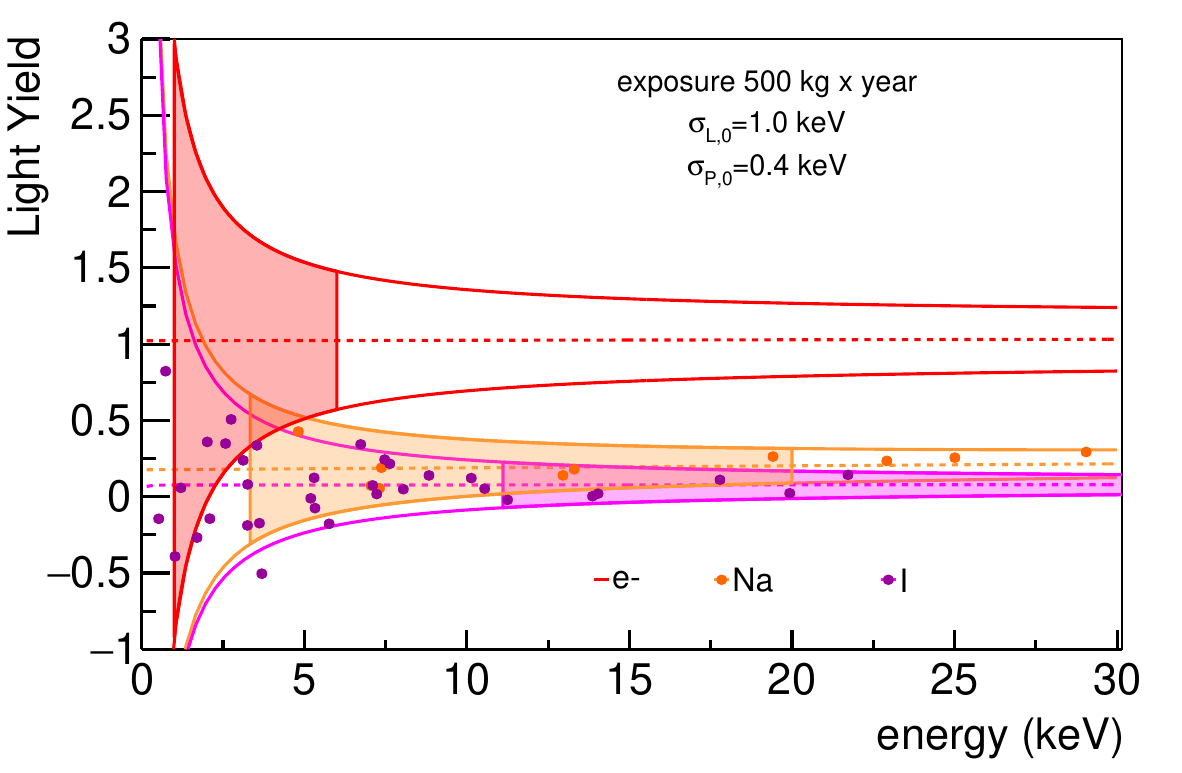}
    \end{minipage}
\caption{\label{LYoficialneutrones} Simulated data in the LY-energy plane for an exposure of 500 kg x year, considering the intrinsic radiogenic neutron background expected in COSINUS. The \textbf{top panel} displays the full spectrum up to 200 keV, while the \textbf{bottom panel} focuses on the low-energy region relevant for DM searches. The elastic scattering events on Na (orange) and I (purple) populate the NR bands accordingly. }

\end{center}
\end{figure}

Figure~\ref{LYoficialneutrones} shows, analogously, the simulated data in the LY-energy plane considering the internal radiogenic neutron background expected in COSINUS for an exposure of 500~kg×~year, a larger exposure chosen, as previously mentioned, to improve statistical significance. In this case, neutron events, producing NRs, populate the NR bands, making them a purely irreducible background in COSINUS and in all DM search experiments under the assumption that WIMPs produce NRs. As seen from the shape of the input spectrum in Figure~\ref{fondoneutrones}, the contribution from iodine is concentrated at low energies, while in the case of sodium it extends across the entire band, with lower probability at low energies.

The presence of false positives in the NR bands, including the DAMA/LIBRA ROI, is non-negligible and must be quantified, which constitutes the primary objective of this study. The rate of false positives in the NR bands depends on several factors, including the contamination levels of the materials, the band width, and the QF values. Additionally, the energy threshold must be considered, as it is related to the phonon channel resolution as 5~$\times$~$\sigma_{\textnormal{P,0}}$. In Figure \ref{LYoficial}, the bands are plotted using typical values from the COSINUS modules performance \cite{PhysRevD.110.043010}. Specifically, COSINUS has achieved a baseline noise of \( \sigma_{\textnormal{L,0}} = 0.988\)~keV in the light channel and \( \sigma_{\textnormal{P,0}} = 0.441 \) keV in the phonon channel.



This study aims to analyze the influence of variations in resolution parameters to evaluate the capability of COSINUS to independently test the DAMA/LIBRA signal and to explore its sensitivity to WIMPs across a broader mass range. For this purpose, the resolution in both channels, \( \sigma_{\textnormal{L,0}}\) and \( \sigma_{\textnormal{P,0}}\), will be varied within reasonable ranges. In particular, \(\sigma_{\textnormal{L,0}}\) will be varied from 0.02 to 2 keV in intervals of 0.2 keV, while \(\sigma_{\textnormal{P,0}}\) will be varied from 0.05 to 0.5 keV in intervals of 0.05 keV. The latter implies an energy threshold ranging from 0.25 to 2.5 keV. Values beyond this range are not considered, as an energy threshold higher than 2.5 keV is outside the scope of the COSINUS current performance plans.

\begin{figure}[t!]
\centering
    \begin{minipage}{0.7\textwidth}
        \centering
        \includegraphics[width=\textwidth]{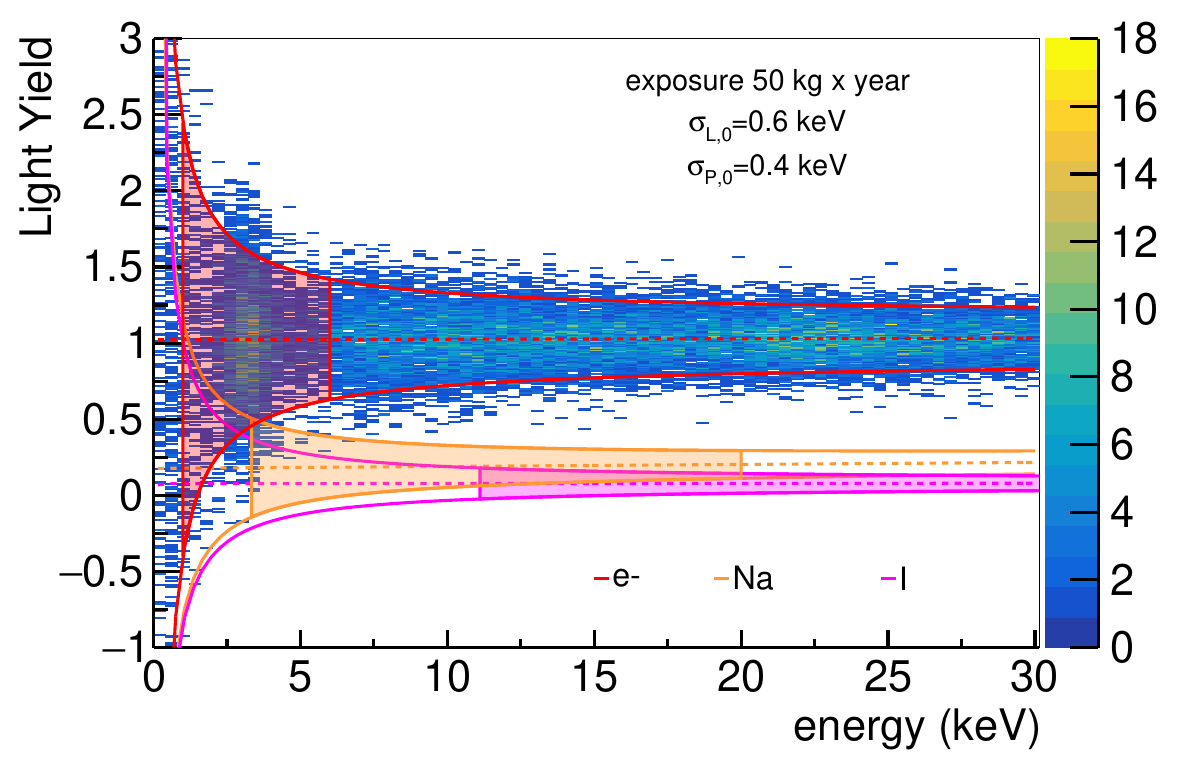}
    \end{minipage}
    \hspace{0.5cm}
    \begin{minipage}{0.7\textwidth}
        \centering
        \includegraphics[width=\textwidth]{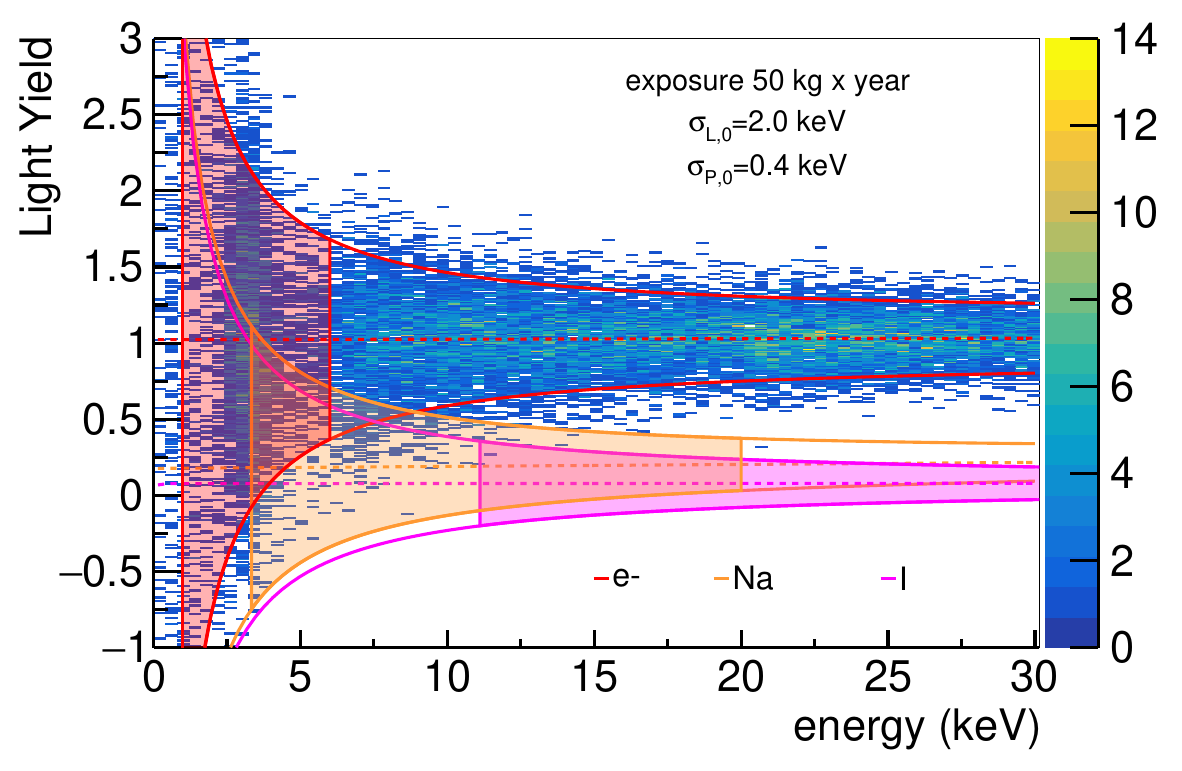}
    \end{minipage}

\caption{\label{LYchangingL} Simulated data in the LY-energy plane for an exposure of 50 kg$\times$year, considering the internal electromagnetic background sources expected in COSINUS. The \textbf{top panel} shows the distribution with a light energy resolution of 0.6 keV, while the \textbf{bottom panel} corresponds to a light detector resolution of 2 keV. The top panel demonstrates a clearer separation between the bands, enabling more reliable particle discrimination.}
\end{figure}

In Figure~\ref{LYchangingL}, simulated data are displayed in the LY-energy plane for two different scenarios regarding the light channel resolution. The top panel corresponds to an energy resolution of 0.6~keV, representing an optimistic yet achievable performance. The bottom panel shows a significantly degraded resolution of 2~keV, considered here as an overly conservative case. As illustrated in the top plot, a better light resolution leads to a clearer separation between bands, particularly at low energies, thereby enhancing the particle discrimination capability. Conversely, a poor light resolution rapidly increases the leakage of electron and gamma events into the NR bands.

False positives are restricted to deposited energies below 15~keV in all cases, and occur first in the sodium band, owing to its higher LY compared to that of iodine recoils. Events with higher LY are more likely to originate from ERs. To mitigate this leakage, particularly at low energies, the region in which WIMP signals are searched for is conservatively limited to events with LY values below the mean of the highest NR band \cite{angloher2016cosinus}, in this case sodium. This selection criterion, originally introduced by the CRESST collaboration for DM searches, is also applied here to quantify the presence of background events in the NR bands. This selection, applied to reject events with large LY, results in a corresponding reduction in signal efficiency, by definition, down to 50\% for sodium recoils, and slightly higher for iodine recoils, as previously calculated in \cite{kahlhoefer2018model}.

Variations in the phonon channel resolution do not significantly affect the width of the recoil bands, which is mainly determined by the resolution of the light detector. However, the phonon resolution governs the energy threshold, which remains relevant. A lower threshold increases the background and may worsen leakage, but it also enhances the sensitivity to the potential DM signal, which is expected to concentrate at low energies. It is also worth noting that an improved phonon resolution enhances the total energy resolution of gamma peaks, which, when leaking into the NR bands, can be more effectively discriminated from a DM signal in the maximum likelihood framework.

To quantify the contribution of background events into the NR bands, two parallel analyses are conducted, taking into account both the leakage of $e^-/\gamma$ events from the internal electromagnetic background and the presence of events from the internal radiogenic neutron background within the NR band, representing an irreducible background that cannot be discriminated. 

First, the ability of COSINUS to test the DAMA/LIBRA signal is evaluated by analyzing the presence of background events in the DAMA/LIBRA ROI. This is done under the assumption that the DAMA signal ([1-6] keV) originates from NRs, and after applying the appropriate QF correction. Second, a more general study is performed by assessing the occurrence of false positives across the full NR band starting from the energy threshold. The number of events falling within these two regions is then counted considering simultaneously the total background contribution in both NR bands; that is, false positives coming from $e^-/\gamma$ internal electromagnetic background and from the elastic scattering on both Na and I are taken into account by integrating over both bands. The 90\%~C.L. upper limit is derived according to poisson statistics for an high exposure and subsequently scaled to a 1~kg~x~year exposure, which is the reference value for reporting results in this work.

\begin{figure}[b!]
\centering
\includegraphics[width=0.7\textwidth]{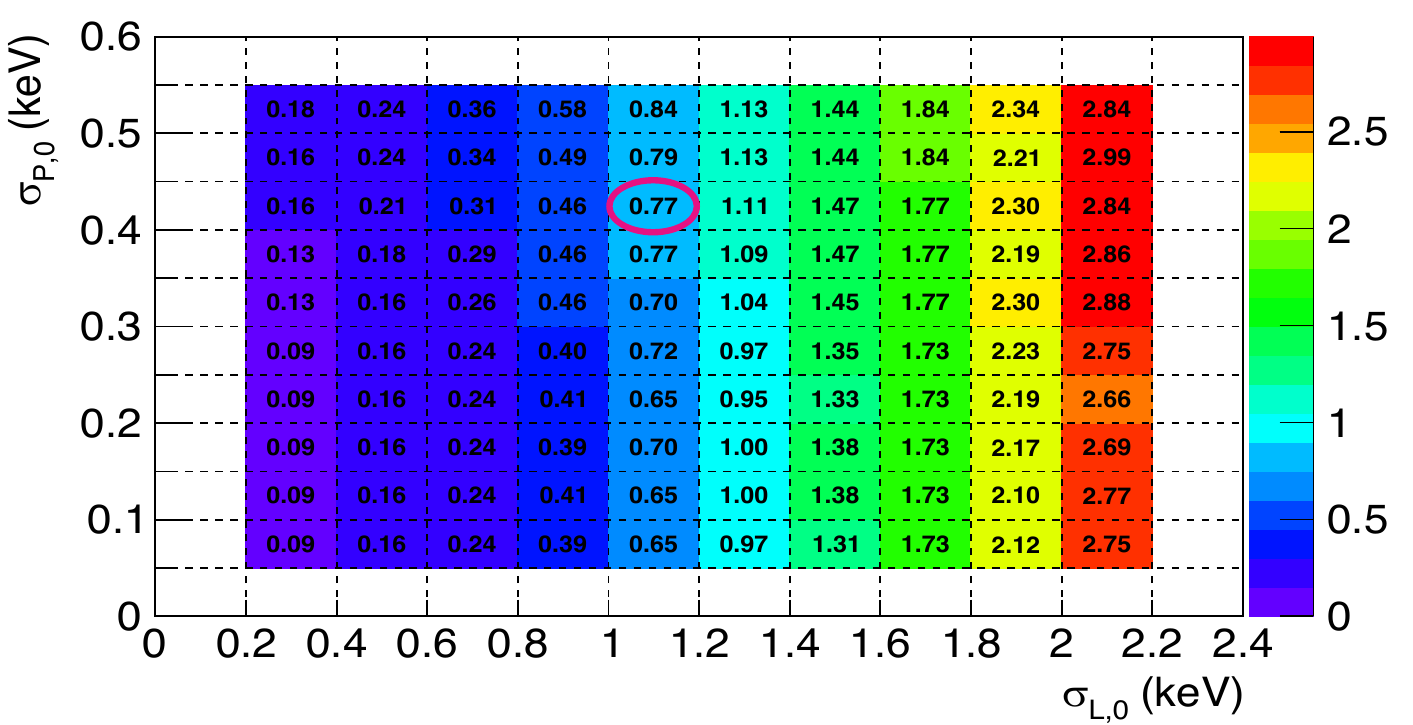}
\caption{\label{resultadoDAMA} Expected rate of false positives in the DAMA/LIBRA ROI of the NR bands ([1--6]~keV, applying the QF correction accordingly) for an exposure of 1 kg $\times$ year, under various energy resolution configurations for the light and phonon channels. The values correspond to the 90\% confidence level upper limit derived according to Poisson statistics and are expressed in c/kg/year units. The figure highlights the expected event rate based on the current performance of the COSINUS detectors ($\sigma_{\textnormal{P,0}} = 0.4$ keV and  $\sigma_{\textnormal{L,0}} = 1$ keV) \cite{PhysRevD.110.043010}.}
\end{figure}

Figure \ref{resultadoDAMA} shows the expected rate of false positives (90\% C.L. upper limits) derived from the study in the DAMA/LIBRA ROI of NR bands for various energy resolution configurations in the light and phonon channels. On the one hand, worsening the resolution of the light channel increases the leakage into the NR bands, as expected given the broader band widths shown in the bottom panel of Figure \ref{LYchangingL}. On the other hand, since the DAMA/LIBRA ROI in the NR bands lies above the energy threshold, it is observed that the resolution of the phonon channel does not have a significant effect on the leakage study in the DAMA/LIBRA ROI.

For a current light channel resolution of 1 keV and a phonon channel resolution of 0.4~keV, as reported in \cite{PhysRevD.110.043010}, the 90\% upper limit is 0.77 c/kg/year. The official COSINUS performance goal is set at an event rate of 0.1 c/kg/day in the DAMA/LIBRA signal region, that is, 36.5 c/kg/year. Thus, it can be concluded that the background leakage predictions are well below the established limit, so with the backgrounds considered in this study and the performance status already achieved, COSINUS is sufficient to exclude a DM interpretation of DAMA/LIBRA signal in this scenario.

\begin{figure}[t!]
\begin{center}
 \centering
     \includegraphics[width=0.7\textwidth]{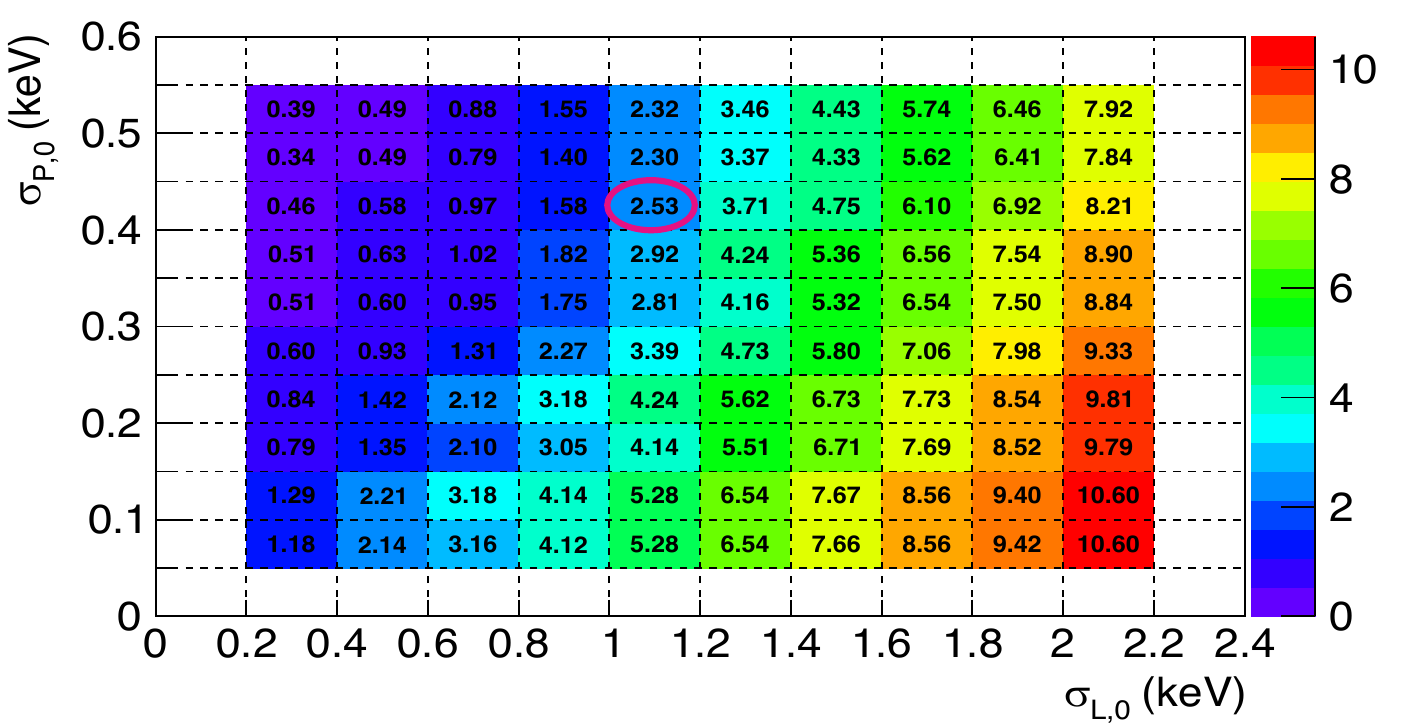}
    
\caption{\label{resultadogeneral} Expected rate of false positives in the full NR bands (from energy threshold upwards) for an exposure of 1 kg × year under various energy resolution configurations of the light and phonon channels. The values correspond to the 90\% confidence level upper limit derived according to poisson statistics and are expressed in c/kg/year units. The figure highlights the expected event rate based on the current performance of the COSINUS detectors (\(\sigma_{\textnormal{P,0}}\) = 0.4 keV and  \(\sigma_{\textnormal{L,0}}\) = 1 keV) \cite{PhysRevD.110.043010}.
\vspace{-0.5cm}
}

\end{center}
\end{figure}

Figure \ref{resultadogeneral} presents the analogous analysis for the full NR bands starting from the energy threshold. In this case, the number of leakage events decreases with increasing \(\sigma_{\textnormal{P,0}}\), i.e. with a worse phonon channel performance. This behavior may appear misleading at first glance. Nevertheless, it is expected when considering the whole band and increasing the phonon threshold/resolution while just counting the number of leakage events. Therefore, in this case, the reduction in leakage comes at the cost of a decreased sensitivity to the DM signal. However, lowering the energy threshold represents a significant advantage for DM searches, since the expected signal is concentrated at low energies.




For the current performance status of the experiment, the resulting 90\% C.L. upper limit on the total rate in the NR band is 2.53 c/kg/year. Results indicate that efforts should primarily focus on improving the resolution of the light channel, as this proves to be the most effective strategy for reducing the rate of non WIMP-like events in the NR bands. Another possible approach to mitigate the background contribution, emphasizing the importance of the development of dedicated simulations, is the modification of the acceptance region of the NR band. If simulations point towards a high number of leakage events in a specific region of the LY-energy plot, such as the 3.2 keV line from the EC $^{40}$K decay (see Figure~\ref{LYoficial}), that region can be excluded from the ROI definition in order to reduce the number of false positives. Nonetheless, it is crucial that such decisions are made prior to data analysis, in compliance with a blind analysis.

Both Figure~\ref{resultadoDAMA} and Figure~\ref{resultadogeneral} account for two distinct sources of background contributing to the NR bands: (i) the leakage of \(e^-/\gamma\) events originating from the internal electromagnetic background, which enter the NR bands due to finite detector resolution, and (ii)~the irreducible internal radiogenic neutron background, whose interactions with the detector nuclei intrinsically produce NRs that are indistinguishable from those induced by DM. It is important to note that the event counts within both defined acceptance regions have been evaluated jointly, and the corresponding 90\% confidence level upper limits have been derived accordingly. Nevertheless, the contribution from intrinsic radiogenic neutrons has been found to be negligible in this study, and the dominant limiting factor in this analysis is clearly the \(e^-/\gamma\) leakage from internal contamination.


\section{Conclusions}

This chapter presents the work carried out by the PhD candidate during her research stay at the Max Planck Institute for Physics in Munich collaborating within the COSINUS experiment. The primary goal of COSINUS is to provide a model-independent cross-check of the annual modulation signal observed by DAMA/LIBRA, using the same target, NaI, but employing a different technology. As a NaI cryogenic scintillating calorimeter, COSINUS offers several advantages over traditional single-channel experiments: in-situ QF measurements, and signal-to-background discrimination based on the dual-readout of light and phonon signals.

Throughout the stay, the internal electromagnetic background model of the COSINUS experiment was developed, which is expected to be the most significant contribution, as observed in other NaI-based DM search experiments \cite{amare2019analysis,adhikari2021background,barberio2023simulation}. The contribution of this background was analyzed in comparison with the background model reported by ANAIS-112. Additionally, the internal radiogenic neutron background simulations previously conducted by the collaboration \cite{privcomMatt} were presented. Both contributions were studied by examining the distribution of the corresponding events in the LY-energy bands for different experimental performance parameters in order to assess COSINUS sensitivity to both the DAMA/LIBRA signal and more general and still viable WIMP candidates, as SD-interacting light mass WIMPs.

It was found that under typical ultra-low background conditions in underground laboratories, the NR bands remain sparsely populated; however, the presence of false positives, defined as those events that fall within the NR bands and cannot be distinguished from a DM signal, is non-negligible. The contribution of background events was estimated both in the DAMA/LIBRA ROI within the NR bands (with the QF correction applied accordingly) and across the entire NR band, starting from an energy threshold determined by the performance of the phonon detector.

In the DAMA/LIBRA NR ROI, the false positive rate (90\% upper limit) for the current performance status of the experiment (\( \sigma_{\textnormal{L,0}} = 0.988\)~keV in the light channel and \( \sigma_{\textnormal{P,0}} = 0.441 \)~keV in the phonon channel) is 0.77 c/kg/year. The COSINUS design goal is to achieve 36.5 c/kg/year in the DAMA/LIBRA ROI, so the results of this study further validate the performance goals of the experiment.


When analyzing the entire NR band, the expected rate (90\% upper limit) with the current performance is 2.53 c/kg/year, with the light channel resolution being the limiting factor. However, the resolution of the phonon channel affects the presence of background signals (and consequently the sensitivity to DM signals) by modifying the energy threshold. In this case, evaluating the signal-to-background ratio requires the selection of a specific WIMP model, making the analysis model-dependent.

The study conducted in this thesis has demonstrated that the contribution of intrinsic radiogenic neutrons to this estimate is negligible, and therefore, the primary concern should be the \(e^-/\gamma\) leakage into the NR bands. However, it should be noted that this analysis only considers the internal electromagnetic background of COSINUS. The NaI crystals for the COSINUS experiment arrived at LNGS in early 2025, and their radiopurity levels will need to be reassessed. Consequently, the final analysis of non-DM-induced leakage events into the NR bands must account for potential cosmogenic activation of the crystals, the actual contamination of \(^{210}\)Pb, specially dangerous if found in the crystal surface, and contributions from external components of the cryostat. 

In addition, the ambient radiogenic neutron background should also be taken into account. These neutrons originate from ($\alpha$,n) and SF reactions occurring in the surrounding rock or construction materials of the underground laboratory. High-energy neutrons may also be produced by interactions of the residual muon flux with the materials surrounding the detectors. However, due to the substantial overburden at LNGS and the implementation of an active water Cherenkov muon veto in COSINUS to reject such events \cite{angloher2024water}, the contribution from muon-induced neutrons is expected to be subdominant compared to that of radiogenic neutrons. These factors are currently being addressed through ongoing screening and Monte Carlo simulations aimed at developing a comprehensive background model for the experiment.



An additional background component expected to play a significant role under real experimental conditions is the occurrence of events induced by mechanical vibrations or originating in or near the temperature sensor (decoupled from the NaI crystal in the remoTES design). In such cases, no scintillation light is detected, but heat is still produced and a phonon signal is recorded.
 These events, which may originate from \(e^-/\gamma\) interactions, could fall within the NR bands, making them indistinguishable from DM signals. While quantifying their contribution is challenging within the current simulation framework, it is essential to recognize their potential impact, as they may represent the dominant limitation to the experiment sensitivity.

Having demonstrated its capability to test the DAMA/LIBRA signal, if COSINUS aims to achieve sensitivity to low-mass WIMPs, several strategies can be pursued. On the one hand, the reduction of the $e^-/\gamma$ background appears feasible in some cases, particularly for natural decay chains. Additionally, it will likely be necessary to improve the resolution of the light channel and to further lower the energy threshold. Regarding the latter, the COSINUS design goal is to achieve a threshold around 1 keV, corresponding to a phonon channel resolution of \(\sigma_{\textnormal{P,0}}\) = 0.2 keV. Achieving this goal will require further improvements in phonon channel performance. Work in these directions is already underway by the collaboration. Significant efforts are currently being directed towards improving energy resolution, with efforts focused on improving the crystal light output, module geometry, and light detector resolution.


In conclusion, this work has validated the sensitivity goals of COSINUS in testing the DAMA/LIBRA signal in a model-independent way and its potential for WIMP searches. Together with ANAIS-112, the COSINUS experiment is poised to contribute significantly to resolving the DAMA/LIBRA puzzle in the coming years. Despite its limited exposure, one of its most notable strengths is its ability to discriminate between NR and ER recoils on an event-by-event basis, providing valuable insights into potential DM signals.

\raggedbottom

\chapter*{} 
\vspace{-2cm}
\label{Chapter:Conclusions}
\addcontentsline{toc}{chapter}{Summary and conclusions} 

{\begin{center}
   \hrule height 0.5pt
    \vspace{1ex}
    \hrule height 0.5pt
    \vspace{2ex}
    {\normalfont\LARGE \bfseries Summary and conclusions}
    \vspace{2ex}
    \hrule height 0.5pt

    \vspace{0.4cm}
\end{center}}

\pagestyle{plain}

Overwhelming astronomical and cosmological evidences across different scales and times of the history of the Universe points to the existence of dark matter (DM). However, to this day, the nature of this elusive component, which constitutes the 27\% of the Universe, still remains one of the most puzzling questions in particle physics, cosmology, and astrophysics.

DM cannot be accounted for by any of the particles included in the Standard Model of particle physics, as none of them exhibit the properties required to reproduce the cosmological and astrophysical observations. DM particles must be massive,  electrically neutral (or at most extremely weakly charged), non-relativistic at the time of galaxy formation, non-baryonic, and stable or very long-lived on cosmological timescales. Moreover, beyond their gravitational interaction, weak couplings with ordinary matter are allowed. This aspect is fundamental for establishing direct detection methods. Among the preferred DM candidates are axions and Weakly Interacting Massive Particles (WIMPs). In the case of WIMPs, the interaction is expected to occur preferably with the nuclei of a convenient detector in the most standard DM interaction scenarios.

Discovering the particle responsible for DM, or simply unveiling more details about its nature, is a complex task that requires complementary strategies: searches for new particles and new physics at large particle accelerators, indirect detection through the products of DM annihilation, and direct detection by identifying energy deposits produced by DM interactions in suitable detectors. Despite significant experimental efforts and remarkable improvements in sensitivity over recent decades, none of these approaches has yet succeeded, although many proposed candidates for DM have been ruled out along the way.

However, a positive result remains noteworthy. For over 20 years, the DAMA/LIBRA
experiment, which uses NaI(Tl) crystal scintillators at the Gran Sasso National Laboratory, in Italy, has provided a long-standing positive result: the observation of a highly statistically significant annual modulation in the detection rate, compatible with that expected for galactic halo DM particles. To independently test the DAMA/LIBRA result, without relying on any specific DM particle or halo model, the primary requirement is to use the same target material, NaI. This is the goal of several experiments within the international context, including ANAIS-112, COSINE-100, SABRE and COSINUS.

This thesis is framed within ANAIS-112, a direct detection DM experiment aiming to confirm or refute the DAMA/LIBRA result in a model-independent way, using 112.5~kg of ultrapure NaI(Tl) scintillators. For this purpose, ANAIS-112 has been operating since August~2017 at the Canfranc Underground Laboratory (LSC) in Spain. Since then, ANAIS-112 has been smoothly taking data, achieving an average live time of 95\% and accumulating nearly eight years of exposure. In 2025, ANAIS has published the results corresponding to six years of data taking. These results, currently the most sensitive using the same target material, NaI(Tl), are incompatible with the DAMA/LIBRA modulation signal at a 4$\sigma$ confidence level (C.L.).

Throughout this thesis, the work carried out within the ANAIS-112 experiment has been presented, with a particular focus on data analysis and the development of Geant4 based simulations, with the goal of reducing systematic effects and increasing the experimental sensitivity.

Firstly, efforts have been directed towards improving the understanding of the response of the ANAIS-112 crystals to nuclear recoils (NRs). The results from ANAIS based on six years of data, strongly challenge the DM interpretation of the DAMA/LIBRA signal. However, there are systematic uncertainties affecting the comparison that need to be addressed. In order to reliably compare the results of ANAIS and DAMA/LIBRA, the scintillation quenching factors for sodium (QF\textsubscript{Na}) and iodine (QF\textsubscript{I}) recoils in NaI(Tl) crystals must be precisely known, particularly if the DM signal is expected to arise from NRs, as expected in the most standard DM interaction scenarios. The QF is defined as the amount of scintillation light produced by a NR, relative to that produced by an electron recoil (ER) depositing the same energy. QFs are thus essential for converting NR energy into electron-equivalent energy (commonly referred to as the visible energy scale), since experiments are typically calibrated using electron/gamma sources.

To date, the QF\textsubscript{Na} and QF\textsubscript{I} in NaI(Tl) remain poorly constrained, with significant experimental uncertainties affecting both their absolute value and their energy dependence. Moreover, no theoretical model currently explains the results satisfactorily. Particularly noteworthy is the difference in the case of the DAMA/LIBRA measurements. While most recent measurements show values for QF\textsubscript{Na} around 20\%, and increasing with energy, DAMA/LIBRA reports a constant value of 30\%. Similarly, for QF\textsubscript{I}, the value reported by DAMA/LIBRA is approximately 50\% higher than recent measurements (9\% vs. 5\%).

Given that this parameter represents the most relevant systematic uncertainty in the comparison with DAMA/LIBRA, ANAIS-112 has conducted a dedicated neutron calibration program. On the one hand, QF measurements of small NaI(Tl) crystals, similar to those used in ANAIS-112, were carried out using a monoenergetic neutron beam at the Triangle
Universities Nuclear Laboratory (TUNL). On the other hand, onsite neutron calibrations have been conducted since 2021 using low-activity \(^{252}\)Cf sources at the LSC. This thesis has focused on the latter approach.

Onsite neutron calibrations serve three primary objectives within the experiment: evaluating the ANAIS-112 event selection efficiency, generating signal-like events in the crystal bulk for machine learning training, and improving the understanding of the QF. In this context, these onsite neutron calibrations provide an important cross-check for the results obtained at TUNL using crystals grown from the same batch as some of the ANAIS crystals. The final goal outlined above, a better understanding of the QF\textsubscript{Na} and QF\textsubscript{I} of the ANAIS crystals, has been pursued in this thesis by comparing the spectra measured during onsite neutron calibrations with those obtained from dedicated Geant4-based neutron simulations.

The calibration is based on the exposure of the full detector array to low-activity \textsuperscript{252}Cf neutron sources, placed outside the anti-radon box and the lead shielding, but inside the muon veto system and the neutron moderator. Neutron interactions have been found to primarily produce bulk scintillation events from elastic scattering off sodium and iodine nuclei, with an even cleaner population when selecting multiple-hit events. Moreover, this analysis has enabled the observation of the
distinct temporal responses of NR and ER by comparing the average pulse shapes corresponding
to each population, yielding an average ratio of the scintillation decay time constant for NRs to that of ERs of $\sim$0.85, consistent with previous results.

The neutron simulations conducted to compare with the neutron calibration data have enabled a critical evaluation of the Geant4 physics databases. Specifically, differences between two Geant4 versions, v9.4.p01 and v11.1.1, regarding cross sections and the presence and intensity of certain inelastic scattering lines have been observed, which impact the agreement between simulation and experimental data. Consequently, Geant4 version v9.4.p01 has been selected for this work due to its better description of the measured data. Additionally, an overestimation of the neutron capture cross section producing \textsuperscript{128}I has been identified in all the versions. 

Several models based on experimental measurements have been proposed and tested. In particular, for QF\textsubscript{Na}, the DAMA/LIBRA result, the three results obtained from measurements at TUNL on crystals similar to those of ANAIS using different calibration procedures, the COSINUS result, and the Tretyak model have been evaluated. Regarding the ANAIS results, to describe the dependence on energy, a modified Lindhard model has been fitted to the ANAIS QF\textsubscript{Na} measurements and compared with a simple linear fit. As for QF\textsubscript{I}, the DAMA/LIBRA result, the measurement obtained for the ANAIS crystals at TUNL, and an energy-dependent QF\textsubscript{I} model proposed in this work, compatible with the ANAIS-112 measurements, have been studied.

 The comparison between data and simulation demonstrates that ANAIS simulation has a very good capability to reproduce the response of the detectors, validating the modelling of the experimental set-up and the DAQ. Although this study does not allow for the extraction of an explicit energy dependence of the QF, as performed in monochromatic source studies involving the measurement of the scattered neutron angle and corresponding NR energy, it enables the comparison of different QF models. In particular, the DAMA/LIBRA QF models are disfavored due to poorer agreement with ANAIS-112 data, as well as QF\textsubscript{Na} models exhibiting a decreasing dependence on energy. 
 
 Regarding the QF\textsubscript{Na} of the ANAIS crystals measured at TUNL, the comparison between data and simulation suggests a preference for increasing with energy QF\textsubscript{Na} models rather than constant ones. Moreover, it has been verified
that the modified Lindhard model provides a slight better description than a linear dependence of QF\textsubscript{Na} with energy. Other energy dependencies cannot
be ruled out, although a stronger reduction of the QF\textsubscript{Na} at lower energies appears to be favored. Concerning iodine, the data-simulation agreement improves significantly when employing an energy-dependent QF\textsubscript{I} compatible with ANAIS-112 results, although the 6\% value reported by ANAIS at 15 keV\textsubscript{NR} and considered here under the hypothesis of being constant with energy also yields a reasonable agreement. Although high-multiplicity events show poorer agreement with the data, a fact that will need to be investigated in future studies, they represent only a small fraction of the total neutron calibration events. The majority of events, specifically the single-hit and m2-hit populations, are very well reproduced by the simulation within the associated uncertainties.

Overall, the preferred QF model for the ANAIS-112 crystals derived from this work is an energy-dependent QF\textsubscript{Na} derived from measurements performed at TUNL using a proportional calibration based on the inelastic peak from the \textsuperscript{127}I(n,n'$\gamma$) process; and an energy-dependent QF\textsubscript{I} consistent with the values obtained for ANAIS crystals. In addition, it has been verified that all the ANAIS-112 crystals exhibit the same
NR spectrum in response to \textsuperscript{252}Cf sources, despite originating from different ingots, though grown following
similar protocols, 
confirming compatible results of the measurements performed at TUNL across several NaI(Tl) crystals built at Alpha Spectra.


The Geant4 version choice has been analyzed as a source of systematic uncertainty affecting the QF determination. The agreement remains satisfactory with the newer version, supporting the selected QFs and disfavoring the DAMA/LIBRA QFs due to their poorer agreement with the ANAIS-112 data. Moreover, the conclusions related to the QFs selected in this thesis have been validated using data from the new ANOD acquisition system, installed at LSC in December 2024, which features a longer acquisition window and operates without dead time. With the ANOD data, the agreement in the low-energy multiple-hit population is remarkably good and improves compared to that obtained with the ANAIS data.


On the other hand, neutron calibrations could be performed using alternative neutron sources. Of particular interest is the use of monoenergetic neutron sources, such as accelerator-based or Y-Be sources, which offer the highest sensitivity to variations in QF values and that should not be affected by the same systematics than \textsuperscript{252}Cf sources simulation. Additionally, these future calibration campaigns will allow the exploration of several sources of uncertainty in the Geant4 modelling of elastic scattering, the neutron multiplicity spectrum from \textsuperscript{252}Cf, and neutron interactions in lead. The latter two may help clarify the observed discrepancies in the multiple-hit spectrum, where agreement with data worsens as multiplicity increases.

Moreover, a particularly relevant future scenario would involve distributing one or more crystals from the DAMA/LIBRA experiment within the ANAIS, COSINE, or SABRE collaborations. Their operation in the ANAIS set-up, in particular, would allow not only for a direct characterization of the DAMA/LIBRA background, but also for an independent determination of their QFs using the same methodology applied to the ANAIS-112 crystals in this work, either with \textsuperscript{252}Cf or alternative neutron sources. Although such a measurement might not unambiguously determine the individual QF\textsubscript{Na} and QF\textsubscript{I} values in NaI(Tl), it could nonetheless shed light, once and for all, on whether the DAMA and ANAIS crystals differ in their response to NRs, thereby removing the major systematic uncertainty from the comparison of both experiments.

In parallel, this thesis has contributed to the revision and improvement of the ANAIS-112 background model through a multiparametric fit of the various contributions, an approach implemented for the first time in the context of ANAIS-112.

Several key aspects were identified while developing the fitting strategy. To start with, an asymmetry in light sharing between photomultipliers (PMTs) was identified, revealing a population of strongly asymmetric single-hit events between $\sim$ [75-350] keV. To investigate their nature, events were classified into symmetric and asymmetric populations, revealing a clear change in the spectral shape of the asymmetric single-hit events around 80~keV, likely due to energy depositions from PMT emissions. Simulations supported this hypothesis, showing position-dependent effects that motivate the implementation in the future of a position-sensitive light collection model. Additionally, assuming bulk contaminations are symmetric, the results suggest the presence of a localized surface contamination by \textsuperscript{210}Pb on the polished ends of the NaI(Tl) crystals.

Subsequently, the PMT modelling has been revised to include a frontal contamination of \textsuperscript{226}Ra and \textsuperscript{232}Th on the photocathode, in addition to the borosilicate glass, as suggested by dedicated HPGe measurements conducted during this thesis. This frontal component is more effective at depositing energy in the crystal. For \textsuperscript{210}Pb, both bulk and surface contaminations within the crystals were considered, as well as surface contamination on both sides of the teflon coating surrounding the crystals. Regarding surface contamination in the crystal, this study has considered exponentially decaying density profiles, rather than fixed-depth layers as assumed in the previous background model, in order to account for the diffusion of the contaminant isotope into varying depths. In particular, the performance of three benchmarks depths (1, 10, and 100 $\mu$m) has been tested. Additionaly, a correction to the $\beta^-$ spectrum for $^{210}$Bi decay has been implemented, incorporating a spectrum derived from experimental measurements, since the modelling included in Geant4 was found not to adequately reproduce the measured data.  This correction improves the agreement between simulation and measurements.

The fit performance at each stage has been evaluated, comparing the fitted activity results with those from the previous background model. A 10 $\mu$m depth has been adopted as the most plausible scenario for the $^{210}$Pb surface contamination. Furthermore, the fitted activity values of bulk versus surface \textsuperscript{210}Pb contamination do not support the earlier assumption made in the ANAIS-112 model regarding the shape of the alpha backgrounds. Two alpha energy deposition peaks are observed in each ANAIS crystal, previously attributed to surface (low-energy peak) and bulk (high-energy peak) \textsuperscript{210}Po decays in a proportion that is not consistent with the results from this work. While all ANAIS crystals exhibit consistent neutron response, they lack the same uniformity for alpha particles. No explanation has yet been found for the structure seen in the alpha energy deposits. Nevertheless, ongoing efforts using the ANOD DAQ system aim to identify Bi-Po sequences and explore possible spatial dependencies that may clarify the behavior observed with $^{210}$Pb.


The new background model yields significant improved agreement with data across all energy regions and populations compared to the previous model. The model's robustness is further supported by its accurate reproduction of coincidence events and spectra obtained from year-to-year differences.

Two additional background sources have also been evaluated. First, the environmental neutron background in hall B of LSC, which has been modeled in this thesis using GEANT4 and compared with the results using FLUKA obtained by HENSA, a neutron spectrometer installed near ANAIS-112 in 2019. Discrepancies between codes are non-negligible, particularly for fast neutrons, but the overall contribution of the neutron background is non relevant relative to the measured background. Second, a population of light noise events that pass the ANAIS-112 event selection filters, referred to as asymmetric events due to their asymmetry in light sharing among both PMTs and identified only with ANOD due to its different acquisition characteristics, has been incorporated as a time-independent component. Their rates align with part of the unexplained excess in the [1–2] keV region in ANAIS data.

In ANAIS-112, the annual modulation search strategy explicitly relies on
the temporal evolution of the background, making accurate and robust background
modelling essential. Thus, the time evolution of the new background model has been developed. While the new model does explain a larger fraction of the measured background, its time evolution remains largely consistent with the previous version, which was already in good agreement with data.

Subsequently, new physics searches with ANAIS-112 data have been presented, making use of the improved background model and the preferred QF models selected in this thesis. Specifically, a reanalysis of the annual modulation signal and a dedicated search for solar axions have been carried out using the full six-year exposure of the experiment.

The annual modulation analysis has been performed in the same energy windows as previous ANAIS-112 studies, showing compatible results and similar sensitivity to DAMA/LIBRA. The updated background model has improved the fit quality in the [1–6]~keV range, increasing the p-value for the null hypothesis, while results in other regions remain comparable. Despite the improved background description, the rate evolution shape, the key input for the analysis, has not changed significantly compared to the previous model. As a result, compatible results have been obtained, further confirming the robustness of the annual modulation analysis.

The result in the [6.7–20] sodium keV\textsubscript{NR} and [22.2–66.7] iodine keV\textsubscript{NR} energy region, corresponding to [2–6] keV in DAMA/LIBRA when assuming QF\textsubscript{Na}=0.3 and QF\textsubscript{I}=0.09, has been reexamined considering ANAIS constant values of QF\textsubscript{Na}=0.2 and QF\textsubscript{I}=0.06, yielding similar results than previous analysis. The energy-dependent QF\textsubscript{Na} selected in this work does not allow the exploration of this region, although this interval can still be probed on the iodine recoil energy scale. The use of the energy-dependent QF\textsubscript{I} results in reduced sensitivity and incompatibility with the DAMA/LIBRA signal. This is primarily due to the larger signal search window in terms of electron-equivalent energy that results from using the energy-dependent QF\textsubscript{I} ([1–4.8] keV, 3.8 keV wide), while covering the same NR energy interval, compared to the narrower window obtained with a constant QF\textsubscript{I} ([1.3–4] keV, 2.7~keV wide). This results in a higher integrated background and, consequently, reduces the statistical significance of a potential DM–induced modulation.

The energy dependence of the modulation amplitude has been examined for both single- and multiple-hit events under two scenarios: assuming identical QFs (i.e., in electron-equivalent energy) and differing QFs (i.e., in NR energy for sodium and iodine recoils). In all cases, the results have consistently favored the null hypothesis. However, when testing the NR energy scale, the current ANAIS-112 threshold of 1 keV limits the analysis to sodium(iodine) recoils above 3(2) keV recoils in the DAMA/LIBRA scale, leaving part of the modulation region unexplored. This limitation motivates future experiments like ANAIS+, which aims to lower the threshold below 0.5 keV by implementing SiPMs for light readout.

The improved background model has enabled, for the first time within ANAIS-112, a search for solar axions considering the expected spectral signatures of various production channels: Primakoff, ABC and the 14.4 keV M1 transition of \textsuperscript{57}Fe. The analysis has explored scenarios considering the simultaneous presence of all three axion components, as well as their individual contributions, while treating the background components determined in the low-energy region as nuisance parameters in the fit.

The analysis revealed correlations between certain background sources and axion signatures. Due to these correlations, only upper limits have been reported for the individual contributions of axions originating from ABC processes and the emission from \textsuperscript{57}Fe. The fit performance favors the absence of an axion signal in the ANAIS-112 data. Upper limits of $g_{\textnormal{Ae}} < 7.40 \times 10^{-12}$ and $g_{\textnormal{AN}}^{\textnormal{eff}} \times g_{\textnormal{Ae}} < 2.03 \times 10^{-17}$ (90\% C.L.) have been set. While not competitive with leading dual-phase xenon TPC experiments because of their reduced background and larger exposures, these limits represent an improvement over other previous results and the most stringent bound obtained to date using NaI(Tl) detectors. This work highlights certain limitations inherent to NaI-based axion searches with comparable contamination levels, particularly the impact of correlated backgrounds. Nevertheless, it demonstrates the capability of ANAIS-112 to probe axion interactions and provides a solid foundation for future searches with increased exposure and improved detector performance.


Finally, this thesis presents the work conducted during a research stay at the Max Planck Institute for Physics in Munich, collaborating within the COSINUS experiment. During this period, the electromagnetic internal background model of COSINUS was developed. Both the latter contribution and the intrinsic radiogenic neutron backgrounds were studied by analyzing the event distributions in the light yield versus energy bands used for NR/ER discrimination, under different experimental conditions. The objective was to assess COSINUS’s sensitivity both to the DAMA/LIBRA signal and to a wider range of viable WIMP candidates. Although this study does not yet include all expected background sources, such as cosmogenic activation, \textsuperscript{210}Pb contamination, or the ambient neutron background, it has validated COSINUS’s sensitivity goals to test DAMA/LIBRA and light WIMPs.

Regarding the future, the eight-year data-taking period of ANAIS-112 will be completed in August 2025. ANAIS will continue operation until the end of the year, by which time, according to the sensitivity prospects, a 5$\sigma$ sensitivity to the DAMA/LIBRA signal will be reached. The analysis of the full dataset will then be carried out using the improved background model developed in this thesis, and the compatibility with the DAMA/LIBRA modulation will be tested taking into account the energy-dependent QF determined in this study for the ANAIS-112 crystals.

For the full-exposure analysis, current efforts within ANAIS are focused on investigating the low-energy anomalous events observed by ANOD, characterized by asymmetric light-sharing. The aim is to improve the performance of the machine learning algorithm using ANOD selected events as training population to enhance their rejection. This could enable a lowering of the energy threshold, leading to improved efficiency and background reduction compared to previous ANAIS analyses. With increased statistics and detailed studies of the ANOD-tagged events, it is expected that these efforts will improve the sensitivity to DAMA/LIBRA’s signal and potentially allow for the exploration of events below the current 1 keV threshold.

After the scheduled ANAIS-112 data-taking ends, a calibration campaign will begin, expected during the first half of 2026, aimed at thoroughly characterizing the detectors before dismantling the experimental set-up. This includes various gamma calibrations (e.g., with $^\text{137}$Cs and $^\text{60}$Co), enabling Compton calibrations to obtain bulk electron populations in the crystals, useful for investigating surface effects, validating efficiency estimations, and probing the non-proportionality of the detector response. Regarding neutron sources, additional neutron calibrations with $^\text{252}$Cf are planned, and, if feasible, monoenergetic neutron sources will be employed to increase sensitivity to QF variations and to study the modelling of elastic scattering in different Geant4 versions, as well as the neutron multiplicity spectrum from \textsuperscript{252}Cf and the effect of lead shielding in the calibrations.

Finally, the possibility remains open that one or more DAMA/LIBRA crystals may be transferred to the LSC. This would allow for a direct characterization of the DAMA/LIBRA background and enable a comparative QF analysis similar to the one developed in this work, helping to assess whether the QF differ between the crystals used in both experiments.

In a long-term future, following the decommissioning of ANAIS-112, the ANAIS program will continue with the ANAIS+ project, whose goal is to lower the energy threshold below 0.5 keV in NaI detectors. Achieving such a low threshold would allow for a more robust test of the DAMA/LIBRA result, enabling the exploration of the full signal region according to the QF estimates derived in this work, and thus significantly reducing the impact of systematics associated with QF uncertainties. Moreover, it would enhance sensitivity to light WIMPs and enable the detection of neutrinos via coherent elastic neutrino–nucleus scattering. 

To reach these objectives, ANAIS+ proposes replacing conventional PMTs with silicon photomultipliers (SiPMs), which offer higher radiopurity and operate at lower voltage, helping to reduce both radioactive backgrounds and non-bulk-scintillation-related events. The detectors would be operated inside a liquid argon tank, which would provide optimal thermal conditions for SiPM performance and act as an active veto system, improving background discrimination and overall sensitivity. Extensive work is already underway in Zaragoza, where the first ANAIS+ prototypes have been developed and tested. Further R\&D efforts are aimed at optimizing and evaluating these prototypes, studying their behavior in liquid argon and pursuing improved crystal radiopurity.

\chapter*{} 
\vspace{-2cm}
\label{Chapter:Conclusiones}
\addcontentsline{toc}{chapter}{Resumen y conclusiones} 

{\begin{center}
   \hrule height 0.5pt
    \vspace{1ex}
    \hrule height 0.5pt
    \vspace{2ex}
    {\normalfont\LARGE \bfseries Resumen y conclusiones}
    \vspace{2ex}
    \hrule height 0.5pt

    \vspace{0.4cm}
\end{center}}

\pagestyle{plain}

Numerosas observaciones astronómicas y cosmológicas correspondientes a distintas escalas y tiempos de la historia del Universo apuntan a la existencia de la materia oscura (MO). Sin embargo, hasta la fecha, la naturaleza de esta componente elusiva, que se estima que constituye el 27\% del Universo, sigue siendo uno de los grandes enigmas de la física de partículas, la cosmología y la astrofísica.

La MO no puede ser explicada por ninguna de las partículas incluidas en el Modelo Estándar de la física de partículas, ya que ninguna presenta las propiedades necesarias para reproducir las observaciones cosmológicas y astrofísicas. Las partículas de MO deben ser masivas, eléctricamente neutras (o poseer una carga extremadamente débil), no relativistas en la época de formación de las galaxias, no bariónicas y estables o con una vida media muy larga en escalas cosmológicas. Además, más allá de su interacción gravitatoria, podrían tener acoplamientos débiles con la materia ordinaria. Este aspecto es fundamental para establecer métodos de detección directa. Entre los candidatos preferidos para la MO se encuentran los axiones y las partículas masivas débilmente interactuantes (WIMPs, por sus siglas en inglés). En el caso de las WIMPs, se espera que la interacción ocurra preferentemente con los núcleos de un detector adecuado en los escenarios más estándar de interacción de la MO.

El descubrimiento de la partícula responsable de la MO, o simplemente la obtención de más detalles sobre su naturaleza, es una tarea compleja que requiere estrategias complementarias: búsquedas de nuevas partículas y física más allá del Modelo Estándar en grandes aceleradores de partículas, detección indirecta mediante la observación de productos de aniquilación de la MO, y detección directa mediante la identificación de los depósitos energéticos producidos por las interacciones de la MO en detectores adecuados. A pesar de los significativos esfuerzos experimentales y de las notables mejoras en sensibilidad durante las últimas décadas, ninguna de estas estrategias ha tenido éxito hasta ahora en la identificación de la MO, aunque muchos candidatos propuestos han sido descartados en el proceso.

No obstante, un resultado positivo merece especial atención. Durante más de 20~años, el experimento DAMA/LIBRA, que utiliza como detectores cristales de NaI(Tl) en el Laboratorio Nacional del Gran Sasso, en Italia, ha reportado un resultado positivo: la observación de una modulación anual en la tasa de detección con alto significado estadístico, compatible con la esperada para partículas de MO del halo galáctico. Para comprobar de forma independiente el resultado de DAMA/LIBRA, sin depender de ningún modelo específico de partícula de MO o de halo galáctico, el requisito principal es emplear el mismo material blanco, yoduro de sodio (NaI). Este es el objetivo de varios experimentos en el ámbito internacional, entre ellos ANAIS-112, COSINE-100, SABRE y COSINUS.

Esta tesis se enmarca dentro del experimento ANAIS-112, un experimento de detección directa de MO que busca confirmar o refutar el resultado de DAMA/LIBRA de manera independiente de modelos, utilizando 112.5 kg de centelleadores ultrapuros de NaI(Tl). Para ello, ANAIS-112 opera desde agosto de 2017 en el Laboratorio Subterráneo de Canfranc (LSC), en España. Desde entonces, ANAIS-112 ha estado adquiriendo datos de forma estable, con un excelente aprovechamiento del tiempo de medida: más del 95\% del tiempo real de adquisición corresponde a tiempo vivo, útil para la extracción de resultados, acumulando así casi ocho años de exposición. En 2025, ANAIS ha publicado los resultados correspondientes a seis años de toma de datos. Estos resultados, actualmente los más sensibles utilizando el mismo material blanco, NaI(Tl), son incompatibles con la señal de modulación de DAMA/LIBRA con un nivel de confianza de 4$\sigma$.

A lo largo de esta tesis, se presenta el trabajo realizado dentro del experimento ANAIS-112, con especial énfasis en el análisis de datos y el desarrollo de simulaciones basadas en Geant4, con el objetivo de mejorar el entendimiento de los efectos sistemáticos que afectan a la comparación entre DAMA/LIBRA y ANAIS-112, y de aumentar la sensibilidad del experimento.

En primer lugar, los esfuerzos se han dirigido a mejorar la comprensión de la respuesta de los cristales de ANAIS-112 frente a retrocesos nucleares (RNs). Los resultados obtenidos por ANAIS basados en seis años de datos ponen fuertemente en duda la interpretación en términos de MO de la señal observada por DAMA/LIBRA. Sin embargo, existen incertidumbres sistemáticas que afectan a la comparación entre ambos resultados y que deben ser abordadas. Para comparar de manera rigurosa los resultados de ANAIS y DAMA/LIBRA, es necesario conocer con precisión los factores de quenching del sodio (QF\textsubscript{Na}) y del yodo (QF\textsubscript{I}) en cristales de NaI(Tl), especialmente si se espera que la señal de MO provenga de RNs, como ocurre en los escenarios más estándar de interacción de MO. El factor de quenching (QF) se define como el cociente entre la cantidad de luz producida por un RN y un retroceso electrónico (RE) que deposita la misma energía. Por tanto, los QFs son esenciales para convertir la energía depositada por RNs a la escala de energía equivalente de electrón (comúnmente denominada escala de energía visible), dado que los experimentos suelen calibrarse con fuentes de electrones o gamma, que generan REs.

Hasta la fecha, el QF en NaI(Tl) no está bien determinado, presentando incertidumbres experimentales significativas tanto en su valor absoluto como en su dependencia energética. Además, ningún modelo teórico actual explica satisfactoriamente los resultados obtenidos. Resulta particularmente relevante la discrepancia observada en los resultados del QF de DAMA/LIBRA. Mientras que la mayoría de las medidas recientes indican valores de QF\textsubscript{Na} alrededor del 20\%, con una tendencia creciente con la energía, DAMA/LIBRA reporta un valor constante del 30\%. De manera similar, para el QF\textsubscript{I}, el valor determinado por DAMA/LIBRA es aproximadamente un 50\% superior a las medidas recientes (9\% frente a 5\%).

Considerando que este parámetro representa la incertidumbre sistemática más relevante en la comparación con DAMA/LIBRA, ANAIS-112 ha llevado a cabo un programa de calibración con neutrones. Por un lado, se realizaron medidas del QF en pequeños cristales de NaI(Tl), similares a los usados en ANAIS-112, mediante un haz monoenergético de neutrones en el Triangle Universities Nuclear Laboratory (TUNL). Por otro lado, desde 2021 se han realizado calibraciones con neutrones in situ utilizando fuentes de $^{252}$Cf de baja actividad en el LSC. Esta tesis se ha centrado en este último enfoque.

Las calibraciones con neutrones in situ cumplen tres objetivos principales dentro del experimento: evaluar la eficiencia de selección de eventos en ANAIS-112, generar eventos con características similares a la señal, distribuidos de forma homogénea en el volumen del cristal, para el entrenamiento de técnicas de aprendizaje automático, y mejorar la comprensión de los QFs. En este contexto, estas calibraciones de neutrones realizadas in situ constituyen una verificación importante de los resultados obtenidos en TUNL para cristales crecidos en el mismo lote que algunos de los cristales de ANAIS. El último objetivo enumerado, una mejor caracterización de los QF\textsubscript{Na} y QF\textsubscript{I} de los cristales de ANAIS, se ha abordado en esta tesis comparando los espectros medidos durante las calibraciones in situ de neutrones con los obtenidos mediante simulaciones de neutrones basadas en Geant4.

La calibración se basa en la exposición de los detectores de ANAIS-112 a fuentes de baja actividad de neutrones de \textsuperscript{252}Cf, colocadas fuera de la caja anti-radón y del blindaje de plomo, pero dentro del veto de muones y el moderador de neutrones. Se ha observado que las interacciones de neutrones producen principalmente eventos de centelleo en el volumen del cristal originados por dispersión elástica con núcleos de sodio y yodo, con una población aún más limpia al seleccionar eventos que interaccionan en múltiples detectores. Además, este análisis ha permitido observar la distinta respuesta temporal de RN y RE mediante la comparación de las formas de pulso promedio de cada población, obteniéndose una razón promedio de las constantes de decaimiento del centelleo para RNs respecto a REs de aproximadamente 0.85, consistente con resultados previos.

Las simulaciones de neutrones realizadas para comparar con los datos de calibración han permitido una evaluación crítica de las bases de datos de Geant4. En particular, se han observado diferencias entre las versiones v9.4.p01 y v11.1.1 de Geant4 en cuanto a secciones eficaces y a la presencia e intensidad de ciertas líneas gamma que se emiten en procesos de dispersión inelástica, las cuales afectan al acuerdo entre simulación y datos. Por consiguiente, se ha seleccionado la versión v9.4.p01 para este trabajo debido a su mejor descripción de los datos. Además, se ha identificado una sobreestimación de la sección eficaz de captura neutrónica para la producción de \textsuperscript{128}I en las dos versiones analizadas.

Se han propuesto y probado varios modelos de QFs basados en medidas experimentales. En particular, para el QF\textsubscript{Na}, se evalúa el resultado de DAMA/LIBRA, los tres resultados obtenidos a partir de medidas realizadas en TUNL sobre cristales similares a los de ANAIS empleando distintos procedimientos de calibración, el resultado de COSINUS y el modelo de Tretyak. Respecto a los resultados de ANAIS, para describir la dependencia con la energía se ha ajustado un modelo de Lindhard modificado a las medidas de QF\textsubscript{Na} y se ha comparado con un ajuste lineal simple. En cuanto a QF\textsubscript{I}, se han estudiado los resultados de DAMA/LIBRA, la medida para cristales de ANAIS en TUNL, y un modelo de QF\textsubscript{I} dependiente de energía propuesto en este trabajo, compatible con las medidas de ANAIS-112.

La comparación entre los datos y la simulación demuestra que la respuesta de los detectores de ANAIS-112 se reproduce de manera adecuada en la simulación. Aunque este estudio no permite extraer una dependencia explícita en energía del QF, a diferencia de otros estudios que hacen uso de fuentes monocromáticas en las que la medida del ángulo de dispersión del neutrón proporciona una medida independiente de la energía del retroceso nuclear, sí permite la comparación de distintos modelos de QF. En particular, los modelos de QF propuestos por DAMA/LIBRA se ven desfavorecidos debido a su menor concordancia con los datos de ANAIS-112, al igual que los modelos de QF\textsubscript{Na} que muestran una dependencia decreciente con la energía.

Respecto al QF\textsubscript{Na} de los cristales de ANAIS medidos en TUNL, la comparación entre datos y simulación sugiere una preferencia por modelos con QF\textsubscript{Na} creciente con la energía, en lugar de modelos de QF constante. Además, se ha verificado que el modelo modificado de Lindhard describe ligeramente mejor los datos que un ajuste lineal. No obstante, no se pueden descartar otras dependencias energéticas, aunque parece preferirse una reducción más pronunciada del QF\textsubscript{Na} a bajas energías. En cuanto al yodo, la concordancia datos-simulación mejora significativamente al emplear un QF\textsubscript{I} dependiente de la energía compatible con los resultados de ANAIS-112, aunque el valor de 6\% reportado por ANAIS a 15 keV\textsubscript{RN} también proporciona un acuerdo razonable. Aunque los eventos de alta multiplicidad muestran un menor acuerdo con los datos, hecho que deberá ser investigado en estudios futuros, estos eventos representan solo una pequeña fracción del total de eventos de la calibración con neutrones. La mayoría de los eventos, en particular las poblaciones de single-hit y m2-hit, es decir, aquellos que interaccionan en uno o en dos detectores, son reproducidos con gran precisión por la simulación dentro de las incertidumbres asociadas.

En conjunto, el modelo preferido para los QF de los cristales de ANAIS-112 derivado de este trabajo es un QF\textsubscript{Na} dependiente de la energía, basado en medidas realizadas en TUNL utilizando una calibración que emplea el pico inelástico del proceso \textsuperscript{127}I(n,n'$\gamma$); y un QF\textsubscript{I} dependiente de la energía compatible con los valores obtenidos para los cristales de ANAIS. Además, se ha verificado que todos los cristales de ANAIS-112 presentan el mismo espectro de retroceso nuclear en respuesta a fuentes de \textsuperscript{252}Cf, a pesar de proceder de distintos lingotes, aunque crecidos siguiendo protocolos similares, lo que confirma la compatibilidad de los resultados obtenidos en las medidas realizadas en TUNL en varios cristales de NaI(Tl) fabricados por Alpha Spectra.



Se ha analizado la elección de la versión de Geant4 como fuente de incertidumbre sistemática que puede afectar a la determinación del QF. El acuerdo es también satisfactorio con la versión más reciente, respaldando los QFs seleccionados en este trabajo y descartando los QFs de DAMA/LIBRA debido a su menor concordancia con los datos de ANAIS-112. Asimismo, las conclusiones relacionadas con los QFs seleccionados en esta tesis se han validado utilizando datos del nuevo sistema de adquisición ANOD, instalado en el LSC en diciembre del 2024, que cuenta con una ventana de adquisición más larga y opera sin tiempo muerto. Con los datos de ANOD, el acuerdo en la población de multiple-hits a baja energía es notablemente bueno y mejora respecto al obtenido con los datos de ANAIS.

Por otro lado, las calibraciones con neutrones podrían realizarse usando fuentes alternativas. De especial interés es el uso de fuentes de neutrones monoenergéticas, como aquellas basadas en aceleradores o fuentes Y-Be, que ofrecen la mayor sensibilidad para detectar variaciones en los valores del QF y que no deberían estar afectadas por los mismos sistemáticos asociados a la simulación de fuentes de $^{252}$Cf. Además, estas futuras campañas de calibración permitirán explorar la incertidumbre de Geant4 en la modelización de la dispersión elástica, del espectro de multiplicidad de neutrones procedentes del \textsuperscript{252}Cf y de las interacciones en plomo. Estas últimas podrían ayudar a aclarar las discrepancias observadas en el espectro de eventos múltiples, donde el acuerdo con los datos empeora a medida que aumenta la multiplicidad.

Además, un escenario futuro particularmente relevante consistiría en distribuir uno o más cristales del experimento DAMA/LIBRA entre las colaboraciones de ANAIS, COSINE o SABRE. Su operación permitiría no solo una caracterización directa del fondo de DAMA/LIBRA, sino también una determinación independiente de sus QFs usando la misma metodología aplicada a los cristales de ANAIS-112 en este trabajo, ya sea con fuentes de \textsuperscript{252}Cf u otras fuentes alternativas de neutrones. Aunque dicha medida podría no determinar de manera inequívoca los valores de QF\textsubscript{Na} y QF\textsubscript{I} en NaI(Tl), sí podría aportar luz, de forma definitiva, sobre si los cristales de DAMA y ANAIS difieren en su respuesta, eliminando así la principal incertidumbre sistemática en la comparación entre ambos experimentos.

Paralelamente, esta tesis ha contribuido a la revisión y mejora del modelo de fondo de ANAIS-112 mediante un ajuste multiparamétrico de las distintas contribuciones, un enfoque implementado por primera vez en el contexto de ANAIS-112.

Durante el desarrollo del ajuste, se identificaron varios aspectos clave. En primer lugar, se detectó una asimetría en la distribución de la luz entre los fotomultiplicadores (PMTs), que reveló una población de eventos que sólo interaccionan con un cristal con fuerte asimetría en la región de $\sim$ [75-350] keV. Para investigar su naturaleza, los eventos fueron clasificados en poblaciones simétricas y asimétricas, evidenciando un cambio claro en la forma espectral de los eventos asimétricos alrededor de 80~keV, probablemente debido a depósitos de energía provenientes de emisiones de los PMTs. Las simulaciones respaldaron esta hipótesis, mostrando efectos dependientes de la posición que motivan la futura implementación de un modelo de recolección de luz sensible a la posición del depósito. Además, asumiendo que las contaminaciones en volumen son simétricas, los resultados sugieren la presencia de una contaminación superficial de \textsuperscript{210}Pb localizada en los extremos pulidos de los cristales de ANAIS-112.

Posteriormente, se revisó la modelización de los PMTs para incluir una contaminación frontal de \textsuperscript{226}Ra y \textsuperscript{232}Th en el fotocátodo, además del vidrio del borosilicato, tal y como sugieren las medidas en un detector de germanio hiperpuro del LSC realizadas durante esta tesis. Esta componente frontal resulta más eficiente en el depósito de energía en el cristal. Para \textsuperscript{210}Pb se consideraron contaminaciones tanto en volumen como en la superficie de los cristales, así como contaminación superficial en ambas caras del recubrimiento de teflón que rodea los cristales. Respecto a la contaminación superficial del cristal, este estudio ha considerado perfiles de contaminación con decaimiento exponencial, en lugar de con profundidad fija como se asumía en el modelo de fondo previo, para tener en cuenta la difusión de los isótopos contaminante a diferentes profundidades. En particular, se evaluó el comportamiento de tres profundidades medias (1, 10 y 100 $\mu$m). Además, se ha implementado una corrección al espectro $\beta^-$ del decaimiento de $^{210}$Bi, incorporando un espectro derivado de medidas experimentales, dado que la modelización incluida en Geant4 no reproduce adecuadamente los datos de ANAIS-112. Esta corrección mejora la concordancia entre la simulación y las mediciones.

El comportamiento del ajuste en cada etapa ha sido evaluado mediante la comparación de los resultados de actividad ajustada en este trabajo con los obtenidos en el modelo de fondo previo. Se ha adoptado una profundidad de 10 $\mu$m como el escenario más plausible para la contaminación superficial de $^{210}$Pb. Además, los valores de actividad no respaldan la suposición previa dentro del modelo ANAIS-112 respecto a la forma de los fondos alfa. En cada cristal de ANAIS se observan dos picos de energía depositada alfa, atribuidos en modelizaciones anteriores a decaimientos de $^{210}$Po en superficie (pico de baja energía) y en volumen (pico de alta energía) en una proporción que no es consistente con los resultados derivados de este trabajo. Aunque todos los cristales de ANAIS presentan una respuesta consistente frente a neutrones, no muestran la misma uniformidad para partículas alfa. En este trabajo no se ha encontrado una explicación para la estructura observada en los depósitos de energía alfa; sin embargo, se están realizando esfuerzos con el sistema de adquisición ANOD para identificar secuencias Bi-Po y explorar posibles dependencias espaciales que puedan aclarar el comportamiento observado con $^{210}$Pb.

El nuevo modelo de fondo muestra un acuerdo significativamente mejorado con los datos en todas las regiones energéticas y poblaciones en comparación con el modelo anterior. La robustez del modelo se refuerza con la reproducción precisa de eventos en coincidencia y de la diferencia entre espectros de fondo correspondientes a diferentes años.

A su vez, se evaluaron dos fuentes adicionales de fondo. Primero, el fondo ambiental de neutrones en el hall B del LSC, simulado en esta tesis con GEANT4 y comparado con resultados basados en FLUKA del experimento HENSA, un espectrómetro de neutrones instalado en las proximidades de ANAIS-112 en 2019. Las discrepancias entre ambos códigos son significativas, particularmente para neutrones rápidos, pero la contribución global del fondo de neutrones resulta no relevante respecto al fondo medido con ambos códigos. En segundo lugar, se ha incorporado una población de eventos anómalos que supera los filtros de selección de eventos de ANAIS-112. Estos eventos, denominados eventos asimétricos debido a la asimetría en el reparto de luz entre los dos PMTs, han podido ser identificados con ANOD debido a las características de su sistema de adquisición. Sus ritmos coinciden con parte del exceso inexplicado en la región [1–2] keV en los datos de ANAIS.

En ANAIS-112, la estrategia para la búsqueda de modulación anual se basa en la evolución temporal del fondo, haciendo esencial un modelo de fondo preciso y robusto. Por ello, se desarrolló la evolución temporal correspondiente al nuevo modelo de fondo, mostrando una descripción notablemente mejorada respecto a la versión anterior. Aunque el nuevo modelo explica una mayor fracción del fondo medido, su evolución temporal se mantiene en gran medida consistente con la versión previa, que ya mostraba un buen acuerdo con los datos.

Posteriormente, se presentan nuevas búsquedas de física utilizando los datos de ANAIS-112, haciendo uso del modelo de fondo mejorado y de los modelos de QFs preferidos seleccionados en esta tesis. En particular, se ha realizado un reanálisis de la señal de modulación anual y una búsqueda dedicada de axiones solares, utilizando la exposición de seis años del experimento.

El análisis de modulación anual se ha llevado a cabo en las mismas ventanas de energía que en estudios previos de ANAIS-112, mostrando resultados compatibles y una sensibilidad similar al resultado de DAMA/LIBRA. El modelo de fondo actualizado ha mejorado la calidad del ajuste en el rango [1–6]~keV, incrementando el valor-p para la hipótesis nula, mientras que los resultados en otras regiones energéticas se mantienen comparables. A pesar de la mejora en la descripción del fondo, la forma de la evolución temporal del ritmo no ha cambiado significativamente con respecto al modelo anterior. En consecuencia, se han obtenido resultados compatibles, lo que refuerza la robustez del análisis de modulación anual de ANAIS-112.

El resultado en la región de energía comprendida entre [6.7–20] keV\textsubscript{NR} para retrocesos de sodio y [22.2–66.7]~keV\textsubscript{NR} para retrocesos de yodo, correspondiente al intervalo [2–6] keV en DAMA/LIBRA al asumir QF\textsubscript{Na} = 0.3 y QF\textsubscript{I} = 0.09, ha sido reanalizado considerando los valores constantes utilizados por ANAIS, QF\textsubscript{Na}=0.2 y QF\textsubscript{I}=0.06, obteniéndose resultados similares a los de análisis anteriores. El QF\textsubscript{Na} dependiente de la energía seleccionado en este trabajo no permite explorar dicha región en términos de retrocesos de sodio, aunque sigue siendo accesible en la escala de energía de retrocesos de yodo. El uso del QF\textsubscript{I} dependiente de la energía considerado en este trabajo ha resultado en una reducción de la sensibilidad e incompatibilidad al resultado de DAMA/LIBRA. Esto se debe principalmente a que la ventana de búsqueda de señal es significativamente más amplia en términos de energía equivalente de electrón al emplear el QF\textsubscript{I} dependiente de la energía ([1-4.8] keV = 3.8 keV) en comparación con el QF\textsubscript{I} constante ([1.3-4] keV = 2.7 keV), aunque corresponde al mismo rango en energía de RN. Esta ampliación implica un fondo integrado mayor, lo que reduce la significancia estadística de una posible modulación inducida por MO.

 Se ha examinado la dependencia de la amplitud de modulación con la energía bajo dos escenarios: asumiendo mismos QFs entre DAMA/LIBRA y ANAIS (es decir, en energía equivalente de electrón) y QFs distintos (es decir, en energía de RN para interacciones con sodio e yodo). En todos los casos, los resultados han favorecido consistentemente la hipótesis nula. Sin embargo, al considerar la escala de energía de RN, el umbral actual de ANAIS-112, de 1 keV, limita el análisis a retrocesos de sodio(yodo) superiores a 3(2) keV en la escala de DAMA/LIBRA, dejando sin explorar parte de la región donde DAMA/LIBRA observa modulación. Esta limitación motiva el desarrollo de futuros experimentos como ANAIS+, que busca reducir el umbral por debajo de 0.5~keV mediante la implementación de fotodetectores de silicio, SiPM, para la lectura de la luz.

El modelo de fondo mejorado ha permitido, por primera vez dentro del marco de ANAIS-112, realizar una búsqueda de axiones solares considerando diversos canales de producción: Primakoff, ABC y la transición M1 de 14.4 keV del \textsuperscript{57}Fe. El análisis ha explorado tanto escenarios con la presencia simultánea de las tres componentes de axiones como sus contribuciones individuales, tratando los componentes del fondo determinados en la región de baja energía como parámetros libres en el ajuste.

El análisis reveló correlaciones entre ciertas fuentes de fondo y las señales de axiones. Debido a estas correlaciones, solo se han reportado límites superiores para las contribuciones individuales de axiones procedentes de procesos ABC y de la emisión del \textsuperscript{57}Fe. El ajuste estadístico favorece la ausencia de señal de axiones en los datos de ANAIS-112. Se han establecido límites superiores de $g_{\textnormal{Ae}} < 7.40 \times 10^{-12}$ y $g_{\textnormal{AN}}^{\textnormal{eff}} \times g_{\textnormal{Ae}} < 2.03 \times 10^{-17}$ (90\%~C.L.). Si bien no son competitivos con los mejores resultados obtenidos por experimentos con cámaras de proyección temporal de xenón en doble fase que ofrecen menores niveles de fondo y mayores exposiciones, estos límites suponen una mejora respecto a resultados anteriores y constituyen el límite más restrictivo obtenido hasta la fecha utilizando detectores de NaI(Tl). Este trabajo pone de relieve ciertas limitaciones inherentes a búsquedas de axiones basadas en NaI con niveles de contaminación similares, en particular el impacto de fondos correlacionados. No obstante, demuestra la capacidad de ANAIS-112 para explorar interacciones de axiones y proporciona una base sólida para futuras búsquedas con mayor exposición y mejoras en el rendimiento del detector.

Finalmente, esta tesis recoge el trabajo realizado durante una estancia de investigación en el Instituto Max Planck de Física en Múnich, en el marco del experimento COSINUS. Durante este período, se desarrolló el modelo de fondo electromagnético interno de COSINUS. Tanto este modelo como el fondo intrínseco de neutrones radiogénicos se estudiaron mediante el análisis de la distribución de eventos en el plano \textit{light yield} vs. energía usado para la discriminación RN/RE, bajo distintas condiciones experimentales. El objetivo fue evaluar la sensibilidad de COSINUS tanto a la señal de DAMA/LIBRA como a un rango más amplio de candidatos viables a WIMPs. Aunque este estudio no incluye aún todas las fuentes de fondo esperadas, como la activación cosmogénica, la contaminación por \textsuperscript{210}Pb o el fondo de neutrones ambientales, ha permitido validar los objetivos de sensibilidad de COSINUS para poner a prueba DAMA/LIBRA y explorar WIMPs ligeros.

Respecto al futuro inmediato, el periodo de toma de datos de ocho años se completará en agosto de 2025. ANAIS continuará operando hasta finales de año, momento en el que, según las proyecciones de sensibilidad, se alcanzará una sensibilidad de 5$\sigma$ a la señal observada por DAMA/LIBRA. El análisis de la exposición completa se llevará a cabo empleando el modelo de fondo mejorado desarrollado en esta tesis, y se evaluará la compatibilidad con la modulación anual de DAMA/LIBRA considerando los QFs dependientes de la energía determinados para los cristales de ANAIS-112 en este trabajo.

Para este análisis final, los esfuerzos actuales de ANAIS se centran en investigar los eventos anómalos de baja energía observados con ANOD, caracterizados por un reparto de luz asimétrico. El objetivo es mejorar el rendimiento del algoritmo de aprendizaje automático utilizando este tipo de señales en su entrenamiento. Esto podría permitir una mejora en el filtrado de estos sucesos, lo que a su vez facilitaría una reducción del umbral energético, incrementando tanto la eficiencia como la capacidad de supresión del fondo con respecto a análisis previos. Con mayor estadística y estudios detallados de los eventos anómalos identificados por ANOD, se espera aumentar la sensibilidad a la señal de DAMA/LIBRA e incluso explorar eventos por debajo del actual umbral de 1~keV.

Tras la finalización del periodo de adquisición de datos de ANAIS-112, se iniciará una campaña de calibración prevista para la primera mitad de 2026, orientada a caracterizar exhaustivamente los detectores antes del desmantelamiento del sistema experimental. Esta campaña incluirá diversas calibraciones gamma (por ejemplo, con \textsuperscript{137}Cs y \textsuperscript{60}Co), permitiendo calibraciones Compton para obtener poblaciones de REs repartidas en el volumen del cristal, útiles para investigar efectos superficiales, validar estimaciones de eficiencia y estudiar la no proporcionalidad en la respuesta del detector. En cuanto a fuentes de neutrones, se prevén nuevas calibraciones con \textsuperscript{252}Cf y, si es viable, el uso de fuentes monoenergéticas para aumentar la sensibilidad a variaciones en los QFs y estudiar la modelización de las interacciones elásticas en las distintas versiones de Geant4, así como el espectro de multiplicidad de neutrones de \textsuperscript{252}Cf y el efecto del blindaje de plomo en las calibraciones.

Además, se mantiene abierta la posibilidad de trasladar uno o más cristales de DAMA/LIBRA al LSC. Esto permitiría una caracterización directa del fondo observado en DAMA/LIBRA, y realizar un análisis comparativo de los QFs con la misma metodología aplicada en este trabajo, lo que contribuiría a determinar si existen diferencias sistemáticas entre la respuesa de los cristales utilizados en ambos experimentos.

En un futuro más lejano, una vez desmontado ANAIS-112, ANAIS continuará con el proyecto ANAIS+, cuyo objetivo es reducir el umbral energético por debajo de 0.5 keV en detectores de NaI. Alcanzar un umbral tan bajo permitiría un análisis más robusto del resultado de DAMA/LIBRA, al permitir la exploración completa de la región de DAMA/LIBRA, según las estimaciones de los QFs obtenidas en este trabajo, reduciendo significativamente el impacto de las incertidumbres sistemáticas asociadas. Además, aumentaría la sensibilidad a WIMPs ligeros y abriría la posibilidad de detectar neutrinos mediante dispersión coherente elástica neutrino-núcleo.

Para alcanzar estos objetivos, ANAIS+ propone sustituir los PMTs convencionales por fotodetectores de silicio (SiPM), que presentan una mayor radiopureza y operan a menor voltaje, lo que contribuye a reducir tanto los fondos radioactivos como los eventos no relacionados con centelleo. Los detectores se operarían en el interior de un tanque de argón líquido, que proporcionaría condiciones térmicas óptimas para el funcionamiento de los SiPM y actuaría como un sistema de veto activo, mejorando la discriminación de fondo y la sensibilidad general. Actualmente, ya se está realizando un gran trabajo en Zaragoza, donde se han desarrollado y probado los primeros prototipos de ANAIS+. Las actividades de I+D continúan enfocadas en la optimización y evaluación de estos prototipos, en el estudio de su comportamiento en argón líquido y en la mejora de la radiopureza de los cristales.

\chapter*{} 
\vspace{-2cm}
\label{Chapter:Agradecimientos}
\phantomsection
\addcontentsline{toc}{chapter}{Agradecimientos} 
{\begin{center}

    {\normalfont\LARGE \bfseries Agradecimientos}

    \vspace{0.4cm}
\end{center}}

\pagestyle{plain}


En primer lugar, quiero agradecer a mis directoras de tesis, Marisa Sarsa y María Martínez, por guiarme en cada paso de este viaje, ayudándome a crecer como persona e investigadora. La verdad es que no podría haber imaginado mejores mentoras. Gracias, Marisa, por saber transmitirme parte de tu gran sabiduría de forma desinteresada, por tu apoyo constante, tu infinita disponibilidad, tu paciencia y perseverancia. Verte trabajar ha sido una inspiración diaria que me motivaba a dar lo mejor de mí, intentando seguir tu ejemplo. Gracias, María, por enseñarme a disfrutar de la física, por animarme a seguir adelante, por tu experiencia, entrega y buen criterio, siempre tan valioso.

 Este primer agradecimiento se extiende también a Iván Coarasa, a quien puedo considerar mi tercer director de tesis. Muchas gracias, Iván, por tu paciencia, por tus consejos, que han sido clave en todo mi trabajo, y por responder a cada una de mis preguntas, que sabemos que han sido muchas.

También quiero agradecer a todos los miembros del experimento ANAIS por crear un ambiente de trabajo tan agradable. En especial, gracias a Alfonso Ortiz de Solórzano por hacerme partícipe de la historia del Laboratorio Subterráneo de Canfranc desde el primer momento, y por su pasión y entusiasmo contagiosos. Gracias por encargarte de las calibraciones de neutrones y por responder a mis infinitas, y muchas veces insignificantes, preguntas sobre los detalles de montaje de ANAIS-112.
También, por supuesto, quiero agradecer a Susana Cebrián por su inmenso trabajo en el modelo de fondo de ANAIS, y por su generosidad, su disponibilidad y su saber hacer siempre que he acudido a ella.
Agradezco a Jorge Puimedón por las medidas de los fotomultiplicadores de ANAIS, y a Eduardo García por sus buenos consejos. Y no me olvido de Javier Mena, por su ayuda siempre fundamental en las tareas informáticas. 

A todos los demás compañeros del departamento, gracias por estar siempre dispuestos a ayudar.
En especial a Gloria Luzón, que me permitió comenzar mi etapa investigadora al darme la oportunidad de realizar prácticas externas en el grupo. Y a Theopisti Dafni, por su interés, ayuda y valiosos consejos. A todos los doctorandos del departamento que compartían mi situación, gracias por hacer que este camino no se sintiera tan solitario, especialmente a Jaime y Sophia, mis compañeros de despacho; a David Diéz, por aportar siempre esa mirada optimista; a Carmen y a Swadheen, que aunque recién llegados ya reflejan lo valioso que es pertenecer a un buen equipo; y a David Cintas, que aunque ya no está físicamente en Zaragoza, me dio un apoyo clave al principio.

I want to give a special thanks to Karo Schäffner and the entire COSINUS team. Thank you for welcoming me into the team from the very beginning, for making my stay so much easier, and for your generosity in sharing your knowledge of cryogenic techniques.

También quiero dar las gracias a todo el equipo del Laboratorio Subterráneo de Canfranc, por su ayuda y colaboración.

Por último, y más importante, gracias a mi familia, en especial a mis padres, a mi hermano y a mi abuela, por su apoyo incondicional y por ser siempre mi lugar seguro e inquebrantable, pase lo que pase.
Quiero dedicar esta tesis a mi madre: mi referente y pilar fundamental. Es ella quien me ha transmitido los valores del esfuerzo, la constancia, la dedicación y la resiliencia. Porque sin ella yo no estaría hoy aquí, y este pequeño logro es, sin duda alguna, de las dos.

También considero parte de mi familia a Elena (con su idea del ascensor a la Luna), a José Miguel, a Clara y a Daniel, a quienes agradezco todo su cariño y su ánimo.
Y a mis amigas, que siguen ahí a pesar de no dedicarles tanto tiempo como merecen.
Finalmente, quiero cerrar estos agradecimientos con Nacho, quien ha sido, es y será mi apoyo fundamental, mi punto fijo y el motor que me impulsa a perseguir mis sueños. Gracias por tu comprensión, por tu paciencia infinita y, en definitiva, por cuidarme tanto y tan bien. A mí y a nuestras hijas gatunas, Anais (en honor al experimento) y Nika, que me han ayudado más de lo que imaginan.
Gracias, Nacho, por elegirnos; sin duda, estar en casa significa estar contigo.

\chapter*{} 
\vspace{-2cm}
\label{Chapter:Agradecimientos}
\phantomsection
\addcontentsline{toc}{chapter}{Acknowledgements} 
{\begin{center}

    {\normalfont\LARGE \bfseries Acknowledgements}

    \vspace{0.4cm}
\end{center}}

\pagestyle{plain}

This work has been financially supported by MCIN/AEI/10.13039/501100011033 under grants PID2022-138357NB-C21 and PID2019-104374GB-I00, as well as by the Consolider-Ingenio 2010 Programme through the MultiDark (CSD2009-00064) and CPAN (CSD2007-00042) projects. Additional support was provided by the LSC Consortium, the Gobierno de Aragón, and the European Social Fund (Group in Nuclear and Astroparticle Physics), as well as funds from the European Union through the NextGenerationEU/PRTR initiative (Planes Complementarios, Programa de Astrofísica y Física de Altas Energías). The author gratefully acknowledge the use of the Servicio General de Apoyo a la Investigación – SAI of the Universidad de Zaragoza, as well as the technical support provided by the staff of LSC and GIFNA.

This PhD dissertation has been possible thanks to the Aid for Predoctoral Contract for the Training of Doctors 2020 of the Ministerio de Ciencia e Innovación-Agencia Estatal de Investigación, reference PRE2020-094910, and the research
stage at the Max Planck Insitute for Physics to the the Ministerio de Ciencia e
Innovación-Agencia Estatal de Investigación Subprograma Estatal de Movilidad.

\clearpage
  \thispagestyle{empty}
  \null
  \clearpage

\pagestyle{rest}

\bibliographystyle{unsrt}
\bibliography{sections/bibliografia}

\addcontentsline{toc}{chapter}{Bibliography}




\end{document}